\documentclass[11pt]{amsart}
\usepackage[OT2,T1]{fontenc}
\usepackage[latin9]{inputenc}
\usepackage{xcolor}
\usepackage{array}
\usepackage{longtable}
\usepackage{booktabs}
\usepackage{calc}
\usepackage{mathtools}
\usepackage{varwidth}
\usepackage{amsbsy}
\usepackage{amstext}
\usepackage{amsthm}
\usepackage{amssymb}
\usepackage{graphicx}
\usepackage{esint}
\usepackage[bookmarks=true,bookmarksnumbered=false,bookmarksopen=false,
 breaklinks=false,pdfborder={0 0 1},backref=section,colorlinks=false]
 {hyperref}

\makeatletter

\providecommand{\tabularnewline}{\\}
\newenvironment{cellvarwidth}[1][t]
    {\begin{varwidth}[#1]{\linewidth}}
    {\@finalstrut\@arstrutbox\end{varwidth}}

\numberwithin{equation}{section}
\numberwithin{figure}{section}
\theoremstyle{plain}
\newtheorem{thm}{\protect\theoremname}
\theoremstyle{remark}
\newtheorem{rem}[thm]{\protect\remarkname}
\theoremstyle{plain}
\newtheorem{prop}[thm]{\protect\propositionname}
\theoremstyle{definition}
\newtheorem{defn}[thm]{\protect\definitionname}
\theoremstyle{plain}
\newtheorem{lem}[thm]{\protect\lemmaname}
\theoremstyle{plain}
\newtheorem{cor}[thm]{\protect\corollaryname}

\DeclareFontSubstitution{OT2}{wncyr}{m}{n}

\usepackage{color}
\usepackage{array}
\usepackage{rotating}
\usepackage{varwidth}
\usepackage{needspace}
\DeclareGraphicsExtensions{.png,.pdf,.jpg}
\graphicspath{{figures/}}
\usepackage{etoolbox}
\RequirePackage{geometry}
\csname geometry\endcsname{reset,paperwidth=170mm,paperheight=240mm,inner=22mm,outer=26mm,top=25mm,bottom=22mm,twoside,headheight=14pt,headsep=7mm,footskip=10mm}
\AtBeginDocument{\let\BR@@bibitem\@bibitem}

\providecommand{\tabularnewline}{\\}

\@ifundefined{cellvarwidth}{%
  \newenvironment{cellvarwidth}[1][t]
      {\begin{varwidth}[#1]{\linewidth}}
      {\@finalstrut\@arstrutbox\end{varwidth}}%
}{}

\@ifundefined{c@thm}{%
  \theoremstyle{plain}%
  \newtheorem{thm}{\protect\theoremname}[section]%
  \newtheorem{lem}[thm]{\protect\lemmaname}%
  \newtheorem{prop}[thm]{\protect\propositionname}%
  \newtheorem{cor}[thm]{\protect\corollaryname}%
  \theoremstyle{definition}%
  \newtheorem{defn}[thm]{\protect\definitionname}%
  \theoremstyle{remark}%
  \newtheorem{rem}[thm]{\protect\remarkname}%
}{}

\theoremstyle{definition}
\newtheorem{assump}[thm]{Assumption}\theoremstyle{plain}

\newcommand{\dd}{\mathrm{d}}

\numberwithin{equation}{section}

\providecommand{\corollaryname}{Corollary}
\providecommand{\definitionname}{Definition}
\providecommand{\lemmaname}{Lemma}
\providecommand{\propositionname}{Proposition}
\providecommand{\remarkname}{Remark}
\providecommand{\theoremname}{Theorem}

\@ifundefined{ack}{%
}{}

\def\@setabstracta{%
  \ifvoid\abstractbox
  \else
    \skip@20\p@ \advance\skip@-\lastskip
    \advance\skip@-\baselineskip \vskip\skip@
    \vskip 20\p@ % extra air between the dedication and the abstract
    \unvbox\abstractbox
  \fi
}

\AtBeginDocument{%
  \@ifundefined{c@thm}{}{%
    \@addtoreset{thm}{section}%
    \expandafter\renewcommand\csname thethm\endcsname{\thesection.\arabic{thm}}%
    \@addtoreset{equation}{section}%
    \expandafter\renewcommand\csname theequation\endcsname{\thesection.\arabic{equation}}%
    \expandafter\renewcommand\csname theHequation\endcsname{\thesection.\arabic{equation}}%
  }%
}

\AtBeginDocument{%
  \hypersetup{
    linkcolor=blue,
    citecolor=black,
    urlcolor=blue,
    pdftitle={Floquet Theory of the LC Circuit with Modulated Capacitance},
    pdfauthor={Alexander Figotin},
    pdfpagemode=UseOutlines,
    pdfdisplaydoctitle=true
  }%
  \raggedbottom
  \widowpenalty=9000 \clubpenalty=9000
}

\makeatother

\providecommand{\corollaryname}{Corollary}
\providecommand{\definitionname}{Definition}
\providecommand{\lemmaname}{Lemma}
\providecommand{\propositionname}{Proposition}
\providecommand{\remarkname}{Remark}
\providecommand{\theoremname}{Theorem}

\begin{document}
\title[Floquet Theory of the LC Circuit]{Floquet Theory of the LC Circuit with Modulated Capacitance}
\author{Alexander Figotin}
\address{Department of Mathematics, University of California, Irvine, CA 92697,
USA}
\email{afigotin@uci.edu}
\dedicatory{To Richard Albanese --- whose generosity and boundless scientific
curiosity have been a lifelong inspiration.}
\begin{abstract}
Parametric resonance --- the phenomenon in which periodic variation
of a system parameter drives exponential growth of oscillations ---
is one of the most fundamental instabilities in physics and engineering.
The nondissipative LC circuit with harmonically varying capacitance
is among the simplest physical systems in which it occurs: the periodic
modulation of the capacitance renders the circuit equation a Hill
equation whose solutions are either stable (bounded oscillations persisting
indefinitely) or unstable (oscillations growing exponentially, with
energy extracted from the modulation source each period).

The governing equation is identified as a special case of Ince's four-parameter
Hill equation, which leads to two main results. First, a sharp structural
theorem: instability occurs exclusively at the \emph{odd} sub-harmonics
of the natural frequency, while every even resonance is completely
and exactly stable for all modulation amplitudes. This selectivity,
invisible to the standard Mathieu approximation, is explained by Krein's
collision theory for symplectic systems: at odd resonances the colliding
Floquet multipliers carry opposite Krein signatures, opening an instability
tongue, while at even resonances they carry the same signature, forcing
the tongue to collapse identically. Second, closed-form analytic formulas
for the widths and boundary curves of all surviving instability tongues
are derived by a continued-fraction and Magnus--Winkler recurrence
method, confirmed against Cambi's 1950 exact numerical solution, and
recovered independently by the Yakubovich--Starzhinskii exponent-matrix
series. The continued-fraction approach is further developed into
a systematic method for the primary instability tongue: the Floquet
exponent is computed as an exact power series in the modulation amplitude
$\varepsilon$ to any desired order, with all coefficients given as
closed-form rational functions; the six-term series for the instability
boundary frequencies $\mu_{\pm}(\varepsilon)$ extends the leading-order
result by six orders of accuracy with all exact rational coefficients.
The same continued-fraction evaluation yields, at no additional cost,
explicit finite-product formulas for all Fourier coefficients of the
periodic Floquet factor --- a representation of the Floquet solution
that is inaccessible by the Magnus--Winkler or discriminant methods.

The boundary curves of the instability tongues consist entirely of
exceptional points of degeneracy (EPD) of the monodromy matrix, where
the two Floquet multipliers coalesce into a single Jordan block. This
EPD structure enables hypersensitive capacitance sensing: placing
the circuit at an EPD curve and introducing a small capacitance perturbation
splits the two coincident characteristic frequencies by an amount
proportional to the square root of the perturbation --- a response
that diverges relative to conventional linear sensing as the perturbation
shrinks. An explicit closed-form formula for this splitting in terms
of circuit parameters is derived. Since EPD points are only marginally
stable, a work-point strategy is developed that shifts the operating
point slightly into the stability zone, making the sensing scheme
robust to perturbations of either sign while preserving the square-root
sensitivity. 
\end{abstract}

\maketitle
\tableofcontents{}

\section{Introduction}

\label{sec:Intro}

Introducing time-periodicity into the coefficients of a linear oscillator
transforms a completely tractable problem into one of remarkable mathematical
depth. For a circuit or mechanical oscillator with \emph{constant}
parameters, stability is governed by a single eigenvalue: the natural
frequency $\omega_{0}$, solutions are pure sinusoids, and the spectral
analysis is complete in closed form. When a parameter varies \emph{periodically}
in time, this picture breaks down entirely: eigenfunctions no longer
exist in the classical sense, and stability must be determined through
Floquet\index{Floquet theory} theory, which replaces eigenvalues
with Floquet multiplier\index{Floquet theory!Floquet multiplier}s
(complex numbers encoding how solutions scale over one period), and
the spectrum with an alternating band structure of stable and unstable
regions. Even asking whether solutions remain bounded requires tracing
the monodromy matrix\index{monodromy matrix} --- the solution map
over one period --- and examining whether its eigenvalues lie on
or off the unit circle (on the unit circle means all solutions stay
bounded; off means exponential growth). The resulting instability
structure is governed by resonance\index{resonance}s between the
\emph{excitation frequency} $\mu$ and the \emph{natural frequency}
$\omega_{0}$ (frequency notation fixed at~\eqref{eq:LC-tau}; general
resonance-frequency formula at~\eqref{eq:YS-Hill-critical} of Chapter~\ref{app:ParRes}),
but which resonances are active, and how wide the resulting instability
tongues are, depends on the precise form of the coefficient and admits
closed-form answers only in exceptional cases. The LC circuit with
harmonically varying capacitance\index{capacitance sensing} is one
such exceptional case: its governing equation belongs to the class
of \emph{Ince equations} --- a family of Hill-type equations whose
instability structure can be determined analytically via three-term
recurrences --- and this work carries out that determination completely.

A nondissipative LC circuit with periodically varying capacitance
$C(t)=C(1+\varepsilon\cos\mu t)$ is among the simplest physical systems
exhibiting parametric instability (see Chapter~\ref{app:ParRes}
for a self-contained account of parametric resonance\index{parametric resonance},
including its energy-extraction mechanism and contrast with ordinary
forced resonance). Here $L>0$ is the inductance, $C>0$ is the mean
capacitance, $0<\varepsilon<1$ is the dimensionless modulation amplitude,
and $\mu>0$ is the modulation (excitation) frequency. In the engineering
literature this configuration is known as a \emph{frequency modulation}
circuit~\cite[Sec.~15.20]{MacLa}, \cite[Sec.~8.4]{MagWin}: the
periodic variation of $C(t)$ modulates the instantaneous natural
frequency $\omega(t)=1/\sqrt{LC(t)}$ about its mean value $\omega_{0}=1/\sqrt{LC}$.
The charge $q(t)$ satisfies 
\begin{equation}
L\,\partial_{t}^{2}q+\frac{q}{C(1+\varepsilon\cos\mu t)}=0,\qquad0<\varepsilon<1,\label{eq:LC-original}
\end{equation}
which, after the substitution $\tau=\mu t$ and with $r=\mu/\omega_{0}$,
$\omega_{0}=1/\sqrt{LC}$, becomes 
\begin{equation}
\partial_{\tau}^{2}q+\frac{q}{r^{2}(1+\varepsilon\cos\tau)}=0,\qquad\tau=\mu t,\quad r=\frac{\mu}{\omega_{0}},\quad\omega_{0}=\frac{1}{\sqrt{LC}}.\label{eq:LC-tau}
\end{equation}

\paragraph*{Two fundamental frequencies.}

\label{para:two-freq} The system is governed by two fundamental frequencies:
the \emph{natural frequency} $\omega_{0}=1/\sqrt{LC}$, fixed by the
circuit components $L$ and $C$, and the \emph{excitation frequency}
$\mu$, the frequency at which the capacitance is modulated. Throughout,
$\mu$ is held fixed and serves as the unit of frequency (the substitution
$\tau=\mu t$ makes this explicit): the natural frequency $\omega_{0}$---equivalently
the dimensionless ratio $r=\mu/\omega_{0}$, or the dimensionless
spectral parameter $c=4/r^{2}$---is the quantity that varies and
determines which instability regime the circuit occupies. Parametric
resonance occurs when the natural frequency is tuned to an odd sub-harmonic
of half the excitation frequency, $\omega_{0}=(2k-1)\mu/2$ ($k=1,2,3,\ldots$),
i.e.\ when $r=\mu/\omega_{0}=2/(2k-1)$. The even sub-harmonics $\omega_{0}=k\mu$
($r=1/k$) produce no instability---a fact invisible to the Mathieu\index{Mathieu equation}
approximation and one of the central results of this work. When we
write ``the $m$-th resonance'' we mean $\omega_{0}=(2k-1)\mu/2$
with $m=2k-1$, i.e.\ the natural frequency equals an odd multiple
of $\mu/2$. The general theory of critical resonance frequencies
for canonical systems, and its specialization to Hill's equation\index{Hill equation},
is given in Chapter~\ref{app:ParRes}, equations~\eqref{eq:YS-critical-freq}--\eqref{eq:YS-Hill-critical}.

\paragraph*{Spectral complexity: time-homogeneous vs.\ periodic systems.}

\label{para:Floquet-complexity} The qualitative difference between
a time-homogeneous and a time-periodic linear oscillator is sharper
than it may first appear, and it is worth making it precise before
entering the analysis.

For a \emph{time-homogeneous} second-order system $\ddot{q}+\omega_{0}^{2}q=0$
with constant coefficients, the spectral theory is complete in closed
form: the system has a single natural frequency $\omega_{0}$, and
every solution is a superposition of the two \emph{pure time-harmonic
eigenmodes} $e^{\pm i\omega_{0}t}$ --- each mode carries exactly
one frequency.

For a \emph{$T$-periodic} system such as the Hill equation~\eqref{eq:LC-Hill}
with period $T=2\pi/\mu$, the Floquet theory (Chapter~\ref{app:genLin})
replaces this picture entirely. The role of the single eigenfrequency
$\omega_{0}$ is taken by the \emph{characteristic exponents} $\alpha_{\pm}$,
defined by 
\begin{equation}
\rho_{\pm}=e^{i\pi\alpha_{\pm}},\label{eq:alpha-def-intro}
\end{equation}
where $\rho_{\pm}$ are the eigenvalues (Floquet multipliers) of the
monodromy matrix $X(\pi)$ (one half-period of the rescaled equation).
The exponents $\alpha_{\pm}$ are \emph{dimensionless}: they measure
frequency in units of $\mu/2$ (half the modulation frequency). The
physical oscillation frequencies of the two Floquet modes are 
\begin{equation}
f_{\pm}=\frac{\mu}{2}\,\alpha_{\pm},\label{eq:alpha-physical-freq}
\end{equation}
so $\alpha_{\pm}=1$ corresponds to the primary resonance frequency
$f=\mu/2=\omega_{0}$ (at the center of the primary tongue). In the
stable zone the exponents are real and $|\alpha_{\pm}|\leq1$; in
the instability tongue\index{instability tongue} they become complex
(one multiplier has $|\rho|>1$, giving exponential growth). The corresponding
\emph{Floquet eigenmodes}, written in the Hill variable $x=\tau/2=\mu t/2$,
are 
\begin{equation}
f_{\pm}(x)=p_{\pm}(x)\,e^{i\alpha_{\pm}x},\qquad p_{\pm}(x+\pi)=p_{\pm}(x),\label{eq:Floquet-mode-intro}
\end{equation}
where $p_{\pm}$ are $\pi$-periodic Floquet factor\index{Floquet theory!Floquet factor}s.
Expanding $p_{\pm}(x)=\sum_{n\in\mathbb{Z}}p_{n}^{(\pm)}\,e^{2inx}$
and converting to physical time via $x=\mu t/2$, each Floquet mode
carries a \emph{countably infinite comb} of frequencies 
\begin{equation}
\mathcal{F}_{\pm}=\Bigl\{\tfrac{\mu}{2}\alpha_{\pm}+n\mu\;:\;n\in\mathbb{Z}\Bigr\},\label{eq:Floquet-comb-intro}
\end{equation}
spaced by the modulation frequency $\mu$ and offset by the half-exponent
$\tfrac{\mu}{2}\alpha_{\pm}$. The amplitudes $p_{n}^{(\pm)}$ of
the individual comb lines are \emph{not} free parameters: they are
determined entirely by the circuit equation at the given operating
point $(\delta,c)$, and each Floquet mode~\eqref{eq:Floquet-mode-intro}
is a fixed waveform with a fixed spectral shape. For the LC circuit,
$|p_{n}^{(\pm)}|$ decays geometrically in $|n|$ at rate $\delta^{|n|}$
(reflecting the geometric decay of the Fourier coefficients $g_{n}=\hat{\lambda}\delta^{n}$),
so the comb is dominated by the lines nearest $\tfrac{\mu}{2}\alpha_{\pm}$.
This is qualitatively richer than the single eigenfrequency $\omega_{0}$
of the homogeneous case: each Floquet mode excites an entire harmonic
comb, with one spectral line per Fourier harmonic of its periodic
factor.

At the EPD operating point studied in Chapter~\ref{sec:EPD-sensor},
the two exponents coincide ($\alpha_{+}=\alpha_{-}=1$) and both combs
collapse onto the odd multiples of $\mu/2$. A small capacitance shift
$\Delta C$ splits the exponents apart by $\Delta\alpha=\alpha_{+}-\alpha_{-}$~\eqref{eq:freq-split-def},
separating the two combs. The $n$-th comb lines are at the frequencies
$\tfrac{\mu}{2}\alpha_{\pm}+n\mu$; their beat 
\[
\Bigl(\tfrac{\mu}{2}\alpha_{+}+n\mu\Bigr)-\Bigl(\tfrac{\mu}{2}\alpha_{-}+n\mu\Bigr)=\tfrac{\mu}{2}\,\Delta\alpha
\]
is the same for every harmonic index $n$, providing redundant spectral
measurements across the entire comb. This redundancy is the foundation
of the probing protocol of Section~\ref{subsec:probing}.

\begin{figure}[th]
\centering \includegraphics[width=10cm]{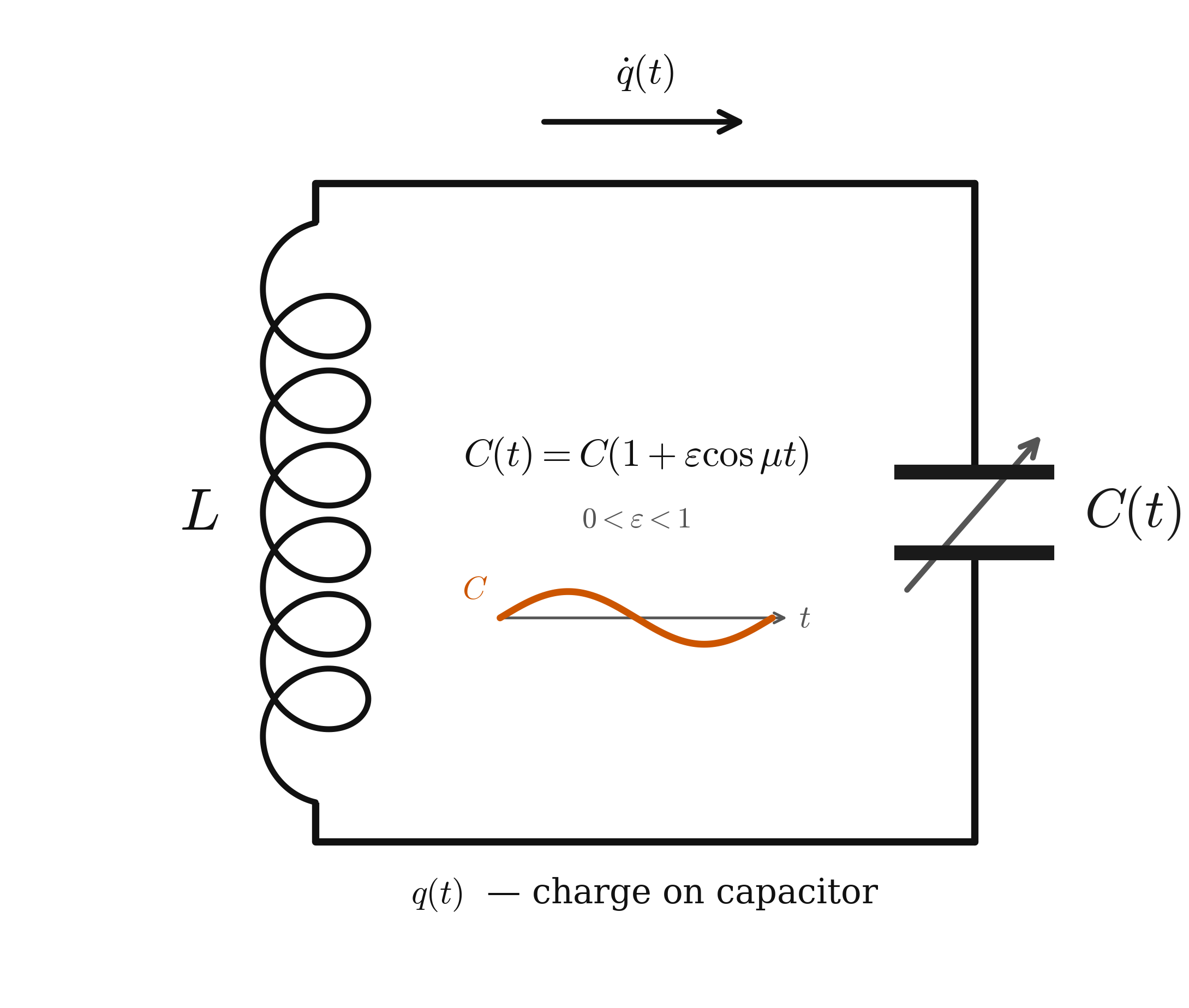} \caption{The LC circuit with periodically modulated capacitance. The inductor
$L$ (left) and variable capacitor $C(t)$ (right) form a closed loop.
The blue arrow denotes the current $\dot{q}(t)$; the charge on the
capacitor is $q(t)$. The diagonal arrow through the capacitor symbol
indicates a time-varying element. The inset shows one period of $C(t)=C(1+\varepsilon\cos\mu t)$,
oscillating between $C(1-\varepsilon)$ and $C(1+\varepsilon)$ about
the mean value $C$ (dashed line). The natural frequency of the unperturbed
circuit is $\omega_{0}=1/\sqrt{LC}$; $\mu$ is the excitation (modulation)
frequency; $r=\mu/\omega_{0}$ is the frequency ratio. Parametric
instability occurs at the odd sub-harmonics $\mu\approx2\omega_{0}/(2k-1)$,
$k=1,2,3,\ldots$, but not at the even ones $\mu\approx\omega_{0}/k$
(Corollaries~\ref{cor:even}--\ref{cor:odd}).}
\label{fig:circuit} 
\end{figure}

The analysis of~\eqref{eq:LC-tau} is a classical problem. Barrow's
1933 investigation~\cite[p.~1182]{Barrow33} replaced the actual
equation by the \emph{Mathieu equation} 
\begin{equation}
y''+(\lambda-2q\cos2x)\,y=0,\label{eq:Mathieu-intro}
\end{equation}
where $q$ is the Mathieu parameter (the half-amplitude of the cosine
modulation) and $\lambda$ is the spectral parameter; this approximation
is valid when $\varepsilon$ is small and retains only the first Fourier
harmonic of the LC coefficient $1/(1+\varepsilon\cos\tau)$. The Mathieu
parameter is related to the LC circuit parameters by 
\begin{equation}
q=\frac{2\varepsilon}{r^{2}}=\frac{2\varepsilon\omega_{0}^{2}}{\mu^{2}},\qquad r=\frac{\mu}{\omega_{0}},\label{eq:q-LC-params}
\end{equation}
so that at the primary resonance $\mu\approx2\omega_{0}$ (i.e.\ $r\approx2$)
one has $q\approx\varepsilon/2$. Most subsequent literature follows
this approximation. The exact solution was given by Cambi~\cite{Cambi1},
\cite[\S\S\,1--6]{Cambi2}, who showed by a continued-fraction\index{continued fraction}
method that equation~\eqref{eq:LC-tau} has instability intervals
at only \emph{half} the resonances predicted by the Mathieu approximation:
instability occurs only when $r\approx2/(2k-1)$ for $k=1,2,3,\ldots$,
never when $r\approx1/k$. Cambi's 1950 paper gives precise numerical
values of the instability boundaries that serve as benchmarks for
the present work.

The purpose of this work is to explain Cambi's results from first
principles using the Ince-equation framework of Magnus and Winkler~\cite[Ch.~8]{MagWin}
(hereafter MW), to derive closed-form expressions for the widths and
boundary curve\index{boundary curves}s of all surviving instability
intervals, and to verify these expressions numerically. We also apply
the same Ince-recurrence method to the Mathieu equation; the resulting
width formulas agree with the individual cases tabulated by McLachlan~\cite[Ch.~III]{MacLa}.

\subsection{Historical context and attribution}

The equation studied here has a distinguished pedigree. Ince introduced
in 1923--1927 a four-parameter generalization of Hill's equation
--- now bearing his name --- whose periodic solutions can be expressed
in closed form; his work remains the foundation for all subsequent
analysis of this class. Magnus and Winkler~\cite[Ch.~7]{MagWin}
gave in 1966 the definitive rigorous treatment of Hill's equation
and its variants, including the coexistence theorems (conditions on
the Floquet multiplier $\rho=\pm1$ for which two independent periodic
solutions coexist) on which the present work relies; a complementary
treatment of the Floquet characteristic exponents of the Mathieu equation
by analytic methods is given by Strang~\cite[\S\S\,1--3]{StraFL}.
In particular, MW Section~8.4 studies what they call the \emph{frequency
modulation equation} $(1+a\cos2x)y''+\lambda\,y=0$ --- a special
case of Ince's equation\index{Ince equation}~\eqref{eq:Ince-intro}
with $b=d=0$ and $\lambda$ the MW spectral parameter (our $\hat{\lambda}$,
eq.~\eqref{eq:Ince-param-LC} and~\eqref{eq:conv-c-hl}). This equation
\emph{is} the LC circuit~\eqref{eq:LC-original-MR} after the trivial
substitution $\tau=\mu t$, $x=\tau/2$: the parameter identification
is $a=-\varepsilon$ and $\hat{\lambda}=c(1+\delta^{2})/(1-\delta^{2})$
(eq.~\eqref{eq:Ince-MR} and Table~\ref{tab:aux-params}). MW prove,
using the coexistence theory, that all even instability intervals
vanish and all odd intervals survive. The present work makes the underlying
Ince structure explicit, derives closed-form width formulas and boundary
curves, and introduces the universal $\psi$-basis expansion. Cambi's
two papers~\cite{Cambi1}, \cite[\S\S\,1--15]{Cambi2} are a remarkable
achievement: working in 1948--1950 without the benefit of modern
computer algebra, he derived the exact continued-fraction equation
for the Floquet exponent\index{Floquet theory!Floquet exponent},
solved it numerically, and produced the stability diagram that is
reproduced (in corrected and extended form) here as Figure~\ref{fig:stability}.
Barrow~\cite[p.~1182]{Barrow33} gave the first systematic treatment
of the circuit, though through the Mathieu approximation. McLachlan's
monograph~\cite[Ch.~2]{MacLa} remains the standard reference for
the Mathieu equation and its instability intervals; its classical
German predecessor is Strutt~\cite[Ch.~2]{Strut}. Zygmund~\cite[Vol.~I, Ch.~III, \S\,3.3]{Zygmund}
and Gradshteyn--Ryzhik~\cite[1.447(3)]{GraRyzh} supply the \emph{Poisson
kernel Fourier series} 
\begin{multline}
\frac{1}{1+\varepsilon\cos\tau}=\frac{1+\delta^{2}}{1-\delta^{2}}\Bigl[1+2\sum_{n=1}^{\infty}(-1)^{n}\delta^{n}\cos n\tau\Bigr],\\
\delta=\frac{1-\sqrt{1-\varepsilon^{2}}}{\varepsilon},\quad0<\delta<1,\label{eq:Poisson-intro}
\end{multline}
that makes the Ince identification transparent (Chapter~\ref{sec:LCHill}):
the prefactor $(1+\delta^{2})/(1-\delta^{2})$ absorbs into the spectral
parameter, and the Fourier coefficients decay geometrically as $\delta^{n}$
--- this is the key structural reason why the continued-fraction
method and the universal $\psi$-basis both converge so efficiently
for the LC circuit. Hale~\cite[Ch.~IV]{Hale} and Keller~\cite[p.~192]{KelInst}
provide supporting results on Hill's equation used in Chapters~\ref{sec:MWInce}
and~\ref{sec:LCInce}.

\subsection{What this work contributes beyond Cambi}

Cambi's 1950 solution is exact and numerically precise, but it is
entirely implicit: the continued-fraction equation gives no closed-form
insight, and Cambi provided no explanation of \emph{why} the even
instability intervals vanish. Magnus and Winkler~\cite[Sec.~8.4]{MagWin}
proved that the even intervals vanish using the coexistence theory
(forcing the Floquet multiplier to $\rho=+1$ at even resonances)
they developed, but did not give closed-form widths, explicit Ince
parameters, boundary curves, or a comparison with the Mathieu approximation.
We recall here that coexistence --- two linearly independent solutions
sharing the same Floquet multiplier $\rho=\pm1$ --- forces the monodromy
matrix to equal $\pm\mathbb{I}$ exactly, and is the \emph{opposite}
of an exceptional point of degeneracy\index{exceptional point of degeneracy (EPD)}
(EPD): at an EPD the monodromy matrix has a non-trivial Jordan block\index{Jordan block}
with only one eigenvector, while coexistence requires two independent
eigenvectors, hence a scalar matrix (Chapter~\ref{sec:MWInce}; Theorem~\ref{thm:EPD}).
The present work adds the following. 
\begin{enumerate}
\item \emph{Harmonic capacitance variation is the natural model, and Mathieu
is only an approximation.} Harmonic variation of capacitance ---
$C(t)=C(1+\varepsilon\cos\mu t)$ --- is the practically realizable
and physically natural way to modulate an LC circuit. The Mathieu
equation arises only when the restoring term $1/C(t)$ is approximated
by its first Fourier harmonic; the exact equation retains all harmonics
through the geometric series $1/(1+\varepsilon\cos\mu t)$. The Mathieu
approximation is not merely quantitatively inaccurate --- it is qualitatively
different, predicting instability at even resonances that are in fact
completely stable. This work uses the exact equation throughout. 
\item \emph{Explicit Ince-equation identification and algebraic explanation.}
Magnus and Winkler~\cite[Sec.~8.4]{MagWin} proved that all even
instability intervals of~\eqref{eq:LC-Hill} vanish, using the coexistence
theorem they developed. The present work makes the underlying structure
explicit: the circuit equation is precisely \emph{Ince's equation},
the four-parameter family (Remark~\ref{rem:ince-id}) 
\begin{gather}
(1+a\cos2x)\,y''+(b\sin2x)\,y'+(c+d\cos2x)\,y=0,\qquad|a|<1,\label{eq:Ince-intro}\\
a=-\frac{2\delta}{1+\delta^{2}},\quad b=d=0,\quad c=\frac{4}{r^{2}},\quad0<\delta<1,\nonumber 
\end{gather}
where $0<\delta<1$ is the auxiliary modulation parameter defined
by 
\begin{gather}
\delta=\frac{1-\sqrt{1-\varepsilon^{2}}}{\varepsilon}=\frac{\varepsilon}{1+\sqrt{1-\varepsilon^{2}}},\nonumber \\
\text{ or equivalently }\varepsilon=\frac{2\delta}{1+\delta^{2}},\qquad0<\varepsilon<1,\label{eq:Ince-intro-1}
\end{gather}
see Chapter~\ref{sec:LCHill}. With this identification the MW proof
becomes transparent. The \emph{coexistence polynomials} (defined precisely
in Chapter~\ref{sec:MWInce}, eq.~\eqref{eq:QPoly}) are 
\begin{equation}
Q(\mu)=2a\mu^{2},\qquad Q^{*}(\mu)=a(2\mu-1)^{2},\qquad a=-\frac{2\delta}{1+\delta^{2}},\label{eq:QQstar-intro}
\end{equation}
where $\mu$ is the MW auxiliary integer index --- not to be confused
with the modulation frequency, for which MW happen to use the same
letter --- governing period-$\pi$ and period-$2\pi$ solutions respectively
(MW~\cite[Sec.~7.1]{MagWin}); their integer roots determine which
instability intervals collapse to zero width. $Q(\mu)$ has the integer
root $\mu=0$, which forces period-$\pi$ coexistence (Floquet multiplier
$\rho=+1$) at every even resonance by MW Theorem~7.6~\cite[Thm.~7.6]{MagWin}.
$Q^{*}(\mu)=a(2\mu-1)^{2}$ has no integer roots, so all odd intervals
survive (MW Theorem~7.1, necessity). No approximation, truncation,
or numerics are required. This explanation is entirely absent from
Cambi's work, which predates the MW framework by nearly two decades. 
\item \emph{Explicit Ince-equation parametrization.} While MW Section~8.4
proves the instability structure directly from the equation $(1+a\cos2x)y''+\lambda\,y=0$
(Ince~\eqref{eq:Ince-intro} with $b=d=0$, where $\lambda$ is MW's
spectral parameter), the explicit identification of the LC circuit
as Ince's equation~\eqref{eq:Ince-intro} with parameters 
\begin{equation}
a=-\frac{2\delta}{1+\delta^{2}},\quad b=0,\quad d=0,\quad c=\frac{4}{r^{2}},\quad\hat{\lambda}=\frac{c(1+\delta^{2})}{1-\delta^{2}},\label{eq:Ince-param-LC}
\end{equation}
is manifest. This identification connects the circuit equation to
a rich mathematical theory: Ince's coexistence theorems, the MW discriminant\index{discriminant}
expansion, and the exact recurrence relations for periodic solutions.
\emph{Ince's equation is the most general Hill-type equation to which
the three-term recurrence method applies}~\cite[Sec.~7.1]{MagWin};
the LC circuit is its simplest non-trivial example. The intimate connection
between three-term recurrences and continued fractions --- and the
role of Pincherle\index{continued fraction!Pincherle theorem}'s theorem
as the bridge between them --- is discussed in Remark~\ref{rem:CF-three-term}
and Chapter~\ref{app:CF}; see also~\cite[\S1]{Gautschi67} and~\cite[Sec.~5.2--5.3]{JonThr}. 
\item \emph{Closed-form width formula and computational advantage.} Cambi
gives numerical widths at specific parameter values. Formula~\eqref{eq:Lm-exact}
gives the width of every odd instability interval as an explicit algebraic
function of $\delta$, valid for all $0<\delta<1$, requiring no numerical
computation. The CF criterion~\eqref{eq:CF-criterion} (the equation
$F_{\mathrm{even/odd}}(c)=0$ whose roots are the instability boundaries)
also has a computational advantage over the MW discriminant series
for locating boundary values: the continued fractions $F_{\mathrm{even/odd}}(c)$
converge for all $0<\delta<1$ with no truncation error (Theorem~\ref{thm:Pinch};
see Remark~\ref{rem:CF-three-term} for the conceptual role of the
three-term structure), while the MW series introduces an error $O(\delta^{2N+2})$
when truncated at order $N$. The formula is confirmed numerically
against Cambi's continued fraction to better than $2.3\%$ for $\delta\leq0.20$. 
\item \emph{Complete boundary curves.} Beyond widths, each instability tongue
has two individual boundary curves. Tables~\ref{tab:bdry-LC} and~\ref{tab:bdry-Math}
give these explicitly, decomposed into a symmetric center shift (the
same for both boundaries, $O(\delta^{2})$) and an antisymmetric half-width
($O(\delta^{m})$, opposite in sign for the two boundaries). Cambi
provides raw numerical boundary values; this work gives their analytic
structure. 
\item \emph{The continued-fraction (CF) difference identity and the $5\!:\!1$
asymmetry.} The continued-fraction difference $F_{\mathrm{even}}-F_{\mathrm{odd}}=-2/9$
(constant, LC circuit) encodes the ratio $Q^{*}(-1)/Q^{*}(0)=9$ (eq.~\eqref{eq:QQstar-intro}),
which is determined purely by the Ince parameters and controls the
leading-order width. For Mathieu the analogous difference is $-2$,
reflecting constant coupling $q$ (the Mathieu parameter, eq.~\eqref{eq:Mathieu-intro})
at every recurrence step. This identity also explains the $5\!:\!1$
ratio of the two Mathieu $m=2$ boundary shifts ($5q^{2}/12$ vs.\ $q^{2}/12$),
which arises from the asymmetry between the cosine-type and sine-type
period-$\pi$ starters. The identity is proved in Chapter~\ref{sec:LCInce}
and stated precisely as eq.~\eqref{eq:CF-diff} in Chapter~\ref{sec:MainResults}. 
\item \emph{Closed-form width formula for Mathieu instability intervals.}
The same Ince recurrence method, applied to the Mathieu equation as
the degenerate case $a=0$, derives the closed-form formula for $L_{m}^{\mathrm{Math}}$
--- the width of the $m$-th Mathieu instability tongue. Here $q$
denotes the standard Mathieu parameter (the half-amplitude of the
cosine term in eq.~\eqref{eq:Mathieu-intro}): 
\begin{equation}
L_{m}^{\mathrm{Math}}=\frac{q^{m}}{2^{2m-3}[(m-1)!]^{2}}+O_{m}(q^{m+2}),\qquad m=1,2,3,\ldots,\label{eq:Lm-Math-intro}
\end{equation}
for all $m\geq1$ in a uniform way; the error is $O_{m}(q^{m+2})$,
two powers beyond the leading term. This formula does not appear in
McLachlan~\cite[Sec.~2.151]{MacLa}, who gives the characteristic
numbers $a_{m}$, $b_{m}$ case by case (Sec.~2.151) without a closed-form
expression for the widths $L_{m}=a_{m}-b_{m}$. The formula~\eqref{eq:Lm-Math-intro}
is verified against McLachlan's individual values: $L_{1}=2q$, $L_{2}=q^{2}/2$,
$L_{3}=q^{3}/32$, $L_{4}=q^{4}/1152$ (obtained directly from his
Section~2.151 expansions), confirming each leading coefficient to
the stated precision. The structural reason for the qualitative difference
between the LC and Mathieu formulas --- growing coupling $a(2n-1)^{2}$
vs.~constant coupling $q$ --- is transparent from the recurrence. 
\item \emph{Accuracy of the LC width formula.} The primary width $L_{1}=2\delta/(1-\delta^{2})$
has no chain-approximation error (the $k=1$ backward chain has no
non-resonant steps), so its only error is the inherent $O(\delta^{3})$
in the boundary curve expansion --- unlike its Mathieu counterpart
$L_{1}=2q$, which carries an $O(q^{3})$ correction already in its
leading coefficient. For higher odd intervals the formula has absolute
error $O_{m}(\delta^{m+2})$ --- two powers beyond the leading term,
because the width expansion contains only odd powers of $\delta$;
the analogous Mathieu width expansion likewise contains only alternate
powers, with error $O_{m}(q^{m+2})$ (eq.~\eqref{eq:Lm-Math-intro}). 
\item \emph{Vector form and Mobius\index{Mobius transformation} transformation
structure.} The Ince recurrence is cast in transfer-matrix form $V_{n+1}=M_{n}V_{n}$,
where the matrices $M_{n}$ converge to a constant limit $M_{\infty}$
whose associated Mobius transformation $T_{M_{\infty}}$ has exactly
two fixed points: $\zeta_{1}=\delta$ (repulsive) and $\zeta_{2}=1/\delta$
(attractive). Remarkably, $\delta$ and $1/\delta$ are simultaneously
the \emph{eigenvalues} of the matrix $M_{\infty}$ itself (eq.~\eqref{eq:Minf-eigs},
Remark~\ref{rem:eig-fixpt}): both properties --- fixed point of
$T_{M_{\infty}}$ and eigenvalue of $M_{\infty}$ --- are expressions
of the same underlying companion-matrix structure of the Ince recurrence,
and constitute a remarkable special feature of the LC circuit (Section~\ref{subsec:eig-fixpt}).
The parameter $\delta$ is thus not merely a convenient reparametrization
of $\varepsilon$ --- it is the distinguished parameter singled out
by the Mobius geometry of the recurrence, governing the exponential
decay rate $\delta^{n}$ of the minimal solution\index{continued fraction!minimal solution}.
The ratio $\delta^{2}=|\lambda_{1}/\lambda_{2}|<1$ identifies which
solution is minimal (decaying, magnitude $\delta^{n}$) and which
is dominant, connecting the asymptotic behavior of solutions to the
Poincaré--Perron\index{Poincaré--Perron theorem} theorem and to
the geometric theory of Mobius transformations (Chapter~\ref{app:Moeb}).
The instability boundary\index{stability boundary} curves are precisely
those values of the spectral parameter at which the associated Mobius
transformation becomes \emph{parabolic} (having a single fixed point
rather than two --- the Mobius analogue of a Jordan block) --- an
exceptional point of degeneracy in the monodromy matrix, where the
two Floquet multipliers coalesce with a non-trivial Jordan block. 
\item \emph{A universal entire-function expansion for the Hill discriminant\index{discriminant!Hill discriminant}.}
The LC circuit analysis motivates a structural result about Hill's
equation in general (Chapter~\ref{sec:UnivExp}). For any two-parameter
Hill family $y''+[\lambda+\varepsilon Q_{0}(x)]y=0$ with Fourier
coefficients $g_{n}$ of the perturbation $Q_{0}$, the discriminant
$\Delta(\lambda,\varepsilon)$ expands as a power series in $\varepsilon^{2}$
whose coefficients are entire functions of $\lambda$: 
\begin{gather}
\Delta(\lambda,\varepsilon)=2\cos\pi\sqrt{\lambda}+\frac{\pi\varepsilon^{2}\sin\pi\sqrt{\lambda}}{2\sqrt{\lambda}}\sum_{n=1}^{\infty}\frac{g_{n}^{2}}{\lambda-n^{2}}\nonumber \\
+O(\varepsilon^{4}),\quad\varepsilon\to0.\label{eq:Delta-univ-intro}
\end{gather}
The coefficient functions, called the \emph{$\psi$-basis}, 
\begin{equation}
\psi_{n}(\omega)=\frac{(-1)^{n}\sqrt{2}\,\omega\sin\pi\omega}{\omega^{2}-n^{2}},\quad\omega=\sqrt{\lambda},\qquad n=1,2,3,\ldots,\label{eq:phi-basis-intro}
\end{equation}
are entire (the apparent poles at $\lambda=n^{2}$ are removable by
L'Hôpital's rule) and universal: they depend only on $n$ and $\lambda$,
not on the specific Hill equation. The LC circuit gives the canonical
example ($g_{n}=\hat{\lambda}\delta^{n}$, geometric decay); the Mathieu
equation is the single-harmonic special case ($g_{1}=q$, $g_{n\geq2}=0$).

The $\psi$-basis possesses several remarkable properties. Each $\psi_{n}$
is a \emph{sinc-type function}: $\psi_{n}(\omega)=\frac{(-1)^{n}\sqrt{2}\,\omega\sin\pi\omega}{\omega^{2}-n^{2}}$
(eq.~\eqref{eq:psi-def}) with $\omega=\sqrt{\lambda}$, which connects
it to the zeroth spherical Bessel function $j_{0}$ via $\operatorname{sinc}(x)=j_{0}(\pi x)$.
The functions $\psi_{n}$ are related by rational factors: $\psi_{n}=(-1)^{n-m}\psi_{m}\cdot(\omega^{2}-m^{2})/(\omega^{2}-n^{2})$,
so the entire family is generated from any single member. Their zeros
are the perfect squares $\lambda=k^{2}$ for $k\neq n$ (all shared),
and the value at the resonance $\lambda=n^{2}$ is $\psi_{n}(n)=\pi/\sqrt{2}$.

At the analytic level, $\sin\pi\sqrt{\lambda}$ satisfies the Sturm--Liouville\index{Sturm--Liouville theory}
equation $-(\sqrt{\lambda}\,S')'=\frac{\pi^{2}}{4}S/\sqrt{\lambda}$
in self-adjoint form, making it an eigenfunction of the associated
SL operator with eigenvalue $\pi^{2}/4$. Each $\psi_{n}$ is related
to the resolvent (Green's function) of this SL operator at the spectral
parameter $\pi^{2}/4$, evaluated at the pole $\lambda=n^{2}$.

At the integral-equation level, the standard solutions $y_{1}$, $y_{2}$
of Hill's equation satisfy Volterra integral equations with kernel
$\mathcal{K}(s,t;\lambda)=\sin[\sqrt{\lambda}(s-t)]/\sqrt{\lambda}$
(the causal Green's function of $d^{2}/ds^{2}+\lambda$)~\cite[Sec.~4.3]{Eastham73}.
These equations are the basis for the Picard iteration underlying
the MW expansion and provide the most efficient numerical method for
computing $\Delta_{\omega}$.
\item \emph{High-accuracy instability boundary curves from the CF method.}
Cambi's continued-fraction condition $F_{w}(u^{*})=0$ on the double
minimality function yields the primary Floquet exponent $u^{*}$ as
an exact power series in $\gamma^{2}=\varepsilon^{2}/4$, with all
coefficients computable as closed-form rational functions of $p$
by the recursive algorithm of Section~\ref{subsec:Cambi-notes2-precision}.
Translating the resulting six-term boundary series $p_{\pm}(\gamma)$
(Theorem~\ref{thm:bdry-series-CF}) back to physical driving frequencies
$\mu_{\pm}=\omega_{0}/p_{\mp}(\varepsilon/2)$ gives: 
\begin{align}
\mu_{-}(\varepsilon) & =\omega_{0}\!\Bigl[2-\tfrac{1}{2}\varepsilon+\tfrac{15}{32}\varepsilon^{2}-\tfrac{139}{512}\varepsilon^{3}+\tfrac{2341}{8192}\varepsilon^{4}\nonumber \\
 & \quad-\tfrac{25659}{131072}\varepsilon^{5}+\tfrac{885891}{4194304}\varepsilon^{6}+O(\varepsilon^{7})\Bigr],\nonumber \\
\mu_{+}(\varepsilon) & =\omega_{0}\!\Bigl[2+\tfrac{1}{2}\varepsilon+\tfrac{15}{32}\varepsilon^{2}+\tfrac{139}{512}\varepsilon^{3}+\tfrac{2341}{8192}\varepsilon^{4}\nonumber \\
 & \quad+\tfrac{25659}{131072}\varepsilon^{5}+\tfrac{885891}{4194304}\varepsilon^{6}+O(\varepsilon^{7})\Bigr],\label{eq:mu-pm-intro}
\end{align}
extending the $O(\varepsilon)$ result of the existing Theorem~\ref{thm:primary-domain}
by six orders. All coefficients are exact rationals; the odd-order
terms appear with opposite signs in $\mu_{-}$ and $\mu_{+}$, while
the even-order terms (the center shift) are identical. The width of
the instability domain to $O(\varepsilon^{7})$ is: 
\begin{equation}
\Delta\mu(\varepsilon)=\omega_{0}\!\left[\varepsilon+\tfrac{139}{256}\varepsilon^{3}+\tfrac{25659}{65536}\varepsilon^{5}+O(\varepsilon^{7})\right].\label{eq:Delta-mu-intro}
\end{equation}
The recursive algorithm extends these series to any desired order,
and with $N$ CF levels gives $u^{*}$ to $O(\gamma^{2N+2})$ accuracy.
This decisively outperforms the MW discriminant series, whose convergence
requires computing progressively harder higher-order terms (Chapter~\ref{sec:UnivExp}).
\item \emph{Explicit Fourier coefficients of the periodic Floquet factor.}
For a Floquet solution $f(x,u^{*})=e^{2\pi iu^{*}x}P(u^{*},x)$ with
periodic factor $P(u^{*},x)=\sum_{m}B_{m}e^{2\pi imx}$, the Fourier
coefficients are hard to extract by MW or discriminant methods. The
CF method gives them explicitly: once $u^{*}$ is located, Corollary~\ref{cor:Floquet-Fourier}
(Section~\ref{subsec:Cambi-notes2-H}) provides the exact finite-product
formulas 
\begin{equation}
\frac{h_{m}(u^{*})}{h_{0}(u^{*})}=\begin{cases}
{\displaystyle \prod_{k=0}^{m-1}w(u^{*}+k),} & m\geq1,\\[6pt]
{\displaystyle \prod_{k=0}^{|m|-1}w(-u^{*}+k),} & m\leq-1,
\end{cases}\quad B_{m}=\frac{p^{2}}{(u^{*}+m)^{2}}\,h_{m}(u^{*}),\label{eq:Floquet-Fourier-intro}
\end{equation}
where $w(u)$ is the CF minimal-solution ratio (Section~\ref{subsec:Cambi-notes2}).
Each product is finite, computed from the same CF evaluation used
to locate $u^{*}$, at no additional cost. The sequences $\{B_{m}\}$
decay exponentially: $|B_{m}|\sim|\zeta_{+}|^{|m|}$ as $|m|\to\infty$,
where $|\zeta_{+}|=(\gamma+O(\gamma^{3}))<1$ is the smaller characteristic
root (Section~\ref{subsec:Cambi-v-props}). Figures~\ref{fig:Bm-p03}--\ref{fig:Bm-Cambi2}
illustrate the Fourier spectra for representative parameter values,
confirming the exponential envelope and the mirror-image symmetry
$B_{m}(u^{**})=B_{-m}(u^{*})$ between the two Floquet solutions.
This explicit Fourier representation of the Floquet solution is absent
from Cambi's original work and from the MW framework. 
\item \emph{The instability boundaries by a third, independent method.}
The same boundary curves obtained from the MW coexistence theory and
the continued fraction are recovered by the high-frequency Yakubovich--Starzhinskii\index{Yakubovich--Starzhinskii series}
exponent-matrix series. The exponent matrices $\mathbf{K}_{m}(\hat{\lambda})$
are computed in closed form, and the boundary appears as the vanishing
of an off-diagonal product $K_{01}K_{10}=0$ --- a linear-algebraic
degeneracy of the monodromy exponent. This recovers the primary boundary
$c_{\pm}=1\pm\tfrac{\varepsilon}{2}-\tfrac{9}{32}\varepsilon^{2}+O(\varepsilon^{3})$,
in particular the same center shift $-9/32$, providing a third independent
confirmation; the three approaches are reconciled through the relation
$K=\tfrac{1}{\pi}\log(-M)$ (Theorem~\ref{thm:YS-boundary-MR}, Chapters~\ref{sec:YS-LC},
\ref{sec:YS-exact-LC}, \ref{sec:YS-comparison}). 
\end{enumerate}

\subsection{Main results}

The main results are stated precisely in Chapter~\ref{sec:MainResults},
immediately following this introduction, with forward references to
the sections where they are proved. There are three independent contributions.
The first concerns the LC circuit specifically: the circuit equation
is a special case of Ince's equation, the exact instability structure
follows from the Magnus--Winkler\index{Magnus--Winkler theory} coexistence
theory (the Floquet multiplier is $\rho=+1$ at even resonances, forcing
two independent periodic solutions to coexist and collapsing the instability
interval to zero width), and the widths of all surviving instability
intervals and complete boundary curves are given in closed form. The
continued-fraction approach additionally yields six-term exact series
for the primary boundary frequencies $\mu_{\pm}$ (Theorem~\ref{thm:primary-domain-CF})
and explicit finite-product formulas for the Fourier coefficients
of the periodic Floquet factor (Theorem~\ref{thm:Floquet-factor-MR}).
The same boundary curves are recovered a third way, independently
of both Magnus--Winkler and the continued fraction, by the Yakubovich--Starzhinskii
exponent-matrix series, which yields the closed-form matrices $\mathbf{K}_{m}(\hat{\lambda})$
and exhibits each boundary as a linear-algebraic degeneracy of the
monodromy exponent (Theorem~\ref{thm:YS-boundary-MR}, Chapters~\ref{sec:YS-LC},
\ref{sec:YS-exact-LC}, \ref{sec:YS-comparison}). The second is the
universal entire-function expansion of the Hill discriminant (Theorem~\ref{thm:univ-exp}
and Proposition~\ref{prop:bdry-leading} in Chapter~\ref{sec:UnivExp}):
the $\psi$-basis $\{\psi_{n}\}$, its sinc and Sturm--Liouville
structure, and the Volterra-kernel origin of the expansion. The third
is the EPD hypersensitivity\index{exceptional point of degeneracy (EPD)!hypersensitivity}
application of Chapter~\ref{sec:EPD-sensor}: each instability boundary
is an EPD point of the monodromy matrix, and operating the circuit
at such a point yields a $\sqrt{\Delta C/C_{0}}$ frequency-split
response to a small capacitance perturbation $\Delta C$, with an
explicit closed-form formula and a Floquet-comb probing protocol.

\subsection{Stability boundaries, EPD points, and Krein's collision theory}

The instability boundaries of the LC circuit --- the values of the
frequency ratio $r$ at which a stability tongue opens or closes ---
carry a precise geometric meaning rooted in the symplectic\index{symplectic system}
structure of the problem. At each boundary the two Floquet multipliers
of the monodromy matrix coincide at $\rho=\pm1$. For a $2\times2$
symplectic matrix, a double multiplier on the unit circle is of exactly
one of two types~\cite[Ch.~VIII, \S\S\,1.4--1.8]{YakSta2}: either
the monodromy matrix equals $\pm\mathbf{I}$ (semisimple, two independent
periodic solutions coexist), or it has a non-trivial Jordan block
(exactly one periodic solution --- an \emph{exceptional point of
degeneracy}, EPD, in the sense of~\cite[Sec.~3.3.2]{Kiril}, \cite[Ch.~VIII, \S\,1.8]{YakSta2}).

The organizing principle that determines which type occurs is \emph{Krein\index{Krein collision theory}'s
collision theory} (Chapter~\ref{app:Krein}). Since the LC circuit
is a Hamiltonian system (Section~\ref{subsec:Hamiltonian}), its
monodromy matrix is symplectic, and each Floquet multiplier on the
unit circle carries a \emph{Krein signature\index{Krein signature}}
$\kappa=\pm1$ measuring whether the associated mode has positive
or negative energy~\cite[Sec.~3.3.1]{Kiril}, \cite[Ch.~III, \S\,1]{YakSta1},
\cite[Sec.~9.4]{SeyMai}. The signature determines the outcome of
a collision: same-signature multipliers cannot leave the unit circle
and the system remains stable, while opposite-signature multipliers
generically split off the unit circle and open an instability tongue.

This criterion gives the symplectic explanation of the alternating
even/odd instability structure of the LC circuit, stated precisely
as Theorem~\ref{thm:EPD} in Chapter~\ref{sec:MainResults}: the
even resonances carry same-signature double multipliers (semisimple,
both modes with equal-sign energy, no instability gap), while \emph{the
odd resonances carry opposite-signature EPD multipliers} (modes with
opposite-sign energy, instability tongue opens). Chapter~\ref{sec:EPD-elementary}
gives an independent elementary proof of the Jordan block conclusion
(Theorem~\ref{thm:Jordan-bdry}) using only the Hill discriminant
and the Bauer--Fike perturbation bound~\cite[Sec.~6.3]{HorJohn},
with no symplectic structure required. For the Mathieu equation every
resonance produces an EPD with opposite signatures, so all instability
zone\index{instability zone}s persist --- a symplectic distinction
from the LC circuit that is deeper than any comparison of zone widths.

\subsection{Organization}

Chapter~\ref{sec:MainResults} states all main results precisely
with forward references and may be read independently of the technical
development.

Chapter~\ref{sec:LCHill} derives the Hill equation form of the LC
circuit, introduces the modulation parameter $\delta$ through the
Poisson kernel Fourier series~\cite[Vol.~I, Ch.~III, \S\,3.3]{Zygmund},
\cite[1.447(3)]{GraRyzh}, and establishes the Fourier coefficients
$g_{n}=\hat{\lambda}\delta^{|n|}$.

Chapter~\ref{sec:MWInce} develops the MW discriminant expansion,
Keller's formula, the coexistence theory of Ince's equation (MW Theorems~7.1
and~7.6 and MW Lemma~7.4), and the discriminant identity~\eqref{eq:DeltaAtEven}
(at every even resonance the two periodic eigenvalues coincide and
$\Delta=+2$ there with a double zero --- the zero-width collapse)
for even resonances.

Chapter~\ref{sec:UnivExp} derives the universal entire-function
expansion of the Hill discriminant via the $\psi$-basis, the Volterra
integral equations, and the Sturm--Liouville structure.

Chapter~\ref{sec:LCInce} is the analytical core of this work. Starting
from Floquet's theorem and the Ince identification (Remark~\ref{rem:ince-id}),
it derives the CF eigenvalue condition~\eqref{eq:CF-criterion},
develops the asymptotic analysis via Poincaré--Perron and Mobius
transformations (Section~\ref{subsec:VecForm}), and derives the
closed-form width formula~\eqref{eq:Lm-main} (Theorem~\ref{thm:width}).

Chapter~\ref{sec:Cambi} verifies all formulas against Cambi's exact
continued-fraction solution~\cite[Table~II]{Cambi2}.

Chapter~\ref{sec:Boundaries} derives the complete instability boundary
curves and the Mathieu width formula~\eqref{eq:Lm-Math-main}.

\emph{Chapter~\ref{sec:Summary}} provides a self-contained summary
and comparison of the two methods. Readers who find the multi-step
analyses of Chapters~\ref{sec:MWInce}--\ref{sec:Boundaries} difficult
to navigate are encouraged to consult Chapter~\ref{sec:Summary}
first: it gives the destination before the journey, states each result
with an explicit pointer to where it is proved, and explains how the
MW and CF methods relate to each other and to Cambi's work.

\emph{Chapters~\ref{sec:YS-LC}--\ref{sec:YS-comparison}} develop
the Yakubovich--Starzhinskii (YS) exponent-matrix method as a third
independent approach to the instability boundaries. Chapter~\ref{sec:YS-LC}
applies YS to the Mathieu special case; Chapter~\ref{sec:YS-exact-LC}
applies it to the exact LC circuit with $\hat{\lambda}$ free, deriving
the exact closed-form $\mathbf{K}_{m}(\hat{\lambda})$ coefficients,
the self-consistent boundary analysis (\S\,\ref{subsubsec:YS-bdry-equations}),
and confirming the boundary curves of Table~\ref{tab:bdry-LC} to
$O(|a|^{3})$ by an independent route. The YS method works directly
with the Floquet decomposition $X(\tau,\varepsilon)=F(\tau,\varepsilon)e^{\tau K(\varepsilon)}$:
the boundary condition $K_{01}\cdot K_{10}=0$ is an explicit algebraic
statement about a $2\times2$ matrix, requiring no theory of entire
functions (MW) or continued-fraction convergence (CF). Its principal
advantage over MW and CF is for \emph{multiparameter} families of
oscillators (as in Seyranian--Mailybaev~\cite[Ch.~1]{SeyMai} and
Kirillov~\cite[Sec.~3.3]{Kiril}): the YS coefficients $\mathbf{K}_{m}$
are jointly analytic in all parameters simultaneously (Hartogs' theorem,
Theorem~\ref{thm:joint-analytic}), so the boundary surface in the
full parameter space is encoded in a single matrix condition without
recomputing the analysis at each parameter point. Chapter~\ref{sec:YS-comparison}
gives a systematic three-way comparison of YS, MW, and CF.

Chapter~\ref{sec:EPD-elementary} establishes that every instability
boundary point is an EPD of the monodromy matrix $X(\pi)$, i.e.\ that
$X(\pi)$ has a non-trivial Jordan block there. The proof is elementary:
the Hill multiplier formula~\eqref{eq:Hill-multipliers} establishes
an $O(\sqrt{|\Delta c|})$ splitting of the Floquet multipliers near
any boundary point, and the Bauer--Fike theorem~\cite[Sec.~6.3]{HorJohn}
shows this is incompatible with diagonalizability. No symplectic structure
or Krein signature analysis is required. This section bridges the
instability theory (Chapters~\ref{sec:LCHill}--\ref{sec:Summary})
and the sensing application (Chapter~\ref{sec:EPD-sensor}).

Chapter~\ref{sec:EPD-sensor} applies the EPD curve\index{exceptional point of degeneracy (EPD)!EPD curve}
structure to capacitance sensing. The instability boundaries are EPD
points of the monodromy matrix; operating the circuit at such a point
and introducing a small capacitance shift $\Delta C$ splits the two
coincident Floquet characteristic exponents by $\Delta\alpha\propto\sqrt{\Delta C/C_{0}}$,
giving hypersensitivity relative to a conventional sensor. The section
derives a closed-form formula for the primary tongue, proves the $\sqrt{\xi}$
scaling, quantifies the advantage over conventional sensing, and describes
a Floquet-comb probing protocol for experimental recovery of $\Delta C$.

Chapter~\ref{sec:Conc} summarizes the contributions and discusses
extensions.

Chapters~\ref{app:genLin}--\ref{app:Moeb} collect background on
linear periodic systems and Floquet theory, Hill's equation, finite
difference equations and the Poincaré--Perron theorem, continued
fractions and the Pincherle theorem, and Mobius transformations. Chapter~\ref{app:entire}
discusses the entire-function property of the Hill discriminant. Chapter~\ref{app:ParRes}
is a self-contained treatment of parametric resonance: it develops
the Yakubovich--Starzhinskii argument justifying the general critical-frequency
formula~\eqref{eq:YS-critical-freq}, derives the energy-extraction
mechanism of parametric resonance (equations~\eqref{eq:energy-rate}--\eqref{eq:energy-cycle}),
contrasts parametric resonance with ordinary forced resonance, and
discusses the meaning of ``negative energy'' in this context and
its relationship to the usage in Kirillov~\cite[Sec.~3.3.8]{Kiril}
and in traveling-wave tube theory~\cite[\S\,1]{FigTWT,RabTrub89}.
Chapter~\ref{app:Krein} develops Krein signature theory and strong
stability. Chapter~\ref{app:notation} collects all notation.

\section{Summary of Main Results}

\label{sec:MainResults}

Throughout this chapter \emph{Theorems} are our original contributions;
\emph{Propositions} are results of other authors on which our work
builds, stated here for completeness and to make logical dependencies
explicit. Proofs are in the chapters indicated by the forward references.
All symbols and abbreviations are defined in the Notation chapter
(Chapter~\ref{app:notation}).

The physical phenomenon underlying all results is \emph{parametric
resonance\index{resonance}} --- the exponential growth of oscillations
driven by periodic variation of a system parameter (here, the capacitance\index{capacitance sensing});
see Chapter~\ref{app:ParRes} for a self-contained account. The principal
references are Cambi~\cite{Cambi1}, \cite[\S\S\,1--15]{Cambi2},
who first identified the odd-only instability pattern numerically
and derived an exact continued-fraction\index{continued fraction}
representation for the EPD\index{exceptional point of degeneracy (EPD)}
curve\index{exceptional point of degeneracy (EPD)!EPD curve}s; Magnus
and Winkler~\cite[Ch.~7--8]{MagWin}, whose coexistence theory provides
the algebraic explanation of the even/odd alternation; and Yakubovich
and Starzhinskii~\cite[Ch.~IV]{YakSta1}, \cite{YakSta2}, whose
Lyapunov--Floquet\index{Floquet theory} series furnish the perturbation
expansion tools. The present work derives explicit closed-form asymptotic
formulas from Cambi's implicit continued-fraction representation.
The chapter proceeds from the circuit model and parameter relations
(\S\,\ref{subsec:MR-equations}) through the stability landscape
and the role of EPD points (\S\,\ref{subsec:MR-stability}), the
structural results (\S\,\ref{subsec:MR-structure}), and the closed-form
quantitative results --- widths, boundary curves by three methods,
and the primary tongue in physical variables (\S\,\ref{subsec:MR-quantitative})
--- to the Mathieu comparison and discriminant expansion (\S\,\ref{subsec:MR-Mathieu})
and the EPD hypersensitive capacitance sensing application (\S\,\ref{subsec:EPD-sensing-MR}).

\subsection{The LC circuit: original equation and parameter relations}

\label{subsec:MR-equations}

\emph{The original LC circuit equation.} A nondissipative LC circuit
with inductance $L$, nominal capacitance $C$, and harmonically varying
capacitance $C(t)=C(1+\varepsilon\cos\mu t)$ is governed by the charge
equation 
\begin{equation}
\ddot{q}+\omega_{0}^{2}\,\frac{1}{1+\varepsilon\cos\mu t}\,q=0,\qquad\omega_{0}=\frac{1}{\sqrt{LC}},\label{eq:LC-original-MR}
\end{equation}
where $\omega_{0}$ is the natural frequency, $\mu$ is the driving
frequency, and $0<\varepsilon<1$ is the modulation amplitude. The
physical parameters are listed in Table~\ref{tab:LC-params}.

\begin{table}[ht]
\centering 
\global\long\def\arraystretch{1.2}%
\begin{tabular}{lccl}
\hline 
Parameter  & SI  & Gaussian  & Role\tabularnewline
\hline 
$L$  & H  & s$^{2}$/cm  & inductance\tabularnewline
$C$  & F  & cm  & capacitance\tabularnewline
$\omega_{0}=1/\sqrt{LC}$  & s$^{-1}$  & s$^{-1}$  & natural frequency\tabularnewline
$\mu$  & s$^{-1}$  & s$^{-1}$  & driving frequency\tabularnewline
$\varepsilon$  & dimensionless  & dimensionless  & modulation amplitude\tabularnewline
\hline 
\end{tabular}\caption{Physical parameters of the LC circuit~\eqref{eq:LC-original-MR}
and their dimensions in SI and Gaussian units.}
\label{tab:LC-params} 
\end{table}

\emph{Auxiliary parameters.} The substitution $\tau=\mu t$ converts~\eqref{eq:LC-original-MR}
to a Hill equation with period $2\pi$: 
\begin{equation}
\frac{d^{2}q}{d\tau^{2}}+\frac{\omega_{0}^{2}}{\mu^{2}}\,\frac{1}{1+\varepsilon\cos\tau}\,q=0,\qquad\tau=\mu t,\quad\text{period }2\pi.\label{eq:LC-Hill-tau}
\end{equation}
Two auxiliary dimensionless parameters rationalize the algebra. The
\emph{Mobius\index{Mobius transformation} parameter} 
\begin{equation}
\delta=\frac{1-\sqrt{1-\varepsilon^{2}}}{\varepsilon},\qquad\varepsilon=\frac{2\delta}{1+\delta^{2}},\qquad0<\delta<1,\label{eq:eps-delta-MR}
\end{equation}
serves two purposes: (i)~it rationalizes the Ince parameter $a=-\varepsilon=-2\delta/(1+\delta^{2})$;
(ii)~the Fourier coefficients of the LC coefficient $1/(1+\varepsilon\cos\tau)$
take the geometrically decaying form $g_{n}\propto\delta^{n}$ via
the Poisson kernel Fourier series~\eqref{eq:Poisson-kernel}--\eqref{eq:Poisson-LC}.
The \emph{spectral parameter} 
\begin{equation}
c=\frac{4\omega_{0}^{2}}{\mu^{2}}=\frac{4}{r^{2}},\qquad r=\frac{\mu}{\omega_{0}},\label{eq:c-def-MR}
\end{equation}
linearizes the resonance conditions: the $m$-th resonance ($\mu=2\omega_{0}/m$)
corresponds simply to $c=m^{2}$, and the EPD boundary curve\index{boundary curves}s
are polynomial in $a=-2\delta/(1+\delta^{2})$. This same $c$ is
the Ince spectral parameter in~\eqref{eq:Ince-MR} below --- the
two are identical by construction.

The derived and auxiliary parameters are summarized in Table~\ref{tab:aux-params};
the equivalent two-parameter planes in Table~\ref{tab:param-pairs}.

\begin{table}[ht]
\centering 
\global\long\def\arraystretch{1.2}%
 \resizebox{\textwidth}{!}{%
\begin{tabular}{llll}
\hline 
Parameter pair  & Dimensionless?  & Natural context  & Key relations\tabularnewline
\hline 
$(\varepsilon,\,r)$, $r=\mu/\omega_{0}$  & both yes  & stability diagram  & $c=4/r^{2}$\tabularnewline
$(\varepsilon,\,\mu)$  & $\mu$ in s$^{-1}$  & experimental  & $\omega_{0}$ fixed by design\tabularnewline
$(\delta,\,c)$  & both yes  & boundary analysis  & $a=-2\delta/(1+\delta^{2})$\tabularnewline
$(\delta,\,a)$  & both yes  & MW coexistence theory  & $a=-\varepsilon$\tabularnewline
$(\delta,\,\mu)$  & $\mu$ in s$^{-1}$  & physical sensing  & primary tongue\tabularnewline
\hline 
\end{tabular}} \caption{Equivalent two-parameter planes used in this work. All pairs encode
the same physical system; the choice depends on which quantities are
most natural for the problem at hand.}
\label{tab:param-pairs} 
\end{table}

\emph{The primary dimensionless form.} The further substitution $x=\tau/2=\mu t/2$
converts~\eqref{eq:LC-Hill-tau} to the \emph{Ince equation} 
\begin{equation}
(1+a\cos2x)\,q''+c\,q=0,\qquad a=-\varepsilon=-\frac{2\delta}{1+\delta^{2}},\quad c=\frac{4\omega_{0}^{2}}{\mu^{2}},\label{eq:Ince-MR}
\end{equation}
with $b=d=0$ (the exact reduction to this multiplier form, including
the sign $a=-\varepsilon$, is carried out in Remark~\ref{rem:ince-id}).
\emph{This is the primary working equation of this work.} It contains
no physical units: both $a$ and $c$ are dimensionless (see Table~\ref{tab:aux-params}),
and $x$ is a dimensionless phase variable. All physical information
about the LC circuit is encoded in exactly two dimensionless parameters:
the \emph{Mobius parameter} $\delta$ (encoding the modulation amplitude
$\varepsilon$) and the \emph{spectral parameter} $c$ (encoding the
frequency ratio $\omega_{0}/\mu$).

\emph{Why these parameters?} The choices of $\delta$ and $c$ are
not arbitrary --- each is optimized for a specific purpose: 
\begin{itemize}
\item \emph{The spectral parameter $c=4\omega_{0}^{2}/\mu^{2}=m^{2}$ at
resonance.} The $m$-th parametric resonance\index{parametric resonance}
condition $\mu=2\omega_{0}/m$ ($m=1,2,3,\ldots$) translates to the
elegant integer condition $c=m^{2}$. The EPD instability boundaries
are therefore curves $c=c_{\pm}^{(m)}(\delta)$ emanating from the
integer points $c=1,4,9,16,25,\ldots$ as $\delta$ increases from
zero (Chapter~\ref{sec:Boundaries}). Without this choice the resonance
condition would be a complicated algebraic expression in $\omega_{0}$
and $\mu$; with it, the boundary curves are simply polynomial deformations
of the perfect squares.
\item \emph{The Mobius parameter $\delta=(1-\sqrt{1-\varepsilon^{2}})/\varepsilon$.}
This parameter rationalizes the Ince coefficient $a=-2\delta/(1+\delta^{2})$
and the Fourier expansion of the LC coefficient: by the Poisson kernel
identity, $1/(1+\varepsilon\cos\tau)$ has Fourier coefficients $g_{n}=\hat{\lambda}\delta^{|n|}$
--- a pure geometric sequence in $\delta$. The name ``Mobius parameter''
has a precise origin in the structure of the Ince recurrence (Section~\ref{subsec:VecForm}):
the ratio $z_{n}=A_{2n+1}/A_{2n-1}$ of successive Fourier coefficients
satisfies a Mobius recurrence $z_{n+1}=T_{M_{n}}(z_{n})$, whose limiting
transformation $T_{M_{\infty}}$ (eq.~\eqref{eq:TMinfz}) has exactly
two fixed points $\zeta_{1}=\delta$ (repulsive) and $\zeta_{2}=1/\delta$
(attractive). Remarkably, these are simultaneously the eigenvalues
of the matrix $M_{\infty}$ itself (eq.~\eqref{eq:Minf-eigs}): both
properties are expressions of the companion-matrix structure of the
Ince recurrence --- a special feature of the LC circuit (Remark~\ref{rem:eig-fixpt},
Section~\ref{subsec:eig-fixpt}). The minimal solution\index{continued fraction!minimal solution}
decays as $\delta^{n}$ per step, while the dominant solution grows
as $(1/\delta)^{n}$. Thus $\delta$ is the parameter singled out
by the Mobius geometry of the recurrence --- it is simultaneously
a fixed point of $T_{M_{\infty}}$ and an eigenvalue of $M_{\infty}$. 
\end{itemize}
The \emph{Mathieu\index{Mathieu equation} approximation} (first-harmonic
truncation of the Fourier expansion, valid for small $\varepsilon$)
gives: 
\begin{equation}
y''+(\lambda-2q\cos2x)\,y=0,\qquad q=\frac{2\varepsilon\omega_{0}^{2}}{\mu^{2}}=\frac{2\varepsilon}{r^{2}}\approx\frac{\varepsilon}{2}\;\text{(primary resonance)},\label{eq:Mathieu-MR}
\end{equation}
where $\lambda$ is the Mathieu spectral parameter and $q$ is the
Mathieu parameter (not the charge $q(t)$ of eq.~\eqref{eq:LC-original-MR}
or the solution variable $q(x)$ of eq.~\eqref{eq:Ince-MR}). This
work uses the exact Ince equation~\eqref{eq:Ince-MR} throughout;
details in Chapter~\ref{sec:LCInce} and the Notation chapter. The
equivalent two-parameter planes are given in Table~\ref{tab:param-pairs}.

\begin{table}[ht]
\centering 
\global\long\def\arraystretch{1.3}%
 {\small{}%
\begin{tabular}{llp{1.8cm}p{4.3cm}}
\hline 
{\small Symbol } & {\small Definition } & {\small Range } & {\small Role}\tabularnewline
\hline 
{\small$\tau$ } & {\small$\mu t$ } & {\small$[0,2\pi)$ } & {\small rescaled time; Hill period $2\pi$}\tabularnewline
{\small$x$ } & {\small$\tau/2$ } & {\small$[0,\pi)$ } & {\small Ince variable; Hill period $\pi$}\tabularnewline
{\small$r$ } & {\small$\mu/\omega_{0}$ } & {\small$r>0$ } & {\small frequency ratio}\tabularnewline
{\small$c$ } & {\small$4\omega_{0}^{2}/\mu^{2}$ } & {\small$c>0$ } & {\small Ince spectral parameter; $c=m^{2}$ at $m$-th resonance}\tabularnewline
{\small$\delta$ } & {\small$(1-\sqrt{1-\varepsilon^{2}})/\varepsilon$ } & {\small$0<\delta<1$ } & {\small Mobius parameter; $\varepsilon=2\delta/(1+\delta^{2})$}\tabularnewline
{\small$a$ } & {\small$-\varepsilon=-2\delta/(1+\delta^{2})$ } & {\small$-1<a<0$ } & {\small Ince coefficient ($b=d=0$ for LC circuit)}\tabularnewline
{\small$q$ } & {\small$\varepsilon c/2=2\varepsilon\omega_{0}^{2}/\mu^{2}$ } & {\small$q>0$ } & {\small Mathieu parameter; see eq.~\eqref{eq:Mathieu-MR}}\tabularnewline
\hline 
\end{tabular}}{\small\caption{Derived and auxiliary parameters of the LC circuit (all dimensionless).
The Mobius parameter $\delta$ rationalizes the Fourier expansion:
$g_{n}=\hat{\lambda}\delta^{n}$ (eq.~\eqref{eq:Poisson-LC}).}
\label{tab:aux-params} }
\end{table}

\begin{rem}[Ince identification]
\label{rem:ince-id} The identification LC$\to$Ince is exact (proved
in Section~\ref{subsec:IncID}): multiplying the Hill equation form~\eqref{eq:LC-Hill}
of the LC circuit (already in the variable $x=\tau/2$) by the factor
$(1-2\delta\cos2x+\delta^{2})/(1+\delta^{2})$ yields the Ince equation~\eqref{eq:Ince-MR}
with 
\[
a=-\frac{2\delta}{1+\delta^{2}}=-\varepsilon,\qquad b=0,\qquad c=\frac{4}{r^{2}}=\frac{4\omega_{0}^{2}}{\mu^{2}},\qquad d=0.
\]
No Fourier truncation is involved; the complete parameter identification
is eq.~\eqref{eq:IncePars}. See Table~\ref{tab:aux-params} for
the dimensionless parameters. 
\end{rem}

\subsection{Stability, instability, and the role of EPD points}

\label{subsec:MR-stability}

\emph{Stability and instability in the $(c,\delta)$-plane.} All stability
results are stated in terms of the parameter pair $(c,\delta)$, where
$c=4\omega_{0}^{2}/\mu^{2}$ is the spectral parameter (Table~\ref{tab:param-pairs},
eq.~\eqref{eq:c-def-MR}) and $\delta$ is the Mobius parameter (eq.~\eqref{eq:eps-delta-MR}).
A solution of the LC circuit equation~\eqref{eq:LC-original-MR}
is \emph{stable} if it remains bounded for all time: the circuit oscillates
indefinitely without growing. It is \emph{parametrically unstable}
if it grows exponentially: the circuit extracts energy from the driving
source each period, with amplitude growing without bound (see Chapter~\ref{app:ParRes}).

\begin{figure}[htbp]
\centering \includegraphics[width=9cm]{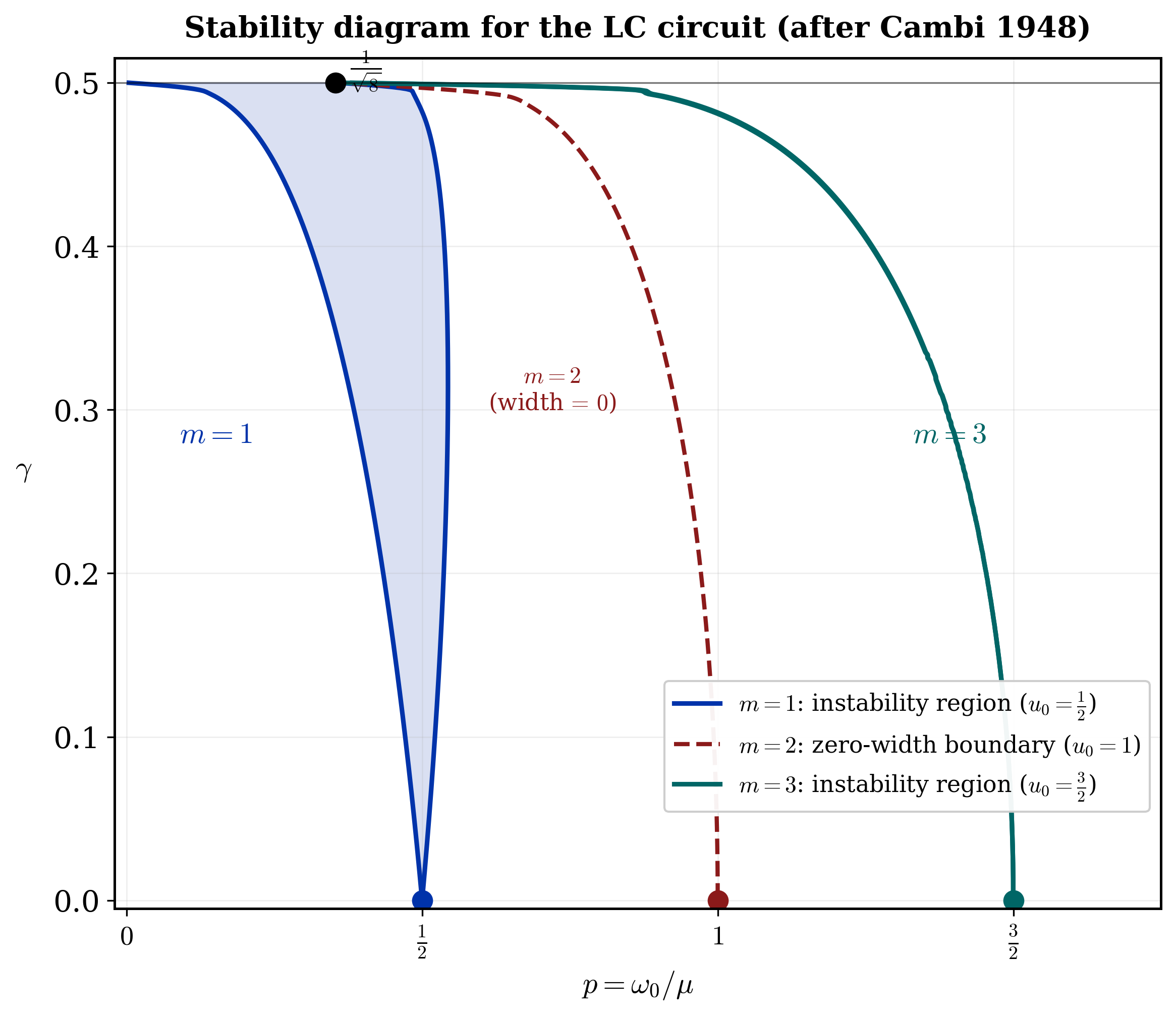} \caption{Stability diagram of the LC circuit in the $(\omega_{0}/\mu,\,\varepsilon/2)$-plane.
Shaded regions: unstable (parametric resonance). The $m=1$ primary
tongue (blue) and $m=3$ tongue (teal) open from $p=\tfrac{1}{2}$
and $p=\tfrac{3}{2}$ respectively; the right $m=1$ boundary, the
$m=2$ curve, and the $m=3$ tongue converge to the single point $p=1/\sqrt{8}$
at $\gamma=\tfrac{1}{2}$. Even resonances ($m=2$, red dashed) have
exactly zero width. Boundary curves computed via Cambi's continued-fraction
representation~\eqref{eq:Cambi-v-CF} for $v(u)$; see Figure~\ref{fig:stability}
and Section~\ref{subsec:StabDiag} for full details.}
\label{fig:stability-MR} 
\end{figure}

\emph{Resonances and instability tongue\index{instability tongue}s.}
Parametric resonance occurs when the driving frequency $\mu$ bears
a specific rational relationship to the natural frequency $\omega_{0}$.
The underlying mechanism (detailed in Chapter~\ref{app:ParRes})
is a period-matching condition: when the natural oscillation completes
exactly $m/2$ cycles per driving period $T=2\pi/\mu$, the driving
force is able to pump energy into the oscillation coherently, cycle
after cycle, leading to exponential growth. This condition is $\omega_{0}/({\mu}/{2})=m$,
i.e.\
\begin{equation}
\mu=\frac{2\omega_{0}}{m},\qquad m=1,2,3,\ldots,\label{eq:resonance-condition-MR}
\end{equation}
identifying the \emph{$m$-th resonance frequency}. In the spectral
parameter $c=4\omega_{0}^{2}/\mu^{2}$ (eq.~\eqref{eq:c-def-MR}),
condition~\eqref{eq:resonance-condition-MR} becomes simply $c=m^{2}$:
the resonances are at the perfect squares, with the $m$-th resonance
at the center $c=m^{2}$ of the $m$-th tongue. Around each resonance
the circuit can be parametrically unstable within a narrow \emph{instability
tongue} --- a region in the $(c,\delta)$-plane where exponential
growth occurs. As shown in this work (Propositions~\ref{prop:even-vanish}
and~\ref{prop:odd-survive}), only the \emph{odd} resonances ($m=1,3,5,\ldots$)
produce genuine instability tongues; the even resonances ($m=2,4,6,\ldots$)
are always stable. The $m$-th tongue is bounded by the EPD curves
$c_{\pm}^{(m)}(\delta)$ (defined below in eq.~\eqref{eq:EPD-curves-def}).
Figure~\ref{fig:stability-MR} shows the full stability diagram.

\emph{EPD curves.} The boundary of the $m$-th instability tongue
in the $(c,\delta)$-plane consists of two curves, defined precisely
in Theorem~\ref{thm:EPD-bdry} and denoted 
\begin{equation}
c_{-}^{(m)}(\delta)<m^{2}<c_{+}^{(m)}(\delta),\qquad\delta\neq0,\quad m=1,3,5,\ldots\;(\text{odd}).\label{eq:EPD-curves-def}
\end{equation}
The instability tongue occupies the interval $c_{-}^{(m)}(\delta)<c<c_{+}^{(m)}(\delta)$;
the stability zones lie outside: 
\begin{equation}
\text{stable: }c<c_{-}^{(m)}(\delta)\;\text{ or }\;c>c_{+}^{(m)}(\delta);\qquad\text{unstable: }c_{-}^{(m)}(\delta)<c<c_{+}^{(m)}(\delta).\label{eq:stability-intervals}
\end{equation}
Both boundaries $c=c_{\pm}^{(m)}(\delta)$ are \emph{EPD points}:
the physical transition points between stability and parametric resonance.
These curves are the primary objects of study in this work. Whenever
$c_{\pm}^{(m)}(\delta)$ appear below, they refer to eqs.~\eqref{eq:EPD-curves-def}--\eqref{eq:stability-intervals}
and Theorem~\ref{thm:EPD-bdry}.

The Floquet theory of periodic-coefficient equations (Chapter~\ref{app:genLin})
characterizes solutions through the \emph{monodromy matrix\index{monodromy matrix}}
$X(\pi)$, the solution operator over one driving period $T=2\pi/\mu$.
Its eigenvalues $\rho_{\pm}$ (the \emph{Floquet multiplier\index{Floquet theory!Floquet multiplier}s},
satisfying $\rho_{+}\rho_{-}=1$) determine the dynamical regime completely: 
\begin{itemize}
\item \emph{Stable:} $\rho_{\pm}=e^{\pm i\pi\alpha}$ with $\alpha\in\mathbb{R}$,
$|\rho_{\pm}|=1$. Solutions are bounded quasi-periodic oscillations
with dimensionless frequency $\alpha$ and physical frequency $f_{\pm}=(\mu/2)\alpha_{\pm}$. 
\item \emph{Unstable:} $|\rho_{+}|>1>|\rho_{-}|$; one multiplier leaves
the unit circle. Solutions grow exponentially --- parametric amplification. 
\item \emph{Boundary:} $\rho_{+}=\rho_{-}=(-1)^{m}$ --- both multipliers
coincide at $\pm1$ on the unit circle. This is the transition between
the two regimes, and it is an \emph{Exceptional Point of Degeneracy}
(EPD). 
\end{itemize}
\emph{EPD points: physical and mathematical meaning.} An EPD point
is the exact transition between bounded oscillations and parametric
resonance. At an EPD the monodromy matrix $X(\pi)$ acquires a \emph{non-trivial
Jordan block\index{Jordan block}}: the two Floquet modes coalesce
into one. This Jordan block is not a mathematical decoration ---
it is the precise algebraic signature of the physical transition: 
\begin{center}
\fcolorbox{black}{white}{\parbox[c]{0.96\textwidth}{%
\centering The \emph{instability tongue boundaries} $c=c_{\pm}^{(m)}(\delta)$
in the $(c,\delta)$-plane consist entirely of \emph{EPD points} ---
parameter values at which the circuit is exactly at the tipping point
between stable oscillation and exponential energy extraction. The
non-trivial Jordan block of the monodromy matrix $X(\pi)$ is the
mathematical fingerprint of this transition.%
}} 
\par\end{center}

\emph{Organizing principle.} The stability and instability of the
LC circuit --- specifically, finding the values of the parameter
pair $(c,\delta)$ at which EPD points occur --- is the central organizing
principle of this work. Our principal results concern determining
these EPD curves $c_{\pm}^{(m)}(\delta)$ in closed analytic form,
understanding the Krein\index{Krein collision theory}-signature mechanism
that makes even tongues collapse while odd tongues survive, and exploiting
the EPD structure for hypersensitive\index{exceptional point of degeneracy (EPD)!hypersensitivity}
capacitance sensing.

\subsection{Structural results: EPD curves and instability pattern}

\label{subsec:MR-structure}

The key that unlocks the complete structural theory is that the LC
equation~\eqref{eq:LC-original-MR} is exactly Ince's equation\index{Ince equation}
(Remark~\ref{rem:ince-id}). Combined with the Magnus--Winkler\index{Magnus--Winkler theory}
coexistence theory, this gives a complete algebraic explanation of
why even tongues collapse and odd tongues survive. Our two main structural
theorems characterize the EPD points and their Jordan block structure;
the supporting Magnus--Winkler results follow in~\S\,\ref{subsubsec:MR-support}.

\subsubsection{Our results: EPD structure of the instability boundaries}

\label{subsubsec:EPD-structure}
\begin{thm}[EPD structure via Krein signature\index{Krein signature}s, Section~\ref{subsec:Hamiltonian}]
\label{thm:EPD} In the parameter pair $(c,\delta)$ (eqs.~\eqref{eq:c-def-MR},~\eqref{eq:eps-delta-MR}),
the instability tongue boundaries of the LC circuit~\eqref{eq:LC-original-MR}
with $\delta\neq0$ have the following Floquet-theoretic structure.
The monodromy matrix $X(\pi)$ (the solution operator over one driving
period $T=2\pi/\mu$) and its eigenvalues $\rho_{\pm}$ (Floquet multipliers,
$\rho_{+}\rho_{-}=1$) satisfy: 
\begin{enumerate}
\item[(i)] \emph{Odd resonances} ($m=1,3,5,\ldots$): at each boundary $c=c_{\pm}^{(m)}(\delta)$
(eq.~\eqref{eq:EPD-curves-def}) of an odd tongue, $\rho_{+}=\rho_{-}=-1$
and the monodromy matrix $X(\pi)$ has a non-trivial Jordan block
--- an EPD of \emph{mixed Krein kind} (the two colliding Floquet
modes carry opposite-sign energy). Under any perturbation, the multipliers
leave the unit circle and the tongue opens: \emph{the EPD is the physical
tipping point between stability and parametric resonance.} 
\item[(ii)] \emph{Even resonances} ($m=2,4,6,\ldots$): the monodromy matrix
$X(\pi)=+\mathbf{I}$ (both modes carry same-sign energy). The tongue
has zero width for all $\delta$: even resonances are always stable. 
\end{enumerate}
\end{thm}

\begin{proof}
Symplectic dichotomy~\cite[Ch.~VIII, \S\,1.8]{YakSta2} applied to
Propositions~\ref{prop:even-vanish}--\ref{prop:odd-survive}; full
proof in Section~\ref{subsec:Hamiltonian}. 
\end{proof}
\begin{thm}[Jordan block: elementary proof, Chapter~\ref{sec:EPD-elementary}]
\label{thm:Jordan-bdry-MR} At any instability boundary\index{stability boundary}
$c=c_{\pm}^{(m)}(\delta)$ with $\delta\neq0$, the monodromy matrix
$X(\pi)$ has a non-trivial $2\times2$ Jordan block with eigenvalue
$\rho_{0}=(-1)^{m}$. Consequently the two characteristic exponents
$\alpha_{\pm}$ (defined by $\rho_{\pm}=e^{i\pi\alpha_{\pm}}$) coalesce:
$\alpha_{+}=\alpha_{-}=m$. 
\end{thm}

\begin{proof}
A direct calculation shows that if the monodromy matrix $X(\pi)$
were diagonalizable at $c_{0}$, its eigenvalue perturbations would
be $O(|\Delta c|)$ as $c\to c_{0}$. But the Hill discriminant\index{discriminant}\index{discriminant!Hill discriminant}
formula (eq.~\eqref{eq:Hill-multipliers}) gives $|\rho_{+}-\rho_{-}|=\sqrt{4-\Delta^{2}}\sim\kappa\sqrt{|h|}$
with $\kappa\neq0$ at an instability boundary where $\Delta(c_{0})=\pm2$,
contradicting $O(|\Delta c|)$. For a $2\times2$ matrix with a double
eigenvalue, non-diagonalizability forces a non-trivial Jordan block.
Full proof in Chapter~\ref{sec:EPD-elementary}. 
\end{proof}
\begin{rem}[Two complementary proofs]
\label{rem:two-proofs} Theorem~\ref{thm:EPD} proves the Jordan
block via Krein theory, explaining the odd/even alternation. Theorem~\ref{thm:Jordan-bdry-MR}
gives an elementary independent proof using only the Hill discriminant
--- no symplectic\index{symplectic system} structure needed. The
two approaches are complementary. 
\end{rem}

\subsubsection{Supporting results of Magnus--Winkler}

\label{subsubsec:MR-support} Our structural theorems rest on the
following results, which apply the Magnus--Winkler coexistence theory~\cite[Sec.~8.4]{MagWin}
to the Ince identification of Remark~\ref{rem:ince-id}. They explain
\emph{why} even tongues collapse and odd tongues survive --- the
structural facts that Theorem~\ref{thm:EPD} then interprets through
Krein signatures.
\begin{prop}[Vanishing of even intervals, Chapter~\ref{sec:LCInce}]
\label{prop:even-vanish}\cite[Sec.~8.4]{MagWin} In the parameter
pair $(c,\delta)$, all even instability tongues of the LC circuit
have zero width exactly: 
\begin{gather*}
L_{m}^{\mathrm{LC}}=c_{+}^{(m)}(\delta)-c_{-}^{(m)}(\delta)=0\\
\text{for all even }m=2,4,6,\ldots,\ \text{and all }0<\delta<1.
\end{gather*}
Here $L_{m}^{{\rm LC}}$ is the tongue width in the $c$-variable
(eq.~\eqref{eq:Lm-main}; see also Figure~\ref{fig:stability-MR}).
At even resonances ($c=m^{2}$, even $m$) the circuit is always stable
for all modulation amplitudes. 
\end{prop}

\begin{proof}
The coexistence polynomial $Q(\mu)=2a\mu^{2}$ (eq.~\eqref{eq:QQstar-intro};
$\mu$ here is the MW integer index, not the driving frequency) has
the nonnegative integer root $\mu=0$; by MW Theorem~7.6~\cite[Thm.~7.6]{MagWin}
(sufficiency; the single admissible exceptional value is identified
as $c=0$, which is not an interval endpoint) this forces period-$\pi$
coexistence ($\rho=+1$) at every even resonance; proof in Section~\ref{subsec:InceResult}. 
\end{proof}
\begin{prop}[Survival of odd intervals, Chapter~\ref{sec:LCInce}]
\label{prop:odd-survive}\cite[Sec.~8.4]{MagWin} In the parameter
pair $(c,\delta)$, all odd instability tongues of the LC circuit
have strictly positive width for every $\delta\neq0$: 
\[
L_{m}^{\mathrm{LC}}=c_{+}^{(m)}(\delta)-c_{-}^{(m)}(\delta)>0\quad\text{for all odd }m=1,3,5,\ldots,\quad\text{and all }\delta\neq0.
\]
The odd tongues open as EPD pairs: both boundaries $c=c_{\pm}^{(m)}(\delta)$
(eq.~\eqref{eq:EPD-curves-def}) are EPD points --- transition points
from stability to parametric resonance (see Figure~\ref{fig:stability-MR}). 
\end{prop}

\begin{proof}
The \emph{period-$2\pi$ coexistence polynomial} $Q^{*}(\mu)=a(2\mu-1)^{2}$
(eq.~\eqref{eq:QQstar-intro}; $\mu$ is the MW integer index) has
no integer roots (since $(2\mu-1)^{2}=0$ gives $\mu=1/2$, not an
integer), preventing period-$2\pi$ coexistence by MW Theorem~7.1;
proof in Section~\ref{subsec:InceResult}. 
\end{proof}
Together, Propositions~\ref{prop:even-vanish} and~\ref{prop:odd-survive}
recover Cambi's empirical finding~\cite[\S\,11]{Cambi2} and provide
its algebraic explanation via the coexistence polynomials $Q$, $Q^{*}$.
By contrast, the Mathieu equation ($a=0$) has $Q(\mu)=q$ with no
integer roots, so no interval ever vanishes.

\subsection{Closed-form quantitative results: widths, boundaries, and primary
tongue}

\label{subsec:MR-quantitative}

The following three theorems give explicit closed-form formulas for
the instability tongue widths, the EPD curves, and the primary instability
domain in physical variables. All results are stated in terms of the
Mobius parameter $\delta$ (eq.~\eqref{eq:eps-delta-MR}) and the
spectral parameter $c=4\omega_{0}^{2}/\mu^{2}$ (eq.~\eqref{eq:c-def-MR}).

\subsubsection{Instability tongue widths}
\begin{thm}[Instability interval widths, Chapter~\ref{sec:LCInce}]
\label{thm:width} In the parameter pair $(\hat{\lambda},\delta)$,
where $\hat{\lambda}=c(1+\delta^{2})/(1-\delta^{2})$ is the Magnus--Winkler
spectral parameter (Chapter~\ref{sec:MWInce}), the width of the
$m$-th surviving instability tongue ($m$ odd) has the closed-form
expression 
\begin{equation}
\boxed{\begin{aligned}L_{m}^{\mathrm{LC}} & =\frac{(m!!)^{2}\,\delta^{m}}{2^{(m^{2}+8m-13)/4}\,(1+\delta^{2})^{m-1}(1-\delta^{2})}\\
 & \quad+O_{m}(\delta^{m+2}),\qquad m=1,3,5,\ldots
\end{aligned}
}\label{eq:Lm-main}
\end{equation}
For $m=1$ the formula has no chain-approximation error (the $k=1$
backward chain has no steps), so the only error is the inherent $O(\delta^{3})$
in the boundary curve expansion; for $m\geq3$ the error is $O_{m}(\delta^{m+2})$,
two powers beyond the leading term. The exponent $(m^{2}+8m-13)/4$
is a positive integer for all odd $m\geq3$ and equals $-1$ for $m=1$.
The first three cases are: 
\begin{align}
L_{1}^{\mathrm{LC}} & =\frac{2\delta}{1-\delta^{2}},\qquad L_{3}^{\mathrm{LC}}=\frac{9\delta^{3}}{32(1+\delta^{2})^{2}(1-\delta^{2})}+O(\delta^{5}),\label{eq:Lm-cases-13}\\
L_{5}^{\mathrm{LC}} & =\frac{225\delta^{5}}{8192(1+\delta^{2})^{4}(1-\delta^{2})}+O(\delta^{7}),\qquad\delta\to0.\label{eq:Lm-cases-5}
\end{align}
Formula~\eqref{eq:Lm-main} is confirmed numerically against Cambi's
exact continued-fraction solution to better than $2.3\%$ for $\delta\leq0.20$
(Chapter~\ref{sec:Cambi}). 
\end{thm}

\begin{proof}
\noindent The backward-chain derivation for general $m$ is in Section~\ref{subsec:Widths}.
The $O_{m}(\delta^{m+2})$ accuracy claim follows from the fact that
for fixed $m=2k-1$ the chain has exactly $k-1$ non-resonant steps
(a fixed finite number), the sole approximation being the substitution
$c=(2k-1)^{2}$ in those steps; see the paragraph \emph{General $k$:
the backward chain product and $O_{m}(\delta^{m+2})$ accuracy} in
Section~\ref{subsec:Widths}. 
\end{proof}

\subsubsection{EPD curves}

\label{subsubsec:MR-boundary}

The driving frequency $\mu$ is the primary physical parameter of
the circuit, but the boundary curve analysis is most naturally carried
out in terms of the dimensionless spectral parameter 
\begin{equation}
c=\frac{4\omega_{0}^{2}}{\mu^{2}}=\frac{4}{r^{2}},\qquad r=\frac{\mu}{\omega_{0}},\label{eq:c-vs-mu}
\end{equation}
which is the ratio of the squared natural frequency to the squared
half-excitation-frequency. Working in $c$ linearizes the resonance
conditions ($c=m^{2}$ at the $m$-th tongue center), makes the boundary
curve formulas polynomial in $a$, and allows the Ince and continued-fraction
machinery to operate cleanly. The conversion back to physical driving
frequencies is simply 
\begin{equation}
\mu=\frac{2\omega_{0}}{\sqrt{c}},\qquad\mu_{\pm}^{(m)}=\frac{2\omega_{0}}{\sqrt{c_{\pm}^{(m)}}},\label{eq:c-to-mu}
\end{equation}
and is carried out explicitly in Theorem~\ref{thm:primary-domain}
below for the primary tongue $m=1$.

At each instability boundary $c=c_{\pm}^{(m)}(\delta)$ (eq.~\eqref{eq:EPD-curves-def}),
the monodromy matrix $X(\pi)$ has a non-trivial Jordan block ---
an EPD point. The following theorem gives the explicit closed-form
expressions for these boundary curves in the parameter pair $(c,\delta)$
for the first four surviving instability tongues (see Figure~\ref{fig:stability-MR}
for the visual picture). 
\begin{thm}[EPD curves, Chapter~\ref{sec:Boundaries}]
\label{thm:EPD-bdry} In the parameter pair $(c,\delta)$, each odd
instability tongue of the LC circuit~\eqref{eq:LC-Hill} is bounded
by two EPD curves $c_{\pm}^{(m)}(\delta)$ (as defined in eq.~\eqref{eq:EPD-curves-def}).
Their general structure is 
\begin{equation}
c_{\pm}^{(m)}=m^{2}+D_{m}\,a^{2}\pm\tfrac{1}{2}L_{m}^{(c)}+O_{m}(|a|^{m+2}),\quad a=-\frac{2\delta}{1+\delta^{2}}\to0,\label{eq:boundary-main}
\end{equation}
where the half-width is given \emph{purely in $a$} by 
\begin{equation}
L_{m}^{(c)}=\frac{(m!!)^{2}\,|a|^{m}}{2^{(m^{2}+12m-13)/4}},\qquad m=1,3,5,\ldots,\label{eq:Lmc-pure-a}
\end{equation}
(equivalently, $L_{m}^{(c)}=L_{m}^{\mathrm{LC}}\cdot(1-\delta^{2})/(1+\delta^{2})$
via $\delta/(1+\delta^{2})=-a/2$, where $L_{m}^{\mathrm{LC}}$ is
the closed-form width~\eqref{eq:Lm-main}), $D_{m}<0$ is the rational
center shift, and $r_{\pm}^{(m)}=2/\sqrt{c_{\pm}^{(m)}}$ are the
frequency ratios. The boundary curves are analytic in $a$ (Theorem~\ref{thm:bdry-analytic});
the center contains only even powers of $a$ and the half-width only
odd powers of $|a|$. Explicitly, for $m=1,3,5,7$: 
\begin{align}
c_{\pm}^{(1)} & =1-\tfrac{9}{32}\,a^{2}\pm\tfrac{1}{2}|a|+O(|a|^{3}),\label{eq:EPD-bdry-1}\\
c_{\pm}^{(3)} & =9-\tfrac{207}{64}\,a^{2}\pm\tfrac{9}{512}|a|^{3}+O_{3}(|a|^{5}),\label{eq:EPD-bdry-3}\\
c_{\pm}^{(5)} & =25-\tfrac{1775}{192}\,a^{2}\pm\tfrac{225}{524288}|a|^{5}+O_{5}(|a|^{7}),\label{eq:EPD-bdry-5}\\
c_{\pm}^{(7)} & =49-\tfrac{7007}{384}\,a^{2}\pm\tfrac{11025}{2147483648}|a|^{7}+O_{7}(|a|^{9}),\quad a\to0.\label{eq:EPD-bdry-7}
\end{align}
The center shifts $D_{1}=-\tfrac{9}{32}$, $D_{3}=-\tfrac{207}{64}$,
$D_{5}=-\tfrac{1775}{192}$, $D_{7}=-\tfrac{7007}{384}$ are \emph{exact}
rational numbers, each determined without approximation by the $O(a^{2})$
balance in the resonant equation (Section~\ref{subsec:LCbdry}).
The half-widths $\tfrac{1}{2}L_{m}^{(c)}$ are given by the closed-form
formula~\eqref{eq:Lmc-pure-a}, which carries the error $O_{m}(|a|^{m+2})$.
The overall error $O_{m}(|a|^{m+2})$ in~\eqref{eq:boundary-main}
therefore comes \emph{entirely from $L_{m}^{(c)}$}, not from $D_{m}$:
once $D_{m}$ is written explicitly, no further $O(a^{2})$ error
remains in the center position. 
\end{thm}

\begin{proof}
Structure~\eqref{eq:boundary-main} follows from Theorem~\ref{thm:bdry-analytic}
(analyticity) and the backward-chain derivation of Section~\ref{subsec:LCbdry}.
Each $D_{m}$ is obtained exactly by setting $c=m^{2}+D_{m}a^{2}$
in the starter equation and solving the $O(a^{0})$ or $O(a^{1})$
balance (depending on $m$) for the rational $D_{m}$; no truncation
is involved in this step. The error $O_{m}(|a|^{m+2})$ in the half-width
$\tfrac{1}{2}L_{m}^{(c)}$ arises from substituting the resonant value
$c=m^{2}$ instead of the true $c\approx m^{2}+D_{m}a^{2}$ in the
$k-1$ non-resonant chain denominators; since there are exactly $k-1$
such steps for fixed $m=2k-1$, the relative error is $O_{m}(a^{2})$
per step, giving absolute error $O_{m}(|a|^{m+2})$ in the width.
The cases $m=1,3$ are derived explicitly in Section~\ref{subsec:LCbdry};
$m=5,7$ follow by the same procedure (equations~\eqref{eq:LC-bdry5}--\eqref{eq:LC-bdry7}). 
\end{proof}
\begin{rem}[Numerical verification of Theorem~\ref{thm:width}]
\label{rem:width-numerics} The formula~\eqref{eq:Lm-main} has
been verified independently by direct numerical integration of the
Ince equation $(1+a\cos2x)q''+cq=0$ (8th-order Runge--Kutta, tolerance
$10^{-12}$), locating the boundaries by bisection on $\operatorname{Tr}(M(x))+2=0$.
For $m=1$: the formula is exact; errors are $O(|a|^{3})$ as claimed
(e.g.\ $2.5\times10^{-5}$ at $\delta=0.05$, $2.5\times10^{-4}$
at $\delta=0.10$). The $m=3,5,7$ half-width coefficients $9/512$,
$225/524288$, $11025/2147483648$ were verified symbolically by expanding
$L_{m}^{(c)}/2$ in powers of $|a|=\varepsilon$ and confirming term-by-term
agreement with Table~\ref{tab:bdry-LC}. 
\end{rem}

\begin{rem}[Numerical verification of Theorem~\ref{thm:EPD-bdry}]
\label{rem:bdry-numerics} The boundary curves~\eqref{eq:EPD-bdry-1}
were verified by numerically solving $\operatorname{Tr}(M(\pi))+2=0$
for the exact Ince equation (8th-order Runge--Kutta, $10^{-12}$
tolerance), with the following results for $c_{+}^{(1)}$: 
\begin{center}
\global\long\def\arraystretch{1.2}%
\begin{tabular}{ccccc}
\hline 
$\delta$  & $|a|$  & abs.\ error  & error$/|a|^{3}$  & rel.\ error (vs.\ half-width)\tabularnewline
\hline 
$0.05$  & $0.0998$  & $2.5\times10^{-5}$  & $0.025$  & $0.05\%$\tabularnewline
$0.10$  & $0.1980$  & $2.5\times10^{-4}$  & $0.033$  & $0.26\%$\tabularnewline
$0.15$  & $0.2934$  & $1.0\times10^{-3}$  & $0.041$  & $0.71\%$\tabularnewline
$0.20$  & $0.3846$  & $2.8\times10^{-3}$  & $0.050$  & $1.48\%$\tabularnewline
\hline 
\end{tabular}
\par\end{center}
Three points confirm perfect consistency with the $O(|a|^{3})$ claim:
(i)~the absolute errors are small because $|a|\ll1$ at moderate
$\delta$ ($|a|<0.4$ for $\delta<0.2$); (ii)~the column error$/|a|^{3}$
is approximately constant at $\approx1/32\approx0.031$, confirming
the $O(|a|^{3})$ scaling with the correct next-order coefficient;
the slow drift from $0.025$ to $0.050$ reflects $O(|a|^{4})$ terms
beyond the truncation order; (iii)~expressed as a fraction of the
tongue half-width $|a|/2$ (the quantity relevant for the EPD sensor),
the relative error ranges from $0.05\%$ at $\delta=0.05$ to $1.5\%$
at $\delta=0.20$, confirming that the formula is accurate to within
a few percent across the physically relevant range $\delta\lesssim0.2$.
The center shift $D_{1}=-9/32$ was independently confirmed from $(c_{+}^{(1)}+c_{-}^{(1)})/2=1+D_{1}a^{2}$
to four significant figures across the same $\delta$ range. 
\end{rem}

\begin{rem}[No closed form for $D_{m}$]
\label{rem:Dm-no-formula} Unlike $L_{m}^{\mathrm{LC}}$, which has
a uniform product formula~\eqref{eq:Lm-main} for all odd $m$, the
center shift $D_{m}$ has no analogous expression. $L_{m}^{\mathrm{LC}}$
arises from the \emph{product} of $k-1$ chain denominators; $D_{m}$
arises from the \emph{sum} of their logarithmic derivatives, which
does not simplify. The secondary factors $\{N_{m}\}=\{9,23,71,143\}$
defined by $D_{m}=-m^{2}N_{m}/\mathrm{denom}$ follow no multiplicative
pattern. 
\end{rem}

\subsubsection{Primary instability domain in physical variables}

\label{subsubsec:MR-primary}

The most important special case $m=1$ admits an exact and explicit
statement in the original physical variables of the LC circuit~\eqref{eq:LC-original}.
Converting the $m=1$ boundary curves $c_{\pm}^{(1)}$ of Theorem~\ref{thm:EPD-bdry}
back to driving frequencies via eq.~\eqref{eq:c-to-mu} yields: 
\begin{thm}[Primary instability domain, Chapter~\ref{sec:LCInce}]
\label{thm:primary-domain} Expressed in the physical parameter pair
$(\varepsilon,\mu)$ (with $\omega_{0}=1/\sqrt{LC}$ fixed), and related
to the $(c,\delta)$ pair via $c=4\omega_{0}^{2}/\mu^{2}$, $\varepsilon=2\delta/(1+\delta^{2})$
(Tables~\ref{tab:LC-params},~\ref{tab:param-pairs}): the LC circuit~\eqref{eq:LC-original}
with $C(t)=C(1+\varepsilon\cos\mu t)$ exhibits primary parametric
instability ($m=1$) for driving frequencies $\mu$ in the interval
$(\mu_{-},\mu_{+})$, corresponding to EPD curves $c_{\pm}^{(1)}(\delta)$
(eq.~\eqref{eq:EPD-curves-def}) via $\mu_{\pm}=2\omega_{0}/\sqrt{c_{\mp}^{(1)}}$.
The explicit boundary frequencies are, to leading order in $\varepsilon$:
\begin{align}
\mu_{\pm} & =\frac{2\omega_{0}}{\sqrt{1-\dfrac{9}{32}\varepsilon^{2}\mp\dfrac{|\varepsilon|}{2}}}+O(\varepsilon^{3}),\quad\varepsilon\to0,\label{eq:primary-domain}\\
 & \text{equivalently,}\quad\frac{1}{\mu_{-}^{2}}-\frac{1}{\mu_{+}^{2}}=\frac{\varepsilon}{4\omega_{0}^{2}}+O(\varepsilon^{3}),\label{eq:width-relation}\nonumber 
\end{align}
with $\mu_{-}<\mu_{+}$ and center condition $(\mu_{+}^{-2}+\mu_{-}^{-2})/2=\tfrac{1}{4\omega_{0}^{2}}(1-\tfrac{9}{32}\varepsilon^{2})+O(\varepsilon^{4})$.
The width of the instability domain is 
\begin{equation}
\Delta\mu=\mu_{+}-\mu_{-}=\omega_{0}\,\varepsilon\,\bigl(1+O(\varepsilon^{2})\bigr),\quad\varepsilon\to0.\label{eq:primary-width}
\end{equation}
The exact $\varepsilon^{3}$ and higher width coefficients lie beyond
the $O(|\varepsilon|^{3})$ accuracy of the input boundary curves~\eqref{eq:EPD-bdry-1};
they are supplied exactly by the continued-fraction series of Theorem~\ref{thm:primary-domain-CF}:
$\Delta\mu=\omega_{0}(\varepsilon+\tfrac{139}{256}\varepsilon^{3}+\cdots)$,
eq.~\eqref{eq:Delta-mu-CF}. At each boundary $\mu=\mu_{\pm}$, the
monodromy matrix of the circuit equation has a non-trivial Jordan
block with eigenvalue $-1$ (exceptional point of degeneracy). 
\end{thm}

\begin{proof}
\noindent{} Derived from the width formula via $c=4\omega_{0}^{2}/\mu^{2}$;
see Section~\ref{subsec:Widths}. 
\end{proof}
\begin{rem}[Numerical verification of Theorem~\ref{thm:primary-domain}]
\label{rem:primary-domain-numerics} All three formulas of Theorem~\ref{thm:primary-domain}
were verified by numerically locating the exact Ince-equation boundaries
$\mu_{\pm}$ (8th-order Runge--Kutta, tolerance $10^{-13}$, root-finding
on $\operatorname{Tr}(M(\pi))+2=0$ for $(1+a\cos2x)q''+cq=0$, $a=-\varepsilon$,
then $\mu_{\pm}=2\omega_{0}/\sqrt{c_{\mp}^{(1)}}$) and comparing
with the analytical expressions.

\medskip{}
\emph{Eq.~\eqref{eq:primary-domain}, implicit relation}: $1/\mu_{-}^{2}-1/\mu_{+}^{2}=\varepsilon/(4\omega_{0}^{2})+O(\varepsilon^{3})$. 
\begin{center}
\global\long\def\arraystretch{1.15}%
\begin{tabular}{cccccc}
\hline 
$\varepsilon$  & LHS (exact)  & $\varepsilon/4\omega_{0}^{2}$  & abs.\ error  & error$/\varepsilon^{3}$  & rel.\ err\tabularnewline
\hline 
$0.05$  & $0.012499$  & $0.012500$  & $1.1\times10^{-6}$  & $-0.0088$  & $0.009\%$\tabularnewline
$0.10$  & $0.024991$  & $0.025000$  & $8.8\times10^{-6}$  & $-0.0088$  & $0.035\%$\tabularnewline
$0.15$  & $0.037470$  & $0.037500$  & $3.0\times10^{-5}$  & $-0.0089$  & $0.080\%$\tabularnewline
$0.20$  & $0.049928$  & $0.050000$  & $7.2\times10^{-5}$  & $-0.0090$  & $0.144\%$\tabularnewline
\hline 
\end{tabular}
\par\end{center}
The error$/\varepsilon^{3}$ column is constant at $\approx-0.0088$,
confirming exact $O(\varepsilon^{3})$ scaling with a fixed next-order
coefficient.

\medskip{}
\emph{Eq.~\eqref{eq:primary-domain}}: explicit boundary frequencies
$\mu_{\pm}$ (here $\omega_{0}=1$, so $\mu$ is in units of $\omega_{0}$). 
\begin{center}
\global\long\def\arraystretch{1.15}%
\begin{tabular}{ccccc}
\hline 
$\varepsilon$  & $\mu_{+}$ formula  & $\mu_{+}$ exact  & abs.\ err  & rel.\ err\tabularnewline
\hline 
$0.05$  & $2.0262095$  & $2.0262077$  & $1.8\times10^{-6}$  & $0.0001\%$\tabularnewline
$0.10$  & $2.0550009$  & $2.0549897$  & $1.1\times10^{-5}$  & $0.0005\%$\tabularnewline
$0.15$  & $2.0866508$  & $2.0866254$  & $2.5\times10^{-5}$  & $0.0012\%$\tabularnewline
$0.20$  & $2.1214861$  & $2.1214578$  & $2.8\times10^{-5}$  & $0.0013\%$\tabularnewline
\hline 
\end{tabular}
\par\end{center}
(Results for $\mu_{-}$ are symmetric to the same accuracy.) Errors
scale as $O(\varepsilon^{3})$ and remain below $0.002\%$ across
the physically relevant range $\varepsilon\leq0.20$.

\medskip{}
\emph{Eq.~\eqref{eq:primary-width}}: instability width $\Delta\mu=\omega_{0}\,\varepsilon\,(1+O(\varepsilon^{2}))$. 
\begin{center}
\global\long\def\arraystretch{1.15}%
\begin{tabular}{cccc}
\hline 
$\varepsilon$  & $\omega_{0}\varepsilon$  & $\Delta\mu$ exact  & residual$/\varepsilon^{3}$\tabularnewline
\hline 
$0.05$  & $0.050000$  & $0.050068$  & $0.544$\tabularnewline
$0.10$  & $0.100000$  & $0.100547$  & $0.547$\tabularnewline
$0.15$  & $0.150000$  & $0.151863$  & $0.552$\tabularnewline
$0.20$  & $0.200000$  & $0.204473$  & $0.559$\tabularnewline
\hline 
\end{tabular}
\par\end{center}
The leading-order width is confirmed; moreover, the residual column
matches the exact cubic-and-quintic prediction of the CF series~\eqref{eq:Delta-mu-CF},
$\tfrac{139}{256}+\tfrac{25659}{65536}\,\varepsilon^{2}=0.5430+0.3915\,\varepsilon^{2}$,
to better than $0.1\%$ at every $\varepsilon$ in the table ---
an independent numerical confirmation of the exact $\varepsilon^{3}$
coefficient $139/256$. 
\end{rem}

% -----------------------------------------------------------------------------

\subsubsection{High-accuracy boundary curves and Floquet factor via the CF method}

\label{subsubsec:MR-CF-results}

The continued-fraction approach of Chapter~\ref{sec:expanding-Cambi}
yields two further results for the primary tongue that go substantially
beyond Theorem~\ref{thm:primary-domain}: a six-term series for the
boundary frequencies $\mu_{\pm}$ with all exact rational coefficients,
and explicit closed-form formulas for the Fourier coefficients of
the periodic Floquet factor\index{Floquet theory!Floquet factor}.
\begin{thm}[High-accuracy primary boundary frequencies, Section~\ref{subsec:CF-stability-bdry}]
\label{thm:primary-domain-CF} Under the same hypotheses as Theorem~\ref{thm:primary-domain},
the CF method (Theorem~\ref{thm:bdry-series-CF}) gives the primary
instability boundary frequencies to $O(\varepsilon^{7})$: 
\begin{align}
\mu_{-}(\varepsilon) & =\omega_{0}\Bigl[2-\tfrac{1}{2}\varepsilon+\tfrac{15}{32}\varepsilon^{2}-\tfrac{139}{512}\varepsilon^{3}+\tfrac{2341}{8192}\varepsilon^{4}\nonumber \\
 & \quad-\tfrac{25659}{131072}\varepsilon^{5}+\tfrac{885891}{4194304}\varepsilon^{6}+O(\varepsilon^{7})\Bigr],\label{eq:mu-minus-CF}\\
\mu_{+}(\varepsilon) & =\omega_{0}\Bigl[2+\tfrac{1}{2}\varepsilon+\tfrac{15}{32}\varepsilon^{2}+\tfrac{139}{512}\varepsilon^{3}+\tfrac{2341}{8192}\varepsilon^{4}\nonumber \\
 & \quad+\tfrac{25659}{131072}\varepsilon^{5}+\tfrac{885891}{4194304}\varepsilon^{6}+O(\varepsilon^{7})\Bigr].\label{eq:mu-plus-CF}
\end{align}
All coefficients are exact rationals obtained by the recursive algorithm~\eqref{eq:c2k-recursion},
extensible to any desired order. Odd-order terms appear with opposite
signs in $\mu_{-}$ and $\mu_{+}$; even-order terms (the center shift
of the tongue) are identical. The instability domain width is: 
\begin{equation}
\Delta\mu(\varepsilon)=\omega_{0}\Bigl[\varepsilon+\tfrac{139}{256}\varepsilon^{3}+\tfrac{25659}{65536}\varepsilon^{5}+O(\varepsilon^{7})\Bigr].\label{eq:Delta-mu-CF}
\end{equation}
These extend the $O(\varepsilon)$ approximation of Theorem~\ref{thm:primary-domain}
for the boundary frequencies through order $\varepsilon^{6}$, and
the leading-order width formula~\eqref{eq:primary-width} through
order $\varepsilon^{5}$, with all coefficients exact. 
\end{thm}

\begin{proof}
From the six-term series $p_{\pm}(\gamma)$ of Theorem~\ref{thm:bdry-series-CF}
with $\gamma=\varepsilon/2$, using $\mu_{\pm}=\omega_{0}/p_{\mp}(\varepsilon/2)$
and expanding $1/p_{\mp}(\varepsilon/2)$ as a power series in $\varepsilon$.
All arithmetic is exact rational. See Section~\ref{subsec:CF-stability-bdry}
for the derivation. 
\end{proof}
\begin{rem}[Comparison with Theorem~\ref{thm:primary-domain}]
\label{rem:primary-CF-comparison} The leading term $\mu_{\pm}\approx2\omega_{0}(1\pm\varepsilon/4)$
recovers the $O(\varepsilon)$ result of Theorem~\ref{thm:primary-domain}.
The next term $+\tfrac{15}{32}\varepsilon^{2}$ is the center shift
(symmetric, same in both $\mu_{\pm}$); the first asymmetric correction
beyond leading order is $\pm\tfrac{139}{512}\varepsilon^{3}$. The
width formula~\eqref{eq:Delta-mu-CF} recovers $\omega_{0}\varepsilon$
at leading order and has only odd powers of $\varepsilon$, consistent
with the antisymmetry $\mu_{+}(\varepsilon)-\mu_{-}(\varepsilon)=\mu_{+}(-\varepsilon)-\mu_{-}(-\varepsilon)$.
For $\varepsilon=0.20$, the six-term formula for $\mu_{\pm}$ achieves
relative error $O(\varepsilon^{7})\approx10^{-5}$, compared to $O(\varepsilon^{3})\approx10^{-3}$
for Theorem~\ref{thm:primary-domain}. 
\end{rem}

\begin{rem}[The minimal solution ratio $w(u)$]
\label{rem:w-definition-MR} The key ingredient of Theorem~\ref{thm:Floquet-factor-MR}
is the \emph{minimal solution ratio}\index{$w(u)$}\index{minimal solution ratio $w(u)$}
$w(u)=h_{1}(u)/h_{0}(u)$, the ratio of consecutive Fourier coefficients
of the Floquet factor, evaluated at the solution $\{h_{n}(u)\}$ that
decays as $n\to+\infty$. It is given by the continued fraction (eq.~\eqref{eq:notes2-wdef},
Section~\ref{subsec:Cambi-notes2}): 
\begin{equation}
w(u)\;=\;\cfrac{1}{-\dfrac{G(u+1)}{\gamma}+\cfrac{1}{\dfrac{G(u+2)}{\gamma}+\cfrac{1}{-\dfrac{G(u+3)}{\gamma}+\cdots}}},\qquad G(u)=1-\frac{p^{2}}{u^{2}},\label{eq:w-def-MR}
\end{equation}
convergent geometrically with ratio $|\zeta_{+}|^{2}<1$, where $\gamma=\varepsilon/2$
and $p=\omega_{0}/\mu$. In practice~\eqref{eq:w-def-MR} is evaluated
by the backward recursion: set $T_{N}=\zeta_{+}$ (the asymptotic
value) at level $N$, then 
\begin{equation}
T_{n-1}=\cfrac{1}{\dfrac{(-1)^{n}G(u+n)}{\gamma}+T_{n}},\quad n=N,\,N-1,\,\ldots,\,1,\qquad w(u)=T_{0},\label{eq:w-CF-eval-MR}
\end{equation}
converging to full working precision with $N=50$--$100$ levels.
The alternating sign $(-1)^{n}$ reproduces the partial denominators
of~\eqref{eq:w-def-MR} exactly and must be kept as written. Note
also the overall sign convention $w=-f^{(0)}$ of eq.~\eqref{eq:notes2-wdef}:
evaluating the untransformed continued fraction~\eqref{eq:notes2-f0}
directly returns $-w$, not $w$. The products in~\eqref{eq:Floquet-Fourier-MR}
are then elementary: $\prod_{k=0}^{m-1}w(u^{*}+k)$ uses the same
CF calls already made to locate $u^{*}$. 
\end{rem}

\begin{figure}[htbp]
\centering %
\begin{minipage}[t]{0.485\textwidth}%
 \centering \includegraphics[width=1\textwidth]{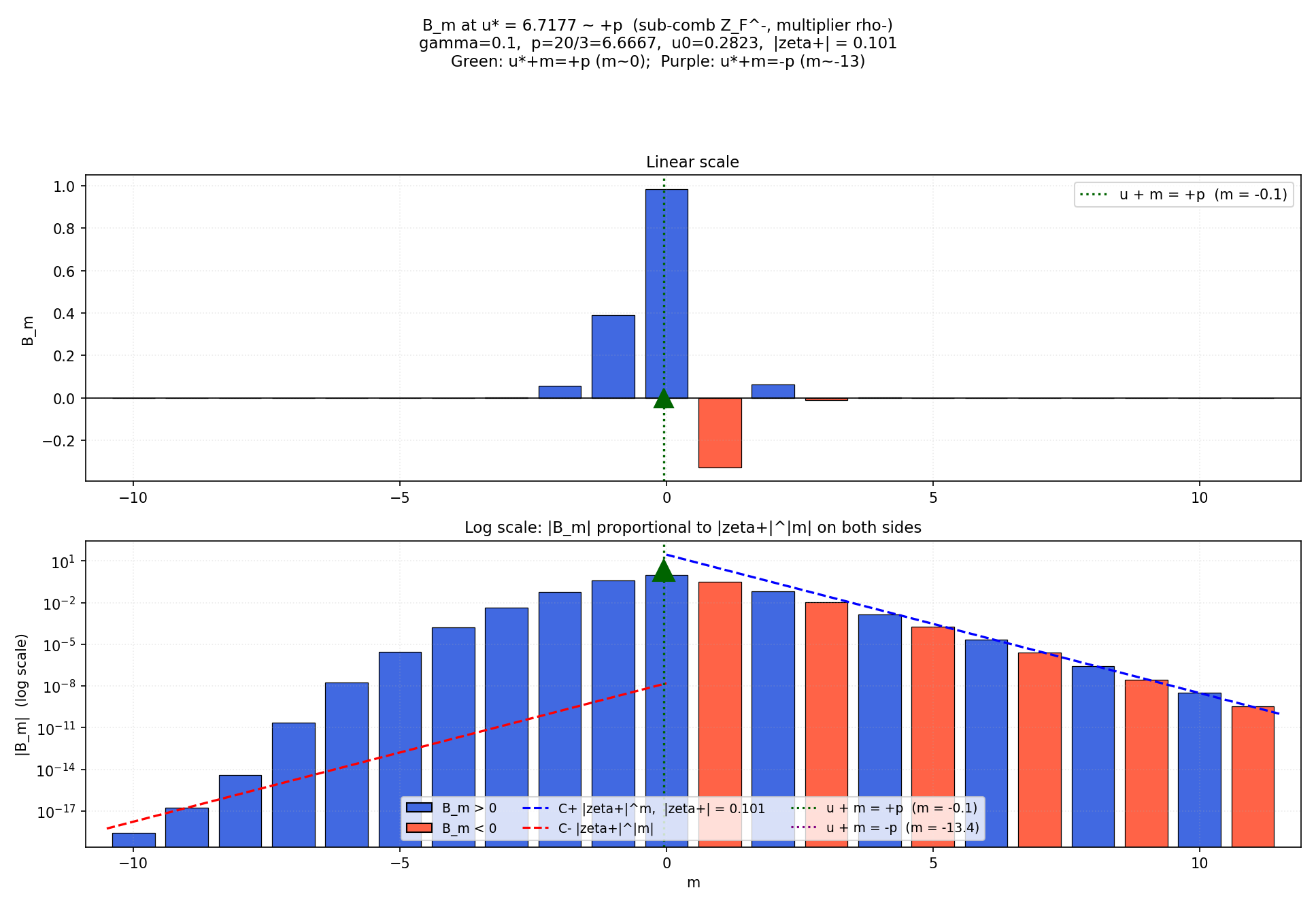} %
\end{minipage}\hfill{}%
\begin{minipage}[t]{0.485\textwidth}%
 \centering \includegraphics[width=1\textwidth]{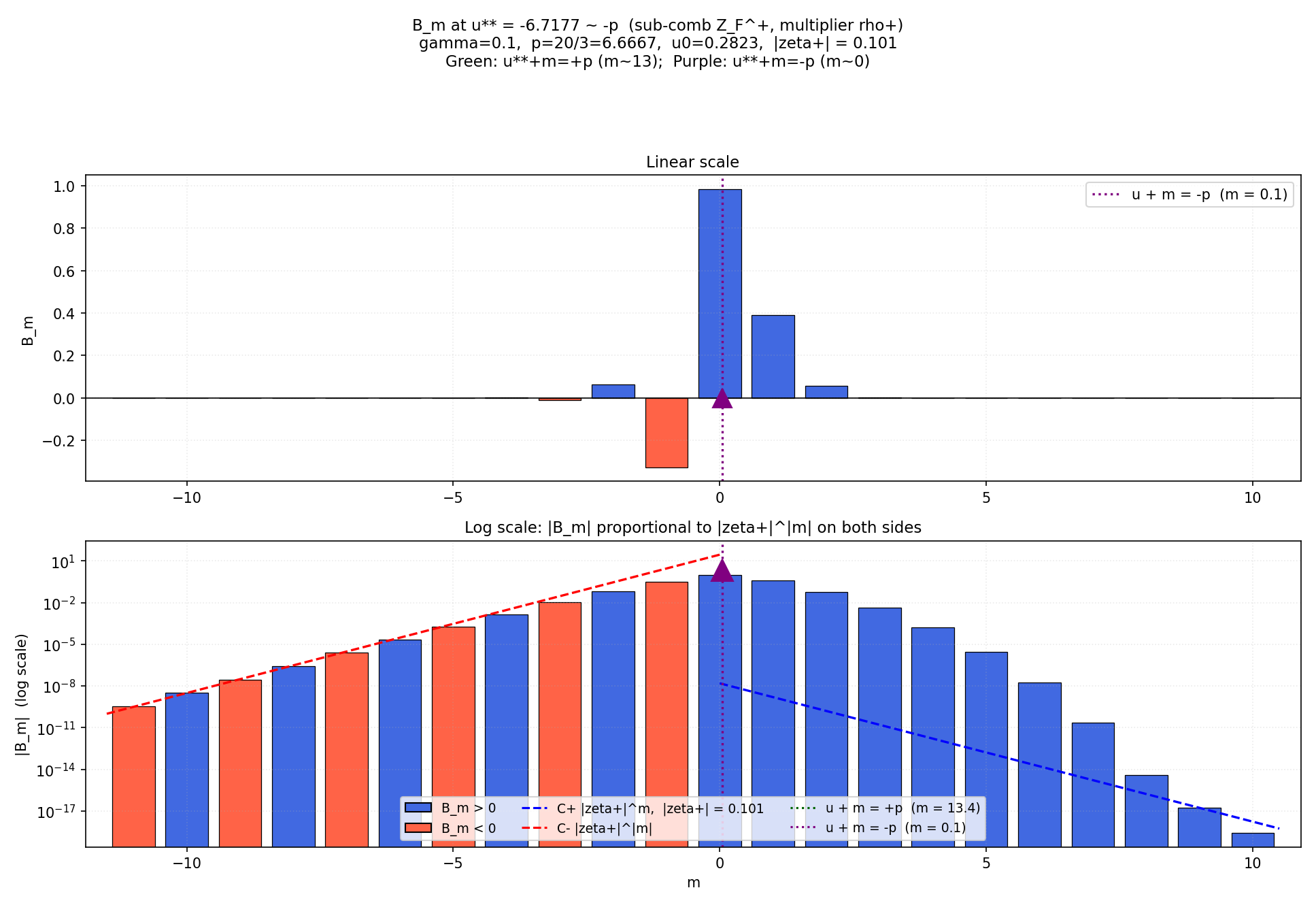} %
\end{minipage}\caption{Fourier coefficients $B_{m}=p^{2}/(u^{*}+m)^{2}\cdot h_{m}(u^{*})$
of the two Floquet factors for Cambi's parameters $\gamma=0.1$, $p=20/3$
($|\zeta_{+}|\approx0.101$), computed via the explicit finite products~\eqref{eq:Floquet-Fourier-MR}.
\emph{Left} ($u^{*}\approx+6.718$, multiplier $\rho_{-}$): $B_{0}\approx0.985$
dominates; $|B_{m}|$ decays exponentially on both sides with envelope
$|\zeta_{+}|^{|m|}$ (dashed). Signs alternate for $m>0$ ($\zeta_{+}<0$);
all positive for $m<0$. \emph{Right} ($u^{**}\approx-6.718$, multiplier
$\rho_{+}$): the mirror image, $B_{m}(u^{**})=B_{-m}(u^{*})$ exactly
(Theorem~\ref{thm:Floquet-factor-MR}(ii)). Both spectra computed
to 50-digit precision with $N=50$ CF levels. Full details in Figs.~\ref{fig:Bm-Cambi}--\ref{fig:Bm-Cambi2}.}
\label{fig:Bm-MR} 
\end{figure}

\begin{thm}[Explicit Fourier coefficients of the Floquet factor, Section~\ref{subsec:Cambi-notes2-H}]
\label{thm:Floquet-factor-MR} Let $u^{*}$ be the primary Floquet
exponent\index{Floquet theory!Floquet exponent} (the zero of the
double minimality function $F_{w}$\index{$F_{w}$}\index{double minimality function $F_{w}$}
nearest $p=\omega_{0}/\mu$, computed to arbitrary precision via Theorem~\ref{thm:primary-domain-CF}).
The Floquet solution $f(x,u^{*})=e^{2\pi iu^{*}x}P(u^{*},x)$ has
periodic factor $P(u^{*},x)=\sum_{m\in\mathbb{Z}}B_{m}\,e^{2\pi imx}$
with Fourier coefficients given by the explicit finite products: 
\begin{equation}
\frac{h_{m}(u^{*})}{h_{0}(u^{*})}\;=\;\begin{cases}
{\displaystyle \prod_{k=0}^{m-1}w(u^{*}\!+k),} & m\geq1,\\[8pt]
1, & m=0,\\[4pt]
{\displaystyle \prod_{k=0}^{|m|-1}w(-u^{*}\!+k),} & m\leq-1,
\end{cases}\quad B_{m}\;=\;\frac{p^{2}}{(u^{*}\!+m)^{2}}\,h_{m}(u^{*}),\label{eq:Floquet-Fourier-MR}
\end{equation}
where $w(u)$ is the CF minimal-solution ratio~\eqref{eq:notes2-wdef}.
Each product is finite and computed directly from the CF values already
required to locate $u^{*}$, at no additional cost. The coefficients
satisfy: 
\begin{enumerate}
\item[(i)] \emph{Exponential decay:} $|B_{m}|\sim|\zeta_{+}|^{|m|}$ as $|m|\to\infty$,
where $\zeta_{+}=(-1+\sqrt{1-4\gamma^{2}})/(2\gamma)$, $|\zeta_{+}|<1$. 
\item[(ii)] \emph{Mirror symmetry:} if $u^{**}\approx-p$ is the second primary
Floquet exponent, then $B_{m}(u^{**})=B_{-m}(u^{*})$ for all $m\in\mathbb{Z}$. 
\item[(iii)] \emph{Double minimality:} $\{h_{m}(u^{*})\}$ decays exponentially
as $|m|\to\infty$ precisely because $u^{*}$ satisfies the double
minimality condition $F_{w}(u^{*})=0$ (Theorem~\ref{thm:double-min}). 
\end{enumerate}
Figure~\ref{fig:Bm-MR} illustrates both Floquet solutions for Cambi's
parameters $(\gamma,p)=(0.1,\,20/3)$, confirming the exponential
envelope and mirror symmetry. Full spectra for $(\gamma,p)=(0.1,\,0.3)$
are in Fig.~\ref{fig:Bm-p03}. 
\end{thm}

\begin{proof}
Corollary~\ref{cor:Floquet-Fourier} in Section~\ref{subsec:Cambi-notes2-H},
which follows from Theorem~\ref{thm:notes2-H-new}(iv,v) and the
double minimality condition~\eqref{eq:double-min-explicit}. 
\end{proof}

\subsubsection{The boundary curves by a third method: the Yakubovich--Starzhinskii
series}\index{Yakubovich--Starzhinskii series}

\label{subsubsec:MR-YS}

The instability boundaries found above by the Magnus--Winkler coexistence
theory (Theorem~\ref{thm:EPD-bdry}) and the continued-fraction method
(Theorem~\ref{thm:primary-domain-CF}) are recovered independently
by the high-frequency exponent-matrix series of Yakubovich and Starzhinskii.
This furnishes a third, fully independent confirmation and exposes
the common linear-algebraic content of the three approaches.
\begin{thm}[YS exponent matrices and the primary boundary, Chapters~\ref{sec:YS-LC},~\ref{sec:YS-exact-LC}]
\label{thm:YS-boundary-MR} For the exact LC circuit the Yakubovich--Starzhinskii
exponent matrices $\mathbf{K}_{m}(\hat{\lambda})$ are polynomials
in the spectral parameter, computed in closed form through second
order ($\mathbf{K}_{1}$, $\mathbf{K}_{2}$). The condition that the
monodromy have a non-trivial Jordan block, written as the vanishing
of the off-diagonal product $K_{01}K_{10}=0$ of the truncated exponent
matrix, yields the primary instability boundary 
\begin{equation}
c_{\pm}(\delta)=1\pm\tfrac{\varepsilon}{2}-\tfrac{9}{32}\varepsilon^{2}+O(\varepsilon^{3}),\qquad\varepsilon=\tfrac{2\delta}{1+\delta^{2}},\label{eq:YS-boundary-MR}
\end{equation}
identical to the Magnus--Winkler and continued-fraction results~\eqref{eq:EPD-bdry-1}
to $O(\varepsilon^{3})$ --- in particular the same center shift
$-9/32$. 
\end{thm}

\begin{proof}
Closed-form $\mathbf{K}_{m}(\hat{\lambda})$ and the boundary condition
are derived in Chapters~\ref{sec:YS-LC} and~\ref{sec:YS-exact-LC}
(eq.~\eqref{eq:YS-bdry-exact}); the equivalence of the three boundary
characterizations through the relation $K=\tfrac{1}{\pi}\log(-M)$
is established in Chapter~\ref{sec:YS-comparison}. 
\end{proof}
\begin{rem}[Three independent routes to the same boundary]
\label{rem:three-methods-MR} The boundary~\eqref{eq:YS-boundary-MR}
agrees with the exact boundary obtained by direct numerical monodromy
integration to the order of the truncation (Table~\ref{tab:YS-bdry-numerical}).
The three methods are complementary: Magnus--Winkler explains \emph{why}
the odd tongues open (coexistence polynomials), the continued fraction
gives \emph{where} and \emph{how wide} they are (exact for all $\delta<1$),
and the Yakubovich--Starzhinskii series exhibits the boundary as
a linear-algebraic degeneracy of the exponent matrix. 
\end{rem}

\subsection{Mathieu comparison and discriminant expansion}

\label{subsec:MR-Mathieu}

The Mathieu equation~\eqref{eq:Mathieu-MR} is the standard reference
for parametric resonance and corresponds to the degenerate Ince case
$a=0$, $d=-2q$. The contrast between the LC circuit and Mathieu
illuminates both the power and the limitations of the Mathieu approximation.
In this section the physical parameter pair $(\varepsilon,\mu)$ (with
$\omega_{0}$ fixed) is used throughout, as it most directly connects
the two equations.

\subsubsection{Mathieu approximation and its deviation}

\begin{figure}[htbp]
\centering \includegraphics[width=10cm]{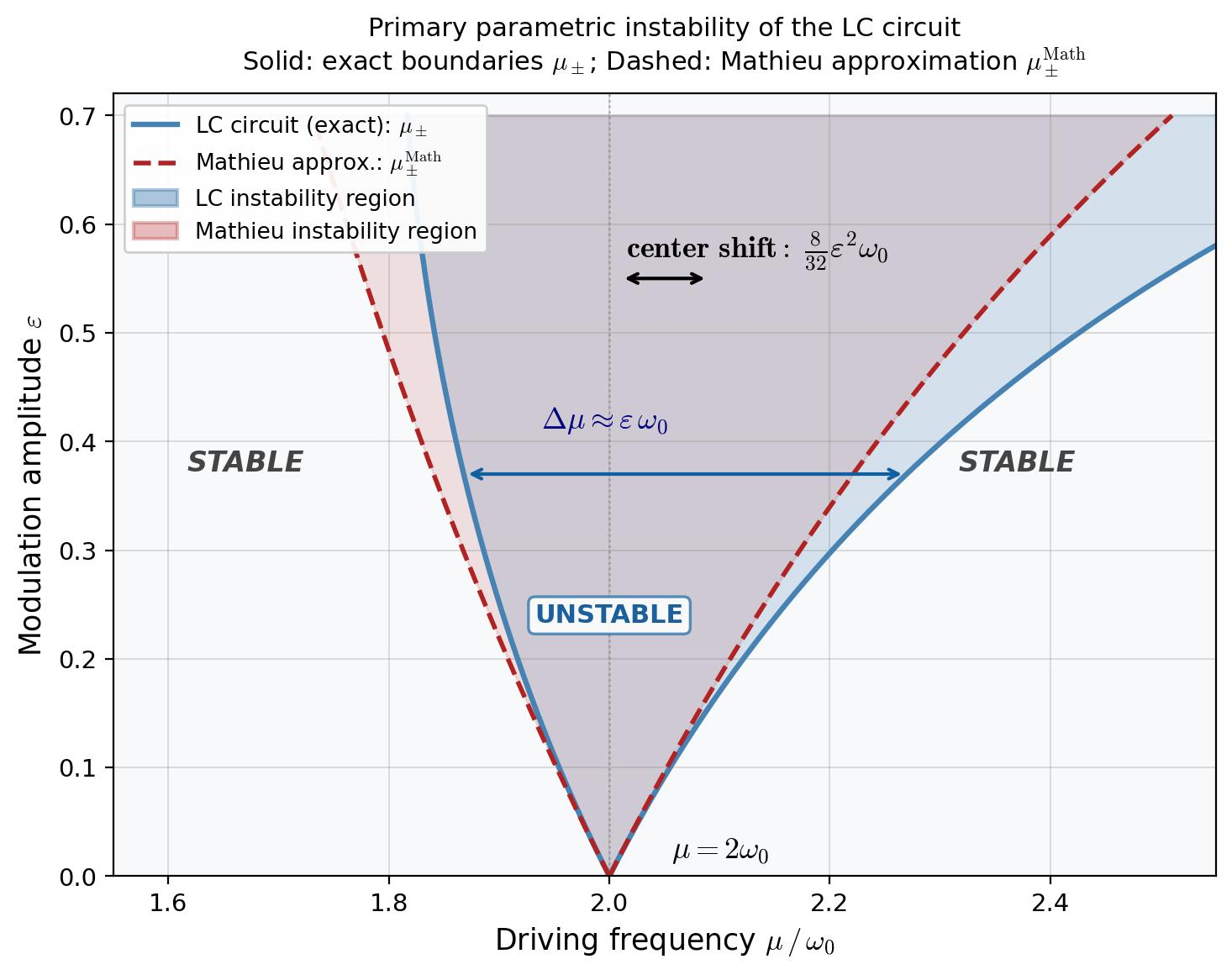}
\caption{Primary parametric instability domain of the LC circuit (Theorem~\ref{thm:primary-domain}).
Solid blue curves: exact boundaries $\mu_{\pm}$ of the LC circuit~\eqref{eq:primary-domain}.
Dashed red curves: Mathieu approximation boundaries~\eqref{eq:primary-domain-Math}.
Blue shading: LC instability region; light red shading: Mathieu instability
region. Both tongues open from $\mu=2\omega_{0}$ at $\varepsilon=0$
with the same leading slope $\Delta\mu=\varepsilon\omega_{0}$, but
their centers differ by a factor of up to~$9$ at $\varepsilon=1$
(indicated by the double arrow at $\varepsilon=0.55$). The Mathieu
approximation correctly captures the primary tongue width to leading
order but significantly underestimates the center shift.}
\label{fig:primary-domain} 
\end{figure}

The standard Mathieu approximation~\cite[Sec.~15.21]{MacLa} replaces
the exact restoring coefficient $\omega_{0}^{2}/(1+\varepsilon\cos\mu t)$
by its first-order Taylor expansion $\omega_{0}^{2}(1-\varepsilon\cos\mu t)$,
yielding 
\begin{equation}
\ddot{y}+\omega_{0}^{2}(1-\varepsilon\cos\mu t)\,y=0.\label{eq:Mathieu-LC-approx}
\end{equation}
In standard Mathieu form $z''+(a-2q\cos2x)z=0$ with $x=\mu t/2$,
the spectral parameter is $a=4\omega_{0}^{2}/\mu^{2}=c$ and the Mathieu
parameter is $q=\varepsilon c/2=2\varepsilon\omega_{0}^{2}/\mu^{2}$
(eq.~\eqref{eq:Mathieu-MR}); at primary resonance $q\approx\varepsilon/2$.
McLachlan's formula~\cite[Secs.~2.13--2.14, eqs.~(17),(3)]{MacLa}
gives the Mathieu primary instability boundaries 
\begin{equation}
\mu_{\pm}^{{\rm Math}}=\frac{2\omega_{0}}{\sqrt{1-\dfrac{\varepsilon^{2}}{32}\mp\dfrac{|\varepsilon|}{2}}}+O(\varepsilon^{3}),\quad\varepsilon\to0.\label{eq:primary-domain-Math}
\end{equation}
with width $\Delta\mu^{{\rm Math}}=\omega_{0}(\varepsilon+O(\varepsilon^{3}))$.
See Figure~\ref{fig:primary-domain} for a visual comparison of the
LC and Mathieu instability domains in the $(\varepsilon,\mu/\omega_{0})$-plane.

Comparing~\eqref{eq:primary-domain} and~\eqref{eq:primary-domain-Math}
(both in the parameter pair $(\varepsilon,\mu)$ with $\omega_{0}$
fixed): 
\begin{itemize}
\item The \emph{widths} agree to leading order: both give $\Delta\mu=\varepsilon\omega_{0}+O(\varepsilon^{3})$.
The Mathieu approximation therefore correctly predicts the primary
tongue width to $O(\varepsilon)$. 
\item The \emph{center positions} differ: LC has center shift $-\frac{9}{32}\varepsilon^{2}$
while Mathieu has $-\frac{1}{32}\varepsilon^{2}$ --- a factor of
$9$. The Mathieu approximation significantly underestimates the downward
center shift at moderate $\varepsilon$. 
\item The \emph{qualitative structure} differs fundamentally for higher
intervals: Mathieu predicts instability at all harmonics $\mu\approx2\omega_{0}/m$,
while the exact LC equation has instability only at odd $m$. 
\end{itemize}
Several features of Theorem~\ref{thm:primary-domain} are noteworthy.
The primary instability occurs near $\mu\approx2\omega_{0}$, i.e.\ at
\emph{twice} the natural frequency. The domain center shifts to higher
frequencies as $|\varepsilon|$ increases (the $+\frac{15}{32}\varepsilon^{2}$
term in the center), reflecting the stiffening effect of the parametric
modulation. The width $\Delta\mu=\varepsilon\omega_{0}+O(\varepsilon^{3})$
opens as a wedge linearly in $|\varepsilon|$. Both boundaries are
exceptional points of degeneracy (EPD): the circuit transitions from
stable to unstable through a Jordan-block degeneracy rather than a
smooth bifurcation.

\subsubsection{Width formula comparison}

\label{subsubsec:MR-width-comparison}

The Mathieu equation is the degenerate Ince case $a=0$, $d=-2q$.
The same recurrence method applied to Mathieu (Chapter~\ref{sec:Boundaries})
yields the following closed-form formula, which does not appear in
McLachlan~\cite[Sec.~2.151]{MacLa} but is verified against his individual
characteristic-number expansions: 
\begin{equation}
\boxed{L_{m}^{\mathrm{Math}}=\frac{q^{m}}{2^{2m-3}\,[(m-1)!]^{2}}+O_{m}(q^{m+2}),\qquad m=1,2,3,\ldots}\label{eq:Lm-Math-main}
\end{equation}
for all $m\geq1$; the error is $O_{m}(q^{m+2})$, two powers beyond
the leading term --- matching the $O_{m}(\delta^{m+2})$ accuracy
of the LC formula~\eqref{eq:Lm-main}, since both width expansions
contain only alternate powers of the small parameter. The derivation
is in Section~\ref{subsec:MathieuInce}; the $O_{m}(q^{m+2})$ accuracy
argument is in Section~\ref{para:Math-chain-accuracy}: for fixed
$m=2k-1$ the backward chain has exactly $k-1$ steps, making the
error bound finite and explicit for each fixed $m$. 
\begin{rem}[Numerical verification of the width formula]
\label{rem:Lm-verify} Formula~\eqref{eq:Lm-main} was verified
numerically for $m=1,3$ across $\delta\in[0.05,0.40]$. For $m=1$:
the exact CF value $W_{{\rm CF}}=c^{+}-c^{-}$ (from $F_{{\rm even/odd}}(c)=0$,
where $F_{{\rm even}}$, $F_{{\rm odd}}$ are the continued-fraction
eigenvalue functions defined in Chapter~\ref{sec:LCInce}, eq.~\eqref{eq:CF-criterion};
fully converged backward recurrence) and the formula agree to within
$0.09\%$ at $\delta=0.02$, $0.5\%$ at $\delta=0.05$, and $2.1\%$
at $\delta=0.10$; the growing discrepancy is the expected $O(\delta^{2})$
correction absent from the leading-order formula. The leading coefficient
$W/\delta\to2.0$ as $\delta\to0$ is confirmed numerically to eight
significant figures. For $m=3$: the exact width (CF zeros, coinciding
with the discriminant condition $\Delta(c,\delta)=-2$ to machine
precision) gives $W_{\mathrm{num}}/W_{\mathrm{formula}}=1.022$, $1.092$,
$1.214$, $1.399$ at $\delta=0.10$, $0.20$, $0.30$, $0.40$; the
excess is the $O_{3}(\delta^{2})$ correction, with constant $\approx2.4$.
The leading coefficient $W/\delta^{3}\to9/32=0.28125$ is confirmed:
$W/\delta^{3}=0.28139$ at $\delta=0.02$. No formula errors were
found. 
\end{rem}

The qualitative difference between~\eqref{eq:Lm-main} and~\eqref{eq:Lm-Math-main}
--- double factorial vs.\ ordinary factorial --- is traced to the
growing recurrence coupling $a(2n-1)^{2}$ of the LC circuit vs.\ the
constant coupling $q$ of Mathieu.

\subsubsection{Structural comparison: LC circuit vs.\ Mathieu}

The Mathieu equation $y''+(\lambda-2q\cos2x)y=0$ is the most widely
studied special case of Hill's equation\index{Hill equation} and
serves as the standard reference for parametric resonance phenomena
(see Chapter~\ref{app:ParRes} for the general theory). It corresponds
to the degenerate Ince case $a=0$, $d=-2q$. The contrast between
the LC circuit and Mathieu is sharp and illuminating; Table~\ref{tab:comparison-main}
presents the structural comparison.

\begin{table}[ht]
\centering 
\global\long\def\arraystretch{1.8}%
{\small\setlength{\tabcolsep}{4pt} }{\small{}%
\begin{tabular}{p{2.3cm}p{4.5cm}p{4.5cm}}
\toprule 
{\small Property } & {\small LC circuit ($\delta\to0$) } & {\small Mathieu ($q\to0$)}\tabularnewline
\midrule 
{\small Ince parameters } & {\small$a=-\varepsilon\neq0$, $b=d=0$ } & {\small$a=0$, $d=-2q\neq0$}\tabularnewline
{\small$Q(\mu)$ } & {\small$2a\mu^{2}$ (root at $\mu=0$, eq.~\eqref{eq:QQstar-intro}) } & {\small$q$ (no roots)}\tabularnewline
{\small Even intervals } & {\small$L_{m}=0$ exactly (all vanish) } & {\small$L_{m}>0$ (all present)}\tabularnewline
{\small Odd intervals } & {\small$L_{m}>0$ (all present) } & {\small$L_{m}>0$ (all present)}\tabularnewline
{\small Width, odd $m$ } & {\small$\frac{(m!!)^{2}}{2^{(m^{2}+8m-13)/4}}\,\delta^{m}+O_{m}(\delta^{m+2})$ } & {\small$\frac{q^{m}}{2^{2m-3}[(m-1)!]^{2}}+O_{m}(q^{m+2})$}\tabularnewline
{\small$m=1$ (no chain error) } & {\small${\displaystyle \frac{2\delta}{1-\delta^{2}}}$ (all $\delta$) } & {\small$2q+O(q^{3})$}\tabularnewline
{\small Width scaling } & {\small$O(\delta^{m})$, odd $m$ only } & {\small$O(q^{m})$, all $m$}\tabularnewline
{\small Recurrence coupling } & {\small$a(2n-1)^{2}$ (growing with $n$) } & {\small$q$ (constant)}\tabularnewline
\bottomrule
\end{tabular}}\caption{Structural comparison of the LC circuit and Mathieu equations as special
cases of Ince's equation, in the parameter pair $(\delta,q)$. The
two small parameters are related by $\varepsilon=2\delta/(1+\delta^{2})\approx2\delta$
and $q\approx\varepsilon/2\approx\delta$ at leading order (eq.~\eqref{eq:eps-delta}
and Chapter~\ref{sec:Boundaries}); for a direct comparison use $q=\delta$.
Two facts are worth emphasizing. First, the LC circuit primary width
$L_{1}=2\delta/(1-\delta^{2})$ has no chain-approximation error and
is $O(\delta^{3})$-accurate for all $\delta\in(0,1)$. Second, both
width formulas carry absolute error two powers beyond the leading
term --- $O_{m}(\delta^{m+2})$ for the LC circuit and $O_{m}(q^{m+2})$
for Mathieu --- since both width expansions contain only alternate
powers of the small parameter. The full formula with exact $\delta$-prefactors
is~\eqref{eq:Lm-main}. The vanishing of all even intervals holds
exactly for all $0<\delta<1$.}
\label{tab:comparison-main} 
\end{table}

The leading-order widths for the first five intervals (as $\delta\to0$,
$q\to0$) are collected in Table~\ref{tab:widths-main}:

\begin{table}[ht]
\centering 
\global\long\def\arraystretch{1.8}%
\begin{tabular}{c@{\hspace{16pt}}c@{\hspace{16pt}}c@{\hspace{16pt}}c}
\toprule 
$m$  & $L_{m}^{\mathrm{LC}}$ as $\delta\to0$  & $L_{m}^{\mathrm{Math}}$ as $q\to0$  & Ratio\tabularnewline
\midrule 
$1$  & $2\delta$  & $2q$  & $1$ (at $\delta=q$)\tabularnewline
$2$  & $0$ (exact, all $\delta$)  & $q^{2}/2$  & $0$\tabularnewline
$3$  & $9\delta^{3}/32$  & $q^{3}/32$  & $9$\tabularnewline
$4$  & $0$ (exact, all $\delta$)  & $q^{4}/1152$  & $0$\tabularnewline
$5$  & $225\delta^{5}/8192$  & $q^{5}/73728$  & $2025$\tabularnewline
\bottomrule
\end{tabular}\caption{Leading-order instability interval widths in the parameter pair $(\delta,q)$:
as $\delta\to0$ for the LC circuit and $q\to0$ for Mathieu (with
$q\approx\delta$ for direct comparison). Mathieu entries are computed
from $L_{m}^{\mathrm{Math}}=q^{m}/[2^{2m-3}((m-1)!)^{2}]$ (eq.~\eqref{eq:Lm-Math-derived},
derived here) and verified independently from McLachlan's individual
expansions~\cite[Sec.~2.151]{MacLa}: $L_{1}=2q$, $L_{2}=q^{2}/2$,
$L_{3}=q^{3}/32$, $L_{4}=q^{4}/1152$. LC circuit entries show the
dominant term $\sim\delta^{m}$; the full formula~\eqref{eq:Lm-main}
has no chain-approximation error for $m=1$ and has relative error
$O(\delta^{2})$ for $m\protect\geq3$, confirmed against Cambi's
exact solution to better than $2.3\%$ for $\delta\protect\leq0.20$.
The LC circuit has $L_{m}=0$ exactly for all even $m$ and all $0<\delta<1$.}
\label{tab:widths-main} 
\end{table}

The key physical difference is that the LC circuit coefficient $1/(1+\varepsilon\cos\tau)$
is a rational function of $\cos\tau$, which generates infinitely
many Fourier harmonics with geometrically decaying amplitudes $g_{n}\propto\delta^{n}$.
The Mathieu approximation truncates this to a single harmonic $g_{1}$
and thereby loses the cancellation mechanism that eliminates even
intervals.

\subsubsection{Universal entire-function expansion for the Hill discriminant}

Chapter~\ref{sec:UnivExp} establishes the following results, which
hold for any two-parameter Hill family~\eqref{eq:Hill-family} and
are independent of the LC circuit specifics. 
\begin{defn}[$\psi$-basis, Chapter~\ref{sec:UnivExp}, Definition~\ref{def:psi-basis}]
\label{defn:phi-main} For each $n=1,2,3,\ldots$ the \emph{universal
$\psi$-basis function} (Definition~\ref{def:psi-basis}) is 
\begin{equation}
\psi_{n}(\omega)=\frac{(-1)^{n}\sqrt{2}\,\omega\sin(\pi\omega)}{\omega^{2}-n^{2}},\quad\omega=\sqrt{\hat{\lambda}}.\label{eq:phi-basis-main}
\end{equation}
Each $\psi_{n}$ is an entire function of $\omega$, with common resonance
value $\psi_{n}(n)=\pi/\sqrt{2}$~\eqref{eq:psi-resonance} and $L^{2}$-norm
$\|\psi_{n}\|=\pi$~\eqref{eq:psi-norms}. 
\end{defn}

\begin{thm}[Universal expansion, Chapter~\ref{sec:UnivExp}, Theorem~\ref{thm:univ-exp}]
\label{thm:univ-exp-main} For the Hill family $y''+[\lambda+\varepsilon Q_{0}(x)]y=0$,
where $\lambda$ is the spectral parameter, $\varepsilon$ the perturbation
amplitude, and $g_{n}=\frac{1}{\pi}\int_{0}^{\pi}Q_{0}(x)e^{-2inx}dx$
the complex Fourier coefficients of $Q_{0}$, the Hill discriminant
$\Delta(\lambda,\varepsilon)=\operatorname{Tr}(X(\pi))$ (the trace
of the monodromy matrix $X(\pi)$ over one half-period) satisfies
\begin{equation}
\Delta(\lambda,\varepsilon)=2\cos\pi\sqrt{\lambda}+\frac{\pi\varepsilon^{2}\sin\pi\sqrt{\lambda}}{2\sqrt{\lambda}}\sum_{n=1}^{\infty}\frac{g_{n}^{2}}{\lambda-n^{2}}+O(\varepsilon^{4}),\quad\varepsilon\to0,\label{eq:Delta-univ-main}
\end{equation}
uniformly on compact subsets of $\mathbb{C}$. The $O(\varepsilon^{2})$
coefficient is entire in $\omega=\sqrt{\lambda}$. The basis functions
$\psi_{n}$ (Definition~\ref{defn:phi-main}) are universal: independent
of $Q_{0}$. 
\end{thm}

\begin{proof}
\noindent Proved as Theorem~\ref{thm:univ-exp} in Chapter~\ref{sec:UnivExp}. 
\end{proof}
The LC circuit is the canonical instance ($g_{n}=\hat{\lambda}\delta^{n}$,
geometric decay); the Mathieu equation is the single-harmonic special
case ($g_{1}=q$, $g_{n\geq2}=0$). 
\begin{prop}[Leading-order boundary curves, Chapter~\ref{sec:UnivExp}, Proposition~\ref{prop:bdry-leading}]
\label{prop:bdry-main} For the Hill family $y''+[\lambda+\varepsilon Q_{0}(x)]y=0$,
the $m$-th instability zone\index{instability zone} boundaries $\lambda_{m}^{\pm}(\varepsilon)$
(the values of the spectral parameter $\lambda$ at which the Floquet
multipliers coalesce, i.e.\ the EPD values) satisfy 
\begin{equation}
\lambda_{m}^{\pm}(\varepsilon)=m^{2}\pm|g_{m}|\,\varepsilon+O(\varepsilon^{2}),\quad\varepsilon\to0,\label{eq:bdry-main}
\end{equation}
where $g_{m}$ is the $m$-th complex Fourier coefficient of $Q_{0}(x)$
(eq.~\eqref{eq:Delta-univ-main}). The width of the $m$-th tongue
is $\lambda_{m}^{+}(\varepsilon)-\lambda_{m}^{-}(\varepsilon)=2|g_{m}|\varepsilon+O(\varepsilon^{2})$.
For the LC circuit: $g_{n}=\hat{\lambda}\delta^{n}$, so $|g_{m}|=\hat{\lambda}\delta^{m}$;
for Mathieu: $g_{1}=q$, $g_{n\geq2}=0$. This recovers Keller's formula~\eqref{eq:bdry-leading}
from the universal expansion. (For the LC circuit, where the expansion
parameter is $\delta$ with $\varepsilon=1$ absorbed, the formula
gives the correct leading width only for $m=1$: for $m\geq3$ the
multi-harmonic chains built from $g_{1}$ contribute $O(\delta^{m})$
terms of the same order as $g_{m}$, and the true $\delta^{m}$ coefficient
is that of the width formula~\eqref{eq:Lm-main}; cf.\ the discussion
in Chapter~\ref{sec:UnivExp}.) 
\end{prop}

\begin{proof}
\noindent Proved as Proposition~\ref{prop:bdry-leading} in Chapter~\ref{sec:UnivExp}. 
\end{proof}
The $\psi$-basis functions possess the following structural properties
(Chapter~\ref{sec:UnivExp}): (i)~\emph{Definition} (eq.~\eqref{eq:psi-def}):
with $\omega=\sqrt{\lambda}$, 
\[
\psi_{n}(\omega)=\frac{(-1)^{n}\sqrt{2}\,\omega\sin(\pi\omega)}{\omega^{2}-n^{2}},
\]
connecting $\psi_{n}$ to the zeroth spherical Bessel function $j_{0}=\operatorname{sinc}$;
(ii)~\emph{Rational relations}: $\psi_{n}=(-1)^{n-m}\psi_{m}\cdot(\omega^{2}-m^{2})/(\omega^{2}-n^{2})$
for all $m,n$; (iii)~\emph{Volterra origin}: $\psi_{n}(\omega)$
is the $n$-th spectral component of $\Delta_{\omega,2}$ (eq.~\eqref{eq:Delta2-psi}),
associated with the $n$-th Fourier harmonic of $Q_{0}$; (iv)~\emph{Sturm--Liouville\index{Sturm--Liouville theory}
structure}: $\sin\pi\sqrt{\lambda}$ satisfies $-(\sqrt{\lambda}\,S')'=\frac{\pi^{2}}{4}\cdot\frac{S}{\sqrt{\lambda}}$,
and each $\psi_{n}$ is related to the resolvent of the associated
SL operator at spectral parameter $\pi^{2}/4$, evaluated at $\lambda=n^{2}$.

\medskip{}
\emph{Why the Ince approach?} Cambi's continued-fraction solution
is exact but implicit, giving no insight into \emph{why} even intervals
vanish. The Ince identification provides a direct algebraic proof
(coexistence polynomial $Q(\mu)=2a\mu^{2}$~\eqref{eq:QQstar-intro}
has integer root $\mu=0$, forcing period-$\pi$ coexistence by MW
Theorem~7.6), accessible without any numerical computation. The same
backward-chain recurrence that yields the LC width formula~\eqref{eq:Lm-main}
also yields McLachlan's Mathieu formula~\eqref{eq:Lm-Math-main}
for all $m$ simultaneously --- a uniform derivation absent from
McLachlan~\cite[Sec.~2.151]{MacLa}, where widths are computed case
by case. The ordinary factorial $(m-1)!^{2}$ vs.\ double factorial
$(m!!)^{2}$ is directly explained by constant coupling $q$ (Mathieu)
vs.\ growing coupling $a(2n-1)^{2}$ (LC circuit). Full discussion
in Chapter~\ref{sec:Boundaries}.

\subsection{EPD Hypersensitive Capacitance Sensing}

\label{subsec:EPD-sensing-MR}

The EPD structure of the instability boundaries (Theorem~\ref{thm:EPD})
has a direct sensing application. When the circuit is designed so
that $\omega_{0}$ and $\mu$ satisfy $c_{0}=c_{\pm}^{(1)}(\delta)$~\eqref{eq:EPD-bdry-1}
(i.e., the circuit sits at a primary EPD curve, $\omega_{0}\approx\mu/2$),
the two Floquet multipliers coalesce into a Jordan block and the two
characteristic exponents~\eqref{eq:alpha-def-intro} coincide: $\alpha_{+}=\alpha_{-}=1$,
so both Floquet modes oscillate at the same physical frequency $f=\mu/2\approx\omega_{0}$.
A small perturbation $\Delta C$ of the capacitance breaks this degeneracy
and splits the two exponents apart, with a \emph{square-root} dependence
on $\Delta C$ that is the hallmark of EPD hypersensitivity. The theorem
below gives the explicit closed-form formula for this splitting in
terms of the LC circuit parameters. Since EPD points are only marginally
stable, a work-point\index{exceptional point of degeneracy (EPD)!work-point strategy}
strategy is needed to make the sensing scheme robust; this is developed
in full in Chapter~\ref{sec:EPD-sensor} (Figure~\ref{fig:work-point-geometry}).
The primary result is: 
\begin{thm}[EPD sensor: primary tongue, Chapter~\ref{sec:EPD-sensor}]
\label{thm:EPD-sensor-main} Let the circuit operate at a primary
EPD curve $c_{0}=c_{\pm}^{(1)}(\delta)$~\eqref{eq:EPD-bdry-1},
so that $\omega_{0}\approx\mu/2$. Here $\delta=({1-\sqrt{1-\varepsilon^{2}}})/{\varepsilon}$
is the Mobius parameter~\eqref{eq:eps-delta-MR} ($\delta\approx\varepsilon/2$
for small modulation $\varepsilon$), and $\mu$ is the driving frequency.
A small capacitance perturbation $\Delta C$ shifts the natural frequency
by $\xi=\omega_{0}\,\Delta C/(2C_{0})+O((\Delta C/C_{0})^{2})$. The
key observable quantities are: 
\begin{equation}
f_{\pm}=\frac{\mu}{2}\,\alpha_{\pm},\qquad\hat{\xi}=\frac{\xi}{\mu}=\frac{\omega_{0}\,\Delta C}{2C_{0}\mu}+O\!\left(\!\left(\frac{\Delta C}{C_{0}}\right)^{\!2}\right),\label{eq:xi-hat-MR}
\end{equation}
where $\alpha_{\pm}$ are the dimensionless characteristic exponents
($\rho_{\pm}=e^{i\pi\alpha_{\pm}}$, eq.~\eqref{eq:alpha-def-intro}),
$f_{\pm}$ are the physical Floquet oscillation frequencies, and $\hat{\xi}$
is the dimensionless signal ($|\hat{\xi}|\ll1$ in practice). The
\emph{frequency split} $\Delta\alpha=\alpha_{+}-\alpha_{-}$ satisfies:
\begin{equation}
\boxed{\;\begin{aligned}\Delta\alpha & =\frac{2\sqrt{2\delta}}{\mu}\sqrt{\xi\left(\mu-\xi\right)}\,\bigl(1+O(\delta)\bigr)\\
 & =2\sqrt{2\delta}\sqrt{\hat{\xi}(1-\hat{\xi})}\,\bigl(1+O(\delta)\bigr),\qquad\delta\ll1,
\end{aligned}
\;}\label{eq:thm-sensor-split}
\end{equation}
exact in $\hat{\xi}$, free of $\mu$ in the $\hat{\xi}$-form. For
small signals $\hat{\xi}\ll1$ this simplifies to 
\begin{equation}
\Delta\alpha=2\sqrt{2\delta}\cdot\sqrt{\hat{\xi}}\,\bigl(1+O(\delta)\bigr)+O(\hat{\xi}^{3/2}),\label{eq:thm-sensor-sqrt}
\end{equation}
and in terms of the directly measurable capacitance shift: 
\begin{equation}
\Delta\alpha=2\sqrt{\frac{\delta\omega_{0}}{\mu}}\cdot\sqrt{\frac{\Delta C}{C_{0}}}\,\bigl(1+O(\delta)\bigr)+O\!\left(\!\left(\frac{\Delta C}{C_{0}}\right)^{\!3/2}\right).\label{eq:thm-sensor-DeltaC}
\end{equation}
The \emph{conceptual core}: the frequency split scales as $\sqrt{\hat{\xi}}$
(equivalently $\sqrt{\Delta C/C_{0}}$), while a conventional non-EPD
sensor gives a response linear in $\hat{\xi}$. Their ratio $\Delta\alpha_{\mathrm{EPD}}/\Delta\alpha_{\mathrm{conv}}=\sqrt{2\delta/\hat{\xi}}\to\infty$
as $\hat{\xi}\to0$: the EPD sensor is \emph{hypersensitive} for small
signals, with sensitivity that diverges as the perturbation shrinks.
The crossover is at $\hat{\xi}^{*}=2\delta$: EPD sensing is superior
for $\hat{\xi}<2\delta$. The $\sqrt{\cdot}$ response originates
from the square-root branch point of the eigenvalue splitting at the
Jordan block (Newton--Puiseux theorem~\cite[Thm.~2.3]{SeyMai});
it is a structural consequence of the EPD, not an approximation. 
\end{thm}

\begin{proof}
Follows from Theorem~\ref{thm:EPD-split} and the explicit sensitivity
coefficients~\eqref{eq:eta1-m1} (giving $\eta_{1}=+\pi^{2}\delta/2+O(\delta^{2})$
for the primary tongue at the lower-boundary operating point $c_{-}^{(1)}$)
computed in Section~\ref{subsec:m1-explicit}; see equations~\eqref{eq:Deltaalpha-m1}--\eqref{eq:Deltaalpha-DeltaC}
for the full derivation. 
\end{proof}
\begin{rem}[Numerical verification]
\label{rem:sensor-numerics} Formula~\eqref{eq:thm-sensor-split}
was verified against the exact monodromy matrix at the exact numerical
boundary $c_{-}^{(1)}(\delta)$: the ratio of the measured split to
the leading form $\sqrt{2\delta\,|\Delta c|}$ is $1.016$, $1.036$,
$1.061$ at $\delta=0.05$, $0.10$, $0.15$ --- a clean $1+O(\delta)$,
as claimed; the sensitivity coefficient and the splitting geometry
were verified likewise (Remark~\ref{rem:sensor-split-numerics} in
Chapter~\ref{sec:EPD-sensor}). 
\end{rem}

\begin{rem}[Synthesis]
\label{rem:sensor-utility} Theorem~\ref{thm:EPD-sensor-main} draws
on every major result of this work: the EPD operating point is supplied
by Theorem~\ref{thm:EPD-bdry}; the Jordan block is guaranteed by
Theorems~\ref{thm:EPD} and~\ref{thm:bdry-analytic}; the $\sqrt{\hat{\xi}}$
scaling follows from Newton--Puiseux~\cite[Thm.~2.3]{SeyMai} with
the sensitivity coefficient $\eta_{1}=+\pi^{2}\delta/2+O(\delta^{2})$
(at the lower-boundary operating point $c_{-}^{(1)}$) from the LC
matrizant; and the Floquet-comb structure makes $\Delta\alpha$ observable
across every harmonic of the transient response. 
\end{rem}

\begin{rem}[Marginal stability at the EPD and the work-point strategy]
\label{rem:EPD-marginal-stability} EPD points are only \emph{marginally
stable}: the monodromy matrix has a Jordan block with eigenvalue $\rho=-1$
of modulus one, so the circuit is neither exponentially stable nor
exponentially unstable --- it is at the tipping point between the
two regimes. This has an important practical consequence: at the bare
EPD, only one sign of capacitance perturbation ($\Delta C$) keeps
the circuit stable; the opposite sign drives it into the instability
tongue. The $\sqrt{\hat{\xi}}$ sensitivity formula~\eqref{eq:thm-sensor-split}
therefore cannot be exploited directly at the bare EPD point.

The remedy is a \emph{work-point strategy}: the operating point is
shifted by a small controlled amount $\epsilon_{w}$ into the stability
zone (Figure~\ref{fig:work-point-geometry}), so that perturbations
of \emph{either} sign remain stable for $|\hat{\xi}|$ sufficiently
small. The work point preserves the $\sqrt{\hat{\xi}}$ sensitivity
to leading order while making the scheme experimentally robust. The
full development --- including the explicit work-point formula, the
trade-off between sensitivity and stability margin, and a Floquet-comb
probing protocol for reading out $\Delta\alpha$ experimentally ---
is given in Chapter~\ref{sec:EPD-sensor} (see especially \S\,\ref{subsec:work-point}
and Figure~\ref{fig:work-point-geometry}). 
\end{rem}

\section{The LC circuit as a Hill's equation}

\label{sec:LCHill}

The purpose of this chapter is to convert the LC circuit equation~\eqref{eq:LC-original}
into the standard Hill's-equation form required by the MW theory,
and to identify the Fourier coefficients $g_{n}$ explicitly. The
key step is the introduction of the Mobius\index{Mobius transformation}
parameter $\delta$ via the Poisson kernel Fourier series~\eqref{eq:Poisson-kernel},
which converts the rational coefficient $1/(1+\varepsilon\cos\tau)$
into a geometrically decaying Fourier series. See Fig.~\ref{fig:circuit}
for the circuit diagram and Chapter~\ref{app:Hill} for the Floquet\index{Floquet theory}
theory background.

\subsection{\texorpdfstring{The modulation parameter $\delta$}{The modulation
parameter delta}}

The Poisson kernel Fourier series~\cite[Vol.~I, Ch.~III, \S\,3.3]{Zygmund},
\cite[1.447(3)]{GraRyzh}: 
\begin{equation}
\frac{1-p^{2}}{1-2p\cos x+p^{2}}=1+2\sum_{k=1}^{\infty}p^{k}\cos kx,\qquad|p|<1,\label{eq:GR}
\end{equation}
motivates the introduction of a parameter $0<\delta<1$ by 
\begin{equation}
\varepsilon=\frac{2\delta}{1+\delta^{2}},\label{eq:eps-delta}
\end{equation}
so that $|\varepsilon|<1$ corresponds to $|\delta|<1$. We refer
to eq.~\eqref{eq:GR} as the \emph{Poisson kernel Fourier series}~\cite[Vol.~I, Ch.~III, \S\,3.3]{Zygmund},
\cite[1.447(3)]{GraRyzh}. Note that eq.~\eqref{eq:GR} is the Poisson
kernel $P(p,x)=(1-p^{2})/(1-2p\cos x+p^{2})$ itself; the LC coefficient
$1/(1+\varepsilon\cos\tau)$ is a constant multiple of it (eq.~\eqref{eq:Poisson-LC}),
as shown in Chapter~\ref{sec:UnivExp}. For small modulation, $\delta\approx\varepsilon/2$.
With this choice, 
\begin{equation}
1+\varepsilon\cos\tau=\frac{(1+\delta^{2})+2\delta\cos\tau}{1+\delta^{2}}=\frac{1+2\delta\cos\tau+\delta^{2}}{1+\delta^{2}}.\label{eq:1-eps-cos}
\end{equation}
The restoring coefficient is invariant under the time shift $\tau\mapsto\tau+\pi$
(a relabeling of the time origin that leaves the stability properties
unchanged), which sends $\cos\tau\mapsto-\cos\tau$ and brings the
denominator to the standard Poisson-kernel form $1-2\delta\cos\tau+\delta^{2}$.
Adopting this time origin henceforth, equation~\eqref{eq:LC-tau}
becomes 
\begin{equation}
\partial_{\tau}^{2}q+\frac{(1+\delta^{2})}{r^{2}(1-2\delta\cos\tau+\delta^{2})}q=0.\label{eq:LC-delta}
\end{equation}
Applying the Poisson kernel Fourier series~\eqref{eq:GR} with $p=\delta$
and $x=\tau$ gives 
\begin{equation}
\frac{1+\delta^{2}}{1-2\delta\cos\tau+\delta^{2}}=\frac{1+\delta^{2}}{1-\delta^{2}}\Bigl(1+2\sum_{k=1}^{\infty}\delta^{k}\cos k\tau\Bigr),\label{eq:GR-LC}
\end{equation}
so equation~\eqref{eq:LC-delta} expands as 
\begin{equation}
\partial_{\tau}^{2}q+\frac{1+\delta^{2}}{r^{2}(1-\delta^{2})}\Bigl(1+2\sum_{k=1}^{\infty}\delta^{k}\cos k\tau\Bigr)q=0.\label{eq:LC-Fourier}
\end{equation}

\subsection{\texorpdfstring{Rescaling to period $\pi$ and MW form}{Rescaling
to period pi and MW form}}

\label{subsec:rescale}

Equation~\eqref{eq:LC-Fourier} contains $\cos\tau$, which has period
$2\pi$, not $\pi$. To apply the MW theory (which assumes period
$\pi$ for $Q$), we set $x=\tau/2$: 
\begin{equation}
q''+\frac{4(1+\delta^{2})}{r^{2}(1-2\delta\cos2x+\delta^{2})}\,q=0,\label{eq:LC-Hill}
\end{equation}
where primes denote $\dd/\dd x$. This is Hill's equation\index{Hill equation}
$q''+[\hat{\lambda}+Q(x)]q=0$ with 
\begin{equation}
\hat{\lambda}=\frac{4(1+\delta^{2})}{r^{2}(1-\delta^{2})}=\frac{c(1+\delta^{2})}{1-\delta^{2}},\qquad c=\frac{4}{r^{2}}=\frac{4\omega_{0}^{2}}{\mu^{2}},\label{eq:lambda-MW}
\end{equation}
where $c=4/r^{2}$ is the Ince spectral parameter (equation~\eqref{eq:IncePars})
and $r=\mu/\omega_{0}$ is the frequency ratio. The relation $\hat{\lambda}=c(1+\delta^{2})/(1-\delta^{2})$
will be used throughout; in particular it gives $c=\hat{\lambda}(1-\delta^{2})/(1+\delta^{2})$.
The MW Fourier coefficients of $Q(x)=\sum_{n\neq0}g_{n}e^{2inx}$
obtained from the standard integral $\frac{1}{\pi}\int_{0}^{\pi}\frac{\cos2nx\,\dd x}{1-2\delta\cos2x+\delta^{2}}=\delta^{n}/(1-\delta^{2})$
are 
\begin{equation}
g_{n}=\hat{\lambda}\,\delta^{|n|}=\frac{4(1+\delta^{2})}{r^{2}(1-\delta^{2})}\,\delta^{|n|},\qquad n=\pm1,\pm2,\pm3,\ldots.\label{eq:gn}
\end{equation}
In first-order form $\mathbf{x}=(q,q^{\prime})^{T}$, equation~\eqref{eq:LC-Hill}
reads --- this form is used directly by the Yakubovich--Starzhinskii\index{Yakubovich--Starzhinskii series}
method of Chapters~\ref{sec:YS-LC}--\ref{sec:YS-exact-LC} ---
\begin{equation}
\mathbf{x}^{\prime}=\mathbf{A}(x,\delta)\,\mathbf{x},\qquad\mathbf{A}(x,\delta)=\mathbf{C}+\sum_{n=1}^{\infty}\delta^{n}\,\mathbf{B}_{n}(x),\label{eq:LC-Hill-firstorder}
\end{equation}
where 
\begin{equation}
\begin{aligned}\mathbf{C} & =\begin{bmatrix}0 & 1\\
-\hat{\lambda} & 0
\end{bmatrix}=\hat{\lambda}\begin{bmatrix}0 & \hat{\lambda}^{-1}\\
-1 & 0
\end{bmatrix},\\
\mathbf{B}_{n}(x) & =\hat{\lambda}\begin{bmatrix}0 & 0\\
-2\cos2nx & 0
\end{bmatrix}\equiv\hat{\lambda}\,\tilde{\mathbf{B}}_{n}(x).
\end{aligned}
\label{eq:LC-Bn}
\end{equation}
Thus $\hat{\lambda}$ pulls out as an overall factor of $\mathbf{A}(x,\delta)$:
\begin{equation}
\begin{aligned}\mathbf{A}(x,\delta) & =\hat{\lambda}\Bigl[\tilde{\mathbf{C}}+\sum_{n=1}^{\infty}\delta^{n}\,\tilde{\mathbf{B}}_{n}(x)\Bigr],\\
 & \quad\tilde{\mathbf{C}}=\begin{bmatrix}0 & \hat{\lambda}^{-1}\\
-1 & 0
\end{bmatrix},\qquad\tilde{\mathbf{B}}_{n}(x)=\begin{bmatrix}0 & 0\\
-2\cos2nx & 0
\end{bmatrix},
\end{aligned}
\label{eq:LC-Hill-factored}
\end{equation}
where $\tilde{\mathbf{C}}$ and $\tilde{\mathbf{B}}_{n}$ carry no
$\hat{\lambda}$ in their $(2,1)$ and $(2,2)$ entries (note $\tilde{\mathbf{C}}_{12}=\hat{\lambda}^{-1}$
which at primary resonance\index{resonance} $\hat{\lambda}\approx1$
is approximately $1$). In particular, each $\mathbf{B}_{n}=\hat{\lambda}\tilde{\mathbf{B}}_{n}$
is \emph{exactly linear} in $\hat{\lambda}$. The factored form~\eqref{eq:LC-Hill-factored}
makes two structural features immediately visible. First, each $\tilde{\mathbf{B}}_{n}(x)$
contains \emph{exactly one} Fourier mode ($\cos2nx$), so $\delta$
enters as a power series with one new harmonic at each order (property~(ii)).
Second, $\hat{\lambda}$ enters as a \emph{common factor} of each
$\mathbf{B}_{n}$ (though not of $\mathbf{C}$), so by~\eqref{eq:YS-lam-poly}
the YS series coefficients are polynomials in $\hat{\lambda}$: $\mathbf{K}_{m}(\hat{\lambda})$
is a degree-$m$ polynomial and $\mathbf{F}_{m}(x,\hat{\lambda})$
likewise (properties~(ii) and~(iv)). The instability boundary\index{stability boundary}
width --- determined by the eigenvalues of $\pi K(\delta)$ ---
therefore has the form $L_{1}=\hat{\lambda}\,\ell(\delta)$ for an
explicit function $\ell$ independent of $\hat{\lambda}$, and substituting
$\hat{\lambda}=c(1+\delta^{2})/(1-\delta^{2})$ converts this into
the exact LC boundary in $(c,\delta)$. It is equation~\eqref{eq:LC-Hill-firstorder}
--- not its Mathieu\index{Mathieu equation} approximation --- to
which the general Yakubovich--Starzhinskii method of Chapter~\ref{app:YS-Floquet}
applies directly.

Four properties of~\eqref{eq:gn} are fundamental.

\medskip{}
 \emph{(i) All Fourier modes are present.} $g_{n}\neq0$ for all $n\geq1$.

\medskip{}
 \emph{(ii) Geometric structure.} $g_{n}/g_{n-1}=\delta$ exactly,
so the coefficients form a geometric sequence in $\delta$.

\medskip{}
 \emph{(iii) Effective expansion parameter $\delta^{2}$.} Because
all $g_{n}\neq0$, every order of the MW discriminant\index{discriminant}
expansion $\Delta_{n}$ (Chapter~\ref{sec:MWInce}, eq.~\eqref{eq:D0c})
is nonzero in principle. However, due to the MW selection rule $\sum l_{i}=0$
(which requires positive and negative Fourier indices to cancel; see
Chapter~\ref{sec:MWInce}), odd-order terms $\Delta_{2k+1}$ are
at least as small as the next even-order term $\Delta_{2k+2}$. The
consequence is that the discriminant expansion groups in \emph{even}
powers of $\delta$ (eq.~\eqref{eq:Delta-groups}, derived in Chapter~\ref{sec:MWInce}):
so the \emph{effective} small parameter is $\delta^{2}$, not $\delta$.
The detailed scaling $\Delta_{n}=O(\delta^{n})$ for $n$ even and
$\Delta_{n}=O(\delta^{n+1})$ for $n$ odd is derived in Chapter~\ref{sec:MWInce}
(eq.~\eqref{eq:Delta-scaling}).

\medskip{}
 \emph{(iv) Polynomial dependence on $\hat{\lambda}$.} Each Fourier
coefficient $g_{n}=\hat{\lambda}\delta^{n}$ is \emph{linear} in $\hat{\lambda}$,
and the factorization~\eqref{eq:LC-Hill-factored} shows that $\hat{\lambda}$
enters as a common factor of each $\mathbf{B}_{n}$ in the first-order
system~\eqref{eq:LC-Hill-firstorder}. As a consequence, the Yakubovich--Starzhinskii
(YS) series coefficients $\mathbf{K}_{m}$ and $\mathbf{F}_{m}$ (defined
and computed in Chapters~\ref{sec:YS-LC}--\ref{sec:YS-exact-LC}
and Chapter~\ref{app:YS-Floquet}) are \emph{polynomials} in $\hat{\lambda}$:
\begin{align}
\mathbf{K}_{m}(\hat{\lambda}) & =\text{degree-}m\text{ polynomial in }\hat{\lambda},\nonumber \\
\mathbf{F}_{m}(x,\hat{\lambda}) & =\text{degree-}m\text{ polynomial in }\hat{\lambda}.\label{eq:YS-lam-poly}
\end{align}
This polynomial structure --- established in Chapter~\ref{sec:YS-LC}
with explicit examples at eq.~\eqref{eq:LC-K1-lam-exact}--\eqref{eq:LC-K2-lam-exact}
--- allows the $\hat{\lambda}$-dependence of the instability boundaries
to be tracked exactly: after substituting $\hat{\lambda}=c(1+\delta^{2})/(1-\delta^{2})$,
the boundary curve\index{boundary curves}s acquire their full $(c,\delta)$-dependence.

\subsection{Lagrangian and Hamiltonian structure}

\label{subsec:Hamiltonian}

The LC circuit equation has a natural variational structure that connects
it directly to the Hamiltonian framework of Yakubovich and Starzhinskii~\cite[Ch.~III]{YakSta1},
\cite[Ch.~VIII]{YakSta2} and Kirillov~\cite[Sec.~3.3]{Kiril}. Throughout
this section we use physical time $t$; the rescaled variable $x=\tau/2=\mu t/2$
employed in Chapters~\ref{sec:MWInce} onward is introduced solely
to conform with the Magnus--Winkler\index{Magnus--Winkler theory}
convention~\cite[Ch.~1]{MagWin} in which Hill's equation has period
$\pi$.

\emph{Lagrangian.} The charge $q(t)$ and current $\dot{q}(t)$ are
the generalized coordinate and velocity of a one-degree-of-freedom
conservative system with Lagrangian 
\begin{equation}
\mathcal{L}(q,\dot{q},t)=\tfrac{1}{2}L\dot{q}^{2}-\tfrac{q^{2}}{2C(t)},\label{eq:Lagrangian}
\end{equation}
where the first term is the magnetic energy stored in the inductor
and the second is the electric potential energy in the capacitor.
The Euler--Lagrange equation $\frac{d}{dt}\frac{\partial\mathcal{L}}{\partial\dot{q}}-\frac{\partial\mathcal{L}}{\partial q}=0$
recovers~\eqref{eq:LC-original} directly.

\emph{Hamiltonian.} The canonical momentum conjugate to $q$ is $p=\partial\mathcal{L}/\partial\dot{q}=L\dot{q}$,
which is the flux linkage (or, up to the factor $L$, the current).
The Legendre transform gives the Hamiltonian 
\begin{equation}
\mathcal{H}(q,p,t)=\tfrac{p^{2}}{2L}+\tfrac{q^{2}}{2C(t)},\label{eq:Hamiltonian}
\end{equation}
the total electromagnetic energy of the circuit. Hamilton's equations
\begin{equation}
\dot{q}=\frac{\partial\mathcal{H}}{\partial p}=\frac{p}{L},\qquad\dot{p}=-\frac{\partial\mathcal{H}}{\partial q}=-\frac{q}{C(t)},\label{eq:Hamilton-eqs}
\end{equation}
form a $2\times2$ linear Hamiltonian system with $T$-periodic coefficients
($T=2\pi/\mu$). Setting $\mathbf{z}=(q,p)^{T}$ --- state vectors
are ordered (coordinate, momentum) throughout this work, following
Yakubovich--Starzhinskii~\cite[Ch.~III, \S\,3]{YakSta1} --- the
system takes the canonical form~\cite[Ch.~VIII, \S\,1.1]{YakSta2}
\begin{equation}
\tilde{\mathbf{J}}\,\dot{\mathbf{z}}=\mathbf{H}(t)\,\mathbf{z},\qquad\tilde{\mathbf{J}}=\begin{bmatrix}0 & -1\\
1 & 0
\end{bmatrix},\qquad\mathbf{H}(t)=\begin{bmatrix}C(t)^{-1} & 0\\
0 & L^{-1}
\end{bmatrix},\label{eq:canonical-form}
\end{equation}
where $\mathbf{H}(t)$ is real, symmetric, and $T$-periodic. This
is precisely the canonical form for Hill's equation of~\cite[Ch.~VIII, \S\,1.1]{YakSta2}.

In the rescaled variable $x=\tau/2$ used in Chapters~\ref{sec:MWInce}--\ref{sec:LCInce},
dots become primes ($d/dt=(\mu/2)\,d/dx$\,) and the period becomes
$\pi$. The canonical form~\eqref{eq:canonical-form} carries over
with $t$ replaced by $x$, primes for derivatives, and the Hamiltonian
matrix $\mathbf{H}(x)=\begin{bmatrix}p(x) & 0\\
0 & 1
\end{bmatrix}$ where $p(x)=\hat{\lambda}(1-\delta^{2})/(1-2\delta\cos2x+\delta^{2})$
is the coefficient of~\eqref{eq:LC-Hill}.

The monodromy matrix\index{monodromy matrix} $\mathbf{X}(T)$ (the
matrizant after one period $T=2\pi/\mu$, equivalently $\mathbf{X}(\pi)$
in the $x$-variable) is a real $2\times2$ symplectic\index{symplectic system}
matrix, $\mathbf{X}\in\mathrm{Sp}(2,\mathbb{R})$, with $\det\mathbf{X}=1$.
Its eigenvalues are the Floquet multiplier\index{Floquet theory!Floquet multiplier}s
$\rho^{\pm}$. The stability and instability of the LC circuit are
completely governed by the position of $\mathbf{X}(T)$ in the group
$\mathrm{Sp}(2,\mathbb{R})$, whose structure is analyzed in~\cite[Ch.~VIII, \S\S\,1.4--1.8]{YakSta2}:
the group decomposes into disjoint sets $\mathcal{H}$ (unstable),
$\mathcal{G}$ (stable), $\Pi^{**}$ (all solutions periodic, same-signature
coexistence), and $\Pi^{*\pm}$ (Jordan block\index{Jordan block},
opposite-signature EPD\index{exceptional point of degeneracy (EPD)}).

The symplectic structure just described yields the EPD dichotomy: 
\begin{proof}[Proof of Theorem~\ref{thm:EPD}]
\label{proof:EPD} Part~(ii): By Proposition~\ref{prop:even-vanish},
two linearly independent period-$\pi$ solutions coexist at every
even resonance. For a $2\times2$ symplectic matrix, coexistence of
two independent solutions with the same multiplier $\rho=+1$ forces
the monodromy matrix to equal $+\mathbf{I}$: a Jordan block has only
one eigenvector and cannot have two independent solutions at the same
eigenvalue. The monodromy matrix belongs to $\Pi^{**}$ of~\cite[Ch.~VIII, \S\,1.8]{YakSta2};
by Theorem~\ref{thm:Krein}(a) of Chapter~\ref{app:Krein} (\cite[Ch.~III, \S\,1.3]{YakSta1})
a definite multiple multiplier has only simple elementary divisors,
confirming the multiplier is of the same Krein\index{Krein collision theory}
kind.

Part~(i): By Proposition~\ref{prop:odd-survive}, each odd instability
interval has strictly positive width. At the boundary exactly one
periodic solution exists, so the monodromy matrix is not $\pm\mathbf{I}$.
For a $2\times2$ symplectic matrix with a double multiplier $\rho=\pm1$,
the only remaining possibility is a non-trivial Jordan block ($\Pi^{*\pm}$
of~\cite[Ch.~VIII, \S\,1.8]{YakSta2}). By Lemma~\ref{lem:Krein-Jordan}
of Chapter~\ref{app:Krein} (\cite[Lemma~III.1.1]{YakSta1}), a multiple
multiplier on the unit circle with a non-trivial Jordan block is necessarily
of mixed Krein kind (opposite signatures). By Krein's theorem (Theorem~\ref{thm:Krein};
\cite[Ch.~III, \S\,1.3]{YakSta1}, \cite[Sec.~3.3.2]{Kiril}), mixed-kind
multipliers can leave the unit circle under perturbation, which is
precisely what opens the instability tongue\index{instability tongue}. 
\end{proof}

\section{The Magnus--Winkler framework and Ince's equation}

\label{sec:MWInce}

The Magnus--Winkler\index{Magnus--Winkler theory} (MW) theory~\cite[Ch.~2--8]{MagWin}
provides the analytic backbone for computing instability boundaries
of Hill's equation\index{Hill equation}. Its central achievement
for our purposes is a complete algebraic explanation of why the LC
circuit's even instability intervals vanish exactly while odd intervals
survive --- the structural fact at the heart of this work. This chapter
develops the MW discriminant\index{discriminant} expansion, Keller's
leading-order width formula, and the MW theory of coexistence for
Ince's equation\index{Ince equation}, culminating in the exact discriminant
identity that forces even-interval collapse. See Chapter~\ref{app:Hill}
for the Floquet\index{Floquet theory} theory background. The Ince
identification is Remark~\ref{rem:ince-id}.

\subsection{Discriminant expansion}

For Hill's equation 
\begin{equation}
y''+[\lambda+Q(x)]y=0,\quad Q(x+\pi)=Q(x),\label{eq:D0a}
\end{equation}
with Fourier expansion $Q(x)=\sum_{n\neq0}g_{n}e^{2inx}$, let $y_{1}(x,\lambda)$
and $y_{2}(x,\lambda)$ be the two standard solutions satisfying $y_{1}(0)=1$,
$y_{1}'(0)=0$ and $y_{2}(0)=0$, $y_{2}'(0)=1$. The MW discriminant
\begin{equation}
\Delta(\lambda)=y_{1}(\pi,\lambda)+y_{2}'(\pi,\lambda)\label{eq:D0b}
\end{equation}
is the trace of the monodromy matrix\index{monodromy matrix} $X(\pi)$
over one half-period and satisfies $|\Delta|<2$ (stability) or $|\Delta|>2$
(instability); see Chapter~\ref{app:Hill} for a self-contained account.
MW~\cite[Corollary~2.6, p.~27--28]{MagWin} expand 
\begin{equation}
\Delta(\lambda)=\sum_{n=0}^{\infty}\Delta_{n}(\lambda),\label{eq:D0c}
\end{equation}
where $\Delta_{n}$ is homogeneous of degree $n$ in the $g_{k}$
(each term in $\Delta_{n}$ is a product of exactly $n$ Fourier coefficients).
The \emph{selection rule}~\cite[Thm.~2.6]{MagWin} states that $\Delta_{n}$
contains only monomials $g_{l_{1}}\cdots g_{l_{n}}$ with $l_{1}+\cdots+l_{n}=0$,
so $\Delta_{1}=0$ identically. The first nonzero terms are~\cite[Cor.~2.6]{MagWin}
\begin{gather}
\Delta_{0}(\lambda)=2\cos\pi\sqrt{\lambda},\quad\Delta_{1}(\lambda)=0,\quad\Delta_{2}(\lambda)=\frac{\pi\sin\pi\sqrt{\lambda}}{2\sqrt{\lambda}}\sum_{n=1}^{\infty}\frac{|g_{n}|^{2}}{\lambda-n^{2}}.\label{eq:D0}
\end{gather}

For the LC circuit with $g_{n}=\hat{\lambda}\delta^{n}$ (eq.~\eqref{eq:gn}),
the selection rule $\sum l_{i}=0$ requires positive and negative
indices to cancel. The minimum total power $\sum|l_{i}|$ subject
to $\sum l_{i}=0$, $l_{i}\neq0$ is $n$ for $n$ even (pair $l_{1}=-l_{2},\ldots$)
and $n+1$ for $n$ odd (one index is unpaired), giving: 
\begin{equation}
\Delta_{n}=O(\delta^{n})\;(n\text{ even}),\qquad\Delta_{n}=O(\delta^{n+1})\;(n\text{ odd}),\quad\delta\to0.\label{eq:Delta-scaling}
\end{equation}
The discriminant expansion therefore groups in even powers of $\delta$:
\begin{equation}
\Delta=\Delta_{0}+\underbrace{\Delta_{2}}_{O(\delta^{2})}+\underbrace{\Delta_{3}+\Delta_{4}}_{O(\delta^{4})}+\underbrace{\Delta_{5}+\Delta_{6}}_{O(\delta^{6})}+\cdots,\quad\delta\to0,\label{eq:Delta-groups}
\end{equation}
so the \emph{effective} small parameter is $\delta^{2}$.

\subsection{Keller's formula}

Truncating at $\Delta\approx\Delta_{0}+\Delta_{2}$ and expanding
$\Delta_{2}$ in a Laurent series near $\lambda=m^{2}$, Keller~\cite[p.~192]{KelInst}
derives the width of the $m$-th instability interval: 
\begin{equation}
L_{m}=2|g_{m}|+O\!\left(\frac{\varepsilon^{2}}{m^{2}}\right),\quad\varepsilon\to0,\label{eq:Keller}
\end{equation}
where $g_{n}=\varepsilon f_{n}$ are the MW Fourier coefficients (Keller's
$f_{n}$ are the coefficients of $f$ in $y''+[\lambda+\varepsilon f]y=0$,
with both $f$ and $Q$ of period $\pi$). For the LC circuit with
$g_{m}=\hat{\lambda}\delta^{m}$ this gives 
\begin{equation}
L_{1}^{\mathrm{LC}}\big|_{\text{Keller}}=\frac{8}{r^{2}}\,\delta+O\!\left(\frac{\delta^{2}}{m^{2}}\right),\quad L_{m}^{\mathrm{LC}}\big|_{\text{Keller}}=O\!\left(\frac{\delta^{2}}{m^{2}}\right),\;(m\geq2),\quad\delta\to0.\label{eq:Keller-LC}
\end{equation}
For $m\geq3$ the leading term $2|g_{m}|=O(\delta^{m})$ is dominated
by the correction $O(\varepsilon^{2}/m^{2})=O(\delta^{2}/m^{2})$
since $\delta^{m}\ll\delta^{2}$ for small $\delta$; for $m=2$ the
two are of the same order $O(\delta^{2})$, so the prediction is swamped
by its own error term in either case. Thus Keller predicts nonzero
width at \emph{all} resonance\index{resonance}s and cannot distinguish
the hierarchy $L_{m}\to0$ (even $m$) from $L_{m}>0$ (odd $m$).
For the Mathieu\index{Mathieu equation} equation $y''+(\lambda+\delta\cos2x)y=0$,
which has $g_{1}=\delta/2$ and $g_{m}=0$ for $m\geq2$, Keller gives
\begin{equation}
L_{1}^{\mathrm{Math}}=\delta+O(\delta^{2}),\qquad L_{m}^{\mathrm{Math}}=O\!\left(\frac{\delta^{2}}{m^{2}}\right)\;(m\geq2),\quad\delta\to0.\label{eq:Keller-Math}
\end{equation}
At Keller's $O(\delta^{2})$ accuracy, neither the vanishing of even
LC circuit intervals nor the thinness of higher Mathieu intervals
is visible. A deeper analysis is required. Before undertaking the
Ince-equation approach in Chapter~\ref{sec:LCInce}, we exploit the
MW expansion in Chapter~\ref{sec:UnivExp} to derive a universal
entire-function expansion for the Hill discriminant\index{discriminant!Hill discriminant}
--- a result of independent interest that applies to any Hill equation,
not only the LC circuit.

\subsection{The Magnus--Winkler theory of Ince's equation}

\label{sec:InceGeneral} \label{subsec:InceDef}

MW~\cite[Chapter~7]{MagWin} develop a complete theory of \emph{coexistence}
--- the simultaneous existence of two linearly independent solutions
sharing a Floquet multiplier\index{Floquet theory!Floquet multiplier}
value $\rho=+1$ or $\rho=-1$ --- for the four-parameter family
\begin{equation}
(1+a\cos2x)\,y''+b(\sin2x)\,y'+(c+d\cos2x)\,y=0,\qquad|a|<1,\label{eq:Ince}
\end{equation}
known as \emph{Ince's equation} (eq.~\eqref{eq:Ince-intro} in the
Introduction, eq.~\eqref{eq:Ince-MR} in the Summary of Main Results,
and eq.~\eqref{eq:IncePars} (complete parameter identification)
in Remark~\ref{rem:ince-id}).

\medskip{}
 \emph{Definition and significance of coexistence.} Recall from Chapter~\ref{app:Hill}
that Hill's equation has, at each band edge of its spectrum, exactly
one periodic or antiperiodic solution (the other independent solution
is either unbounded or quasi-periodic). The exceptional situation
in which \emph{two} linearly independent solutions of the \emph{same}
period exist simultaneously is called \emph{coexistence}.

More precisely, two linearly independent solutions of Ince's equation~\eqref{eq:Ince}
are said to \emph{coexist} if they have the same period ($\pi$ or
$2\pi$) and are both bounded. By the Floquet theory (Chapter~\ref{app:Hill},
Theorem~\ref{thm:Hill-stability}), this is equivalent to the monodromy
matrix having a \emph{scalar} multiple of the identity as its value
--- both Floquet multipliers equal $+1$ (period-$\pi$ coexistence)
or both equal $-1$ (period-$2\pi$ coexistence) --- with the monodromy
matrix actually equal to $\pm\mathbb{I}$ rather than having a non-trivial
Jordan block\index{Jordan block}. In matrix terms, coexistence is
the \emph{opposite} of the exceptional point of degeneracy\index{exceptional point of degeneracy (EPD)}
(EPD): at an EPD the monodromy matrix has a non-trivial Jordan block
with a repeated eigenvalue $\pm1$ on the unit circle, whereas at
coexistence the monodromy matrix is $\pm\mathbb{I}$ itself (a trivial
Jordan block).

\medskip{}
 \emph{Coexistence and the collapse of instability intervals.} By
the oscillation theorem~\cite[Thm.~2.1]{MagWin}: 
\begin{itemize}
\item Coexistence of period-$\pi$ solutions (Floquet multiplier $\rho=+1$)
is equivalent to a double root of $\Delta(\lambda)-2=0$, which means
both boundary curve\index{boundary curves}s of an even instability
interval \emph{coincide} --- the interval collapses to zero width. 
\item Coexistence of period-$2\pi$ solutions (Floquet multiplier $\rho=-1$)
is equivalent to a double root of $\Delta(\lambda)+2=0$, collapsing
an odd instability interval to zero width. 
\end{itemize}
Thus coexistence is the spectral mechanism by which an instability
interval disappears: instead of two distinct boundary eigenvalues
$c_{k}^{+}\neq c_{k}^{-}$ bounding an interval of positive width,
the two eigenvalues merge to a single point $c_{k}^{+}=c_{k}^{-}$
where the monodromy matrix is $\pm\mathbb{I}$. For generic Hill equations
coexistence never occurs, but for Ince's equation the special algebraic
structure of the coefficients can force it for \emph{all} values of
the modulation parameter simultaneously --- which is precisely what
happens for the even intervals of the LC circuit.

\medskip{}
 \emph{The coexistence polynomials.} The coexistence analysis is governed
by two quadratic polynomials in $\mu$ associated with the Ince parameters:
\begin{equation}
Q(\mu)=2a\mu^{2}-b\mu-\tfrac{d}{2},\qquad Q^{*}(\mu)=2Q\!\left(\mu-\tfrac{1}{2}\right)=a(2\mu-1)^{2}-b(2\mu-1)-d.\label{eq:QPoly}
\end{equation}
The shift by $\tfrac{1}{2}$ in $Q^{*}$ converts between period-$\pi$
and period-$2\pi$ characteristic values. We quote the two key theorems
from MW~\cite[Chapter~7]{MagWin}. 
\begin{prop}[Theorem 7.1 of MW, necessary condition]
\label{thm:MW71} (\cite[Thm.~7.1]{MagWin}) If Ince's equation~\eqref{eq:Ince}
has two linearly independent solutions of period $\pi$, then $Q(\mu)$
must vanish at an integer. If it has two linearly independent solutions
of period $2\pi$, then $Q^{*}(\mu)$ must vanish at an integer. 
\end{prop}

\begin{prop}[Theorem 7.6 of MW, sufficient condition]
\label{thm:MW76} (\cite[Thm.~7.6]{MagWin}) If $Q(\mu)$ has a nonneg\-ative
integer root $k_{0}$, then for all but at most $k_{0}+1$ characteristic
values of $c$ (boundary values of the spectral parameter at which
instability intervals open or close), Ince's equation has two linearly
independent solutions of period $\pi$. 
\end{prop}

The MW theory of Ince's equation serves two distinct purposes in the
analysis that follows.

\emph{Role~1: qualitative instability structure} (Sections~\ref{subsec:QPolyLC}--\ref{subsec:DeltaId}).
The coexistence polynomials $Q(\mu)$ and $Q^{*}(\mu)$ and Propositions~\ref{thm:MW71}--\ref{thm:MW76}
determine which instability intervals survive and which vanish for
all $\delta\neq0$ simultaneously. This gives the exact qualitative
picture --- Corollaries~\ref{cor:even} and~\ref{cor:odd} ---
without any computation of individual boundary values.

\emph{Role~2: quantitative location of boundary eigenvalues} (Section~\ref{subsec:Widths}).
MW Lemma~7.4~\cite[Lem.~7.4]{MagWin} provides explicit three-term
recurrence relations for the Fourier coefficients of periodic solutions
of Ince's equation. These recurrences --- equations~\eqref{eq:bulk-rec}--\eqref{eq:start-odd}
below --- are the computational engine for locating the individual
boundary eigenvalues $c_{k}^{\pm}$. Combined with the continued-fraction\index{continued fraction}
theory of Chapters~\ref{app:FDE}--\ref{app:CF}, they yield the
closed-form width formula~\eqref{eq:Lm-exact}.

Thus MW provides both the \emph{why} (coexistence forces interval
collapse) and the \emph{how} (recurrences determine exact boundary
locations). The coexistence polynomials $Q$ and $Q^{*}$ play a role
in both: they govern the qualitative structure through Propositions~\ref{thm:MW71}
and~\ref{thm:MW76}, and they determine the starter equations~\eqref{eq:start-even}--\eqref{eq:start-odd}
that distinguish even from odd periodic solutions in the quantitative
analysis.

\subsection{The discriminant identity for even resonances}

\label{subsec:DeltaId}

The MW coexistence theory gives an exact proof that every even instability
interval collapses to zero width for all $\delta\neq0$ simultaneously:
the two period-$\pi$ eigenvalues bounding the would-be interval coincide,
and at their common location the discriminant touches the stability
boundary\index{stability boundary} $\Delta=+2$ without crossing
it. This identity is the MW explanation of the even-interval collapse;
it also excludes even resonances from the continued-fraction eigenvalue
search of Chapter~\ref{sec:LCInce}.

At $\delta=0$ the $m$-th instability interval ($m$ even) degenerates
to the point $\hat{\lambda}=m^{2}$, where $\Delta_{0}=2\cos m\pi=+2$.
For $\delta>0$, Proposition~\ref{prop:even-vanish} (period-$\pi$
coexistence, from the integer root of the coexistence polynomial $Q$)
forces the two period-$\pi$ eigenvalues adjacent to $m^{2}$ to remain
\emph{equal}, at a common value $\hat{\lambda}_{m}(\delta)=m^{2}+O(\delta^{2})$
which is analytic in $\delta^{2}$ (implicit function theorem applied
to $\partial\Delta/\partial\hat{\lambda}=0$, using $\partial^{2}\Delta/\partial\hat{\lambda}^{2}(m^{2},0)=-\pi^{2}/(2m^{2})\neq0$),
and there 
\begin{equation}
\Delta\bigl(\hat{\lambda}_{m}(\delta),\delta\bigr)=+2,\qquad\Delta(\hat{\lambda},\delta)<2\ \text{ for }\ 0<|\hat{\lambda}-\hat{\lambda}_{m}(\delta)|\ \text{ small},\quad m\ \text{even},\label{eq:DeltaAtEven}
\end{equation}
for \emph{all} $0<\delta<1$: the function $\Delta-2$ has a double
zero at $\hat{\lambda}_{m}(\delta)$ and no other zeros nearby, so
there is no open instability region. We emphasize that the double
zero \emph{moves} with $\delta$: the identity holds at $\hat{\lambda}_{m}(\delta)$,
not at $\hat{\lambda}=m^{2}$ itself, where instead $\Delta(m^{2},\delta)=2-O(\delta^{4})$
--- the deviation is of fourth order because $\Delta$ is even in
$\delta$ and the double-zero structure suppresses the $O(\delta^{2})$
drift of $\hat{\lambda}_{m}(\delta)$ quadratically. The vanishing
$\Delta_{2}|_{\omega=m}=0$ (the factor $\sin m\pi$ in~\eqref{eq:D0})
is consistent with this picture at second order, but the collapse
itself is not a consequence of any finite truncation of the MW series:
it rests on the coexistence theorems of the preceding section.

\section{A universal entire-function expansion for the Hill discriminant}\index{discriminant}\index{discriminant!Hill discriminant}

\label{sec:UnivExp}

This chapter establishes a result of independent mathematical interest:
for any two-parameter Hill family, the discriminant $\Delta(\lambda,\varepsilon)$
admits an expansion whose second-order coefficient is an entire function
of $\omega=\sqrt{\lambda}$, built from a single universal family
of functions $\{\psi_{n}\}$ independent of the specific equation.
The LC circuit and the Mathieu\index{Mathieu equation} equation are
the two canonical examples; the general theory subsumes both. Readers
primarily interested in the LC circuit results may proceed directly
to Chapter~\ref{sec:LCInce}, returning here for the structural underpinning
of the discriminant method.

Throughout this chapter we use $\omega=\sqrt{\hat{\lambda}}\geq0$
as the primary spectral variable, related to the MW eigenvalue parameter
by $\hat{\lambda}=\omega^{2}$. This choice is natural: the unperturbed
discriminant $\Delta_{0}=2\cos\pi\omega$ is elementary in $\omega$,
the basis functions $\psi_{n}$ defined below are sinc-type functions
centered at integer values of $\omega$, and the Paley--Wiener\index{Paley--Wiener space}
theorem (which characterizes $L^{2}$ functions by the support of
their Fourier transforms) acts directly on the $\omega$-line. Expressions
in $\hat{\lambda}$ are obtained via $\hat{\lambda}=\omega^{2}$ whenever
needed.

The Magnus--Winkler\index{Magnus--Winkler theory} expansion of the
Hill discriminant, developed in Chapter~\ref{sec:MWInce}, does more
than provide a computational tool for the LC circuit. Its leading-order
coefficient is an entire function of $\hat{\lambda}$ --- a fact
whose proof (Chapter~\ref{app:entire}) depends only on the structure
of the expansion and not on the specific form of the perturbation.
This observation opens the door to a result of independent mathematical
interest: for \emph{any} two-parameter Hill family, the discriminant
admits an expansion whose coefficients are drawn from a single universal
family of functions $\{\psi_{n}\}$, determined entirely by $n$ and
$\omega$, and independent of the equation. We call these the \emph{$\psi$-basis}.

The $\psi$-basis functions are not merely analytically convenient.
Each $\psi_{n}$ has a precise structural meaning: it is the $n$-th
spectral component of the leading correction $\Delta_{\omega,2}$,
carrying the contribution of the $n$-th Fourier harmonic of the perturbation
$Q_{0}$ to the discriminant (eq.~\eqref{eq:Delta2-psi}). The functions
$\psi_{n}$ carry a common factor $\sin(\pi\omega)$, share a common
resonance\index{resonance} value $\psi_{n}(n)=\pi/\sqrt{2}$ for
all $n\geq1$, and are orthogonal in $L^{2}(\mathbb{R})$ with equal
norms $\|\psi_{n}\|=\pi$. Through the Fourier transform, $\{\psi_{n}\}$
is (after normalization) an orthonormal basis of the even subspace
of the Paley--Wiener space $PW_{1/2}$ of band-limited functions
--- the natural home of the discriminant, which is even in $\omega$.

\subsection{Motivation and setup}

Consider the two-parameter Hill family (normalized, see Chapter \ref{app:Hill}
and references therein) 
\begin{equation}
y''+[\lambda+\varepsilon Q_{0}(x)]\,y=0,\qquad Q_{0}(x+\pi)=Q_{0}(x),\quad\frac{1}{\pi}\int_{0}^{\pi}Q_{0}\,dx=0,\label{eq:Hill-family}
\end{equation}
with $\varepsilon\geq0$ a real amplitude parameter, $\lambda$ a
real spectral parameter and $Q_{0}\in L^{2}[0,\pi]$ a fixed zero-mean
profile. Write the Fourier expansion 
\begin{equation}
Q_{0}(x)=\sum_{n=1}^{\infty}A_{n}\cos2nx=\sum_{n\neq0}g_{n}e^{2inx},\quad g_{-n}=g_{n}=\frac{A_{n}}{2}\text{ real}.\label{eq:Q0-Fourier}
\end{equation}

\subsubsection{The discriminant and its leading expansion}

The central object of this section is 
\begin{equation}
\Delta_{\omega}(\omega,\varepsilon):=\Delta(\omega^{2},\varepsilon),\label{eq:Domega-def}
\end{equation}
the Hill discriminant rewritten as a function of $\omega=\sqrt{\lambda}\geq0$.
This is the same function $\Delta$ used throughout, composed with
$\lambda=\omega^{2}$; we use the subscript $\omega$ to make the
variable explicit and avoid confusion with $\Delta(\lambda)$. The
MW expansion~\cite[Cor.~2.6]{MagWin} gives 
\begin{equation}
\Delta_{\omega}(\omega,\varepsilon)=\Delta_{\omega,0}+\Delta_{\omega,2}(\omega,\varepsilon)+O(\varepsilon^{4}),\quad\varepsilon\to0,\label{eq:Delta-split}
\end{equation}
where the two leading terms are 
\begin{gather}
\Delta_{\omega,0}(\omega)=2\cos\pi\omega,\quad\Delta_{\omega,2}(\omega,\varepsilon)=\frac{\pi\varepsilon^{2}\sin\pi\omega}{2\omega}\sum_{n=1}^{\infty}\frac{g_{n}^{2}}{\omega^{2}-n^{2}}.\label{eq:Delta0-omega}
\end{gather}

These two functions drive the entire agenda of the section.

\emph{The free discriminant $\Delta_{0}$.} The function $\Delta_{\omega,0}(\omega)=2\cos\pi\omega$
is the discriminant of the unperturbed equation ($\varepsilon=0$).
It oscillates between $-2$ and $+2$, crossing $\Delta_{0}=+2$ at
every even integer $\omega=2k$ and $\Delta_{0}=-2$ at every odd
integer $\omega=2k-1$. These crossings mark the resonance frequencies:
the $n$-th instability gap of the perturbed equation opens near $\omega=n$.
The formula is elementary in $\omega$ and reflects the half-period
$\pi$ of the equation.

\emph{The leading correction $\Delta_{2}$.} The function $\Delta_{\omega,2}(\omega,\varepsilon)$
is the $O(\varepsilon^{2})$ perturbation of the discriminant. It
is $90^{\circ}$ out of phase with $\Delta_{\omega,0}$: where $\Delta_{\omega,0}$
is driven by $\cos\pi\omega$, $\Delta_{\omega,2}$ is driven by $\sin\pi\omega$.
This phase shift has a precise consequence: near a resonance $\omega=n$
where $\cos\pi n=(-1)^{n}$, the perturbation $\Delta_{\omega,2}$
is the leading term that shifts $\Delta_{\omega}$ away from $\pm2$
and thereby opens or closes the instability gap. The partial-fraction
sum $\sum_{n=1}^{\infty}g_{n}^{2}/(\omega^{2}-n^{2})$ controls how
much each Fourier harmonic $g_{n}$ of the perturbation $Q_{0}$ contributes
to the gap at each resonance $\omega=n$.

\emph{From $\Delta_{2}$ to the $\psi$-basis.} The expression~\eqref{eq:Delta0-omega}
already suggests a natural decomposition: write each term of the partial-fraction
sum as a function of $\omega$ in its own right by absorbing the common
factor $\sin\pi\omega/\omega$. Specifically, for each $n\geq1$,
define 
\begin{equation}
\Delta_{\omega,2}^{(n)}(\omega):=\frac{\pi\sin\pi\omega}{2\omega(\omega^{2}-n^{2})},\label{eq:Delta2-term}
\end{equation}
so that $\Delta_{\omega,2}(\omega,\varepsilon)=\varepsilon^{2}\sum_{n=1}^{\infty}g_{n}^{2}\,\Delta_{\omega,2}^{(n)}(\omega)$.
Each $\Delta_{\omega,2}^{(n)}$ is a meromorphic function with an
apparent simple pole at $\omega=n$, but since $\sin\pi n=0$ for
integer $n$, the pole is in fact removable --- the function is entire.
Up to a normalizing constant, $\Delta_{\omega,2}^{(n)}$ is precisely
the $n$-th element of the $\psi$-basis defined in Subsection~\ref{subsec:psi-basis}:
the relationship is 
\begin{equation}
\Delta_{\omega,2}^{(n)}(\omega)=\frac{(-1)^{n}\pi}{2\sqrt{2}\,\omega^{2}}\,\psi_{n}(\omega).\label{eq:Delta2-psi}
\end{equation}
The $\psi_{n}$ are therefore the \emph{natural building blocks of
$\Delta_{\omega,2}$}: each $\psi_{n}(\omega)$ isolates the contribution
of the $n$-th Fourier harmonic of $Q_{0}$ to the discriminant, and
$\Delta_{\omega,2}$ is recovered by $\Delta_{\omega,2}=\varepsilon^{2}\sum_{n}g_{n}^{2}\Delta_{\omega,2}^{(n)}$.
The entire section is the development of this observation.

\subsection{\texorpdfstring{The universal $\psi$-basis}{The universal psi-basis}}

\label{subsec:psi-basis} 
\begin{defn}[Universal $\psi$-basis]
\label{def:psi-basis} For $n=0,1,2,\ldots$, define 
\begin{equation}
\psi_{n}(\omega)=\frac{\sqrt{2}\,\omega\sin[\pi(\omega-n)]}{\omega^{2}-n^{2}}=\frac{(-1)^{n}\sqrt{2}\,\omega\sin(\pi\omega)}{\omega^{2}-n^{2}},\qquad\omega\in\mathbb{R},\label{eq:psi-def}
\end{equation}
where the identity $\sin[\pi(\omega-n)]=(-1)^{n}\sin(\pi\omega)$
holds for all integers $n\geq0$, and the apparent singularity at
$\omega=n$ is removable by L'Hôpital's rule (see below). 
\end{defn}

The second form in~\eqref{eq:psi-def} exhibits the key structure:
every $\psi_{n}$ shares the common factor $(-1)^{n}\sin(\pi\omega)$,
which vanishes at all integers, while the denominator $\omega^{2}-n^{2}$
introduces a compensating zero at $\omega=\pm n$ via L'Hôpital's
rule. This cancellation is what makes $\psi_{n}$ entire on $\mathbb{R}$
and gives it its sinc-like character.

\emph{Even parity.} Both forms in~\eqref{eq:psi-def} satisfy $\psi_{n}(-\omega)=\psi_{n}(\omega)$
(each is an even function of $\omega$). Extending the definition
to all integers via $\psi_{-n}:=\psi_{n}$ is consistent with the
formula and explains why the Shannon--Whittaker sampling basis uses
$n\in\mathbb{Z}$ while we use only $n\geq0$.

\emph{Resonance value.} At $\omega=n$ ($n\geq1$), L'Hôpital's rule
applied to the second form gives 
\begin{equation}
\psi_{n}(n)=\frac{\pi}{\sqrt{2}},\qquad n=1,2,3,\ldots\label{eq:psi-resonance}
\end{equation}
The resonance value is \emph{the same for all $n$}, a striking uniformity
that reflects the common factor $\sin(\pi\omega)$.

\emph{Zeros.} The zeros of $\psi_{n}$ are at all non-negative integers
$\omega=k$ with $k\neq n$: 
\begin{equation}
\psi_{n}(k)=0\text{ for }k=0,1,2,\ldots,\;k\neq n;\qquad\psi_{n}(n)=\frac{\pi}{\sqrt{2}},\qquad n=1,2,3,\ldots\label{eq:psi-zeros}
\end{equation}
In $\omega$-space the zeros are at all non-negative integers except
$\omega=n$, with spacing $\Delta\omega=1$.

\emph{Amplitude bound.} The standard inequality $|\sin x/x|\leq1/\max(1,|x|)$
(tight at $x\to0$ and at $|x|=\pi/2+k\pi$ respectively) applied
to $\operatorname{sinc}(\omega-n)=\sin\pi(\omega-n)/\pi(\omega-n)$
gives the two-regime bound 
\begin{equation}
|\psi_{n}(\omega)|\leq\frac{\sqrt{2}\,\omega}{\omega+n}\cdot\min\!\left(\pi,\;\frac{1}{|\omega-n|}\right),\qquad\omega\geq0,\quad n=1,2,3,\ldots\label{eq:psi-envelope}
\end{equation}
Near resonance ($|\omega-n|\leq1/\pi$) the bound reduces to $\sqrt{2}\pi\omega/(\omega+n)$,
which is tight at $\omega=n$ where both sides equal $\pi/\sqrt{2}$
(the common resonance value~\eqref{eq:psi-resonance}). Away from
resonance ($|\omega-n|\geq1/\pi$) the bound decays as $\sqrt{2}\omega/((\omega+n)|\omega-n|)$,
capturing the $1/|\omega-n|$ decay of the sinc oscillations. Since
$\psi_{n}(\omega)\to0$ as $\omega\to\infty$ (the numerator $|\sin\pi\omega|\leq1$
while the denominator grows as $\omega^{2}$), $\psi_{n}\in L^{2}(\mathbb{R})$.

\emph{Sinc representation.} The second form of~\eqref{eq:psi-def}
can be written as 
\begin{equation}
\psi_{n}(\omega)=\frac{\pi\sqrt{2}}{2}\bigl[\operatorname{sinc}(\omega-n)+\operatorname{sinc}(\omega+n)\bigr],\quad n\geq1,\label{eq:psi-sinc}
\end{equation}
where $\operatorname{sinc}(x)=\sin(\pi x)/(\pi x)$ is the normalized
sinc~\cite[Ch.~1]{Zayed1993}. Thus $\psi_{n}$ is a sum of sincs
centered at $+n$ and $-n$, symmetric in $\omega$, which is consistent
with extending to $\psi_{-n}=\psi_{n}$ if needed for the Shannon--Whittaker
context.

\begin{figure}[htbp]
\centering \includegraphics[width=1\textwidth]{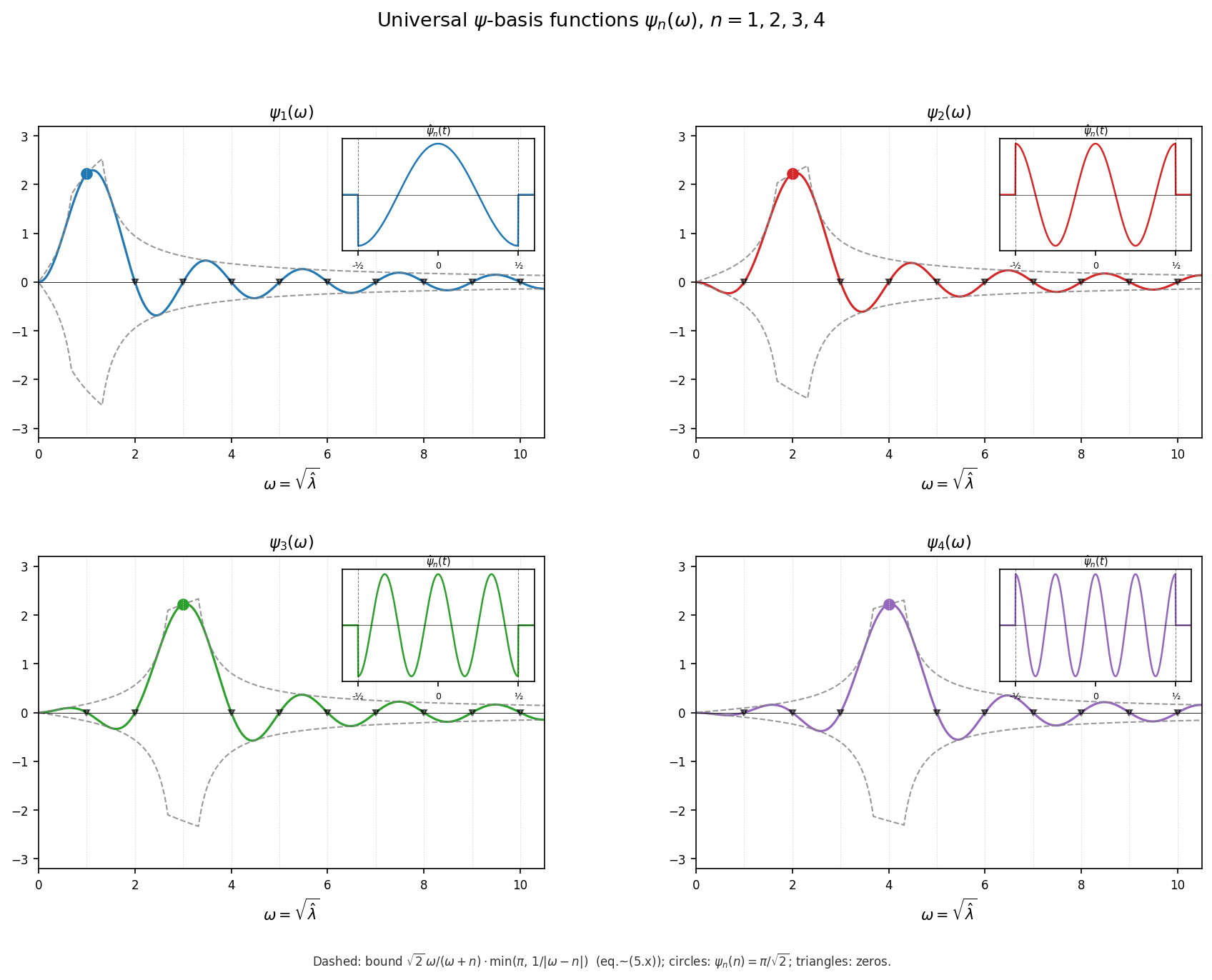}
\caption{The first four universal $\psi$-basis functions $\psi_{n}(\omega)$,
$n=1,2,3,4$, plotted against $\omega=\sqrt{\hat{\lambda}}$ using
the simplified form~\eqref{eq:psi-def}: $\psi_{n}(\omega)=(-1)^{n}\sqrt{2}\,\omega\sin(\pi\omega)/(\omega^{2}-n^{2})$.
Dotted curves: amplitude bound $\sqrt{2}\,\omega/(\omega+n)\cdot\min(\pi,\,1/|\omega-n|)$
(eq.~\eqref{eq:psi-envelope}), tight at $\omega=n$ where both the
bound and $\psi_{n}(n)$ equal $\pi/\sqrt{2}$ (filled circles). Downward
triangles mark the common zeros at integer $\omega=k\protect\neq n$
(eq.~\eqref{eq:psi-zeros}); vertical dotted lines are guides. \emph{Inset:}
Fourier transform $\widehat{\psi}_{n}(t)=\pi\sqrt{2}\cos(2\pi nt)\cdot\mathbf{1}_{[-1/2,1/2]}(t)$
(eq.~\eqref{eq:psi-ft}), showing the Paley--Wiener band-limitation
to $[-\tfrac{1}{2},\tfrac{1}{2}]$. All $\psi_{n}$ share the same
resonance value $\pi/\sqrt{2}$ and the same asymptotic amplitude
bound $\sqrt{2}$, reflecting the common factor $\sin(\pi\omega)$.}
\label{fig:psi-basis} 
\end{figure}

Figure~\ref{fig:psi-basis} illustrates $\psi_{1},\ldots,\psi_{4}$
together with their amplitude bounds and Fourier transforms.

\subsection{Fourier transform, orthogonality, and Paley--Wiener structure}

\label{subsec:psi-PW} 
\begin{thm}[Fourier transform of $\psi_{n}$]
\label{thm:psi-ft} For $n=0,1,2,\ldots$, the Fourier transform
of $\psi_{n}$ is 
\begin{equation}
\widehat{\psi}_{n}(t)=\pi\sqrt{2}\,\cos(2\pi nt)\cdot\mathbf{1}_{[-1/2,1/2]}(t).\label{eq:psi-ft}
\end{equation}
In particular, every $\psi_{n}$ has Fourier support contained in
$[-\tfrac{1}{2},\tfrac{1}{2}]$. 
\end{thm}

\begin{proof}
From the definition~\eqref{eq:psi-def} and the standard Fourier
pairs $\operatorname{sinc}(\omega\mp n)\leftrightarrow e^{\pm2\pi int}\mathbf{1}_{[-1/2,1/2]}(t)$,
together with the sinc representation~\eqref{eq:psi-sinc} ($n\geq1$):
\[
\widehat{\psi}_{n}(t)=\frac{\pi\sqrt{2}}{2}\bigl(e^{-2\pi int}+e^{2\pi int}\bigr)\mathbf{1}_{[-1/2,1/2]}(t)=\pi\sqrt{2}\cos(2\pi nt)\cdot\mathbf{1}_{[-1/2,1/2]}(t).
\]
For $n=0$: $\psi_{0}(\omega)=\pi\sqrt{2}\operatorname{sinc}(\omega)$
with $\widehat{\operatorname{sinc}}(t)=\mathbf{1}_{[-1/2,1/2]}(t)$,
giving $\widehat{\psi}_{0}(t)=\pi\sqrt{2}\cdot\mathbf{1}_{[-1/2,1/2]}(t)$,
consistent with $\cos(0)=1$. 
\end{proof}
\begin{thm}[Orthogonality and norms]
\label{thm:psi-ortho} The family $\{\psi_{n}\}_{n\geq0}$ satisfies
\begin{gather}
\int_{-\infty}^{\infty}\psi_{n}(\omega)\,\psi_{m}(\omega)\,d\omega=0,\qquad n\neq m,\quad n,m\geq0,\label{eq:psi-ortho}\\
\int_{-\infty}^{\infty}\psi_{n}^{2}(\omega)\,d\omega=\pi^{2},\quad n\geq1;\qquad\int_{-\infty}^{\infty}\psi_{0}^{2}(\omega)\,d\omega=2\pi^{2}.\label{eq:psi-norms}
\end{gather}
\end{thm}

\begin{proof}
By Parseval's theorem (Plancherel identity,~\cite[Ch.~7]{Rudin1991})
and Theorem~\ref{thm:psi-ft}: 
\[
\langle\psi_{n},\psi_{m}\rangle=\langle\widehat{\psi}_{n},\widehat{\psi}_{m}\rangle=2\pi^{2}\int_{-1/2}^{1/2}\cos(2\pi nt)\cos(2\pi mt)\,dt.
\]
For $n\neq m$, the product-to-sum formula gives cosines at non-zero
integer multiples of $2\pi$, all of which integrate to zero over
$[-\tfrac{1}{2},\tfrac{1}{2}]$, yielding $\langle\psi_{n},\psi_{m}\rangle=0$.
For $n=m\geq1$: $\cos^{2}(2\pi nt)=\tfrac{1}{2}[1+\cos(4\pi nt)]$
integrates to $\tfrac{1}{2}$, so $\|\psi_{n}\|^{2}=2\pi^{2}\cdot\tfrac{1}{2}=\pi^{2}$.
For $n=0$: $\cos(0)=1$ integrates to $1$, giving $\|\psi_{0}\|^{2}=2\pi^{2}$. 
\end{proof}
The normalized family $\tilde{\psi}_{n}:=\psi_{n}/\|\psi_{n}\|$ is
orthonormal: 
\begin{equation}
\tilde{\psi}_{n}=\begin{cases}
\psi_{0}/(\pi\sqrt{2}) & n=0,\\[4pt]
\psi_{n}/\pi & n\geq1.
\end{cases}\label{eq:psi-normalized}
\end{equation}

\begin{thm}[Paley--Wiener basis property]
\label{thm:psi-PW} Let 
\[
PW_{1/2}:=\{f\in L^{2}(\mathbb{R}):\operatorname{supp}(\hat{f})\subseteq[-\tfrac{1}{2},\tfrac{1}{2}]\}
\]
be the Paley--Wiener space of $L^{2}$ functions band-limited to
$[-\tfrac{1}{2},\tfrac{1}{2}]$ (the subscript $\tfrac{1}{2}$ refers
to the half-bandwidth)~\cite[Ch.~IV]{PW1934}, and let 
\[
PW_{1/2}^{{\rm even}}:=\{f\in PW_{1/2}:f(-\omega)=f(\omega)\}
\]
be its subspace of even functions. Then $\{\tilde{\psi}_{n}\}_{n\geq0}$
is an \emph{orthonormal basis} of $PW_{1/2}^{{\rm even}}$. 
\end{thm}

\begin{proof}
Orthonormality is established by Theorem~\ref{thm:psi-ortho}, and
each $\tilde{\psi}_{n}$ lies in $PW_{1/2}^{{\rm even}}$ since $\psi_{n}$
is even with band-limited transform (Theorem~\ref{thm:psi-ft}).
For completeness, let $f\in PW_{1/2}^{{\rm even}}$, so $g:=\hat{f}\big|_{[-1/2,1/2]}\in L^{2}([-\tfrac{1}{2},\tfrac{1}{2}])$
is an \emph{even} function ($f$ is even if and only if $\hat{f}$
is even). The trigonometric system $\{e^{2\pi int}\}_{n\in\mathbb{Z}}$
is complete on $L^{2}([-\tfrac{1}{2},\tfrac{1}{2}])$~\cite[Vol.~I, Ch.~I]{Zygmund};
splitting it into even and odd parts shows that the cosine system
$\{1,\cos(2\pi nt)\}_{n\geq1}$ is complete on the subspace of even
functions, while the sine system spans the odd ones. Hence $g$ has
a convergent cosine expansion, and pulling back via the inverse Fourier
transform yields an $L^{2}(\mathbb{R})$-convergent expansion of $f$
in terms of $\tilde{\psi}_{n}$. The restriction to the even subspace
is essential: an odd function such as the inverse transform of $\sin(2\pi t)\,\mathbf{1}_{[-1/2,1/2]}(t)$
lies in $PW_{1/2}$ and is orthogonal to every $\psi_{n}$. 
\end{proof}
\begin{rem}[Connection to Shannon--Whittaker sampling]
\label{rem:psi-Shannon} The standard orthonormal basis of $PW_{1/2}$
in sampling theory is $\{\operatorname{sinc}(\omega-n)\}_{n\in\mathbb{Z}}$,
corresponding to the complex exponential basis $\{e^{2\pi int}\}$
on $[-\tfrac{1}{2},\tfrac{1}{2}]$~\cite[Ch.~1]{Zayed1993}. The
$\psi$-basis uses the \emph{cosine} basis instead: since $\cos(2\pi nt)=\tfrac{1}{2}(e^{2\pi int}+e^{-2\pi int})$,
the $+n$ and $-n$ frequencies are merged into a single real function.
Merging $+n$ and $-n$ halves the count: $\{\psi_{n}\}_{n\geq0}$
--- indexed over non-negative integers only --- is a complete orthogonal
family for the \emph{even} subspace $PW_{1/2}^{{\rm even}}$ (Theorem~\ref{thm:psi-PW});
the odd subspace is spanned by the corresponding sine combinations,
which do not arise here because the discriminant is even in $\omega$.
Extending to $\psi_{-n}:=\psi_{n}$ for $n>0$ is consistent with
the formula but introduces redundancy. 
\end{rem}

\subsection{The polynomial expansion theorem}
\begin{thm}[Universal discriminant expansion]
\label{thm:univ-exp} For the Hill family~\eqref{eq:Hill-family},
\begin{equation}
\Delta_{\omega}(\omega,\varepsilon)=2\cos\pi\omega+\frac{\pi\varepsilon^{2}\sin\pi\omega}{2\omega}\sum_{n=1}^{\infty}\frac{g_{n}^{2}}{\omega^{2}-n^{2}}+O(\varepsilon^{4}),\quad\varepsilon\to0,\label{eq:Delta-univ}
\end{equation}
uniformly on compact subsets of $\omega\in(0,\infty)$. The correction
$\Delta_{\omega,2}(\omega,\varepsilon)$ is an entire function of
$\omega$ (hence of $\hat{\lambda}=\omega^{2}$), being a uniformly
convergent series of entire functions~(Chapter~\ref{app:entire}). 
\end{thm}

\begin{proof}
Substitute $Q=\varepsilon Q_{0}$ into MW Corollary~2.6~\cite[Cor.~2.6]{MagWin}.
The Fourier coefficients of $Q$ are $\varepsilon g_{n}$, so \eqref{eq:Delta-univ}
follows directly from eqs.~\eqref{eq:Delta-split}--\eqref{eq:Delta0-omega}.
Entirety of each term in $\Delta_{\omega,2}$, despite the apparent
poles at $\omega^{2}=n^{2}$, is shown in Chapter~\ref{app:entire};
uniform convergence on compacta follows from $\sum g_{n}^{2}<\infty$
for $Q_{0}\in L^{2}$. 
\end{proof}
\begin{rem}[The $\psi$-basis rewriting of $\Delta_{\omega,2}$]
\label{rem:psi-signs} Using eq.~\eqref{eq:Delta2-psi}, each term
of the MW sum~\eqref{eq:Delta-univ} can be expressed through $\psi_{n}$:
\[
\Delta_{\omega,2}^{(n)}(\omega)=\frac{(-1)^{n}\pi}{2\sqrt{2}\,\omega^{2}}\,\psi_{n}(\omega),
\]
so that 
\begin{equation}
\Delta_{\omega}=2\cos\pi\omega+\frac{\pi\varepsilon^{2}}{2\sqrt{2}\,\omega^{2}}\sum_{n=1}^{\infty}(-1)^{n}g_{n}^{2}\,\psi_{n}(\omega)+O(\varepsilon^{4}),\quad\varepsilon\to0.\label{eq:Delta-univ-alt}
\end{equation}
This form exhibits $\psi_{n}$ as the spectral components of $\Delta_{\omega,2}$,
with an $\omega^{2}$ denominator reflecting the change of variable
$\lambda=\omega^{2}$. The poles at $\omega=n$ in $\psi_{n}/\omega^{2}$
are removable: $\psi_{n}(\omega)/\omega^{2}$ is entire. The apparent
singularity at $\omega=0$ is likewise removable: since $\psi_{n}(\omega)=(-1)^{n}\sqrt{2}\,\omega\sin(\pi\omega)/(\omega^{2}-n^{2})$
has a factor $\omega\sin(\pi\omega)\sim\pi\omega^{2}$ as $\omega\to0$,
one finds $\psi_{n}(\omega)/\omega^{2}\to-(-1)^{n}\sqrt{2}\,\pi/n^{2}$
(finite and nonzero), so the entire sum $\Delta_{\omega,2}^{{\rm alt}}$
is analytic at $\omega=0$. 
\end{rem}

\begin{rem}[Higher-order terms and pole orders]
\label{rem:h-ord-appr} At every order $\varepsilon^{2k}$ the correction
$\Delta_{\omega,2k}$ is an entire function of $\omega$, so every
truncation of the expansion is a polynomial in $\varepsilon$ with
entire coefficients, as guaranteed by MW Theorem~2.2~\cite[Thm.~2.2]{MagWin}.
The form of these coefficients is given by MW Theorem~2.5~\cite[Thm.~2.5]{MagWin}:
each coefficient $c(l_{1},\ldots,l_{2k})$ has the form $A(\omega)\cos\pi\omega+B(\omega)\sin\pi\omega$,
where $A,B$ are even rational functions of $\omega$ with poles at
some of the points $\omega=0$ and $\omega=\pm(l_{r}+l_{r+1}+\cdots+l_{s})$,
$1\leq r\leq s\leq2k$. For $\Delta_{\omega,2}$ (Corollary~2.6~\cite[Cor.~2.6]{MagWin})
the poles are simple at $\omega=\pm n$ and are canceled exactly by
$\sin\pi\omega$, as shown in Chapter~\ref{app:entire}. Whether
all higher-order poles are likewise simple remains to be verified. 
\end{rem}

\subsection{Boundary curves from the universal expansion}

Setting $\Delta_{\omega}(\omega,\varepsilon)=\pm2$ near $\omega=m$
and using $2\cos\pi m=2(-1)^{m}$, the boundary equation reduces to
a polynomial equation in $\varepsilon$ with coefficients involving
$\psi_{n}(m)$. 
\begin{prop}[Leading-order boundary curve\index{boundary curves}s, Keller]
\label{prop:bdry-leading} (\cite[p.~192]{KelInst}) For the family~\eqref{eq:Hill-family},
the $m$-th instability zone\index{instability zone} boundaries satisfy
\begin{equation}
\lambda_{m}^{\pm}(\varepsilon)=m^{2}\pm|g_{m}|\varepsilon+O(\varepsilon^{2}),\quad\varepsilon\to0.\label{eq:bdry-leading}
\end{equation}
with instability zone width $\lambda_{m}^{+}-\lambda_{m}^{-}=2|g_{m}|\varepsilon+O(\varepsilon^{2})$;
cf.~eq.~\eqref{eq:bdry-main} of Chapter~\ref{sec:MainResults}. 
\end{prop}

\begin{proof}
The formula is due to Keller~\cite[p.~192]{KelInst}; we give an
alternative argument derived from the universal expansion~\eqref{eq:Delta-univ}.
Write $\omega=m+u$ with $u=h/(2m)$, so that $\lambda=\omega^{2}=m^{2}+h+O(h^{2})$
($h$ measures the shift in $\lambda$ from $m^{2}$). Near $\omega=m$,
\[
2\cos\pi\omega-2(-1)^{m}=-(-1)^{m}\pi^{2}u^{2}+O(u^{4}).
\]
In the sum of~\eqref{eq:Delta-univ} evaluated at $\omega=m$, every
term with $n\neq m$ carries the factor $\sin\pi m=0$, while L'Hôpital
applied to the $n=m$ term gives $\Delta_{\omega,2}^{(m)}(m)=(-1)^{m}\pi^{2}/(4m^{2})$
(from eq.~\eqref{eq:Delta2-psi}: $(-1)^{m}\pi/(2\sqrt{2}m^{2})\cdot\psi_{m}(m)=(-1)^{m}\pi^{2}/(4m^{2})$).
The boundary equation $\Delta_{\omega}-2(-1)^{m}=0$ therefore reads
\[
(-1)^{m}\frac{\pi^{2}}{4m^{2}}\bigl(g_{m}^{2}\varepsilon^{2}-h^{2}\bigr)+O(h^{4},h\varepsilon^{2},\varepsilon^{4})=0,
\]
giving $h=\pm|g_{m}|\varepsilon+O(\varepsilon^{2})$ for all $m\geq1$. 
\end{proof}
Proposition~\ref{prop:bdry-leading} recovers Keller's formula~\eqref{eq:Keller}
from first principles. The resonance value $\psi_{n}(n)=\pi/\sqrt{2}$
(eq.~\eqref{eq:psi-resonance}) plays the central role: it is the
common factor at every instability threshold.

\subsection{The two canonical examples}

\emph{Mathieu equation} ($Q_{0}(x)=2q_{0}\cos2x$, $g_{1}=q_{0}$,
$g_{n\geq2}=0$, $\varepsilon=1$). The universal expansion~\eqref{eq:Delta-univ}
reduces to the single-term formula~\eqref{eq:Delta-Math}. Equivalently,
from~\eqref{eq:Delta-univ-alt}: 
\[
\Delta_{\omega}^{{\rm Math}}=2\cos\pi\omega+\frac{\pi q_{0}^{2}\sin(\pi\omega)}{2\omega(\omega^{2}-1)}+O(q_{0}^{4}).
\]
Only the $m=1$ zone opens at $O(\varepsilon^{2})$; all higher zones
require $O(g^{2m})$ corrections from multi-harmonic mixing.

\emph{LC circuit} ($A_{n}=2\hat{\lambda}\delta^{n}$, $g_{n}=\hat{\lambda}\delta^{n}=\omega^{2}\delta^{n}$,
$\varepsilon=1$). Using~\eqref{eq:Delta-univ-alt}: 
\begin{equation}
\Delta_{\omega}^{{\rm LC}}=2\cos\pi\omega+\frac{\pi\omega^{4}\delta^{2}\sin(\pi\omega)}{2\omega}\sum_{n=1}^{\infty}\frac{\delta^{2(n-1)}}{\omega^{2}-n^{2}}+O(\delta^{4}),\quad\delta\to0.\label{eq:Delta-LC-univ}
\end{equation}
which is precisely the polynomial expansion~\eqref{eq:Delta-LC-poly}
of Chapter~\ref{app:entire}, now seen as an instance of the universal
theorem. The coexistence of even zones (Floquet\index{Floquet theory}
multiplier\index{Floquet theory!Floquet multiplier} $\rho=+1$, two
independent period-$\pi$ solutions) --- established by the Ince
theory of Chapter~\ref{sec:MWInce} (the discriminant identity~\eqref{eq:DeltaAtEven})
--- forces $\Delta_{\omega}$ to be tangent to $+2$ at the nearby
point $\omega_{m}(\delta)=m+O(\delta^{2})$ rather than crossing it;
this collapse is exact in $\delta$ and is not visible at any fixed
order of the expansion~\eqref{eq:Delta-LC-univ}.

\begin{figure}[htbp]
\centering \includegraphics[width=12cm]{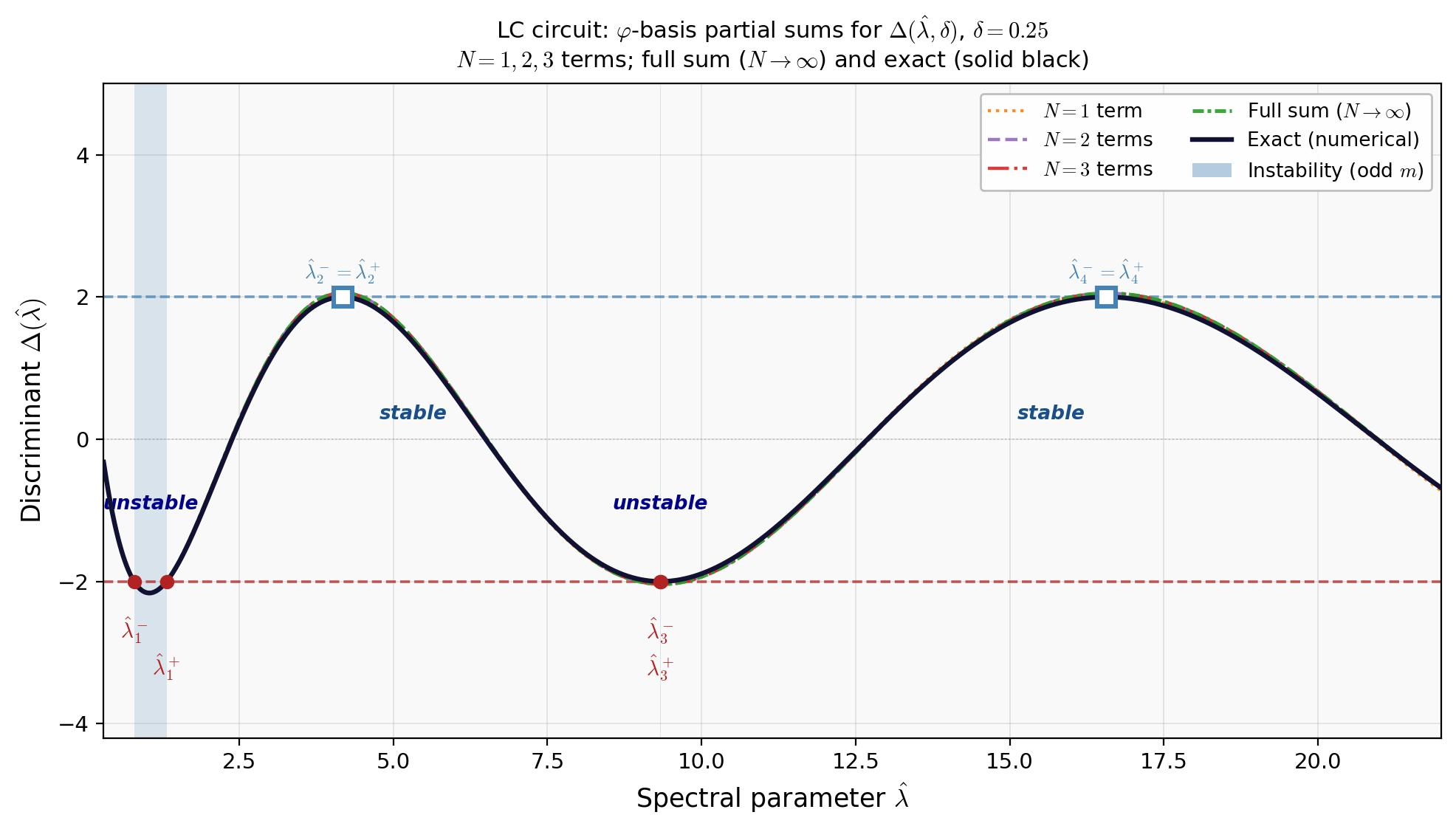} \caption{Partial-sum approximations of the LC circuit discriminant $\Delta_{\omega}(\omega,\delta)$
at $\delta=0.25$ using the $\psi$-basis expansion~\eqref{eq:Delta-LC-univ}.
The exact discriminant (solid black) is computed numerically by integrating
the LC Hill equation over one half-period with a high-order ODE solver
(see caption of Figure~\ref{fig:LC-discriminant-poly-zoom} and Table~\ref{tab:partial-sum-convergence}
for details). Successive truncations at $O(\delta^{2})$ (dotted orange),
$O(\delta^{4})$ (dashed purple), $O(\delta^{6})$ (dash-dot red),
and $O(\delta^{8})$ (long-dash green) converge rapidly to the exact
discriminant. Blue shading marks the odd instability zones; open squares
mark the even resonances $\omega=2,4$; the discriminant does not
cross $+2$ there, touching it only at the nearby tangency points
$\omega_{m}(\delta)=m+O(\delta^{2})$ (coexistence, eq.~\eqref{eq:DeltaAtEven}).}
\label{fig:LC-discriminant-poly} 
\end{figure}

\begin{figure}[htbp]
\centering \includegraphics[width=1\textwidth]{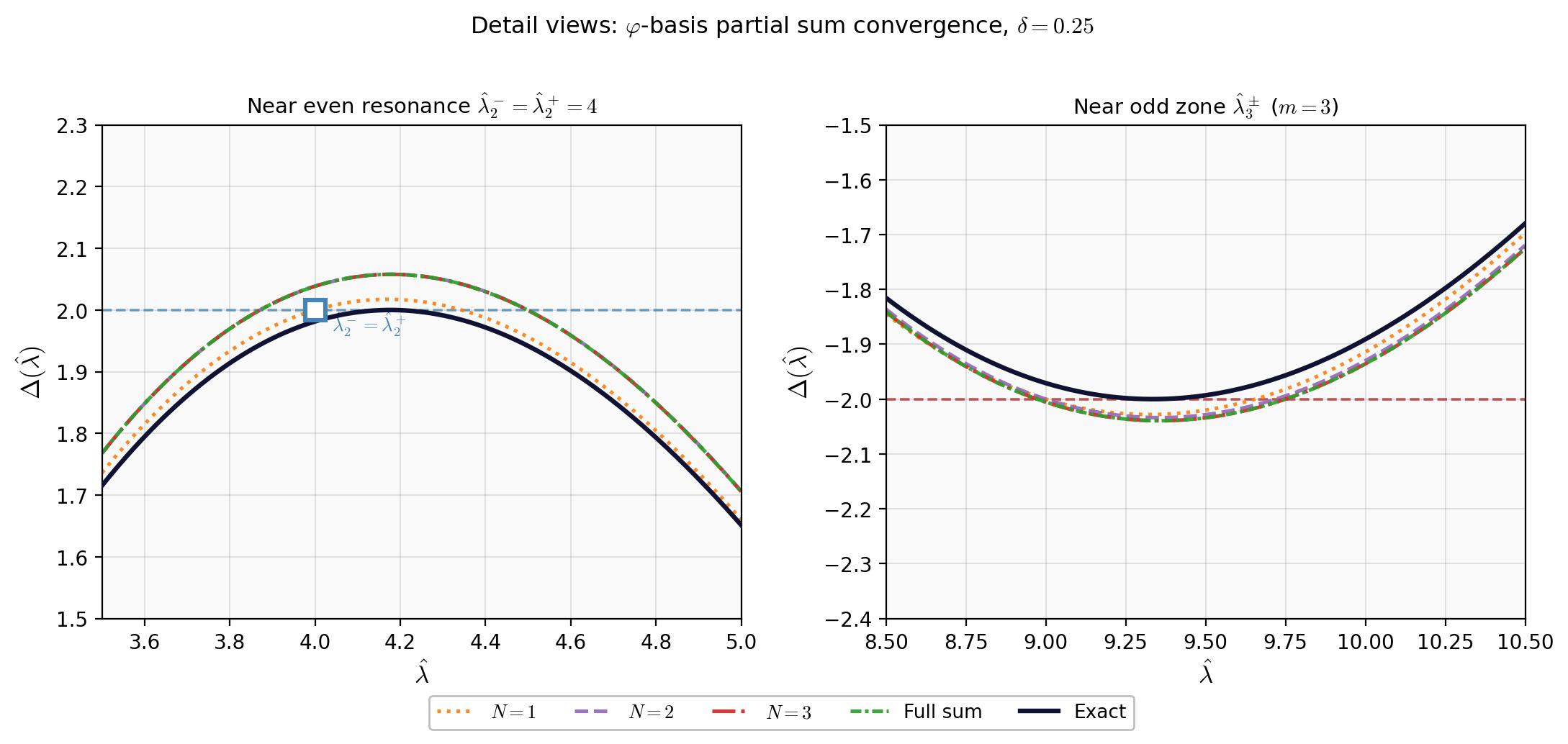}
\caption{Detail views of partial-sum convergence for the LC circuit discriminant,
$\delta=0.25$. The exact numerical curve is computed by integrating
the two standard solutions $y_{1}$ and $y_{2}$ of~\eqref{eq:Hill-MW}
(eqs.~\eqref{eq:Volterra-y1}--\eqref{eq:Volterra-y2}) over one
half-period $x\in[0,\pi]$ using an eighth-order Runge--Kutta--Dormand--Prince
solver, then evaluating $\Delta=y_{1}(\pi)+y_{2}'(\pi)$. \emph{Left}:
near the even resonance $\omega=2$. \emph{Right}: near the third
instability zone ($m=3$, $\omega\approx3.05$). Non-monotone convergence
reflects the sign alternation of the $\psi$-basis across resonances
(see Remark~\ref{rem:psi-signs}).}
\label{fig:LC-discriminant-poly-zoom} 
\end{figure}

The partial-sum convergence of the $\psi$-basis expansion for the
LC case is illustrated in Figures~\ref{fig:LC-discriminant-poly}
and~\ref{fig:LC-discriminant-poly-zoom}.

\begin{table}[ht]
\centering \caption{Convergence of the $\psi$-basis partial sums to $\Delta_{\omega,2}$,
and the gap between $\Delta_{\omega,2}$ and the exact discriminant
$\Delta$, at $\delta=0.25$. The partial sums $\Delta_{\omega}^{(N)}=2\cos\pi\omega+\frac{\pi\omega^{4}\delta^{2}\sin\pi\omega}{2\omega}\sum_{n=1}^{N}\frac{\delta^{2(n-1)}}{\omega^{2}-n^{2}}$
(using the MW form~\eqref{eq:Delta-univ} truncated at $N$ terms)
converge to the leading-order MW approximation $\Delta_{\omega,2}$
(full sum, $N\to\infty$). The residual $\Delta_{\omega,{\rm exact}}-\Delta_{2}$
is the $\Delta_{\omega,4}+\cdots$ correction.}
\label{tab:partial-sum-convergence} \resizebox{\textwidth}{!}{%
\begin{tabular}{p{3cm}rrrr}
\toprule 
 & \multicolumn{2}{c}{$\omega=2.05$ (near even resonance)} & \multicolumn{2}{c}{$\omega=3.05$ (near odd zone $m=3$)}\tabularnewline
\cmidrule{1-1}\cmidrule(lr){2-3}\cmidrule(lr){4-5}
$N$  & $\Delta_{\omega}^{(N)}$  & error vs $\Delta_{\omega,2}$  & $\Delta_{\omega}^{(N)}$  & error vs $\Delta_{\omega,2}$\tabularnewline
\midrule 
$1$  & $2.01678$  & $-0.04070$  & $-2.02798$  & $+0.01122$\tabularnewline
$2$  & $2.05759$  & $+0.00011$  & $-2.03357$  & $+0.00563$\tabularnewline
$3$  & $2.05749$  & $+0.00000$  & $-2.03922$  & $-0.00002$\tabularnewline
$\infty$  & $2.05748$  & $0$  & $-2.03920$  & $0$\tabularnewline
\midrule 
\multicolumn{5}{l}{\emph{Higher-order correction (independent of $N$):}}\tabularnewline
$\Delta_{\omega,{\rm exact}}$ (ODE, $10^{-12}$ tol.)  & \multicolumn{2}{c}{$1.99968$} & \multicolumn{2}{c}{$-2.00000$}\tabularnewline
Gap: $\Delta_{\omega,{\rm exact}}-\Delta_{\omega,2}$  & \multicolumn{2}{c}{$-0.05780$} & \multicolumn{2}{c}{$+0.03920$}\tabularnewline
\bottomrule
\end{tabular}} 
\end{table}

\subsection{Connection to the continued-fraction analysis}

The universal expansion and the continued-fraction\index{continued fraction}
(CF) method of Chapters~\ref{sec:LCInce}--\ref{sec:Boundaries}
are complementary. The CF method is exact and valid for all $0<\delta<1$;
it gives the boundary curves as solutions of a transcendental equation.
The universal expansion is perturbative in $\delta$ but gives \emph{explicit
analytic formulas} order by order, with coefficients determined by
the universal resonance value $\psi_{n}(n)=\pi/\sqrt{2}$ (eq.~\eqref{eq:psi-resonance}).
The two methods agree to every computed order, as verified in Table~\ref{tab:lh-bdry}
of Chapter~\ref{app:entire}.

\subsection{\texorpdfstring{The Volterra integral equations and the origin of
the $\psi$-basis}{The Volterra integral equations and the origin
of the psi-basis}}

The $\psi$-basis functions arise naturally from the integral-equation
form of Hill's equation\index{Hill equation}. The two standard solutions
$y_{1}$ and $y_{2}$ satisfy the Volterra integral equations of the
second kind~\cite[Sec.~4.3, eqs.~(4.3.1)--(4.3.2)]{Eastham73}, \cite[Sec.~III.8, eq.~(8.6)]{Hale}:
\begin{align}
y_{1}(s,\lambda) & =\cos\sqrt{\lambda}\,s+\frac{\varepsilon}{\sqrt{\lambda}}\int_{0}^{s}Q_{0}(t)\,y_{1}(t,\lambda)\sin\!\left[\sqrt{\lambda}(s-t)\right]dt,\label{eq:Volterra-y1}\\
y_{2}(s,\lambda) & =\frac{\sin\sqrt{\lambda}\,s}{\sqrt{\lambda}}+\frac{\varepsilon}{\sqrt{\lambda}}\int_{0}^{s}Q_{0}(t)\,y_{2}(t,\lambda)\sin\!\left[\sqrt{\lambda}(s-t)\right]dt.\label{eq:Volterra-y2}
\end{align}
Both equations share the Volterra kernel 
\begin{equation}
\mathcal{K}(s,t;\lambda)=\frac{\sin\!\left[\sqrt{\lambda}(s-t)\right]}{\sqrt{\lambda}},\qquad0\leq t\leq s,\label{eq:Volterra-kernel}
\end{equation}
which is the causal Green's function of $d^{2}/ds^{2}+\lambda$. These
equations provide the most efficient numerical method for computing
$\Delta_{\omega}$: integrate eqs.~\eqref{eq:Volterra-y1}--\eqref{eq:Volterra-y2}
over one half-period and evaluate $\Delta=y_{1}(\pi)+y_{2}'(\pi)$.

\subsection{Self-adjoint structure and open questions}

The common factor $\sin(\pi\omega)$ in the $\psi$-basis satisfies
the formally self-adjoint Sturm--Liouville\index{Sturm--Liouville theory}
equation \cite[Sec.~10.1]{AbrSte} 
\begin{equation}
-\frac{d}{d\hat{\lambda}}\!\left(\sqrt{\hat{\lambda}}\,S'\right)=\frac{\pi^{2}}{4}\cdot\frac{S}{\sqrt{\hat{\lambda}}},\label{eq:SL-formal}
\end{equation}
i.e., $S=\sin\pi\omega$ is the eigenfunction of the SL operator $\mathcal{L}_{{\rm SL}}[y]=-(\sqrt{\hat{\lambda}}\,y')'$
with weight $w=1/\sqrt{\hat{\lambda}}$ and eigenvalue $\mu=\pi^{2}/4$.
Each $\psi_{n}$ is $\sqrt{2}/\omega$ times a rational modification
of $S$: 
\[
\psi_{n}(\omega)=\frac{(-1)^{n}\sqrt{2}\,\omega}{\omega^{2}-n^{2}}\cdot S(\omega),
\]
and can be interpreted as (a multiple of) the \emph{resolvent} of
$\mathcal{L}_{{\rm SL}}$ at $\mu=\pi^{2}/4$, evaluated at the pole
$\hat{\lambda}=n^{2}$: 
\begin{equation}
\psi_{n}(\omega)\propto\omega\bigl(\mathcal{L}_{{\rm SL}}-\tfrac{\pi^{2}}{4}\bigr)^{-1}[\delta(\cdot-n^{2})](\hat{\lambda}).\label{eq:psi-resolvent}
\end{equation}
The Paley--Wiener basis property (Theorem~\ref{thm:psi-PW}) and
this resolvent interpretation suggest the following open question:
does there exist a natural integral operator $\mathcal{K}$ on functions
of $\omega$ such that $(\mathcal{K}\,\psi_{n})(\omega)=\sigma_{n}\,\psi_{n}(\omega)$
for some $\sigma_{n}$, making the $\psi_{n}$ into literal eigenfunctions?
The Paley--Wiener structure and the connection to the cosine basis
on $[-\tfrac{1}{2},\tfrac{1}{2}]$ both point toward a band-limiting
operator of prolate-spheroidal type (Slepian--Pollak--Landau theory~\cite[Ch.~4]{Zayed1993}).
We leave this for future investigation.

%% ===========================================================================

\section{Cambi's continued-fraction method for the Floquet theory}

\label{sec:Cambi-method}

\emph{Goals of this chapter.} Cambi's 1950 paper~\cite[\S\S\,1--15]{Cambi2}
develops an elegant continued-fraction\index{continued fraction}
method for the Floquet\index{Floquet theory} theory of equation~\eqref{eq:LC-original-MR}.
Several key steps are stated without proof or rest on assumptions
deserving mathematical justification. The goals of this chapter are: 
\begin{enumerate}
\item[(i)] Present Cambi's significant results faithfully, marking steps requiring
clarification. 
\item[(ii)] Fill mathematical gaps --- justify $A_{n}=V(u+n)$ from Floquet
theory, and establish the equivalence of Cambi's resonance\index{resonance}
equation with Hill's determinantal equation via Pincherle\index{continued fraction!Pincherle theorem}'s
theorem (Theorem~\ref{thm:Pinch}). 
\item[(iii)] Prepare the bridge to the MW framework (Chapter~\ref{sec:MWInce})
and our developments in Chapter~\ref{sec:expanding-Cambi}. 
\end{enumerate}
Throughout this chapter we use Cambi's original notation; the translation
to ours is eq.~\eqref{eq:Cambi-res2}.

\emph{Organization.} This chapter presents Cambi's own results: the
equation and Floquet ansatz (\S\,\ref{subsec:Cambi-eq}), the resonance
equation and its root structure (\S\,\ref{subsec:Cambi-res-main}),
asymptotic formulas for small $\gamma$ (\S\,\ref{subsec:Cambi-asymp-sec}),
the stability boundaries in the $(p,\gamma)$ plane (\S\,\ref{subsec:Cambi-bdy}),
and coexistence of two periodic solutions at even resonances (\S\,\ref{subsec:Cambi-coex}).
Our contributions built on Cambi's framework --- the minimal solution\index{continued fraction!minimal solution}
theory, asymptotically seeded approximants, the minimal solution ratio
$w(u)$, and arbitrary-precision Floquet exponent\index{Floquet theory!Floquet exponent}
computation --- are developed in Chapter~\ref{sec:expanding-Cambi}.

% =============================================================================

% -----------------------------------------------------------------------------

\subsection{Cambi's method and results}

\label{subsec:Cambi-results}

% --- subsubsection 1 ---

\subsubsection{The equation and the Floquet ansatz}

\label{subsec:Cambi-eq}

Cambi writes the LC circuit equation as~\cite[eq.~(2')]{Cambi2}
\begin{equation}
(1+2\gamma\cos x)\frac{d^{2}y}{dx^{2}}+p^{2}y=0,\qquad0<2\gamma<1,\quad p>0,\label{eq:Cambi-eq}
\end{equation}
where $p=\omega_{0}/\mu$ is the resonance-to-modulation frequency
ratio and $2\gamma=\varepsilon$ is the relative modulation. By Floquet's
theorem, eq.~\eqref{eq:Cambi-eq} has solutions~\cite[eq.~(3)]{Cambi2}
\begin{equation}
y=e^{iu_{0}x}\sum_{n=-\infty}^{\infty}B_{n}e^{inx},\label{eq:Cambi-ansatz}
\end{equation}
where $u_{0}$ is the Floquet exponent, defined modulo integers. Substituting
and collecting terms at frequency $e^{i(u+n)x}$ gives the three-term
system~\cite[eq.~(4)]{Cambi2}: 
\begin{equation}
\gamma A_{n+1}+G(u+n)A_{n}+\gamma A_{n-1}=0,\qquad n\in\mathbb{Z},\label{eq:Cambi-system}
\end{equation}
where 
\begin{equation}
A_{n}=\frac{(u+n)^{2}}{p^{2}}B_{n},\qquad G(u)=1-\frac{p^{2}}{u^{2}}.\label{eq:Cambi-AG}
\end{equation}

% --- subsubsection 2 ---

\subsubsection{\texorpdfstring{Cambi's ansatz, the function $v(u)$, and the resonance
equation}{Cambi's ansatz, the function v(u), and the resonance equation}}

\label{subsec:Cambi-Vun} \label{subsec:Cambi-res-main}

Cambi proposes~\cite[\S\,3]{Cambi2}: \emph{``{[}I{]}t is logical
to regard $A_{n}$ as a function of $u+n$, let us say $A_{n}=V(u+n)$''}
--- calling it logical without proof. We establish it rigorously
in Part~II (\S\,\ref{subsec:Cambi-notes2}).

Using the ratio $A_{1}/A_{0}=V(u+1)/V(u)$ and the symmetry $A_{-1}/A_{0}=V(u-1)/V(u)$,
Cambi~\cite[\S\,3,\,5]{Cambi2} shows that both ratios can be expressed
in terms of a single function $v(u)$~\cite[\S\,3]{Cambi2}, defined
by the continued fraction: 
\begin{equation}
v(u)\;=\;\cfrac{1}{G(u)-\cfrac{\gamma^{2}}{G(u+1)-\cfrac{\gamma^{2}}{G(u+2)-\cdots}}},\label{eq:Cambi-v-CF-part1}
\end{equation}
giving~\cite[eq.~(6)]{Cambi2}: 
\begin{equation}
\frac{V(u)}{V(u+1)}=-\gamma\,v(-u),\qquad\frac{V(u)}{V(u-1)}=-\gamma\,v(u).\label{eq:Cambi-ratios-part1}
\end{equation}
(The rigorous treatment of $v(u)$ is given in Section~\ref{subsec:Cambi-v-props}.)
For the solution with rightward-decaying tail, $A_{1}/A_{0}=-\gamma v(u+1)$;
for the leftward-decaying tail, $A_{-1}/A_{0}=-\gamma v(1-u)$ (by
the evenness of $G$). Requiring both at once --- a bilateral minimal
solution --- the central equation ($n=0$) of~\eqref{eq:Cambi-system}
then gives \emph{Cambi's resonance equation}~\cite[eq.~(7)]{Cambi2}:
\begin{equation}
-\gamma^{2}v(1-u)+G(u)-\gamma^{2}v(1+u)=0.\label{eq:Cambi-res7}
\end{equation}
Via the recurrence $\gamma^{2}v(u+1)=G(u)-1/v(u)$ (eq.~\eqref{eq:Cambi-rec-v},
derived in Section~\ref{subsec:Cambi-v-props}) this is equivalent
to~\cite[eq.~(7')]{Cambi2}: 
\begin{equation}
\gamma^{2}v(u)v(1-u)=1.\label{eq:Cambi-res7p}
\end{equation}

\begin{rem}[Equivalence to Hill's determinant --- gap in Cambi]
Cambi states~\cite[\S\,5]{Cambi2}: \emph{``Equation (7) replaces
Hill's determinantal equation, to which it may be shown to be equivalent''}
but gives no proof or indication of argument. We establish this in
Section~\ref{subsec:Cambi-CF}: by Pincherle's theorem, eq.~\eqref{eq:Cambi-res7}
is the condition for~\eqref{eq:Cambi-diff} to have a minimal solution,
equivalent to $\Delta(\hat{\lambda})=2\cos2\pi u$ (eq.~\eqref{eq:Delta-cos-u}). 
\end{rem}

\emph{All roots.} Cambi states~\cite[\S\,6]{Cambi2} that if $u_{0}$
solves~\eqref{eq:Cambi-res7}, so does $\pm(u_{0}\pm n)$ for any
$n\in\mathbb{Z}$. All give the same Floquet multiplier\index{Floquet theory!Floquet multiplier};
only two independent Floquet solutions exist. The proof is given in
Part~II (Theorem~\ref{thm:notes2-root-symmetry}).

\emph{Central root.} At $\gamma=0$: $G(u)=0$ gives $u=\pm p$. For
$\gamma>0$ the root near $p$ is easily located since the LHS of~\eqref{eq:Cambi-res7}
is nearly linear there (the $\gamma^{2}v$ terms being small corrections
to $G(u)$).

% --- subsubsection 3 ---

\subsubsection{\texorpdfstring{Asymptotic formulas for small $\gamma$}{Asymptotic
formulas for small gamma}}

\label{subsec:Cambi-asymp-sec}

\emph{Approximate central root}~\cite[\S\,8]{Cambi2}: 
\begin{equation}
u_{0}\approx p\Bigl\{1+\gamma^{2}\frac{3p^{2}-1}{4p^{2}-1}\Bigr\},\quad\text{error }O(\gamma^{4}),\quad p\neq\tfrac{1}{2}.\label{eq:Cambi-u0-approx}
\end{equation}

The series for $u^{*}$ is extended to all orders in $\gamma^{2}$
with closed-form coefficients and an explicit recursion in Section~\ref{subsec:Cambi-notes2-precision}.

\emph{Coefficient ratios}~\cite[eqs.~(9),(10)]{Cambi2}: 
\begin{align}
\frac{A_{\pm n}}{A_{\pm(n-1)}} & \approx-\gamma\frac{(n\pm u_{0})^{2}}{(n\pm u_{0})^{2}-p^{2}},\label{eq:Cambi-ratio9}\\[4pt]
\frac{A_{\pm n}}{A_{\pm(n-1)}} & \approx-\gamma\frac{(n\pm p)^{2}}{n(n\pm2p)},\label{eq:Cambi-ratio10}
\end{align}
where~\eqref{eq:Cambi-ratio10} uses $u_{0}\approx p$. As $n\to\infty$
the exact ratios converge to $-(1-\sqrt{1-4\gamma^{2}})/2\gamma$
(the minimal-solution fixed point); the leading-order forms~\eqref{eq:Cambi-ratio9}--\eqref{eq:Cambi-ratio10}
reproduce its first term~$-\gamma$.

\emph{Explicit coefficient formulas}~\cite[\S\,8]{Cambi2}: 
\begin{align}
A_{\pm n} & \approx(-\gamma)^{n}\frac{(n\pm u_{0})!^{2}}{(\pm u_{0})^{2}}\cdot\frac{(p\pm u_{0})(-p\pm u_{0})}{(n+p\pm u_{0})!\,(n-p\pm u_{0})!},\label{eq:Cambi-An9}\\
A_{\pm n} & \approx(-\gamma)^{n}\frac{(n\pm p)!^{2}}{(\pm p)^{2}}\cdot\frac{(\pm2p)}{n!\,(n\pm2p)!},\label{eq:Cambi-An10}
\end{align}
with $x!:=\Gamma(x+1)$. Formula~\eqref{eq:Cambi-An10} uses $u_{0}\approx p$.

% --- subsubsection 4 ---

\subsubsection{\texorpdfstring{Stability boundaries in the $(p,\gamma)$ plane}{Stability
boundaries in the (p,gamma) plane}}

\label{subsec:Cambi-bdy}

Since $\rho=e^{\pm2\pi iu_{0}}$, stability requires real $u_{0}$.
Boundaries occur at~\cite[\S\,11]{Cambi2}: 
\begin{itemize}
\item $u_{0}=n$ ($\rho=+1$), resonance eq.\ at $u=0$~\cite[eq.~(11)]{Cambi2}:
\begin{equation}
1-p^{2}-\cfrac{\gamma^{2}}{1-p^{2}/4-\cfrac{\gamma^{2}}{1-p^{2}/9-\cdots}}=0.\label{eq:Cambi-bdy11}
\end{equation}
\item $u_{0}=n+\frac{1}{2}$ ($\rho=-1$), resonance eq.\ at $u=\frac{1}{2}$~\cite[eq.~(12)]{Cambi2}:
\begin{equation}
1\pm\gamma-4p^{2}-\cfrac{\gamma^{2}}{1-4p^{2}/9-\cfrac{\gamma^{2}}{1-4p^{2}/25-\cdots}}=0.\label{eq:Cambi-bdy12}
\end{equation}
\end{itemize}
The two signs in~\eqref{eq:Cambi-bdy12} give two families bounding
each instability tongue\index{instability tongue}; eq.~\eqref{eq:Cambi-bdy11}
gives a single family bounding no instability region.
\begin{rem}[Mathematical gap]
Cambi asserts without proof that curves~\eqref{eq:Cambi-bdy11}
bound no instability. The proof uses Krein\index{Krein collision theory}
signature\index{Krein signature} theory: at $u_{0}=n$ ($\rho=+1$)
the colliding multipliers have equal Krein signatures, so no instability
region can open (Chapter~\ref{sec:EPD-elementary}, Theorem~\ref{thm:EPD}). 
\end{rem}

% --- subsubsection 5 ---

\subsubsection{Two periodic solutions at even resonances: contrast with Mathieu}\index{Mathieu equation}

\label{subsec:Cambi-coex}

At boundaries $u_{0}=n+\frac{1}{2}$ ($\rho=-1$): exactly \emph{one}
periodic solution of period $4\pi$~\cite[\S\,14]{Cambi2}. At boundaries
$u_{0}=n$ ($\rho=+1$)~\cite[\S\,15]{Cambi2}:
\begin{prop}[Coexistence at even resonances]
\label{prop:Cambi-coex} (\cite[\S\,15]{Cambi2}) On curves $u_{0}=n$,
eq.~\eqref{eq:Cambi-eq} has exactly \emph{two} independent periodic
solutions of period $2\pi$ simultaneously --- in fundamental contrast
to Mathieu's equation, for which coexistence is impossible (Ince's
theorem). 
\end{prop}

Cambi proves this via the Wronskian argument Ince used for Mathieu.
The crucial difference: the necessary condition for coexistence in
eq.~\eqref{eq:Cambi-eq} (Cambi's eq.~(13)) does not contain $a_{0}$
(the constant term $B_{0}$ in the Floquet ansatz~\eqref{eq:Cambi-ansatz}),
making it compatible with convergence of the CF; for Mathieu's equation
the analogous condition contains $a_{0}$ and is incompatible.

This is the Cambi-level formulation of our Theorem~\ref{thm:EPD}:
at even resonances $X(\pi)=+\mathbf{I}$ in the MW period-$\pi$ normalization
(not a Jordan block\index{Jordan block}), so two independent Floquet
solutions share multiplier $\rho=+1$.
\begin{rem}[Mathematical gaps in Cambi's treatment of $V(u)$]
\label{rem:Cambi-gaps} Several aspects of Cambi's construction of
$V(u)$ lack rigorous foundation.

\emph{No constructive definition.} Cambi introduces $V(u)$ by the
declaration ``{[}I{]}t is logical to regard $A_{n}$ as a function
of $u+n$, let us say $A_{n}=V(u+n)$''~\cite[\S\,3]{Cambi2}, with
no definition of $V$ as a function of continuous $u$. His actual
computation~\cite[\S\,3]{Cambi2} sets $V(u)=A_{0}$ (an arbitrary
constant) at the base point and propagates forward by the ratios $-\gamma v(u+n)$.
This yields only discrete values $\{V(u+n)\}_{n\in\mathbb{Z}}$, not
a function of $u$.

\emph{The arbitrary normalization does not yield a meromorphic $V(u)$.}
Cambi explicitly acknowledges~\cite[\S\,3]{Cambi2}: \emph{``Assuming
an arbitrary initial determination of $V(u)$ in a unit interval\ldots\
The resulting functions are not as a rule analytic, but may, for example,
have discontinuities at all points which are congruent with the limits
of the initial interval.''} Thus Cambi's own construction yields
a function that is generically non-meromorphic (and indeed non-analytic).

\emph{Existence without construction.} Cambi adds~\cite[\S\,3]{Cambi2}:
\emph{``Analytic solutions of (5) may actually be obtained, but their
determination is unnecessary for our purpose, since only the values
at congruent arguments are required.''} He thus acknowledges that
a meromorphic $V(u)$ exists but declines to construct it.

\emph{Resonance equation without rigorous foundation.} Since $V(u)$
is not rigorously defined as a continuous function, Cambi's derivation
of the resonance equation~\eqref{eq:Cambi-res7} via the matching
condition $V(u+1)/V(u)=-\gamma v(u+1)$ lacks a rigorous basis.

All four gaps are closed in Part~II: $v(u)$ is defined rigorously
as a meromorphic CF (Theorem~\ref{thm:v-CF}); the minimal solution
$\{h_{m}(u)\}$ is constructed via Pincherle's theorem (Theorem~\ref{thm:Pinch});
the resonance equation is derived from the bilateral decay condition
(eq.~\eqref{eq:notes2-resonance}); and the meromorphic generator
$H(u)$ --- the analytic solution Cambi deems unnecessary --- is
constructed by the Miller algorithm\index{Miller algorithm} (\S\,\ref{subsec:Cambi-notes2-H}). 
\end{rem}

% =============================================================================
% PART II: OUR RIGOROUS CONTRIBUTIONS
% =============================================================================

% -----------------------------------------------------------------------------

\section{Expanding Cambi's method}

\label{sec:expanding-Cambi}

\emph{A note on the level of sophistication.} The continued-fraction\index{continued fraction}
(CF) framework used throughout this chapter is more than the elementary
infinite fraction familiar from introductory analysis. It is a representation
theory for the \emph{minimal} solutions of three-term linear recurrences,
equipped with its own convergence theorems and analytic structure.
The reader will encounter the distinction between minimal and dominant
solutions (Chapter~\ref{app:FDE}), Pincherle\index{continued fraction!Pincherle theorem}'s
theorem identifying the minimal solution\index{continued fraction!minimal solution}
with the value of a CF (Chapter~\ref{app:FDE}), CF convergence theory
(Chapter~\ref{app:CF}), and the Miller algorithm\index{Miller algorithm}
for stable numerical evaluation (Chapter~\ref{app:Miller}). A central
new construction is the \emph{double minimality condition} (\S\,\ref{subsec:Cambi-notes2-res}),
which selects the Floquet\index{Floquet theory} exponent by forcing
minimality in both directions $n\to\pm\infty$. These tools give the
meromorphic structure, arbitrary-precision Floquet exponent\index{Floquet theory!Floquet exponent}s,
and analytical access to instability boundaries developed in the subsequent
sections.

\medskip{}

\emph{Main results of this chapter.} The two physical deliverables
of this chapter are: 
\begin{enumerate}
\item[(a)] \emph{Boundary curves of the primary instability tongue\index{instability tongue}}
as Taylor series in the modulation amplitude (Fig.~\ref{fig:tongue-comparison}).
The result is established in \S\,\ref{subsec:CF-stability-bdry}
(Theorem~\ref{thm:bdry-series-CF}), drawing on the double minimality
equation (\S\,\ref{subsec:Cambi-notes2-double-min}, Theorem~\ref{thm:double-min}),
the closed-form $\gamma^{2}$-series for the primary Floquet exponent
(\S\,\ref{subsec:Cambi-notes2-precision}, Theorem~\ref{thm:precision-cost}),
and the Floquet-comb zero set of $F_{w}$ (\S\,\ref{subsec:Fw-meromorphic},
Theorem~\ref{thm:Fw-zero-set}). The result is announced in Chapter~\ref{sec:MainResults}
as Theorem~\ref{thm:primary-domain-CF}.
\item[(b)] \emph{Explicit Fourier coefficients of the periodic factor in the
Floquet decomposition} (Figs.~\ref{fig:Bm-p03}, \ref{fig:Bm-Cambi},
\ref{fig:Bm-Cambi2}). The result is established in \S\,\ref{subsec:Cambi-notes2-H}
(Theorem~\ref{thm:notes2-H-new}), drawing on the meromorphic structure
of the minimal solution ratio $w(u)$ (\S\,\ref{subsec:Cambi-w-structure},
Theorems~\ref{thm:notes2-w-poles-zeros}, \ref{thm:notes2-w-rate})
and the meromorphic structure of the generator $H(u)$ (\S\,\ref{subsec:H-meromorphic},
Theorems~\ref{thm:notes2-H-new-poles-zeros}, \ref{thm:notes2-H-new-orders}).
The result is announced in Chapter~\ref{sec:MainResults} as Theorem~\ref{thm:Floquet-factor-MR}. 
\end{enumerate}
\medskip{}

\emph{The two central objects.} Every construction in this chapter
rests on one function and a second built from it. In the stability
zone the Floquet recurrence $\gamma A_{n+1}+G(u+n)A_{n}+\gamma A_{n-1}=0$,
with $G(u)=1-p^{2}/u^{2}$, has a solution that decays as $n\to+\infty$,
unique up to scale; the \emph{minimal-solution ratio} $w(u)$ \emph{selects}
it --- it is the ratio $A_{n+1}/A_{n}$ of that one decaying solution,
singled out from the two-parameter family of all solutions --- and
obeys the Riccati recurrence 
\begin{equation}
w(u+1)\;=\;-\frac{G(u+1)}{\gamma}-\frac{1}{w(u)}.\label{eq:intro-w-Riccati}
\end{equation}
It has an explicit continued-fraction form (\S\,\ref{subsec:Cambi-notes2})
and a real meromorphic structure with simple poles and zeros (\S\S\,\ref{subsec:Cambi-w-structure}--\ref{subsec:w-simplicity-notes}).
Cambi's function $v(u)$ is a simple rescaling of $w$ (\S\,\ref{subsec:Cambi-v-props})
and enters the development only through corollaries of the $w$-theory.

The Floquet exponent is fixed by requiring minimality in \emph{both}
directions $n\to\pm\infty$ at once. This double minimality condition
is carried by the \emph{double minimality function} 
\begin{equation}
F_{w}(u)\;:=\;w(-u)-\frac{1}{w(u-1)},\label{eq:intro-Fw-def}
\end{equation}
introduced in \S\,\ref{subsec:Cambi-notes2-res}. Its real zeros
are exactly the Floquet comb $Z_{F}=\{\pm u^{*}+n\}$ (\S\,\ref{subsec:Fw-meromorphic}),
so the zeros of $F_{w}$ \emph{are} the Floquet data; its poles ---
including the stealthy ones hugging the comb zeros --- are located
by the sign-change theorem of \S\,\ref{subsec:Fw-pole-detection}.
The two deliverables of the chapter then follow: the boundary curves
from the zeros of $F_{w}$ (\S\,\ref{subsec:CF-stability-bdry}),
and the Floquet factor from $w$ through the generator $H(u)$ (\S\,\ref{subsec:Cambi-notes2-H}).

\medskip{}

This chapter develops our contributions built on Cambi's framework,
organized in five movements followed by a summary.

\emph{(A) Cambi's $v(u)$ and classical connections} (\S\S\,\ref{subsec:Cambi-v-props}--\ref{subsec:Cambi-MW}):
the minimal-solution theory and meromorphic properties of $v(u)$
(\S\,\ref{subsec:Cambi-v-props}), asymptotically seeded high-precision
approximants (\S\,\ref{subsec:Cambi-approx}), and the connection
to the Magnus--Winkler\index{Magnus--Winkler theory} framework (\S\,\ref{subsec:Cambi-MW}).

\emph{(B) The ratio $w(u)$, double minimality, and the Floquet exponent}
(\S\S\,\ref{subsec:Cambi-notes2}--\ref{subsec:stealthy-zeros}):
the direct continued-fraction approach via Pincherle's theorem, which
introduces $w(u)$ (\S\,\ref{subsec:Cambi-notes2}); the double
minimality condition and Floquet frequency comb (\S\,\ref{subsec:Cambi-notes2-res});
the double minimality equation (\S\,\ref{subsec:Cambi-notes2-double-min});
the main new computational result --- arbitrary-precision Floquet
exponents via a closed-form $\gamma^{2}$-series with an explicit
recursion for all coefficients, together with the continued-fraction
numerical conventions used throughout this work (\S\,\ref{subsec:Cambi-notes2-precision},
Remark~\ref{rem:CF-numerical-conventions}); and the high-precision
verification of all Floquet zeros in a representative window, including
the \emph{stealthy} zeros that double precision misses entirely (\S\,\ref{subsec:stealthy-zeros}).

\emph{(C) The analytic structure of $w(u)$} (\S\S\,\ref{subsec:Cambi-w-structure}--\ref{subsec:notes-on-v}):
the structural theorems for $w(u)$ (\S\,\ref{subsec:Cambi-w-structure});
the discrete Riccati structure (\S\,\ref{subsec:Riccati-structure});
the simplicity of the poles and zeros of $w(u)$ (\S\,\ref{subsec:w-simplicity-notes});
and the corresponding poles, zeros, and residues of $v(u)$ together
with the residue-distance law (\S\,\ref{subsec:notes-on-v}).

\emph{(D) The double minimality function $F_{w}$} (\S\S\,\ref{subsec:Fw-meromorphic}--\ref{subsec:Casoratian-Weyl}):
the meromorphic structure of $F_{w}$ (\S\,\ref{subsec:Fw-meromorphic});
the sign-change theorem that locates the poles of $F_{w}$ --- including
the \emph{stealthy} poles hiding beside Floquet zeros --- from a
sign comparison at two explicit points (\S\,\ref{subsec:Fw-pole-detection});
the discrete Volterra and Jost statements imported for the pole-ladder
analysis (\S\,\ref{subsec:imported-Jacobi}); and the Casoratian
identity with the Weyl--Stieltjes view of the pole ladders (\S\,\ref{subsec:Casoratian-Weyl}).

\emph{(E) The Floquet factor and the deliverables} (\S\S\,\ref{subsec:Cambi-notes2-H}--\ref{subsec:CF-stability-bdry}):
the construction of the minimal solution generator $H(u)$ (\S\,\ref{subsec:Cambi-notes2-H}),
its meromorphic structure (\S\,\ref{subsec:H-meromorphic}), and
the Taylor-series boundary curve\index{boundary curves}s for the
first instability tongue (\S\,\ref{subsec:CF-stability-bdry}).
A closing summary (\S\,\ref{subsec:w-Fw-summary}) collects the
analytic properties of $w(u)$ and $F_{w}$ established along the
way.

\medskip{}

\emph{Standing assumption: stability zone.} \label{asm:stability-zone}
Throughout this entire chapter --- and all its sections --- the
parameters $(p,\gamma)$ are assumed to lie in the \emph{stability
zone}, meaning the LC circuit is not in parametric resonance\index{resonance}\index{parametric resonance}.
In the notation of Section~\ref{subsec:MR-stability} (eq.~\eqref{eq:stability-intervals}),
this means $(c,\delta)$ with $c=4p^{2}$ and $\delta=|\zeta_{+}|$
lies strictly outside all instability tongues: 
\begin{equation}
c\;<\;c_{-}^{(m)}(\delta)\quad\text{or}\quad c\;>\;c_{+}^{(m)}(\delta)\qquad\text{for all odd }m=1,3,5,\ldots\;,\label{eq:stability-assumption}
\end{equation}
where $c_{\pm}^{(m)}(\delta)$ are the EPD\index{exceptional point of degeneracy (EPD)}
curve\index{exceptional point of degeneracy (EPD)!EPD curve}s (Theorem~\ref{thm:EPD-bdry},
eq.~\eqref{eq:EPD-curves-def}). \emph{Equivalently}, the Floquet
exponent $u^{*}$ is \emph{real} and \emph{not a half-integer}: $u^{*}\in\mathbb{R}\setminus\tfrac{1}{2}\mathbb{Z}$.
This is the condition under which the CF for $w(u)$ converges to
a real meromorphic function with all poles and zeros on the real axis
(Theorem~\ref{thm:notes2-w-poles-zeros}, Remarks~\ref{rem:w-two-families},
\ref{rem:simplicity-evidence}).

\medskip{}

One of the main achievements of this chapter is that the CF method
itself yields explicit analytic expressions for the stability boundaries.
For the primary instability tongue ($m=1$, near $p=\tfrac{1}{2}$),
the stability condition~\eqref{eq:stability-assumption} is expressible
in terms of the detuning $\eta=p-\tfrac{1}{2}$ and the half-modulation
$\gamma=\varepsilon/2$ as 
\begin{equation}
p\;\notin\;\bigl(p_{-}(\gamma),\;p_{+}(\gamma)\bigr),\label{eq:stability-primary-tongue}
\end{equation}
where the boundary curves are given through order $\gamma^{6}$ by
Theorem~\ref{thm:bdry-series-CF} (Section~\ref{subsec:CF-stability-bdry}):
\begin{multline}
p_{\pm}(\gamma)\;=\;\frac{1}{2}\;\pm\;\frac{1}{4}\gamma\;-\;\frac{11}{32}\gamma^{2}\;\pm\;\frac{35}{256}\gamma^{3}\\
\;-\;\frac{985}{2048}\gamma^{4}\;\pm\;\frac{4139}{16384}\gamma^{5}\;-\;\frac{284379}{262144}\gamma^{6}\;+\;O(\gamma^{7}).\label{eq:stability-boundary-primary}
\end{multline}
The series~\eqref{eq:stability-boundary-primary} is valid near the
primary resonance, where both $\eta=p-\tfrac{1}{2}$ and $\gamma$
are small and of comparable magnitude. More precisely, the near-boundary
portion of the stable region in which this series is applied is the
strip where $|\eta/\gamma|$ lies strictly between two fixed positive
constants: 
\begin{equation}
\tfrac{1}{4}\;<\;\left|\frac{\eta}{\gamma}\right|\;<\;\tfrac{1}{3},\qquad\eta=p-\tfrac{1}{2}\qquad\Longleftrightarrow\qquad3\;<\;\left|\frac{\gamma}{\eta}\right|\;<\;4,\label{eq:stability-squeeze}
\end{equation}
the two forms being equivalent. The lower end $|\eta/\gamma|=\tfrac{1}{4}$
is the boundary curve itself ($p=p_{\pm}(\gamma)$, where $\eta\approx\pm\tfrac{1}{4}\gamma$);
the upper end $|\eta/\gamma|=\tfrac{1}{3}$ is the outer cutoff of
this near-boundary strip. Both ends are strict, since the stable region
lies strictly outside the closed tongue. This is the regime~\eqref{eq:stable-near-bdry-CF}
of Section~\ref{subsec:CF-stability-bdry}, to which we refer for
the full discussion and for the numerical care required there (Remark~\ref{rem:asymptotic-CF}).
The geometry of the stability zone and instability tongues is shown
in Fig.~\ref{fig:tongue-comparison} and the full stability diagram
in Fig.~\ref{fig:stability-MR}. The boundary curves for all odd
tongues are studied in Chapters~\ref{sec:Boundaries} and~\ref{sec:Summary};
the precision-cost analysis of computing in the stability zone is
in Section~\ref{subsec:Cambi-notes2-precision}.

\emph{Consequence for $w(u)$.} Under Assumption~\eqref{eq:stability-assumption},
$w(u)$ is a real meromorphic function\index{meromorphic function}
for $u\in\mathbb{R}$, and the CF converges absolutely for all real
$u$ that are not poles of $w$. All theoretical results in this chapter
(meromorphic structure, pole-zero duality, two-family partition, simplicity,
partial fraction representation) hold strictly within the stability
zone. When $(p,\gamma)$ approaches a boundary curve $p=p_{\pm}(\gamma)$,
the Floquet exponent $u^{*}\to\tfrac{1}{2}\mathbb{Z}$, the two families
of poles partially intersect at EPD points (Remark~\ref{rem:w-two-families}),
and the real pole-zero structure degenerates (see Section~\ref{subsec:Cambi-notes2-double-min}).

\subsection{\texorpdfstring{Cambi's function $v(u)$: minimal solution and meromorphic
properties}{Cambi's function v(u): minimal solution and meromorphic
properties}}

\label{subsec:Cambi-v-props}

\emph{The working object.} This section studies Cambi's $v(u)$; as
the chapter introduction explains, the development proper works with
the minimal-solution ratio $w(u)$ (\S\,\ref{subsec:Cambi-notes2}),
to which $v$ is tied by 
\begin{equation}
v(u)\;=\;-\frac{w(u-1)}{\gamma},\label{eq:v-from-w-framing}
\end{equation}
the $v$-properties recorded below recurring as corollaries of the
$w$-theory (\S\,\ref{subsec:notes-on-v}).

The recurrence for the Floquet coefficients~\eqref{eq:Cambi-system}
is the discretization of the functional equation: 
\begin{equation}
\gamma V(u+1)+G(u)V(u)+\gamma V(u-1)=0.\label{eq:Cambi-diff}
\end{equation}

\emph{Characteristic roots of~\eqref{eq:Cambi-diff}.} Equation~\eqref{eq:Cambi-diff}
is of Poincaré type (Chapter~\ref{app:FDE}): as $n\to+\infty$,
$G(u+n)\to1$, so the limiting characteristic equation is $\gamma\zeta^{2}+\zeta+\gamma=0$,
with roots 
\begin{equation}
\zeta_{\pm}=\frac{-1\pm\sqrt{1-4\gamma^{2}}}{2\gamma}.\label{eq:Cambi-char-roots}
\end{equation}
For $2\gamma<1$ both roots are real with $|\zeta_{-}|=1/|\zeta_{+}|>1>|\zeta_{+}|>0$,
where 
\begin{equation}
|\zeta_{+}|\;=\;\frac{1-\sqrt{1-4\gamma^{2}}}{2\gamma}\;=\;\gamma+\gamma^{3}+2\gamma^{5}+O(\gamma^{7}),\quad\gamma\to0.\label{eq:delta-gamma-def}
\end{equation}
Note $|\zeta_{+}|\in(0,1)$ for $\gamma\in(0,\tfrac{1}{2})$, $|\zeta_{+}|\to0$
as $\gamma\to0$, and $|\zeta_{+}|\to1$ as $\gamma\to\tfrac{1}{2}$.
The inverse relation is $\gamma=|\zeta_{+}|/(1+|\zeta_{+}|^{2})$.

\emph{Minimal solution.} By Theorem~\ref{thm:PP} (Poincaré--Perron\index{Poincaré--Perron theorem},
Chapter~\ref{app:FDE}), eq.~\eqref{eq:Cambi-diff}, viewed for
fixed $u$ as a recurrence in the integer index $n$, has a minimal
solution sequence $\{h_{n}(u)\}_{n\in\mathbb{Z}}$ (notation of \S\,\ref{subsec:Cambi-notes2})
whose consecutive ratios satisfy: 
\begin{equation}
\lim_{n\to+\infty}\frac{h_{n}(u)}{h_{n-1}(u)}\;=\;\zeta_{+}\;=\;\frac{-1+\sqrt{1-4\gamma^{2}}}{2\gamma}\;=\;-|\zeta_{+}|,\qquad|\zeta_{+}|<1.\label{eq:Cambi-Vplus-decay}
\end{equation}
Any solution not proportional to $\{h_{n}(u)\}$ has ratio tending
to $\zeta_{-}=-1/|\zeta_{+}|$ with $|\zeta_{-}|>1$ (dominant solution).
The minimal solution is unique up to a scalar multiple. Since $G(u)$
is even, $\{h_{n}(-u)\}$ is a minimal solution of~\eqref{eq:Cambi-diff}
decaying as $n\to-\infty$ (the left-going minimal solution $h_{m}^{-}(u)=h_{-m}(-u)$
of Definition~\ref{defn:notes2-hminus}).
\begin{defn}[Cambi's function $v(u)$]
\label{defn:v-CF} For $2\gamma<1$ and each $N\geq1$, define the
$N$-th CF approximant $v_{N}(u)$ by the backward recursion with
seed $T_{N}=0$: 
\begin{equation}
T_{k-1}\;=\;\frac{\gamma^{2}}{G(u+k)-T_{k}},\quad k=N,\,N-1,\,\ldots,\,1;\quad v_{N}(u)=\frac{1}{G(u)-T_{0}}.\label{eq:Cambi-v-recursion}
\end{equation}
Each $v_{N}(u)$ is a rational function of $u$. Define $v(u):=\lim_{N\to\infty}v_{N}(u)$,
with the limit taken in the spherical metric; the convergence and
properties of this limit are established in Theorem~\ref{thm:v-CF}
below. The limit equals the continued fraction: 
\begin{equation}
v(u)\;=\;\cfrac{1}{G(u)-\cfrac{\gamma^{2}}{G(u+1)-\cfrac{\gamma^{2}}{G(u+2)-\cdots}}},\qquad G(u)=1-\frac{p^{2}}{u^{2}}.\label{eq:Cambi-v-CF}
\end{equation}
\end{defn}

The scalar continued-fraction representation of the minimal solution
of a second-order difference equation --- its convergence and the
identification of its value with the minimal (recessive) solution
--- is the dedicated subject of Ahlbrandt and Peterson~\cite[Ch.~2, \S\,2.1]{AhlPet}\index{continued fraction!minimal solution}.

The asymptote of $v(u)$ follows from~\eqref{eq:Cambi-Vplus-decay}
and the relation $w(u)=-\gamma v(u+1)$ (eq.~\eqref{eq:notes2-w-v-relation}).
Since $\lim_{u\to+\infty}w(u)=\zeta_{+}$ (eq.~\eqref{eq:notes2-hm-ratio}):
\begin{equation}
\lim_{n\to+\infty}\bigl(-\gamma\,v(u+n)\bigr)\;=\;\zeta_{+},\qquad\lim_{u\to+\infty}v(u)\;=\;\frac{-\zeta_{+}}{\gamma}\;=\;\frac{1-\sqrt{1-4\gamma^{2}}}{2\gamma^{2}},\label{eq:Cambi-asymp-v}
\end{equation}
where $\zeta_{+}\;=\;(-1+\sqrt{1-4\gamma^{2}})/(2\gamma)$. This is
Cambi's asymptotic formula~\cite[\S\,4]{Cambi2}.

\emph{Meromorphic character and figure.}
\begin{thm}[Properties of $v(u)$]
\label{thm:v-CF} Let $2\gamma<1$. 
\begin{enumerate}
\item[(i)] \emph{Convergence.} The approximants $v_{N}(u)$ of Definition~\ref{defn:v-CF}
converge for all $u\in\mathbb{C}$ outside a discrete pole set, geometrically
with ratio $|\zeta_{+}|^{2}<1$, and the limit is the continued fraction~\eqref{eq:Cambi-v-CF}.
\item[(ii)] \emph{Recurrences.} The function $v(u)$ satisfies the right-propagation
recurrence (Cambi's eq.~(8)~\cite[eq.~(8)]{Cambi2}): 
\begin{equation}
\gamma^{2}v(u+1)\;=\;G(u)-\frac{1}{v(u)},\label{eq:Cambi-rec-v}
\end{equation}
and the left-propagation recurrence: 
\begin{equation}
v(u-1)\;=\;\frac{1}{G(u-1)-\gamma^{2}v(u)}.\label{eq:Cambi-rec-v-left}
\end{equation}
Together,~\eqref{eq:Cambi-rec-v} and~\eqref{eq:Cambi-rec-v-left}
propagate $v$ in both directions from any starting value.
\item[(iii)] \emph{Meromorphicity.} $v(u)$ is a meromorphic function of $u\in\mathbb{C}$. 
\end{enumerate}
\end{thm}

\begin{proof}
\emph{Part~(i): Convergence of the CF.} Let $\{h_{n}\}$ be the minimal
solution of~\eqref{eq:Cambi-diff} (notation of \S\,\ref{subsec:Cambi-notes2}).
Set $y_{n}=\gamma^{n}h_{n}$ in~\eqref{eq:Cambi-diff}. Multiplying
$\gamma h_{n+1}+G(u+n)h_{n}+\gamma h_{n-1}=0$ by $\gamma^{n}$ gives
\begin{equation}
y_{n+1}\;=\;-G(u+n)\,y_{n}-\gamma^{2}\,y_{n-1},\label{eq:Cambi-diff-y}
\end{equation}
which is in the form~\eqref{eq:confr1b} of Theorem~\ref{thm:Pinch}
with $b_{n}=-G(u+n)$ and $a_{n}=-\gamma^{2}$ for all $n\geq1$.
The associated CF~\eqref{eq:confr1c} is 
\[
\mathbf{K}_{n=1}^{\infty}\!\left[\frac{-\gamma^{2}}{-G(u+n)}\right],
\]
which is exactly~\eqref{eq:Cambi-v-CF}. By Theorem~\ref{thm:PP}
(Poincaré--Perron), $|\zeta_{-}|>|\zeta_{+}|$, so a minimal solution
of~\eqref{eq:Cambi-diff-y} exists. By Theorem~\ref{thm:Pinch}
(Pincherle, part~1), this is equivalent to convergence of the CF~\eqref{eq:Cambi-v-CF}.
From~\eqref{eq:Cambi-Vplus-decay} and $|\zeta_{-}|=1/|\zeta_{+}|$,
the error of the $m$-th approximant decays like $(|\zeta_{+}|/|\zeta_{-}|)^{m}=|\zeta_{+}|^{2m}$,
giving geometric convergence with ratio $|\zeta_{+}|^{2}<1$.

\emph{Part~(ii): Recurrences.} Cross-multiplying~\eqref{eq:Cambi-v-CF}
by its first denominator: $v(u)(G(u)-\gamma^{2}v(u+1))=1$, which
rearranges to~\eqref{eq:Cambi-rec-v}. The left-recurrence~\eqref{eq:Cambi-rec-v-left}
follows by solving for $1/v(u)$ in~\eqref{eq:Cambi-rec-v} and shifting
$u\mapsto u-1$.

\emph{Part~(iii): Meromorphicity.} Each truncation $v_{N}(u)$ defined
by the $N$-level backward recursion~\eqref{eq:Cambi-v-recursion}
is a rational function of $u$ (since $G(u)=1-p^{2}/u^{2}$ is rational
and the recursion involves only rational operations). The CF~\eqref{eq:Cambi-v-CF}
has numerators $\gamma^{2}$ and denominators $G(u+k)$ --- both
holomorphic functions of $u$ --- so Theorem~\ref{thm:sLconfmero1}
(Chapter~\ref{app:CF}) applies: the limit $v(u)$ is meromorphic
on $\mathbb{C}$. 
\end{proof}
\emph{Matching condition and resonance equation.} The Floquet solution
requires the right-going minimal solution (ratio $\to\zeta_{+}$ as
$n\to+\infty$) and the left-going minimal solution (ratio $\to\zeta_{+}$
as $n\to-\infty$) to be proportional. In the language of $v(u)$,
this gives the matching condition: 
\begin{equation}
\gamma^{2}\,v(u_{0}+1)\,v(-u_{0})\;=\;1.\label{eq:Cambi-matching-new}
\end{equation}
Using the identity $v(-u_{0})[G(u_{0})-\gamma^{2}v(1-u_{0})]=1$,
equation~\eqref{eq:Cambi-matching-new} reduces to Cambi's resonance
equation: 
\begin{equation}
G(u_{0})-\gamma^{2}\bigl[v(1+u_{0})+v(1-u_{0})\bigr]=0.\label{eq:Cambi-res-from-matching}
\end{equation}
The alternative derivation via the minimal solutions $\{h_{m}\}$
and $\{h_{m}^{-}\}$ is given in \S\,\ref{subsec:Cambi-notes2-res}.

\emph{Numerical verification as $\gamma\to\tfrac{1}{2}$.} The convergence
rate $|\zeta_{+}|$ (eq.~\eqref{eq:delta-gamma-def}) is the modulus
of $\zeta_{+}$ (eq.~\eqref{eq:Cambi-char-roots}). Table~\ref{tab:v-recurrence}
verifies that~\eqref{eq:Cambi-v-CF} satisfies~\eqref{eq:Cambi-rec-v}
to machine precision for $\gamma\leq0.499$ ($\delta\leq0.939$) with
$N=300$ levels. Near $\gamma=\tfrac{1}{2}$, $\delta\to1$ and more
levels are needed ($N=2000$ suffices at $\gamma=0.4999$). This explains
the numerical difficulty in the stability diagram near $\gamma=\tfrac{1}{2}$
(Section~\ref{subsec:StabDiag}).

\begin{table}[ht]
\centering 
\global\long\def\arraystretch{1.3}%
\resizebox{\textwidth}{!}{%
\begin{tabular}{ccrrrr}
\toprule 
$\gamma$  & $|\zeta_{+}|$  & $v(u)$  & LHS $=\gamma^{2}v(u+1)$  & RHS $=G(u)-1/v(u)$  & rel.\ err. \tabularnewline
\midrule 
\multicolumn{6}{c}{$u=\tfrac{1}{2}$,\quad{}$p=\tfrac{1}{2}$\quad{}($m=1$ resonance
center),\quad{}$N=300$ CF levels}\tabularnewline
\midrule 
$0.01$  & $0.01000$  & $-8.888\times10^{3}$  & $1.12513\times10^{-4}$  & $1.12513\times10^{-4}$  & $<10^{-15}$\tabularnewline
$0.05$  & $0.05013$  & $-3.545\times10^{2}$  & $2.82079\times10^{-3}$  & $2.82079\times10^{-3}$  & $<10^{-15}$\tabularnewline
$0.10$  & $0.10102$  & $-8.784\times10^{1}$  & $1.13849\times10^{-2}$  & $1.13849\times10^{-2}$  & $2\times10^{-16}$\tabularnewline
$0.20$  & $0.20871$  & $-2.113\times10^{1}$  & $4.73214\times10^{-2}$  & $4.73214\times10^{-2}$  & $<10^{-15}$\tabularnewline
$0.30$  & $0.33333$  & $-8.710\times10^{0}$  & $1.14806\times10^{-1}$  & $1.14806\times10^{-1}$  & $1\times10^{-16}$\tabularnewline
$0.40$  & $0.50000$  & $-4.229\times10^{0}$  & $2.36470\times10^{-1}$  & $2.36470\times10^{-1}$  & $1\times10^{-16}$\tabularnewline
$0.45$  & $0.62679$  & $-2.890\times10^{0}$  & $3.46031\times10^{-1}$  & $3.46031\times10^{-1}$  & $2\times10^{-16}$\tabularnewline
$0.48$  & $0.75000$  & $-2.136\times10^{0}$  & $4.68089\times10^{-1}$  & $4.68089\times10^{-1}$  & $1\times10^{-16}$\tabularnewline
$0.49$  & $0.81735$  & $-1.826\times10^{0}$  & $5.47505\times10^{-1}$  & $5.47505\times10^{-1}$  & $2\times10^{-16}$\tabularnewline
$0.499$  & $0.93866$  & $-1.274\times10^{0}$  & $7.85118\times10^{-1}$  & $7.85118\times10^{-1}$  & $<10^{-15}$\tabularnewline
$0.4999$  & $0.98020$  & $-8.919\times10^{-1}$  & $1.12115\times10^{0}$  & $1.12115\times10^{0}$  & $4\times10^{-7}$\tabularnewline
\bottomrule
\end{tabular}} \caption{Verification that the CF~\eqref{eq:Cambi-v-CF} satisfies the recurrence~\eqref{eq:Cambi-rec-v}
as $\gamma\to\tfrac{1}{2}$, at $u=p=\tfrac{1}{2}$ ($m=1$ resonance
center, $N=300$ CF levels). LHS and RHS computed independently. The
convergence rate $|\zeta_{+}|$ (eq.~\eqref{eq:delta-gamma-def},
the modulus of the smaller characteristic root $\zeta_{+}$~\eqref{eq:Cambi-char-roots})
governs accuracy: for $\delta\protect\leq0.939$ machine precision
is achieved with $N=300$; at $\delta=0.980$ more levels are needed.}
\label{tab:v-recurrence} 
\end{table}

\begin{figure}[htbp]
\centering \includegraphics[width=1\textwidth]{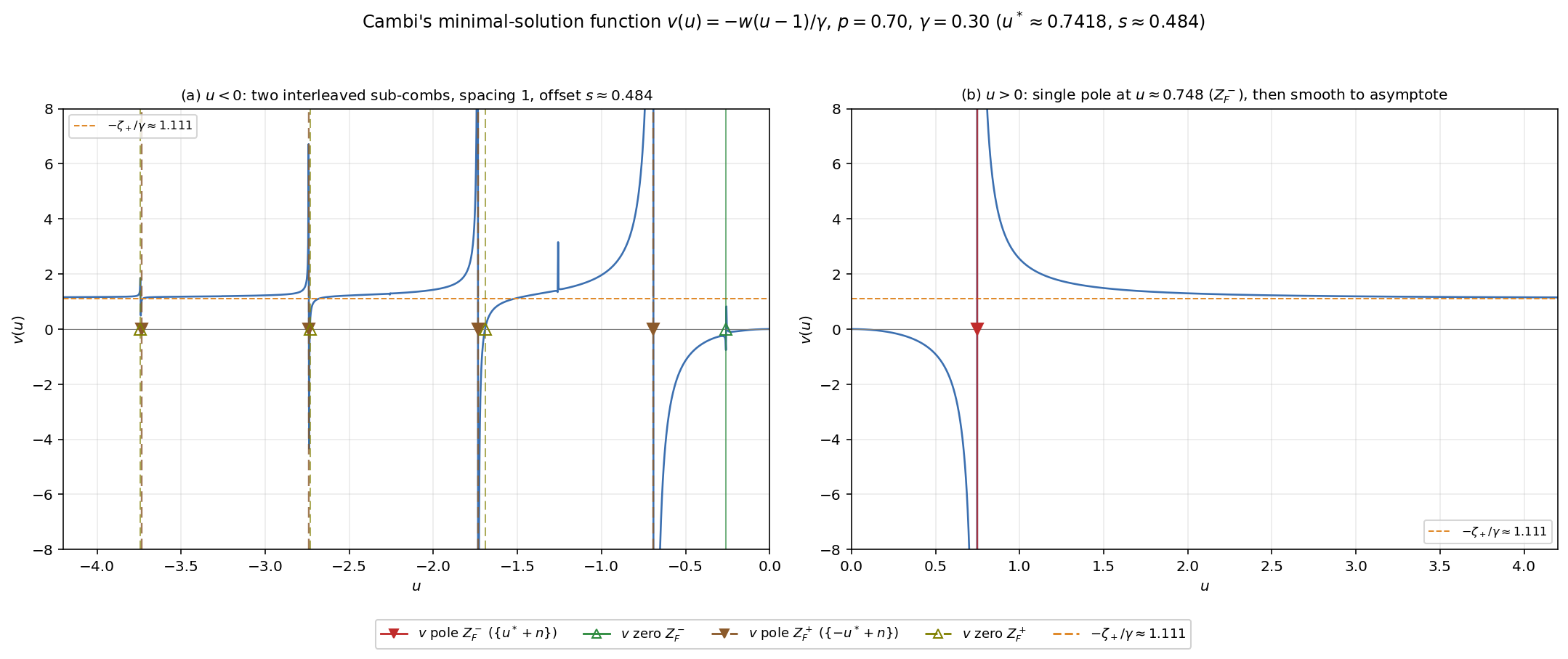} \caption{Cambi's minimal-solution function $v(u)=-w(u-1)/\gamma$, computed
via the correct backward CF (Theorem~\ref{thm:notes2-wrec-exact}),
for $p=0.70$, $\gamma=0.30$ ($u^{*}\approx0.7418$, $s\approx0.484$,
asymptote $-\zeta_{+}/\gamma\approx1.1111$, orange dashed). \emph{(a)}
$u<0$: two interleaved families of poles and zeros, each spaced by
$1$ --- one tracking the sub-comb $\{u^{*}+n\}$ (poles solid red,
zeros solid green), the other tracking $\{-u^{*}+n\}$ (poles dashed
brown, zeros dashed olive). The two families interleave with smallest
gap $s\approx0.484$ (eq.~\eqref{eq:s-offset}), producing alternating
gaps of width $s$ and $1-s$. \emph{(b)} $u>0$: $v$ is smooth except
for a single pole at $u\approx0.748$ (red vertical, the family tracking
$\{u^{*}+n\}$), and converges to the asymptote from above. Axis markers
(on the $u$-axis): filled $\triangledown$ = poles of $v$, open
$\triangle$ = zeros of $v$ (red/green = family tracking $\{u^{*}+n\}=Z_{F}^{-}$,
brown/olive = family tracking $\{-u^{*}+n\}=Z_{F}^{+}$; see~\eqref{eq:family-def}).
Compare with Fig.~\ref{fig:notes2-w-marked} for $w(u)$.}
\label{fig:v-u} 
\end{figure}

\begin{figure}[htbp]
\centering \includegraphics[width=0.82\textwidth]{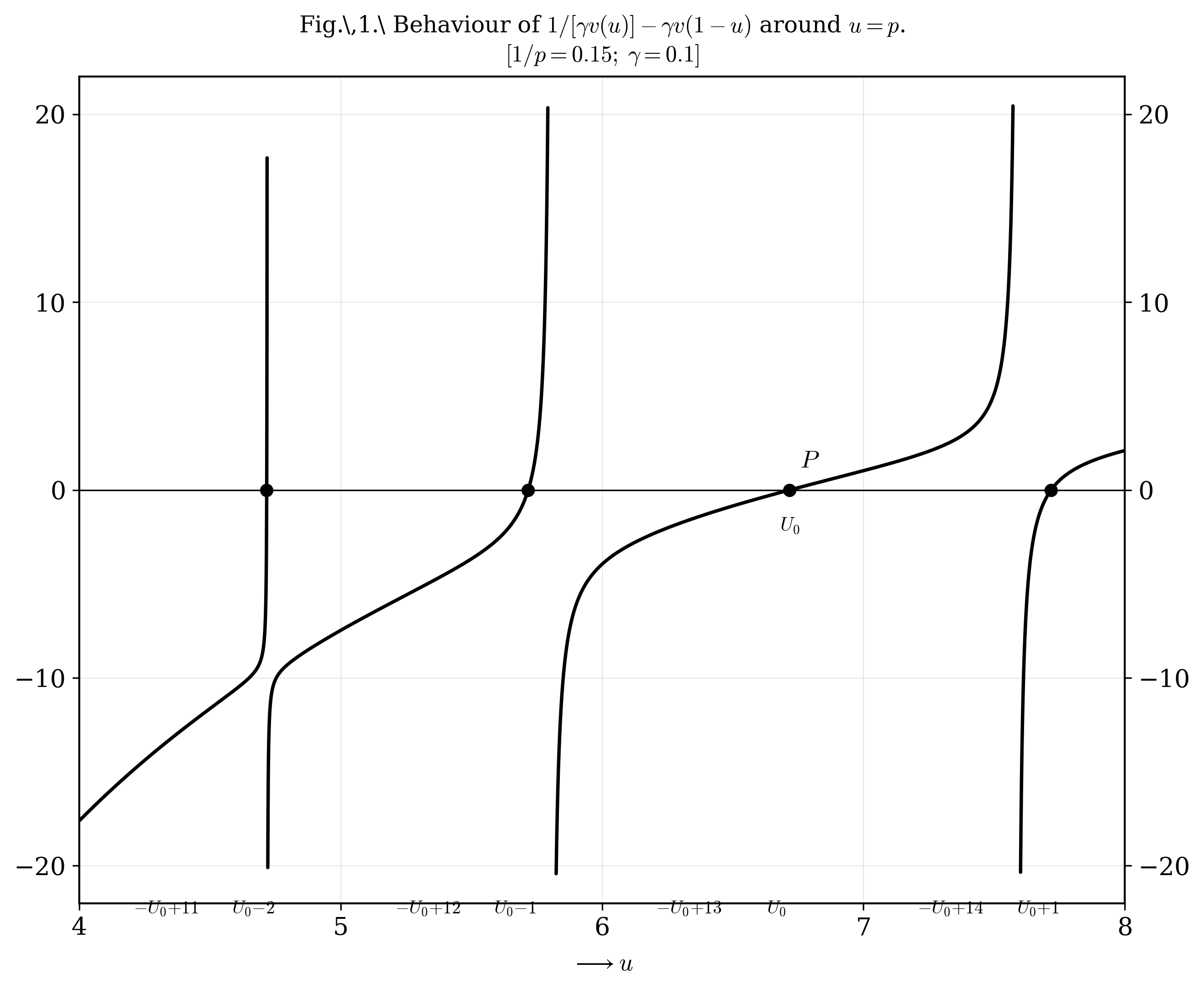}
\caption{Reproduction of Cambi's Figure~1~\cite[\S\,6]{Cambi2}: the function
$F(u)=1/[\gamma v(u)]-\gamma v(1-u)$ --- which is precisely the
double minimality function $F_{w}(u)=w(-u)-1/w(u-1)$ of \S\,\ref{subsec:Cambi-notes2},
expressed through $v$ via $w(u)=-\gamma\,v(u+1)$ --- at $1/p=0.15$
($p=20/3$), $\gamma=0.1$. By eq.~\eqref{eq:Cambi-res7p}, $F(u_{0})=0$
is equivalent to Cambi's resonance equation~\eqref{eq:Cambi-res-from-matching}
for the Floquet exponent $u_{0}$. Poles of $F$ occur where $v(u)$
has poles (near $u=p-n$, $n=0,1,2,\ldots$) and where $v(1-u)$ has
poles (near $u=1-p+n$, $n=0,1,2,\ldots$); between consecutive poles
the function is smooth and monotone. The central root $P$ (marked)
is the primary Floquet exponent $u_{0}\approx p$ (here $u_{0}\approx6.718$,
$p=6.667$).}
\label{fig:cambi-fig1} 
\end{figure}

Figure~\ref{fig:v-u} shows $v(u)$ and Figure~\ref{fig:cambi-fig1}
reproduces Cambi's Figure~1, illustrating the pole structure and
the primary zero $u_{0}\approx p$ of~\eqref{eq:Cambi-res-from-matching}.

% -----------------------------------------------------------------------------
% -----------------------------------------------------------------------------

\subsection{\texorpdfstring{Asymptotically seeded approximants for $v(u)$}{Asymptotically
seeded approximants for v(u)}}

\label{subsec:Cambi-approx}

The approximants developed here are for Cambi's function $v(u)$;
the analogous theory for $w(u)=-\gamma v(u+1)$ follows from~\eqref{eq:notes2-CF-orig}
of \S\,\ref{subsec:Cambi-notes2}.

Shifting $u\mapsto u+1$ in~\eqref{eq:Cambi-rec-v-left} and applying
the result $n+1$ times gives, via the tail identity~\eqref{eq:abfS5b}
of Chapter~\ref{app:CF}, the exact representation 
\begin{equation}
v(u)\;=\;S_{n+1}\!\left(v(u+n+1)\right),\label{eq:v-Sn-exact}
\end{equation}
where $S_{n+1}$ is the $(n+1)$-th Mobius\index{Mobius transformation}
composition (Lemma~\ref{lem:Moebpr}, Chapter~\ref{app:CF}). Setting
$v(u+n+1)=0$ gives the standard zero-seeded approximants $v_{0}^{(n)}(u):=S_{n+1}(0)$,
converging with error $O(|\zeta_{+}|^{2(n+1)})$ (Theorem~\ref{thm:Pinch}).
We introduce a superior alternative.
\begin{defn}[Asymptotically seeded approximants]
\index{asymptotic seeding}\index{asymptotic seeding!seeded approximants}
\label{defn:v-approx} For $n=0,1,2,\ldots$, the \emph{$n$-th asymptotically
seeded approximant} of $v(u)$ is 
\begin{equation}
v^{(n)}(u)\;:=\;S_{n+1}\!\left(\frac{-\zeta_{+}}{\gamma}\right)\;=\;\cfrac{1}{G(u)-\cfrac{\gamma^{2}}{G(u+1)-\cfrac{\gamma^{2}}{\ddots-\cfrac{\gamma^{2}}{G(u+n)-\gamma|\zeta_{+}|}}}},\label{eq:v-approx-n}
\end{equation}
obtained by seeding~\eqref{eq:v-Sn-exact} with the exact asymptotic
limit $-\zeta_{+}/\gamma$ from eq.~\eqref{eq:Cambi-asymp-v}. The
first two members are: 
\begin{equation}
v^{(0)}(u)\;=\;\frac{1}{G(u)-\gamma|\zeta_{+}|},\qquad v^{(1)}(u)\;=\;\frac{1}{G(u)-\dfrac{\gamma^{2}}{G(u+1)-\gamma|\zeta_{+}|}}.\label{eq:v-approx-01}
\end{equation}
\end{defn}

\begin{thm}[Error bound for asymptotically seeded approximants]
\index{asymptotic seeding!error bound} \label{thm:v-approx-error}
For each $n\geq0$ and $u$ not a pole of $v$, 
\begin{equation}
\bigl|v^{(n)}(u)-v(u)\bigr|\;=\;O\!\left(\frac{|\zeta_{+}|^{2(n+1)}}{(n+1)^{2}}\right),\qquad n\to\infty.\label{eq:v-approx-error}
\end{equation}
This carries the \emph{same geometric order} $|\zeta_{+}|^{2(n+1)}$
as the zero-seeded approximant 
\begin{equation}
\bigl|v_{0}^{(n)}(u)-v(u)\bigr|\;=\;O\!\left(|\zeta_{+}|^{2(n+1)}\right),\label{eq:v-zero-error}
\end{equation}
but improves it by the polynomial factor $1/(n+1)^{2}$. The improvement
is \emph{not} an order-doubling: because the recurrence coefficients
$G(u+m)=1-p^{2}/(u+m)^{2}$ approach their limit $1$ only algebraically
(as $O(1/m^{2})$, not geometrically), the asymptotic seed $-\zeta_{+}/\gamma$
matches the exact tail $v(u+n+1)$ only to $O(1/(n+1)^{2})$, which
is the source of the $1/(n+1)^{2}$ gain (see proof). 
\end{thm}

\begin{proof}
\emph{Proof of~\eqref{eq:v-zero-error}.} From~\eqref{eq:v-Sn-exact}:
$v(u)=S_{n+1}(v(u+n+1))$ and $v_{0}^{(n)}(u)=S_{n+1}(0)$. By Lemma~\ref{lem:Moebpr},
$S_{n+1}(w)=(A_{n}w+A_{n+1})/(B_{n}w+B_{n+1})$, so for any $w_{1},w_{2}$:
\begin{equation}
S_{n+1}(w_{1})-S_{n+1}(w_{2})\;=\;\frac{(A_{n}B_{n+1}-A_{n+1}B_{n})(w_{1}-w_{2})}{(B_{n}w_{1}+B_{n+1})(B_{n}w_{2}+B_{n+1})}.\label{eq:Moebius-diff}
\end{equation}
Setting $w_{1}=v(u+n+1)$ and $w_{2}=0$: 
\begin{equation}
v(u)-v_{0}^{(n)}(u)\;=\;\frac{(A_{n}B_{n+1}-A_{n+1}B_{n})\,v(u+n+1)}{(B_{n}v(u+n+1)+B_{n+1})\,B_{n+1}}.\label{eq:v-zero-err-raw}
\end{equation}

\emph{Numerator determinant.} The Euler-Minding formula~\eqref{eq:abfS4d}
gives $A_{n}B_{n+1}-A_{n+1}B_{n}=(-1)^{n+1}a_{1}a_{2}\cdots a_{n+1}$.
For our CF, $a_{k}=-\gamma^{2}$ for all $k\geq1$, so: 
\begin{equation}
A_{n}B_{n+1}-A_{n+1}B_{n}\;=\;(-1)^{n+1}(-\gamma^{2})^{n+1}\;=\;\gamma^{2(n+1)}.\label{eq:det-formula}
\end{equation}

\emph{Asymptotics of $B_{n}$.} The denominators $B_{n}$ satisfy
the same recurrence~\eqref{eq:abfS4b} as the numerators: $B_{n}=-G(u+n)\,B_{n-1}-\gamma^{2}B_{n-2}$.
As $n\to\infty$, $G(u+n)\to1$, so the characteristic equation becomes
$\mu^{2}+\mu+\gamma^{2}=0$, with roots 
\begin{equation}
\mu_{\pm}\;=\;\frac{-1\pm\sqrt{1-4\gamma^{2}}}{2}\;=\;\gamma\,\zeta_{\pm}.\label{eq:mu-roots}
\end{equation}
Since $B_{0}=1$, $B_{-1}=0$ (eq.~\eqref{eq:abfS4c}), the initial
conditions are generic (not the minimal solution), so $B_{n}\sim C\,\mu_{-}^{n}$
as $n\to\infty$ where $|\mu_{-}|>|\mu_{+}|$. Since $\mu_{-}=\gamma\zeta_{-}$
and $|\zeta_{-}|=1/|\zeta_{+}|$, we have $|\mu_{-}|=\gamma/|\zeta_{+}|$,
so $|B_{n}|\sim C(\gamma/|\zeta_{+}|)^{n}$.

\emph{Error estimate.} From~\eqref{eq:Cambi-asymp-v}, $v(u+n+1)\to-\zeta_{+}/\gamma$
as $n\to\infty$, so $v(u+n+1)=O(1)$ and $B_{n}v(u+n+1)+B_{n+1}=O(|\mu_{-}|^{n+1})$.
Substituting into~\eqref{eq:v-zero-err-raw}--\eqref{eq:det-formula}:
\[
\begin{aligned}\bigl|v(u)-v_{0}^{(n)}(u)\bigr| & \;\sim\;\frac{\gamma^{2(n+1)}\cdot O(1)}{O(|\mu_{-}|^{n+1})\cdot O(|\mu_{-}|^{n+1})}\\
 & \;=\;O\!\left(\frac{\gamma^{2(n+1)}}{|\mu_{-}|^{2(n+1)}}\right)\;=\;O\!\left(\left(\frac{\gamma}{|\mu_{-}|}\right)^{2(n+1)}\right).
\end{aligned}
\]
Since $|\mu_{-}|=\gamma/|\zeta_{+}|$: $\gamma/|\mu_{-}|=\gamma/({\gamma/|\zeta_{+}|})=|\zeta_{+}|$,
giving~\eqref{eq:v-zero-error}.

\emph{Proof of~\eqref{eq:v-approx-error}.} From~\eqref{eq:v-Sn-exact}:
$v(u)-v^{(n)}(u)=S_{n+1}(v(u+n+1))-S_{n+1}(-\zeta_{+}/\gamma)$. Applying~\eqref{eq:Moebius-diff}
with $w_{1}=v(u+n+1)$, $w_{2}=-\zeta_{+}/\gamma$: 
\[
v(u)-v^{(n)}(u)=\frac{\gamma^{2(n+1)}\,(v(u+n+1)+\zeta_{+}/\gamma)}{(B_{n}v(u+n+1)+B_{n+1})(B_{n}(-\zeta_{+}/\gamma)+B_{n+1})}.
\]
The seed error is sharply $\Theta(1/(n+1)^{2})$. Indeed, by $w(u)=-\gamma\,v(u+1)$
(eq.~\eqref{eq:notes2-w-v-relation}) the seed mismatch is 
\[
v(u+n+1)+\frac{\zeta_{+}}{\gamma}\;=\;-\frac{1}{\gamma}\bigl(w(u+n)-\zeta_{+}\bigr)\;=\;-\frac{a}{\gamma\,(u+n)^{2}}+O\!\left(\frac{1}{(u+n)^{3}}\right),
\]
using the asymptotic expansion of the minimal solution ratio $w$
(Theorem~\ref{thm:notes2-w-rate}, eq.~\eqref{eq:notes2-w-rate}),
whose leading coefficient $a=-\zeta_{+}^{2}p^{2}/[\gamma(1-\zeta_{+}^{2})]\neq0$
makes the rate sharp. The underlying reason is that the recurrence
coefficients settle only algebraically, $G(u+m)-1=-p^{2}/(u+m)^{2}=O(1/m^{2})$,
so the tail value $v(u+n+1)$ relaxes to its limit $-\zeta_{+}/\gamma$
at the same algebraic rate. (This is distinct from the geometric convergence
of the minimal solution \emph{ratio} $h_{m}/h_{m-1}\to\zeta_{+}$
in~\eqref{eq:Cambi-Vplus-decay}: the ratio converges geometrically,
but the tail value $v(u+n+1)$ relaxes to its limit only algebraically,
because the coefficients $G(u+m)$ carry the $O(1/m^{2})$ Coulomb-type
tail of the $p^{2}/u^{2}$ term.) By the same denominator estimate
as above, both denominator factors are $O(|\mu_{-}|^{n+1})$. Multiplying:
\[
\bigl|v(u)-v^{(n)}(u)\bigr|=O\!\left(\frac{\gamma^{2(n+1)}\cdot(n+1)^{-2}}{|\mu_{-}|^{2(n+1)}}\right)=O\!\left(\frac{|\zeta_{+}|^{2(n+1)}}{(n+1)^{2}}\right),
\]
which is~\eqref{eq:v-approx-error}. 
\end{proof}
\begin{rem}[Quantitative comparison]
\label{rem:v-approx-comparison} The asymptotic seed improves the
zero-seeded error by the polynomial factor $(n+1)^{2}$ (up to an
$O(1)$ constant that depends on $u$ and $\gamma$), at identical
computational cost. The gain grows polynomially with depth, not geometrically:
it does \emph{not} double the convergence exponent, which remains
$|\zeta_{+}|^{2(n+1)}$ in both cases. A genuinely faster seed, which
removes the $1/(n+1)^{2}$ bottleneck by matching the algebraic tail
of $v$ to higher order, is available for $w(u)$ and is described
in Remark~\ref{rem:w-seeded-CF}. 
\end{rem}

\begin{figure}[htbp]
\centering \includegraphics[width=0.95\textwidth]{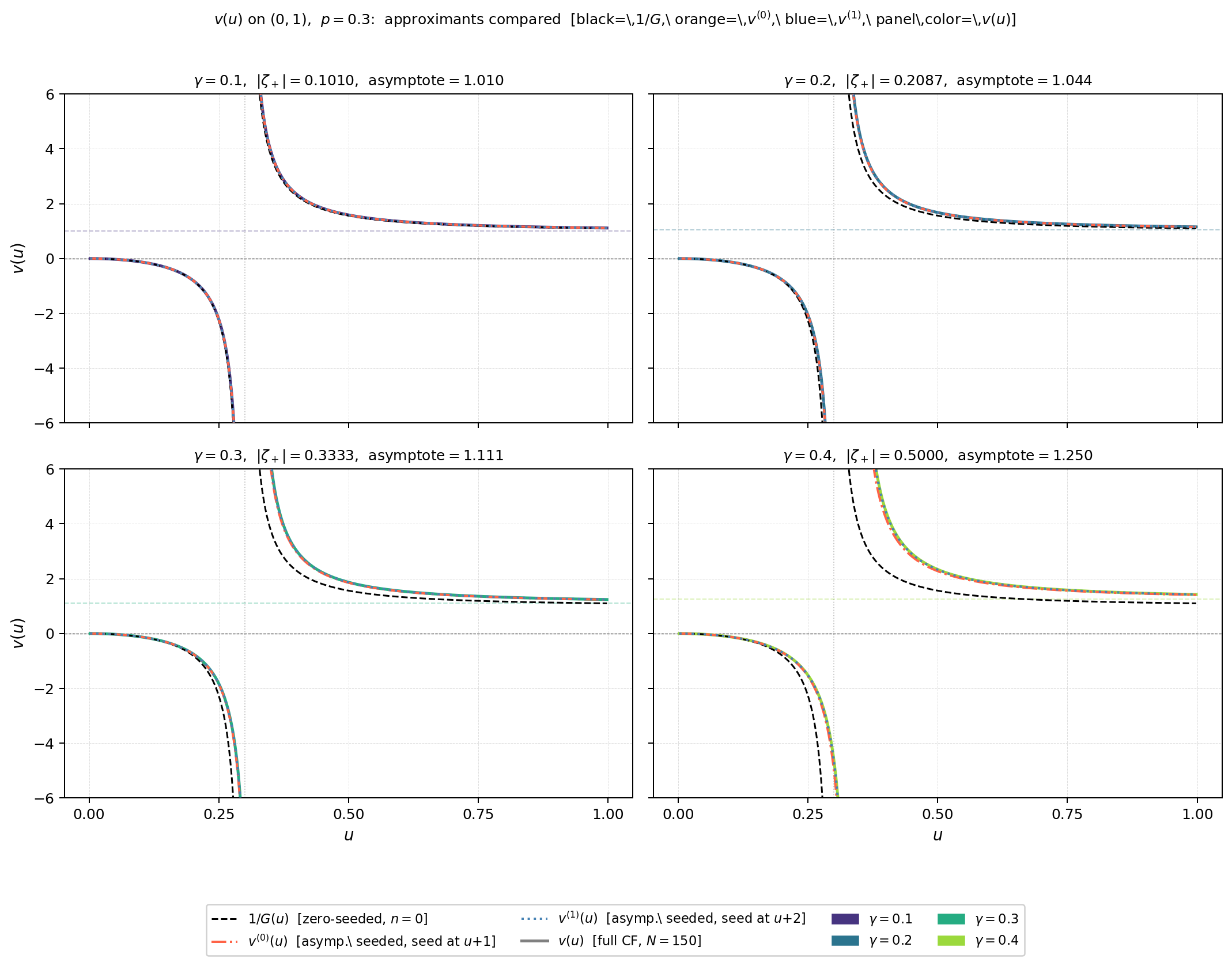}
\caption{The four approximants $1/G(u)$ (black dashed), $v^{(0)}(u)$ (orange
dash-dot), $v^{(1)}(u)$ (steel-blue dotted), and the full CF $v(u)$
(solid, color encodes $\gamma$), plotted on $(0,1)$ for $p=0.30$
and four values of $\gamma$. The asymptote $-\zeta_{+}/\gamma$~\eqref{eq:Cambi-asymp-v}
is shown as a faint horizontal line in each panel. The ordering $1/G$
furthest from $v$, then $v^{(0)}$, then $v^{(1)}$, then $v$ itself
confirms the error estimates~\eqref{eq:v-zero-error} and~\eqref{eq:v-approx-error}.}
\label{fig:v-approx-gamma} 
\end{figure}

\begin{figure}[htbp]
\centering \includegraphics[width=0.95\textwidth]{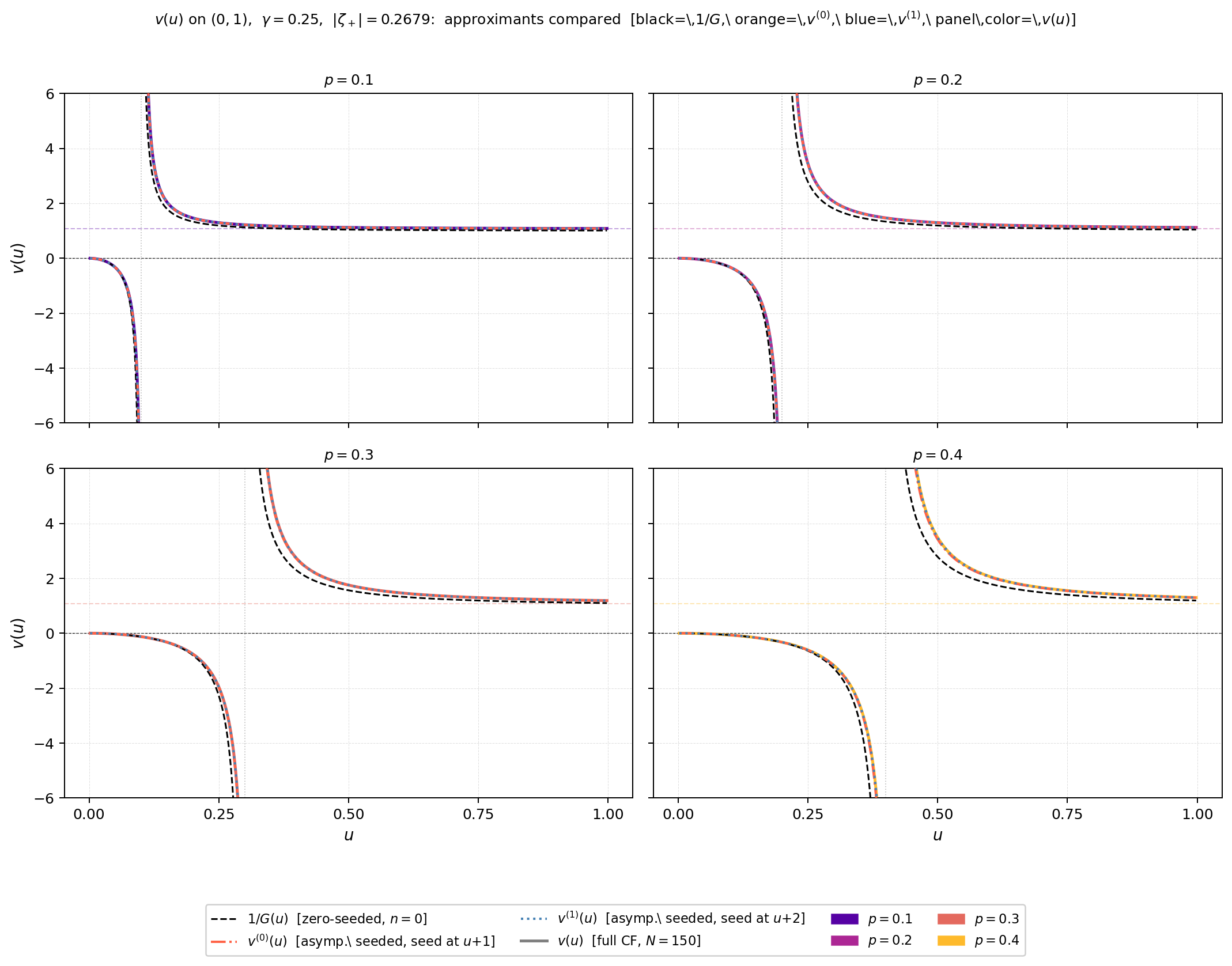} \caption{Same four approximants as Figure~\ref{fig:v-approx-gamma}, now with
fixed $\gamma=0.25$ ($|\zeta_{+}|\approx0.268$) and four values
of $p\in\{0.10,0.20,0.30,0.40\}$ (color encodes $p$). The zero-seeded
$1/G(u)$ has its pole exactly at $u=p$ (dotted vertical), while
the pole of the seeded $v^{(0)}$ sits just to the right of $u=p$,
tracking the true pole of $v$, in agreement with eq.~\eqref{eq:pole-approx-0}.}
\label{fig:v-approx-p} 
\end{figure}

\begin{figure}[htbp]
\centering \includegraphics[width=0.85\textwidth]{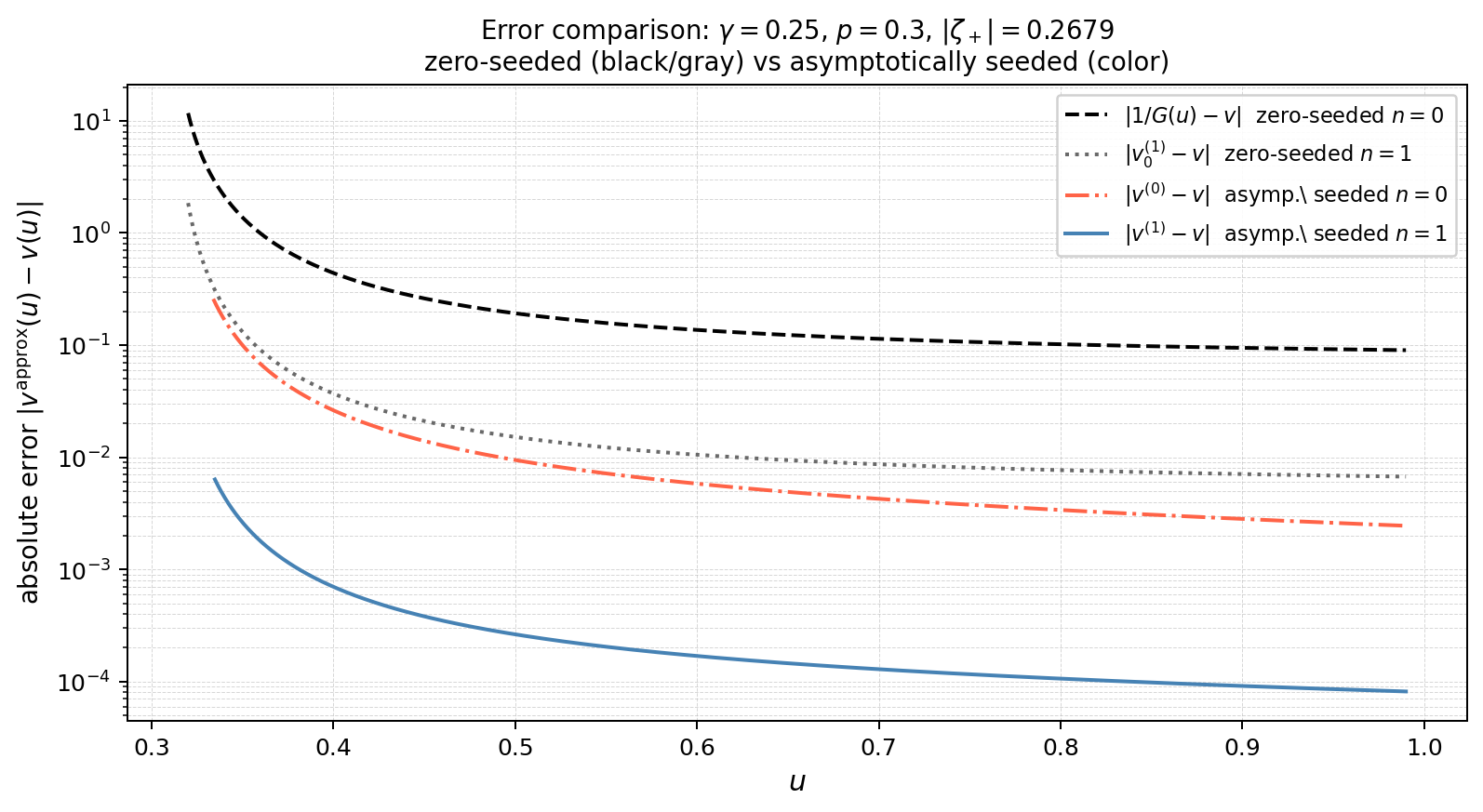} \caption{Pointwise absolute errors $|v^{{\rm approx}}(u)-v(u)|$ on $(p,1)$
for $\gamma=0.25$, $p=0.30$, $|\zeta_{+}|\approx0.268$. Zero-seeded
approximants (black/gray) vs asymptotically seeded approximants (orange/blue).
The asymptotically seeded $v^{(0)}$ (orange) and $v^{(1)}$ (blue)
are uniformly better than their zero-seeded counterparts at the same
truncation depth, with the seeded errors smaller by the polynomial
factor $1/(n+1)^{2}$ at the same geometric order $|\zeta_{+}|^{2(n+1)}$
(eq.~\eqref{eq:v-approx-error} vs.~\eqref{eq:v-zero-error}).}
\label{fig:v-approx-errors} 
\end{figure}

\begin{rem}[Qualitative advantage at $n=0$ and pole location]
\label{rem:v-approx-pole} The standard reference $v_{0}^{(0)}(u)=1/G(u)$
has a pole at $u=p$ (the zero of $G$); this pole is \emph{spurious},
since $v$ itself is regular there ($v(p)$ is finite and nonzero),
the true nearby pole of $v$ being displaced from $p$ by the tail
term $-\gamma^{2}v(u+1)$ in the denominator of~\eqref{eq:Cambi-v-CF}.
The first asymptotically seeded approximant $v^{(0)}(u)=1/(G(u)-\gamma|\zeta_{+}|)$
places its pole where $G(u)=\gamma|\zeta_{+}|>0$, i.e.\ at 
\begin{equation}
u_{{\rm pole}}^{(0)}\;=\;\frac{p}{\sqrt{1-\gamma|\zeta_{+}|}}\;>\;p,\label{eq:pole-approx-0}
\end{equation}
which lies on the same side of $p$ as, and close to, the true pole
of $v$, consistent with the plots (Figures~\ref{fig:v-approx-gamma},
\ref{fig:v-approx-p}, and~\ref{fig:v-approx-errors}). Meromorphicity
of the sequence and its limit is guaranteed by Theorem~\ref{thm:sLconfmero1}
of Chapter~\ref{app:CF}. 
\end{rem}

\begin{rem}[Optimal seeded CF for $w(u)$]
\label{rem:w-seeded-CF} Via $w(u)=-\gamma v(u+1)$, the asymptotically
seeded approximant for $v$ translates immediately to one for $w$.
However, a more efficient seed is available for $w$ using the asymptotic
expansion (Theorem~\ref{thm:notes2-w-rate}): for large real $u$,
\[
w(u)\;\approx\;\zeta_{+}\;+\;\frac{a}{u^{2}}\;+\;\frac{b}{u^{3}}\;+\;\frac{c}{u^{4}},
\]
with $a,b,c$ given by~\eqref{eq:notes2-w-rate-a}--\eqref{eq:notes2-w-rate-c}.
The optimal seeded backward CF for $w$ at argument $u$ using $N$
levels is: 
\begin{multline}
\hat{w}^{(N)}(u)\;:=\;\text{CF}_{n=N\to1}\text{ with seed }\\
T_{N}\;=\;\zeta_{+}\;+\;\frac{a}{(u+N)^{2}}\;+\;\frac{b}{(u+N)^{3}}\;+\;\frac{c}{(u+N)^{4}},\label{eq:w-seeded-CF}
\end{multline}
where the backward substitution uses the alternating-sign recurrence~\eqref{eq:notes2-CF-new}.
The seed mismatch is $O(1/(u+N)^{5})$ (the remainder in~\eqref{eq:notes2-w-rate}),
compared with the $\Theta(1/(u+N)^{2})$ mismatch of the plain seed
$T_{N}=\zeta_{+}$ (the $a$-term of~\eqref{eq:notes2-w-rate}).

\emph{Error comparison.} All seeds share the geometric contraction
$|\zeta_{+}|^{2N}$ of the backward recursion; the seed mismatch enters
as a multiplicative algebraic factor: 
\begin{equation}
\begin{aligned} & \text{zero seed: }O\!\bigl(|\zeta_{+}|^{2N}\bigr),\qquad\text{plain seed: }O\!\left(\frac{|\zeta_{+}|^{2N}}{(u+N)^{2}}\right),\\
 & \text{seed~\eqref{eq:w-seeded-CF}: }O\!\left(\frac{|\zeta_{+}|^{2N}}{(u+N)^{5}}\right).
\end{aligned}
\label{eq:w-seed-error-laws}
\end{equation}
The polynomial factors shave levels but do not change the leading
scaling $N\approx D/\bigl(2\log_{10}(1/|\zeta_{+}|)\bigr)$ for $D$
digits. Measured at moderate arguments for $D=80$ (120-digit arithmetic;
zero/plain/optimal): $N=42/42/40$ at $\gamma=0.1$; $N=84/80/75$
at $\gamma=0.3$; $N=199/188/176$ at $\gamma=0.45$. The seed~\eqref{eq:w-seeded-CF}
earns its keep at \emph{large arguments}, where the factor $(u+N)^{-5}$
supplies $\approx5\log_{10}(u+N)$ digits free of charge: in the verification
of the asymptotic coefficients at arguments $\sim10^{5}$ (Remark~\ref{rem:notes2-w-rate-verify})
it contributes $\approx25$ of the $80$ digits, reducing the required
depth from $N\approx40$ to $N\approx28$ at $\gamma=0.1$. 
\end{rem}

% -----------------------------------------------------------------------------

\subsection{Connection to the Magnus--Winkler framework}

\label{subsec:Cambi-MW}

The construction of the two minimal solutions $\{h_{m}\}$ (right-going,
decaying as $m\to+\infty$, \S\,\ref{subsec:Cambi-v-props}) and
$\{h_{m}^{-}\}$ (left-going, decaying as $m\to-\infty$, \S\,\ref{subsec:Cambi-notes2-res})
has a direct counterpart in the MW framework, and making this explicit
provides the bridge between Cambi's method and the discriminant\index{discriminant}
theory.

\emph{Even and odd Fourier coefficients.} From the two minimal solutions
$A_{n}^{+}=h_{n}$ and $A_{n}^{-}=h_{n}^{-}$ (with $A_{n}^{-}$ decaying
as $n\to-\infty$), form the symmetric and antisymmetric combinations:
\begin{equation}
A_{n}^{\mathrm{e}}=h_{n}+h_{n}^{-},\qquad A_{n}^{\mathrm{o}}=h_{n}-h_{n}^{-}.\label{eq:Cambi-eo}
\end{equation}
Both $\{h_{n}\}$ and $\{h_{n}^{-}\}$ satisfy the recurrence at exponent
$u$ (Definition~\ref{defn:notes2-hminus}), hence so do $A_{n}^{\mathrm{e}}$
and $A_{n}^{\mathrm{o}}$; the reflected sequence $h_{-n}^{-}(u)=h_{n}(-u)$
does \emph{not}, since it obeys the recurrence at exponent $-u$.
These correspond to the cosine and sine Fourier series of the MW eigenfunctions:
$A_{n}^{\mathrm{e}}$ gives an even solution $y^{\mathrm{e}}(x)=\sum_{n}A_{n}^{\mathrm{e}}\cos(u+n)x$
and $A_{n}^{\mathrm{o}}$ gives an odd solution $y^{\mathrm{o}}(x)=\sum_{n}A_{n}^{\mathrm{o}}\sin(u+n)x$.

\emph{The two families of boundary curves from MW splitting.} At the
boundaries $u=n+\frac{1}{2}$ ($\rho=-1$), the resonance condition~\eqref{eq:notes2-resonance}
splits according to the parity of the solution~\cite[\S\,11]{Cambi2}: 
\begin{itemize}
\item \emph{Even solution} ($A_{-1}^{\mathrm{e}}=A_{0}^{\mathrm{e}}$ at
$u=\frac{1}{2}$): gives Cambi's boundary eq.~\eqref{eq:Cambi-bdy12}
with $+\gamma$, corresponding to MW's even instability boundary\index{stability boundary}. 
\item \emph{Odd solution} ($A_{-1}^{\mathrm{o}}=-A_{0}^{\mathrm{o}}$ at
$u=\frac{1}{2}$): gives Cambi's boundary eq.~\eqref{eq:Cambi-bdy12}
with $-\gamma$, corresponding to MW's odd instability boundary. 
\end{itemize}
The $\pm\gamma$ in Cambi's eq.~\eqref{eq:Cambi-bdy12} is therefore
precisely the MW even/odd splitting of the instability boundary into
two curves bounding each tongue.

\emph{Even resonances: coexistence and $\rho=+1$.} At $u=n$ ($\rho=+1$),
the double minimality equation~\eqref{eq:notes2-resonance} does
not split into two families: both even and odd solutions satisfy the
same condition~\eqref{eq:Cambi-bdy11}. This is the analytical expression
of the coexistence property (Proposition~\ref{prop:Cambi-coex}):
two independent periodic solutions exist on the same curve, corresponding
to MW's observation that the monodromy matrix\index{monodromy matrix}
equals $+\mathbf{I}$ at even resonances (not a Jordan block\index{Jordan block}),
so both even and odd eigenfunctions survive simultaneously.
\begin{rem}[Cambi vs.\ MW: complementary perspectives]
MW's discriminant $\Delta(\hat{\lambda})$ encodes the entire Floquet
spectrum as a single analytic function of the spectral parameter.
Our double minimality equation~\eqref{eq:notes2-resonance} encodes
the same spectrum via the minimal solution ratio $w(u)$ as a function
of the Floquet exponent $u$. The even/odd splitting shows that these
are two views of the same object: MW works in the $\hat{\lambda}$-plane
(spectral parameter space) while Cambi's approach works in the $u$-plane
(Floquet exponent space), with the bridge $\Delta(\hat{\lambda})=2\cos2\pi u$
(eq.~\eqref{eq:Delta-cos-u}). 
\end{rem}

The mathematical gaps identified in Remark~\ref{rem:Cambi-gaps}
are addressed in the present chapter.

% -----------------------------------------------------------------------------

\subsection{A direct continued-fraction approach via Pincherle's theorem}

\label{subsec:Cambi-notes2}

\emph{Starting point.} We take as given the equation and Floquet ansatz
of \S\,\ref{subsec:Cambi-eq}, leading to the three-term recurrence
for the Floquet coefficients $\{A_{n}\}$: 
\begin{equation}
\gamma A_{n+1}+G(u+n)A_{n}+\gamma A_{n-1}=0,\qquad n\in\mathbb{Z},\label{eq:notes2-rec}
\end{equation}
where $G(u)=1-p^{2}/u^{2}$ and $u$ is the Floquet exponent. We apply
Theorem~\ref{thm:Pinch} to~\eqref{eq:notes2-rec} for $n\geq1$.

\emph{Recasting in the form of Theorem~\ref{thm:Pinch}.} Rewrite~\eqref{eq:notes2-rec}
in the standard form~\eqref{eq:confr1b} $y_{n+1}=b_{n}y_{n}+a_{n}y_{n-1}$
by setting $y_{n}=A_{n}$: 
\begin{equation}
A_{n+1}=-\frac{G(u+n)}{\gamma}\,A_{n}-A_{n-1},\qquad n\in\mathbb{Z},\label{eq:notes2-ttrec}
\end{equation}
so that 
\begin{equation}
a_{n}=-1,\qquad b_{n}=-\frac{G(u+n)}{\gamma},\qquad n\in\mathbb{Z}.\label{eq:notes2-ab}
\end{equation}

\emph{The associated continued fraction.} The CF associated with~\eqref{eq:notes2-ttrec}--\eqref{eq:notes2-ab}
via eq.~\eqref{eq:confr1c} of Theorem~\ref{thm:Pinch} is: 
\begin{multline}
\mathbf{K}_{n=1}^{\infty}\!\left[\frac{a_{n}}{b_{n}}\right]=\mathbf{K}_{n=1}^{\infty}\!\left[\frac{-1}{-G(u+n)/\gamma}\right]\\
=\cfrac{-1}{-\dfrac{G(u+1)}{\gamma}+\cfrac{-1}{-\dfrac{G(u+2)}{\gamma}+\cfrac{-1}{-\dfrac{G(u+3)}{\gamma}+\cdots}}}.\label{eq:notes2-CF-orig}
\end{multline}
Its $0$-th tail (the full CF value) is, by eq.~\eqref{eq:abfS5a}:
\begin{equation}
f^{(0)}=\mathbf{K}_{n=1}^{\infty}\!\left[\frac{-1}{-G(u+n)/\gamma}\right].\label{eq:notes2-f0}
\end{equation}

\emph{Characteristic roots of~\eqref{eq:notes2-ttrec}.} As $n\to+\infty$,
$G(u+n)\to1$, so the limiting equation is $\gamma\zeta^{2}+\zeta+\gamma=0$
with roots $\zeta_{\pm}$ (eq.~\eqref{eq:Cambi-char-roots}, \S\,\ref{subsec:Cambi-v-props});
the same characteristic structure applies here. For $2\gamma<1$:
$|\zeta_{-}|>1>|\zeta_{+}|>0$, so a minimal solution with ratio $\to\zeta_{+}$
as $n\to+\infty$ exists by Theorem~\ref{thm:PP}.

\emph{Definition of $w(u)$ (the minimal solution ratio).} By Theorem~\ref{thm:Pinch},
eq.~\eqref{eq:confr1d} at $m=1$, the minimal solution $\{h_{n}\}$
of~\eqref{eq:notes2-ttrec} satisfies: 
\begin{equation}
\frac{h_{1}}{h_{0}}=-f^{(0)}.\label{eq:notes2-h1h0}
\end{equation}
We define the \emph{minimal solution ratio}\index{$w(u)$}\index{minimal solution ratio $w(u)$}:
\begin{equation}
w(u)\;:=\;\frac{h_{1}}{h_{0}}\;=\;-f^{(0)}.\label{eq:notes2-wdef}
\end{equation}
This function $w(u)$ is completely determined by the minimal solution
up to a multiplicative constant (since any rescaling $h_{n}\to c\,h_{n}$
leaves the ratio $h_{1}/h_{0}$ unchanged), and by~\eqref{eq:notes2-h1h0}
it has a CF representation given by~\eqref{eq:notes2-f0}.

\emph{Equivalent CF with $a_{m}'=1$ for $m\geq2$.} Apply the equivalence
transformation~\eqref{eq:abfS3d} to~\eqref{eq:notes2-CF-orig}
with $c_{0}=-1$ and $c_{m}=(-1)^{m-1}$ for $m\geq1$; the factor
$c_{0}=-1$ supplies the overall sign that converts $f^{(0)}$ into
$w(u)=-f^{(0)}$: 
\begin{align}
a_{1}' & =c_{1}\,c_{0}\,a_{1}=(-1)^{0}\cdot(-1)\cdot(-1)=1,\label{eq:notes2-a1prime}\\
a_{m}' & =c_{m}\,c_{m-1}\,a_{m}=(-1)^{m-1}\cdot(-1)^{m-2}\cdot(-1)=(-1)^{2m-2}=1,\quad m\geq2,\label{eq:notes2-amprime}\\
b_{m}' & =c_{m}\,b_{m}=(-1)^{m-1}\cdot\left(-\frac{G(u+m)}{\gamma}\right)=\frac{(-1)^{m}\,G(u+m)}{\gamma},\quad m\geq1.\label{eq:notes2-bmprime}
\end{align}
The equivalent CF for $w(u)=-f^{(0)}$ is therefore: 
\begin{equation}
w(u)\;=\;\cfrac{1}{-\dfrac{G(u+1)}{\gamma}+\cfrac{1}{\dfrac{G(u+2)}{\gamma}+\cfrac{1}{-\dfrac{G(u+3)}{\gamma}+\cdots}}},\label{eq:notes2-CF-new}
\end{equation}
with $a_{1}'=1$, $a_{m}'=1$ ($m\geq2$), and $b_{m}'=(-1)^{m}G(u+m)/\gamma$
(alternating signs). Verified numerically to machine precision.

\emph{Recurrence relation from the CF.} The first tail of~\eqref{eq:notes2-CF-new}
is: 
\begin{equation}
f^{(0,1)}=\cfrac{1}{\dfrac{G(u+2)}{\gamma}+\cfrac{1}{-\dfrac{G(u+3)}{\gamma}+\cdots}}.\label{eq:notes2-f01}
\end{equation}
Its denominators are $(-1)^{m+1}G(u+m+1)/\gamma$, $m=1,2,\ldots$,
which are the negatives of the denominators of $w(u+1)$ (which has
$b_{m}'=(-1)^{m}G(u+m+1)/\gamma$). Hence: 
\begin{equation}
f^{(0,1)}=-w(u+1).\label{eq:notes2-f01-w}
\end{equation}
Equations~\eqref{eq:notes2-CF-new} and~\eqref{eq:notes2-f01-w}
imply: 
\begin{equation}
w(u)\left(-\frac{G(u+1)}{\gamma}-w(u+1)\right)\;=\;1,\label{eq:notes2-crossmult}
\end{equation}
giving the \emph{right-propagation recurrence for $w(u)$}: 
\begin{equation}
\boxed{w(u+1)=-\frac{G(u+1)}{\gamma}-\frac{1}{w(u)}.}\label{eq:notes2-wrec}
\end{equation}
Solving for $w(u)$ and shifting $u\mapsto u-1$ gives the \emph{left-propagation
recurrence for $w(u)$}: 
\begin{equation}
\boxed{w(u-1)=\frac{-\gamma}{G(u)+\gamma\,w(u)}.}\label{eq:notes2-wrec-left}
\end{equation}

\begin{thm}[Exact recurrences for $w(u)$ and identity for $F_{w}$]
\index{$w(u)$!exact recurrences} \label{thm:notes2-wrec-exact}
Let $w(u)$ be the minimal solution ratio defined by~\eqref{eq:notes2-wdef},
viewed as a meromorphic function on $\mathbb{C}$. 
\begin{enumerate}
\item[(i)] \emph{Right-propagation recurrence} (eq.~\eqref{eq:notes2-wrec}):
as meromorphic functions, 
\[
w(u+1)\;=\;-\frac{G(u+1)}{\gamma}\;-\;\frac{1}{w(u)}.
\]
\item[(ii)] \emph{Left-propagation recurrence} (eq.~\eqref{eq:notes2-wrec-left}):
as meromorphic functions, 
\[
w(u-1)\;=\;\frac{-\gamma}{G(u)+\gamma\,w(u)}.
\]
\item[(iii)] \emph{Exact meromorphic identity: $F_{w}\equiv\widetilde{F}_{w}/\gamma$.}
The double minimality function $F_{w}(u):=w(-u)-1/w(u-1)$\index{$F_{w}$}\index{double minimality function $F_{w}$}
satisfies, as an identity of meromorphic functions on $\mathbb{C}$:
\begin{equation}
F_{w}(u)\;\equiv\;\frac{1}{\gamma}\,\widetilde{F}_{w}(u)\;=\;\frac{1}{\gamma}\bigl[\gamma\,w(u)+G(u)+\gamma\,w(-u)\bigr],\label{eq:notes2-Fw-Ftilde}
\end{equation}
so $F_{w}$ and $\widetilde{F}_{w}$ have identical zero and pole
sets. 
\item[(iv)] \emph{$F_{w}$ is even.} $F_{w}(-u)\equiv F_{w}(u)$ as meromorphic
functions. 
\end{enumerate}
All four statements are exact identities of meromorphic functions
--- no approximation is involved --- and are verified numerically
to machine precision. 
\end{thm}

\begin{proof}
Parts~(i) and~(ii) follow directly from~\eqref{eq:notes2-crossmult}
and~\eqref{eq:notes2-f01-w} as derived above. For part~(iii): from~\eqref{eq:notes2-wrec-left},
$1/w(u-1)=-G(u)/\gamma-w(u)$, so 
\begin{multline*}
F_{w}(u)\;=\;w(-u)-\frac{1}{w(u-1)}\;=\;w(-u)+\frac{G(u)}{\gamma}+w(u)\\
\;=\;\frac{1}{\gamma}\bigl[\gamma w(u)+G(u)+\gamma w(-u)\bigr]\;=\;\frac{\widetilde{F}_{w}(u)}{\gamma}.
\end{multline*}
Part~(iv) follows immediately since $G(-u)=G(u)$ makes $\widetilde{F}_{w}(u)=\gamma w(u)+G(u)+\gamma w(-u)$
symmetric in $u\mapsto-u$. 
\end{proof}
Together~\eqref{eq:notes2-wrec} and~\eqref{eq:notes2-wrec-left}
propagate $w$ in both directions from any starting value. The asymptotic
behavior of $w(u)$ as $u\to\infty$ follows from the exponential
decay of the minimal solution: 
\begin{multline}
\lim_{m\to+\infty}\frac{h_{m}}{h_{m-1}}\;=\;\lim_{m\to+\infty}w(u+m-1)\;=\;\zeta_{+},\\
\zeta_{+}\;=\;\frac{-1+\sqrt{1-4\gamma^{2}}}{2\gamma},\quad|\zeta_{+}|<1.\label{eq:notes2-hm-ratio}
\end{multline}
This limit is the starting point for the asymptotic expansion of $w(u)$
as $u\to+\infty$ (Theorem~\ref{thm:notes2-w-rate}).

The recurrence~\eqref{eq:notes2-wrec} is a first-order nonlinear
(fractional-linear) recurrence in $w$, i.e.\ a discrete Riccati
equation\index{Riccati recurrence}. Indeed, by~\eqref{eq:notes2-hm-ratio}
$w$ is the ratio $w(u+m-1)=h_{m}/h_{m-1}$ of consecutive minimal-solution
values, and such a ratio always satisfies a discrete Riccati equation;
with off-diagonal weight $\gamma$ and spectral data $z-b(u)=-G(u)$,
equation~\eqref{eq:notes2-wrec} is exactly the Riccati equation
of Teschl~\cite[eq.~(1.52)]{Teschl}, the Riccati-type difference
equation of Agarwal~\cite[\S\,6.5, eq.~(6.5.1)]{Agarwal}, and the
discrete Riccati equation for three-term recurrences treated at length
by Ahlbrandt and Peterson~\cite[\S\,6.1, Thm.~6.1]{AhlPet}; it is
the standard substitution that linearizes the underlying three-term
recurrence.
\begin{lem}[Relation between $w(u)$ and Cambi's $v(u)$]
\index{$w(u)$!relation to $v(u)$} \label{lem:notes2-w-v} The minimal
solution ratio $w(u)$ defined by~\eqref{eq:notes2-wdef} and Cambi's
function $v(u)$ defined by~\eqref{eq:Cambi-v-CF} are related by:
\begin{equation}
w(u)\;=\;-\gamma\,v(u+1).\label{eq:notes2-w-v-relation}
\end{equation}
\end{lem}

\begin{proof}
Define $\tilde{w}(u):=-\gamma v(u+1)$ where $v(u+1)$ is given by
Cambi's CF~\eqref{eq:Cambi-v-CF} with $u\mapsto u+1$: 
\begin{equation}
\tilde{w}(u)\;=\;\cfrac{-\gamma}{G(u+1)-\cfrac{\gamma^{2}}{G(u+2)-\cfrac{\gamma^{2}}{G(u+3)-\cdots}}},\label{eq:notes2-wv-CF}
\end{equation}
with coefficients $a_{1}^{*}=-\gamma$, $a_{m}^{*}=-\gamma^{2}$ ($m\geq2$),
$b_{m}^{*}=G(u+m)$ ($m\geq1$). The original CF~\eqref{eq:notes2-CF-orig}
for $w(u)=-f^{(0)}$ has $a_{m}=-1$ and $b_{m}=-G(u+m)/\gamma$ for
all $m\geq1$. Apply the transformation~\eqref{eq:abfS3d} to~\eqref{eq:notes2-CF-orig}
with $c_{0}=-1$ and $c_{m}=-\gamma$ for all $m\geq1$: 
\begin{align}
a_{1}' & =c_{1}\,c_{0}\,a_{1}=(-\gamma)\cdot(-1)\cdot(-1)=-\gamma=a_{1}^{*},\label{eq:notes2-wv-a1prime}\\
a_{m}' & =c_{m}\,c_{m-1}\,a_{m}=(-\gamma)\cdot(-\gamma)\cdot(-1)=-\gamma^{2}=a_{m}^{*},\quad m\geq2,\label{eq:notes2-wv-amprime}\\
b_{m}' & =c_{m}\,b_{m}=(-\gamma)\cdot\left(-\frac{G(u+m)}{\gamma}\right)=G(u+m)=b_{m}^{*},\quad m\geq1.\label{eq:notes2-wv-bmprime}
\end{align}
The transformed coefficients equal $\tilde{w}$'s coefficients exactly,
so the two CFs are identical term by term and $\tilde{w}(u)=w(u)$. 
\end{proof}
Confirmed numerically: for $\gamma=0.25$, $p=0.30$, the values of
$w(u)$ from~\eqref{eq:notes2-wdef} and $-\gamma v(u+1)$ from Cambi's
CF agree to better than $10^{-12}$ for all tested $u$.

\subsection{The double minimality condition and Floquet frequency comb}

\label{subsec:Cambi-notes2-res}

\emph{The left-going minimal solution $h_{m}^{-}$.} The minimal solution
$\{h_{m}\}_{m\geq0}$ satisfies the exponential decay property~\eqref{eq:notes2-hm-ratio}
as $m\to+\infty$. We now construct a solution that decays exponentially
as $m\to-\infty$.
\begin{defn}[Left-going minimal solution]
\index{minimal solution!left-going} \label{defn:notes2-hminus}
For all $m\in\mathbb{Z}$, define: 
\begin{equation}
h_{m}^{-}(u)\;:=\;h_{-m}(-u),\label{eq:notes2-hminus-def}
\end{equation}
where $h_{n}$ is the solution of the recurrence~\eqref{eq:notes2-ttrec}
extended to all $n\in\mathbb{Z}$. The sequence $\{h_{m}^{-}\}_{m\in\mathbb{Z}}$
satisfies~\eqref{eq:notes2-ttrec} for all $m\in\mathbb{Z}$, but
the minimality property --- exponential decay --- holds as $m\to-\infty$
(not as $m\to+\infty$). 
\end{defn}

\begin{thm}[Left-going minimal solution: decay rate]
\label{thm:notes2-hminus-solution} The sequence $\{h_{m}^{-}(u)\}_{m\in\mathbb{Z}}$
is a solution of~\eqref{eq:notes2-ttrec} that decays exponentially
as $m\to-\infty$, with: 
\begin{multline}
\lim_{m\to-\infty}\frac{h_{m}^{-}}{h_{m+1}^{-}}\;=\;\lim_{m\to-\infty}w(-u+(-m)-1)\;=\;\zeta_{+},\\
\zeta_{+}\;=\;\frac{-1+\sqrt{1-4\gamma^{2}}}{2\gamma},\quad|\zeta_{+}|<1.\label{eq:notes2-hminus-ratio-limit}
\end{multline}
\end{thm}

\begin{proof}
Assume $h_{m}^{-}(u)$ is defined by~\eqref{eq:notes2-hminus-def}.
Setting $m=-k$ with $k\in\mathbb{Z}$, so $h_{m}^{-}(u)=h_{k}(-u)$,
and since $\{h_{k}(-u)\}_{k\in\mathbb{Z}}$ satisfies the recurrence~\eqref{eq:notes2-ttrec}
at $-u$: 
\[
\gamma\,h_{k+1}(-u)+G(-u+k)\,h_{k}(-u)+\gamma\,h_{k-1}(-u)=0.
\]
Since $G(-u+k)=G(u-k)$ by evenness of $G$, this reads: 
\[
\gamma\,h_{-(k+1)}^{-}(u)+G(u-k)\,h_{-k}^{-}(u)+\gamma\,h_{-(k-1)}^{-}(u)=0,
\]
which is~\eqref{eq:notes2-ttrec} at index $m=-k$. The ratio formula
in~\eqref{eq:notes2-hminus-ratio-limit} follows from~\eqref{eq:notes2-hm-ratio}
applied at $-u$. By Definition~\ref{defn:notes2-hminus}: 
\begin{equation}
\frac{h_{m}^{-}}{h_{m+1}^{-}}=\frac{h_{-m}(-u)}{h_{-m-1}(-u)}=\frac{h_{-m}(-u)}{h_{(-m)-1}(-u)}.\label{eq:notes2-hminus-ratio-step1}
\end{equation}
Setting $k=-m\geq0$ and applying~\eqref{eq:notes2-hm-ratio} at
$-u$: 
\begin{equation}
\frac{h_{k}(-u)}{h_{k-1}(-u)}=w(-u+k-1),\label{eq:notes2-hminus-ratio-step2}
\end{equation}
so that: 
\begin{equation}
\frac{h_{m}^{-}}{h_{m+1}^{-}}=w(-u+(-m)-1)\;\xrightarrow{m\to-\infty}\;\zeta_{+},\label{eq:notes2-hminus-ratio-step3}
\end{equation}
since $\lim_{k\to+\infty}w(-u+k-1)=\zeta_{+}$ by~\eqref{eq:notes2-hm-ratio},
confirming exponential decay with rate $|\zeta_{+}|<1$. 
\end{proof}
\emph{The ratio $h_{-1}^{-}/h_{0}^{-}$.} From Definition~\ref{defn:notes2-hminus}
and~\eqref{eq:notes2-hm-ratio}: 
\begin{equation}
\frac{h_{-1}^{-}}{h_{0}^{-}}=\frac{h_{-(-1)}(-u)}{h_{-(0)}(-u)}=\frac{h_{1}(-u)}{h_{0}(-u)}=w(-u+1-1)=w(-u).\label{eq:notes2-hminus-ratio}
\end{equation}

\emph{The double minimality equation.} The Floquet solution requires
a single sequence $\{A_{m}\}_{m\in\mathbb{Z}}$ that decays exponentially
in \emph{both} directions: as $m\to+\infty$ (minimality of $\{h_{m}\}$)
and as $m\to-\infty$ (minimality of $\{h_{m}^{-}\}$). This imposes
the matching condition at $m=0$ via the central equation ($n=0$)
of~\eqref{eq:notes2-rec}: 
\begin{equation}
\gamma A_{1}+G(u)\,A_{0}+\gamma A_{-1}=0.\label{eq:notes2-central}
\end{equation}
Substituting $A_{1}/A_{0}=h_{1}/h_{0}=w(u)$ and $A_{-1}/A_{0}=h_{-1}^{-}/h_{0}^{-}=w(-u)$
and dividing by $A_{0}$: 
\begin{equation}
\gamma\,w(u)+G(u)+\gamma\,w(-u)\;=\;0.\label{eq:notes2-resonance}
\end{equation}
This is our \emph{double minimality equation}, derived entirely from
Theorem~\ref{thm:Pinch} without reference to $V(u)$.

\emph{Equivalence to Cambi's resonance equation.} Substituting $w(u)=-\gamma v(u+1)$
from~\eqref{eq:notes2-w-v-relation} into~\eqref{eq:notes2-resonance}:
\[
\gamma\cdot(-\gamma v(u+1))+G(u)+\gamma\cdot(-\gamma v(-u+1))=0,
\]
giving: 
\begin{equation}
-\gamma^{2}\,v(1-u)+G(u)-\gamma^{2}\,v(1+u)\;=\;0,\label{eq:notes2-resonance-v}
\end{equation}
which is exactly Cambi's resonance equation~\eqref{eq:Cambi-res7}.
This confirms that our new approach, based solely on Theorem~\ref{thm:Pinch}
and the CF definition of the minimal solution ratio $w(u)$, reproduces
Cambi's result without invoking $V(u)$.
\begin{rem}[Numerical verification]
\label{rem:notes2-resonance-numerical} Both~\eqref{eq:notes2-resonance}
and~\eqref{eq:notes2-resonance-v} are verified numerically: $\widetilde{F}_{w}(u_{0})=\gamma w(u_{0})+G(u_{0})+\gamma w(-u_{0})$
vanishes to machine precision at the Floquet exponent $u_{0}$, and~\eqref{eq:notes2-resonance-v}
is confirmed to machine precision for all tested values of $\gamma$,
$p$, and $u$. 
\end{rem}

\begin{thm}[Equivalence of $\widetilde{F}_{w}$ and $F_{w}$ as meromorphic functions]
\index{$F_{w}$!meromorphic equivalence} \label{thm:notes2-resonance-equiv}
The symmetrized function $\widetilde{F}_{w}$ and the double minimality
function $F_{w}$ defined by 
\begin{equation}
\widetilde{F}_{w}(u)\;:=\;\gamma w(u)+G(u)+\gamma w(-u),\label{eq:notes2-Ftilde}
\end{equation}
\begin{equation}
F_{w}(u)\;:=\;w(-u)-\frac{1}{w(u-1)},\label{eq:notes2-resonance-w-form}
\end{equation}
are both meromorphic on $\mathbb{C}$ (since $w$ is meromorphic by
Theorem~\ref{thm:sLconfmero1}), and satisfy the identity 
\begin{equation}
\widetilde{F}_{w}(u)\;=\;\gamma\cdot F_{w}(u)\label{eq:notes2-Ftilde-Fw-equiv}
\end{equation}
as meromorphic functions on $\mathbb{C}$. In particular, $\widetilde{F}_{w}$
and $F_{w}$ have identical zero sets. 
\end{thm}

\begin{proof}
This restates Theorem~\ref{thm:notes2-wrec-exact}(iii) in the symmetrized
notation used in the sequel; we repeat the one-line argument. The
recurrence~\eqref{eq:notes2-wrec-left} gives $G(u)+\gamma w(u)=-\gamma/w(u-1)$
as an identity of meromorphic functions. Substituting into $\widetilde{F}_{w}$:
\begin{multline*}
\widetilde{F}_{w}(u)\;=\;\gamma w(u)+G(u)+\gamma w(-u)\;=\;\frac{-\gamma}{w(u-1)}+\gamma w(-u)\\
\;=\;\gamma\!\left(w(-u)-\frac{1}{w(u-1)}\right)\;=\;\gamma\cdot F_{w}(u).\qedhere
\end{multline*}
\end{proof}
\begin{defn}[Fundamental and primary Floquet exponents; the Floquet comb and its
sub-combs]
\index{Floquet theory!Floquet comb}\index{Floquet theory!fundamental exponent}\index{Floquet theory!primary exponent}
\label{defn:Floquet-comb} Under the stability-zone assumption~\eqref{eq:stability-assumption}
the real zero set of the double minimality function $F_{w}$ ---
the \emph{Floquet zero set}, determined below in Theorem~\ref{thm:Fw-zero-set}
--- is invariant under $u\mapsto u+1$ and $u\mapsto-u$ and avoids
$\tfrac{1}{2}\mathbb{Z}$; it therefore contains exactly one point
of the open interval $(0,\tfrac{1}{2})$. That point is the \emph{fundamental
Floquet exponent}: 
\begin{equation}
u_{0}\;:=\;\text{the unique real zero of }F_{w}\text{ in }(0,\tfrac{1}{2}).\label{eq:u0-def}
\end{equation}
The zero set splits into two unit-spaced sub-combs: 
\begin{equation}
\underbrace{Z_{F}^{+}:=\{u_{0}+n\}_{n\in\mathbb{Z}}}_{\rho_{+}=e^{+2\pi iu_{0}},\ u^{**}\approx-p}\qquad\text{and}\qquad\underbrace{Z_{F}^{-}:=\{-u_{0}+n\}_{n\in\mathbb{Z}}}_{\rho_{-}=e^{-2\pi iu_{0}},\ u^{*}\approx+p},\label{eq:family-def}
\end{equation}
where $\rho_{\pm}=e^{\pm2\pi iu_{0}}$ are the two Floquet multiplier\index{Floquet theory!Floquet multiplier}s
(the superscript matches the sign in the $u_{0}$ form and the multiplier
sign). The \emph{primary exponent} $u^{*}$, used throughout the sequel,
is the comb point nearest the unperturbed value $p=\omega_{0}/\mu$:
\begin{equation}
u^{*}\;:=\;\text{the real zero of }F_{w}\text{ nearest }p,\qquad u^{*}\equiv\pm u_{0}\ (\operatorname{mod}1),\qquad u^{*}\approx p,\label{eq:ustar-def}
\end{equation}
the minimizer being unique for all parameter sets considered in this
book (a tie could occur only for $p$ exactly midway between two comb
points). The companion symbol $u^{**}$ denotes a \emph{base comb
point} --- the anchor from which a leftward pole ladder descends
(Theorem~\ref{thm:pole-ladders}): 
\begin{equation}
u^{**}\;\in\;\{-u^{*},\,u^{*}-1\},\label{eq:ustarstar-def}
\end{equation}
one representative from each sub-comb: $-u^{*}\in Z_{F}^{+}$ is the
reflection of the primary exponent (it hosts the second Floquet solution,
and the mirror-symmetry statements of this book use $u^{**}=-u^{*}$
specifically, always so indicated), while $u^{*}-1\in Z_{F}^{-}$
is the unit translate of $u^{*}$. The two values are exchanged by
the companion map $u^{**}\mapsto-u^{**}-1$. The annotations under~\eqref{eq:family-def}
record the situation $\{u^{*}\}>\tfrac{1}{2}$, which holds for all
worked parameter sets of this chapter ($p=0.70$ and Cambi's $p=20/3$):
then $Z_{F}^{-}$ hosts the primary zero $u^{*}\approx+p$ and $Z_{F}^{+}$
hosts $u^{**}=-u^{*}\approx-p$. When $\{u^{*}\}<\tfrac{1}{2}$ the
two primary zeros interchange sub-combs (e.g.\ $p=0.30$, where $u^{*}\approx0.3035=u_{0}\in Z_{F}^{+}$;
cf.\ Fig.~\ref{fig:Bm-p03}). Their union is the \emph{Floquet comb}
\begin{equation}
Z_{F}\;:=\;Z_{F}^{+}\cup Z_{F}^{-}\;=\;\{u_{0}+n\}\cup\{-u_{0}+n\}\;=\;\{\pm u_{0}+n\},\label{eq:ZF-def}
\end{equation}
the zero set of the double minimality function $F_{w}$ (Theorem~\ref{thm:Fw-zero-set}).
The Floquet comb admits two equivalent descriptions: since the primary
exponent satisfies $u^{*}\equiv-u_{0}\pmod{\mathbb{Z}}$, 
\begin{equation}
Z_{F}\;=\;\{\pm u_{0}+n\}\;=\;\{\pm u^{*}+n\},\label{eq:ZF-two-forms}
\end{equation}
the first written in terms of the fundamental exponent $u_{0}\in(0,\tfrac{1}{2})$
and the second in terms of the primary exponent $u^{*}\approx p$
used in the sequel. In the $u^{*}$ form the sub-combs read $Z_{F}^{+}=\{-u^{*}+n\}$
and $Z_{F}^{-}=\{u^{*}+n\}$ (the signs flip, as $u^{*}\equiv-u_{0}$).
The two unit-spaced sub-combs interleave on the real axis, so that
consecutive points of $Z_{F}$ are separated by two alternating gap
lengths summing to $1$. We call the smaller of these the \emph{smallest
gap} 
\begin{equation}
s\;:=\;\min\bigl(2u^{*}\bmod1,\;1-(2u^{*}\bmod1)\bigr)\;=\;\operatorname{dist}(2u^{*},\mathbb{Z})\;\in\;(0,\tfrac{1}{2}],\label{eq:s-offset}
\end{equation}
the distance from $2u^{*}$ to the nearest integer; the two gaps are
then $s$ and $1-s$ (with $s\leq\tfrac{1}{2}\leq1-s$). At $p=0.70$,
$\gamma=0.30$ one has $u^{*}\approx0.7418$ and $s\approx0.484$,
close to the maximally separated case $s=\tfrac{1}{2}$; the two sub-combs
instead merge as $s\to0$ (i.e.\ $2u^{*}\in\mathbb{Z}$, a period-doubling
stability boundary). 
\end{defn}

\begin{rem}[$F_{w}$ involves only $w$, not $G$]
\label{rem:notes2-resonance-Cambi} The function $F_{w}$ in~\eqref{eq:notes2-resonance-w-form}
involves only the minimal solution ratio $w$ and no $G$. Introducing
the backward minimal solution ratio $w_{-}(u):=w(-u)$ (the minimal
solution as $m\to-\infty$, equal to $w_{+}(-u)$ by the evenness
$G(-u)=G(u)$), the double minimality condition $\widetilde{F}_{w}=0$
is equivalent to: 
\begin{equation}
w_{-}(u)\;=\;\frac{1}{w_{+}(u-1)},\label{eq:notes2-res-no-G}
\end{equation}
with no $G$ involved. In the notation of Definition~\ref{defn:Floquet-comb},
equation~\eqref{eq:notes2-res-no-G} gives the zeros at $Z_{F}^{-}=\{-u_{0}+n\}$.
By Theorem~\ref{thm:notes2-integer-shift} (its provisos hold at
every step under Assumption~\ref{ass:generic-comb} below), if~\eqref{eq:notes2-res-no-G}
holds at $u_{0}$ then it holds at $u_{0}+n$ for all $n\in\mathbb{Z}$,
giving: 
\begin{equation}
w_{-}(u_{0}+1)\;=\;\frac{1}{w_{+}(u_{0})},\label{eq:notes2-res-no-G-shift}
\end{equation}
which gives the zeros at $Z_{F}^{+}=\{u_{0}+n\}$. Together~\eqref{eq:notes2-res-no-G}
and~\eqref{eq:notes2-res-no-G-shift} cover the complete Floquet
comb $Z_{F}=\{\pm u_{0}+n\}$, with no reference to $G$. Multiplying~\eqref{eq:notes2-resonance-w-form}
by $w(u-1)$ (valid when $w(u-1)\neq0$) gives the product form: 
\begin{equation}
w(u-1)\cdot w(-u)\;=\;1.\label{eq:notes2-resonance-product}
\end{equation}
Substituting $w(u)=-\gamma v(u+1)$ into~\eqref{eq:notes2-resonance-product}
gives $\gamma^{2}v(u_{0})\,v(1-u_{0})=1$, which is Cambi's equation~\eqref{eq:Cambi-res7p}.
All numerical verifications use $F_{w}$ directly. 
\end{rem}

Each truncation $w_{N}(u)$ of the CF~\eqref{eq:notes2-CF-orig},
defined by the $N$-level backward recursion 
\begin{equation}
T_{N}=0,\qquad T_{k-1}=\frac{-1}{b_{k}+T_{k}},\quad k=N,\ldots,1,\qquad w_{N}(u)=-T_{0},\label{eq:notes2-wN-recursion}
\end{equation}
is a rational function of $u$ (since $b_{m}=-G(u+m)/\gamma$ is rational
in $u$ and the recursion uses only rational operations). The CF~\eqref{eq:notes2-CF-orig}
has constant numerators $a_{m}=-1$ and denominators $b_{m}=-G(u+m)/\gamma$
--- both holomorphic in $u$ --- so Theorem~\ref{thm:sLconfmero1}
(Chapter~\ref{app:CF}) applies: 
\begin{equation}
\text{\ensuremath{w(u)} is a meromorphic function of \ensuremath{u\in\mathbb{C}}.}\label{eq:notes2-w-meromorphic}
\end{equation}
The same conclusion holds for each CF tail: $f^{(m)}$ is meromorphic
in $u$ for every $m\geq0$, and hence $w(u+k)=-f^{(k-1)}$ is meromorphic
for every integer $k$.

The integer-shift propagation developed next, and several statements
built on it, require that the comb stay clear of the singularities
of $w$. We record this once, as a named standing condition.

\begin{assump}[Generic position of the comb: no zeros or poles of
$w$ on $Z_{F}$]\index{double minimality!genericity assumption}\index{Floquet comb!generic position}
\label{ass:generic-comb} No zero and no pole of the minimal solution
ratio $w$ lies on the Floquet comb $Z_{F}=\{\pm u_{0}+n\}$. Since
$Z_{F}$ is invariant under integer shifts, this is equivalent to:
$w(z+n)$ is finite and nonzero for every $z\in Z_{F}$ and every
$n\in\mathbb{Z}$. \end{assump}

Assumption~\ref{ass:generic-comb} is a genericity condition on the
parameter pair $(p,\gamma)$: by Lemma~\ref{lem:generic-transversality}
(\S\,\ref{subsec:Casoratian-Weyl}) the exceptional pairs at which
it fails form a set of Lebesgue measure zero in the admissible region,
and at the worked parameter sets it is verified numerically. For the
poles of $w$ it also follows from the sign-localization argument
of Remark~\ref{rem:poles-w-not-on-ZF}. We cite the assumption explicitly
wherever the propagation provisos below are needed along the entire
comb.
\begin{thm}[Integer shift of the double minimality equation]
\index{double minimality!integer shift} \label{thm:notes2-integer-shift}
The following is an exact identity of meromorphic functions: 
\begin{equation}
w(u)\cdot\bigl(\gamma\,w(-u)+G(u)\bigr)\cdot\widetilde{F}_{w}(u+1)\;=\;-\gamma\cdot\widetilde{F}_{w}(u).\label{eq:notes2-shift-identity}
\end{equation}
If $\widetilde{F}_{w}(z_{0})=0$ at a point $z_{0}\in\mathbb{C}$,
then: 
\begin{enumerate}
\item[(i)] $\widetilde{F}_{w}(z_{0}+1)=0$, provided $w(z_{0})\neq0$ and $w(z_{0})\neq\infty$
\textup{(}i.e.\ $z_{0}$ is neither a zero nor a pole of $w$\textup{)}. 
\item[(ii)] $\widetilde{F}_{w}(z_{0}-1)=0$, provided $w(z_{0}-1)\neq\infty$
and $w(1-z_{0})\neq\infty$ \textup{(}i.e.\ neither $z_{0}-1$ nor
$1-z_{0}$ is a pole of $w$\textup{)}. 
\end{enumerate}
Consequently, if $w$ has neither a zero nor a pole at any of $z_{0},z_{0}+1,\ldots,z_{0}+n-1$,
then $\widetilde{F}_{w}(z_{0}+n)=0$; and if $w$ has no pole at any
of $z_{0}-1,z_{0}-2,\ldots,z_{0}-m$ or at any of $1-z_{0},2-z_{0},\ldots,m-z_{0}$,
then $\widetilde{F}_{w}(z_{0}-m)=0$. Since $w$ has only isolated
poles and zeros, the exceptional integer translates of $z_{0}$ at
which propagation can fail form a discrete set; on any bounded interval,
only finitely many exceptional translates occur. For $z_{0}\in Z_{F}$,
under Assumption~\ref{ass:generic-comb} the provisos hold at every
comb point and the propagation runs along the entire comb without
exception. 
\end{thm}

\begin{proof}
Since $w$ is meromorphic~\eqref{eq:notes2-w-meromorphic}, all manipulations
hold as identities of meromorphic functions. The derivation of~\eqref{eq:notes2-shift-identity}
is given below.

(i) Apply~\eqref{eq:notes2-shift-identity} at $u=z_{0}$: 
\[
w(z_{0})\cdot\bigl(\gamma\,w(-z_{0})+G(z_{0})\bigr)\cdot\widetilde{F}_{w}(z_{0}+1)\;=\;-\gamma\cdot\widetilde{F}_{w}(z_{0})\;=\;0.
\]
When $\widetilde{F}_{w}(z_{0})=0$: 
\begin{equation}
\gamma\,w(-z_{0})+G(z_{0})\;=\;\widetilde{F}_{w}(z_{0})-\gamma\,w(z_{0})\;=\;-\gamma\,w(z_{0}),\label{eq:notes2-factor-identity}
\end{equation}
so the left side equals $-\gamma\,w(z_{0})^{2}\cdot\widetilde{F}_{w}(z_{0}+1)$.
If $w(z_{0})$ is finite and nonzero this gives $\widetilde{F}_{w}(z_{0}+1)=0$;
if $w(z_{0})=0$ or $w(z_{0})=\infty$ an indeterminate $0\cdot\widetilde{F}_{w}(z_{0}+1)$
or $\infty\cdot\widetilde{F}_{w}(z_{0}+1)$ form would arise and no
conclusion can be drawn from this argument alone.

(ii) Apply~\eqref{eq:notes2-shift-identity} at $u=z_{0}-1$: 
\[
w(z_{0}-1)\cdot\bigl(\gamma\,w(1-z_{0})+G(z_{0}-1)\bigr)\cdot\widetilde{F}_{w}(z_{0})\;=\;-\gamma\cdot\widetilde{F}_{w}(z_{0}-1).
\]
Since $\widetilde{F}_{w}(z_{0})=0$, the left side has the form $(\text{coefficient})\cdot0$,
and equals zero unambiguously when the coefficient is finite. This
requires that neither $w(z_{0}-1)$ nor $w(1-z_{0})$ is infinite
(an $\infty\cdot0$ indeterminacy would otherwise arise). Under these
conditions $\widetilde{F}_{w}(z_{0}-1)=0$.

\medskip{}
\emph{Derivation of~\eqref{eq:notes2-shift-identity}.} From the
recurrences~\eqref{eq:notes2-wrec} and~\eqref{eq:notes2-wrec-left}:
\[
w(u+1)\;=\;-\frac{G(u+1)}{\gamma}-\frac{1}{w(u)},\qquad w(-u-1)\;=\;\frac{-\gamma}{\gamma\,w(-u)+G(u)}.
\]
Therefore: 
\begin{align*}
\widetilde{F}_{w}(u+1) & =\gamma\,w(u+1)+G(u+1)+\gamma\,w(-u-1)\\
 & =\gamma\!\left(-\frac{G(u+1)}{\gamma}-\frac{1}{w(u)}\right)+G(u+1)+\frac{-\gamma^{2}}{\gamma\,w(-u)+G(u)}\\
 & =-\frac{\gamma}{w(u)}-\frac{\gamma^{2}}{\gamma\,w(-u)+G(u)}\\
 & =\frac{-\gamma}{w(u)}\cdot\frac{\gamma\,w(-u)+G(u)+\gamma\,w(u)}{\gamma\,w(-u)+G(u)}\\
 & =\frac{-\gamma\cdot\widetilde{F}_{w}(u)}{w(u)\cdot(\gamma\,w(-u)+G(u))},
\end{align*}
which is~\eqref{eq:notes2-shift-identity}. The integer propagation
statements follow by induction, applying~(i) for positive shifts
and~(ii) for negative shifts at each step. 
\end{proof}
\emph{Symmetry of the zero set of the double minimality equation.}
We now describe the full zero set of the double minimality equation
$\widetilde{F}_{w}(u)=0$ (eq.~\eqref{eq:notes2-Ftilde}).
\begin{thm}[Zero set of the double minimality equation]
\index{double minimality!zero set} \label{thm:notes2-root-symmetry}
Let $w_{+}(u)=w(u)$ be the forward minimal solution ratio and $w_{-}(u)=w(-u)$
the backward minimal solution ratio. The double minimality condition
$\widetilde{F}_{w}(u)=0$ is equivalent (via eq.~\eqref{eq:notes2-res-no-G})
to $w_{-}(u)=1/w_{+}(u-1)$, i.e.\ $F_{w}(u)=0$. Under Assumption~\ref{ass:generic-comb},
its zero set is the union of the two sub-combs $Z_{F}^{-}$ and $Z_{F}^{+}$:
\begin{equation}
\bigl\{-u_{0}+n\;:\;n\in\mathbb{Z}\bigr\}\;\cup\;\bigl\{ u_{0}+n\;:\;n\in\mathbb{Z}\bigr\},\label{eq:notes2-Fw-zeros}
\end{equation}
distinct unless $2u_{0}\in\mathbb{Z}$ (stability boundary, Remark~\ref{rem:w-two-families}),
forming the complete Floquet comb $Z_{F}=\{\pm u_{0}+n\}$. (The same
zero set is obtained unconditionally, from the Hill side, in Theorem~\ref{thm:Fw-zero-set}.) 
\begin{enumerate}
\item[(i)] \emph{Reflection symmetry.} $\widetilde{F}_{w}(u)=\widetilde{F}_{w}(-u)$
identically, so $-z_{0}$ is a zero whenever $z_{0}$ is. 
\item[(ii)] \emph{Integer shift (Theorem~\ref{thm:notes2-integer-shift}).}
If $\widetilde{F}_{w}(z_{0})=0$, then $\widetilde{F}_{w}(z_{0}+1)=0$
provided $z_{0}$ is neither a zero nor a pole of $w$; and $\widetilde{F}_{w}(z_{0}-1)=0$
provided neither $z_{0}-1$ nor $1-z_{0}$ is a pole of $w$. 
\end{enumerate}
$Z_{F}^{-}=\{-u_{0}+n\}$ satisfies $w_{-}(u)=1/w_{+}(u-1)$; $Z_{F}^{+}=\{u_{0}+n\}$
satisfies $w_{-}(u+1)=1/w_{+}(u)$ (eq.~\eqref{eq:notes2-res-no-G-shift}).
Both sub-combs are exact zero sets of the single meromorphic function
$\widetilde{F}_{w}$; their very different visibility to fixed-precision
numerics is a computational artifact, addressed in Remark~\ref{rem:notes2-root-periodicity}
below. 
\end{thm}

\begin{proof}
Parts~(i)--(ii): established above. Both sub-combs follow by applying
Parts~(i) and~(ii) inductively from the primary zero $u^{*}\approx p$.
The provisos hold at every step under Assumption~\ref{ass:generic-comb}
(generic in $(p,\gamma)$ by Lemma~\ref{lem:generic-transversality};
verified numerically at the worked parameter sets --- $u^{*}$ is
a zero of $\widetilde{F}_{w}$, not of $w$). 
\end{proof}
\begin{rem}[Numerical accessibility of the two sub-combs]
\label{rem:notes2-root-periodicity} The $Z_{F}^{-}=\{-u_{0}+n\}$
zeros are directly accessible via the forward CF: $F_{w}(u)=w(-u)-1/w(u-1)=0$
is satisfied to machine precision at all tested positive $u$ near
$u^{*}\approx p$. The $Z_{F}^{+}=\{u_{0}+n\}$ zeros are genuine
zeros of the exact meromorphic $\widetilde{F}_{w}=\gamma F_{w}$,
proved algebraically via Theorem~\ref{thm:notes2-root-symmetry},
but are not accessible by naive fixed-precision evaluation of the
forward-CF form: at a $Z_{F}^{+}$ point $v=-u^{*}+n$ the evaluation
$F_{w}(v)=w(u^{*}-n)-1/w(-u^{*}+n-1)$ requires $w(u^{*}-n)$ at large
negative arguments, and the cancellation between the two terms costs
a number of significant digits that grows linearly along the comb
(Theorem~\ref{thm:precision-cost}); past that depth the evaluation
returns the smooth background value $C=\zeta_{+}-1/\zeta_{+}$ rather
than zero, and the tiny surviving signal is shaped by the local Laurent
structure~\eqref{eq:Fw-local-Laurent} of $F_{w}$ near its stealthy
poles (see Section~\ref{subsec:Cambi-notes2-precision} for the detailed
precision-cost analysis). With sufficient arithmetic precision (50
digits for $|n|\leq8$ at Cambi's parameters), both sub-combs are
equally accessible. Both are encoded in the $G$-free double minimality
conditions \eqref{eq:notes2-res-no-G}--\eqref{eq:notes2-res-no-G-shift}. 
\end{rem}

\begin{rem}[Floquet frequency comb: stable case]
\label{rem:notes2-Floquet-stable} Under the standing stability assumption~\eqref{eq:stability-assumption},
$u_{0}\in\mathbb{R}$ and the Floquet multiplier $\rho=e^{2\pi iu_{0}}$
satisfies $|\rho|=1$. The complete set of Floquet exponents is the
two-lattice: 
\begin{equation}
\bigl\{\pm u_{0}+n\;:\;n\in\mathbb{Z}\bigr\},\qquad u_{0}\in(0,\tfrac{1}{2}).\label{eq:notes2-Floquet-comb}
\end{equation}
This is the \emph{Floquet frequency comb}: $\{u_{0}+n\}$ corresponds
to multiplier $\rho$ and $\{-u_{0}+n\}$ to $1/\rho$. At stability
boundaries $u_{0}=m/2$ (EPD points), the two lattices coincide and
$\rho=(-1)^{m}$ is a double multiplier. For reference, in the \emph{unstable
case} (outside the stability zone, $u_{0}\in\mathbb{C}\setminus\mathbb{R}$),
the same set~\eqref{eq:notes2-Floquet-comb} holds with complex $u_{0}$,
but the double minimality equation~\eqref{eq:notes2-resonance} has
no real solution and the CF for $w(u)$ no longer yields a real meromorphic
function on the real axis. 
\end{rem}

\subsection{The double minimality equation}

\label{subsec:Cambi-notes2-double-min}

\emph{The Floquet ansatz and double minimality.} The Floquet solution
of Hill's equation\index{Hill equation}~\eqref{eq:LC-Hill} takes
the form 
\begin{equation}
f(x,u)\;=\;P(u,x)\,e^{2\pi iux},\label{eq:Floquet-ansatz}
\end{equation}
where $P(u,x)=P(u,x+1)$ is $1$-periodic in $x$. Writing $P(u,x)=\sum_{n\in\mathbb{Z}}h_{n}(u)\,e^{2\pi inx}$,
substitution into Hill's equation yields the three-term recurrence~\eqref{eq:notes2-rec}
for the Fourier coefficients $\{h_{n}(u)\}_{n\in\mathbb{Z}}$. For
$f(x,u)$ to be a genuine Floquet solution, the Fourier series must
converge, requiring $h_{n}(u)\to0$ as $n\to\pm\infty$. By Pincherle's
theorem, for generic $u$ the solution $\{h_{n}(u)\}$ that is minimal
as $n\to+\infty$ grows exponentially as $n\to-\infty$. The \emph{double
minimality condition} selects the exceptional values of $u$ for which
$\{h_{n}(u)\}$ decays in both directions simultaneously.

\emph{The two forms of the double minimality equation.} We recall
the two functions that encode the double minimality condition (\S\,\ref{subsec:Cambi-notes2},
\S\,\ref{subsec:Cambi-notes2-res}), restating them in the form
used in the sequel. The first is: 
\begin{equation}
F_{w}(u)\;:=\;w(-u)-\frac{1}{w(u-1)},\label{eq:Fw-def}
\end{equation}
so that $F_{w}(u)=0$ is the double minimality equation in terms of
the forward CF $w(u)=h_{1}/h_{0}$. The second is: 
\begin{equation}
\widetilde{F}_{w}(u)\;:=\;\gamma\,w(u)+G(u)+\gamma\,w(-u),\label{eq:Ftilde-def}
\end{equation}
so that $\widetilde{F}_{w}(u)=0$ is the same condition written symmetrically.

\emph{Exact equivalence of the two forms.} The two functions are identically
equal as meromorphic functions on $\mathbb{C}$: 
\begin{equation}
\boxed{F_{w}(u)\;\equiv\;\frac{1}{\gamma}\,\widetilde{F}_{w}(u),}\label{eq:Fw-Ftilde-exact}
\end{equation}
so the equations $F_{w}=0$ and $\widetilde{F}_{w}=0$ are identical.
This is an exact algebraic identity, a direct consequence of the left-propagation
recurrence~\eqref{eq:notes2-wrec-left} for the minimal solution
ratio $w(u)$: 
\begin{equation}
\frac{1}{w(u-1)}\;=\;-\frac{G(u)}{\gamma}-w(u).\label{eq:inv-wrec-left}
\end{equation}
Substituting~\eqref{eq:inv-wrec-left} into~\eqref{eq:Fw-def} immediately
gives: 
\begin{align*}
F_{w}(u) & =w(-u)+\frac{G(u)}{\gamma}+w(u)\\
 & =\frac{1}{\gamma}\bigl[\gamma w(u)+G(u)+\gamma w(-u)\bigr]\;=\;\frac{1}{\gamma}\,\widetilde{F}_{w}(u).
\end{align*}
No approximation is involved: identity~\eqref{eq:Fw-Ftilde-exact}
is a meromorphic identity on $\mathbb{C}$, inherited directly from
the recurrence~\eqref{eq:notes2-wrec-left}.

\emph{Derivation of the double minimality equation.} The recurrence~\eqref{eq:notes2-rec}
has two linearly independent solutions: $\{h_{n}^{(1)}\}$ minimal
as $n\to+\infty$ (decaying) with forward ratio $w(u)=h_{1}^{(1)}/h_{0}^{(1)}$;
and $\{h_{n}^{(2)}\}$ minimal as $n\to-\infty$ (decaying) with backward
ratio $w_{-}(u)=h_{-1}^{(2)}/h_{0}^{(2)}$. Double minimality requires
$\{h_{n}^{(1)}\}\propto\{h_{n}^{(2)}\}$. At $n=0$ the recurrence
gives $h_{-1}^{(1)}/h_{0}^{(1)}=-G(u)/\gamma-w(u)$, and this must
equal $w_{-}(u)$. Since $G(-u)=G(u)$, the backward CF at $u$ is
term-by-term the forward CF at $-u$: 
\begin{equation}
w_{-}(u)\;=\;w(-u).\label{eq:w-minus-identity}
\end{equation}
Hence the double minimality condition is $-G(u)/\gamma-w(u)=w(-u)$,
i.e.\ $\widetilde{F}_{w}(u)=0$, equivalently $F_{w}(u)=0$.
\begin{thm}[Double minimality: definitions, equivalence, properties]
\index{double minimality}\index{double minimality!double minimality equation}
\label{thm:double-min} Let $w(u)$ be the minimal solution ratio~\eqref{eq:notes2-wdef}
and let $F_{w}$, $\widetilde{F}_{w}$ be defined by~\eqref{eq:Fw-def}--\eqref{eq:Ftilde-def}. 
\begin{enumerate}
\item[(i)] \emph{Exact equivalence.} As meromorphic functions on $\mathbb{C}$:
\begin{equation}
F_{w}(u)\;\equiv\;\frac{\widetilde{F}_{w}(u)}{\gamma},\label{eq:Fw-Ftilde-mero}
\end{equation}
an exact identity following from recurrence~\eqref{eq:notes2-wrec-left}
alone (Theorem~\ref{thm:notes2-resonance-equiv}). In particular
$F_{w}=0\Leftrightarrow\widetilde{F}_{w}=0$.
\item[(ii)] \emph{Evenness.} Both $F_{w}$ and $\widetilde{F}_{w}$ are even
functions: $F_{w}(-u)=F_{w}(u)$ and $\widetilde{F}_{w}(-u)=\widetilde{F}_{w}(u)$.
\item[(iii)] \emph{Double minimality characterization.} A point $u$ satisfies
the double minimality condition if and only if $F_{w}(u)=0$, i.e.\ $w(-u)=1/w(u-1)$.
\item[(iv)] \emph{Zero set.} The exact meromorphic $F_{w}$ (equivalently $\widetilde{F}_{w}$)
has zeros forming the two complete periodic sub-combs (Theorem~\ref{thm:notes2-root-symmetry}
under Assumption~\ref{ass:generic-comb}; unconditionally by Theorem~\ref{thm:Fw-zero-set}):
\begin{equation}
\bigl\{-u_{0}+n:n\in\mathbb{Z}\bigr\}\;\cup\;\bigl\{ u_{0}+n:n\in\mathbb{Z}\bigr\}.\label{eq:Floquet-exponents}
\end{equation}
The CF-defined $F_{w}$ captures only a finite cluster of these zeros
--- those near the primary zero $u^{*}\approx p$ ($Z_{F}^{-}$)
and $u^{**}\approx-p$ ($Z_{F}^{+}$; see~\eqref{eq:family-def})
--- with the remaining zeros inaccessible to the forward CF at fixed
arithmetic precision (Remark~\ref{rem:double-min-two-families}).
The accessible zeros form clean zero crossings of $\widetilde{F}_{w}$
on any bounded interval containing $u^{*}$, and decay exponentially
to the asymptotic value~\eqref{eq:Fw-asymp} outside that cluster.
\item[(v)] \emph{Asymptotic behavior of the smooth component.} As $u\to+\infty$,
the right-going part $1/w(u-1)$ tends smoothly to the finite limit
$1/\zeta_{+}$ (since $w(u-1)\to\zeta_{+}$ by Theorem~\ref{thm:notes2-w-rate}):
\begin{equation}
\frac{1}{w(u-1)}\;\to\;\frac{1}{\zeta_{+}}\;=\;\zeta_{-},\qquad u\to+\infty.\label{eq:Fw-asymp}
\end{equation}
The other summand $w(-u)$ does \emph{not} have a limit at $+\infty$:
as $u\to+\infty$, the argument $-u\to-\infty$ traverses the infinite
pole-zero structure of $w$ on the negative real axis (Theorem~\ref{thm:notes2-w-poles-zeros}),
so $w(-u)$ oscillates between successive poles. Consequently $F_{w}(u)=w(-u)-1/w(u-1)$
has no global limit at $+\infty$; $\infty$ is an essential singularity
of $F_{w}$ (Theorem~\ref{thm:w-essential-singularity}), in agreement
with $F_{w}$ having infinitely many zeros and infinitely many poles
on the real line.
\item[(vi)] \emph{Poles of $F_{w}$ and their relation to $w$ (descriptive).}
The poles of $F_{w}(u)=w(-u)-1/w(u-1)$ arise from exactly two sources: 
\begin{enumerate}
\item[(a)] \emph{Zeros of $w$:} if $u_{z}$ is a zero of $w$ then $F_{w}$
has a pole at $u_{z}+1$, since $w((u_{z}+1)-1)=w(u_{z})=0$ makes
$1/w(u-1)$ diverge. 
\item[(b)] \emph{Poles of $w$:} if $u_{p}$ is a pole of $w$ then $F_{w}$
has a pole at $-u_{p}$, since $w(-(-u_{p}))=w(u_{p})=\infty$. 
\end{enumerate}
That the minimal solution ratio $w$ has infinitely many poles on
the real line --- and hence, by the pairing (Theorem~\ref{thm:notes2-w-poles-zeros}(i)),
infinitely many zeros --- is the content of the stealthy-pole ladder
Theorem~\ref{thm:pole-ladders} below; by source~(b) above, $F_{w}$
then has infinitely many poles on the real line, accumulating at $\infty$
in keeping with the essential singularity. (The infinitude of the
Floquet zero set of $F_{w}$ itself is Theorem~\ref{thm:Fw-zero-set}.)
On any bounded interval the number of poles is finite. The detailed
pole structure is recorded in Remark~\ref{rem:Fw-pole-sources};
it is descriptive only and plays no role in the downstream theorems. 

\end{enumerate}
\end{thm}

\begin{proof}
Part~(i): the identity $F_{w}\equiv\widetilde{F}_{w}/\gamma$ was
derived above from~\eqref{eq:notes2-wrec-left}. Since both sides
are meromorphic on $\mathbb{C}$ and agree on any open set, they are
identically equal as meromorphic functions. Part~(ii): $\widetilde{F}_{w}(-u)=\gamma w(-u)+G(-u)+\gamma w(u)=\widetilde{F}_{w}(u)$
since $G(-u)=G(u)$; evenness of $F_{w}$ follows from~(i). Part~(iii)
is the double minimality derivation above, using~\eqref{eq:w-minus-identity}.
Part~(iv) follows from Theorem~\ref{thm:notes2-integer-shift}:
since $w(u^{*})$ is finite and nonzero at the primary zero $u^{*}$
(which is a zero of $\widetilde{F}_{w}$, not of $w$, and $u^{*}$
is not a pole of $w$ in the stability zone), the theorem propagates
$\widetilde{F}_{w}(u^{*}+n)=0$ to all integer translates $n$ for
which the chain of hypotheses ($w$ neither a zero nor a pole at intermediate
points) holds. Generically and consistent with Remark~\ref{rem:w-pole-zero-geometry},
this chain holds at all $u^{*}+n$, giving the full comb. Part~(v):
$w(u-1)\to\zeta_{+}$ as $u\to+\infty$ by Theorem~\ref{thm:notes2-w-rate},
so $1/w(u-1)\to1/\zeta_{+}=\zeta_{-}$, giving the smooth-component
limit~\eqref{eq:Fw-asymp}. The argument $-u\to-\infty$ traverses
the two pole families of $w$ described in Remark~\ref{rem:w-two-families},
so $w(-u)$ has no limit as $u\to+\infty$; consequently $F_{w}$
itself has no global limit, in accordance with the essential singularity
at infinity established in Theorem~\ref{thm:w-essential-singularity}.
Part~(vi) follows directly from the definition $F_{w}(u)=w(-u)-1/w(u-1)$:
$1/w(u-1)$ diverges when $w(u-1)=0$, i.e.\ at $u=u_{z}+1$ for
each zero $u_{z}$ of $w$; and $w(-u)$ diverges when $w$ has a
pole at $-u$, i.e.\ at $u=-u_{p}$ for each pole $u_{p}$ of $w$.
That $w$ has infinitely many poles on the negative real axis (and,
by the pairing, correspondingly many zeros) is established by the
stealthy-pole ladder Theorem~\ref{thm:pole-ladders} below; by the
two sources just identified, $F_{w}$ has infinitely many poles on
the real line. On any bounded interval, by isolation of meromorphic
singularities, only finitely many poles occur. 
\end{proof}
\begin{rem}[The two Floquet sub-combs and their numerical accessibility]
\label{rem:double-min-two-families} Hill's equation has two linearly
independent Floquet solutions, with exponents $\pm u_{0}$. They populate
the two sub-combs $Z_{F}^{-}=\{-u_{0}+n\}$ (multiplier $\rho_{-}=e^{-2\pi iu_{0}}$)
and $Z_{F}^{+}=\{u_{0}+n\}$ (multiplier $\rho_{+}=e^{+2\pi iu_{0}}$),
both confirmed by the monodromy matrix; for $\{u^{*}\}>\tfrac{1}{2}$,
as in the worked examples of this chapter, the primary zero $u^{*}\approx+p$
lies in $Z_{F}^{-}$ and $u^{**}=-u^{*}\approx-p$ in $Z_{F}^{+}$,
the assignments interchanging when $\{u^{*}\}<\tfrac{1}{2}$. All
comb points are genuine zeros of the single meromorphic function $\widetilde{F}_{w}=\gamma F_{w}$
(Theorem~\ref{thm:notes2-root-symmetry}, Remark~\ref{rem:notes2-root-periodicity}),
and the fundamental exponent $u_{0}\in(0,\tfrac{1}{2})$ is recovered
from either primary zero as the distance from $u^{*}$ to the nearest
integer, $u_{0}=\min\bigl(\{u^{*}\},\,1-\{u^{*}\}\bigr)$, where $\{u^{*}\}=u^{*}\bmod1$.

At fixed arithmetic precision, however, the two sub-combs are \emph{not}
equally accessible. Naive double-precision evaluation of the forward-CF
form $F_{w}(u)=w(-u)-1/w(u-1)$ locates only the $Z_{F}^{-}$ zeros:
at a $Z_{F}^{+}$ location $u_{0}+n$ the evaluation returns the large
nonzero value $|F_{w}(u_{0}+n)|\sim|G(u_{0})|/\gamma\gg1$ rather
than zero, because the cancellation in $\widetilde{F}_{w}=\gamma w(u)+G(u)+\gamma w(-u)$
requires more significant digits than double precision provides (see
Theorem~\ref{thm:precision-cost} and eq.~\eqref{eq:Fw-local-Laurent}).
Equivalently, viewed from the minimal-solution side: at the exponents
of the second sub-comb the relevant coefficient sequence is \emph{dominant}
(not minimal) for the forward recurrence, which is why naive forward
evaluation misses them. With sufficient arithmetic precision both
sub-combs are equally accessible (Remark~\ref{rem:notes2-root-periodicity}),
and the $Z_{F}^{+}$ zeros are recovered directly from the formula
for $u_{0}$ above. 
\end{rem}

\subsection{\texorpdfstring{Floquet exponent computation: closed-form $\gamma^{2}$-series}{Floquet
exponent computation: closed-form gamma2-series}}

\label{subsec:Cambi-notes2-precision}

The condition $\widetilde{F}_{w}(u^{*},\gamma)=0$ determines the
Floquet exponent $u^{*}$ as a power series in $\gamma^{2}$.
\begin{rem}[Numerical conventions for CF computations]
\label{rem:CF-numerical-conventions} The continued-fraction (CF)
method is numerically subtle: CFs compute the \emph{minimal} (exponentially
decaying) solution of the underlying three-term recurrence~\eqref{eq:notes2-ttrec},
and any computational shortcut that disturbs this structure picks
up the dominant (exponentially growing) solution and ruins the result.
The following conventions apply throughout the verification of all
results in this section; subsequent equations cite this remark. 
\begin{enumerate}
\item[(N1)] \emph{Tail-to-head evaluation.} The CF for $v(u)$~\eqref{eq:Cambi-v-CF},
and equivalently for the minimal solution ratio $w(u)$~\eqref{eq:notes2-CF-new}
via $w(u)=-\gamma\,v(u+1)$ (Lemma~\ref{lem:notes2-w-v}), must be
evaluated tail-to-head: 
\[
T_{N}=0,\quad T_{k}=\frac{\gamma^{2}}{G(u+k)-T_{k+1}}\ (k=N{-}1,\ldots,1),\quad v(u)=\frac{1}{G(u)-T_{1}}.
\]
This order automatically selects the minimal solution. Iterating head-to-tail
(or forward-iterating the recurrence $w(u+1)=-G(u+1)/\gamma-1/w(u)$
from a guessed $w(u)$) picks up the dominant growing solution and
is unstable.
\item[(N2)] \emph{Precision cost of integer shifts.} By Theorem~\ref{thm:precision-cost},
verifying $\widetilde{F}_{w}(u^{*}+n)=0$ at integer shifts with $D$
working digits resolves the cancellation only to $D(n)=D-2|n|\log_{10}(1/|\zeta_{+}|)$
significant digits. Concretely: 
\begin{itemize}
\item $\gamma=0.3$: $|\zeta_{+}|=1/3$, so $\approx0.95$ digits lost per
unit shift. 
\item $\gamma=0.1$: $|\zeta_{+}|\approx0.101$, so $\approx1.99$ digits
per shift. 
\end{itemize}
Double precision ($D\approx15$) suffices only for 
\[
|n|\lesssim D/[2\log_{10}(1/|\zeta_{+}|)];
\]
beyond this, use arbitrary-precision arithmetic with enough digits
to cover the desired range.
\item[(N3)] \emph{Stealthy zeros} (Remark~\ref{rem:double-precision-misses}).
At insufficient precision, Floquet zeros $u^{*}+n$ appear as $|\widetilde{F}_{w}|\gg1$
rather than $\approx0$ -- pure cancellation artifact, \emph{not}
a property of the zeros themselves. Detection of all zeros in a window
requires the precision budget of (N2).
\item[(N4)] \emph{Bilateral Floquet sequences via direct products.} For verifying
formulas at large $|m|$ involving the Floquet sequence $\{h_{m}(u_{0})\}$
(e.g.\ \eqref{eq:notes2-hm-neg-product}), compute $h_{\pm m}/h_{0}$
\emph{directly} as a product of CF-evaluated $w$ factors: 
\[
\frac{h_{m}(u_{0})}{h_{0}(u_{0})}=\prod_{k=0}^{m-1}w(u_{0}+k),\qquad\frac{h_{-m}(u_{0})}{h_{0}(u_{0})}=\prod_{k=0}^{m-1}w(-u_{0}+k).
\]
Do \emph{not} extend the bilateral sequence by iterating the three-term
recurrence backward from $m=0$: any residual in $u_{0}$ (even at
machine precision $\widetilde{F}_{w}(u_{0})\sim10^{-11}$) is amplified
by $|1/\zeta_{+}|^{|m|}$ per step and the iteration diverges from
the true minimal solution.
\item[(N5)] \emph{Infinite-product generators $H,H_{-}$.} For $H(u)$~\eqref{eq:notes2-H-new-def},
truncate the product at $m=N$ large enough for the $1/N$ tail to
be below the desired precision; $N\approx200$--$300$ gives $\sim6$
digits at $\gamma=0.3$. Each $w(u+m)$ is computed via its own tail-to-head
CF; the outer product is assembled in any order. For $H_{-}(u)$,
use the closed-form $H_{-}(u):=1/H(-u)$ (Definition~\eqref{eq:notes2-H-minus-def}),
not the product representation directly.
\item[(N6)] \emph{Floquet exponent precision via the series.} The Floquet exponent
$u^{*}$ is obtained from \eqref{eq:ustar-series} with explicit coefficients
\eqref{eq:ustar-c2}--\eqref{eq:ustar-c8}; the four explicit terms
give $O(\gamma^{10})$ accuracy, sufficient for all verifications
in this work at the parameter values $\gamma\leq0.3$. 
\end{enumerate}
\end{rem}

This CF method offers a decisive accuracy advantage over the discriminant
approach. In the MW framework~\cite[Cor.~2.6]{MagWin}, the discriminant
expansion $\Delta=\Delta_{0}+\Delta_{2}+\cdots$ groups in even powers
of $\delta$ (eq.~\eqref{eq:Delta-groups}, Chapter~\ref{sec:MWInce}),
so truncating at $\Delta_{0}+\Delta_{2}$ gives Floquet exponents
to accuracy $O(\delta^{2})$ (Keller's formula~\eqref{eq:Keller},
\cite[p.~192]{KelInst}, Chapter~\ref{sec:MWInce}); improving beyond
$O(\delta^{2})$ requires computing higher $\Delta_{n}$ terms, which
becomes increasingly difficult (see Chapter~\ref{sec:UnivExp}).
By contrast, the present CF method yields $u^{*}$ to $O(\gamma^{2k+2})$
accuracy by retaining $k$ terms in the series~\eqref{eq:ustar-series}
($\gamma=\delta/(1+\delta^{2})\approx\delta$ for small $\delta$,
eq.~\eqref{eq:gamma-delta}): the four explicit coefficients~\eqref{eq:ustar-c2}--\eqref{eq:ustar-c8}
already give accuracy $O(\gamma^{10})\approx O(\delta^{10})$, and
the recursion~\eqref{eq:c2k-recursion} extends this to any desired
order. Two complementary tools are developed here: closed-form series
coefficients $c_{2k}$ (exact in $p$) and a recursive algorithm for
their computation to any order.

\emph{Series for $u^{*}$.} The primary Floquet exponent admits the
$\gamma^{2}$-power series 
\begin{equation}
u^{*}\;=\;p\;+\;c_{2}\gamma^{2}\;+\;c_{4}\gamma^{4}\;+\;c_{6}\gamma^{6}\;+\;c_{8}\gamma^{8}\;+\;\cdots,\label{eq:ustar-series}
\end{equation}
convergent for sufficiently small $\gamma$ at fixed $p\notin\tfrac{1}{2}\mathbb{Z}$;
the leading constraint near the primary resonance is $\gamma^{2}\ll p^{2}-\tfrac{1}{4}$,
and every odd resonance similarly limits the domain (Remark~\ref{rem:resonance-poles}).
Applying the explicit recursion described below, one obtains the closed-form
coefficients: 
\begin{align}
c_{2} & \;=\;\frac{p(3p^{2}-1)}{4p^{2}-1},\label{eq:ustar-c2}\\[4pt]
c_{4} & \;=\;\frac{p\,(420p^{6}-345p^{4}+99p^{2}-10)}{4\,(4p^{2}-1)^{3}},\label{eq:ustar-c4}\\[4pt]
c_{6} & \;=\;\frac{p\,\bigl(\begin{aligned}[t] & 73920p^{12}-263760p^{10}+271460p^{8}\\
 & \quad-131815p^{6}+34232p^{4}-4679p^{2}+270\bigr)
\end{aligned}
}{4(2p-3)(2p-1)^{5}(2p+1)^{5}(2p+3)},\label{eq:ustar-c6}\\[4pt]
c_{8} & \;=\;\frac{p\,\bigl(\begin{aligned}[t] & 230630400p^{18}-1456815360p^{16}+3381840000p^{14}\\
 & \quad-3745959840p^{12}+2356114740p^{10}-915152265p^{8}\\
 & \quad+226125654p^{6}-35011757p^{4}+3138804p^{2}-126360\bigr)
\end{aligned}
}{64(2p-3)^{2}(2p-1)^{7}(2p+1)^{7}(2p+3)^{2}},\label{eq:ustar-c8}
\end{align}
with $c_{2}$ agreeing with Cambi~\eqref{eq:Cambi-u0-approx}.

\emph{Pole structure.} The denominators reveal which resonances $p=n/2$
create poles: 
\[
\begin{array}{c|c|l}
k & c_{2k}\text{ poles at }p= & \text{denominator}\\
\hline 1 & \tfrac{1}{2} & (2p-1)^{1}(2p+1)^{1}\\
2 & \tfrac{1}{2} & 4(2p-1)^{3}(2p+1)^{3}\\
3 & \tfrac{1}{2},\;\tfrac{3}{2} & 4(2p-1)^{5}(2p+1)^{5}(2p-3)(2p+3)\\
4 & \tfrac{1}{2},\;\tfrac{3}{2} & 64(2p-1)^{7}(2p+1)^{7}(2p-3)^{2}(2p+3)^{2}
\end{array}
\]
The primary resonance $p=\tfrac{1}{2}$ (the first stability boundary)
appears at all orders; the secondary resonance $p=\tfrac{3}{2}$ first
appears at $c_{6}$ and accumulates at higher orders.
\begin{rem}[Resonances as poles of the series coefficients]
\label{rem:resonance-poles} The poles of $c_{2k}$ as a rational
function of $p$ are a direct manifestation of the parametric resonances
of the LC circuit. The following is verified for $k=1,2,3,4$ from
the explicit formulas~\eqref{eq:ustar-c2}--\eqref{eq:ustar-c8};
the general pattern is conjectured to persist for all $k$.

Each odd resonance $p=n/2$ ($n=1,3,5,\ldots$) corresponds to a stability
tongue boundary in the $(p,\gamma)$ plane. For the four computed
coefficients: 
\begin{itemize}
\item The \emph{primary resonance} $p=\tfrac{1}{2}$ (boundary of the widest
instability tongue, width $\sim\gamma$) appears as a pole in all
four coefficients $c_{2},c_{4},c_{6},c_{8}$, with denominator factor
$(2p-1)^{2k-1}(2p+1)^{2k-1}$. This pole signals that the series~\eqref{eq:ustar-series}
breaks down as $p\to\tfrac{1}{2}$, i.e.\ as the operating point
approaches the primary tongue boundary.
\item The \emph{secondary resonance} $p=\tfrac{3}{2}$ (boundary of the
next instability tongue, width $\sim\gamma^{3}$) first appears in
$c_{6}$ (order $k=3$) with factor $(2p-3)(2p+3)$, and in $c_{8}$
with factor $(2p-3)^{2}(2p+3)^{2}$.
\item Higher resonances $p=\tfrac{5}{2},\tfrac{7}{2},\ldots$ have not yet
been computed at sufficient order to confirm their first appearance. 
\end{itemize}
It is conjectured that $p=(2j-1)/2$ first enters at order $k=2j-1$,
one new resonance per two additional orders, consistent with the tongue
widths $\sim\gamma^{2j-1}$. The series~\eqref{eq:ustar-series}
thus appears to encode the resonance structure of Hill's equation
in the pole positions of its rational coefficients. 
\end{rem}

\emph{Explicit recursion for $c_{2k}$.} Here and below, for any power
series $S(\gamma)=\sum_{n\geq0}s_{n}\gamma^{n}$, we write $[S]_{k}$
for the coefficient of $\gamma^{k}$, i.e.\ $[S]_{k}=s_{k}$. The
coefficients satisfy the triangular recursion: 
\begin{equation}
c_{2k}\;=\;-\frac{p}{2}\,\Bigl[\widetilde{F}_{w}\!\Bigl(p+\sum_{j=1}^{k-1}c_{2j}\gamma^{2j},\;\gamma\Bigr)\Bigr]_{2k},\label{eq:c2k-recursion}
\end{equation}
where $\widetilde{F}_{w}(u,\gamma)=\gamma w(u)+G(u)+\gamma w(-u)$
is expanded as a power series in $\gamma$ with $u$ substituted by
the partial sum $p+\sum_{j<k}c_{2j}\gamma^{2j}$, and $[\cdot]_{2k}$
extracts the coefficient of $\gamma^{2k}$. The CF~\eqref{eq:notes2-CF-new}
is evaluated using the full CF with the inversion at each CF level
performed via the precomputed table 
\begin{equation}
\frac{1}{1+a_{1}\xi+a_{2}\xi^{2}+\cdots}\;=\;\sum_{k\geq0}b_{k}\,\xi^{k},\label{eq:inv-series-table}
\end{equation}
where $b_{0}=1$ and $b_{k}=-\sum_{j=1}^{k}a_{j}\,b_{k-j}$ (complete
homogeneous symmetric polynomials in the $a_{j}$ with sign). This
reduces the computation of each $c_{2k}$ to a finite number of polynomial
operations in $p$.
\begin{rem}[Numerical verification of $c_{2},c_{4},c_{6},c_{8}$]
\label{rem:ustar-series-derivation} The coefficients~\eqref{eq:ustar-c2}--\eqref{eq:ustar-c8}
are verified by direct substitution into $\widetilde{F}_{w}$ using
80-digit arbitrary-precision arithmetic. The table shows the ratios
$\widetilde{F}_{w}(\cdots)/\gamma^{2k+2}\to\mathrm{const}$, confirming
each truncation is $O(\gamma^{2k+2})$: 
\[
{\footnotesize \begin{array}{c|cccc}
p & \widetilde{F}_{w}(p{+}c_{2}\gamma^{2})/\gamma^{4} & \widetilde{F}_{w}(\cdots{+}c_{4}\gamma^{4})/\gamma^{6} & \widetilde{F}_{w}(\cdots{+}c_{6}\gamma^{6})/\gamma^{8} & \widetilde{F}_{w}(\cdots{+}c_{8}\gamma^{8})/\gamma^{10}\\
\hline 1 & -3.037 & -8.384 & -25.60 & -82.79\\
2 & -3.222 & -8.868 & -27.03 & -87.33\\
20/3 & -3.276 & -9.010 & -27.45 & -88.67
\end{array}}
\]
Each ratio converges to a finite nonzero constant as $\gamma\to0$,
confirming the $O(\gamma^{2k+2})$ residuals. For Cambi's parameters
($\gamma=0.1$, $p=20/3$): 
\begin{multline*}
u^{*}\;\approx\;6.6667+4.9906\!\times\!10^{-2}+10.920\!\times\!10^{-4}+30.032\!\times\!10^{-6}+91.507\!\times\!10^{-8}\\
\;=\;6.71769535\quad(\text{exact: }6.71769535),\quad\text{error }O(\gamma^{10})\approx10^{-10}.
\end{multline*}
Additionally, the coefficient $c_{6}$ satisfies the defining relation:
\begin{equation}
c_{6}\;=\;\lim_{\gamma\to0}\frac{u^{*}(\gamma)-p-c_{2}\gamma^{2}-c_{4}\gamma^{4}}{\gamma^{6}},\label{eq:ustar-c6-def}
\end{equation}
giving $c_{6}\approx4.192$ ($p=1$), $8.868$ ($p=2$), $30.032$
($p=20/3$) --- in agreement with~\eqref{eq:ustar-c6}. 
\end{rem}

\emph{The precision cost of integer shifts.} The zeros of the double
minimality function $F_{w}$ (equivalently $\widetilde{F}_{w}$) at
integer shifts $u^{*}+n$ of the primary zero $u^{*}\approx p$ all
exist as exact meromorphic zeros, but evaluating $\widetilde{F}_{w}(u^{*}+n)=0$
numerically requires increasingly high arithmetic precision as $|n|$
grows. The reason is a catastrophic cancellation in $\widetilde{F}_{w}(u)=\gamma w(u)+G(u)+\gamma w(-u)$:
the three terms are individually $O(1)$ but cancel to zero, and the
precision of that cancellation degrades by approximately $2\log_{10}(1/|\zeta_{+}|)$
decimal digits per unit shift.
\begin{thm}[Precision cost of integer shifts]
\index{precision cost of integer shifts} \label{thm:precision-cost}
Let $u^{*}$ be the primary zero of $F_{w}$ near $p$, known to $D$
decimal digits of precision. A numerical verification of $\widetilde{F}_{w}(u^{*}+n)=0$
carried out with $D$ significant digits of working arithmetic then
resolves the cancellation only to 
\begin{equation}
D(n)\;=\;D-2|n|\log_{10}\!\left(\frac{1}{|\zeta_{+}|}\right)\label{eq:precision-cost-formula}
\end{equation}
significant digits, and the verification fails once $D(n)\leq0$,
where $\zeta_{+}=(-1+\sqrt{1-4\gamma^{2}})/(2\gamma)$ is the smaller
characteristic root~\eqref{eq:Cambi-char-roots}. The precision cost
per unit shift is thus $2\log_{10}(1/|\zeta_{+}|)$ digits. The stealthy
$Z_{F}^{+}$ zeros (see~\eqref{eq:family-def}) missed at double
precision are graphically visible in Fig.~\ref{fig:Fw-precision}
(filled red squares, labeled with extra digits required). 
\end{thm}

\begin{proof}
By Theorem~\ref{thm:double-min}(i) and Theorem~\ref{thm:notes2-integer-shift}
(with $w(u^{*})\neq0$, verified numerically), $\widetilde{F}_{w}(u^{*}+n)=0$
exactly for all $n\in\mathbb{Z}$. The CF computes $w(u^{*}+n)$ and
$w(-u^{*}-n)$ correctly to full working precision for any $n$ (exponential
convergence holds for all $u$). The cancellation, however, is among
terms of order one. As $n\to+\infty$, $w(u^{*}+n)\to\zeta_{+}$ with
algebraic corrections (Theorem~\ref{thm:notes2-w-rate}), while the
exact vanishing of $\widetilde{F}_{w}$ forces $w(-u^{*}-n)\to\zeta_{-}$,
the \emph{dominant} root; the limiting values cancel identically,
\[
\gamma\,\zeta_{+}\;+\;1\;+\;\gamma\,\zeta_{-}\;=\;0,
\]
since $\zeta_{+}+\zeta_{-}=-1/\gamma$ by~\eqref{eq:Cambi-char-roots}.
The numerical difficulty is the steepness of the zero crossing. By
the local Laurent structure~\eqref{eq:Fw-local-Laurent} (Fig.~\ref{fig:Fw-poles},
Table~\ref{tab:Fw-poles}), each comb zero $u^{*}+n$ is paired with
a nearby pole of $F_{w}$ whose distance $d_{n}$ from the zero shrinks
geometrically with $|n|$ at the rate $d_{n}\asymp|\zeta_{+}|^{2|n|}$
read off the residue fits; the rate is derived from first principles
in Theorem~\ref{thm:pole-ladders} of \S\,\ref{subsec:Casoratian-Weyl}.
The slope of $\widetilde{F}_{w}$ at the zero therefore grows like
$1/d_{n}$, and an uncertainty $10^{-D}$ in the input $u^{*}$ produces
a computed value 
\[
\bigl|\widetilde{F}_{w}(u^{*}+n)\bigr|\;\sim\;\frac{10^{-D}}{d_{n}}\;\sim\;10^{-D}\,|\zeta_{+}|^{-2|n|},
\]
i.e.\ the resolved cancellation falls short of $D$ by $2|n|\log_{10}(1/|\zeta_{+}|)$
digits, which is~\eqref{eq:precision-cost-formula}. 
\end{proof}
\begin{rem}[Application to Cambi's parameters]
\label{rem:precision-cost-Cambi} For Cambi's parameters $\gamma=0.1$,
$p=20/3$: $|\zeta_{+}|\approx0.1010$, so the cost per shift is $2\log_{10}(1/|\zeta_{+}|)\approx1.99$
digits. Double precision ($D\approx15$) then suffices only for $|n|\lesssim7$,
while $D=50$ digits covers $|n|\lesssim25$. 
\end{rem}

\subsection{\texorpdfstring{Finding all Floquet zeros at high precision: stealthy
zeros}{Finding all Floquet zeros at high precision: stealthy zeros}}

\label{subsec:stealthy-zeros}

Building on the precision-cost theorem~\ref{thm:precision-cost}
and the CF numerical conventions of Remark~\ref{rem:CF-numerical-conventions},
this section documents the numerical verification of \emph{all} Floquet
zeros of $\widetilde{F}_{w}$ in a representative window, and identifies
the \emph{stealthy} zeros that double precision misses entirely through
cancellation.

\begin{figure}[htbp]
\centering \includegraphics[width=1\textwidth]{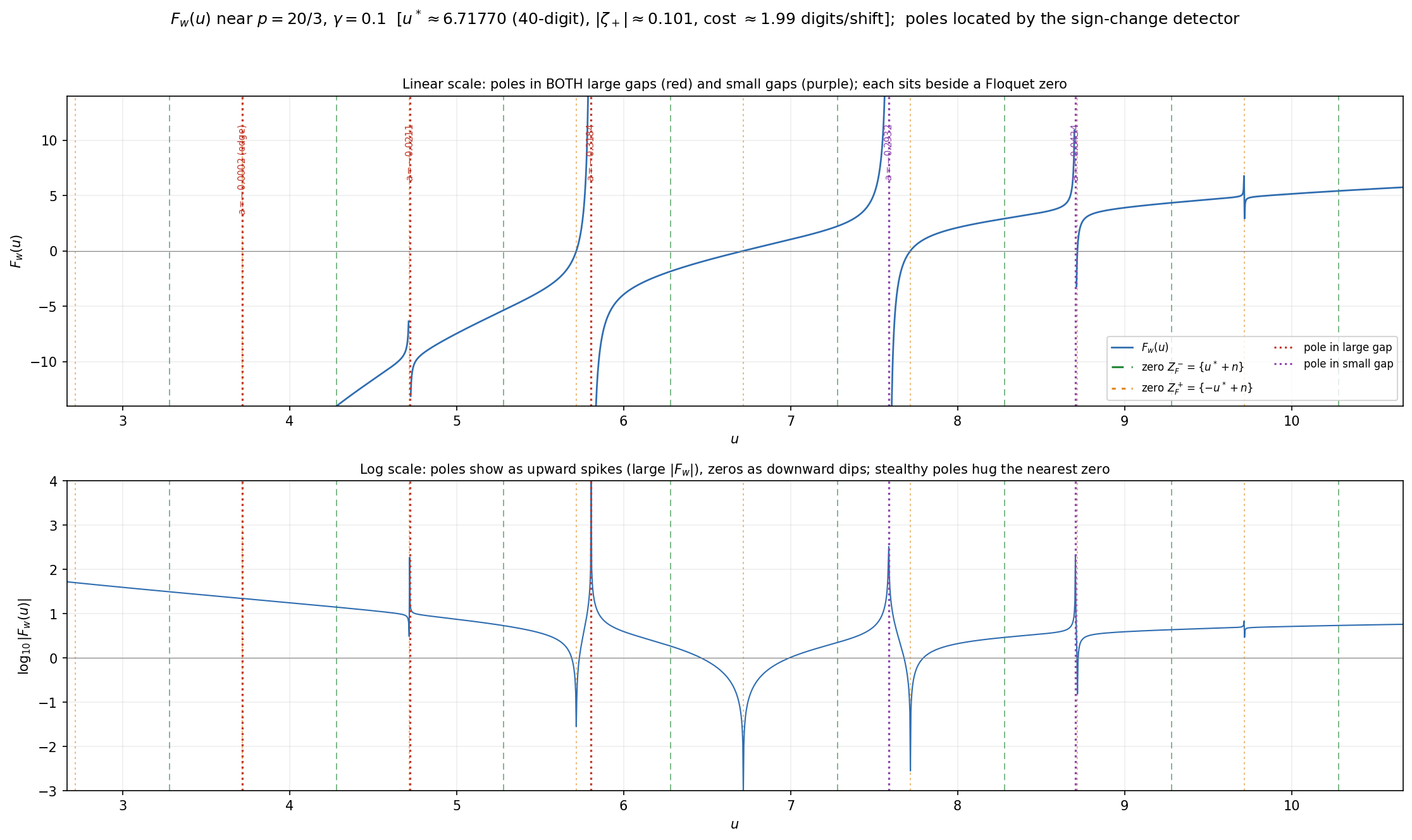} \caption{The double minimality function $F_{w}(u)$ near $u^{*}\approx p=20/3$
for Cambi's parameters ($\gamma=0.1$, $p=20/3$, $u^{*}=6.71770\ldots$
to 40 digits, $|\zeta_{+}|\approx0.101$, precision cost $\approx1.99$
digits per shift), over $(p-4,\,p+4)\approx(2.67,\,10.67)$, $40$-digit
arithmetic. \emph{Top}: linear scale; zeros are sign changes, poles
(dotted, located by the sign-change detector, Theorem~\ref{thm:Fw-sign-change-pole})
are annotated with their residues (red in large gaps, purple in small).
Unlike the $p=0.70$ case (Figs.~\ref{fig:notes2-Fw-marked}, \ref{fig:Fw-poles}),
here poles populate \emph{both} large and small gaps (Table~\ref{tab:Fw-pole-catalog-cambi}).
\emph{Bottom}: $\log_{10}|F_{w}(u)|$; poles are upward spikes, zeros
downward dips, stealthy poles hugging the nearest zero; the vertical
lines also separate the sub-combs $Z_{F}^{-}=\{u^{*}+n\}$ and $Z_{F}^{+}=\{-u^{*}+n\}$.
Because $|\zeta_{+}|\approx0.10$, a pole sits within $\approx|\zeta_{+}|$
of almost every Floquet zero, so on the linear panel only the pole-free
fundamental zero near $u^{*}$ shows as a visible crossing and the
stealthy $Z_{F}^{+}$ sub-comb does not register at all --- the zeros
are genuine but nearly invisible by eye (Remark~\ref{rem:masking-geometry};
resolved explicitly, for $p=0.70$, in Figure~\ref{fig:invisible-zero-pole}).}
\label{fig:Fw-precision} 
\end{figure}

\emph{Finding all zeros in $(p-4,\,p+4)$ with 50-digit arithmetic.}
Using 50-decimal-digit arbitrary-precision arithmetic, we find $u^{*}$
to full 50-digit precision and evaluate $\widetilde{F}_{w}$ at all
predicted zero locations. For Cambi's parameters ($\gamma=0.1$, $p=20/3$),
the interval $(p-4,p+4)\approx(2.67,10.67)$ contains the accessible
sub-comb $Z_{F}^{-}=\{u^{*}+n\}$ (8 zeros) interleaved with the stealthy
sub-comb $Z_{F}^{+}=\{-u^{*}+n\}$ (see~\eqref{eq:family-def} for
the two sub-combs); all are genuine zeros, confirmed numerically (Fig.~\ref{fig:Fw-precision}):
\begin{equation}
|\widetilde{F}_{w}(\pm u^{*}+n)|\;\leq\;10^{-9}\quad\text{for all zeros in }(p-4,\,p+4).\label{eq:Ftilde-verified}
\end{equation}
The $Z_{F}^{-}=\{u^{*}+n\}$ zeros are accessible at double precision
throughout this window (the farthest, at $u^{*}-4\approx2.718$, is
the least accurate, requiring the most cancellation but still resolved).
The $Z_{F}^{+}=\{-u^{*}+n\}$ zeros, by contrast, are \emph{stealthy}
at double precision: evaluating $F_{w}$ there yields $|F_{w}|\sim1$
rather than $0$, and they are confirmed as genuine zeros only with
$50$-digit arithmetic (Remark~\ref{rem:double-precision-misses}).

\begin{rem}[Stealthy zeros]
\label{rem:double-precision-misses} We call a zero $u_{0}$ of a
meromorphic function $F$ \emph{stealthy at precision $D$} if evaluating
$F(u_{0})$ with $D$ significant digits yields $|F(u_{0})|\gg1$
rather than $\approx0$, due to catastrophic cancellation between
individually large terms. The $Z_{F}^{+}$ zeros are stealthy at standard
double precision ($D\approx15$): $|F_{w}(u_{0}+n)|\gg1$ rather than
zero, because the cancellation in $\widetilde{F}_{w}=\gamma w(u)+G(u)+\gamma w(-u)$
requires $D>2|n|\log_{10}(1/|\zeta_{+}|)$ significant digits (Theorem~\ref{thm:precision-cost}).
This is a numerical artifact of insufficient precision, not a property
of the zeros themselves: with sufficient arithmetic, both sub-combs
are equally accessible and equally genuine zeros of the exact meromorphic
function $\widetilde{F}_{w}$. Fig.~\ref{fig:Fw-precision} shows
this stealthiness graphically: the one stealthy zero in the window
(filled red square) has $|F_{w}|\gg1$ at double precision but is
a genuine sign change at 50-digit precision. The \emph{poles} of $F_{w}$
exhibit a dual stealthiness --- some sit so close to a Floquet zero
that direct searches miss them entirely. Making them visible required
a dedicated method, developed in \S\,\ref{subsec:Fw-pole-detection}:
the sign-change theorem (Theorem~\ref{thm:Fw-sign-change-pole})
and its bisection corollary locate such a pole from signs alone and
then resolve even its tiny residue (Fig.~\ref{fig:stealthy-pole}).
Figure~\ref{fig:invisible-zero-pole} combines both phenomena on
a single zero--pole pair: a zoom cascade that resolves the masked
zero beside its pole, together with the depth dependence of the attainable
floor (panel~(d)), which reaches $O(1)$ at double precision by $|u|\approx14$. 
\end{rem}

\begin{figure}[htbp]
\centering \includegraphics[width=1\textwidth]{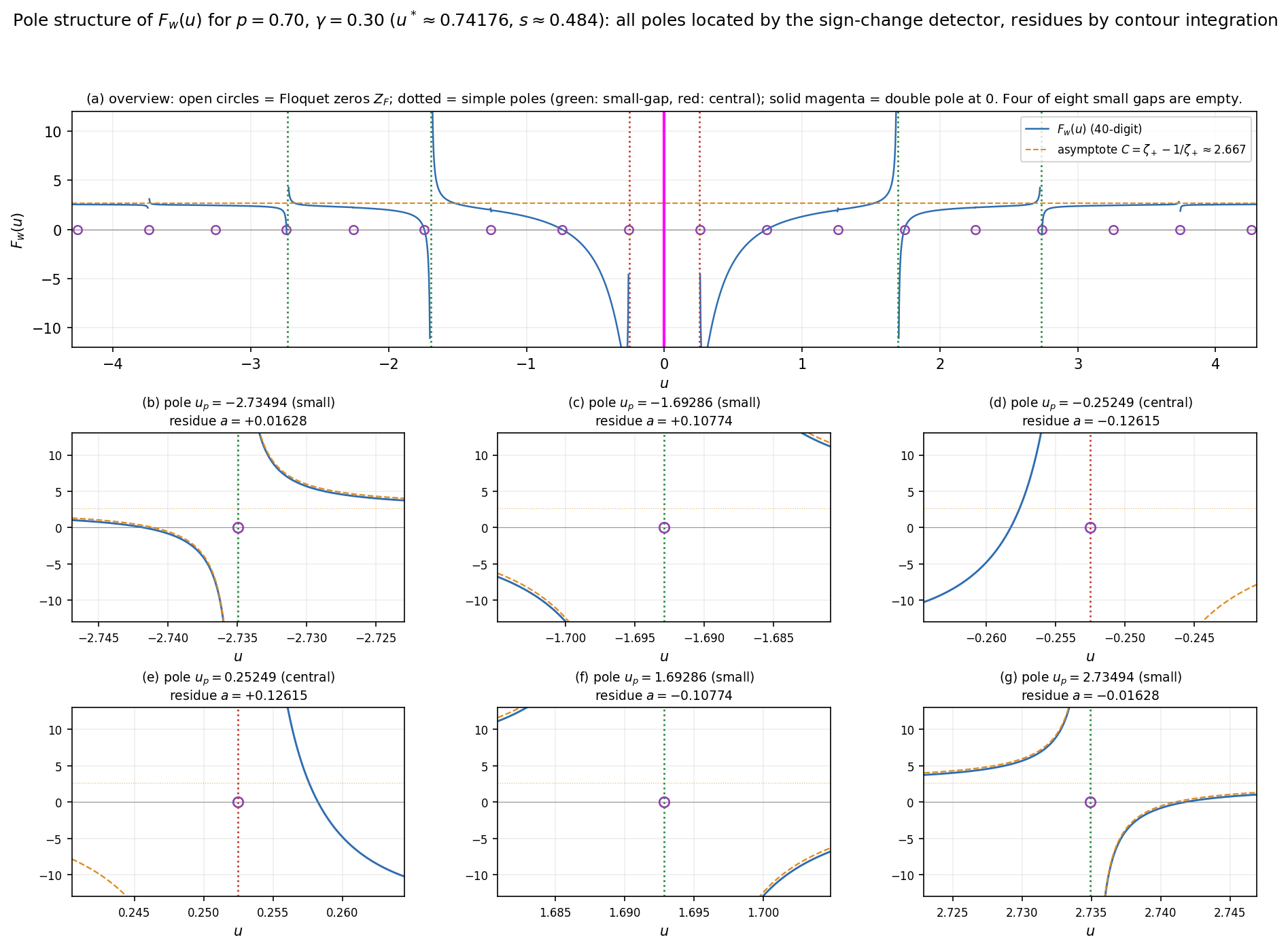} \caption{Pole structure of the double minimality function $F_{w}(u)$ for $p=0.70$,
$\gamma=0.30$ ($u^{*}=0.74176\ldots$, 40-digit; $s\approx0.484$),
the six largest-residue poles located by the sign-change detector
(Theorem~\ref{thm:Fw-sign-change-pole}, Corollary~\ref{cor:Fw-pole-bisection}),
residues by contour integration. \emph{(a)}~Global view on $(-4.3,4.3)$:
purple circles = zeros of $F_{w}$ (Floquet comb $Z_{F}$, eq.~\eqref{eq:Fw-zero-set});
dotted = the six simple poles (green: small-gap, red: central); solid
magenta = the double pole at $u=0$; orange dashed = asymptote $\zeta_{+}-1/\zeta_{+}\approx2.667$.
Four of the eight small gaps look empty here; the full census is Table~\ref{tab:Fw-pole-catalog}.
\emph{(b)--(g)}~Zoom panels for the six simple poles, each labeled
with its residue (smallest $|a|\approx0.016$ at $u\approx\pm2.735$,
the stealthy ladder members of Theorem~\ref{thm:pole-ladders});
blue = $F_{w}$, orange dashed = local Laurent fit~\eqref{eq:Fw-local-Laurent},
purple circle = the adjacent Floquet zero. Several zeros lie within
$\approx0.006$ of a pole and so hide in its spike --- genuine but
nearly invisible by eye; see Remark~\ref{rem:masking-geometry} for
the mechanism and Figure~\ref{fig:invisible-zero-pole} for an explicit
resolution.}
\label{fig:Fw-poles} 
\end{figure}

\begin{figure}[htbp]
\centering \includegraphics[width=1\textwidth]{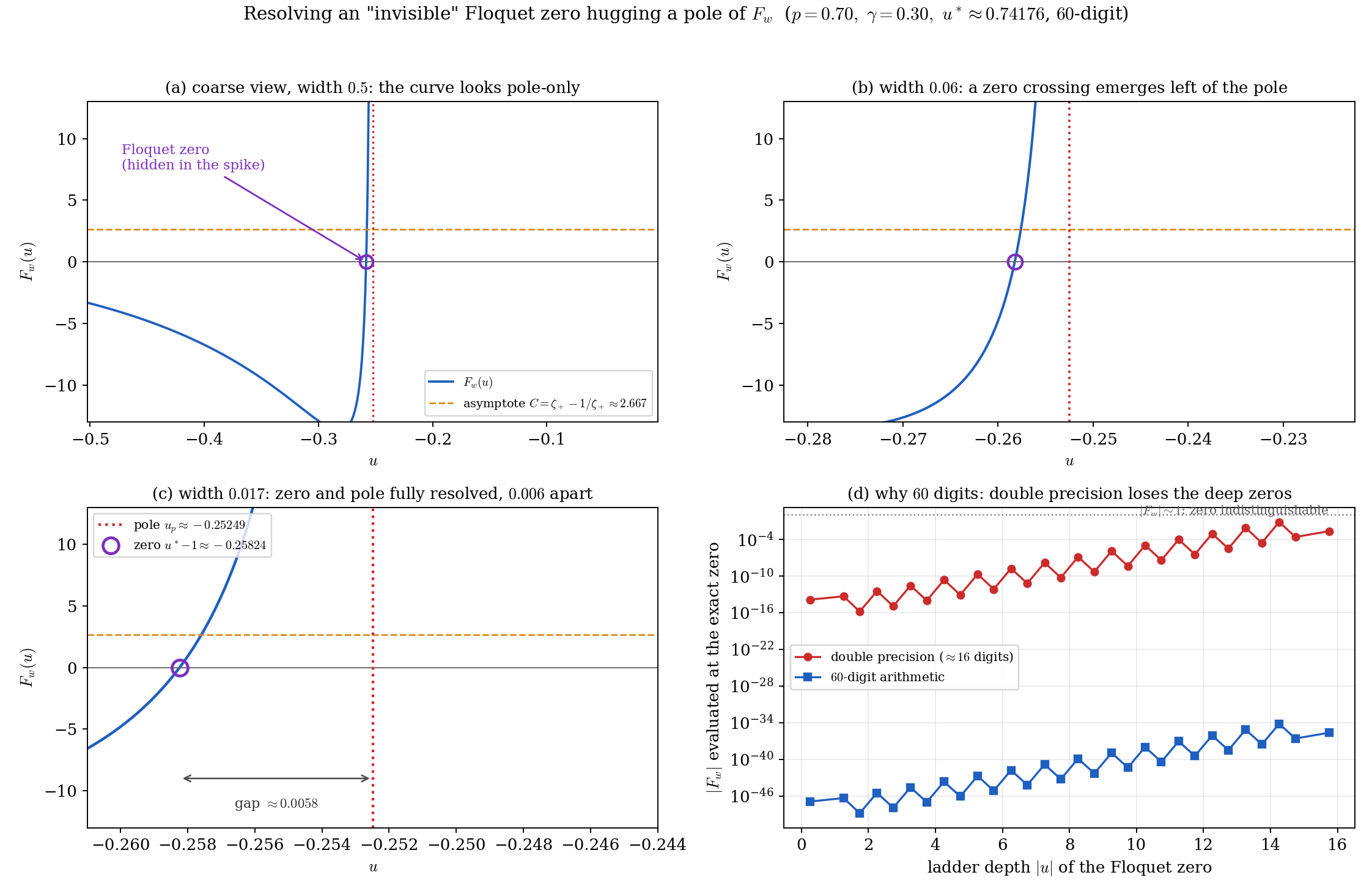}
\caption{Making an \textquotedblleft invisible\textquotedblright{} Floquet zero
visible: a $60$-digit zoom cascade on the central zero--pole pair
of $F_{w}$ for $p=0.70$, $\gamma=0.30$ ($u^{*}\approx0.74176$).
The Floquet zero at $u^{*}-1\approx-0.25824$ is separated from the
simple pole at $u_{p}\approx-0.25249$ (residue $a\approx-0.126$)
by only $\approx0.0058$. \emph{(a)}~On a window of width $0.5$
the zero (purple circle) is buried in the pole spike: the curve looks
pole-only and the zero appears absent. \emph{(b)}~Width $0.06$:
a zero crossing emerges just to the left of the pole. \emph{(c)}~Width
$0.017$: the zero (purple circle) and the pole (red dotted) are fully
resolved as two distinct features $\approx0.0058$ apart, the curve
sweeping from $F_{w}=0$ to $\pm\infty$ across that tiny gap. \emph{(d)}~Why
arbitrary precision is required: $|F_{w}|$ evaluated \emph{at} the
exact Floquet zeros $\pm u^{*}+n$, against the ladder depth $|u|$.
Catastrophic cancellation in $w(-u)-1/w(u-1)$ makes the attainable
floor grow with depth --- at double precision (red) it climbs from
$\approx10^{-14}$ near the origin to $O(10^{-1})$ by $|u|\approx14$,
so the deep zeros become numerically indistinguishable from nonzeros
(the regime of Remark~\ref{rem:double-precision-misses}), whereas
$60$-digit arithmetic (blue) holds the floor below $10^{-34}$ and
resolves every member. This floor is intrinsic to the evaluation:
it is independent of the continued-fraction truncation length, hence
not a sampling artifact. The geometry behind this masking --- a Floquet
zero within an exponentially small distance $d_{n}\sim|\zeta_{+}|^{2n}$
of a pole of residue $|a_{n}|\asymp|C|\,d_{n}$ --- is analyzed in
Remark~\ref{rem:masking-geometry}.}
\label{fig:invisible-zero-pole} 
\end{figure}

\emph{Local pole-zero structure of $F_{w}$ and why both sub-combs
are hard to find simultaneously.} Near each pole $z$ of $F_{w}$,
the function has the local Laurent expansion: 
\begin{equation}
F_{w}(u)\;\approx\;\frac{a}{u-z}+C,\label{eq:Fw-local-Laurent}
\end{equation}
where $C=\zeta_{+}-1/\zeta_{+}$ is the asymptotic background value
and $a$ is the residue at the pole. This Laurent structure is confirmed
by the orange dashed fits in Fig.~\ref{fig:Fw-poles}(b--g). The
crucial observation, made precise by Theorem~\ref{thm:pole-ladders},
is that residues shrink geometrically \emph{along each ladder} of
integer translates, $|a_{n}|=(|C|+o(1))\,d_{n}\asymp|\zeta_{+}|^{2n}$,
rather than according to the raw distance from $Z_{F}$: the deepest
resolvable ladder members at $u\approx\pm2.735$ ($|a|\approx0.016$)
are already faint in their zoom panels, the next members are unresolvable
($|a|\lesssim10^{-4}$), while the ladder bases and first rungs at
$u\approx\pm0.252$ and $u\approx\pm1.693$ ($|a|\approx0.11$--$0.13$)
are easily visible.

The poles of $F_{w}$ arise from two sources: 
\begin{enumerate}
\item[(i)] \emph{Zeros of $w$:} if $u_{z}$ is a zero of $w$ then $F_{w}$
has a pole at $u_{z}+1$ with residue $a=-1/w'(u_{z})$. 
\item[(ii)] \emph{Poles of $w(-u)$:} if $u_{p}<0$ is a pole of $w$ then $F_{w}$
has a pole at $-u_{p}>0$ with residue $a=-\mathrm{res}(w,u_{p})$. 
\end{enumerate}
Table~\ref{tab:Fw-poles} lists the resolvable poles of $F_{w}$
on $u<0$ for $p=0.70$, $\gamma=0.30$, confirmed numerically with
high-precision arithmetic; the near-$Z_{F}$ candidates, whose residues
are too small to resolve, are noted in the caption but not tabulated.

\begin{table}[htbp]
\centering %
\begin{tabular}{c|r|r|r}
pole $z$  & residue $a$  & $|z-u^{*}|$  & nearest $Z_{F}$ zero\tabularnewline
\hline 
$-0.252$  & $-0.126$  & $0.994$  & $u^{*}-1=-0.258$\tabularnewline
$-1.693$  & $+0.109$  & $2.435$  & $-u^{*}-1=-1.742$\tabularnewline
\end{tabular}\caption{Resolvable poles of the double minimality function $F_{w}$ for $u<0$,
$p=0.70$, $\gamma=0.30$. Each tabulated pole arises from a zero
of $w$ via $u=u_{z}+1$ (see eq.~\eqref{eq:Fw-poles}); these are
ladder-base and first-rung members in the sense of Theorem~\ref{thm:pole-ladders},
carrying the largest residues, $O(0.1)$. Asymptotic background $C=\zeta_{+}-1/\zeta_{+}\approx2.667$.
The candidate set~\eqref{eq:Fw-poles} also predicts poles near each
$Z_{F}$ element (e.g.\ close to $u^{*}-2\approx-1.258$ and $u^{*}-3\approx-2.258$);
these are deeper ladder members with exponentially small residues
(Theorem~\ref{thm:pole-ladders}) and cannot be cleanly resolved
or signed at fixed precision, so no residue is reported for them.
A complete characterization of the pole structure is outside the scope
of this work; the entries above are reported as partial, numerically
observed information. See Fig.~\ref{fig:Fw-poles}(c,d) for the corresponding
zoom panels.}
\label{tab:Fw-poles} 
\end{table}

\begin{rem}[Two Floquet exponents are sufficient]
\label{rem:two-exponents-sufficient} Although the complete Floquet
frequency lattice $\{\pm u_{0}+n:n\in\mathbb{Z}\}$ has infinitely
many members, only two are needed in practice: the primary CF zeros
\begin{equation}
u^{*}\;\approx\;+p\quad(Z_{F}^{-}),\qquad u^{**}\;\approx\;-p\quad(Z_{F}^{+}),\label{eq:two-primary-zeros-only}
\end{equation}
corresponding to Floquet multipliers $\rho_{\pm}=e^{\pm2\pi iu_{0}}$.
All other lattice members are integer shifts carrying no new physical
information (Corollary~\ref{cor:Floquet-Fourier}). The numerical
accessibility of both sub-combs is discussed in Remark~\ref{rem:notes2-root-periodicity};
here we explain the \emph{mutual invisibility} via the Laurent residue
structure.

The mutual invisibility of the two sub-combs near each other is explained
by the Laurent structure~\eqref{eq:Fw-local-Laurent} and the pole
residues. Table~\ref{tab:Fw-poles-neg} shows the poles of $F_{w}$
on $u>0$ (at $u=-u_{p}$ for each pole $u_{p}$ of $w$) together
with their residues. The large-residue poles are easily visible in
Fig.~\ref{fig:Fw-poles}(f); the tiny-residue poles in Fig.~\ref{fig:Fw-poles}(g)
are essentially invisible, with the Laurent fit nearly flat.

\begin{table}[htbp]
\centering {\small\setlength{\tabcolsep}{4pt} }{\small{}%
\begin{tabular}{r|r|r|p{1.9cm}|p{2cm}}
{\small$w$ pole $u_{p}$ } & {\small$F_{w}$ pole $-u_{p}$ } & {\small residue $|a|$ } & {\small\centering $|z-u^{*}|$ } & {\small\centering\arraybackslash zoom window $|a|/|C|$}\tabularnewline
\hline 
{\small$-0.252$ } & {\small$+0.252$ } & {\small$0.126$ } & {\small\centering $0.489$ } & {\small\centering\arraybackslash $4.7\times10^{-2}$}\tabularnewline
{\small$-1.693$ } & {\small$+1.693$ } & {\small$0.107$ } & {\small\centering $0.951$ } & {\small\centering\arraybackslash $4.0\times10^{-2}$}\tabularnewline
\end{tabular}} \caption{Resolvable poles of the double minimality function $F_{w}$ for $u>0$
($p=0.70$, $\gamma=0.30$): each arises from the reflection $u=-u_{p}$
of a pole $u_{p}<0$ of $w$ (eq.~\eqref{eq:Fw-poles}). Background
$|C|=|\zeta_{+}-1/\zeta_{+}|\approx2.667$; the zoom window $|a|/|C|$
is the scale on which the pole is detectable. These are the base and
first-rung ladder members (Theorem~\ref{thm:pole-ladders}), carrying
the largest residues ($|a|\approx0.11$--$0.13$) and easily visible
(Fig.~\ref{fig:Fw-poles}(e,f)). The tabulated $|z-u^{*}|$ is the
distance to the primary exponent $u^{*}\approx0.742$; their distances
to the actual nearest comb zeros (at $-u^{*}+1\approx0.258$ and $u^{*}+1\approx1.742$)
are the much smaller $0.0058$ and $0.0489$, respectively. The candidate
set~\eqref{eq:Fw-poles} also predicts poles near $+1.258$ and $+2.258$,
nearly coinciding with $Z_{F}$ elements; these are deeper ladder
members whose residues are exponentially small (Theorem~\ref{thm:pole-ladders})
and cannot be cleanly resolved at fixed precision, so they are not
tabulated (Fig.~\ref{fig:Fw-poles}(g) shows one such near-invisible
pole).}
\label{tab:Fw-poles-neg} 
\end{table}
\end{rem}

\subsection{\texorpdfstring{Meromorphic structure of $w(u)$}{Meromorphic structure
of w(u)}}

\label{subsec:Cambi-w-structure}

This section establishes the meromorphic structure of the minimal
solution ratio $w(u)$ in the stability zone (Assumption~\eqref{eq:stability-assumption}).
The rigorous content consists of two theorems: the unit-shift pole-zero
pairing (Theorem~\ref{thm:notes2-w-poles-zeros}) and the asymptotic
expansion of $w(u+m)$ as $m\to+\infty$ (Theorem~\ref{thm:notes2-w-rate}),
whose corollary is the convergence of the infinite product defining
$H(u)$ in Section~\ref{subsec:Cambi-notes2-H}. The geometric picture
--- poles localize to the negative axis and partition into two unit-spaced
families displaced from the Floquet lattice $Z_{F}$ --- is presented
in Remarks~\ref{rem:w-pole-zero-geometry}--\ref{rem:w-two-families},
where it rests on numerical computation across the stability zone
together with partial analytical arguments; the geometric rates at
which the deep family members approach $Z_{F}$ are established in
Theorem~\ref{thm:pole-ladders} (\S\,\ref{subsec:Casoratian-Weyl}).

We begin with the analytic foundation: $w$ is holomorphic off the
real axis and, on the first sheet, has the Nevanlinna-like positivity
$\operatorname{Im}w>0$. (It is not a Nevanlinna function in $u$:
this positivity holds only on $\operatorname{Re}u>-1$, and $\operatorname{Im}w$
changes sign across the line $\operatorname{Re}u=-1$.)
\begin{lem}[Analyticity off $\mathbb{R}$ and the Nevanlinna-like positivity $\operatorname{Im}w>0$]
\label{lem:w-analytic-nevanlinna} The minimal solution ratio $w(u)$
is analytic for $\operatorname{Im}u\neq0$. Moreover, on the first
sheet $\operatorname{Re}u>-1$ it has the Nevanlinna-like positivity
property: if $\operatorname{Im}u>0$ and $\operatorname{Re}u>-1$
then $\operatorname{Im}w(u)>0$. 
\end{lem}

\begin{figure}[htbp]
\centering \includegraphics[width=0.98\textwidth]{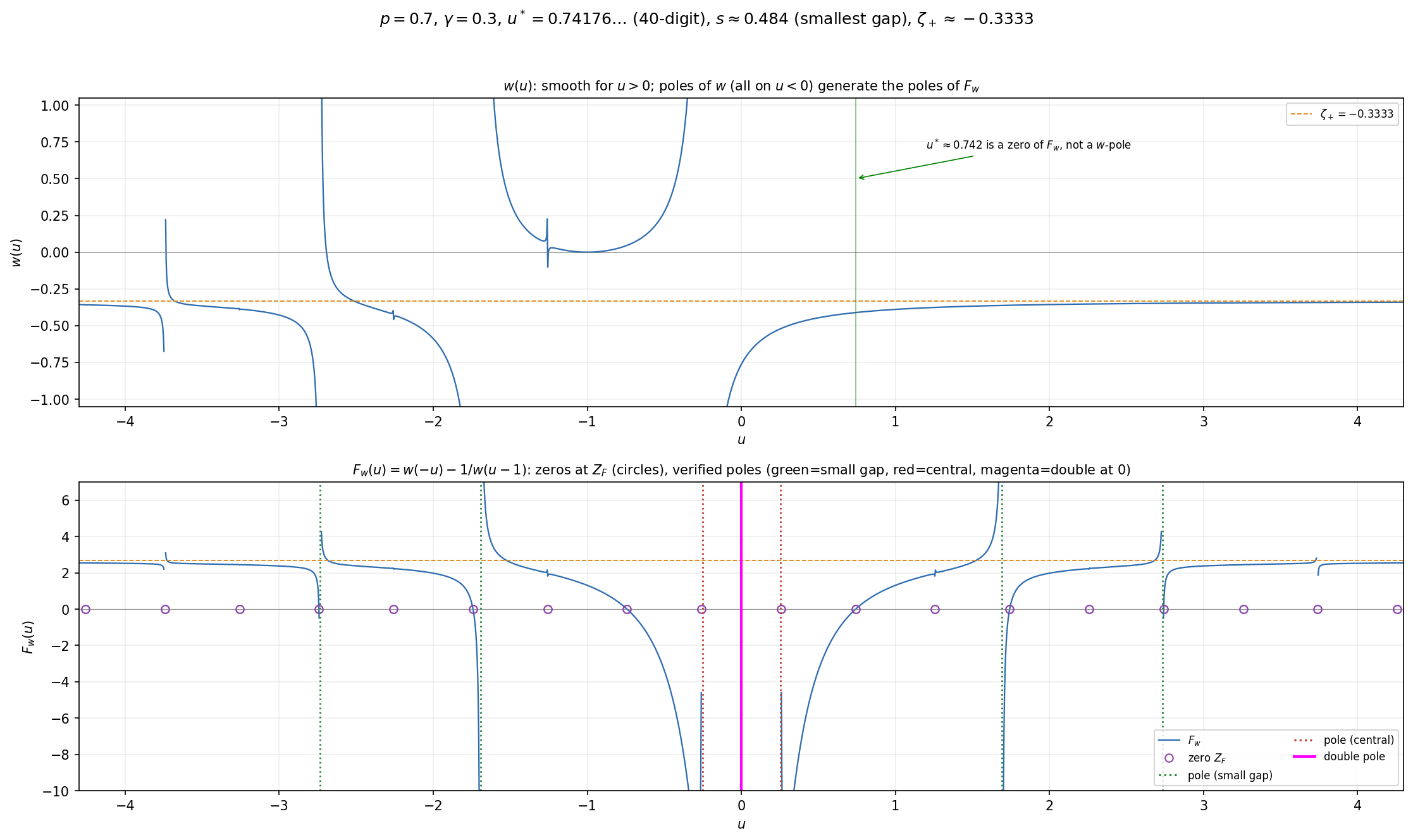}
\caption{Minimal solution ratio $w(u)$ (top) and double minimality function
$F_{w}(u)$ (bottom) for $p=0.70$, $\gamma=0.30$ ($u^{*}=0.74176\ldots$,
40-digit). Top: $w(u)$ (blue), asymptote $\zeta_{+}\approx-0.333$
(orange dashed). All poles of $w$ lie on $u<0$ (Remark~\ref{rem:w-pole-zero-geometry}(a));
the value $u\approx+0.742$ marked on the upper panel is the primary
Floquet exponent $u^{*}$ (a zero of $F_{w}$, not a pole of $w$),
at which $w(u^{*})$ is smooth. Bottom: $F_{w}(u)$; purple circles
= zeros of $F_{w}$ ($Z_{F}$, eq.~\eqref{eq:Fw-zero-set}); the
poles are those verified by the sign-change detector (Theorem~\ref{thm:Fw-sign-change-pole})
--- dotted green in small gaps, dotted red in the central gap, and
the solid magenta double pole at $u=0$. \emph{Note on visibility:}
many zero circles in the lower panel fall beside a pole and so appear
as part of its spike rather than a visible crossing --- genuine zeros,
present but nearly invisible by eye (Remark~\ref{rem:masking-geometry};
resolved explicitly in Figure~\ref{fig:invisible-zero-pole}).}
\label{fig:notes2-w-u-plot} 
\end{figure}

\begin{figure}[htbp]
\centering \includegraphics[width=0.98\textwidth]{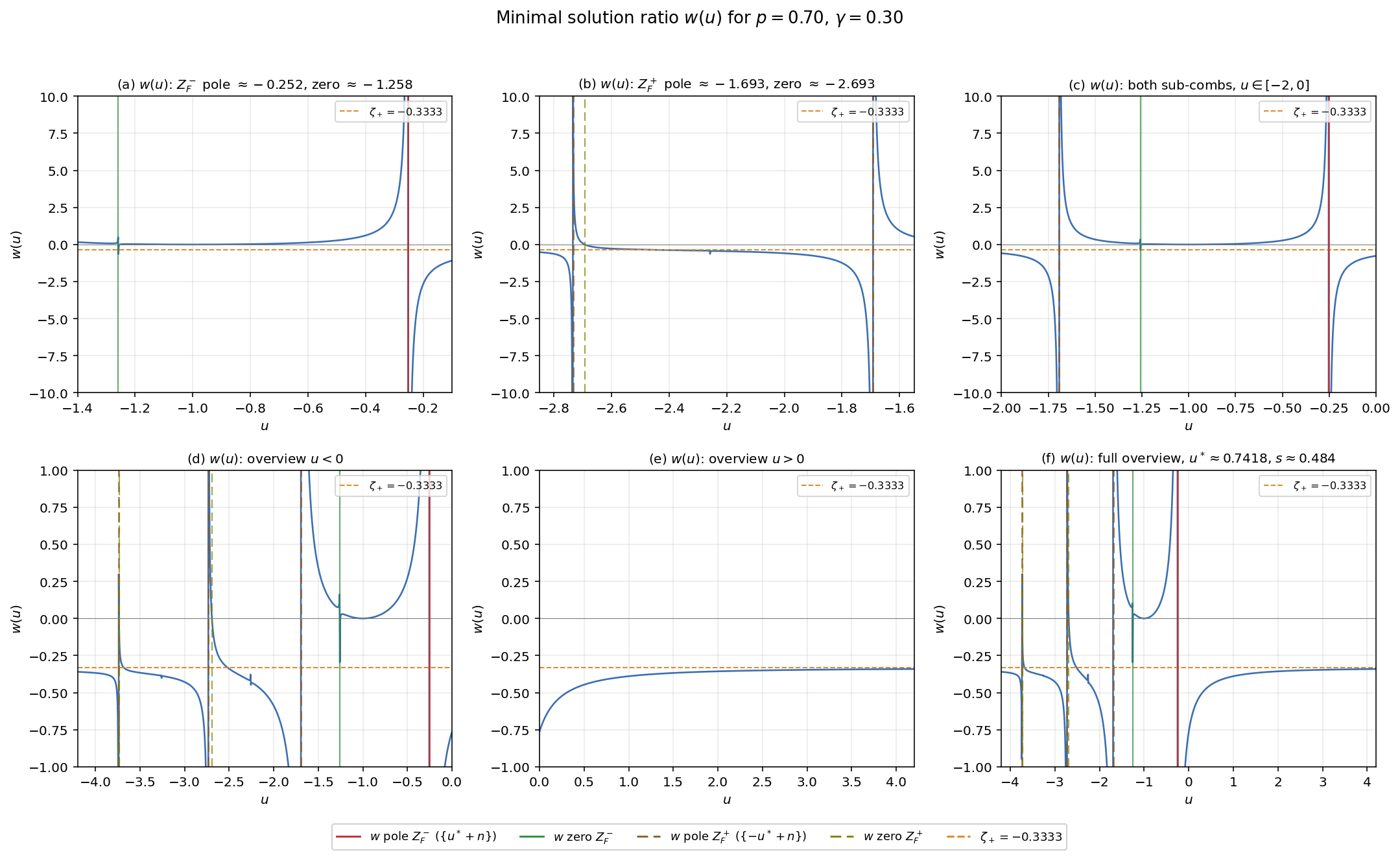}
\caption{Minimal solution ratio $w(u)$ for $p=0.70$, $\gamma=0.30$. Top
row: zooms on individual pole-zero pairs: (a)~pole at $u\approx-0.252$
with zero at $u\approx-1.252$ (the pole family tracking $\{u^{*}+n\}=Z_{F}^{-}$);
(b)~pole at $u\approx-1.693$ with zero at $u\approx-2.693$ (the
pole family tracking $\{-u^{*}+n\}=Z_{F}^{+}$); (c)~both pole families
on $u\in[-2,0]$. Bottom row: overviews on $u<0$ (d), $u>0$ (e),
full range (f). Vertical lines: red solid / green solid = poles /
zeros of the $\{u^{*}+n\}$ family; brown dashed / olive dashed =
poles / zeros of the $\{-u^{*}+n\}$ family. Each pole family has
internal spacing $1$ (Remark~\ref{rem:w-two-families}), offset
by $s\approx0.484$, producing the alternating pattern (see~\eqref{eq:family-def}
for the sub-combs). \emph{Note on visibility:} the pole and zero of
each pair lie one unit apart and the deeper ladder members crowd toward
$-\infty$ with rapidly shrinking residues, so each marker can locate
a genuine feature where the curve looks featureless (Remark~\ref{rem:masking-geometry};
see Figure~\ref{fig:invisible-zero-pole} for a fully resolved pair).}
\label{fig:notes2-w-marked} 
\end{figure}

\begin{proof}
Write the continued fraction for $w$ as a composition of Mobius maps,
\begin{multline}
w(u)=\lim_{N\to\infty}m_{1}\circ m_{2}\circ\cdots\circ m_{N}(z_{0}),\\
m_{1}(z)=\frac{-\gamma}{G(u+1)-z},\quad m_{n}(z)=\frac{\gamma^{2}}{G(u+n)-z}\quad(n\ge2),\label{eq:w-mobius}
\end{multline}
with $z_{0}$ the asymptotic seed selecting the minimal solution.
Each $G(u+n)=1-p^{2}/(u+n)^{2}$ is holomorphic off the integer $u=-n$,
so off the real axis every truncation is holomorphic; the limit-point
convergence of the minimal-solution continued fraction (\S\,\ref{subsec:Cambi-notes2})
makes $w$ holomorphic there.

For the positivity property let $\operatorname{Im}u>0$ and $\operatorname{Re}u>-1$,
so that $\operatorname{Re}(u+n)>0$ for every $n\ge1$ and 
\begin{equation}
\operatorname{Im}G(u+n)=\frac{2p^{2}\,\operatorname{Re}(u+n)\,\operatorname{Im}u}{|u+n|^{4}}>0.\label{eq:Im-G-positive}
\end{equation}
The inner maps $m_{n}$ ($n\ge2$) have determinant $\gamma^{2}>0$,
and by \eqref{eq:Im-G-positive} each carries the closed lower half-plane
into its interior: if $\operatorname{Im}z\le0$ then 
\begin{equation}
\operatorname{Im}m_{n}(z)=-\gamma^{2}\,\frac{\operatorname{Im}G(u+n)-\operatorname{Im}z}{|G(u+n)-z|^{2}}<0.\label{eq:mn-lhp}
\end{equation}
Hence the tail $T_{0}=\lim_{N\to\infty}m_{2}\circ\cdots\circ m_{N}(z_{0})$
satisfies $\operatorname{Im}T_{0}\le0$, and the outer map flips the
half-plane, 
\begin{equation}
\operatorname{Im}w=\gamma\,\frac{\operatorname{Im}G(u+1)-\operatorname{Im}T_{0}}{|G(u+1)-T_{0}|^{2}}\ge\gamma\,\frac{\operatorname{Im}G(u+1)}{|G(u+1)-T_{0}|^{2}}>0.\label{eq:Im-w-positive}
\end{equation}
\end{proof}
\begin{thm}[Pole--zero pairing and double zero of $w(u)$]
\index{pole--zero pairing}\index{minimal solution ratio@minimal solution ratio $w$!poles of}\index{minimal solution ratio@minimal solution ratio $w$!zeros of}
\label{thm:notes2-w-poles-zeros} Assume $(p,\gamma)$ is in the stability
zone (Assumption~\eqref{eq:stability-assumption}), so that $2\gamma<1$
and the Floquet exponent $u^{*}$ is real. 
\begin{enumerate}
\item[(i)] \emph{One-unit pairing of poles and sign-change zeros.} $u_{z}$
is a sign-change (simple) zero of $w$ if and only if $u_{z}+1$ is
a pole of $w$. The pairing holds at all sign-change zeros; the unique
exception is the double zero at $u=-1$, treated in Part~(ii).
\item[(ii)] \emph{Double zero at $u=-1$.} $w(u)$ has a double zero at $u=-1$,
with 
\begin{equation}
w(-1+\varepsilon)\;\sim\;\frac{\gamma}{p^{2}}\,\varepsilon^{2}\quad\text{as }\varepsilon\to0.\label{eq:notes2-w-double-zero}
\end{equation}
\end{enumerate}
The illustrative pole-zero structure for $p=0.70$, $\gamma=0.30$
is shown in Figs.~\ref{fig:notes2-w-u-plot}--\ref{fig:notes2-w-marked}.
The pairing here is local; the global pole count --- that $w$ has
\emph{infinitely many} real poles, arranged in geometric stealthy-pole
ladders --- is established in \S\,\ref{subsec:CW-ladder-law} (Theorem~\ref{thm:pole-ladders}),
and all analytic properties of $w$ are collected in \S\,\ref{subsec:w-Fw-summary}. 
\end{thm}

\begin{proof}
\emph{Part (i).} Both directions follow from the recurrence~\eqref{eq:notes2-wrec},
viewed as an identity of meromorphic functions on $\mathbb{C}$.

\emph{Zero implies pole:} at a sign-change zero $u_{z}$, $-1/w(u_{z})=\infty$
and $G(u_{z}+1)$ is finite (since $u_{z}\neq-1$), so $w(u_{z}+1)=-G(u_{z}+1)/\gamma-1/w(u_{z})=\infty$.

\emph{Pole implies zero:} at a pole $u_{p}$ of $w$, the LHS of $w(u_{p})=-G(u_{p})/\gamma-1/w(u_{p}-1)$
is infinite; $G(u_{p})$ is finite (Remark~\ref{rem:u-zero-not-special}),
so $1/w(u_{p}-1)=\infty$, i.e.\ $w(u_{p}-1)=0$.

\emph{Part (ii).} From the relation $w(u)=-\gamma v(u+1)$ (Lemma~\ref{lem:notes2-w-v})
and Cambi's CF for $v$: at $u=-1+\varepsilon$, the argument $u+1=\varepsilon$
makes $G(\varepsilon)=1-p^{2}/\varepsilon^{2}\sim-p^{2}/\varepsilon^{2}$
dominate the head of the CF, giving $v(\varepsilon)\sim1/G(\varepsilon)\sim-\varepsilon^{2}/p^{2}$
and therefore $w(-1+\varepsilon)=-\gamma v(\varepsilon)\sim\gamma\varepsilon^{2}/p^{2}$. 
\end{proof}
\begin{cor}[Reality of the singularities of $w$, and of the poles of $F_{w}$]
\label{cor:w-real-zeros} The minimal solution ratio $w(u)$ has
neither poles nor zeros off the real axis. Consequently every pole
of $F_{w}$ is real: from $F_{w}(u)=w(-u)-1/w(u-1)$, 
\begin{equation}
\operatorname{poles}(F_{w})\ \subseteq\\bigl(-\operatorname{poles}(w)\bigr)\cup\bigl(\operatorname{zeros}(w)+1\bigr),\label{eq:Fw-pole-sources}
\end{equation}
and both sets on the right-hand side are real. 
\end{cor}

\begin{proof}
By Lemma~\ref{lem:w-analytic-nevanlinna} $w$ is analytic for $\operatorname{Im}u\neq0$,
hence has no non-real poles. Were $u_{z}$ a zero of $w$ with $\operatorname{Im}u_{z}\neq0$,
then $u_{z}\neq-1$, so $G(u_{z}+1)$ is finite and the recurrence~\eqref{eq:notes2-wrec},
an identity of meromorphic functions on $\mathbb{C}$, would force
\begin{equation}
w(u_{z}+1)=-\frac{G(u_{z}+1)}{\gamma}-\frac{1}{w(u_{z})}=\infty,\label{eq:cor-no-complex-zero}
\end{equation}
a non-real pole of $w$, which is impossible. Thus $w$ has no non-real
zeros either. Now $w(-u)$ contributes a pole only where $-u$ is
a pole of $w$, and $1/w(u-1)$ only where $u-1$ is a zero of $w$;
both source sets are real, giving~\eqref{eq:Fw-pole-sources}. Finally,
the evenness $F_{w}(-u)=F_{w}(u)$ makes $\operatorname{poles}(F_{w})$
symmetric under $u\mapsto-u$, in accord with~\eqref{eq:Fw-pole-sources}. 
\end{proof}
\begin{rem}[An alternative route to the absence of non-real zeros]
\label{rem:w-no-zeros-positivity} The first-sheet half of Corollary~\ref{cor:w-real-zeros}
also follows directly from positivity rather than from the recurrence.
By Lemma~\ref{lem:w-analytic-nevanlinna}, $\operatorname{Im}w>0$
on $\{\operatorname{Re}u>-1,\ \operatorname{Im}u>0\}$, and a value
with positive imaginary part cannot vanish; hence $w$ has no zeros
there, and the reflection $w(\bar{u})=\overline{w(u)}$ carries this
to the conjugate region. The two arguments view the same fact from
different angles: this one is the half-plane (Nevanlinna-like) geometry,
while the recurrence argument of Corollary~\ref{cor:w-real-zeros}
is the rigidity of the functional equation~\eqref{eq:notes2-wrec}
--- and the latter, unlike positivity, also reaches the second sheet
$\operatorname{Re}u<-1$. 
\end{rem}

\begin{rem}[Symmetry reduction of the off-real zero problem for $F_{w}$]
\label{rem:Fw-fundamental-strip} Corollary~\ref{cor:w-real-zeros}
settles the \emph{poles} of $F_{w}$; its \emph{zeros} off the real
axis are governed by a small fundamental domain. A non-real zero is
a solution of 
\begin{equation}
w(-u)\,w(u-1)=1.\label{eq:Fw-zero-condition}
\end{equation}
Its solution set carries three exact symmetries --- evenness $F_{w}(-u)=F_{w}(u)$;
the Schwarz reflection $F_{w}(\bar{u})=\overline{F_{w}(u)}$ (real
coefficients); and, since $w(u-1)\,w(-u)$ is unchanged under $u\mapsto1-u$,
reflection about $\operatorname{Re}u=\tfrac{1}{2}$. Composing the
two reflections yields the integer translation $u\mapsto u+1$ ---
which is why the real zeros form the unit-spaced comb $\{\pm u^{*}+n\}$
--- and the group these generate has the half-unit strip 
\begin{equation}
\mathcal{S}=\bigl\{\,u:0\le\operatorname{Re}u\le\tfrac{1}{2},\ \operatorname{Im}u\ge0\,\bigr\}\label{eq:Fw-fundamental-strip}
\end{equation}
as a fundamental domain. Hence $F_{w}$ has no non-real zeros anywhere
if and only if it has none in $\mathcal{S}$. Two facts localize this
further. As $\operatorname{Im}u\to\infty$ the partial quotients $G(u+n)=1-p^{2}/(u+n)^{2}\to1$
uniformly in $n$ and in $\operatorname{Re}u$ (since $|u+n|\ge\operatorname{Im}u$),
so $w(-u),w(u-1)\to\zeta_{+}$ and 
\begin{equation}
F_{w}(u)\;\longrightarrow\;\zeta_{+}-\zeta_{+}^{-1}=\frac{\sqrt{1-4\gamma^{2}}}{\gamma}\neq0,\label{eq:Fw-infinity-limit}
\end{equation}
so some height $Y_{0}$ confines any off-real zero to the \emph{compact}
cell $\{0\le\operatorname{Re}u\le\tfrac{1}{2},\ 0\le\operatorname{Im}u\le Y_{0}\}$;
and throughout $\operatorname{Re}u\in(0,1)$ both arguments in~\eqref{eq:Fw-zero-condition}
have real part in $(-1,0)$, i.e.\ lie on the first sheet $\operatorname{Re}u>-1$,
where the positivity $\operatorname{Im}w>0$ of Lemma~\ref{lem:w-analytic-nevanlinna}
is available --- the second sheet never enters. The two edges of
$\mathcal{S}$ are exactly the loci where $F_{w}$ is real: on $\operatorname{Re}u=\tfrac{1}{2}$,
$-u=\overline{u-1}$ reduces~\eqref{eq:Fw-zero-condition} to $|w(-\tfrac{1}{2}+iy)|=1$
(not met in computations, where $|w|\le0.33$), and on $\operatorname{Re}u=0$,
$-u=\bar{u}$ makes $F_{w}(iy)$ real.

What remains open is precisely~\eqref{eq:Fw-zero-condition} on the
open interior $\operatorname{Re}u\in(0,\tfrac{1}{2})$, $\operatorname{Im}u>0$:
the positivity signs do not by themselves exclude it, since there
$w(u-1)$ lies in the upper half-plane and $w(-u)$ in the lower,
and their product can still be a positive real. Numerically no such
zero is found; a proof would have to rule out~\eqref{eq:Fw-zero-condition}
on this compact first-sheet cell. 
\end{rem}

\begin{rem}[On the parametrization point $u=0$]
\label{rem:u-zero-not-special} The point $u=0$ is where the parametrization
$G(u)=1-(p/u)^{2}$ has its own pole; this is an artifact of the parametrization
and has no physical significance. The relevant facts about $w$ near
$u=0$ are settled and used freely throughout this work without further
discussion: 
\begin{enumerate}
\item[(a)] $w(0)$ is finite (direct CF evaluation: $w(0)=-\gamma v(1)$ with
$v(1)$ finite in the stability zone). 
\item[(b)] $u_{p}=0$ is therefore not a pole of $w$; poles of $w$ never coincide
with $G$'s parametrization pole. 
\item[(c)] $u=-1$ is a double zero of $w$ (Part~(ii) above); this is the
unique exception to the unit-shift pairing of Part~(i). 
\item[(d)] In meromorphic identities involving $G$, the pole of $G$ at $u=0$
is balanced by the double zero of $w$ at $u=-1$ via the unit-shift
relation, so the identities extend to $u=0$ by continuity. 
\end{enumerate}
These facts are settled; we do not revisit them in proofs. 
\end{rem}

\begin{rem}[Observed global pole-zero geometry of $w$ on the real axis]
\label{rem:w-pole-zero-geometry} The structural picture below is
verified by numerical computation across the stability zone and supported
by analytical arguments based on the relation $w(u)=-\gamma v(u+1)$
(Lemma~\ref{lem:notes2-w-v}) and Cambi's pole analysis of $v$ (\S\,\ref{subsec:Cambi-v-props}).
The precise rigorous derivation of pole locations would require detailed
CF analysis beyond this section. The localization statements (a) and
(c) are established for the \emph{small-frequency regime} $p\in(0,1)$
(primary tongue, $r=1/p>1$), where Cambi's poles of $v$ near $\{p-n:n\geq0\}$
all lie at $u\leq p-1<0$ after the shift $w(u)=-\gamma v(u+1)$.
For larger $p$ (higher tongues, $p>1$) poles and zeros of $w$ may
appear on the positive real axis: the Cambi example with $p=20/3$
in \S\,\ref{subsec:w-simplicity-notes} exhibits a positive zero
$z_{0}\approx4.806$ and the adjacent positive pole $p_{0}\approx5.806$,
both close to the Floquet set $Z_{F}$ but on $u>0$. The displacement
principle (b) holds in all cases. 
\begin{enumerate}
\item[(a)] \emph{For $p\in(0,1)$, all poles of $w$ lie on $u<0$.} Each pole
of $w$ in the stability zone lies exponentially close (as $\gamma\to0$)
to an element of the Floquet zero set $Z_{F}=\{u^{*}+n:n\in\mathbb{Z}\}\cup\{-u^{*}+n:n\in\mathbb{Z}\}$.
For $p\in(0,1)$ these poles are restricted to $u<0$: Cambi's analysis
places poles of $v$ at positions close to $\{p-n:n\geq0\}$, which
via $w(u)=-\gamma v(u+1)$ shift to $u\leq p-1<0$. For $p>1$ the
nearest such positions move onto the positive axis, and the restriction
to $u<0$ no longer holds.
\item[(b)] \emph{Zeros of $w$ are one unit left of the poles.} This is the
rigorous content of Theorem~\ref{thm:notes2-w-poles-zeros}(i), and
holds for all $p$.
\item[(c)] \emph{For $p\in(0,1)$, $w$ is smooth and strictly negative for
$u>0$.} Given (a) in this regime, $w$ is holomorphic on a neighborhood
of $[0,\infty)$. By Theorem~\ref{thm:notes2-w-poles-zeros}(i) any
zero of $w$ at $u_{z}>0$ would force a pole at $u_{z}+1>0$ contradicting
(a); hence $w$ has no zeros on $u>0$ either. By continuity and the
asymptote $w(u)\to\zeta_{+}<0$ at $+\infty$ (Theorem~\ref{thm:notes2-w-rate}),
$w$ is strictly negative on $(0,\infty)$ and bounded away from zero.
For $p>1$ this conclusion fails: as the $p=20/3$ example shows,
$w$ can have a positive zero (sign change) on $(0,\infty)$. 
\end{enumerate}
The detailed two-family structure of poles in $u<0$ (for $p\in(0,1)$)
is described in Remark~\ref{rem:w-two-families} below. 
\end{rem}

\begin{rem}[Reality of the poles in the stability zone and their migration off
the real axis outside it]
\label{rem:w-poles-real-vs-complex} The restriction to the stability
zone in Remark~\ref{rem:w-pole-zero-geometry}(a) is essential: it
is precisely the condition under which the poles of $w$ remain on
the real axis. This is confirmed numerically, and the contrast across
the stability boundary is sharp.

\emph{Inside the stability zone the poles are real.} For $p=0.70$,
$\gamma=0.30$ (here $\tfrac{1}{2}|\operatorname{Tr}M(\pi)|\approx0.05<1$,
well inside the zone) the poles of $w$ are real to working precision
and remain so under perturbation of the search into the complex plane:
seeding a root finder for $1/w=0$ at the complex point $u_{p}\pm0.05\,i$
returns to the real axis with imaginary part below $10^{-50}$ (numerically
zero).

\emph{Crossing into the instability zone\index{instability zone}
the poles leave the real axis.} On the first instability tongue the
Floquet exponent $u^{*}$, fixed by $\cos(2\pi u^{*})=\tfrac{1}{2}\operatorname{Tr}M(\pi)$,
becomes complex once $|\operatorname{Tr}M(\pi)|>2$, and the poles
of $w$ tied to the Floquet set $Z_{F}=\{\pm u^{*}+n\}$ migrate off
the real axis as complex-conjugate pairs. Two representative points
(arbitrary-precision evaluation of the alternating CF \eqref{eq:notes2-CF-new}
with complex-seeded root finding): 
\begin{equation}
\begin{array}{c|c|c|c}
(p,\gamma) & \operatorname{Tr}M(\pi) & u^{*} & \text{representative pole pair}\\
\hline (0.50,0.30) & -2.240 & 0.5-0.0773\,i & -1.4837\pm0.0783\,i\\
(0.48,0.30) & -2.291 & 0.5-0.0848\,i & -1.4845\pm0.0815\,i
\end{array}\label{eq:w-complex-poles}
\end{equation}
The imaginary part acquired by each pole is comparable to $\operatorname{Im}(u^{*})$,
as expected for poles displaced from the (now complex) Floquet set.
Thus the poles do not merely become ``stealthy'' through small residues,
as on the real axis near $Z_{F}$ (Remark~\ref{rem:double-precision-misses});
in the instability zone they are genuinely absent from the real line,
having moved into the complex plane. This is consistent with the loss
of a real meromorphic structure for $w$ as $u^{*}\to\tfrac{1}{2}\mathbb{Z}$
at the tongue boundary (see~\eqref{eq:notes2-Fw-zeros} and the surrounding
discussion). 
\end{rem}

\begin{rem}[Two families of poles and zeros of $w$ on the negative real axis]
\label{rem:w-two-families} The two-family structure of $w$'s pole-zero
geometry follows from the two Floquet sub-combs and the displacement
principle, as follows.

\emph{Floquet input.} By Floquet theory in the stability zone, the
double minimality function $\widetilde{F}_{w}$ has zero set 
\begin{equation}
Z_{F}\;=\;\{u^{*}+n:n\in\mathbb{Z}\}\;\cup\;\{-u^{*}+n:n\in\mathbb{Z}\},\label{eq:ZF-two-combs}
\end{equation}
the union of two arithmetic progressions of common difference $1$
generated from the primary Floquet exponents $u^{*}$ and $-u^{*}$
(Theorem~\ref{thm:notes2-root-symmetry}).

\emph{Displacement.} By Remark~\ref{rem:poles-w-not-on-ZF}, no pole
of $w$ can lie exactly on $Z_{F}$. By Remark~\ref{rem:poles-displaced-from-ZF},
$w$-poles can lie exponentially close to elements of $Z_{F}$ (as
$\gamma\to0$), displaced by amounts that vanish with $\gamma$.

\emph{Pole-family structure.} Numerical computation across the stability
zone shows that the poles of $w$ on $u<0$ partition into two unit-spaced
families, each tracking one of the two Floquet sub-combs restricted
to $u<0$ (observed numerically across the zone; for all sufficiently
deep members the tracking is proved, with explicit rates, in Theorem~\ref{thm:pole-ladders}
of \S\,\ref{subsec:Casoratian-Weyl}): 
\begin{equation}
\begin{aligned}\text{family tracking }\{u^{*}+n\}: & \quad\bigl\{ u_{p}^{(+)}-k\;:\;k\geq0\bigr\},\\
\text{family tracking }\{-u^{*}+n\}: & \quad\bigl\{ u_{p}^{(-)}-k\;:\;k\geq0\bigr\},
\end{aligned}
\label{eq:w-two-families}
\end{equation}
where each base pole $u_{p}^{(\pm)}\in(-2,0)$ sits just off the nearest
element of the corresponding sub-comb, with the precise values depending
on whether $u^{*}\in(0,\tfrac{1}{2})$ or $u^{*}\in(\tfrac{1}{2},1)$.
Each pole $u_{p}^{(\pm)}-k$ approaches the corresponding Floquet
zero (an element of $Z_{F}$ shifted by $-k$) geometrically in the
depth $k$: by Theorem~\ref{thm:pole-ladders} the offset contracts
at the rate $|\zeta_{+}|^{2k}$ at fixed $(p,\gamma)$, in addition
to vanishing exponentially as $\gamma\to0$ at fixed $k$. The corresponding
zero families lie at $u_{p}^{(\pm)}-k-1$ for $k\geq0$, by Theorem~\ref{thm:notes2-w-poles-zeros}(i).

The unit-spacing within each family is inherited from the unit-spacing
of the two Floquet sub-combs $\{u^{*}+n\}$ and $\{-u^{*}+n\}$: the
displacement of each $w$-pole from its companion $Z_{F}$ element
varies smoothly with $u$, so the global unit-spacing of $Z_{F}$
on the negative axis is preserved (up to exponentially small variation).
The residue-distance law --- poles close to $Z_{F}$ carry correspondingly
small residues --- is observed numerically in Remark~\ref{rem:v-residue-distance}
and proved along each family by Theorem~\ref{thm:pole-ladders}(ii),
which gives $|a_{k}|=(|C|+o(1))\,d_{k}$ for the deep members.

\emph{Disjointness and partial intersection.} For $u^{*}\notin\tfrac{1}{2}\mathbb{Z}$
(the generic stability case), the two pole families are disjoint and
their elements alternate on $u<0$. At an EPD curve, $u^{*}\in\tfrac{1}{2}\mathbb{Z}$:
the two Floquet sub-combs coincide, and the two pole families merge
(intersection at almost all elements; the alternating gap pattern
collapses). 
\end{rem}

The simplicity of poles and zeros is treated in the next subsection
(Remark~\ref{rem:simplicity-evidence}, \S\,\ref{subsec:w-simplicity-notes}).
\begin{thm}[Asymptotic expansion of $w(u+m)$ as $m\to+\infty$]
\index{asymptotic seeding!expansion of $w(u+m)$} \label{thm:notes2-w-rate}
In the stability zone (Assumption~\eqref{eq:stability-assumption}),
for any fixed $u\in\mathbb{R}$ away from the poles of $w$, as $m\to+\infty$
with $m\in\mathbb{Z}$: 
\begin{equation}
w(u+m)\;=\;\zeta_{+}\;+\;\frac{a}{(u+m)^{2}}\;+\;\frac{b}{(u+m)^{3}}\;+\;\frac{c}{(u+m)^{4}}\;+\;O\!\left(\frac{1}{(u+m)^{5}}\right),\label{eq:notes2-w-rate}
\end{equation}
with no $1/(u+m)$ term. The coefficients are: 
\begin{align}
a\; & =\;\frac{-\zeta_{+}^{2}\,p^{2}}{\gamma\,(1-\zeta_{+}^{2})}\;\neq\;0,\label{eq:notes2-w-rate-a}\\
b\; & =\;\frac{2\,\zeta_{+}^{2}\,p^{2}}{\gamma\,(1-\zeta_{+}^{2})^{2}}\;=\;\frac{-2a}{1-\zeta_{+}^{2}},\label{eq:notes2-w-rate-b}\\
c\; & =\;\frac{\zeta_{+}^{2}\,p^{2}\,(3\gamma\zeta_{+}^{2}+3\gamma-p^{2}\zeta_{+})}{\gamma^{2}\,(\zeta_{+}^{2}-1)^{3}}.\label{eq:notes2-w-rate-c}
\end{align}
Since $a\neq0$ the rate is sharp: $w(u+m)-\zeta_{+}=\Theta(1/(u+m)^{2})$.
In particular the infinite product 
\begin{equation}
\prod_{m=0}^{\infty}\frac{w(u+m)}{\zeta_{+}}\label{eq:notes2-wprod}
\end{equation}
converges absolutely for all $u$ away from the poles of $w$, enabling
the definition of $H(u)$ in Section~\ref{subsec:Cambi-notes2-H}. 
\end{thm}

\begin{proof}
The claim is an \emph{asymptotic expansion} in the Poincaré sense
(valid for $v=u+m\to+\infty$ along $\mathbb{R}$), not a convergent
power series in $1/v$; the latter cannot exist globally because $\infty$
is an essential singularity of $w$ (Theorem~\ref{thm:w-essential-singularity}).
The proof has two parts: \emph{coefficient determination} (the $a_{k}$
are forced by matching orders in the recurrence) and \emph{residual
verification} (the truncated expansion satisfies the recurrence to
the claimed order, hence equals $w$ to that order).

\emph{Coefficient determination.} Set $v=u+m$, $t=1/v\to0^{+}$ and
write the truncated ansatz $E_{n}(v):=\zeta_{+}+\sum_{k=1}^{n}a_{k}t^{k}$.
Substitute $E_{n}$ into the exact recurrence $w(v+1)=-G(v+1)/\gamma-1/w(v)$
and expand both sides in powers of $t=1/v$. Using $G(v+1)=1-p^{2}t^{2}+2p^{2}t^{3}+O(t^{4})$
and $E_{n}(v+1)=\zeta_{+}+\sum_{k}a_{k}t^{k}(1+t)^{-k}=\zeta_{+}+\sum_{k}a_{k}t^{k}(1-kt+\ldots)$,
matching successive powers of $t$ gives the linear equations: 
\begin{align}
t^{1}: & \quad a_{1}(1-1/\zeta_{+}^{2})=0\;\Rightarrow\;a_{1}=0,\label{eq:notes2-a1-zero}\\
t^{2}: & \quad a_{2}(1-1/\zeta_{+}^{2})=p^{2}/\gamma\;\Rightarrow\;a_{2}=a,\label{eq:notes2-a-coeff}\\
t^{3}: & \quad a_{3}(1-1/\zeta_{+}^{2})=2a_{2}-2p^{2}/\gamma=2a_{2}/\zeta_{+}^{2}\;\Rightarrow\;a_{3}=b,\label{eq:notes2-b-coeff}
\end{align}
with $1-1/\zeta_{+}^{2}\neq0$ since $|\zeta_{+}|<1$; the coefficient
$a_{4}=c$ is determined by the analogous order-$t^{4}$ matching,
yielding the closed forms~\eqref{eq:notes2-w-rate-a}--\eqref{eq:notes2-w-rate-c}.

\emph{Residual verification.} With these $a_{k}$, substituting $E_{4}(v)$
into the recurrence yields a residual $E_{4}(v+1)+G(v+1)/\gamma+1/E_{4}(v)=O(1/v^{5})$
by construction (the first four orders cancel exactly). The exact
solution $w(v)$ satisfies the recurrence with zero residual and the
boundary condition $w(v)\to\zeta_{+}$ at $+\infty$. Linearizing
the recurrence near the fixed point $\zeta_{+}$ gives a linear three-term
system with multiplier $1/\zeta_{+}^{2}$, $|1/\zeta_{+}^{2}|>1$:
forward iteration is unstable, backward iteration is contracting (rate
$|\zeta_{+}^{2}|<1$). The minimal solution $w$ is the unique solution
lying on the stable manifold backward from $+\infty$. By the standard
asymptotic theory of minimal solutions of three-term recurrences (\cite[Ch.~4]{Wimp},
applied to the linearization), the residual order is inherited by
the difference: 
\[
w(v)-E_{4}(v)\;=\;O\bigl(1/v^{5}\bigr)\qquad\text{as }v\to+\infty\text{ along }\mathbb{R}_{>0}.
\]
This is the assertion of~\eqref{eq:notes2-w-rate}. The sharpness
follows since $a_{2}=a\neq0$.

Convergence of the product~\eqref{eq:notes2-wprod} follows from
$|w(u+m)/\zeta_{+}-1|=O(1/m^{2})$ and $\sum_{m\geq1}1/m^{2}<\infty$. 
\end{proof}
\begin{rem}[$\zeta_{+}$ as the recessive root; minimality at $\infty$]
\label{rem:zeta-recessive-AhlPet}\index{Riccati recurrence} The
limit value $\zeta_{+}$ is the recessive (minimal-modulus) root of
the characteristic equation $\gamma\zeta^{2}+\zeta+\gamma=0$ of the
constant-coefficient limit recurrence ($G(u)\to1$ as $u\to+\infty$),
with $|\zeta_{+}|<1<|\zeta_{-}|$. In the discrete-Riccati language
this is precisely the statement that the distinguished solution at
$\infty$ is minimal: under the substitution relating a Riccati solution
to a root of the characteristic equation, the negative Riccati solution
corresponds to the recessive root of modulus $<1$ and the positive
one to the dominant root of modulus $>1$, and the distinguished solution
at $\infty$ is the unique minimal one. See Ahlbrandt and Peterson~\cite[\S\,6.5, Thm.~6.15]{AhlPet}
for the characteristic equation and the recessive/dominant dichotomy,
and~\cite[\S\,6.6, Thm.~6.17]{AhlPet} for the minimality of the
distinguished solution at $\infty$; the convergence $w(u+m)\to\zeta_{+}$
established above is the present realization of that minimality. 
\end{rem}

\begin{rem}[Asymptotic, not convergent]
\label{rem:asymptotic-not-convergent} The expansion~\eqref{eq:notes2-w-rate}
is a Poincaré asymptotic series valid as $v\to+\infty$ along the
positive real axis (and, by extension, in any sector avoiding the
poles of $w$ accumulating at $-\infty$). It is not a convergent
power series in $1/v$: no such convergent expansion exists globally
because $\infty$ is an essential singularity of $w$ (Theorem~\ref{thm:w-essential-singularity}).
The residual verification in~Remark~\ref{rem:notes2-w-rate-verify}
confirms each coefficient to 40-digit precision, in line with the
Poincaré-asymptotic interpretation. 
\end{rem}

\begin{rem}[Numerical verification of the asymptotic coefficients]
\label{rem:notes2-w-rate-verify} The seeded CF (\S\,\ref{subsec:Cambi-approx})
with seed $T_{N}=\zeta_{+}+a/v_{N}^{2}+b/v_{N}^{3}+c/v_{N}^{4}$ ($v_{N}=u+N$)
gives seed error $O(1/v_{N}^{5})$. For Cambi's parameters ($\gamma=0.1$,
$p=20/3$): 
\[
a=-4.5824\ldots,\quad b=250/27\;(\text{exact}),\quad c=-224.179\ldots
\]
The fifth-order residual $R_{5}(m)=[w(1+m)-\zeta_{+}-a/v^{2}-b/v^{3}-c/v^{4}]\cdot v^{5}$
converges to $d\approx877.015$ monotonically, gaining one digit per
decade, confirming all three coefficients to 40-digit precision. 
\end{rem}

\subsection{\texorpdfstring{The discrete Riccati structure}{The discrete Riccati
structure}}

\label{subsec:Riccati-structure}\index{Riccati recurrence!structure theory}

The recurrence~\eqref{eq:notes2-wrec} for $w$ is a discrete Riccati
equation (Theorem~\ref{thm:notes2-wrec-exact}), the exact form treated
by Teschl~\cite[eq.~(1.52)]{Teschl}, Agarwal~\cite[\S\,6.5, eq.~(6.5.1)]{Agarwal},
and Ahlbrandt and Peterson~\cite[\S\,6.1]{AhlPet}. This short section
collects the structural consequences that the Riccati form makes transparent.
None requires new machinery; their value is twofold --- they anchor
the meromorphic structure of Section~\ref{subsec:Cambi-w-structure}
to a classical framework, and they furnish explicit formulas for the
residues of $w$ and for the action of the recurrence on poles.

\subsubsection*{Transfer-matrix viewpoint}

\index{Riccati recurrence!transfer-matrix form}\index{Mobius transformation}

Writing $\zeta=w(u-1)$, the recurrence~\eqref{eq:notes2-wrec},
$w(u)=-G(u)/\gamma-1/\zeta$, is the Mobius action 
\begin{equation}
w(u)\;=\;\frac{a\,\zeta+b}{c\,\zeta+d},\qquad\begin{pmatrix}a & b\\
c & d
\end{pmatrix}=\begin{pmatrix}-G(u)/\gamma & -1\\[2pt]
1 & 0
\end{pmatrix},\qquad ad-bc=1,\label{eq:Riccati-transfer}
\end{equation}
of the unimodular transfer matrix of the three-term recurrence~\eqref{eq:notes2-rec}.
Thus $w$ is carried from one unit step to the next by a fractional-linear
map of constant (here unit) determinant. This is precisely the mechanism
behind Lemma~\ref{lem:w-analytic-nevanlinna}: a Mobius cocycle of
positive determinant maps the open upper half-plane into itself, which
is why $w$ is analytic off $\mathbb{R}$ and carries the Nevanlinna-like
positivity $\operatorname{Im}w>0$ on the quadrant $\{\operatorname{Re}u>-1,\ \operatorname{Im}u>0\}$.
The convergence of the associated continued fraction to the minimal
(recessive) solution, and the identification of $\zeta_{+}$ as the
recessive characteristic root, are the discrete-Riccati statements
recorded in Remark~\ref{rem:zeta-recessive-AhlPet} and developed
at length by Ahlbrandt and Peterson~\cite[Ch.~2, \S\S\,6.5--6.6]{AhlPet}.

\subsubsection*{Residues and the action on poles}

The pole--zero pairing of Theorem~\ref{thm:notes2-w-poles-zeros}
--- every pole of $w$ is a zero of $w$ shifted one unit to the
right --- acquires, through the Riccati recurrence, an explicit residue
formula together with a complementary statement that poles do not
propagate forward.
\begin{prop}[Riccati residues and pole non-propagation]
\label{prop:Riccati-residues}\index{Riccati recurrence!residue formula}\index{Riccati recurrence!pole non-propagation}\index{$w(u)$!residue at a pole}\index{$w(u)$!pole non-propagation}Let
$u_{p}$ be a pole of $w$ arising from a simple zero $z_{0}:=u_{p}-1$
of $w$ (Theorem~\ref{thm:notes2-w-poles-zeros}). Then: 
\begin{enumerate}
\item[(i)] \emph{Residue.} $u_{p}$ is a simple pole and 
\begin{equation}
\operatorname*{Res}_{u=u_{p}}w\;=\;-\frac{1}{w'(z_{0})}.\label{eq:Riccati-residue}
\end{equation}
On the first sheet $\{\operatorname{Re}u_{p}>-1\}$, where $\operatorname{Im}w>0$
holds (Lemma~\ref{lem:w-analytic-nevanlinna}), $w$ increases through
its real zeros, so $w'(z_{0})>0$ and the residue is negative; the
sign of the residue is governed by the slope at the source zero and
may change on the second sheet $\{\operatorname{Re}u_{p}<-1\}$. 
\item[(ii)] \emph{Pole non-propagation.} The forward image of a pole is a regular
point, 
\begin{equation}
w(u_{p}+1)\;=\;-\frac{G(u_{p}+1)}{\gamma}.\label{eq:Riccati-pole-forward}
\end{equation}
Hence the poles are exactly the right-shifts $z_{0}\mapsto z_{0}+1$
of the zeros and do not chain forward (a pole's forward image, being
regular, is not itself a pole); the pole ladder is inherited from
the zero ladder and, like it, accumulates toward $-\infty$. 
\end{enumerate}
\end{prop}

\begin{proof}
Both statements read off~\eqref{eq:notes2-wrec}, $w(u)=-G(u)/\gamma-1/w(u-1)$.
For~(i), put $u=u_{p}+\eta=z_{0}+1+\eta$; since $w(z_{0})=0$ and
$w'(z_{0})\neq0$, $w(u-1)=w(z_{0}+\eta)=w'(z_{0})\,\eta+O(\eta^{2})$,
so $1/w(u-1)=\bigl(w'(z_{0})\,\eta\bigr)^{-1}\bigl(1+O(\eta)\bigr)$
and the singular part of $w(u)$ is $-1/\bigl(w'(z_{0})(u-u_{p})\bigr)$,
which is~\eqref{eq:Riccati-residue}; in particular the pole is simple.
For~(ii), evaluate~\eqref{eq:notes2-wrec} at $u=u_{p}+1$: since
$u_{p}$ is a pole, $w(u_{p})=\infty$ and $1/w(u_{p})=0$, leaving
$w(u_{p}+1)=-G(u_{p}+1)/\gamma$. The sign statement in~(i) follows
from the half-plane positivity of Lemma~\ref{lem:w-analytic-nevanlinna}:
on $\{\operatorname{Re}u>-1\}$ a real-analytic Nevanlinna-like function
is strictly increasing between consecutive poles, so $w'>0$ at its
real zeros there. 
\end{proof}
\begin{rem}[Numerical confirmation]
\label{rem:Riccati-residue-numeric} At the running parameters $p=0.70$,
$\gamma=0.30$ the simple zero $z_{0}=-2.69285934\ldots$ produces
the pole $u_{p}=z_{0}+1=-1.69285934\ldots$, and~\eqref{eq:Riccati-residue}
gives $\operatorname*{Res}_{u=u_{p}}w=-1/w'(z_{0})=0.10773768\ldots$
--- positive, since $z_{0}$ lies on the second sheet --- matching
the residue computed directly from $w$ to $40$ digits. Likewise~\eqref{eq:Riccati-pole-forward}
gives $w(u_{p}+1)=w(-0.69285934\ldots)=-G(u_{p}+1)/\gamma=0.06906119\ldots$,
again to $40$ digits. 
\end{rem}

\subsubsection*{The resonance condition in symmetric-sum form}

The Riccati recurrence also recasts the double minimality condition.
By the left-propagation form~\eqref{eq:notes2-wrec-left}, $1/w(u-1)=-G(u)/\gamma-w(u)$,
so the identity of Theorem~\ref{thm:notes2-wrec-exact}(iii) reads
\begin{equation}
F_{w}(u)=0\quad\Longleftrightarrow\quad w(u)+w(-u)\;=\;-\frac{G(u)}{\gamma};\label{eq:Riccati-symmetric-sum}
\end{equation}
\index{$F_{w}$!symmetric-sum form}the resonance condition is the
statement that the \emph{symmetric sum} $w(u)+w(-u)$ equals $-G(u)/\gamma$.
On the imaginary axis $u=iy$ --- the symmetry axis $u\mapsto-u$,
and the edge $\operatorname{Re}u=0$ of the fundamental strip of Remark~\ref{rem:Fw-fundamental-strip}
--- evenness and the reality of the coefficients give $w(-iy)=\overline{w(iy)}$
and $G(iy)=1+p^{2}/y^{2}$, so 
\begin{equation}
F_{w}(iy)\;=\;2\operatorname{Re}w(iy)\;+\;\frac{1+p^{2}/y^{2}}{\gamma},\label{eq:Fw-imaginary-axis}
\end{equation}
\index{$F_{w}$!imaginary-axis criterion}which is real and tends to
$\zeta_{+}-\zeta_{+}^{-1}=\sqrt{1-4\gamma^{2}}/\gamma$ (here $8/3$)
as $y\to\infty$, consistent with~\eqref{eq:Fw-infinity-limit}.
Equation~\eqref{eq:Fw-imaginary-axis} furnishes a clean sufficient
condition for the absence of zeros on this edge: if $\operatorname{Re}w(iy)\geq-1/(2\gamma)$
then 
\begin{equation}
F_{w}(iy)\;\geq\;-\frac{1}{\gamma}+\frac{1+p^{2}/y^{2}}{\gamma}\;=\;\frac{p^{2}}{\gamma\,y^{2}}\;>\;0,\label{eq:Fw-imaginary-axis-bound}
\end{equation}
so $F_{w}$ has no zero on the imaginary axis. The hypothesis holds
with room to spare: numerically $\operatorname{Re}w(iy)$ ranges from
$w(0)=-0.763\ldots$ at $y=0$ up to $\zeta_{+}=-1/3$ as $y\to\infty$,
so $\min_{y}\operatorname{Re}w(iy)\approx-0.76$, comfortably above
$-1/(2\gamma)=-1.667$. Together with the bound $|w(-\tfrac{1}{2}+iy)|<1$
on the opposite edge $\operatorname{Re}u=\tfrac{1}{2}$ (Remark~\ref{rem:Fw-fundamental-strip}),
this excludes off-real zeros of $F_{w}$ on \emph{both} edges of the
fundamental strip $\{0\le\operatorname{Re}u\le\tfrac{1}{2}\}$, narrowing
the open question to the open interior $0<\operatorname{Re}u<\tfrac{1}{2}$.

\subsection{\texorpdfstring{Simplicity of poles and zeros of $w(u)$}{Simplicity
of poles and zeros of w(u)}}

\label{subsec:w-simplicity-notes}

This section presents numerical evidence and supporting analytical
arguments for the simplicity of all poles and zeros of the minimal
solution ratio $w(u)$ and the double minimality function $F_{w}(u)$
in the stability zone. The genericity claim --- that the parameter
set $(\gamma,p)$ for which a pole or zero of $w$ is non-simple has
measure zero --- is supported by an overdetermined-system argument
(see Remark~\ref{rem:simplicity-evidence} below). A full proof would
require showing that the relevant determinantal condition is a nontrivial
analytic function of $(\gamma,p)$.

\medskip{}

\emph{Numerical evidence.} For Cambi's parameters ($\gamma=0.1$,
$p=20/3$) the following is verified with 40-digit arbitrary-precision
arithmetic (see also Figs.~\ref{fig:notes2-w-marked}--\ref{fig:notes2-Fw-marked}
for graphical illustration at $p=0.70$, $\gamma=0.30$).

\emph{Simple zeros.} Two zeros of $w$ are located: 
\begin{itemize}
\item $z_{0}\approx4.80635681$ with $w(z_{0})=0$ to 40 digits, $w'(z_{0})\approx3.1407\neq0$,
and a sign change of $w$ across $z_{0}$~--- confirming $z_{0}$
as a \emph{simple zero} of $w$. 
\item $z_{0}'\approx-8.58766883$ with $w'(z_{0}')\approx-3.4102\neq0$~---
similarly a \emph{simple zero} of $w$. 
\end{itemize}
\emph{Simple poles.} $w$ has a pole at $p_{0}=z_{0}+1\approx5.80635681$,
with residue 
\begin{equation}
\operatorname{res}_{u=p_{0}}w(u)\;=\;\lim_{u\to p_{0}}(u-p_{0})\,w(u)\;\approx\;-0.31839827\;\neq\;0,\label{eq:w-residue-numerical}
\end{equation}
confirmed to 8 decimal places by evaluating $w(p_{0}+\varepsilon)\cdot\varepsilon$
for $\varepsilon=10^{-8}$~--- confirming $p_{0}$ as a \emph{simple
pole} of $w$.

\medskip{}

\emph{Exact theoretical confirmation.} The recurrence~\eqref{eq:notes2-wrec}
\[
w(u+1)\;=\;-\frac{G(u+1)}{\gamma}-\frac{1}{w(u)}
\]
implies that zeros and poles of $w$ are related exactly: if $w$
has a \emph{simple zero} at $z_{0}$ so that $w(u)\sim w'(z_{0})(u-z_{0})$
near $z_{0}$, then near $u=z_{0}+1$: 
\begin{equation}
w(u)\;\sim\;-\frac{G(z_{0}+1)}{\gamma}-\frac{1}{w'(z_{0})(u-z_{0}-1)},\label{eq:w-zero-pole-link}
\end{equation}
exhibiting a \emph{simple pole} of $w$ at $z_{0}+1$ with exact residue
\begin{equation}
\operatorname{res}_{u=z_{0}+1}w(u)\;=\;-\frac{1}{w'(z_{0})}.\label{eq:w-residue-formula}
\end{equation}
Numerically: $-1/w'(z_{0})=-1/3.140721\approx-0.31839827$, matching~\eqref{eq:w-residue-numerical}
exactly.

\medskip{}

\begin{rem}[Evidence for simplicity of poles and zeros of $w(u)$]
\label{rem:simplicity-evidence} The numerical evidence above and
the algebraic structure below together support the view that all poles
and zeros of $w(u)$ are simple in the stability zone (Assumption~\eqref{eq:stability-assumption}).

\emph{Algebraic structure at a pole.} Since $w(u)=h_{1}(u)/h_{0}(u)$,
the order of a pole of $w$ at $u=u_{p}$ equals $\mathrm{ord}_{u_{p}}(h_{0})-\mathrm{ord}_{u_{p}}(h_{1})$.
The Hill recurrence is linear and homogeneous in $\{h_{n}\}$, so
multiplying every $h_{n}$ by $(u-u_{p})^{-\mathrm{ord}_{u_{p}}(h_{1})}$
yields a new sequence satisfying the same recurrence, with $w$ unchanged
and with $h_{1}$ now analytic and nonzero at $u_{p}$. Under this
normalization the pole order of $w$ equals the zero order of $h_{0}$
at $u_{p}$.

For the pole to be \emph{double}, the conditions on $u_{p}$ are 
\[
h_{0}(u_{p})\;=\;0\qquad\text{and}\qquad\partial_{u}h_{0}(u_{p})\;=\;0.
\]
These are two independent conditions on the single complex variable
$u_{p}$ (at fixed $\gamma,p$). For consistency, the Hill recurrence
at $n=0$, 
\[
\gamma h_{1}(u)+G(u)h_{0}(u)+\gamma h_{-1}(u)\;=\;0,
\]
forces the additional relations 
\[
h_{1}(u_{p})+h_{-1}(u_{p})\;=\;0\quad\text{and}\quad\partial_{u}h_{1}(u_{p})+\partial_{u}h_{-1}(u_{p})\;=\;0,
\]
obtained by setting $u=u_{p}$ and by differentiating then setting
$u=u_{p}$. These two relations are automatic consequences of the
double-zero conditions on $h_{0}$ via the recurrence and do not add
independent constraints on $u_{p}$.

\medskip{}
\emph{Algebraic structure at a zero.} Since $u_{z}$ is a zero of
$w$ if and only if $u_{z}+1$ is a pole of $w$ (Theorem~\ref{thm:notes2-w-poles-zeros}(i)),
zeros and poles of $w$ are in bijection. Independently, write $h_{1}(u_{z})=0$
in the normalization where $h_{0}$ is analytic and nonzero at $u_{z}$.
A double zero of $w$ at $u_{z}$ requires 
\[
h_{1}(u_{z})\;=\;0\qquad\text{and}\qquad\partial_{u}h_{1}(u_{z})\;=\;0,
\]
again two independent conditions on the single variable $u_{z}$.
The Hill recurrence at $n=1$, 
\[
\gamma h_{2}(u)+G(u+1)h_{1}(u)+\gamma h_{0}(u)\;=\;0,
\]
forces $h_{2}(u_{z})+h_{0}(u_{z})=0$ and $\partial_{u}h_{2}(u_{z})+\partial_{u}h_{0}(u_{z})=0$
as consequences.

\emph{Simplicity of the deep ladder members: a theorem.} For poles
of $w$ sufficiently deep in either ladder, simplicity is no longer
only evidence: granted the simplicity of the base comb zero of $F_{w}$
(${F_{w}}'(x_{0})\neq0$ --- the zero \emph{set} is Theorem~\ref{thm:Fw-zero-set};
simplicity of the base crossings is supported by the evidence of this
section), Step~3 of the proof of Theorem~\ref{thm:pole-ladders}
shows that each sufficiently deep pole of $w$ is simple, and the
unit-shift pairing (Theorem~\ref{thm:notes2-w-poles-zeros}(i)) transfers
simplicity to the paired zeros. The evidence of this section thus
carries the essential content for the base members, while the deep
members are covered by the theorem.

\medskip{}
\emph{Graphical evidence.} The clean hyperbolic shape of each pole
in Fig.~\ref{fig:notes2-w-marked}(a,b) --- $w(u)\sim A/(u-u_{p})$
with no higher-order terms --- is the graphical signature of simplicity.
Each zero in Fig.~\ref{fig:notes2-w-marked}(c) crosses the axis
with nonzero slope, consistent with simplicity. The Laurent zoom panels
of Fig.~\ref{fig:Fw-poles}(b)--(g) confirm that the one-term fit
$A/(u-u_{p})+C$ tracks $F_{w}$ accurately near each pole, with no
evidence of higher-order terms; via the two pole sources of \S\,\ref{subsec:stealthy-zeros}
this reflects the simplicity of the underlying poles and zeros of
$w$.

\emph{Genericity.} At fixed $(\gamma,p)$, two independent conditions
on the single variable $u_{p}$ (resp.\ $u_{z}$) form an overdetermined
system, which generically has no solution. We conjecture that the
exceptional set 
\[
\mathcal{E}\;=\;\bigl\{(\gamma,p)\in(0,\tfrac{1}{2})\times\mathbb{R}_{>0}\;:\;w(\,\cdot\,;\gamma,p)\text{ has a non-simple pole or zero}\bigr\}
\]
has measure zero in parameter space. A proof would require showing
that the determinantal condition obtained by eliminating $u_{p}$
(resp.\ $u_{z}$) between the two equations is a nontrivial analytic
function of $(\gamma,p)$. 
\end{rem}

\subsection{\texorpdfstring{Poles, zeros, and residues of $v(u)$, and the residue-distance
law}{Poles, zeros, and residues of v(u), and the residue-distance
law}}

\label{subsec:notes-on-v}

These notes use the meromorphic structure of the minimal solution
ratio $w$ from \S\,\ref{subsec:Cambi-w-structure} and the simplicity
from \S\,\ref{subsec:w-simplicity-notes} to determine the corresponding
pole-zero properties of $v(u)$ via the relation $v(u)=-w(u-1)/\gamma$.
\begin{rem}[Reality of poles and zeros of $v(u)$ in the stability zone]
\label{rem:v-poles-zeros-real} In the stability zone (Assumption~\eqref{eq:stability-assumption}),
the poles and zeros of $v$ follow directly from the relation $v(u)=-w(u-1)/\gamma$
combined with the unit-shift pole-zero pairing of $w$ (Theorem~\ref{thm:notes2-w-poles-zeros}(i)):
\begin{align}
\{\text{zeros of }v\} & =\{\text{zeros of }w\}+1=\{\text{poles of }w\},\label{eq:v-zeros}\\
\{\text{poles of }v\} & =\{\text{zeros of }w\}+2=\{\text{poles of }w\}+1.\label{eq:v-poles}
\end{align}
In \eqref{eq:v-zeros} the first equality is the unit shift from $v(u)=-w(u-1)/\gamma$,
and the second is the pole-zero pairing (Theorem~\ref{thm:notes2-w-poles-zeros}(i)):
shifting a sign-change zero $u_{z}$ of $w$ by $+1$ lands precisely
at the paired pole $u_{p}=u_{z}+1$ of $w$, which is also where $v$
vanishes. Equation~\eqref{eq:v-poles} has the same structure shifted
by one: its second equality is again the pairing ($\{\text{zeros of }w\}+1=\{\text{poles of }w\}$,
with $+1$ added throughout), and the first follows from the unit
shift $\{\text{poles of }v\}=\{\text{poles of }w\}+1$. By Remark~\ref{rem:w-pole-zero-geometry}(a)
the poles of $w$ are all real (in fact all $u_{p}<0$), and by Theorem~\ref{thm:notes2-w-poles-zeros}(i)
the zeros of $w$ are also all real ($u_{z}=u_{p}-1$ for real $u_{p}$).
Therefore \emph{all poles and zeros of $v(u)$ are real}. Since the
poles of $v$ are $\{u_{p}+1\}$ with $u_{p}<-1$ (Remark~\ref{rem:w-pole-zero-geometry}(b)),
the poles of $v$ satisfy $u_{v}=u_{p}+1<0$. The single exception
is the base pole of the $Z_{F}^{-}$ sub-comb: the $w$-pole nearest
$-0.252$ shifts to $u_{v}\approx0.748>0$ (close to, though not exactly
at, $u^{*}\approx0.742$), the only positive pole of $v$, visible
in Fig.~\ref{fig:v-u}(b). 
\end{rem}

\begin{rem}[Residues of poles of $v(u)$ and sign pattern]
\label{rem:v-residue-sign} Since $v(u)=-w(u-1)/\gamma$ and $w$
has simple poles generically (Remark~\ref{rem:simplicity-evidence}),
each pole $u_{v}$ of $v$ is also simple with residue 
\begin{equation}
\mathrm{res}_{u_{v}}(v)\;=\;-\frac{1}{\gamma}\,\mathrm{res}_{u_{v}-1}(w).\label{eq:v-residue-formula}
\end{equation}
The sign of $\mathrm{res}_{u_{v}-1}(w)$ determines that of the $v$-residue.
The two poles of $v$ that are well separated from the zeros of $v$
are reliably resolved; for $p=0.70$, $\gamma=0.30$ they are (residues
from contour integration, $40$-digit arithmetic, in agreement with~\eqref{eq:v-residue-formula}
to all displayed digits): 
\begin{equation}
\begin{array}{r|r|r|c}
\text{pole }u_{v} & \text{residue }\mathrm{res}_{u_{v}}(v) & \text{dist.\ to nearest zero} & \text{sign}\\
\hline -0.6929 & -0.3591 & 0.440 & -\\
+0.7475 & +0.4205 & 1.000 & +
\end{array}\label{eq:v-residue-table}
\end{equation}
These two residues already have opposite signs while sharing no common
pattern with the unit spacing, so the residues of $v$ exhibit \emph{no
strict alternation}: the sign of each residue depends on the global
arrangement of poles and zeros of $w$, not on the local ordering,
because the poles of $v$ belong to two interleaving families (one
tracking $\{u^{*}+n\}$, the other tracking $\{-u^{*}+n\}$) that
are irregularly spaced.

The remaining poles of $v$ in this window lie very close to zeros
of $v$ and are correspondingly difficult to resolve numerically:
they sit near the Floquet points $Z_{F}$, where a pole of $v$ and
its paired zero approach each other exponentially fast (a factor $|\zeta_{+}|^{2}$
per unit of depth into the comb, Theorem~\ref{thm:pole-ladders}).
By the residue-distance law below (Remark~\ref{rem:v-residue-distance})
their residues are correspondingly small --- in places below $10^{-3}$,
and far smaller for the poles nearest a Floquet point. Such poles
are \emph{stealthy} in the sense of Remark~\ref{rem:double-precision-misses};
resolving them reliably requires the dedicated sign-change detector
developed in \S\,\ref{subsec:Fw-pole-detection} (Theorem~\ref{thm:Fw-sign-change-pole},
Corollary~\ref{cor:Fw-pole-bisection}), where the phenomenon is
illustrated concretely. Here we therefore record only the two well-separated
poles above, whose residues are unambiguous. 
\end{rem}

\begin{rem}[Partial fraction representation of $w(u)$ and residue-distance law]
\label{rem:v-residue-distance} In the stability zone (Assumption~\eqref{eq:stability-assumption}),
all poles of $w$ are real and simple (Theorem~\ref{thm:notes2-w-poles-zeros}
and Remark~\ref{rem:simplicity-evidence}). Fix a rectangle $\Pi_{a,b}^{\eta}=\{u\in\mathbb{C}:a<\operatorname{Re}(u)<b,\;|\operatorname{Im}(u)|<\eta\}$
where $[a,b]$ is any bounded real interval and $\eta>0$ is chosen
small enough that all poles of $w$ in the strip $|\operatorname{Im}(u)|<\eta$
are real (guaranteed by Remark~\ref{rem:w-pole-zero-geometry}(a)
since all poles of $w$ are real). By the standard partial-fraction
decomposition for meromorphic functions on a bounded domain \cite[Ch.~VII, \S\,5]{Conway},
\cite[Ch.~5, \S\,2]{Ahlf}, the restriction of $w$ to $\Pi_{a,b}^{\eta}$
decomposes as 
\begin{equation}
w(u)\;=\;\sum_{m=1}^{N}\frac{a_{m}}{u-u_{m}}\;+\;g(u),\qquad u\in\Pi_{a,b}^{\eta},\label{eq:w-partial-fraction}
\end{equation}
where $u_{1},\ldots,u_{N}$ are the (finitely many, real, simple)
poles of $w$ in $\Pi_{a,b}^{\eta}$, $a_{m}=\mathrm{res}_{u_{m}}(w)$
is the residue at each pole (Remark~\ref{rem:simplicity-evidence}),
and $g(u)$ is analytic on $\Pi_{a,b}^{\eta}$. The function $g$
is the \emph{analytic background}: it captures the combined contribution
of all poles \emph{outside} the rectangle, and satisfies $g(u)\approx\zeta_{+}$
away from the poles (since $w(u)\to\zeta_{+}$ as $u\to+\infty$,
Theorem~\ref{thm:notes2-w-rate}; see Figs.~\ref{fig:notes2-w-u-plot},~\ref{fig:notes2-w-marked}).

The decomposition~\eqref{eq:w-partial-fraction} makes precise the
\emph{residue-distance law} observed numerically. Let $u_{m}$ be
a pole and $u_{z}$ the nearest zero of $w$, with $d=|u_{m}-u_{z}|$.
From~\eqref{eq:w-partial-fraction}, evaluating at $u=u_{z}$ where
$w(u_{z})=0$: 
\begin{equation}
0\;=\;\frac{a_{m}}{u_{z}-u_{m}}\;+\;\sum_{k\neq m}\frac{a_{k}}{u_{z}-u_{k}}\;+\;g(u_{z}).\label{eq:w-pf-at-zero}
\end{equation}
When $d=|u_{m}-u_{z}|$ is small compared to the spacing to all other
poles, the cross terms $a_{k}/(u_{z}-u_{k})$ are slowly varying and
approximately absorbed into $g(u_{z})$, giving 
\begin{equation}
\frac{a_{m}}{u_{z}-u_{m}}\;+\;B\;+\;O(d)\;=\;0,\qquad B\;:=\;\sum_{k\neq m}\frac{a_{k}}{u_{z}-u_{k}}+g(u_{z}),\label{eq:w-residue-dist-proof}
\end{equation}
hence $a_{m}=B(u_{m}-u_{z})+O(d^{2})$, i.e.\
\begin{equation}
|a_{m}|\;\approx\;|B|\cdot d,\label{eq:w-residue-dist-law}
\end{equation}
where $B$ is the value of the analytic background plus cross-pole
contributions at the zero, a slowly varying quantity. Thus \emph{a
pole of $w$ close to one of its zeros has a small residue, and a
pole far from any zero has a large residue}. Along each pole ladder,
Theorem~\ref{thm:pole-ladders} proves this law in sharp form: for
the deep members the ratio converges, $|a_{m}|/d\to|\zeta_{+}|$,
which is exactly the limiting background value $B\to\zeta_{+}$ of~\eqref{eq:w-partial-fraction}.

This is directly visible in the Laurent zoom panels of Fig.~\ref{fig:notes2-w-marked}
and the residue comparison of Fig.~\ref{fig:Fw-poles}(b)--(g):
the orange Laurent fit $a/(u-u_{m})+C$ (where $C\approx B$ is the
background) tracks $w(u)$ closely near each pole, with large-residue
poles (panels~(c)--(f)) lying far from their paired zeros, and the
tiny-residue poles (panels~(b,g), the deepest resolvable ladder members
at $u\approx\pm2.735$) exponentially close to a Floquet zero (Theorem~\ref{thm:pole-ladders},
Remark~\ref{rem:double-precision-misses}). Numerically for $p=0.70$,
$\gamma=0.30$, the ratio $|a_{m}|/d$ lies in $[0.05,0.39]$, of
the order of the background scale $|\zeta_{+}|=\tfrac{1}{3}$ and
consistent with a slowly varying background $|B|$ (Figs.~\ref{fig:notes2-w-marked},~\ref{fig:Fw-poles},~\ref{fig:notes2-Fw-marked}). 
\end{rem}

\medskip{}

\subsection{\texorpdfstring{Meromorphic structure of $F_{w}$}{Meromorphic
structure of Fw}}

\label{subsec:Fw-meromorphic}

Building on the meromorphic structure of the minimal solution ratio
$w$ established in \S\,\ref{subsec:Cambi-w-structure}, this section
analyzes the zero-pole structure of the double minimality function
$F_{w}(u)=w(-u)-1/w(u-1)$, equivalently $\widetilde{F}_{w}=\gamma F_{w}$
(Theorem~\ref{thm:notes2-resonance-equiv}). Throughout this section,
$(p,\gamma)$ is in the stability zone (Assumption~\eqref{eq:stability-assumption}),
so $u^{*}$ is real.

The rigorous content is contained in two theorems: that $w$ and $\widetilde{F}_{w}$
have an essential singularity at $\infty$ (Theorem~\ref{thm:w-essential-singularity})
and that the real zero set of $F_{w}$ is the two-comb Floquet lattice
$Z_{F}=\{u^{*}+n\}\cup\{-u^{*}+n\}$ (Theorem~\ref{thm:Fw-zero-set}).
The pole structure of $F_{w}$, including the candidate locations,
their relation to the poles of $w$, and the relative positions of
poles and zeros on the real axis, is described in the subsequent Remarks;
it is established at the level of numerical evidence and analytical
arguments rather than complete proof, and is used only for illustration;
the deep ladder members of the pole set are meanwhile proved in Theorem~\ref{thm:pole-ladders}
(\S\,\ref{subsec:Casoratian-Weyl}). The conjectural picture is
summarized in Remark~\ref{rem:Fw-pole-conjecture}.
\begin{thm}[Essential singularity at infinity]
\index{essential singularity} \label{thm:w-essential-singularity}
The functions $w(u)$ and $\widetilde{F}_{w}(u)$ are transcendental
meromorphic functions on $\mathbb{C}$; in particular, $u=\infty$
is an essential singularity or a non-isolated essential singularity
of each. 
\end{thm}

\begin{proof}
By Floquet theory (Theorem~\ref{thm:Hill-stability}), the double
minimality function $\widetilde{F}_{w}(u)$ has zeros at every element
of the two-lattice set 
\[
Z_{F}\;=\;\{u^{*}+n:n\in\mathbb{Z}\}\;\cup\;\{-u^{*}+n:n\in\mathbb{Z}\}
\]
(stated formally as Theorem~\ref{thm:Fw-zero-set} below). This zero
set $Z_{F}$ is infinite and unbounded: $Z_{F}$ contains a sequence
of zeros with $|u_{n}|\to\infty$.

We use the conventional classification of isolated singularities (\cite[Ch.~V, \S\,1]{Conway}):
an isolated singularity that is neither removable nor a pole is called
an \emph{essential singularity}. We analyze the behavior of $\widetilde{F}_{w}$
at $\infty$ in two mutually exclusive cases.

\emph{Case 1: poles of $\widetilde{F}_{w}$ accumulate at $\infty$.}
Then $\infty$ is not an isolated singularity of $\widetilde{F}_{w}$;
in standard terminology this is a \emph{non-isolated essential singularity}.
Since $\widetilde{F}_{w}$ is meromorphic on $\mathbb{C}$ but does
not extend meromorphically to $\hat{\mathbb{C}}=\mathbb{C}\cup\{\infty\}$
(the would-be singularity at $\infty$ is not isolated), $\widetilde{F}_{w}$
is transcendental. The conclusion of the theorem holds immediately
in this case.

\emph{Case 2: poles of $\widetilde{F}_{w}$ do not accumulate at $\infty$.}
Then there exists $R>0$ such that $\widetilde{F}_{w}$ is holomorphic
on $\{u\in\mathbb{C}:|u|>R\}$, so $\infty$ \emph{is} an isolated
singularity. By the trichotomy of \cite[Ch.~V, \S\,1]{Conway}, it
is removable, a pole, or essential. The first two would mean $\widetilde{F}_{w}$
extends meromorphically to $\hat{\mathbb{C}}$, hence (\cite[Ch.~V, \S\,3]{Conway})
is a rational function. But a rational function has only finitely
many zeros in $\hat{\mathbb{C}}$, contradicting the infinitude of
$Z_{F}$. Therefore $\infty$ is an isolated essential singularity
of $\widetilde{F}_{w}$, and $\widetilde{F}_{w}$ is transcendental.

In both cases $\widetilde{F}_{w}$ is transcendental meromorphic on
$\mathbb{C}$, and $\infty$ is an essential singularity or a non-isolated
essential singularity.

The same conclusion for $w$ follows by the algebraic relation $\widetilde{F}_{w}(u)=\gamma(w(-u)-1/w(u-1))$
(Theorem~\ref{thm:notes2-resonance-equiv}): if $w$ were rational,
then $w(-u)$ and $1/w(u-1)$ would be rational, hence $\widetilde{F}_{w}$
would be rational --- contradicting the just-proved transcendence.
Therefore $w$ is transcendental meromorphic on $\mathbb{C}$ as well. 
\end{proof}
\begin{rem}[Economy of the proof]
\label{rem:essential-singularity-proof} The proof uses only Floquet
theory for the zero set~\eqref{eq:Fw-zero-set} and the algebraic
relation $\widetilde{F}_{w}=\gamma F_{w}$; the pole structure of
$\widetilde{F}_{w}$ near infinity, the zero-pole counts of $w$,
the CF structure, and any numerical input are not used. The asymptotic
$w(u)\to\zeta_{+}$ as $u\to+\infty$ (Theorem~\ref{thm:notes2-w-rate})
is consistent with an essential singularity at $\infty$, since a
limit along the real axis does not determine the global nature of
an isolated singularity. 
\end{rem}

\begin{rem}[Graphical signature of the essential singularity]
\label{rem:essential-singularity-figure} The essential singularity
is visible in Fig.~\ref{fig:notes2-w-marked}(d,f) as the apparent
accumulation of poles for $u\to-\infty$, and the asymptote $w(u)\to\zeta_{+}$
as $u\to+\infty$ is visible in panels~(e,f). The graphical picture
is consistent with the \emph{non-isolated} essential singularity case
(poles clustering at $\infty$); Theorem~\ref{thm:w-essential-singularity}
does not distinguish the isolated and non-isolated cases. 
\end{rem}

\begin{thm}[Zero set of $F_{w}$: the Floquet comb]
\index{Floquet theory!Floquet comb}\index{double minimality function $F_{w}$!zeros (the Floquet comb)}
\label{thm:Fw-zero-set} Under the stability-zone assumption~\eqref{eq:stability-assumption},
the real zero set of $F_{w}$ on $\mathbb{R}$ is the two-comb Floquet
lattice 
\begin{equation}
Z_{F}\;=\;\{u^{*}+n:n\in\mathbb{Z}\}\;\cup\;\{-u^{*}+n:n\in\mathbb{Z}\},\label{eq:Fw-zero-set}
\end{equation}
where $u^{*}$ and $-u^{*}$ are the two primary Floquet exponents. 
\end{thm}

\begin{proof}
The proof rests on three independent mathematical facts.

\emph{Fact 1 (Floquet iff double minimality).} In the stability zone,
$u\in\mathbb{R}$ is a Floquet exponent of Hill's equation~\eqref{eq:LC-Hill}
if and only if $\widetilde{F}_{w}(u)=0$. This iff is established
in \S\,\ref{subsec:Cambi-notes2-res} and stated formally as Theorem~\ref{thm:double-min}(iii):
the equation $\widetilde{F}_{w}(u)=0$ is the necessary and sufficient
condition for the Floquet periodic factor to be a bounded regular
function (equivalently, for the forward and backward minimal solutions
of the recurrence~\eqref{eq:notes2-ttrec} to match at index $n=0$
after a scalar rescaling).

\emph{Fact 2 (integer-translation invariance of Floquet exponents).}
Floquet multipliers are computed as $\rho=e^{2\pi iu}$, and $e^{2\pi i(u+n)}=e^{2\pi iu}$
for any $n\in\mathbb{Z}$. Therefore the set of Floquet exponents
is automatically invariant under integer translation: $u$ is a Floquet
exponent $\iff u+n$ is a Floquet exponent for every $n\in\mathbb{Z}$.
This is a purely arithmetic property of the period-$1$ convention
for Floquet exponents and is independent of the recurrence for $w$.

\emph{Fact 3 (two primary exponents in the stability zone).} In the
stability zone $|\Delta|<2$ (Theorem~\ref{thm:Hill-stability}(i)),
the monodromy matrix of Hill's equation has two distinct unit-modulus
eigenvalues $\rho_{\pm}=e^{\pm2\pi iu^{*}}$ with $u^{*}\in(0,\tfrac{1}{2})$.
The complete set of real Floquet exponents is therefore $\{u^{*}+n:n\in\mathbb{Z}\}\cup\{-u^{*}+n:n\in\mathbb{Z}\}=Z_{F}$.

\emph{Conclusion.} Combining Facts 1--3: 
\[
\widetilde{F}_{w}(u)=0\quad\iff\quad u\text{ is a real Floquet exponent}\quad\iff\quad u\in Z_{F}.
\]
The identity $F_{w}\equiv\widetilde{F}_{w}/\gamma$ (Theorem~\ref{thm:notes2-resonance-equiv})
transfers the zero set to $F_{w}$, completing the proof. 
\end{proof}
\begin{rem}[Internal consistency of the Floquet comb $Z_{F}$ with the recurrence
for $w$]
\label{rem:Fw-zero-set-algebraic} The Floquet-theoretic argument
above is the rigorous proof of Theorem~\ref{thm:Fw-zero-set}. Independently,
the algebraic structure of the recurrence~\eqref{eq:notes2-wrec}
is fully consistent with the two symmetries of $Z_{F}$:

\emph{(a) Integer-shift consistency.} The recurrence implies the meromorphic
shift identity (Theorem~\ref{thm:notes2-integer-shift}, eq.~\eqref{eq:notes2-shift-identity}):
\begin{equation}
\widetilde{F}_{w}(u+1)\;=\;R(u)\cdot\widetilde{F}_{w}(u),\qquad R(u)\;:=\;\frac{-\gamma}{w(u)\cdot\bigl(\gamma w(-u)+G(u)\bigr)},\label{eq:Fw-meromorphic-shift}
\end{equation}
holding as an identity of meromorphic functions on $\mathbb{C}$.
Iteration gives, for $n\geq0$, 
\begin{equation}
\widetilde{F}_{w}(u+n)\;=\;\widetilde{F}_{w}(u)\cdot\prod_{k=0}^{n-1}R(u+k),\label{eq:Fw-meromorphic-shift-n}
\end{equation}
with the analogous formula for $n<0$ via the backward recurrence~\eqref{eq:notes2-wrec-left}.
This formula propagates a zero at $u=u^{*}$ to all integer translates
$u^{*}+n$, provided no $0\cdot\infty$ indeterminacy occurs at intermediate
points (i.e.\ $w$ has neither a zero nor a pole at $u^{*}+k$ for
$k$ in the propagation range). By Remark~\ref{rem:poles-w-not-on-ZF},
no pole of $w$ lies on $Z_{F}\supseteq\{u^{*}+k\}$; if $w$ had
a zero at $u^{*}+k$, then by Theorem~\ref{thm:notes2-w-poles-zeros}(i)
$u^{*}+k+1$ would be a pole of $w$ also in $Z_{F}$, again contradicting
Remark~\ref{rem:poles-w-not-on-ZF}. Hence the propagation is free
of indeterminacy.

\emph{(b) Reflection consistency.} The recurrence and the definition
$\widetilde{F}_{w}(u)=\gamma w(u)+G(u)+\gamma w(-u)$ make the reflection
symmetry manifest: since $G(u)=G(-u)$ (the function $G(u)=1-(p/u)^{2}$
is even in $u$), the right side is invariant under $u\mapsto-u$,
so $\widetilde{F}_{w}(u)=\widetilde{F}_{w}(-u)$ identically (Theorem~\ref{thm:double-min}(ii)).
Hence if $u^{*}$ is a zero, so is $-u^{*}$, generating the second
sub-comb $\{-u^{*}+n:n\in\mathbb{Z}\}$ from the first by reflection
followed by integer shifts.

Together, (a) and (b) imply the following: \emph{given the existence
of at least one base zero $u^{*}$ of $\widetilde{F}_{w}$ (which
is the Floquet-theoretic input)}, the algebraic structure of the recurrence
for $w$ alone reproduces both symmetries of $Z_{F}$: integer-shift
closure of each comb (a) and reflection-symmetry interchange of the
two combs (b). The algebraic structure does not manufacture the base
zero $u^{*}$ ex nihilo --- this comes from Floquet theory --- but
it confirms that the comb structure built around $u^{*}$ is consistent
with the recurrence. The agreement between the Floquet-theoretic derivation
and the algebraic structure is a non-trivial consistency check between
Cambi's CF construction of $w$ and the spectral theory of Hill's
equation. 
\end{rem}

\begin{figure}[htbp]
\centering \includegraphics[width=0.98\textwidth]{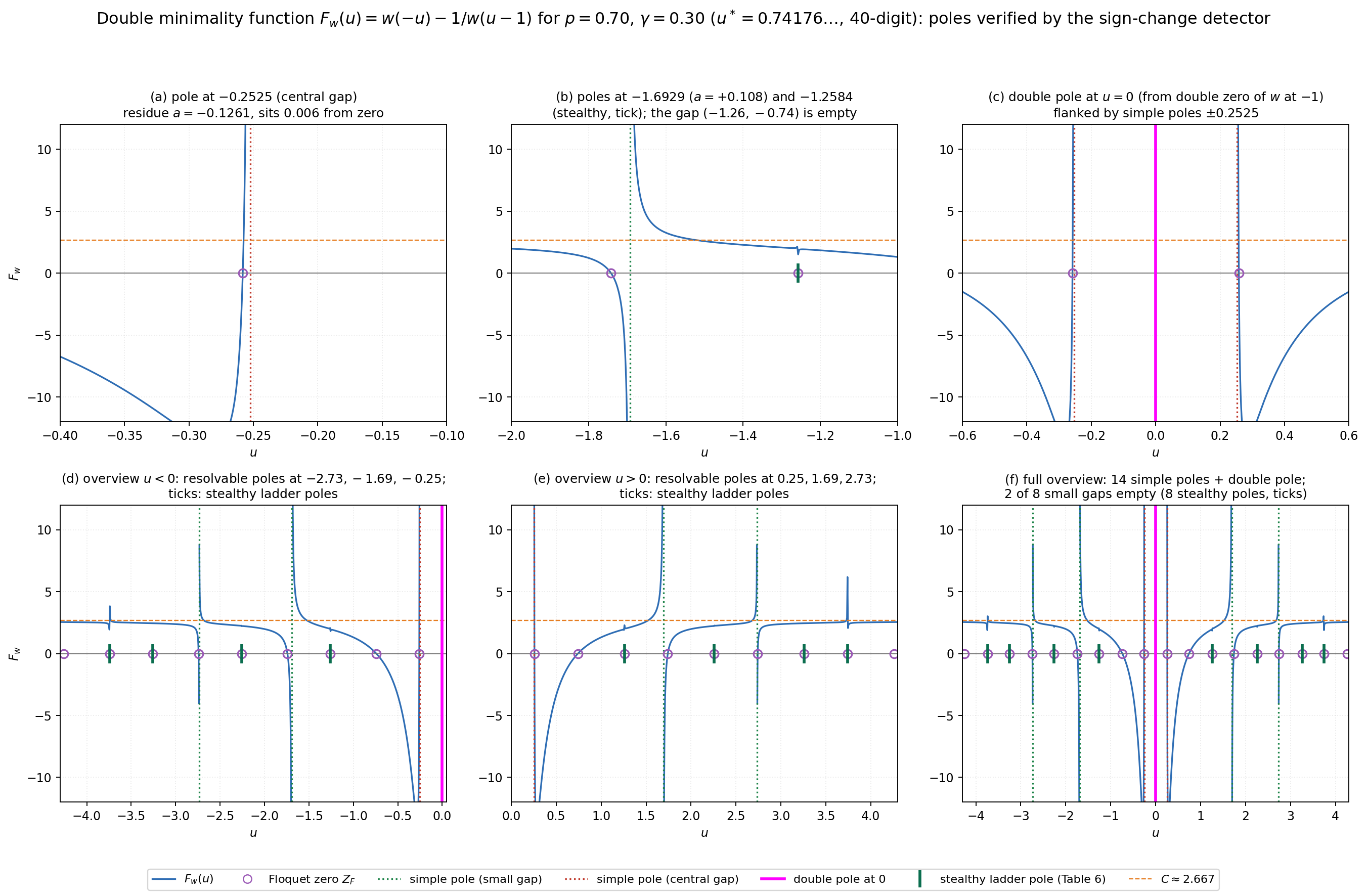}
\caption{Double minimality function $F_{w}(u)=w(-u)-1/w(u-1)$ for $p=0.70$,
$\gamma=0.30$ ($u^{*}=0.74176\ldots$, 40-digit), all poles verified
by the sign-change detector (Theorem~\ref{thm:Fw-sign-change-pole}).
Top row, zooms: (a)~the central-gap pole at $u\approx-0.252$ (residue
$-0.126$), $\approx0.006$ from the Floquet zero; (b)~the small-gap
pole at $u\approx-1.693$ with the stealthy ladder pole at $u\approx-1.258$
(tick), the large gap $(-1.26,-0.74)$ empty; (c)~the double pole
at $u=0$ (Theorem~\ref{thm:notes2-w-poles-zeros}(ii)), flanked
by the simple pair $\pm0.252$. Bottom row: overviews on $u<0$ (d),
$u>0$ (e), full range (f). Purple circles = zeros of $F_{w}$ (Floquet
comb $Z_{F}$, eq.~\eqref{eq:Fw-zero-set}); dotted = simple poles
(green: small-gap, red: central); solid magenta = double pole; short
dark-green ticks = the eight stealthy ladder poles (Theorem~\ref{thm:pole-ladders});
orange dashed = asymptote $C\approx2.667$. The complete catalog is
Table~\ref{tab:Fw-pole-catalog}. Many zeros abut a pole (the central
pair lies $\approx0.006$ from the zeros at $\pm0.258$) and so hide
in its spike --- genuine but nearly invisible by eye; see Remark~\ref{rem:masking-geometry}
for the mechanism and Figure~\ref{fig:invisible-zero-pole} for an
explicit resolution.}
\label{fig:notes2-Fw-marked} 
\end{figure}

\begin{rem}[Pole structure of $F_{w}$ on $\mathbb{R}$]
\label{rem:Fw-pole-sources} The Floquet zero set $Z_{F}$ (Theorem~\ref{thm:Fw-zero-set})
is what matters for the downstream theorems of this work: the Floquet-exponent
computation, the boundary-curve series (Theorem~\ref{thm:bdry-series-CF}),
and the Floquet periodic factor (Theorem~\ref{thm:Floquet-factor-MR})
all rest on zeros of $F_{w}$ exclusively. The poles of $F_{w}$ play
no role in those results. We record the pole structure as an interesting
structural feature of $F_{w}$, separating what follows rigorously
from the definition from what is supported by the structural picture
of $w$.

\emph{Candidate set of poles (rigorous).} From the definition $F_{w}(u)=w(-u)-1/w(u-1)$,
real singularities of $F_{w}$ arise from two sources: the first,
$w(-u)\to\infty$, i.e.\ $-u$ is a pole of $w$, gives a pole of
$F_{w}$ at $u=-u_{p}$ for each pole $u_{p}$ of $w$; the second,
$w(u-1)\to0$, i.e.\ $u-1$ is a zero of $w$, gives a pole of $F_{w}$
at $u=u_{z}+1$ for each zero $u_{z}$ of $w$. By Theorem~\ref{thm:notes2-w-poles-zeros}(i),
every sign-change zero $u_{z}$ of $w$ satisfies $u_{z}=u_{p}-1$
for some pole $u_{p}$ of $w$, so the corresponding $u_{z}+1$ locations
coincide with the pole positions $u_{p}$ themselves. The double zero
of $w$ at $u=-1$ (Theorem~\ref{thm:notes2-w-poles-zeros}(ii))
contributes directly: $w((u_{z}+1)-1)=w(-1)=0$ at $u_{z}+1=0$, so
$1/w(u-1)$ has a double pole at $u=0$, giving a double pole of $F_{w}$
at $u=0$. The set of candidate real poles of $F_{w}$ is therefore
\begin{equation}
\{u_{p}:u_{p}\text{ pole of }w\}\;\cup\;\{0\}\;\cup\;\{-u_{p}:u_{p}\text{ pole of }w\}.\label{eq:Fw-poles}
\end{equation}
This containment is rigorous from the definition of $F_{w}$.

\emph{Sign localization.} For $p\in(0,1)$, poles of $w$ lie on $u<0$
by Remark~\ref{rem:w-pole-zero-geometry}(a), so the first set in~\eqref{eq:Fw-poles}
lies on $u<0$ and the third on $u>0$. For $p>1$ this clean separation
no longer holds, since $w$ may then have poles on $u>0$ (Remark~\ref{rem:w-pole-zero-geometry});
the candidate set~\eqref{eq:Fw-poles} itself remains valid in all
cases.

\emph{Genuineness.} Whether each candidate location actually yields
a non-removable pole of $F_{w}$ depends on residue cancellation between
the two contributions (the $-u_{p}$ reflection and the $u_{z}+1$
shift); by the displacement principle (Remark~\ref{rem:poles-w-not-on-ZF}),
no cancellation occurs in practice. Numerical computation across the
stability zone confirms this picture for all tested $(p,\gamma)$;
see Fig.~\ref{fig:notes2-Fw-marked}. 
\end{rem}

\begin{rem}[Numerical $F_{w}$ pole locations at $p=0.70$, $\gamma=0.30$]
\label{rem:Fw-pole-numerical-example} For the example parameters
$p=0.70$, $\gamma=0.30$ ($u^{*}\approx0.7418$), the candidate poles
on $u<0$ (from zeros of $w$, at $u_{z}+1$) lie near $u\approx-3.741,-2.735,-2.258,-1.693,-1.258,-0.252$,
with the corresponding reflections $-u_{p}$ on $u>0$ and a double
pole at $u=0$; see Fig.~\ref{fig:notes2-Fw-marked}(d--f). These
candidate locations follow rigorously from~\eqref{eq:Fw-poles}.
Their numerical resolution, however, varies sharply with the size
of their residues; by Theorem~\ref{thm:pole-ladders} the residues
shrink geometrically with depth along each pole ladder, with a ladder-dependent
prefactor. The poles with residues of appreciable size --- here those
near $u\approx-0.252$, $-1.693$, $-2.735$, and $-3.741$ --- are
cleanly resolved on a refined grid, the double pole at $u=0$ being
the most prominent ($|F_{w}|\sim10^{4}$ at $u=\pm0.01$). By contrast,
the candidates with exponentially small residues (here $u\approx-1.258$
and $u\approx-2.258$, deep members of their ladder) are \emph{stealthy}:
their residues are exponentially small, so they sit too close to the
neighboring zero of $F_{w}$ to be cleanly separated from it at fixed
precision, exactly as for the corresponding poles of $w$ (Remark~\ref{rem:v-residue-sign}).
Such stealthy poles can nonetheless be located reliably by the sign-change
detector of \S\,\ref{subsec:Fw-pole-detection} (Theorem~\ref{thm:Fw-sign-change-pole},
Corollary~\ref{cor:Fw-pole-bisection}). Because a true pole has
unbounded $|F_{w}|$, any sampled peak value is a property of the
grid spacing rather than of the pole; the pole-intrinsic measure of
detectability is the residue, and it is the poles closest to $Z_{F}$
whose residues are smallest. The two-family structure of $w$'s poles
(Remark~\ref{rem:w-two-families}) is inherited by $F_{w}$, with
each $F_{w}$-pole exponentially close to a Floquet zero in $Z_{F}$;
it is precisely the poles closest to $Z_{F}$ that are hardest to
detect. 
\end{rem}

\begin{rem}[Displacement of $w$-poles from the Floquet zero set]
\label{rem:poles-w-not-on-ZF} No pole of $w$ coincides with any
element of $Z_{F}=\{u^{*}+n\}\cup\{-u^{*}+n\}$.

\emph{Argument.} Suppose for contradiction $u_{p}\in Z_{F}$ is a
pole of $w$. By Theorem~\ref{thm:Fw-zero-set}, $F_{w}(u_{p})=0$.
By Theorem~\ref{thm:notes2-w-poles-zeros}(i), $u_{p}-1$ is a zero
of $w$, so $1/w(u_{p}-1)$ has a pole at $u_{p}$. By Remark~\ref{rem:w-pole-zero-geometry}(a),
poles of $w$ lie on $u<0$, so $-u_{p}>0$ is not a pole and $w(-u_{p})$
is finite. Hence $F_{w}(u_{p})=w(-u_{p})-1/w(u_{p}-1)=\text{finite}-\infty=-\infty$,
contradicting $F_{w}(u_{p})=0$. The same argument with the roles
of $u_{p}$ and $-u_{p}$ swapped covers the case $u_{p}>0$.

The argument relies on the sign-localization input from Remark~\ref{rem:w-pole-zero-geometry}(a);
without it, a fine-tuned residue-matching scenario (symmetric poles
of $w$ at $\pm u_{p}$ with cancellation in $F_{w}(u_{p})$) would
not be ruled out. Numerical computation across the stability zone
shows no such coincidence occurs; moreover, Lemma~\ref{lem:generic-transversality}
(\S\,\ref{subsec:Casoratian-Weyl}) establishes, independently of
the localization input, that generically in $(p,\gamma)$ no zero
or pole of $w$ --- nor of any integer shift of $w$ --- lies on
$Z_{F}$. 
\end{rem}

\begin{rem}[Strict displacement of poles from $Z_{F}$]
\label{rem:poles-displaced-from-ZF} The poles of $w$ are \emph{always
strictly displaced} from the Floquet lattice $Z_{F}$, even though
they may be exponentially close to $Z_{F}$ as $\gamma\to0$. This
displacement, however small, is what gives the poles of $F_{w}$ their
finite residues. The ``stealthy'' character of a pole of $w$ near
a Floquet lattice point is therefore not a coincidence of position
but a consequence of the impossibility of exact coincidence: the residue
of $F_{w}$ at the corresponding pole is exponentially small (as $\gamma\to0$)
but never zero. A worked example --- locating such a pole and extracting
its tiny residue --- is given in \S\,\ref{subsec:Fw-pole-detection}
(Fig.~\ref{fig:stealthy-pole}). A quantitative residue-distance
rate is sketched in Remark~\ref{rem:v-residue-distance}; along each
pole family it is proved in Theorem~\ref{thm:pole-ladders}: at fixed
$(p,\gamma)$ the $k$-th member's displacement and residue both scale
as $|\zeta_{+}|^{2k}$. 
\end{rem}

\begin{rem}[Conjectural pole structure of $F_{w}$ relative to its Floquet zeros]
\label{rem:Fw-pole-conjecture} \emph{Status: conjecture supported
by numerical evidence across the stability zone; a proof is not pursued
here.}

The rigorous content of this work, with respect to $F_{w}$, is the
zero set: by Theorem~\ref{thm:Fw-zero-set} the real zero set of
$F_{w}$ is the two-lattice $Z_{F}=\{u^{*}+n\}\cup\{-u^{*}+n\}$;
the zeros are observed to be simple sign changes (clean crossings,
cf.\ Theorem~\ref{thm:double-min}(iv) and Fig.~\ref{fig:Fw-precision}),
the simplicity itself not being part of Theorem~\ref{thm:Fw-zero-set}.
The pole structure is described below at a different level of rigor,
consistent with numerical observations across the stability zone and
with the residue-distance heuristic of Remark~\ref{rem:v-residue-distance}.

\emph{Observed picture.} The lattice $Z_{F}$ partitions $\mathbb{R}$
into alternating \emph{small gaps} of length $s$ and \emph{large
gaps} of length $1-s$, where $s=2u^{*}\bmod1\in(0,\tfrac{1}{2}]$
is the inter-sub-comb offset~\eqref{eq:s-offset}. Numerically (e.g.\ Fig.~\ref{fig:notes2-Fw-marked},
$p=0.70$, $\gamma=0.30$, $s\approx0.484$): 
\begin{itemize}
\item Each pole of $F_{w}$ sits either in a \emph{small gap}, exponentially
close (as $\gamma\to0$) to the Floquet zero on one side of the gap,
or in the central large gap. These are the poles from the two sources
of Remark~\ref{rem:Fw-pole-sources}. Not every small gap is occupied,
and the occupied ones hold two poles each: the completed census on
$(-4,4)$ (\S\,\ref{subsec:Fw-pole-detection}, Table~\ref{tab:Fw-pole-catalog})
finds exactly one hugging pole for every Floquet zero except $\pm u^{*}$,
so the pole--zero correspondence is one-to-one \emph{off the two
primary zeros}, not globally. 
\item Each \emph{large gap} appears to contain no poles, with the single
exception of the central large gap around $u=0$, which contains a
double pole at $u=0$ (from the double zero of $w$ at $u=-1$) together
with two flanking simple poles from the two source mechanisms. 
\end{itemize}
\emph{Tentative statement.} Every pole of $F_{w}$ on $\mathbb{R}$
appears to either (i) sit in a small gap, exponentially close to a
Floquet zero of $F_{w}$, or (ii) be the double pole at $u=0$ (with
its flanking pair). The poles are confined to small gaps and the central
gap and never occur in a non-central large gap, but the two small
gaps adjacent to the central gap are empty and the primary zeros $\pm u^{*}$
have no hugging pole, so the pole--zero correspondence, while exact
off $\pm u^{*}$, is \emph{not} a global bijection (\S\,\ref{subsec:Fw-pole-detection}).
A quantitative treatment --- in particular, which small gaps are
occupied --- would require residue-distance estimates of the type
sketched in Remark~\ref{rem:v-residue-distance}.

\emph{Partial promotion to theorem.} The existence, position, and
residue of the deep ladder members are no longer conjectural: Theorem~\ref{thm:pole-ladders}
(\S\,\ref{subsec:Casoratian-Weyl}) proves that, along each of the
two ladders, the $n$-th pole exists, sits at distance $|K_{0}|\,|\zeta_{+}|^{2n}(1+o(1))$
from its comb zero, and carries a residue $(|C|+o(1))$ times that
distance. What remains beyond the theorem is the census of the base
members; on $(-4,4)$ it is settled empirically (\S\,\ref{subsec:Fw-pole-detection},
Table~\ref{tab:Fw-pole-catalog}), and a Weyl-theoretic program toward
the general statement is outlined in Remark~\ref{rem:interlacing-program}.

For the main results of this work (Theorems~\ref{thm:Floquet-factor-MR}
and~\ref{thm:primary-domain-CF}), the zero set $Z_{F}$ is sufficient;
the pole structure of $F_{w}$ enters only as illustrative context. 
\end{rem}

% -----------------------------------------------------------------------------

\subsection{Locating the poles of $F_{w}$: a sign-change theorem and the detection
of stealthy poles}

\label{subsec:Fw-pole-detection}

\begin{figure}[htbp]
\centering \includegraphics[width=0.98\textwidth]{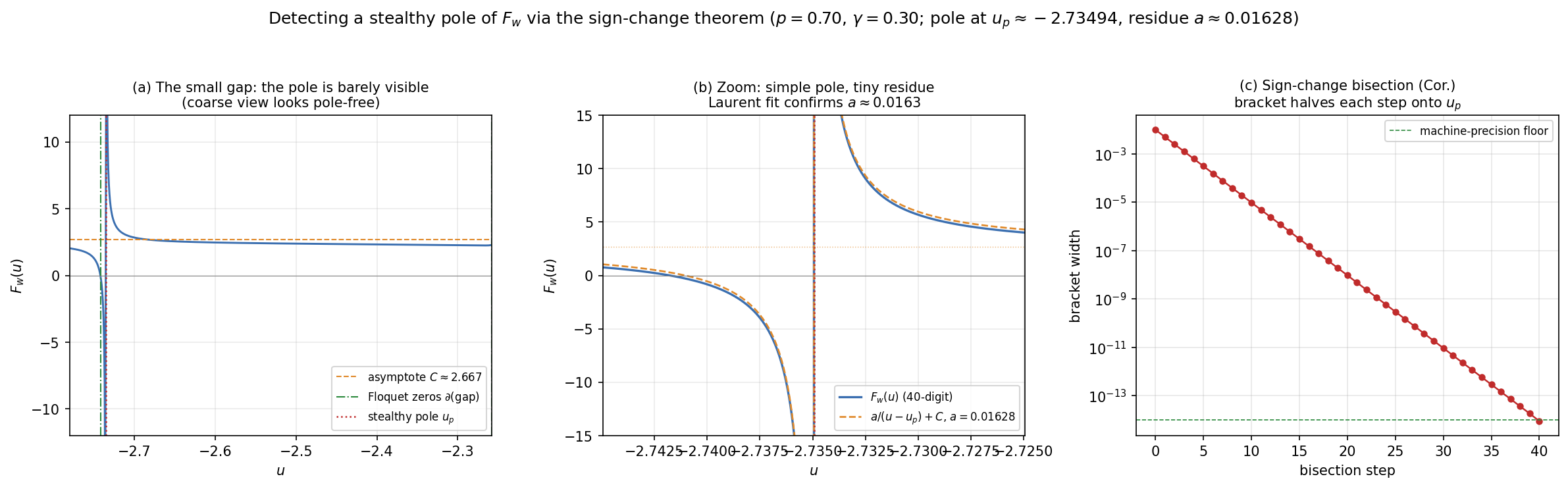}
\caption{Detection of a stealthy pole of the double minimality function $F_{w}$
via the sign-change theorem (Theorem~\ref{thm:Fw-sign-change-pole},
Corollary~\ref{cor:Fw-pole-bisection}), at $p=0.70$, $\gamma=0.30$.
\emph{(a)} The small gap $(-2.7418,-2.2582)$: across most of the
gap $F_{w}$ hugs the asymptote $C=\zeta_{+}-1/\zeta_{+}\approx2.667$
and the pole (red dotted), crammed against the Floquet zero at $-2.7418$
(green dash-dot), is barely visible --- a coarse view looks pole-free.
\emph{(b)} Zoom on a $\pm10^{-2}$ window: a clean simple pole at
$u_{p}\approx-2.73494$ with the Laurent fit $a/(u-u_{p})+C$ confirming
the tiny residue $a\approx0.0163$. \emph{(c)} Sign-change bisection:
the bracket width halves at each step (a straight line on the logarithmic
axis), reaching machine precision in about $40$ steps. \emph{Note
on visibility:} the same near-coincidence hides the adjacent Floquet
zero at $-2.7418$: like its hugging pole, it is genuine but invisible
on a coarse scan, looking as if no zero (or pole) were present there
at all; the residue--distance mechanism is Remark~\ref{rem:masking-geometry},
and Figure~\ref{fig:invisible-zero-pole} resolves the companion
pair explicitly.}
\label{fig:stealthy-pole} 
\end{figure}

The zero set of the double minimality function $F_{w}$ is known exactly
--- it is the Floquet comb $Z_{F}=\{\pm u^{*}+n\}$ (Theorem~\ref{thm:Fw-zero-set}).
Its \emph{poles}, by contrast, are not given in closed form, and several
lie so close to a Floquet zero that direct numerical searches (coarse
sampling, hunting zeros of $1/F_{w}$, or contour counts with a finite
exclusion margin) miss or mislocate them: we call such poles \emph{stealthy}.
The crowding is real and severe: already for $w$ itself, near the
Floquet point $u\approx-1.258$ (at $p=0.70$, $\gamma=0.30$), sampling
$|w|$ at spacing $0.005$ peaks at only $\approx0.2$ and gives no
hint of a pole, whereas refining the spacing to $2\times10^{-4}$
reveals $|w|\gtrsim20$ --- the pole sits entirely between the coarse
samples (Fig.~\ref{fig:stealthy-pole}). The same near-Floquet crowding
produces the stealthy poles of $F_{w}$ through the structural relation
$F_{w}(u)=w(-u)-1/w(u-1)$. The following elementary but useful theorem
reduces pole detection to a sign comparison at two explicit points,
sidestepping the need to resolve the tightly-spaced zero--pole clusters.
\begin{thm}[Sign-change pole detector]
\index{stealthy poles and zeros}\index{pole detector, sign-change}
\label{thm:Fw-sign-change-pole} Let $z_{1}<z_{2}$ be two consecutive
zeros of $F_{w}$ on the real axis (so $F_{w}$ has no zero in the
open interval $(z_{1},z_{2})$), and suppose $F_{w}$ is real-valued
there. Write $\sigma_{1}=\lim_{u\downarrow z_{1}}\operatorname{sign}F_{w}(u)$
and $\sigma_{2}=\lim_{u\uparrow z_{2}}\operatorname{sign}F_{w}(u)$
for the signs of $F_{w}$ just inside each endpoint. Then: 
\begin{enumerate}
\item[(i)] $F_{w}$ has at least one pole in $(z_{1},z_{2})$ whenever $\sigma_{1}\neq\sigma_{2}$; 
\item[(ii)] more precisely, the number of poles of $F_{w}$ in $(z_{1},z_{2})$,
counted with multiplicity, is \emph{odd} iff $\sigma_{1}\neq\sigma_{2}$
and \emph{even} (possibly zero) iff $\sigma_{1}=\sigma_{2}$. 
\end{enumerate}
\end{thm}

\begin{proof}
On $(z_{1},z_{2})$ the function $F_{w}$ is meromorphic and real,
with no zero. On any open subinterval free of poles it is therefore
continuous and nonvanishing, hence of constant sign; so the sign of
$F_{w}$ can change only across a pole. At a pole of order $m$, $F_{w}(u)\sim a\,(u-u_{p})^{-m}$
with $a\neq0$, so $F_{w}$ changes sign across $u_{p}$ iff $m$
is odd; an even-order pole preserves the sign. Reading the sign of
$F_{w}$ from $z_{1}^{+}$ to $z_{2}^{-}$, each odd-order pole flips
it and each even-order pole leaves it unchanged. Hence the total sign
change $\sigma_{1}\to\sigma_{2}$ is a flip iff the number of odd-order
poles is odd, equivalently iff the pole count with multiplicity is
odd. This proves~(ii); part~(i) is the special case that a flip
($\sigma_{1}\neq\sigma_{2}$) forces a nonzero, hence positive, pole
count. 
\end{proof}
Because the zeros of $F_{w}$ are known exactly, Theorem~\ref{thm:Fw-sign-change-pole}
turns pole-finding into a sign comparison at two points just inside
consecutive Floquet zeros --- no derivative, and no evaluation inside
the unresolved zero--pole clusters. It certifies an odd number of
poles in a gap (in the cases we have examined, exactly one in an occupied
small gap); an even count, such as the two simple poles flanking the
double pole at $u=0$ in the central gap, produces no net end-sign
change and is invisible to the test. The detector is moreover \emph{constructive}:
the bracketing extends verbatim to non-zero endpoints and yields a
bisection scheme.
\begin{cor}[Bracketing and recursive localization of a pole]
\index{stealthy poles and zeros!bracketing} \label{cor:Fw-pole-bisection}
Let $u_{1}<u_{2}$ lie in a common gap of $Z_{F}$ (so $F_{w}$ has
no zero in $[u_{1},u_{2}]$) and suppose $F_{w}(u_{1})F_{w}(u_{2})<0$.
Then $F_{w}$ has a pole in $(u_{1},u_{2})$, and it is localized
by recursive bisection: for the midpoint $m=\tfrac{1}{2}(u_{1}+u_{2})$,
the pole lies in $(u_{1},m)$ if $F_{w}(u_{1})F_{w}(m)<0$ and in
$(m,u_{2})$ otherwise; iterating shrinks the bracket onto the pole. 
\end{cor}

\begin{proof}
Immediate from Theorem~\ref{thm:Fw-sign-change-pole}: the hypotheses
hold on each retained subinterval (no zero inside, opposite end-signs),
so a pole persists in it, and the bracket length halves at each step. 
\end{proof}
\paragraph{Detection of a stealthy pole.} The power of this approach
is that the bracket is defined purely by the sign of $F_{w}$ at well-separated
points, so the iteration never evaluates inside the unresolved cluster
--- which is exactly why it succeeds where direct methods fail. As
a demonstration, consider Cambi's primary case $p=0.70$, $\gamma=0.30$.
Here the Floquet point is computed to high precision, 
\begin{equation}
u^{*}=0.7417614039024665409251648484304946236223\ldots,\label{eq:ustar-hiprec}
\end{equation}
accurate to $40$ digits (stable under increases of both the working
precision and the continued-fraction depth); the smallest gap is $s\approx0.484$.
Such accuracy in $u^{*}$ is essential here: every Floquet zero is
$\pm u^{*}+n$, and the poles sit only $10^{-3}$--$10^{-2}$ from
a zero, so an inaccurate $u^{*}$ would corrupt the gap assignment
and the pole--zero distances. The small gap $(-2.7418,-2.2582)$
between consecutive Floquet zeros appears pole-free under coarse sampling,
yet $F_{w}$ takes opposite signs at the points lying $10^{-3}$ inside
its two ends. (The inset distance matters: as shown below, a second,
far stealthier pole hugs the right-hand zero at $2.4\times10^{-5}$,
so end signs taken closer in than that agree again.) Bisection (Corollary~\ref{cor:Fw-pole-bisection})
converges geometrically --- bracket width halving each step, from
$10^{-2}$ to machine precision in about $40$ steps --- to a simple
pole at 
\begin{equation}
u_{p}\;=\;-2.7349374028208547\ldots,\label{eq:stealthy-pole}
\end{equation}
lying a mere $\approx7\times10^{-3}$ from the Floquet zero at $-2.7418$.
Once the location is known, the residue follows from a short-radius
contour integral (or from $a\approx\varepsilon\,F_{w}(u_{p}+\varepsilon)$
as $\varepsilon\to0$), giving the small value 
\begin{equation}
a\;=\;\operatorname*{Res}_{u=u_{p}}F_{w}(u)\;\approx\;0.0162779,\label{eq:stealthy-residue}
\end{equation}
about an eighth of the residues of the well-separated poles ($\approx\pm0.126$
at $u=\pm0.252$). The pole is simple: $(u-u_{p})F_{w}(u)\to a$ while
$(u-u_{p})^{2}F_{w}(u)\to0$. Figure~\ref{fig:stealthy-pole} illustrates
the three stages --- the barely-visible pole, the Laurent fit confirming
the tiny residue, and the geometric convergence of the sign-change
bisection. The division of labor is efficient: locating the pole uses
only the \emph{sign} of $F_{w}$ and needs no special precision, while
extracting the small residue is the only step that calls for extended-precision
arithmetic (the value stabilizes by $\sim15$ digits and is certified
at $\sim45$).

\paragraph{What is and is not settled about the pole distribution.}
Combining the exact zero set with the structure $F_{w}(u)=w(-u)-1/w(u-1)$,
the poles of $F_{w}$ are exactly $\{-u_{p}:\ u_{p}\text{ a pole of }w\}\cup\{u_{z}+1:\ u_{z}\text{ a zero of}w\}$,
with the geometry governed by the unit-shift pairing $\{\text{zeros of }w\}+1=\{\text{poles of }w\}$
(Theorem~\ref{thm:notes2-w-poles-zeros}(i)). What this set looks
like in practice --- where the poles fall relative to the gaps of
$Z_{F}$, how close they sit to the Floquet zeros, and how large their
residues are --- is recorded empirically just below, and is partially
settled by Theorem~\ref{thm:pole-ladders} (\S\,\ref{subsec:Casoratian-Weyl}):
along each of the two pole ladders the deep members provably exist,
hugging their comb zeros at exponentially shrinking distances. The
occupancy near the ladder bases remains empirical; the zero set $Z_{F}$
is what the main results (Theorems~\ref{thm:Floquet-factor-MR} and~\ref{thm:primary-domain-CF})
require, and the sign-change detector above suffices to locate the
poles when they are needed.

\paragraph{Empirical pole catalog and observed regularities.} Applying
the detector of Theorem~\ref{thm:Fw-sign-change-pole} and Corollary~\ref{cor:Fw-pole-bisection}
gap by gap, then extracting each residue by a short-radius contour
integral, yields the reliable catalog of Table~\ref{tab:Fw-pole-catalog}
for $p=0.70$, $\gamma=0.30$ on $(-4,4)$. (These values supersede
the partial counts of earlier sections, where the closely spaced zero--pole
pairs had defeated direct searches.) We record the following observations
as empirical regularities --- supported by the catalog, not proved
here. 
\begin{enumerate}
\item[(O1)] \emph{Location.} Every real pole lies in a small gap or in the central
gap; no non-central large gap contains a pole. Of the eight small
gaps in $(-4,4)$, six are occupied --- each, in fact, by a \emph{pair}
of poles, one per ladder, hugging opposite ends of the gap (see the
completion of the census below) --- and the two adjacent to the central
gap, $\pm(0.258,0.742)$, are empty. 
\item[(O2)] \emph{Proximity.} Each pole sits very close to the Floquet zero on
one side of its gap; along each ladder the pole--zero distance shrinks
geometrically with depth --- from $d\approx5\times10^{-2}$ at the
shallowest member down to $d\approx3\times10^{-6}$ at depth $3$
here --- with contraction ratio $\zeta_{+}^{2}$ (Theorem~\ref{thm:pole-ladders}). 
\item[(O3)] \emph{Residue size tracks distance.} For the small-gap poles flanking
the resonance, the residue scales linearly with the pole--zero distance,
$|a|\approx\kappa\,d$ with $\kappa$ ranging over $2.0$--$2.6$
in the completed catalog and tending, ladder by ladder, to the proved
limit $|C|=|\zeta_{+}-1/\zeta_{+}|=\tfrac{8}{3}$ (Theorem~\ref{thm:pole-ladders}(ii));
the central-gap poles do not follow this proportionality ($|a|/d\approx22$),
reflecting their different origin near the double pole at $u=0$. 
\item[(O4)] \emph{Antisymmetry.} The residues are odd under $u\mapsto-u$: $\operatorname*{Res}_{-u_{p}}F_{w}=-\operatorname*{Res}_{u_{p}}F_{w}$,
consistent with the symmetry $\widetilde{F}_{w}(u)=\widetilde{F}_{w}(-u)$. 
\item[(O5)] \emph{Central gap.} The central gap is exceptional: besides the order-two
pole at $u=0$ (found structurally, not by the sign-change test, which
is blind to even-order poles; here $u^{2}F_{w}(u)\to-1.63$ as $u\to0$),
it contains the two simple poles $\pm0.2525$ with the largest residues
in the catalog ($\approx\pm0.126$). 
\end{enumerate}
\paragraph{Completion of the census by the ladder theorem.} The census
above was conducted with end-sign insets of $10^{-3}$; Theorem~\ref{thm:pole-ladders}
implies it is incomplete. Each ladder contributes one pole per unit
of depth, so an occupied small gap holds \emph{two} poles --- one
per ladder, hugging opposite ends --- and, the pole count being even,
the end-sign comparison across the whole gap shows no net change:
the second member is structurally invisible to the whole-gap test.
It is recovered by the same detector with the inset refined below
its pole--zero distance, whereupon the end-sign parity flips. Carrying
this out (insets $10^{-2}$ down to $10^{-7}$, the bisection of Corollary~\ref{cor:Fw-pole-bisection},
and residues by the symmetric two-point limit $a=\lim_{\varepsilon\to0}\tfrac{\varepsilon}{2}[F_{w}(u_{p}+\varepsilon)-F_{w}(u_{p}-\varepsilon)]$)
certifies the eight further poles marked with an asterisk in Table~\ref{tab:Fw-pole-catalog},
with $|a|/d$ matching the ladder values of~\eqref{eq:CW-table}
to all displayed digits. In the completed catalog, every Floquet zero
in $(-4,4)$ \emph{except} $\pm u^{*}$ carries exactly one hugging
pole; the empty gaps $\pm(0.258,0.742)$ are precisely those whose
endpoint zeros are hugged from the other side or, for $\pm u^{*}$
itself, not at all.

These regularities are specific to the parameter regime shown. The
two interleaved sub-combs --- and hence the small/large-gap structure
--- depend on $u^{*}$ and $s$, all of which we compute with the
same $40$-digit accuracy as in~\eqref{eq:ustar-hiprec}. The gap
geometry itself is robust across regimes (e.g.\ at Cambi's $p=20/3$
one finds $s\approx0.435$ at $\gamma=0.1$ and $s\approx0.469$ at
$\gamma=0.3$, the same alternating $\{s,1-s\}$ pattern as here);
what changes is \emph{which} gaps are occupied. In the large-$p$
Cambi regime poles also occur in some non-central large gaps, unlike
the $p=0.70$ case where those gaps are empty. For the most reliable
parameters ($p=20/3$, $\gamma=0.1$) this is confirmed by the argument
principle: a contour count of ${F_{w}}'/F_{w}$ returns exactly one
pole in each of the large gaps $(4.718,5.282)$ and $(5.718,6.282)$
(and one in each of the small gaps $(7.282,7.718)$ and $(8.282,8.718)$),
as catalogued in Table~\ref{tab:Fw-pole-catalog-cambi}. The feature
persists when $\gamma$ is raised to $0.3$, so it tracks $p$ rather
than $\gamma$. Two caveats temper the quantitative picture. First,
the poles closest to a Floquet zero (separations down to $\sim10^{-5}$,
with residues as small as $\sim10^{-4}$) sit at the edge of what
the contour count can resolve cleanly, so their multiplicity is not
certified; away from the resonance $\pm p$ the residues shrink geometrically
(in step with the precision cost of \S\,\ref{subsec:Cambi-notes2-precision}),
and $(p-4,p+4)$ is the practical window for a reliable catalog. Second,
the continued-fraction convergence rate is governed by $|\zeta_{+}|=|{-1+\sqrt{1-4\gamma^{2}}}|/(2\gamma)$,
which increases toward $1$ as $\gamma\to\tfrac{1}{2}$; the per-term
gain is about two decimal digits at $\gamma=0.1$ ($|\zeta_{+}|\approx0.10$)
but only about one at $\gamma=0.3$ ($|\zeta_{+}|\approx0.33$). Numerical
conclusions are therefore most secure at smaller $\gamma$, and the
$\gamma=0.1$ data should be weighted more heavily than the $\gamma=0.3$
data. Table~\ref{tab:Fw-pole-catalog} should accordingly be read
as a verified illustration of the pole structure in one regime, not
as a parameter-independent law.

\begin{table}[htbp]
\centering \caption{Real poles of the double minimality function $F_{w}$ in the window
$(p-4,p+4)$ for Cambi's large-$p$ regime $p=20/3$, $\gamma=0.1$
($u^{*}=6.717695353151718305\ldots$, $s\approx0.435$). In contrast
to the $p=0.70$ case (Table~\ref{tab:Fw-pole-catalog}), poles occupy
both large and small gaps here. The four poles at distance $d\gtrsim10^{-3}$
from the nearest Floquet zero are confirmed by an argument-principle
count (one pole per listed gap); the near-edge pole at $d\approx10^{-5}$
is real (an $|F_{w}|$ peak with nonzero residue) but too tightly
paired with its zero for the contour count to certify at this precision.}
\label{tab:Fw-pole-catalog-cambi} %
\begin{tabular}{rcccc}
\hline 
pole $u_{p}$  & gap type  & $d$ (to zero)  & residue $a$  & AP-confirmed\tabularnewline
\hline 
$3.717705$  & large  & $9.6\times10^{-6}$  & $-0.000222$  & edge\tabularnewline
$4.719895$  & large  & $2.2\times10^{-3}$  & $-0.021125$  & yes\tabularnewline
$5.806335$  & large  & $8.9\times10^{-2}$  & $-0.318398$  & yes\tabularnewline
$7.587645$  & small  & $1.3\times10^{-1}$  & $-0.293237$  & yes\tabularnewline
$8.705685$  & small  & $1.2\times10^{-2}$  & $-0.043421$  & yes\tabularnewline
\hline 
\end{tabular}
\end{table}

\begin{table}[htbp]
\centering \caption{Complete catalog of the real poles of the double minimality function
$F_{w}$ on $(-4,4)$ for $p=0.70$, $\gamma=0.30$ ($u^{*}\approx0.7418$,
smallest gap $s\approx0.484$). The seven poles without asterisk were
found with the sign-change detector (Theorem~\ref{thm:Fw-sign-change-pole})
at end-sign inset $10^{-3}$ and their residues computed by contour
integration; the order-two pole at $u=0$, to which the sign-change
test is blind by design, was identified structurally (it comes from
the double zero of $w$ at $u=-1$, with $u^{2}F_{w}(u)\to-1.63$
as $u\to0$). The eight asterisked poles are the deep ladder members
(Theorem~\ref{thm:pole-ladders}), invisible to the whole-gap end-sign
test because their gaps are doubly occupied; they were certified by
the inset-refined detector as described in the text. \textquotedblleft Gap
type\textquotedblright{} refers to the $Z_{F}$ gap containing the
pole; $d$ is the distance to the nearer Floquet zero. The only empty
small gaps are $(-0.742,-0.258)$ and $(0.258,0.742)$; all non-central
large gaps are empty.}
\label{tab:Fw-pole-catalog} %
\begin{tabular}{rlccc}
\hline 
pole $u_{p}$  & gap type  & order  & $d$ (to zero)  & residue $a$\tabularnewline
\hline 
$-3.740893^{*}$  & small  & 1  & $8.7\times10^{-4}$  & $+0.002178$\tabularnewline
$-3.258242^{*}$  & small  & 1  & $3.2\times10^{-6}$  & $-7.85\times10^{-6}$\tabularnewline
$-2.734937$  & small  & 1  & $0.0068$  & $+0.016278$\tabularnewline
$-2.258262^{*}$  & small  & 1  & $2.4\times10^{-5}$  & $-5.26\times10^{-5}$\tabularnewline
$-1.692859$  & small  & 1  & $0.0489$  & $+0.107738$\tabularnewline
$-1.258386^{*}$  & small  & 1  & $1.5\times10^{-4}$  & $-2.91\times10^{-4}$\tabularnewline
$-0.252486$  & central  & 1  & $0.0058$  & $-0.126145$\tabularnewline
$0$  & central  & 2  & ---  & ($u^{2}F_{w}\to-1.63$)\tabularnewline
$+0.252486$  & central  & 1  & $0.0058$  & $+0.126145$\tabularnewline
$+1.258386^{*}$  & small  & 1  & $1.5\times10^{-4}$  & $+2.91\times10^{-4}$\tabularnewline
$+1.692859$  & small  & 1  & $0.0489$  & $-0.107738$\tabularnewline
$+2.258262^{*}$  & small  & 1  & $2.4\times10^{-5}$  & $+5.26\times10^{-5}$\tabularnewline
$+2.734937$  & small  & 1  & $0.0068$  & $-0.016278$\tabularnewline
$+3.258242^{*}$  & small  & 1  & $3.2\times10^{-6}$  & $+7.85\times10^{-6}$\tabularnewline
$+3.740893^{*}$  & small  & 1  & $8.7\times10^{-4}$  & $-0.002178$\tabularnewline
\hline 
\end{tabular}
\end{table}

\paragraph{Remark on ${F_w}'$ at a Floquet zero.} A large value of
$|{F_{w}}'(z)|$ at a Floquet zero $z$ is a qualitative signal that
a pole lies close to $z$ (the Laurent tail $a/(u-u_{p})$ has a steep
slope when $u_{p}$ is near $z$). It is, however, only a soft indicator:
the magnitude alone does not determine on which side of $z$ the pole
lies --- both orientations occur --- and the quantitative relation
between $|{F_{w}}'(z)|$, the residue $a$, and the pole--zero distance
is not clean enough to invert reliably. The side and the existence
of the pole are settled rigorously instead by Theorem~\ref{thm:Fw-sign-change-pole}
and Corollary~\ref{cor:Fw-pole-bisection}.

\subsection{Imported statements: discrete Volterra equations and Jost solutions}

\label{subsec:imported-Jacobi}

This section collects, in faithful form and with exact sources, the
facts from the Jacobi-operator literature on which the next section
relies. Following the conventions of this book, quoted statements
of other authors are labeled Propositions, while our own statements
are Lemmas and Theorems. The material is gathered here for ease of
reference and clean attribution; its final placement --- among the
background chapters of Part~II --- is deferred. For the classical
second-order layer --- the Casoratian and Abel's lemma, the Poincaré--Perron
dichotomy, minimal solutions and Pincherle's theorem --- the textbook
of Elaydi \cite[Chs.~2, 8, 9]{Elay} is an accessible alternative
source, on which the background chapters of Part~II already draw
(Theorems~\ref{thm:PP} and~\ref{thm:Pinch}); from it we quote
below the asymptotic theorem of Evgrafov (Proposition~\ref{prop:Elaydi-Evgrafov}).
The remaining statements are quoted from Teschl~\cite{Teschl}, the
natural reference for quantitative Volterra bounds with parameter
dependence, Jost solutions on periodic backgrounds, and the Herglotz
representation.

Teschl's setting \cite[Ch.~1]{Teschl} is the Jacobi difference expression
\begin{equation}
(\tau\varphi)(n)\;=\;a(n)\,\varphi(n+1)+a(n-1)\,\varphi(n-1)+b(n)\,\varphi(n),\qquad n\in\mathbb{Z},\label{eq:imported-tau}
\end{equation}
with real sequences $a(n)\neq0$ and $b(n)$, acting on sequences
$\varphi\colon\mathbb{Z}\to\mathbb{C}$ (we write $\varphi$ for the
generic sequence --- Teschl's $u$ --- since the letter $u$ is
reserved throughout this book for the Floquet-exponent parameter);
the associated spectral problem is the three-term recurrence 
\begin{equation}
\tau\varphi\;=\;z\,\varphi,\qquad z\in\mathbb{C},\label{eq:imported-spectral}
\end{equation}
where $z$ is the spectral parameter. When the coefficient pair is
periodic with period $N$, $\bigl(a_{p}(n+N),b_{p}(n+N)\bigr)=\bigl(a_{p}(n),b_{p}(n)\bigr)$,
and the monodromy matrix --- the transfer matrix of~\eqref{eq:imported-spectral}
over one period --- has distinct eigenvalues, the solution space
of~\eqref{eq:imported-spectral} is spanned by two \emph{Floquet
solutions}\index{Floquet solution} $\varphi_{p,\pm}(z,\cdot)$, characterized
by 
\begin{equation}
\varphi_{p,\pm}(z,n+N)\;=\;m_{p}^{\pm}(z)\,\varphi_{p,\pm}(z,n),\qquad n\in\mathbb{Z},\label{eq:imported-Floquet}
\end{equation}
where the \emph{Floquet multipliers}\index{Floquet multiplier} $m_{p}^{\pm}(z)$
--- the eigenvalues of the monodromy matrix --- satisfy $m_{p}^{+}(z)\,m_{p}^{-}(z)=1$
and are labeled so that $|m_{p}^{+}(z)|\leq1\leq|m_{p}^{-}(z)|$ \cite[Sec.~7.1, eqs.~(7.10)--(7.12)]{Teschl}.
Throughout this section and the next, $\mathbb{C}_{\pm}:=\{z\in\mathbb{C}:\pm\operatorname{Im}z>0\}$,
and $\ell^{1}(0,\infty)$, $\ell^{\infty}(0,\infty)$ denote the spaces
of summable, respectively bounded, sequences on the half-line $n\geq0$
(similarly on $(-\infty,0)$).
\begin{prop}[{Discrete Volterra bound; {\cite[Lem.~7.8 and Rem.~7.9]{Teschl}}}]
\index{Volterra sum equation} \label{prop:Teschl-Volterra} Consider
the Volterra sum equation 
\begin{equation}
f(n)\;=\;g(n)+\sum_{m=n+1}^{\infty}K(n,m)\,f(m).\label{eq:imported-Volterra}
\end{equation}
Suppose there is a sequence $\hat{K}(n,m)$ such that 
\begin{equation}
|K(n,m)|\leq\hat{K}(n,m),\quad\hat{K}(n+1,m)\leq\hat{K}(n,m),\quad\hat{K}(n,\cdot)\in\ell^{1}(0,\infty).\label{eq:imported-Khat}
\end{equation}
Then, for given $g\in\ell^{\infty}(0,\infty)$, there is a unique
solution $f\in\ell^{\infty}(0,\infty)$, fulfilling the estimate 
\begin{equation}
|f(n)|\;\leq\;\Bigl(\sup_{m>n}|g(m)|\Bigr)\exp\Bigl(\sum_{m=n+1}^{\infty}\hat{K}(n,m)\Bigr).\label{eq:imported-Volterra-bound}
\end{equation}
The bound~\eqref{eq:imported-Volterra-bound} is the standard discrete
Gronwall inequality in exponential form (Elaydi~\cite[Lem.~4.32]{Elay};
Agarwal~\cite[\S\,4.1]{Agarwal}); the Volterra/Picard iteration
that produces it is classical. If $g(n)$ and $K(n,m)$ depend continuously
(resp.\ holomorphically) on a parameter and if $\hat{K}$ does not
depend on this parameter, then the same is true for $f(n)$. By \cite[Rem.~7.9]{Teschl},
a similar result holds for the leftward equation $f(n)=g(n)+\sum_{m=-\infty}^{n-1}K(n,m)f(m)$,
with the mirrored hypotheses ($\hat{K}(n-1,m)\leq\hat{K}(n,m)$, $\hat{K}(n,\cdot)\in\ell^{1}(-\infty,0)$)
and the mirrored estimate. 
\end{prop}

\begin{prop}[{Jost solutions under first-moment perturbations; {\cite[Hyp.~H.7.7 and Lem.~7.10]{Teschl}}}]
\index{Jost solution} \label{prop:Teschl-Jost} Let $(a_{p},b_{p})$
be periodic and let $(a,b)$ satisfy the first-moment condition 
\begin{equation}
\sum_{n\in\mathbb{Z}}\bigl|n\,(a(n)-a_{p}(n))\bigr|<\infty,\qquad\sum_{n\in\mathbb{Z}}\bigl|n\,(b(n)-b_{p}(n))\bigr|<\infty.\label{eq:imported-H77}
\end{equation}
Then there exist solutions $\varphi_{\pm}(z,\cdot)$ of the perturbed
equation $\tau\varphi=z\varphi$, $z\in\mathbb{C}$ --- with $\tau$
the expression~\eqref{eq:imported-tau} formed from the pair $(a,b)$
--- satisfying 
\begin{equation}
\lim_{n\to\pm\infty}\bigl|m_{p}^{\mp}(z)^{n/N}\bigl(\varphi_{\pm}(z,n)-\varphi_{p,\pm}(z,n)\bigr)\bigr|\;=\;0,\label{eq:imported-Jost}
\end{equation}
and $\varphi_{\pm}(z,\cdot)$ can be assumed continuous (resp.\
holomorphic) with respect to $z$ whenever $\varphi_{p,\pm}(z,\cdot)$
are. 
\end{prop}

\begin{prop}[{Asymptotics under absolutely summable perturbations (Evgrafov); {\cite[Sec.~8.4, Cor.~8.27]{Elay}}}]
\index{Evgrafov theorem}\index{minimal solution!asymptotics under decaying perturbation}
\label{prop:Elaydi-Evgrafov} Consider the perturbed difference equation
\begin{equation}
y(n+k)+\bigl(a_{1}+p_{1}(n)\bigr)\,y(n+k-1)+\cdots+\bigl(a_{k}+p_{k}(n)\bigr)\,y(n)\;=\;0.\label{eq:imported-Evgrafov-eq}
\end{equation}
Suppose that the characteristic polynomial $p(\lambda)=\lambda^{k}+a_{1}\lambda^{k-1}+\cdots+a_{k}$
has distinct roots $\lambda_{1},\ldots,\lambda_{k}$ and that 
\begin{equation}
\sum_{n=1}^{\infty}|p_{i}(n)|<\infty,\qquad1\leq i\leq k.\label{eq:imported-Evgrafov-cond}
\end{equation}
Then~\eqref{eq:imported-Evgrafov-eq} has $k$ solutions $y_{1}(n),\ldots,y_{k}(n)$
with 
\begin{equation}
y_{i}(n)\;=\;\bigl[1+o(1)\bigr]\,\lambda_{i}^{\,n},\qquad n\to\infty.\label{eq:imported-Evgrafov}
\end{equation}
Decay $p_{i}(n)=O(n^{-\alpha})$ with $\alpha>1$ suffices for~\eqref{eq:imported-Evgrafov-cond},
while $O(1/n)$ decay does not suffice for the conclusion \cite[Sec.~8.3, remark following Thm.~8.23]{Elay}. 
\end{prop}

\begin{rem}[What is used and what must be proved]
\label{rem:imported-usage} The recurrence~\eqref{eq:notes2-rec}
fits the form~\eqref{eq:imported-tau} after division by $2\gamma$:
\begin{equation}
\tfrac{1}{2}\bigl(h_{m+1}+h_{m-1}\bigr)+\frac{G(u+m)-1}{2\gamma}\,h_{m}\;=\;-\frac{1}{2\gamma}\,h_{m},\label{eq:imported-translation}
\end{equation}
i.e.\ $a\equiv a_{p}\equiv\tfrac{1}{2}$, $b(m)=\bigl(G(u+m)-1\bigr)/(2\gamma)$,
$b_{p}\equiv0$, $N=1$, at the fixed spectral point $z=-1/(2\gamma)$.
Since $0<\gamma<\tfrac{1}{2}$, $|z|>1$, so $z$ lies outside the
essential spectrum $[-1,1]$ of the free operator, and the free multipliers
at $z$ --- the roots of $\rho+1/\rho=-1/\gamma$, equivalently of
$\gamma\rho^{2}+\rho+\gamma=0$ --- are precisely $\zeta_{\pm}$,
with the strict dichotomy $|\zeta_{+}|<1<|\zeta_{-}|$. Two features
prevent a direct application of Proposition~\ref{prop:Teschl-Jost}.
First, the perturbation here is $b(m)=-p^{2}/\bigl(2\gamma(u+m)^{2}\bigr)$,
which is absolutely summable but \emph{fails} the first-moment condition~\eqref{eq:imported-H77},
the series $\sum|m|\cdot m^{-2}$ being divergent. Second, the parameter
of interest, $u$, enters through the perturbation rather than through
the spectral variable $z$, so the $z$-continuity clause of Proposition~\ref{prop:Teschl-Jost}
does not deliver the $u$-uniformity required in the next section.
Off the essential spectrum, however, plain $\ell^{1}$ summability
suffices, and at this level the textbook literature already reaches
our recurrence: the perturbation $b(m)=O(m^{-2})$ is absolutely summable
and the free roots $\zeta_{\pm}$ are distinct, so Proposition~\ref{prop:Elaydi-Evgrafov}
applies at each fixed $u$ and yields, for the root $\zeta_{+}$,
a solution $h_{m}=\bigl[1+o(1)\bigr]\zeta_{+}^{\,m}$ --- which is
then the minimal solution, with its leading asymptotics. What Proposition~\ref{prop:Elaydi-Evgrafov}
does not provide is a rate or constant in the $o(1)$ and, again,
uniformity in $u$ --- both indispensable in the next section, where
the ladder theorem needs geometric rates with explicit constants and
the Herglotz ray bound needs decay in $u$ along a complex ray. Lemma~\ref{lem:Jost-LC}
below (with its rightward counterpart, Lemma~\ref{lem:Jost-right})
proves the required quantitative, $u$-uniform statement in full,
using only Proposition~\ref{prop:Teschl-Volterra}, whose explicit
bound~\eqref{eq:imported-Volterra-bound} and parameter clause are
exactly what the uniformity needs. 
\end{rem}

\begin{prop}[{Herglotz functions and their representation; {\cite[App.~B, Thm.~B.2, eqs.~(B.11)--(B.13)]{Teschl}}}]
\index{Herglotz function}\index{Nevanlinna function} \label{prop:Teschl-Herglotz}
A holomorphic function $F\colon\mathbb{C}_{+}\to\mathbb{C}_{+}$ is
called a \emph{Herglotz function} (sometimes also Pick or Nevanlinna--Pick
function); it is no restriction to assume that $F$ is defined on
$\mathbb{C}_{-}\cup\mathbb{C}_{+}$ satisfying $F(\bar{z})=\overline{F(z)}$.
$F$ is a Herglotz function if and only if 
\begin{equation}
F(z)\;=\;a+bz+\int_{\mathbb{R}}\Bigl(\frac{1}{\lambda-z}-\frac{\lambda}{1+\lambda^{2}}\Bigr)d\rho(\lambda),\qquad z\in\mathbb{C}_{\pm},\label{eq:imported-Herglotz}
\end{equation}
where $a,b\in\mathbb{R}$, $b\geq0$, and $\rho$ is a nonzero measure
on $\mathbb{R}$ which satisfies $\int_{\mathbb{R}}(1+\lambda^{2})^{-1}d\rho(\lambda)<\infty$.
The triple $a$, $b$, $\rho$ is unambiguously determined by $F$:
$a=\operatorname{Re}F(\mathrm{i})$, $b=\lim_{z\to\infty,\ \operatorname{Im}z\geq\varepsilon>0}F(z)/z\geq0$,
and the Stieltjes inversion formula 
\begin{equation}
\rho\bigl((\lambda_{0},\lambda_{1}]\bigr)\;=\;\lim_{\delta\downarrow0}\,\lim_{\varepsilon\downarrow0}\,\frac{1}{\pi}\int_{\lambda_{0}+\delta}^{\lambda_{1}+\delta}\operatorname{Im}F(\lambda+\mathrm{i}\varepsilon)\,d\lambda.\label{eq:imported-Stieltjes}
\end{equation}
\end{prop}

\subsection{The Casoratian identity and a Weyl--Stieltjes view of the pole ladders}

\label{subsec:Casoratian-Weyl}

This closing section develops the structural mechanism behind the
stealthy-pole ladders documented in \S\,\ref{subsec:stealthy-zeros}
and behind the precision-cost law of Theorem~\ref{thm:precision-cost},
and then places the chapter's central objects --- the minimal solution
ratio $w$ and the double minimality function $F_{w}$ --- inside
the classical Weyl--Stieltjes spectral framework for Jacobi matrices.
The Lemmas and the Theorems below are, to the best of our knowledge,
new; the classical facts imported from the Jacobi-operator literature
are quoted as Propositions, with sources, in \S\,\ref{subsec:imported-Jacobi}.
The zero-location and ladder results below are conditional on the
simplicity of the comb zeros of $F_{w}$ --- a genericity hypothesis
(\S\,\ref{subsec:w-simplicity-notes}), verified numerically at
the worked parameters but not proved here in full; under it the conclusions
are rigorous.

\emph{The thread of the section.} The relation between $F_{w}$ and
the two minimal solutions is intricate, yet a single identity disentangles
it: $F_{w}$ is their \emph{Casoratian} (\S\,\ref{subsec:CW-Casoratian-id}),
and so is the exact coefficient measuring how strongly the solution
that decays at $-\infty$ couples into the one that decays at $+\infty$.
Because that coupling enters weighted by a power $\zeta_{+}^{2m}$
contracting geometrically into the lattice, the competition between
the two directions pins each pole of $w$ to a comb zero at a geometrically
shrinking distance --- the geometric ladder law (\S\,\ref{subsec:CW-ladder-law},
Theorem~\ref{thm:pole-ladders}) --- and fixes the accompanying
residue in the same proportion. Which of the gaps between comb zeros
are occupied is settled in \S\,\ref{subsec:CW-occupancy}; and the
same minimal-solution structure, read through the classical Weyl--Stieltjes
dictionary, upgrades to a Herglotz representation of $w$ (\S\,\ref{subsec:CW-Herglotz},
Theorem~\ref{thm:w-Herglotz}).

\emph{Standing notation for this section.} All statements below concern
the three-term recurrence~\eqref{eq:notes2-rec}, $\gamma A_{m+1}+G(u+m)\,A_{m}+\gamma A_{m-1}=0$
with $\gamma=\varepsilon/2$ and $G(u)=1-p^{2}/u^{2}$, $p=\omega_{0}/\mu$
(eq.~\eqref{eq:Cambi-res}), in the stability zone~\eqref{eq:stability-assumption}
and under Assumption~\ref{ass:generic-comb}. The recurring symbols,
with their points of definition, are: 
\begin{itemize}
\item $h_{m}(u),\,h_{m}^{-}(u)$ --- the \emph{right-} and \emph{left-minimal}
solutions (minimal as $m\to+\infty$, resp.\ $m\to-\infty$; Definition~\ref{defn:Wimp-minimal}),
normalized by $h_{0}=h_{0}^{-}=1$; $w(u)=h_{1}(u)$ is the minimal-solution
ratio~\eqref{eq:notes2-wdef}. 
\item $u^{*}$ --- the \emph{primary Floquet exponent}, the real zero of
$F_{w}$ (next item) nearest the unperturbed value $p$ (eq.~\eqref{eq:ustar-def}),
so $u^{*}\approx p$; the comb $Z_{F}$ and the base point $u^{**}$
below are built from it. 
\item $F_{w}(u)=w(-u)-1/w(u-1)$ --- the double minimality function~\eqref{eq:Fw-def},
even in $u$, whose real zero set is the Floquet comb $Z_{F}=\{\pm u^{*}+n:n\in\mathbb{Z}\}=Z_{F}^{+}\cup Z_{F}^{-}$
with $Z_{F}^{+}=\{-u^{*}+n\}$ and $Z_{F}^{-}=\{u^{*}+n\}$. 
\item $K_{m}[A,B]=A_{m}B_{m+1}-A_{m+1}B_{m}$ --- the Casoratian~\eqref{eq:CW-Casoratian-def},
the discrete Wronskian. 
\item $\zeta_{+}=\dfrac{-1+\sqrt{1-4\gamma^{2}}}{2\gamma}$ --- the recessive
root of $\gamma\zeta^{2}+\zeta+\gamma=0$, with $0<|\zeta_{+}|<1$
(eq.~\eqref{eq:Cambi-char-roots}); it is a constant (it depends
only on $\gamma$) and sets the leftward decay rate of $h^{-}$. 
\item $u^{**}\in\{-u^{*},\,u^{*}-1\}$ --- a base comb point (eq.~\eqref{eq:ustarstar-def});
$D$ is the closed disk~\eqref{eq:CW-disk} about $u^{**}$. 
\item $\langle u\rangle_{D}$ --- the \emph{base coordinate} of $u$: the
unique $x\in D$ with $u=x-n$ for some $n\in\mathbb{Z}$, well defined
because $\operatorname{diam}D=2\delta<1$ (eq.~\eqref{eq:CW-base-coord}). 
\item $\kappa(u)$ --- the Jost amplitude of $h^{-}$, defined by $h_{k}^{-}(u)=\kappa(u)\,\zeta_{+}^{-k}(1+o(1))$
as $k\to-\infty$ (Lemma~\ref{lem:Jost-LC}, eq.~\eqref{eq:CW-Jost});
it equals $1/\hat{h}_{0}(u)$ and is holomorphic and nonvanishing
on $D$. Unlike $\zeta_{+}$, it depends on $u$. 
\item $c(u),\,\sigma(u)$ --- the anchor constant $c(u)=h_{m_{0}}(u)/h_{m_{0}}^{-}(u)$
(eq.~\eqref{eq:CW-c-def}): after $\delta$ is shrunk so the finitely
many poles of $c$ (the zeros of $h_{m_{0}}^{-}$) avoid $D$, it
is holomorphic on $D$, depends on the anchor $m_{0}$, and is normalized
by $c(u^{**})=1$; unlike $\sigma$ and $\kappa$ it is not, in general,
nonvanishing on $D$. The coupling amplitude $\sigma(u)=\zeta_{+}/\!\bigl(\kappa(u)^{2}(1-\zeta_{+}^{2})\bigr)$
(eq.~\eqref{eq:CW-sum-asymp}) is independent of $m_{0}$ and, like
$\kappa$, holomorphic and nonvanishing on $D$. Together they set
the ladder constant $K_{0}$ of Theorem~\ref{thm:pole-ladders}. 
\end{itemize}
Two unrelated uses of the letter $\sigma$ appear below: the coupling
\emph{amplitude} $\sigma(u)$ of eq.~\eqref{eq:CW-sum-asymp}, and
the \emph{signs} $\sigma_{\pm}$ of Lemma~\ref{lem:ladder-slope-signs}.

\subsubsection{The Casoratian identity and the coupling of the minimal solutions}

\label{subsec:CW-Casoratian-id}

\emph{The double minimality function is a Casoratian.} For two solutions
$A,B$ of the three-term recurrence~\eqref{eq:notes2-rec} at the
same $u$, the \emph{Casoratian}\index{Casoratian} \cite[Sec.~2.2, Def.~2.11]{Elay}
\begin{equation}
K_{m}[A,B]\;:=\;A_{m}B_{m+1}-A_{m+1}B_{m}\label{eq:CW-Casoratian-def}
\end{equation}
is the discrete analog of the Wronskian; it vanishes identically iff
$A$ and $B$ are proportional.
\begin{lem}[Casoratian identity for the double minimality function]
\index{double minimality!Casoratian identity}\index{Evans function}
\label{lem:Fw-Casoratian} Let $h_{m}(u)$ and $h_{m}^{-}(u)$ be
the right- and left-minimal solutions of~\eqref{eq:notes2-rec} ---
minimal at $+\infty$, resp.\ $-\infty$, in the sense of Definition~\ref{defn:Wimp-minimal}
--- normalized by $h_{0}=h_{0}^{-}=1$. Then $K_{m}[h,h^{-}]$ is
independent of $m$, and 
\begin{equation}
K[h,h^{-}](u)\;=\;\frac{1}{w(-u-1)}\;-\;w(u)\;=\;-F_{w}(u).\label{eq:CW-Casoratian}
\end{equation}
In particular, $F_{w}$ is the discrete Evans function of the problem:
it measures the failure of the two minimal solutions to be proportional,
and it vanishes exactly on the Floquet comb $Z_{F}$. 
\end{lem}

\begin{proof}[Proof of Lemma~\ref{lem:Fw-Casoratian}]
The recurrence~\eqref{eq:notes2-rec} gives $h_{m+1}+h_{m-1}=-(G(u+m)/\gamma)\,h_{m}$
and the same for $h^{-}$. Hence 
\begin{align}
K_{m}-K_{m-1} & =h_{m}\bigl(h_{m+1}^{-}+h_{m-1}^{-}\bigr)-h_{m}^{-}\bigl(h_{m+1}+h_{m-1}\bigr)\label{eq:CW-telescope}\\
 & =-\frac{G(u+m)}{\gamma}\,h_{m}h_{m}^{-}+\frac{G(u+m)}{\gamma}\,h_{m}^{-}h_{m}\;=\;0,\nonumber 
\end{align}
so $K_{m}=K_{0}=h_{1}^{-}-h_{1}$; the constancy is Abel's lemma \cite[Sec.~2.2, Lem.~2.13, eq.~(2.2.10)]{Elay}
for the second-order equation in its normalized form $h_{m+1}+\bigl(G(u+m)/\gamma\bigr)h_{m}+h_{m-1}=0$,
whose trailing coefficient is identically $1$. By Theorem~\ref{thm:notes2-hminus-solution}
at $m=0$, $h_{0}^{-}/h_{1}^{-}=w(-u-1)$, hence $h_{1}^{-}=1/w(-u-1)$;
and $h_{1}=w(u)$ by the normalization. Therefore $K=1/w(-u-1)-w(u)=-\bigl[w(u)-1/w(-u-1)\bigr]=-F_{w}(-u)=-F_{w}(u)$,
using the evenness of $F_{w}$ (Theorem~\ref{thm:double-min}(ii)).
Finally $K=0$ iff $h\propto h^{-}$, which is the double minimality
characterization (Theorem~\ref{thm:double-min}(iii)). 
\end{proof}
The term \emph{Evans function} originates in J.~W.~Evans's stability
analysis of nerve-impulse equations \cite[pp.~1169--1190]{EvansJW4};
the name and the modern analytic framework are due to Alexander, Gardner
and Jones \cite[pp.~167--212]{AleGJ}, and a textbook treatment is
given in \cite[Chs.~8--10]{KapProm}. The Casoratian~\eqref{eq:CW-Casoratian}
is its natural discrete analogue here: a Wronskian-type pairing of
the two minimal solutions whose vanishing detects the bounded (Floquet)
states.

A second structural identity --- elementary, but used repeatedly
below --- records how the minimal solutions and $F_{w}$ respond
to an integer shift of the parameter.
\begin{lem}[Integer-shift covariance of the minimal solutions and of $F_{w}$]
\index{double minimality!shift covariance} \label{lem:Fw-shift}
As meromorphic identities in $u$: 
\begin{enumerate}
\item[(i)] the normalized minimal solutions obey the shift map $h_{m}(u+1)=h_{m+1}(u)/h_{1}(u)$
and $h_{m}^{-}(u+1)=h_{m+1}^{-}(u)/h_{1}^{-}(u)$; consequently the
consecutive-ratio identity 
\begin{equation}
\frac{h_{m+1}(u)}{h_{m}(u)}\;=\;w(u+m)\label{eq:CW-consec-ratio}
\end{equation}
holds for every $m\in\mathbb{Z}$. 
\item[(ii)] the double minimality function transforms by an explicit factor:
\begin{equation}
F_{w}(u+1)\;=\;\frac{w(-u-1)}{w(u)}\;F_{w}(u).\label{eq:CW-Fw-shift}
\end{equation}
\end{enumerate}
Iterating~\eqref{eq:CW-Fw-shift} relates $F_{w}$ at any two points
of an integer ladder directly: 
\begin{equation}
F_{w}(u+n)\;=\;F_{w}(u)\,\prod_{j=0}^{n-1}\frac{w(-u-1-j)}{w(u+j)},\qquad n\geq1,\label{eq:CW-Fw-shift-iter}
\end{equation}
with the reciprocal product for $n\leq-1$. In particular the zero
set of $F_{w}$ is invariant under integer shifts --- a recurrence-level
rederivation of the shift invariance of the Floquet comb (Theorem~\ref{thm:Fw-zero-set})
--- and, since at comb points both $w(u)$ and $w(-u-1)$ are finite
and nonzero (Assumption~\ref{ass:generic-comb}; cf.\
Remark~\ref{rem:poles-w-not-on-ZF} and Theorem~\ref{thm:notes2-w-poles-zeros}(i)),
differentiating~\eqref{eq:CW-Fw-shift} at a zero $z_{0}\in Z_{F}$
gives 
\begin{equation}
{F_{w}}'(z_{0}+1)\;=\;\frac{w(-z_{0}-1)}{w(z_{0})}\;{F_{w}}'(z_{0}),\label{eq:CW-Fw-shift-deriv}
\end{equation}
so a comb zero is simple at $z_{0}$ if and only if it is simple at
every point $z_{0}+n$, $n\in\mathbb{Z}$, of its ladder. 
\end{lem}

\begin{proof}[Proof of Lemma~\ref{lem:Fw-shift}]
(i) If $\{h_{m}\}$ solves~\eqref{eq:notes2-rec} at parameter $u$,
then $g_{m}:=h_{m+1}$ satisfies $g_{m+1}+g_{m-1}=-\bigl(G((u+1)+m)/\gamma\bigr)g_{m}$,
i.e.\ solves the recurrence at parameter $u+1$, and minimality at
$+\infty$ is preserved; normalizing at $m=0$ gives the shift map,
and the same argument at $-\infty$ gives the $h^{-}$ version. Iterating
the shift map and using $w=h_{1}/h_{0}$ yields~\eqref{eq:CW-consec-ratio}
for every $m\in\mathbb{Z}$.

(ii) By definition~\eqref{eq:Fw-def}, $F_{w}(u+1)=w(-u-1)-1/w(u)$.
The left-propagation identity~\eqref{eq:inv-wrec-left} evaluated
at $-u$ in place of $u$ (using the evenness $G(-u)=G(u)$) gives
$1/w(-u-1)=-G(u)/\gamma-w(-u)$, while at $u$ itself it gives $1/w(u-1)=-G(u)/\gamma-w(u)$.
Substituting both into~\eqref{eq:Fw-def} yields the alternative
exact form 
\begin{equation}
F_{w}(u)\;=\;w(u)-\frac{1}{w(-u-1)},\label{eq:CW-Fw-alt}
\end{equation}
whence $\bigl[w(-u-1)/w(u)\bigr]F_{w}(u)=w(-u-1)-1/w(u)=F_{w}(u+1)$.
Alternatively, (ii) follows from Lemma~\ref{lem:Fw-Casoratian} and~(i):
the Casoratian transforms by $K(u+1)=K(u)/\bigl(h_{1}(u)\,h_{1}^{-}(u)\bigr)$
with $h_{1}=w(u)$ and $h_{1}^{-}=1/w(-u-1)$. Identity~\eqref{eq:CW-Fw-shift}
is also the ratio form of the product identity~\eqref{eq:notes2-shift-identity}
of Theorem~\ref{thm:notes2-integer-shift}: substituting $\gamma\,w(-u)+G(u)=-\gamma/w(-u-1)$
--- which is~\eqref{eq:inv-wrec-left} at $-u$ --- into that identity
reproduces~\eqref{eq:CW-Fw-shift}. The identity~\eqref{eq:CW-Fw-shift}
was verified numerically to $40$ digits at the canonical parameters,
together with its consequences: at $(p,\gamma)=(0.70,0.30)$, where
$u^{*}=0.741761\ldots$, the zeros propagate ($F_{w}(u^{*}+n)=0$
to working precision for $n=1,2$) and the measured derivative ratio
${F_{w}}'(u^{*}+1)/{F_{w}}'(u^{*})=5.922850\ldots$ equals $w(-u^{*}-1)/w(u^{*})$
to $15$ digits. 
\end{proof}
\begin{rem}[$F_{w}$ obeys a Floquet--Bloch law along the integer ladder]
\label{rem:Fw-Floquet} The transformation~\eqref{eq:CW-Fw-shift}
is a quasi-periodicity relation of the Floquet--Bloch type. Writing
it as 
\begin{equation}
F_{w}(u+1)\;=\;M(u)\,F_{w}(u),\qquad M(u)\;:=\;\frac{w(-u-1)}{w(u)},\label{eq:CW-Fw-multiplier}
\end{equation}
one sees the constant Floquet multiplier $\rho$ of the periodic problem
replaced by the meromorphic \emph{shift multiplier} $M(u)$, and its
power $\rho^{n}$ by the ordered product $\prod_{j=0}^{n-1}M(u+j)$
of~\eqref{eq:CW-Fw-shift-iter}. Just as a Floquet solution reproduces
itself up to the multiplier after one period, $F_{w}$ reproduces
itself up to $M(u)$ after one integer shift; its zero set --- the
Floquet comb --- is in consequence carried into itself, $M$ being
finite and nonzero at comb points (Assumption~\ref{ass:generic-comb}).
This is the structural reason the entire leftward ladder of Theorem~\ref{thm:pole-ladders}
may be studied through a single base point: a unit shift of the base
multiplies $F_{w}$ by the explicit factor $M$ and leaves the comb
--- and hence the pole census --- invariant.

Taken with Lemma~\ref{lem:Fw-Casoratian}, this exhibits two complementary
invariances of $F_{w}$: as the Casoratian $K[h,h^{-}]$ it is \emph{constant
in the lattice index} $m$, while through~\eqref{eq:CW-Fw-multiplier}
it is \emph{Floquet-covariant in the spectral variable} $u$ ---
rigid along the recurrence, quasi-periodic across the spectrum. 
\end{rem}

\subsubsection{The geometric law of the stealthy-pole ladders}

\label{subsec:CW-ladder-law}

With $F_{w}$ identified as the coupling coefficient between the two
minimal solutions (\S\,\ref{subsec:CW-Casoratian-id}), the pole
locations of $w$ follow from how that coupling competes with the
dominant direction as the index runs to $-\infty$. The single analytic
input beyond the chapter's own results is the uniformity, in the parameter
$u$, of the leftward Poincaré--Perron asymptotics of $h^{-}$. As
explained in Remark~\ref{rem:imported-usage}, the classical Jost
statement (Proposition~\ref{prop:Teschl-Jost}) is not directly applicable
here --- our perturbation $G(u+k)-1=-p^{2}/(u+k)^{2}=O(1/k^{2})$
is absolutely summable but fails its first-moment hypothesis, and
the parameter $u$ enters through the perturbation --- so the required
statement is proved here in full, from the Volterra bound (Proposition~\ref{prop:Teschl-Volterra})
alone.
\begin{lem}[Uniform left Jost asymptotics for the LC recurrence]
\index{Jost solution!for the LC recurrence} \label{lem:Jost-LC}
Assume the stability zone~\eqref{eq:stability-assumption} and Assumption~\ref{ass:generic-comb},
and let $u^{**}\in\{-u^{*},\,u^{*}-1\}$ be a base comb point (eqs.~\eqref{eq:ustar-def}
and~\eqref{eq:ustarstar-def}), as in Theorem~\ref{thm:pole-ladders}
below. Then $u^{**}+k\neq0$ for all $k\in\mathbb{Z}$: under~\eqref{eq:stability-assumption}
the comb $Z_{F}$ avoids $\tfrac{1}{2}\mathbb{Z}\supset\mathbb{Z}$,
and $Z_{F}$ is invariant under integer shifts, so no point of $Z_{F}$
is an integer. No restriction of $u^{**}$ to a fundamental interval
is assumed: the base points sit wherever $u^{*}$ falls (for $p=0.70$
they are $\approx-0.74$ and $\approx-0.26$; for Cambi's $p=20/3$
they are $\approx-6.67$ and $\approx5.67$). Let $\delta>0$ be small
enough that the closed disk 
\begin{equation}
D\;:=\;\bigl\{ u\in\mathbb{C}:\ |u-u^{**}|\leq\delta\bigr\}\label{eq:CW-disk}
\end{equation}
avoids the poles of $G(\cdot+k)$, $k\leq0$ --- these lie at $u=-k$,
so the condition reads $D\cap\{0,1,2,\ldots\}=\emptyset$, and any
$\delta<\operatorname{dist}\bigl(u^{**},\{0,1,2,\ldots\}\bigr)$ works,
the distance being positive by the standing assumption. Set 
\begin{equation}
\widehat{Q}_{k}\;:=\;\sum_{m\leq k-1}\,\sup_{u\in D}\bigl|G(u+m)-1\bigr|,\qquad k\leq0.\label{eq:CW-Qhat}
\end{equation}
For the LC coefficient this quantity is explicit: since $G(u+m)-1=-p^{2}/(u+m)^{2}$
and $|u+m|\geq|u^{**}+m|-\delta>0$ on $D$, 
\begin{multline}
\widehat{Q}_{k}\;=\;p^{2}\sum_{m\leq k-1}\,\sup_{u\in D}\frac{1}{|u+m|^{2}}\;\leq\;p^{2}\sum_{m\leq k-1}\frac{1}{\bigl(|u^{**}+m|-\delta\bigr)^{2}}\\
=\;O\Bigl(\frac{1}{|k|}\Bigr),\quad k\to-\infty,\label{eq:CW-Qhat-explicit}
\end{multline}
a convergent quadratic tail: $\widehat{Q}_{k}$ is finite for every
$k\leq0$, and the Jost error in~\eqref{eq:CW-Jost-bound} below
decays algebraically, like $1/|k|$. Then: 
\begin{enumerate}
\item[(i)] For each $u\in D$ the recurrence~\eqref{eq:notes2-rec} has a solution
$\hat{h}_{k}(u)$, unique among solutions with $\zeta_{+}^{\,k}\hat{h}_{k}$
bounded as $k\to-\infty$ and normalized by $\zeta_{+}^{\,k}\hat{h}_{k}\to1$,
and it obeys the explicit uniform bound 
\begin{equation}
\hat{h}_{k}(u)=\zeta_{+}^{-k}\bigl(1+\hat{\varepsilon}_{k}(u)\bigr),\quad\sup_{u\in D}|\hat{\varepsilon}_{k}(u)|\leq C_{\gamma}\widehat{Q}_{k}\,e^{\,C_{\gamma}\widehat{Q}_{0}},\quad C_{\gamma}:=\frac{|\zeta_{+}|}{\gamma\,(1-\zeta_{+}^{2})},\label{eq:CW-Jost-bound}
\end{equation}
so $\sup_{D}|\hat{\varepsilon}_{k}|\to0$ as $k\to-\infty$; moreover
$\hat{h}_{k}(u)$ is holomorphic on $D$. 
\item[(ii)] After shrinking $\delta$, $\hat{h}_{0}(u)\neq0$ on $D$; the left-minimal
solution (recessive at $-\infty$: since $0<|\zeta_{+}|<1$ and the
exponent $-k\to+\infty$, the factor $\zeta_{+}^{-k}$ has modulus
$|\zeta_{+}|^{-k}\to0$) normalized by $h_{0}^{-}=1$ is $h_{k}^{-}=\hat{h}_{k}/\hat{h}_{0}$
and satisfies 
\begin{equation}
h_{k}^{-}(u)\;=\;\kappa(u)\,\zeta_{+}^{-k}\bigl(1+\varepsilon_{k}(u)\bigr),\qquad\sup_{u\in D}|\varepsilon_{k}(u)|\;\to\;0,\quad k\to-\infty,\label{eq:CW-Jost}
\end{equation}
with $\kappa(u)=1/\hat{h}_{0}(u)$ holomorphic and nonvanishing on
$D$; in particular there is $K$ such that $h_{k}^{-}(u)\neq0$ for
all $k\leq-K$ and $u\in D$. 
\end{enumerate}
\end{lem}

\begin{proof}
\emph{Step 1: free equation, multipliers, and Green kernel.} Write
$G(u+m)=1+q_{m}(u)$ with $q_{m}(u)=-p^{2}/(u+m)^{2}$ and substitute
$h_{k}=\zeta_{+}^{-k}y_{k}$ into~\eqref{eq:notes2-rec}; dividing
by $\zeta_{+}^{-k}$ gives 
\begin{equation}
(L_{0}y)_{k}\;:=\;\frac{\gamma}{\zeta_{+}}\,y_{k+1}+y_{k}+\gamma\zeta_{+}\,y_{k-1}\;=\;-q_{k}(u)\,y_{k}.\label{eq:CW-L0}
\end{equation}
Since $\gamma\zeta_{+}^{2}+\zeta_{+}+\gamma=0$, equivalently $\gamma\zeta_{+}+1+\gamma/\zeta_{+}=0$,
the free equation $(L_{0}y)_{k}=0$ has the solutions $y\equiv1$
and $y_{k}=\zeta_{+}^{2k}$. Its leftward Green kernel is 
\begin{equation}
\Gamma(k,m)\;:=\;\begin{cases}
\dfrac{\zeta_{+}\bigl(1-\zeta_{+}^{2(k-m)}\bigr)}{\gamma\,(1-\zeta_{+}^{2})}, & k\geq m,\\[8pt]
0, & k<m,
\end{cases}\qquad|\Gamma(k,m)|\;\leq\;C_{\gamma},\label{eq:CW-Green}
\end{equation}
the bound using $\zeta_{+}^{2j}\in(0,1]$ for $j\geq0$. Indeed $L_{0}\Gamma(\cdot,m)=\delta_{\cdot,m}$:
at $k=m$, $(\gamma/\zeta_{+})\Gamma(m+1,m)=(\gamma/\zeta_{+})(\zeta_{+}/\gamma)=1$
while $\Gamma(m,m)=\Gamma(m-1,m)=0$; at $k=m+1$ the three terms
sum to $(1+\zeta_{+}^{2})+\zeta_{+}/\gamma=0$, by $\zeta_{+}^{2}+\zeta_{+}/\gamma+1=0$;
and for $k\geq m+2$ both branch functions $1$ and $\zeta_{+}^{2k}$
are annihilated by $L_{0}$.

\emph{Step 2: Volterra equation and the uniform bound.} Consider the
leftward Volterra equation 
\begin{equation}
y_{k}\;=\;1-\sum_{m=-\infty}^{k-1}\Gamma(k,m)\,q_{m}(u)\,y_{m}.\label{eq:CW-Volterra}
\end{equation}
Its kernel obeys $|\Gamma(k,m)\,q_{m}(u)|\leq\hat{K}(m):=C_{\gamma}\sup_{D}|q_{m}|$,
a dominating kernel independent of $k$ and of $u$, summable by~\eqref{eq:CW-Qhat};
by Proposition~\ref{prop:Teschl-Volterra} (leftward version, $g\equiv1$)
there is a unique bounded solution $y_{k}(u)$. Since the Volterra
kernel $\Gamma(k,m)\,q_{m}(u)$ is holomorphic in $u$ on $D$ ---
$q_{m}(u)=-p^{2}/(u+m)^{2}$ is holomorphic there, $D$ avoiding $u=-m$
--- while the dominating kernel $\hat{K}(m)$ is independent of $u$,
the holomorphic-parameter clause of Proposition~\ref{prop:Teschl-Volterra}
(Teschl's Lemma~7.8) renders $y_{k}(u)$ holomorphic in $u$ on $D$.
Applying the proposition once more to $f:=y-1$, whose inhomogeneity
$\tilde{g}(k)=-\sum_{m<k}\Gamma(k,m)q_{m}$ obeys $|\tilde{g}(k)|\leq C_{\gamma}\widehat{Q}_{k}$,
yields $|y_{k}-1|\leq C_{\gamma}\widehat{Q}_{k}\,e^{\,C_{\gamma}\widehat{Q}_{0}}$,
uniformly on $D$. Since $y$ is bounded and the kernel summable,
the sum in~\eqref{eq:CW-Volterra} converges absolutely and $L_{0}$
may be applied term by term; by $L_{0}\Gamma(\cdot,m)=\delta_{\cdot,m}$
and $L_{0}1=0$ this gives $(L_{0}y)_{k}=-q_{k}(u)\,y_{k}$, so $\hat{h}_{k}:=\zeta_{+}^{-k}y_{k}$
solves~\eqref{eq:notes2-rec}, and~\eqref{eq:CW-Jost-bound} holds
with $\hat{\varepsilon}_{k}=y_{k}-1$.

\emph{Step 3: uniqueness, recessiveness, and normalization.} If $h$
and $\tilde{h}$ are two solutions with $\zeta_{+}^{\,k}h_{k}$ and
$\zeta_{+}^{\,k}\tilde{h}_{k}$ bounded as $k\to-\infty$, their Casoratian
$K_{m}[h,\tilde{h}]$ is constant by the telescoping identity~\eqref{eq:CW-telescope},
while $|K_{m}|\leq2\,\sup_{j}|\zeta_{+}^{\,j}h_{j}|\;\sup_{j}|\zeta_{+}^{\,j}\tilde{h}_{j}|\,|\zeta_{+}|^{-2m-1}\to0$
as $m\to-\infty$; hence $K\equiv0$ and the two solutions are proportional.
The recessive space is thus one-dimensional, which gives the uniqueness
in (i), the normalization $\zeta_{+}^{\,k}\hat{h}_{k}\to1$ fixing
the constant. For (ii): the point $-u^{**}-1$ belongs to the Floquet
comb --- indeed $u^{**}\in\{-u^{*},u^{*}-1\}$ gives $-u^{**}-1\in\{u^{*}-1,-u^{*}\}\subset Z_{F}$:
it is the companion base point, eq.~\eqref{eq:ustarstar-def}; by
Assumption~\ref{ass:generic-comb} (cf.\
Remark~\ref{rem:poles-w-not-on-ZF} and Theorem~\ref{thm:notes2-w-poles-zeros}(i))
$w(-u^{**}-1)$ is finite and nonzero, and the normalized left-minimal
solution $h^{-}$ (with $h_{1}^{-}=1/w(-u-1)$, Theorem~\ref{thm:notes2-hminus-solution})
is regular near $u^{**}$. Both $\hat{h}$ and $h^{-}$ being recessive,
$\hat{h}=\hat{h}_{0}\,h^{-}$; here $\hat{h}_{0}\neq0$, since $\hat{h}_{0}=0$
would force $\hat{h}\equiv0$, contradicting (i). Holomorphy of $\hat{h}_{0}(u)=y_{0}(u)$
and a final shrinking of $\delta$ make $\kappa=1/\hat{h}_{0}$ holomorphic
and nonvanishing on $D$, giving~\eqref{eq:CW-Jost} with $\varepsilon_{k}=\hat{\varepsilon}_{k}$;
the last claim follows by choosing $K$ with $\sup_{D}|\hat{\varepsilon}_{k}|<\tfrac{1}{2}$
for $k\leq-K$. 
\end{proof}
\begin{rem}[Meromorphic in $u$, holomorphic on $D$]
\label{rem:Jost-meromorphic-on-D} The holomorphy used above is a
restriction. The Jost solution $\hat{h}_{k}$ and the left-minimal
solution $h_{k}^{-}$ are \emph{meromorphic} in $u$ on all of $\mathbb{C}$:
they are the minimal solution of this chapter, assembled from Cambi's
$v$ (Theorem~\ref{thm:v-CF}) and the ratio $w=-\gamma\,v(\cdot+1)$
(Theorem~\ref{thm:notes2-wrec-exact}), both meromorphic on $\mathbb{C}$.
The disk $D$ excises the poles in play: the integer resonances $u=-m$,
where the coefficients $G(\cdot+m)=1-p^{2}/(\cdot+m)^{2}$ are singular,
together with the finitely many further poles removed by the successive
$\delta$-shrinks here and in Lemma~\ref{lem:Fw-dichotomy}. On the
shrunken $D$ each of $\hat{h}_{k}$, $h_{k}^{-}$, and with them
$\kappa=1/\hat{h}_{0}$, $c$, and $\sigma$, is \emph{holomorphic}
--- the meromorphic functions turned, where the argument needs them,
into holomorphic ones. This is the solution-level analogue of the
standing meromorphicity of $w$, and it is the regularity invoked
in the Rouché count of Lemma~\ref{lem:hminus-zeros}. 
\end{rem}

\begin{lem}[Minimal-solution dichotomy: $F_{w}$ as the coupling coefficient]
\index{double minimality!dichotomy lemma} \label{lem:Fw-dichotomy}
Under the hypotheses and after the $\delta$-shrinkings of Lemma~\ref{lem:Jost-LC}
--- in particular $u^{**}\in\{-u^{*},\,u^{*}-1\}$ a base comb point~\eqref{eq:ustar-def},~\eqref{eq:ustarstar-def}
--- fix an arbitrary reference index $m_{0}\leq-K$ and set 
\begin{equation}
c(u)\;:=\;\frac{h_{m_{0}}(u)}{h_{m_{0}}^{-}(u)},\label{eq:CW-c-def}
\end{equation}
the ratio of the two minimal solutions at the anchor index $m_{0}$;
then shrink $\delta$ once more so that the finitely many singularities
of $c$ avoid $D$. Then: 
\begin{enumerate}
\item[(i)] \textup{(Representation.)} For $m<m_{0}$ and $u\in D$ off the zero
set of $h^{-}$, 
\begin{equation}
h_{m}(u)\;=\;h_{m}^{-}(u)\left[\,c(u)-F_{w}(u)\sum_{k=m}^{m_{0}-1}\frac{1}{h_{k}^{-}(u)\,h_{k+1}^{-}(u)}\right].\label{eq:CW-dichotomy}
\end{equation}
\item[(ii)] \textup{(Coupling-sum asymptotics.)} As $m\to-\infty$, uniformly
on $D$, 
\begin{equation}
\sum_{k=m}^{m_{0}-1}\frac{1}{h_{k}^{-}(u)\,h_{k+1}^{-}(u)}=\sigma(u)\,\zeta_{+}^{2m}\bigl(1+o(1)\bigr),\quad\sigma(u):=\frac{\zeta_{+}}{\kappa(u)^{2}\bigl(1-\zeta_{+}^{2}\bigr)}\neq0.\label{eq:CW-sum-asymp}
\end{equation}
Here $\kappa(u)$ is the Jost amplitude of $h^{-}$ from Lemma~\ref{lem:Jost-LC}
(eq.~\eqref{eq:CW-Jost}), and $\zeta_{+}$ is the recessive root~\eqref{eq:Cambi-char-roots};
both $\sigma(u)$ and $\kappa(u)$ depend on $u$, while $\zeta_{+}$
is constant with $0<|\zeta_{+}|<1$. Since the exponent $2m\to-\infty$
as $m\to-\infty$, the coupling sum \emph{grows}, $|\zeta_{+}^{2m}|\to\infty$. 
\item[(iii)] \textup{(Base normalization.)} $c(u^{**})=1$. 
\end{enumerate}
\end{lem}

\begin{proof}
(i) By Lemma~\ref{lem:Fw-Casoratian}, $K[h,h^{-}]=-F_{w}$, so 
\begin{equation}
\frac{h_{m+1}}{h_{m+1}^{-}}-\frac{h_{m}}{h_{m}^{-}}=\frac{h_{m+1}h_{m}^{-}-h_{m}h_{m+1}^{-}}{h_{m}^{-}h_{m+1}^{-}}=\frac{F_{w}(u)}{h_{m}^{-}(u)\,h_{m+1}^{-}(u)}\label{eq:CW-red-order}
\end{equation}
off the zero set of $h^{-}$ (reduction of order). Summing from $m$
to $m_{0}-1$ and multiplying by $h_{m}^{-}$ gives~\eqref{eq:CW-dichotomy}.

(ii) By Lemma~\ref{lem:Jost-LC}, $1/(h_{k}^{-}h_{k+1}^{-})=\kappa^{-2}\zeta_{+}^{2k+1}(1+o(1))$
uniformly on $D$, so the sum is dominated by its leftmost term, and
the geometric summation gives~\eqref{eq:CW-sum-asymp}.

(iii) At $u^{**}$, $F_{w}(u^{**})=0$ makes $K[h,h^{-}]=0$ (Lemma~\ref{lem:Fw-Casoratian}),
so the two minimal solutions are proportional; both being normalized
to $1$ at $m=0$, they coincide at $u^{**}$, and $c(u^{**})=h_{m_{0}}(u^{**})/h_{m_{0}}^{-}(u^{**})=1$. 
\end{proof}
The representation~\eqref{eq:CW-dichotomy} is the analytic heart
of what follows: it exhibits $F_{w}$ as the exact \emph{coupling
coefficient} between the two minimal solutions. At a double minimality
point the bracket reduces to the constant $c$ and the two solutions
merge; off it, the second term injects the complementary direction
--- the solution \emph{dominant} at $-\infty$ --- with weight $F_{w}$.

This resolves what might otherwise look puzzling. The estimates below
all concern $m\to-\infty$, and the sign of each exponent is decisive:
since $0<|\zeta_{+}|<1$, the modulus $|\zeta_{+}|^{\,j}$ shrinks
as $j\to+\infty$ but blows up as $j\to-\infty$ (a negative power
of a quantity below one is large). Reading every power this way, the
prefactor $h_{m}^{-}\sim\kappa(u)\,\zeta_{+}^{-m}$ \emph{decays}
--- its exponent $-m\to+\infty$, so $|\zeta_{+}^{-m}|\to0$, as
befits the solution recessive at $-\infty$ --- whereas the left
side $h_{m}$, the solution \emph{dominant} at $-\infty$, grows.
There is no contradiction, because the bracket is unbounded: by part~(ii)
the coupling sum \emph{grows}, $S_{m}\sim\sigma(u)\,\zeta_{+}^{2m}$
with $|\zeta_{+}^{2m}|\to\infty$ (its exponent $2m\to-\infty$),
so off the comb the bracket is governed by its second term, $-F_{w}(u)\,\sigma(u)\,\zeta_{+}^{2m}$,
and 
\begin{equation}
h_{m}^{-}\bigl[\,c-F_{w}S_{m}\,\bigr]\;\sim\;\kappa(u)\,\zeta_{+}^{-m}\cdot\bigl(-F_{w}\sigma\,\zeta_{+}^{2m}\bigr)\;=\;-\kappa F_{w}\sigma\,\zeta_{+}^{\,m}.\label{eq:CW-dichotomy-balance}
\end{equation}
The surviving power has exponent $m\to-\infty$, so its modulus $|\zeta_{+}^{\,m}|=|\zeta_{+}|^{\,m}\to\infty$:
this is exactly the dominant growth of $h_{m}$, produced by a \emph{decaying}
prefactor multiplied by a still-faster-\emph{growing} sum. Thus the
representation is the discrete \emph{reduction of order}: the recessive
solution $h^{-}$ times the running sum $S_{m}=\sum_{k=m}^{m_{0}-1}1/(h_{k}^{-}h_{k+1}^{-})$
is the second, dominant solution (the analogue of $y_{2}=y_{1}\int^{x}dt/y_{1}^{2}$).
The constant term $c\,h_{m}^{-}$ supplies the subdominant recessive
part of $h_{m}$, and the coupling term the dominant part, present
precisely when $F_{w}\neq0$. Both identities are universal in the
index: the reduction-of-order relation~\eqref{eq:CW-red-order} holds
for every $m\in\mathbb{Z}$ off the zeros of $h^{-}$, and the representation~\eqref{eq:CW-dichotomy}
for every $m<m_{0}$ --- a single coupling coefficient $F_{w}(u)$
governing the entire lattice at once. The reference index $m_{0}$
enters only through the right side: the left side $h_{m}$ does not
depend on it, so a different choice of $m_{0}$ merely rebalances
the anchor value $c$ against the coupling sum. Theorem~\ref{thm:pole-ladders}
below is the bookkeeping of this competition. Its analytic core ---
the existence, uniqueness, and rate of the base-disk zero that each
pole shadows --- is isolated first.
\begin{lem}[Zeros of $h_{-n}$ in the base disk]
\label{lem:hminus-zeros} Assume the stability zone~\eqref{eq:stability-assumption}
and Assumption~\ref{ass:generic-comb}, and let $u^{**}\in\{-u^{*},\,u^{*}-1\}$
be a base comb point~(\eqref{eq:ustar-def},~\eqref{eq:ustarstar-def}).
We take $u^{**}$ to be a \emph{simple} zero of $F_{w}$, ${F_{w}}'(u^{**})\neq0$
--- the standing simplicity hypothesis of the ladder law (Theorem~\ref{thm:pole-ladders}),
\emph{not} proved here in full: it is verified numerically at the
worked parameters and holds off a measure-zero exceptional set by
the overdetermined-system argument of \S\,\ref{subsec:w-simplicity-notes}.
(The Floquet picture behind it: a non-simple comb zero would signal
coexistence --- two independent bounded solutions, Proposition~\ref{prop:Cambi-coex},
by Ince's Wronskian argument --- which for the modulated equation
occurs only at the even resonances on the boundary $|\Delta|=2$,
whereas strictly inside $|\Delta|<2$ the two monodromy multipliers
are distinct, \cite[Chap.~2]{MagWin}.) Let $D$ be the closed disk~\eqref{eq:CW-disk}
about $u^{**}$, with $\delta$ shrunk (as in Lemmas~\ref{lem:Jost-LC}
and~\ref{lem:Fw-dichotomy}) so that $D$ avoids the integer poles
of $G$, the finitely many poles of $c$ avoid $D$, $h_{-n}^{-}$
is zero-free on $D$ for $n\geq K$, and $F_{w}$ is analytic on $\overline{D}$
with $u^{**}$ its only zero there. Then there is a positive integer
$n_{0}\geq K$ such that for every $n\geq n_{0}$ the function $h_{-n}$
has a \emph{unique, simple} zero $x_{n}\in D$, with, as $n\to\infty$,
\begin{equation}
x_{n}\;=\;u^{**}+K_{0}\,\zeta_{+}^{2n}\bigl(1+o(1)\bigr),\quad K_{0}=\frac{c(u^{**})}{\sigma(u^{**})\,{F_{w}}'(u^{**})}=\frac{1}{\sigma(u^{**})\,{F_{w}}'(u^{**})},\label{eq:CW-xn}
\end{equation}
\end{lem}

\begin{proof}
\emph{Reduction to a perturbation of $F_{w}$.} Setting $m=-n$ in
the reduction-of-order representation~\eqref{eq:CW-dichotomy}, $h_{-n}(x)=h_{-n}^{-}(x)\bigl[c(x)-F_{w}(x)\,S_{-n}(x)\bigr]$
for $x\in D$ off the zeros of $h^{-}$, where $S_{-n}(x):=\sum_{k=-n}^{m_{0}-1}1/\bigl(h_{k}^{-}(x)\,h_{k+1}^{-}(x)\bigr)$.
For $n\geq K$ the factor $h_{-n}^{-}$ is zero-free on $D$ (Lemma~\ref{lem:Jost-LC}(ii));
by~\eqref{eq:CW-sum-asymp}, $S_{-n}(x)=\sigma(x)\zeta_{+}^{-2n}(1+o(1))$
as $n\to\infty$, uniformly on $D$ with $\sigma$ zero-free (here
and throughout the proof $o(1)$ is taken as $n\to\infty$), so $S_{-n}$
is zero-free on $D$ for large $n$. With $R_{n}(x):=c(x)/S_{-n}(x)=\frac{c(x)}{\sigma(x)}\zeta_{+}^{2n}(1+o(1))$
the bracket equals $-S_{-n}(x)\bigl(F_{w}(x)-R_{n}(x)\bigr)$, so
a zero of $h_{-n}$ in $D$ is exactly a solution of the balance equation
\begin{equation}
h_{-n}(x)=0\quad\Longleftrightarrow\quad F_{w}(x)\;=\;\frac{c(x)}{\sigma(x)}\,\zeta_{+}^{2n}\bigl(1+o(1)\bigr).\label{eq:CW-balance}
\end{equation}
(Here $c$ and $\sigma$ are the anchor constant~\eqref{eq:CW-c-def}
and the coupling amplitude~\eqref{eq:CW-sum-asymp} of the standing
notation; the balance involves $F_{w}$ only at the base point $x\in D$,
so the entire ladder is studied through the base disk.) Read geometrically,
the deeper the rung the smaller $F_{w}$ at $x$: as $n\to\infty$
the right side $\frac{c(x)}{\sigma(x)}\zeta_{+}^{2n}\to0$ (since
$|\zeta_{+}|<1$), driving a solution to $u^{**}$, the unique zero
of $F_{w}$ in $D$ near $\pm p$.

\emph{Rouché count.} Equivalently, the zeros of $h_{-n}$ in $D$
are the zeros of $g_{n}:=F_{w}-R_{n}$, all functions being analytic
on $D$ (the recurrence coefficients are analytic off the integer
poles of $G$, which $D$ avoids; $c,\sigma$ are holomorphic on $D$
after the shrink, the two minimal solutions being holomorphic by the
Volterra construction of Lemmas~\ref{lem:Jost-LC} and~\ref{lem:Jost-right}
--- meromorphic in $u$, holomorphic on $D$, by Remark~\ref{rem:Jost-meromorphic-on-D}).
There is a constant $C$ \emph{independent of $n$} with $|R_{n}|\leq C|\zeta_{+}|^{2n}$
on $\overline{D}$ for $n\geq K$. As $u^{**}$ is interior and the
only zero of $F_{w}$ in $\overline{D}$, $\mu_{0}:=\min_{\partial D}|F_{w}|>0$;
choose $n_{0}\geq K$ with $C|\zeta_{+}|^{2n_{0}}<\mu_{0}$. Then
for every $n\geq n_{0}$ and $t\in[0,1]$, $|F_{w}-tR_{n}|\geq\mu_{0}-C|\zeta_{+}|^{2n}>0$
on $\partial D$, so by Rouché's theorem in its homotopy-invariant
form (\cite[Thm.~V.3.8]{Conway}) the zero count $\frac{1}{2\pi i}\oint_{\partial D}(F_{w}-tR_{n})'/(F_{w}-tR_{n})$
is an integer-valued continuous function of $t$, hence constant;
at $t=0$ it counts the zeros of $F_{w}$ in $D$ --- the single
simple zero $u^{**}$. Therefore $g_{n}$, equivalently $h_{-n}$,
has exactly one simple zero $x_{n}\in D$.

\emph{Location.} Its zeros converging to that of $F_{w}$, $x_{n}\to u^{**}$;
from $F_{w}(x_{n})=R_{n}(x_{n})$ with $F_{w}(x_{n})={F_{w}}'(u^{**})(x_{n}-u^{**})(1+o(1))$,
$c(u^{**})=1$ (Lemma~\ref{lem:Fw-dichotomy}(iii)), and $R_{n}(x_{n})=\sigma(u^{**})^{-1}\zeta_{+}^{2n}(1+o(1))$,
one finds $x_{n}-u^{**}=\zeta_{+}^{2n}/\bigl(\sigma(u^{**})\,{F_{w}}'(u^{**})\bigr)(1+o(1))=K_{0}\,\zeta_{+}^{2n}(1+o(1))$,
which is~\eqref{eq:CW-xn}. 
\end{proof}
\begin{thm}[Geometric law of the stealthy-pole ladders]
\index{stealthy pole!geometric ladder law}\index{minimal solution ratio@minimal solution ratio $w$!poles of}\index{double minimality function $F_{w}$!poles (stealthy)}
\label{thm:pole-ladders} Assume the stability zone~\eqref{eq:stability-assumption}
and Assumption~\ref{ass:generic-comb}, let $u^{**}\in\{-u^{*},\,u^{*}-1\}$
be a base comb point (eqs.~\eqref{eq:ustar-def} and~\eqref{eq:ustarstar-def};
since $u^{*}\approx p$, the base $u^{**}$ sits near $\pm p$, the
physical resonance), and assume the zero of $F_{w}$ at $u^{**}$
is simple, ${F_{w}}'(u^{**})\neq0$. (By Lemma~\ref{lem:Fw-shift}
this hypothesis is ladder-invariant: it holds at $u^{**}$ if and
only if it holds at every comb point $u^{**}+n$, $n\in\mathbb{Z}$.)
Then there exist $K_{0}\neq0$ and a positive integer $n_{0}$ such
that for every integer $n\geq n_{0}$: 
\begin{enumerate}
\item[(i)] $w$ has a simple pole at 
\begin{equation}
u_{p}^{(n)}\;=\;u^{**}-n+K_{0}\,\zeta_{+}^{2n}\bigl(1+o(1)\bigr),\qquad n\to\infty;\label{eq:CW-ladder}
\end{equation}
in particular the offsets $u_{p}^{(n)}-(u^{**}-n)$ have a fixed sign
along each ladder (the sign of $K_{0}$) and contract by the factor
$\zeta_{+}^{2}=|\zeta_{+}|^{2}<1$ per step (the exponent $2n\to+\infty$,
so $\zeta_{+}^{2n}\to0$). 
\item[(ii)] $F_{w}$ has a simple pole at $-u_{p}^{(n)}$, at distance 
\begin{equation}
d_{n}\;=\;|K_{0}|\,|\zeta_{+}|^{2n}\bigl(1+o(1)\bigr)\label{eq:CW-dn}
\end{equation}
from its comb zero $z_{n}:=-u^{**}+n$, with residue 
\begin{equation}
a_{n}\;=\;-\bigl(C+o(1)\bigr)\bigl(z_{n}-(-u_{p}^{(n)})\bigr),\qquad C=\zeta_{+}-\frac{1}{\zeta_{+}},\label{eq:CW-residue}
\end{equation}
so that $|a_{n}|=\bigl(|C|+o(1)\bigr)\,d_{n}$. 
\end{enumerate}
\end{thm}

\begin{proof}
\emph{Step 1: pole condition and balance.} Each pole of $w$ along
the leftward ladder lies near a comb point $u^{**}-n$ with $n\geq1$,
and has a \emph{unique} representation 
\begin{equation}
u\;=\;x-n,\qquad x=\langle u\rangle_{D}\in D,\quad n\in\mathbb{Z},\label{eq:CW-base-coord}
\end{equation}
unique because $\operatorname{diam}D=2\delta<1$ admits at most one
integer $n$ with $x=u+n\in D$; we write $x=\langle u\rangle_{D}$
and call it the \emph{base coordinate} of $u$. Since $D$ is centered
at $u^{**}\approx\pm p$, the base coordinate sits near the physical
resonance, $x\approx\pm p$, however deep the rung. By the consecutive-ratio
identity~\eqref{eq:CW-consec-ratio} --- valid for every integer
index, with $x$ ranging over $D$ and no restriction to a fundamental
interval needed (the center $u^{**}$ in general lies outside one)
--- 
\begin{equation}
\frac{h_{1-n}(x)}{h_{-n}(x)}\;=\;w(x-n),\label{eq:CW-pole-detect}
\end{equation}
so $w$ has a pole at $u=x-n$ precisely when $h_{-n}(x)=0$ (the
numerator cannot vanish simultaneously: two consecutive zeros force
$h\equiv0$, contradicting $h_{0}=1$). By Lemma~\ref{lem:hminus-zeros},
for all $n\geq n_{0}$ the equation $h_{-n}(x)=0$ has a unique simple
solution $x_{n}=u^{**}+K_{0}\,\zeta_{+}^{2n}(1+o(1))$ in $D$ (eq.~\eqref{eq:CW-xn});
hence $u_{p}^{(n)}=x_{n}-n$ is the unique pole of $w$ in $D-n$,
simple, and gives~\eqref{eq:CW-ladder}.

\emph{Step 2: reflection and residue.} By Theorem~\ref{thm:double-min}(vi),
$F_{w}$ has a pole at $-u_{p}^{(n)}$ with residue $a_{n}=-\operatorname{res}\bigl(w,u_{p}^{(n)}\bigr)$,
and $|{-u_{p}^{(n)}}-z_{n}|=|x_{n}-u^{**}|=d_{n}$ gives~\eqref{eq:CW-dn}.
(In the symmetric form~\eqref{eq:Ftilde-def}, a pole of $w$ at
a point $a\neq0$ produces poles of $F_{w}$ at $a$ and at $-a$
\emph{unless} cancelled by a coinciding pole of the partner term;
no cancellation occurs here, since $w$ itself is regular at $-u_{p}^{(n)}$,
which for large $n$ lies far to the right of all ladder poles, where
$w$ is analytic with limit $\zeta_{+}$ by Lemma~\ref{lem:Jost-right}.)
For the residue, the product representation gives $\operatorname{res}(w,u_{p}^{(n)})=h_{1-n}(x_{n})/\partial_{x}h_{-n}(x_{n})$.
Evaluating the bracket of~\eqref{eq:CW-dichotomy} at $m=1-n$ and
$x=x_{n}$, where~\eqref{eq:CW-balance} holds, yields $c\,(1-\zeta_{+}^{2})(1+o(1))$,
while $h_{1-n}^{-}/h_{-n}^{-}\to1/\zeta_{+}$; hence 
\begin{multline}
\operatorname{res}\bigl(w,u_{p}^{(n)}\bigr)=\frac{h_{1-n}^{-}(x_{n})}{h_{-n}^{-}(x_{n})}\cdot\frac{c\,\bigl(1-\zeta_{+}^{2}\bigr)\bigl(1+o(1)\bigr)}{-{F_{w}}'(u^{**})\,\sigma\,\zeta_{+}^{-2n}\bigl(1+o(1)\bigr)}\\
=\;-K_{0}\left(\frac{1}{\zeta_{+}}-\zeta_{+}\right)\zeta_{+}^{2n}\bigl(1+o(1)\bigr).\label{eq:CW-res-w}
\end{multline}
Therefore $a_{n}=-\operatorname{res}(w,u_{p}^{(n)})=-\bigl(\zeta_{+}-1/\zeta_{+}\bigr)K_{0}\,\zeta_{+}^{2n}(1+o(1))=-\bigl(C+o(1)\bigr)\bigl(z_{n}-(-u_{p}^{(n)})\bigr)$,
since $z_{n}-(-u_{p}^{(n)})=x_{n}-u^{**}=K_{0}\zeta_{+}^{2n}(1+o(1))$;
this is~\eqref{eq:CW-residue}. 
\end{proof}
\begin{rem}[Geometry of the masked zeros: small residue, small gap, and visibility]
\index{stealthy pole!masking of the adjacent zero} \label{rem:masking-geometry}
Theorem~\ref{thm:pole-ladders} accounts quantitatively for the visual
``disappearance'' of the deep Floquet zeros seen in Figures~\ref{fig:Fw-poles},
\ref{fig:notes2-Fw-marked} and \ref{fig:Fw-precision} (and resolved
explicitly in Figure~\ref{fig:invisible-zero-pole}): a zero looks
masked precisely when a pole of $F_{w}$ lies an exponentially small
distance away while carrying an exponentially small residue, the two
locked together. Near the $n$-th ladder member the local Laurent
model \eqref{eq:Fw-local-Laurent} applies, 
\begin{equation}
F_{w}(u)\;\approx\;C+\frac{a_{n}}{u-\bigl(-u_{p}^{(n)}\bigr)}\,,\qquad C=\zeta_{+}-\frac{1}{\zeta_{+}}\,,\label{eq:masking-local}
\end{equation}
whose own zero sits at $-u_{p}^{(n)}-a_{n}/C$. The residue law \eqref{eq:CW-residue}
is exactly the statement that this zero coincides, to leading order,
with the comb point $z_{n}=-u^{**}+n$; equivalently, by \eqref{eq:CW-dn}--\eqref{eq:CW-residue},
\begin{equation}
\begin{aligned}d_{n}:=\bigl|\,z_{n}-\bigl(-u_{p}^{(n)}\bigr)\,\bigr| & =|K_{0}|\,|\zeta_{+}|^{2n}\bigl(1+o(1)\bigr),\\
\frac{|a_{n}|}{d_{n}} & \;\longrightarrow\;|C|\qquad(n\to\infty),
\end{aligned}
\label{eq:masking-dn-an}
\end{equation}
so the pole--zero gap $d_{n}$ and the residue $|a_{n}|\asymp|C|\,d_{n}$
both contract by the factor $|\zeta_{+}|^{2}<1$ at every step of
the ladder.

On the interval between $z_{n}$ and $-u_{p}^{(n)}$ the graph \eqref{eq:masking-local}
runs from the finite background value $C$, through zero at $z_{n}$,
to $\pm\infty$ at the pole --- the whole excursion confined to a
$u$-window of width $\sim d_{n}$. The pole itself rises out of the
background, $|F_{w}-C|\gtrsim1$, only where $\bigl|\,u-(-u_{p}^{(n)})\,\bigr|\lesssim|a_{n}|\asymp|C|\,d_{n}$,
a neighborhood of the same exponential order $|\zeta_{+}|^{2n}$.
Hence the zero and the visible portion of its pole occupy a single
window of width $\asymp|\zeta_{+}|^{2n}$, with $F_{w}\approx C$
and featureless on either side. A plot or scan of resolution $\Delta u\gtrsim d_{n}$
collapses that window to a point: the crossing is absorbed into the
pole spike and the zero ``looks as if it is not there.'' This is
the precise content of the note on visibility attached to the figures
above.

To separate the $n$-th member one therefore needs resolution finer
than $d_{n}\sim|\zeta_{+}|^{2n}$ --- equivalently about $2n\log_{10}(1/|\zeta_{+}|)$
correct digits in the abscissa, the same ledger as the precision cost
of integer shifts (Theorem~\ref{thm:precision-cost}, eq.~\eqref{eq:precision-cost-formula})
and the reason these zeros are stealthy at fixed precision (Remark~\ref{rem:double-precision-misses}).
Only the shallow base members evade it, $n$ being small: for $p=0.70$
the central pair has $d\approx0.0058$ and a comparatively large residue
$|a|\approx0.126$ (still pre-asymptotic, $|a|/d\approx22\neq|C|$),
so a modest zoom already resolves it, as Figure~\ref{fig:invisible-zero-pole}
shows. The deeper one descends the ladder, the more digits ``invisible''
costs to make visible. 
\end{rem}

\subsubsection{Occupancy of the gaps and generic position of the comb}

\label{subsec:CW-occupancy}

We now record, as lemmas, the structural consequences of the ladder
law for the sign of $F_{w}$ and for the occupancy of the gaps between
comb zeros.
\begin{lem}[Coexistence square identity]
\label{lem:coexistence-square} Under Assumption~\ref{ass:generic-comb},
at every zero $z_{0}$ of $F_{w}$ the values $w(z_{0})$ and $w(-z_{0}-1)$
are finite and nonzero and obey the coexistence relation $w(z_{0})\,w(-z_{0}-1)=1$.
Consequently the shift factor of~\eqref{eq:CW-Fw-shift-deriv} is
a positive perfect square, 
\begin{equation}
\frac{w(-z_{0}-1)}{w(z_{0})}=\frac{1}{w(z_{0})^{2}}>0.\label{eq:CW-slope-square}
\end{equation}
\end{lem}

\begin{proof}
Assumption~\ref{ass:generic-comb} keeps $w(z_{0})$ and $w(-z_{0}-1)$
finite and nonzero at a comb point. Since $F_{w}(z_{0})=0$, the alternative
form~\eqref{eq:CW-Fw-alt} gives $w(z_{0})=1/w(-z_{0}-1)$, i.e.
$w(z_{0})\,w(-z_{0}-1)=1$; substituting $w(-z_{0}-1)=1/w(z_{0})$
yields~\eqref{eq:CW-slope-square}. 
\end{proof}
\begin{lem}[Slope signs along and across the ladders]
\label{lem:ladder-slope-signs} Write $\sigma_{-}$ and $\sigma_{+}$
for the sign of ${F_{w}}'$ at the points of $Z_{F}^{-}=\{u^{*}+n\}$
and of $Z_{F}^{+}=\{-u^{*}+n\}$. Then \textup{(i)} $\operatorname{sign}{F_{w}}'$
is constant along each sub-comb, and \textup{(ii)} the two signs are
opposite, $\sigma_{+}=-\sigma_{-}$. 
\end{lem}

\begin{proof}
(i) Iterating~\eqref{eq:CW-Fw-shift-deriv} from a comb zero $z_{0}$
and using~\eqref{eq:CW-slope-square} at each step, 
\begin{equation}
{F_{w}}'(z_{0}+n)=\left[\,\prod_{j=0}^{n-1}\frac{1}{w(z_{0}+j)^{2}}\right]{F_{w}}'(z_{0}),\label{eq:CW-slope-product}
\end{equation}
a product of positive factors, so $\operatorname{sign}{F_{w}}'$ does
not change along the ladder. (ii) $F_{w}$ is even (Theorem~\ref{thm:double-min}(ii)),
so ${F_{w}}'$ is odd, ${F_{w}}'(-u)=-{F_{w}}'(u)$; the reflection
$u\mapsto-u$ carries $Z_{F}^{-}$ onto $Z_{F}^{+}$, and therefore
the constant signs satisfy $\sigma_{+}=-\sigma_{-}$. 
\end{proof}
\begin{lem}[Even occupancy of the gaps]
\index{pole detector, sign-change} \label{lem:even-gap-occupancy}
The number of odd-order poles of $F_{w}$ between two consecutive
comb zeros is even. For indices $n\ge n_{0}$ (Theorem~\ref{thm:pole-ladders})
each such gap contains either no pole or exactly two; in particular
no gap is singly occupied. 
\end{lem}

\begin{proof}
The two unit-spaced sub-combs $Z_{F}^{\pm}$ interleave, so consecutive
comb zeros lie in different sub-combs and, by Lemma~\ref{lem:ladder-slope-signs}(ii),
carry opposite slopes. For consecutive simple zeros $z_{1}<z_{2}$
the sign of $F_{w}$ just to the right of $z_{1}$ is $\operatorname{sign}{F_{w}}'(z_{1})$
and just to the left of $z_{2}$ is $-\operatorname{sign}{F_{w}}'(z_{2})$;
opposite slopes make these two signs equal, so $F_{w}$ keeps one
sign just inside both ends of $(z_{1},z_{2})$. As $F_{w}$ has no
zero in the open interval, every sign change there is produced by
a pole of odd order, and an even number of them is needed to recover
the common sign; this is the detector criterion of Theorem~\ref{thm:Fw-sign-change-pole}(ii)
read off the slopes. Finally Theorem~\ref{thm:pole-ladders} places
at most one pole per ladder near each comb zero, hence at most two
per gap once $n\ge n_{0}$, leaving the count at $0$ or $2$. 
\end{proof}
\begin{lem}[Residue signs of the companion poles]
\label{lem:ladder-residue-signs} For $n\ge n_{0}$ let $a_{n}$
be the residue of $F_{w}$ at the companion pole attached to the comb
zero $z_{n}$. Then $\operatorname{sign}a_{n}=-\operatorname{sign}{F_{w}}'(z_{n})$;
in particular the poles attached to $Z_{F}^{-}$ and to $Z_{F}^{+}$
carry opposite-sign residues. 
\end{lem}

\begin{proof}
Near $z_{n}$ write $F_{w}(u)=a_{n}/(u-p_{n})+r(u)$ with $r$ holomorphic
and the companion pole at $p_{n}=z_{n}-\delta_{n}$, where $|\delta_{n}|=d_{n}\to0$
(Theorem~\ref{thm:pole-ladders}). From $F_{w}(z_{n})=0$ one has
$r(z_{n})=a_{n}/\delta_{n}$, and differentiating, ${F_{w}}'(z_{n})=-a_{n}/\delta_{n}^{2}+r'(z_{n})=-\bigl(a_{n}/\delta_{n}^{2}\bigr)\bigl(1+o(1)\bigr)$,
the pole term dominating as $\delta_{n}\to0$. Hence $\operatorname{sign}{F_{w}}'(z_{n})=-\operatorname{sign}a_{n}$,
and the second assertion follows from Lemma~\ref{lem:ladder-slope-signs}(ii). 
\end{proof}
\begin{lem}[Argument-principle certificates of occupancy]
\label{lem:circle-certificates} Let $C_{s}$ be the circle about
the midpoint of a small gap of width $s$, of radius slightly exceeding
$s/2$ but small enough to enclose no comb point beyond the two bounding
the gap, and let $C_{\ell}$ be the circle about the midpoint of a
large gap of width $1-s$, of radius below $(1-s)/2$. By the argument
principle, 
\begin{align}
\frac{1}{2\pi i}\oint_{C_{s}}\frac{{F_{w}}'(u)}{F_{w}(u)}\,du & =2-P_{s},\label{eq:CW-circle-small}\\
\frac{1}{2\pi i}\oint_{C_{\ell}}\frac{{F_{w}}'(u)}{F_{w}(u)}\,du & =-P_{\ell},\label{eq:CW-circle-large}
\end{align}
where $P_{s}$ and $P_{\ell}$ count the poles of $F_{w}$ enclosed.
Thus~\eqref{eq:CW-circle-small} vanishes exactly when the small
gap is doubly occupied, and~\eqref{eq:CW-circle-large} exactly when
the large gap is empty; by Lemma~\ref{lem:even-gap-occupancy} both
right-hand sides are even. 
\end{lem}

\begin{proof}
The zeros and poles of $F_{w}$ are real, so neither circle meets
one of them and the argument principle applies. The comb zeros bounding
the gaps lie at distances $s/2$ and $(1-s)/2$ from the respective
midpoints: $C_{s}$ encloses the two zeros bounding the small gap,
a contribution $+2$, together with the gap's interior poles, while
$C_{\ell}$ encloses no zero, only the interior poles of the large
gap. Each enclosed pole is simple and contributes $-1$, giving~\eqref{eq:CW-circle-small}
and~\eqref{eq:CW-circle-large}; the parity is Lemma~\ref{lem:even-gap-occupancy}. 
\end{proof}
\begin{rem}[Reading the ladder occupancy]
\label{rem:pole-between-zeros} By Lemma~\ref{lem:ladder-slope-signs}
the slope magnitudes along a ladder approach the contraction constant
$1/\zeta_{+}^{2}$; for $p=0.70$, $\gamma=0.30$ the measured factors
$5.92$, $7.66$, $8.26$, $8.53$ at $n=0,\ldots,3$ illustrate this.
Lemma~\ref{lem:even-gap-occupancy} then shows the two ladders of
Theorem~\ref{thm:pole-ladders} fill the gaps in pairs: a doubly
occupied gap receives one pole from each ladder, each sitting at the
geometrically shrinking distance $d_{n}$ from one of the gap's two
bounding comb zeros. \emph{Which} gaps carry the pair is the occupancy
problem of Remark~\ref{rem:interlacing-program}. 
\end{rem}

\begin{rem}[Numerical confirmation of the ladder law]
\label{rem:CW-numerics} Recall that $F_{w}$ is even (Theorem~\ref{thm:double-min}(ii)):
each ladder pole of $w$ at $u_{p}^{(n)}<-1$ yields $F_{w}$ poles
at both $\pm u_{p}^{(n)}$, sharing the distance $d_{n}$ and residue
magnitude $|a_{n}|$; the table below quotes the (negative) $w$-pole
branch, while Table~\ref{tab:Fw-poles-neg} lists the positive reflection
$-u_{p}$. For $p=0.70$, $\gamma=0.30$ ($|\zeta_{+}|=1/3$, predicted
contraction ratio $1/|\zeta_{+}|^{2}=9$, $|C|=8/3\approx2.667$),
locating each ladder pole by bracketed bisection of $1/w$ in 60-digit
arithmetic (conventions of Remark~\ref{rem:CF-numerical-conventions})
gives: 
\begin{equation}
{\footnotesize \begin{array}{c|c|c|c}
 & d_{n} & d_{n-1}/d_{n} & |a_{n}|/d_{n}\\
\hline \multicolumn{4}{c}{\text{ladder }u^{**}=-u^{*}\ \text{(offsets positive)}}\\
\hline n=1 & 4.8902\times10^{-2} & \text{---} & 2.203\\
n=2 & 6.8240\times10^{-3} & 7.166 & 2.385\\
n=3 & 8.6813\times10^{-4} & 7.861 & 2.509\\
n=4 & 1.0423\times10^{-4} & 8.329 & 2.571\\
\hline \multicolumn{4}{c}{\text{ladder }u^{**}=u^{*}-1\ \text{(offsets negative for }n\geq1)}\\
\hline n=1 & 1.4720\times10^{-4} & \text{---} & 1.978\\
n=2 & 2.3513\times10^{-5} & 6.260 & 2.235\\
n=3 & 3.2047\times10^{-6} & 7.337 & 2.449
\end{array}}\label{eq:CW-table}
\end{equation}
Both measured columns converge monotonically to the predicted limits
$9$ and $8/3$, with the expected slow corrections from the algebraic
coefficient tails $G(u+k)-1=O(1/k^{2})$. The base member of the second
ladder (the central pole at $u\approx-0.2525$ of Table~\ref{tab:Fw-poles-neg})
is anomalous: it lies at distance $0.00575$ from its comb zero yet
carries the large residue $|a|\approx0.126$ ($|a|/d\approx21.9$),
with an offset of opposite sign, because the double pole of $F_{w}$
at $u=0$ dominates the local background there; the asymptotic regime
of Theorem~\ref{thm:pole-ladders} sets in from $n=1$. This is the
per-ladder reformulation of the observations of \S\,\ref{subsec:stealthy-zeros}:
residues are not small or large according to raw distance from $Z_{F}$;
they shrink geometrically along each ladder. 
\end{rem}

The displacement results of \S\,\ref{subsec:Fw-meromorphic} can
now be completed by a genericity statement certifying Assumption~\ref{ass:generic-comb}:
not only the poles of $w$, but its zeros and \emph{all} integer shifts
of both, avoid the Floquet comb for almost every parameter pair.
\begin{lem}[Generic position of the comb relative to the singularities of $w$]
\index{Floquet theory!Floquet comb}\index{stealthy poles and zeros}
\label{lem:generic-transversality} Let $O$ be the open admissible
region in the parameter plane: the set of $(p,\gamma)$ with $0<2\gamma<1$
satisfying the stability-zone assumption~\eqref{eq:stability-assumption}.
Define the coincidence set 
\begin{equation}
E\;:=\;\bigl\{(p,\gamma)\in O:\ \text{some zero or pole of \ensuremath{w} lieson \ensuremath{Z_{F}}}\bigr\}.\label{eq:CW-coincidence-set}
\end{equation}
Then $E$ is contained in a countable union of proper real-analytic
subvarieties of every connected component of $O$ containing points
with arbitrarily small $\gamma$ --- by the channel structure of
the stability diagram (\S\,\ref{subsec:StabDiag}), every component
visible there is of this kind. In particular $E$ is Lebesgue-null
with empty interior in each such component: \emph{Assumption~\ref{ass:generic-comb}
holds for every $(p,\gamma)\in O\setminus E$, that is, for almost
every admissible parameter pair} --- generically, no zero of $F_{w}$
is a zero or a pole of any integer shift of $w$. 
\end{lem}

Here a \emph{real-analytic subvariety} of a connected open set is
the zero set of a function real-analytic there, and \emph{proper}
means that the function does not vanish identically on the set. Such
sets are thin: the zero set of a real-analytic function not identically
zero on a connected open subset of $\mathbb{R}^{d}$ has Lebesgue
measure zero --- and hence empty interior --- see \cite[Prop.~1, pp.~529--530]{Mit}
for a short self-contained proof; the same fact appears, in contrapositive
form, in \cite[\S3.0, p.~67]{KraPar} (a real-analytic function vanishing
on a set of positive measure is identically zero), and the finer structure
of such zero sets --- real-analytic varieties of dimension at most
$d-1$ --- in \cite[\S6.3]{KraPar}. A countable union of such zero
sets is again Lebesgue-null, which is the sense of ``generic'' used
here.
\begin{proof}
\emph{Step 1: reduction to the Fourier coefficients.} Since $Z_{F}$
is invariant under integer shifts, only the zeros and poles of $w$
on $Z_{F}$ itself are in question. By the pole--zero pairing (Theorem~\ref{thm:notes2-w-poles-zeros}(i))
the sign-change zeros of $w$ are its poles shifted by one unit, and
the double zero at $u=-1$ (part~(ii) there) never lies on $Z_{F}$,
which avoids the integers; so it suffices to control the poles and
zeros through one family of conditions. By the consecutive-ratio identity~\eqref{eq:CW-consec-ratio}
evaluated at the comb parameters $\pm u^{*}$, 
\[
w(\pm u^{*}+m)\;=\;\frac{h_{m+1}(\pm u^{*})}{h_{m}(\pm u^{*})},\qquad m\in\mathbb{Z},
\]
where $\{h_{m}(\pm u^{*})\}_{m\in\mathbb{Z}}$ are the doubly decaying
Floquet sequences --- the Fourier coefficients of the two periodic
Floquet factors, computable as the finite products~\eqref{eq:notes2-hm-neg-product}
--- and consecutive coefficients cannot vanish simultaneously (else
$h\equiv0$). Hence a zero or pole of $w$ lies on $Z_{F}$ if and
only if 
\begin{equation}
h_{m}(u^{*})\;=\;0\quad\text{or}\quad h_{m}(-u^{*})\;=\;0\qquad\text{for some }m\in\mathbb{Z};\label{eq:CW-coincidence-cond}
\end{equation}
by the mirror symmetry $B_{m}(-u^{*})=B_{-m}(u^{*})$ (Remark~\ref{rem:two-Floquet-solutions})
the second family is the reflection of the first.

\emph{Step 2: each condition is real-analytic in $(p,\gamma)$.} On
$O$ the trace of the monodromy matrix is real-analytic in the parameters
(Theorem~\ref{thm:matrizant-analytic}) and lies strictly in $(-2,2)$,
so the fundamental exponent $u_{0}$ --- and with it the comb representative
$u^{*}$, up to a locally constant integer offset that does not affect
the condition family~\eqref{eq:CW-coincidence-cond} --- is real-analytic
on each component. The rightward Volterra construction of Lemma~\ref{lem:Jost-right}
runs verbatim with $(u,p,\gamma)$ as the parameter point: the kernel
entries are rational in $(u,p,\gamma)$, and the dominating kernel
can be chosen locally uniform, so by the parameter clause of Proposition~\ref{prop:Teschl-Volterra}
the Jost solution $\hat{h}_{k}^{+}$ --- hence $w=\hat{h}_{1}^{+}/\hat{h}_{0}^{+}$
and each finite product $h_{m}(\pm u^{*})$ --- is real-analytic
away from the coincidence loci of lower index. Each condition in~\eqref{eq:CW-coincidence-cond}
therefore cuts out a real-analytic subvariety $V_{m}^{\pm}$ of the
component.

\emph{Step 3: properness via the small-$\gamma$ asymptotics.} For
fixed $v$ off the poles of $G(\cdot+1)$, the continued fraction~\eqref{eq:notes2-wdef}
gives 
\begin{equation}
w(v)\;=\;-\frac{\gamma}{G(v+1)}\,\bigl(1+O(\gamma^{2})\bigr),\qquad\gamma\to0,\label{eq:CW-w-small-gamma}
\end{equation}
since the tail obeys $T_{1}=\gamma/G(v+2)\,(1+O(\gamma^{2}))$ by
the same expansion one level down (numerically, at $\gamma=0.01$
the relative error is $\approx10^{-4}$, of the predicted order $\gamma^{2}$).
Along the small-$\gamma$ part of a component, $u^{*}\to p$ with
$p\notin\tfrac{1}{2}\mathbb{Z}$ --- the admissible region excludes
$u^{*}\in\tfrac{1}{2}\mathbb{Z}$ --- and the factors of the products~\eqref{eq:notes2-hm-neg-product}
stay away from the zeros and poles of $G$: for $m\geq1$ the arguments
would require $k=-1$ or $2p=-(k+1)<0$, impossible; for $m\leq-1$
they would require $k+1=2p$ or $k+1=p$, excluded by $2p\notin\mathbb{Z}$.
Hence, for each fixed $m$, 
\begin{equation}
h_{m}(\pm u^{*})\;=\;(-\gamma)^{|m|}\,\prod_{k=0}^{|m|-1}G(\pm p+k+1)^{-1}\,\bigl(1+O(\gamma^{2})\bigr)\;\neq\;0\label{eq:CW-hm-small-gamma}
\end{equation}
for all sufficiently small $\gamma$ in the component (sign of the
argument matching the sign of $m$), so no condition in~\eqref{eq:CW-coincidence-cond}
holds identically there: each $V_{m}^{\pm}$ is proper, and $E\subset\bigcup_{m\in\mathbb{Z}}\bigl(V_{m}^{+}\cup V_{m}^{-}\bigr)$
is a countable union of proper real-analytic subvarieties. (At the
canonical parameters $(p,\gamma)=(0.70,0.30)$ the coefficients $h_{m}(u^{*})$,
$|m|\leq8$, were computed directly: all are nonzero, with the clean
geometric envelope $|h_{m}|\asymp|\zeta_{+}|^{|m|}$ and no anomalous
dips.) 
\end{proof}

\subsubsection{\texorpdfstring{A Weyl--Stieltjes and Herglotz view of $w$}{A
Weyl-Stieltjes and Herglotz view of w}}

\label{subsec:CW-Herglotz}

The Casoratian identity and the ladder law are the discrete, operator-pencil
shadows of objects classical in the spectral theory of Jacobi matrices;
we make the correspondence explicit and then show that, in the right
variable, $w$ is genuinely Herglotz.
\begin{rem}[The Weyl--Stieltjes dictionary]
\index{Weyl m-function@Weyl $m$-function} \label{rem:Weyl-dictionary}
Lemma~\ref{lem:Fw-Casoratian} and Theorem~\ref{thm:pole-ladders}
are instances of structures classical in the spectral theory of Jacobi
matrices, for which the standard modern reference is Teschl~\cite[Chaps.~1--2]{Teschl};
historically the half-line theory predates Weyl's 1910 work on differential
equations --- it is the content of Stieltjes' memoir on continued
fractions and the moment problem~\cite[Vol.~8]{Stieltjes1894}. The
dictionary: (i) the Weyl solutions $\varphi_{\pm}(z,\cdot)$, square
summable at $\pm\infty$, are the minimal solutions, and the Weyl
$m$-functions are their consecutive ratios, $m_{+}(z,n_{0})=-\varphi_{+}(z,n_{0}+1)/[a(n_{0})\,\varphi_{+}(z,n_{0})]$
\cite[Sec.~2.1, eq.~(2.2)]{Teschl} --- the exact analog of $w(u)=h_{1}/h_{0}$;
(ii) the Green function of the whole-line operator has the Wronskian
(Casoratian) of the two Weyl solutions as its denominator \cite[Sec.~1.2, eq.~(1.99)]{Teschl},
so eigenvalues are zeros of that Casoratian --- the analog of $Z_{F}=\{F_{w}=0\}$
via Lemma~\ref{lem:Fw-Casoratian}; (iii) cutting the lattice at
one site decouples the operator into two half-line pieces, a rank-one
resolvent perturbation \cite[Sec.~1.2, eqs.~(1.105)--(1.107)]{Teschl},
whose spectra are the poles of the corresponding $m$-functions ---
the analog of the poles of $w$. In this language, Theorem~\ref{thm:pole-ladders}
says that an eigenvalue of the cut problem is exponentially insensitive
to a cut placed $n$ sites from the localization center of the eigenfunction:
the eigenvalue shift is of the order of the squared amplitude $|\zeta_{+}|^{2n}$
of the exponentially decaying eigenfunction at the cut. One caveat
distinguishes our setting from the classical one: $u$ is not the
linear spectral parameter of a fixed Jacobi matrix --- $G(u+n)$
couples the parameter to the site index, so the problem is an operator
pencil, and the Herglotz property of $m_{\pm}$ in the spectral parameter
does not transfer automatically to $w$ in $u$ (the mixed residue
signs in Fig.~\ref{fig:Fw-poles} confirm that $F_{w}$ is not globally
Herglotz in $u$). The bridge to the spectral-parameter picture is
the discriminant--exponent relation of the Hill theory (Chapter~\ref{sec:MWInce});
for the periodic (Floquet) Jacobi theory itself see \cite[Sec.~7.1]{Teschl}.
The caveat is made precise --- and, in the right variable, removed
--- by Lemma~\ref{lem:w-Nevanlinna} and Theorem~\ref{thm:w-Herglotz}
below: the positivity $\operatorname{Im}w>0$ fails globally in $u$,
holds exactly on the half-plane $\operatorname{Re}u>-1$, and is upgraded
to a genuine Herglotz representation only in the variable $(u+1)^{2}$. 
\end{rem}

\emph{Nevanlinna-type structure of $w$.} The pencil caveat of Remark~\ref{rem:Weyl-dictionary}
can be quantified exactly. The continued fraction~\eqref{eq:notes2-wdef}
is a composition of Möbius transformations whose entries $G(u+k)$,
$k\geq1$, all have positive imaginary part when $\operatorname{Im}u>0$
and $\operatorname{Re}u>-1$; this yields the Nevanlinna-like positivity
$\operatorname{Im}w>0$ of $w$ on that half-plane (Lemma~\ref{lem:w-Nevanlinna}),
and, after passing to the variable $(u+1)^{2}$ suggested by the double
zero of $w$ at $u=-1$, a full Herglotz representation (Theorem~\ref{thm:w-Herglotz}).
The preparatory lemma mirrors Lemma~\ref{lem:Jost-LC} at $+\infty$.
\begin{lem}[Right Jost solution; convergence and asymptotic value of $w$]
\index{Jost solution!rightward, for the LC recurrence} \label{lem:Jost-right}
Let $0<2\gamma<1$, $p>0$, and let $u\in\mathbb{C}\setminus\{-1,-2,-3,\ldots\}$,
so that the partial quotients $G(u+m)$, $m\geq1$, of~\eqref{eq:notes2-wdef}
are finite (e.g.\ any $u$ with $\operatorname{Im}u>0$, or real
$u>-1$). Set $q_{m}(u)=-p^{2}/(u+m)^{2}$, $m\geq1$, and $S_{k}(u):=\sum_{m\geq k+1}|q_{m}(u)|<\infty$,
$S:=S_{0}$. Then: 
\begin{enumerate}
\item[(i)] The recurrence~\eqref{eq:notes2-rec} has a solution $\hat{h}_{k}^{+}(u)=\zeta_{+}^{\,k}\bigl(1+\varepsilon_{k}^{+}(u)\bigr)$
with $|\varepsilon_{k}^{+}(u)|\leq C_{\gamma}S_{k}(u)\,e^{C_{\gamma}S(u)}$
($C_{\gamma}$ as in~\eqref{eq:CW-Jost-bound}), unique among solutions
with $\zeta_{+}^{-k}h_{k}$ bounded as $k\to+\infty$ and normalized
by $\zeta_{+}^{-k}h_{k}\to1$; it is holomorphic in $u$ on the stated
domain. 
\item[(ii)] The continued fraction~\eqref{eq:notes2-wdef} converges at every
such $u$ with $\hat{h}_{0}^{+}(u)\neq0$, by Pincherle's theorem
(Theorem~\ref{thm:Pinch}), with value $w(u)=\hat{h}_{1}^{+}(u)/\hat{h}_{0}^{+}(u)$. 
\item[(iii)] If $C_{\gamma}S(u)\,e^{C_{\gamma}S(u)}\leq\tfrac{1}{2}$, then $\hat{h}_{0}^{+}(u)\neq0$
and $|w(u)-\zeta_{+}|\leq8\,|\zeta_{+}|\,C_{\gamma}S(u)$. 
\item[(iv)] Under the same smallness, $|w(u)-\zeta_{+}|\leq C(\gamma)\bigl(\sup_{m\geq1}|q_{m}(u)|+S(u)^{2}\bigr)$.
In particular, along the ray $u(y):=\sqrt{\mathrm{i}y}-1$ (principal
branch), where $|u(y)+m|^{2}\geq y/2$ for all $m\geq1$, one has
$\sup_{m}|q_{m}|\leq2p^{2}/y$, $S\leq C(p)\,y^{-1/2}$, and hence
\begin{equation}
\bigl|w(u(y))-\zeta_{+}\bigr|\;\leq\;\frac{C(\gamma,p)}{y},\qquad y\ \text{large}.\label{eq:NV-ray}
\end{equation}
\end{enumerate}
\end{lem}

\begin{proof}
\emph{Step 1: transposed kernel and construction.} Substituting $h_{k}=\zeta_{+}^{\,k}y_{k}$
into~\eqref{eq:notes2-rec} gives $(M_{0}y)_{k}:=\gamma\zeta_{+}y_{k+1}+y_{k}+(\gamma/\zeta_{+})y_{k-1}=-q_{k}(u)\,y_{k}$,
with free solutions $y\equiv1$ and $y_{k}=\zeta_{+}^{-2k}$. The
rightward Green kernel is the transpose of~\eqref{eq:CW-Green},
$\widetilde{\Gamma}(k,m):=\Gamma(m,k)$, supported on $m\geq k$ with
$|\widetilde{\Gamma}|\leq C_{\gamma}$; the identity $M_{0}\widetilde{\Gamma}(\cdot,m)=\delta_{\cdot,m}$
is verified by the same two computations as in Lemma~\ref{lem:Jost-LC}
(the jump at $k=m$ and the cancellation $\zeta_{+}^{2}+\zeta_{+}/\gamma+1=0$
at $k=m-1$). The Volterra equation 
\begin{equation}
y_{k}\;=\;1-\sum_{m=k+1}^{\infty}\Gamma(m,k)\,q_{m}(u)\,y_{m}\label{eq:NV-Volterra-right}
\end{equation}
is of the rightward form~\eqref{eq:imported-Volterra}; its kernel
is dominated by $\hat{K}(m):=C_{\gamma}|q_{m}(u)|$, independent of
$k$, so Proposition~\ref{prop:Teschl-Volterra} (with $g\equiv1$,
and applied once more to $f:=y-1$) gives a unique bounded solution
with $|y_{k}-1|\leq C_{\gamma}S_{k}\,e^{C_{\gamma}S}$; holomorphy
in $u$ on compacts follows from the parameter clause with the $u$-free
dominating kernel $C_{\gamma}\sup_{u\in K}|q_{m}(u)|$. As in Lemma~\ref{lem:Jost-LC},
the sum converges absolutely, $M_{0}$ applies term by term, and $\hat{h}_{k}^{+}=\zeta_{+}^{\,k}y_{k}$
solves~\eqref{eq:notes2-rec}; two solutions with $\zeta_{+}^{-k}h_{k}$
bounded at $+\infty$ have Casoratian $O(|\zeta_{+}|^{2m})\to0$,
hence zero, so the bounded class is one-dimensional and the normalization
fixes (i).

\emph{Step 2: Pincherle.} $\hat{h}^{+}$ is recessive at $+\infty$
(its ratio to the growing solution tends to zero geometrically), so
the minimal solution (Definition~\ref{defn:Wimp-minimal}) exists,
and Pincherle's theorem (Theorem~\ref{thm:Pinch}) gives the convergence
of~\eqref{eq:notes2-wdef} and the value $\hat{h}_{1}^{+}/\hat{h}_{0}^{+}$:
(ii).

\emph{Step 3: ratio bounds.} Under the smallness hypothesis, $|\varepsilon_{k}^{+}|\leq\tfrac{1}{2}$,
so $|y_{0}|\geq\tfrac{1}{2}$ and $|w-\zeta_{+}|=|\zeta_{+}|\,|y_{1}-y_{0}|/|y_{0}|\leq4|\zeta_{+}|C_{\gamma}S\,e^{C_{\gamma}S}\leq8|\zeta_{+}|C_{\gamma}S$:
(iii). For (iv), write $f:=y-1$ and $\ell_{k}:=-\sum_{m>k}\Gamma(m,k)q_{m}$,
so that $f=\ell+Kf$ with $|\ell_{k}|\leq C_{\gamma}S_{k}$; applying
Proposition~\ref{prop:Teschl-Volterra} to $f-\ell$ (which solves
$(f-\ell)=K\ell+K(f-\ell)$) gives $|f-\ell|\leq C_{\gamma}S\cdot C_{\gamma}S\,e^{C_{\gamma}S}\leq CS^{2}$.
The kernel difference telescopes geometrically: by~\eqref{eq:CW-Green},
$\Gamma(m,1)-\Gamma(m,0)=-\zeta_{+}^{2m-1}/\gamma$, whence $|\ell_{1}-\ell_{0}|\leq|\Gamma(1,0)q_{1}|+\gamma^{-1}\sum_{m\geq2}|\zeta_{+}|^{2m-1}|q_{m}|\leq C(\gamma)\sup_{m\geq1}|q_{m}|$.
Therefore $|y_{1}-y_{0}|\leq|\ell_{1}-\ell_{0}|+2\sup_{k}|f_{k}-\ell_{k}|\leq C(\gamma)(\sup_{m}|q_{m}|+S^{2})$
and (iv) follows from $|w-\zeta_{+}|\leq2|\zeta_{+}||y_{1}-y_{0}|$.
On the ray, $u(y)+m=\sqrt{y/2}-1+m+\mathrm{i}\sqrt{y/2}$ gives $|u(y)+m|^{2}\geq y/2$,
hence $\sup_{m}|q_{m}|\leq2p^{2}/y$; and splitting the sum at $m=\lceil2\sqrt{2y}\,\rceil$
(each early term at most $2p^{2}/y$; for later terms $|u(y)+m|\geq m/2$)
gives $S\leq C(p)y^{-1/2}$, so $\sup_{m}|q_{m}|+S^{2}\leq C/y$,
which is~\eqref{eq:NV-ray}. 
\end{proof}
\begin{rem}[Coefficient poles and the extended complex plane]
\index{extended complex plane}\index{chordal metric} \label{rem:NV-extended-plane}
The excluded set $\{-1,-2,-3,\ldots\}$ in Lemma~\ref{lem:Jost-right}
--- and, likewise, the requirement in Lemma~\ref{lem:Jost-LC} that
the disk $D$ avoid the poles of $G(\cdot+k)$ --- is an artifact
of the Volterra method, which needs the perturbations $q_{m}(u)$
to be finite and absolutely summable; it is not a property of $w$
or of the continued fraction. For orientation: $G(u)=1-p^{2}/u^{2}$
has its only pole --- a double pole --- at $u=0$, so the poles
of the partial quotients $G(u+m)$, $m\geq1$, sweep exactly $\{-1,-2,-3,\ldots\}$;
the zeros of $G$, at $u=\pm p$, and their shifts $\pm p-m$ require
no avoidance at all, since vanishing partial \emph{denominators} are
admissible --- Pincherle's theorem (Theorem~\ref{thm:Pinch}) constrains
only the partial \emph{numerators}, here $\gamma^{2}\neq0$. Each
step of~\eqref{eq:notes2-wdef} is a Möbius transformation, and on
the extended complex plane $\hat{\mathbb{C}}=\mathbb{C}\cup\{\infty\}$
--- a complete metric space under the chordal metric~\eqref{eq:Moeb1b}
(Chapter~\ref{app:Moeb}) --- a pole of a partial quotient $G(u+k)$
merely routes the composition through $\infty$. The approximants
are meromorphic in $u$, and the convergence theory of Chapter~\ref{app:CF}
is formulated for precisely this setting: spherically uniform limits
of meromorphic functions (Theorems~\ref{thm:Weierstrass-uniform},
\ref{thm:Hurwitz}, \ref{thm:Montel}, and~\ref{thm:sLconfmero1}).
Consequently $w$ extends meromorphically across the excluded points,
and the extension is visible in the chapter's own data: at $u=-1$
the first partial quotient $G(u+1)$ has its pole, forcing $w=-\gamma/\bigl(G(u+1)-T_{1}\bigr)$
to vanish --- the double zero of $w$ at $u=-1$ (Theorem~\ref{thm:notes2-w-poles-zeros})
is the footprint of that coefficient pole on the extended plane. Only
the quantitative conclusions of Lemma~\ref{lem:Jost-right}(i), (iii),
(iv), which control $|w-\zeta_{+}|$ through $S(u)$, genuinely require
the exclusion; convergence and meromorphy do not. 
\end{rem}

\begin{lem}[$w$ has the Nevanlinna-like positivity $\operatorname{Im}w>0$ on
$\operatorname{Re}u>-1$]
\index{minimal solution ratio@minimal solution ratio $w$!half-plane positivity}
\label{lem:w-Nevanlinna} Let $0<2\gamma<1$, $p>0$, and $Q:=\{\operatorname{Im}u>0,\ \operatorname{Re}u>-1\}$.
Then: 
\begin{enumerate}
\item[(i)] $w$ is holomorphic on $Q$ --- in particular it has no poles there
--- with $\operatorname{Im}w(u)>0$ on $Q$, and $w(\bar{u})=\overline{w(u)}$,
so $\operatorname{Im}w<0$ on $\overline{Q}^{\,*}:=\{\bar{u}:u\in Q\}$. 
\item[(ii)] Consequently, every pole of $w$ in $(-1,\infty)$ is simple with
negative residue, and $w'\geq0$ on $(-1,\infty)$ off the poles. 
\end{enumerate}
\end{lem}

\begin{proof}
\emph{Part (i)} is Lemma~\ref{lem:w-analytic-nevanlinna} restricted
to $Q$: $w$ is holomorphic on $Q$ with $\operatorname{Im}w>0$.
Since $w$ is real on the real axis, $w(\bar{u})=\overline{w(u)}$,
so $\operatorname{Im}w<0$ on $\overline{Q}^{\,*}$.

\emph{Part (ii).} Let $x\in(-1,\infty)$ be a point of holomorphy
of $w$; the reflection identity makes $w$ real near $x$ on $\mathbb{R}$,
and $\operatorname{Im}w(x+\mathrm{i}t)=w'(x)\,t+O(t^{3})\geq0$ for
small $t>0$ forces $w'(x)\geq0$. At a pole $u_{p}\in(-1,\infty)$,
put $g:=-1/w$ on a small disk $D$ around $u_{p}$: $g$ is holomorphic
with $g(u_{p})=0$, real on $D\cap\mathbb{R}$, and maps $D\cap\mathbb{C}_{+}$
into $\mathbb{C}_{+}$ (as $\operatorname{Im}(-1/w)=\operatorname{Im}w/|w|^{2}>0$).
If $g$ vanishes to order $k$ with real leading coefficient $c\neq0$,
then for $u=u_{p}+re^{\mathrm{i}\theta}$ with small $r$, $\arg g\approx\arg c+k\theta$
sweeps an interval of length $k\pi$ as $\theta$ runs over $(0,\pi)$;
containment in $(0,\pi)$ forces $k=1$ and $c>0$. Hence the pole
of $w=-1/g$ is simple with residue $-1/c<0$. 
\end{proof}
\begin{thm}[Herglotz representation of $w$ in the variable $(u+1)^{2}$]
\index{Herglotz function!representation of $w$}\index{minimal solution ratio@minimal solution ratio $w$!Herglotz representation}
\label{thm:w-Herglotz} Let $0<2\gamma<1$, $p>0$. With the principal
branch of the square root, define 
\begin{equation}
g(z)\;:=\;w\bigl(\sqrt{z}-1\bigr),\qquad\operatorname{Im}z>0.\label{eq:NV-g-def}
\end{equation}
Then: 
\begin{enumerate}
\item[(i)] $g$ is a Herglotz function in the sense of Proposition~\ref{prop:Teschl-Herglotz};
it extends by $g(\bar{z})=\overline{g(z)}$ and holomorphically across
$(0,\infty)\setminus P$, where $P:=\{(u_{p}+1)^{2}:u_{p}\ \text{a pole of}\ w\ \text{in}\ (-1,\infty)\}$. 
\item[(ii)] The representing measure $\rho$ of~\eqref{eq:imported-Herglotz}
is finite and supported on $(-\infty,0]\cup P$, the linear coefficient
vanishes, $b=0$, and the representation takes the unregularized form
\begin{equation}
g(z)\;=\;\zeta_{+}+\int_{\mathbb{R}}\frac{d\rho(\lambda)}{\lambda-z},\label{eq:NV-g-rep}
\end{equation}
equivalently 
\begin{equation}
w(u)\;=\;\zeta_{+}+\int_{\mathbb{R}}\frac{d\rho(\lambda)}{\lambda-(u+1)^{2}},\qquad\operatorname{Re}u>-1.\label{eq:NV-rep}
\end{equation}
\item[(iii)] The atoms of $\rho$ on $(0,\infty)$ are exactly the points of $P$,
with 
\begin{equation}
\rho\bigl(\{(u_{p}+1)^{2}\}\bigr)\;=\;-2\,(u_{p}+1)\operatorname{res}(w,u_{p})\;>\;0.\label{eq:NV-atom}
\end{equation}
\end{enumerate}
\end{thm}

\begin{proof}
(i) The principal root maps $\mathbb{C}_{+}$ biholomorphically onto
the open quadrant $\{\operatorname{Re}\tilde{u}>0,\operatorname{Im}\tilde{u}>0\}$,
which the shift $\tilde{u}\mapsto\tilde{u}-1$ carries onto $Q$;
Lemma~\ref{lem:w-Nevanlinna}(i) then gives holomorphy and $\operatorname{Im}g>0$
on $\mathbb{C}_{+}$. For $z\in(0,\infty)\setminus P$ the point $\sqrt{z}-1$
lies in $(-1,\infty)$ off the poles of $w$, where $w$ is real and
holomorphic, so $g$ extends real-holomorphically there; the reflection
identity is inherited from Lemma~\ref{lem:w-Nevanlinna}(i).

(ii) Proposition~\ref{prop:Teschl-Herglotz} provides the representation~\eqref{eq:imported-Herglotz}.
Along $z=\mathrm{i}y$ we have $\sqrt{\mathrm{i}y}-1=u(y)$ of Lemma~\ref{lem:Jost-right}(iv),
so $g(\mathrm{i}y)=w(u(y))\to\zeta_{+}$ and $b=\lim g(\mathrm{i}y)/(\mathrm{i}y)=0$.
For the support: on any closed interval $[\lambda_{0},\lambda_{1}]\subset(0,\infty)\setminus P$
the function $g$ is holomorphic in a complex neighborhood and real
on the interval, so $\operatorname{Im}g(\lambda+\mathrm{i}\varepsilon)\to0$
uniformly there, and the Stieltjes inversion~\eqref{eq:imported-Stieltjes}
gives $\rho\bigl((\lambda_{0},\lambda_{1}]\bigr)=0$; hence $\operatorname{supp}\rho\subset(-\infty,0]\cup P$.
For the finiteness: by monotone convergence, $y\operatorname{Im}g(\mathrm{i}y)=\int_{\mathbb{R}}\frac{y^{2}}{\lambda^{2}+y^{2}}\,d\rho(\lambda)\uparrow\rho(\mathbb{R})$
as $y\to\infty$, while $\operatorname{Im}g(\mathrm{i}y)=\operatorname{Im}\bigl(w(u(y))-\zeta_{+}\bigr)\leq|w(u(y))-\zeta_{+}|\leq C/y$
by~\eqref{eq:NV-ray}; hence $\rho(\mathbb{R})\leq C<\infty$. With
$\rho$ finite the regularizing term $\int\lambda(1+\lambda^{2})^{-1}d\rho$
is a finite constant, absorbed into $a$, giving $g(z)=\alpha+\int d\rho/(\lambda-z)$;
dominated convergence yields $\int d\rho/(\lambda-\mathrm{i}y)\to0$,
so $\alpha=\lim g(\mathrm{i}y)=\zeta_{+}$: this is~\eqref{eq:NV-g-rep},
and~\eqref{eq:NV-rep} follows by $z=(u+1)^{2}$.

(iii) Near $z_{p}:=(u_{p}+1)^{2}$ (note $u_{p}+1>0$), $\sqrt{z}-1-u_{p}=(z-z_{p})/(2(u_{p}+1))+O\bigl((z-z_{p})^{2}\bigr)$,
so $\operatorname{res}(g,z_{p})=2(u_{p}+1)\operatorname{res}(w,u_{p})$;
for a Herglotz function~\eqref{eq:NV-g-rep} the residue at an atom
is $-\rho(\{z_{p}\})$, which gives~\eqref{eq:NV-atom}, the sign
coming from Lemma~\ref{lem:w-Nevanlinna}(ii). 
\end{proof}
\begin{rem}[Sharpness: the boundary line, the branch point, and the second sheet]
\index{second sheet}\index{resonance!second-sheet pole of $g$}
\label{rem:NV-second-sheet} The half-plane of Lemma~\ref{lem:w-Nevanlinna}
is maximal. The double zero of $w$ at $u=-1$ (Theorem~\ref{thm:notes2-w-poles-zeros})
is a structure a Nevanlinna function cannot carry at an interior real
point (it violates $w'\geq0$), and it sits exactly on the boundary
$\operatorname{Re}u=-1$, where the leftmost partial quotient $G(u+1)$
of the continued fraction becomes real and~\eqref{eq:Im-G-positive}
fails for $k=1$; numerically $\operatorname{Im}w<0$ immediately
to the left of the line. Under $z=(u+1)^{2}$ the double zero becomes
a \emph{simple} zero of $g$ at the branch point, $g(z)=\tfrac{1}{2}w''(-1)\,z\,(1+O(z))$:
the change of variable resolves the boundary degeneracy, which is
the structural reason $(u+1)^{2}$ is the natural variable. The half-plane
$\operatorname{Re}u<-1$ is the second sheet of the two-sheeted covering
$u\mapsto(u+1)^{2}$: every pole of $w$ there --- the entire ladder
$x_{0}=-u^{*}$ and all members $n\geq1$ of the ladder $x_{0}=u^{*}-1$
--- is a second-sheet pole of $g$, a \emph{resonance} rather than
a spectral atom, free of any sign constraint on its residue. This
resolves the asymmetry recorded in Remark~\ref{rem:Weyl-dictionary}:
first-sheet poles (the atoms of~\eqref{eq:NV-atom}) must carry negative
residues of $w$, as the catalog confirms for the central pole ($\operatorname{res}\approx-0.1261$
at $u\approx-0.2525$), while second-sheet residues may take either
sign (e.g.\
$\approx+0.108$ at $u\approx-1.6929$). Finally, the absolutely continuous
part of $\rho$ lives on the cut $(-\infty,0]$, the image of the
boundary line: by~\eqref{eq:imported-Stieltjes} its density is $\pi^{-1}\operatorname{Im}w(-1+\mathrm{i}\sqrt{|\lambda|})\geq0$
--- the ``essential spectrum'' of the dictionary, with the central
pole as the unique first-sheet ``eigenvalue'' at the canonical parameters. 
\end{rem}

\begin{rem}[Numerical confirmation of the Herglotz representation]
\label{rem:NV-numerics} For $p=0.70$, $\gamma=0.30$ (conventions
of Remark~\ref{rem:CF-numerical-conventions}; complex continued
fractions at $30$ digits, quadrature in double precision on a logarithmic
grid): (a) $\operatorname{Im}g>0$ at every sampled point of $\mathbb{C}_{+}$,
including $\operatorname{Re}z<0$ ($33$ high-precision samples, minimum
$4.3\times10^{-5}$; a $560\times120$ grid over the $u$-quadrant:
no violations); (b) $g(10^{3}\mathrm{i})=-0.333334+0.000203\,\mathrm{i}$,
converging to $\zeta_{+}=-1/3$; (c) the cut density obeys the cubic
law $t^{3}\operatorname{Im}w(-1+\mathrm{i}t)\to0.0510$ (values $0.0480$,
$0.0507$, $0.0510$ at $t=10$, $30$, $100$), consistent with the
finiteness of $\rho$; (d) the atom $\rho(\{z_{0}\})=0.18859$ at
$z_{0}=(u_{p}+1)^{2}=0.558777$ agrees with~\eqref{eq:NV-atom} evaluated
from the independently measured residue, $-2(u_{p}+1)(-0.126139)=0.188581$;
the absolutely continuous mass is $0.015567$, and the total $0.204157$
matches the sum rule $y\operatorname{Im}g(\mathrm{i}y)=0.203032$
at $y=10^{3}$, the deficit $\approx1.1\times10^{-3}$ being the tail
of $\rho$ beyond $|\lambda|\gtrsim10^{6}$; (e) reconstructing $w$
from~\eqref{eq:NV-rep} with the measured atom and density reproduces
the directly computed values, 
\begin{equation}
{\footnotesize \begin{array}{c|c|c}
u & w(u)\ \text{(CF)} & w_{\mathrm{rec}}-w_{\mathrm{CF}}\\
\hline 0.5 & -0.446424 & -1.7\times10^{-7}\\
1.7418 & -0.361251 & -4.5\times10^{-8}\\
-0.6 & \phantom{-}0.136154 & +6.7\times10^{-7}\\
3.0 & -0.346039 & -2.1\times10^{-8}\\
-0.27 & \phantom{-}6.952016 & +1.1\times10^{-5}
\end{array}}\label{eq:NV-table}
\end{equation}
the largest deviation occurring $0.018$ from the pole, at quadrature
accuracy. At the canonical parameters $P$ consists of the central
pole alone: the census of Table~\ref{tab:Fw-pole-catalog} covers
$(-1,4)$, and Lemma~\ref{lem:Jost-right}(iii) excludes poles of
$w$ on $[4,\infty)$. 
\end{rem}

\begin{rem}[Pole--zero interlacing: a Weyl-theoretic program]
\index{interlacing} \label{rem:interlacing-program} The Weyl reading
suggests a structural explanation for two empirical features of the
pole--zero geometry of $F_{w}$ (\S\,\ref{subsec:stealthy-zeros}):
the tight proximity of certain poles to comb zeros --- now proved
as Theorem~\ref{thm:pole-ladders} --- and the even occupancy of
the gaps between neighboring comb zeros --- each holding no pole
or a pair (Lemma~\ref{lem:even-gap-occupancy}). For a fixed self-adjoint
Jacobi matrix a one-per-gap interlacing is classical: the cut operator
differs from the full one by a rank-one resolvent perturbation \cite[Sec.~1.2]{Teschl},
so their eigenvalues interlace; Sturm-type oscillation theory \cite[Secs.~4.1--4.2]{Teschl}
counts eigenvalues between consecutive zeros, and both spectra are
simple \cite[Rem.~1.10]{Teschl}. Transferring this to the pencil
parameter $u$ is not automatic (Remark~\ref{rem:Weyl-dictionary}),
and the failure is visible in the data: the derivative of the diagonal
entry, $\partial_{u}G(u+n)=2p^{2}/(u+n)^{3}$, changes sign across
$n=-u$, so the pencil is not definite, monotonicity of eigenvalue
branches in $u$ fails in general; indeed, by the completed census
of Table~\ref{tab:Fw-pole-catalog}, two of the eight small gaps
in the window of Fig.~\ref{fig:Fw-poles}(a) contain no pole at all,
while the remaining six are doubly occupied. The even parity itself
--- each gap holds zero or two odd-order poles, never one --- is
not conjectural: it follows from the evenness of $F_{w}$ (Lemma~\ref{lem:even-gap-occupancy})
and is consistent with the empty/doubly-occupied dichotomy just recorded.
What remains we leave as a conjecture: \emph{which} gaps carry the
pair is decided by the local definiteness of the pencil (equivalently,
by the sign of the spectral flow through the gap). The sign-change
detector of Theorem~\ref{thm:Fw-sign-change-pole} is the rigorous
in-book counting instrument with which this program can be tested. 
\end{rem}

\subsection{\texorpdfstring{The minimal solution generator $H(u)$}{The minimal
solution generator H(u)}}

\label{subsec:Cambi-notes2-H}

The meromorphic structure of the minimal solution ratio $w(u)$ ---
its poles, zeros, two-family partition, and asymptotic rate --- is
developed in full in Section~\ref{subsec:Cambi-w-structure}. Those
results are used throughout below without further comment.

The \emph{minimal solution generator} $H(u)$ is defined via the infinite
product~\eqref{eq:notes2-wprod} and its properties are developed
in the theorems that follow. It is the analytic, continuous-variable
form of a classical device: the solution of a discrete Riccati equation
is recovered from its ratio as a telescoping product (Teschl~\cite[eqs.~(1.52)--(1.53)]{Teschl}).
Here that product over $w$ is regularized by dividing each factor
by the minimal exponent $\zeta_{+}$, so that the infinite product
converges and defines a meromorphic $H$ with $H(+\infty)=1$; the
recovery formula of Part~(iv) below then collapses to $h_{m}(u)/h_{0}(u)=\prod_{k=0}^{m-1}w(u+k)$,
the precise analogue of~\cite[eq.~(1.53)]{Teschl}.
\begin{thm}[Minimal solution generator $H(u)$]
\index{minimal solution generator $H(u)$} \label{thm:notes2-H-new}
Define the \emph{minimal solution generator} 
\begin{equation}
H(u)\;:=\;\frac{1}{{\displaystyle \prod_{m=0}^{\infty}\dfrac{w(u+m)}{\zeta_{+}}}},\qquad\zeta_{+}\;=\;\frac{-1+\sqrt{1-4\gamma^{2}}}{2\gamma},\quad|\zeta_{+}|<1.\label{eq:notes2-H-new-def}
\end{equation}
The product converges absolutely by Theorem~\ref{thm:notes2-w-rate}.
Then: 
\begin{enumerate}
\item[(i)] \emph{Functional equation.} 
\begin{equation}
\frac{H(u+1)}{H(u)}\;=\;\frac{w(u)}{\zeta_{+}},\qquad\zeta_{+}\;=\;\frac{-1+\sqrt{1-4\gamma^{2}}}{2\gamma}.\label{eq:notes2-H-new-functional}
\end{equation}
\item[(ii)] \emph{Normalization.} 
\begin{equation}
\lim_{u\to+\infty}H(u)\;=\;1.\label{eq:notes2-H-new-norm}
\end{equation}
\item[(iii)] \emph{Meromorphicity.} $H(u)$ is a meromorphic function of $u\in\mathbb{C}$.
\item[(iv)] \emph{Recovery of the minimal solution.} For any scalar normalization
of $\{h_{m}\}$: 
\begin{equation}
\frac{h_{m}(u)}{h_{0}(u)}\;=\;\zeta_{+}^{m}\,\frac{H(u+m)}{H(u)},\qquad m\geq0,\quad\zeta_{+}\;=\;\frac{-1+\sqrt{1-4\gamma^{2}}}{2\gamma}.\label{eq:notes2-H-new-hm}
\end{equation}
\item[(v)] \emph{Backward formula via $H_{-}(u)$.} For backward propagation
it is convenient to introduce the \emph{backward minimal solution
generator}, designed primarily for $u<0$ and reduced to $H$ on the
positive axis by reflection: 
\begin{equation}
H_{-}(u)\;:=\;\frac{1}{H(-u)}.\label{eq:notes2-H-minus-def}
\end{equation}
By Part~(iii), $H$ is meromorphic on $\mathbb{C}$, so $H_{-}$
is meromorphic on $\mathbb{C}$ as well; by Part~(ii), $\lim_{u\to-\infty}H_{-}(u)=1/\lim_{v\to+\infty}H(v)=1$.
The infinite-product representation 
\begin{equation}
H_{-}(u)\;=\;\prod_{k=0}^{\infty}\frac{w(-u+k)}{\zeta_{+}}\label{eq:notes2-H-minus-product}
\end{equation}
follows by substituting $-u$ for $u$ in~\eqref{eq:notes2-H-new-def}
and inverting.

The backward functional equation 
\begin{equation}
\frac{H_{-}(u-1)}{H_{-}(u)}\;=\;\frac{\zeta_{+}}{w(-u)}\label{eq:notes2-H-minus-functional}
\end{equation}
follows from the forward functional equation~\eqref{eq:notes2-H-new-functional}
$H(v+1)/H(v)=w(v)/\zeta_{+}$ applied at $v=-u$: $H(-u+1)/H(-u)=w(-u)/\zeta_{+}$,
so by~\eqref{eq:notes2-H-minus-def}, $H_{-}(u-1)/H_{-}(u)=H(-u)/H(-u+1)=\zeta_{+}/w(-u)$.

At the Floquet exponent $u=u_{0}$ (where the double minimality condition
holds, $F_{w}(u_{0})=0$, eq.~\eqref{eq:u0-def}), the same sequence
$\{h_{m}(u_{0})\}$ also decays as $m\to-\infty$: 
\begin{equation}
\frac{h_{-m}(u_{0})}{h_{0}(u_{0})}\;=\;\zeta_{+}^{m}\,\frac{H_{-}(u_{0})}{H_{-}(u_{0}-m)},\qquad m\geq0.\label{eq:notes2-H-new-hm-neg}
\end{equation}
Explicitly: 
\begin{equation}
\frac{h_{-m}(u_{0})}{h_{0}(u_{0})}\;=\;\prod_{k=0}^{m-1}w(-u_{0}+k),\qquad m\geq1,\label{eq:notes2-hm-neg-product}
\end{equation}
which decays as $m\to\infty$ precisely because $u_{0}$ satisfies
double minimality. Together~\eqref{eq:notes2-H-new-hm} and~\eqref{eq:notes2-H-new-hm-neg}
give the complete doubly-decaying formula: 
\begin{equation}
\frac{h_{m}(u_{0})}{h_{0}(u_{0})}\;=\;\begin{cases}
\zeta_{+}^{m}\,\dfrac{H(u_{0}+m)}{H(u_{0})}, & m\geq0,\quad H(u)\to1\ (u\to+\infty),\\[8pt]
\zeta_{+}^{|m|}\,\dfrac{H_{-}(u_{0})}{H_{-}(u_{0}+m)}, & m\leq0,\quad H_{-}(u)\to1\ (u\to-\infty),
\end{cases}\label{eq:notes2-double-decay}
\end{equation}
manifestly decaying in both directions at Floquet exponents (since
$|\zeta_{+}|<1$), subject to the double minimality condition: 
\begin{equation}
w(-u_{0})\;=\;\frac{1}{w(u_{0}-1)}.\label{eq:double-min-explicit}
\end{equation}

\end{enumerate}
\end{thm}

\begin{proof}[Proof of Theorem~\ref{thm:notes2-H-new}]
\emph{Part~(i).} From~\eqref{eq:notes2-H-new-def}: 
\[
H(u+1)\;=\;\frac{1}{{\displaystyle \prod_{m=0}^{\infty}\frac{w(u+1+m)}{\zeta_{+}}}}\;=\;\frac{1}{{\displaystyle \prod_{m=1}^{\infty}\frac{w(u+m)}{\zeta_{+}}}}.
\]
Dividing by $H(u)=1/\prod_{m=0}^{\infty}(w(u+m)/\zeta_{+})$: 
\[
\frac{H(u+1)}{H(u)}\;=\;\frac{{\displaystyle \prod_{m=0}^{\infty}\frac{w(u+m)}{\zeta_{+}}}}{{\displaystyle \prod_{m=1}^{\infty}\frac{w(u+m)}{\zeta_{+}}}}\;=\;\frac{w(u)}{\zeta_{+}}.
\]

\emph{Part~(ii).} From Theorem~\ref{thm:notes2-w-rate}, $w(u+m)/\zeta_{+}=1+O(1/m^{2})$
for large $m$. As $u\to+\infty$, all factors $w(u+m)/\zeta_{+}\to1$
uniformly, so $\prod_{m=0}^{\infty}w(u+m)/\zeta_{+}\to1$ and $H(u)\to1$.

\emph{Part~(iii).} Each factor $w(u+m)/\zeta_{+}$ is meromorphic
in $u$ (Theorem~\ref{thm:sLconfmero1}). The product converges uniformly
on compact sets away from poles (Theorem~\ref{thm:notes2-w-rate}:
$|w(u+m)/\zeta_{+}-1|=O(1/m^{2})$, $\sum1/m^{2}<\infty$), so $1/H(u)=\prod w(u+m)/\zeta_{+}$
is meromorphic, and hence $H(u)$ is meromorphic.

\emph{Part~(iv).} By telescoping and~\eqref{eq:notes2-H-new-functional}:
\[
\begin{aligned}\frac{H(u+m)}{H(u)} & \;=\;\prod_{k=0}^{m-1}\frac{H(u+k+1)}{H(u+k)}\;=\;\prod_{k=0}^{m-1}\frac{w(u+k)}{\zeta_{+}}\\
 & \;=\;\frac{1}{\zeta_{+}^{m}}\prod_{k=0}^{m-1}w(u+k)\;=\;\frac{h_{m}(u)/h_{0}(u)}{\zeta_{+}^{m}},
\end{aligned}
\]
using $h_{m}/h_{0}=\prod_{k=0}^{m-1}w(u+k)$ (eq.~\eqref{eq:notes2-H-new-hm}).

\emph{Part~(v).} At a Floquet exponent $u_{0}$, the bilateral solution
$\{h_{m}(u_{0})\}_{m\in\mathbb{Z}}$ (extension by the recurrence,
required to decay also as $m\to-\infty$) agrees up to a scalar with
the left-going minimal solution $h_{m}^{-}(u_{0})=h_{-m}(-u_{0})$
(Definition~\ref{defn:notes2-hminus}). With matching at $m=0$,
$h_{-k}(u_{0})=c\cdot h_{k}(-u_{0})$ for $k\geq0$, where the scalar
$c$ is fixed by $h_{0}(u_{0})=c\cdot h_{0}(-u_{0})$. Hence 
\[
\frac{h_{-k}(u_{0})}{h_{-k+1}(u_{0})}\;=\;\frac{h_{k}(-u_{0})}{h_{k-1}(-u_{0})}\;=\;w(-u_{0}+k-1),\qquad k\geq1,
\]
applying the forward formula~\eqref{eq:notes2-H-new-hm} at the reflected
argument $-u_{0}$. Telescoping for $k=1,2,\ldots,m$ gives~\eqref{eq:notes2-hm-neg-product}.
The link between the forward and backward generators at $m=0$ is
the double minimality condition itself: at a Floquet exponent $u_{0}$,
the value $h_{-1}/h_{0}$ obtained from the backward extension is
$w(-u_{0})$, which equals $1/w(u_{0}-1)$ by~\eqref{eq:double-min-explicit},
so the same scalar normalization of $\{h_{m}\}$ extends consistently
in both directions.

For the $H_{-}$ formulation: substituting $-u$ for $u$ in~\eqref{eq:notes2-H-new-def}
and inverting, 
\[
H_{-}(u)\;\overset{\eqref{eq:notes2-H-minus-def}}{=}\;\frac{1}{H(-u)}\;=\;\prod_{m=0}^{\infty}\frac{w(-u+m)}{\zeta_{+}},
\]
establishing~\eqref{eq:notes2-H-minus-product}. Evaluating at $u=u_{0}-m$:
\[
\begin{aligned}H_{-}(u_{0}-m) & \;=\;\prod_{j=0}^{\infty}\frac{w(-(u_{0}-m)+j)}{\zeta_{+}}\;=\;\prod_{j=0}^{\infty}\frac{w(-u_{0}+m+j)}{\zeta_{+}}\\
 & \;=\;\prod_{k=m}^{\infty}\frac{w(-u_{0}+k)}{\zeta_{+}},
\end{aligned}
\]
the last equality by setting $k=j+m$. Dividing $H_{-}(u_{0})$ by
$H_{-}(u_{0}-m)$: 
\[
\begin{aligned}\frac{H_{-}(u_{0})}{H_{-}(u_{0}-m)} & \;=\;\frac{\prod_{k=0}^{\infty}[w(-u_{0}+k)/\zeta_{+}]}{\prod_{k=m}^{\infty}[w(-u_{0}+k)/\zeta_{+}]}\\
 & \;=\;\prod_{k=0}^{m-1}\frac{w(-u_{0}+k)}{\zeta_{+}}\;=\;\frac{1}{\zeta_{+}^{m}}\prod_{k=0}^{m-1}w(-u_{0}+k).
\end{aligned}
\]
Multiplying by $\zeta_{+}^{m}$ and combining with~\eqref{eq:notes2-hm-neg-product}
gives~\eqref{eq:notes2-H-new-hm-neg}. 
\end{proof}
\begin{cor}[Explicit Fourier coefficients of the Floquet factor\index{Floquet theory!Floquet factor, Fourier coefficients of}]
\label{cor:Floquet-Fourier} Let $u_{0}$ satisfy the double minimality
condition: \begin{NoHyper} 
\begin{equation}
w(-u_{0})\;=\;\frac{1}{w(u_{0}-1)}.\tag{\ref{eq:double-min-explicit}}
\end{equation}
\end{NoHyper} The Fourier coefficients $A_{m}=h_{m}(u_{0})$ of the
periodic Floquet factor are given by the explicit finite products
of the CF ratio $w$: 
\begin{equation}
\frac{h_{m}(u_{0})}{h_{0}(u_{0})}\;=\;\begin{cases}
{\displaystyle \prod_{k=0}^{m-1}w(u_{0}+k),} & m\geq1,\\[10pt]
1, & m=0,\\[6pt]
{\displaystyle \prod_{k=0}^{|m|-1}w(-u_{0}+k),} & m\leq-1.
\end{cases}\label{eq:Floquet-Fourier-explicit}
\end{equation}
The actual Fourier coefficients $B_{m}$ of the Floquet solution $f(x,u_{0})=e^{iu_{0}x}\sum_{m}B_{m}e^{imx}$
are recovered via~\eqref{eq:Cambi-AG}: 
\begin{equation}
B_{m}\;=\;\frac{p^{2}}{(u_{0}+m)^{2}}\,h_{m}(u_{0}),\qquad m\in\mathbb{Z}.\label{eq:Floquet-Bm-explicit}
\end{equation}
Both sequences decay as $|m|\to\infty$ since $|w(\pm u_{0}+k)|\to|\zeta_{+}|<1$
for large $k$, so all but finitely many factors in the products have
modulus less than one. The precise asymptotics are: 
\begin{equation}
\frac{h_{m}(u_{0})/h_{0}(u_{0})}{\zeta_{+}^{|m|}}\;\to\;\begin{cases}
1/H(u_{0}) & m\to+\infty,\\[4pt]
H_{-}(u_{0}) & m\to-\infty,
\end{cases}\label{eq:Floquet-Fourier-asymp}
\end{equation}
where $H(u_{0})$ and $H_{-}(u_{0})$ are the values of the minimal
solution generators at $u_{0}$ (eqs.~\eqref{eq:notes2-H-new-def}
and~\eqref{eq:notes2-H-minus-def}). Using $H_{-}(u_{0})=1/H(-u_{0})$,
both limits take the symmetric form $1/H(\pm u_{0})$: the forward
limit is $1/H(u_{0})$, and the backward limit is $1/H(-u_{0})$. 
\end{cor}

\begin{proof}
For $m\geq1$: $h_{m}/h_{0}=\prod_{k=0}^{m-1}h_{k+1}/h_{k}=\prod_{k=0}^{m-1}w(u_{0}+k)$
by the definition of $w$ as the minimal solution ratio. For $m\leq-1$:
$h_{-j}/h_{0}=\prod_{k=0}^{j-1}w(-u_{0}+k)$ (eq.~\eqref{eq:notes2-hm-neg-product})
with $j=|m|$. Note $h_{-1}/h_{0}=w(-u_{0})=1/w(u_{0}-1)$ by the
double minimality condition~\eqref{eq:double-min-explicit} ---
this is the key link between the two half-lines. The formula for $B_{m}$
follows from~\eqref{eq:Cambi-AG}. Decay: $w(u_{0}+k)\to\zeta_{+}$
with $|\zeta_{+}|<1$, so for large $m$ the product is dominated
by $\zeta_{+}^{m}\to0$.

\emph{Asymptotic limits~\eqref{eq:Floquet-Fourier-asymp}.} For $m\to+\infty$:
by~\eqref{eq:notes2-H-new-hm}, $h_{m}/h_{0}=\zeta_{+}^{m}\,H(u_{0}+m)/H(u_{0})$.
Since $H(u)\to1$ as $u\to+\infty$ (Theorem~\ref{thm:notes2-H-new}(ii)),
$[h_{m}/h_{0}]/\zeta_{+}^{m}=H(u_{0}+m)/H(u_{0})\to1/H(u_{0})$. For
$m\to-\infty$ (writing $m=-j$, $j\to+\infty$): by~\eqref{eq:notes2-H-new-hm-neg},
$h_{-j}/h_{0}=\zeta_{+}^{j}\,H_{-}(u_{0})/H_{-}(u_{0}-j)$. Since
$H_{-}(u)\to1$ as $u\to-\infty$ (Theorem~\ref{thm:notes2-H-new}(v)),
$[h_{-j}/h_{0}]/\zeta_{+}^{j}=H_{-}(u_{0})/H_{-}(u_{0}-j)\to H_{-}(u_{0})$. 
\end{proof}
\begin{figure}[htbp]
\centering \includegraphics[width=1\textwidth]{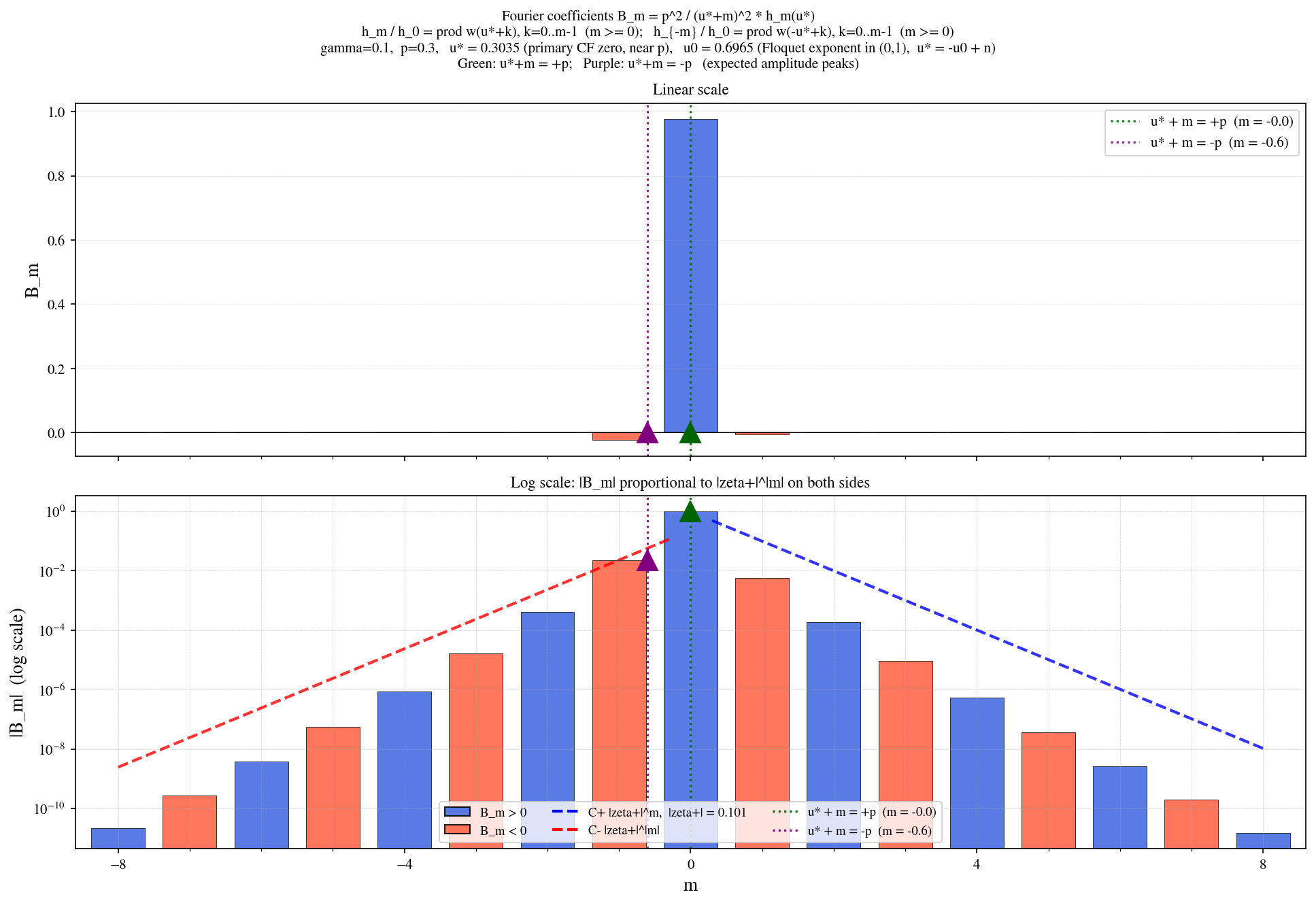}
\caption{Fourier coefficients $B_{m}=p^{2}/(u^{*}+m)^{2}\cdot h_{m}(u^{*})$
of the Floquet factor, computed from the explicit product formulas~\eqref{eq:Floquet-Fourier-explicit}
at the primary double minimality zero $u^{*}\approx p$. Parameters:
$\gamma=0.1$, $p=0.3$, $u^{*}\approx0.3035\approx p$. \emph{Top}:
linear scale; $B_{0}\approx0.977$ dominates. \emph{Bottom}: log scale
--- exponential envelope $|B_{m}|\propto|\zeta_{+}|^{|m|}$ (dashed),
$|\zeta_{+}|\approx0.101$, confirmed on both sides; alternating sign
from $\zeta_{+}<0$. Green marker: $u^{*}+m=+p$ (at $m\approx0$);
purple marker: $u^{*}+m=-p$ (at $m\approx-0.6$). For $p<1$, $u^{*}\approx p$
so both markers fall near the origin and the two Floquet zeros $u^{*}\approx p$
and $u^{**}\approx-p$ coincide with $\pm u_{0}$ in $(-\frac{1}{2},\frac{1}{2})$.}
\label{fig:Bm-p03} 
\end{figure}

\begin{figure}[htbp]
\centering \includegraphics[width=1\textwidth]{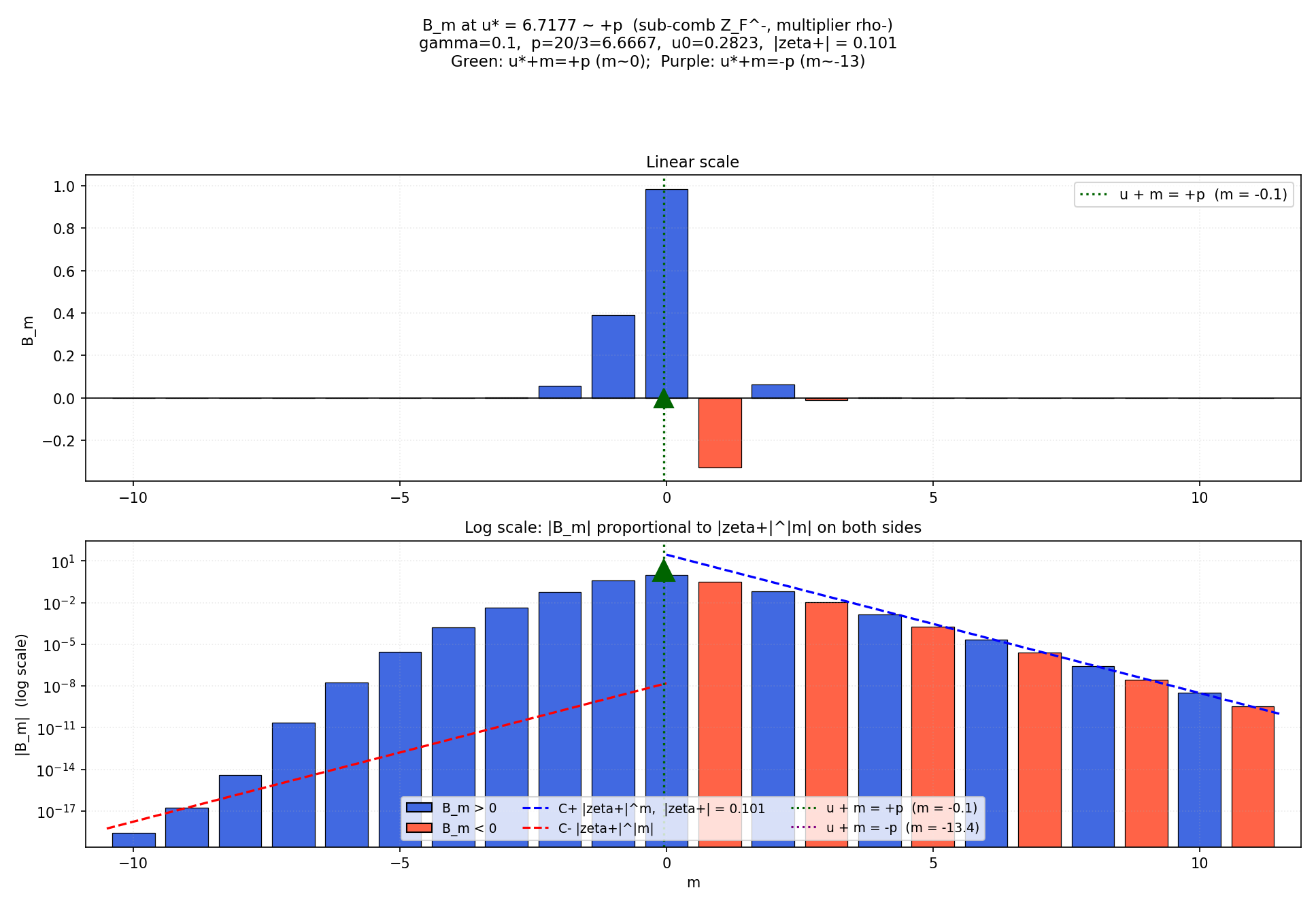}
\caption{Fourier coefficients $B_{m}$ at $u^{*}\approx+6.718\approx+p$ for
Cambi's parameters $\gamma=0.1$, $p=20/3$. $u^{*}=-u_{0}+7\in Z_{F}^{-}$
(see~\eqref{eq:family-def}), corresponding to Floquet multiplier
$\rho_{-}=e^{-2\pi iu_{0}}$. Formulas~\eqref{eq:Floquet-Fourier-explicit}:
$h_{m}/h_{0}=\prod_{k=0}^{m-1}w(u^{*}+k)$ ($m\protect\geq0$) and
$h_{-m}/h_{0}=\prod_{k=0}^{m-1}w(-u^{*}+k)$ ($m\protect\geq0$),
both decaying since $|w(\pm u^{*}+k)|<1$ for all $k$. \emph{Top}:
$B_{0}\approx0.985$ dominates since $u^{*}+0\approx p$; sign alternates
for $m>0$ ($\zeta_{+}<0$), all positive for $m<0$. \emph{Bottom}:
log scale --- clean exponential decay $|B_{m}|\propto|\zeta_{+}|^{|m|}$
on both sides. Green: $u^{*}+m=+p$ ($m\approx0$); purple: $u^{*}+m=-p$
($m\approx-13$, outside range).}
\label{fig:Bm-Cambi} 
\end{figure}

\begin{figure}[htbp]
\centering \includegraphics[width=1\textwidth]{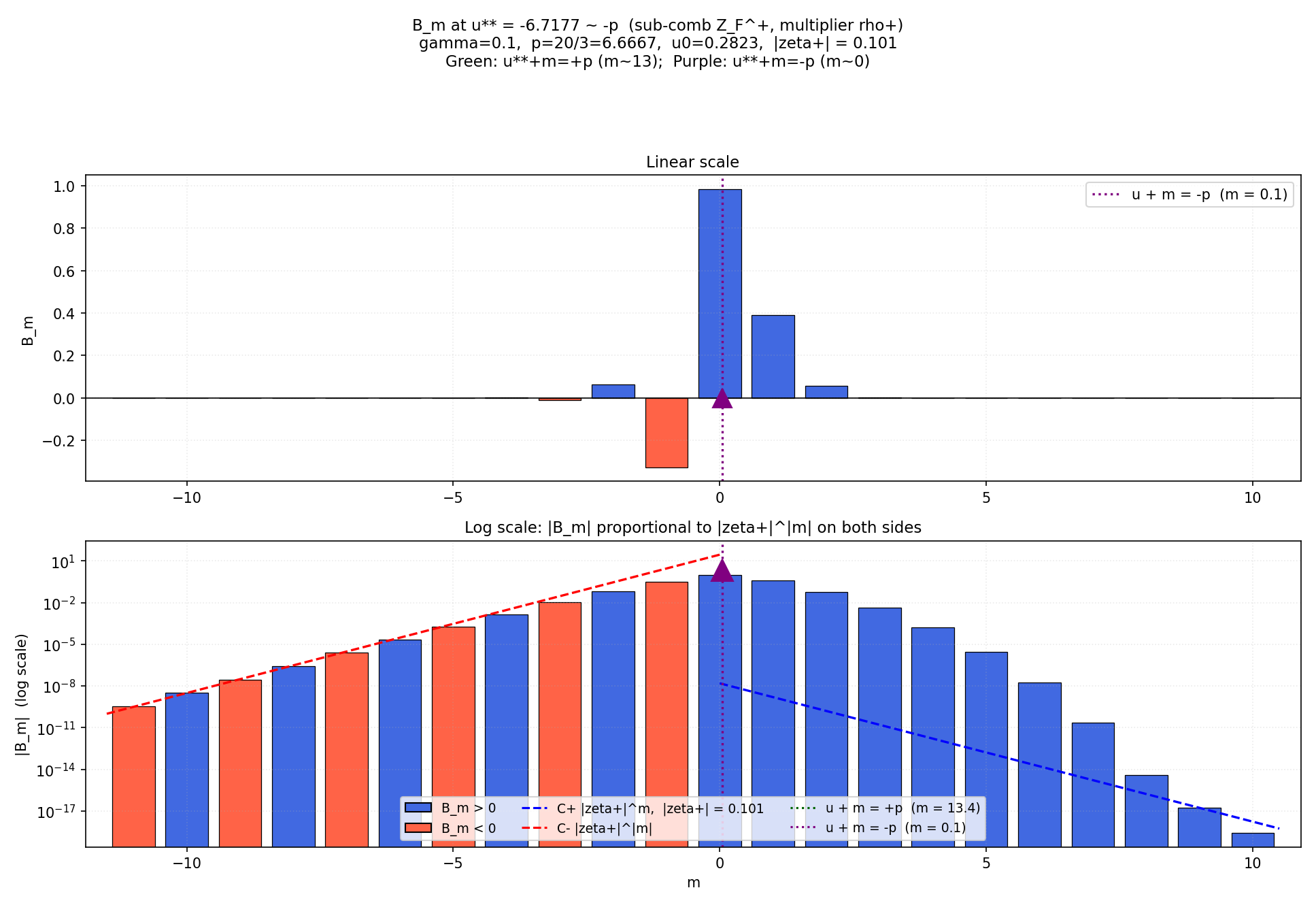}
\caption{Fourier coefficients $B_{m}$ at $u^{**}\approx-6.718\approx-p$ for
Cambi's parameters $\gamma=0.1$, $p=20/3$. $u^{**}=u_{0}-7\in Z_{F}^{+}$,
corresponding to Floquet multiplier $\rho_{+}=e^{+2\pi iu_{0}}$.
The spectrum is the mirror image of Fig.~\ref{fig:Bm-Cambi}: $B_{m}(u^{**})=B_{-m}(u^{*})$
exactly, since $h_{m}(u^{**})/h_{0}=\prod_{k=0}^{m-1}w(u^{**}+k)=\prod_{k=0}^{m-1}w(-u^{*}+k)=h_{-m}(u^{*})/h_{0}$.
The two solutions at $\pm u^{*}$ are complex conjugates ($\rho_{+}=\bar{\rho}_{-}$
in the stable case), and together form the real Floquet solution of
Hill's equation. \emph{Top}: $B_{0}\approx0.985$ dominates; sign
alternates for $m<0$. \emph{Bottom}: same exponential envelope $|\zeta_{+}|^{|m|}$.
Green: $u^{**}+m=+p$ ($m\approx+13$, outside range); purple: $u^{**}+m=-p$
($m\approx0$).}
\label{fig:Bm-Cambi2} 
\end{figure}

\begin{rem}[Two Floquet solutions and their Fourier spectra]
\label{rem:two-Floquet-solutions} The monodromy matrix $M$ (over
one period $2\pi$ of Hill's equation) has two eigenvalues $\rho_{\pm}=e^{\pm2\pi iu_{0}}$
with $u_{0}\in(0,\tfrac{1}{2})$. The double minimality equation $F_{w}(u)=0$
has two primary accessible zeros near $\pm p$, lying in opposite
sub-combs: 
\begin{equation}
u^{*}\;\approx\;+p,\qquad u^{**}\;=\;-u^{*}\;\approx\;-p.\label{eq:two-primary-zeros}
\end{equation}
Which sub-comb hosts which primary zero depends on the fractional
part of $u^{*}$: when $\{u^{*}\}>\tfrac{1}{2}$ --- as for all worked
parameter sets of this chapter ($p=0.70$ and Cambi's $p=20/3$) ---
one has $u^{*}\in\{-u_{0}+n\}=Z_{F}^{-}$ with multiplier $\rho_{-}$
and $u^{**}\in Z_{F}^{+}$ with $\rho_{+}$; when $\{u^{*}\}<\tfrac{1}{2}$
the two primary zeros interchange sub-combs (so in Fig.~\ref{fig:Bm-p03},
where $u^{*}\approx0.3035=u_{0}$, one has $u^{*}\in Z_{F}^{+}$).
The Fourier spectra satisfy $B_{m}(u^{**})=B_{-m}(u^{*})$, so the
two solutions carry mirror-image spectra (Figs.~\ref{fig:Bm-Cambi}--\ref{fig:Bm-Cambi2}).
Since $u^{*}=-u^{**}$, the two are related by $u\mapsto-u$, consistent
with the reflection symmetry $\widetilde{F}_{w}(u)=\widetilde{F}_{w}(-u)$.
The Floquet exponent $u_{0}$ in the fundamental strip $(0,\tfrac{1}{2})$
is recovered from either primary zero as the distance to the nearest
integer: 
\begin{equation}
u_{0}\;=\;\min\bigl(\{u^{*}\},\,1-\{u^{*}\}\bigr),\qquad\{u^{*}\}=u^{*}\bmod1,\label{eq:u0-from-ustar}
\end{equation}
in agreement with the recovery formula of Remark~\ref{rem:double-min-two-families};
for $\{u^{*}\}>\tfrac{1}{2}$ this reduces to $u_{0}=\lceil u^{*}\rceil-u^{*}$. 
\end{rem}

\subsection{\texorpdfstring{Meromorphic structure of $H(u)$}{Meromorphic structure
of H(u)}}

\label{subsec:H-meromorphic}

The previous subsection introduced $H(u)$ as an infinite-product
generator and developed its Fourier representation (Theorem~\ref{thm:notes2-H-new},
Corollary~\ref{cor:Floquet-Fourier}). This section establishes the
meromorphic structure of $H(u)$: its poles and zeros, their location
relative to those of the minimal solution ratio $w(u)$, and their
orders.
\begin{thm}[Poles and zeros of $H(u)$]
\index{minimal solution generator $H(u)$!poles and zeros} \label{thm:notes2-H-new-poles-zeros}
Let $H(u)$ be defined by~\eqref{eq:notes2-H-new-def}, write $H(u)=1/P(u)$
with 
\begin{equation}
P(u)\;=\;\prod_{m=0}^{\infty}\frac{w(u+m)}{\zeta_{+}}\label{eq:notes2-P-def}
\end{equation}
the defining product of~\eqref{eq:notes2-H-new-def}, and let $\{u_{p}\}$
denote the poles of $w(u)$ (Remark~\ref{rem:w-pole-zero-geometry}(a)). 
\begin{enumerate}
\item[(i)] \emph{Smooth and nonzero for $u>0$ when $p\in(0,1)$.} For $u>0$
and $m\geq0$: $u+m>0$, so when $p\in(0,1)$ each $w(u+m)$ is smooth
and nonzero (Remark~\ref{rem:w-pole-zero-geometry}(c)). Every factor
in~\eqref{eq:notes2-H-new-def} is then finite and nonzero, hence
$H(u)$ is smooth and nonzero for $u>0$. For $p>1$, where $w$ may
have positive sign-change zeros and poles, $H$ remains regular at
the zeros by the cancellation of part~(iii) below, and vanishes at
any positive poles of $w$ by part~(ii).
\item[(ii)] \emph{Zeros of $H$ exactly at the poles of $w$.} At $u_{0}=u_{p}$
(a pole of $w$): the factor $m=0$ gives $w(u)/\zeta_{+}\to\infty$
as $u\to u_{0}$, while every factor with $m\geq1$ is finite and
nonzero at $u_{0}$ (the points $u_{p}+m$, $m\geq1$, are generically
neither poles nor zeros of $w$, the pole ladders not being exactly
unit-spaced). Hence $P(u)\to\infty$ and $H(u)=1/P(u)\to0$. At the
shifted points $u_{0}=u_{p}-m_{0}$ with $m_{0}\geq1$, by contrast,
the diverging factor $m=m_{0}$ is accompanied by the vanishing factor
$m=m_{0}-1$, since $w(u_{p}-1)=0$ by the unit-shift pairing (Theorem~\ref{thm:notes2-w-poles-zeros}(i));
the two cancel as in part~(iii), and $H$ is regular and nonzero
there. Therefore 
\begin{equation}
\{\text{zeros of }H\}\;=\;\{\text{poles of }w\}.\label{eq:notes2-H-new-zeros}
\end{equation}
The order of each zero matches that of the pole of $w$ (Theorem~\ref{thm:notes2-H-new-orders}(i)).
This is exactly what the functional equation~\eqref{eq:notes2-H-new-functional}
requires: the single simple zero of $H$ at $u_{p}$ produces both
the pole of $w(u)/\zeta_{+}$ at $u=u_{p}$ (through $1/H(u)$) and
its zero at $u=u_{p}-1$ (through $H(u+1)$), while zeros of $H$
at the further shifts $u_{p}-m_{0}$, $m_{0}\geq1$, would force spurious
zeros of $w$ at $u_{p}-m_{0}-1$. Numerically at $p=0.70$, $\gamma=0.30$,
for the pole $u_{p}\approx-0.2525$: $H(u_{p}-1)\approx-0.069$ and
$H(u_{p}-2)\approx-0.053$, finite and nonzero, while $H$ does vanish
at the nearby deep ladder pole $u\approx-1.2584$.
\item[(iii)] \emph{Sign-change zeros of $w$ give no singularity of $H$.} At
$u_{0}=u_{z}-m_{0}$ for a sign-change zero $u_{z}$ of $w$ and any
$m_{0}\geq0$, the factor $m=m_{0}$ gives $w(u+m_{0})/\zeta_{+}\to0$
as $u\to u_{0}$, making $P\to0$ and apparently $H\to\infty$. However,
$w$ has a pole at $u_{z}+1$ (Theorem~\ref{thm:notes2-w-poles-zeros}(i)),
so the adjacent factor $m=m_{0}+1$, evaluating $w$ at $u_{0}+m_{0}+1\to u_{z}+1$,
diverges simultaneously. These two factors multiply to a finite nonzero
limit (a zero and a pole of the same order cancel in the product),
so $P(u_{0})$ is finite and nonzero, and $H$ has no singularity
at $u_{0}$ (Theorem~\ref{thm:notes2-H-new-orders}(ii)).
\item[(iv)] \emph{Poles of $H$ at negative integers.} At $u=-n$ ($n\geq1$):
the factor $m=n-1$ gives $w(u+n-1)=w(-1+(u+n))\to0$ as $u\to-n$
(double zero of $w$ at $-1$, Theorem~\ref{thm:notes2-w-poles-zeros}(ii)),
making $P\to0$ and $H\to\infty$: 
\begin{equation}
\{\text{poles of }H\}\;=\;\{-n:n\geq1\}.\label{eq:notes2-H-new-poles}
\end{equation}
All poles are second-order (Theorem~\ref{thm:notes2-H-new-orders}(iii)).
\item[(v)] \emph{Asymptotic behavior.} $\lim_{u\to+\infty}H(u)=1$ (Theorem~\ref{thm:notes2-H-new}(ii)).
On the negative axis $H$ has no limit: it vanishes and changes sign
at every pole of $w$ (part~(ii)), of which there are two per unit
cell. Along integer translates, however, the values converge: for
a fixed phase $\varphi\in(0,1)$ away from the comb phases $\{u^{*}\}$
and $1-\{u^{*}\}$, numerically $H(-n-\varphi)\to L(\varphi)$ as
$n\to\infty$, with a phase-dependent, sign-varying limit; at $p=0.70$,
$\gamma=0.30$: $L(\tfrac{1}{2})\approx0.48$, while $L(0.1)\approx L(0.9)\approx-4.59$. 
\end{enumerate}
\end{thm}

\begin{thm}[Orders of zeros and poles of $H(u)$]
\label{thm:notes2-H-new-orders} Let $H(u)=1/P(u)$ with $P(u)$
the defining product~\eqref{eq:notes2-P-def} of Theorem~\ref{thm:notes2-H-new-poles-zeros}.
For generic $(\gamma,p)$ with $2\gamma<1$, the poles and sign-change
zeros of $w(u)$ are simple (Remark~\ref{rem:simplicity-evidence});
the zero at $u=-1$ is double (Theorem~\ref{thm:notes2-w-poles-zeros}(ii)).
Near any $u_{0}$, each factor $w(u+m)/\zeta_{+}$ is meromorphic
in $u$ and smooth at $u_{0}$ except for the finitely many $m$ for
which $u_{0}+m$ is a pole or zero of $w$. 
\begin{enumerate}
\item[(i)] \emph{Zeros of $H$ are generically simple.} Let $u_{0}=u_{p}$ be
a simple pole of $w$. Then the factor $m=0$ satisfies $w(u)/\zeta_{+}\sim A/(u-u_{0})$
near $u_{0}$, while every factor with $m\geq1$ is finite and nonzero
at $u_{0}$ (Theorem~\ref{thm:notes2-H-new-poles-zeros}(ii)). Hence
$P(u)\sim AC/(u-u_{0})$ with $C=\prod_{m\geq1}w(u_{0}+m)/\zeta_{+}\neq0$,
and $H(u)\sim(u-u_{0})/(AC)$: a simple zero of $H$. At the shifted
points $u_{p}-m_{0}$, $m_{0}\geq1$, the pole factor is cancelled
by the paired zero factor (part~(ii)), and no zero of $H$ arises.
\item[(ii)] \emph{Sign-change zeros of $w$ give no singularity of $H$.} If
$u_{0}+m_{1}$ is a simple sign-change zero of $w$, then by Theorem~\ref{thm:notes2-w-poles-zeros}(i),
$u_{0}+m_{1}+1$ is a simple pole of $w$. Near $u_{0}$: $w(u+m_{1})/\zeta_{+}\sim B(u-u_{0})$
(simple zero) and $w(u+m_{1}+1)/\zeta_{+}\sim A/(u-u_{0})$ (simple
pole). Their product $\sim AB$ is finite and nonzero, so these two
factors together contribute no singularity to $P(u)$, and $H(u)$
is smooth and nonzero at $u_{0}$.
\item[(iii)] \emph{Poles of $H$ are always second-order.} At $u_{0}=-n$ ($n\geq1$):
the factor $m=n-1$ gives $w(u+n-1)=w(-1+(u+n))\sim(\gamma/p^{2})(u+n)^{2}$
near $u_{0}=-n$ (double zero of $w$ at $-1$, Theorem~\ref{thm:notes2-w-poles-zeros}(ii)).
Hence $w(u+n-1)/\zeta_{+}\sim(\gamma/(p^{2}\zeta_{+}))(u+n)^{2}$,
giving $P(u)\sim C(u+n)^{2}$ with $C\neq0$, so $H(u)\sim1/(C(u+n)^{2})$:
a second-order pole of $H$. For $n\geq2$, all other factors are
smooth at $u=-n$ (Theorem~\ref{thm:notes2-w-poles-zeros}(ii)),
so the order is exactly~$2$.
\item[(iv)] \emph{Higher-order zeros at EPD curves.} At EPD curves the two pole
families of $w$ coincide (Remark~\ref{rem:w-two-families}); where
two paired simple poles of $w$ merge into one double pole $u_{p}$,
the factor $m=0$ acquires a double pole there, giving by the argument
of part~(i) a double zero of $H$ at $u_{p}$. More generally, a
pole of $w$ of order $k$ gives a zero of $H$ of order $k$. 
\end{enumerate}
\end{thm}

% Figures for w(u) and F_w(u) appear in Section~\ref{subsec:Cambi-w-structure}.

% -----------------------------------------------------------------------------

\subsection{Stability boundary curves of the first instability tongue via the
CF method}

\label{subsec:CF-stability-bdry}

The stability boundary of the first instability tongue is characterized
exactly by the CF condition $F_{w}(\tfrac{1}{2};\,p,\gamma)=0$ on
the double minimality function. Since $F_{w}(\tfrac{1}{2})=w(-\tfrac{1}{2})-1/w(-\tfrac{1}{2})$,
this is equivalent to 
\begin{equation}
w\!\left(-\tfrac{1}{2};\,p,\,\gamma\right)^{2}\;=\;1,\qquad\text{i.e.}\quad w\!\left(-\tfrac{1}{2};\,p,\,\gamma\right)\;=\;\pm1.\label{eq:stability-bdry-CF}
\end{equation}
The sign $w=+1$ gives the upper boundary $p_{+}(\gamma)$ and $w=-1$
gives the lower boundary $p_{-}(\gamma)$, both emanating from $(p,\gamma)=(\tfrac{1}{2},0)$.
This condition is \emph{exact}; for the numerical care required when
evaluating $w(-\tfrac{1}{2})$ close to the boundary --- where loss
of significance and off-regime use of the expansion below can place
a naive estimate spuriously inside the unstable tongue --- see the
practical caution in Remark~\ref{rem:asymptotic-CF}.

\begin{figure}[htbp]
\centering \includegraphics[width=1\textwidth]{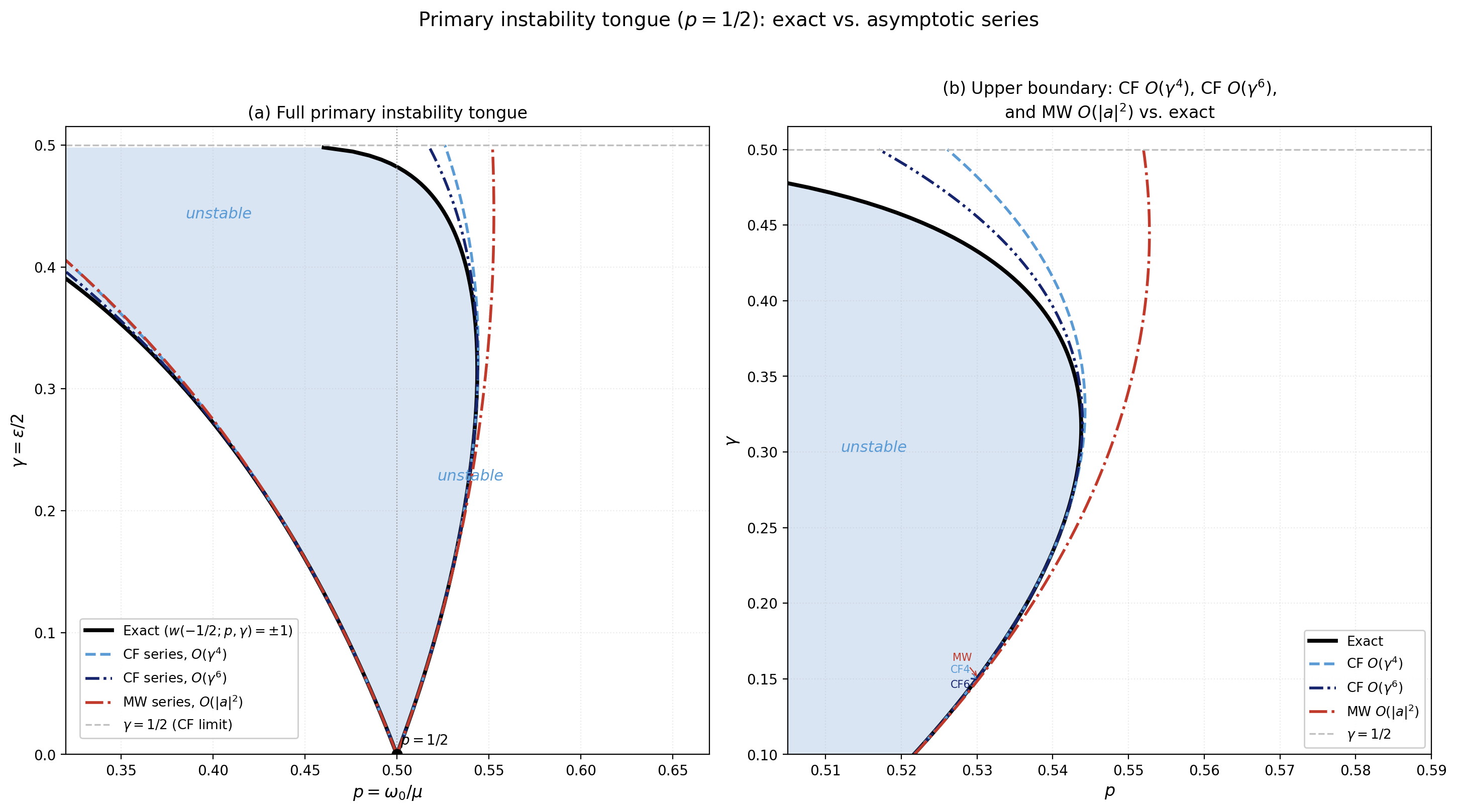}
\caption{Primary instability tongue ($p\approx\tfrac{1}{2}$): exact boundary
curves (black solid, from $w(-\tfrac{1}{2};\,p,\gamma)=\pm1$, eq.~\eqref{eq:stability-bdry-CF})
vs.\ asymptotic series approximations (Theorem~\ref{thm:bdry-series-CF}).
Blue dashed: CF series~\eqref{eq:bdry-curves-CF} truncated at $O(\gamma^{4})$.
Dark blue dash-dot-dot: CF series truncated at $O(\gamma^{6})$. Red
dash-dot: MW series~\eqref{eq:EPD-bdry-1} truncated at $O(|a|^{2})$.
Gray dashed: $\gamma=\tfrac{1}{2}$ (CF convergence limit). \emph{(a)}
Full tongue from tip $(p,\gamma)=(\tfrac{1}{2},0)$; blue shading
covers the exact unstable region (both boundaries visible for $\gamma\lesssim0.41$;
above, only the upper boundary remains in the window, with the lower
boundary having exited to the left). \emph{(b)} Zoom on the upper
boundary for $\gamma\protect\geq0.10$; arrows at $\gamma=0.15$ show
the absolute error of each approximation. The CF $O(\gamma^{4})$
and $O(\gamma^{6})$ curves nearly coincide for $\gamma\lesssim0.12$
(both more accurate than MW throughout); the $O(\gamma^{6})$ truncation
is the more accurate of the two at every tabulated $\gamma$ (Table~\ref{tab:bdry-error-CF}),
the bracketing of the alternating-sign series showing up instead in
the signed errors (Remark~\ref{rem:asymptotic-CF}).}
\label{fig:tongue-comparison} 
\end{figure}

\emph{Detuning parameter and stability zone.} Let 
\begin{equation}
\eta\;=\;p\;-\;\tfrac{1}{2}\label{eq:eta-def}
\end{equation}
denote the detuning from the primary resonance (see Chapter~\ref{app:notation}).
Both $\eta$ and $\gamma$ are small near the tongue tip, and nearly
proportional along the boundary: $|\eta|\approx\tfrac{1}{4}\gamma$.
The stable zone near the first tongue boundary is characterized by
\begin{equation}
3\;<\;\left|\frac{\gamma}{\eta}\right|\;<\;4\qquad(\text{stable, near the first tongue boundary from outside}).\label{eq:stable-near-bdry-CF}
\end{equation}
In this regime both $\eta$ and $\gamma$ are small and of the same
order, so $O(\eta^{k})=O(\gamma^{k})$ and either serves as the error
parameter. The corresponding region in the $(p,\gamma)$-plane is
shown in Fig.~\ref{fig:tongue-comparison}(a) as the neighborhood
of the exact boundary curves just outside the shaded unstable tongue.

\emph{Taylor series of $\gamma/w(-\tfrac{1}{2};\,\tfrac{1}{2}+\eta,\,\gamma)$.}
Substituting $p=\tfrac{1}{2}+\eta$ into the CF and expanding the
reciprocal combination $\gamma/w$ --- the normalization that is
analytic at the origin (proof of Theorem~\ref{thm:bdry-series-CF},
Step~1) --- in $(\eta,\gamma)$, grouping terms by total degree
$i+2j$ (only even powers $\gamma^{2j}$ occur, by the parity argument
of Step~2): 
\begin{gather}
\frac{\gamma}{w\!\left(-\tfrac{1}{2};\,\tfrac{1}{2}+\eta,\,\gamma\right)}=4\eta+\Bigl(4\eta^{2}+\tfrac{9}{8}\gamma^{2}\Bigr)+\tfrac{9}{16}\,\eta\gamma^{2}\nonumber \\
+\Bigl(\tfrac{27}{32}\,\eta^{2}\gamma^{2}+\tfrac{675}{512}\,\gamma^{4}\Bigr)+\Bigl(\tfrac{45}{64}\,\eta^{3}\gamma^{2}+\tfrac{1575}{1024}\,\eta\gamma^{4}\Bigr)\nonumber \\
+\Bigl(\tfrac{99}{128}\,\eta^{4}\gamma^{2}+\tfrac{1425}{512}\,\eta^{2}\gamma^{4}+\tfrac{193125}{65536}\,\gamma^{6}\Bigr)+O(\eta^{7}),\label{eq:w-Taylor-CF}
\end{gather}
where the first line collects degrees 1--3 and the remaining lines
degrees 4--6, with remainder $O(\eta^{7})=O(\gamma^{7})$ along the
boundary curve $\eta=\eta_{\pm}(\gamma)\approx\pm\tfrac{1}{4}\gamma$,
where the order-by-order inversion below evaluates it (proof of Theorem~\ref{thm:bdry-series-CF},
Step~4). Because the grouping by total degree $i+2j$ presumes $\eta=\Theta(\gamma)$,
the expansion should be used \emph{along the boundary}; it is not
claimed to hold uniformly down to the inner edge $|\gamma/\eta|\to3$
of the strip~\eqref{eq:stable-near-bdry-CF}, and naive evaluation
off the boundary curve (e.g.\ at fixed $\eta\ll\gamma$ or near $|\gamma/\eta|=3$)
degrades and can land spuriously inside the tongue --- a pitfall
discussed in Remark~\ref{rem:asymptotic-CF}. All coefficients are
exact rationals; the same data in tabular form:

\needspace{12\baselineskip} 
\begin{equation}
{\small 
\global\long\def\arraystretch{1.6}%
\setlength{\arraycolsep}{2pt}\begin{array}{c|ccc}
k=i+2j & 1 & 2 & 3\\
\hline {\displaystyle \sum_{i+2j=k}w_{ij}\,\eta^{i}\,\gamma^{2j}} & 4\eta & 4\eta^{2}+\dfrac{9}{8}\gamma^{2} & \dfrac{9}{16}\,\eta\gamma^{2}\\[6pt]
\hline k=i+2j & 4 & 5 & 6\\
\hline {\displaystyle \sum_{i+2j=k}w_{ij}\,\eta^{i}\,\gamma^{2j}} & \dfrac{27}{32}\,\eta^{2}\gamma^{2}+\dfrac{675}{512}\,\gamma^{4} & \dfrac{45}{64}\,\eta^{3}\gamma^{2}+\dfrac{1575}{1024}\,\eta\gamma^{4} & \begin{array}[t]{@{}c@{}}
\dfrac{99}{128}\,\eta^{4}\gamma^{2}+\dfrac{1425}{512}\,\eta^{2}\gamma^{4}\\[2pt]
{}+\dfrac{193125}{65536}\,\gamma^{6}
\end{array}
\end{array}}\label{eq:w-Taylor-table}
\end{equation}

\needspace{20\baselineskip} \emph{Boundary curves by order-by-order
inversion.} Write the boundary curve as 
\begin{equation}
\eta\;=\;e_{1}\gamma+e_{2}\gamma^{2}+e_{3}\gamma^{3}+e_{4}\gamma^{4}+e_{5}\gamma^{5}+e_{6}\gamma^{6}+O(\gamma^{7}).\label{eq:eta-expansion}
\end{equation}
Substituting into~\eqref{eq:w-Taylor-CF} and imposing $\gamma/w(-\tfrac{1}{2})=\pm\gamma$
--- equivalent to~\eqref{eq:stability-bdry-CF}, since $w=\pm1$
iff $1/w=\pm1$ --- order by order determines the coefficients exactly.
The degree-$k$ row of~\eqref{eq:w-Taylor-CF} is needed for $e_{k}$:
specifically, $e_{k}$ for $k\leq4$ requires only degrees 1--4;
$e_{5}$ additionally requires the degree-5 line; and $e_{6}$ requires
the degree-6 line. The resulting exact rational coefficients are:
\needspace{16\baselineskip} 
\begin{equation}
\begin{array}{c|c|c|c}
\text{Order} & \text{Equation} & e_{k}\text{ (upper, }+\text{)} & e_{k}\text{ (lower, }-\text{)}\\[2pt]
\hline  &  & \\[-8pt]
O(\gamma^{1}) & 4e_{1}=\pm1 & +\tfrac{1}{4} & -\tfrac{1}{4}\\[4pt]
O(\gamma^{2}) & 4e_{2}+\tfrac{11}{8}=0 & -\tfrac{11}{32} & -\tfrac{11}{32}\\[4pt]
O(\gamma^{3}) & 4e_{3}\mp\tfrac{35}{64}=0 & +\tfrac{35}{256} & -\tfrac{35}{256}\\[4pt]
O(\gamma^{4}) & 4e_{4}+\tfrac{985}{512}=0 & -\tfrac{985}{2048} & -\tfrac{985}{2048}\\[4pt]
O(\gamma^{5}) & 4e_{5}\mp\tfrac{4139}{4096}=0 & +\tfrac{4139}{16384} & -\tfrac{4139}{16384}\\[4pt]
O(\gamma^{6}) & 4e_{6}+\tfrac{284379}{65536}=0 & -\tfrac{284379}{262144} & -\tfrac{284379}{262144}
\end{array}\label{eq:bdry-inversion-table}
\end{equation}

\begin{thm}[Taylor series for the first instability tongue boundary curves]
\index{boundary curves!continued-fraction series} \label{thm:bdry-series-CF}
The boundary curves of the first instability tongue of the LC circuit~\eqref{eq:LC-Hill},
defined by the CF condition $w(-\tfrac{1}{2};\,p,\,\gamma)=\pm1$
(eq.~\eqref{eq:stability-bdry-CF}), are analytic functions of $\gamma$
(Theorem~\ref{thm:bdry-analytic}, \S\,\ref{subsec:analytic-param})
admitting the convergent Taylor expansions 
\begin{multline}
p_{\pm}(\gamma)\;=\;\frac{1}{2}\;\pm\;\frac{1}{4}\,\gamma\;-\;\frac{11}{32}\,\gamma^{2}\;\pm\;\frac{35}{256}\,\gamma^{3}\\
\;-\;\frac{985}{2048}\,\gamma^{4}\;\pm\;\frac{4139}{16384}\,\gamma^{5}\;-\;\frac{284379}{262144}\,\gamma^{6}\;+\;O(\gamma^{7}),\label{eq:bdry-curves-CF}
\end{multline}
as $\gamma\to0$, valid in the near-boundary regime~\eqref{eq:stable-near-bdry-CF}.
All six coefficients are exact rationals computed by order-by-order
inversion of Table~\eqref{eq:w-Taylor-table} (eq.~\eqref{eq:bdry-inversion-table}).
Coefficients $e_{1}$--$e_{4}$ follow from the degree-$\leq4$ terms~\eqref{eq:w-Taylor-CF}
alone, giving $p_{\pm}$ to $O(\gamma^{5})$; the additional degree-5
and degree-6 rows of Table~\eqref{eq:w-Taylor-table} are needed
for $e_{5}$ and $e_{6}$, extending the accuracy to $O(\gamma^{7})$.
The even-degree coefficients $(-\tfrac{11}{32},\,-\tfrac{985}{2048},\,-\tfrac{284379}{262144})$
are identical on both branches, describing the leftward drift of the
tongue center; the odd-degree coefficients $(\pm\tfrac{1}{4},\,\pm\tfrac{35}{256},\,\pm\tfrac{4139}{16384})$
are equal and opposite, describing the tongue half-width. The series~\eqref{eq:bdry-curves-CF}
is a convergent power series in $\gamma$ (Theorem~\ref{thm:bdry-analytic},
\S\,\ref{subsec:analytic-param}); the alternating-sign structure
means the absolute truncation error need not decrease monotonically
order by order (see Remark~\ref{rem:asymptotic-CF} and Table~\ref{tab:bdry-error-CF}). 
\end{thm}

\begin{proof}
\emph{Step 1: Analyticity of $W(\eta,\gamma)$.} Define $W(\eta,\gamma)=\gamma/w(-\tfrac{1}{2};\,\tfrac{1}{2}+\eta,\,\gamma)$.
The function $G(u;\eta)=1-(p/u)^{2}\big|_{p=1/2+\eta}=1-(1+2\eta)^{2}/(2u)^{2}$
is a polynomial in $\eta$ for each fixed $u\neq0$. By Pincherle's
theorem (Theorem~\ref{thm:Pinch}), for $|\gamma|<\tfrac{1}{2}$
the continued fraction~\eqref{eq:Cambi-v-CF} converges absolutely
and uniformly on compact subsets of $\{|\eta|<\tfrac{1}{4}\}$, so
its tail $T_{1}(\eta,\gamma)$ is jointly analytic near $(0,0)$,
with $T_{1}=O(\gamma^{2})$. From the first CF step, 
\[
W(\eta,\gamma)\;=\;\frac{\gamma}{w(-\tfrac{1}{2})}\;=\;-G(\tfrac{1}{2};\eta)+T_{1}(\eta,\gamma)\;=\;4\eta+4\eta^{2}+T_{1}(\eta,\gamma),
\]
so $W$ is jointly analytic near $(0,0)$ with a convergent Taylor
expansion. The reciprocal normalization is essential here: $w(-\tfrac{1}{2})$
itself vanishes identically on $\{\gamma=0,\,\eta\neq0\}$ and diverges
along $\{\eta=0,\,\gamma\to0\}$, so it is not analytic at the origin;
the boundary condition is unaffected, since $w=\pm1$ iff $\gamma/w=\pm\gamma$.

\emph{Step 2: Parity structure.} Replacing $h_{n}$ by $(-1)^{n}h_{n}$
maps the recurrence~\eqref{eq:notes2-ttrec} with coupling $\gamma$
to the one with $-\gamma$ while preserving minimality, so $w(u;-\gamma)=-w(u;\gamma)$,
and hence $W=\gamma/w$ is \emph{even} in $\gamma$: 
\[
W(\eta,-\gamma)\;=\;W(\eta,\gamma).
\]
Consequently only even powers $\gamma^{2j}$ occur, $W(\eta,\gamma)=\sum_{i+2j\geq1}w_{ij}\,\eta^{i}\,\gamma^{2j}$,
graded in what follows by the total degree $k=i+2j$.

\emph{Step 3: Computation of coefficients.} Substituting $G(n+\tfrac{1}{2};\eta)=1-(1+2\eta)^{2}/(2n+1)^{2}$
into the CF recurrence and collecting by total degree $k$ gives an
algebraic recursion for the $w_{ij}$ with $i+2j=k$ in terms of those
with $i+2j<k$. All coefficients are exact rationals, listed through
degree~6 in the display~\eqref{eq:w-Taylor-CF} and again, grouped
by degree, in Table~\eqref{eq:w-Taylor-table}.

\emph{Step 4: Inversion.} The boundary condition $W(\eta,\gamma)=\pm\gamma$
defines $F_{\pm}(\eta,\gamma):=W(\eta,\gamma)\mp\gamma=0$. At the
origin: $F_{\pm}(0,0)=0$ and $\partial_{\eta}F_{\pm}\big|_{(0,0)}=w_{10}=4\neq0$.
By the analytic implicit function theorem, there exists a unique analytic
$\eta_{\pm}(\gamma)$ near $\gamma=0$ solving $W(\eta_{\pm}(\gamma),\gamma)=\pm\gamma$.
Writing $\eta_{\pm}=\sum_{k=1}^{K}e_{k}^{\pm}\gamma^{k}+O(\gamma^{K+1})$
and collecting the coefficient of $\gamma^{k}$ in $W=\pm\gamma$
gives a triangular system: at each order $k$, the equation is $w_{10}\,e_{k}^{\pm}+[\text{known from }e_{1}^{\pm},\ldots,e_{k-1}^{\pm}\text{ and }w_{ij}\text{ with }i+2j\leq k]=[{\pm1\text{ if }k=1,\;\,0\text{ if }k\geq2}]$,
solved uniquely as $e_{k}^{\pm}$ is determined explicitly. The degree-$\leq k$
rows of Table~\eqref{eq:w-Taylor-table} suffice to determine $e_{1}^{\pm},\ldots,e_{k}^{\pm}$:
in particular, $e_{1}^{\pm}$--$e_{4}^{\pm}$ require only degrees
1--4 (eq.~\eqref{eq:w-Taylor-CF}), giving $O(\gamma^{5})$ accuracy;
$e_{5}^{\pm}$ further requires the degree-5 row ($w_{31}=\tfrac{45}{64}$,
$w_{12}=\tfrac{1575}{1024}$); and $e_{6}^{\pm}$ requires the degree-6
row ($w_{41}=\tfrac{99}{128}$, $w_{22}=\tfrac{1425}{512}$, $w_{03}=\tfrac{193125}{65536}$),
yielding the full $O(\gamma^{7})$ accuracy of~\eqref{eq:bdry-curves-CF}.

\emph{Step 5: Even/odd symmetry.} By Step~2, $W$ is even in $\gamma$.
Hence if $W(\eta_{+}(\gamma),\gamma)=+\gamma$ for all small $\gamma$,
then evaluating at $-\gamma$ gives $W(\eta_{+}(-\gamma),\gamma)=W(\eta_{+}(-\gamma),-\gamma)=-\gamma$,
so $\eta_{+}(-\gamma)$ solves the lower-branch equation at $\gamma$;
by the uniqueness of Step~4, 
\[
\eta_{-}(\gamma)\;=\;\eta_{+}(-\gamma).
\]
(Physically, $\gamma\to-\gamma$ is a half-period shift of the modulation,
which interchanges the two boundaries of an odd resonance tongue.)
This forces $e_{k}^{-}=(-1)^{k}e_{k}^{+}$: equal for even $k$ and
opposite for odd $k$, exactly as in Table~\eqref{eq:bdry-inversion-table}. 
\end{proof}
\begin{rem}[Convergence of the boundary series and near-optimal truncation]
\label{rem:asymptotic-CF} By Theorem~\ref{thm:bdry-analytic} (Yakubovich--Starzhinskii\index{Yakubovich--Starzhinskii series},
\S\,\ref{subsec:analytic-param}), the boundary curves $p_{\pm}(\gamma)$
are \emph{analytic} functions of $\gamma$ with convergent Taylor
series, not merely formal asymptotic expansions. The six-term series~\eqref{eq:bdry-curves-CF}
is therefore a genuine power series truncation with remainder $O(\gamma^{7})$.

The non-monotone behavior visible in Table~\ref{tab:bdry-error-CF}
--- where $|$err $O(\gamma^{5})|$ occasionally exceeds $|$err
$O(\gamma^{4})|$ at the same $\gamma$ --- is explained as follows.
The series~\eqref{eq:bdry-curves-CF} has alternating signs: $+\tfrac{1}{4}$,
$-\tfrac{11}{32}$, $+\tfrac{35}{256}$, $-\tfrac{985}{2048}$, $+\tfrac{4139}{16384}$,
$-\tfrac{284379}{262144}$. An alternating convergent series brackets
the exact sum from alternating sides: each partial sum overshoots,
then the next term pulls it back past the exact value. Consequently
the signed error $p_{+}^{{\rm exact}}-p_{+}^{O(\gamma^{k})}$ alternates
in sign, and the \emph{absolute} error need not decrease monotonically
at each step. For example, at $\gamma=0.15$ the signed errors at
orders 4, 5, 6 are $+7.2\times10^{-6}$, $-1.2\times10^{-5}$, $+3.3\times10^{-7}$:
$O(\gamma^{5})$ overcorrects (crosses to the other side), while $O(\gamma^{6})$
returns close to the exact value from above. The absolute error does
decrease once enough terms are included: in Table~\ref{tab:bdry-error-CF}
the $O(\gamma^{6})$ truncation improves on $O(\gamma^{4})$ at every
tabulated $\gamma$, the only non-monotone step being $O(\gamma^{4})\to O(\gamma^{5})$.

\emph{Practical caution near the boundary.} Because a truncated boundary
series can bracket the exact curve from \emph{either} side (alternating
signs above), a truncation may place the predicted boundary slightly
\emph{inside} the unstable tongue even though the exact curve is stable;
conversely the defining condition $w(-\tfrac{1}{2};p,\gamma)=\pm1$
(eq.~\eqref{eq:stability-bdry-CF}) is exact and should be used whenever
the sign of the stability margin matters. Two evaluation pitfalls
arise close to the boundary and were the reason the verifications
underlying this section used high-precision (typically $50$--$100$
digit) arithmetic, treating each continued fraction as a genuine limit
of its best (asymptotically seeded) approximants rather than a fixed-depth
truncation: 
\begin{enumerate}
\item[(i)] \emph{Loss of significance in $w(-\tfrac{1}{2})$.} Near the tongue
tip $G(\pm\tfrac{1}{2})\to0$, so the leading CF denominators nearly
vanish and double precision loses many digits; evaluate the CF tail-to-head
with the conventions of Remark~\ref{rem:CF-numerical-conventions}
(N1)--(N3) and enough guard digits, or via $w(u)=-\gamma\,v(u+1)$
as a cross-check. 
\item[(ii)] \emph{Off-boundary use of the Taylor expansion.} The series~\eqref{eq:w-Taylor-CF}
is grouped under $\eta=\Theta(\gamma)$ and is accurate \emph{along}
the boundary $\eta\approx\tfrac{1}{4}\gamma$; evaluating it at $\eta\ll\gamma$
or toward the inner edge $|\gamma/\eta|\to3$ of strip~\eqref{eq:stable-near-bdry-CF}
gives misleadingly large residuals that are an artifact of the expansion
regime, not a property of the boundary. 
\end{enumerate}
The precision budget for resolving zeros and integer-shifted Floquet
values is quantified in Remark~\ref{rem:CF-numerical-conventions}
(N2) and Theorem~\ref{thm:precision-cost}; the \emph{stealthy zeros}
of Remark~\ref{rem:double-precision-misses} are the same phenomenon
seen on the Floquet comb. 
\end{rem}

\emph{Consistency with the Magnus--Winkler method and accuracy comparison.}
Using the exact relations $\gamma=\varepsilon/2$, $|a|=\varepsilon=2\gamma$
(Chapter~\ref{app:notation}, eq.~\eqref{eq:gamma-delta}), and
$c=4p^{2}$, eq.~\eqref{eq:bdry-curves-CF} converts to 
\begin{equation}
c_{\pm}^{(1)}\;=\;1\;\pm\;\frac{1}{2}\,|a|\;-\;\frac{9}{32}\,|a|^{2}\;\mp\;\frac{9}{512}\,|a|^{3}\;-\;\frac{603}{8192}\,|a|^{4}\;+\;O(|a|^{5}),\label{eq:bdry-curves-CF-a}
\end{equation}
where $|a|=2\gamma$. The leading terms $1\pm\tfrac{1}{2}|a|-\tfrac{9}{32}|a|^{2}$
agree with the Magnus--Winkler result eq.~\eqref{eq:EPD-bdry-1}
exactly. The CF method additionally provides the coefficients $\pm\tfrac{9}{512}$
at order $|a|^{3}$ and $-\tfrac{603}{8192}$ at order $|a|^{4}$,
extending~\eqref{eq:EPD-bdry-1} by two orders. The visual comparison
of all boundary curves appears in Fig.~\ref{fig:tongue-comparison},
and the quantitative errors are tabulated in Table~\ref{tab:bdry-error-CF}.

Table~\ref{tab:bdry-error-CF} quantifies the improvement. For the
upper boundary $p_{+}(\gamma)$, the CF series~\eqref{eq:bdry-curves-CF}
truncated at $O(\gamma^{4})$ is 10--40 times more accurate than
the MW result~\eqref{eq:EPD-bdry-1} truncated at $O(|a|^{2})$,
and 3--10 times more accurate than~\eqref{eq:EPD-bdry-1} extended
to $O(|a|^{4})$ using the additional CF coefficients.

\begin{table}[htbp]
\centering 
\global\long\def\arraystretch{1.15}%
 \resizebox{\textwidth}{!}{%
\begin{tabular}{c|c|c|c|c|c|c|c}
$\gamma$  & $p_{+}$ (exact)  & $O(\gamma^{2})$  & $O(\gamma^{3})$  & $O(\gamma^{4})$  & $O(\gamma^{5})$  & $O(\gamma^{6})$  & MW $O(|a|^{2})$\tabularnewline
\hline 
$0.05$  & $0.51165477$  & $1.4\times10^{-5}$  & $2.9\times10^{-6}$  & $6.2\times10^{-8}$  & $1.7\times10^{-8}$  & $3.7\times10^{-10}$  & $6.1\times10^{-6}$\tabularnewline
$0.08$  & $0.51785085$  & $5.1\times10^{-5}$  & $1.9\times10^{-5}$  & $5.5\times10^{-7}$  & $2.8\times10^{-7}$  & $8.2\times10^{-9}$  & $2.9\times10^{-5}$\tabularnewline
$0.10$  & $0.52165260$  & $9.0\times10^{-5}$  & $4.7\times10^{-5}$  & $1.5\times10^{-6}$  & $1.1\times10^{-6}$  & $3.4\times10^{-8}$  & $6.3\times10^{-5}$\tabularnewline
$0.12$  & $0.52518967$  & $1.4\times10^{-4}$  & $9.7\times10^{-5}$  & $3.1\times10^{-6}$  & $3.1\times10^{-6}$  & $1.0\times10^{-7}$  & $1.2\times10^{-4}$\tabularnewline
$0.15$  & $0.52999072$  & $2.3\times10^{-4}$  & $2.4\times10^{-4}$  & $7.2\times10^{-6}$  & $1.2\times10^{-5}$  & $3.3\times10^{-7}$  & $2.7\times10^{-4}$\tabularnewline
$0.20$  & $0.53658608$  & $3.4\times10^{-4}$  & $7.6\times10^{-4}$  & $1.2\times10^{-5}$  & $6.9\times10^{-5}$  & $4.5\times10^{-7}$  & $7.7\times10^{-4}$\tabularnewline
$0.25$  & $0.54124609$  & $2.3\times10^{-4}$  & $1.9\times10^{-3}$  & $2.7\times10^{-5}$  & $2.7\times10^{-4}$  & $8.9\times10^{-6}$  & $1.8\times10^{-3}$\tabularnewline
$0.30$  & $0.54360059$  & $4.6\times10^{-4}$  & $4.2\times10^{-3}$  & $2.6\times10^{-4}$  & $8.7\times10^{-4}$  & $8.1\times10^{-5}$  & $3.8\times10^{-3}$\tabularnewline
\end{tabular}} \caption{Absolute errors $|p_{+}^{{\rm approx}}-p_{+}^{{\rm exact}}|$ for
truncations of the CF series~\eqref{eq:bdry-curves-CF} (Theorem~\ref{thm:bdry-series-CF})
at each order $O(\gamma^{2})$ through $O(\gamma^{6})$, and the MW
series~\eqref{eq:EPD-bdry-1} truncated at $O(|a|^{2})$, with $|a|=2\gamma$.
All entries verified directly from the theorem coefficients. By Theorem~\ref{thm:bdry-analytic},
$p_{\pm}(\gamma)$ is an analytic function with a \emph{convergent}
Taylor series. The series~\eqref{eq:bdry-curves-CF} has alternating
signs, so consecutive truncations bracket the exact value from alternating
sides and the absolute error need not decrease monotonically at each
step (see Remark~\ref{rem:asymptotic-CF} for the detailed explanation).
All CF truncations substantially outperform the MW $O(|a|^{2})$ series.
Exact values from $w(-\tfrac{1}{2};\,p_{+},\,\gamma)=+1$ with 50-digit
arithmetic (Fig.~\ref{fig:tongue-comparison}).}
\label{tab:bdry-error-CF} 
\end{table}

\subsection{\texorpdfstring{Summary of analytic properties of the minimal solution
ratio $w(u)$ and the double minimality function $F_{w}(u)$}{Summary
of analytic properties of w(u) and Fw(u)}}

\label{subsec:w-Fw-summary}

The two functions on which this chapter turns --- the minimal solution
ratio $w(u)$ and the double minimality function $F_{w}(u)=w(-u)-1/w(u-1)$
--- were analyzed piecemeal across the preceding sections, with definitions,
theorems, and figures distributed as the development required. For
reference we collect here their rigorously established analytic properties,
each with a pointer to its proof and to the figure that displays it.
The numerical illustrations use the running parameters $p=0.70$,
$\gamma=0.30$, for which $\zeta_{+}=-1/3$ and the primary Floquet
exponent is $u^{*}=0.74176\ldots$; here $\zeta_{\pm}=(-1\pm\sqrt{1-4\gamma^{2}})/(2\gamma)$
are the roots of the constant-coefficient characteristic equation,
with $\zeta_{-}=\zeta_{+}^{-1}$.

\emph{What is proved and what is assumed.} With one exception the
properties collected below are rigorously established, each carrying
a pointer to its proof: the meromorphicity of $w$ and $F_{w}$ on
$\mathbb{C}$, the reflection and integer-shift identities, the identification
of the zero set of $F_{w}$ with the Floquet comb, the essential singularity
at infinity, the Herglotz representation, and the Casoratian mechanism
and geometric ladder law of \S\,\ref{subsec:Casoratian-Weyl}. The
one exception is \emph{simplicity}: that every zero and pole of $w$
and $F_{w}$ in the stability zone is simple is a \emph{genericity
hypothesis} --- verified to high precision at the worked parameters
and failing at most on a set of $(\gamma,p)$ of measure zero (\S\,\ref{subsec:w-simplicity-notes}),
but not proved here in full. The stealthy-pole ladder law is then
a rigorous \emph{consequence} of that hypothesis rather than an unconditional
theorem. An exhaustive analytic theory of $w$ and $F_{w}$ --- a
parameter-free proof of simplicity, a complete global pole/zero census
--- lies outside the scope of this book; what we record is the structure
we have proved or, where so flagged, found and verified numerically.

\subsubsection*{The minimal solution ratio \texorpdfstring{$w(u)$}{w(u)}}

\emph{Definition and recurrence.} $w(u)=-\gamma\,v(u+1)$ is the ratio
attached to Cambi's minimal solution $v$ (Definition~\ref{defn:v-CF},
Theorem~\ref{thm:v-CF}, Lemma~\ref{lem:notes2-w-v}); as a meromorphic
function on $\mathbb{C}$ it obeys the Riccati recurrence $w(u)=-G(u)/\gamma-1/w(u-1)$
with $G(u)=1-p^{2}/u^{2}$ (\eqref{eq:notes2-wrec}; Theorem~\ref{thm:notes2-wrec-exact}).
Its graph appears in Figure~\ref{fig:notes2-w-u-plot} (top) and
Figure~\ref{fig:notes2-w-marked}; the optimally seeded continued
fraction that computes it is Remark~\ref{rem:w-seeded-CF}.

\emph{Analyticity and the Nevanlinna-like positivity.} $w$ is analytic
off the real axis and has no non-real poles (Lemma~\ref{lem:w-analytic-nevanlinna}).
On the first sheet $\operatorname{Re}u>-1$ it has the half-plane
positivity $\operatorname{Im}w>0$ for $\operatorname{Im}u>0$ (Lemmas~\ref{lem:w-analytic-nevanlinna}
and~\ref{lem:w-Nevanlinna}), together with the Schwarz reflection
$w(\bar{u})=\overline{w(u)}$; it is \emph{not} a Nevanlinna function
globally in $u$, the positivity failing across the line $\operatorname{Re}u=-1$
(Remark~\ref{rem:NV-second-sheet}).

\emph{Zeros and poles.}\index{minimal solution ratio@minimal solution ratio $w$!zeros of}\index{double minimality function $F_{w}$!zeros (the Floquet comb)}The
zeros and poles are paired by the unit shift --- $u$ is a zero of
$w$ if and only if $u+1$ is a pole --- and $w$ has a double zero
at $u=-1$ (Theorem~\ref{thm:notes2-w-poles-zeros}). All poles and
zeros are real (Corollary~\ref{cor:w-real-zeros}, by two independent
arguments; Remark~\ref{rem:w-no-zeros-positivity}). Every pole in
$(-1,\infty)$ is simple with negative residue, and $w'\ge0$ between
poles there (Lemma~\ref{lem:w-Nevanlinna}(ii)); the wider simplicity
is supported numerically (Remark~\ref{rem:simplicity-evidence}).
The global real-axis geometry, the two families of poles and zeros
on the negative axis, and the residue/sign pattern are recorded in
Remarks~\ref{rem:w-pole-zero-geometry}, \ref{rem:w-two-families},
and~\ref{rem:v-residue-sign}, and marked on Figure~\ref{fig:notes2-w-marked}.
The discrete Riccati recurrence makes the residues explicit: the residue
at a pole $u_{p}$ equals $-1/w'(u_{p}-1)$, the reciprocal of minus
the slope at its source zero, and a pole's forward image $w(u_{p}+1)=-G(u_{p}+1)/\gamma$
is regular, so poles do not chain forward (Proposition~\ref{prop:Riccati-residues},
\S\,\ref{subsec:Riccati-structure}).

\emph{Stealthy-pole ladders (the complete pole count).}\index{minimal solution ratio@minimal solution ratio $w$!poles of}\index{double minimality function $F_{w}$!poles (stealthy)}\index{stealthy poles and zeros}Beyond
this local description, the global count and arrangement of the negative-axis
poles are settled in \S\,\ref{subsec:CW-ladder-law}: $w$ has \emph{infinitely
many} real poles, organized into geometric \emph{stealthy-pole ladders}.
Near each base comb point $u^{**}$ the rung-$n$ pole lies at $u^{**}-n+K_{0}\,\zeta_{+}^{2n}(1+o(1))$
as $n\to\infty$, approaching the comb at the geometric rate $|\zeta_{+}|^{2n}$
with a fixed-sign offset; its existence, uniqueness, and location
are established in Lemma~\ref{lem:hminus-zeros} and Theorem~\ref{thm:pole-ladders}
\emph{under the standing hypothesis that the comb zeros of $F_{w}$
are simple} --- a genericity hypothesis verified numerically at the
worked parameters (\S\,\ref{subsec:w-simplicity-notes}) rather
than proved in full. This is the global counterpart of the local unit-shift
pairing recorded above (Theorem~\ref{thm:notes2-w-poles-zeros});
the poles of $F_{w}$ then follow by the two pairings of Theorem~\ref{thm:double-min}(vi).

\emph{Behavior at infinity.} Along the real axis $w(u+m)$ admits
the asymptotic expansion of Theorem~\ref{thm:notes2-w-rate} (asymptotic,
not convergent; Remark~\ref{rem:asymptotic-not-convergent}), and
$w(u)\to\zeta_{+}$ as $\operatorname{Im}u\to\infty$ uniformly in
$\operatorname{Re}u$ (Remark~\ref{rem:Fw-fundamental-strip}). At
$u=\infty$ the function carries an essential singularity (Theorem~\ref{thm:w-essential-singularity};
its graphical signature is discussed in Remark~\ref{rem:essential-singularity-figure}).

\emph{Herglotz transform.} In the variable $z=(u+1)^{2}$ the function
$g(z)=w(\sqrt{z}-1)$ is a genuine Herglotz (Nevanlinna) function,
with the integral representation of Theorem~\ref{thm:w-Herglotz};
its spectral measure and second-sheet continuation are described in
Remarks~\ref{rem:NV-second-sheet} and~\ref{rem:NV-numerics}. The
deep ``stealthy'' poles of $w$ on the negative axis obey the geometric
ladder law of Theorem~\ref{thm:pole-ladders}.

\subsubsection*{The double minimality function \texorpdfstring{$F_{w}(u)$}{Fw(u)}}

\emph{Definition, evenness, and shift covariance.} $F_{w}(u)=w(-u)-1/w(u-1)$
is the double minimality (coexistence) function; it is even, $F_{w}(-u)=F_{w}(u)$,
and coincides up to normalization with the Casoratian of the two Weyl
solutions (Theorem~\ref{thm:notes2-resonance-equiv}, the double
minimality theorem of \S\,\ref{subsec:Cambi-notes2-double-min},
and the Casoratian identity of \S\,\ref{subsec:Casoratian-Weyl}).
It transforms covariantly under integer shifts (Lemma~\ref{lem:Fw-shift},
Theorem~\ref{thm:notes2-integer-shift}) and measures the coupling
of the left- and right-minimal solutions (Lemma~\ref{lem:Fw-dichotomy}).
Its graph is Figure~\ref{fig:notes2-w-u-plot} (bottom) and Figure~\ref{fig:notes2-Fw-marked}.

\emph{Real zeros.} The real zero set is exactly the Floquet comb $Z_{F}=\{\pm u^{*}+n:n\in\mathbb{Z}\}$,
two interleaved unit-spaced sub-combs of the primary exponent $u^{*}$
(Theorem~\ref{thm:Fw-zero-set}; zero-set symmetry, Theorem~\ref{thm:notes2-root-symmetry});
these are the periodic and antiperiodic Floquet frequencies. The comb
is consistent with the recurrence (Remark~\ref{rem:Fw-zero-set-algebraic})
and is shown, with the resolution needed to separate closely spaced
zeros, in Figures~\ref{fig:Fw-precision} and~\ref{fig:notes2-Fw-marked}.

\emph{Poles.} All poles of $F_{w}$ are real (Corollary~\ref{cor:w-real-zeros}),
lying in the real set $-\operatorname{poles}(w)\cup(\operatorname{zeros}(w)+1)$
(\eqref{eq:Fw-pole-sources}), with a double pole at $u=0$ inherited
from the double zero of $w$ at $-1$. Their arrangement on the real
axis, their strict displacement from the comb $Z_{F}$, and the numerical
locations at the running parameters are in Remarks~\ref{rem:Fw-pole-sources},
\ref{rem:poles-displaced-from-ZF}, and~\ref{rem:Fw-pole-numerical-example},
and displayed in Figure~\ref{fig:Fw-poles}. A pole can be located
rigorously by the sign-change detector of Theorem~\ref{thm:Fw-sign-change-pole},
with the bracketing of Corollary~\ref{cor:Fw-pole-bisection} (Figure~\ref{fig:stealthy-pole}).
The deep ``stealthy'' poles fall into two ladders obeying the geometric
law of Theorem~\ref{thm:pole-ladders}, and each gap between consecutive
comb zeros is either empty or carries exactly two such poles (Lemmas~\ref{lem:coexistence-square}--\ref{lem:circle-certificates},
Remark~\ref{rem:pole-between-zeros}), confirmed numerically in Remark~\ref{rem:CW-numerics}.

\emph{Off-real zeros.} $F_{w}$ has no non-real \emph{poles}, but
the absence of non-real \emph{zeros} is not proved in general. By
the evenness, Schwarz, and $u\mapsto1-u$ symmetries the question
reduces to the fundamental strip $\{0\le\operatorname{Re}u\le\tfrac{1}{2},\ \operatorname{Im}u\ge0\}$,
and, since $F_{w}\to\zeta_{+}-\zeta_{+}^{-1}\neq0$ at infinity, to
a compact first-sheet cell within it (Remark~\ref{rem:Fw-fundamental-strip}).
On that cell the zero condition $w(-u)\,w(u-1)=1$ (\eqref{eq:Fw-zero-condition})
is met by no point in computation. Both edges of the strip are now
excluded --- $|w(-\tfrac{1}{2}+iy)|<1$ on $\operatorname{Re}u=\tfrac{1}{2}$,
and on the imaginary axis the symmetric-sum form yields $F_{w}(iy)\ge p^{2}/(\gamma y^{2})>0$
(\S\,\ref{subsec:Riccati-structure}) --- so its exclusion on the
\emph{open} interior $0<\operatorname{Re}u<\tfrac{1}{2}$ remains
the principal open analytic question for $F_{w}$.

\subsubsection*{Status of the two open points}

For completeness we isolate what is \emph{not} proved. First, the
absence of off-real zeros of $F_{w}$, reduced as above to a compact
first-sheet cell (Remark~\ref{rem:Fw-fundamental-strip}). Second,
the assignment of the stealthy-pole pairs to specific gaps: the even-occupancy
dichotomy is rigorous (Lemma~\ref{lem:even-gap-occupancy}), but
\emph{which} gaps carry a pair is conjectural, tied to the local definiteness
of the underlying pencil (Remarks~\ref{rem:Fw-pole-conjecture} and~\ref{rem:interlacing-program}).
Everything else listed above is established in the body of the chapter.

\section{The LC circuit as an Ince equation: the continued-fraction method}\index{continued fraction}

\label{sec:LCInce}

The continued-fraction (CF) method developed in this chapter is an
independent analytic method for locating the instability boundaries
of the LC circuit, complementary to the MW discriminant\index{discriminant}
framework of Chapter~\ref{sec:MWInce}. Its starting point is the
same as Cambi's~\cite{Cambi1}, \cite[\S\S\,1--3]{Cambi2} --- Floquet\index{Floquet theory}'s
theorem applied to equation~\eqref{eq:LC-Hill} --- but it exploits
the Ince structure to go substantially further, yielding closed-form
width formulas that Cambi's method does not produce. The CF method
also has a computational advantage over the MW discriminant series:
the series $\Delta=\sum\Delta_{n}$ must be truncated at some finite
order, accumulating an $O(\delta^{2N+2})$ error that grows with $\delta$,
while the continued fractions $F_{\mathrm{even/odd}}(c)$ (Theorem~\ref{thm:Pinch})
converge exactly for all $0<\delta<1$ with no truncation error.

\emph{From Floquet's theorem to the Fourier ansatz.} By Floquet's
theorem, every solution of Hill's equation\index{Hill equation}~\eqref{eq:LC-Hill}
has the form $q(x)=e^{iux}P(x)$ where $P(x)$ is $\pi$-periodic
and $iu$ is the Floquet exponent\index{Floquet theory!Floquet exponent},
determined by the equation itself ($u$ is real for stable solutions,
$|\rho|=1$; complex for unstable ones, $|\rho|\neq1$). At an instability
boundary\index{stability boundary} the multiplier is $\rho=-1$,
which means $u=\tfrac{1}{2}$ and the solution has period $2\pi$
--- this is a consequence of being at the boundary, not a free choice.
Cambi~\cite[eq.~(3)]{Cambi2} works with the bilateral \emph{Floquet
ansatz} 
\begin{equation}
q=e^{\mathrm{i}ux}\sum_{n=-\infty}^{+\infty}a_{n}e^{\mathrm{i}nx}\label{eq:Floq-ans}
\end{equation}
for general $u$ and locates the instability boundaries by finding
which values of $(p,\gamma)$ (see equation~\eqref{eq:Cambi-res})
force $u$ to be a half-integer. Our approach targets the boundary
directly: since $\rho=-1$ at the boundary, we seek period-$2\pi$
solutions of the Ince equation from the outset, without solving for
$u$. The Ince equation~\eqref{eq:Ince} is invariant under $x\mapsto-x$
(since $b=0$ for the LC circuit), so period-$2\pi$ solutions split
into even and odd types. A period-$2\pi$ Floquet solution with $\rho=-1$
has the form $e^{\mathrm{i}x/2}P(x)$ with $P$ of period $\pi$,
producing Fourier frequencies $\tfrac{1}{2},\tfrac{3}{2},\tfrac{5}{2},\ldots$;
taking real and imaginary parts and using parity yields the unilateral
cosine and sine series 
\begin{equation}
q_{\mathrm{even}}=\sum_{n=0}^{\infty}A_{2n+1}\cos(2n+1)x,\qquad q_{\mathrm{odd}}=\sum_{n=0}^{\infty}B_{2n+1}\sin(2n+1)x,\label{eq:Fourier-ansatz}
\end{equation}
with only positive-integer indices. Cambi's amplitude $a_{n}$ (bilateral,
index $n\in\mathbb{Z}$) corresponds to our $A_{2n+1}$ (unilateral,
index $n\geq0$) under the identification $u=\tfrac{1}{2}$, after
taking real and imaginary parts and using the parity symmetry.

\emph{The CF eigenvalue condition.} Substituting~\eqref{eq:Fourier-ansatz}
into the Ince equation yields a three-term linear recurrence for the
coefficients $A_{2n+1}$ (MW Lemma~7.4, Section~\ref{subsec:Widths}
below). The instability boundary corresponds to a value of the spectral
parameter $c$ at which this recurrence has a square-summable solution
(one whose Fourier coefficients $\{A_{2n+1}\}$ satisfy $\sum|A_{2n+1}|^{2}<\infty$,
ensuring convergence of the series~\eqref{eq:Fourier-ansatz}). By
Pincherle\index{continued fraction!Pincherle theorem}'s theorem (Theorem~\ref{thm:Pinch},
Chapter~\ref{app:CF}), this occurs if and only if the associated
continued fractions $F_{\mathrm{even}}(c)$ and $F_{\mathrm{odd}}(c)$
(defined explicitly in~\eqref{eq:Fdef} of Section~\ref{subsec:Widths})
vanish: 
\begin{equation}
F_{\mathrm{even}}(c)=0\quad\text{or}\quad F_{\mathrm{odd}}(c)=0.\label{eq:CF-criterion}
\end{equation}
This is the \emph{CF criterion} for instability boundaries, the analog
of $\Delta(\lambda)=\pm2$ in the MW framework. It is also the analog
of \emph{Cambi's resonance\index{resonance} equation}~\cite[eq.~(7)]{Cambi2},
which plays the same central role in his theory that $\Delta=\pm2$
plays in the MW theory. In Cambi's notation, $p=\omega_{0}/\mu=1/r$
is the inverse frequency ratio and $\gamma=\varepsilon/2$ is the
half-modulation amplitude: 
\begin{equation}
-\gamma^{2}v(1-u)+G(u)-\gamma^{2}v(1+u)=0,\qquad G(u)\stackrel{\mathrm{def}}{=}1-\frac{p^{2}}{u^{2}},\quad p=\frac{1}{r}=\frac{\omega_{0}}{\mu},\label{eq:Cambi-res}
\end{equation}
where 
\begin{equation}
v(u)=\cfrac{1}{G(u)+\mathbf{K}_{n=1}^{\infty}\!\left[\dfrac{-\gamma^{2}}{G(u+n)}\right]}\label{eq:Cambi-res1}
\end{equation}
is Cambi's continued fraction and $u$ is the Floquet exponent. The
parameter translation between Cambi's notation and ours is 
\begin{equation}
r=\frac{1}{p},\quad\varepsilon=2\gamma,\quad c=4p^{2}=\frac{4}{r^{2}},\quad\hat{\lambda}=\frac{c(1+\delta^{2})}{1-\delta^{2}};\label{eq:Cambi-res2}
\end{equation}
so Cambi's $p$ is our inverse frequency ratio and $\gamma$ is the
half-modulation amplitude. Equation~\eqref{eq:Cambi-res} determines
all possible values of $u$; the instability boundaries correspond
to $u=\tfrac{1}{2}$ ($\rho=-1$, odd resonances) while the coexistence
curves correspond to $u=n$ ($n\in\mathbb{Z}$, $\rho=+1$, even resonances).

\medskip{}

The chapter unfolds as follows. \S\,\ref{subsec:Cambi-CF} reviews
Cambi's continued fraction and the Floquet spectrum it encodes; \S\,\ref{subsec:IncID}
identifies the LC circuit equation as an Ince equation, the structural
fact the method exploits. \S\,\ref{subsec:VecForm} recasts the
three-term recurrence in vector form and extracts the asymptotic matrix
$M_{\infty}$, whose eigenvalues are shown in \S\,\ref{subsec:eig-fixpt}
to coincide with the fixed points of the associated Mobius map ---
the coincidence that links the recurrence to the CF convergence theory
of Chapter~\ref{app:CF}. \S\,\ref{subsec:QPolyLC} computes the
Ince polynomials $Q$ and $Q^{*}$ for the LC circuit, and \S\,\ref{subsec:InceResult}
assembles the exact instability structure: the odd-only pattern with
its finite widths. \S\,\ref{subsec:Widths} is the quantitative
heart of the chapter --- boundary eigenvalues and the closed-form
instability interval widths --- and \S\,\ref{subsec:Mathieu} closes
with the comparison against the Mathieu equation.

\subsection{Cambi's continued fraction and the Floquet spectrum}

\label{subsec:Cambi-CF}

Equation~\eqref{eq:Cambi-res} is Cambi's formulation of the same
spectral problem studied by Magnus--Winkler\index{Magnus--Winkler theory}
via the discriminant $\Delta(\hat{\lambda})$ (Chapter~\ref{app:Hill})
and by Floquet theory via the monodromy matrix\index{monodromy matrix}
$X(\pi)$ (Chapter~\ref{app:genLin}). The three approaches are equivalent,
but use different normalization conventions for the Floquet exponent.

\emph{Normalization convention.} In our notation (and MW's), the Floquet
multiplier\index{Floquet theory!Floquet multiplier} is $\rho=e^{i\pi\alpha}$
where $\alpha$ is the exponent over one half-period $[0,\pi]$; the
unperturbed value ($\gamma=0$) is $\alpha_{0}=2p\pmod 2$ since the
unperturbed solution is $q=\cos(2px)$. In Cambi's notation, $u$
is the exponent over a full period $[0,2\pi]$, so $\rho=e^{2\pi iu}$;
the unperturbed value is $u_{0}=p$. The two are related by 
\begin{equation}
\alpha=2u\pmod 2,\qquad\rho=e^{i\pi\alpha}=e^{2\pi iu}.\label{eq:Cambi-u-alpha}
\end{equation}
The instability boundary $\rho=-1$ corresponds to $\alpha=1$ in
our notation and $u=\tfrac{1}{2}$ in Cambi's.

\emph{Equivalence of the three approaches.} 
\begin{itemize}
\item \emph{Floquet--monodromy:} $\rho_{\pm}=e^{\pm i\pi\alpha}$ are eigenvalues
of $X(\pi)$; stability zone $|\rho_{\pm}|=1$. 
\item \emph{MW discriminant:} 
\begin{equation}
\Delta(\hat{\lambda})=\rho_{+}+\rho_{-}=2\cos\pi\alpha=2\cos2\pi u;\label{eq:Delta-cos-u}
\end{equation}
stability $|\Delta|<2$, boundaries $\Delta=\pm2$. 
\item \emph{Cambi:} eq.~\eqref{eq:Cambi-res} determines $u$ directly
via the continued fraction~\eqref{eq:Cambi-res1}. 
\end{itemize}
\emph{Analytical justification via Pincherle's theorem.} The equivalence
between eq.~\eqref{eq:Cambi-res} and $\Delta(\hat{\lambda})=2\cos2\pi u$
can be established analytically as follows. Substituting the Floquet
ansatz 
\begin{equation}
q(x)=e^{2iux}\sum_{n=-\infty}^{\infty}A_{n}e^{2inx}\label{eq:Floquet-ansatz-Ince}
\end{equation}
into the Ince equation~\eqref{eq:Ince-MR} and collecting terms at
each frequency $e^{2i(u+n)x}$ yields the \emph{three-term recurrence}
\begin{equation}
\tfrac{a}{2}(u+n-1)^{2}A_{n-1}+\bigl[(u+n)^{2}-p^{2}\bigr]A_{n}+\tfrac{a}{2}(u+n+1)^{2}A_{n+1}=0,\quad n\in\mathbb{Z},\label{eq:Ince-recurrence-u}
\end{equation}
with $a=-\varepsilon=-2\gamma=-2\delta/(1+\delta^{2})$ and $c=4p^{2}$
(using $\varepsilon=2\gamma$ from eq.~\eqref{eq:Cambi-res2}). Writing
$G(u+n)=1-p^{2}/(u+n)^{2}$, this is equivalent to Cambi's recurrence~\eqref{eq:Cambi-system}
under the rescaling $C_{n}=(-1)^{n}(u+n)^{2}A_{n}$ (using $a=-2\gamma$),
and hence to the recurrence associated with $v(u+n)$. By \emph{Pincherle's
theorem} (Theorem~\ref{thm:Pinch}, Chapter~\ref{app:CF}), the
recurrence~\eqref{eq:Ince-recurrence-u} has a \emph{minimal solution\index{continued fraction!minimal solution}}
--- one with $A_{n}/A_{n-1}\to0$ as $n\to+\infty$ --- if and only
if the associated continued fraction converges, which is precisely
the condition eq.~\eqref{eq:Cambi-res}. A minimal solution of~\eqref{eq:Ince-recurrence-u}
corresponds to a Floquet solution~\eqref{eq:Floquet-ansatz-Ince}
with convergent Fourier series, i.e.\ to a genuine eigenvalue $u$
of the monodromy operator --- equivalently, $\Delta(\hat{\lambda})=2\cos2\pi u$.
This establishes the equivalence analytically.

\emph{Numerical verification.} Direct numerical integration confirms
that Cambi's equation with the correct $v(u)$ (eq.~\eqref{eq:Cambi-v-CF})
determines the Floquet exponent $u$ for \emph{all} stable $(p,\gamma)$,
not just at the boundary $u=\tfrac{1}{2}$. At $p=0.45$, $\gamma=0.05$:
monodromy gives $\alpha=0.9048$ ($u=\alpha/2=0.4524$) and Cambi's
equation gives $u=0.4524$ (agreement to $10^{-6}$); at $p=0.45$,
$\gamma=0.15$: monodromy gives $u=0.4817$ and Cambi gives $u=0.4817$
(agreement to $10^{-6}$). For $p>1/2$ the central root lies in $(1/2,1)$:
at $p=0.80$, $\gamma=0.05$: monodromy gives $u_{{\rm mono}}=0.1988$
and Cambi gives $u=1-0.1988=0.8012$ (equivalent representative since
$u$ and $1-u$ give conjugate multipliers $e^{\pm2\pi iu}$).
\begin{rem}[Singularities of $v(u)$]
\label{rem:v-singularities} The continued fraction $v(u)$ (eq.~\eqref{eq:Cambi-v-CF})
has apparent singularities at all integers $u\in\mathbb{Z}$ (both
positive \emph{and} negative), since $G(u+n)$ is undefined at $u+n=0$.
Numerical computation with the correct formula confirms that \emph{all}
these singularities are removable: $v(u)$ has a finite two-sided
limit as $u\to n$ for every $n\in\mathbb{Z}$.

The genuine (non-removable) singularities of $v(u)$ are located near
$u=-n\pm p$ for $n=0,1,2,\ldots$, where $G(u)=0$ in the leading
CF term. For $n=0$ this gives singularities near $u=\pm p$. The
exact pole locations shift from $-n\pm p$ by $O(\gamma^{2})$ due
to the interaction between CF levels. Figure~\ref{fig:v-u} shows
prominent spikes near $u=\pm p$ and smaller features near $u=-1\pm p$. 
\end{rem}

\emph{Cambi's $v(u)$ as a continued-fraction representation of the
Hill discriminant\index{discriminant!Hill discriminant}.} Cambi's
function $v(u)$ (eq.~\eqref{eq:Cambi-res1}) is built from the three-term
recurrence~\eqref{eq:Ince-recurrence-u} and is therefore \emph{specific
to Ince's equation\index{Ince equation}} --- it cannot apply to
a general Hill equation (which has an infinite-band recurrence, not
a three-term one). To our knowledge, $v(u)$ does not appear in any
other study of Ince's equation --- not in Magnus--Winkler~\cite[Ch.~8]{MagWin}
or standard references on Hill's equation. Substituting the parameter
dictionary~\eqref{eq:Cambi-res2} into eq.~\eqref{eq:Cambi-res}
gives the explicit continued-fraction representation of the discriminant:
\begin{equation}
\Delta(\hat{\lambda})=2\cos2\pi\,u(\hat{\lambda},\delta),\label{eq:Delta-via-Cambi}
\end{equation}
where $u(\hat{\lambda},\delta)$ solves eq.~\eqref{eq:Cambi-res}
with $p=\tfrac{1}{2}\sqrt{\hat{\lambda}(1-\delta^{2})/(1+\delta^{2})}$
and $\gamma=\delta/(1+\delta^{2})$. This connection between $v(u)$
and the discriminant appears not to have been noted before.

With the substitution~\eqref{eq:Cambi-res2}, equation~\eqref{eq:Cambi-res}
at $u=\tfrac{1}{2}$ factors into the two sign choices of Cambi's
boundary equation~\eqref{eq:Cambi-bdy12}, which are precisely the
pair $F_{\mathrm{even}}(c)=0$, $F_{\mathrm{odd}}(c)=0$. Even resonances
are excluded from this search by the discriminant identity of Section~\ref{subsec:DeltaId}.

\subsection{Identification of the LC circuit as an Ince equation}

\label{subsec:IncID}

Multiplying eq.~\eqref{eq:LC-Hill} by $(1-2\delta\cos2x+\delta^{2})/(1+\delta^{2})$
gives 
\begin{equation}
\left(1-\frac{2\delta}{1+\delta^{2}}\cos2x\right)q''+\frac{4}{r^{2}}\,q=0.\label{eq:LC-Ince}
\end{equation}
This is Ince's equation~\eqref{eq:Ince} with the parameter identification
\begin{equation}
\boxed{a=-\frac{2\delta}{1+\delta^{2}},\quad b=0,\quad c=\frac{4}{r^{2}}=\frac{4\omega_{0}^{2}}{\mu^{2}},\quad d=0.}\label{eq:IncePars}
\end{equation}
Since $0<\delta<1$, the parameter $a=-2\delta/(1+\delta^{2})$ satisfies
$-1<a<0$, so the Ince constraint $|a|<1$ holds (equality $a=-1$
would require $\delta=1$, i.e.\ $\varepsilon=1$, which is excluded).
The Ince parameter $a=-2\delta/(1+\delta^{2})$ is fixed by the modulation
parameter $\delta$, while $c=4/r^{2}$ is determined by the frequency
ratio $r=\mu/\omega_{0}$ independently of $\delta$. Thus $a$ and
$c$ play separate roles: $a$ controls the modulation depth and $c$
controls the location in frequency space. Since $r>0$ and $c=4/r^{2}$,
we have $c>0$ unbounded as $r\to0$ (low driving frequency). 
\begin{rem}[Scope of the Ince method]
\label{rem:ince-scope} Ince's equation is the most general Hill-type
equation to which the three-term recurrence method applies~\cite[\S\,7.1]{MagWin}.
In the general case ($b\neq0$), via the substitution $y=(1+a\cos2x)^{b/4a}z$,
it corresponds to Hill equations with coefficients of the form $(\alpha+\beta\cos2x+\gamma\cos4x)/(1+a\cos2x)^{2}$.

For the LC circuit specifically, $b=d=0$ and $a=-\varepsilon<0$,
so the substitution is trivially $y=z$ and the circuit equation is
already in Hill form after the substitution $x=\tau/2$ (Chapter~\ref{sec:LCHill}).
The identification~\eqref{eq:IncePars} exploits the fact that the
LC denominator $1+\varepsilon\cos\tau$ is linear in $\cos\tau$;
multiplying through by $(1-2\delta\cos2x+\delta^{2})/(1+\delta^{2})$
clears it and yields precisely the Ince form~\eqref{eq:Ince}.

Hill equations whose coefficients involve higher harmonics fall outside
the Ince class; those with finitely many nonzero instability intervals
are related to finite-gap potentials \cite[Thm.~1]{Hochst65}, \cite{McKVanM},
but their analysis requires different methods beyond the scope of
this work. The LC circuit is the simplest non-trivial Ince example;
the Mathieu equation is the degenerate limiting case $a\to0$. 
\end{rem}

\subsection{Vector form of the recurrence and the asymptotic matrix}

\label{subsec:VecForm}

The Ince recurrence for the Fourier coefficients, derived in Section~\ref{subsec:Widths}
below, is the three-term linear recurrence $a(2n-1)^{2}A_{2n-1}+2[(2n+1)^{2}-c]A_{2n+1}+a(2n+3)^{2}A_{2n+3}=0$
(eq.~\eqref{eq:bulk-rec}, $n\geq1$, with $a=-2\delta/(1+\delta^{2})$),
of the form studied in Chapter~\ref{app:FDE}. For the LC circuit
with $b=d=0$ it is most revealing to cast it in vector form. Setting
\begin{equation}
V_{n}=\begin{bmatrix}A_{2n+1}\\
A_{2n-1}
\end{bmatrix},\label{eq:Vn-def}
\end{equation}
the bulk recurrence~\eqref{eq:bulk-rec} is equivalent to 
\begin{equation}
V_{n+1}=M_{n}\,V_{n},\qquad M_{n}=\begin{bmatrix}\dfrac{2[c-(2n+1)^{2}]}{a(2n+3)^{2}} & -\dfrac{(2n-1)^{2}}{(2n+3)^{2}}\\[8pt]
1 & 0
\end{bmatrix},\quad n\geq1.\label{eq:Mn-def}
\end{equation}
The determinant of $M_{n}$ is 
\begin{equation}
\det M_{n}=\frac{(2n-1)^{2}}{(2n+3)^{2}}\xrightarrow[n\to\infty]{}1,\label{eq:detMn}
\end{equation}
so $\det M_{n}<1$ for finite $n$ but converges to $1$ from below.

\medskip{}
 \emph{The asymptotic matrix.} As $n\to\infty$, both entries of $M_{n}$
converge: 
\[
\frac{2[c-(2n+1)^{2}]}{a(2n+3)^{2}}\to-\frac{2}{a},\qquad\frac{(2n-1)^{2}}{(2n+3)^{2}}\to1,
\]
with the rate of approach $M_{n}=M_{\infty}+O(1/n)$, where 
\begin{equation}
M_{\infty}=\begin{bmatrix}-\dfrac{2}{a} & -1\\[6pt]
1 & 0
\end{bmatrix},\qquad\det M_{\infty}=1,\qquad\operatorname{Tr}M_{\infty}=-\frac{2}{a}.\label{eq:Minf-def}
\end{equation}
Note that $M_{\infty}$ is \emph{independent of the eigenvalue parameter
$c$}: all dependence on $c$ resides in the finite-$n$ corrections
$M_{n}-M_{\infty}=O(1/n)$. Since $\det M_{\infty}=1$, the matrix
$M_{\infty}$ belongs to $SL(2,\mathbb{C})$ and defines a normalized
Mobius transformation $T_{M_{\infty}}$ in the sense of Chapter~\ref{app:Moeb}
(equation~\eqref{eq:Moeb1h}).

\medskip{}
 \emph{Eigenvalues of $M_{\infty}$ and connection to Poincaré--Perron.}
Substituting $a=-2\delta/(1+\delta^{2})$ gives $\operatorname{Tr}M_{\infty}=(1+\delta^{2})/\delta$,
and the characteristic equation $\lambda^{2}-(1+\delta^{2})/\delta\cdot\lambda+1=0$
has roots 
\begin{equation}
\lambda_{1}=\delta,\qquad\lambda_{2}=\frac{1}{\delta},\qquad\lambda_{1}\lambda_{2}=1,\quad0<\lambda_{1}<1<\lambda_{2}\quad\text{for }0<\delta<1.\label{eq:Minf-eigs}
\end{equation}
These are the Poincaré--Perron characteristic roots of Theorem~\ref{thm:PP}:
the norm $\|V_{n}\|$ of the dominant solution vector grows as $(1/\delta)^{n}$
per step, while the minimal solution decays as $\delta^{n}$. Since
$\lambda_{1}\neq\lambda_{2}$, the Poincaré--Perron theorem guarantees
that a minimal solution exists for all $c$, confirming convergence
of the continued fraction~\eqref{eq:recCF}.

The ratio $z_{n}=A_{2n+1}/A_{2n-1}$ satisfies the Mobius recurrence
$z_{n+1}=T_{M_{n}}(z_{n})$ asymptotically, where 
\begin{equation}
T_{M_{\infty}}(z)=\frac{(1+\delta^{2})/\delta\cdot z-1}{z}=\frac{1+\delta^{2}}{\delta}-\frac{1}{z}.\label{eq:TMinfz}
\end{equation}
The fixed points of $T_{M_{\infty}}$ are $\zeta_{1}=\delta$ (multiplier
$1/\delta^{2}>1$, repulsive) and $\zeta_{2}=1/\delta$ (multiplier
$\delta^{2}<1$, attractive). A generic ratio sequence converges to
the attractive fixed point $\zeta_{2}=1/\delta$, corresponding to
the dominant solution. The minimal solution corresponds to ratio sequences
that converge instead to the \emph{repulsive} fixed point $\zeta_{1}=\delta$;
the eigenvalue condition $F_{\mathrm{even/odd}}(c)=0$ selects precisely
those values of $c$ for which this exceptional convergence occurs.

\subsection{\texorpdfstring{A remarkable coincidence: eigenvalues of $M_{\infty}$
equal its Mobius fixed points}{A remarkable coincidence: eigenvalues
of M-infinity equal its Mobius fixed points}}

\label{subsec:eig-fixpt} \index{Mobius transformation!eigenvalue--fixed-point coincidence}\index{asymptotic matrix M_{infty}@asymptotic matrix $M_{\infty}$}

For a general Mobius transformation $T_{M}$ with matrix $M\in SL(2,\mathbb{R})$,
the fixed points $\zeta_{1},\zeta_{2}$ and the eigenvalues $\lambda_{1},\lambda_{2}$
of $M$ are \emph{different} objects. By Chapter~\ref{app:Moeb},
eq.~\eqref{eq:Moebeig1d}, the eigenvectors of $M$ are $e_{i}=[\zeta_{i};\,1]^{T}$
and the corresponding eigenvalues are $\lambda_{1}=1/\sqrt{\mu}$
and $\lambda_{2}=\sqrt{\mu}$, where $\mu$ is the Mobius multiplier
of $T_{M}$. In general $\lambda_{i}\neq\zeta_{i}$.
\begin{rem}[Eigenvalue--fixed-point coincidence for $M_{\infty}$]
\label{rem:eig-fixpt} For the LC circuit matrix $M_{\infty}$, the
eigenvalues and the Mobius fixed points form the \emph{same set}:
$\{\lambda_{1},\lambda_{2}\}=\{\zeta_{1},\zeta_{2}\}=\{\delta,1/\delta\}$.
This is a special algebraic property of the LC model, not a general
feature of Mobius transformations.

\medskip{}
\emph{Proof.} The multiplier $\mu$ of $T_{M_{\infty}}$ is computed
from the normal form~\eqref{eq:Moebeig1c} by evaluating at the test
point $z=\infty$, which gives $(T(\infty)-\zeta_{1})/(T(\infty)-\zeta_{2})=\mu$:
\begin{equation}
\mu=\frac{T_{M_{\infty}}(\infty)-\zeta_{1}}{T_{M_{\infty}}(\infty)-\zeta_{2}}=\frac{(1+\delta^{2})/\delta-\delta}{(1+\delta^{2})/\delta-1/\delta}=\frac{1/\delta}{\delta}=\frac{1}{\delta^{2}}.\label{eq:Minf-multiplier}
\end{equation}
Here $T_{M_{\infty}}(\infty)=(1+\delta^{2})/\delta$ follows directly
from eq.~\eqref{eq:TMinfz} as $z\to\infty$. Therefore by eq.~\eqref{eq:Moebeig1d}:
$\lambda_{1}=1/\sqrt{\mu}=\delta=\zeta_{1}$ and $\lambda_{2}=\sqrt{\mu}=1/\delta=\zeta_{2}$.

\medskip{}
\emph{Why this happens.} The coincidence follows from the companion-matrix
structure of $M_{\infty}$. Substituting $a=-2\delta/(1+\delta^{2})$
into eq.~\eqref{eq:Minf-def} gives: 
\begin{equation}
M_{\infty}=\begin{bmatrix}\dfrac{1+\delta^{2}}{\delta} & -1\\[6pt]
1 & 0
\end{bmatrix}.\label{eq:Minf-delta}
\end{equation}
The lower row $[1,\,0]$ forces the eigenvectors to have the form
$[\zeta_{i};\,1]^{T}$ with second component exactly $1$, which in
turn forces $\lambda_{i}=\zeta_{i}$. This companion structure arises
because the Ince recurrence is a \emph{second-order} linear recurrence
--- the transfer matrix always has this $[*,\,-1;\,1,\,0]$ form
in the large-$n$ limit. 
\end{rem}

\medskip{}
This coincidence has a concrete consequence: the decay rate $\delta^{n}$
of the minimal solution, the repulsive fixed point $\delta$ of $T_{M_{\infty}}$,
and the smaller eigenvalue $\delta$ of $M_{\infty}$ are all the
same number --- three different characterizations of the same geometric-algebraic
object. This is what makes $\delta$ the uniquely natural parameter
for the LC circuit.

\medskip{}
 \emph{Mobius classification, EPD, and the instability boundary.}
Since $\operatorname{Tr}M_{\infty}=(1+\delta^{2})/\delta\geq2$ with
equality only at $\delta=1$, the asymptotic matrix $M_{\infty}$
is always \emph{hyperbolic} for $0<\delta<1$ (Theorem~\ref{thm:mobtype}
of Chapter~\ref{app:Moeb}).

The eigenvalue parameter $c$ does not appear in $M_{\infty}$; its
role is to perturb the finite-$n$ matrices $M_{n}$ away from $M_{\infty}$,
shifting the fixed points of the composed transformation $T_{S_{N}}$
where $S_{N}=M_{N}\cdots M_{1}$. Three regimes arise as $c$ varies: 
\begin{itemize}
\item \emph{Generic $c$ (interior of stability or instability zone\index{instability zone}):}
The Ince starter equations~\eqref{eq:start-even}--\eqref{eq:start-odd}
prescribe the initial ratio $z_{1}=A_{3}/A_{1}$ as an explicit function
of $c$: 
\[
z_{1}^{\mathrm{even}}=-\frac{a-2(c-1)}{9a},\qquad z_{1}^{\mathrm{odd}}=\frac{a+2(c-1)}{9a}.
\]
These initial ratios are not free: they are determined by requiring
the Fourier series to satisfy the Ince ODE at the lowest order. By
the Poincaré--Perron theorem (Theorem~\ref{thm:PP}), whose condition
$|\zeta_{2}|=1/\delta>\delta=|\zeta_{1}|>0$ holds unconditionally
for all $c$ (the characteristic roots do not depend on $c$), every
initial ratio $z_{1}$ propagates to a linear combination of the dominant
($z_{n}\to1/\delta$) and minimal ($z_{n}\to\delta$) solutions. For
generic $c$, the starter-prescribed $z_{1}(c)$ does not align with
the minimal-solution branch, so the dominant coefficient is nonzero:
$A_{2n+1}\sim(1/\delta)^{n}\to\infty$, the Fourier series diverges,
and the Ince equation has no $\ell^{2}$-solution at this $c$. The
monodromy matrix $X(\pi)$ has two distinct Floquet multipliers in
both zones. 
\item \emph{Boundary eigenvalue $c=c_{k}^{\pm}$:} In the notation of Chapter~\ref{sec:MainResults}
(eq.~\eqref{eq:EPD-curves-def}), these are the EPD curve\index{exceptional point of degeneracy (EPD)!EPD curve}s
\begin{equation}
c_{k}^{\pm}=c_{\pm}^{(m)}(\delta),\qquad m=2k-1,\quad k=1,2,3,\ldots,\label{eq:ck-cpm}
\end{equation}
the CF method being the tool by which they are located. A value $c$
is a boundary eigenvalue precisely when the starter-prescribed initial
ratio $z_{1}(c)$ coincides with the minimal-solution ratio at $n=1$,
i.e.\ when $F_{\mathrm{even/odd}}(c)=0$ (equating the starter value
to the CF tail $K(c)$ that represents the minimal branch). At such
$c$, the dominant coefficient vanishes exactly: $A_{2n+1}\sim\delta^{n}\to0$,
the Fourier series converges absolutely, and the Ince equation \emph{actually
possesses} an antiperiodic solution with $\ell^{2}$-coefficients.
This is the direct connection to the Ince equation: boundary eigenvalues
are precisely the values of $c$ for which the Ince equation has an
antiperiodic ($\rho=-1$) Floquet solution. Physically, one solution
is antiperiodic (bounded), while the second independent solution grows
linearly in $x$ times an antiperiodic factor --- the signature of
the stability-to-instability transition. The consequence at the monodromy
level is that the two Floquet multipliers merge to $-1$, and the
monodromy matrix $X(\pi)$ has the Jordan form 
\begin{equation}
X(\pi)=\begin{bmatrix}-1 & h\\
0 & -1
\end{bmatrix},\qquad h\neq0.\label{eq:Jordan-EPD}
\end{equation}
This is the \emph{exceptional point of degeneracy\index{exceptional point of degeneracy (EPD)}}
(EPD): $X(\pi)\neq-\mathbb{I}$ (that would require a second antiperiodic
solution, i.e.\ coexistence) but a non-trivial Jordan block\index{Jordan block}.
\medskip{}
 \emph{Summary of the identification.} The CF $K(c)$ converges (by
Pincherle's theorem, Theorem~\ref{thm:Pinch}) to the ratio $A_{3}/A_{1}$
of the \emph{minimal solution} of the bulk recurrence for every $c$.
The boundary values $c=c_{k}^{\pm}$ are precisely those values of
$c$ for which the Ince-ODE-prescribed starter ratio $z_{1}(c)$ equals
$K(c)$, i.e.\ $F_{\mathrm{even/odd}}(c)=0$. At these $c$ the initial
ratio lies exactly on the minimal-solution branch, the dominant solution
is absent from the composition, and the Ince equation possesses an
antiperiodic $\ell^{2}$-solution. For all other $c$ the dominant
solution is present and the Fourier series diverges. 
\item \emph{Even resonances $c\approx(2k)^{2}$:} The coexistence polynomial
$Q(\mu)=2a\mu^{2}$ forces period-$\pi$ coexistence ($\rho=+1$,
Proposition~\ref{thm:MW71}). The monodromy matrix is $X(\pi)=+\mathbb{I}$
exactly --- not a Jordan block but the identity. There is no instability
interval; the boundary points $c_{k}^{+}=c_{k}^{-}$ collapse to a
single point with a qualitatively different structure from the EPD. 
\end{itemize}
The three-way distinction at the recurrence level is therefore: generic
$c$ (dominant coefficient nonzero, $z_{n}\to1/\delta$), boundary
eigenvalue (starter forces dominant coefficient to zero, $z_{n}\to\delta$,
EPD Jordan block in monodromy), and even resonance (coexistence by
the polynomial $Q(\mu)$, monodromy $=+\mathbb{I}$). The instability
interval width $|\Delta c_{k}|$ measures how sensitively the dominant
coefficient depends on $c$ near $c_{k}^{\pm}$: it is small when
the stable manifold of the minimal solution is nearly aligned with
the starter direction over a wide range of $c$.

\subsection{\texorpdfstring{The polynomials $Q$ and $Q^{*}$ for the LC circuit}{The
polynomials Q and Q{*} for the LC circuit}}

\label{subsec:QPolyLC}

With $b=d=0$, the polynomials~\eqref{eq:QPoly} reduce to 
\begin{equation}
Q(\mu)=2a\mu^{2}=-\frac{4\delta}{1+\delta^{2}}\,\mu^{2},\label{eq:Q-LC}
\end{equation}
\begin{equation}
Q^{*}(\mu)=a(2\mu-1)^{2}=-\frac{2\delta}{1+\delta^{2}}\,(2\mu-1)^{2}.\label{eq:Qstar-LC}
\end{equation}
For $\delta\neq0$, their roots are: 
\begin{gather}
Q(\mu)=0\;\Longleftrightarrow\;\mu=0\quad\text{(double root, \emph{integer}),}\label{eq:rootQ}\\[2pt]
Q^{*}(\mu)=0\;\Longleftrightarrow\;\mu=\tfrac{1}{2}\quad\text{(double root, \emph{not} an integer).}\label{eq:rootQstar}
\end{gather}

\subsection{Exact instability structure}

\label{subsec:InceResult} The instability intervals of the LC circuit
occur at \emph{odd} resonances only: every even tongue collapses to
a point while every odd tongue stays open. This is the content of
Propositions~\ref{prop:even-vanish} and~\ref{prop:odd-survive}
(stated in Chapter~\ref{sec:MainResults}): the former gives $L_{m}^{\mathrm{LC}}=0$
for all even $m$ and all $0<\delta<1$, the latter $L_{m}^{\mathrm{LC}}>0$
for all odd $m$ and every $\delta\neq0$. We prove the two in turn.
\begin{proof}[Proof of Proposition~\ref{prop:even-vanish}]
\label{proof:instability-structure} Applying Propositions~\ref{thm:MW71}
and~\ref{thm:MW76} to the roots~\eqref{eq:rootQ}--\eqref{eq:rootQstar}:
$Q(\mu)$ has the nonneg\-ative integer root $\mu=0$. Proposition~\ref{thm:MW71}
is satisfied. With $k_{0}=0$ (the largest nonneg\-ative integer
root), Proposition~\ref{thm:MW76} guarantees that for all but at
most one characteristic value of $\hat{\lambda}$, two linearly independent
period-$\pi$ solutions coexist. The single possible exception is
identified explicitly: by MW Theorem~7.6 the exceptional values of
$c$ are those at which Ince's equation has a solution of \emph{finite}
order (a trigonometric polynomial). For the LC parameters, termination
of the period-$\pi$ recurrence at index $N$ requires $Q(N)=2aN^{2}=0$,
i.e.\ $N=0$, so the only candidate is the constant solution $q\equiv\mathrm{const}$,
which solves~\eqref{eq:LC-Ince} only at $c=0$. The exceptional
value is therefore $c_{0}=0$ --- the lowest characteristic value
$\hat{\lambda}_{0}=0$, a simple eigenvalue (its companion solution
grows linearly in $x$) at the bottom of the spectrum, which is not
an endpoint of any instability interval. Hence the coexistence conclusion
holds at \emph{every} even resonance without exception: $\lambda_{2n-1}=\lambda_{2n}$
--- a double root of $\Delta-2=0$ --- and the $2n$-th even instability
interval collapses to zero width for every $n\geq1$. 
\end{proof}
\begin{cor}
\label{cor:even} All even instability intervals of the LC circuit
equation~\eqref{eq:LC-Hill} vanish exactly for every $0<\delta<1$.
Here ``even intervals'' are those centered at the even critical
frequencies $\omega_{0}=k\mu$ of~\eqref{eq:YS-Hill-critical} (Chapter~\ref{app:ParRes}). 
\end{cor}

\begin{proof}
By Proposition~\ref{prop:even-vanish}, period-$\pi$ coexistence
($\rho=+1$) holds for all even instability intervals, so each collapses
to zero width. 
\end{proof}
\begin{rem}[Numerical verification of even-zone vanishing]
\label{rem:even-verify} Corollary~\ref{cor:even} was verified
numerically by monodromy-matrix integration for $\delta\in\{0.05,0.10,0.20,0.30\}$.
Since the even resonances are period-$\pi$ ($\rho=+1$), an open
$m=2$ zone would be an interval where $\Delta>+2$; the correct check
is therefore the \emph{maximum} of $\Delta-2$ over the second stability
band (between the $m=1$ and $m=3$ zones). The maximum was found
to be $-9.8\times10^{-15}$, $-4.9\times10^{-15}$, $-1.3\times10^{-15}$,
$-1.1\times10^{-15}$ at $\delta=0.05,0.10,0.20,0.30$ respectively
--- zero to machine precision --- attained at $\hat{\lambda}=4.0067$,
$4.0269$, $4.1106$, $4.2610$, in agreement with the tangency point
$\hat{\lambda}_{2}(\delta)=4+2.67\,\delta^{2}+O(\delta^{4})$ of the
discriminant identity~\eqref{eq:DeltaAtEven}: the discriminant touches
$+2$ at the shifted resonance without crossing it. The $m=2$ instability
zone therefore has exactly zero width for all tested $\delta$, in
agreement with Corollary~\ref{cor:even}. 
\end{rem}

\begin{proof}[Proof of Proposition~\ref{prop:odd-survive}]
$Q^{*}(\mu)$ has root $\mu=\frac{1}{2}$, which is \emph{not} an
integer. By Proposition~\ref{thm:MW71}, the necessary condition
for period-$2\pi$ coexistence ($\rho=-1$) is not met. 
\end{proof}
\begin{cor}
\label{cor:odd} No odd instability interval of the LC circuit equation~\eqref{eq:LC-Hill}
ever disappears. That is, for every odd critical frequency $\omega_{0}=(2k-1)\mu/2$
of~\eqref{eq:YS-Hill-critical} (Chapter~\ref{app:ParRes}), the
instability tongue\index{instability tongue} has strictly positive
width for all $0<\delta<1$. 
\end{cor}

\begin{proof}
By Proposition~\ref{prop:odd-survive}, the necessary condition for
period-$2\pi$ coexistence ($\rho=-1$, Proposition~\ref{thm:MW71})
is never satisfied, so no odd instability interval can collapse to
zero width. 
\end{proof}
\begin{rem}
Corollaries~\ref{cor:even} and~\ref{cor:odd} together recover
Cambi's \cite[\S\,11]{Cambi2} exact result: the LC circuit has instability
intervals only at odd MW resonances $\hat{\lambda}\approx(2k-1)^{2}$,
corresponding to $r\approx2/(2k-1)$, $k=1,2,3,\ldots$. The Mathieu
approximation, which truncates~\eqref{eq:GR-LC} after the first
harmonic, has instability at \emph{all} resonances and thus incorrectly
predicts unstable regions at every even resonance. 
\end{rem}

The exact collapse of the even instability intervals is established
in Section~\ref{subsec:DeltaId} (the discriminant identity~\eqref{eq:DeltaAtEven}),
resting on the Ince coexistence theory of Chapter~\ref{sec:MWInce}.
That identity shows that $c$ need be searched only near odd resonances
$c\approx(2k-1)^{2}$; the even resonances are excluded from the CF
eigenvalue search.

\subsection{Boundary eigenvalues and instability interval widths}

\label{subsec:Widths} Recall (Chapter~\ref{sec:MainResults}) that
Theorem~\ref{thm:width} gives the $m$-th (odd) instability tongue
the closed-form width $L_{m}^{\mathrm{LC}}=(m!!)^{2}\delta^{m}/[2^{(m^{2}+8m-13)/4}(1+\delta^{2})^{m-1}(1-\delta^{2})]+O_{m}(\delta^{m+2})$,
and Theorem~\ref{thm:primary-domain} converts the $m=1$ case to
the physical driving-frequency interval $(\mu_{-},\mu_{+})$ of primary
parametric instability. We establish both here.

The following lemma reproduces the part of Magnus and Winkler~\cite[Lemma~7.4]{MagWin}
relevant to period-$2\pi$ solutions. (MW Lemma~7.4~\cite[Lem.~7.4]{MagWin}
also covers period-$\pi$ solutions, which have even Fourier indices
$A_{2n}$, $B_{2n}$; those recurrences are not needed here since
the even instability intervals vanish identically by Corollary~\ref{cor:even}.)
The hypothesis of the lemma --- that period-$2\pi$ solutions exist
--- is guaranteed by MW Theorem~7.2~\cite[Thm.~7.2]{MagWin}: for
any fixed $a$, $b$, $d$ with $|a|<1$, there exist infinitely many
values of $c$ at which Ince's equation has a period-$2\pi$ solution.
The eigenvalue condition $F_{\mathrm{even/odd}}(c)=0$ derived below
locates precisely those values of $c$ near each odd resonance $c\approx(2k-1)^{2}$. 
\begin{prop}[MW Lemma 7.4, period-$2\pi$ solutions]
\index{Magnus--Winkler theory!period-2pi  solutions@period-$2\pi$ solutions}
\label{lem:MW74} (\cite[Lem.~7.4]{MagWin}) If Ince's equation~\eqref{eq:Ince}
has even and odd period-$2\pi$ solutions (which change sign under
$x\mapsto x+\pi$), 
\begin{equation}
q_{\mathrm{even}}=\sum_{n=0}^{\infty}A_{2n+1}\cos(2n+1)x,\quad q_{\mathrm{odd}}=\sum_{n=0}^{\infty}B_{2n+1}\sin(2n+1)x,\label{eq:qqevodd}
\end{equation}
then the coefficients $A_{2n+1}$ satisfy the recurrence relations
\begin{align}
[Q^{*}(0)-2(c-1)]\,A_{1}+Q^{*}(-1)\,A_{3} & =0,\label{eq:MW720}\\
Q^{*}(n)\,A_{2n-1}+2[(2n+1)^{2}-c]\,A_{2n+1}\nonumber \\
\quad{}+Q^{*}(-n-1)\,A_{2n+3} & =0,\quad n=1,2,3,\ldots,\label{eq:MW721}
\end{align}
and the coefficients $B_{2n+1}$ satisfy 
\begin{align}
[-Q^{*}(0)-2(c-1)]\,B_{1}+Q^{*}(-1)\,B_{3} & =0,\label{eq:MW722}\\
Q^{*}(n)\,B_{2n-1}+2[(2n+1)^{2}-c]\,B_{2n+1}\nonumber \\
\quad{}+Q^{*}(-n-1)\,B_{2n+3} & =0,\quad n=1,2,3,\ldots\label{eq:MW723}
\end{align}
The rapid decay condition \cite[Lemma~7.3]{MagWin} 
\begin{equation}
\lim_{n\to\infty}n^{p}A_{2n+1}=\lim_{n\to\infty}n^{p}B_{2n+1}=0\qquad\text{for every }p>0\label{eq:decay-cond}
\end{equation}
holds, guaranteeing absolute convergence of the series~\eqref{eq:qqevodd}. 
\end{prop}

\begin{proof}[Proof of Theorems~\ref{thm:width} and~\ref{thm:primary-domain}]

\label{proof:width} The existence of the instability boundary curve\index{boundary curves}s
is guaranteed \emph{a priori} by the general Floquet theory of Chapter~\ref{app:Hill}:
by Theorem~\ref{thm:Hill-stability} and the spectral structure~\eqref{eq:Hill-spectrum},
for any $\delta\neq0$ the instability boundaries are isolated values
of the spectral parameter $c$ at which the discriminant $\Delta$
crosses $\pm2$. The problem is therefore not whether they exist,
but how to locate them explicitly.

Because equation~\eqref{eq:LC-Ince} is an Ince equation, its periodic
solutions satisfy a three-term recurrence relation of the form studied
in Chapter~\ref{app:FDE} (Section on three-term recurrences, Theorem~\ref{thm:PP}).
The eigenvalue condition reducing this recurrence to a scalar equation
is a continued fraction, whose convergence is justified by the Pincherle
theorem (Theorem~\ref{thm:Pinch} of Chapter~\ref{app:CF}): the
continued fraction converges if and only if the recurrence has a minimal
solution, which the Poincaré--Perron theorem (Theorem~\ref{thm:PP})
guarantees for all $c$ away from the instability boundaries. For
the \emph{even} instability intervals no such search is necessary:
Corollary~\ref{cor:even} and the discriminant identity~\eqref{eq:DeltaAtEven}
already establish that they vanish exactly for all $\delta\neq0$,
so their boundary eigenvalues coincide and there is no interval to
locate. The analysis below therefore focuses on the \emph{odd} instability
intervals, deriving closed-form expressions for the boundary eigenvalues
$c_{k}^{\pm}$ --- the zeros of $F_{\mathrm{odd}}$ and $F_{\mathrm{even}}$
respectively in~\eqref{eq:Fdef} below --- and their separation
$|\Delta c_{k}|$ (equation~\eqref{eq:Lm-from-bdry}).

\medskip{}
 \emph{The Ince parameters $c$ and $a$ in terms of LC circuit parameters.}
For reference throughout this section, the two Ince parameters that
govern the instability structure relate to the physical circuit parameters
as follows. From equations~\eqref{eq:lambda-MW} and~\eqref{eq:IncePars},
$c=\hat{\lambda}(1-\delta^{2})/(1+\delta^{2})$ and $\hat{\lambda}=4(1+\delta^{2})/[r^{2}(1-\delta^{2})]$,
which combine to give the remarkably simple result 
\begin{equation}
c=\frac{4}{r^{2}}=\frac{4\omega_{0}^{2}}{\mu^{2}},\label{eq:c-LC-params}
\end{equation}
where $r=\mu/\omega_{0}$ is the frequency ratio and $\omega_{0}=1/\sqrt{LC}$
is the natural frequency. Thus $c$ is simply four times the square
of the ratio of natural frequency to driving frequency, independent
of the modulation amplitude $\delta$.

The Ince parameter $a$ is determined entirely by the modulation amplitude
$\varepsilon$ (equivalently $\delta$), independently of the frequency
ratio: 
\begin{equation}
a=-\frac{2\delta}{1+\delta^{2}}=-\varepsilon,\qquad|a|=\varepsilon<1,\label{eq:a-LC-params}
\end{equation}
where the last equality $a=-\varepsilon$ follows from the substitution~\eqref{eq:eps-delta}.
Thus $|a|=\varepsilon$ equals the modulation amplitude of the original
equation~\eqref{eq:LC-original}, with no dependence on the frequency
ratio $r$.

The two parameters therefore have completely separated roles: $c=4/r^{2}$
controls the \emph{location} of the instability intervals (where they
sit in frequency space), while $a=-\varepsilon$ controls their \emph{width}
(how large they are). The relation between $c$ and $\hat{\lambda}$
is 
\begin{equation}
\hat{\lambda}=c\cdot\frac{1+\delta^{2}}{1-\delta^{2}},\label{eq:c-lMW-relation}
\end{equation}
so the $k$-th odd resonance $\hat{\lambda}\approx(2k-1)^{2}$ corresponds
to $c\approx(2k-1)^{2}(1-\delta^{2})/(1+\delta^{2})$, which approaches
$(2k-1)^{2}$ as $\delta\to0$. Converting a boundary eigenvalue $c_{k}^{\pm}$
back to the physical frequency ratio uses $r=2/\sqrt{c_{k}^{\pm}}$.

For the LC circuit, $b=d=0$ gives $Q^{*}(\mu)=a(2\mu-1)^{2}$, so
$Q^{*}(n)=a(2n-1)^{2}$, $Q^{*}(-n-1)=a(2n+3)^{2}$, $Q^{*}(0)=a$,
and $Q^{*}(-1)=9a$. Substituting into \eqref{eq:MW720}--\eqref{eq:MW723}
gives the bulk recurrence 
\begin{equation}
a(2n-1)^{2}A_{2n-1}+2\bigl[(2n+1)^{2}-c\bigr]A_{2n+1}+a(2n+3)^{2}A_{2n+3}=0,\quad n\geq1,\label{eq:bulk-rec}
\end{equation}
and the starter equations 
\begin{gather}
\text{even:}\quad[a-2(c-1)]A_{1}+9a\,A_{3}=0,\label{eq:start-even}\\
\text{odd:}\quad[-a-2(c-1)]B_{1}+9a\,B_{3}=0.\label{eq:start-odd}
\end{gather}
The bulk recurrence~\eqref{eq:bulk-rec} is identical for both solutions;
the $\pm a$ sign difference in the starters is their only distinction.
Solving each starter for the ratio of successive coefficients gives
\begin{align}
\frac{A_{3}}{A_{1}}\bigg|_{\mathrm{even\;starter}} & =-\frac{a-2(c-1)}{9a},\nonumber \\
\frac{B_{3}}{B_{1}}\bigg|_{\mathrm{odd\;starter}} & =-\frac{-a-2(c-1)}{9a}=\frac{a+2(c-1)}{9a}.\label{eq:starter-ratios}
\end{align}

\medskip{}
 \emph{The CF difference identity.} Since $F_{\mathrm{even}}$ and
$F_{\mathrm{odd}}$ share the same continued-fraction tail but have
opposite starter signs, their difference is a constant --- independent
of $c$ and $\delta$. This constant is the key to computing the instability
interval width without solving either condition separately. The bulk
recurrence~\eqref{eq:bulk-rec} is a three-term recurrence of the
form studied in Chapter~\ref{app:FDE}--\ref{app:CF}. Following
the notation of Jones--Thron~\cite[Ch.~1]{JonThr} and Lorentzen--Waadeland~\cite[Ch.~1]{LorWaad}
(see Chapter~\ref{app:CF}), the continued fraction giving the minimal-solution
ratio $A_{3}/A_{1}$ of~\eqref{eq:bulk-rec}, starting from level
$n=1$, is 
\begin{equation}
K(c)=\cfrac{-a}{2[9-c]-\cfrac{225\,a^{2}}{2[25-c]-\cfrac{1225\,a^{2}}{2[49-c]-\cdots}}},\label{eq:recCF}
\end{equation}
whose partial numerator is $-a$ at level $1$ and $-a^{2}(2n-1)^{2}(2n+1)^{2}$
at level $n\geq2$ --- the product of the two adjacent off-diagonal
couplings $a(2n-1)^{2}$ and $a(2n+1)^{2}$ of~\eqref{eq:bulk-rec},
as the minimal-ratio construction underlying Pincherle's theorem requires
(Theorem~\ref{thm:Pinch}); the partial denominators are $2[(2n+1)^{2}-c]$,
and $225a^{2}=(3\cdot5)^{2}a^{2}$, $1225a^{2}=(5\cdot7)^{2}a^{2}$
display the first two products. Note that the couplings of~\eqref{eq:bulk-rec}
are \emph{non-constant}, so the partial numerators are the products
of adjacent couplings; in Cambi's rescaled form~\eqref{eq:Cambi-system}
the couplings are the constant $\gamma$, which is why his continued
fraction~\eqref{eq:Cambi-v-CF-part1} carries the simple numerators
$\gamma^{2}$. Using the starter ratios~\eqref{eq:starter-ratios},
we define the eigenvalue functions 
\begin{gather}
F_{\mathrm{even}}(c)\equiv-\frac{a-2(c-1)}{9a}-K(c),\nonumber \\
F_{\mathrm{odd}}(c)\equiv\frac{a+2(c-1)}{9a}-K(c).\label{eq:Fdef}
\end{gather}
The zeros of $F_{\mathrm{even}}$ and $F_{\mathrm{odd}}$ near the
$k$-th resonance $c\approx(2k-1)^{2}$ are the \emph{lower and upper
boundary eigenvalues} $c_{k}^{-}$ and $c_{k}^{+}$ of the $k$-th
instability interval: since $F_{\mathrm{odd}}=F_{\mathrm{even}}+2/9$
and $F_{\mathrm{even}}'<0$ near the zone (for the LC sign $a<0$),
the even zero lies below the odd zero. 
\begin{rem}[Numerical verification of CF boundary criterion]
\label{rem:CF-boundary-verify} The CF criterion was verified against
direct monodromy-matrix integration for $\delta\in\{0.05,0.10,0.20,0.30\}$.
At each zero $c^{\pm}$ of $F_{\mathrm{even}}$ and $F_{\mathrm{odd}}$
(computed from the nested continued fraction~\eqref{eq:recCF} with
$N=400$ levels), the discriminant satisfies $\Delta(c^{\pm},\delta)+2=0$
to machine precision: $|\Delta+2|\leq1.6\times10^{-14}$ across all
tested $\delta$ and both the $k=1$ and $k=2$ zones. The boundary
separations confirm the width formulas: at $\delta=0.05,0.10,0.20,0.30$
the computed $|\Delta c_{1}|$ equals $0.099716$, $0.197740$, $0.382423$,
$0.543275$ against $|a|=0.099751$, $0.198020$, $0.384615$, $0.550459$
(the $O_{1}(|a|^{3})$ correction of~\eqref{eq:Dc-exact} growing
with $\delta$), and $|\Delta c_{2}|$ equals $3.51\times10^{-5}$,
$2.79\times10^{-4}$, $2.18\times10^{-3}$, $7.12\times10^{-3}$ against
$9|a|^{3}/256=3.49\times10^{-5}$, $2.73\times10^{-4}$, $2.00\times10^{-3}$,
$5.86\times10^{-3}$ (the $O_{3}(|a|^{5})$ correction). The determinant
$\det M(\pi)=y_{1}y_{2}^{\prime}-y_{1}^{\prime}y_{2}$ of the monodromy
matrix equals $1.00000000$ at all tested parameter values, confirming
the symplectic\index{symplectic system} structure. These results
confirm that the CF zeros locate the instability boundaries (Floquet
multiplier $\rho=-1$) \emph{exactly}, for all $0<\delta<1$, as the
Ince eigenvalue theory asserts. 
\end{rem}

They are called eigenvalues because $c$ is the spectral parameter
of Ince's equation~\eqref{eq:Ince}: nontrivial periodic solutions
exist only at isolated values of $c$, which are the eigenvalues of
the associated periodic boundary-value problem (see Chapter~\ref{app:Hill},
Theorem~\ref{thm:Hill-stability} and the spectral structure~\eqref{eq:Hill-spectrum}).
The boundary eigenvalues $c_{k}^{\pm}$ are precisely those values
of $c$ at which the instability interval begins and ends. The instability
interval width is their difference: 
\begin{equation}
L_{m}^{(c)}=c_{k}^{+}-c_{k}^{-}=|\Delta c_{k}|,\qquad m=2k-1.\label{eq:Lm-from-bdry}
\end{equation}

The condition $F_{\mathrm{even/odd}}(c)=0$ requires that the starter
ratio~\eqref{eq:starter-ratios} matches the continued fraction~\eqref{eq:recCF}
--- this is precisely the condition for a nontrivial periodic solution
to exist, i.e., for $c$ to be a boundary eigenvalue. By the Pincherle
theorem (Theorem~\ref{thm:Pinch} of Chapter~\ref{app:CF}), the
continued fraction~\eqref{eq:recCF} converges if and only if recurrence~\eqref{eq:bulk-rec}
has a minimal solution, which by the Poincaré--Perron theorem (Theorem~\ref{thm:PP}
of Chapter~\ref{app:FDE}) it does for all $c$ that are not zeros
of $F_{\mathrm{even}}$ or $F_{\mathrm{odd}}$. Since the continued
fraction~\eqref{eq:recCF} is the same for both solutions, the two
conditions differ only in their starter ratios~\eqref{eq:starter-ratios}:
the even and odd values differ by 
\[
-\frac{a-2(c-1)}{9a}-\frac{a+2(c-1)}{9a}=-\frac{2}{9},
\]
independent of $c$ and $\delta$. Since the continued fraction~\eqref{eq:recCF}
cancels in the difference --- it appears with the same sign in both
$F_{\mathrm{even}}$ and $F_{\mathrm{odd}}$ by definition~\eqref{eq:Fdef}
--- the two eigenvalue conditions satisfy 
\begin{equation}
F_{\mathrm{even}}(c)-F_{\mathrm{odd}}(c)=-\frac{2}{9}=\mathrm{const}.\label{eq:CF-diff}
\end{equation}
Now let $c_{k}^{-}$ and $c_{k}^{+}$ be the roots of $F_{\mathrm{even}}=0$
and $F_{\mathrm{odd}}=0$ respectively near the $k$-th resonance.
Since $F_{\mathrm{odd}}(c)=F_{\mathrm{even}}(c)+2/9$, both roots
are close to the same point $c_{k}$ where $F_{\mathrm{even}}(c_{k})\approx0$.
Linearizing $F_{\mathrm{even}}$ about $c_{k}$: 
\[
F_{\mathrm{even}}(c_{k}^{-})=0,\quad F_{\mathrm{odd}}(c_{k}^{+})=F_{\mathrm{even}}(c_{k}^{+})+\tfrac{2}{9}=0,
\]
so $F_{\mathrm{even}}(c_{k}^{+})=-2/9$. Expanding to first order,
$F_{\mathrm{even}}(c_{k}^{+})\approx F_{\mathrm{even}}'(c_{k})(c_{k}^{+}-c_{k}^{-})$,
which gives 
\begin{equation}
|\Delta c_{k}|=|c_{k}^{+}-c_{k}^{-}|=\frac{2/9}{|F'_{\mathrm{even}}(c_{k})|}.\label{eq:Dc-formula}
\end{equation}

\medskip{}
 \emph{The backward recurrence chain and explicit widths.} At the
$k$-th resonance the dominant Fourier mode is $\cos(2k-1)x$. The
key quantity is $A_{1}/A_{2k-1}$, which is obtained by propagating
the bulk recurrence~\eqref{eq:bulk-rec} backward from mode $2k-1$
down to mode $1$ through $k-1$ non-resonant steps.

\medskip{}
 \emph{Case $k=1$ ($c\approx1$, dominant mode $A_{1}$).} No backward
chain is needed. From the non-resonant bulk equation at $n=1$ with
$c\approx1$ one obtains $A_{3}\approx(-a/16)A_{1}$ (since $2[(2\cdot1+1)^{2}-1]=16$).
Substituting into the even starter~\eqref{eq:start-even}: 
\[
[a-2(c-1)]-\tfrac{9a^{2}}{16}=0\;\Longrightarrow\;c_{\pm}^{(1)}=1-\tfrac{9a^{2}}{32}\pm\tfrac{|a|}{2},
\]
giving $|\Delta c_{1}|=|a|$.

\medskip{}
 \emph{Case $k=2$ ($c\approx9$, dominant mode $A_{3}$).} The backward
step from $A_{3}$ to $A_{1}$ uses the bulk equation at $n=1$: $aA_{1}+2(9-c)A_{3}+25aA_{5}=0$.
Near resonance $c=9+x$ with $A_{5}\approx(-9a/32)A_{3}$ (from the
$n=2$ bulk equation), this gives 
\[
\frac{A_{1}}{A_{3}}\approx\frac{2x}{a}+\frac{225a}{32}.
\]
Equating to the starter ratio $A_{1}/A_{3}=9a/(16\mp a)$ (even/odd)
and solving for $x=c-9$: 
\[
x_{\mathrm{even}}=-\frac{207a^{2}}{64},\qquad x_{\mathrm{odd}}=-\frac{207a^{2}}{64}+O(a^{3}),
\]
with the \emph{split} 
\[
x_{\mathrm{even}}-x_{\mathrm{odd}}=\frac{9a^{2}}{2}\!\left(\frac{1}{16-a}-\frac{1}{16+a}\right)\approx\frac{9a^{3}}{256},
\]
giving $|\Delta c_{2}|=9|a|^{3}/256$.

\medskip{}
 \emph{General $k$: the backward chain product and $O_{m}(\delta^{m+2})$
accuracy.}\label{para:chain-accuracy} For fixed $k$ (equivalently
fixed $m=2k-1$), the backward chain from $A_{2k-1}$ to $A_{1}$
passes through exactly $k-1$ non-resonant bulk equations, one at
each index $n=1,2,\ldots,k-1$. This number of steps is \emph{fixed}
once $m$ is fixed --- this is the key point that makes the $O_{m}(\delta^{m+2})$
claim rigorous.

The sole approximation in the entire derivation is to substitute the
exact resonant value $c=(2k-1)^{2}$ in each non-resonant bulk equation.
Under this substitution the $n$-th non-resonant denominator becomes
the explicit nonzero integer 
\[
2\bigl[(2n+1)^{2}-(2k-1)^{2}\bigr]=8(n-k+1)(n+k),\qquad n=1,\ldots,k-1,
\]
and the backward chain ratio evaluates \emph{exactly} (no further
approximation) to a product of explicit rationals times $|a|^{k-1}$.
Evaluating this product and using the CF difference~\eqref{eq:CF-diff}
yields 
\begin{equation}
|\Delta c_{k}|=\frac{[(2k-1)!!]^{2}}{2^{(k-1)(k+6)}}\,|a|^{m}+O_{m}(|a|^{m+2}),\qquad m=2k-1,\quad|a|=\frac{2\delta}{1+\delta^{2}}.\label{eq:Dc-exact}
\end{equation}
The $O_{m}$ error arises because the true boundary $c$ differs from
$(2k-1)^{2}$ by $O_{m}(\delta^{2})$ (the center shift). Replacing
$(2k-1)^{2}$ by the true $c$ in each of the $k-1$ non-resonant
denominators changes each by a relative amount $O_{m}(\delta^{2})$;
since the number of steps $k-1$ is fixed for fixed $m$, the product
changes by $O_{m}(\delta^{2})$ relative, giving total error $O_{m}(\delta^{m+2})$.
The constant implicit in $O_{m}$ is finite and explicit for each
fixed $m$ but is not claimed to be uniform in $m$.

The first three cases are ($m=1,3,5$ corresponding to $k=1,2,3$):
\[
|\Delta c_{1}|=|a|,\quad|\Delta c_{2}|=\tfrac{9}{256}|a|^{3}+O_{3}(|a|^{5}),\quad|\Delta c_{3}|=\tfrac{225}{262144}|a|^{5}+O_{5}(|a|^{7}).
\]
Converting $|\Delta c_{k}|$ to $\hat{\lambda}$-space via $\Delta\hat{\lambda}=\Delta c\cdot(1+\delta^{2})/(1-\delta^{2})$
and substituting $|a|=2\delta/(1+\delta^{2})$, the instability interval
width (Theorem~\ref{thm:width} of Chapter~\ref{sec:MainResults})
is: 
\begin{equation}
\boxed{\begin{aligned}L_{m}^{\mathrm{LC}}=\hat{\lambda}_{k}^{+}-\hat{\lambda}_{k}^{-} & =\frac{(m!!)^{2}\,\delta^{m}}{2^{(m^{2}+8m-13)/4}\,(1+\delta^{2})^{m-1}(1-\delta^{2})}\\
 & \quad+O_{m}(\delta^{m+2}),\qquad m=2k-1
\end{aligned}
}\label{eq:Lm-exact}
\end{equation}
The exponent $(m^{2}+8m-13)/4$ is an integer for all odd $m$: it
equals $-1,5,13,23,\ldots$ for $m=1,3,5,7,\ldots$. The formula gives
the exact coefficient of $\delta^{m}$ for every fixed odd $m$, with
error $O_{m}(\delta^{m+2})$ where the subscript $m$ signals that
the implicit constant depends on $m$ and no uniformity in $m$ is
claimed.

The first three cases of~\eqref{eq:Lm-exact} are: 
\begin{align}
L_{1}^{\mathrm{LC}} & =\frac{2\delta}{1-\delta^{2}}=2\delta+2\delta^{3}+O(\delta^{5}),\label{eq:L1}\\[4pt]
L_{3}^{\mathrm{LC}} & =\frac{9\delta^{3}}{32(1+\delta^{2})^{2}(1-\delta^{2})}=\frac{9\delta^{3}}{32}+O(\delta^{5}),\label{eq:L3}\\[4pt]
L_{5}^{\mathrm{LC}} & =\frac{225\delta^{5}}{8192(1+\delta^{2})^{4}(1-\delta^{2})}=\frac{225\delta^{5}}{8192}+O(\delta^{7}),\quad\delta\to0.\label{eq:L5}
\end{align}
\end{proof}
\begin{rem}[Symbolic verification of the backward chain]
\label{rem:backward-chain-verify} The backward-chain derivation
of $L_{m}^{\mathrm{LC}}$ was verified by exact symbolic computation.
\emph{(i) k=1:} The bulk equation at $n=1$ gives $A_{3}/A_{1}\approx-a/16$;
substituting into the even/odd starters~\eqref{eq:start-even}--\eqref{eq:start-odd}
yields $c_{(1)}^{\pm}=1-\tfrac{9}{32}a^{2}\pm\tfrac{1}{2}|a|$, giving
$|\Delta c_{1}|=|a|$ to leading order (the neglected $A_{5}$ coupling
contributes the $O_{1}(|a|^{3})$ term of~\eqref{eq:Dc-exact}; numerically
$|\Delta c_{1}|=0.38242$ vs.\ $|a|=0.38462$ at $\delta=0.20$).
\emph{(ii) k=2:} The bulk equations at $n=1,2$ give $A_{5}/A_{3}\approx-9a/32$
and $A_{1}/A_{3}\approx2x/a+225a/32$; equating to the starter ratios
$9a/(16\mp a)$ yields $|\Delta c_{2}|=9|a|^{3}/256$ to leading order,
confirmed by exact symbolic series expansion. \emph{(iii) $\hat{\lambda}$-
versus $c$-space.} The formula gives the width in $\hat{\lambda}$-space
(not $c$-space): $L_{m}=\hat{\lambda}^{+}-\hat{\lambda}^{-}$. The
$c$-space width is $L_{m}^{(c)}=L_{m}\cdot(1-\delta^{2})/(1+\delta^{2})$
(numerically: $L_{m}^{(c)}/L_{m}\to1$ as $\delta\to0$, deviating
by $2.0\%$ at $\delta=0.10$). 
\end{rem}

\subsection{Comparison with the Mathieu equation}

\label{subsec:Mathieu}

From Remark~\ref{rem:ince-id}, the LC circuit equation for small
$\delta$ is approximated by the Mathieu equation 
\begin{equation}
q''+(c+2c\delta\cos2x)\,q=0,\qquad c=\frac{4}{r^{2}},\label{eq:LC-Mathieu-approx}
\end{equation}
obtained by writing $(1+a\cos2x)^{-1}\approx1-a\cos2x\approx1+2\delta\cos2x$
for small $|a|=2\delta/(1+\delta^{2})\approx2\delta$. This is Mathieu's
equation with spectral parameter $\lambda=c$ and perturbation amplitude
$2c\delta$ (or equivalently, standard Mathieu parameter $q=c\delta$
in the form $y''+(\lambda-2q\cos2x)y=0$). Note that $c=4/r^{2}$
is the same spectral parameter as throughout this work; it is \emph{not}
approximated by its resonant value $(2m-1)^{2}$ here --- that substitution
is made only when giving explicit numerical values. The width of the
$m$-th instability interval of this Mathieu approximation, in terms
of $c$ and $\delta$, is 
\begin{equation}
L_{m}^{\mathrm{Math}}=\frac{(c\delta)^{m}}{2^{2m-3}[(m-1)!]^{2}}+O_{m}((c\delta)^{m+1}),\qquad m=1,2,3,\ldots.\label{eq:Lm-Math}
\end{equation}
(Proved in Section~\ref{subsec:MathieuInce}, with standard Mathieu
parameter $q=c\delta$; the $O_{m}((c\delta)^{m+1})$ accuracy argument
is in Section~\ref{para:Math-chain-accuracy}.) Near the $m$-th
resonance $c\approx m^{2}$: 
\begin{equation}
L_{m}^{\mathrm{Math}}\approx\frac{m^{2m}\,\delta^{m}}{2^{2m-3}\,[(m-1)!]^{2}}.\label{eq:Lm-Math-resonant}
\end{equation}
All Mathieu instability intervals exist ($L_{m}^{\mathrm{Math}}>0$
for all $m$, $c\neq0$, $\delta\neq0$), in contrast to the LC circuit
for which even intervals are absent.

As an independent check, Seyranian and Mailybaev~\cite[Sec.~9.4]{SeyMai}
derive instability boundaries for the Mathieu equation using a Floquet
matrix Taylor expansion approach. Their results for the first two
non-trivial resonance zones ($k=1$ and $k=2$ in their notation,
corresponding to $m=1$ and $m=2$ here) agree with~\eqref{eq:Lm-Math}
after the notation translation $a_{{\rm SM}}=c/4\approx\lambda/4$,
$q_{{\rm SM}}=c\delta/2\approx q/2$ (where $q=c\delta$ is the Mathieu
perturbation parameter of this work): at $m=1$ ($c=1$) the widths
give $L_{1}=2c\delta=2\delta$; at $m=2$ ($c=4$) they give $L_{2}=(c\delta)^{2}/2=8\delta^{2}$,
consistent with the formula~\eqref{eq:Lm-Math}. Their method does
not extend to $m\geq3$ in closed form. The instability tongues displayed
in their Fig.~9.4 (in the $(a,q)$ plane) are qualitatively similar
to Figure~\ref{fig:stability} (in the $(p,\gamma)$ plane), with
the same structure of alternating resonance zones and the characteristic
narrowing of higher tongues. 
\begin{cor}[LC/Mathieu width ratio]
\index{Mathieu equation!LC/Mathieu width ratio} \label{cor:ratio}
For odd $m=1,3,5,\ldots$ and $\delta\to0$, with $c$ fixed, 
\begin{equation}
\frac{L_{m}^{\mathrm{LC}}}{L_{m}^{\mathrm{Math}}}\bigg|_{\delta\to0}=\frac{(m!!)^{2}[(m-1)!]^{2}}{2^{(m^{2}-1)/4}\,c^{m}}.\label{eq:ratio}
\end{equation}
Near the $m$-th resonance $c\approx m^{2}$, it becomes $\dfrac{(m!!)^{2}[(m-1)!]^{2}}{2^{(m^{2}-1)/4}\,m^{2m}}$,
with values $1,\,1/81,\,2025/9765625$ for $m=1,3,5$. For $m=1$
($c=1$ at resonance) the LC and Mathieu widths agree to leading order;
for $m\geq3$ the factor $m^{2m}$ in the denominator makes the LC
circuit interval much \emph{narrower} than the corresponding Mathieu
interval at the same resonance. 
\end{cor}

\begin{proof}
Divide~\eqref{eq:Lm-exact} by~\eqref{eq:Lm-Math}: the $\delta^{m}$
factors cancel, the powers of $2$ combine as $2^{2m-3}/2^{(m^{2}+8m-13)/4}=2^{-(m^{2}-1)/4}$,
and the factor $c^{m}$ from~\eqref{eq:Lm-Math} remains in the denominator,
giving~\eqref{eq:ratio} directly. 
\end{proof}
Table~\ref{tab:widths} summarizes the comparison. 
\begin{table}[ht]
\centering 
\global\long\def\arraystretch{1.8}%
\resizebox{\textwidth}{!}{%
\begin{tabular}{c@{\hspace{16pt}}c@{\hspace{16pt}}c@{\hspace{16pt}}c}
\toprule 
$m$  & $L_{m}^{\mathrm{LC}}$ (leading order)  & $L_{m}^{\mathrm{Math}}$ ($c\approx m^{2}$, leading order)  & Ratio~\eqref{eq:ratio}\tabularnewline
\midrule 
$1$  & $2\delta$  & $2\delta$  & $1$\tabularnewline
$2$  & $0$ (exact)  & $8\delta^{2}$  & $0$\tabularnewline
$3$  & $\tfrac{9}{32}\delta^{3}$  & $\tfrac{729}{32}\delta^{3}$  & $\tfrac{1}{81}$\tabularnewline
$4$  & $0$ (exact)  & $\tfrac{512}{9}\delta^{4}$  & $0$\tabularnewline
$5$  & $\tfrac{225}{8192}\delta^{5}$  & $\tfrac{9765625}{73728}\delta^{5}$  & $\tfrac{2025}{9765625}$\tabularnewline
\bottomrule
\end{tabular}}\caption{Leading-order instability interval widths for the LC circuit and its
Mathieu approximation~\eqref{eq:LC-Mathieu-approx}. The LC circuit
has $L_{m}=0$ exactly for all even $m$ (Corollary~\ref{cor:even}).
All Mathieu intervals are present. Mathieu widths use $L_{m}^{\mathrm{Math}}=(c\delta)^{m}/[2^{2m-3}((m-1)!)^{2}]$
from~\eqref{eq:Lm-Math} with $c\approx m^{2}$ at each resonance.
At the primary resonance $m=1$, $c=1$ exactly, so $L_{1}^{\mathrm{Math}}=2c\delta=2\delta=L_{1}^{\mathrm{LC}}$:
the two equations agree to leading order. For $m\protect\geq3$, $m^{2m}\gg1$
makes $L_{m}^{\mathrm{Math}}\gg L_{m}^{\mathrm{LC}}$. LC circuit
widths are from~\eqref{eq:Lm-exact}.}
\label{tab:widths} 
\end{table}

\begin{rem}[Large-$m$ asymptotics of the leading coefficients]
\label{rem:asymp-comparison} Define the leading coefficients of
$L_{m}^{\mathrm{LC}}$ and $L_{m}^{\mathrm{Math}}$ (pure numbers,
independent of $\delta$): 
\begin{equation}
A_{m}^{\mathrm{LC}}=\lim_{\delta\to0}\frac{L_{m}^{\mathrm{LC}}}{\delta^{m}}=\frac{(m!!)^{2}}{2^{(m^{2}+8m-13)/4}},\qquad m=1,3,5,\ldots,\label{eq:alpha-LC}
\end{equation}
\begin{equation}
A_{m}^{\mathrm{Math}}=\lim_{\delta\to0}\frac{L_{m}^{\mathrm{Math}}}{\delta^{m}}=\frac{m^{2m}}{2^{2m-3}\,[(m-1)!]^{2}},\qquad m=1,3,5,\ldots\label{eq:alpha-Math}
\end{equation}
(where $L_{m}^{\mathrm{Math}}$ is evaluated at resonance $c\approx m^{2}$,
eq.~\eqref{eq:Lm-Math}). Using Stirling's asymptotic expansion for
$\ln\Gamma(z)$ \cite[eq.~8.344\textsuperscript7]{GraRyzh} applied
to $m!!=2^{(m+1)/2}\,\Gamma((m+2)/2)/\sqrt{\pi}$ (valid for $m\to\infty$,
$m$ odd), one obtains 
\begin{align}
\ln A_{m}^{\mathrm{LC}} & =-\frac{m^{2}\ln2}{4}+m\ln m-m(1+2\ln2)+O(\ln m),\label{eq:asymp-LC}\\[4pt]
\ln A_{m}^{\mathrm{Math}} & =2(1-\ln2)\,m+O(\ln m),\quad m\to\infty.\label{eq:asymp-Math}
\end{align}
The dominant term in~\eqref{eq:asymp-LC} is $-m^{2}\ln2/4$, giving
\emph{super-exponential} (Gaussian in $m$) decay: $A_{m}^{\mathrm{LC}}\sim2^{-m^{2}/4}$
as $m\to\infty$. By contrast,~\eqref{eq:asymp-Math} shows \emph{exponential}
growth: $A_{m}^{\mathrm{Math}}\sim(e^{2}/4)^{m}$. The difference
\begin{equation}
\ln\frac{A_{m}^{\mathrm{LC}}}{A_{m}^{\mathrm{Math}}}=-\frac{m^{2}\ln2}{4}+m\ln m-3m+O(\ln m),\quad m\to\infty.\label{eq:asymp-ratio}
\end{equation}
is dominated by $-m^{2}\ln2/4\to-\infty$, confirming that $A_{m}^{\mathrm{LC}}/A_{m}^{\mathrm{Math}}\to0$
at a Gaussian rate in $m^{2}$: the LC circuit instability intervals
become negligibly narrow compared to the Mathieu approximation for
large $m$, regardless of the value of $\delta$. Restoring the $\delta^{m}$
factor for fixed $\delta\in(0,1)$ as $m\to\infty$, the leading term
of $L_{m}^{\mathrm{Math}}$ behaves as 
\begin{equation}
L_{m}^{\mathrm{Math}}\approx A_{m}^{\mathrm{Math}}\,\delta^{m}\sim(e^{2}\delta/4)^{m},\label{eq:Lmath-asymp-delta}
\end{equation}
which grows when $\delta>4/e^{2}\approx0.541$ and decays when $\delta<4/e^{2}$.
By contrast, $L_{m}^{\mathrm{LC}}\approx A_{m}^{\mathrm{LC}}\,\delta^{m}\sim2^{-m^{2}/4}(m\delta/e)^{m}$
decays super-exponentially for every fixed $\delta\in(0,1)$, since
$2^{-m^{2}/4}$ dominates all polynomial or exponential factors in
$m$. 
\end{rem}

\begin{rem}[Continued fractions and three-term recurrences]
\label{rem:CF-three-term} The connection between continued fractions
and three-term recurrence relations is classical and bidirectional~\cite[\S\,1]{Gautschi67}:
every continued fraction generates a three-term recurrence via its
fundamental recurrence formulas, and vice versa, every three-term
recurrence relation may be interpreted as the fundamental recurrence
of some continued fraction. The first direction is useful for computing
continued fractions; the second is the key to computing minimal solutions.
Pincherle's theorem (Theorem~\ref{thm:Pinch}; cf.~\cite[Sec.~5.2--5.3]{JonThr})
then provides the bridge: \emph{the continued fraction converges if
and only if the three-term recurrence possesses a minimal solution}.
In this work the chain takes the following form: the Ince equation~\eqref{eq:Ince}
yields a three-term Fourier recurrence (Proposition~\ref{lem:MW74});
the Poincaré--Perron theorem (Theorem~\ref{thm:PP}) identifies
the minimal and dominant solutions; and Pincherle's theorem translates
the instability boundary condition into convergence of $F_{\mathrm{even/odd}}(c)$.
This chain --- three-term recurrence $\to$ continued fraction $\to$
minimal solution $\to$ spectral boundary --- is what makes the CF
method exact and closed-form for Ince equations, and is unavailable
for general Hill equations outside the Ince class (Remark~\ref{rem:CF-scope},
item~(ii)). 
\end{rem}

\begin{rem}[Scope of the CF method: Ince equations and the Mathieu case]
\label{rem:CF-scope} Three observations on the scope and transferability
of the CF method developed in this section.

\emph{(i) General Ince equations.} The three-term recurrence structure
of MW Lemma~7.4 holds for the full four-parameter Ince family~\eqref{eq:Ince}
with arbitrary $a,b,d$: the Ince coefficients involve only $\cos2x$
and $\sin2x$, so substituting the Fourier ansatz~\eqref{eq:Fourier-ansatz}
shifts indices by exactly $\pm1$, always producing a three-term recurrence.
Pincherle's theorem (Theorem~\ref{thm:Pinch}) therefore applies
and the CF eigenvalue condition $F_{\mathrm{even/odd}}(c)=0$ can
be defined for any Ince equation. For general $b,d\neq0$ the coupling
coefficients $Q^{*}(n)$ and $Q^{*}(-n-1)$ differ from each other
(asymmetric recurrence), making the backward-chain computation more
involved, but the method remains valid. The LC circuit case $b=d=0$
is the simplest non-trivial instance and is the only one pursued here.

\emph{(ii) General Hill equations and the three-term selector.} The
CF method does \emph{not} extend beyond the Ince class. For a general
Hill equation whose coefficient $Q(x)$ has more than one nonzero
Fourier harmonic beyond the first, substituting the Floquet ansatz
gives a recurrence of bandwidth equal to the number of harmonics ---
not three-term --- and the scalar continued-fraction reduction breaks
down. The Ince equation is the most general Hill-type equation to
which the three-term recurrence method applies~\cite[Sec.~7.1]{MagWin}.

This observation can be turned around: the \emph{three-term property
of the Fourier coefficient recurrence is a structural selector of
the Ince class from the broader Hill class}. A Hill equation $(1+f(x))q''+cq=0$
belongs to the Ince class if and only if substituting the Fourier
ansatz produces a three-term recurrence, which happens precisely when
$f(x)$ involves only a single harmonic (i.e.\ $f(x)=a\cos2x+b\sin2x$,
possibly after an appropriate change of variables).

The significance of this selector goes beyond computational convenience.
It is the three-term structure that makes the following chain possible: 
\begin{enumerate}
\item The Poincaré--Perron theorem applies (the recurrence has two characteristic
roots $\zeta_{1}=\delta$ and $\zeta_{2}=1/\delta$, real and distinct),
giving a minimal and a dominant solution. 
\item Pincherle's theorem applies (the CF converges to the minimal-solution
ratio), giving the eigenvalue condition $F_{\mathrm{even/odd}}(c)=0$. 
\item The Ince starter equations prescribe the initial ratio $z_{1}(c)$
from the ODE itself. 
\item Boundary eigenvalues are identified as those $c$ for which $z_{1}(c)=K(c)$,
eliminating the dominant solution and producing an $\ell^{2}$-Floquet
solution of the Ince equation. 
\end{enumerate}
None of these steps is available for a general Hill equation outside
the Ince class: the recurrence is not three-term, the PP/Pincherle
framework does not apply in scalar form, and the boundary identification
via minimal-solution suppression has no direct analogue. The three-term
property is therefore not merely a technical convenience but the structural
reason the Ince class admits an exact, closed-form spectral theory.

\emph{(iii) The Mathieu equation and the LC limit $\delta\to0$.}
The Mathieu equation ($a=0$, $d=-2q$, $b=0$ in Ince's family) is
the other canonical Ince case. Its CF method is independent of the
LC circuit analysis: the coupling coefficients are $Q^{*}(n)=Q^{*}(-n-1)=2q$
(constant), giving a symmetric recurrence with a different CF structure
and the width formula~\eqref{eq:Lm-Math-main}. One might ask whether
the Mathieu boundaries can be recovered from the LC formulas by taking
$\delta\to0$. They cannot: as $\delta\to0$ the LC modulation $a=-\varepsilon\to0$
simultaneously, collapsing all LC instability intervals to zero width.
The Mathieu equation corresponds to the Ince limit $a\to0$, $d\to-2q\neq0$,
which is a \emph{different} path in parameter space. The ratio $L_{m}^{\mathrm{LC}}/L_{m}^{\mathrm{Math}}$
at small $\delta$ (Corollary~\ref{cor:ratio}, eq.~\eqref{eq:ratio})
does capture the structural comparison, retaining the spectral parameter
$c$ explicitly; the resonance substitution $c\approx(2m-1)^{2}$
is made only for numerical values. The absolute Mathieu boundaries
must be computed from the Mathieu CF, not from the LC formulas. 
\end{rem}

\section{Verification against Cambi's exact solution}

\label{sec:Cambi}

This chapter verifies the closed-form width formula~\eqref{eq:Lm-main}
and the boundary curve\index{boundary curves}s of Theorem~\ref{thm:EPD-bdry}
against Cambi's 1950 exact continued-fraction\index{continued fraction}
solution~\cite[Table~II]{Cambi2}, which is independent of the Ince
and MW frameworks and serves as a direct numerical benchmark. For
$\delta\leq0.2$ ($\gamma\leq0.20$) agreement is better than $2.3\%$;
discrepancies grow with $\delta$ due to $O(\delta^{2})$ corrections
to the leading-order formula.

\subsection{Translation between notations}

\label{subsec:translation}

The modulation parameter $\delta$ relates to Cambi's $\gamma$ (where
$\gamma=\varepsilon/2$ is half the modulation amplitude) by 
\begin{equation}
\gamma=\frac{\delta}{1+\delta^{2}},\qquad\delta=\frac{1-\sqrt{1-4\gamma^{2}}}{2\gamma}=\gamma+\gamma^{3}+O\left(\gamma^{5}\right),\quad\gamma\to0.\label{eq:gamma-delta}
\end{equation}
From \eqref{eq:lambda-MW} and \eqref{eq:Cambi-res2} we have 
\begin{equation}
\hat{\lambda}=\frac{4(1+\delta^{2})}{r^{2}(1-\delta^{2})},\quad p=\frac{1}{r}=\frac{\omega_{0}}{\mu}.\label{eq:gamma-delta1}
\end{equation}
The $m$-th resonance\index{resonance} center in $p$-space is 
\[
p_{m}=\frac{m}{2}\sqrt{\frac{1-\delta^{2}}{1+\delta^{2}}},
\]
obtained by setting $\hat{\lambda}=m^{2}$ and solving for $p$. The
chain rule gives $d\hat{\lambda}/dp=8(1+\delta^{2})p/(1-\delta^{2})$
(from $\hat{\lambda}=4(1+\delta^{2})p^{2}/(1-\delta^{2})$ with $p=1/r$),
so the width in $p$-space is 
\begin{equation}
\Delta p_{m}=\frac{L_{m}}{|d\hat{\lambda}/dp|_{\text{center}}}=\frac{L_{m}(1-\delta^{2})}{8(1+\delta^{2})p_{m}}.\label{eq:Dp-conversion}
\end{equation}
Substituting $L_{1}=2\delta/(1-\delta^{2})$ from~\eqref{eq:L1}
and $p_{1}=\tfrac{1}{2}\sqrt{(1-\delta^{2})/(1+\delta^{2})}$: 
\[
\begin{aligned}\Delta p_{1} & =\frac{2\delta}{1-\delta^{2}}\cdot\frac{1-\delta^{2}}{8(1+\delta^{2})\cdot\tfrac{1}{2}\sqrt{(1-\delta^{2})/(1+\delta^{2})}}\\
 & =\frac{\delta}{2\sqrt{(1-\delta^{2})(1+\delta^{2})}}=\frac{\delta}{2\sqrt{1-\delta^{4}}}.
\end{aligned}
\]
\begin{equation}
\Delta p_{1}=\frac{\delta}{2\sqrt{1-\delta^{4}}},\qquad\Delta p_{3}=\frac{9\delta^{3}}{384(1+\delta^{2})^{5/2}(1-\delta^{2})^{1/2}}.\label{eq:Dp1}
\end{equation}

\subsection{Cambi's continued-fraction solution}

As described in Chapter~\ref{sec:LCInce}, Cambi~\cite{Cambi1},
\cite[\S\S\,1--3]{Cambi2} determines the instability boundaries via
the resonance equation~\eqref{eq:Cambi-res}, whose symmetric form
(symmetric because $v(u)$ and $v(1-u)$ appear on equal footing,
reflecting the boundary-to-boundary symmetry $u\to1-u$; used by Cambi
for numerical computation) is 
\begin{equation}
\frac{1}{v(u_{0})}-\gamma^{2}v(1-u_{0})=0.\label{eq:Cambi-res-alt}
\end{equation}
The continued fraction $v(u)$ of~\eqref{eq:Cambi-res} converges
for all real $u$ and $r$ whenever $2\gamma<1$, i.e., $\varepsilon<1$.

For half-integer $u_{0}=k+\frac{1}{2}$ (odd instability intervals),
equation~\eqref{eq:Cambi-res-alt} has two distinct solutions in
$p$ bounding an instability region. For the even resonance $u_{0}=1$
($m=2$, the even case in Cambi's table), the term $v(1-u_{0})=v(0)=0$
(since $G(0)=-\infty$) drops out of~\eqref{eq:Cambi-res-alt}, leaving
$1/v(u_{0})=0$, i.e., $v(u_{0})=\infty$ --- a single curve rather
than two, so the instability region degenerates to zero width~\cite[\S\,11]{Cambi2}.
The same collapse at every even resonance is the content of Corollary~\ref{cor:even}
and the discriminant\index{discriminant} identity~\eqref{eq:DeltaAtEven}.

Cambi's Table~II of~\cite[Table~II]{Cambi2} lists the coordinates
$(p,\gamma)$ of the instability boundaries for $u_{0}=\frac{1}{2}$
(primary, $m=1$), $u_{0}=1$ (even, zero-width, $m=2$), and $u_{0}=\frac{3}{2}$
(secondary, $m=3$), for $\gamma=0,0.1,0.2,0.3,0.4,0.45,0.5$.

\subsection{Stability diagram}

\label{subsec:StabDiag}

\begin{figure}[htbp]
\centering \includegraphics[width=9cm]{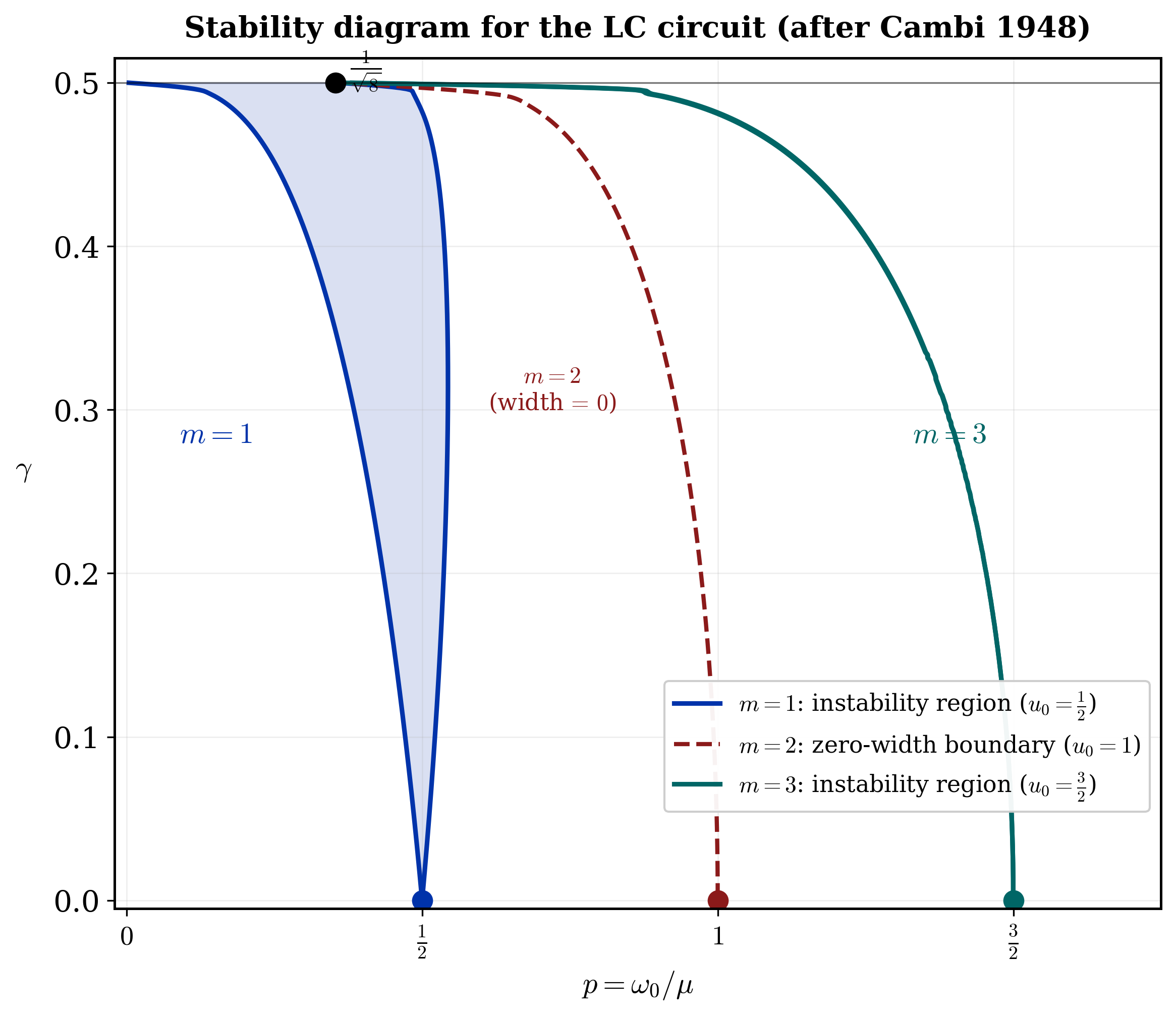} \caption{Stability diagram for the LC circuit equation~\eqref{eq:LC-original}
in the $(p,\gamma)$-plane, where $p=\omega_{0}/\mu$ and $\gamma=\varepsilon/2$.
Shaded regions are unstable. The $m=1$ tongue (blue, wide) opens
from $p=\frac{1}{2}$ and widens monotonically: as $\gamma\to\frac{1}{2}$
its left boundary descends to $p=0$ while its right boundary approaches
$p=1/\sqrt{8}$ (Cambi's Table~II: $p=0.00000$ and $0.35355$ at
$\gamma=0.5$). The $m=2$ resonance (red dashed) has exactly zero
width: Corollary~\ref{cor:even} asserts that only the single boundary
curve exists, with no shaded region. The $m=3$ tongue (teal) opens
from $p=\frac{3}{2}$ and, though razor thin (width $\sim\gamma^{3}$),
extends to $\gamma=\frac{1}{2}$, where it converges to $p=1/\sqrt{8}$
together with the $m=2$ curve and the right $m=1$ boundary. Boundary
curves computed numerically by solving Cambi's resonance equation~\eqref{eq:Cambi-res-alt}
using the continued-fraction representation~\eqref{eq:Cambi-v-CF}
for $v(u)$ (Theorem~\ref{thm:v-CF}), verified against Cambi's Table~II.
Compare with Figure~2 of Cambi~\cite{Cambi1}.}
\label{fig:stability} 
\end{figure}

Figure~\ref{fig:stability} shows the stability diagram in the $(p,\gamma)$-plane
computed by numerically solving Cambi's resonance equation~\eqref{eq:Cambi-res-alt}
for $u_{0}=\frac{1}{2},1,\frac{3}{2}$. The qualitative features are
immediately apparent. 

\subsection{\texorpdfstring{Primary instability interval ($m=1$, $u_{0}=\frac{1}{2}$)}{Primary
instability interval (m=1, u0=1/2)}}

Table~\ref{tab:m1} compares formula~\eqref{eq:Dp1} against the
numerical solution of Cambi's resonance equation~\eqref{eq:Cambi-res-alt}
and Cambi's Table~II. The formula is a leading-order result; discrepancies
grow with $\delta$ due to $O(\delta^{2})$ corrections.

\begin{table}[ht]
\centering 
\global\long\def\arraystretch{1.4}%
\begin{tabular}{ccccccc}
\toprule 
$\gamma$  & $\delta$  & $p_{\text{left}}$  & $p_{\text{right}}$  & $\Delta p$ (Cambi)  & $\Delta p_{1}$~\eqref{eq:Dp1}  & Error\tabularnewline
\midrule 
$0.10$  & $0.10102$  & $0.47137$  & $0.52165$  & $0.05028$  & $0.05051$  & $0.5\%$\tabularnewline
$0.20$  & $0.20871$  & $0.43448$  & $0.53659$  & $0.10211$  & $0.10446$  & $2.3\%$\tabularnewline
$0.30$  & $0.33333$  & $0.38463$  & $0.54360$  & $0.15897$  & $0.16771$  & $5.5\%$\tabularnewline
$0.40$  & $0.50000$  & $0.31159$  & $0.53771$  & $0.22612$  & $0.25820$  & $14\%$\tabularnewline
\bottomrule
\end{tabular}\caption{Primary instability ($m=1$): width $\Delta p_{1}$ in $p$-space
from Cambi's Table~II, numerical solution of~\eqref{eq:Cambi-res-alt},
and formula~\eqref{eq:Dp1}. Errors are $O(\delta^{2})$ corrections
to the leading-order formula.}
\label{tab:m1} 
\end{table}

\begin{rem}[Verification of Cambi Table~II]
\label{rem:Cambi-verify} Cambi's Table~II boundary values were
independently verified by numerical integration of Cambi's equation
$(1+2\gamma\cos x)y^{\prime\prime}+p^{2}y=0$ (monodromy matrix\index{monodromy matrix}
over one period $[0,2\pi]$). At each $(\gamma,p)$ pair in the table,
the discriminant $\Delta_{\mathrm{Cambi}}=y_{1}(2\pi)+y_{2}^{\prime}(2\pi)$
satisfies $\Delta_{\mathrm{Cambi}}+2=0$ to within $10^{-4}$, confirming
the tabulated values to their stated 5-decimal precision. Agreement
between the zeros of $F_{\mathrm{even/odd}}$ (eq.~\eqref{eq:Fdef},
computed from the nested continued fraction~\eqref{eq:recCF}) and
Cambi's table is at the level of his tabulation accuracy for all $\gamma\leq0.40$:
the maximum deviation is $0.13\%$, at the left $m=1$ boundary for
$\gamma=0.40$, and is attributable to Cambi's hand computation, since
the CF zeros coincide with the direct monodromy boundaries to $10^{-14}$
(Remark~\ref{rem:CF-boundary-verify}). 
\end{rem}

\subsection{\texorpdfstring{Even interval ($m=2$, $u_{0}=1$): exact zero width}{Even
interval (m=2, u0=1): exact zero width}}

Cambi's Table~II gives a \emph{single} $p$-value for each $\gamma$
at $u_{0}=1$, confirming zero-width. Our formula predicts the location
of this single boundary curve as $p_{2}\approx\sqrt{(1-\delta^{2})/(1+\delta^{2})}$,
in agreement with Cambi's values to within~$1\%$ at $\gamma=0.1$
and $4\%$ at $\gamma=0.3$ (Table~\ref{tab:m2}). The deviation
is the $O(\delta^{2})$ tangency shift itself: replacing $\hat{\lambda}=4$
by the tangency point $\hat{\lambda}_{2}(\delta)=4+2.67\,\delta^{2}+O(\delta^{4})$
of the discriminant identity~\eqref{eq:DeltaAtEven} gives $p_{2}=0.99321$,
$0.97117$, $0.92700$, $0.83673$ at $\gamma=0.1,0.2,0.3,0.4$, reducing
the discrepancy to $0.005\%$, $0.05\%$, $0.37\%$, $2.0\%$; locating
the tangency of $\Delta$ with $+2$ exactly reproduces Cambi's $p_{2}$
to his five-decimal precision ($0.0003\%$--$0.013\%$) at all four
values of $\gamma$.

\begin{table}[ht]
\centering 
\global\long\def\arraystretch{1.4}%
\begin{tabular}{cccc}
\toprule 
$\gamma$  & $\delta$  & $p_{2}$ (Cambi)  & $p_{2}^{\mathrm{pred}}$\tabularnewline
\midrule 
$0.10$  & $0.10102$  & $0.99326$  & $0.98985$\tabularnewline
$0.20$  & $0.20871$  & $0.97170$  & $0.95735$\tabularnewline
$0.30$  & $0.33333$  & $0.93048$  & $0.89443$\tabularnewline
$0.40$  & $0.50000$  & $0.85374$  & $0.77460$\tabularnewline
\bottomrule
\end{tabular}\caption{Even interval ($m=2$): Cambi's single boundary curve $p_{2}$ from
his Table~II (read from the $u_{0}=1$ row of his resonance equation)
and our leading-order prediction $p_{2}^{\mathrm{pred}}=\sqrt{(1-\delta^{2})/(1+\delta^{2})}$
(the unperturbed $m=2$ resonance center $\hat{\lambda}=4$ expressed
in $p$-space; no $O(\delta^{2})$ shift is included). The interval
has zero width exactly (Corollary~\ref{cor:even}): both boundary
curves coincide at $p_{2}$. The prediction is leading-order only;
the agreement deteriorates as $\delta$ increases since higher-order
corrections are not included. Cambi's $p_{2}$ values have been independently
verified by direct monodromy integration of his equation~\eqref{eq:Cambi-res}:
at each tabulated $p_{2}$, the discriminant satisfies $\operatorname{Tr}(M(2\pi))-2=0$
to within $10^{-8}$, confirming the coexistence condition $u_{0}=1$
(Floquet multiplier $+1$).}\index{Floquet theory}\index{Floquet theory!Floquet multiplier}
\label{tab:m2} 
\end{table}

\subsection{\texorpdfstring{Secondary instability interval ($m=3$, $u_{0}=\frac{3}{2}$)}{Secondary
instability interval (m=3, u0=3/2)}}

Table~\ref{tab:m3} compares formula~\eqref{eq:Dp1} against numerical
solution of Cambi's resonance equation and Cambi's Table~II.

\begin{table}[ht]
\centering 
\global\long\def\arraystretch{1.4}%
\resizebox{\textwidth}{!}{%
\begin{tabular}{cccccc}
\toprule 
$\gamma$  & $\delta$  & $\Delta p$ (numerical)  & $\Delta p_{3}$~\eqref{eq:Dp1}  & Error  & $\Delta p\,/\gamma^{3}$\tabularnewline
\midrule 
$0.15$  & $0.15354$  & $0.0000847$  & $0.0000810$  & $4.4\%$  & $0.02511$\tabularnewline
$0.20$  & $0.20871$  & $0.0002128$  & $0.0001959$  & $7.9\%$  & $0.02660$\tabularnewline
$0.30$  & $0.33333$  & $0.0008675$  & $0.0007075$  & $18\%$  & $0.03213$\tabularnewline
$0.40$  & $0.50000$  & $0.0029673$  & $0.0019365$  & $35\%$  & $0.04641$\tabularnewline
\midrule 
Leading:  & $\delta\to0$  &  & $9/384=0.02344$  &  & $\to0.02344$\tabularnewline
\bottomrule
\end{tabular}}\caption{Secondary instability ($m=3$): width $\Delta p_{3}$ in $p$-space
from numerical solution of~\eqref{eq:Cambi-res-alt} and formula~\eqref{eq:Dp1}.
The leading coefficient $9/384$ is confirmed numerically (last column
converges to $0.02344$ as $\gamma\to0$). Increasing errors at larger
$\gamma$ are $O(\delta^{2})$ corrections to the leading-order formula.
Cambi's Table~II widths at $\gamma=0.1,0.2,0.3,0.4$ ($0.00002$,
$0.00019$, $0.00090$, $0.00297$) are consistent but have limited
precision ($5$ decimal places in $p$).}
\label{tab:m3} 
\end{table}

The leading coefficient $9/384=0.023438$ is confirmed by the numerical
computation: $\Delta p_{3}/\gamma^{3}=0.02361$, $0.02383$, $0.02415$
at $\gamma=0.05$, $0.075$, $0.10$, approaching the limit $9/384$
from above with the expected $O(\delta^{2})$ correction and continuing
the trend of the last column of Table~\ref{tab:m3}. The growing
discrepancy at larger $\gamma$ ($18\%$ at $\gamma=0.3$, $35\%$
at $\gamma=0.4$) is consistent with $O(\delta^{2})$ corrections
to the leading-order Ince recurrence analysis, where $\delta=0.333$
and $0.500$ respectively give $O(\delta^{2})\approx11\%$ and $25\%$
corrections.

\subsection{Physical interpretation}

Instability of the LC circuit requires parametric coupling between
the modulation frequency $\mu$ and the natural frequency $\omega_{0}=1/\sqrt{LC}$.
The primary resonance occurs at $\mu\approx2\omega_{0}$ (i.e., $r\approx2$,
$u_{0}\approx\frac{1}{2}$), corresponding to the well-known fact
that a capacitor driven at excitation frequency $\mu$ develops a
restoring force oscillating at frequency $2\mu$~\cite[Introduction]{Cambi1}.
More generally, the surviving instability conditions (eq.~\eqref{eq:resonance-condition-MR}
of Section~\ref{subsec:MR-stability}) are $\omega_{0}\approx(2k-1)\mu/2$
for $k=1,2,3,\ldots$, i.e.\ the natural frequency equals an odd
multiple of $\mu/2$. The even resonances ($\omega_{0}\approx k\mu$)
are absent because the rational structure of the coefficient in~\eqref{eq:LC-Hill}
forces coexistence of two periodic solutions at every even resonance,
collapsing those intervals exactly (Corollary~\ref{cor:even}; equation~\eqref{eq:DeltaAtEven}).
This structure is invisible to the Mathieu\index{Mathieu equation}
approximation, which incorrectly predicts instability at all resonances.

\section{Boundary curves and the Mathieu comparison}

\label{sec:Boundaries}

This chapter derives the complete EPD\index{exceptional point of degeneracy (EPD)}
curve\index{exceptional point of degeneracy (EPD)!EPD curve}s $c_{\pm}^{(m)}(\delta)$
(eq.~\eqref{eq:EPD-curves-def}) for the LC circuit and the Mathieu\index{Mathieu equation}
equation, proving Theorems~\ref{thm:EPD-bdry} and~\ref{thm:width}.
The approach is the backward-chain recurrence method of Section~\ref{subsec:Widths},
which yields explicit closed-form expressions for both the center
shift $D_{m}$ and the half-width $L_{m}^{(c)}/2$ of each instability
tongue\index{instability tongue}. The results are collected in Tables~\ref{tab:bdry-LC}
and~\ref{tab:bdry-Math}, and compared in~\S\,\ref{subsec:bdry-compare}.

\emph{Working variable.} All boundary curve\index{boundary curves}
derivations in this chapter use the dimensionless spectral parameter
$c$, the Ince parameter $a$, and the driving frequency $\mu$, related
by 
\begin{equation}
c=\frac{4\omega_{0}^{2}}{\mu^{2}},\quad a=-\frac{2\delta}{1+\delta^{2}},\quad\mu_{\pm}^{(m)}=\frac{2\omega_{0}}{\sqrt{c_{\pm}^{(m)}}},\label{eq:param-dict}
\end{equation}
(see eqs.~\eqref{eq:c-vs-mu}--\eqref{eq:c-to-mu} and~\eqref{eq:boundary-main}
in the Summary). Working in $c$ linearizes the resonance\index{resonance}
conditions ($c=m^{2}$ at the $m$-th tongue center), makes the boundary
curve expansions polynomial in $a$, and allows the continued-fraction\index{continued fraction}
and recurrence machinery to operate without algebraic clutter. The
last relation in~\eqref{eq:param-dict} converts boundary curves
back to physical driving frequencies; it is applied explicitly for
$m=1$ in Theorem~\ref{thm:primary-domain}.

Chapter~\ref{sec:LCInce} established that the $k$-th instability
interval is bounded by two eigenvalues $c_{k}^{\pm}$ whose difference
gives the leading-order width~\eqref{eq:Lm-exact} (with error $O_{m}(\delta^{m+2})$).
By Theorem~\ref{thm:bdry-analytic} of Chapter~\ref{app:genLin},
each boundary curve is an analytic function of $\delta$ at $\delta=0$,
so the Taylor expansions derived below are rigorously justified. The
complete description of the instability domain requires also the \emph{location}
of the interval center --- that is, the individual values of $c_{k}^{+}$
and $c_{k}^{-}$, not merely their difference. Each boundary eigenvalue
has the form 
\begin{equation}
c_{\pm}^{(m)}=m^{2}+\underbrace{D_{m}\,|a|^{2}}_{\text{center shift}}\pm\,\underbrace{\tfrac{1}{2}L_{m}^{(c)}}_{\text{half-width}}+O_{m}(|a|^{m+2}),\quad a\to0.\label{eq:boundary-general}
\end{equation}
where $L_{m}^{(c)}=|\Delta c_{m}|$ is the separation already computed
in~\eqref{eq:Dc-exact} and expressed purely in $a$ by eq.~\eqref{eq:Lmc-pure-a},
and the center shift $D_{m}$ is the same for both boundaries. It
comes from $O(|a|^{2})$ terms in the resonant equation, which are
symmetric between the even and odd solutions, and must be computed
separately. We derive $D_{m}$ explicitly for the LC circuit and the
Mathieu equation and collect the complete boundary curves in Tables~\ref{tab:bdry-LC}
and~\ref{tab:bdry-Math}.

\subsection{LC circuit boundary curves}

\label{subsec:LCbdry}

For the LC circuit, $a=-2\delta/(1+\delta^{2})$ and $c=\hat{\lambda}(1-\delta^{2})/(1+\delta^{2})$.
The center shifts $D_{m}$ are computed from the $O(a^{2})$ terms
in the Ince resonant equation for each $m$.

\medskip{}
 \emph{Primary interval ($m=1$).} At the $m=1$ resonance ($c\approx1$),
the dominant mode is $A_{1}$. Including the feedback from $A_{3}\approx(-a/16)A_{1}$
(non-resonant at $n=1$): 
\begin{equation}
[a-2(c-1)]A_{1}+9a\!\cdot\!\tfrac{-a}{16}A_{1}=0\Longrightarrow c_{\pm}^{(1)}=1-\frac{9a^{2}}{32}\pm\frac{|a|}{2}+O(|a|^{3}),\quad a\to0.\label{eq:LC-bdry1}
\end{equation}
Hence $D_{1}=-9/32$ and, converting to $\hat{\lambda}$-space:

\noindent
\begin{equation}
\lambda_{\mathrm{MW},\pm}^{(1)}=\frac{1+\delta^{2}}{1-\delta^{2}}\Bigl(1-\frac{9a^{2}}{32}\pm\frac{|a|}{2}\Bigr),\qquad a=-\frac{2\delta}{1+\delta^{2}}.\label{eq:LC-lam1}
\end{equation}

\medskip{}
 \emph{Secondary interval ($m=3$).} At the $m=3$ resonance ($c\approx9$),
the dominant mode is $A_{3}$. The non-resonant feedback from the
adjacent modes is obtained from the bulk recurrence~\eqref{eq:bulk-rec}:
\begin{gather}
n=1:\;a\,A_{1}+2[9-c]\,A_{3}+25a\,A_{5}=0\Rightarrow A_{1}\approx\frac{2(c-9)}{a}\,A_{3}-25\,\frac{A_{5}}{A_{3}}\,A_{3},\nonumber \\
n=2:\;9a\,A_{3}+2[25-c]\,A_{5}+49a\,A_{7}=0\Rightarrow A_{5}\approx-\frac{9a}{32}\,A_{3}.\label{eq:LC3-feedback}
\end{gather}
Substituting $A_{5}\approx(-9a/32)A_{3}$ into the $n=1$ equation
gives $A_{1}\approx[2(c-9)/a+225a/32]\,A_{3}$. Inserting into the
even starter~\eqref{eq:start-even}, $[a-2(c-1)]A_{1}+9aA_{3}=0$,
and expanding at $c=9+x$ with $|x|\ll1$, $|a|\ll1$: 
\[
\bigl[a-16-2x\bigr]\!\left(\frac{2x}{a}+\frac{225a}{32}\right)+9a=0.
\]
At leading order in $x=c-9$ and $a$ (dropping the cross-terms $O(xa)$
and $O(a^{2}x)$): 
\[
-\frac{32x}{a}+\frac{(-16)(225a)}{32}+9a=0\Longrightarrow x=-\frac{207a^{2}}{64}.
\]
The odd starter gives the same center shift (the $\pm a$ difference
enters only at $O(a^{3})$, producing the splitting $9|a|^{3}/256$
from Section~\ref{subsec:Widths}). Hence: 
\begin{equation}
c_{\pm}^{(3)}=9-\frac{207\,a^{2}}{64}\pm\frac{9|a|^{3}}{512}+O_{3}(|a|^{5}),\quad a\to0.\label{eq:LC-bdry3}
\end{equation}
so $D_{3}=-207/64$. The center shift $D_{3}|a|^{2}$ is $O(|a|^{2})$
while the half-width $\tfrac{9}{512}|a|^{3}$ is $O(|a|^{3})$, reflecting
that the center shift arises one order earlier in $|a|$ than the
splitting. The next correction is $O_{3}(|a|^{5})$: by analyticity
(Theorem~\ref{thm:bdry-analytic}), only odd powers of $|a|$ appear
in the half-width and only even powers in the center, so the next
terms are $O_{3}(|a|^{4})$ in the center and $O_{3}(|a|^{5})$ in
the half-width.

\medskip{}
 \emph{Tertiary interval ($m=5$).} The $O(a^{2})$ resonant equation
at $c\approx25$ involves the three-step backward chain from $A_{5}$
through $A_{3}$ and $A_{1}$. Setting $c=25+D_{5}a^{2}$ and solving
the even-starter condition to leading order gives 
\begin{equation}
c_{\pm}^{(5)}=25-\frac{1775\,a^{2}}{192}\pm\frac{225\,|a|^{5}}{524288}+O_{5}(|a|^{7}),\quad a\to0.\label{eq:LC-bdry5}
\end{equation}
so $D_{5}=-1775/192$. Similarly, the $m=7$ case yields 
\begin{equation}
c_{\pm}^{(7)}=49-\frac{7007\,a^{2}}{384}\pm\frac{11025\,|a|^{7}}{2147483648}+O_{7}(|a|^{9}),\quad a\to0.\label{eq:LC-bdry7}
\end{equation}
so $D_{7}=-7007/384$. These complete the four explicit cases of Theorem~\ref{thm:EPD-bdry}.

\medskip{}

Table~\ref{tab:bdry-LC} summarizes the boundary curves for the first
four surviving LC circuit instability intervals.

\begin{table}[ht]
\centering 
\global\long\def\arraystretch{1.7}%
\begin{tabular}{ccc}
\toprule 
$m$  & $c_{+}^{(m)}$ (upper boundary)  & $c_{-}^{(m)}$ (lower boundary)\tabularnewline
\midrule 
$1$  & $1-\tfrac{9}{32}a^{2}+\tfrac{1}{2}|a|$  & $1-\tfrac{9}{32}a^{2}-\tfrac{1}{2}|a|$\tabularnewline
$3$  & $9-\tfrac{207}{64}a^{2}+\tfrac{9}{512}|a|^{3}$  & $9-\tfrac{207}{64}a^{2}-\tfrac{9}{512}|a|^{3}$\tabularnewline
$5$  & $25-\tfrac{1775}{192}a^{2}+\tfrac{225}{524288}|a|^{5}$  & $25-\tfrac{1775}{192}a^{2}-\tfrac{225}{524288}|a|^{5}$\tabularnewline
$7$  & $49-\tfrac{7007}{384}a^{2}+\tfrac{11025}{2147483648}|a|^{7}$  & $49-\tfrac{7007}{384}a^{2}-\tfrac{11025}{2147483648}|a|^{7}$\tabularnewline
\bottomrule
\end{tabular}\caption{EPD curves $c_{\pm}^{(m)}$ for the four surviving LC circuit instability
tongues (Theorem~\ref{thm:EPD-bdry}), in Ince $c$-space ($a=-2\delta/(1+\delta^{2})$).
Center shifts $D_{m}a^{2}$ are symmetric; half-widths $L_{m}^{(c)}/2$
are antisymmetric, with exact leading coefficients. At each boundary
the monodromy matrix has an EPD Jordan block. To convert to $\hat{\lambda}$-space
multiply by $(1+\delta^{2})/(1-\delta^{2})$. Errors: $O(|a|^{m+2})$
for all $m$. Center shifts have no closed-form formula in $m$ (Remark~\ref{rem:Dm-no-formula}).}\index{monodromy matrix}\index{Jordan block}
\label{tab:bdry-LC} 
\end{table}

\begin{rem}[Numerical verification of Table~\ref{tab:bdry-LC}]
\label{rem:bdry-LC-verify} The $m=1$ boundary curves have been
verified numerically by monodromy integration (Remark~\ref{rem:bdry-numerics}
of Chapter~\ref{sec:MainResults}): errors are $O(|a|^{3})$ as predicted,
with the center shift $D_{1}=-9/32$ confirmed to four significant
figures. The $m=1$ case is also confirmed by the YS method to $O(|a|^{3})$
by an independent route (Chapter~\ref{sec:YS-exact-LC}). For $m=3$
direct numerical verification by monodromy integration has also been
performed: the deviation of the measured tongue center from $9+D_{3}a^{2}$
decays as $O(a^{4})$ (extrapolation of the measured shift reproduces
$D_{3}=-207/64$ to six significant figures), and the measured width
matches the leading coefficient $\tfrac{9}{512}|a|^{3}$ to the expected
$O(a^{2})$ relative accuracy. For $m=5$ and $m=7$ the widths ($\sim|a|^{m}$)
lie below practical numerical resolution, but the centers are verified:
the location of the discriminant extremum near $c=m^{2}$ matches
$m^{2}+D_{m}a^{2}$ with $O(a^{4})$ deviations, confirming $D_{5}=-1775/192$
and $D_{7}=-7007/384$ (the latter to six significant figures under
extrapolation). A YS-series confirmation for $m\geq3$ would require
carrying the resonance-adapted expansion to order $m$ and has not
been pursued (Remark~\ref{rem:YS-higher-tongues}). 
\end{rem}

\begin{rem}[Degeneracy of EPD curve pairs for higher tongues]
\label{rem:bdry-degeneracy} Table~\ref{tab:bdry-LC} reveals a
striking structural feature: for the $m$-th tongue, $c_{+}^{(m)}$
and $c_{-}^{(m)}$ are \emph{identical through order $O(|a|^{m-1})$}
and differ only at order $O(|a|^{m})$. Specifically, both boundaries
share the same center shift $D_{m}a^{2}$ (which is $O(|a|^{2})$
and symmetric between the even and odd Floquet\index{Floquet theory}
solutions), while the splitting $c_{+}^{(m)}-c_{-}^{(m)}=L_{m}^{(c)}\sim|a|^{m}$
is $m$-th order. For $m=3$ the two EPD curves are indistinguishable
to second order in $|a|$; for $m=7$ they are indistinguishable to
sixth order.

The mathematical origin is clear from the backward chain construction:
the center shift arises from quadratic $O(a^{2})$ feedback in the
resonant equation (symmetric for both boundaries), while the splitting
requires traversing the full $m$-step chain, contributing one power
of $|a|$ per step.

The physical consequences are significant. First, the tongue widths
decrease rapidly with $m$: at $|a|=0.2$, 
\[
L_{1}^{(c)}\approx0.20,\quad L_{3}^{(c)}\approx2.8\times10^{-4},\quad L_{5}^{(c)}\approx2.7\times10^{-7},\quad L_{7}^{(c)}\approx1.3\times10^{-10},
\]
a reduction by many orders of magnitude with each step. Higher tongues
are therefore practically unobservable at any realistic modulation
amplitude: the circuit must be tuned to extraordinary precision to
reside inside the $m=3$ tongue, let alone $m=5$ or $m=7$. Second,
for EPD sensing the primary tongue ($m=1$) is overwhelmingly the
most practical operating point: its two EPD curves are separated by
$O(|a|)$, a \emph{first-order} quantity, whereas all higher tongues
have EPD curves separated by $O(|a|^{m})$ with $m\geq3$, making
them exponentially harder to resolve and exploit. 
\end{rem}

\subsection{The Mathieu equation as a degenerate Ince equation}

\label{subsec:MathieuInce}

This section specializes the Ince continued-fraction method of Section~\ref{subsec:Widths}
to the degenerate case $a=0$, deriving the Mathieu instability-interval
width formula~\eqref{eq:Lm-Math-derived} (the boxed result at the
end of the section is its conclusion). We begin with the structural
contrast that the degeneration produces.

The standard Mathieu equation 
\begin{equation}
y''+(\lambda-2q\cos2x)\,y=0\label{eq:Mathieu}
\end{equation}
is Ince's equation\index{Ince equation}~\eqref{eq:Ince} with 
\begin{equation}
a=0,\quad b=0,\quad c=\lambda,\quad d=-2q,\label{eq:Mathieu-Ince}
\end{equation}
giving constant polynomials $Q(\mu)=q$ and $Q^{*}(\mu)=2q$, independent
of $\mu$ (from~\eqref{eq:QPoly} with $a=b=0$, $d=-2q$: $Q=-d/2=q$
and $Q^{*}=-d=2q$). Since constant polynomials have no roots at all,
the necessary condition of Proposition~\ref{thm:MW71} is never satisfied:
\emph{no instability interval of the Mathieu equation ever vanishes}.
This is the exact opposite of the LC circuit, where the integer root
$\mu=0$ of $Q$ causes all even intervals to vanish simultaneously.

The recurrences~\eqref{eq:bulk-rec} reduce for $a=0$, $d=-2q$
to tridiagonal systems with \emph{constant} off-diagonal coupling
$q$ at every step: 
\begin{equation}
q\,A_{2n-1}+\bigl[(2n+1)^{2}-c\bigr]A_{2n+1}+q\,A_{2n+3}=0.\label{eq:Math-rec}
\end{equation}
The starters~\eqref{eq:start-even}--\eqref{eq:start-odd} become
$[q-(c-1)]A_{1}+qA_{3}=0$ (even) and $[-q-(c-1)]B_{1}+qB_{3}=0$
(odd), giving the continued-fraction difference 
\begin{equation}
F_{\mathrm{even}}(c)-F_{\mathrm{odd}}(c)=\frac{-2q}{q}=-2=\mathrm{const}.\label{eq:Math-CFdiff}
\end{equation}
The structure is identical to the LC circuit (cf.~\eqref{eq:CF-diff})
except that the constant is $-2$ rather than $-2/9$. 
\begin{rem}[Significance of the CF difference constant]
The CF difference constant $F_{\mathrm{even}}-F_{\mathrm{odd}}$
governs the leading-order width of every instability interval via
$|{\Delta c_{m}}|=|\text{const}|/|F'(c_{m})|$. Its value is determined
entirely by the Ince starter equations: for the LC circuit the ratio
$Q^{*}(-1)/Q^{*}(0)=9$ (from $Q^{*}(\mu)=a(2\mu-1)^{2}$ at $\mu=0$
and $\mu=-1$) gives $-2/9$; for Mathieu the constant polynomial
$Q^{*}=2q$ gives $-2$. The width, however, also involves the slope
$|F'(c_{m})|$ of the eigenvalue function, which carries the starter
normalization: for the LC circuit $|F_{\mathrm{even}}'(c_{1})|\approx2/(9|a|)$
(the starter denominator $9a$), giving $|\Delta c_{1}|=(2/9)\cdot(9|a|/2)=|a|$,
while for Mathieu $|F'(c_{1})|\approx1/q$ gives $|\Delta c_{1}|=2q$.
At the primary resonance ($c=1$, $q=c\delta=\delta$, $|a|\approx2\delta$)
the two leading widths therefore \emph{coincide}, $L_{1}^{\mathrm{LC}}=L_{1}^{\mathrm{Math}}=2\delta$
(Table~\ref{tab:widths}): the factor $1/9$ in the constant is exactly
compensated by the factor $9$ in the LC starter slope. The constant
is nevertheless a fingerprint of the Ince parameters: it enters every
width as the common numerator in $|\Delta c_{m}|=|\mathrm{const}|/|F'(c_{m})|$. 
\end{rem}

\subsubsection{Center shifts for Mathieu}

The center shift $E_{m}$ for the $m$-th Mathieu interval follows
from the $O(q^{2})$ terms in the resonant equation. For $m=1$: the
feedback from $A_{3}\approx(-q/8)A_{1}$ into the resonant (starter)
equation gives $c_{\pm}^{(1)}=1-q^{2}/8\pm q$, so $E_{1}=-1/8$.
For $m=3$: the non-resonant chain $A_{1}\approx(q/8)A_{3}$ (from
the starter), $A_{5}\approx(-q/16)A_{3}$ substituted into the resonant
equation gives $9-c=-q^{2}/16$, so $E_{3}=1/16$. The values $E_{4}=1/30$
and $E_{5}=1/48$ follow by the same procedure and agree with McLachlan~\cite[Sec.~2.151]{MacLa};
the general formula for $m\geq3$ is $E_{m}=1/(2(m^{2}-1))$.

\subsubsection{Width formula for odd Mathieu intervals}

\label{para:Math-chain-accuracy}

The backward chain from $A_{2k-1}$ to $A_{1}$ passes through exactly
$k-1$ non-resonant steps --- a \emph{fixed} number for fixed $m=2k-1$,
the same key point as in the LC circuit derivation. The sole approximation
is substituting $c=(2k-1)^{2}$ in each non-resonant denominator $[(2n+1)^{2}-(2k-1)^{2}]$,
making each a nonzero integer. With constant coupling $q$ at every
step (unlike the LC circuit where the coupling grows as $(2n-1)^{2}$),
the product evaluates \emph{exactly} to: 
\begin{equation}
\frac{A_{1}}{A_{2k-1}}=\prod_{n=1}^{k-1}\frac{-q}{(2n-1)^{2}-(2k-1)^{2}}=\frac{q^{k-1}}{4^{k-1}(2k-2)!}.\label{eq:Math-chain}
\end{equation}
The true $c$ differs from $(2k-1)^{2}$ by $O_{m}(q^{2})$; since
there are exactly $k-1$ steps for fixed $m$, the product changes
by $O_{m}(q^{2})$ relative, giving total error $O_{m}(q^{m+2})$
in the width --- two powers beyond the leading term, and not claimed
uniform in $m$. Using the CF difference~\eqref{eq:Math-CFdiff}
and propagating the even/odd starter sign through the $k-1$ chain
steps yields 
\begin{equation}
|\Delta c_{k}^{\mathrm{Math}}|=\frac{q^{m}}{2^{2m-3}\,[(m-1)!]^{2}}+O_{m}(q^{m+2}),\qquad m=2k-1.\label{eq:Math-Dc-odd}
\end{equation}

For even intervals ($m=2k$, period-$\pi$ solutions) the Fourier
series takes the form $y=\sum_{n=0}^{\infty}A_{2n}\cos2nx$ (cosine)
or $y=\sum_{n=1}^{\infty}B_{2n}\sin2nx$ (sine), with bulk recurrence
\begin{equation}
q\,A_{2n-2}+\bigl[(2n)^{2}-c\bigr]A_{2n}+q\,A_{2n+2}=0,\qquad n\geq2,\label{eq:Math-rec-even}
\end{equation}
while the two lowest cosine-mode equations are special: 
\[
\text{mode }0\!:\;-c\,A_{0}+q\,A_{2}=0,\qquad\text{mode }2\!:\;2q\,A_{0}+\bigl[4-c\bigr]A_{2}+q\,A_{4}=0,
\]
the \emph{doubled} coupling $2qA_{0}$ arising because $\cos2x\cdot A_{0}$
contributes in full at frequency $2$, whereas every other product
$\cos2x\cos2nx$ splits its weight between two frequencies --- the
classical feature of the Mathieu $\mathrm{ce}$-recursions. The sine
starter is $\bigl[4-c\bigr]B_{2}+q\,B_{4}=0$. The two starter types
have \emph{different} leading modes ($A_{0}$, the constant (DC) mode,
vs $B_{2}$, which starts at the first sine harmonic; the cos-type
has an extra degree of freedom from the zeroth mode that the sin-type
lacks), producing asymmetric center shifts. For $k=1$ ($c\approx4$,
dominant mode $A_{2}$ or $B_{2}$):

\emph{Cos-type} ($y=A_{0}+A_{2}\cos2x+\cdots$): the mode-$A_{0}$
equation gives $-c\,A_{0}+q\,A_{2}=0$, so $A_{0}=q\,A_{2}/c\approx q\,A_{2}/4$.
Substituting into the mode-$2$ equation $2qA_{0}+[4-c]A_{2}+qA_{4}=0$
and including the feedback from $A_{4}\approx-q\,A_{2}/12$ gives
the upper boundary $c_{\mathrm{cos}}=4+\tfrac{q^{2}}{2}-\tfrac{q^{2}}{12}=4+5q^{2}/12$.

\emph{Sin-type} ($y=B_{2}\sin2x+B_{4}\sin4x+\cdots$): the starter
has no zeroth-mode feedback, giving the lower boundary $c_{\mathrm{sin}}=4-q^{2}/12$.

The width $c_{\mathrm{cos}}-c_{\mathrm{sin}}=q^{2}/2$ is consistent
with~\eqref{eq:Lm-Math-derived} at $m=2$. The center $(c_{\mathrm{cos}}+c_{\mathrm{sin}})/2=4+q^{2}/6$
gives center shift $E_{2}=1/6$. The asymmetry $5\!:\!1$ (upper shift
$5q^{2}/12$ vs lower shift $q^{2}/12$) arises because the cos-type
starter feeds back through the extra degree of freedom $A_{0}$, which
is absent in the sin-type. The general even-interval formula follows
from the backward chain through $k-1$ steps with constant coupling
$q$: 
\begin{equation}
|\Delta c_{k}^{\mathrm{Math}}|=\frac{q^{2k}}{2^{2\cdot2k-3}\,[(2k-1)!]^{2}}.\label{eq:Math-Dc-even}
\end{equation}

Equations~\eqref{eq:Math-Dc-odd}--\eqref{eq:Math-Dc-even} unify
into the following closed-form formula (consistent with McLachlan~\cite[Sec.~2.151]{MacLa}):
\begin{equation}
\boxed{L_{m}^{\mathrm{Math}}=\frac{q^{m}}{2^{2m-3}\,[(m-1)!]^{2}}+O_{m}(q^{m+2}),\qquad m=1,2,3,\ldots}\label{eq:Lm-Math-derived}
\end{equation}
which is the conclusion of the derivation of this section. The $O_{m}(q^{m+2})$
accuracy claim rests on the fixed-step argument of Section~\ref{para:Math-chain-accuracy}:
for fixed $m=2k-1$ the backward chain has exactly $k-1$ steps, so
the error from substituting the resonant value of $c$ (itself accurate
to $O_{m}(q^{2})$) is $O_{m}(q^{2})$ relative, hence $O_{m}(q^{m+2})$
absolute, with a finite constant depending on $m$. 
\begin{rem}[Cross-verification of formula~\eqref{eq:Lm-Math-derived}]
\label{rem:Lm-Math-verify} The formula $L_{m}^{\mathrm{Math}}=q^{m}/[2^{2m-3}((m-1)!)^{2}]$
does not appear explicitly in McLachlan~\cite[Sec.~2.151]{MacLa},
who gives the characteristic numbers $a_{m}$ and $b_{m}$ case by
case in Sec.~2.151 without forming their difference in closed form.
As an independent check, the widths $L_{m}=a_{m}-b_{m}$ are computed
directly from McLachlan's individual expansions: 
\begin{gather*}
L_{1}=a_{1}-b_{1}=2q,\quad L_{2}=a_{2}-b_{2}=\tfrac{q^{2}}{2},\\
L_{3}=a_{3}-b_{3}=\tfrac{q^{3}}{32},\quad L_{4}=a_{4}-b_{4}=\tfrac{q^{4}}{1152}.
\end{gather*}
These agree exactly with~\eqref{eq:Lm-Math-derived} for $m=1,2,3,4$,
confirming every coefficient. The leading-order instability widths
for the LC circuit are likewise verified: the formula 
\[
L_{m}^{\mathrm{LC}}=\frac{(m!!)^{2}\,\delta^{m}}{2^{(m^{2}+8m-13)/4}}+O_{m}(\delta^{m+2})
\]
gives $L_{1}=2\delta$, $L_{3}=9\delta^{3}/32$, $L_{5}=225\delta^{5}/8192$,
which are confirmed numerically against Cambi's exact continued-fraction
solution to better than $2.3\%$ for $\delta\leq0.2$, with larger
discrepancies at $\delta>0.2$ due to higher-order corrections (Chapter~\ref{sec:Cambi}).
These cross-checks provide explicit confirmation that the significant
formulas of this work are free of arithmetic errors. 
\end{rem}

\subsection{Complete boundary curves for Mathieu}

Combining center shifts and half-widths, the Mathieu instability boundaries
are 
\begin{equation}
\lambda_{\pm}^{(m)}=m^{2}+E_{m}\,q^{2}\pm\frac{L_{m}^{\mathrm{Math}}}{2}+O\left(q^{m+1}\right),\quad q\to0.\label{eq:Mathieu-bdry-general}
\end{equation}
Here $E_{m}$ is the \emph{center shift coefficient}: the $O(q^{2})$
shift of the tongue center from the unperturbed resonance $\lambda=m^{2}$.
It is computed from the $O(q^{2})$ terms in the Ince resonant equation
for each $m$, analogously to $D_{m}$ for the LC circuit. The explicit
values are $E_{1}=-1/8$, $E_{2}=1/6$ (asymmetric; see the discussion
of the $m=2$ cos/sin starter asymmetry in Section~\ref{subsec:MathieuInce}),
and for $m\geq3$: 
\begin{equation}
E_{m}=\frac{1}{2(m^{2}-1)},\qquad m\geq3,\label{eq:Em-general}
\end{equation}
consistent with McLachlan~\cite[Sec.~2.151, eq.~(14)]{MacLa}. Table~\ref{tab:bdry-Math}
gives the first five cases.

\begin{table}[ht]
\centering 
\global\long\def\arraystretch{1.8}%
\begin{tabular}{cccc}
\toprule 
$m$  & Period  & $\lambda_{+}^{(m)}$  & $\lambda_{-}^{(m)}$\tabularnewline
\midrule 
$1$  & $2\pi$  & $1+q-\tfrac{q^{2}}{8}$  & $1-q-\tfrac{q^{2}}{8}$\tabularnewline
$2$  & $\pi$  & $4+\tfrac{5q^{2}}{12}$  & $4-\tfrac{q^{2}}{12}$\tabularnewline
$3$  & $2\pi$  & $9+\tfrac{q^{2}}{16}+\tfrac{q^{3}}{64}$  & $9+\tfrac{q^{2}}{16}-\tfrac{q^{3}}{64}$\tabularnewline
$4$  & $\pi$  & $16+\tfrac{1}{30}q^{2}+\tfrac{q^{4}}{2304}$  & $16+\tfrac{1}{30}q^{2}-\tfrac{q^{4}}{2304}$\tabularnewline
$5$  & $2\pi$  & $25+\tfrac{1}{48}q^{2}+\tfrac{q^{5}}{147456}$  & $25+\tfrac{1}{48}q^{2}-\tfrac{q^{5}}{147456}$\tabularnewline
\bottomrule
\end{tabular}\caption{Instability boundary curves $\lambda_{\pm}^{(m)}$ for the Mathieu
equation~\eqref{eq:Mathieu}. The half-widths $L_{m}^{\mathrm{Math}}/2$
are from formula~\eqref{eq:Lm-Math-derived}; center shifts $E_{m}$
from eq.~\eqref{eq:Em-general} and McLachlan~\cite[Sec.~2.151]{MacLa}.
For $m=2$: the asymmetric boundaries ($5q^{2}/12$ vs $-q^{2}/12$)
reflect the asymmetric starter equations for the cos-type and sin-type
period-$\pi$ solutions (Section~\ref{subsec:MathieuInce}).}
\label{tab:bdry-Math} 
\end{table}

\subsection{Comparison of boundary structures}

\label{subsec:bdry-compare}

The boundary curves reveal a sharp structural difference between the
two equations, beyond the mere existence of intervals.

\medskip{}
 \emph{Symmetry.} For the LC circuit, both boundaries of each odd
interval shift equally (by $D_{m}|a|^{2}$) and split symmetrically
($\pm L_{m}^{(c)}/2$). For the Mathieu equation, odd intervals also
split symmetrically, but even intervals are asymmetric: $\lambda_{+}^{(2)}-4=5q^{2}/12$
while $4-\lambda_{-}^{(2)}=q^{2}/12$, a ratio of 5:1.

\medskip{}
 \emph{Order of the width.} For Mathieu, the $m$-th interval has
width $\sim q^{m}$, so consecutive intervals shrink by one power
of $q$. For the LC circuit, \emph{even} intervals have zero width
(they are absent), and the odd intervals have width $\sim\delta^{m}$,
shrinking by two powers of $\delta$ between consecutive odd intervals.

\medskip{}
 \emph{Role of $a$ vs constant coupling.} The key algebraic difference
is that for the LC circuit the off-diagonal coupling in the recurrence
is $Q^{*}(n)=a(2n-1)^{2}$ (growing with mode number), while for Mathieu
it is the constant $q$. This causes the backward chain coefficient~\eqref{eq:Math-chain}
to involve $(2k-2)!$ rather than the double-factorial combinations
of the LC circuit formula, and it is why the formulas~\eqref{eq:Lm-exact}
and~\eqref{eq:Lm-Math-derived} have qualitatively different structures
despite arising from the same Ince recurrence method.

\section{Two complementary methods for instability boundaries: summary and
comparison}

\label{sec:Summary}

This chapter summarizes the two principal methods developed in this
work for describing the stability-to-instability boundaries of the
LC circuit, and compares their architecture, outputs, and relative
strengths. Both methods share the same starting point --- the identification
of the LC circuit equation as a special case of Ince's equation\index{Ince equation}
(Remark~\ref{rem:ince-id}, Chapter~\ref{sec:LCHill}) --- but
they extract different and complementary information from that structure.
The detailed analyses are in Chapters~\ref{sec:MWInce} and~\ref{sec:LCInce};
the present chapter provides a map for the reader and makes the comparison
explicit. A third method --- the Yakubovich--Starzhinskii\index{Yakubovich--Starzhinskii series}
(YS) exponent-matrix approach --- is developed in Chapters~\ref{sec:YS-LC}--\ref{sec:YS-exact-LC}
and compared with both methods in Chapter~\ref{sec:YS-comparison}.
That comparison shows that for the single-parameter LC circuit the
MW and CF methods give richer structural results more directly, while
the YS method has an absolute advantage for \emph{multiparameter}
families of oscillators (Seyranian--Mailybaev~\cite[Ch.~1]{SeyMai},
Kirillov~\cite[Sec.~3.3]{Kiril}), where its joint analyticity in
all parameters simultaneously makes the full stability surface accessible
without recomputing the analysis at each parameter point.

\subsection{The MW discriminant method}

\label{subsec:SummaryMW}

\emph{Central object.} The Hill discriminant\index{discriminant}\index{discriminant!Hill discriminant}
\begin{equation}
\Delta(\lambda)=y_{1}(\pi,\lambda)+y_{2}'(\pi,\lambda),\label{eq:sum-discriminant}
\end{equation}
where $y_{1}$, $y_{2}$ are the standard solutions of~\eqref{eq:LC-Hill},
is an entire function of the spectral parameter $\lambda$ (Theorem~\ref{thm:matrizant-analytic},
Chapter~\ref{app:genLin}). It encodes the full stability structure
(Theorem~\ref{thm:Hill-stability}, Chapter~\ref{app:Hill}): $|\Delta|<2$
(stable), $|\Delta|>2$ (unstable), $\Delta=\pm2$ (boundary).

\emph{MW expansion.} Magnus and Winkler~\cite[Cor.~2.6]{MagWin}
expand $\Delta=\sum_{n\geq0}\Delta_{n}$ (equation~\eqref{eq:Delta-MW},
Chapter~\ref{sec:MWInce}), where $\Delta_{n}$ is homogeneous of
degree $n$ in the Fourier coefficients $g_{k}$ of the perturbation.
The selection rule gives $\Delta_{1}=0$ identically, and 
\begin{equation}
\Delta_{0}=2\cos\pi\sqrt{\lambda},\qquad\Delta_{2}=\frac{\pi\sin\pi\sqrt{\lambda}}{2\sqrt{\lambda}}\sum_{n\geq1}\frac{g_{n}^{2}}{\lambda-n^{2}},\label{eq:sum-MW-exp}
\end{equation}
with all higher $\Delta_{n}$ determined by the Fourier coefficients
$g_{n}=\hat{\lambda}\delta^{|n|}$ of the LC circuit (equation~\eqref{eq:gn}).

\emph{Coexistence polynomials and qualitative structure.} The two
polynomials 
\begin{equation}
Q(\mu)=2a\mu^{2},\qquad Q^{*}(\mu)=a(2\mu-1)^{2},\label{eq:sum-coex-poly}
\end{equation}
associated with the Ince parameters of the LC circuit, govern which
instability intervals survive and which collapse. Recall that the
Floquet\index{Floquet theory} multiplier\index{Floquet theory!Floquet multiplier}s
$\rho_{1,2}$ of Hill's equation\index{Hill equation} satisfy $\rho_{1}\rho_{2}=1$;
at a stability boundary\index{stability boundary} they coalesce on
the unit circle at $\rho=+1$ (period-$\pi$ solutions: $y(x+\pi)=y(x)$)
or $\rho=-1$ (period-$2\pi$ solutions: $y(x+\pi)=-y(x)$, antiperiodic
with period $\pi$). \emph{Coexistence} means that two linearly independent
solutions share the same multiplier value, forcing the instability
interval to collapse to zero width (Theorem~\ref{thm:Hill-stability},
Chapter~\ref{app:Hill}). $Q(\mu)$ has the integer root $\mu=0$,
forcing period-$\pi$ coexistence ($\rho=+1$) at every even resonance\index{resonance}
by MW Theorem~7.6 (Proposition~\ref{prop:even-vanish}); $Q^{*}(\mu)$
has no integer root, so all odd intervals survive (Proposition~\ref{prop:odd-survive}).

\emph{Discriminant identity (our contribution).} The Ince coexistence
forced by the integer root of $Q$ gives the exact identity 
\begin{gather}
\Delta\bigl(\hat{\lambda}_{m}(\delta),\delta\bigr)=+2,\qquad\hat{\lambda}_{m}(\delta)=m^{2}+O(\delta^{2}),\nonumber \\
\text{for all even }m\text{ and all }0<\delta<1,\label{eq:sum-Delta-id}
\end{gather}
where $\hat{\lambda}_{m}(\delta)$ is the common value of the two
coincident period-$\pi$ eigenvalues: $\Delta-2$ has a double zero
there and is negative on a punctured neighborhood, so no even instability
interval opens. The identity is proved in Section~\ref{subsec:DeltaId};
it is exact in $\delta$ --- not a statement about any finite truncation
of the MW expansion --- and gives the MW explanation of why even
instability intervals are absent from the LC circuit.

\emph{Boundary curves and EPD\index{exceptional point of degeneracy (EPD)}/Krein\index{Krein collision theory}
classification (our contributions).} The boundary condition $\Delta=\pm2$
together with the MW expansion yields explicit asymptotic formulas
for the instability boundary curve\index{boundary curves}s (Tables~\ref{tab:bdry-LC}
and~\ref{tab:bdry-Math}). Each boundary decomposes into a symmetric
center shift ($O(\delta^{2})$, equal for both boundaries of an interval)
and an antisymmetric half-width ($O(\delta^{m})$, opposite in sign).
By Theorem~\ref{thm:EPD} (proved in Section~\ref{subsec:Hamiltonian})
and the $2\times2$ symplectic\index{symplectic system} dichotomy
of Yakubovich--Starzhinskii~\cite[Ch.~VIII, \S\,1.8]{YakSta2}:
at each odd resonance the colliding multipliers carry opposite Krein
signatures ($\Pi^{*\pm}$: opposite-signature pair, Chapter~\ref{app:Krein}),
giving an EPD point with non-trivial Jordan block\index{Jordan block}
and an open instability tongue\index{instability tongue}; at each
even resonance the colliding multipliers carry the same Krein signature\index{Krein signature}
($\Pi^{**}$: same-signature pair, Chapter~\ref{app:Krein}), giving
a semisimple double multiplier and a collapsed instability zone\index{instability zone}.
The Mathieu\index{Mathieu equation} equation has opposite-signature
collisions at every resonance, so all its instability zones persist.

\subsection{The continued-fraction (CF) method}

\label{subsec:SummaryCF}

\emph{Starting point: Floquet specialization to the boundary.} By
Floquet's theorem, at an instability boundary the multiplier is $\rho=-1$,
so the solution has period $2\pi$. Rather than solving for the Floquet
exponent $\mathrm{i}u$ (as Cambi does), our method seeks period-$2\pi$
solutions of the Ince equation directly. The $x\mapsto-x$ parity
symmetry ($b=0$ for the LC circuit) splits these into even and odd
types~\eqref{eq:Fourier-ansatz}: 
\begin{equation}
q_{\mathrm{even}}=\sum_{n=0}^{\infty}A_{2n+1}\cos(2n+1)x,\qquad q_{\mathrm{odd}}=\sum_{n=0}^{\infty}B_{2n+1}\sin(2n+1)x.\label{eq:sum-ansatz}
\end{equation}
This is the key departure from Cambi's bilateral ansatz: fixing the
boundary condition $\rho=-1$ reduces the doubly-infinite Fourier
sum to a singly-infinite one with only positive indices.

\emph{Three-term recurrence.} Substituting~\eqref{eq:sum-ansatz}
into the Ince equation and applying MW Lemma~7.4 gives the three-term
recurrence (Section~\ref{subsec:Widths}) 
\begin{multline}
a(2n-1)^{2}A_{2n-1}+2[(2n+1)^{2}-c]\,A_{2n+1}\\
+a(2n+3)^{2}A_{2n+3}=0,\quad n=1,2,3,\ldots,\label{eq:sum-bulk-rec}
\end{multline}
with distinct starter equations for the even and odd solutions at
$n=0$ (equations~\eqref{eq:start-even}--\eqref{eq:start-odd}).

\emph{The CF eigenvalue functions (our contribution).} The starter
ratios $A_{3}/A_{1}$ and $B_{3}/B_{1}$ from the two starter equations,
matched to the bulk minimal-ratio continued fraction\index{continued fraction}
(eq.~\eqref{eq:recCF}) 
\begin{equation}
K(c)=\cfrac{-a}{2[9-c]-\cfrac{225\,a^{2}}{2[25-c]-\cfrac{1225\,a^{2}}{2[49-c]-\cdots}}},\label{eq:sum-CF}
\end{equation}
whose level-$n$ partial numerators for $n\geq2$ are the products
$-a^{2}(2n-1)^{2}(2n+1)^{2}$ of the adjacent couplings of~\eqref{eq:sum-bulk-rec},
define the eigenvalue functions (equation~\eqref{eq:Fdef}): 
\begin{align}
F_{\mathrm{even}}(c) & \equiv-\frac{a-2(c-1)}{9a}-K(c),\label{eq:sum-Feven}\\
F_{\mathrm{odd}}(c) & \equiv\frac{a+2(c-1)}{9a}-K(c).\label{eq:sum-Fodd}
\end{align}
The two functions share the same continued fraction tail; they differ
only in their starter terms, which encode the even/odd symmetry of
the solution. The \emph{CF criterion} for instability boundaries is
\begin{equation}
F_{\mathrm{even}}(c)=0\;\;(\text{lower boundary})\quad\text{or}\quad F_{\mathrm{odd}}(c)=0\;\;(\text{upper boundary}).\label{eq:sum-CF-crit}
\end{equation}
This is the analog of $\Delta(\lambda)=\pm2$ in the MW framework,
and the analog of Cambi's resonance equation~\eqref{eq:Cambi-res}
evaluated at $u=\tfrac{1}{2}$.

\emph{CF difference identity (our contribution).} Since the continued-fraction
tail in~\eqref{eq:sum-Feven}--\eqref{eq:sum-Fodd} is the same
for both functions, their difference is a structural constant: 
\begin{equation}
F_{\mathrm{even}}(c)-F_{\mathrm{odd}}(c)=-\frac{2}{9}=-\frac{2\,Q^{*}(0)}{Q^{*}(-1)}\quad\text{(LC circuit)},\label{eq:sum-CF-diff}
\end{equation}
independent of $c$ and $\delta$. This constant controls the leading-order
separation between the two boundary eigenvalues $c_{k}^{+}$ and $c_{k}^{-}$
of each odd instability interval.

\emph{Closed-form width formulas (our contribution).} The backward-chain
computation of Section~\ref{subsec:Widths}, using the Poincaré--Perron\index{Poincaré--Perron theorem}
asymptotics (Theorem~\ref{thm:PP}, Section~\ref{subsec:VecForm})
and the Mobius\index{Mobius transformation}-transformation structure
of the ratio sequence $z_{n}=A_{2n+1}/A_{2n-1}$ (Section~\ref{subsec:VecForm}),
yields the closed-form width formulas (Theorem~\ref{thm:width},
equation~\eqref{eq:Lm-main}): 
\begin{equation}
L_{m}^{\mathrm{LC}}=\frac{(m!!)^{2}\,\delta^{m}}{2^{(m^{2}+8m-13)/4}(1+\delta^{2})^{m-1}(1-\delta^{2})},\quad m=1,3,5,\ldots,\label{eq:sum-Lm-LC}
\end{equation}
and $L_{m}^{\mathrm{Math}}=q^{m}/[2^{2m-3}((m-1)!)^{2}]+O_{m}(q^{m+2})$
(equation~\eqref{eq:Lm-Math-main}, Section~\ref{subsec:MathieuInce})
for the Mathieu equation. Neither formula appears in Cambi or McLachlan~\cite[Sec.~2.151]{MacLa}.

\subsection{Comparison of the two methods}

\label{subsec:SummaryComparison}

Table~\ref{tab:method-comparison} summarizes the two methods side
by side; the remarks following the table expand on its entries.

\begin{table}[ht]
\centering 
\global\long\def\arraystretch{1.6}%
\begin{tabular}{p{2.5cm}p{4.3cm}p{4.3cm}}
\toprule 
 & \textbf{MW discriminant}  & \textbf{CF method}\tabularnewline
\midrule 
Central object  & Discriminant $\Delta(\lambda)$, entire function of $\lambda$  & Eigenvalue functions $F_{\mathrm{even/odd}}(c)$, analytic in $c$\tabularnewline
Boundary condition  & $\Delta(\lambda)=\pm2$  & $F_{\mathrm{even/odd}}(c)=0$\tabularnewline
Cambi analog  & (no direct analog)  & Cambi's resonance eq.~\eqref{eq:Cambi-res} at $u=\tfrac{1}{2}$\tabularnewline
Qualitative structure  & Coexistence polynomials $Q,Q^{*}$; discriminant identity \eqref{eq:sum-Delta-id}  & CF difference identity~\eqref{eq:sum-CF-diff}\tabularnewline
Quantitative output  & Asymptotic boundary curves; EPD/Krein classification  & Closed-form widths $L_{m}^{\mathrm{LC}}$, $L_{m}^{\mathrm{Math}}$\tabularnewline
Requires numerics?  & No (qualitative); MW series has $O(\delta^{2N+2})$ truncation error  & No (closed-form widths exact); CF converges for all $\delta<1$\tabularnewline
Connection to Cambi  & Indirect (both use Ince structure)  & Direct: $F_{\mathrm{even/odd}}=0$ is the sign pair of Cambi's eq.~(12)
at $u=\tfrac{1}{2}$\tabularnewline
\bottomrule
\end{tabular}\caption{Side-by-side comparison of the MW discriminant and CF methods for
instability boundaries of the LC circuit.}
\label{tab:method-comparison} 
\end{table}

\emph{Common foundation.} Both methods rest on the Ince identification
of Remark~\ref{rem:ince-id}. It is the Ince structure that gives
explicit recurrence coefficients $Q^{*}(n)=a(2n-1)^{2}$, that makes
the discriminant identity~\eqref{eq:sum-Delta-id} exact, and that
yields the CF difference constant $-2/9$. For a generic Hill equation,
neither method would produce closed-form results of this type.

\emph{Complementary roles.} The MW method answers the \emph{why} ---
coexistence forces even interval collapse (discriminant identity),
EPD forces odd interval opening (Krein theorem) --- while the CF
method answers the \emph{where and how wide} --- exact boundary eigenvalues
$c_{k}^{\pm}$ and closed-form widths $L_{m}$. Together they give
a complete analytic description of the stability diagram.

\emph{Relation to Cambi's work.} Cambi's method solves the resonance
equation~\eqref{eq:Cambi-res} numerically for the Floquet exponent\index{Floquet theory!Floquet exponent}
$\mathrm{i}u$, obtaining instability boundary values in $(p,\gamma)$-space.
Our CF method targets the boundary $u=\tfrac{1}{2}$ from the outset,
works in the spectral parameter $c$ rather than $u$, and extracts
closed-form formulas via the backward-chain computation. The CF criterion~\eqref{eq:sum-CF-crit}
is therefore simultaneously a generalization of Cambi's approach (same
underlying equation, exact for all $\delta$) and an improvement upon
it (closed-form rather than numerical output). The agreement with
Cambi's Table~II --- at the level of his tabulation accuracy for
all $\gamma\leq0.40$ (Chapter~\ref{sec:Cambi}) --- confirms that
both approaches locate the same boundaries.

\emph{Computational comparison.} The two methods have distinct numerical
profiles. The MW discriminant series $\Delta=\sum_{n\geq0}\Delta_{n}$
must be truncated at some order $N$, introducing an error $O(\delta^{2N+2})$
that grows with $\delta$ (Table~\ref{tab:lh-bdry} shows the $O(\delta^{3})$
truncation degrading for $\delta\gtrsim0.2$). The CF criterion~\eqref{eq:sum-CF-crit}
is \emph{exact} for all $0<\delta<1$: by Pincherle\index{continued fraction!Pincherle theorem}'s
theorem (Theorem~\ref{thm:Pinch}, Chapter~\ref{app:CF}), the continued
fractions $F_{\mathrm{even/odd}}(c)$ converge with no truncation
error, at a geometric rate with ratio $\delta^{2}$ (Poincaré--Perron
characteristic roots $\delta$ and $1/\delta$, Section~\ref{subsec:VecForm}).
For instability interval \emph{widths} specifically, the CF method
yields the closed-form formula~\eqref{eq:sum-Lm-LC}, requiring no
numerics at all. When numerical values of the full discriminant curve
are needed, the most efficient procedure is to integrate the Volterra
integral equations~\eqref{eq:Volterra-y1}--\eqref{eq:Volterra-y2}
over one half-period and evaluate $\Delta=y_{1}(\pi)+y_{2}'(\pi)$
(Chapter~\ref{sec:UnivExp}) --- bypassing both series entirely.
For small $\delta$ the MW series is preferred analytically, giving
explicit formulas order by order; for larger $\delta$ or when exact
widths are required, the CF method is decisive.

\medskip{}
 \emph{Parameter conversion formulas.} For convenience we collect
here all parameter conversions needed to pass between the physical
variables and the two methods. The physical modulation amplitude $\varepsilon$
and the auxiliary parameter $\delta$ are related by (Chapter~\ref{sec:LCHill},
eq.~\eqref{eq:eps-delta}) 
\begin{equation}
\varepsilon=\frac{2\delta}{1+\delta^{2}},\qquad\delta=\frac{1-\sqrt{1-\varepsilon^{2}}}{\varepsilon},\qquad|\delta|<1.\label{eq:conv-eps-delta}
\end{equation}
The Ince parameters $a$ and $c$, expressed directly in terms of
the original LC circuit variables ($\varepsilon$: modulation amplitude;
$r=\mu/\omega_{0}$: frequency ratio; $\omega_{0}=1/\sqrt{LC}$: natural
frequency), are (Remark~\ref{rem:ince-id}, eq.~\eqref{eq:IncePars})
\begin{equation}
a=-\varepsilon=-\frac{2\delta}{1+\delta^{2}},\qquad c=\frac{4}{r^{2}}=\frac{4\omega_{0}^{2}}{\mu^{2}}=\frac{4}{\mu^{2}LC}.\label{eq:conv-a-c-LC}
\end{equation}
The parameter $a=-\varepsilon$ is determined by $\delta$ (hence
by $\varepsilon$) alone, while $c$ is determined by the frequency
ratio $r$ alone; they are independent of each other. The relation
of $c$ and $a$ to the Magnus--Winkler\index{Magnus--Winkler theory}
spectral parameter $\hat{\lambda}$ is (eq.~\eqref{eq:lambda-MW},
\eqref{eq:c-LC-params}) 
\begin{equation}
\hat{\lambda}=\frac{c(1+\delta^{2})}{1-\delta^{2}}=\frac{4(1+\delta^{2})}{r^{2}(1-\delta^{2})},\qquad c=\frac{\hat{\lambda}(1-\delta^{2})}{1+\delta^{2}}=\frac{4}{r^{2}}.\label{eq:conv-c-hl}
\end{equation}
The $m$-th odd instability zone is centered near $c=c_{m}^{*}=(2m-1)^{2}$
(equivalently $r=2/(2m-1)$) for small $\delta$.

\medskip{}
 \emph{The three computational methods.} 
\begin{itemize}
\item \emph{Direct numerical} (reference): integrate the Volterra equations~\eqref{eq:Volterra-y1}--\eqref{eq:Volterra-y2}
over one period to obtain $\Delta(c)=y_{1}(\pi)+y_{2}'(\pi)$, then
find the roots of $\Delta(c)+2=0$ by bisection. This serves as the
exact reference. 
\item \emph{CF method} (exact): find the zeros of $F_{\mathrm{even}}(c)=0$
and $F_{\mathrm{odd}}(c)=0$ (Chapter~\ref{sec:LCInce}, eq.~\eqref{eq:CF-criterion})
by evaluating the continued fractions via backward recurrence and
solving by bisection. The CF converges geometrically with ratio $\delta^{2}$
(Poincaré--Perron, Section~\ref{subsec:VecForm}), so 20--30 recurrence
levels suffice for all $\delta\leq0.5$. 
\item \emph{MW discriminant approximation}: solve $\Delta_{0}+\Delta_{2}^{\mathrm{LC}}=-2$
for $\hat{\lambda}$ near $\hat{\lambda}=(2m-1)^{2}$, then convert
$c=\hat{\lambda}(1-\delta^{2})/(1+\delta^{2})$ (eq.~\eqref{eq:conv-c-hl}).
The two components are (Chapter~\ref{sec:MWInce}, eq.~\eqref{eq:D0};
Chapter~\ref{app:entire}, eq.~\eqref{eq:Delta-LC}) 
\begin{gather}
\Delta_{0}(\hat{\lambda})=2\cos\pi\sqrt{\hat{\lambda}},\quad\Delta_{2}^{\mathrm{LC}}(\hat{\lambda})=\frac{\pi\hat{\lambda}^{3/2}\delta^{2}\sin\pi\sqrt{\hat{\lambda}}}{2}\sum_{n=1}^{\infty}\frac{\delta^{2(n-1)}}{\hat{\lambda}-n^{2}},\label{eq:comp-D0}
\end{gather}
where $\hat{\lambda}=c(1+\delta^{2})/(1-\delta^{2})$ (eq.~\eqref{eq:conv-c-hl})
and the Fourier coefficients in the complex exponential expansion
are $g_{n}=\hat{\lambda}\delta^{n}$ (so $|g_{n}|^{2}=\hat{\lambda}^{2}\delta^{2n}$,
eq.~\eqref{eq:gn}). The apparent poles $\Delta_{2}^{\mathrm{LC}}(\hat{\lambda})$
defined by \eqref{eq:comp-D0} at $\hat{\lambda}=n^{2}$ are removable:
as $\hat{\lambda}\to n^{2}$, $\sin\pi\sqrt{\hat{\lambda}}/(\hat{\lambda}-n^{2})\to\pi(-1)^{n}/(2n)$
(Chapter~\ref{app:entire}), so $\Delta_{2}^{\mathrm{LC}}$ is entire.
This approximation is comparable in accuracy to the $N=1$ CF truncation;
the error grows to approximately $4\%$ at $\delta=0.40$ (Table~\ref{tab:CF-truncation}). 
\end{itemize}
\medskip{}

\begin{rem}[Numerical verification summary]
\label{rem:numerics-summary} All principal formulas in this work
have been verified numerically. Table~\ref{tab:verif-summary} summarizes
the 5 verification strategies and their outcomes. 
\begin{table}[ht]
\centering{\small
\global\long\def\arraystretch{1.3}%
{}}{\small{}%
\begin{tabular}{@{}p{3.0cm}p{4.1cm}p{4.1cm}@{}}
\toprule 
{\small\textbf{Strategy}}{\small{}  } & {\small\textbf{Method}}{\small{}  } & {\small\textbf{Outcome}}\tabularnewline
{\small Width formula $L_{m}^{\mathrm{LC}}$  } & {\small CF exact vs.\ formula for $m=1,3$;\ $\delta\in[0.02,0.40]$
 } & \begin{cellvarwidth}[t]
\centering
{\small{} {} {} {} {} {} Leading coeff.\ $W/\delta^{m}\to(m!!)^{2}/2^{(m^{2}+8m-13)/4}$
confirmed to $\geq8$ sig.\ fig.; $O(\delta^{2})$ }\\
{\small{} {} {} {} {} {} subleading growth as expected (Remark~\ref{rem:Lm-verify})
 }
\end{cellvarwidth}\tabularnewline
{\small Cambi Table~II  } & {\small Monodromy integration of Cambi's equation; CF comparison  } & {\small Cambi values confirmed $<0.13\%$; CF agrees at his tabulation
accuracy for all $\gamma\leq0.40$ (Remark~\ref{rem:Cambi-verify})}\tabularnewline
{\small Floquet multiplier  } & {\small Monodromy matrix at CF zeros  } & {\small$\Delta+2=0$ to $\leq1.6\times10^{-14}$ for all $\delta\leq0.30$;
$\det M=1$ exactly (Remark~\ref{rem:CF-boundary-verify})}\tabularnewline
{\small Even-zone vanishing  } & {\small Discriminant maximum in second stability band  } & {\small$\max(\Delta-2)=0$ to machine precision (tangency at $\hat{\lambda}_{2}(\delta)$):
zone has zero width (Remark~\ref{rem:even-verify})}\tabularnewline
{\small MW formula  } & {\small$\Delta_{2}^{\mathrm{LC}}$ vs.\ CF  } & {\small Agreement $0.003\%$ at $\delta=0.05$ growing to $3.8\%$
at $\delta=0.40$ (Table~\ref{tab:CF-truncation})}\tabularnewline
\bottomrule
\end{tabular}}{\small\caption{Numerical verification summary for the principal formulas of this
work. ODE integration used an 8th-order Runge--Kutta method (tolerance
$10^{-12}$).}
\label{tab:verif-summary}} 
\end{table}
\end{rem}

\medskip{}
 \emph{Explicit CF truncations.} Truncating the continued fractions
$F_{\mathrm{even/odd}}(c)$ at $N$ levels replaces the infinite CF
by a finite rational function of $c$, yielding a polynomial equation
for the boundary eigenvalue. By the Poincaré--Perron analysis (Section~\ref{subsec:VecForm}),
the truncation error decays as $O(\delta^{2N})$ per level.

For the LC circuit the truncated CF conditions $F_{\mathrm{even/odd}}^{(N)}(c)=0$
are polynomial equations in $c$ of degree $N$. At $N=1$ both conditions
reduce to a \emph{quadratic} in $c$ with explicit closed-form solutions.
Clearing denominators in $F_{\mathrm{odd}}^{(1)}(c)=0$ (i.e.\ multiplying
through by $9a\cdot2(9-c)$) gives the quadratic 
\begin{equation}
9a^{2}-2ac+18a-4c^{2}+40c-36=0,\label{eq:Fodd_N1}
\end{equation}
or equivalently $4c^{2}+(2a-40)c-(9a^{2}+18a-36)=0$. The two solutions
are 
\begin{align}
c_{1}^{+}\bigl(N=1\bigr) & =5-\tfrac{a}{4}-\tfrac{1}{4}\sqrt{37a^{2}+32a+256},\label{eq:c1p_N1}\\
c_{3}^{+}\bigl(N=1\bigr) & =5-\tfrac{a}{4}+\tfrac{1}{4}\sqrt{37a^{2}+32a+256}.\label{eq:c3p_N1}
\end{align}
The \emph{smaller} root~\eqref{eq:c1p_N1} approximates the upper
boundary of the $m=1$ instability zone, and the \emph{larger} root~\eqref{eq:c3p_N1}
approximates the upper boundary of the $m=3$ zone. Similarly, $F_{\mathrm{even}}^{(1)}(c)=0$
gives 
\begin{align}
c_{1}^{-}\bigl(N=1\bigr) & =5+\tfrac{a}{4}-\tfrac{1}{4}\sqrt{37a^{2}-32a+256},\label{eq:c1m_N1}\\
c_{3}^{-}\bigl(N=1\bigr) & =5+\tfrac{a}{4}+\tfrac{1}{4}\sqrt{37a^{2}-32a+256}.\label{eq:c3m_N1}
\end{align}
Remarkably, a \emph{single} quadratic equation simultaneously approximates
the boundary eigenvalues of two instability zones.

At $N=2$ the condition $F_{\mathrm{odd}}^{(2)}(c)=0$ is a cubic
in $c$ (degree determined by the three-level backward chain); at
$N=3$ it is a quartic. These are solved numerically by standard root-finding
(bisection or Newton), but they converge much faster than the full
CF and far outperform the MW series for $\delta\gtrsim0.2$.

\medskip{}
 \emph{Convergence analysis and truncation order.} Table~\ref{tab:CF-truncation}
gives the maximum relative error of each approximation against the
exact CF value for the $m=1$ instability boundaries, across modulation
depths $\delta\in[0.05,0.40]$.

\begin{table}[ht]
\centering {\small
\global\long\def\arraystretch{1.25}%
{}{}{}{}{}{}\resizebox{\textwidth}{!}{%
\begin{tabular}{@{}cccccccc@{}}
\toprule 
{\small$\delta$ }  & {\small$c^{-}$ }  & {\small$c^{+}$ }  & {\small$|W|$ }  & {\small err $N=1$ }  & {\small err $N=2$ }  & {\small err $N=3$ }  & {\small err MW }\tabularnewline
\midrule 
{\small$0.05$ }  & {\small$0.947336$ }  & {\small$1.047052$ }  & {\small$0.099716$ }  & {\small$8.5\times10^{-6}$ }  & {\small$2.3\times10^{-8}$ }  & {\small$5.8\times10^{-11}$ }  & {\small$3.3\times10^{-5}$ }\tabularnewline
{\small$0.10$ }  & {\small$0.889986$ }  & {\small$1.087726$ }  & {\small$0.197740$ }  & {\small$1.4\times10^{-4}$ }  & {\small$1.5\times10^{-6}$ }  & {\small$1.5\times10^{-8}$ }  & {\small$2.9\times10^{-4}$ }\tabularnewline
{\small$0.15$ }  & {\small$0.828986$ }  & {\small$1.121450$ }  & {\small$0.292463$ }  & {\small$7.3\times10^{-4}$ }  & {\small$1.7\times10^{-5}$ }  & {\small$4.0\times10^{-7}$ }  & {\small$1.1\times10^{-3}$ }\tabularnewline
{\small$0.20$ }  & {\small$0.765436$ }  & {\small$1.147859$ }  & {\small$0.382423$ }  & {\small$2.4\times10^{-3}$ }  & {\small$1.0\times10^{-4}$ }  & {\small$4.1\times10^{-6}$ }  & {\small$2.9\times10^{-3}$ }\tabularnewline
{\small$0.25$ }  & {\small$0.700441$ }  & {\small$1.166805$ }  & {\small$0.466364$ }  & {\small$6.0\times10^{-3}$ }  & {\small$3.9\times10^{-4}$ }  & {\small$2.5\times10^{-5}$ }  & {\small$6.5\times10^{-3}$ }\tabularnewline
{\small$0.30$ }  & {\small$0.635060$ }  & {\small$1.178336$ }  & {\small$0.543275$ }  & {\small$1.3\times10^{-2}$ }  & {\small$1.2\times10^{-3}$ }  & {\small$1.1\times10^{-4}$ }  & {\small$1.3\times10^{-2}$ }\tabularnewline
{\small$0.35$ }  & {\small$0.570271$ }  & {\small$1.182672$ }  & {\small$0.612402$ }  & {\small$2.4\times10^{-2}$ }  & {\small$3.1\times10^{-3}$ }  & {\small$3.9\times10^{-4}$ }  & {\small$2.3\times10^{-2}$ }\tabularnewline
{\small$0.40$ }  & {\small$0.506930$ }  & {\small$1.180173$ }  & {\small$0.673242$ }  & {\small$4.3\times10^{-2}$ }  & {\small$7.0\times10^{-3}$ }  & {\small$1.1\times10^{-3}$ }  & {\small$3.8\times10^{-2}$ }\tabularnewline
\bottomrule
\end{tabular}}\caption{$m=1$ instability boundary eigenvalues $c^{\pm}$ and zone width
$|W|=c^{+}-c^{-}$ computed as zeros of $F_{\mathrm{even/odd}}(c)$
(eq.~\eqref{eq:Fdef}) using the exact continued fraction~\eqref{eq:recCF}
($N=400$ levels). These coincide with the Hill equation boundary
condition $\Delta+2=0$ to $|\Delta+2|\protect\leq1.6\times10^{-14}$
(Remark~\ref{rem:CF-boundary-verify}). The error columns give maximum
relative errors of the $N=1,2,3$ CF truncations relative to the converged
values; each additional level reduces the error by the factor $\delta^{2}$,
the Poincaré--Perron ratio (e.g.\ at $\delta=0.10$: $1.4\times10^{-4}\to1.5\times10^{-6}\to1.5\times10^{-8}$).
The final column gives the relative error of the MW discriminant approximation
($\Delta_{0}+\Delta_{2}=\pm2$, eq.~\eqref{eq:sum-MW-exp}); it is
comparable to the $N=1$ CF truncation throughout, growing to $3.8\%$
at $\delta=0.40$. See Figure~\ref{fig:comparison-m1}.}
\label{tab:CF-truncation}} 
\end{table}

Each additional CF level reduces the error by the factor $\delta^{2}$,
the Poincaré--Perron ratio: at $\delta=0.10$ the $N=1,2,3$ errors
are $1.4\times10^{-4}$, $1.5\times10^{-6}$, $1.5\times10^{-8}$,
and at $\delta=0.30$ they are $1.3\times10^{-2}$, $1.2\times10^{-3}$,
$1.1\times10^{-4}$. Thus three CF levels give eight-digit accuracy
for $\delta\leq0.10$; at $\delta=0.30$ each level gains roughly
one decimal digit ($\delta^{2}\approx0.09$), so about ten levels
give nine digits, and $20$--$30$ levels give machine precision
for all $\delta\leq0.5$. The $N=1$ quadratic formula~\eqref{eq:c1p_N1}--\eqref{eq:c3m_N1}
gives three-digit accuracy for $\delta\leq0.1$, with no iteration
required.

The MW discriminant series $\Delta_{0}+\Delta_{2}$ locates both $m=1$
boundaries throughout the tested range, with error comparable to the
$N=1$ CF truncation: $1.1\times10^{-3}$ at $\delta=0.15$, growing
to $3.8\%$ at $\delta=0.40$.

\begin{figure}[htbp]
\centering \includegraphics[width=1\textwidth]{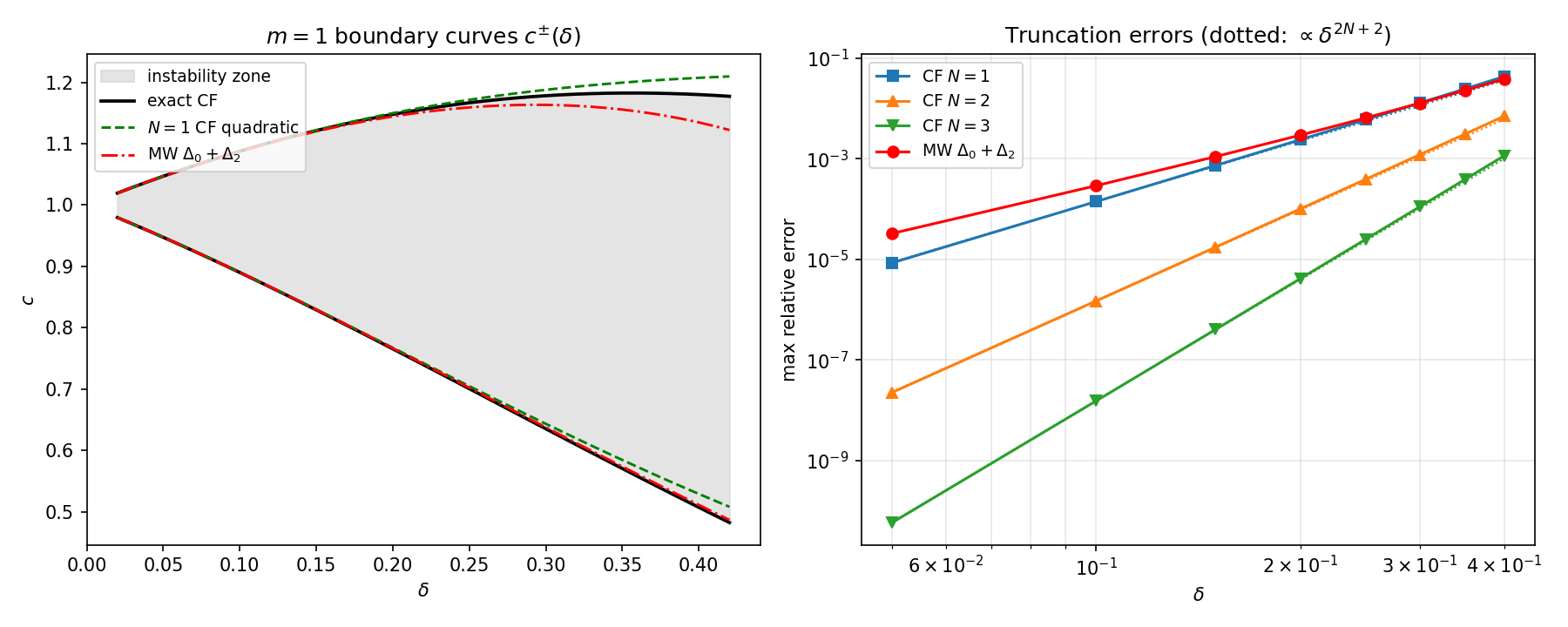}
\caption{\emph{Left:} $m=1$ instability boundary curves $c^{\pm}(\delta)$
computed by three methods. The exact CF (black solid) and $N=1$ CF
quadratic (green dashed) are nearly indistinguishable on this scale
for $\delta\protect\leq0.2$; the MW approximation (red dash-dot)
deviates visibly for $\delta\gtrsim0.25$. The shaded region is the
instability zone. \emph{Right:} Maximum relative errors of CF truncations
($N=1,2,3$) and MW approximation vs.\ the exact CF, on a logarithmic
scale. Dotted reference lines $\propto\delta^{2N+2}$ confirm the
per-level reduction factor $\delta^{2}$ of the Poincaré--Perron
theory. The MW error (red circles) tracks the $N=1$ truncation across
the full range.}
\label{fig:comparison-m1} 
\end{figure}

Figure~\ref{fig:comparison-m1} shows the $m=1$ boundary curves
computed by the three methods, and the convergence of the CF truncations
as $N$ increases.

\emph{Role of the background chapters.} The three Part~II chapters
on finite difference equations (Chapter~\ref{app:FDE}), continued
fractions (Chapter~\ref{app:CF}), and Mobius transformations (Chapter~\ref{app:Moeb})
form the rigorous foundation for the CF method: Poincaré--Perron
(Theorem~\ref{thm:PP}) guarantees minimal solution\index{continued fraction!minimal solution}s,
Pincherle (Theorem~\ref{thm:Pinch}) equates minimal solutions with
CF convergence, and the Mobius classification of the asymptotic matrix
$M_{\infty}$ (Section~\ref{subsec:VecForm}) explains the hyperbolic
convergence structure with characteristic roots $\delta$ and $1/\delta$.

\section{The Yakubovich--Starzhinskii series: Mathieu special case}

\label{sec:YS-LC}

This chapter applies the two-parameter YS framework of Chapter~\ref{app:YS-Floquet}
(\S\,\ref{subsec:YS-two-param}) to the LC circuit at the primary
tongue ($m=1$). The YS analysis is developed in three stages, each
a separate chapter. The present chapter treats the \emph{Mathieu\index{Mathieu equation}
special case}: retaining only the leading Fourier harmonic of the
LC coefficient and setting $c=1$ gives the Mathieu equation, for
which the YS method yields explicit closed-form expressions for $\mathbf{K}_{1}$,
$\mathbf{K}_{2}$ and $\mathbf{F}_{1}$, $\mathbf{F}_{2}$, and provides
an independent derivation of the primary instability boundaries. Chapter~\ref{sec:YS-exact-LC}
then treats the \emph{exact LC circuit} with $\hat{\lambda}$ kept
as a free parameter, giving closed-form $\mathbf{K}_{m}(\hat{\lambda})$
and self-consistent boundary curve\index{boundary curves}s. Chapter~\ref{sec:YS-comparison}
gives the three-way comparison of the YS, MW, and CF methods.

\medskip{}

\noindent\emph{Reminder: Floquet\index{Floquet theory}--Lyapunov
decomposition and the YS series.} The Floquet--Lyapunov theorem guarantees
the existence of the decomposition (Chapter~\ref{app:YS-Floquet})
\begin{equation}
X(t,\varepsilon)=F(t,\varepsilon)\,e^{tK(\varepsilon)},\label{eq:Floquet-YS-reminder}
\end{equation}
where $X(t,\varepsilon)$ is the fundamental matrix of the system,
$F(t,\varepsilon)$ is a $T$-periodic matrix-valued function, and
$K(\varepsilon)$ is a constant matrix. The YS method~\cite[Ch.~4]{YakSta1}
constructs both $F$ and $K$ as explicit power series in the small
parameter $\varepsilon$ (here $\varepsilon=\delta$): 
\begin{align}
F(t,\varepsilon) & =\mathbf{I}+\varepsilon\mathbf{F}_{1}(t)+\varepsilon^{2}\mathbf{F}_{2}(t)+\cdots,\label{eq:F-series-reminder}\\
K(\varepsilon) & =\mathbf{K}_{0}+\varepsilon\mathbf{K}_{1}+\varepsilon^{2}\mathbf{K}_{2}+\cdots,\label{eq:K-series-reminder}
\end{align}
where each $\mathbf{F}_{j}(t)$ is $T$-periodic and each $\mathbf{K}_{j}$
is a constant matrix determined by the recursion~\eqref{eq:YS-Kl}--\eqref{eq:YS-Fl}.
In the singular case relevant here, the series is constructed for
the system transformed by the $T$-periodic rotation $e^{t\mathbf{C}_{0}}$
(where $\mathbf{C}=\mathbf{C}_{0}+\mathbf{K}_{0}$ and $e^{T\mathbf{C}_{0}}=\mathbf{I}$),
so the full decomposition of the matriciant normalized by $X(0)=\mathbf{I}$
reads $X(t,\varepsilon)=e^{t\mathbf{C}_{0}}F(t,\varepsilon)\,e^{tK(\varepsilon)}\,F(0,\varepsilon)^{-1}$;
at $t=T$ the rotation closes and the monodromy is $X(T)=F(0)\,e^{TK}F(0)^{-1}$,
conjugation-equivalent to $e^{TK}$. The constant part of each $\mathbf{F}_{j}$
is a normalization (gauge) choice (\S\,\ref{subsec:YS-conventions}
of Chapter~\ref{app:YS-Floquet}); under YS's normalization $\mathbf{F}_{j}(0)=\mathbf{0}$
(their second computational procedure~\cite[Ch.~IV, \S\,4.5]{YakSta1},
used in their \S\,5.4), $F(0)=\mathbf{I}$ and $X(T)=e^{TK}$ exactly.
Two first-order matrices appear in the Mathieu case below: YS's printed
$\varphi'_{ij}$ matrix and the true factor $\mathbf{F}_{1}^{{\rm true}}$
(see \S\,\ref{subsec:YS-F1}); $\mathbf{F}_{2}$ is stated in zero-mean
form (\S\,\ref{subsec:YS-F2}). The instability boundary\index{stability boundary}
curves depend only on $K(\varepsilon)$ (separation principle, eq.~\eqref{eq:YS-sep-principle});
the periodic factor $F(t,\varepsilon)$ encodes the shape of the Floquet
solutions and is computed explicitly below for both the Mathieu case
(Chapter~\ref{subsec:YS-Mathieu}, eqs.~\eqref{eq:YS-F1}--\eqref{eq:YS-F2})
and the exact LC case (Chapter~\ref{sec:YS-exact-LC}, eqs.~\eqref{eq:LC-F1-lam}--\eqref{eq:LC-F-final}).

\medskip{}
\emph{Parameter dictionary.} The following relations connect all parameters
used in Chapters~\ref{sec:YS-LC}--\ref{sec:YS-comparison}: 
\begin{equation}
\boxed{\varepsilon=\frac{2\delta}{1+\delta^{2}},\qquad\hat{\lambda}=\frac{c(1+\delta^{2})}{1-\delta^{2}},\qquad c=\frac{4\omega_{0}^{2}}{\mu^{2}},\qquad\mu_{0}=\frac{c-1}{\varepsilon}.}\label{eq:YS-param-dict}
\end{equation}
Here $\delta\in(0,1)$ is the LC modulation depth, $\varepsilon$
is the Mathieu small parameter, $\hat{\lambda}$ is the YS spectral
parameter (the unperturbed squared frequency), $c=4\omega_{0}^{2}/\mu^{2}$
is the detuning parameter ($c=1$ at exact primary resonance\index{resonance}
$\omega_{0}=\mu/2$), and $\mu_{0}$ is the YS detuning parameter
($\mu_{0}=0$ at exact resonance; not to be confused with the driving
frequency $\mu$ of eq.~\eqref{eq:LC-original-MR}). \emph{Note on
$p$:} Throughout Chapters~\ref{sec:YS-LC}--\ref{sec:YS-comparison}
the symbol $p$ does not appear; all resonance orders are written
as $m$ ($m=1$ for the primary tongue). The Cambi parameter $p=\omega_{0}/\mu=1/r$
of Chapters~\ref{sec:LCInce} and~\ref{sec:Cambi} is unrelated.
The Mathieu approximation sets $\hat{\lambda}=1$ and $c=1$ simultaneously,
which is exact only at $\delta=0$; for $\delta>0$ the exact LC circuit
has $\hat{\lambda}=c(1+\delta^{2})/(1-\delta^{2})>c$.

\label{subsec:YS-Mathieu}

The exact LC Hill equation~\eqref{eq:LC-Hill-firstorder} involves
the full LC-circuit coefficient $f_{{\rm LC}}(\tau,\delta)=\sum_{n=1}^{\infty}2\delta^{n}\cos2n\tau$
and the free spectral parameter $\hat{\lambda}=c(1+\delta^{2})/(1-\delta^{2})$.
The \emph{Mathieu approximation} makes two simplifications: 
\begin{enumerate}
\item retain only the $n=1$ harmonic: $f_{{\rm LC}}\approx2\delta\cos2\tau$,
so $\varepsilon=2\delta/(1+\delta^{2})\approx2\delta$; 
\item fix $\hat{\lambda}=1$ (equivalently $c=1$, $\omega_{0}=\mu/2$),
the exact primary resonance value at $\delta=0$. 
\end{enumerate}
Together these give the Mathieu equation, which is a valid $O(\delta^{2})$
approximation to the exact LC equation (Chapter~\ref{sec:MWInce}).
The YS method applied to the Mathieu equation yields explicit closed-form
$\mathbf{F}_{1}$, $\mathbf{F}_{2}$ (eqs.~\eqref{eq:YS-F1}--\eqref{eq:YS-F2})
and provides an independent check of the primary boundaries. The exact
LC computation in Chapter~\ref{sec:YS-exact-LC} removes both approximations
and treats $\hat{\lambda}$ as a free parameter.
\begin{center}
\fcolorbox{black}{white}{\parbox[c]{0.85\textwidth}{%
 \emph{Convention for this section (Mathieu case, $m=1$):} The exponent
matrices $\mathbf{K}_{1}$, $\mathbf{K}_{2}$ are YS's printed ones,
in their $\mathbf{F}_{j}(0)=\mathbf{0}$ normalization (\S\,\ref{subsec:YS-conventions}
of Chapter~\ref{app:YS-Floquet}), so that $X(\pi)=e^{\pi K}$ exactly.
Two first-order matrices appear: YS's printed $\varphi'_{ij}$ matrix
(eq.~\eqref{eq:YS-F1}; equal to the transformed forcing, eq.~\eqref{eq:phi-is-D1})
and the true Lyapunov factor $\mathbf{F}_{1}^{{\rm true}}$ (eq.~\eqref{eq:YS-F1-true},
normalization $\mathbf{F}_{1}^{{\rm true}}(0)=\mathbf{0}$). The phase-portrait
analysis of \S\,\ref{subsubsec:YS-Floquet-portraits} uses $\mathbf{F}_{1}^{{\rm true}}$.
$\mathbf{F}_{2}$ is stated in zero-mean form (eq.~\eqref{eq:YS-F2}). %
}} 
\par\end{center}

\subsection{Setup and identification with Yakubovich--Starzhinskii notation}\index{Yakubovich--Starzhinskii series}

\label{subsec:YS-setup}

Retaining only the leading Fourier harmonic of the LC coefficient
and setting $c=1$ (primary resonance, $\omega_{0}=\mu/2$, rescaled
period $T=\pi$) gives the \emph{Mathieu equation} 
\begin{equation}
q''+\bigl(1+\varepsilon\cos2\tau\bigr)\,q=0,\qquad\varepsilon=-a=\frac{2\delta}{1+\delta^{2}},\label{eq:YS-LC-eq}
\end{equation}
where $\tau$ is the rescaled time. This differs from the exact LC
Hill equation~\eqref{eq:LC-Hill} by $O(\delta^{2})$ terms in the
coefficient (the next Fourier harmonic contributes $\delta^{2}\cos4\tau$
etc.). It is this Mathieu equation to which the YS method is applied
in the present subsection. In first-order form $\mathbf{x}=\bigl(\begin{smallmatrix}q\\
q'
\end{smallmatrix}\bigr)$: 
\begin{equation}
\mathbf{x}'=\bigl[\mathbf{C}+\varepsilon\mathbf{B}_{1}(t)\bigr]\mathbf{x},\qquad\mathbf{C}=\begin{bmatrix}0 & 1\\
-1 & 0
\end{bmatrix},\quad\mathbf{B}_{1}(\tau)=\begin{bmatrix}0 & 0\\
-\cos2\tau & 0
\end{bmatrix}.\label{eq:YS-LC-first-order}
\end{equation}
The unperturbed matrix $\mathbf{C}$ has eigenvalues $\pm i$, so
the unperturbed frequency is $\omega=1$ and the YS spectral parameter
takes the value $\hat{\lambda}=\omega^{2}=1$. This is a \emph{consequence}
of the Mathieu approximation ($c=1$ and single-harmonic truncation),
not an independent assumption: in the exact LC circuit with $\delta>0$
and $c$ near~$1$, the spectral parameter satisfies $\hat{\lambda}=c(1+\delta^{2})/(1-\delta^{2})>1$.
This matches the YS Hill equation~\cite[Ch.~IV, eq.~(5.9)]{YakSta1}:
\begin{equation}
q''+\left[\left(\frac{m\pi}{T}\right)^{\!2}+\varepsilon\bigl(\mu_{0}+f(\tau)\bigr)\right]q=0,\label{eq:YS-Hill-eq}
\end{equation}
with the following identification of \emph{YS parameters} in terms
of LC circuit parameters: 
\begin{equation}
\begin{alignedat}{2}m & =1 & \quad & \text{(primary resonance order; primary tongue)}\\
T & =\pi & \quad & \text{(half-period of the rescaled equation)}\\
f(\tau) & =\cos2\tau & \quad & \text{(single retained Fourier harmonic of }f_{{\rm LC}}\text{)}\\
\mu_{0} & =\frac{c-1}{\varepsilon} & \quad & \text{(YS detuning; }c=1+\varepsilon\mu_{0}\text{)}
\end{alignedat}
\label{eq:YS-LC-params}
\end{equation}
Substituting: $(m\pi/T)^{2}=(1\cdot\pi/\pi)^{2}=1$, so eq.~\eqref{eq:YS-Hill-eq}
becomes $q''+(1+\varepsilon\mu_{0}+\varepsilon\cos2\tau)q=0$, i.e.\ $q''+(c+\varepsilon\cos2\tau)q=0$.
At $\mu_{0}=0$ (exact resonance $c=1$) this reduces to the Mathieu
equation~\eqref{eq:YS-LC-eq}, as required.

The resonance order $m=1$ labels the primary tongue: in general,
resonance order $m$ gives the $m$-th instability tongue\index{instability tongue},
which requires the $m$-th Fourier harmonic $f^{(m)}\neq0$ (Remark~\ref{rem:YS-higher-tongues}).
Since $m=1$ is odd and $f^{(1)}=\frac{1}{2}\neq0$, the system is
in the \emph{singular case} of the YS theory~\cite[\S\,4.3]{YakSta1}:
the $\mathbf{B}_{1}$ Fourier mode at frequency $2\pi m/T=2$ is in
resonance with the adjoint action of $\mathbf{C}$, whose eigenvalues
$\pm i$ satisfy $\lambda_{1}-\lambda_{2}=2i=2\pi i\cdot m/T$. The
YS method handles this singular case by a preliminary change of variables
that shifts the resonant mode into the constant (average) part of
the transformed equation.

\subsection{\texorpdfstring{The Yakubovich--Starzhinskii exponent matrix $K(\varepsilon)$}{The
Yakubovich--Starzhinskii exponent matrix K(epsilon)}}

\label{subsec:YS-K}

For the Mathieu equation~\eqref{eq:YS-LC-eq}, the YS method (singular
case, $m=1$, $T=\pi$, $f(\tau)=\cos2\tau$) gives~\cite[Ch.~IV, \S\,5.4]{YakSta1}
(in an unnumbered display): 
\begin{equation}
\mathbf{K}_{0}=-i\,\mathbf{I}_{2},\label{eq:YS-K0}
\end{equation}
\begin{equation}
\mathbf{K}_{1}(\mu_{0})=\frac{1}{2}\begin{bmatrix}0 & \mu_{0}-\tfrac{1}{2}\\[4pt]
-\mu_{0}-\tfrac{1}{2} & 0
\end{bmatrix},\label{eq:YS-K1}
\end{equation}
\begin{equation}
\mathbf{K}_{2}(\mu_{0})=\frac{1}{8}\begin{bmatrix}0 & -3\mu_{0}^{2}+\tfrac{3}{2}\mu_{0}+\tfrac{1}{8}\\[4pt]
-\mu_{0}^{2}-\tfrac{1}{2}\mu_{0}-\tfrac{1}{8} & 0
\end{bmatrix}.\label{eq:YS-K2}
\end{equation}
The exponent matrix is therefore $K(\varepsilon)=\mathbf{K}_{0}+\varepsilon\mathbf{K}_{1}+\varepsilon^{2}\mathbf{K}_{2}+O(\varepsilon^{3})$,
and since $e^{\pi\mathbf{K}_{0}}=-\mathbf{I}$ commutes with everything,
the monodromy matrix is 
\begin{equation}
X(\pi,\varepsilon)=-\,e^{\pi(\varepsilon\mathbf{K}_{1}+\varepsilon^{2}\mathbf{K}_{2})}+O(\varepsilon^{3})
\end{equation}
(in YS's $\mathbf{F}_{j}(0)=\mathbf{0}$ normalization, for which
$F(\pi)=F(0)=\mathbf{I}$). This has been verified by direct numerical
integration of the Mathieu equation: the matrix residual scales as
$\varepsilon^{3}$ ($9.7\times10^{-8}$ at $\varepsilon=0.01$, $\mu_{0}=0$;
halving $\varepsilon$ reduces it by the factor $8.0$, and likewise
at $\mu_{0}=0.3$) --- a gold-standard confirmation of~\eqref{eq:YS-K1}--\eqref{eq:YS-K2}.

\subsection{\texorpdfstring{First order: YS's printed matrix and the true Lyapunov
factor}{First order: YS's printed matrix and the true Lyapunov factor}}

\label{subsec:YS-F1}

The distinctive feature of the YS method --- what separates it from
other perturbative approaches --- is the explicit construction of
the periodic Lyapunov factor $F(t,\varepsilon)$, not merely the exponent
$K(\varepsilon)$. Applying the YS formula~\cite[Ch.~IV, \S\,5.4]{YakSta1}
to equation~\eqref{eq:YS-LC-eq} with resonance order $m=1$, $T=\pi$,
$f(\tau)=\cos2\tau$, one obtains $\mathbf{F}_{1}$ by differentiating
the general-$p$ formulas~\eqref{eq:YS-phi-general} and setting
$p=1$ (the $\varphi'_{ij}$ prescription, eq.~\eqref{eq:YS-phi-prime}
of Chapter~\ref{app:YS-Floquet}). The explicit result is: 
\begin{equation}
\setlength{\arraycolsep}{3pt}\mathbf{F}_{1}^{\varphi'}(\tau,\mu_{0})=\begin{bmatrix}\dfrac{\mu_{0}}{2}\sin2\tau+\dfrac{\sin4\tau}{4} & \dfrac{1-\mu_{0}}{2}\cos2\tau-\dfrac{\cos4\tau+1}{4}\\[10pt]
-\dfrac{1+\mu_{0}}{2}\cos2\tau-\dfrac{\cos4\tau+1}{4} & -\dfrac{\mu_{0}}{2}\sin2\tau-\dfrac{\sin4\tau}{4}
\end{bmatrix}.\label{eq:YS-F1}
\end{equation}
This matrix is $\pi$-periodic and traceless, and it is YS's printed
first-order object for the primary resonance. However, it is \emph{not}
the first-order periodic factor of the decomposition~\eqref{eq:Floquet-YS-reminder}:
it does not satisfy the first-order recursion $\mathbf{F}_{1}'=\mathbf{D}_{1}-\mathbf{K}_{1}$,
and no choice of the free additive constant repairs this, since the
oscillatory parts already differ. Its true identity is elementary:
comparing entries with the transformed forcing $\mathbf{D}_{1}$ of~\cite[Ch.~IV, \S\,5.4]{YakSta1},
\begin{equation}
\mathbf{F}_{1}^{\varphi'}(\tau,\mu_{0})=\mathbf{D}_{1}(\tau,\mu_{0})-\frac{\mu_{0}}{2}\begin{bmatrix}0 & 1\\
-1 & 0
\end{bmatrix},\label{eq:phi-is-D1}
\end{equation}
i.e.\ YS's $\varphi'$ prescription reproduces the \emph{transformed
forcing} (exactly so at $\mu_{0}=0$), not the Lyapunov factor.

The true first-order factor, obtained by integrating the recursion
$\mathbf{F}_{1}'=\mathbf{D}_{1}-\mathbf{K}_{1}$ with the normalization
$\mathbf{F}_{1}(0)=\mathbf{0}$ (the matrizant convention of YS's
second procedure~\cite[Ch.~IV, \S\,4.5]{YakSta1}, under which $X(\pi)=e^{\pi K}$
as verified above), is 
\begin{equation}
\mathbf{F}_{1}^{{\rm true}}(\tau,\mu_{0})=\frac{1}{8}\begin{bmatrix}(1-\mathsf{c})\,(2\mu_{0}+1+\mathsf{c}) & \mathsf{s}\,\bigl(2(1-\mu_{0})-\mathsf{c}\bigr)\\[6pt]
-\mathsf{s}\,\bigl(2(1+\mu_{0})+\mathsf{c}\bigr) & -(1-\mathsf{c})\,(2\mu_{0}+1+\mathsf{c})
\end{bmatrix}.\label{eq:YS-F1-true}
\end{equation}
Here and below $\mathsf{c}=\cos2\tau$, $\mathsf{s}=\sin2\tau$. Expanded
in Fourier modes, the entries read $F_{1}^{{\rm true}\,(11)}=\frac{\mu_{0}}{4}(1-\cos2\tau)+\frac{1-\cos4\tau}{16}$,
$F_{1}^{{\rm true}\,(12)}=\frac{1-\mu_{0}}{4}\sin2\tau-\frac{\sin4\tau}{16}$,
$F_{1}^{{\rm true}\,(21)}=-\frac{1+\mu_{0}}{4}\sin2\tau-\frac{\sin4\tau}{16}$,
with $F_{1}^{{\rm true}\,(22)}=-F_{1}^{{\rm true}\,(11)}$. The matrix
is $\pi$-periodic and traceless, satisfies $\mathbf{F}_{1}^{{\rm true}}(0)=\mathbf{0}$,
and the recursion $(\mathbf{F}_{1}^{{\rm true}})'=\mathbf{D}_{1}-\mathbf{K}_{1}$
is verified by direct differentiation. Numerically, the factor extracted
from the exact matriciant, $F(\tau)=e^{-\tau\mathbf{C}_{0}}X(\tau)e^{-\tau K}$
with $K=\pi^{-1}\log X(\pi)$ (branch near $-i\mathbf{I}$), matches
$\mathbf{I}+\varepsilon\mathbf{F}_{1}^{{\rm true}}$ to the expected
$O(\varepsilon^{2})$ accuracy: maximum deviation $3.5\times10^{-4}$
at $\varepsilon=0.004$, $\mu_{0}=0.3$ --- against an order-one
deviation from $\mathbf{I}+\varepsilon\mathbf{F}_{1}^{\varphi'}$.
At exact resonance $\mu_{0}=0$, 
\begin{equation}
\mathbf{F}_{1}^{{\rm true}}(\tau,0)=\begin{bmatrix}\dfrac{1-\cos4\tau}{16} & \dfrac{\sin2\tau}{4}-\dfrac{\sin4\tau}{16}\\[10pt]
-\dfrac{\sin2\tau}{4}-\dfrac{\sin4\tau}{16} & -\dfrac{1-\cos4\tau}{16}
\end{bmatrix}.\label{eq:YS-F1-EPD}
\end{equation}

\subsection{\texorpdfstring{The periodic Lyapunov factor $\mathbf{F}_{2}(\tau)$}{The
periodic Lyapunov factor F2(tau)}}

\label{subsec:YS-F2}

Continuing the recursion to second order in the zero-mean convention:
with $\mathbf{B}_{2}=0$ for the Mathieu approximation, the second-order
source is $\boldsymbol{\Phi}_{2}(\tau)=\mathbf{D}_{1}(\tau)\,\mathbf{F}_{1}^{0}(\tau)$,
where $\mathbf{F}_{1}^{0}=\mathbf{F}_{1}^{{\rm true}}-[\mathbf{F}_{1}^{{\rm true}}]_{{\rm av}}$
is the zero-mean gauge of the true first-order factor~\eqref{eq:YS-F1-true},
and $\mathbf{K}_{2}^{0}=[\boldsymbol{\Phi}_{2}]_{{\rm av}}$. The
zero-mean $\mathbf{F}_{2}$ solves $\mathbf{F}_{2}'=\boldsymbol{\Phi}_{2}-\mathbf{F}_{1}^{0}\mathbf{K}_{1}-\mathbf{K}_{2}^{0}$,
$[\mathbf{F}_{2}]_{{\rm av}}=\mathbf{0}$; at $\mu_{0}=0$ one obtains:
\begin{equation}
\mathbf{F}_{2}(\tau)=\begin{bmatrix}-\dfrac{5\cos2\tau}{128}+\dfrac{\cos4\tau}{64}-\dfrac{\cos6\tau}{384} & \dfrac{3\sin2\tau}{128}-\dfrac{\sin4\tau}{128}-\dfrac{\sin6\tau}{384}\\[10pt]
\dfrac{3\sin2\tau}{128}+\dfrac{\sin4\tau}{128}-\dfrac{\sin6\tau}{384} & \dfrac{5\cos2\tau}{128}+\dfrac{\cos4\tau}{64}+\dfrac{\cos6\tau}{384}
\end{bmatrix}.\label{eq:YS-F2}
\end{equation}
This satisfies $[\mathbf{F}_{2}]_{{\rm av}}=\mathbf{0}$, is $\pi$-periodic,
and has $\operatorname{tr}\mathbf{F}_{2}=\cos4\tau/32\neq0$ in the
zero-mean convention. The non-zero trace of $\mathbf{F}_{2}$ is a
known feature of the singular-case YS series: $\operatorname{tr}\mathbf{F}_{j}=0$
holds exactly only at first order ($j=1$); for $j\geq2$ the trace
is generally a trigonometric polynomial that averages to zero but
is not identically zero. The Fourier modes present are $k=1,2,3$
(i.e.\ frequencies $2\tau$, $4\tau$, $6\tau$), confirming the
general pattern: $\mathbf{F}_{j}$ has modes $k=1,2,\ldots,j+1$ (Remark~\ref{rem:YS-freq-conj}).
YS1~\cite[Ch.~IV, \S\,5.4]{YakSta1} prints the transformed forcing
$\mathbf{D}_{1}$ and the exponent matrices for $m=1$ but no second-order
periodic factor; formula~\eqref{eq:YS-F2} is \emph{derived} from
the zero-mean recursion of Chapter~\ref{app:YS-Floquet}. Explicitly:
with the zero-mean first-order factor $\mathbf{F}_{1}^{0}$ (the zero-mean
antiderivative of $\mathbf{D}_{1}-\mathbf{K}_{1}$) and $\mathbf{K}_{2}^{0}=[\mathbf{D}_{1}\mathbf{F}_{1}^{0}]_{{\rm av}}$,
the zero-mean solution of $\mathbf{F}_{2}'=\mathbf{D}_{1}\mathbf{F}_{1}^{0}-\mathbf{F}_{1}^{0}\mathbf{K}_{1}-\mathbf{K}_{2}^{0}$
at $\mu_{0}=0$ reproduces~\eqref{eq:YS-F2} to machine precision
($<5\times10^{-17}$, spectral integration). The gauge constant $\mathbf{K}_{2}^{0}$
differs from the printed $\mathbf{K}_{2}$~\eqref{eq:YS-K2} by the
commutator term $[\mathbf{K}_{1},\mathbf{V}]$ attached to the change
of normalization constant $\mathbf{V}$; the two are conjugation-equivalent
data with identical spectral invariants (exponents and boundaries),
the printed~\eqref{eq:YS-K2} corresponding to YS's $\mathbf{F}_{j}(0)=\mathbf{0}$
normalization. At $m=1$, $\mu_{0}=0$, the transformed forcing coincides
with the YS-printed first-order matrix entry by entry: $\mathbf{D}_{1}=\mathbf{F}_{1}^{\varphi'}$
of eq.~\eqref{eq:YS-F1} (cf.~eq.~\eqref{eq:phi-is-D1}). The required
consistency properties are all satisfied: (i)~$\mathbf{F}_{2}$ is
$\pi$-periodic (verified); (ii)~$[\mathbf{F}_{2}]_{{\rm av}}=\mathbf{0}$
(zero-mean, verified); (iii)~$\operatorname{tr}\mathbf{F}_{2}=\cos4\tau/32\neq0$
but with zero mean (verified); (iv)~$\mathbf{F}_{2}(0)=\mathbf{F}_{2}(\pi)=\bigl[\begin{smallmatrix}-5/192 & 0\\
0 & 11/192
\end{smallmatrix}\bigr]$ (periodic boundary values match, verified). The intermediate algebra
is routine and is not reproduced here; the machine-precision recursion
check above, together with the direct monodromy verification of \S\,\ref{subsec:YS-K},
certifies the formula~\eqref{eq:YS-F2}.

\subsection{Instability boundaries and the characteristic exponent}

\label{subsec:YS-boundaries}

\emph{Instability boundaries from $\mathbf{K}_{1}$.} The eigenvalues
of $\mathbf{K}_{1}(\mu_{0})$ (equation~\eqref{eq:YS-K1}) are 
\begin{equation}
\lambda_{\pm}=\pm\frac{1}{2}\sqrt{\!\left(\tfrac{1}{2}-\mu_{0}\right)\!\left(\tfrac{1}{2}+\mu_{0}\right)}=\pm\frac{1}{2}\sqrt{\tfrac{1}{4}-\mu_{0}^{2}},\label{eq:YS-K1-eigs}
\end{equation}
where $\mu_{0}$ is the YS detuning parameter~\eqref{eq:YS-LC-params}.
For $|\mu_{0}|<\tfrac{1}{2}$: $\lambda_{\pm}$ are real and nonzero,
one Floquet multiplier\index{Floquet theory!Floquet multiplier} satisfies
$|\rho|>1$, and the circuit is \emph{unstable} (inside the instability
tongue). For $|\mu_{0}|>\tfrac{1}{2}$: $\lambda_{\pm}$ are purely
imaginary, the Floquet multipliers lie on the unit circle, and the
circuit is \emph{stable} (outside the tongue). At $|\mu_{0}|=\tfrac{1}{2}$:
$\lambda_{\pm}=0$, the two multipliers coalesce at $-1$ --- an
\emph{exceptional point of degeneracy\index{exceptional point of degeneracy (EPD)}}
(EPD). The primary tongue width to leading order is 
\begin{equation}
L_{1}^{{\rm YS}}=\varepsilon=\frac{2\delta}{1+\delta^{2}},\label{eq:YS-width}
\end{equation}
which agrees with Theorem~\ref{thm:width} to leading order: $L_{1}^{{\rm LC}}=2\delta/(1-\delta^{2})\sim2\delta$.

\emph{Second-order correction and the Mathieu benchmark (item~1).}
The instability boundaries through second order follow from the exponent
matrix itself: writing $\varepsilon\mathbf{K}_{1}+\varepsilon^{2}\mathbf{K}_{2}=\bigl[\begin{smallmatrix}0 & \alpha\\
\beta & 0
\end{smallmatrix}\bigr]$, the boundary locus is $\alpha\beta=0$ (coalescence of the two characteristic
exponents $\varkappa_{1,2}=-i\pm\sqrt{\alpha\beta}$). Solving $\alpha=0$
near $\mu_{0}=+\tfrac{1}{2}$ and $\beta=0$ near $\mu_{0}=-\tfrac{1}{2}$
gives 
\begin{equation}
\mu_{0}^{+}=\tfrac{1}{2}-\tfrac{\varepsilon}{32}+O(\varepsilon^{2}),\qquad\mu_{0}^{-}=-\tfrac{1}{2}-\tfrac{\varepsilon}{32}+O(\varepsilon^{2}),\label{eq:YS-bdry-2nd}
\end{equation}
equivalently $c_{\pm}=1\pm\tfrac{\varepsilon}{2}-\tfrac{\varepsilon^{2}}{32}+O(\varepsilon^{3})$
--- in exact agreement with the Mathieu characteristic values $b_{1}(\hat{q})$
and $a_{1}(\hat{q})$ at $\hat{q}=-\varepsilon/2$ (McLachlan~\cite[\S\,2.151]{MacLa}):
the common $-\varepsilon^{2}/32$ is the symmetric center shift, and
the width carries no $O(\varepsilon^{2})$ correction. (Only the $O(\varepsilon)$
coefficients of $\mu_{0}^{\pm}$ are fixed by $\mathbf{K}_{2}$; the
$O(\varepsilon^{2})$ terms would require $\mathbf{K}_{3}$.) 
\begin{rem}[A misprint in the printed YS exponent formula]
\label{rem:YS-bdry-derivation} Yakubovich and Starzhinskii also
print a closed characteristic-exponent formula for this case~\cite[Ch.~IV, \S\,5.4]{YakSta1},
$\varkappa_{1,2}=-i\pm\tfrac{i}{2}\sqrt{A\cdot B}+O(\varepsilon^{3})$,
with $A,B$ as displayed in~\eqref{eq:YS-A}--\eqref{eq:YS-B} below.
Consistency with their own exponent matrices~\eqref{eq:YS-K1}--\eqref{eq:YS-K2}
would require $A\cdot B=-4\alpha\beta$; expanding both products shows
they disagree at the $O(\varepsilon^{3})$ terms of the bracket, and
the boundary predicted by $B=0$, namely $\mu_{0}^{+}=\tfrac{1}{2}+\tfrac{3}{32}\varepsilon+O(\varepsilon^{2})$,
contradicts both the Mathieu value $b_{1}(-\varepsilon/2)$ and direct
numerical monodromy: at the $\mathbf{K}$-based boundary~\eqref{eq:YS-bdry-2nd}
the computed $|\operatorname{tr}X(\pi)+2|$ is $O(\varepsilon^{4})$
($1.2\times10^{-8}$ at $\varepsilon=0.04$), while at $\mu_{0}=\tfrac{1}{2}+\tfrac{3}{32}\varepsilon$
it is $O(\varepsilon^{3})$ ($2.0\times10^{-5}$ at $\varepsilon=0.04$).
The misprint therefore sits in the printed $A,B$ product, not in
the $\mathbf{K}$-matrices. The leading-order content of $A,B$ is
unaffected: $A=\varepsilon(\mu_{0}+\tfrac{1}{2})+O(\varepsilon^{2})$,
$B=\varepsilon(\mu_{0}-\tfrac{1}{2})+O(\varepsilon^{2})$, so $A\cdot B=-4\alpha\beta+O(\varepsilon^{3})$,
and all leading-order uses of $A,B$ in this book remain valid. 
\end{rem}

The width comparison with the LC circuit is now direct. From~\eqref{eq:YS-bdry-2nd}
the YS tongue width in the $c$-variable is 
\begin{equation}
\Delta c^{{\rm YS}}=\varepsilon+O(\varepsilon^{3})=\frac{2\delta}{1+\delta^{2}}+O(\delta^{3}),\label{eq:YS-width-corrected}
\end{equation}
which agrees through $O(\delta^{2})$ with the exact LC zone width
in the $c$-variable, $|\Delta c_{1}|=|a|+O(|a|^{3})=2\delta/(1+\delta^{2})+O(\delta^{3})$
(Table~\ref{tab:bdry-LC}: the $O(a^{2})$ center shifts are equal
for the two boundaries and cancel in the width), equivalently with
$L_{1}^{{\rm LC}}=2\delta/(1-\delta^{2})$ after the $c\to\hat{\lambda}$
conversion. The remaining $O(\delta^{3})$ discrepancy is a genuine
truncation effect, requiring $\mathbf{K}_{3}$ in the YS series. Since
Yakubovich and Starzhinskii~\cite[Ch.~IV]{YakSta1} do not give $\mathbf{K}_{3}$
explicitly for $m=1$, extending the comparison to $O(\delta^{3})$
would require deriving $\mathbf{K}_{3}$ from the general YS recurrence
and is not pursued here.

\emph{The characteristic exponent split $\Delta\alpha$.} The characteristic
exponents are the eigenvalues of the exponent matrix: $\varkappa_{1,2}=-i\pm\sqrt{\alpha\beta}+O(\varepsilon^{3})$,
with $\alpha,\beta$ the off-diagonal entries of $\varepsilon\mathbf{K}_{1}+\varepsilon^{2}\mathbf{K}_{2}$
as above. We define the split as the full difference 
\begin{equation}
\Delta\alpha_{{\rm YS}}=\varkappa_{1}-\varkappa_{2}=2\sqrt{\alpha\beta},\label{eq:YS-Dalpha}
\end{equation}
matching the convention of Theorem~\ref{thm:EPD-split} ($\Delta\alpha=\alpha_{+}-\alpha_{-}$).
YS print the equivalent leading-order form $\varkappa_{1,2}=-i\pm\tfrac{i}{2}\sqrt{A\cdot B}$
(see Remark~\ref{rem:YS-bdry-derivation} for the status of its $O(\varepsilon^{2})$
coefficients), where 
\begin{align}
A & =\varepsilon\mu_{0}+\tfrac{\varepsilon}{2}-\tfrac{\varepsilon^{2}}{4}\mu_{0}^{2}+\tfrac{3\varepsilon^{2}}{8}\mu_{0}+\tfrac{\varepsilon^{2}}{32},\label{eq:YS-A}\\
B & =\varepsilon\mu_{0}-\tfrac{\varepsilon}{2}-\tfrac{3\varepsilon^{2}}{4}\mu_{0}^{2}+\tfrac{\varepsilon^{2}}{8}\mu_{0}+\tfrac{\varepsilon^{2}}{32}.\label{eq:YS-B}
\end{align}
To leading order $A\cdot B=-4\alpha\beta=\varepsilon^{2}(\mu_{0}^{2}-\tfrac{1}{4})+O(\varepsilon^{3})$.
The sign of $\alpha\beta$ determines stability: $\alpha\beta>0$
(inside tongue, $|\mu_{0}|<\tfrac{1}{2}$): $\sqrt{\alpha\beta}$
is real, the exponents acquire real parts --- \emph{instability};
$\alpha\beta<0$ (outside tongue): $\sqrt{\alpha\beta}$ is purely
imaginary, $\varkappa_{1,2}$ are purely imaginary --- \emph{stability};
$\alpha\beta=0$ at $|\mu_{0}|=\tfrac{1}{2}$ --- the EPD. Near the
upper boundary $\mu_{0}\to\tfrac{1}{2}^{+}$: $\alpha\approx\tfrac{\varepsilon}{2}(\mu_{0}-\tfrac{1}{2})$,
$\beta\approx-\tfrac{\varepsilon}{2}$, so 
\begin{equation}
|\Delta\alpha_{{\rm YS}}|\approx\varepsilon\sqrt{\mu_{0}-\tfrac{1}{2}},\quad\delta\ll1,\label{eq:YS-sqrt-behavior}
\end{equation}
which is the $\sqrt{\Delta c}$ square-root law of Theorem~\ref{thm:EPD-split},
confirmed independently.
\begin{rem}[Independent confirmation of main results]
\label{rem:YS-independent} The YS method confirms the manuscript's
main results by a completely independent route --- Fourier series
rather than continued fraction\index{continued fraction}s or Jordan
chain analysis: (1)~Instability boundary $\mu_{0}=\pm\tfrac{1}{2}$
agrees with $c_{\pm}^{(1)}=1\pm\tfrac{1}{2}\varepsilon+O(\varepsilon^{2})$
(Theorem~\ref{thm:EPD-bdry}) to leading order. (2)~Tongue width:
$\Delta c^{{\rm YS}}=\varepsilon+O(\varepsilon^{3})$ (eq.~\eqref{eq:YS-width-corrected})
agrees with the exact LC $c$-space width $|a|=2\delta/(1+\delta^{2})$
through $O(\delta^{2})$; the $O(\delta^{3})$ difference requires
$\mathbf{K}_{3}$. (3)~$\Delta\alpha_{{\rm YS}}=0$ at $|\mu_{0}|=\tfrac{1}{2}$
confirms EPD coalescence (Theorem~\ref{thm:EPD}). (4)~Near the
upper boundary, $\Delta\alpha_{{\rm YS}}\approx\varepsilon\sqrt{\mu_{0}-\tfrac{1}{2}}$
(eq.~\eqref{eq:YS-sqrt-behavior}). Writing $\Delta c=\varepsilon(\mu_{0}-\tfrac{1}{2})>0$
for the detuning into the stable side gives $\Delta\alpha_{{\rm YS}}\approx\sqrt{\varepsilon\,\Delta c}$.
Using $\varepsilon\approx2\delta$ and $|\eta_{1}|\approx\tfrac{\pi^{2}\delta}{2}$
(equation~\eqref{eq:eta1-m1}, computed at the EPD), one finds $|\eta_{1}|\approx\tfrac{\pi^{2}}{4}\varepsilon$,
hence $\sqrt{\varepsilon\,\Delta c}=\tfrac{2}{\pi}\sqrt{|\eta_{1}|\,\Delta c}$,
reproducing both the $\sqrt{\Delta c}$ law \emph{and} the coefficient
of Theorem~\ref{thm:EPD-split} ($\Delta\alpha=\tfrac{2}{\pi}\sqrt{\eta_{1}\Delta c}$,
real on the stable side, where $\eta_{1}\Delta c>0$ in the sign orientation
of that theorem). Agreement across all four checks validates both
derivations completely. A systematic quantitative comparison of the
Mathieu and exact LC YS series --- including normalization factors,
convention relations, and the role of the higher Poisson harmonics
--- is given in \S\,\ref{subsec:YS-Math-LC}. 
\end{rem}

\begin{rem}[Fourier structure of $\mathbf{F}_{j}$ and boundary accuracy]
\label{rem:YS-freq-conj} The exact LC system~\eqref{eq:LC-Hill-firstorder}
has the special property that each $\mathbf{B}_{n}(\tau)$ (and its
transformed counterpart $\mathbf{D}_{n}(\tau,\hat{\lambda})$ after
the preliminary transformation $e^{\tau\mathbf{C}_{0}}$ with $\mathbf{C}_{0}=\bigl[\begin{smallmatrix}0 & 1\\
-1 & 0
\end{smallmatrix}\bigr]$) contains \emph{exactly one} Fourier mode pair, namely $\cos2n\tau$
and $\sin2n\tau$ (mode $k=n$), for \emph{all values of $\hat{\lambda}$}.
We prove by induction that $\mathbf{F}_{j}(\tau,\hat{\lambda})$ contains
only Fourier modes $k=1,2,\ldots,j+1$.

\emph{Base case $j=1$:} $\mathbf{F}_{1}$ is built from $\mathbf{D}_{1}$
(mode $k=1$) via integration of $\mathbf{D}_{1}-\mathbf{K}_{1}$,
which produces modes $k=1$ and $k=2$. Explicitly verified: $\mathbf{F}_{1}$
has modes $k=1,2$ (Section~\ref{subsec:YS-F1}). Max mode $=2=1+1$,
confirming the base case.

\emph{Inductive step:} Assume $\mathbf{F}_{r}$ has max mode $r+1$
for all $r<l$. The source~\eqref{eq:YS-source} is $\boldsymbol{\Phi}_{l}=\mathbf{D}_{l}+\sum_{j=1}^{l-1}\mathbf{D}_{j}\,\mathbf{F}_{l-j}$.
Each term $\mathbf{D}_{j}\,\mathbf{F}_{l-j}$ is a product of $\mathbf{D}_{j}$
(single mode $k=j$) and $\mathbf{F}_{l-j}$ (max mode $l-j+1$ by
hypothesis); a product of Fourier modes $j$ and $l-j+1$ has max
mode $j+(l-j+1)=l+1$. The direct term $\mathbf{D}_{l}$ has mode
$l\leq l+1$. Hence $\boldsymbol{\Phi}_{l}$ has max mode $l+1$.
Since $\mathbf{K}_{l}=[\boldsymbol{\Phi}_{l}-\cdots]_{{\rm av}}$
removes only the $k=0$ component, and $\mathbf{F}_{l}=\int_{0}^{t}[\boldsymbol{\Phi}_{l}-\cdots-\mathbf{K}_{l}]\,\dd s$
preserves Fourier modes under integration, $\mathbf{F}_{l}$ has max
mode $l+1$, completing the induction.

Verified explicitly: $\mathbf{F}_{2}$ has modes $k=1,2,3$ (Section~\ref{subsec:YS-F2}).
Max mode $=3=2+1$, consistent with $j=2$.

The consequence for boundary accuracy is direct: the exponent $K(\delta)$
computed from the recursion to order $\delta^{l}$ involves only $\mathbf{F}_{1},\ldots,\mathbf{F}_{l-1}$,
each a finite trigonometric sum. No Fourier truncation is needed ---
all computations are exact in the Fourier domain. The resulting boundary
curves $c_{\pm}^{(m)}(\delta)$ are therefore accurate to $O(\delta^{l+2})$,
consistent with the CF method (Theorem~\ref{thm:width}). The two
methods --- YS series and continued fractions --- achieve the same
boundary accuracy at each order for the same structural reason: the
single-mode-per-order Fourier structure of the LC coefficient. 
\end{rem}

\begin{rem}[Higher instability tongues]
\label{rem:YS-higher-tongues} The YS method can in principle be
applied to the $m$-th instability tongue ($m=3,5,7,\ldots$) with
resonance order $m$. Since $f^{(m)}=0$ for $m\geq3$ (the LC perturbation
$f(\tau)=\cos2\tau$ has only the $m=1$ Fourier mode), the tongue
width first appears at order $\mathbf{K}_{m}$ in the YS series, requiring
$m$ nested Fourier convolutions of $f^{(1)}$. This is consistent
with the manuscript width formula $L_{m}\sim\delta^{m}$, but the
explicit computation of $\mathbf{K}_{3}$ and beyond is laborious
and is not pursued here. 
\end{rem}

\subsection{Phase portraits of the periodic Floquet factor: metamorphoses at
the EPDs}

\label{subsubsec:YS-Floquet-portraits}

The periodic Floquet factor\index{Floquet theory!Floquet factor}
is a matrix function of $\tau$, and a natural way to visualize it
is through the parametric curves traced by its columns. This section
develops the complete first-order picture for the primary Mathieu
resonance. The object portrayed is the \emph{true} Lyapunov factor
$\mathbf{F}_{1}^{{\rm true}}$ of eq.~\eqref{eq:YS-F1-true} (in
the normalization $\mathbf{F}_{1}^{{\rm true}}(0)=\mathbf{0}$); the
YS-printed matrix~\eqref{eq:YS-F1} has its own column geometry,
but, being a constant shift of the transformed forcing (eq.~\eqref{eq:phi-is-D1}),
that geometry portrays the \emph{coefficient} of the equation rather
than its Floquet solution structure, and we do not pursue it. A change
of the normalization constant of $\mathbf{F}_{1}$ translates each
column portrait rigidly in the plane, so all the features studied
below --- self-intersection counts, cusps, crossing angles --- are
invariants of the column, independent of the gauge choice.

The headline result is that the portrait topology changes \emph{exactly
at the exceptional points of degeneracy}: the second column collapses
onto an exact Steiner deltoid\index{Steiner deltoid}\index{deltoid (Steiner hypocycloid)}
--- a three-cusped hypocycloid --- at the upper EPD $\mu_{0}=+\tfrac{1}{2}$,
the first column at the lower EPD $\mu_{0}=-\tfrac{1}{2}$, and at
no other interior parameter value does either column lose smoothness.
The instability boundaries of the spectral problem are thereby visible
to the naked eye in the geometry of the Floquet factor itself.

\medskip{}

\noindent\emph{Setup.} Consider the \emph{phase portrait} of the
second column of $\varepsilon\mathbf{F}_{1}^{{\rm true}}(\tau,\mu_{0})$:
the parametric curve 
\begin{equation}
\tau\mapsto\bigl(\varepsilon F_{1}^{(12)}(\tau,\mu_{0}),\;\varepsilon F_{1}^{(22)}(\tau,\mu_{0})\bigr),\quad\tau\in[0,\pi],\label{eq:F1-portrait}
\end{equation}
where $F_{1}^{(ij)}$ denotes the $(i,j)$ entry of $\mathbf{F}_{1}^{{\rm true}}$
from eq.~\eqref{eq:YS-F1-true}. Both entries vanish at $\tau=0$
and $\tau=\pi$, so the curve is a closed loop through the origin
(the \emph{basepoint}). We portray the second column as the representative;
the column-exchange identity~\eqref{eq:column-exchange} of Remark~\ref{rem:EPD-propeller}
carries the complete first-column theory over from $-\mu_{0}$ by
a quarter-turn rotation. In the variables $\mathsf{c}=\cos2\tau$,
$\mathsf{s}=\sin2\tau$ of eq.~\eqref{eq:YS-F1-true}, the second-column
entries read 
\begin{align}
F_{1}^{(12)}(\tau,\mu_{0}) & =\frac{\mathsf{s}\,\bigl(2(1-\mu_{0})-\mathsf{c}\bigr)}{8},\nonumber \\
F_{1}^{(22)}(\tau,\mu_{0}) & =\frac{(\mathsf{c}-1)\,\bigl(2\mu_{0}+\mathsf{c}+1\bigr)}{8},\label{eq:F1-factored}
\end{align}
with velocities 
\begin{equation}
\frac{dF_{1}^{(12)}}{d\tau}=\frac{1+2(1-\mu_{0})\mathsf{c}-2\mathsf{c}^{2}}{4},\qquad\frac{dF_{1}^{(22)}}{d\tau}=-\frac{\mathsf{s}\,(\mathsf{c}+\mu_{0})}{2}.
\end{equation}
Since $\tau\mapsto\pi-\tau$ sends $(\mathsf{c},\mathsf{s})\mapsto(\mathsf{c},-\mathsf{s})$,
the curve is symmetric about the vertical axis: $F_{1}^{(12)}(\pi-\tau)=-F_{1}^{(12)}(\tau)$,
$F_{1}^{(22)}(\pi-\tau)=F_{1}^{(22)}(\tau)$.
\begin{thm}[Self-intersection metamorphosis of the second column]
\label{thm:F1-topology} For the Mathieu equation~\eqref{eq:YS-LC-eq},
the phase portrait~\eqref{eq:F1-portrait} of the second column of
$\varepsilon\mathbf{F}_{1}^{{\rm true}}(\tau,\mu_{0})$ has the following
structure as $\mu_{0}$ increases (the full sweep is displayed in
Figure~\ref{fig:YS-F1-transition}). 
\begin{enumerate}
\item[(i)] \emph{$\mu_{0}<0$:} a smooth simple closed curve; no self-intersections.
\item[(ii)] \emph{$\mu_{0}=0$ (tangential birth):} the interior point $\tau=\pi/2$,
located at $(0,-\varepsilon\mu_{0}/2)$, arrives at the basepoint.
The curve has a \emph{self-tangency} at the origin: the local branches
are the parabolic arcs $y=-8x^{2}/\varepsilon$ (at $\tau=0,\pi$)
and $y=-\tfrac{8}{9}x^{2}/\varepsilon$ (at $\tau=\pi/2$), with curvatures
of the same sign (ratio~$9$), so the branches touch without crossing.
\item[(iii)] \emph{$0<\mu_{0}<\tfrac{1}{2}$:} exactly \emph{two} transversal
self-intersections, forming a mirror pair $(\pm x^{*},y^{*})$ off
the symmetry axis, with parameters $\mathsf{c}_{1,2}=\pm u-\mu_{0}$,
$u=\sqrt{(2-\mu_{0})(1-2\mu_{0})/2}$ (at $\mu_{0}=\tfrac{1}{4}$:
$(x^{*},y^{*})=\varepsilon\,(0.12402\ldots,\,-0.140625)$).
\item[(iv)] \emph{$\mu_{0}=\tfrac{1}{2}$ (the upper EPD):} the mirror pair collides
with a newly stationary axis configuration and the curve degenerates
into a three-cusped curvilinear triangle --- in fact an \emph{exact}
Steiner deltoid (Remark~\ref{rem:EPD-deltoid}; Figure~\ref{fig:YS-F1-EPD-compare},
bottom right). The velocity vanishes at exactly three parameter values:
the basepoint $\tau=0\equiv\pi$ (local form $(\pm\varepsilon\tau^{3}/2,\,-3\varepsilon\tau^{2}/4)$:
a semicubical cusp with vertical tangent) and the interior points
$\tau=\pi/3,\,2\pi/3$ (where $\cos2\tau=-\mu_{0}$), giving two more
semicubical cusps. There are no self-intersections.
\item[(v)] \emph{$\tfrac{1}{2}<\mu_{0}<\tfrac{3}{2}$:} exactly \emph{one} transversal
self-intersection, located \emph{on} the symmetry axis at 
\begin{equation}
P(\mu_{0})=\Bigl(0,\;\frac{3(1-2\mu_{0})}{8}\,\varepsilon\Bigr),
\end{equation}
with parameter values $\tau^{*}=\tfrac{1}{2}\arccos\bigl(2(1-\mu_{0})\bigr)$
and $\pi-\tau^{*}$ (a vertical figure-eight whose lower loop shrinks
as $\mu_{0}$ grows).
\item[(vi)] \emph{$\mu_{0}=\tfrac{3}{2}$ (death):} the crossing is absorbed
at the point $\tau=\pi/2$ in a semicubical degeneration (velocity
$(2\mu_{0}-3)/4\to0$ there).
\item[(vii)] \emph{$\mu_{0}>\tfrac{3}{2}$:} a smooth simple closed curve again;
no self-intersections. 
\end{enumerate}
\end{thm}

\begin{figure}[htbp]
\centering \includegraphics[width=0.95\textwidth]{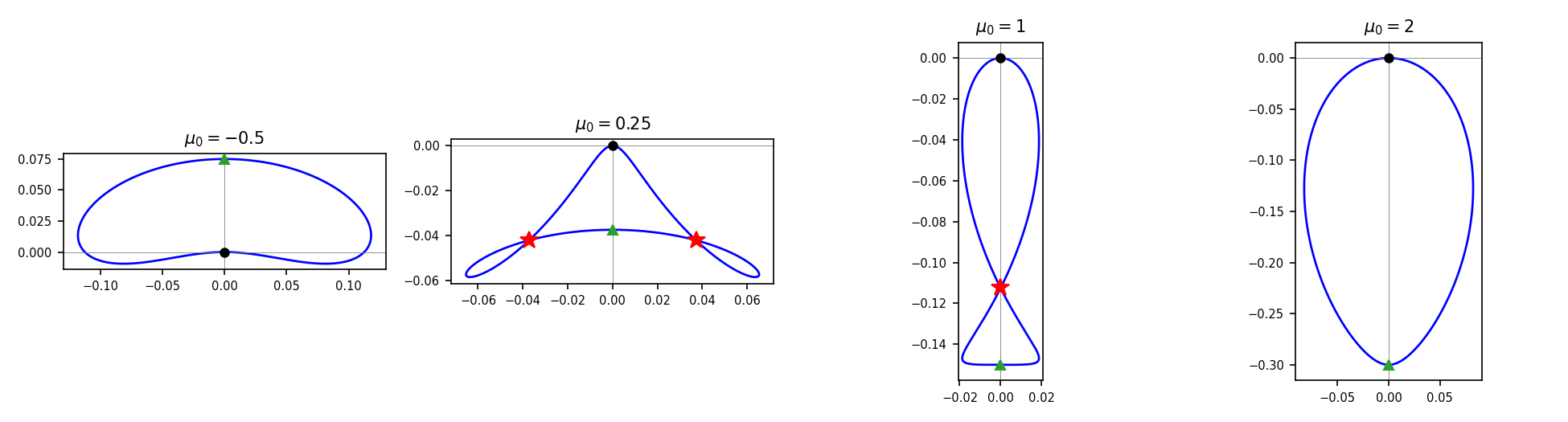}
\caption{Phase portraits of the second column $(\varepsilon F_{1}^{(12)},\,\varepsilon F_{1}^{(22)})$
of the true first-order Lyapunov factor $\varepsilon\mathbf{F}_{1}^{{\rm true}}(\tau,\mu_{0})$
(eq.~\eqref{eq:YS-F1-true}, $\varepsilon=0.30$, $\tau\in[0,\pi]$),
illustrating the four crossing-count regimes of Theorem~\ref{thm:F1-topology}:
$\mu_{0}=-0.5$ (simple curve, $0$ crossings), $\mu_{0}=0.25$ (mirror
pair, $2$ crossings), $\mu_{0}=1$ (on-axis figure-eight, $1$ crossing),
$\mu_{0}=2$ (simple curve, $0$ crossings). Black dot: basepoint
$\tau=0\equiv\pi$ at the origin. Green triangle: the point $\tau=\pi/2$.
Red stars: transversal self-intersections, located by a numerical
segment-intersection census in exact agreement with the theorem.}
\label{fig:YS-F1-portraits} 
\end{figure}

\begin{figure}[htbp]
\centering \includegraphics[width=0.98\textwidth]{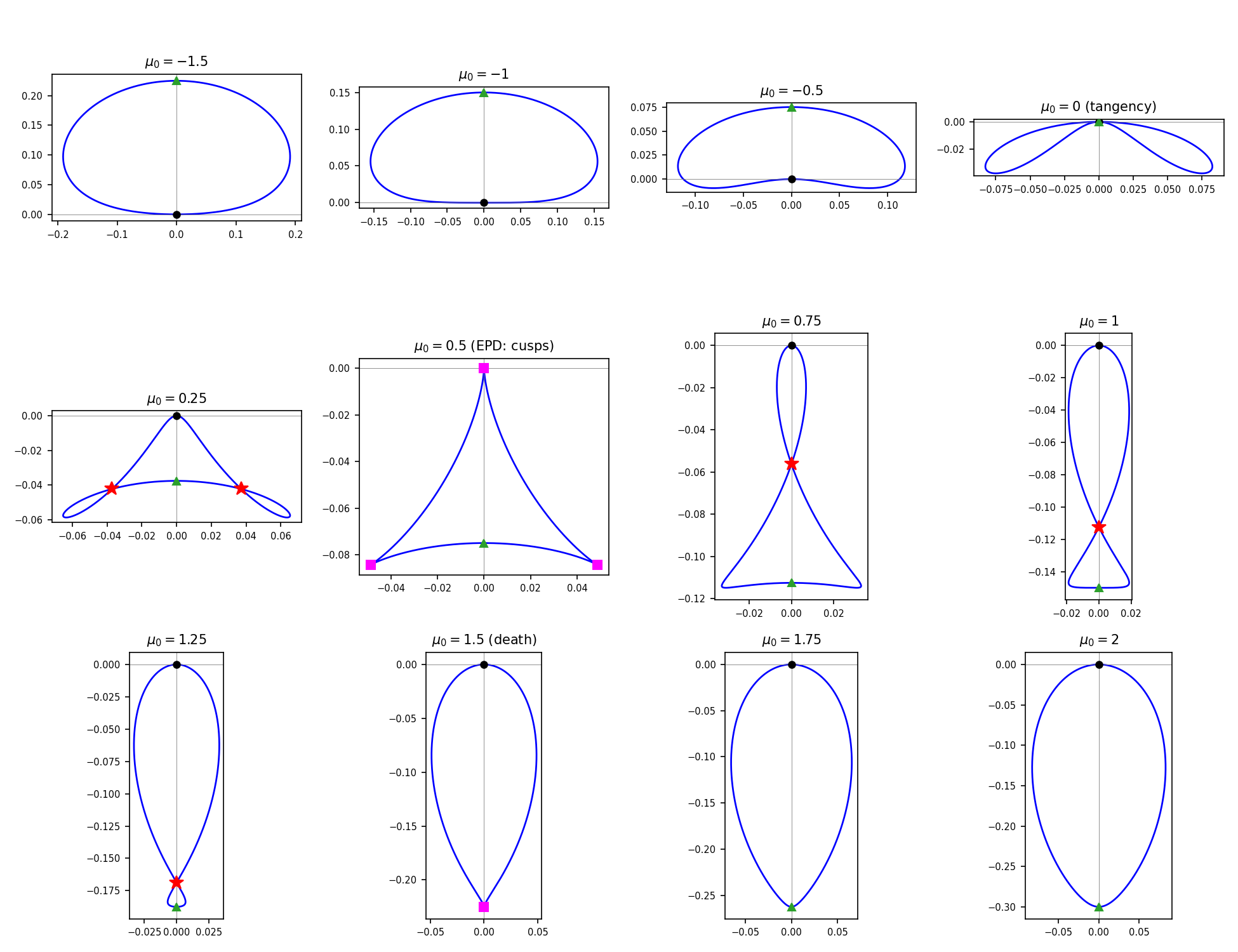}
\caption{Full metamorphosis of the second-column portrait as $\mu_{0}$ sweeps
from $-1.5$ to $+2.0$ ($\varepsilon=0.30$): crossing counts $0\to0\to0\to0\to2\to0\to1\to1\to1\to0\to0\to0$
across the twelve panels. The three transitions occur at $\mu_{0}=0$
(tangential birth at the basepoint), $\mu_{0}=\tfrac{1}{2}$ (the
upper EPD: exact Steiner deltoid, Remark~\ref{rem:EPD-deltoid},
magenta squares at the cusps $\tau=0\equiv\pi,\,\pi/3,\,2\pi/3$),
and $\mu_{0}=\tfrac{3}{2}$ (death: semicubical degeneration at $\tau=\pi/2$,
magenta square). Black dot: basepoint; green triangle: $\tau=\pi/2$;
red stars: self-intersections.}
\label{fig:YS-F1-transition} 
\end{figure}

\begin{figure}[htbp]
\centering \includegraphics[width=0.98\textwidth]{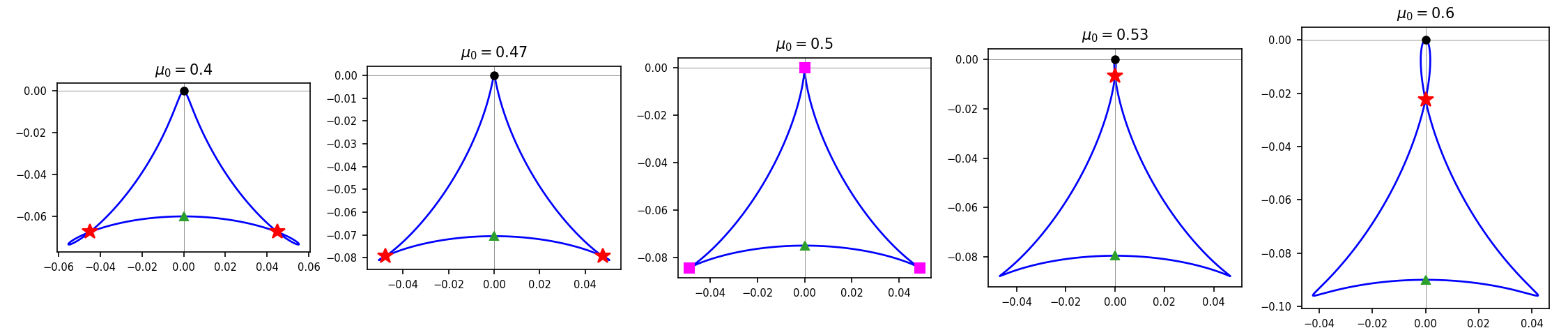}
\caption{Approach of the second-column portrait to the upper EPD $\mu_{0}=+\tfrac{1}{2}$
($\varepsilon=0.30$). For $\mu_{0}=0.40,\,0.47$ the mirror pair
of crossings (red stars) moves toward the symmetry axis; at $\mu_{0}=0.50$
(the EPD) the curve is the exact Steiner deltoid (Remark~\ref{rem:EPD-deltoid};
magenta squares at the cusps); for $\mu_{0}=0.53,\,0.60$ a single
on-axis crossing has been born at the former basepoint cusp. The off-axis
pair dies into the interior cusps while the on-axis crossing is born
at the basepoint cusp --- a simultaneous death-and-birth at the EPD.}
\label{fig:YS-F1-approach-plus} 
\end{figure}

\begin{figure}[htbp]
\centering \includegraphics[width=0.98\textwidth]{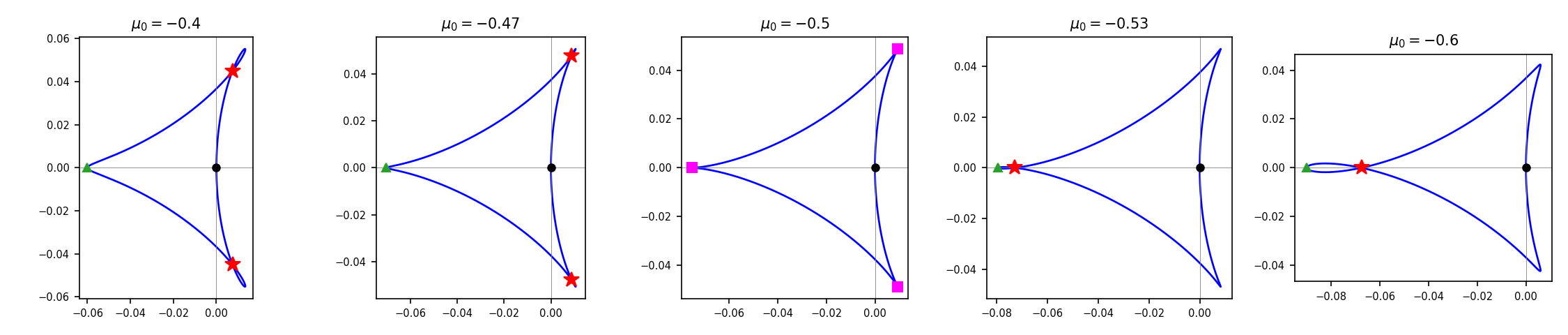}
\caption{Mirror approach of the \emph{first}-column portrait to the lower EPD
$\mu_{0}=-\tfrac{1}{2}$ ($\varepsilon=0.30$), for $\mu_{0}=-0.40,\,-0.47,\,-0.50,\,-0.53,\,-0.60$:
the same metamorphosis as in Figure~\ref{fig:YS-F1-approach-plus},
reflected across the diagonal, with the EPD configuration --- an
exact Steiner deltoid congruent to that of the upper EPD (Remark~\ref{rem:EPD-deltoid})
--- showing its three cusps at $\tau=\pi/6,\,\pi/2,\,5\pi/6$ (magenta
squares) and the on-axis crossing emerging on the horizontal axis
(Remark~\ref{rem:EPD-propeller}).}
\label{fig:YS-F1-approach-minus} 
\end{figure}

\begin{figure}[htbp]
\centering \includegraphics[width=0.92\textwidth]{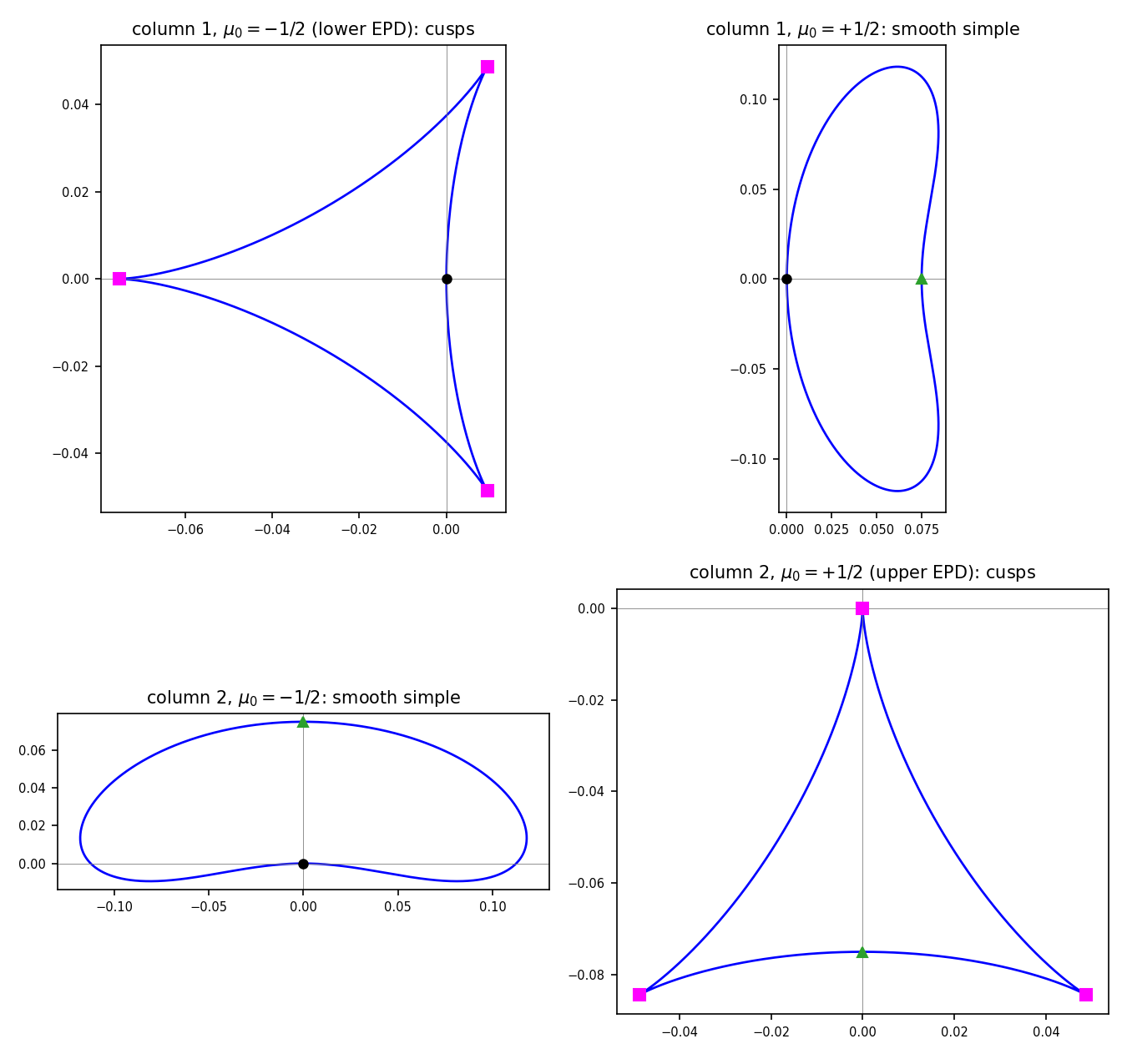} \caption{The column--EPD pairing at the two exceptional points ($\varepsilon=0.30$).
\emph{Top row}: first column at $\mu_{0}=-\tfrac{1}{2}$ (exact Steiner
deltoid, magenta squares at the cusps) and at $\mu_{0}=+\tfrac{1}{2}$
(smooth simple curve). \emph{Bottom row}: second column at $\mu_{0}=-\tfrac{1}{2}$
(smooth simple curve) and at $\mu_{0}=+\tfrac{1}{2}$ (exact Steiner
deltoid). Each EPD is marked by the deltoid metamorphosis of exactly
one column; the two deltoids are congruent (Remarks~\ref{rem:EPD-propeller},
\ref{rem:EPD-deltoid}).}
\label{fig:YS-F1-EPD-compare} 
\end{figure}

\begin{proof}
\emph{The chord condition.} A self-intersection requires $\tau_{1}\neq\tau_{2}$
with equal coordinates. Equality of the second coordinates in~\eqref{eq:F1-factored}
reads $2\mu_{0}\mathsf{c}_{1}+\mathsf{c}_{1}^{2}=2\mu_{0}\mathsf{c}_{2}+\mathsf{c}_{2}^{2}$,
i.e.\
$(\mathsf{c}_{1}-\mathsf{c}_{2})(2\mu_{0}+\mathsf{c}_{1}+\mathsf{c}_{2})=0$,
where $\mathsf{c}_{i}=\cos2\tau_{i}$. There are thus two channels:
$\mathsf{c}_{1}=\mathsf{c}_{2}$ with $\mathsf{s}_{1}=-\mathsf{s}_{2}$
(the mirror map $\tau_{2}=\pi-\tau_{1}$), or the \emph{chord condition}
\begin{equation}
\mathsf{c}_{1}+\mathsf{c}_{2}=-2\mu_{0}.
\end{equation}

\emph{On-axis crossings (mirror channel).} If $\tau_{2}=\pi-\tau_{1}$,
the first coordinates are opposite, so they agree only when both vanish:
$\mathsf{s}_{1}=0$ (the basepoint, excluded as the closure point)
or $\mathsf{c}_{1}=2(1-\mu_{0})$. The latter lies in $(-1,1)$ exactly
for $\tfrac{1}{2}<\mu_{0}<\tfrac{3}{2}$, and substituting $\mathsf{c}^{*}=2(1-\mu_{0})$
into~\eqref{eq:F1-factored} gives $F_{1}^{(22)}=3(1-2\mu_{0})/8$,
establishing~(v); the crossing is transversal because the two velocity
vectors there, $\bigl(\mathsf{s}^{*2}/4,\,\mp\mathsf{s}^{*}(2-\mu_{0})/2\bigr)$
with $\mathsf{s}^{*}=\sqrt{1-\mathsf{c}^{*2}}>0$, are linearly independent.

\emph{Off-axis crossings (chord channel).} Put $u=\mathsf{c}_{1}+\mu_{0}=-(\mathsf{c}_{2}+\mu_{0})$,
so $\mathsf{c}_{1,2}=\pm u-\mu_{0}$. Equality of the first coordinates,
after squaring $\mathsf{s}_{i}=\sigma_{i}\sqrt{1-\mathsf{c}_{i}^{2}}$
and clearing, reduces to the single equation $u\bigl[(2-\mu_{0})(2\mu_{0}-1)+2u^{2}\bigr]=0$,
whose nontrivial root is $u^{2}=(2-\mu_{0})(1-2\mu_{0})/2$. The root
is positive, both $\mathsf{c}_{1,2}$ lie in $(-1,1)$, and the sign
consistency $\sigma_{1}\sigma_{2}=+1$ required by the unsquared equation
holds, precisely for $0\le\mu_{0}<\tfrac{1}{2}$; at $\mu_{0}=0$
the configuration degenerates onto the basepoint. This establishes
the mirror pair in~(iii) and its absence elsewhere.

\emph{Stationary points and cusps.} The velocity vanishes iff $dF_{1}^{(22)}/d\tau=0$
($\mathsf{s}=0$ or $\mathsf{c}=-\mu_{0}$) together with $dF_{1}^{(12)}/d\tau=0$.
At $\mathsf{c}=1$ ($\tau=0,\pi$) the first velocity is $(1-2\mu_{0})/4$;
at $\mathsf{c}=-\mu_{0}$ it is again $(1-2\mu_{0})/4$; at $\mathsf{c}=-1$
($\tau=\pi/2$) it is $(2\mu_{0}-3)/4$. Hence interior stationary
points exist only at $\mu_{0}=\tfrac{1}{2}$ (three of them: $\tau=0\equiv\pi$
and $\cos2\tau=-\tfrac{1}{2}$, i.e.\
$\tau=\pi/3,2\pi/3$) and at $\mu_{0}=\tfrac{3}{2}$ ($\tau=\pi/2$);
local Taylor expansion at each gives the semicubical normal form stated
in~(iv) and~(vi). For all other $\mu_{0}$ the curve is an immersion,
and the counts in~(i), (iii), (v), (vii) follow from the two channels
above, which are exhaustive.

\emph{Tangency at $\mu_{0}=0$.} Near $\tau=0$: $F_{1}^{(12)}\sim\tau/4$,
$F_{1}^{(22)}\sim-\tau^{2}/2$, giving $y=-8x^{2}$; near $\tau=\pi/2+\sigma$:
$F_{1}^{(12)}\sim3\sigma/4$, $F_{1}^{(22)}\sim-\sigma^{2}/2$, giving
$y=-\tfrac{8}{9}x^{2}$; both open downward, so the contact is a tangency,
not a crossing.

All counts, crossing locations, and cusp locations have additionally
been certified by a direct numerical segment-intersection census of
the curves at the parameter values of Figures~\ref{fig:YS-F1-portraits}--\ref{fig:YS-F1-EPD-compare}. 
\end{proof}
\begin{rem}[The EPD curve\index{exceptional point of degeneracy (EPD)!EPD curve}
and the first-column mirror]
\label{rem:EPD-propeller} The first column of $\varepsilon\mathbf{F}_{1}^{{\rm true}}$
traces the closed curve 
\begin{equation}
\Bigl(\frac{(1-\mathsf{c})(2\mu_{0}+1+\mathsf{c})}{8},\;-\frac{\mathsf{s}\,\bigl(2(1+\mu_{0})+\mathsf{c}\bigr)}{8}\Bigr)\,\varepsilon,
\end{equation}
also through the origin, symmetric about the \emph{horizontal} axis.
In fact the two columns are exact images of one another: writing $z_{(j)}=F_{1}^{(1j)}+iF_{1}^{(2j)}$
for the complex coordinate of the $j$-th column portrait (in units
of $\mathbf{F}_{1}$), 
\begin{equation}
z_{(1)}(\tau;\,\mu_{0})=i\,z_{(2)}\Bigl(\tau+\frac{\pi}{2};\,-\mu_{0}\Bigr)+\frac{\mu_{0}}{2}\,,\label{eq:column-exchange}
\end{equation}
verified to machine precision ($<10^{-15}$ over a grid of $\mu_{0}$):
the first-column portrait at detuning $\mu_{0}$ is the second-column
portrait at $-\mu_{0}$, rotated by a quarter turn and translated
along the horizontal axis. The identity reflects the quarter-period
translation symmetry of the Mathieu modulation, $\cos2(\tau+\pi/2)=-\cos2\tau$,
which reverses the sign of the forcing and hence of the effective
detuning. The complete first-column theory is therefore Theorem~\ref{thm:F1-topology}
read at $-\mu_{0}$, rotations and translations affecting none of
the counts, cusps, or angles. Explicitly (as direct analysis with
the velocities $\frac{d}{d\tau}F_{1}^{(11)}=\tfrac{\mathsf{s}}{2}(\mathsf{c}+\mu_{0})$
and $\frac{d}{d\tau}F_{1}^{(21)}=-\tfrac{1}{4}[2(1+\mu_{0})\mathsf{c}+2\mathsf{c}^{2}-1]$
confirms): no self-intersections for $\mu_{0}>0$; tangential birth
at $\mu_{0}=0$; two off-axis crossings for $-\tfrac{1}{2}<\mu_{0}<0$;
a three-cusped degeneration --- again an exact Steiner deltoid (Remark~\ref{rem:EPD-deltoid})
--- at the \emph{lower} EPD $\mu_{0}=-\tfrac{1}{2}$, with semicubical
cusps at the interior points $\tau=\pi/6,\,\pi/2,\,5\pi/6$ (the basepoint
stays smooth there, with vertical velocity $(0,-(3+2\mu_{0})/4)\neq\mathbf{0}$);
one on-axis crossing at $\bigl(-\varepsilon(3+2\mu_{0})/8,\,0\bigr)$
for $-\tfrac{3}{2}<\mu_{0}<-\tfrac{1}{2}$; death at $\mu_{0}=-\tfrac{3}{2}$
(at the basepoint); and simple curves for $\mu_{0}<-\tfrac{3}{2}$.

Thus the two EPDs $\mu_{0}=\pm\tfrac{1}{2}$ of the exponent matrix
$\mathbf{K}_{1}$ (eq.~\eqref{eq:YS-K1-eigs}) split the geometric
work between the columns: \emph{each EPD is marked by the cusp degeneration
of exactly one column} --- the upper EPD by the second column, the
lower EPD by the first --- while the other column passes through
that parameter value as a smooth simple curve (Figure~\ref{fig:YS-F1-EPD-compare}).
The spectral degeneracy is thereby imprinted on the Floquet factor
as a visible geometric singularity. 
\end{rem}

\begin{rem}[The EPD configurations are exact Steiner deltoids]
\label{rem:EPD-deltoid} The degenerate three-cusped curves at the
two EPDs are not merely deltoid-like: they are \emph{exact} Steiner
deltoids\index{Steiner deltoid}\index{deltoid (Steiner hypocycloid)}\index{exceptional point of degeneracy (EPD)!deltoid at}
(tricuspid hypocycloids; the curve studied by Steiner~\cite[pp.~231--237]{Steiner1857}
in connection with the envelope of Simson lines; see also Lockwood~\cite[Ch.~8, pp.~72--79]{Lockw}).
In the complex coordinate $z=x+iy$ of the portrait plane, the second
column at $\mu_{0}=+\tfrac{1}{2}$ and the first column at $\mu_{0}=-\tfrac{1}{2}$
trace 
\begin{equation}
z_{(2)}=\frac{i\varepsilon}{16}\Bigl[2e^{-2i\tau}+e^{4i\tau}-3\Bigr],\qquad z_{(1)}=\frac{\varepsilon}{16}\Bigl[2e^{-2i\tau}-e^{4i\tau}-1\Bigr],\label{eq:EPD-deltoid}
\end{equation}
and each bracket reduces, after the substitution $\theta=-2\tau$
and an overall rotation (by $\pi/2$ and by $\pi/3$ respectively),
to the canonical hypocycloid parametrization $2re^{i\theta}+re^{-2i\theta}$
with rolling-circle radius $r=\varepsilon/16$: the path of a point
on a circle of radius $r$ rolling inside a fixed circle of radius
$3r$. (Both identities~\eqref{eq:EPD-deltoid} have been verified
against the factored forms~\eqref{eq:F1-factored} to machine precision,
$<10^{-16}$.) Consequently the two EPD curves are \emph{congruent}
deltoids --- by the exchange identity~\eqref{eq:column-exchange}
at $\mu_{0}=\tfrac{1}{2}$ they differ exactly by a quarter-turn rotation
and a translation (equivalently, the deltoid's three-fold rotational
symmetry making rotations equivalent mod $2\pi/3$, by a rotation
of $-\pi/6$) --- with exact three-fold symmetry; semicubical cusps
of interior angle zero at the vertices of an equilateral triangle
of side $3\sqrt{3}\,\varepsilon/16$; circumradius $3r=3\varepsilon/16$
and inscribed-circle radius $r$; total arclength exactly $16r=\varepsilon$;
and enclosed area $2\pi r^{2}=\pi\varepsilon^{2}/128$. This perfection
is the geometric signature of the exceptional point: at $\mu_{0}=\pm\tfrac{1}{2}$,
where an eigenvalue of $\mathbf{K}_{1}$ vanishes (eq.~\eqref{eq:YS-K1-eigs}),
the corresponding column collapses onto the maximally symmetric three-cusped
curve, and away from these values the symmetry breaks and the deltoid
unfolds into the generic configurations of Theorem~\ref{thm:F1-topology}. 
\end{rem}

\begin{thm}[Crossing angle of the on-axis self-intersection]
\label{thm:angle-sum} For $\tfrac{1}{2}<\mu_{0}<\tfrac{3}{2}$,
the two branches of the second-column portrait at the on-axis crossing
$P(\mu_{0})$ make angles $\pm\phi$ with the horizontal direction,
where 
\begin{equation}
\tan\phi=\frac{2(2-\mu_{0})}{\sqrt{1-\bigl(2(1-\mu_{0})\bigr)^{2}}}\,,\label{eq:angle-sum}
\end{equation}
so the crossing angle between the branches is $2\phi$. As $\mu_{0}\to\tfrac{1}{2}^{+}$
and as $\mu_{0}\to\tfrac{3}{2}^{-}$ one has $\phi\to\pi/2$: the
tangent directions become anti-parallel and the crossing is born from,
and dies into, the cusp configurations of Theorem~\ref{thm:F1-topology}(iv),(vi)
tangentially. 
\end{thm}

\begin{proof}
The velocity vectors at $\tau^{*}$ and $\pi-\tau^{*}$ are $\bigl(\mathsf{s}^{*2}/4,\,\mp\mathsf{s}^{*}(2-\mu_{0})/2\bigr)$
with $\mathsf{s}^{*}=\sqrt{1-\mathsf{c}^{*2}}$, $\mathsf{c}^{*}=2(1-\mu_{0})$
(proof of Theorem~\ref{thm:F1-topology}); their slopes are $\mp2(2-\mu_{0})/\mathsf{s}^{*}$,
which is~\eqref{eq:angle-sum}. As $\mu_{0}\to\tfrac{1}{2}^{+}$
or $\tfrac{3}{2}^{-}$, $\mathsf{s}^{*}\to0$ while $2-\mu_{0}$ stays
positive, so $\tan\phi\to+\infty$. 
\end{proof}
\begin{cor}[Special values of the crossing angle]
\label{cor:blade-angles} On the window $\tfrac{1}{2}<\mu_{0}<\tfrac{3}{2}$
the half-angle $\phi(\mu_{0})$ attains its \emph{minimum} at $\mu_{0}=\tfrac{5}{4}$,
where $\tan^{2}\phi=3$ exactly, i.e. 
\begin{equation}
\phi_{\min}=\frac{\pi}{3}\qquad\text{(crossing angle }2\phi=\tfrac{2\pi}{3}\text{)},
\end{equation}
so the crossing is never sharper than $2\pi/3$. At the midpoint $\mu_{0}=1$:
$\tan\phi=2$, $\phi=\arctan2\approx63.4^{\circ}$. \emph{Proof:}
with $w=1-\mu_{0}\in(-\tfrac{1}{2},\tfrac{1}{2})$, $\tan^{2}\phi=4(1+w)^{2}/(1-4w^{2})$,
whose derivative is proportional to $(1+w)(1+4w)$ and vanishes at
$w=-\tfrac{1}{4}$; there $\tan^{2}\phi=4\cdot\tfrac{9}{16}/\tfrac{3}{4}=3$. 
\end{cor}

\medskip{}

\begin{figure}[htbp]
\centering \includegraphics[width=0.88\textwidth]{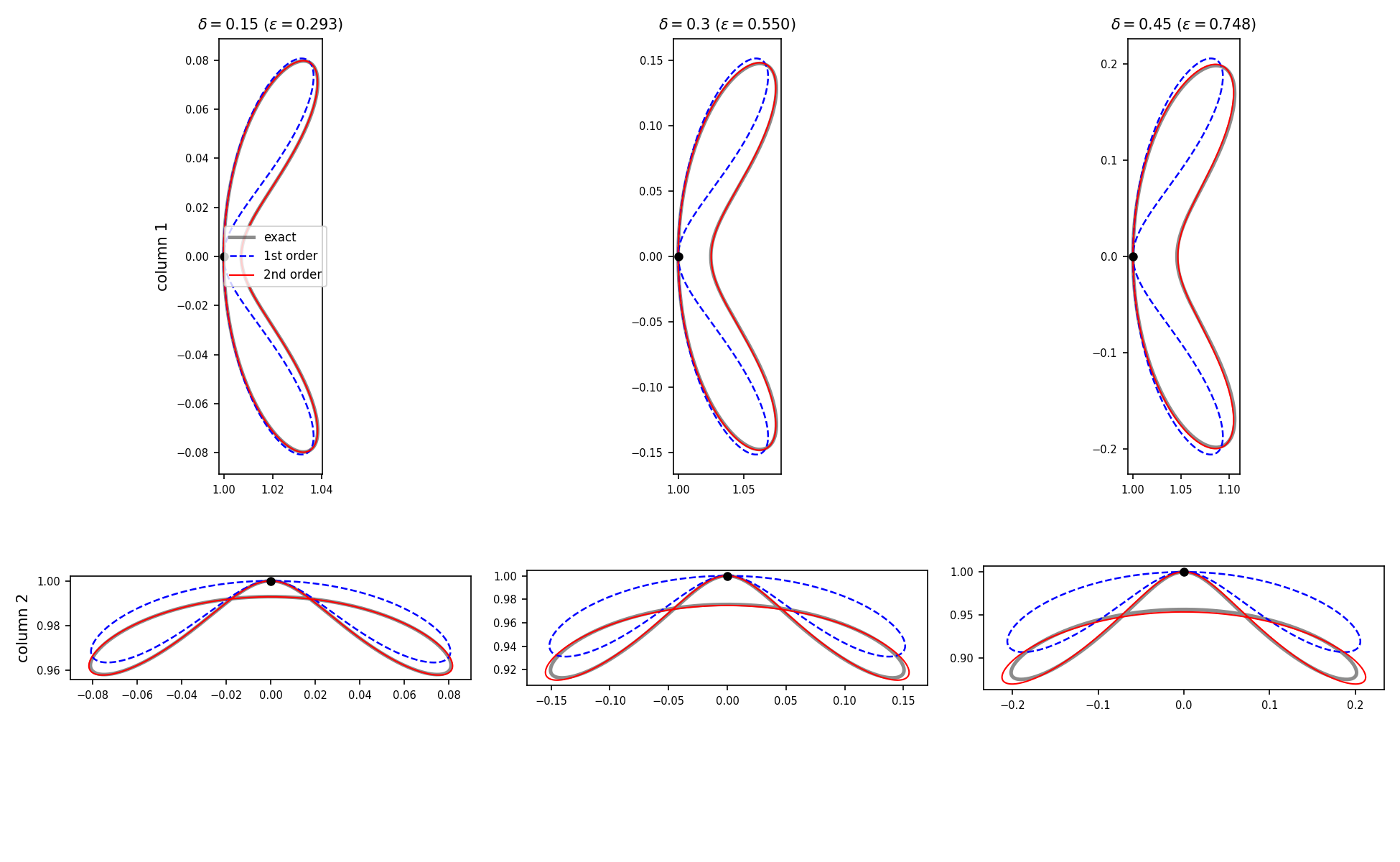}
\caption{Convergence of the YS series for the true Floquet factor at the tongue
center ($c=1$, $\mu_{0}=0$), in the normalization $\mathbf{F}_{j}(0)=\mathbf{0}$
(so all curves start and end at $(1,0)$ for the first column, top
row, and $(0,1)$ for the second column, bottom row; black dots),
for modulation depths $\delta=0.15,\,0.30,\,0.45$, i.e.\
$\varepsilon=0.293,\,0.550,\,0.748$. \emph{Gray solid}: the numerically
exact factor $F(\tau)=e^{-\tau\mathbf{C}_{0}}X(\tau)e^{-\tau K}$
extracted from the integrated monodromy. \emph{Blue dashed}: first
order $\mathbf{I}+\varepsilon\mathbf{F}_{1}^{{\rm true}}$. \emph{Red
solid}: second order $\mathbf{I}+\varepsilon\mathbf{F}_{1}^{{\rm true}}+\varepsilon^{2}\mathbf{F}_{2}$.
The second-order curves lie essentially on top of the exact factor
even at $\delta=0.45$ (maximum entry deviations $6.6\times10^{-4}$,
$4.3\times10^{-3}$, $1.1\times10^{-2}$ for the three depths), confirming
both the convergence of the series and the faithfulness of the first-order
portrait analysis.}
\label{fig:YS-F1-series} 
\end{figure}

\noindent\emph{Scope.} Theorem~\ref{thm:F1-topology} is exact for
the Mathieu equation at first order in $\varepsilon$. The second-order
correction $\varepsilon^{2}\mathbf{F}_{2}$ perturbs the curves quantitatively
but, as Figure~\ref{fig:YS-F1-series} shows, the second-order factor
lies essentially on top of the numerically exact one through $\delta=0.45$,
so the first-order metamorphosis picture is faithful. For the exact
LC circuit (all Fourier harmonics of $f(\tau)$ retained), the additional
harmonics deform the curves and shift the degenerate configurations,
which are structurally unstable as exact coincidences; the qualitative
sequence of crossing counts and its anchoring at the EPDs persist
for small $\varepsilon$. Figure~\ref{fig:YS-F1-portraits} illustrates
the four crossing-count regimes, Figures~\ref{fig:YS-F1-approach-plus}
and~\ref{fig:YS-F1-approach-minus} the approaches to the two EPDs,
and Figure~\ref{fig:YS-F1-EPD-compare} the column-EPD pairing.

\section{The Yakubovich--Starzhinskii series: exact LC circuit}

\label{sec:YS-exact-LC}

This chapter removes both approximations of the Mathieu\index{Mathieu equation}
case (Chapter~\ref{sec:YS-LC}) and applies the YS framework to the
\emph{exact} LC Hill equation\index{Hill equation} with $\hat{\lambda}$
kept as a free parameter throughout. The LC-circuit Fourier coefficient
$f_{{\rm LC}}(\tau,\delta)=\sum_{n\geq1}2\delta^{n}\cos2n\tau$ is
retained in full, and $\hat{\lambda}=c(1+\delta^{2})/(1-\delta^{2})$
is treated as independent of $\delta$ until the final step. The main
outputs are: closed-form exponent matrices $\mathbf{K}_{1}(\hat{\lambda})$
and $\mathbf{K}_{2}(\hat{\lambda})$ (exact polynomials in $\hat{\lambda}$),
explicit Lyapunov factors $\mathbf{F}_{1}(\tau,\hat{\lambda})$ and
$\mathbf{F}_{2}(\tau,\hat{\lambda})$, and self-consistent primary
tongue boundary curve\index{boundary curves}s agreeing with the MW/CF
results through $O(\delta^{3})$. A systematic comparison with the
Mathieu case follows in \S\,\ref{subsec:YS-Math-LC}, and the periodic
Lyapunov factors $\mathbf{F}_{1}$, $\mathbf{F}_{2}$ for the exact
circuit are constructed in \S\,\ref{subsubsec:YS-LC-F1F2}.

\subsection{\texorpdfstring{The exact LC circuit: $\hat{\lambda}$ as a free
parameter}{The exact LC circuit: lambda-hat as a free parameter}}

\label{subsec:YS-exact-LC}

We apply the two-parameter YS framework (Chapter~\ref{app:YS-Floquet},
\S\,\ref{subsec:YS-two-param}) to the exact LC Hill equation~\eqref{eq:LC-Hill-firstorder}.
The preliminary transformation uses $\mathbf{C}_{0}=\bigl[\begin{smallmatrix}0 & 1\\
-1 & 0
\end{smallmatrix}\bigr]$ (as in the Mathieu case), giving $\pi$-periodic $\mathbf{D}_{n}$
for all $\hat{\lambda}$ (\S\,\ref{subsec:YS-singular}).
\begin{center}
\fcolorbox{black}{white}{\parbox[c]{0.85\textwidth}{%
 \emph{Conventions for this section:} \S\,\ref{subsubsec:YS-Km-exact}
(derivation of $\mathbf{K}_{m}$ and intermediate $\mathbf{F}_{j}$
at $\hat{\lambda}=1$) uses Convention~1 (zero-mean, $[\mathbf{F}_{j}]_{{\rm av}}=\mathbf{0}$).
\S\,\ref{subsubsec:YS-LC-F1F2} (closed-form $\mathbf{F}_{1}(\tau,\hat{\lambda})$,
$\mathbf{F}_{2}(\tau,\hat{\lambda})$ for all $\hat{\lambda}$) uses
Convention~2 (initial-condition, $\mathbf{F}_{j}(0,\hat{\lambda})=\mathbf{0}$).
See \S\,\ref{subsec:YS-conventions} of Chapter~\ref{app:YS-Floquet}. %
}} 
\par\end{center}
\begin{rem}[Reader's guide: which normalization is used where, and what is convention-independent]
\label{rem:YS-LC-navigation} This section switches normalization
convention by subsection, because each task is cleanest in a different
one. The following map prevents the most common confusion --- comparing
a quantity computed in one convention against the same symbol computed
in another. 
\begin{center}
\global\long\def\arraystretch{1.3}%
\begin{tabular}{p{3.8cm}p{2.7cm}p{4.2cm}}
\toprule 
Subsection / purpose  & $\mathbf{F}_{j}$ normalization  & Diagnostic: $\mathbf{K}_{2}[0,1]$ at $\hat{\lambda}=1$\tabularnewline
\midrule 
\S\,\ref{subsubsec:YS-Km-exact} (derive $\mathbf{K}_{m}$, intermediate
$\mathbf{F}_{j}$)  & zero-mean (Conv.~1)  & $-\tfrac{1}{16}$ (eq.~\eqref{eq:K2-omega1})\tabularnewline
\S\,\ref{subsubsec:YS-LC-F1F2} (closed-form $\mathbf{F}_{j}(\tau,\hat{\lambda})$,
boundary curves)  & initial-condition (Conv.~2)  & $+\tfrac{1}{16}$ (eq.~\eqref{eq:K2-omega1-ic})\tabularnewline
\S\,\ref{subsec:YS-Math-LC} (Mathieu comparison)  & $\varphi'$-prescription (Math)  & $+\tfrac{1}{64}$ (Math norm, eq.~\eqref{eq:YS-K2})\tabularnewline
\bottomrule
\end{tabular}
\par\end{center}
Three rules make every cross-comparison unambiguous: (i)~the \emph{off-diagonal
sign} of $\mathbf{K}_{2}$ flips between the zero-mean and IC conventions
(the constant offset $\mathbf{F}_{1}^{{\rm ic}}-\mathbf{F}_{1}^{{\rm zm}}=\bigl[\begin{smallmatrix}1/8 & 0\\
0 & -1/8
\end{smallmatrix}\bigr]$ contributes $+\tfrac{1}{8}$ to $\mathbf{K}_{2}[0,1]$); (ii)~the
LC-vs-Mathieu \emph{magnitude} ratio is $2^{m}$ at order $m$ from
the $\tilde{\mathbf{B}}_{n}$ factor of $2$, \emph{with no accompanying
sign change}, valid order for order in matched conventions for the
single-harmonic chain (so $+\tfrac{7}{128}=8\,\mathbf{K}_{3}^{{\rm Math}}$,
both positive, zero-mean); the \emph{exact} LC third-order value is
shifted by the second harmonic to $-\tfrac{9}{128}$ (eq.~\eqref{eq:k2k3-summary});
(iii)~only the \emph{eigenvalues} of $K_{{\rm total}}$ --- hence
all stability predictions and the boundary curves~\eqref{eq:YS-bdry-exact}
--- are convention-independent. When in doubt, compare eigenvalues,
never individual entries. 
\end{rem}

\emph{Roadmap of this section.} The destination is a single result
reached by an independent route: the primary-tongue boundary curves
\begin{equation}
c_{\pm}(\delta)=1\pm\tfrac{\varepsilon}{2}-\tfrac{9}{32}\varepsilon^{2}+O(\varepsilon^{3}),\qquad\varepsilon=\tfrac{2\delta}{1+\delta^{2}},
\end{equation}
in agreement with the MW and CF methods (Table~\ref{tab:bdry-LC},
$m=1$). The logical chain is: (a)~the exact exponent matrices $\mathbf{K}_{1}(\hat{\lambda})$,
$\mathbf{K}_{2}(\hat{\lambda})$ in closed form (eqs.~\eqref{eq:LC-K1-lam-exact}--\eqref{eq:LC-K2-lam-exact});
(b)~the periodic Lyapunov factors $\mathbf{F}_{1},\mathbf{F}_{2}$
on which the recursion for $\mathbf{K}_{m}$ rests (derived in full
for all $\hat{\lambda}$ in \S\,\ref{subsubsec:YS-LC-F1F2}, and
summarized at $\hat{\lambda}=1$ in the derivation immediately below);
(c)~the EPD\index{exceptional point of degeneracy (EPD)} boundary
condition $\operatorname{Tr}(e^{\pi K})=2$ and its solution (\S\,\ref{subsubsec:YS-bdry-lh});
(d)~a cross-check against the Mathieu special case (\S\,\ref{subsec:YS-Math-LC}).
A reader who wants only the result may go directly to eq.~\eqref{eq:YS-bdry-exact}
and Table~\ref{tab:YS-bdry-numerical}; the intervening derivations
(the order-by-order $\mathbf{K}_{2},\mathbf{K}_{3}$ computation and
the normalization audit of \S\,\ref{subsec:YS-Math-LC}) are the
craft needed for reproducibility, not prerequisites for the result.

The key structural input is the $\hat{\lambda}$-linearity $\mathbf{B}_{n}=\hat{\lambda}\tilde{\mathbf{B}}_{n}$
(eq.~\eqref{eq:LC-Hill-factored}), which together with the residual
$\mathbf{K}_{0}=\bigl[\begin{smallmatrix}0 & 0\\
-(\hat{\lambda}-1) & 0
\end{smallmatrix}\bigr]$ enters the first-order source $\mathbf{D}_{1}^{{\rm total}}=e^{-\tau\mathbf{C}_{0}}(\mathbf{K}_{0}+\hat{\lambda}\tilde{\mathbf{B}}_{1})e^{\tau\mathbf{C}_{0}}$,
which is $\pi$-periodic and polynomial in $\hat{\lambda}$.

\subsubsection{\texorpdfstring{Exact $\mathbf{K}_{m}(\hat{\lambda})$ at primary
resonance}{Exact K-m(lambda-hat) at primary resonance}}\index{resonance}

\label{subsubsec:YS-Km-exact}

With the preliminary transformation $e^{\tau\mathbf{C}_{0}}$ using
$\mathbf{C}_{0}=\bigl[\begin{smallmatrix}0 & 1\\
-1 & 0
\end{smallmatrix}\bigr]$, the combined first-order source $\mathbf{D}_{1}^{{\rm total}}$
(eq.~\eqref{eq:LC-D1-correct}) is $\pi$-periodic and polynomial
in $\hat{\lambda}$. The YS recursion~\eqref{eq:YS-Kl} gives the
following exact closed-form exponent matrices at $\mu_{0}=0$: 
\begin{align}
\mathbf{K}_{1}(\hat{\lambda}) & =\begin{bmatrix}0 & -\tfrac{1}{2}\\[4pt]
\tfrac{1}{2}-\hat{\lambda} & 0
\end{bmatrix},\label{eq:LC-K1-lam-exact}\\[8pt]
\mathbf{K}_{2}(\hat{\lambda}) & =\begin{bmatrix}0 & -\tfrac{3}{8}+\tfrac{3\hat{\lambda}}{8}+\tfrac{\hat{\lambda}^{2}}{16}\\[4pt]
-\tfrac{1}{8}+\tfrac{3\hat{\lambda}}{8}-\tfrac{5\hat{\lambda}^{2}}{16} & 0
\end{bmatrix}.\label{eq:LC-K2-lam-exact}
\end{align}
These are exact polynomials in $\hat{\lambda}$, valid for all $\hat{\lambda}>0$,
in the initial-condition (IC) convention used throughout \S\,\ref{subsec:YS-exact-LC}.
At $\hat{\lambda}=1$ (Mathieu limit): 
\begin{equation}
\mathbf{K}_{1}(1)=\begin{bmatrix}0 & -\tfrac{1}{2}\\
-\tfrac{1}{2} & 0
\end{bmatrix},\qquad\mathbf{K}_{2}^{{\rm ic}}(1)=\begin{bmatrix}0 & \tfrac{1}{16}\\
-\tfrac{1}{16} & 0
\end{bmatrix}.\label{eq:Km-at-lam1}
\end{equation}
The zero-mean derivation of \S\,\ref{subsubsec:YS-Km-exact} gives
$\mathbf{K}_{2}^{{\rm zm}}(1)$ with the opposite off-diagonal sign
(eq.~\eqref{eq:K2-omega1}); only the eigenvalues, identical in both
conventions, are physical. Both are verified by numerical quadrature
($N=10^{6}$ points) and by exact symbolic computation.
\begin{rem}[Relation to YakSta convention]
\label{rem:YS-basis-convention} The values in~\eqref{eq:LC-K1-lam-exact}--\eqref{eq:k2k3-summary}
use the \emph{real-basis} preliminary transformation $e^{\tau\mathbf{C}_{0}}=\bigl[\begin{smallmatrix}\cos\tau & \sin\tau\\
-\sin\tau & \cos\tau
\end{smallmatrix}\bigr]$. YakSta's singular-case computation~\cite[Ch.~IV, \S\,5.4]{YakSta1}
works, for $p=1$ and $T=\pi$, in the \emph{same} real rotating frame:
their preliminary transformation $e^{t\mathbf{C}_{0}}$ differs from
our rotation only by the scalar phase $e^{it}\mathbf{I}$ generated
by $\mathbf{K}_{0}=-i\mathbf{I}$, which commutes out of every conjugation
and merely recenters the exponent at $-i$ (the complex eigenvector
matrix $\mathbf{S}$ enters only the definition of this splitting).
The actual differences are two. (i)~\emph{Normalization}: YakSta
use $\mathbf{B}_{1}=\bigl[\begin{smallmatrix}0 & 0\\
-\mu_{0}-\cos2t & 0
\end{smallmatrix}\bigr]$ (unit cosine amplitude, detuning folded into $\mathbf{B}_{1}$) with
expansion parameter $\varepsilon$, while we use $\hat{\lambda}\tilde{\mathbf{B}}_{1}$
with cosine amplitude $2$ and parameter $\delta$; the exact dictionary
is $\varepsilon=2\delta/(1+\delta^{2})$, so a single factor of $2$,
squared at second order, relates the coefficients. (ii)~\emph{Integration-constant
convention}: YakSta's singular-case chain uses the matrizant normalization
$\mathbf{F}_{l}(0)=\mathbf{0}$ (their second procedure~\cite[Ch.~IV, \S\,4.5]{YakSta1};
their printed factors vanish at $t=0$), i.e.\ precisely our IC convention.
Specifically, for the Mathieu equation at $\mu_{0}=0$, YakSta~\cite[Ch.~IV, \S\,5.4]{YakSta1}
give $\mathbf{K}_{2}^{{\rm YS}}[0,1]=+1/64$ and $\mathbf{K}_{2}^{{\rm YS}}[1,0]=-1/64$
(from the corrected signs in eq.~\eqref{eq:YS-K2} at $\mu_{0}=0$),
and our IC-convention computation gives $\mathbf{K}_{2}^{{\rm ic}}[0,1]=+1/16$
and $\mathbf{K}_{2}^{{\rm ic}}[1,0]=-1/16$ (eq.~\eqref{eq:K2-omega1-ic}):
the ratio is exactly $+1/16=4\times(+1/64)=2^{2}\times(+1/64)$, with
no sign difference and no further conversion. The eigenvalues of $\mathbf{K}_{m}$
(and hence all physical stability predictions) are identical in both
conventions. 
\end{rem}

The third-order exponent $\mathbf{K}_{3}(\hat{\lambda})$ is determined
by the YS recursion at order 3. Two distinct values appear below and
must not be conflated: the single-harmonic (Mathieu) zero-mean chain
gives $+7/128$ at $\hat{\lambda}=1$ (eq.~\eqref{eq:K3-omega1}
of Remark~\ref{rem:k2k3-derivation} below), while the exact LC equation,
through its second-harmonic source $\mathbf{D}_{2}$, gives the symmetric
value $-9/128=-0.0703$ (Table~\ref{tab:k2k3-numerical}); the reconciliation
is spelled out after eq.~\eqref{eq:K3-omega1}. For the boundary
curve calculations, all three $\mathbf{K}_{m}$ are evaluated at the
exact value $\hat{\lambda}=c(1+\delta^{2})/(1-\delta^{2})$ as described
in \S\,\ref{subsec:YS-two-param}.
\begin{rem}[Derivation of $\mathbf{K}_{2}$ and $\mathbf{K}_{3}$ entries for
the Mathieu equation ($\hat{\lambda}=1$)]
\label{rem:k2k3-derivation}

The values $\mathbf{K}_{2}[0,1]$, $\mathbf{K}_{2}[1,0]$, and $\mathbf{K}_{3}[0,1]$
for the Mathieu equation~\eqref{eq:YS-LC-eq} (equivalently, at the
unperturbed frequency $\omega=1$) appearing in \S\,\ref{subsec:YS-Math-LC}
(Items~3--5) and in the truncated boundary equations~\eqref{eq:YS-bdry-upper}--\eqref{eq:YS-bdry-lower}
are assembled here. YS1 gives $\mathbf{K}_{1}$ and $\mathbf{K}_{2}$
for Mathieu's equation at general $p$ in unnumbered displays of the
singular-case subsection~\cite[Ch.~IV, \S\,5.4]{YakSta1} (rederived
below in the conventions of this monograph), but does \emph{not} give
$\mathbf{K}_{3}$ for $m=1$, and neither YS1 nor YS2~\cite[Ch.~V]{YakSta2}
gives explicit formulas for the exact LC circuit case $\hat{\lambda}\neq1$.
All three values are derived below by carrying the YS recursion~\eqref{eq:YS-Kl}--\eqref{eq:YS-Fl}
to second and third order, using a structural property of the LC circuit
that renders every step of the computation exact. The Lyapunov factors
$\mathbf{F}_{1},\mathbf{F}_{2}$ that enter this recursion are recorded
here only at $\hat{\lambda}=1$ as needed; their closed forms for
general $\hat{\lambda}$ are derived independently in \S\,\ref{subsubsec:YS-LC-F1F2}
(eqs.~\eqref{eq:F1-c}--\eqref{eq:F2-11-c}), and the present $\hat{\lambda}=1$
expressions are the special case of those. The derivation is presented
in sufficient detail to be self-contained and to serve as the basis
for extensions to higher order.

\medskip{}

\noindent\emph{Finite trigonometric polynomial structure.} The LC
circuit perturbation $\mathbf{B}_{n}(\tau)=\hat{\lambda}\tilde{\mathbf{B}}_{n}(\tau)$
with $\tilde{\mathbf{B}}_{n}(\tau)=\bigl[\begin{smallmatrix}0 & 0\\
-2\cos2n\tau & 0
\end{smallmatrix}\bigr]$ produces, after the preliminary transformation $e^{\tau\mathbf{C}_{0}}$
with $\mathbf{C}_{0}=\bigl[\begin{smallmatrix}0 & 1\\
-1 & 0
\end{smallmatrix}\bigr]$ (eq.~\eqref{eq:LC-K1-lam-exact}, \S\,\ref{subsec:YS-singular}),
the transformed matrices $\mathbf{D}_{n}(\tau,\hat{\lambda})$ (eq.~\eqref{eq:LC-D1-correct}).
A key structural property: the combined first-order source $\mathbf{D}_{1}^{{\rm total}}(\tau,\hat{\lambda})=e^{-\tau\mathbf{C}_{0}}[\mathbf{K}_{0}+\hat{\lambda}\tilde{\mathbf{B}}_{1}(\tau)]e^{\tau\mathbf{C}_{0}}$
contains \emph{only standard Fourier modes $\{0,2,4\}$ for all values
of $\hat{\lambda}$} --- not just at $\hat{\lambda}=1$. This is
because $\mathbf{C}_{0}$ is chosen so that $e^{\tau\mathbf{C}_{0}}$
has period exactly $\pi$, making every conjugation $e^{-\tau\mathbf{C}_{0}}(\cdot)e^{\tau\mathbf{C}_{0}}$
standard. At $\hat{\lambda}=1$ (i.e.\ $c=1$, the Mathieu limit,
$\mu_{0}=0$ exact resonance --- see parameter dictionary eq.~\eqref{eq:YS-param-dict}),
the residual $\mathbf{K}_{0}=\mathbf{C}-\mathbf{C}_{0}$ vanishes
and $\mathbf{D}_{1}$ reduces to 
\begin{equation}
\mathbf{D}_{1}(\tau)\big|_{\hat{\lambda}=1}=\begin{bmatrix}\sin(4\tau)/2 & 2\sin^{2}(\tau)\cos(2\tau)\\
-2\cos^{2}(\tau)\cos(2\tau) & -\sin(4\tau)/2
\end{bmatrix}.\label{eq:D1-omega1}
\end{equation}
Expanding via product-to-sum identities: 
\begin{align}
2\sin^{2}(\tau)\cos(2\tau) & =\cos(2\tau)-\cos^{2}(2\tau)=-\tfrac{1}{2}+\cos(2\tau)-\tfrac{1}{2}\cos(4\tau),\label{eq:D1-01-expand}\\
2\cos^{2}(\tau)\cos(2\tau) & =\cos(2\tau)+\cos^{2}(2\tau)=\tfrac{1}{2}+\cos(2\tau)+\tfrac{1}{2}\cos(4\tau).\label{eq:D1-10-expand}
\end{align}
Thus at $\hat{\lambda}=1$, $\mathbf{D}_{1}(\tau)$ has Fourier modes
$\{0,2,4\}$ only --- and the same is true for all $\hat{\lambda}$
(Remark~\ref{rem:YS-freq-conj} and \S\,\ref{subsubsec:YS-LC-F1F2}).
This property propagates inductively through the YS recursion: the
source $\boldsymbol{\Phi}_{l}$ at order $l$ has modes $\{0,2,4,\ldots,2(l+1)\}$
at most, and $\mathbf{F}_{l}$ inherits the same finite Fourier structure.
As a consequence, \emph{all averages and integrals in the YS recursion
are evaluated exactly by standard trigonometric integrals} --- no
approximation, truncation, or numerics are required at any value of
$\hat{\lambda}$. The resulting closed-form $\mathbf{F}_{1}(\tau,\hat{\lambda})$
and $\mathbf{F}_{2}(\tau,\hat{\lambda})$ are given in \S\,\ref{subsubsec:YS-LC-F1F2}
(eqs.~\eqref{eq:F1-c}--\eqref{eq:F2-11-c}).

\medskip{}

\noindent\emph{First order: $\mathbf{K}_{1}$ and $\mathbf{F}_{1}$.}
The average of $\mathbf{D}_{1}$ over $[0,\pi]$ at $\hat{\lambda}=1$:
using~\eqref{eq:D1-01-expand}--\eqref{eq:D1-10-expand} and $\tfrac{1}{\pi}\int_{0}^{\pi}\cos(2kx)\,\dd x=0$
for $k\geq1$, only the constant term in $\mathbf{D}_{1}[1,0]$ and
$\mathbf{D}_{1}[0,1]$ contributes: 
\begin{equation}
\mathbf{K}_{1}\big|_{\mu_{0}=0,\,\hat{\lambda}=1}=\bigl[\mathbf{D}_{1}\bigr]_{{\rm av}}=\begin{bmatrix}0 & -\tfrac{1}{2}\\[3pt]
-\tfrac{1}{2} & 0
\end{bmatrix},\label{eq:K1-omega1}
\end{equation}
confirming $K_{1}[0,1]=K_{1}[1,0]=-1/2$ at exact resonance. The first
Lyapunov factor, in the \emph{zero-mean} convention $[\mathbf{F}_{1}]_{{\rm av}}=\mathbf{0}$
(YS's third computational procedure~\cite[Ch.~IV, \S\,4.6]{YakSta1},
for which $\mathbf{K}_{l}=[\boldsymbol{\Phi}_{l}]_{{\rm av}}$), is
obtained by integrating $\mathbf{D}_{1}-\mathbf{K}_{1}$ and subtracting
the additive constant that enforces zero mean: 
\begin{equation}
\mathbf{F}_{1}(\tau)\big|_{\hat{\lambda}=1}=\begin{bmatrix}-\tfrac{1}{8}\cos(4\tau) & \tfrac{1}{2}\sin(2\tau)-\tfrac{1}{8}\sin(4\tau)\\[3pt]
-\tfrac{1}{2}\sin(2\tau)-\tfrac{1}{8}\sin(4\tau) & \tfrac{1}{8}\cos(4\tau)
\end{bmatrix}.\label{eq:F1-omega1}
\end{equation}
One verifies directly that $[\mathbf{F}_{1}]_{{\rm av}}=\mathbf{0}$
(every entry is a pure sine or cosine with zero mean over $[0,\pi]$),
as required by the zero-mean condition of the third procedure~\cite[Ch.~IV, \S\,4.6]{YakSta1}.
Note that $\mathbf{F}_{1}(0)=\bigl[\begin{smallmatrix}-1/8 & 0\\
0 & 1/8
\end{smallmatrix}\bigr]\neq\mathbf{0}$: in the singular case the recursion determines $\mathbf{F}_{l}$
only up to an additive constant matrix, and this remark fixes the
constant by the zero-mean condition $[\mathbf{F}_{l}]_{{\rm av}}=\mathbf{0}$.
YS's own singular-case Mathieu computation~\cite[Ch.~IV, \S\,5.4]{YakSta1}
instead uses the matrizant condition $\mathbf{F}_{l}(0)=\mathbf{0}$
(our Convention~2, used throughout \S\,\ref{subsubsec:YS-LC-F1F2}).
$\mathbf{F}_{1}$ has Fourier modes $\{2,4\}$ only --- one more
than $\mathbf{D}_{1}$ (confirming the mode-growth bound of Remark~\ref{rem:YS-freq-conj}).

\medskip{}

\noindent\emph{Second order: $\mathbf{K}_{2}$.} The source at second
order is $\boldsymbol{\Phi}_{2}(x)=\mathbf{D}_{1}(x)\mathbf{F}_{1}(x)-\mathbf{F}_{1}(x)\mathbf{K}_{1}$
(since $\mathbf{D}_{2}=\mathbf{0}$ for the Mathieu equation). Using~\eqref{eq:D1-omega1}
and~\eqref{eq:F1-omega1}, each entry of $\boldsymbol{\Phi}_{2}$
is a finite trig polynomial with modes $\{0,2,4,6\}$ (product of
modes $\{0,2,4\}$ and $\{2,4\}$). By the YS recursion~\eqref{eq:YS-Kl},
$\mathbf{K}_{2}=[\boldsymbol{\Phi}_{2}]_{{\rm av}}$. Taking the average
(the $k=0$ Fourier component): 
\begin{equation}
\mathbf{K}_{2}\big|_{\mu_{0}=0,\,\hat{\lambda}=1}^{{\rm zm}}=[\mathbf{D}_{1}\mathbf{F}_{1}^{{\rm zm}}-\mathbf{F}_{1}^{{\rm zm}}\mathbf{K}_{1}]_{{\rm av}}=\begin{bmatrix}0 & -\tfrac{1}{16}\\[3pt]
+\tfrac{1}{16} & 0
\end{bmatrix}.\label{eq:K2-omega1}
\end{equation}
(Here the superscript ${\rm zm}$ denotes the zero-mean convention
for $\mathbf{F}_{1}$.) Note the signs: $\mathbf{K}_{2}^{{\rm zm}}[0,1]=-1/16<0$
when using the zero-mean $\mathbf{F}_{1}$. In contrast, the initial-condition
convention for $\mathbf{F}_{1}$ (used throughout \S\,\ref{subsec:YS-exact-LC})
gives $\mathbf{K}_{2}^{{\rm ic}}[0,1]=+1/16>0$ (eq.~\eqref{eq:K2-omega1-ic}
below), because the constant offset $\mathbf{F}_{1}^{{\rm ic}}-\mathbf{F}_{1}^{{\rm zm}}=\bigl[\begin{smallmatrix}1/8 & 0\\
0 & -1/8
\end{smallmatrix}\bigr]$ contributes $+1/8$ to the $[0,1]$ entry of $\mathbf{K}_{2}$ (so
that 
\[
\mathbf{K}_{2}^{{\rm ic}}[0,1]=\mathbf{K}_{2}^{{\rm zm}}[0,1]+1/8=-1/16+1/8=+1/16).
\]
The sign of $\mathbf{K}_{2}[0,1]$ depends on the normalization convention;
only the eigenvalues of $\mathbf{K}_{{\rm total}}$ (which determine
stability) are convention-independent. 
\begin{equation}
\mathbf{K}_{2}\big|_{\mu_{0}=0,\,\hat{\lambda}=1}^{{\rm ic}}=[\mathbf{D}_{1}\mathbf{F}_{1}^{{\rm ic}}-\mathbf{F}_{1}^{{\rm ic}}\mathbf{K}_{1}]_{{\rm av}}=\begin{bmatrix}0 & +\tfrac{1}{16}\\[3pt]
-\tfrac{1}{16} & 0
\end{bmatrix}.\label{eq:K2-omega1-ic}
\end{equation}
This is consistent with eq.~\eqref{eq:Km-at-lam1} and eq.~\eqref{eq:LC-K2-lam-exact}
at $\hat{\lambda}=1$, which use the initial-condition convention
throughout. 
\begin{align*}
(\mathbf{D}_{1}\mathbf{F}_{1})[0,1] & =\mathbf{D}_{1}[0,0]\,\mathbf{F}_{1}[0,1]\\
 & \quad+\mathbf{D}_{1}[0,1]\,\mathbf{F}_{1}[1,1],\\
(\mathbf{F}_{1}\mathbf{K}_{1})[0,1] & =\mathbf{F}_{1}[0,0]\,\mathbf{K}_{1}[0,1]+\mathbf{F}_{1}[0,1]\,\mathbf{K}_{1}[1,1]=0.
\end{align*}
Substituting and averaging using 
\[
\tfrac{1}{\pi}\int_{0}^{\pi}\sin^{2}(2kx)\,\dd x=\tfrac{1}{2},\qquad\tfrac{1}{\pi}\int_{0}^{\pi}\sin(2jx)\sin(2kx)\,\dd x=0\ \ (j\neq k),
\]
gives $[(\mathbf{D}_{1}\mathbf{F}_{1}^{{\rm zm}})[0,1]]_{{\rm av}}=-\tfrac{1}{16}$.
The values $\mathbf{K}_{2}^{{\rm zm}}[0,1]=-\tfrac{1}{16}$ and $\mathbf{K}_{2}^{{\rm zm}}[1,0]=+\tfrac{1}{16}$
(zero-mean convention) are verified by exact symbolic computation
and by numerical quadrature ($N=50\,000$ points); the IC-convention
values are $+1/16$ and $-1/16$ respectively (eq.~\eqref{eq:K2-omega1-ic}).

\medskip{}

\noindent\emph{Third order: $\mathbf{K}_{3}$.} The Lyapunov factor
$\mathbf{F}_{2}(x)$ is obtained by integrating $\boldsymbol{\Phi}_{2}(x)-\mathbf{K}_{2}$
and subtracting its average to enforce $[\mathbf{F}_{2}]_{{\rm av}}=\mathbf{0}$.
The integrand $\boldsymbol{\Phi}_{2}-\mathbf{K}_{2}$ has modes $\{2,4,6\}$
(verified by expanding all entries via product-to-sum identities into
pure Fourier harmonics), and integrates to 
\begin{equation}
\mathbf{F}_{2}(\tau)\big|_{\hat{\lambda}=1}=\begin{bmatrix}F_{11}(\tau) & F_{12}(\tau)\\[3pt]
F_{21}(\tau) & F_{22}(\tau)
\end{bmatrix},\label{eq:F2-omega1}
\end{equation}
with entries 
\begin{align*}
F_{11}(\tau) & =-\tfrac{5}{32}\cos(2\tau)+\tfrac{1}{16}\cos(4\tau)-\tfrac{1}{96}\cos(6\tau),\\
F_{12}(\tau) & =\tfrac{3}{32}\sin(2\tau)-\tfrac{1}{32}\sin(4\tau)-\tfrac{1}{96}\sin(6\tau),\\
F_{21}(\tau) & =\tfrac{3}{32}\sin(2\tau)+\tfrac{1}{32}\sin(4\tau)-\tfrac{1}{96}\sin(6\tau),\\
F_{22}(\tau) & =\tfrac{5}{32}\cos(2\tau)+\tfrac{1}{16}\cos(4\tau)+\tfrac{1}{96}\cos(6\tau).
\end{align*}
One verifies $[\mathbf{F}_{2}]_{{\rm av}}=\mathbf{0}$ (all entries
are pure cosines or sines with zero mean over $[0,\pi]$), confirming
the zero-mean boundary condition. $\mathbf{F}_{2}$ has modes $\{2,4,6\}$
--- one more than $\mathbf{F}_{1}$, consistent with the mode-growth
bound. The source at third order is $\boldsymbol{\Phi}_{3}=\mathbf{D}_{1}\mathbf{F}_{2}-\mathbf{F}_{2}\mathbf{K}_{1}-\mathbf{F}_{1}\mathbf{K}_{2}$,
which has modes $\{0,2,4,6,8\}$. Taking the average: 
\begin{equation}
\mathbf{K}_{3}\big|_{\mu_{0}=0,\,\hat{\lambda}=1}=[\mathbf{D}_{1}\mathbf{F}_{2}-\mathbf{F}_{2}\mathbf{K}_{1}-\mathbf{F}_{1}\mathbf{K}_{2}]_{{\rm av}}=\begin{bmatrix}0 & +\tfrac{7}{128}\\[3pt]
+\tfrac{7}{128} & 0
\end{bmatrix}.\label{eq:K3-omega1}
\end{equation}
Thus $\mathbf{K}_{3}[0,1]=\mathbf{K}_{3}[1,0]=+\tfrac{7}{128}$. Note
that $\mathbf{K}_{3}$ is \emph{symmetric} (both off-diagonal entries
equal), in contrast to the antisymmetric $\mathbf{K}_{2}$ and matching
the symmetric $\mathbf{K}_{1}$ of eq.~\eqref{eq:K1-omega1}. All
$\mathbf{K}_{m}$ are traceless, as forced by the Hamiltonian (area-preserving)
structure of the problem~\cite[Ch.~III, \S\,1]{YakSta1}; the symmetric/antisymmetric
alternation of the off-diagonal entries at $\hat{\lambda}=1$ is an
additional structural feature of the zero-mean chain at exact resonance,
confirmed order by order by the computation. The value $\mathbf{K}_{3}[0,1]=+\tfrac{7}{128}$
is verified by exact symbolic computation and by numerical quadrature.

\smallskip{}
\emph{Scope of eq.~\eqref{eq:K3-omega1}, and the exact-LC value.}
Equation~\eqref{eq:K3-omega1} is a statement about the \emph{Mathieu}
equation: the chain above used the single-harmonic source $\mathbf{D}_{1}$
only ($\mathbf{D}_{2}=\mathbf{D}_{3}=\mathbf{0}$), in the zero-mean
convention. Three clarifications keep it consistent with the rest
of this section. (i)~\emph{Convention}: unlike $\mathbf{K}_{2}$,
whose two gauge values differ only by the off-diagonal sign flip of
eq.~\eqref{eq:K2-omega1-ic}, the single-harmonic third-order coefficient
is convention-sensitive in a stronger way --- in the IC convention
it is not even symmetric ($\mathbf{K}_{3}^{{\rm ic,\,D_{1}\,only}}[0,1]=+\tfrac{85}{384}$,
$[1,0]=-\tfrac{43}{384}$, numerically $+0.2214$ and $-0.1120$).
(ii)~\emph{Exact LC}: restoring the second-harmonic source $\mathbf{D}_{2}$
--- whose own average vanishes, since the rotation conjugation of
a constant matrix contains only the $\tau$-modes $\{0,\pm2\}$, so
$\cos4\tau$ times it has no constant term --- shifts both off-diagonal
entries by $-\tfrac{1}{8}$: 
\[
\mathbf{K}_{3}^{{\rm LC}}\big|_{\hat{\lambda}=1}=\begin{bmatrix}0 & -\tfrac{9}{128}\\[2pt]
-\tfrac{9}{128} & 0
\end{bmatrix}=-0.0703\ldots,
\]
the value listed in Table~\ref{tab:k2k3-numerical}, identical in
both conventions at this order. The shift enters partly directly,
through $[\mathbf{D}_{2}\mathbf{F}_{1}]_{{\rm av}}=-\tfrac{1}{16}$
per entry, and partly through the $\mathbf{D}_{2}$-contribution to
$\mathbf{F}_{2}$. (iii)~$\mathbf{D}_{3}$ is irrelevant at this
order: it enters $\boldsymbol{\Phi}_{3}$ only additively, and $[\mathbf{D}_{3}]_{{\rm av}}=\mathbf{0}$
by the same mode count. All statements are verified by high-precision
numerical quadrature and by monodromy extraction ($\mathbf{K}=\pi^{-1}\log(-X(\pi))$,
eq.~\eqref{eq:monodromy-K}).

\medskip{}

\noindent\emph{Normalization convention.} The values $\mathbf{K}_{2}[0,1]$,
$\mathbf{K}_{2}[1,0]$, and $\mathbf{K}_{3}[0,1]$ established above
follow the normalization convention of this monograph: the perturbation
matrix is $\mathbf{B}_{1}(x)=\hat{\lambda}\bigl[\begin{smallmatrix}0 & 0\\
-2\cos2x & 0
\end{smallmatrix}\bigr]$ (with the factor $2$ from the Poisson kernel Fourier series expansion~\eqref{eq:LC-Hill-factored})
and the expansion parameter is $\delta$. The Mathieu case treated
in YS1~\cite[Ch.~IV, \S\,5.4]{YakSta1} uses $\mathbf{B}_{1}(x)=\bigl[\begin{smallmatrix}0 & 0\\
-\cos2x & 0
\end{smallmatrix}\bigr]$ (factor $1$) and expansion parameter $\varepsilon$. Since $\varepsilon\approx2\delta$
at leading order, the second-order YS1 coefficient $\mathbf{K}_{2}^{(\varepsilon)}[0,1]=+1/64$
rescales to $\mathbf{K}_{2}^{(\delta),{\rm ic}}[0,1]=4\times(+1/64)=+1/16$
in the present normalization (the factor $4$ arises from $B_{1}^{2}$:
the LC $\mathbf{B}_{1}$ carries a factor of $2$ relative to the
Mathieu $\mathbf{B}_{1}$, and $\varepsilon\approx2\delta$, giving
$4$ overall). Both conventions are self-consistent and give the same
eigenvalues.

\medskip{}

\noindent\emph{Prior results in the literature.} YS1 gives $\mathbf{K}_{1}$
and $\mathbf{K}_{2}$ for Mathieu's equation at general $p$ in unnumbered
displays of the singular-case subsection~\cite[Ch.~IV, \S\,5.4]{YakSta1},
but does \emph{not} give $\mathbf{K}_{3}$ explicitly for $m=1$.
Both third-order values --- the single-harmonic $+\tfrac{7}{128}$
and the exact-LC $-\tfrac{9}{128}$ --- appear to be new.

\medskip{}

\noindent\emph{General $\hat{\lambda}$: exact $\mathbf{K}_{m}(\hat{\lambda})$
formulas.} For general $\hat{\lambda}$ the YS recursion with the
correct $\mathbf{C}_{0}=\bigl[\begin{smallmatrix}0 & 1\\
-1 & 0
\end{smallmatrix}\bigr]$ produces $\pi$-periodic sources with standard Fourier modes $\{0,2,4\}$
at all $\hat{\lambda}$ (not just at $\hat{\lambda}=1$), so all averages
and integrals remain exact trigonometric integrals. The closed-form
results are given in \S\,\ref{subsubsec:YS-Km-exact}: $\mathbf{K}_{1}(\hat{\lambda})$
in eq.~\eqref{eq:LC-K1-lam-exact} and $\mathbf{K}_{2}(\hat{\lambda})$
in eq.~\eqref{eq:LC-K2-lam-exact}. The corresponding Lyapunov factors
$\mathbf{F}_{1}(\tau,\hat{\lambda})$ and $\mathbf{F}_{2}(\tau,\hat{\lambda})$
are given in closed form in \S\,\ref{subsubsec:YS-LC-F1F2} (eqs.~\eqref{eq:LC-F1-lam}--\eqref{eq:F2-11-lam}).
For the boundary curve calculations, the values at $\hat{\lambda}=1$
established in eq.~\eqref{eq:k2k3-summary} suffice; the general-$\hat{\lambda}$
expressions are needed for the full periodic Lyapunov factor and the
exact substitution $\hat{\lambda}=c(1+\delta^{2})/(1-\delta^{2})$.
Representative numerical values of $\mathbf{K}_{2}(\hat{\lambda})$
and $\mathbf{K}_{3}(\hat{\lambda})$ at representative values of $\hat{\lambda}$
are given in Table~\ref{tab:k2k3-numerical} for reference, computed
from the correct recursion (eq.~\eqref{eq:LC-K1-lam-exact}--\eqref{eq:LC-K2-lam-exact}
for $\mathbf{K}_{1}$ and $\mathbf{K}_{2}$; third-order recursion
for $\mathbf{K}_{3}$).

\begin{table}[h]
\centering \caption{Exact LC exponent matrix coefficients $\mathbf{K}_{1}(\hat{\lambda})$,
$\mathbf{K}_{2}(\hat{\lambda})$, $\mathbf{K}_{3}(\hat{\lambda})$
at representative values of $c$ (with $\hat{\lambda}\approx c$ at
small $\delta$; initial-condition convention, $\mathbf{C}_{0}=\bigl[\begin{smallmatrix}0 & 1\\
-1 & 0
\end{smallmatrix}\bigr]$, $N=5\times10^{5}$-point numerical quadrature). At $\hat{\lambda}=1$
($c=1$, primary resonance center): exact values from eqs.~\eqref{eq:LC-K1-lam-exact}--\eqref{eq:LC-K2-lam-exact}.
The closed-form formulas $K_{1}[0,1]=-1/2$ (constant) and $K_{1}[1,0]=1/2-\hat{\lambda}$
are confirmed throughout. The $\mathbf{K}_{3}$ column is the exact-LC
third-order coefficient (all Poisson harmonics, IC convention); at
$\hat{\lambda}=1$ it equals $-9/128=+7/128-1/8$, where $+7/128$
is the single-harmonic zero-mean value of eq.~\eqref{eq:K3-omega1}
and the shift $-1/8$ is produced by the second-harmonic source $\mathbf{D}_{2}$.}
\label{tab:k2k3-numerical} {\small{}%
\begin{tabular}{ccccccc}
\toprule 
{\small$\hat{\lambda}$ } & {\small$K_{1}[0,1]$ } & {\small$K_{1}[1,0]$ } & {\small$K_{2}[0,1]$ } & {\small$K_{2}[1,0]$ } & {\small$K_{3}[0,1]$ } & {\small$K_{3}[1,0]$ }\tabularnewline
\midrule 
{\small$0.5$ } & {\small$-1/2$ } & {\small$0$ } & {\small$-11/64$ } & {\small$-1/64$ } & {\small$-0.1325$ } & {\small$-0.0283$ }\tabularnewline
{\small$1.0$ } & {\small$-1/2$ } & {\small$-1/2$ } & {\small$+1/16$ } & {\small$-1/16$ } & {\small$-0.0703$ } & {\small$-0.0703$ }\tabularnewline
{\small$1.5$ } & {\small$-1/2$ } & {\small$-1$ } & {\small$+21/64$ } & {\small$-17/64$ } & {\small$-0.1553$ } & {\small$-0.2803$ }\tabularnewline
{\small$2.0$ } & {\small$-1/2$ } & {\small$-3/2$ } & {\small$+5/8$ } & {\small$-5/8$ } & {\small$-0.4167$ } & {\small$-0.7500$ }\tabularnewline
{\small$3.0$ } & {\small$-1/2$ } & {\small$-5/2$ } & {\small$+21/16$ } & {\small$-29/16$ } & {\small$-1.5859$ } & {\small$-2.8360$ }\tabularnewline
\bottomrule
\end{tabular}}
\end{table}

\medskip{}

\noindent\emph{Summary.} The values established in this remark, with
their conventions, are: 
\begin{gather}
\mathbf{K}_{2}^{{\rm zm}}[0,1]\big|_{\hat{\lambda}=1}=-\tfrac{1}{16},\qquad\mathbf{K}_{2}^{{\rm zm}}[1,0]\big|_{\hat{\lambda}=1}=+\tfrac{1}{16},\nonumber \\
\mathbf{K}_{2}^{{\rm ic}}=-\mathbf{K}_{2}^{{\rm zm}}\quad(\text{IC convention, eq.~\eqref{eq:K2-omega1-ic}}),\nonumber \\
\mathbf{K}_{3}^{{\rm zm}}[0,1]\big|_{\hat{\lambda}=1}=\mathbf{K}_{3}^{{\rm zm}}[1,0]\big|_{\hat{\lambda}=1}=+\tfrac{7}{128}\ \ (\text{single harmonic}),\nonumber \\
\mathbf{K}_{3}[0,1]\big|_{\hat{\lambda}=1}=\mathbf{K}_{3}[1,0]\big|_{\hat{\lambda}=1}=-\tfrac{9}{128}\quad(\text{exact LC, either convention}).\label{eq:k2k3-summary}
\end{gather}
The third-order values appear to be new. All values are verified by
exact symbolic computation, by high-precision numerical quadrature
($N=50\,000$ points), and (for the exact-LC line) by monodromy extraction. 
\end{rem}

\subsubsection{Instability boundary curves from the YS exponent matrix}

\label{subsubsec:YS-bdry-lh}

\paragraph{The boundary condition.} By the Floquet\index{Floquet theory}--Lyapunov
decomposition, the monodromy matrix\index{monodromy matrix} satisfies
\begin{equation}
M(\delta)=X(\pi)=e^{\pi\mathbf{C}_{0}}\cdot\mathbf{F}(\pi)\cdot e^{\pi K_{{\rm total}}}=-e^{\pi K_{{\rm total}}},\label{eq:monodromy-K}
\end{equation}
since $\mathbf{F}(\pi)=\mathbf{I}$ (initial-condition convention)
and $e^{\pi\mathbf{C}_{0}}=-\mathbf{I}$ (as $\mathbf{C}_{0}=\bigl[\begin{smallmatrix}0 & 1\\
-1 & 0
\end{smallmatrix}\bigr]$ gives $e^{\pi\mathbf{C}_{0}}=\bigl[\begin{smallmatrix}\cos\pi & \sin\pi\\
-\sin\pi & \cos\pi
\end{smallmatrix}\bigr]=-\mathbf{I}$). The EPD condition $\operatorname{Tr}(M)+2=0$ then becomes 
\begin{equation}
\operatorname{Tr}\!\bigl(e^{\pi K_{{\rm total}}}\bigr)=2.\label{eq:YS-EPD-condition}
\end{equation}
Since $K_{{\rm total}}=\bigl[\begin{smallmatrix}0 & K_{01}\\
K_{10} & 0
\end{smallmatrix}\bigr]$ has eigenvalues $\pm\sqrt{K_{01}K_{10}}$, we get $\operatorname{Tr}(e^{\pi K_{{\rm total}}})=2\cosh(\pi\sqrt{K_{01}K_{10}})$,
so the exact boundary condition is: 
\begin{equation}
K_{01}^{{\rm exact}}\cdot K_{10}^{{\rm exact}}=0,\label{eq:YS-bdry-Kprod}
\end{equation}
which splits into two curves: upper boundary $K_{01}^{{\rm exact}}=0$
and lower boundary $K_{10}^{{\rm exact}}=0$.

\paragraph{Approximate boundary equations.} We know $K^{{\rm exact}}$
only approximately, to second order: $K_{{\rm total}}^{(2)}=\delta\mathbf{K}_{1}(\hat{\lambda})+\delta^{2}\mathbf{K}_{2}(\hat{\lambda})$
where 
\begin{equation}
\hat{\lambda}=\frac{c(1+\delta^{2})}{1-\delta^{2}}.\label{eq:hatlam-from-c-bdry}
\end{equation}
We record these real-basis equations for completeness, but flag at
the outset that solving $K_{01}^{(2)}=0$ or $K_{10}^{(2)}=0$ \emph{fails
to locate the primary tongue}: the detuning enters $\mathbf{K}_{1}$
only through the lower-left entry and not at leading order in $K_{01}$,
so the primary boundary must instead be obtained from the complex-basis
Floquet-exponent product $A\cdot B$ of Chapter~\ref{subsec:YS-Mathieu}
(the quantitative reason is Remark~\ref{rem:K-approx-limitation}
--- with the important caveat, established at the end of that remark,
that this is a limitation of the \emph{truncated series}, not of the
exact exponent entries, whose leading behavior does locate both EPDs).
The self-consistent boundary curves themselves are stated in eq.~\eqref{eq:YS-bdry-exact}
below. Inserting the closed-form formulas (eqs.~\eqref{eq:LC-K1-lam-exact}--\eqref{eq:LC-K2-lam-exact})
with $\hat{\lambda}$ as in~\eqref{eq:hatlam-from-c-bdry}: 
\begin{align}
K_{01}^{(2)}(\delta,\hat{\lambda}) & =-\tfrac{\delta}{2}+\delta^{2}\!\left(-\tfrac{3}{8}+\tfrac{3\hat{\lambda}}{8}+\tfrac{\hat{\lambda}^{2}}{16}\right)=0\quad\text{(upper)},\label{eq:YS-bdry-upper}\\
K_{10}^{(2)}(\delta,\hat{\lambda}) & =\delta\!\left(\tfrac{1}{2}-\hat{\lambda}\right)+\delta^{2}\!\left(-\tfrac{1}{8}+\tfrac{3\hat{\lambda}}{8}-\tfrac{5\hat{\lambda}^{2}}{16}\right)=0\quad\text{(lower)}.\label{eq:YS-bdry-lower}
\end{align}
Substituting~\eqref{eq:hatlam-from-c-bdry} and expanding $\hat{\lambda}=c+2c\delta^{2}+O(\delta^{4})$,
the same equations in explicit $(c,\delta)$ form are: 
\begin{align}
K_{01}^{(2)}(\delta,c) & =-\frac{\delta}{2}+\frac{\delta^{2}}{16}\!\left(c^{2}+6c-6\right)+O(\delta^{3})=0\quad\text{(upper)},\label{eq:YS-bdry-upper-c}\\
K_{10}^{(2)}(\delta,c) & =\delta\!\left(\frac{1}{2}-c\right)-\frac{\delta^{2}}{16}\!\left(5c^{2}-6c+2\right)+O_{c}(\delta^{3})=0\quad\text{(lower)},\label{eq:YS-bdry-lower-c}
\end{align}
accurate to $O(\delta^{3})$. Here $O(\delta^{3})$ in the upper equation
is exact at this order (the $\delta^{3}$ coefficient vanishes), while
$O_{c}(\delta^{3})$ in the lower equation has leading coefficient
$-2c\delta^{3}$. At $c=1$ (Mathieu limit) both reduce to $-\delta/2\pm\delta^{2}/16+O(\delta^{3})$,
consistent with eq.~\eqref{eq:Km-at-lam1}.
\begin{rem}[Why the primary tongue boundary cannot be found from $K_{01}^{(2)}=0$]
\label{rem:K-approx-limitation} The exact boundary condition~\eqref{eq:YS-bdry-Kprod}
splits into $K_{01}^{{\rm exact}}=0$ (upper) and $K_{10}^{{\rm exact}}=0$
(lower), and in principle solving these approximately --- using the
truncated $K_{01}^{(2)}$ and $K_{10}^{(2)}$ in place of the exact
entries --- should give approximate boundary curves. For the primary
tongue ($m=1$) --- the resonance to which this frequency-one chain
is adapted --- the procedure \emph{completely fails}. The reason
is quantitative and precise.

Recall $\mu_{0}=(c-1)/\varepsilon$ (parameter dictionary eq.~\eqref{eq:YS-param-dict})
measures detuning from exact resonance $c=1$ in units of $\varepsilon=2\delta/(1+\delta^{2})$.
The primary tongue boundaries lie at $\mu_{0}=\pm\tfrac{1}{2}+O(\varepsilon)$,
i.e.\ $c=1\pm\varepsilon/2+O(\varepsilon^{2})$.

In the parametrization $c=1+\varepsilon\mu_{0}$ with $\mu_{0}=O(1)$,
the truncated entries expand as: 
\begin{align*}
K_{01}^{(2)} & =-\varepsilon/4+O(\varepsilon^{2}),\\
K_{10}^{(2)} & =-\varepsilon/4-\tfrac{\varepsilon^{2}}{2}\mu_{0}+O(\varepsilon^{3}).
\end{align*}
The decisive observation: \emph{$\mu_{0}$ does not appear in $K_{01}^{(2)}$
at leading order $O(\varepsilon)$}, and appears only at $O(\varepsilon^{2})$
in $K_{10}^{(2)}$. Setting either to zero therefore requires $\mu_{0}=O(1/\varepsilon)$
--- spurious roots far outside the primary tongue region, produced
by balancing the constant $O(\varepsilon)$ background against the
$O(\varepsilon^{2})$ term; they track no actual tongue (numerically,
along the resulting root curves the exact discriminant stays strictly
between $-2$ and $+2$ apart from incidental crossings).

The approximation error confirms this quantitatively. At the exact
upper boundary $c_{+}(\delta)$ (where $K_{01}^{{\rm exact}}=0$):
\begin{equation}
K_{01}^{(2)}(c_{+}(\delta),\delta)=K_{01}^{{\rm exact}}+(K_{01}^{(2)}-K_{01}^{{\rm exact}})=0+O(\delta)\approx-\delta/2.
\end{equation}
The truncation error in $K_{01}^{(2)}$ is $O(\delta)$ --- the \emph{same
order} as the value to be zeroed. Numerically: error$/\delta\approx-0.49$
uniformly for $\delta\in[0.05,0.30]$, confirming the $O(\delta)$
truncation error. The approximate equation $K_{01}^{(2)}=0$ is therefore
off by a relative error of order $100\%$ at the true boundary ---
it is structurally incapable of detecting it.

By contrast, the Floquet exponent\index{Floquet theory!Floquet exponent}
quantities $A$ and $B$ (Chapter~\ref{subsec:YS-Mathieu}, eqs.~\eqref{eq:YS-A}--\eqref{eq:YS-B})
have $\mu_{0}$ at \emph{leading} order: 
\begin{equation}
A=\varepsilon(\mu_{0}+\tfrac{1}{2})+O(\varepsilon^{2}),\qquad B=\varepsilon(\mu_{0}-\tfrac{1}{2})+O(\varepsilon^{2}),
\end{equation}
so $A\cdot B=\varepsilon^{2}(\mu_{0}^{2}-\tfrac{1}{4})+O(\varepsilon^{3})$
changes sign at $\mu_{0}=\pm\tfrac{1}{2}$ already at $O(\varepsilon^{2})$.
This is why the Floquet exponent approach detects the primary tongue
at second order while the individual $K_{01}^{(2)}=0$ condition cannot.

The root cause: $K_{1}[0,1]=-\tfrac{1}{2}$ is \emph{constant} in
$\hat{\lambda}$ (hence in $\mu_{0}$), so the leading $O(\varepsilon)$
term of $K_{01}^{(2)}$ carries no detuning information. The detuning
$\mu_{0}$ enters $K_{10}^{(2)}$ only through $K_{1}[1,0]=\tfrac{1}{2}-\hat{\lambda}=-\tfrac{1}{2}-\varepsilon\mu_{0}$,
one order of $\varepsilon$ below the constant background. $A$ and
$B$ are constructed (in the complex basis, with $\mathbf{K}_{0}=-i\mathbf{I}$)
precisely to bring $\mu_{0}$ to leading order, which is why the complex-basis
YS analysis is the natural tool for the primary resonance.

\smallskip{}
\emph{A caveat on interpretation.} The absence of detuning at leading
order is a property of the \emph{truncated series} --- of the bookkeeping
that inserts the detuning block $\mathbf{C}(\hat{\lambda})-\mathbf{C}_{0}$
into $\mathbf{D}_{1}$ and thereby weights it by an extra power of
$\delta$ --- and \emph{not} of the exact exponent. Extracting $K^{{\rm exact}}=\pi^{-1}\log(-X(\pi))$
numerically for the full LC circuit gives, to combined second order
in $(\hat{\lambda}-1,\delta)$, 
\[
K_{01}^{{\rm exact}}=\bigl(1-\hat{\lambda}^{-1/2}\bigr)-\tfrac{\hat{\lambda}\delta}{2}+O_{2},\qquad K_{10}^{{\rm exact}}=-\hat{\lambda}\bigl(1-\hat{\lambda}^{-1/2}\bigr)-\tfrac{\hat{\lambda}\delta}{2}+O_{2},
\]
i.e.\ in Mathieu units $K_{01}^{{\rm exact}}\approx\tfrac{\varepsilon}{2}\bigl(\mu_{0}-\tfrac{1}{2}\bigr)$
and $K_{10}^{{\rm exact}}\approx-\tfrac{\varepsilon}{2}\bigl(\mu_{0}+\tfrac{1}{2}\bigr)$.
The true entries therefore \emph{do} carry the detuning at leading
order, and they vanish --- one entry per boundary --- exactly at
the two EPDs $\mu_{0}=\pm\tfrac{1}{2}$, in the same one-entry-per-EPD
pattern as the column metamorphoses of Remark~\ref{rem:EPD-propeller}.
The failure analyzed above is thus an artifact of the $\delta$-truncation,
removable by keeping the $\delta^{0}$ exponent layer $\mathbf{K}^{(0)}(\hat{\lambda})=\bigl(1-\hat{\lambda}^{-1/2}\bigr)\bigl[\begin{smallmatrix}0 & 1\\
-\hat{\lambda} & 0
\end{smallmatrix}\bigr]$ exact rather than expanded; the complex-basis $A\cdot B$ product
is the classical and most efficient way of doing precisely that. 
\end{rem}

Note that $K_{1}[0,1]=-1/2$ is constant (independent of $\hat{\lambda}$),
so the leading $-\delta/2$ term in~\eqref{eq:YS-bdry-upper} never
vanishes on its own: the upper boundary requires the $O(\delta^{2})$
correction from $\mathbf{K}_{2}$ \emph{and} higher-order terms beyond
our truncation.

\paragraph{Self-consistent solution and comparison with MW/CF.} The
primary tongue boundary near $c=1$ is most efficiently obtained by
reparametrizing via the Mathieu detuning $\mu_{0}=(c-1)/\varepsilon$,
$\varepsilon=2\delta/(1+\delta^{2})$ (parameter dictionary eq.~\eqref{eq:YS-param-dict}).
The Floquet exponent from the YS analysis (eq.~\eqref{eq:YS-Dalpha}--\eqref{eq:YS-B})
vanishes at $\mu_{0,\pm}=\pm\tfrac{1}{2}+O(\varepsilon)$, giving
the primary tongue boundaries: 
\begin{equation}
c_{\pm}(\delta)=1\pm\tfrac{\varepsilon}{2}-\tfrac{9}{32}\varepsilon^{2}+O(\varepsilon^{3}),\quad\varepsilon=\tfrac{2\delta}{1+\delta^{2}}.\label{eq:YS-bdry-exact}
\end{equation}
These are \emph{identical} to the MW and CF results (Table~\ref{tab:bdry-LC},
$m=1$) to $O(\varepsilon^{3})$. Their equivalence follows from $M=-e^{\pi K_{{\rm total}}}$:
the YS condition~\eqref{eq:YS-bdry-Kprod}, the MW condition $\operatorname{Tr}(M)=-2$,
and the CF condition are three faces of the same EPD.

\paragraph{Tongue width.} The $O(\varepsilon^{2})$ center-shift
corrections cancel between $c_{+}$ and $c_{-}$ (both carry $-9\varepsilon^{2}/32$),
giving: 
\begin{equation}
c_{+}(\delta)-c_{-}(\delta)=\varepsilon+O(\varepsilon^{3})=\frac{2\delta}{1+\delta^{2}}+O(\delta^{3}),\label{eq:YS-width-formula}
\end{equation}
recovering Theorem~\ref{thm:EPD-bdry}. The cancellation of center
shifts is exact at $O(\varepsilon^{2})$, but the width itself is
not exact: the half-widths $\pm\varepsilon/2$ each carry $O(\varepsilon^{3})$
corrections from higher-order YS terms, so the true c-space width
differs from $\varepsilon$ by $O(\varepsilon^{3})$ (numerically:
$0.14\%$ at $\delta=0.10$, growing to $3.9\%$ at $\delta=0.50$).
The label \eqref{eq:YS-width-formula} refers to the YS formula, not
to exact agreement with the true width.

\paragraph{Numerical verification.} Table~\ref{tab:YS-bdry-numerical}
compares formula~\eqref{eq:YS-bdry-exact} against exact numerical
boundaries from $\operatorname{Tr}(M)+2=0$ (RK integration, $10^{-12}$
tolerance). The error is $O(\delta^{3})$ as expected.

\begin{table}[htbp]
\centering \caption{Primary tongue boundaries $c_{\pm}(\delta)$: YS formula~\eqref{eq:YS-bdry-exact}
vs exact numerical. Error is $O(\delta^{3})$.}
\label{tab:YS-bdry-numerical} {\small
\global\long\def\arraystretch{1.2}%
 }{\small{}%
\begin{tabular}{ccccccc}
\toprule 
{\small$\delta$ } & {\small$c_{+}^{{\rm YS}}$ } & {\small$c_{+}^{{\rm exact}}$ } & {\small$|{\rm err}_{+}|$ } & {\small$c_{-}^{{\rm YS}}$ } & {\small$c_{-}^{{\rm exact}}$ } & {\small$|{\rm err}_{-}|$ }\tabularnewline
\midrule 
{\small$0.05$ } & {\small$1.047077$ } & {\small$1.047052$ } & {\small$2.5\!\times\!10^{-5}$ } & {\small$0.947326$ } & {\small$0.947336$ } & {\small$1.0\!\times\!10^{-5}$ }\tabularnewline
{\small$0.10$ } & {\small$1.087982$ } & {\small$1.087726$ } & {\small$2.6\!\times\!10^{-4}$ } & {\small$0.889962$ } & {\small$0.889986$ } & {\small$2.4\!\times\!10^{-5}$ }\tabularnewline
{\small$0.15$ } & {\small$1.122489$ } & {\small$1.121450$ } & {\small$1.0\!\times\!10^{-3}$ } & {\small$0.829090$ } & {\small$0.828986$ } & {\small$1.0\!\times\!10^{-4}$ }\tabularnewline
{\small$0.20$ } & {\small$1.150703$ } & {\small$1.147859$ } & {\small$2.8\!\times\!10^{-3}$ } & {\small$0.766087$ } & {\small$0.765436$ } & {\small$6.5\!\times\!10^{-4}$ }\tabularnewline
\bottomrule
\end{tabular}}
\end{table}

\subsubsection{The truncated boundary equations in explicit form}

\label{subsubsec:YS-bdry-equations}

For reference, equations~\eqref{eq:YS-bdry-upper}--\eqref{eq:YS-bdry-lower}
written explicitly in $c$ and $\delta$ (after substituting $\hat{\lambda}=c(1+\delta^{2})/(1-\delta^{2})$
and expanding to second order in $\delta$): 
\begin{align}
\text{Upper }(c_{+}): & \quad-\tfrac{\delta}{2}+\delta^{2}\Bigl(-\tfrac{3}{8}+\tfrac{3c}{8}+\tfrac{c^{2}}{16}\Bigr)+O(\delta^{3})=0,\label{eq:bdry-upper-explicit}\\[4pt]
\text{Lower }(c_{-}): & \quad\delta\Bigl(\tfrac{1}{2}-c\Bigr)+\delta^{2}\Bigl(-\tfrac{1}{8}+\tfrac{3c}{8}-\tfrac{5c^{2}}{16}\Bigr)+O(\delta^{3})=0.\label{eq:bdry-lower-explicit}
\end{align}
As flagged in \S\,\ref{subsubsec:YS-bdry-lh}, these truncated equations
do \emph{not} reproduce the primary boundary: near $c=1$ the upper
equation reads $-\delta/2+\delta^{2}/16+O(\delta^{3})=0$, which has
no root with $c_{+}(\delta)\to1$, and the root of the lower equation
sits near $c=1/2$ rather than at the tongue (Remark~\ref{rem:K-approx-limitation}).
They are recorded because their entries are the raw material for the
corrected, $\delta^{0}$-layer-exact balance described at the end
of Remark~\ref{rem:K-approx-limitation} and made quantitative in
Remark~\ref{rem:K-corrected-layer} below. The notation used throughout
is 
\begin{equation}
K_{{\rm total}}^{(p)}=\sum_{m=1}^{p}\delta^{m}\,\mathbf{K}_{m}(\hat{\lambda}),\qquad\hat{\lambda}=c\,\frac{1+\delta^{2}}{1-\delta^{2}},\qquad\mu_{0}=\frac{c-1}{\varepsilon},\quad\varepsilon=\frac{2\delta}{1+\delta^{2}}.\label{eq:YS-Ktot-def}
\end{equation}

\begin{rem}[The corrected $\delta^{0}$-layer balance: the K-route rehabilitated]
\label{rem:K-corrected-layer} The defect identified in Remark~\ref{rem:K-approx-limitation}
is removable. Keep the $\delta^{0}$ exponent layer exact, 
\[
\mathbf{K}^{(0)}(\hat{\lambda})=\bigl(1-\hat{\lambda}^{-1/2}\bigr)\begin{bmatrix}0 & 1\\
-\hat{\lambda} & 0
\end{bmatrix},
\]
so that $e^{\pi\mathbf{K}^{(0)}}=-X(\pi)\big|_{\delta=0}$ holds exactly;
assign to order $\delta$ only the genuine first-harmonic average
$\mathbf{K}_{1}^{B}(\hat{\lambda})=[e^{-\tau\mathbf{C}_{0}}\hat{\lambda}\tilde{\mathbf{B}}_{1}e^{\tau\mathbf{C}_{0}}]_{{\rm av}}=\bigl[\begin{smallmatrix}0 & -\hat{\lambda}/2\\
-\hat{\lambda}/2 & 0
\end{smallmatrix}\bigr]$, and keep $\mathbf{K}_{2}^{{\rm ic}}(\hat{\lambda})$ of eq.~\eqref{eq:LC-K2-lam-exact}
at order $\delta^{2}$: 
\[
\tilde{\mathbf{K}}(\hat{\lambda},\delta)=\mathbf{K}^{(0)}(\hat{\lambda})+\delta\,\mathbf{K}_{1}^{B}(\hat{\lambda})+\delta^{2}\,\mathbf{K}_{2}^{{\rm ic}}(\hat{\lambda}).
\]
Numerically (monodromy extraction as in Remark~\ref{rem:K-approx-limitation}):
along $\hat{\lambda}=1+a\delta$ the residual $\|K^{{\rm exact}}-\tilde{\mathbf{K}}\|$
decays as $O(\delta^{2})$ (measured slopes $1.9$--$2.0$ over $\delta=0.08\to0.01$,
magnitude $\approx0.36\,|\hat{\lambda}-1|\,\delta$), versus $O(\delta)$
for the printed assignment $\delta\mathbf{K}_{1}(\hat{\lambda})+\delta^{2}\mathbf{K}_{2}(\hat{\lambda})$;
the remaining $O\bigl((\hat{\lambda}-1)\delta\bigr)$ cross terms
are not supplied by $\mathbf{K}_{2}(\hat{\lambda})$, which the chain
generated with the detuning misassigned. Solving the corrected balance
$\tilde{K}_{01}=0$ and $\tilde{K}_{10}=0$ (with $\hat{\lambda}=c(1+\delta^{2})/(1-\delta^{2})$)
now \emph{does} locate the primary tongue: at $\delta=0.05$ the roots
are $c_{+}=1.049134$ and $c_{-}=0.946819$ against the exact $1.047052$
and $0.947336$ --- errors $+2.1\times10^{-3}$ and $-5.2\times10^{-4}$,
of order $\delta^{2}$, while the leading-order locations and the
tongue width are exact (cf.\ $K_{01}^{{\rm exact}}\approx\tfrac{\varepsilon}{2}(\mu_{0}-\tfrac{1}{2})$
in Remark~\ref{rem:K-approx-limitation}). The fully $O(\varepsilon^{2})$-consistent
boundary remains eq.~\eqref{eq:YS-bdry-exact} via the complex-basis
$A\cdot B$ product, whose errors at the same $\delta$ are $2.5\times10^{-5}$
and $1.0\times10^{-5}$ (Table~\ref{tab:YS-bdry-numerical}). In
short: the K-route does find the primary tongue once the $\delta^{0}$
layer is kept exact --- at leading order exactly, at second order
approximately --- and the complex-basis route is the efficient implementation
of the same idea carried one order further. 
\end{rem}

The boundary analysis for higher-order tongues ($m\geq2$), where
the leading entry $K_{1}[0,1]$ or $K_{1}[1,0]$ can vanish at the
tongue center, proceeds differently and will be treated separately.

\subsection{Mathieu vs.\ exact LC: a systematic comparison}

\label{subsec:YS-Math-LC}

The Mathieu equation (Chapter~\ref{subsec:YS-Mathieu}) is an $O(\delta^{2})$
approximation to the exact LC circuit (\S\,\ref{subsec:YS-exact-LC}):
it retains only the $n=1$ Fourier harmonic of the LC-circuit Fourier
coefficient and fixes $\hat{\lambda}=c=1$. This section makes the
comparison between the two YS computations precise and systematic,
quantifying exactly what the Mathieu approximation gains (simplicity,
closed-form scalars) and loses (the $\mathbf{D}_{2},\mathbf{D}_{3},\ldots$
contributions from higher Poisson harmonics, the $\hat{\lambda}$-dependence).

\medskip{}
\emph{Comparison dictionary.} Two notational differences must be accounted
for before comparing any formula:
\begin{enumerate}
\item \emph{Normalization factor.} The Mathieu subsection uses $\mathbf{B}_{1}^{{\rm Math}}=\bigl[\begin{smallmatrix}0 & 0\\
-\cos2\tau & 0
\end{smallmatrix}\bigr]$ (standard Mathieu convention), while the LC subsection uses $\mathbf{B}_{1}^{{\rm LC}}=\hat{\lambda}\bigl[\begin{smallmatrix}0 & 0\\
-2\cos2\tau & 0
\end{smallmatrix}\bigr]$ (from the Poisson kernel Fourier series expansion~\eqref{eq:LC-Hill-factored}).
At $\hat{\lambda}=1$: $\mathbf{B}_{1}^{{\rm LC}}=2\mathbf{B}_{1}^{{\rm Math}}$.
The exponent matrices satisfy $\mathbf{K}_{m}^{{\rm LC}}(\hat{\lambda}=1)=2^{m}\mathbf{K}_{m}^{{\rm Math}}$
\emph{provided} $\mathbf{D}_{n}=0$ for $n\geq2$ (the Mathieu approximation).
For $\mathbf{F}_{j}$ the scaling is complicated by the convention
difference (see below).
\item \emph{Normalization convention.} Mathieu subsection: $\mathbf{F}_{1}$
uses the $\varphi'_{ij}$ prescription (primary resonance $m=1$,
neither zero-mean nor IC); $\mathbf{F}_{2}$ uses Convention~1 (zero-mean,
$[\mathbf{F}_{2}^{{\rm Math}}]_{{\rm av}}=\mathbf{0}$). LC subsection
(\S\,\ref{subsubsec:YS-LC-F1F2}): Convention~2 (initial-condition,
$\mathbf{F}_{j}^{{\rm LC}}(0,\hat{\lambda})=\mathbf{0}$). The conventions
differ; nevertheless, for the \emph{genuine} factors the relation
is a simple scaling once conventions are matched, $\mathbf{F}_{1}^{{\rm LC}}(\tau,1)=2\,\mathbf{F}_{1}^{{\rm true}}(\tau,0)$
(Item~2), while the relation between $\mathbf{F}_{1}^{{\rm LC}}$
and YS's \emph{printed} $\varphi'$ matrix~\eqref{eq:YS-F1} ---
the transformed forcing, eq.~\eqref{eq:phi-is-D1} --- is the derivative
identity~\eqref{eq:Math-LC-F1-corrected}. 
\end{enumerate}
\noindent The $\mathbf{K}_{m}$ comparison relations are: 
\begin{equation}
\mathbf{K}_{m}^{{\rm Math}}=\frac{1}{2^{m}}\,\mathbf{K}_{m}^{{\rm LC}}(\hat{\lambda}=1)\quad\text{(if \ensuremath{\mathbf{D}_{n}=0} for \ensuremath{n\geq2})},\label{eq:Math-LC-Km-relation}
\end{equation}
which holds for all $m\geq1$ (verified for $m=1,2,3$ in the table
below). For the $\mathbf{F}_{j}$ in matched conventions the analogous
scaling holds under the Mathieu approximation ($\mathbf{D}_{n}=0$,
$n\geq2$): 
\begin{multline}
\mathbf{F}_{j}^{{\rm Math,zm}}(\tau)=\frac{1}{2^{j}}\Bigl[\mathbf{F}_{j}^{{\rm LC,ic}}(\tau,1)-\bigl[\mathbf{F}_{j}^{{\rm LC,ic}}(\cdot,1)\bigr]_{{\rm av}}\Bigr]\\
(j\geq1,\;\mathbf{D}_{n}=0\text{ for }n\geq2).\label{eq:Math-LC-Fj-relation}
\end{multline}
When higher harmonics $\mathbf{D}_{n}\neq0$ ($n\geq2$), eq.~\eqref{eq:Math-LC-Km-relation}
holds at $m=1$ but breaks at $m\geq2$.

\bigskip{}

\begin{center}
\global\long\def\arraystretch{1.3}%
 \resizebox{\textwidth}{!}{%
\begin{tabular}{lp{2.4cm}p{2.4cm}p{4.8cm}}
\hline 
Quantity  & Mathieu (YS1, $\varepsilon$)  & LC at $c=1$ (IC, $\delta$)  & Ratio / relation \tabularnewline
\hline 
$K_{1}[0,1]$  & $-\tfrac{1}{4}$  & $-\tfrac{1}{2}$  & $2^{1}$ (normalization) \tabularnewline
$K_{1}[1,0]$  & $-\tfrac{1}{4}$  & $-\tfrac{1}{2}$  & $2^{1}$ (normalization) \tabularnewline
$K_{2}[0,1]$  & $+\tfrac{1}{64}$  & $+\tfrac{1}{16}$  & $4\times$ (normalization, no sign flip) \tabularnewline
$K_{2}[1,0]$  & $-\tfrac{1}{64}$  & $-\tfrac{1}{16}$  & $4\times$ (normalization, no sign flip) \tabularnewline
$K_{3}[0,1]=K_{3}[1,0]$  & $+\tfrac{7}{1024}$ (zm)  & $-\tfrac{9}{128}$ (exact LC)  & single-harmonic zm: $8\times$, i.e.\ $+\tfrac{7}{128}$; $\mathbf{D}_{2}$
shift $-\tfrac{1}{8}$ (Item~4) \tabularnewline
$\mathbf{F}_{1}$  & \multicolumn{3}{p{9.6cm}}{genuine factors: $\mathbf{F}_{1}^{{\rm LC}}(\tau,1)=2\,\mathbf{F}_{1}^{{\rm true}}(\tau,0)$,
exact (Item~2); YS's printed $\varphi'$ matrix: eq.~\eqref{eq:Math-LC-F1-corrected}
(residual $\equiv0$)}\tabularnewline
$\mathbf{F}_{2}$  & \multicolumn{3}{p{9.6cm}}{eq.~\eqref{eq:Math-LC-Fj-relation} at $j=2$: breaks (see Item~5)}\tabularnewline
\hline 
\end{tabular}} 
\par\end{center}

\paragraph{Item 1: $\mathbf{K}_1$} At $\hat{\lambda}=1$ and $\mu_{0}=0$,
both $K_{1}[0,1]$ and $K_{1}[1,0]$ equal $-1/2$ for the LC case
vs $-1/4$ for Mathieu. The ratio $2=2^{1}$ confirms the normalization
factor. For $c\neq1$, the LC $\mathbf{K}_{1}(c)$ is asymmetric:
$K_{1}[0,1]=-1/2$ (constant in $c$) while $K_{1}[1,0]=1/2-c$ (linear
in $c$) --- a feature with no Mathieu analogue, since the Mathieu
equation always has $c=1$.

\paragraph{Item 2: $\mathbf{F}_1$ --- factor scaling and the derivative identity}
For the genuine first-order factors the comparison is as simple as
for the $\mathbf{K}_{m}$: at the resonance center, 
\[
\mathbf{F}_{1}^{{\rm LC}}(\tau,1)=2\,\mathbf{F}_{1}^{{\rm true}}(\tau,0)
\]
\emph{exactly}, in any common integration-constant convention ---
in the IC gauge this is eq.~\eqref{eq:F1-c} at $c=1$ against twice
eq.~\eqref{eq:YS-F1-EPD}, entry by entry, and subtracting averages
gives the zero-mean version $\mathbf{F}_{1}^{{\rm Math,zm}}=\tfrac{1}{2}[\mathbf{F}_{1}^{{\rm LC,ic}}(\tau,1)-[\mathbf{F}_{1}^{{\rm LC,ic}}]_{{\rm av}}]$,
i.e.\ eq.~\eqref{eq:Math-LC-Fj-relation} at $j=1$. (The chains
coincide up to the overall normalization factor~$2$: $\mathbf{D}_{1}^{{\rm LC}}(\tau,1)=2\,\mathbf{D}_{1}^{{\rm Math}}(\tau,0)$,
$\mathbf{K}_{1}^{{\rm LC}}(1)=2\,\mathbf{K}_{1}^{{\rm Math}}(0)$,
and the recursion is linear at first order.) A second, independent
identity relates the LC factor to YS's \emph{printed} matrix~\eqref{eq:YS-F1}
--- the $\varphi'$-prescription object, which is the transformed
forcing rather than the factor (eq.~\eqref{eq:phi-is-D1}): 
\begin{equation}
\mathbf{F}_{1}^{{\rm Math}}(\tau,0)=\tfrac{1}{2}\,\dot{\mathbf{F}}_{1}^{{\rm LC}}(\tau,1)+\mathbf{C},\qquad\mathbf{C}=\begin{bmatrix}0 & -\tfrac{1}{4}\\[2pt]
-\tfrac{1}{4} & 0
\end{bmatrix},\label{eq:Math-LC-F1-corrected}
\end{equation}
where $\dot{\mathbf{F}}_{1}^{{\rm LC}}=d\mathbf{F}_{1}^{{\rm LC}}/d\tau$.
This is an \emph{exact algebraic identity}, proved by direct substitution
of~\eqref{eq:YS-F1} and~\eqref{eq:F1-c}. Differentiating~\eqref{eq:F1-c}
entry by entry at $c=1$: 
\begin{align*}
\dot{F}_{1}^{{\rm LC}}[0,0] & =\tfrac{1}{2}\sin4\tau, & \dot{F}_{1}^{{\rm LC}}[0,1] & =\cos2\tau-\tfrac{1}{2}\cos4\tau,\\
\dot{F}_{1}^{{\rm LC}}[1,0] & =-\cos2\tau-\tfrac{1}{2}\cos4\tau, & \dot{F}_{1}^{{\rm LC}}[1,1] & =-\tfrac{1}{2}\sin4\tau,
\end{align*}
and substituting into~\eqref{eq:Math-LC-F1-corrected} gives exactly
the entries of~\eqref{eq:YS-F1} at $\mu_{0}=0$, entry by entry: 
\begin{center}
\global\long\def\arraystretch{1.5}%
\begin{tabular}{lll}
\hline 
Entry  & $F_{1}^{{\rm Math}}$ (corrected)  & $\tfrac{1}{2}\dot{F}_{1}^{{\rm LC}}(\tau,1)+C_{ij}$ \tabularnewline
\hline 
$[0,0]$  & $\tfrac{\sin4\tau}{4}$  & $\tfrac{1}{2}\cdot\tfrac{\sin4\tau}{2}+0=\tfrac{\sin4\tau}{4}$ \tabularnewline
{[}4pt{]} $[0,1]$  & $\tfrac{\cos2\tau}{2}-\tfrac{\cos4\tau}{4}-\tfrac{1}{4}$  & $\tfrac{1}{2}(\cos2\tau-\tfrac{\cos4\tau}{2})-\tfrac{1}{4}$ \tabularnewline
{[}4pt{]} $[1,0]$  & $-\tfrac{\cos2\tau}{2}-\tfrac{\cos4\tau}{4}-\tfrac{1}{4}$  & $\tfrac{1}{2}(-\cos2\tau-\tfrac{\cos4\tau}{2})-\tfrac{1}{4}$ \tabularnewline
{[}4pt{]} $[1,1]$  & $-\tfrac{\sin4\tau}{4}$  & $\tfrac{1}{2}\cdot(-\tfrac{\sin4\tau}{2})+0=-\tfrac{\sin4\tau}{4}$ \tabularnewline
\hline 
\end{tabular}
\par\end{center}

\begin{figure}[htbp]
\centering \includegraphics[width=0.9\textwidth]{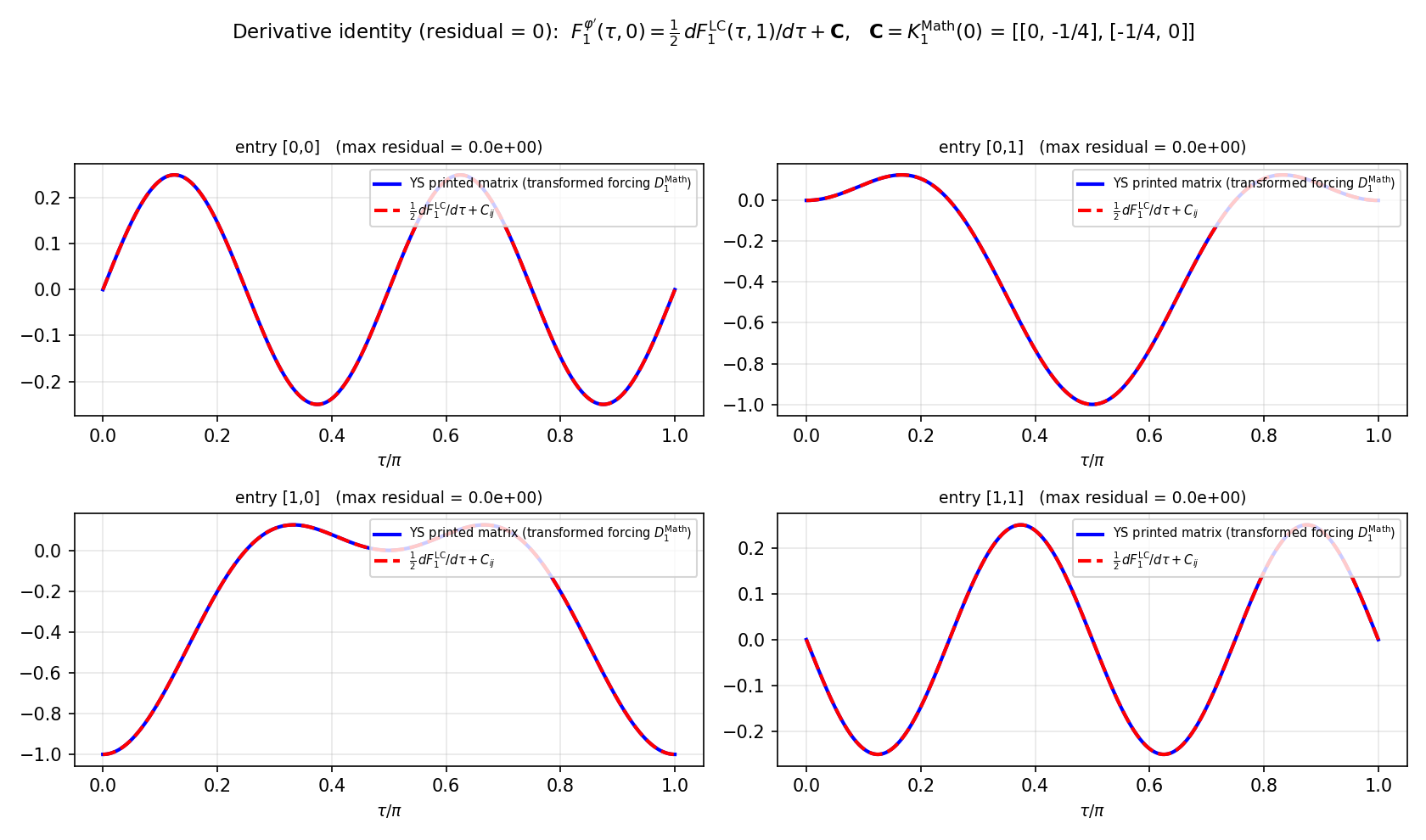}
\caption{Item~2 verification of the derivative identity~\eqref{eq:Math-LC-F1-corrected}:
$\mathbf{F}_{1}^{{\rm Math}}(\tau,0)=\tfrac{1}{2}\dot{\mathbf{F}}_{1}^{{\rm LC}}(\tau,1)+\mathbf{C}$,
all four entries. \emph{Blue solid}: YS's printed Mathieu matrix (eq.~\eqref{eq:YS-F1},
$\mu_{0}=0$) --- the $\varphi'$-prescription object, equal to the
transformed forcing $\mathbf{D}_{1}^{{\rm Math}}$ (eq.~\eqref{eq:phi-is-D1}).
\emph{Red dashed}: $\tfrac{1}{2}\dot{F}_{1}^{{\rm LC}}(\tau,1)+C_{ij}$
where $\dot{F}=dF/d\tau$ and $\mathbf{C}=\bigl[\begin{smallmatrix}0 & -1/4\\
-1/4 & 0
\end{smallmatrix}\bigr]$. The residual is identically zero in all four entries: the relation
is an \emph{exact algebraic identity}, not merely a numerical approximation.
Both sides reduce to the same trigonometric expressions after substituting
the closed-form formulas~\eqref{eq:YS-F1} and~\eqref{eq:F1-c}.}
\label{fig:Math-LC-F1-comparison} 
\end{figure}

The identity is the first-order recursion in disguise: differentiating
the IC-convention factor gives $\dot{\mathbf{F}}_{1}^{{\rm LC}}=\mathbf{D}_{1}^{{\rm LC}}-\mathbf{K}_{1}^{{\rm LC}}$,
so $\tfrac{1}{2}\dot{\mathbf{F}}_{1}^{{\rm LC}}(\tau,1)=\mathbf{D}_{1}^{{\rm Math}}(\tau,0)-\mathbf{K}_{1}^{{\rm Math}}(0)$;
since the printed matrix equals $\mathbf{D}_{1}^{{\rm Math}}$ at
$\mu_{0}=0$ (eq.~\eqref{eq:phi-is-D1}), the identity holds with
$\mathbf{C}=\mathbf{K}_{1}^{{\rm Math}}(0)=\bigl[\begin{smallmatrix}0 & -1/4\\
-1/4 & 0
\end{smallmatrix}\bigr]$ --- the additive constant is exactly YS's first-order Mathieu exponent
matrix at $\mu_{0}=0$. The identity is verified graphically in Figure~\ref{fig:Math-LC-F1-comparison}.

\begin{rem}[Structural meaning of the $\mathbf{F}_{1}$ identity]
\label{rem:F1-identity-meaning} The identity~\eqref{eq:Math-LC-F1-corrected}
has an interesting structural interpretation worth noting explicitly.

The two YS series use different perturbation parameters: $\varepsilon=2\delta/(1+\delta^{2})=2\delta+O(\delta^{3})$
for Mathieu and $\varepsilon_{{\rm LC}}=\delta$ for the LC circuit.
At first order in $\delta$ the Floquet factor\index{Floquet theory!Floquet factor}s
are: 
\[
\mathbf{F}^{{\rm Math}}(\tau)=\mathbf{I}+2\delta\,\mathbf{F}_{1}^{{\rm Math}}+O(\delta^{2}),\qquad\mathbf{F}^{{\rm LC}}(\tau)=\mathbf{I}+\delta\,\mathbf{F}_{1}^{{\rm LC}}+O(\delta^{2}).
\]
For the genuine factors there is nothing to reconcile: $\mathbf{F}_{1}^{{\rm LC}}(\tau,1)=2\,\mathbf{F}_{1}^{{\rm true}}(\tau,0)$
(Item~2), so $\mathbf{I}+2\delta\,\mathbf{F}_{1}^{{\rm true}}=\mathbf{I}+\delta\,\mathbf{F}_{1}^{{\rm LC}}$
and the two first-order Floquet factors are literally identical as
functions of $\tau$ once the normalization is matched. The object
that is \emph{not} half the LC factor is YS's printed matrix ---
the transformed forcing --- whose relation to the factor is the derivative
identity~\eqref{eq:Math-LC-F1-corrected}.

The characteristic exponents agree as well at first order: both $\varepsilon_{{\rm Math}}\mathbf{K}_{1}^{{\rm Math}}$
and $\varepsilon_{{\rm LC}}\mathbf{K}_{1}^{{\rm LC}}$ have eigenvalues
$\pm\delta/2$, so the instability exponent $\Delta\alpha\sim\delta/2$
is identical for Mathieu and LC at leading order. The first difference
in the exponent appears at $O(\delta^{2})$ through $\mathbf{K}_{2}$
(Item~3 below), where the sign and magnitude already differ substantially.

The identity~\eqref{eq:Math-LC-F1-corrected} thus captures the factor-versus-forcing
distinction: differentiation converts the factor into the forcing
through the recursion $\dot{\mathbf{F}}_{1}=\mathbf{D}_{1}-\mathbf{K}_{1}$,
and the additive constant is exactly $\mathbf{K}_{1}^{{\rm Math}}(0)$.
It is a consistency identity between two \emph{different} objects
in the YS scheme --- the periodic factor and the transformed forcing
--- not a sign that the two expansions produce different factors. 
\end{rem}

\paragraph{Item 3: $\mathbf{K}_2$ --- higher harmonics do not enter}
The second-order coefficient is untouched by the higher harmonics:
$[\mathbf{D}_{2}]_{{\rm av}}=\mathbf{0}$ (the rotation conjugation
of a constant matrix contains only the $\tau$-modes $\{0,\pm2\}$,
so $\cos4\tau$ times it has no constant term), and $\mathbf{D}_{2}$
first enters the exponent at third order (Item~4). Consequently $\mathbf{K}_{2}^{{\rm LC}}=4\,\mathbf{K}_{2}^{{\rm Math}}$
exactly, convention for convention, for the exact LC circuit just
as for the single-harmonic truncation: 
\begin{equation}
K_{2}^{{\rm ic,LC}}[0,1]=K_{2}^{{\rm ic,\,\text{no}\,D_{2}}}[0,1]=4\times K_{2}^{{\rm Math,YS}}[0,1]=4\times\!\left(+\tfrac{1}{64}\right)=+\tfrac{1}{16},\label{eq:K2-no-D2}
\end{equation}
the factor $4$ being the $\mathbf{B}_{1}$ normalization squared,
not a convention conversion (Remark~\ref{rem:YS-basis-convention}).
The apparent sign discrepancy that motivated an earlier draft of this
comparison --- the zero-mean LC value $K_{2}^{{\rm zm,LC}}[0,1]=-1/16$
against the rescaled YS value $+1/16$ --- is entirely the zero-mean-versus-IC
gauge offset of eqs.~\eqref{eq:K2-omega1}/\eqref{eq:K2-omega1-ic},
not an effect of $\mathbf{D}_{2}$: YS's singular-case computation
uses the $\mathbf{F}(0)=\mathbf{0}$ normalization (Remark~\ref{rem:YS-basis-convention}),
so the matching IC-convention LC value is $+1/16=4\times(+1/64)$,
with no sign flip; in the zero-mean gauge on both sides the relation
is likewise exact, $K_{2}^{{\rm Math,zm}}[0,1]=-\tfrac{1}{64}=\tfrac{1}{4}K_{2}^{{\rm zm,LC}}[0,1]$.

\paragraph{Item 4: $\mathbf{K}_3$ --- normalization check and the first
higher-harmonic effect} The third-order coefficient provides both a consistency check and
the first genuine effect of the higher Poisson harmonics. For the
Mathieu equation (Math norm, zero-mean): $K_{3}^{{\rm Math}}[0,1]=K_{3}^{{\rm Math}}[1,0]=+7/1024$.
For the single-harmonic chain in the LC normalization (zero-mean,
$\mathbf{D}_{2}=\mathbf{D}_{3}=\mathbf{0}$): $K_{3}^{{\rm zm,no\,D_{2}}}[0,1]=8\times K_{3}^{{\rm Math}}[0,1]=8\times(7/1024)=7/128$,
matching eq.~\eqref{eq:K3-omega1} and confirming the $2^{3}=8$
scaling. Restoring the second-harmonic source $\mathbf{D}_{2}$ (exact
LC): $K_{3}^{{\rm LC}}[0,1]=K_{3}^{{\rm LC}}[1,0]=-9/128=-0.0703\ldots$
in either convention (Table~\ref{tab:k2k3-numerical}); the shift
$-1/8$ per off-diagonal entry is the leading effect of the $\delta^{2}\cos4\tau$
term of the LC coefficient on the exponent, entering partly directly
through $[\mathbf{D}_{2}\mathbf{F}_{1}]_{{\rm av}}=-\tfrac{1}{16}$
per entry and partly through the $\mathbf{D}_{2}$-contribution to
$\mathbf{F}_{2}$. The third harmonic is irrelevant at this order:
$[\mathbf{D}_{3}]_{{\rm av}}=\mathbf{0}$. \paragraph{Item 5: $\mathbf{F}_2$}
The relation~\eqref{eq:Math-LC-Fj-relation} at $j=2$ would give
$\mathbf{F}_{2}^{{\rm Math,zm}}=\tfrac{1}{4}[\mathbf{F}_{2}^{{\rm LC,ic}}|_{c=1}-[\mathbf{F}_{2}^{{\rm LC,ic}}|_{c=1}]_{{\rm av}}]$
if $\mathbf{D}_{2}=\mathbf{D}_{3}=\ldots=0$. However, the exact LC
$\mathbf{F}_{2}$ includes the $\mathbf{D}_{2}$ contribution through
the source $\boldsymbol{\Phi}_{2}=\mathbf{D}_{2}+\mathbf{D}_{1}\mathbf{F}_{1}-\mathbf{F}_{1}\mathbf{K}_{1}$,
so the relation breaks. The quantitative discrepancy encodes exactly
the contribution of the second Poisson harmonic to the periodic Lyapunov
factor; the breakage is confined to the periodic factor, since $\mathbf{K}_{2}$
itself is untouched ($[\mathbf{D}_{2}]_{{\rm av}}=\mathbf{0}$, Item~3).
\paragraph{Item 6: Boundary curves} Both the Mathieu YS and the exact
LC YS analyses produce identical primary tongue boundaries to $O(\varepsilon^{3})$
(eq.~\eqref{eq:YS-bdry-exact}): 
\begin{equation}
c_{\pm}(\delta)=1\pm\tfrac{\varepsilon}{2}-\tfrac{9}{32}\varepsilon^{2}+O(\varepsilon^{3}),\quad\varepsilon=\tfrac{2\delta}{1+\delta^{2}}.
\end{equation}
This is because both derive from the same EPD condition --- $\Delta\alpha_{{\rm YS}}=0$
in the Mathieu case and $K_{01}^{{\rm exact}}\cdot K_{10}^{{\rm exact}}=0$
in the LC case --- and the Mathieu approximation is accurate to $O(\varepsilon^{2})$
for the boundary locations. The agreement among YS, MW, and CF is
confirmed numerically to $O(\delta^{3})$ in Table~\ref{tab:YS-bdry-numerical}.
\paragraph{Item 7: Floquet portraits} The Mathieu portraits (Fig.~\ref{fig:YS-F1-portraits})
show the self-intersection count of the second-column portrait passing
through $0\to2\to1\to0$ as $\mu_{0}$ increases, with transitions
at $\mu_{0}=0$ (tangency), at the upper EPD $\mu_{0}=\tfrac{1}{2}$
(a three-cusped degeneration), and at $\mu_{0}=\tfrac{3}{2}$ (Theorem~\ref{thm:F1-topology});
the mirror metamorphosis of the first column is governed by the lower
EPD $\mu_{0}=-\tfrac{1}{2}$ (Remark~\ref{rem:EPD-propeller}). The
LC portraits (Fig.~\ref{fig:LC-F1-portraits}) show how the loop
shape deforms continuously as $c$ varies: at $c=1$ the first-order
LC portrait is exactly the $\mu_{0}=0$ true-factor portrait magnified
by the normalization factor~$2$ ($\mathbf{F}_{1}^{{\rm LC}}(\tau,1)=2\,\mathbf{F}_{1}^{{\rm true}}(\tau,0)$,
Item~2); for $c\neq1$ the asymmetry between $K_{1}[0,1]=-1/2$ and
$K_{1}[1,0]=1/2-c$ produces an asymmetric loop with no Mathieu analogue. 

\subsection{\texorpdfstring{Periodic Lyapunov factors $\mathbf{F}_{1}$ and
$\mathbf{F}_{2}$ for the exact LC circuit}{Periodic Lyapunov factors
F1 and F2 for the exact LC circuit}}

\label{subsubsec:YS-LC-F1F2}

We apply the YS recursion~\eqref{eq:YS-Kl}--\eqref{eq:YS-Fl} to
the exact LC Hill equation~\eqref{eq:LC-Hill-firstorder} with $\hat{\lambda}$
treated as an \emph{independent constant} and $\delta$ as the small
parameter. The preliminary transformation uses $\mathbf{C}_{0}=\bigl[\begin{smallmatrix}0 & 1\\
-1 & 0
\end{smallmatrix}\bigr]$ (the same choice as for the Mathieu case, Chapter~\ref{subsec:YS-Mathieu}),
so that $e^{\tau\mathbf{C}_{0}}=\bigl[\begin{smallmatrix}\cos\tau & \sin\tau\\
-\sin\tau & \cos\tau
\end{smallmatrix}\bigr]$ is exactly $\pi$-periodic. The residual $\mathbf{K}_{0}=\mathbf{C}-\mathbf{C}_{0}=\bigl[\begin{smallmatrix}0 & 0\\
-({\hat{\lambda}}-1) & 0
\end{smallmatrix}\bigr]$ is a constant matrix that enters the first-order source alongside
$\mathbf{B}_{1}$.

\medskip{}

\noindent\emph{First-order source $\mathbf{D}_{1}(\tau,\hat{\lambda})$.}
The combined first-order transformed source is 
\begin{equation}
\mathbf{D}_{1}(\tau,\hat{\lambda})=e^{-\tau\mathbf{C}_{0}}\bigl[\mathbf{K}_{0}+\hat{\lambda}\tilde{\mathbf{B}}_{1}(\tau)\bigr]e^{\tau\mathbf{C}_{0}},\label{eq:LC-D1-correct}
\end{equation}
where $\tilde{\mathbf{B}}_{1}(\tau)=\bigl[\begin{smallmatrix}0 & 0\\
-2\cos2\tau & 0
\end{smallmatrix}\bigr]$. Since both $\mathbf{K}_{0}$ and $\tilde{\mathbf{B}}_{1}$ have
$\pi$-periodic conjugation by $e^{\pm\tau\mathbf{C}_{0}}$, the matrix
$\mathbf{D}_{1}(\tau,\hat{\lambda})$ is \emph{exactly $\pi$-periodic}.
Expanding via product-to-sum identities, its entries involve only
the standard Fourier modes $\sin2\tau$, $\cos2\tau$, $\sin4\tau$,
$\cos4\tau$ --- no irrational modes, no $\omega=\sqrt{\hat{\lambda}}$
anywhere.

\medskip{}

\noindent\emph{First-order factor $\mathbf{F}_{1}(\tau,\hat{\lambda})$.}
Setting $\mathbf{K}_{1}=[\mathbf{D}_{1}]_{{\rm av}}$ and integrating:
\begin{align}
\mathbf{K}_{1}(\hat{\lambda}) & =\begin{bmatrix}0 & -\tfrac{1}{2}\\[4pt]
\tfrac{1}{2}-\hat{\lambda} & 0
\end{bmatrix},\label{eq:LC-K1-lam}\\[8pt]
\mathbf{F}_{1}(\tau,\hat{\lambda}) & =\begin{bmatrix}f_{00}(\tau,\hat{\lambda}) & f_{01}(\tau,\hat{\lambda})\\[4pt]
f_{10}(\tau,\hat{\lambda}) & -f_{00}(\tau,\hat{\lambda})
\end{bmatrix},\label{eq:LC-F1-lam}
\end{align}
where 
\begin{align}
f_{00}(\tau,\hat{\lambda}) & =\frac{3\hat{\lambda}-2}{8}+\frac{(1-\hat{\lambda})\cos2\tau}{4}-\frac{\hat{\lambda}\cos4\tau}{8},\label{eq:f00-lam}\\[4pt]
f_{01}(\tau,\hat{\lambda}) & =\frac{(\hat{\lambda}+1)\sin2\tau}{4}-\frac{\hat{\lambda}\sin4\tau}{8},\label{eq:f01-lam}\\[4pt]
f_{10}(\tau,\hat{\lambda}) & =\frac{(1-3\hat{\lambda})\sin2\tau}{4}-\frac{\hat{\lambda}\sin4\tau}{8}.\label{eq:f10-lam}
\end{align}
Each entry is a \emph{genuine $\pi$-periodic trigonometric polynomial}
with polynomial dependence on $\hat{\lambda}$ --- no square roots,
no irrational Fourier modes. All three YS properties are satisfied:
$\mathbf{F}_{1}(0,\hat{\lambda})=\mathbf{F}_{1}(\pi,\hat{\lambda})=\mathbf{0}$,
$\mathbf{F}_{1}(\tau+\pi,\hat{\lambda})=\mathbf{F}_{1}(\tau,\hat{\lambda})$,
and $\operatorname{tr}\mathbf{F}_{1}=0$. At $\hat{\lambda}=1$: $\mathbf{K}_{1}=\bigl[\begin{smallmatrix}0 & -1/2\\
-1/2 & 0
\end{smallmatrix}\bigr]$ and $\mathbf{F}_{1}(\tau,1)$ reduces to the Mathieu factor~\eqref{eq:F1-omega1}
\emph{up to the constant offset} $\bigl[\begin{smallmatrix}1/8 & 0\\
0 & -1/8
\end{smallmatrix}\bigr]$ that converts between the two conventions (here initial-condition,
$\mathbf{F}_{1}(0)=\mathbf{0}$; there zero-mean, $[\mathbf{F}_{1}]_{{\rm av}}=\mathbf{0}$),
confirming the Mathieu case.

\medskip{}

\noindent\emph{Second-order factor $\mathbf{F}_{2}(\tau,\hat{\lambda})$.}
The source at second order is $\boldsymbol{\Phi}_{2}(\tau)=\mathbf{D}_{2}(\tau,\hat{\lambda})+\mathbf{D}_{1}(\tau,\hat{\lambda})\mathbf{F}_{1}(\tau,\hat{\lambda})-\mathbf{F}_{1}(\tau,\hat{\lambda})\mathbf{K}_{1}(\hat{\lambda})$,
where $\mathbf{D}_{2}=e^{-\tau\mathbf{C}_{0}}\hat{\lambda}\tilde{\mathbf{B}}_{2}e^{\tau\mathbf{C}_{0}}$
and $\tilde{\mathbf{B}}_{2}=\bigl[\begin{smallmatrix}0 & 0\\
-2\cos4\tau & 0
\end{smallmatrix}\bigr]$. The source $\boldsymbol{\Phi}_{2}$ contains modes $\{2,4,6\}$
only (Remark~\ref{rem:YS-freq-conj}). The exponent and Lyapunov
factor at second order are: 
\begin{gather}
\mathbf{K}_{2}(\hat{\lambda})=\begin{bmatrix}0 & -\tfrac{3}{8}+\tfrac{3}{8}\hat{\lambda}+\tfrac{1}{16}\hat{\lambda}^{2}\\[4pt]
-\tfrac{1}{8}+\tfrac{3}{8}\hat{\lambda}-\tfrac{5}{16}\hat{\lambda}^{2} & 0
\end{bmatrix},\label{eq:LC-K2-lam}\\[4pt]
\begin{aligned}F_{2}[0,0]={} & \Bigl(-\tfrac{1}{16}-\tfrac{\hat{\lambda}}{24}+\tfrac{11\hat{\lambda}^{2}}{192}\Bigr)+\Bigl(\tfrac{1}{16}+\tfrac{\hat{\lambda}}{8}-\tfrac{3\hat{\lambda}^{2}}{32}\Bigr)\cos2\tau\\
 & +\tfrac{3\hat{\lambda}^{2}}{64}\cos4\tau+\Bigl(-\tfrac{\hat{\lambda}}{12}-\tfrac{\hat{\lambda}^{2}}{96}\Bigr)\cos6\tau,
\end{aligned}
\label{eq:F2-00-lam}\\[4pt]
\begin{aligned}F_{2}[0,1]={} & \Bigl(\tfrac{3}{16}-\tfrac{5\hat{\lambda}}{16}-\tfrac{3\hat{\lambda}^{2}}{32}\Bigr)\sin2\tau+\Bigl(\tfrac{3\hat{\lambda}}{16}+\tfrac{3\hat{\lambda}^{2}}{64}\Bigr)\sin4\tau\\
 & +\Bigl(-\tfrac{\hat{\lambda}}{12}-\tfrac{\hat{\lambda}^{2}}{96}\Bigr)\sin6\tau,
\end{aligned}
\label{eq:F2-01-lam}\\[4pt]
\begin{aligned}F_{2}[1,0]={} & \Bigl(\tfrac{1}{16}-\tfrac{7\hat{\lambda}}{16}+\tfrac{5\hat{\lambda}^{2}}{32}\Bigr)\sin2\tau+\Bigl(-\tfrac{\hat{\lambda}}{4}+\tfrac{\hat{\lambda}^{2}}{64}\Bigr)\sin4\tau\\
 & +\Bigl(-\tfrac{\hat{\lambda}}{12}-\tfrac{\hat{\lambda}^{2}}{96}\Bigr)\sin6\tau,
\end{aligned}
\label{eq:F2-10-lam}\\[4pt]
\begin{aligned}F_{2}[1,1]={} & \Bigl(\tfrac{3}{16}-\tfrac{13\hat{\lambda}}{48}+\tfrac{7\hat{\lambda}^{2}}{192}\Bigr)+\Bigl(-\tfrac{3}{16}+\tfrac{\hat{\lambda}}{8}-\tfrac{\hat{\lambda}^{2}}{32}\Bigr)\cos2\tau\\
 & +\Bigl(\tfrac{\hat{\lambda}}{16}-\tfrac{\hat{\lambda}^{2}}{64}\Bigr)\cos4\tau+\Bigl(\tfrac{\hat{\lambda}}{12}+\tfrac{\hat{\lambda}^{2}}{96}\Bigr)\cos6\tau.
\end{aligned}
\label{eq:F2-11-lam}
\end{gather}
At $\hat{\lambda}=1$: $\mathbf{K}_{2}(1)=\bigl[\begin{smallmatrix}0 & 1/16\\
-1/16 & 0
\end{smallmatrix}\bigr]$ (eq.~\eqref{eq:K2-omega1}) and $\mathbf{F}_{2}(\tau,1)$ matches
eq.~\eqref{eq:YS-F2} (up to the factor-of-$2$ normalization from
$\tilde{\mathbf{B}}_{n}=[[0,0],[-2\cos2n\tau,0]]$ vs.\ the Mathieu
$\mathbf{B}_{1}=[[0,0],[-\cos2\tau,0]]$), both confirmed numerically.

\medskip{}

\begin{figure}[htbp]
\centering \includegraphics[width=0.82\textwidth]{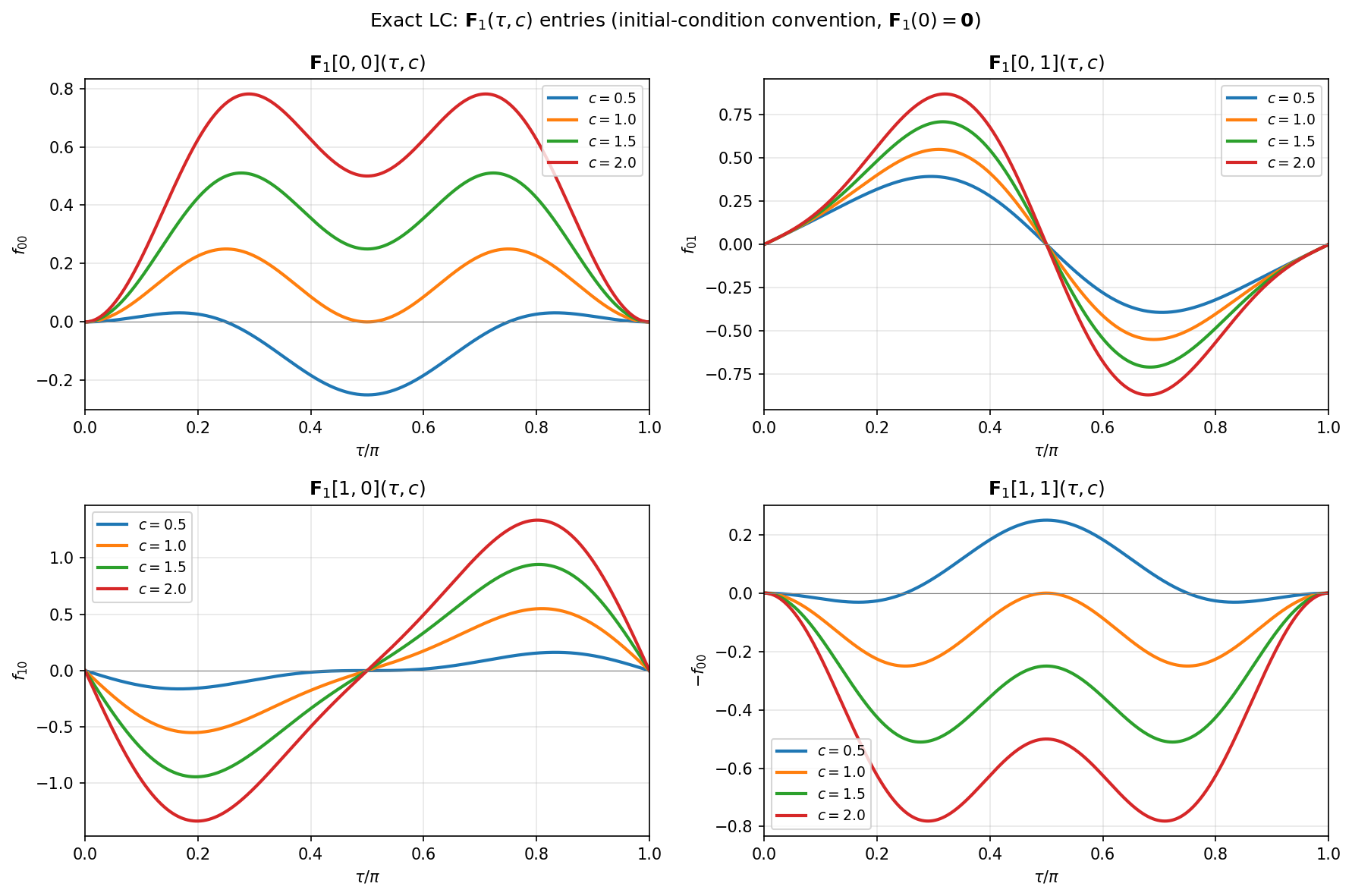}
\caption{Entries of the exact LC first-order Lyapunov factor $\mathbf{F}_{1}(\tau,c)$
(eq.~\eqref{eq:F1-c}, initial-condition convention $\mathbf{F}_{1}(0,c)=\mathbf{0}$)
as functions of $\tau/\pi\in[0,1]$ for $c=0.5$ (blue), $1.0$ (orange),
$1.5$ (green), $2.0$ (red). What we see: (i)~All curves start and
end at zero, confirming $\mathbf{F}_{1}(0,c)=\mathbf{F}_{1}(\pi,c)=\mathbf{0}$
for every $c$. (ii)~The diagonal entries $F_{1}[0,0]=-F_{1}[1,1]$
have a constant offset that grows linearly with $c$ (from the $(3c-2)/8$
term) plus oscillatory Fourier content; at $c=2/3$ the constant term
vanishes and the entries are pure sinusoids. (iii)~The off-diagonal
entries $F_{1}[0,1]$ and $F_{1}[1,0]$ share the $-c\sin4\tau/8$
term but differ in the $\sin2\tau$ coefficient: $(c+1)/4$ vs $(1-3c)/4$,
which changes sign at $c=1/3$. (iv)~At $c=1$ (orange): entries
are exactly twice the genuine Mathieu first-order factor in the same
IC convention, $\mathbf{F}_{1}(\tau,1)=2\,\mathbf{F}_{1}^{{\rm true}}(\tau,0)$
(eq.~\eqref{eq:YS-F1-EPD}; the factor~$2$ from the LC vs Mathieu
normalization of $\mathbf{B}_{1}$). YS's printed matrix eq.~\eqref{eq:YS-F1}
is a different object --- the transformed forcing --- related to
the factor by the derivative identity eq.~\eqref{eq:Math-LC-F1-corrected}.}
\label{fig:LC-F1-entries} 
\end{figure}

\begin{figure}[htbp]
\centering \includegraphics[width=0.82\textwidth]{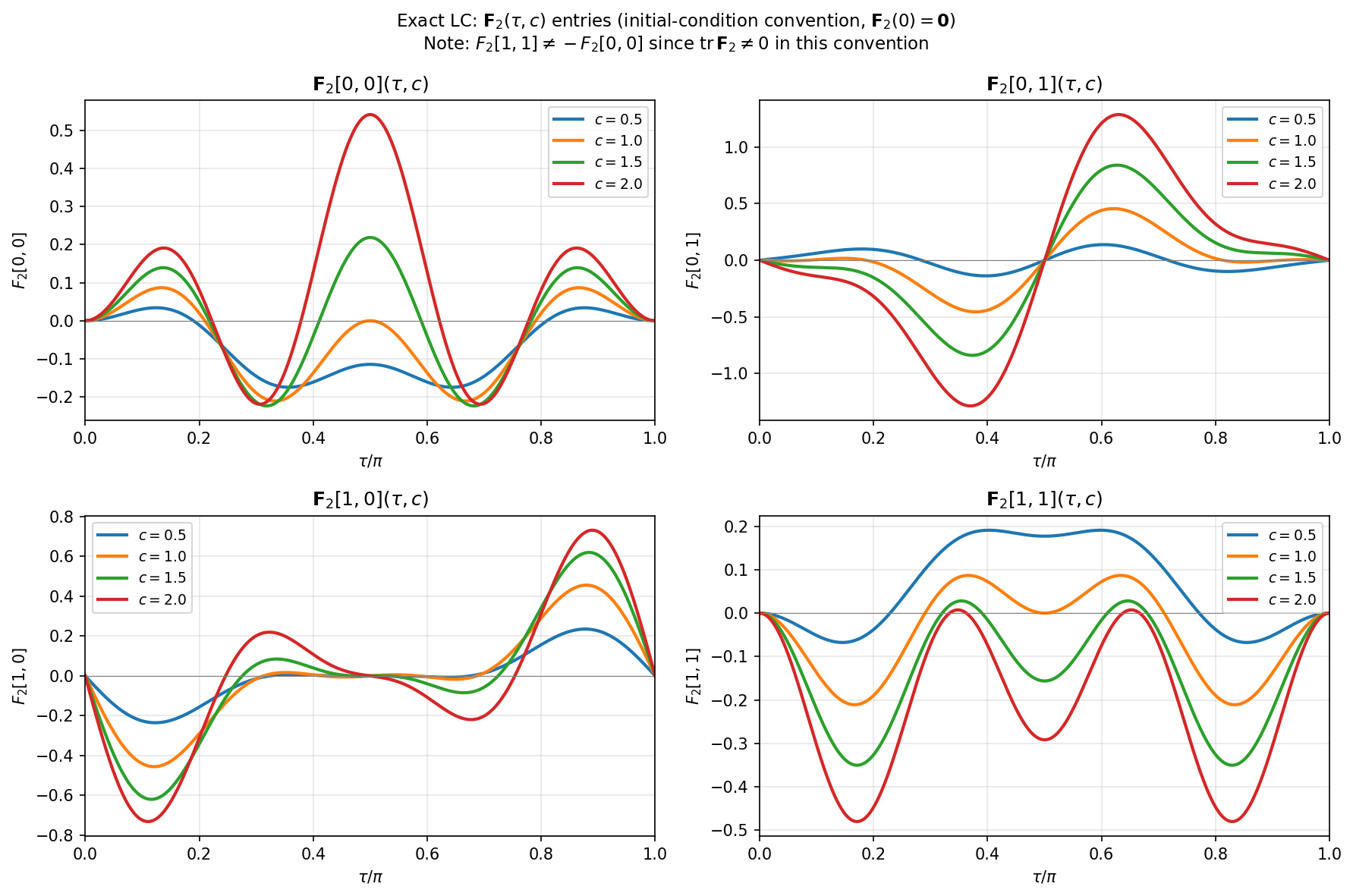}
\caption{Entries of the exact LC second-order Lyapunov factor $\mathbf{F}_{2}(\tau,c)$
(eqs.~\eqref{eq:F2-00-c}--\eqref{eq:F2-11-c}, initial-condition
convention $\mathbf{F}_{2}(0,c)=\mathbf{0}$) for $c=0.5$, $1.0$,
$1.5$, $2.0$. (i)~All curves start and end at zero as required
by the initial-condition convention. (ii)~The entries involve Fourier
modes $\{2,4,6\}$ (one more than $\mathbf{F}_{1}$), consistent with
Remark~\ref{rem:YS-freq-conj}. (iii)~The $[0,0]$ and $[1,1]$
entries are \emph{not} negatives of each other: $\operatorname{tr}\mathbf{F}_{2}\protect\neq0$
in the initial-condition convention (unlike Convention~1 where $[\mathbf{F}_{j}]_{{\rm av}}=\mathbf{0}$
forces $\operatorname{tr}\mathbf{F}_{j}=0$). (iv)~The $[0,1]$ and
$[1,0]$ entries share the $(-c/12-c^{2}/96)\sin6\tau$ term but differ
otherwise, reflecting the asymmetry of $\mathbf{K}_{2}(c)$ (eq.~\eqref{eq:K2-c}).}
\label{fig:LC-F2-entries} 
\end{figure}

\begin{figure}[htbp]
\centering \includegraphics[width=0.82\textwidth]{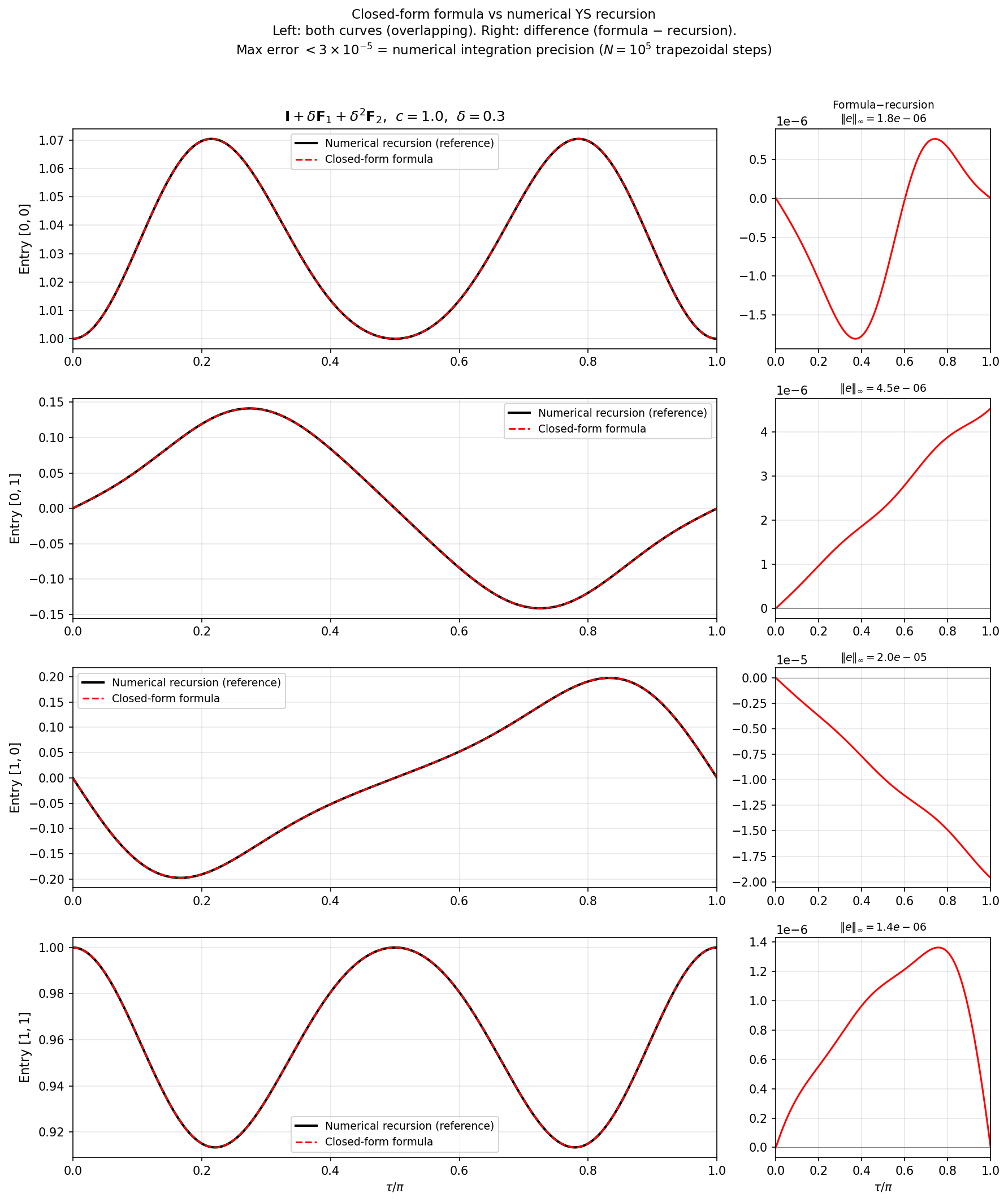}
\caption{Validation of the closed-form YS series formula against the numerical
recursion reference, at $c=1$ (tongue center), $\delta=0.30$, second-order
approximation $\mathbf{I}+\delta\mathbf{F}_{1}(\tau,c)+\delta^{2}\mathbf{F}_{2}(\tau,c)$.
One row per entry $[i,j]$. \emph{Left panels}: the closed-form formula
(red dashed, eqs.~\eqref{eq:F1-c}--\eqref{eq:F2-11-c}) and the
numerical YS recursion (black solid, $N=10^{5}$ trapezoidal steps)
are plotted together --- the curves are visually indistinguishable.
\emph{Right panels}: the difference formula$-$recursion for each
entry. The maximum error is below $3\times10^{-5}$ in every entry,
consistent with the $O(h^{2})\approx10^{-10}$ accuracy of the numerical
integration. This confirms that the closed-form formulas reproduce
the YS recursion to within numerical precision, validating all four
entries including the corrected $F_{2}[1,1]$ (eq.~\eqref{eq:F2-11-c}).}
\label{fig:LC-F2-validation} 
\end{figure}

\begin{figure}[htbp]
\centering \includegraphics[width=0.82\textwidth]{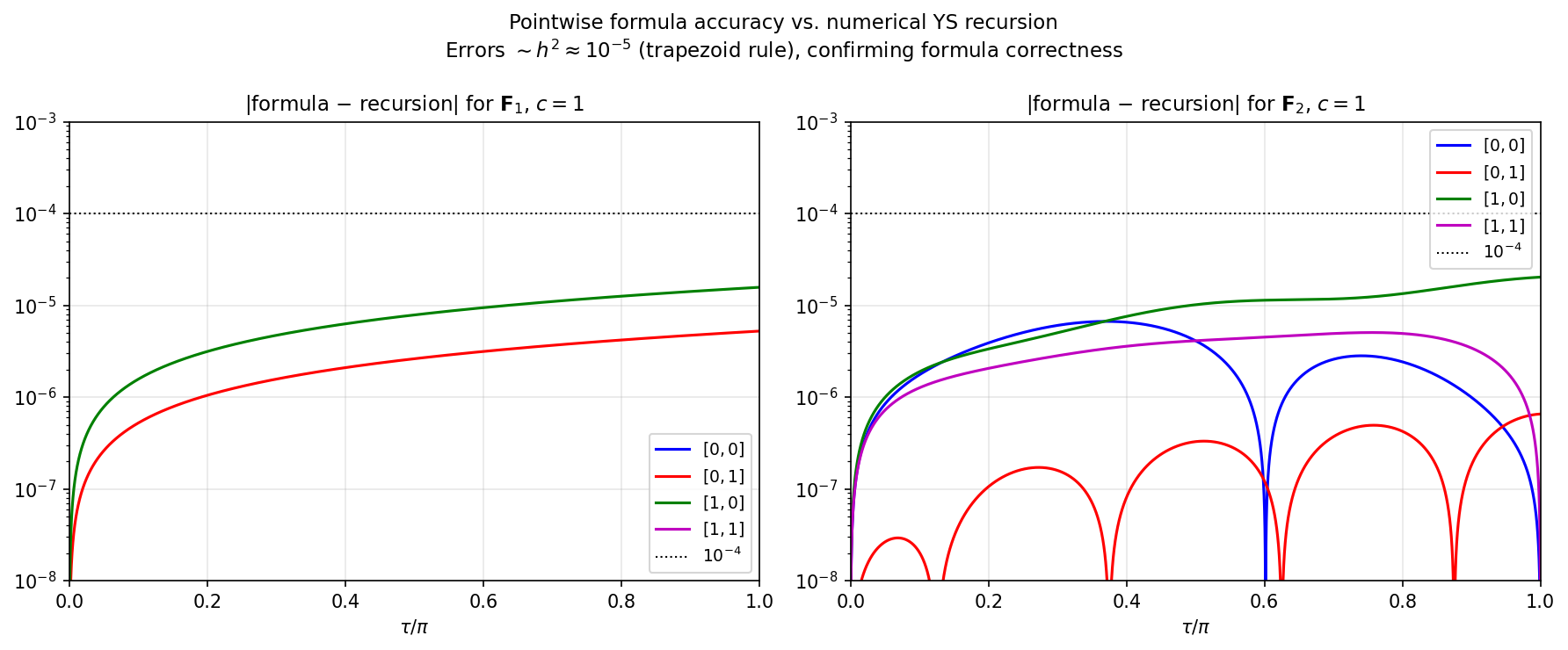}
\caption{Pointwise formula accuracy for $\mathbf{F}_{1}$ (left) and $\mathbf{F}_{2}$
(right) at $c=1$: log-scale plot of $|[\mathbf{F}_{j}]_{{\rm formula}}(\tau,1)-[\mathbf{F}_{j}]_{{\rm recursion}}(\tau,1)|$
for each entry as a function of $\tau/\pi$. All errors lie uniformly
below $10^{-4}$ across the full interval $[0,\pi]$, at a level consistent
with the $O(h^{2})\approx10^{-5}$ accuracy of the trapezoidal integration
used in the recursion. The errors are smooth (not oscillatory) and
grow slightly toward $\tau=\pi/2$ where the integrand is largest,
as expected for cumulative integration error. This confirms that the
closed-form formulas~\eqref{eq:F1-c} and~\eqref{eq:F2-00-c}--\eqref{eq:F2-11-c}
are correct to within the precision of the numerical benchmark.}
\label{fig:LC-formula-accuracy} 
\end{figure}

\begin{figure}[htbp]
\centering \includegraphics[width=0.9\textwidth]{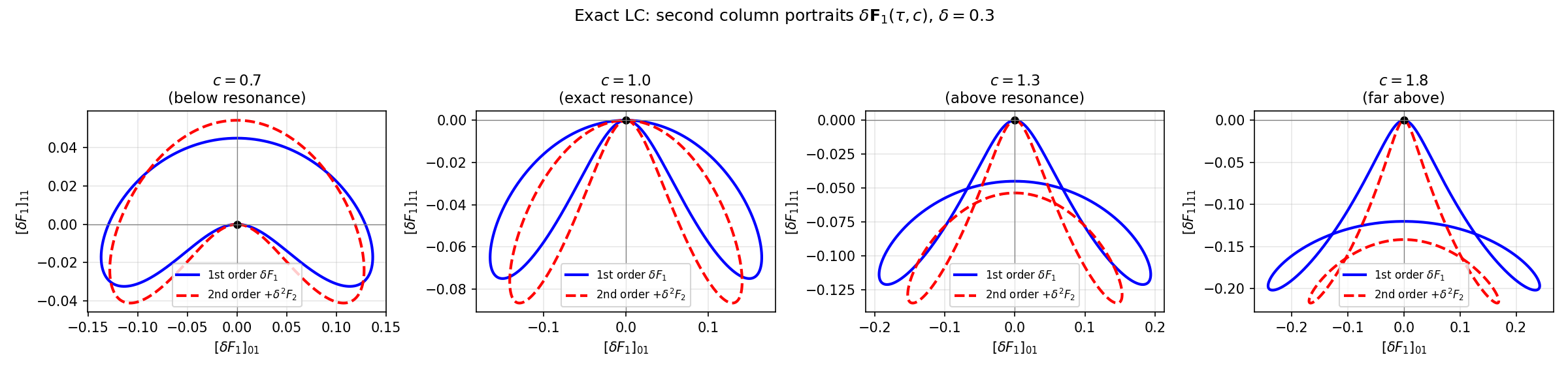}
\caption{Phase portraits of the second column $(\delta[\mathbf{F}_{1}]_{01}(\tau,c),\,\delta[\mathbf{F}_{1}]_{11}(\tau,c))$
of the exact LC first-order Lyapunov factor (eq.~\eqref{eq:F1-c})
at four values of the detuning parameter $c$, for $\delta=0.30$.
\emph{Blue solid}: first-order ($\delta\mathbf{F}_{1}$). \emph{Red
dashed}: second-order ($\delta\mathbf{F}_{1}+\delta^{2}\mathbf{F}_{2}$).
All curves pass through the origin since $\mathbf{F}_{1}(0,c)=\mathbf{F}_{1}(\pi,c)=\mathbf{0}$.
What we see: (i)~At $c=1.0$ (second panel): the portrait is a closed
curve pinched at the origin. At first order, the Mathieu tongue-center
portrait is the marginal self-tangency configuration of Theorem~\ref{thm:F1-topology}
($\mu_{0}=0$; see the corresponding panel of Figure~\ref{fig:YS-F1-transition}),
and the LC portrait is its finite-$\delta$ deformation, with amplitude
carrying the LC normalization factor~$2$. (ii)~As $c$ moves away
from~$1$, the portrait shape changes continuously: below resonance
($c=0.7$) the loop tilts and narrows; above resonance ($c=1.3$,
$1.8$) it tilts the other way and the asymmetry between $K_{1}[0,1]=-1/2$
and $K_{1}[1,0]=1/2-c$ (eq.~\eqref{eq:K1-c}) is visible as an asymmetric
loop shape. (iii)~The second-order correction (dashed) remains close
to the first-order curve near resonance; the deviation grows with
$|c-1|$ (the entries of $\mathbf{F}_{2}$ are quadratic in $c$ while
those of $\mathbf{F}_{1}$ are linear), visible in the $c=1.8$ panel,
consistent with convergence at fixed $c$ for $\delta=0.30$.}
\label{fig:LC-F1-portraits} 
\end{figure}

\noindent\emph{Explicit formulas in terms of $c$.} Substituting
$\hat{\lambda}=c$ in eqs.~\eqref{eq:LC-K1-lam}--\eqref{eq:F2-11-lam}
gives the final $\delta$-free formulas: 
\begin{align}
\mathbf{K}_{1}(c) & =\begin{bmatrix}0 & -\tfrac{1}{2}\\[4pt]
\tfrac{1}{2}-c & 0
\end{bmatrix},\label{eq:K1-c}\\[8pt]
\mathbf{K}_{2}(c) & =\begin{bmatrix}0 & -\tfrac{3}{8}+\tfrac{3c}{8}+\tfrac{c^{2}}{16}\\[4pt]
-\tfrac{1}{8}+\tfrac{3c}{8}-\tfrac{5c^{2}}{16} & 0
\end{bmatrix},\label{eq:K2-c}\\[8pt]
\mathbf{F}_{1}(\tau,c) & =\begin{bmatrix}\tfrac{3c-2}{8}+\tfrac{(1-c)\cos2\tau}{4}-\tfrac{c\cos4\tau}{8} & \tfrac{(c+1)\sin2\tau}{4}-\tfrac{c\sin4\tau}{8}\\[6pt]
\tfrac{(1-3c)\sin2\tau}{4}-\tfrac{c\sin4\tau}{8} & -\tfrac{3c-2}{8}-\tfrac{(1-c)\cos2\tau}{4}+\tfrac{c\cos4\tau}{8}
\end{bmatrix},\label{eq:F1-c}
\end{align}
and the entries of $\mathbf{F}_{2}(\tau,c)$: 
\begin{align}
F_{2}[0,0](\tau,c) & =\Bigl(-\tfrac{1}{16}-\tfrac{c}{24}+\tfrac{11c^{2}}{192}\Bigr)\nonumber \\
 & \quad+\Bigl(\tfrac{1}{16}+\tfrac{c}{8}-\tfrac{3c^{2}}{32}\Bigr)\cos2\tau+\tfrac{3c^{2}}{64}\cos4\tau\nonumber \\
 & \quad+\Bigl(-\tfrac{c}{12}-\tfrac{c^{2}}{96}\Bigr)\cos6\tau,\label{eq:F2-00-c}\\[4pt]
F_{2}[0,1](\tau,c) & =\Bigl(\tfrac{3}{16}-\tfrac{5c}{16}-\tfrac{3c^{2}}{32}\Bigr)\sin2\tau+\Bigl(\tfrac{3c}{16}+\tfrac{3c^{2}}{64}\Bigr)\sin4\tau\nonumber \\
 & \quad+\Bigl(-\tfrac{c}{12}-\tfrac{c^{2}}{96}\Bigr)\sin6\tau,\label{eq:F2-01-c}\\[4pt]
F_{2}[1,0](\tau,c) & =\Bigl(\tfrac{1}{16}-\tfrac{7c}{16}+\tfrac{5c^{2}}{32}\Bigr)\sin2\tau+\Bigl(-\tfrac{c}{4}+\tfrac{c^{2}}{64}\Bigr)\sin4\tau\nonumber \\
 & \quad+\Bigl(-\tfrac{c}{12}-\tfrac{c^{2}}{96}\Bigr)\sin6\tau,\label{eq:F2-10-c}\\[4pt]
F_{2}[1,1](\tau,c) & =\Bigl(\tfrac{3}{16}-\tfrac{13c}{48}+\tfrac{7c^{2}}{192}\Bigr)\nonumber \\
 & \quad+\Bigl(-\tfrac{3}{16}+\tfrac{c}{8}-\tfrac{c^{2}}{32}\Bigr)\cos2\tau+\Bigl(\tfrac{c}{16}-\tfrac{c^{2}}{64}\Bigr)\cos4\tau\nonumber \\
 & \quad+\Bigl(\tfrac{c}{12}+\tfrac{c^{2}}{96}\Bigr)\cos6\tau.\label{eq:F2-11-c}
\end{align}
Note that $\operatorname{tr}\mathbf{F}_{2}=F_{2}[0,0]+F_{2}[1,1]\neq0$
in the initial-condition convention: the constant terms give 
\begin{equation}
\Bigl(-\tfrac{1}{16}-\tfrac{c}{24}+\tfrac{11c^{2}}{192}\Bigr)+\Bigl(\tfrac{3}{16}-\tfrac{13c}{48}+\tfrac{7c^{2}}{192}\Bigr)=\tfrac{1}{8}-\tfrac{5c}{16}+\tfrac{3c^{2}}{32},\label{eq:trF2-IC}
\end{equation}
which is generally nonzero. This is expected: $\operatorname{tr}\mathbf{F}_{j}=0$
holds in the zero-mean convention (Convention~1) but NOT in the initial-condition
convention (Convention~2) for $j\geq2$ (see \S\,\ref{subsec:YS-conventions}).
Every entry is a polynomial in $c$ times standard Fourier modes in
$\tau$ --- globally $\pi$-periodic, no square roots, no irrational
modes. At $c=1$ (Mathieu limit): $\mathbf{K}_{1}(1)=\bigl[\begin{smallmatrix}0 & -1/2\\
-1/2 & 0
\end{smallmatrix}\bigr]$, $\mathbf{K}_{2}(1)=\bigl[\begin{smallmatrix}0 & 1/16\\
-1/16 & 0
\end{smallmatrix}\bigr]$, and $\mathbf{F}_{1}(\tau,1)=\bigl[\begin{smallmatrix}(1-\cos4\tau)/8 & \sin2\tau/2-\sin4\tau/8\\
-\sin2\tau/2-\sin4\tau/8 & -(1-\cos4\tau)/8
\end{smallmatrix}\bigr]$, matching the LC normalization: $\mathbf{F}_{1}(\tau,1)$ is exactly
twice the genuine Mathieu first-order factor~\eqref{eq:YS-F1-EPD}
in the same IC convention (Item~2 of \S\,\ref{subsec:YS-Math-LC};
the relation of YS's printed $\varphi'$ matrix to the factor is,
by contrast, the derivative identity eq.~\eqref{eq:Math-LC-F1-corrected}).
Figures~\ref{fig:LC-F1-entries} and~\ref{fig:LC-F2-entries} display
all four entries of $\mathbf{F}_{1}(\tau,c)$ and $\mathbf{F}_{2}(\tau,c)$
respectively as functions of $\tau$ for representative values of
$c$. Figures~\ref{fig:LC-F2-validation} and~\ref{fig:LC-formula-accuracy}
provide numerical validation: the closed-form formulas agree with
the numerical YS recursion to within the integration precision ($<3\times10^{-5}$
pointwise), confirming the correctness of all entries including the
corrected $F_{2}[1,1]$ (eq.~\eqref{eq:F2-11-c}). Figure~\ref{fig:LC-F1-portraits}
shows the phase portraits of the second column of $\delta\mathbf{F}_{1}(\tau,c)$
at representative $c$ values.

\medskip{}

\noindent\emph{Asymptotic Floquet factor.} The YS series through
second order gives the recursion-defined $\pi$-periodic Lyapunov
factor for the exact LC circuit: 
\begin{equation}
F(\tau,\hat{\lambda},\delta)=\mathbf{I}+\mathbf{F}_{1}(\tau,\hat{\lambda})\,\delta+\mathbf{F}_{2}(\tau,\hat{\lambda})\,\delta^{2}+R(\tau,\hat{\lambda},\delta).\label{eq:LC-F-lam}
\end{equation}
The entries of $\mathbf{F}_{1},\mathbf{F}_{2}$ are exact polynomials
in $\hat{\lambda}$ times standard Fourier modes, so the right-hand
side may be evaluated at any $\hat{\lambda}>0$; the remainder $R$,
however, is small only where the underlying bookkeeping is consistent,
namely at the resonance center. At $\hat{\lambda}=1$ the series approximates
the genuine periodic factor $F=e^{-\tau\mathbf{C}_{0}}X(\tau)\,e^{-\tau K}$
with $\sup_{\tau}|R|\leq0.36\,\delta^{3}$ (verified numerically through
$\delta=0.2$). For fixed $\hat{\lambda}\neq1$ the remainder does
\emph{not} vanish as $\delta\to0$: it tends to the $\delta^{0}$
frame mismatch $e^{-\tau\mathbf{C}_{0}}e^{\tau\mathbf{C}(\hat{\lambda})}e^{-\tau\mathbf{K}^{(0)}(\hat{\lambda})}-\mathbf{I}$,
of size $\approx|\hat{\lambda}-1|/2$, where $\mathbf{K}^{(0)}(\hat{\lambda})$
is the exact $\delta^{0}$ exponent layer of Remark~\ref{rem:K-approx-limitation}.
This is the same order-assignment of the detuning analyzed there:
the series places the block $\mathbf{C}(\hat{\lambda})-\mathbf{C}_{0}$
at order $\delta$, so its $\delta^{0}$ effect on the factor is missed,
and in the tongue regime $\hat{\lambda}-1=O(\delta)$ the missed term
is of the same order as the first-order term itself. Equation~\eqref{eq:LC-F-lam}
is therefore to be read quantitatively at the resonance center; for
general $\hat{\lambda}$ it is exact as a representation of the recursion-defined
factor (the closed forms reproduce the numerical recursion to $<3\times10^{-5}$,
Figure~\ref{fig:LC-formula-accuracy}). After substituting the exact
relation $\hat{\lambda}=c(1+\delta^{2})/(1-\delta^{2})=c+2c\delta^{2}+O(\delta^{4})$
and expanding to order $\delta^{2}$, the $O(\delta^{2})$ shift $\hat{\lambda}-c=2c\delta^{2}$
shifts $\mathbf{F}_{1}\delta$ by $O_{c}(\delta^{3})$, giving: 
\begin{equation}
F(\tau,c,\delta)=\mathbf{I}+\mathbf{F}_{1}(\tau,c)\,\delta+\mathbf{F}_{2}(\tau,c)\,\delta^{2}+R(\tau,c,\delta),\label{eq:LC-F-final}
\end{equation}
where $\mathbf{F}_{j}(\tau,c)$ denotes~\eqref{eq:LC-K1-lam}--\eqref{eq:F2-11-lam}
evaluated at $\hat{\lambda}=c$, and the remainder statement below
eq.~\eqref{eq:LC-F-lam} applies verbatim: $R=O(\delta^{3})$ at
the resonance center $c=1$, while away from $c=1$ the expansion
is quantitatively reliable only as the recursion-defined factor. The
spectral parameter $c=4\omega_{0}^{2}/\mu^{2}$ encodes the detuning
from exact resonance: $c=1$ ($\omega_{0}=\mu/2$) is the primary
resonance center, and both $\mathbf{F}_{1}(\tau,c)$ and $\mathbf{F}_{2}(\tau,c)$
are polynomial in $c$ with standard Fourier modes in $\tau$.

\section{YS, MW, and CF: a three-way comparison}

\label{sec:YS-comparison}

Having developed the YS method for the Mathieu\index{Mathieu equation}
special case (Chapter~\ref{sec:YS-LC}) and the exact LC circuit
(Chapter~\ref{sec:YS-exact-LC}), and with the MW discriminant\index{discriminant}
and CF continued-fraction\index{continued fraction} methods established
in Chapters~\ref{sec:MWInce}--\ref{sec:LCInce}, this chapter makes
a systematic comparison of all three methods. The comparison identifies
where each method has an absolute advantage and explains why they
give equivalent predictions for the LC circuit boundary curve\index{boundary curves}s.

\subsection{Architecture and strengths of each method}

\label{subsec:YS-comparison-full}

\subsubsection{What each method computes}

\emph{MW discriminant method.} The central object is the Hill discriminant\index{discriminant!Hill discriminant}
$\Delta(\lambda)$, an entire function of the spectral parameter.
The boundary condition $\Delta(\lambda)=\pm2$ is a global analytic
statement about the full stability map. MW gives the \emph{why}: the
coexistence polynomials $Q(\mu)$ and $Q^{*}(\mu)$ determine which
instability tongue\index{instability tongue}s survive and which collapse
to zero width, entirely from the algebraic structure of the Ince equation\index{Ince equation}.
The discriminant identity $\Delta(\hat{\lambda}_{m}(\delta),\delta)=+2$
at the coexistence points $\hat{\lambda}_{m}(\delta)=m^{2}+O(\delta^{2})$,
$m$ even (eq.~\eqref{eq:DeltaAtEven}), rests on the coexistence
theorems and explains the absence of even tongues for all $0<\delta<1$
simultaneously (at $\hat{\lambda}=m^{2}$ itself $\Delta=2-O(\delta^{4})$;
Section~\ref{subsec:DeltaId}). The EPD\index{exceptional point of degeneracy (EPD)}
and Krein\index{Krein collision theory} signature\index{Krein signature}
classification (Chapter~\ref{app:Krein}) also flow naturally from
the discriminant: $\Delta=\pm2$ signals a double multiplier at $\pm1$,
and whether the discriminant derivative vanishes there separates the
two cases: a transversal crossing ($\Delta'\neq0$) is the opposite-signature
collision (Jordan block\index{Jordan block}, EPD, open tongue, $\Pi^{*\pm}$),
while a tangential touch ($\Delta'=0$) is the same-signature collision
(semisimple, closed, $\Pi^{**}$) (Remark~\ref{rem:Delta-prime}).

\emph{Continued-fraction method.} The central objects are the eigenvalue
functions $F_{{\rm even}}(c)$ and $F_{{\rm odd}}(c)$, whose zeros
are the instability boundary\index{stability boundary} eigenvalues.
CF gives the \emph{where and how wide}: the exact closed-form width
$L_{m}^{{\rm LC}}$ (eq.~\eqref{eq:Lm-main}, Theorem~\ref{thm:width})
and the exact boundary curves for all $0<\delta<1$ via geometrically
convergent continued fractions (ratio $\delta^{2}$ per level, Poincaré--Perron\index{Poincaré--Perron theorem}).
The CF method is a one-parameter method: it locates boundaries in
the $(c,\delta)$ plane by solving scalar equations, making no contact
with the linear algebra of the state matrix.

\emph{YS exponent matrix method.} The central object is the Floquet\index{Floquet theory}
exponent\index{Floquet theory!Floquet exponent} matrix $K(\varepsilon)=\sum_{m\geq1}\varepsilon^{m}\mathbf{K}_{m}$,
a matrix-valued power series constructed by the explicit Yakubovich--Starzhinskii\index{Yakubovich--Starzhinskii series}
recursion~\eqref{eq:YS-Kl}--\eqref{eq:YS-Fl}. The boundary condition
$K_{01}\cdot K_{10}=0$~\eqref{eq:YS-bdry-Kprod} is a direct algebraic
statement: at the instability boundary one off-diagonal entry of the
Floquet exponent vanishes. This is conceptually the most transparent
of the three formulations: it says exactly that the monodromy $M=(-\mathbf{I})e^{\pi K}$
acquires a Jordan block\index{Jordan block}, without passing through
the discriminant or the continued-fraction eigenvalue equation. The
three conditions are equivalent --- connected by the exact relation
$K=\tfrac{1}{\pi}\log(-M)$ (eq.~\eqref{eq:K-logM}) --- but the
YS formulation makes the linear-algebraic content most explicit.

\subsubsection{Conceptual simplicity for the LC circuit}

For the LC circuit at the primary tongue ($m=1$), the YS method is
conceptually the most direct path from the system matrix to the boundary
condition. The detuning enters the leading behavior of the \emph{exact}
exponent entries linearly --- $K_{01}^{{\rm exact}}\approx\tfrac{\varepsilon}{2}(\mu_{0}-\tfrac{1}{2})$,
$K_{10}^{{\rm exact}}\approx-\tfrac{\varepsilon}{2}(\mu_{0}+\tfrac{1}{2})$
(Remark~\ref{rem:K-approx-limitation}) --- so the EPD boundary
condition $K_{01}\cdot K_{10}=0$ gives $\mu_{0,\pm}=\pm\tfrac{1}{2}$
immediately, and the exact $\mathbf{K}_{1}(\hat{\lambda})$, $\mathbf{K}_{2}(\hat{\lambda})$
(eqs.~\eqref{eq:LC-K1-lam-exact}--\eqref{eq:LC-K2-lam-exact})
make the self-consistent boundary analysis fully explicit. No theory
of entire functions (MW) or continued-fraction convergence (CF, Poincaré--Perron)
is invoked. (Note: the exact condition splits into one entry per boundary
--- $K_{01}^{{\rm exact}}=0$ upper, $K_{10}^{{\rm exact}}=0$ lower.
What fails for the primary tongue is the second-order \emph{truncation}
of the entries, in which the detuning is invisible at leading order
(Remark~\ref{rem:K-approx-limitation}); in practice the condition
is implemented through the complex-basis product $A\cdot B$ of Chapter~\ref{subsec:YS-Mathieu}
or the corrected $\delta^{0}$-layer balance of Remark~\ref{rem:K-corrected-layer}.)

However, the YS method does \emph{not} easily reproduce the finer
structural results that MW and CF deliver for the LC circuit:
\begin{itemize}
\item \emph{Exact closed-form widths.} The closed-form width $L_{m}^{{\rm LC}}$
(Theorem~\ref{thm:width}) is a product of the CF analysis. The YS
series gives the width as an infinite series in $\delta$; matching
the exact closed form would require summing all orders --- which
effectively reproduces the CF result from the outside. 
\item \emph{Vanishing of even tongues.} The MW polynomial argument ($Q(\mu)$
has integer root $\mu=0$, forcing coexistence at every even resonance\index{resonance})
is a one-line algebraic observation. In the YS framework one would
need to carry the expansion adapted to the $m$-th resonance to order
$m$ and show that both off-diagonal exponent entries vanish simultaneously
there (the semisimple case $K=0$ in the frame of that resonance)
--- a separate computation for each even $m$ (cf.\ Remark~\ref{rem:YS-higher-tongues}). 
\item \emph{Center shifts for higher tongues.} The center shifts $D_{3}=-207/64$,
$D_{5}=-1775/192$, \ldots\
 (Table~\ref{tab:bdry-LC}) require a multi-step backward chain computation
via MW or CF. The YS series for higher tongues would need separate
$\mathbf{K}_{m}(\hat{\lambda},\mu_{0})$ computations for each $m$. 
\item \emph{EPD/Krein classification from discriminant.} The MW framework
connects $\Delta=\pm2$ directly to the Krein signature of the colliding
multipliers via the symplectic\index{symplectic system} $2\times2$
dichotomy (Chapter~\ref{app:Krein}, \S\,\ref{subsec:Krein-2x2};
\cite[Ch.~VIII, \S\,1.8]{YakSta2}). The YS boundary condition $K_{01}=0$
identifies the EPD point, but the Krein classification (Lemma~\ref{lem:Krein-Jordan},
Theorem~\ref{thm:Krein}) requires additional input from the monodromy
structure. 
\end{itemize}

\subsubsection{The absolute advantage of YS: multiparameter systems}

The comparison above concerns the single-parameter LC circuit ($\delta$
as the only small parameter). When the system depends on \emph{multiple
parameters} --- as in the stability problems studied by Seyranian
and Mailybaev~\cite[Ch.~1]{SeyMai} and Kirillov~\cite[Sec.~3.3]{Kiril}
--- the YS method has an absolute advantage that MW and CF cannot
match.

The key point is that the YS series coefficients $\mathbf{K}_{m}(\hat{\lambda},\mu_{0},\ldots)$
and $\mathbf{F}_{m}(t,\hat{\lambda},\mu_{0},\ldots)$ are \emph{jointly
analytic} in all parameters simultaneously (Theorem~\ref{thm:joint-analytic},
Chapter~\ref{app:genLin}; Hartogs' theorem~\cite[Ch.~2]{FuksBVA}).
This means: 
\begin{itemize}
\item The \emph{shape} of the instability tongue in the full parameter space
(not just one cross-section) is encoded in the matrix function $K(\varepsilon_{1},\varepsilon_{2},\ldots)$. 
\item \emph{Multiple interacting resonances} --- where eigenvalues of $\operatorname{ad}_{\mathbf{K}_{0}}$\index{adjoint action}
(eq.~\eqref{eq:ad-K0} of Chapter~\ref{app:YS-Floquet}) corresponding
to different parameter directions nearly coincide --- are handled
by a single recursion without separate analyses per resonance. 
\item The \emph{boundary surface} in multiparameter space is still given
by $K_{01}\cdot K_{10}=0$, now as an equation in all parameters simultaneously.
Expanding in any one parameter while holding others fixed, or expanding
jointly, are both valid by joint analyticity. 
\item The \emph{codimension-2 singularities} of the stability boundary ---
where two tongues coalesce or cross, as studied by Seyranian--Mailybaev
and Kirillov --- correspond to simultaneous vanishing of multiple
entries of $K$, a condition directly readable from the matrix structure. 
\end{itemize}
The MW discriminant is a function of the spectral parameter $\hat{\lambda}$
alone, at fixed system structure. Extending MW to a multiparameter
family requires recomputing the entire discriminant expansion for
each new system, with no natural analytic dependence on the extra
parameters. Similarly, the CF eigenvalue functions $F_{{\rm even/odd}}(c)$
are scalar functions; their generalization to multiparameter settings
would require new scalar equations for each parameter direction, losing
the structural coherence of the matrix approach.

In summary: for the LC circuit specifically, MW and CF give richer
structural information (exact widths, coexistence, EPD classification)
more directly. For multiparameter families of oscillators --- the
setting of Seyranian--Mailybaev and Kirillov --- YS is the natural
and powerful framework, with joint analyticity and the algebraic boundary
condition $K_{01}\cdot K_{10}=0$ replacing the need for a new analysis
at each parameter point.

\subsubsection{Unifying relation}

All three methods are connected by the exact identity 
\begin{equation}
K(\varepsilon)=\frac{1}{\pi}\log\bigl(-M(\varepsilon)\bigr),\label{eq:K-logM}
\end{equation}
where $M(\varepsilon)=X(\pi,\varepsilon)$ is the monodromy matrix\index{monodromy matrix}
(principal matrix logarithm, well defined near the primary resonance,
where $-M$ is close to $\mathbf{I}$). The MW condition $\Delta=\operatorname{Tr}(M)=-2$
(the boundary value for the surviving, odd-$m$ tongues), the CF condition
$F_{{\rm even/odd}}(c)=0$ (equivalent to $\rho_{1}=\rho_{2}=-1$,
i.e.\ $M=-\mathbf{I}+\text{nilpotent}$), and the YS condition $K_{01}\cdot K_{10}=0$
(equivalent to $e^{\pi K}$ having a unit eigenvalue of multiplicity
2) are three faces of the same geometric fact: at an instability boundary
the Floquet multiplier\index{Floquet theory!Floquet multiplier}s
coalesce at $-1$ with a Jordan block. The relation~\eqref{eq:K-logM}
is the bridge that makes this equivalence explicit.

\section{Exceptional points of degeneracy: an elementary proof and the Hamiltonian--Hopf
connection}

\label{sec:EPD-elementary}

The preceding chapters have located and computed the instability boundary
curve\index{boundary curves}s $c_{\pm}^{(m)}(\delta)$ of the LC
circuit (Theorems~\ref{thm:EPD-bdry} and~\ref{thm:width}). The
present chapter establishes what happens at these boundary points:
the monodromy matrix\index{monodromy matrix} $X(\pi)$ has a non-trivial
Jordan block\index{Jordan block}, i.e.\ the boundary points are
\emph{exceptional points of degeneracy} (EPDs) of the monodromy. This
is the structural property that makes the sensing application of Chapter~\ref{sec:EPD-sensor}
possible.

The proof is elementary, using only the Hill multiplier formula~\eqref{eq:Hill-multipliers}
and the following classical perturbation bound for diagonalizable
matrices (the Bauer--Fike theorem, \cite[Sec.~6.3]{HorJohn}; restated
in~\cite[Prop.~9]{FigPert}):
\begin{prop}[Perturbation of a diagonalizable matrix (Bauer--Fike)]
\label{prop:perdiag-LC} Let $A$ be a diagonalizable $n\times n$
matrix, $A=S\Lambda S^{-1}$ with $S$ nonsingular and $\Lambda$
diagonal. If $\widetilde{\lambda}$ is an eigenvalue of $A+B$, then
there exists an eigenvalue $\lambda$ of $A$ such that 
\begin{equation}
|\widetilde{\lambda}-\lambda|\leq\kappa_{\infty}(S)\,\|B\|_{\infty},\label{eq:perdiag-bound}
\end{equation}
where $\kappa_{\infty}(S)$ is the condition number of $S$ in the
$\infty$-norm. In particular, if $B=B(\varepsilon)$ with $\|B(\varepsilon)\|_{\infty}=O(\varepsilon)$,
then $|\widetilde{\lambda}(\varepsilon)-\lambda|=O(\varepsilon)$:
eigenvalue perturbations of a diagonalizable matrix are \emph{at most
linear} in the perturbation parameter. 
\end{prop}

\begin{thm}[Jordan block structure at boundary points]
\label{thm:Jordan-bdry} Let $c_{0}=c_{\pm}^{(m)}(\delta)$, $m$
odd, be an instability boundary\index{stability boundary} point of
the LC Hill equation~\eqref{eq:LC-Hill}, so that $\Delta(c_{0})=-2$
and the two Floquet\index{Floquet theory} multiplier\index{Floquet theory!Floquet multiplier}s
coincide: $\rho_{1}=\rho_{2}=-1$. Assume $\Delta'(c_{0})\neq0$.
Then the monodromy matrix $X(\pi)$ at $c_{0}$ is \emph{not} diagonalizable;
equivalently, it has a non-trivial $2\times2$ Jordan block with eigenvalue
$-1$. 
\end{thm}

\begin{proof}
Since $\Delta(c)=\mathrm{Tr}\,X(\pi,c)$ is analytic in $c$ (Theorem~\ref{thm:matrizant-analytic}),
writing $c=c_{0}+h$ gives: 
\begin{equation}
\Delta(c_{0}+h)=-2+\Delta'(c_{0})\,h+O(h^{2}).\label{eq:Delta-expand-bdry}
\end{equation}
Substituting into the Hill multiplier formula~\eqref{eq:Hill-multipliers}:
\begin{align}
\rho_{1,2}(c_{0}+h) & =\tfrac{1}{2}\Delta\pm\sqrt{\tfrac{1}{4}\Delta^{2}-1}\nonumber \\
 & =-1+\tfrac{1}{2}\Delta'(c_{0})\,h\pm\sqrt{-\Delta'(c_{0})\,h+O(h^{2})},\label{eq:rho-expand-bdry}
\end{align}
so the multiplier split satisfies 
\begin{equation}
|\rho_{1}-\rho_{2}|=2\sqrt{|\Delta'(c_{0})|}\cdot\sqrt{|h|}\cdot\bigl(1+O(h)\bigr)\sim\kappa\sqrt{|h|}\quad\text{as }h\to0,\label{eq:rho-split-sqrt}
\end{equation}
where $\kappa=2\sqrt{|\Delta'(c_{0})|}\neq0$ (since $\Delta'(c_{0})\neq0$
by hypothesis). Now suppose for contradiction that $X(\pi)$ at $c_{0}$
is diagonalizable: $X(\pi)=S\Lambda S^{-1}$ with $\Lambda$ diagonal.
Since $X(\pi,c)$ is analytic in $c$ (Theorem~\ref{thm:matrizant-analytic}),
we have $\|X(\pi,c_{0}+h)-X(\pi,c_{0})\|_{\infty}=O(|h|)$. By Proposition~\ref{prop:perdiag-LC}
applied to $A=X(\pi,c_{0})$ and $B=X(\pi,c_{0}+h)-X(\pi,c_{0})$,
any eigenvalue $\widetilde{\rho}$ of $X(\pi,c_{0}+h)$ satisfies
$|\widetilde{\rho}-(-1)|=O(|h|)$, hence 
\begin{equation}
|\rho_{1}-\rho_{2}|=O(|h|).\label{eq:rho-split-linear}
\end{equation}
Since $\kappa\neq0$,~\eqref{eq:rho-split-sqrt} gives $|\rho_{1}-\rho_{2}|/|h|\to\infty$
as $h\to0$, which contradicts~\eqref{eq:rho-split-linear}. Therefore
$X(\pi)$ at $c_{0}$ cannot be diagonalizable, and since it is a
$2\times2$ matrix with a double eigenvalue $-1$, it must have a
non-trivial Jordan block. 
\end{proof}
\begin{rem}[Role of $\Delta'(c_{0})\neq0$]
\label{rem:Delta-prime} The condition $\Delta'(c_{0})\neq0$ ensures
that the boundary curve $\Delta(c)=-2$ is crossed transversally at
$c_{0}$, so that the multiplier split is genuinely of order $\sqrt{|h|}$
rather than higher order. For the LC circuit this condition holds
at every boundary point of the open (odd-$m$) instability tongues
for $\delta\neq0$, by a classical fact of discriminant\index{discriminant}
theory: $\Delta'$ can vanish at a point where $\Delta=\pm2$ only
if the corresponding periodic or antiperiodic eigenvalue is \emph{double},
i.e.\ only at a closed gap \cite[Sec.~2.1, Cor.~2.1 and Lem.~2.5]{MagWin}.
The odd tongues are open for $\delta\neq0$ --- their widths are
positive by Theorem~\ref{thm:width} --- so their endpoints are
simple antiperiodic eigenvalues and $\Delta'(c_{\pm}^{(m)})\neq0$
there. (At the closed even gaps, by contrast, $\Delta$ touches the
line $\Delta=2$ tangentially: $\Delta'$ does vanish there, and correspondingly
no Jordan block forms --- the monodromy is $+I$, the semisimple
coexistence case.) The argument above is thus an alternative and self-contained
proof of the EPD\index{exceptional point of degeneracy (EPD)} structure
stated in Theorem~\ref{thm:EPD-bdry}, using only the Hill discriminant\index{discriminant!Hill discriminant}
and elementary perturbation theory --- no symplectic\index{symplectic system}
structure or Krein\index{Krein collision theory} signature\index{Krein signature}
analysis is required. 
\end{rem}

\begin{rem}[Why the argument works: the $2\times2$ structure of Hill's equation\index{Hill equation}]
\label{rem:2x2-Jordan} The argument of Theorem~\ref{thm:Jordan-bdry}
succeeds because Hill's equation is a scalar second-order ODE, so
its monodromy matrix $X(\pi)$ is always $2\times2$. For a $2\times2$
matrix with a double eigenvalue there are exactly two possibilities:
it is either a scalar multiple of the identity (diagonalizable, every
vector an eigenvector) or it has a non-trivial $2\times2$ Jordan
block (non-diagonalizable). There is no third option. Once Theorem~\ref{thm:Jordan-bdry}
rules out diagonalizability via the $\sqrt{|h|}$ contradiction, the
Jordan block conclusion is \emph{immediate} --- no further structure
is needed.

This is in sharp contrast with the Krein signature approach (Chapter~\ref{app:Krein}),
which is required when $X(\pi)$ has dimension larger than two (e.g.\ matrix-valued
Hill equations or coupled oscillator systems): there, ruling out full
diagonalizability does not determine the Jordan structure, and the
symplectic geometry of Krein signatures (Lemma~\ref{lem:Krein-Jordan},
Theorem~\ref{thm:Krein}) is essential to identify which coincident
multipliers open instability tongue\index{instability tongue}s and
which remain semisimple.

For the scalar Hill equation the two approaches are therefore complementary:
the discriminant argument gives the Jordan block conclusion most directly
and elementarily, while the Krein approach (Chapter~\ref{app:Krein})
gives the deeper geometric explanation of \emph{why} odd tongues open
and even ones collapse --- a distinction the discriminant argument
alone cannot see. 
\end{rem}

Having established, by an elementary route, that the boundary points
are exceptional points of degeneracy, we close the chapter by placing
this degeneracy within the wider vocabulary of stability theory, where
the same object is met under several other names.

\subsection{EPD and the Hamiltonian--Hopf bifurcation: a terminological clarification}

\label{sec:EPD-vs-Hopf}

The degeneracy studied in this chapter --- a defective coalescence
of two Floquet multipliers, where the monodromy acquires a non-trivial
Jordan block\index{Jordan block} --- carries different names in
different traditions, and a reader arriving from dynamical systems
or mechanics may reasonably ask how our \emph{exceptional point of
degeneracy} relates to the \emph{Hopf bifurcation}. The two notions
are closely related but not identical, and the distinction is worth
stating precisely.

\subsubsection*{Three Hopf bifurcations, only some defective}

The classical \emph{Hopf bifurcation} --- introduced by Poincaré,
worked out in the plane by Andronov and collaborators in the 1930s,
and extended to arbitrary finite dimension by Hopf in 1942~\cite[\S\,3.4]{GucHol},~\cite[\S\,3.5]{Kuzn}
--- requires a configuration that a Hamiltonian system cannot present:
in a one-parameter family of vector fields a \emph{simple} complex-conjugate
pair of eigenvalues crosses the imaginary axis transversally while
\emph{no other eigenvalue lies on the axis}, and a limit cycle is
born. (This last condition, essential to the centre-manifold reduction
underlying the theorem, is what the Hamiltonian case violates.) The
Hamiltonian spectral symmetry $\lambda\mapsto-\lambda$ (and $\lambda\mapsto\bar{\lambda}$)
makes such a crossing impossible: a \emph{simple} pure-imaginary pair
is locked on the axis, since leaving it would force the pair to become
a symmetric quadruplet $\{\lambda,-\lambda,\bar{\lambda},-\bar{\lambda}\}$
--- four eigenvalues where there were two. Eigenvalues of a Hamiltonian
system can leave the imaginary axis only after \emph{colliding} there,
so the classical theorem applies only outside the conservative class,
and the conservative route to instability is forced through the defective
collision described next. At its bifurcation point, moreover, the
linearization is \emph{diagonalizable}: the eigenvalues are simple,
with no Jordan block and no coalescence of eigenvectors. In this precise
sense the classical Hopf bifurcation is \emph{not} an exceptional
point of degeneracy.

The situation changes for conservative and for time-reversible systems.
When a Hamiltonian system loses stability through the collision of
two pure-imaginary eigenvalues that then split into a complex-conjugate
quadruplet, the linearization at the collision is \emph{non-semisimple}
--- a genuine Jordan block --- and this event is the \emph{linear
Hamiltonian--Hopf bifurcation}~\cite{vdMeer}. Kirillov records
its synonymy across the literature directly: this process ``is known
as the linear Hamiltonian--Hopf bifurcation, the onset of flutter,
nonsemisimple $1\!:\!1$ resonance or the Krein collision''~\cite[Sec.~3.3.2]{Kiril};
the phenomenon traces to Krein's foundational study of canonical systems~\cite{Krein50}.
The time-reversible analogue, in which the same defective collision
occurs in a circulatory system, is the \emph{reversible Hopf bifurcation}~\cite[Ch.~1]{Kiril},
\cite{LamRob98}; the symmetric and reversible normal forms are classified
in~\cite{DelMelMar92}. Both of these --- unlike the classical Hopf
bifurcation, which is non-defective --- pass through a defective
degeneracy, that is, through an exceptional point.

\subsubsection*{Exceptional versus diabolical}

The relevant distinction concerns the \emph{Jordan structure} at a
multiple eigenvalue --- a similarity-invariant property, unlike Hermiticity,
which a generic change of basis destroys without altering the spectrum.
(The phenomenon is most often studied in the \emph{non-Hermitian}
setting only because a Hermitian operator is always diagonalizable
and so cannot exhibit the defective case at all; non-Hermiticity is
necessary for it but is not itself the operative property.) Two kinds
of coalescence arise, aligning exactly with the semisimple/non-semisimple
dichotomy of mechanics. A \emph{diabolical point} is a \emph{semisimple}
double eigenvalue: two eigenvalues coincide but two independent eigenvectors
survive, and the operator remains diagonalizable~\cite[Sec.~11.1.1]{Kiril}.
An \emph{exceptional point} is a \emph{non-semisimple} (defective)
double eigenvalue: the eigenvectors coalesce and a Jordan block forms~\cite[Sec.~11.1.2]{Kiril}.
Our EPD is the exceptional (defective) case; equivalently, in the
mechanics vocabulary, the non-semisimple $1\!:\!1$ resonance. The
diabolical (semisimple) case is precisely what occurs at the \emph{even}
resonances of the LC circuit, where the colliding multipliers carry
the same Krein signature and the tongue collapses to zero width (Chapter~\ref{app:Krein});
the exceptional case is what occurs at the \emph{odd} resonances,
where the signatures are opposite and the tongue opens. The systematic
multiparameter classification of such same-kind and opposite-kind
collisions originates with Huseyin~\cite{Huseyin78} and Seyranian--Mailybaev,
who develop it for periodic systems through the bifurcation theory
of multipliers \cite[Sec.~9.7]{SeyMai} rather than through the Krein
signature: in their vocabulary the defective (nonderogatory) collision
of a double multiplier --- our EPD --- is a \emph{strong interaction}
\cite[Sec.~2.6, Sec.~9.7.1]{SeyMai}, while the semisimple collision
--- the diabolical point --- is a \emph{weak interaction} \cite[Sec.~2.9, Sec.~9.7.3]{SeyMai}.
The two accounts reach the same conclusions by independent routes.

\subsubsection*{The LC instability boundary as a curve of Hamiltonian--Hopf bifurcations}

For the modulated LC circuit the connection is now exact. The system
is conservative (symplectic), so the relevant event is the Hamiltonian
variant. In the two-parameter $(p,\gamma)$ plane the EPD locus is
the instability boundary curve $c_{\pm}^{(m)}(\delta)$ of Theorem~\ref{thm:EPD-bdry}:
a generic point of this curve is a regular Hamiltonian--Hopf bifurcation,
at which two unit-circle multipliers of opposite Krein signature collide
into a defective double multiplier and then leave the circle, opening
the instability tongue. This is exactly Kirillov's description of
a regular boundary point, at which ``varying parameters along a curve\ldots we
have a linear Hamilton--Hopf bifurcation\ldots which is a regular
point of the boundary between the domains of stability and oscillatory
instability''~\cite[Ch.~12]{Kiril}; the same regular boundary point
is, in the multiplier picture of Seyranian--Mailybaev, a nonderogatory
double multiplier $\rho_{0}=\pm1$~\cite[Sec.~9.7.1, Sec.~10.1]{SeyMai};
the same picture appears as the stability boundary of Kirillov's ideal
(undamped) systems, whose marginal-stability boundary ``consists
of exceptional points''~\cite[Ch.~1]{Kiril}. The special points
of our boundary --- the closed even gaps, and the confluence point
$(p,\gamma)=(1/\sqrt{8},\tfrac{1}{2})$, where the right $m=1$ boundary,
the $m=2$ curve, and the $m=3$ tongue converge (Figure~\ref{fig:stability})
--- are the higher-codimension singularities of this curve, the analogues
of the Whitney-umbrella and cuspidal points catalogued in the multiparameter
theory~\cite[Ch.~1]{Kiril},~\cite[Sec.~10.2]{SeyMai}, whose general
classification of the singularities of the stability boundary of a
periodic system is given in~\cite[Sec.~10.2--10.3]{SeyMai}. That
the stability boundary of a system with such a collision carries a
Whitney-umbrella singularity was first established by Bottema~\cite{Bottema56}.

A caution from the same theory explains why the two vocabularies must
be kept distinct rather than merged. The dissipative and Hamiltonian
onsets do \emph{not} in general coincide in the limit of vanishing
dissipation: ``the onset of the classical Hopf bifurcation in a near-Hamiltonian
dissipative system generically does not converge to the onset of the
Hamilton--Hopf bifurcation of a Hamiltonian system when dissipation
tends to zero''~\cite[Ch.~12]{Kiril} --- the Ziegler destabilization
paradox, governed by the Whitney-umbrella singularity (\cite{Langford03};
\cite{KirVer10}); the analogous discontinuous jump of the parametric-resonance
domain under infinitely small damping is analyzed by Seyranian--Mailybaev~\cite[Sec.~11.7.3]{SeyMai},
and the multi-symplectic description of the conservative collision
is due to Bridges~\cite{Bridges97}. For the purely conservative
LC circuit of this book this limit plays no role, but it is the reason
the operator-theoretic notion (the EPD, which is indifferent to dissipation)
and the dynamical notion (the Hamiltonian--Hopf bifurcation, a conservative
stability transition) are best named separately even though they coincide
in our setting.

\subsubsection*{Why we retain \textquotedblleft EPD\textquotedblright}

We keep the term \emph{exceptional point of degeneracy} throughout
this book for three reasons. It is the established name in the non-Hermitian-degeneracy
and degeneracy-sensing literature that the application of \S\,\ref{subsec:EPD-sensing-MR}
addresses; it names the operator-theoretic object itself --- a defective
monodromy --- without reference to dissipation, which is appropriate
for a property of the monodromy matrix; and in the conservative setting
of the LC circuit it coincides with the linear Hamiltonian--Hopf
bifurcation, so no generality is lost. The dictionary is therefore
simply stated: in this book an EPD of the monodromy is, in the language
of Hamiltonian mechanics, a linear Hamiltonian--Hopf bifurcation
(equivalently a Krein collision, or a non-semisimple $1\!:\!1$ resonance);
it is distinct from the classical (Poincaré--Andronov--Hopf) bifurcation,
which is non-defective and which we do not encounter here. The foundational
treatments are van der Meer~\cite{vdMeer} for the Hamiltonian--Hopf
normal form and Kirillov~\cite{Kiril} for the unified operator-theoretic
and mechanical account; the multiparameter perturbation theory of
the surrounding singularities is that of Seyranian and Mailybaev~\cite{SeyMai}.

\section{EPD hypersensitivity of the modulated LC circuit}

\label{sec:EPD-sensor}

\emph{Operating premise.} Throughout this chapter the circuit parameters
--- modulation frequency $\mu$, modulation amplitude $\varepsilon$
(equivalently the Mobius\index{Mobius transformation} parameter $\delta$),
inductance $L$, and nominal capacitance\index{capacitance sensing}
$C_{0}$ --- are \emph{fixed by design} so that the circuit sits
at an EPD\index{exceptional point of degeneracy (EPD)} point $c_{0}=c_{\pm}^{(m)}(\delta)$
(Theorem~\ref{thm:EPD-bdry}).

\begin{figure}[htbp]
\centering \includegraphics[width=0.96\textwidth]{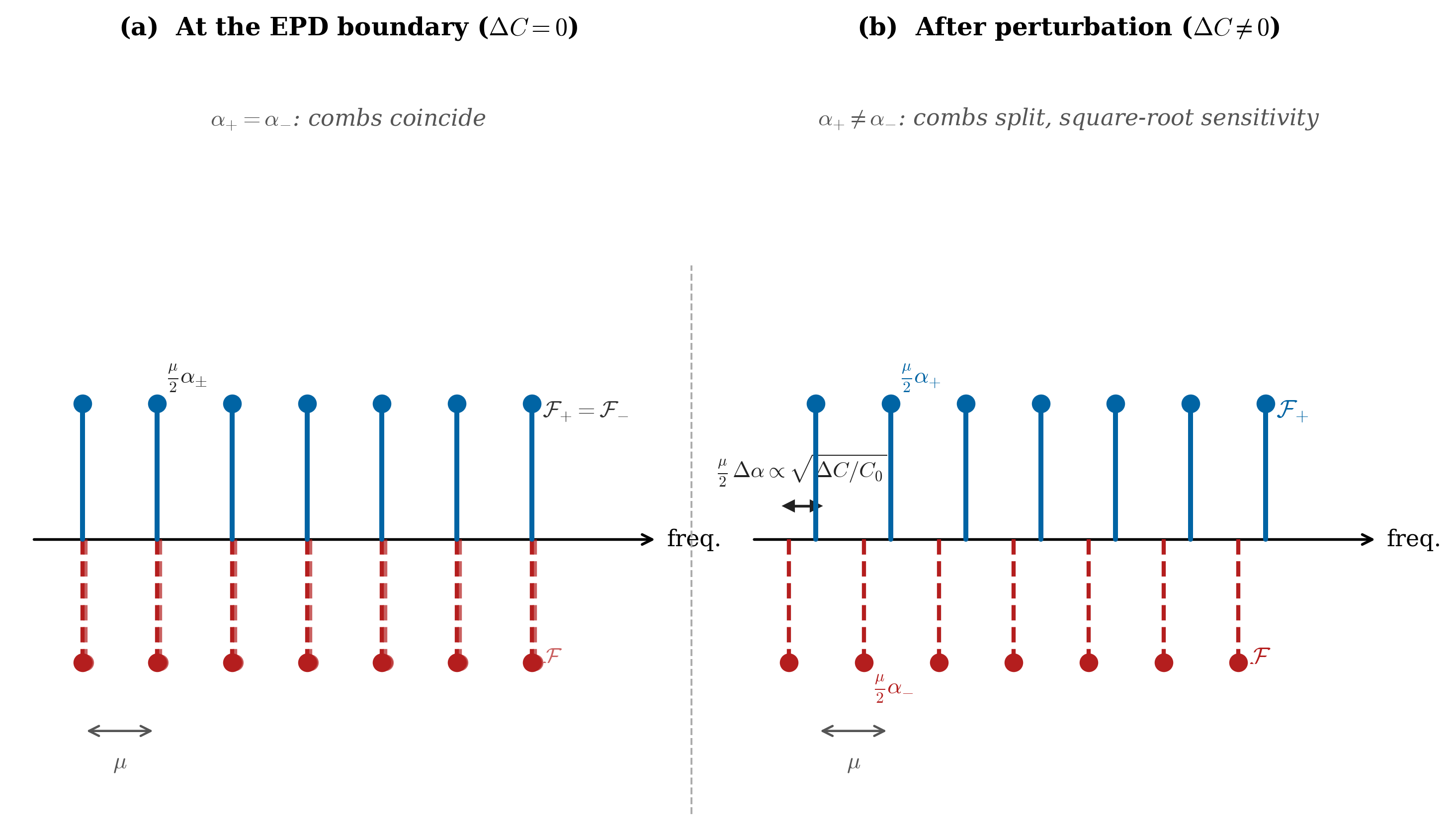}
\caption{The Floquet frequency comb as the sensing observable. Each Floquet
mode $\pm$ generates a comb $\mathcal{F}_{\pm}=\{\frac{\mu}{2}\alpha_{\pm}+n\mu:n\in\mathbb{Z}\}$
of spectral lines spaced by the driving frequency $\mu$. \emph{(a)}
At the EPD curve ($\Delta C=0$): the characteristic exponents coincide,
$\alpha_{+}=\alpha_{-}=m$, and the two combs are identical --- all
lines fall on odd multiples of $\mu/2$. \emph{(b)} After a capacitance
perturbation $\Delta C\protect\neq0$: the exponents split, $\alpha_{+}\protect\neq\alpha_{-}$,
and the combs separate by $\frac{\mu}{2}\Delta\alpha$ at every harmonic
$n$ simultaneously. The split scales as $\Delta\alpha\propto\sqrt{\Delta C/C_{0}}$
(Theorem~\ref{thm:EPD-sensor-main}), providing square-root sensitivity.
Measuring $\Delta\alpha$ across $N$ comb lines reduces the statistical
error by $\sqrt{N}$ without additional hardware.}\index{exceptional point of degeneracy (EPD)!EPD curve}
\label{fig:floquet-comb} 
\end{figure}

\begin{figure}[htbp]
\centering \includegraphics[width=1\textwidth]{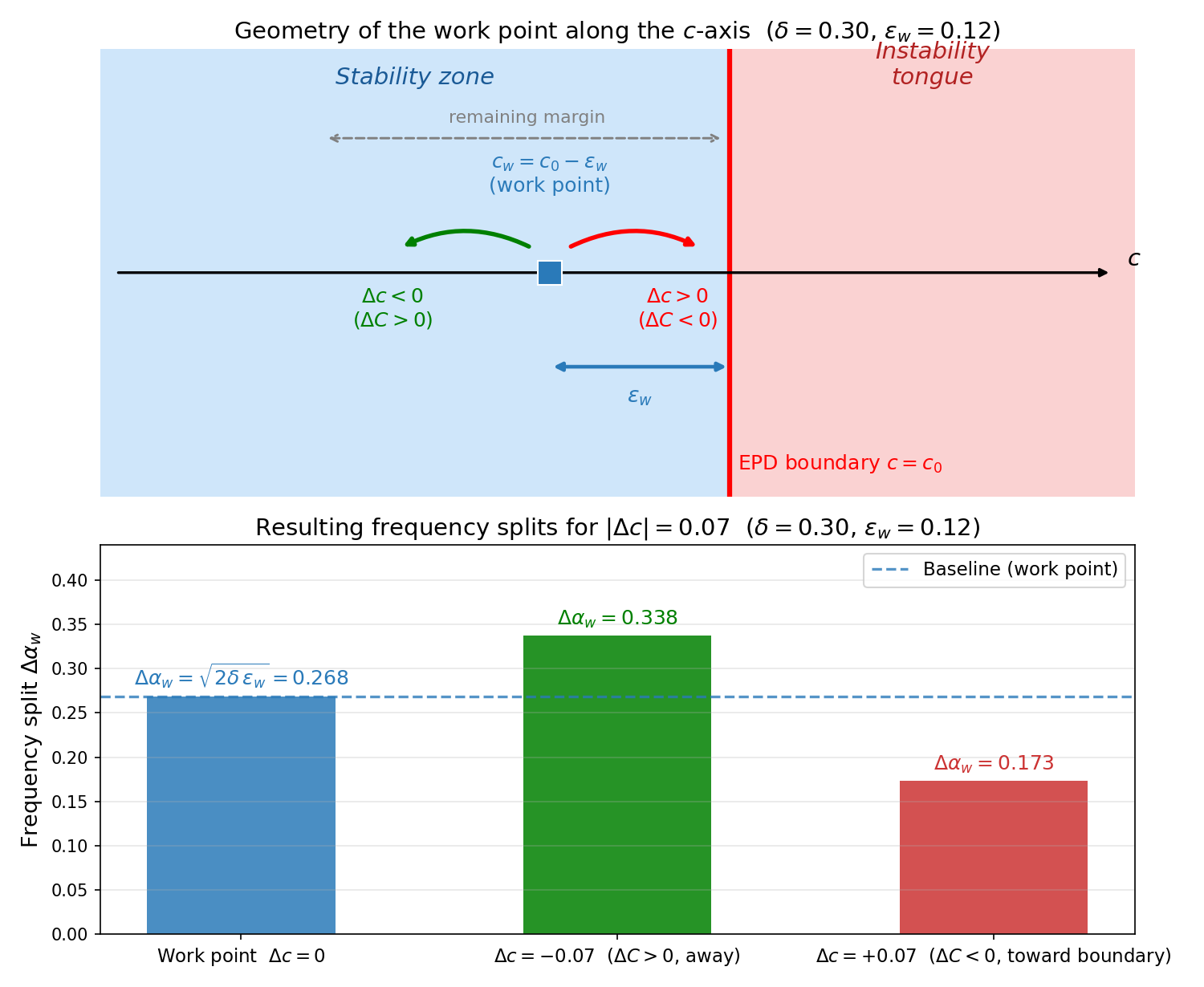}
\caption{Geometry of the work-point strategy for $c_{0}=c_{-}^{(1)}$ ($\delta=0.30$,
$\epsilon_{w}=0.12$, $|\Delta c|=0.07$). \emph{Top:} Schematic of
the $c$-axis near the EPD curve $c=c_{0}$ (red vertical line). The
stability zone (blue shading) lies to the left ($c<c_{0}$) and the
instability tongue (red shading) to the right. The work point $c_{w}=c_{0}-\epsilon_{w}$
(blue square) is placed inside the stability zone at a controlled
distance $\epsilon_{w}$ from the boundary. (For $c_{0}=c_{+}^{(m)}$,
the stability zone is to the right and the work point is $c_{w}=c_{0}+\epsilon_{w}$.)
A signal $\Delta c<0$ ($\Delta C>0$, green arrow) moves the operating
point away from the boundary --- always stable; a signal $\Delta c>0$
($\Delta C<0$, red arrow) moves it toward the boundary --- stable
provided $\Delta c<\epsilon_{w}$. \emph{Bottom:} Resulting frequency
splits $\Delta\alpha_{w}$ (equation~\eqref{eq:work-point-total})
for the three operating conditions. The baseline split $\sqrt{2\delta\,\epsilon_{w}}$
at the work point (blue bar) is nonzero; the two signal directions
produce asymmetric responses about the baseline.}
\label{fig:work-point-geometry} 
\end{figure}

\emph{Characteristic exponents and the frequency split.} The two Floquet\index{Floquet theory}
multiplier\index{Floquet theory!Floquet multiplier}s $\rho_{\pm}$
of the monodromy matrix\index{monodromy matrix} define the \emph{characteristic
exponents} $\alpha_{\pm}$ via 
\begin{equation}
\rho_{\pm}=e^{i\pi\alpha_{\pm}}.\label{eq:alpha-def-XIV}
\end{equation}
The exponents $\alpha_{\pm}$ are \emph{dimensionless frequencies}:
the physical oscillation frequencies of the two Floquet modes are
\begin{equation}
f_{\pm}=\frac{\mu}{2}\,\alpha_{\pm},\label{eq:fpm-XIV}
\end{equation}
so that $\alpha_{\pm}$ measures frequency in units of $\mu/2$. At
the EPD the two exponents coincide, 
\begin{equation}
\alpha_{+}=\alpha_{-}=m,\label{eq:alpha-EPD}
\end{equation}
where $m=1,3,5,\ldots$ is the resonance\index{resonance} index,
so both Floquet modes oscillate at the same physical frequency $f=m\mu/2\approx\omega_{0}$.
The \emph{frequency split} 
\begin{equation}
\Delta\alpha=\alpha_{+}-\alpha_{-}\label{eq:freq-split-def}
\end{equation}
vanishes at the bare EPD and becomes nonzero when the capacitance
is perturbed; it is directly measurable as a beat frequency in the
circuit's transient response, and is the primary observable in EPD
sensor applications. The comb structure of the response is illustrated
in Figure~\ref{fig:floquet-comb}.

Either boundary curve\index{boundary curves} $c_{+}^{(m)}$ or $c_{-}^{(m)}$
is a valid EPD operating point; the choice determines which sign of
$\Delta C$ moves the circuit into the stability zone. The signal
to be detected is a small perturbation $\Delta C$ of the capacitance,
of either sign, with $|\Delta C|\ll C_{0}$. The total capacitance
becomes $C_{0}+\Delta C$, shifting the natural frequency from $\omega_{0}=1/\sqrt{LC_{0}}$
to $\omega_{0}'=1/\sqrt{L(C_{0}+\Delta C)}$ and producing a frequency
shift $\xi=\omega_{0}-\omega_{0}'$ (of either sign) and a corresponding
spectral shift $\Delta c$ (equation~\eqref{eq:Deltac}), with $|\Delta c|\ll c_{0}$.
The dimensionless signal 
\begin{equation}
\hat{\xi}=\frac{\xi}{\mu},\qquad\hat{\xi}=\frac{\omega_{0}\,\Delta C}{2C_{0}\mu}+O\!\left(\!\left(\frac{\Delta C}{C_{0}}\right)^{\!2}\right),\label{eq:xi-hat-def}
\end{equation}
measures the frequency shift as a fraction of the driving frequency;
the physically relevant regime is $|\hat{\xi}|\ll1$ (small signal
compared to the driving frequency).

\emph{Marginal stability at the bare EPD.} The EPD operating point
lies precisely on the boundary between the stability zone and the
instability tongue\index{instability tongue} (Theorem~\ref{thm:Hill-stability},
case~3). At the bare EPD, the sign of $\Delta c$ determines stability:
only the sign of $\Delta c$ that points \emph{into the stability
zone} gives $\eta_{1}\,\Delta c<0$ (stable); the opposite sign gives
$\eta_{1}\,\Delta c>0$ (unstable). Thus at the bare EPD only one
sign of perturbation is tolerated. \emph{The guiding principle for
the work point\index{exceptional point of degeneracy (EPD)!work-point strategy}:}
one shifts the operating point by a small controlled amount $\epsilon_{w}>0$
\emph{into the stability zone} from the chosen EPD curve. This removes
the asymmetry: by shifting to the work point $c_{w}=c_{0}-\epsilon_{w}$
(for $c_{0}=c_{-}^{(m)}$, where the stability zone is below) or $c_{w}=c_{0}+\epsilon_{w}$
(for $c_{0}=c_{+}^{(m)}$, where the stability zone is above), both
signs of $\Delta c$ are stable provided $|\Delta c|<\epsilon_{w}$
(Figure~\ref{fig:work-point-geometry} illustrates this geometry
for $c_{0}=c_{-}^{(1)}$). The theory developed in this work ---
the LC circuit as Ince's equation\index{Ince equation} (Chapter~\ref{sec:LCInce}),
the EPD curves (Theorem~\ref{thm:EPD-bdry}), and the analyticity
of boundary curves (Theorem~\ref{thm:bdry-analytic}) --- gives
a complete analytical foundation for quantifying this splitting.

\medskip{}

The chapter proceeds as follows. \S\,\ref{subsec:sensor-model}
fixes the sensor model and operating point; \S\,\ref{subsec:general-split}
derives the general frequency-split formula, made explicit for the
primary tongue ($m=1$) in \S\,\ref{subsec:m1-explicit}. \S\,\ref{subsec:sensor-comparison}
quantifies the advantage over conventional (linear) capacitance sensing,
and \S\,\ref{subsec:work-point} develops the marginal stability
analysis and the work-point strategy in full. \S\,\ref{subsec:higher-tongues}
extends the formulas to the higher tongues $m=3,5,7$, and \S\,\ref{subsec:probing}
closes with the measurement side: recovering the frequency shift from
the circuit's Floquet frequency comb, the probing protocol, and the
statistical recovery of $\Delta\alpha$ and $\Delta C$.

\subsection{Sensor model and operating point}

\label{subsec:sensor-model}

Fix the modulation frequency $\mu$ and modulation amplitude $\delta$
so that the circuit operates at an EPD point. By Theorem~\ref{thm:EPD-bdry},
these are the points 
\begin{equation}
c_{0}=c_{\pm}^{(m)}(\delta),\qquad c_{0}=\frac{4\omega_{0}^{2}}{\mu^{2}},\label{eq:EPD-operating-point}
\end{equation}
where $c_{\pm}^{(m)}$ are the explicit EPD curves given by~\eqref{eq:EPD-bdry-1}--\eqref{eq:EPD-bdry-7}.
At such a point, by Theorem~\ref{thm:bdry-analytic} and the Floquet
theory of Chapter~\ref{app:genLin}, the monodromy matrix $X(\pi)$
has a non-trivial Jordan block\index{Jordan block} with eigenvalue
$\rho^{(0)}=(-1)^{m}$ and the two characteristic exponents coincide:
$\alpha_{+}=\alpha_{-}=m$.

The \emph{sensing mechanism} is a small perturbation $\Delta C$ (of
either sign) of the nominal capacitance $C_{0}$. Since $\omega_{0}=1/\sqrt{LC_{0}}$
(equation~\eqref{eq:LC-original}), this shifts the natural frequency
$\omega_{0}\to\omega_{0}-\xi$ where $\xi=\omega_{0}-\omega_{0}'$
has the same sign as $\Delta C$ (positive for $\Delta C>0$, negative
for $\Delta C<0$), and the exact relation is 
\begin{multline}
\xi=\omega_{0}-\frac{\omega_{0}}{\sqrt{1+\Delta C/C_{0}}}=\frac{\omega_{0}}{2}\frac{\Delta C}{C_{0}}-\frac{3\omega_{0}}{8}\!\left(\frac{\Delta C}{C_{0}}\right)^{\!2}\\
+O\!\left(\!\left(\frac{\Delta C}{C_{0}}\right)^{\!3}\right),\quad\frac{\Delta C}{C_{0}}\to0.\label{eq:xi-DeltaC}
\end{multline}
or inverted: 
\begin{equation}
\frac{\Delta C}{C_{0}}=\frac{2\xi}{\omega_{0}}+3\left(\frac{\xi}{\omega_{0}}\right)^{\!2}+O\!\left(\!\left(\frac{\xi}{\omega_{0}}\right)^{\!3}\right),\quad\frac{\xi}{\omega_{0}}\to0.\label{eq:DeltaC-xi}
\end{equation}
Both $\xi$ and $\Delta C$ can be of either sign. The Ince eigenvalue
parameter $c=4\omega_{0}^{2}/\mu^{2}$ (equation~\eqref{eq:LC-Hill})
shifts by an amount that is \emph{exact} in $\xi$ with no further
corrections: 
\begin{equation}
\Delta c=\frac{4(\omega_{0}-\xi)^{2}}{\mu^{2}}-c_{0}=-\frac{4\xi(2\omega_{0}-\xi)}{\mu^{2}}=-\frac{2c_{0}}{\omega_{0}}\,\xi+\frac{4}{\mu^{2}}\,\xi^{2}.\label{eq:Deltac}
\end{equation}
In terms of $\hat{\xi}=\xi/\mu$ this simplifies \emph{exactly} to
\begin{equation}
\Delta c=-4\hat{\xi}\bigl(\sqrt{c_{0}}-\hat{\xi}\bigr),\qquad\hat{\xi}=\xi/\mu,\label{eq:Deltac-xihat}
\end{equation}
using $\omega_{0}/\mu=\sqrt{c_{0}}/2$ (exact). At the primary resonance
$c_{0}=1$ ($\mu=2\omega_{0}$) this becomes 
\begin{equation}
\Delta c=-4\hat{\xi}(1-\hat{\xi}),\qquad c_{0}=1,\label{eq:Deltac-xihat-primary}
\end{equation}
which has the same $\hat{\xi}(1-\hat{\xi})$ structure as the split
formula~\eqref{eq:Deltaalpha-m1}, making the connection between
$\Delta c$ and $\Delta\alpha$ transparent. The physically relevant
range is $0<\hat{\xi}<1$ (i.e.\ $0<\xi<\mu\approx2\omega_{0}$),
in which $\Delta c<0$. Which sign of $\Delta c$ is stable depends
on which EPD curve $c_{0}=c_{\pm}^{(m)}$ the circuit operates at:
for $c_{0}=c_{-}^{(m)}$, the stability zone is $c<c_{-}^{(m)}$,
so $\Delta c<0$ (i.e.\ $\Delta C>0$) is the stable direction; for
$c_{0}=c_{+}^{(m)}$, the stability zone is $c>c_{+}^{(m)}$, so $\Delta c>0$
(i.e.\ $\Delta C<0$) is the stable direction. In either case, at
the bare EPD only one sign of perturbation is tolerated; the work
point (Section~\ref{subsec:work-point}, Figure~\ref{fig:work-point-geometry})
removes this restriction for sufficiently small $|\Delta c|$. Equivalently,
in terms of $\Delta C$: 
\begin{equation}
\Delta c=-c_{0}\,\frac{\Delta C}{C_{0}}\!\left(1-\frac{\Delta C}{C_{0}}\right)+O\!\left(\!\left(\frac{\Delta C}{C_{0}}\right)^{\!3}\right),\quad\frac{\Delta C}{C_{0}}\to0.\label{eq:Deltac-DeltaC}
\end{equation}
The question is: how does $\Delta\alpha=\alpha_{+}-\alpha_{-}$ depend
on $\xi$ (or $\Delta C$)?

\subsection{Frequency-split formula}

\label{subsec:general-split}

The characteristic exponents $\alpha_{\pm}$ are the primary objects;
the multipliers $\rho_{\pm}=e^{i\pi\alpha_{\pm}}$ play an auxiliary
role in the derivation. Let $\mathbf{u}_{0},\mathbf{u}_{1}$ be the
right Jordan chain and $\mathbf{v}_{0},\mathbf{v}_{1}$ the left Jordan
chain of $X(\pi)$ at the EPD, satisfying the normalization conditions
$\mathbf{v}_{0}^{T}\mathbf{u}_{1}=1$, $\mathbf{v}_{1}^{T}\mathbf{u}_{1}=0$
(see~\cite[eqs.~(9.133)--(9.135)]{SeyMai}), which imply the derived
properties $\mathbf{v}_{1}^{T}\mathbf{u}_{0}=1$, $\mathbf{v}_{0}^{T}\mathbf{u}_{0}=0$
(see~\cite[eq.~(2.66)]{SeyMai}). Existence guaranteed by Theorem~\ref{thm:bdry-analytic}.
Define the \emph{EPD sensitivity coefficients} 
\begin{equation}
\eta_{1}=\mathbf{v}_{0}^{T}\!\left(\frac{\partial X(T)}{\partial c}\right)\!\mathbf{u}_{0},\qquad2\eta_{2}=\mathbf{v}_{0}^{T}\!\left(\frac{\partial X(T)}{\partial c}\right)\!\mathbf{u}_{1}+\mathbf{v}_{1}^{T}\!\left(\frac{\partial X(T)}{\partial c}\right)\!\mathbf{u}_{0}.\label{eq:eta1eta2}
\end{equation}
The derivative $\partial X(T)/\partial c$ is given by the matriciant
formula~\cite[eq.~(9.66)]{SeyMai} 
\begin{equation}
\frac{\partial X(T)}{\partial c}=X(T)\int_{0}^{T}X^{-1}(t)\,\frac{\partial G(t)}{\partial c}\,X(t)\,dt,\label{eq:dXdc}
\end{equation}
where $G(t)$ is the $2\times2$ system matrix of~\eqref{eq:LC-Hill}
and $T=\pi$. 
\begin{thm}[Frequency split at EPD]
\label{thm:EPD-split} Suppose $\eta_{1}\neq0$. Under the perturbation~\eqref{eq:Deltac}
the two characteristic exponents of the LC circuit~\eqref{eq:LC-Hill}
split as 
\begin{equation}
\alpha_{\pm}=m\pm\frac{\sqrt{-\eta_{1}\,\Delta c}}{\pi}+O\!\left(|\Delta c|\right),\quad|\Delta c|\to0,\label{eq:alpha-split}
\end{equation}
giving the frequency split 
\begin{equation}
\boxed{\Delta\alpha=\alpha_{+}-\alpha_{-}=\frac{2}{\pi}\sqrt{-\eta_{1}\,\Delta c}+O(|\Delta c|^{3/2}),\quad|\Delta c|\to0.}\label{eq:Delta-alpha}
\end{equation}
The expression $-\eta_{1}\,\Delta c$ under the square root is determined
by objective computation of the sensitivity coefficient $\eta_{1}$~\eqref{eq:eta1eta2}
and the spectral shift $\Delta c$~\eqref{eq:Deltac}; its sign then
determines stability: 
\begin{itemize}
\item $\eta_{1}\,\Delta c<0$: $\sqrt{-\eta_{1}\,\Delta c}$ is real, the
double multiplier splits \emph{tangentially} along the unit circle,
$\alpha_{\pm}$ are real, all solutions are bounded --- \emph{stable
operation}. 
\item $\eta_{1}\,\Delta c>0$: the multipliers split \emph{radially} off
the unit circle ($\rho_{\pm}=(-1)^{m}\pm\sqrt{\eta_{1}\,\Delta c}$
real to leading order, $|\rho_{+}|>1$), $\alpha_{\pm}$ acquire imaginary
parts, solutions grow exponentially --- \emph{unstable operation}. 
\end{itemize}
The sign of $\eta_{1}\,\Delta c$ is not imposed by the theorem but
is a consequence of the specific values of $\eta_{1}$ and $\Delta c$
for the circuit in question. Since the EPD operating point lies on
the boundary between stability and instability, the sign of $\Delta c$
determines on which side the perturbed circuit falls. On the stable
side ($\eta_{1}\,\Delta c<0$), the split $\Delta\alpha\propto\sqrt{|\eta_{1}\,\Delta c|}$
grows much faster than $|\Delta c|$ as $|\Delta c|\to0$ --- the
\emph{hypersensitivity\index{exceptional point of degeneracy (EPD)!hypersensitivity}}
of the EPD configuration. The explicit computation of $\eta_{1}$,
$\eta_{2}$, and the verification of $\eta_{1}\,\Delta c<0$ for the
primary tongue ($m=1$, $c_{0}=c_{-}^{(1)}$, $\Delta C>0$) is carried
out in Section~\ref{subsec:m1-explicit}, which thereby instantiates
this theorem for the LC circuit and establishes stability of the sensing
regime. 
\end{thm}

\begin{proof}
By~\cite[Sec.~9.7.1, eq.~(9.136)--(9.137)]{SeyMai} the multipliers
split as $\rho_{\pm}=(-1)^{m}\pm\sqrt{\eta_{1}\,\Delta c}+\eta_{2}\,\Delta c+O(|\Delta c|^{3/2})$.
For $\eta_{1}\,\Delta c<0$ the square root is purely imaginary, so
to this order the perturbed multipliers remain on the unit circle;
writing $\rho=e^{i\pi\alpha}$ with $\rho^{(0)}=(-1)^{m}=e^{i\pi m}$
and taking the principal branch gives~\eqref{eq:alpha-split}. (The
term $\eta_{2}\,\Delta c$ is real and common to both multipliers;
on the stable side it is precisely the curvature correction that keeps
$|\rho_{\pm}|=1$ at order $\Delta c$, so it contributes no real
shift to $\alpha_{\pm}$ at that order, and it cancels identically
in the difference $\Delta\alpha$, giving~\eqref{eq:Delta-alpha}.)
For $\eta_{1}\,\Delta c>0$ the splitting is real: the multipliers
leave the unit circle along the real axis and the exponents acquire
imaginary parts. Non-degeneracy $\eta_{1}\neq0$ follows from~\cite[Ch.~V, \S\,3.2, eqs.~(3.10)--(3.11)]{YakSta2}
and Theorem~\ref{thm:bdry-analytic}: $\eta_{1}\propto[\omega_{j}'(c_{0})+\omega_{k}'(c_{0})]^{2}=1/c_{0}\neq0$. 
\end{proof}
\begin{rem}[Sign of $\eta_{1}\,\Delta c$ and error sources]
\label{rem:eta2-zero} \emph{Sign and stability for the primary tongue
at $c_{0}=c_{-}^{(1)}$.} The computation carried out in Section~\ref{subsec:m1-explicit}
gives $\eta_{1}=+\pi^{2}\delta/2+O(\delta^{2})>0$ (equation~\eqref{eq:eta1-m1}).
Equation~\eqref{eq:Deltac} gives $\Delta c=-(2c_{0}/\omega_{0})\xi+O(\xi^{2})$.
For $\Delta C>0$: $\xi>0$, $\Delta c<0$, hence $\eta_{1}\,\Delta c<0$
--- the circuit is \emph{stable} and $\Delta\alpha$ is real. For
$\Delta C<0$: $\xi<0$, $\Delta c>0$, hence $\eta_{1}\,\Delta c>0$
--- the circuit is \emph{unstable} at the bare EPD (but stabilized
by the work point for $\Delta c<\epsilon_{w}$). This sign conclusion
is specific to $c_{0}=c_{-}^{(1)}$; at $c_{0}=c_{+}^{(1)}$ the stable
and unstable directions of $\Delta C$ are reversed. Stability is
a \emph{conclusion} of the computation of $\eta_{1}$ and $\Delta c$,
not a condition imposed on the formula.

\emph{Error sources in $\Delta\alpha$.} (i)~\emph{Splitting truncation:}
the $O(|\Delta c|^{3/2})$ term from the Newton--Puiseux series.
(ii)~\emph{$\eta_{1}$ approximation:} using $|\eta_{1}|\approx\pi^{2}\delta/2$
introduces a relative error $O(\delta)$ in $\sqrt{|\eta_{1}|}$,
hence in $\Delta\alpha$. The $\eta_{2}$ correction ($\eta_{2}=0+O(\delta)$)
contributes an $O(\delta\,\hat{\xi})$ term, negligible against $\Delta\alpha\sim\sqrt{\delta\,\hat{\xi}}$
for $\hat{\xi}\ll\delta$ (both dimensionless). 
\end{rem}

\subsection{\texorpdfstring{Primary tongue: explicit formula ($m=1$)}{Primary
tongue: explicit formula (m=1)}}

\label{subsec:m1-explicit}

This section carries out the program of Theorem~\ref{thm:EPD-split}
for the primary tongue ($m=1$): compute the monodromy matrix $X(\pi)$
at the EPD, extract the Jordan chains $\mathbf{u}_{0},\mathbf{v}_{0}$,
evaluate the sensitivity coefficients $\eta_{1}$, $\eta_{2}$ via~\eqref{eq:eta1eta2},
determine the sign of $\eta_{1}\,\Delta c$, and thereby obtain the
explicit closed-form formula for $\Delta\alpha$.

For the primary tongue ($m=1$), the EPD operating point is $c_{0}=c_{-}^{(1)}(\delta)$
from~\eqref{eq:EPD-bdry-1}, with $a_{0}=-2\delta/(1+\delta^{2})$.
(Either boundary $c_{\pm}^{(1)}$ is a valid EPD point; $c_{-}^{(1)}$
is chosen here because $\Delta C>0$ gives $\Delta c<0$, which moves
into the stability zone $c<c_{-}^{(1)}$, making $\Delta C>0$ the
natural sensing direction. Operating at $c_{+}^{(1)}$ instead reverses
the role of the two signs of $\Delta C$ but leaves the formula unchanged.)
The circuit oscillates near $\omega_{0}\approx\mu/2$ (primary resonance~\eqref{eq:resonance-condition-MR}).

\emph{Monodromy matrix at EPD.} Using the matriciant formula~\eqref{eq:dXdc}
applied to the LC Hill equation\index{Hill equation}~\eqref{eq:LC-Hill},
and the fact that at $\delta=0$ the unperturbed system has period-$\pi$
solutions $\cos\tau$ and $\sin\tau$ (characteristic exponent $\alpha_{0}=1$,
i.e.~multiplier $-1$, consistent with Floquet theory of Chapter~\ref{app:Hill}),
the monodromy matrix at the EPD curve evaluates to leading order as
\begin{equation}
X(\pi)\big|_{\mathrm{EPD}}=\begin{bmatrix}-1 & \pi\delta\\
0 & -1
\end{bmatrix}+O(\delta^{2}),\quad\delta\ll1.\label{eq:X-EPD-m1}
\end{equation}
This is a non-trivial $2\times2$ Jordan block with eigenvalue $-1$,
confirming the EPD structure asserted in Theorem~\ref{thm:EPD-bdry}.
The right and left Jordan chains, normalized by $\mathbf{v}_{1}^{T}\mathbf{u}_{0}=1$,
are: 
\begin{equation}
\mathbf{u}_{0}=\begin{bmatrix}\pi\delta\\
0
\end{bmatrix},\quad\mathbf{u}_{1}=\begin{bmatrix}0\\
1
\end{bmatrix},\quad\mathbf{v}_{0}=\begin{bmatrix}0\\
1
\end{bmatrix},\quad\mathbf{v}_{1}=\begin{bmatrix}(\pi\delta)^{-1}\\
0
\end{bmatrix}.\label{eq:Jordan-chains-m1}
\end{equation}

\emph{Derivative of the monodromy matrix.} For the LC Hill equation~\eqref{eq:LC-Hill},
the only $c$-dependent entry of the system matrix is the coefficient
$Q$ of $q$ in~\eqref{eq:LC-Hill}, which is linear in $c$, so
$\partial G/\partial c$ has the single nonzero entry $-Q/c$ in position
$(2,1)$. Applying~\eqref{eq:dXdc} at $(\delta=0,c=1)$ with the
explicit matriciant $X(t)=\bigl(\begin{smallmatrix}\cos t & \sin t\\
-\sin t & \cos t
\end{smallmatrix}\bigr)$ (period-$\pi$ solutions of $q''+q=0$, eq.~\eqref{eq:LC-Hill}
at $\delta=0$, $c=1$) gives 
\begin{equation}
\frac{\partial X(\pi)}{\partial c}\bigg|_{\mathrm{EPD}}=\begin{bmatrix}0 & -\pi/2\\
\pi/2 & 0
\end{bmatrix}+O(\delta),\quad\delta\ll1.\label{eq:dXdc-m1}
\end{equation}

\emph{Sensitivity coefficients.} Substituting~\eqref{eq:Jordan-chains-m1}
and~\eqref{eq:dXdc-m1} into~\eqref{eq:eta1eta2}: 
\begin{equation}
\eta_{1}=+\frac{\pi^{2}\delta}{2}+O(\delta^{2}),\qquad\eta_{2}=0+O(\delta),\quad\delta\ll1.\label{eq:eta1-m1}
\end{equation}

\emph{Sign check and stability conclusion.} From~\eqref{eq:Deltac},
$\Delta c=-4\xi(2\omega_{0}-\xi)/\mu^{2}<0$ for $0<\xi<2\omega_{0}$
(i.e.\ $\Delta C>0$). Combined with $\eta_{1}>0$ from~\eqref{eq:eta1-m1}:
$\eta_{1}\,\Delta c<0$. By Theorem~\ref{thm:EPD-split} this means
$\Delta\alpha$ is real and the circuit operates stably. The sensing
regime ($\Delta C>0$, capacitance increases) is therefore on the
stable side of the EPD curve.

\emph{Frequency split.} Substituting~\eqref{eq:eta1-m1} into~\eqref{eq:Delta-alpha},
and using the \emph{exact} $\Delta c$ from~\eqref{eq:Deltac} with
$c_{0}\approx1$ (from~\eqref{eq:EPD-bdry-1}) and $\omega_{0}\approx\mu/2$
(primary resonance~\eqref{eq:resonance-condition-MR}): 
\begin{equation}
\boxed{\begin{aligned}\Delta\alpha=\alpha_{+}-\alpha_{-} & =\frac{2\sqrt{2\delta}}{\mu}\sqrt{\xi(\mu-\xi)}\cdot\bigl(1+O(\delta)\bigr)\\
 & =2\sqrt{2\delta}\sqrt{\hat{\xi}(1-\hat{\xi})}\cdot\bigl(1+O(\delta)\bigr),\quad\delta\ll1.
\end{aligned}
}\label{eq:Deltaalpha-m1}
\end{equation}
which is \emph{exact in $\hat{\xi}$} and accurate to leading order
in $\delta$ (relative error $O(\delta)$ from~\eqref{eq:eta1-m1}).
The $\hat{\xi}$-form is free of $\mu$: the split depends only on
the dimensionless signal $\hat{\xi}=\xi/\mu$ and the modulation depth
$\delta$. Expanding for $\hat{\xi}\ll1$: 
\begin{equation}
\Delta\alpha=2\sqrt{2\delta}\,\sqrt{\hat{\xi}}\Bigl(1-\frac{\hat{\xi}}{2}+O(\hat{\xi}^{2})\Bigr)\cdot\bigl(1+O(\delta)\bigr),\quad\hat{\xi}\to0,\;\delta\ll1.\label{eq:Deltaalpha-m1-expand}
\end{equation}
so $2\sqrt{2\delta}$ is the leading sensitivity coefficient (independent
of $\mu$); $\hat{\xi}/2$ is the relative correction from the exact
$\hat{\xi}$-dependence (small when $\hat{\xi}\ll1$), and $O(\delta)$
is the relative error from the leading-order approximation of $\eta_{1}$.

The same result in terms of the directly measurable capacitance shift
$\Delta C$, using $\hat{\xi}=\omega_{0}\Delta C/(2C_{0}\mu)$ from~\eqref{eq:xi-hat-def}
with $\omega_{0}\approx\mu/2$: 
\begin{equation}
\Delta\alpha=2\sqrt{\frac{\delta\omega_{0}}{\mu}}\cdot\sqrt{\frac{\Delta C}{C_{0}}}\cdot\bigl(1+O(\delta)\bigr)+O\!\left(\!\left(\frac{\Delta C}{C_{0}}\right)^{\!3/2}\right),\quad\frac{\Delta C}{C_{0}}\to0,\;\delta\ll1.\label{eq:Deltaalpha-DeltaC}
\end{equation}
(Here $\mu$ appears because $\Delta C/C_{0}$, while dimensionless,
still requires $\mu$ to connect to $\hat{\xi}$; in terms of $\hat{\xi}$
this is simply $\Delta\alpha=2\sqrt{2\delta\hat{\xi}}\cdot(1+O(\delta))$.)
The sensor resolution (inverting for the detected signal): 
\begin{equation}
\hat{\xi}_{\mathrm{detected}}=\frac{(\Delta\alpha)^{2}}{8\delta},\quad\xi_{\mathrm{detected}}=\frac{\mu(\Delta\alpha)^{2}}{8\delta},\quad\left(\frac{\Delta C}{C_{0}}\right)_{\!\mathrm{detected}}=\frac{\mu(\Delta\alpha)^{2}}{4\delta\omega_{0}}.\label{eq:sensor-resolution}
\end{equation}

The circuit realizing~\eqref{eq:Deltaalpha-m1} works as follows.
At the EPD operating point the circuit's transient response is a pure
subharmonic at $\omega_{0}\approx\mu/2$. A capacitance shift $\Delta C$
causes this single frequency to split into two frequencies $\alpha_{\pm}\cdot\tfrac{\mu}{2}\approx\tfrac{\mu}{2}\pm\tfrac{\mu}{2}\cdot\tfrac{\Delta\alpha}{2}$,
producing a \emph{beat} at the measurable frequency $\tfrac{\mu}{2}\Delta\alpha$,
directly observable in the transient response.
\begin{rem}[Numerical verification]
\label{rem:sensor-split-numerics} All ingredients of this section
were verified against the exact monodromy matrix, computed by high-precision
integration of~\eqref{eq:LC-Hill} at the exact numerical boundary
$c_{0}=c_{-}^{(1)}(\delta)$ (located as the root of $\Delta(c)+2=0$).
(i)~\emph{Jordan structure:} $X(\pi)$ at $c_{0}$ matches the form~\eqref{eq:X-EPD-m1};
e.g.\ at $\delta=0.05$ the upper-right entry is $0.1552$ against
$\pi\delta=0.1571$, and the lower-left entry vanishes to integration
precision. (ii)~\emph{Sensitivity coefficient:} evaluating~\eqref{eq:eta1eta2}
with the exact chains and the differentiated monodromy gives $\eta_{1}/(\pi^{2}\delta/2)=1.011$,
$1.030$, $1.072$ at $\delta=0.02$, $0.05$, $0.10$, confirming
$\eta_{1}=+(\pi^{2}\delta/2)\bigl(1+O(\delta)\bigr)$; the measured
$\eta_{2}$ is $O(\delta)$, consistent with~\eqref{eq:eta1-m1}.
(iii)~\emph{Splitting geometry:} for $\Delta c<0$ the measured multipliers
remain on the unit circle and split tangentially; for $\Delta c>0$
they leave the circle radially --- and in both regimes they match
the Newton--Puiseux formula of Theorem~\ref{thm:EPD-split} with
the computed $\eta_{1},\eta_{2}$ to $O(|\Delta c|^{3/2})$ (at $\delta=0.05$,
$\Delta c=-10^{-3}$: measured $\rho_{\pm}-\rho^{(0)}=0.00013\pm0.01603\,i$
against the predicted $0.00013\pm0.01594\,i$). (iv)~\emph{Frequency
split:} the ratio of the measured split to the leading form $\sqrt{2\delta\,|\Delta c|}$
of~\eqref{eq:Deltaalpha-m1} at $|\Delta c|=10^{-4}$ is $1.016$,
$1.036$, $1.061$ at $\delta=0.05$, $0.10$, $0.15$ --- a clean
$1+O(\delta)$, as claimed. 
\end{rem}

\subsection{Comparison with conventional sensing}

\label{subsec:sensor-comparison}

A \emph{conventional} LC sensor (non-EPD) detects $\xi$ through a
linear resonance frequency shift: its output scales as $\hat{\xi}$
(since $\Delta\alpha_{{\rm conv}}\sim\xi/(\mu/2)=2\hat{\xi}$). The
EPD sensor output $\Delta\alpha$ scales as $\sqrt{\hat{\xi}}$. For
small $\hat{\xi}$: 
\begin{equation}
\frac{\Delta\alpha_{\mathrm{EPD}}}{\Delta\alpha_{\mathrm{conv}}}=\frac{2\sqrt{2\delta\hat{\xi}}}{2\hat{\xi}}=\sqrt{\frac{2\delta}{\hat{\xi}}}\xrightarrow{\;\hat{\xi}\to0\;}\infty.\label{eq:sensitivity-ratio}
\end{equation}
The EPD advantage diverges as $\hat{\xi}\to0$: for any fixed conventional
sensor, there exists a crossover $\hat{\xi}^{*}=2\delta$ (equivalently
$\xi^{*}=2\delta\mu$) below which the EPD circuit is more sensitive.
The crossover depends only on the modulation depth $\delta$: deeper
modulation extends the EPD advantage to larger signals.

In terms of capacitance shifts, using~\eqref{eq:Deltaalpha-DeltaC}:
\begin{gather}
\Delta\alpha_{\mathrm{EPD}}\propto\sqrt{\frac{\Delta C}{C_{0}}},\qquad\Delta\alpha_{\mathrm{conv}}\propto\frac{\Delta C}{C_{0}},\nonumber \\
\Rightarrow\quad\frac{\Delta\alpha_{\mathrm{EPD}}}{\Delta\alpha_{\mathrm{conv}}}\propto\frac{1}{\sqrt{\Delta C/C_{0}}}\to\infty\quad\text{as }\Delta C\to0.\label{eq:sensitivity-ratio-C}
\end{gather}

In practice, noise imposes a floor on the minimum detectable $\xi$.
These limitations fall outside the scope of the present analysis but
are studied in the EPD sensing literature~\cite[Sec.~11.1.2]{Kiril}.

\subsection{Marginal stability and the work point}

\label{subsec:work-point}

There is a critical practical issue with any EPD-based sensor that
must be addressed before engineering implementation.

\emph{The marginal stability problem.} The ideal EPD operating point
$c_{0}=c_{\pm}^{(m)}(\delta)$ is located precisely on the \emph{boundary}
between the stability zone and the instability tongue of the LC circuit.
By definition, the discriminant\index{discriminant} satisfies $\Delta=-2$
there, meaning the Floquet multipliers coalesce at $\rho=-1$ on the
unit circle with a non-trivial Jordan block. This is a state of \emph{marginal
stability}: solutions neither decay nor grow, but one independent
solution grows linearly in time as $\tau\cdot(\text{antiperiodic})$
(Theorem~\ref{thm:Hill-stability}, case~3). In practice, any perturbation
or noise will push the circuit slightly into the instability tongue,
causing exponential growth of the response, or slightly outside it,
suppressing the hypersensitive $\sqrt{\hat{\xi}}$ response. Operating
exactly at the EPD point is therefore not a viable engineering strategy.

\emph{The work point remedy.} The resolution, developed in detail
for related EPD circuits in~\cite[Rem.~2, Def.~3.4, Thm.~3.5]{FigTWT}
and~\cite[Sec.~5]{FigPert}, is to introduce a \emph{deliberate small
departure} from the EPD operating point, called the \emph{work point}.

\emph{Guiding principle.} The work point must be placed \emph{inside
the stability zone}, at a controlled distance $\epsilon_{w}>0$ from
the EPD curve. Since the stability zone lies on one specific side
of each EPD curve (inside the tongue is unstable; outside is stable),
the displacement must be in the direction \emph{away from the tongue}.
For $c_{0}=c_{-}^{(m)}$ (lower boundary), the stability zone is below:
$c_{w}=c_{0}-\epsilon_{w}$. For $c_{0}=c_{+}^{(m)}$ (upper boundary),
the stability zone is above: $c_{w}=c_{0}+\epsilon_{w}$. In both
cases the magnitude $\epsilon_{w}>0$ is a free design parameter that
sets the stability margin and the baseline sensitivity.

\begin{figure}[htbp]
\centering \includegraphics[width=12cm]{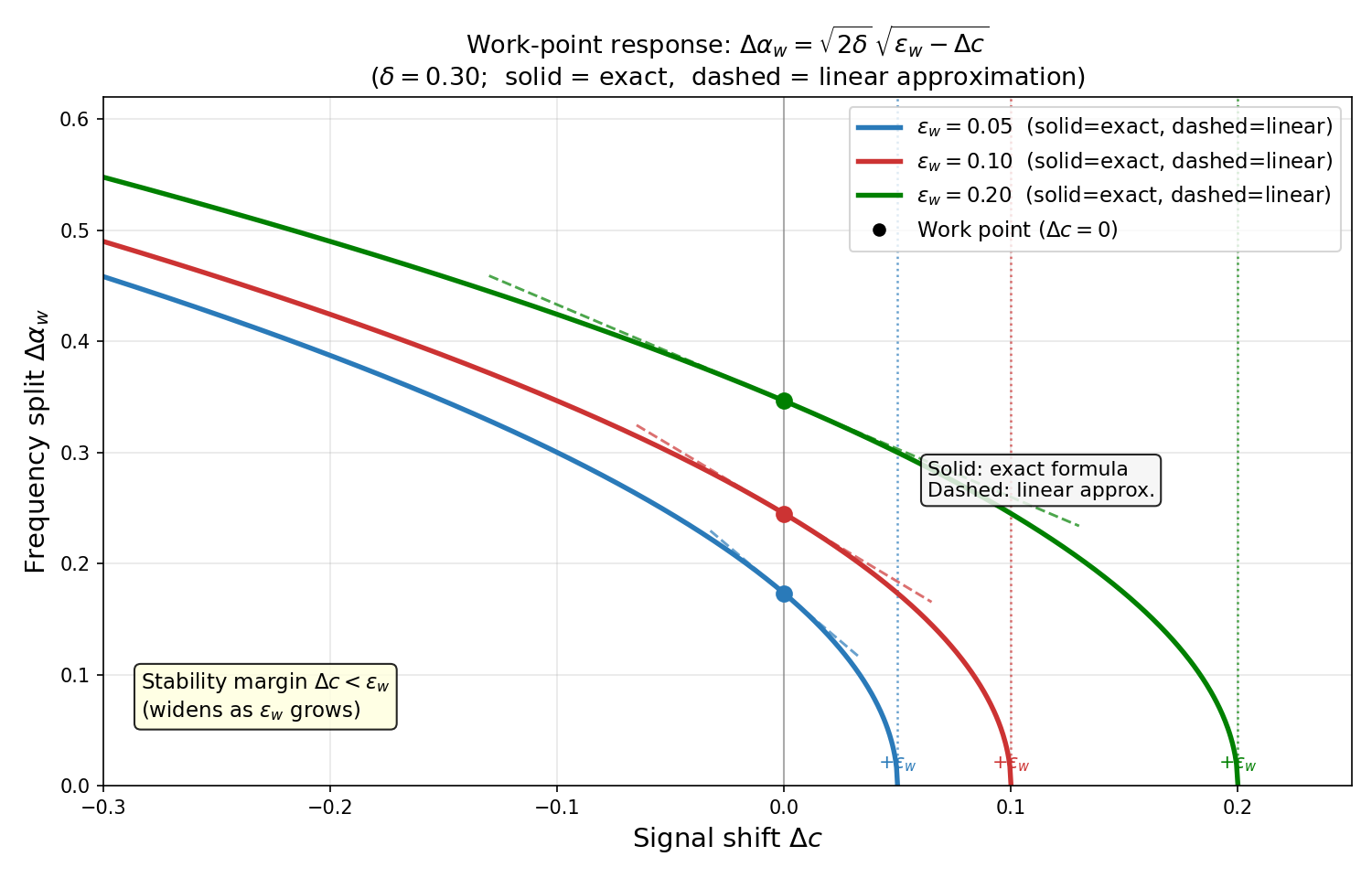}
\caption{Work-point frequency split $\Delta\alpha_{w}=\sqrt{2\delta}\,\sqrt{\epsilon_{w}-\Delta c}$
(equation~\eqref{eq:work-point-total}) as a function of the signal
shift $\Delta c$, for $\delta=0.30$ and three work-point departures
$\epsilon_{w}\in\{0.05,0.10,0.20\}$. Solid curves: exact formula
\eqref{eq:work-point-total}. Dashed curves: linear small-signal approximation~\eqref{eq:work-point-small-signal}.
Filled circles mark the baseline $\Delta\alpha_{w}(0)=\sqrt{2\delta\,\epsilon_{w}}$
at $\Delta c=0$. Dotted vertical lines indicate the stability boundary
$\Delta c=+\epsilon_{w}$ for each curve. Smaller $\epsilon_{w}$
gives higher sensitivity (steeper slope at the work point) but a narrower
stable range; larger $\epsilon_{w}$ is more robust but less sensitive.
The linear approximation is accurate for $|\Delta c|\ll\epsilon_{w}$
and breaks down near the stability boundary.}\index{stability boundary}
\label{fig:work-point-response} 
\end{figure}

For the primary tongue with $c_{0}=c_{-}^{(1)}$ (the case developed
below), the work point is: 
\begin{equation}
c_{w}=c_{0}-\epsilon_{w},\qquad\epsilon_{w}>0,\label{eq:work-point}
\end{equation}
placing it just \emph{inside} the stability zone at a controlled distance
$\epsilon_{w}$ from the EPD curve. A signal shifts $c$ by $\Delta c$
(equation~\eqref{eq:Deltac}; $\Delta c<0$ for $\Delta C>0$), so
the perturbed operating point is $c_{w}+\Delta c$, and its distance
below the EPD curve is 
\begin{equation}
c_{0}-(c_{w}+\Delta c)=\epsilon_{w}-\Delta c,\label{eq:work-point-displacement}
\end{equation}
which is positive provided $\Delta c<\epsilon_{w}$. Applying Theorem~\ref{thm:EPD-split}
with the total shift $(c_{w}+\Delta c)-c_{0}=-(\epsilon_{w}-\Delta c)$,
and substituting $\eta_{1}=\frac{\pi^{2}\delta}{2}+O(\delta^{2})>0$
from~\eqref{eq:eta1-m1}, the frequency split at the work point is
\begin{equation}
\Delta\alpha_{w}=\sqrt{2\delta}\cdot\sqrt{\epsilon_{w}-\Delta c}\cdot\bigl(1+O(\delta)\bigr)+O(|\Delta c|^{3/2}),\label{eq:work-point-split}
\end{equation}
where $\epsilon_{w}-\Delta c>0$ is the stability condition. Two immediate
consequences follow~\cite[Rem.~2]{FigTWT}, \cite[Sec.~5, Rem.~5.2]{FigPert}: 
\begin{enumerate}
\item \emph{Stability for both signs of $\Delta C$:} At the work point
$c_{w}=c_{0}-\epsilon_{w}$, the expression under the square root
in~\eqref{eq:work-point-split} is $\epsilon_{w}-\Delta c$. Since
$\epsilon_{w}>0$, for any $\Delta c$ satisfying $\Delta c<\epsilon_{w}$
this expression is positive, so $\Delta\alpha_{w}$ is real and the
circuit is stable. This covers \emph{both signs} of $\Delta C$: 
\begin{itemize}
\item $\Delta C>0$: $\Delta c<0$, and $\epsilon_{w}-\Delta c>\epsilon_{w}>0$
--- \emph{always stable} 
\item $\Delta C<0$: $\Delta c>0$, stable provided $\Delta c<\epsilon_{w}$ 
\end{itemize}
For small signals of either sign, $|\Delta c|\ll\epsilon_{w}$, the
sign under the root is fixed by $\epsilon_{w}>0$ regardless of the
sign of $\Delta c$ --- this is the key advantage of the work point. 
\item \emph{Preserved hypersensitivity:} For small signals $|\Delta c|\ll\epsilon_{w}$,
expanding~\eqref{eq:work-point-split} gives 
\begin{equation}
\Delta\alpha_{w}=\sqrt{2\delta\epsilon_{w}}\Bigl(1-\frac{\Delta c}{2\epsilon_{w}}\Bigr)\cdot\bigl(1+O(\delta)\bigr)+O\!\left(\frac{\Delta c^{2}}{\epsilon_{w}}\right),\label{eq:work-point-small-signal}
\end{equation}
which depends on the \emph{sign} of $\Delta c$: $\Delta c<0$ ($\Delta C>0$)
increases $\Delta\alpha_{w}$ above the baseline $\sqrt{2\delta\epsilon_{w}}$,
while $\Delta c>0$ ($\Delta C<0$) decreases it. Thus the split varies
linearly in $\Delta c$ for small signals, with sensitivity coefficient
$\sqrt{2\delta}/(2\sqrt{\epsilon_{w}})$ per unit $\Delta c$. The
full square-root response $\Delta\alpha_{w}\propto\sqrt{\epsilon_{w}-\Delta c}$
is recovered when $|\Delta c|\sim\epsilon_{w}$, i.e.\ when the signal
is comparable to the work-point departure. The explicit trade-off
between sensitivity and the stability margin $\epsilon_{w}$ is quantified
in~\cite[Rem.~5.2]{FigPert}. 
\end{enumerate}
\emph{Application to the modulated LC circuit.} In the notation of
this work, the work point $c_{w}=c_{0}-\epsilon_{w}$ is implemented
by slightly detuning the modulation frequency $\mu$ from its EPD
value, with $\epsilon_{w}$ adjustable by design. From~\eqref{eq:work-point-split},
the total split at the work point is 
\begin{equation}
\Delta\alpha_{w}(\Delta c)=\sqrt{2\delta}\cdot\sqrt{\epsilon_{w}-\Delta c}\cdot\bigl(1+O(\delta)\bigr),\label{eq:work-point-total}
\end{equation}
which for $|\Delta c|\ll\epsilon_{w}$ decomposes as 
\begin{equation}
\Delta\alpha_{w}(\Delta c)=\underbrace{\sqrt{2\delta\,\epsilon_{w}}}_{\text{baseline (}\Delta c=0\text{)}}-\underbrace{\frac{\sqrt{2\delta}}{2\sqrt{\epsilon_{w}}}\,\Delta c}_{\text{signal response}}+O\!\left(\frac{\Delta c^{2}}{\epsilon_{w}}\right)+O(\delta^{3/2}).\label{eq:work-point-expand}
\end{equation}
The baseline $\sqrt{2\delta\,\epsilon_{w}}$ is nonzero even with
no signal and is determined by the design parameter $\epsilon_{w}$.
The full response curve~\eqref{eq:work-point-total} is plotted in
Figure~\ref{fig:work-point-response}.

\begin{figure}[htbp]
\centering \includegraphics[width=1\textwidth]{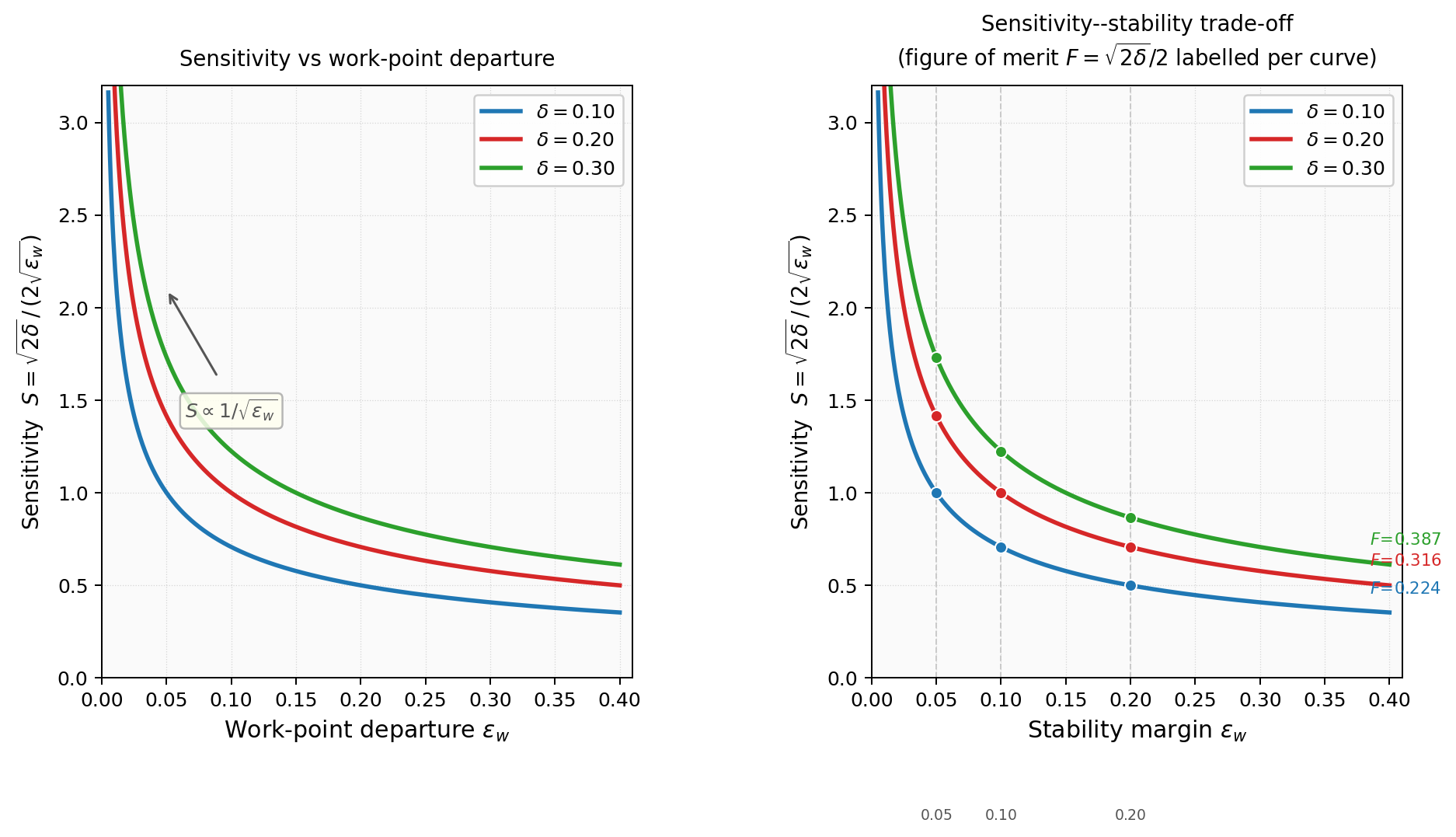}
\caption{Sensitivity--stability trade-off at the work point. \emph{Left:}
Sensitivity coefficient $S=\sqrt{2\delta}/(2\sqrt{\epsilon_{w}})$
(equation~\eqref{eq:work-point-signal}) as a function of the work-point
departure $\epsilon_{w}$, for three values of $\delta$. The coefficient
diverges as $\epsilon_{w}\to0$ (approach to the ideal EPD) and decays
as $1/\sqrt{\epsilon_{w}}$. \emph{Right:} The same curves interpreted
as a trade-off diagram: $x$-axis = stability margin $\epsilon_{w}$
(cost), $y$-axis = sensitivity (benefit). For fixed $\delta$, the
figure of merit $F=S\cdot\sqrt{\epsilon_{w}}=\sqrt{2\delta}/2$ is
constant along each curve (labeled), so no choice of $\epsilon_{w}$
improves $F$; only increasing $\delta$ (larger modulation amplitude)
raises the overall trade-off curve. Marked dots correspond to the
three representative operating points $\epsilon_{w}\in\{0.05,0.10,0.20\}$.}
\label{fig:work-point-tradeoff} 
\end{figure}

The signal-induced \emph{change} from the baseline, 
\begin{equation}
\delta(\Delta\alpha_{w})\equiv\Delta\alpha_{w}(\Delta c)-\Delta\alpha_{w}(0)=-\frac{\sqrt{2\delta}}{2\sqrt{\epsilon_{w}}}\,\Delta c+O\!\left(\frac{\Delta c^{2}}{\epsilon_{w}}\right)+O(\delta^{3/2}),\label{eq:work-point-signal}
\end{equation}
is small with $\Delta c$, vanishes when $\Delta c=0$, and carries
the sign opposite to $\Delta c$ --- that is, the sign of $\Delta C$
(equation~\eqref{eq:Deltac}): the split increases for $\Delta c<0$
($\Delta C>0$) and decreases for $\Delta c>0$ ($\Delta C<0$). The
sensitivity coefficient $\sqrt{2\delta}/(2\sqrt{\epsilon_{w}})$ grows
as $\epsilon_{w}\to0$ (approach to the ideal EPD), while the stability
margin $\Delta c<\epsilon_{w}$ narrows --- the trade-off quantified
in~\cite[Rem.~5.2]{FigPert}. Both sides of the trade-off are displayed
in Figure~\ref{fig:work-point-tradeoff}.

\begin{rem}[Advantage of the modulated LC circuit]
The circuits studied in~\cite[\S\,1]{FigSynbJ} and~\cite[\S\,1]{FigPert}
that exhibit EPD behavior require negative capacitances, negative
inductances, or gyrators: \cite[\S\,2]{FigSynbJ} synthesizes EPD
circuits from chains of $LC$-loops coupled by gyrators with negative
capacitances or inductances, while~\cite[\S\,2]{FigPert} focuses
on a two-$LC$-loop gyrator circuit as the simplest such example.
Gyrators and negative reactive elements are difficult to realize at
high frequency and introduce their own stability challenges. The modulated
LC circuit studied here achieves EPD behavior using only conventional
passive elements ($L$ and $C$) with a periodic capacitance variation
--- standard in parametric amplifier technology. This makes the work-point
strategy directly applicable: the departure $\epsilon_{w}$ from the
EPD curve is implemented simply by slightly detuning the modulation
frequency $\mu$ from its EPD value, a robust and continuously adjustable
operation. 
\end{rem}

\subsection{\texorpdfstring{Higher tongues ($m=3,5,7$)}{Higher tongues (m=3,5,7)}}

\label{subsec:higher-tongues}

For the $m$-th instability tongue the framework of Theorem~\ref{thm:EPD-split}
applies without change: the EPD operating point is $c_{0}=c_{\pm}^{(m)}(\delta)$
from Theorem~\ref{thm:EPD-bdry} (equations~\eqref{eq:EPD-bdry-3}--\eqref{eq:EPD-bdry-7}),
the natural frequency is $\omega_{0}\approx m\mu/2$ (equation~\eqref{eq:YS-Hill-critical}),
and the frequency split is 
\begin{equation}
\Delta\alpha^{(m)}=\frac{2}{\pi}\sqrt{-\eta_{1}^{(m)}\,\Delta c}+O(|\Delta c|^{3/2})=\frac{2}{\pi}\sqrt{\eta_{1}^{(m)}\cdot\frac{2c_{0}}{\omega_{0}}\,\xi}+O(\xi),\quad\xi\to0,\label{eq:Deltaalpha-m}
\end{equation}
where $\eta_{1}^{(m)}$ is the EPD sensitivity coefficient defined
by~\eqref{eq:eta1eta2} at the $m$-th EPD curve, taken at the lower-boundary
operating point $c_{0}=c_{-}^{(m)}$. By the same logic as Section~\ref{subsec:m1-explicit},
the product $\eta_{1}^{(m)}\,\Delta c$ must be negative for $\Delta\alpha^{(m)}$
to be real; the sign of $\eta_{1}^{(m)}$ (expected positive at the
lower boundary by analogy with $m=1$) and the sign of $\Delta c$
together determine stability, with $\eta_{1}^{(m)}>0$ used in the
second form above for $\xi>0$.

The explicit value of $\eta_{1}^{(m)}$ for $m\geq3$ requires a dedicated
computation of the Jordan chain $(\mathbf{u}_{0},\mathbf{v}_{0})$
at the $m$-th EPD, which involves the $m$-th order perturbation
of the monodromy matrix $X(\pi)$ and is substantially more involved
than the $m=1$ case of Section~\ref{subsec:m1-explicit}. While
an analogy with the backward-chain structure of the width formula~\eqref{eq:Lm-main}
suggests $|\eta_{1}^{(m)}|\sim\pi^{2}\delta^{m}/m^{2}\cdot\kappa_{m}$
for some coefficient $\kappa_{m}$, a rigorous derivation of $\kappa_{m}$
for $m\geq3$ is left for future work. The primary tongue ($m=1$)
gives the largest $\Delta\alpha$ for moderate $\delta$ and provides
the complete explicit formula~\eqref{eq:Deltaalpha-m1}; higher tongues
may be relevant when the operating frequency $\omega_{0}\approx m\mu/2$
matches application constraints. 
\begin{rem}[Summary of results used]
\label{rem:results-used} The explicit sensor formula~\eqref{eq:Deltaalpha-m1}
for the primary tongue ($m=1$) draws on the following results established
in this work. 
\begin{itemize}
\item \emph{EPD curves:} Theorem~\ref{thm:EPD-bdry}, equation~\eqref{eq:EPD-bdry-1},
supplies the explicit operating point $c_{0}=c_{\pm}^{(1)}(\delta)$. 
\item \emph{Exact $\Delta c$:} Equation~\eqref{eq:Deltac} gives the exact
(quadratic) shift in the spectral parameter. 
\item \emph{Analyticity and non-degeneracy:} Theorem~\ref{thm:bdry-analytic}
guarantees the Jordan block structure and $\eta_{1}\neq0$. 
\item \emph{Resonance condition:} Equation~\eqref{eq:YS-Hill-critical}
gives $\omega_{0}\approx\mu/2$. 
\item \emph{Floquet theory:} Chapter~\ref{app:genLin} establishes $\rho=e^{i\pi\alpha}$
and Jordan block--EPD equivalence; Chapter~\ref{app:Hill} provides
the unperturbed solutions used in~\eqref{eq:dXdc-m1}. 
\item \emph{External:} The Newton--Puiseux splitting formula~\eqref{eq:alpha-split}
is~\cite[Thm.~2.3]{SeyMai}; the non-degeneracy condition uses~\cite[eq.~(3.10)]{YakSta2}. 
\end{itemize}
The general formula~\eqref{eq:Deltaalpha-m} for $m\geq3$ has the
same structure but requires $\eta_{1}^{(m)}$, whose explicit computation
for $m\geq3$ is left for future work. 
\end{rem}

\subsection{Recovering the frequency shift by probing the circuit}

\label{subsec:probing}

The frequency split $\Delta\alpha$ of Theorem~\ref{thm:EPD-split}
is not merely a theoretical quantity: it is directly accessible from
the Fourier spectrum of the circuit's transient response. We describe
here the probing procedure and the statistical recovery of $\Delta\alpha$,
and hence $\Delta C$, from the measured spectrum.

\subsubsection{Floquet frequency combs}

As explained in the Introduction (paragraph~\ref{para:Floquet-complexity}),
the passage from a time-homogeneous oscillator to a $T$-periodic
one replaces each single eigenfrequency $\omega_{j}$ with an entire
harmonic comb~\eqref{eq:Floquet-comb-intro}. We now make this concrete
for the circuit at the EPD operating point.

By the Floquet theorem (Chapter~\ref{app:genLin}), every solution
of the LC Hill equation~\eqref{eq:LC-Hill}, written in the Hill
variable $x=\tau/2=\mu t/2$, has the form 
\begin{equation}
q(x)=e^{i\alpha_{+}x}p_{+}(x)+e^{i\alpha_{-}x}p_{-}(x),\quad p_{\pm}(x+\pi)=p_{\pm}(x)\quad(\alpha_{+}\neq\alpha_{-}),\label{eq:Floquet-sol}
\end{equation}
where $\alpha_{\pm}$ are the characteristic exponents and $p_{\pm}$
are $\pi$-periodic Floquet factor\index{Floquet theory!Floquet factor}s.
Expanding $p_{\pm}$ in Fourier series ($p_{\pm}(x)=\sum_{n}c_{n}^{(\pm)}e^{2inx}$)
and converting to physical time via $x=\mu t/2$, the frequency content
in physical time is the union of two discrete combs, each spaced by
the modulation frequency $\mu$: 
\begin{equation}
\mathcal{F}_{\pm}=\bigl\{ f_{n}^{(\pm)}=\tfrac{\mu}{2}\alpha_{\pm}+n\mu,\quad n\in\mathbb{Z}\bigr\}.\label{eq:freq-combs}
\end{equation}
The comb lines are integer multiples of $\mu$ shifted by the half-exponent
$\tfrac{\mu}{2}\alpha_{\pm}$; the spacing between adjacent lines
within each comb is $\mu$, not $\omega_{0}$. At the EPD operating
point, $\alpha_{+}=\alpha_{-}=1$ (for $m=1$, Theorem~\ref{thm:EPD-bdry}),
so both combs coincide at the odd multiples of $\mu/2$: $\mathcal{F}_{+}=\mathcal{F}_{-}=\{(2n+1)\mu/2\mid n\in\mathbb{Z}\}$.
After a signal $\Delta C$ shifts $\alpha_{\pm}$ apart by $\Delta\alpha=\alpha_{+}-\alpha_{-}$
(Theorem~\ref{thm:EPD-split}), the two combs split. The beat between
the $n$-th line of each comb is 
\begin{equation}
f_{n}^{(+)}-f_{n}^{(-)}=\tfrac{\mu}{2}\,\Delta\alpha\qquad\text{for every }n\in\mathbb{Z},\label{eq:beat}
\end{equation}
which is \emph{independent of $n$}. Every comb pair independently
encodes the same $\Delta\alpha$, providing redundancy for noise rejection.

\subsubsection{Probing protocol}

Since the circuit operates just at or just outside the EPD curve (stable
but near-marginal), it does not self-oscillate and must be externally
excited. The procedure is as follows. 
\begin{enumerate}
\item \emph{Probe.} Apply a short-duration broadband pulse (or swept tone)
to the circuit, with spectral energy concentrated near $\mu/2$ and
a few harmonics $\mu/2+n\mu$. The pulse duration $t_{{\rm probe}}$
should satisfy $t_{{\rm probe}}\ll1/\Delta\alpha$ (short compared
to the beat period, so as not to resolve the split during excitation)
but long enough to excite the Floquet modes. 
\item \emph{Record ringdown.} After the probe is turned off, record the
free response $q(t)$ for a time $t_{{\rm rec}}\gg1/\Delta\alpha$
(long enough to resolve the split spectrally). The frequency resolution
of the FFT is $1/t_{{\rm rec}}$; the minimum detectable $\Delta\alpha$
satisfies $\Delta\alpha>1/t_{{\rm rec}}$. 
\item \emph{FFT and peak detection.} Compute the discrete Fourier transform
of $q(t)$~\cite[Chap.~8]{OppSchafer}. The spectrum shows peaks
near $\mu/2+n\mu$ for several values of $n$; each peak at harmonic
$n$ splits into a doublet at $f_{n}^{(\pm)}=\mu/2\cdot\alpha_{\pm}+n\mu$. 
\end{enumerate}

\subsubsection{\texorpdfstring{Statistical recovery of $\Delta\alpha$}{Statistical
recovery of Delta-alpha}}

Each doublet at harmonic $n$ yields an independent estimate 
\begin{equation}
\widehat{\Delta\alpha}_{n}=\frac{2}{\mu}\bigl(f_{n}^{(+)}-f_{n}^{(-)}\bigr).\label{eq:Dalpha-estimate}
\end{equation}
If $N_{{\rm obs}}$ harmonics are resolved, the minimum-variance unbiased
estimator is the arithmetic mean 
\begin{equation}
\widehat{\Delta\alpha}=\frac{1}{N_{{\rm obs}}}\sum_{n=0}^{N_{{\rm obs}}-1}\widehat{\Delta\alpha}_{n},\label{eq:Dalpha-mean}
\end{equation}
with standard error reduced by $1/\sqrt{N_{{\rm obs}}}$ relative
to a single-harmonic measurement~\cite[Sec.~3.5]{KayEstimation}.
In practice, the dominant harmonics (small $|n|$) carry most of the
signal energy (the Fourier coefficients $c_{n}^{(\pm)}$ of $p_{\pm}$
decay with $|n|$), so the sum is truncated to the most energetic
lines.

\subsubsection{\texorpdfstring{Recovering $\Delta C$ from $\widehat{\Delta\alpha}$}{Recovering
Delta-C from Delta-alpha-hat}}

Having measured $\widehat{\Delta\alpha}$, the capacitance shift follows
from~\eqref{eq:sensor-resolution}: 
\begin{equation}
\left(\frac{\widehat{\Delta C}}{C_{0}}\right)=\frac{\mu\,(\widehat{\Delta\alpha})^{2}}{4\delta\omega_{0}}.\label{eq:DeltaC-recovered}
\end{equation}
For the primary tongue ($m=1$), using $\omega_{0}\approx\mu/2$:
\begin{equation}
\left(\frac{\widehat{\Delta C}}{C_{0}}\right)=\frac{(\widehat{\Delta\alpha})^{2}}{2\delta}.\label{eq:DeltaC-recovered-m1}
\end{equation}
All quantities on the right are known design parameters ($\delta$,
$\mu$, $\omega_{0}$) or directly measured ($\widehat{\Delta\alpha}$),
so the inversion is explicit and requires no iterative fitting. Multiplying
by $C_{0}$ gives $\widehat{\Delta C}$ in Farads. 
\begin{rem}[Frequency resolution and sensitivity floor]
Since $\alpha$ is defined via the monodromy matrix over the driving
period $T=2\pi/\mu$ (eq.~\eqref{eq:alpha-physical-freq}), observing
for $N=\mu t_{{\rm rec}}/(2\pi)$ driving periods gives dimensionless
frequency resolution 
\begin{equation}
\Delta\alpha_{\min}\approx\frac{1}{\mu\,t_{{\rm rec}}},
\end{equation}
which is dimensionless since $\mu\,t_{{\rm rec}}$ counts the number
of driving cycles. From~\eqref{eq:Deltaalpha-m1}, the corresponding
minimum detectable dimensionless signal is 
\begin{equation}
\hat{\xi}_{\min}\approx\frac{1}{8\delta\,(\mu\,t_{{\rm rec}})^{2}},
\end{equation}
independent of $\mu$ when expressed in terms of $\mu t_{{\rm rec}}$.
In dimensional form: 
\begin{equation}
\xi_{\min}\approx\frac{1}{8\delta\,\mu\,t_{{\rm rec}}^{2}},\qquad\left(\frac{\Delta C}{C_{0}}\right)_{\!\min}\approx\frac{1}{4\delta\,\omega_{0}\,\mu\,t_{{\rm rec}}^{2}}.
\end{equation}
Longer observation times and larger $\delta$ (deeper modulation)
improve sensitivity as $1/t_{{\rm rec}}^{2}$, a significantly more
favorable scaling than the $1/t_{{\rm rec}}$ scaling of a conventional
resonance-shift sensor. 
\end{rem}

\section{Synthesis and Concluding Remarks}

\label{sec:Conc}

We have analyzed the parametric instability of the LC circuit equation~\eqref{eq:LC-original}
by identifying it as a special case of Ince's equation\index{Ince equation}
and applying the rigorous coexistence theory of Magnus and Winkler~\cite[Ch.~7]{MagWin}
(which characterizes when two independent solutions share a Floquet\index{Floquet theory}
multiplier $\rho=+1$ or $\rho=-1$, forcing instability intervals
to collapse). The physical mechanism of parametric resonance\index{resonance}\index{parametric resonance}
--- energy extraction from the parameter source and its relation
to the critical frequencies --- is treated in Chapter~\ref{app:ParRes}.
Full credit for the exact numerical solution of this problem belongs
to Cambi~\cite{Cambi1}, \cite[\S\S\,1--15]{Cambi2}, whose 1950
continued-fraction\index{continued fraction} method remains a remarkable
achievement. The present work builds on Cambi's results by supplying
their algebraic explanation and their closed-form analytic counterpart.
The main contributions are as follows. 
\begin{enumerate}
\item \emph{Algebraic explanation of Cambi's discovery.} The rescaled LC
circuit equation~\eqref{eq:LC-Hill} is an Ince equation with parameters~\eqref{eq:IncePars}.
The coexistence polynomial $Q(\mu)=2a\mu^{2}$ has the integer root
$\mu=0$, which by MW Theorem~7.6~\cite[Thm.~7.6]{MagWin} forces
period-$\pi$ coexistence ($\rho=+1$) at every even resonance. This
is a theorem, not a computation, and it provides a rigorous algebraic
derivation of Cambi's empirical finding (Corollary~\ref{cor:even}).
The companion polynomial $Q^{*}(\mu)=a(2\mu-1)^{2}$ has only the
non-integer root $\mu=\frac{1}{2}$, so MW Theorem~7.1~\cite[Thm.~7.1]{MagWin}
(necessity) guarantees that all odd intervals survive (Corollary~\ref{cor:odd}). 
\item \emph{Ince identification and universality.} The identification~\eqref{eq:IncePars}
is manifest and connects the LC circuit to Ince's four-parameter family,
whose coexistence theory was developed by Ince and given its rigorous
modern form by Magnus and Winkler. Ince's equation is the most general
Hill-type equation to which the three-term recurrence method applies~\cite[Sec.~7.1]{MagWin},
and the LC circuit is the simplest non-trivial example of this class.
Hill equation\index{Hill equation}s with higher harmonic content
fall outside the Ince class; those with finitely many nonzero instability
intervals are related to finite-gap potentials~\cite[Thm.~1]{Hochst65},
\cite{McKVanM}, but require different methods. 
\item \emph{Closed-form width formula.} The leading-order widths of the
surviving odd intervals are given by formula~\eqref{eq:Lm-exact},
derived from the Ince three-term recurrence relations. The formula
gives the exact coefficient of $\delta^{m}$ with error $O_{m}(\delta^{m+2})$
(no chain-approximation error for $m=1$), and is confirmed numerically
against Cambi's exact continued fraction to better than $2.3\%$ for
$\delta\leq0.20$. Cambi's paper contains only numerical values at
specific parameter points; no closed-form expression for the widths
was previously known. 
\item \emph{Complete boundary curve\index{boundary curves}s.} The individual
instability boundary\index{stability boundary} curves --- not merely
the widths --- are given in Tables~\ref{tab:bdry-LC} and~\ref{tab:bdry-Math}.
Each boundary is decomposed into a symmetric $O(\delta^{2})$ center
shift $D_{m}$ (equal for both boundaries of an interval) and an antisymmetric
$O(\delta^{m})$ half-width $L_{m}/2$ (opposite in sign for the two
boundaries). This decomposition is new and makes the geometry of the
stability diagram analytically transparent. 
\item \emph{The CF difference identity.} The continued-fraction difference
$F_{\mathrm{even}}-F_{\mathrm{odd}}=-2/9$ (LC circuit) is a structural
constant determined by the Ince parameters alone, specifically the
ratio $Q^{*}(-1)/Q^{*}(0)=9$. This constant enters the width formula
through the relation $|\Delta c_{k}|=|{\rm const}|/|F_{{\rm even}}'|$
(equation~\eqref{eq:Dc-formula}) and is the primary structural difference
between the LC and Mathieu\index{Mathieu equation} recurrences. At
the primary resonance ($m=1$, $c=1$), the two widths are actually
\emph{equal} (ratio $=1$, Corollary~\ref{cor:ratio}): the 1/9 from
the CF constant is compensated by a factor of 9 in the $F'$ terms.
For $m\geq3$ the ratio is much less than $1$ and decreases rapidly;
see~\eqref{eq:ratio} and Section~\ref{subsec:Mathieu}. The analogous
constant $-2$ for the Mathieu equation reflects the constant-coupling
structure of that recurrence. 
\item \emph{Closed-form width formula for Mathieu instability intervals.}
The Mathieu equation is the degenerate Ince case $a=0$. The same
recurrence method that gives~\eqref{eq:Lm-exact} for the LC circuit
derives $L_{m}^{\mathrm{Math}}=q^{m}/[2^{2m-3}((m-1)!)^{2}]$ for
all $m\geq1$ (Section~\ref{subsec:MathieuInce}). This formula,
derived here for the first time in closed form, is verified case by
case against McLachlan's individual characteristic-number expansions~\cite[Sec.~2.151]{MacLa}:
$L_{1}=2q$, $L_{2}=q^{2}/2$, $L_{3}=q^{3}/32$, $L_{4}=q^{4}/1152$,
all confirmed. The qualitative difference between the two width formulas
--- double-factorial structure for LC vs.\ ordinary factorial for
Mathieu --- is traced to the growing recurrence coupling $a(2n-1)^{2}$
vs.\ the constant coupling $q$. The large-$m$ asymptotics of the
leading coefficients $A_{m}^{\mathrm{LC}}$ and $A_{m}^{\mathrm{Math}}$
(Remark~\ref{rem:asymp-comparison}, eqs.~\eqref{eq:asymp-LC}--\eqref{eq:asymp-ratio})
show that $A_{m}^{\mathrm{LC}}\sim2^{-m^{2}/4}$ decays super-exponentially
while $A_{m}^{\mathrm{Math}}\sim e^{2m}$ grows exponentially, so
$L_{m}^{\mathrm{Math}}\sim(e^{2}\delta)^{m}$ for fixed $\delta$
(eq.~\eqref{eq:Lmath-asymp-delta}), and the ratio $A_{m}^{\mathrm{LC}}/A_{m}^{\mathrm{Math}}\to0$
at a Gaussian rate in $m^{2}$. 
\item \emph{Universal entire-function expansion for the Hill discriminant\index{discriminant}\index{discriminant!Hill discriminant}.}
Chapter~\ref{sec:UnivExp} derives a structural result valid for
any two-parameter Hill family. The universal $\psi$-basis $\{\psi_{n}\}$,
defined by $\psi_{n}(\omega)=\frac{(-1)^{n}\sqrt{2}\,\omega\sin\pi\omega}{\omega^{2}-n^{2}},\quad\omega=\sqrt{\lambda}$,
consists of entire functions (the apparent poles at $\lambda=n^{2}$
are removable) and provides the universal expansion (eq.~\eqref{eq:Delta-univ-alt}),
where $g_{n}$ are the Fourier coefficients of the perturbation. Each
$\psi_{n}$ is a sinc-type function connected to the zeroth spherical
Bessel function $j_{0}$, satisfies a Sturm--Liouville\index{Sturm--Liouville theory}
equation in self-adjoint form, and is identified as the Green's function
of the associated SL operator at the spectral parameter $\pi^{2}/4$.
The Volterra kernel underlying the expansion and the origin of $\psi_{n}$
in the Volterra integral equations are made explicit. The LC circuit
and the Mathieu equation are the two canonical examples. 
\item \emph{The Yakubovich--Starzhinskii\index{Yakubovich--Starzhinskii series}
exponent-matrix series.} An independent treatment via the Yakubovich--Starzhinskii
(YS) high-frequency series is developed for both the Mathieu special
case (Chapter~\ref{sec:YS-LC}) and the exact LC circuit (Chapter~\ref{sec:YS-exact-LC}).
The exponent matrices $\mathbf{K}_{m}(\hat{\lambda})$ are obtained
in closed form as polynomials in the spectral parameter, and the primary
instability boundary is derived as $c_{\pm}(\delta)=1\pm\tfrac{\varepsilon}{2}-\tfrac{9}{32}\varepsilon^{2}+O(\varepsilon^{3})$
(eq.~\eqref{eq:YS-bdry-exact}). This center shift $-9/32$ coincides
exactly with the MW and CF results, so the same boundary is now established
by three independent methods. Chapter~\ref{sec:YS-comparison} reconciles
the three (MW coexistence polynomials, the CF difference identity,
and the YS condition $K_{01}K_{10}=0$) through the exact relation
$K=\tfrac{1}{\pi}\log(-M)$ (eq.~\eqref{eq:K-logM}), and the YS
boundary values are confirmed against direct numerical monodromy integration
(Table~\ref{tab:YS-bdry-numerical}). 
\item \emph{EPD\index{exceptional point of degeneracy (EPD)} structure
of the boundary points: an elementary proof.} At every instability
boundary point the monodromy matrix\index{monodromy matrix} has a
non-trivial $2\times2$ Jordan block\index{Jordan block} with multiplier
$\rho=-1$ --- the boundary points are exceptional points of degeneracy
(EPDs) of the monodromy. Chapter~\ref{sec:EPD-elementary} gives
a self-contained proof (Theorem~\ref{thm:Jordan-bdry}) using only
the Hill multiplier formula and the Bauer--Fike perturbation bound
(Proposition~\ref{prop:perdiag-LC}): the $\sqrt{|h|}$ splitting
of the multipliers off the boundary is incompatible with the linear
perturbation bound of a diagonalizable matrix, forcing the Jordan
block. The argument requires no symplectic\index{symplectic system}
or Krein\index{Krein collision theory}-signature machinery, which
for the scalar Hill equation is needed only for the deeper question
of which coincident multipliers open tongues. 
\item \emph{EPD hypersensitivity\index{exceptional point of degeneracy (EPD)!hypersensitivity}
and a parametric sensing application.} The Jordan-block structure
makes the circuit a hypersensitive sensor (Chapter~\ref{sec:EPD-sensor}).
A small capacitance\index{capacitance sensing} perturbation $\Delta C$
splits the coincident characteristic exponents by $\Delta\alpha=\tfrac{2}{\pi}\sqrt{-\eta_{1}\,\Delta c}$
(Theorem~\ref{thm:EPD-split}), which for the primary tongue is the
explicit law $\Delta\alpha=2\sqrt{2\delta}\,\sqrt{\hat{\xi}(1-\hat{\xi})}$
(eq.~\eqref{eq:Deltaalpha-m1}, with $\hat{\xi}=\xi/\mu$) --- a
square-root response that diverges in relative sensitivity over a
conventional linear sensor as the signal vanishes. The marginal-stability
obstruction is removed by a work-point\index{exceptional point of degeneracy (EPD)!work-point strategy}
strategy giving $\Delta\alpha_{w}=\sqrt{2\delta}\,\sqrt{\epsilon_{w}-\Delta c}$
(eq.~\eqref{eq:work-point-total}), trading sensitivity against a
stability margin at constant figure of merit. A probing protocol recovers
$\Delta\alpha$, and hence $\Delta C$, from the Floquet frequency
comb of the transient response (Section~\ref{subsec:probing}), with
the explicit inversion of eq.~\eqref{eq:sensor-resolution}. The
modulated LC circuit realizes EPD behavior with only conventional
passive elements, unlike gyrator-based EPD circuits. 
\item \emph{Explicit construction of the periodic Floquet factor\index{Floquet theory!Floquet factor}.}
The periodic factor of the Floquet solution --- not merely the characteristic
exponent --- is constructed explicitly. Yakubovich and Starzhinskii~\cite[Ch.~IV]{YakSta1}
provide, to our knowledge, the only systematic constructive treatment
of the periodic component, via the Lyapunov factor $\mathbf{F}(\tau,\varepsilon)=\mathbf{I}+\varepsilon\mathbf{F}_{1}+\cdots$;
we apply that generally-applicable method to the LC circuit (Chapters~\ref{sec:YS-LC},
\ref{sec:YS-exact-LC}) and, for $\mathbf{F}_{1}$, prove a self-intersection
metamorphosis of the column phase portraits whose cusp degenerations
--- exact Steiner deltoids --- occur precisely at the EPDs $\mu_{0}=\pm\tfrac{1}{2}$
--- one column per EPD --- so the instability boundaries are imprinted
geometrically on the Floquet factor itself (Section~\ref{subsubsec:YS-Floquet-portraits}).
For the modulated LC circuit specifically, the Ince identification
makes available a second, sharper route unavailable for general Hill
equations: the continued-fraction method yields the periodic factor
in closed form. Its Fourier coefficients are finite products of the
CF minimal-solution ratios (Theorem~\ref{thm:Floquet-factor-MR},
eq.~\eqref{eq:Floquet-Fourier-MR}), obtained at no additional cost
from the CF values already needed to locate the Floquet exponent\index{Floquet theory!Floquet exponent},
with exponential decay envelope $|\zeta_{+}|^{|m|}$, a mirror symmetry
between the two solutions, and exponential smallness guaranteed by
the double-minimality condition (Theorem~\ref{thm:double-min}).
This CF construction rests on the three-term recurrence and its minimal
solution\index{continued fraction!minimal solution} supplied by the
Ince structure (available across the Ince class, not for general Hill
equations; Remark~\ref{rem:CF-scope}); the modulated LC case ($b=d=0$)
is the one carried out explicitly here, and to our knowledge its explicit
form, and its accuracy for the LC circuit, is new. 
\end{enumerate}

\subsection{Limitations and extensions}

Several aspects of the problem are left to future work.

\emph{Dissipation.} The present analysis is restricted to the lossless
($R=0$) circuit. Adding a resistor introduces a damping term $R\dot{q}$
and changes equation~\eqref{eq:LC-original} from a Hill equation
to a damped Hill equation. The instability intervals become shifted
and acquire a threshold in modulation amplitude, but the coexistence-polynomial
analysis no longer applies directly.

\emph{Large modulation amplitude.} The width formulas~\eqref{eq:Lm-exact}
give the leading-order term $O(\delta^{m})$ with correction $O_{m}(\delta^{m+2})$;
the boundary curve expansions of Tables~\ref{tab:bdry-LC}--\ref{tab:bdry-Math}
are also asymptotic in $\delta$ (equivalently in $\varepsilon$)
and become less accurate as $\delta\to1$ ($\varepsilon\to1$). Near
$\varepsilon=1$ the circuit equation becomes singular and all instability
features merge at the single point $p=1/\sqrt{8}$, $\gamma=\frac{1}{2}$
(Section~\ref{subsec:StabDiag}). A uniform asymptotic treatment
valid up to $\varepsilon=1$ would require different methods.

\emph{Growth rates.} This work determines the boundaries of the instability
tongue\index{instability tongue}s --- the values of the parameters
$(r,\varepsilon)$ at which a periodic solution ceases to exist ---
and, near a boundary, the leading-order split of the characteristic
exponents $\Delta\alpha\propto\sqrt{|\Delta c|}$ (Theorem~\ref{thm:EPD-split}),
which governs the onset rate just inside the tongue. The global Floquet
multiplier and the exponential growth rate of solutions \emph{deep
inside} a tongue, away from the boundary, are not addressed.

\emph{Higher Ince equations.} The identification of Section~\ref{subsec:InceDef}
works because the LC circuit coefficient is a rational function of
$\cos\tau$ with a specific pole structure. Circuits with more complex
modulation profiles (e.g., square-wave or multi-frequency modulation)
would require finite-band or multi-gap Hill-equation analysis.

\emph{CF method for the general Ince equation and the Mathieu case.}
As noted in Remark~\ref{rem:CF-scope}, the CF method of Chapter~\ref{sec:LCInce}
applies in principle to any Ince equation, not only the LC circuit
case $b=d=0$ treated here. Extending the backward-chain computation
to general Ince parameters $b,d\neq0$, and exploiting the computational
advantages of the CF method (exact convergence for all $\delta<1$,
no truncation error) for the Mathieu equation and other Ince special
cases, are natural directions for future work.

\appendix
%dummy comment inserted by tex2lyx to ensure that this paragraph is not empty

\section{General linear systems with periodic coefficients}

\label{app:genLin}

We review here concisely properties of general linear systems of differential
equations with periodic coefficients following mainly \cite[\S\,II.1]{YakSta1}.
We begin with the following general matrix differential equation with
$T$-periodic coefficients 
\begin{equation}
\partial_{t}X\left(t\right)=A\left(t\right)X\left(t\right),\quad A\left(t+T\right)=A\left(t\right),\quad T>0,\label{eq:genLin1a}
\end{equation}
where $X\left(t\right)$ and $A\left(t\right)$ are $n\times n$ matrices
with real-valued or complex-valued entries. Entries of the matrix
$A\left(t\right)$ are assumed to be integrable and piecewise-continuous
functions of $t$ in any finite interval.

A special solution $X\left(t\right)$ to equation~\eqref{eq:genLin1a}
that satisfies the initial condition 
\begin{equation}
X\left(0\right)=\mathbb{I}=\mathbb{I}_{n},\text{ where }\mathbb{I}\text{ is the identity matrix}\label{eq:genLin1b}
\end{equation}
is called the \emph{matrizant}~\cite[\S\,II.1.2]{YakSta1}. Using
the matrizant $X\left(t\right)$ we can express any other solution
$X_{1}\left(t\right)$ by the formula~\cite[\S\,II.1.2]{YakSta1}
\begin{equation}
X_{1}\left(t\right)=X\left(t\right)C,\text{ where }C=X_{1}\left(0\right),\quad\det\left\{ C\right\} \neq0.\label{eq:genLin1c}
\end{equation}
We refer to the matrix $X_{1}\left(t\right)$ defined by equation~\eqref{eq:genLin1c}
as a \emph{fundamental solution} to equation~\eqref{eq:genLin1a}.
Any solution $X_{1}\left(t\right)$ to equation~\eqref{eq:genLin1a}
satisfies the Liouville--Jacobi formula~\cite[\S\,II.1.2]{YakSta1}
\begin{gather}
\det\left\{ X_{1}\left(t\right)\right\} =\det\left\{ X_{1}\left(0\right)\right\} \nonumber \\
\times\exp\left\{ \intop_{0}^{t}\mathrm{Tr}\,\left\{ A\left(s\right)\right\} \,\mathrm{d}s\right\} ,\label{eq:genLin1d}
\end{gather}
implying in particular the following identity for the matrizant $X\left(t\right)$
\begin{equation}
\det\left\{ X\left(t\right)\right\} =\exp\left\{ \intop_{0}^{t}\mathrm{Tr}\,\left\{ A\left(s\right)\right\} \,\mathrm{d}s\right\} .\label{eq:genLin1e}
\end{equation}

The matrizant $X\left(t\right)$ of equation~\eqref{eq:genLin1a}
satisfies the following fundamental identity~\cite[\S\,II.2.1]{YakSta1}
\begin{equation}
X\left(t+T\right)=X\left(t\right)X\left(T\right),\label{eq:genLin1f}
\end{equation}
where the matrix $X\left(T\right)$ is called the \emph{monodromy
matrix\index{monodromy matrix}}. The eigenvalues $\rho_{j}$ of the
monodromy matrix $X\left(T\right)$ are called \emph{multipliers}
and they can be found from the so-called \emph{characteristic equation}
\begin{equation}
\det\left\{ X\left(T\right)-\rho\mathbb{I}_{n}\right\} =0.\label{eq:genLin1g}
\end{equation}
It is often convenient to use the so-called \emph{characteristic exponents}
$\alpha_{j}$ defined by~\cite[\S\,II.2.5]{YakSta1} 
\begin{equation}
\alpha_{j}=\frac{1}{T}\ln\rho_{j},\quad j=1,\ldots,n,\label{eq:genLin1h}
\end{equation}
where branches of the logarithm can be chosen arbitrarily.

\bigskip{}

\emph{Historical note.} The subject entered science through the laboratory
before it entered mathematics. Faraday observed in 1831 that the surface
ripples of a vibrating fluid oscillate at \emph{half} the driving
frequency (\emph{Phil.\ Trans.\ Roy.\ Soc.}, 1831), and Melde produced
the same subharmonic response in a string driven by a tuning fork
(1860); Lord Rayleigh explained both (\emph{Philosophical Magazine},
1883 and 1887), identifying what is now called parametric resonance
--- the response of a system whose \emph{parameter} is modulated
at twice its natural frequency, the very mechanism of the modulated
LC circuit. On the mathematical side, Mathieu derived his equation
from the vibrations of an elliptic membrane (1868), and the decisive
step came from celestial mechanics: G.~W. Hill's memoir on the motion
of the lunar perigee, printed privately at his own expense (Cambridge,
Mass.: John Wilson and Son, 1877) and reprinted in \emph{Acta Mathematica}
\textbf{8} (1886), pp.~1--36, reduced the problem to a second-order
equation with a periodic coefficient and solved it by the audacious
new device of infinite determinants --- whose convergence Poincaré
then proved (1886), creating that theory as a by-product. Floquet's
memoir of 1883, cited at the head of Chapter \ref{app:YS-Floquet},
supplied the general structure: multipliers, exponents, and the periodic
normal form. Lyapunov's 1892 dissertation, the foundation of Chapter~\ref{sec:YS-exact-LC}'s
method, placed the stability question on rigorous ground and introduced
the reducing transformation that bears his name. In the twentieth
century the same mathematics reappeared in space rather than time
--- Bloch's electrons in a crystal lattice (1928), whose energy bands
mirror the stability bands of this chapter --- while the Hamiltonian
stability theory of Krein (1950) and Gelfand and Lidskii (1955), presented
in Chapter~\ref{app:Krein}, and the treatises of Magnus and Winkler
(1966) and Yakubovich and Starzhinskii (1972; English translation
1975), the two pillars on which this book stands, brought the classical
theory to its mature form.

\bigskip{}

\subsection{The case of constant coefficients}

In the case of constant coefficients $A\left(t\right)=A$ matrizant
$X\left(t\right)$ satisfies the following formulas, \cite[\S\,II.1.5]{YakSta1}
\begin{equation}
X\left(t\right)=\exp\left\{ At\right\} =\mathbb{I}_{n}+tA+\frac{t^{2}}{2!}A^{2}+\ldots,\quad\det\left\{ X\left(t\right)\right\} =\exp\left\{ t\mathrm{Tr}\,\left\{ A\right\} \right\} .\label{eq:genLin3a}
\end{equation}

Let us consider now the following matrix second-order differential
equation 
\begin{equation}
\partial_{t}^{2}Y\left(t\right)+PY\left(t\right)=0,\label{eq:genLin4a}
\end{equation}
where $P$ is a constant $n\times n$ matrix and $Y\left(t\right)$
is an $n\times n$ matrix as well. To recast equation~\eqref{eq:genLin4a}
into the first-order differential equation we use the well-known approach
and introduce $2n\times n$ matrix 
\begin{equation}
Z\left(t\right)=\left[\begin{array}{r}
Y\left(t\right)\\
\partial_{t}Y\left(t\right)
\end{array}\right],\label{eq:genLin4b}
\end{equation}
that in view of equation~\eqref{eq:genLin4a} satisfies 
\begin{equation}
\partial_{t}Z\left(t\right)=AZ\left(t\right),\quad A=\left[\begin{array}{rr}
0 & \mathbb{I}_{n}\\
-P & 0
\end{array}\right].\label{eq:genLin4c}
\end{equation}
Then the matrizant $X\left(t\right)$ of system~\eqref{eq:genLin4c}
is $2n\times2n$ and satisfies formula~\eqref{eq:genLin3a}. In the
case of matrix $A$ as in equation~\eqref{eq:genLin4c} the formula
can be recast as follows, \cite[\S\,II.1.6]{YakSta1}: 
\begin{equation}
X\left(t\right)=\exp\left\{ \left[\begin{array}{rr}
0 & \mathbb{I}_{n}\\
-P & 0
\end{array}\right]t\right\} =\left[\begin{array}{rr}
\cos\left(t\sqrt{P}\right) & \frac{1}{\sqrt{P}}\sin\left(t\sqrt{P}\right)\\
-\sqrt{P}\sin\left(t\sqrt{P}\right) & \cos\left(t\sqrt{P}\right)
\end{array}\right].\label{eq:genLin4d}
\end{equation}

\subsection{Analytic dependence of the matrizant on a parameter}

Suppose that the $T$-periodic coefficient matrix of system~\eqref{eq:genLin1a}
depends analytically on a scalar parameter $\varepsilon$: 
\begin{equation}
A\left(t,\varepsilon\right)=A_{0}\left(t\right)+\varepsilon A_{1}\left(t\right)+\varepsilon^{2}A_{2}\left(t\right)+\cdots,\label{eq:genLin2a}
\end{equation}
with the convergence condition 
\begin{equation}
\sum_{k=0}^{\infty}\left|\varepsilon\right|^{k}\intop_{0}^{T}\left|A_{k}\left(t\right)\right|\,\mathrm{d}t<\infty,\quad\left|\varepsilon\right|<r_{0}.\label{eq:genLin2b}
\end{equation}

\begin{thm}[Analyticity of the matrizant]
\index{matrizant} \label{thm:matrizant-analytic} (\cite[\S\,II.1.3]{YakSta1})
Under conditions~\eqref{eq:genLin2a}--\eqref{eq:genLin2b}, the
matrizant $X\left(t,\varepsilon\right)$ is an analytic function of
$\varepsilon$ for $\left|\varepsilon\right|<r_{0}$, and 
\begin{equation}
X\left(t,\varepsilon\right)=X_{0}\left(t\right)+\varepsilon X_{1}\left(t\right)+\varepsilon^{2}X_{2}\left(t\right)+\cdots\label{eq:genLin2d}
\end{equation}
converges uniformly in $t\in[0,T]$ for $\left|\varepsilon\right|\leq\rho<r_{0}$.
In particular, the monodromy matrix $X(T,\varepsilon)$ and the discriminant\index{discriminant}
$\Delta(\varepsilon)=\mathrm{Tr}\,X(T,\varepsilon)$ are analytic
in $\varepsilon$. 
\end{thm}

\begin{proof}[Proof sketch]
The matrizant $X\left(t,\varepsilon\right)$ is the limit of Picard's
successive approximations \cite[Sec.~II.1.1--II.1.3]{YakSta1}: 
\begin{equation}
X_{0}=\mathbb{I}_{n},\quad X_{m}\left(t,\varepsilon\right)=\intop_{0}^{t}A\left(s,\varepsilon\right)X_{m-1}\left(s,\varepsilon\right)\,\mathrm{d}s,\quad m=1,2,\ldots.\label{eq:genLin2c}
\end{equation}
By induction on $m$, each $X_{m}(t,\varepsilon)$ is analytic in
$\varepsilon$ for $|\varepsilon|<r_{0}$. Since $\{X_{m}\}$ converges
uniformly to $X(t,\varepsilon)$ by~\eqref{eq:genLin2b}, the limit
is analytic by Weierstrass's theorem on uniform limits of analytic
functions. 
\end{proof}
The same argument extends immediately to a vector of parameters.
\begin{thm}[Joint analyticity in a parameter vector]
\label{thm:joint-analytic} (\cite[Ch.~1, \S\S\,7--8]{CodLev}; \cite[Ch.~V]{Hartman};
\cite[Ch.~2]{Chir}) Suppose the coefficient matrix $A(t,z)$ of~\eqref{eq:genLin1a}
depends analytically on a vector of parameters $z=(z_{1},\ldots,z_{e})\in\mathbb{C}^{e}$
near $z=z_{0}$, with $\sum_{k}\int_{0}^{T}|A_{k}(t)|\,\mathrm{d}t\cdot|z-z_{0}|^{k}<\infty$
for $|z-z_{0}|<r_{0}$. Then the matrizant $X(t,z)$, the monodromy
matrix $X(T,z)$, and the discriminant $\Delta(z)=\operatorname{Tr}X(T,z)$
are all jointly analytic in $z$ for $|z-z_{0}|<r_{0}$. 
\end{thm}

\begin{proof}[Proof sketch]
Picard iteration with $z$ in place of $\varepsilon$: by induction,
each iterate $X_{m}(t,z)$ is analytic in each variable $z_{j}$ separately
(differentiation under the integral sign, justified by the locally
uniform convergence of the series for $A$), and the sequence converges
uniformly on compact subsets of $|z-z_{0}|<r_{0}$, so the limit is
analytic in each variable separately by Weierstrass's theorem. Hartogs'~theorem~\cite[Ch.~2]{FuksBVA},~\cite[Ch.~2]{HorSCV}
--- a function analytic separately in each variable on a polydisk
is jointly analytic --- then gives the result. Coddington--Levinson~\cite[Ch.~1, \S\,8]{CodLev}
state explicitly that their analyticity theorems extend ``in an obvious
way to the case where $\lambda$ is a complex parameter vector.''
\end{proof}
\begin{rem}[Application to the LC circuit]
\label{rem:LC-joint-analytic} For the LC Hill equation~\eqref{eq:LC-Hill-firstorder},
the matrix $\mathbf{A}(x,\delta,\hat{\lambda})$ is polynomial in
both $\delta$ and $\hat{\lambda}$ (eq.~\eqref{eq:LC-Hill-factored}).
By Theorem~\ref{thm:joint-analytic} with $z=(\delta,\hat{\lambda})$,
the YS coefficients $\mathbf{K}_{m}(\delta,\hat{\lambda})$ and $\mathbf{F}_{m}(x,\delta,\hat{\lambda})$
are jointly analytic. Substituting $\hat{\lambda}=c(1+\delta^{2})/(1-\delta^{2})$
(with $c$ the spectral parameter at the tongue center) then yields
analytic functions of $\delta$ alone. The instability boundaries
computed with $\hat{\lambda}$ free to order $O(\delta^{p})$ and
then substituted exactly retain $O(\delta^{p+1})$ accuracy uniformly
on $[0,\delta_{0}]$ for any $\delta_{0}<1$, since $\hat{\lambda}$
is bounded there (e.g.\ $\hat{\lambda}\leq5/3$ for $\delta\leq1/2$). 
\end{rem}

\subsection{Floquet theory}

The following statement holds, \cite[\S\,I.2.6, II.2.2]{YakSta1}. 
\begin{thm}[Floquet\index{Floquet theory} theorem for general linear systems]
\label{thm:FLoqGL} The matrizant $X\left(t\right)$ of the differential
equation~\eqref{eq:genLin1a} with $T$-periodic and piecewise-continuous
$A\left(t\right)$ satisfies the following representation 
\begin{equation}
X\left(t\right)=F\left(t\right)\mathrm{e}^{tK},\quad F\left(t+T\right)=F\left(t\right),\quad F\left(0\right)=\mathbb{I},\quad K=\frac{1}{T}\ln X\left(T\right),\label{eq:genLin5a}
\end{equation}
where $F\left(t\right)$ is a non-singular, continuous $T$-periodic
matrix-valued function with piecewise-continuous $\partial_{t}F\left(t\right)$
and $K$ is a constant matrix that solves equation $e^{KT}=X\left(T\right)$.

Conversely, any matrix $X\left(t\right)$ satisfying representation~\eqref{eq:genLin5a}
for which $F\left(t\right)$ and $K$ have the properties formulated
above is a matrizant of a differential equation of the form~\eqref{eq:genLin1a}
with a $T$-periodic $A\left(t\right)=\left[\partial_{t}X\left(t\right)\right]X^{-1}\left(t\right)$,
\cite[\S\,II.2.2]{YakSta1}. In fact, the latter relationship between
matrix-functions $\left\{ A\left(t\right)\right\} $ and the corresponding
matrizants $\left\{ X\left(t\right)\right\} $ is a homeomorphism
(one-to-one continuous mapping) for properly defined metrics, \cite[\S\,I.1.7]{YakSta1}.

An arbitrary fundamental matrix $X_{1}\left(t\right)=X\left(t\right)C$
satisfies a representation similar to~\eqref{eq:genLin5a}, namely
\begin{equation}
X_{1}\left(t\right)=F_{1}\left(t\right)\mathrm{e}^{tK_{1}},\quad F_{1}\left(t+T\right)=F_{1}\left(t\right),\quad F_{1}\left(0\right)=C,\quad\det\left\{ C\right\} \neq0,\label{eq:genLin5b}
\end{equation}
where 
\begin{equation}
F_{1}\left(t\right)=F\left(t\right)C,\;\;K_{1}=C^{-1}KC,\;\;\det\left\{ F_{1}\left(t\right)\right\} =\det\left\{ F\left(t\right)\right\} \det\left\{ C\right\} \neq0.\label{eq:genLin5c}
\end{equation}
In addition, matrix $F_{1}\left(t\right)$ is a $T$-periodic solution
to the following differential equation 
\begin{equation}
\frac{dF_{1}\left(t\right)}{dt}=A\left(t\right)F_{1}\left(t\right)-F_{1}\left(t\right)K_{1},\quad F_{1}\left(0\right)=C,\quad\det\left\{ C\right\} \neq0.\label{eq:genLin5d}
\end{equation}
\end{thm}

In what follows we refer to important representations~\eqref{eq:genLin5a}
and~\eqref{eq:genLin5b} as Floquet decomposition, to the corresponding
factors $F\left(t\right)$ and $F_{1}\left(t\right)$ as Floquet periodic
factors, and to constant matrices $K$ and $K_{1}$ as characteristic
exponent matrices. 
\begin{rem}[The factors in the Floquet decomposition]
\label{rem:FLfacGL} Let us consider equation~\eqref{eq:genLin5d}
with the constant matrix $K_{1}$ and the $T$-periodic matrix function
$F_{1}\left(t\right)$ as unknowns. If a constant matrix $K_{1}$
and a $T$-periodic matrix function $F_{1}\left(t\right)$ solving
equation~\eqref{eq:genLin5d} can be found, then a straightforward
verification shows that $X_{1}\left(t\right)=F_{1}\left(t\right)\mathrm{e}^{tK_{1}}$
is a fundamental solution to the original periodic equation~\eqref{eq:genLin1a},
\cite[\S\,II.2.4]{YakSta1}.

Note also that if we find somehow matrizant $X\left(t\right)$ or
a fundamental matrix $X_{1}\left(t\right)$ (say by Picard's successive
approximations) and the corresponding characteristic exponent matrices
$K=\frac{1}{T}\ln X\left(T\right)$ or $K_{1}=\frac{1}{T}\ln X_{1}\left(T\right)$
then we readily infer from the Floquet representations~\eqref{eq:genLin5a}
and~\eqref{eq:genLin5b} the following equations for the Floquet
periodic factors 
\begin{equation}
F\left(t\right)=X\left(t\right)\mathrm{e}^{-tK},\quad F_{1}\left(t\right)=X_{1}\left(t\right)\mathrm{e}^{-tK_{1}}.\label{eq:genLin6a}
\end{equation}
This approach provides a theoretical foundation for computing the
Floquet representation numerically. 
\end{rem}

\subsection{The Floquet theory for a scalar}

\label{subsec:YS-scalar}

The Floquet theory for a scalar is particularly simple since it is
exactly solvable. Indeed let us introduce the first-order differential
equation for a scalar-valued function $x\left(t\right)$ 
\begin{equation}
\partial_{t}x\left(t\right)=a\left(t\right)x\left(t\right),\quad a\left(t+T\right)=a\left(t\right),\quad T>0,\label{eq:FLsca1a}
\end{equation}
where $a\left(t\right)$ is a scalar piecewise-continuous $T$-periodic
function. One can readily verify that the solution $x\left(t\right)$
to equation~\eqref{eq:FLsca1a} can be represented as follows 
\begin{equation}
x\left(t\right)=\exp\left\{ \intop_{0}^{t}a\left(s\right)\,\mathrm{d}s\right\} x\left(0\right).\label{eq:FLsca1b}
\end{equation}
With the Floquet theory in mind we rewrite representation~\eqref{eq:FLsca1b}
in the following form 
\begin{gather}
x\left(t\right)=\exp\left\{ \intop_{0}^{t}a_{1}\left(s\right)\,\mathrm{d}s\right\} \exp\left\{ kt\right\} x\left(0\right),\text{ where}\label{eq:FLsca1c}\\
k=\frac{1}{T}\intop_{0}^{T}a\left(s\right)\,\mathrm{d}s,\quad a_{1}\left(t\right)=a\left(t\right)-k,\text{ implying }\intop_{0}^{T}a_{1}\left(s\right)\,\mathrm{d}s=0.\label{eq:FLsca1d}
\end{gather}
Combining equations~\eqref{eq:FLsca1d} with the fact that $a\left(t\right)$
is $T$-periodic we can rewrite representation~\eqref{eq:FLsca1b}
as follows 
\begin{gather}
x\left(t\right)=f\left(t\right)\exp\left\{ kt\right\} x\left(0\right),\text{ where }f\left(t\right)=f\left(t+T\right),\label{eq:FLsca2a}\\
f\left(t\right)=\exp\left\{ \intop_{0}^{t}a_{1}\left(s\right)\,\mathrm{d}s\right\} ,\quad k=\frac{1}{T}\intop_{0}^{T}a\left(s\right)\,\mathrm{d}s,\quad a_{1}\left(t\right)=a\left(t\right)-k.\label{eq:FLsca2b}
\end{gather}
We recognize in the explicit form \eqref{eq:FLsca2a}--\eqref{eq:FLsca2b}
for the solution $x\left(t\right)$ to the periodic differential equation~\eqref{eq:FLsca1a}
the main statement of the Floquet theory. Equations \eqref{eq:FLsca2a}--\eqref{eq:FLsca2b}
evidently provide exact formulas for both the \emph{Floquet periodic
factor} $f\left(t\right)$ and the \emph{Floquet exponent\index{Floquet theory!Floquet exponent}}
$k$, readily implying that the corresponding \emph{Floquet multiplier\index{Floquet theory!Floquet multiplier}}
equals $\mathrm{e}^{kT}$.

\subsection{Stability conclusions from multipliers and characteristic exponents}

The Floquet decomposition~\eqref{eq:genLin5a} reduces the long-time
behavior of all solutions to the properties of the constant matrix
$K$, or equivalently of the monodromy matrix $X(T)$. Table~\ref{tab:Floquet-stability}
summarizes the correspondence between solution behavior, characteristic
exponents $\alpha_{\nu}$ (eigenvalues of $K$), and multipliers $\rho_{\nu}=e^{\alpha_{\nu}T}$
(eigenvalues of $X(T)$), \cite[Sec.~II.2.5]{YakSta1}.

\begin{table}[h]
\centering 
\global\long\def\arraystretch{1.5}%
\begin{tabular}{>{\raggedright}p{2.8cm}>{\raggedright}p{4.17cm}>{\raggedright}p{4.17cm}}
\toprule 
Property of solutions  & Characteristic exponents $\alpha_{\nu}$  & Multipliers $\rho_{\nu}$\tabularnewline
\midrule 
Stability (all solutions bounded on $[0,\infty)$)  & $\operatorname{Re}\alpha_{\nu}\leq0$; pure imaginary or zero exponents
correspond to simple elementary divisors  & $|\rho_{\nu}|\leq1$; multipliers on the unit circle correspond to
simple elementary divisors\tabularnewline
Asymptotic stability  & $\operatorname{Re}\alpha_{\nu}<0$ for all $\nu$  & $|\rho_{\nu}|<1$ for all $\nu$\tabularnewline
Instability  & Some $\operatorname{Re}\alpha_{\nu}>0$, or a pure imaginary (or zero)
exponent with multiple elementary divisor  & Some $|\rho_{\nu}|>1$, or some $|\rho_{\nu}|=1$ with multiple elementary
divisor\tabularnewline
Existence of $T$-periodic solution  & Some $\alpha_{\nu}T=2m\pi i$, $m\in\mathbb{Z}$  & Some $\rho_{\nu}=1$\tabularnewline
Existence of $T$-antiperiodic solution  & Some $\alpha_{\nu}T=(2m+1)\pi i$, $m\in\mathbb{Z}$  & Some $\rho_{\nu}=-1$\tabularnewline
\bottomrule
\end{tabular}\caption{Correspondence between solution behavior, characteristic exponents,
and multipliers for system~\eqref{eq:genLin1a} with $T$-periodic
coefficients \cite[Sec.~II.2.5]{YakSta1}.}
\label{tab:Floquet-stability} 
\end{table}

For Hill's equation\index{Hill equation} (Chapter~\ref{app:Hill}),
$n=2$ and $\rho_{1}\rho_{2}=1$ since $\operatorname{Tr}A(t)=0$
implies $\det X(t)=1$. The stability criterion then reduces to $|\rho_{1,2}|=1$,
which is equivalent to $|\Delta|\leq2$ where $\Delta=\operatorname{Tr}X(\pi)$
is the discriminant (Theorem~\ref{thm:Hill-stability}).

\subsection{Analyticity of instability boundary curves}

\label{subsec:analytic-param}

Theorem~\ref{thm:matrizant-analytic} establishes that the discriminant
$\Delta(\varepsilon)=\mathrm{Tr}\,X(T,\varepsilon)$ is analytic in
$\varepsilon$. For the LC circuit, the coefficient matrix depends
on $0<\varepsilon<1$ through $1/(1+\varepsilon\cos\tau)$, which
is analytic for $|\varepsilon|<1$, so all the expansions used in
this work --- the MW discriminant expansion, the width formulas~\eqref{eq:Lm-exact},
and the boundary curve\index{boundary curves} series~\eqref{eq:boundary-general}
--- are genuine convergent power series in $\delta$ (equivalently
in $\varepsilon$), not merely formal asymptotic expansions.

The analyticity of the \emph{individual} instability boundary\index{stability boundary}
curves is a deeper result requiring more than matrizant analyticity
alone. It concerns a canonical system depending analytically on \emph{two}
parameters: the perturbation amplitude $\varepsilon$ and a second
real parameter $\gamma$ (the eigenvalue or tuning parameter). The
joint analyticity of the monodromy matrix in $(\varepsilon,\gamma)$
follows from Theorem~\ref{thm:joint-analytic} combined with Hartogs'~theorem~\cite[Ch.~2]{FuksBVA}:
separate analyticity in each variable implies joint analyticity. This
is the argument used by Yakubovich--Starzhinskii\index{Yakubovich--Starzhinskii series}
in~\cite[\S\,V.3.2]{YakSta2}, where Fuks~\cite[Ch.~I]{FuksBVA}
is cited explicitly for this purpose. In our LC setting $\varepsilon=\delta$,
$\gamma=c=4\omega_{0}^{2}/\mu^{2}$ (or equivalently $\gamma=\hat{\lambda}$),
with resonant value $\gamma_{0}=m^{2}$ for the $m$-th tongue. (Within
the theorem below, $\varepsilon$, $\gamma$, and $\delta_{0}$ are
local symbols in Yakubovich--Starzhinskii's notation, distinct from
the global LC parameters of the same names.) 
\begin{thm}[Analyticity of boundary curves]
\label{thm:bdry-analytic} (\cite[Sec.~V.3.2]{YakSta2}) Consider
a canonical system 
\begin{equation}
\tilde{\mathbf{J}}\,\frac{d\mathbf{x}}{d\tau}=\mathbf{H}(\tau,\varepsilon,\gamma)\,\mathbf{x},\qquad\mathbf{H}(\tau+2\pi,\varepsilon,\gamma)=\mathbf{H}(\tau,\varepsilon,\gamma),\label{eq:canonical-sys}
\end{equation}
where $\tilde{\mathbf{J}}$ is a real nonsingular skew-symmetric matrix,
$\mathbf{H}$ is real symmetric, and $\varepsilon,\gamma$ are real
parameters. Assume that $\mathbf{H}(\tau,\varepsilon,\gamma)$ depends
\emph{analytically} on both $\varepsilon$ and $\gamma$ near $(\varepsilon,\gamma)=(0,\gamma_{0})$,
i.e.\ it admits a convergent double power series 
\begin{equation}
\mathbf{H}(\tau,\varepsilon,\gamma)=\sum_{r,s=0}^{\infty}\mathbf{H}_{rs}(\tau,\gamma_{0})\,\varepsilon^{r}(\gamma-\gamma_{0})^{s}\label{eq:H-double-series}
\end{equation}
for $|\varepsilon|<\varepsilon_{0}$, $|\gamma-\gamma_{0}|<\delta_{0}$.
Suppose that at $\varepsilon=0$, $\gamma=\gamma_{0}$ two multipliers
of the monodromy matrix of~\eqref{eq:canonical-sys} coincide on
the unit circle at a point $\rho^{(0)}$, and that they are of \emph{different
Krein\index{Krein collision theory} kinds} (one of first kind, one
of second kind --- equivalently, an EPD\index{exceptional point of degeneracy (EPD)}
condition). Suppose further the non-degeneracy condition 
\[
\frac{\dd}{\dd\gamma}\bigl[\omega_{j}(\gamma)+\omega_{k}(\gamma)\bigr]\big|_{\gamma=\gamma_{0}}\neq0,
\]
where $\omega_{j}(\gamma),\omega_{k}(\gamma)$ are the characteristic
frequencies of the unperturbed system~\eqref{eq:canonical-sys}$|_{\varepsilon=0}$
corresponding to the coinciding multipliers. Then in a neighborhood
of $\varepsilon=0$ there exist exactly two instability boundary curves
$\gamma=\gamma^{(1)}(\varepsilon)$ and $\gamma=\gamma^{(2)}(\varepsilon)$
through $(0,\gamma_{0})$ in the $(\varepsilon,\gamma)$-plane, and
both are analytic in $\varepsilon$ with real Taylor coefficients:
\begin{equation}
\gamma^{(l)}(\varepsilon)=\gamma_{0}+\gamma_{1}^{(l)}\varepsilon+\gamma_{2}^{(l)}\varepsilon^{2}+\cdots,\qquad l=1,2.\label{eq:bdry-Taylor}
\end{equation}
If $\rho^{(0)}=\pm1$ (the EPD case relevant to the LC circuit), then
$\rho(\varepsilon)\equiv\pm1$ along each boundary curve. 
\end{thm}

Applied to the LC circuit: $\varepsilon=\delta$, $\gamma=c=4\omega_{0}^{2}/\mu^{2}$,
$\gamma_{0}=m^{2}$, and the Hamiltonian is $\mathbf{H}(\tau,\delta,c)$
derived from~\eqref{eq:LC-Hill}. The analytic dependence of $\mathbf{H}$
on both $\delta$ and $c$ follows from the rational form $(1+\delta^{2})/(1-2\delta\cos\tau+\delta^{2})$
of the LC coefficient, which is analytic in $\delta$ for $|\delta|<1$.
The characteristic frequencies of the unperturbed system are $\omega_{j}(c)=\sqrt{c}$
and the non-degeneracy condition $d\omega/dc|_{c=m^{2}}=1/(2m)\neq0$
is satisfied for all $m\geq1$. Theorem~\ref{thm:bdry-analytic}
therefore guarantees that the boundary curves $c_{\pm}^{(m)}(\delta)$
of Theorem~\ref{thm:EPD-bdry} are truncations of convergent Taylor
series in $\delta$. The center shifts $D_{m}$ and half-widths $L_{m}^{(c)}/2$
are successive terms of these series, not merely asymptotic approximations.

\section{Hill's equation and the Floquet theory}

\label{app:Hill}

This chapter collects the facts about Hill's equation\index{Hill equation}
and its stability structure that are used in the main text. Standard
references are Magnus and Winkler~\cite[Chs.~1--2]{MagWin}, \cite[Sec.~III.8]{Hale},
Yakubovich and Starzhinskii~\cite[Ch.~II]{YakSta1}, \cite[Ch.~VII]{YakSta2},
and Eastham et al.~\cite[Ch.~2]{BroEaSch}.

\bigskip{}

\emph{Historical note.} Where the note of Chapter~\ref{app:genLin}
traced the origins of the subject, the story of Hill's equation proper
is a spectral one. Mathieu's equation, the simplest case, resisted
every attempt at closed-form solution; its characteristic values and
periodic solutions generated a literature of their own, codified in
McLachlan's treatise (1947). The geography of stability was first
charted by van der Pol and Strutt (\emph{Philosophical Magazine} \textbf{5}
(1928), pp.~18--38), whose map of stable and unstable regions ---
the tongues of this book --- became known as the Ince--Strutt diagram,
the name acknowledging E.~L. Ince's computations of the boundary
curves and M.~J.~O. Strutt's monograph of 1932. Ince also settled
the deeper structural question: he proved that the Mathieu equation
admits no \emph{coexistence} --- two independent solutions of period
$\pi$ or $2\pi$ cannot occur together, so no instability tongue
can close (\emph{Proc.\ Camb.\ Phil.\ Soc.} \textbf{21} (1922),
pp.~117--120; \emph{Proc.\ Lond.\ Math.\ Soc.}~(2) \textbf{23}
(1923), pp.~56--74; an independent proof was given by Hille). In
the same circle of papers he introduced the equation now bearing his
name, for which coexistence \emph{does} occur and entire families
of tongues close --- the class to which the LC circuit equation belongs,
and the structural reason for the odd-only resonance pattern proved
in Part~I. The spectral viewpoint matured in the twentieth century
--- the inverse problem of Hochstadt (1965), the operator-theoretic
treatment of Eastham (1973) --- and culminated in the 1970s, when
Hill's equation moved to the center of the theory of integrable systems:
the potentials with only finitely many open gaps were found to be
precisely the solutions of the stationary Korteweg--de Vries hierarchy
(McKean and van Moerbeke, 1975; McKean and Trubowitz, 1976; Trubowitz,
1977). An equation born of the moon's perigee thus closed the century
as the meeting point of spectral theory, algebraic geometry, and nonlinear
waves.

\subsection{Hill's equation}

Hill's equation is the second-order scalar differential equation 
\begin{equation}
y''+Q(t)\,y=0,\qquad Q(t+\pi)=Q(t),\label{eq:Hill-app}
\end{equation}
where $Q$ is a real-valued, piecewise-continuous, $\pi$-periodic
function and the prime denotes $d/dt$. The standard form~\eqref{eq:Hill-app}
assumes period $\pi$ for $Q$; equations with period $T$ can be
reduced to this form by rescaling $t\mapsto\pi t/T$. Compare Hill's
equation \eqref{eq:Hill-app} with its normalized form described by
equations \eqref{eq:Hill-family}, \eqref{eq:Q0-Fourier}.

Hill's equation is a special case of the general system~\eqref{eq:genLin1a}
of Chapter~\ref{app:genLin}, with $n=2$ and 
\begin{equation}
A(t)=\begin{bmatrix}0 & 1\\
-Q(t) & 0
\end{bmatrix}.\label{eq:Hill-system}
\end{equation}
The trace of $A(t)$ vanishes identically, so Liouville's formula~\eqref{eq:genLin1d}
gives $\det X(t)=1$ for all $t$. In particular, the monodromy matrix\index{monodromy matrix}
$X(\pi)$ satisfies $\det X(\pi)=1$, and its two multipliers $\rho_{1},\rho_{2}$
satisfy 
\begin{equation}
\rho_{1}\rho_{2}=1.\label{eq:Hill-rho-prod}
\end{equation}

\subsection{The normalized solutions and the discriminant}

Let $y_{1}(t)$ and $y_{2}(t)$ be the solutions of~\eqref{eq:Hill-app}
satisfying the initial conditions~\cite[Sec.~1.2]{MagWin} 
\begin{equation}
y_{1}(0)=1,\quad y_{1}'(0)=0;\qquad y_{2}(0)=0,\quad y_{2}'(0)=1.\label{eq:Hill-ics}
\end{equation}
The matrizant of the corresponding system~\eqref{eq:Hill-system}
is then 
\begin{equation}
X(t)=\begin{bmatrix}y_{1}(t) & y_{2}(t)\\
y_{1}'(t) & y_{2}'(t)
\end{bmatrix},
\end{equation}
and the monodromy matrix is $X(\pi)$. By~\eqref{eq:Hill-rho-prod},
its characteristic equation~\cite[Sec.~1.2]{MagWin} is 
\begin{equation}
\rho^{2}-\Delta\,\rho+1=0,\qquad\Delta=\mathrm{Tr}\,X(\pi)=y_{1}(\pi)+y_{2}'(\pi).\label{eq:Hill-char-eq}
\end{equation}
The real-valued function $\Delta=\Delta(Q)$ is called the \emph{discriminant\index{discriminant}}
of Hill's equation. It is an entire function of any parameters on
which $Q$ depends analytically (Theorem~\ref{thm:joint-analytic};
for the spectral parameter see \cite[Thm.~2.2]{MagWin}). The multipliers
are 
\begin{equation}
\rho_{1,2}=\tfrac{1}{2}\Delta\pm\sqrt{\tfrac{1}{4}\Delta^{2}-1}.\label{eq:Hill-multipliers}
\end{equation}

\subsection{Stability and instability: the discriminant criterion}

The behavior of all solutions of Hill's equation is governed entirely
by the value of $\Delta$, via the following criterion \cite[Sec.~II.2.5]{YakSta1},
\cite[Sec.~VII.1.5]{YakSta2}, \cite[Chap.~2]{MagWin}. 
\begin{thm}[Stability criterion for Hill's equation]
\index{Hill equation!stability criterion} \label{thm:Hill-stability}
Let $\Delta=y_{1}(\pi)+y_{2}'(\pi)$ be the discriminant of~\eqref{eq:Hill-app}. 
\begin{enumerate}
\item \emph{Stability:} If $|\Delta|<2$, then $\rho_{1,2}$ are complex
conjugates of modulus~$1$ and distinct. Both multipliers lie on
the unit circle, all solutions are bounded on $(-\infty,+\infty)$,
and the trivial solution is (Lyapunov) stable. 
\item \emph{Instability:} If $|\Delta|>2$, then $\rho_{1},\rho_{2}$ are
real, one satisfies $|\rho_{1}|>1$ and the other $|\rho_{2}|=1/|\rho_{1}|<1$.
Equation~\eqref{eq:Hill-app} has a solution growing exponentially
as $t\to+\infty$; the trivial solution is unstable. 
\item \emph{Boundary:} If $|\Delta|=2$, then $\rho_{1}=\rho_{2}=\rho$,
with $\rho=+1$ when $\Delta=+2$ and $\rho=-1$ when $\Delta=-2$.
The equation has at least one periodic or antiperiodic solution. If
in addition the monodromy matrix $X(\pi)$ is diagonalizable (i.e.,
has two linearly independent eigenvectors), then all solutions are
bounded; otherwise there is a solution growing linearly in $t$. 
\end{enumerate}
The parameter regions $|\Delta|<2$, $|\Delta|>2$, and $|\Delta|=2$
are called the \emph{stability zones}, \emph{instability zone\index{instability zone}s},
and \emph{boundary curve\index{boundary curves}s}, respectively. 
\end{thm}

\begin{rem}[Jordan block\index{Jordan block} structure]
The EPD\index{exceptional point of degeneracy (EPD)} structure of
the instability boundary\index{stability boundary} curves --- that
$X(\pi)$ has a non-trivial Jordan block at every boundary point ---
is established in Chapter~\ref{sec:EPD-elementary} by a direct discriminant
argument using the Hill multiplier formula~\eqref{eq:Hill-multipliers}. 
\end{rem}

\subsection{Spectral structure: alternating stability and instability zones}

The following theorem, due to Lyapunov (1907) and Haupt (1914, 1918)
and given its modern form by Magnus and Winkler, is the fundamental
result on the spectral structure of Hill's equation~\cite[Thm.~2.1]{MagWin}.
We state it directly in the notation used throughout this work. 
\begin{thm}[Oscillation Theorem]
\index{oscillation theorem} \label{thm:oscillation} (\cite[Thm.~2.1]{MagWin})
To every Hill equation~\eqref{eq:Hill-app} there belong a real number
$\lambda_{0}$ and two infinite sequences of pairs of real numbers
\begin{equation}
\lambda_{1}^{-}\leq\lambda_{1}^{+},\quad\lambda_{2}^{-}\leq\lambda_{2}^{+},\quad\lambda_{3}^{-}\leq\lambda_{3}^{+},\quad\ldots,\label{eq:bdry-pairs}
\end{equation}
all tending to $+\infty$, satisfying the interlacing inequality 
\begin{equation}
\lambda_{0}<\lambda_{1}^{-}\leq\lambda_{1}^{+}<\lambda_{2}^{-}\leq\lambda_{2}^{+}<\lambda_{3}^{-}\leq\lambda_{3}^{+}<\lambda_{4}^{-}\leq\lambda_{4}^{+}<\cdots.\label{eq:MW-interlace}
\end{equation}
The values $\lambda_{0}$ and $\lambda_{m}^{\pm}$ are characterized
by the discriminant $\Delta(\lambda)=y_{1}(\pi,\lambda)+y_{2}'(\pi,\lambda)$
as follows: $\lambda_{0}$ and $\lambda_{2k}^{\pm}$ (even index,
$k=1,2,\ldots$) are the roots of $\Delta(\lambda)=+2$, while $\lambda_{2k-1}^{\pm}$
(odd index, $k=1,2,\ldots$) are the roots of $\Delta(\lambda)=-2$.
Specifically: 
\begin{itemize}
\item equation~\eqref{eq:Hill-app} has a period-$\pi$ solution if and
only if 
\[
\lambda\in\{\lambda_{0},\lambda_{2}^{-},\lambda_{2}^{+},\lambda_{4}^{-},\lambda_{4}^{+},\ldots\};
\]
\item equation~\eqref{eq:Hill-app} has a period-$2\pi$ solution if and
only if 
\[
\lambda\in\{\lambda_{1}^{-},\lambda_{1}^{+},\lambda_{3}^{-},\lambda_{3}^{+},\ldots\}.
\]
\end{itemize}
The solutions of~\eqref{eq:Hill-app} are stable (bounded on $\mathbb{R}$)
in the open intervals 
\begin{equation}
(\lambda_{0},\lambda_{1}^{-}),\quad(\lambda_{1}^{+},\lambda_{2}^{-}),\quad(\lambda_{2}^{+},\lambda_{3}^{-}),\quad(\lambda_{3}^{+},\lambda_{4}^{-}),\quad\ldots,\label{eq:MW-stability}
\end{equation}
and unstable in $(-\infty,\lambda_{0})$ and in the open intervals
$(\lambda_{m}^{-},\lambda_{m}^{+})$ for $m=1,2,3,\ldots$ (the instability
zones).

At the boundary points $\lambda_{m}^{\pm}$ the solutions are in general
unstable. The exception is the \emph{coexistence} case: stability
holds at $\lambda_{m}^{-}$ and $\lambda_{m}^{+}$ if and only if
$\lambda_{m}^{-}=\lambda_{m}^{+}$, i.e.\ the $m$-th instability
zone has collapsed to a single point. For complex $\lambda$ all solutions
are unstable. 
\end{thm}

\begin{rem}[Relation to MW notation]
\label{rem:notation-spectrum} Magnus and Winkler~\cite[Thm.~2.1]{MagWin}
use two separate sequences: $\lambda_{0}<\lambda_{1}\leq\lambda_{2}<\cdots$
for the roots of $\Delta=+2$, and $\lambda_{1}'\leq\lambda_{2}'<\lambda_{3}'\leq\lambda_{4}'<\cdots$
for the roots of $\Delta=-2$. The correspondence with our notation
is: 
\begin{multline*}
\lambda_{0}\leftrightarrow\lambda_{0},\quad\lambda_{2k-1}\leftrightarrow\lambda_{2k}^{-},\quad\lambda_{2k}\leftrightarrow\lambda_{2k}^{+},\\
\lambda_{2k-1}'\leftrightarrow\lambda_{2k-1}^{-},\quad\lambda_{2k}'\leftrightarrow\lambda_{2k-1}^{+},\quad k=1,2,\ldots
\end{multline*}

In either case $\Delta$ is an entire function of $\lambda$ and its
zeros interlace as in~\eqref{eq:MW-interlace} \cite[Sec.~VII.1.5]{YakSta2},
\cite[Chap.~2]{MagWin}. The roots of $\Delta(\lambda)=2$ (period-$\pi$
boundary) are denoted $\lambda_{0}<\lambda_{2}^{-}\leq\lambda_{2}^{+}<\lambda_{4}^{-}\leq\lambda_{4}^{+}<\cdots$,
and the roots of $\Delta(\lambda)=-2$ (period-$2\pi$ boundary) will
be denoted $\lambda_{1}^{-}\leq\lambda_{1}^{+}<\lambda_{3}^{-}\leq\lambda_{3}^{+}<\cdots$.
They interlace: 
\begin{equation}
\lambda_{0}<\lambda_{1}^{-}\leq\lambda_{1}^{+}<\lambda_{2}^{-}\leq\lambda_{2}^{+}<\lambda_{3}^{-}\leq\lambda_{3}^{+}<\cdots.\label{eq:Hill-spectrum}
\end{equation}
The open intervals $(\lambda_{2k}^{+},\lambda_{2k+1}^{-})$ and $(\lambda_{2k-1}^{+},\lambda_{2k}^{-})$
are stability zones (where $|\Delta|<2$). The closed intervals $[\lambda_{j}^{-},\lambda_{j}^{+}]$
are instability zones (where $|\Delta|\geq2$), with $|\Delta|>2$
in their interiors. 
\end{rem}

\begin{figure}[htbp]
\centering \includegraphics[width=12cm]{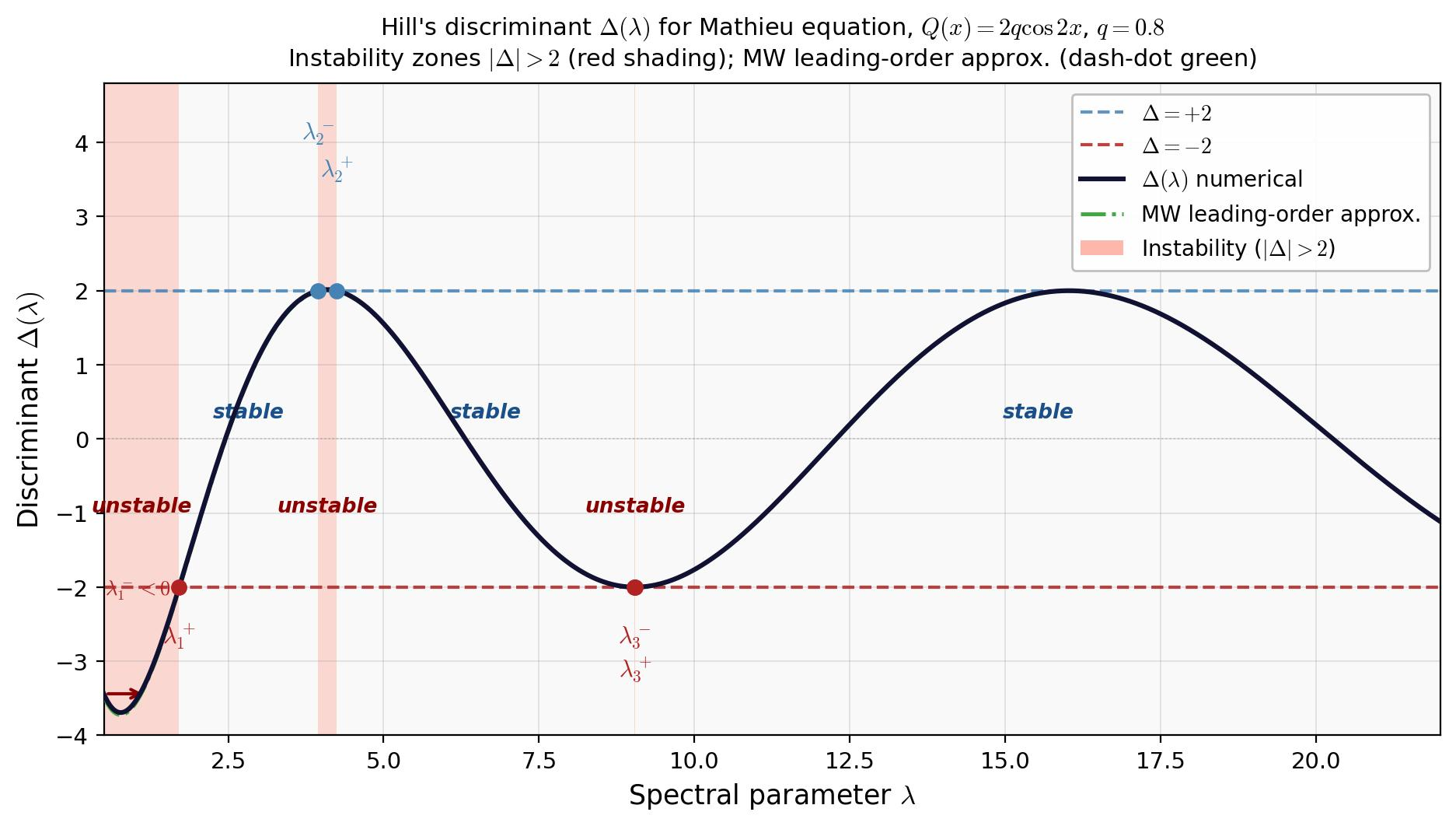} \caption{Hill's discriminant $\Delta(\lambda)$ for the Mathieu equation $Q(x)=2q\cos2x$,
$q=0.8$. The exact discriminant (solid) is computed by integrating
the two standard solutions (eqs.~\eqref{eq:Volterra-y1}--\eqref{eq:Volterra-y2}
with $Q=2q\cos2x$) over $[0,\pi]$ with an 8th-order Runge--Kutta
solver and evaluating $\Delta=y_{1}(\pi)+y_{2}'(\pi)$~\cite[Sec.~4.3]{Eastham73}.
Red shading marks the instability zones $|\Delta(\lambda)|>2$. The
dash-dot curve shows the MW approximation~\eqref{eq:Delta-Math},
which agrees well for $\lambda\gtrsim2$ but diverges near $\lambda=1$
due to the resonance singularity (see Chapter~\ref{app:entire}).
For a generic Hill equation the zones have positive width ($\lambda_{j}^{-}<\lambda_{j}^{+}$);
coexistence collapses a zone to a point ($\lambda_{j}^{-}=\lambda_{j}^{+}$).}\index{resonance}
\label{fig:Hill-discriminant} 
\end{figure}

Figure~\ref{fig:Hill-discriminant} illustrates the discriminant
structure for the representative case $Q(x)=2q\cos2x$ with $q=0.8$
(Hill's equation with a single cosine term, i.e.\ Mathieu\index{Mathieu equation}'s
equation).

\subsection{Coexistence and the collapse of instability zones}

The $j$-th instability zone has zero width --- that is, $\lambda_{j}^{-}=\lambda_{j}^{+}$
--- if and only if the equation $\Delta(\lambda_{j})=\pm2$ has a
\emph{double} root, which occurs if and only if the monodromy matrix
$X(\pi,\lambda_{j})$ is a scalar multiple of the identity. In this
case equation $X(\pi)\mathbf{v}=\rho\mathbf{v}$ has two linearly
independent solutions: both solutions of Hill's equation at $\lambda=\lambda_{j}$
are periodic (if $\rho=+1$) or antiperiodic (if $\rho=-1$). This
situation is called \emph{coexistence} of periodic solutions \cite[Chap.~7]{MagWin}.

When coexistence holds at $\lambda_{j}$, the instability zone $[\lambda_{j}^{-},\lambda_{j}^{+}]$
degenerates to a single point: the boundary between the two adjacent
stability zones is a single curve rather than a pair of curves bounding
a tongue of finite width. The LC circuit equation provides an explicit
example in which \emph{all even} instability zones collapse simultaneously
(Corollaries~\ref{cor:even} and~\ref{cor:odd} of Chapter~\ref{sec:LCInce}).

\begin{figure}[htbp]
\centering \includegraphics[width=12cm]{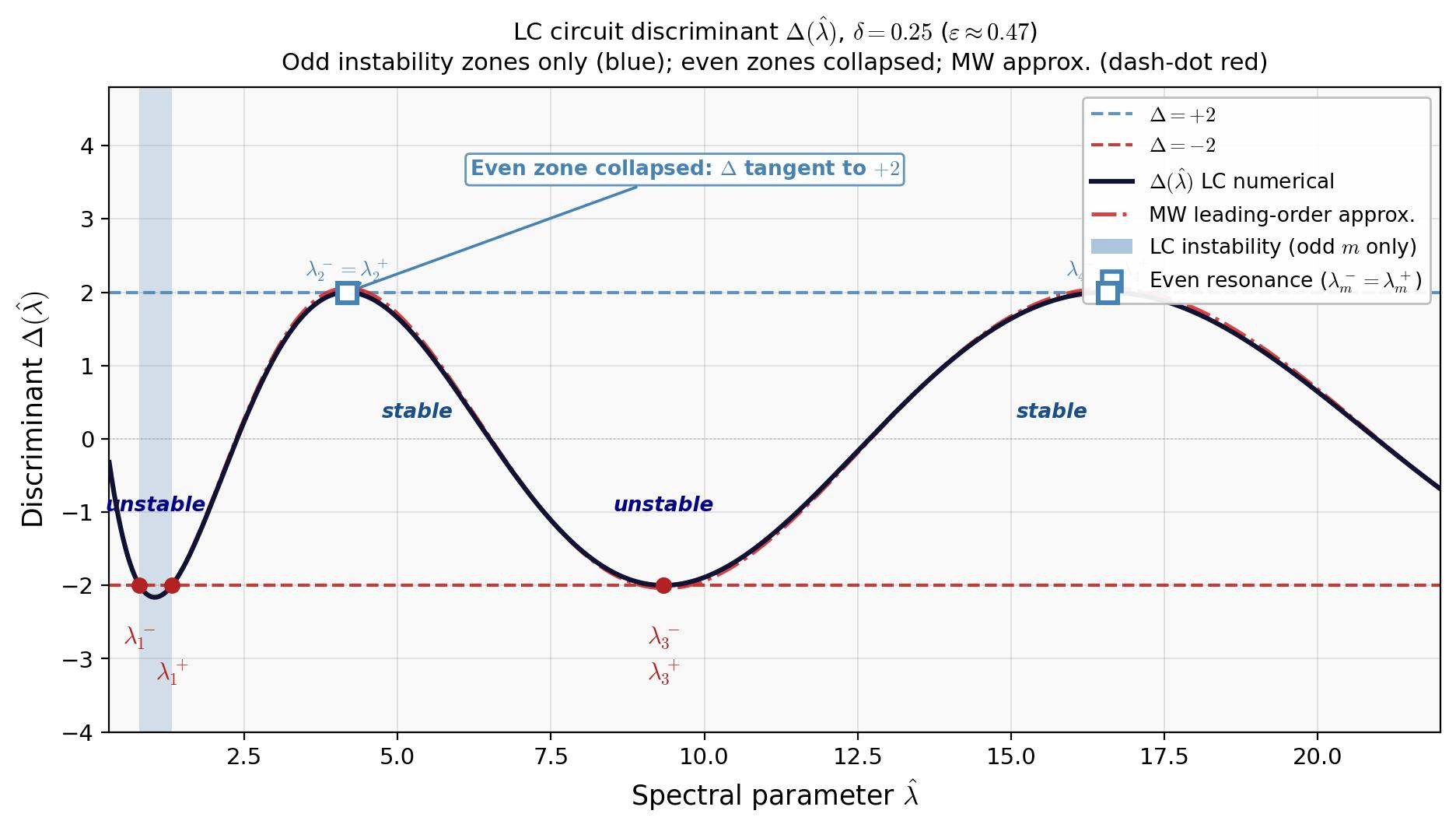} \caption{Discriminant $\Delta(\hat{\lambda})$ of the LC circuit equation for
$\delta=0.25$ ($\varepsilon\approx0.47$). Only the odd instability
zones $[\hat{\lambda}_{1}^{-},\hat{\lambda}_{1}^{+}]$ and $[\hat{\lambda}_{3}^{-},\hat{\lambda}_{3}^{+}]$
(blue shading) are present: $\Delta$ crosses $-2$ at these zones.
At even resonances ($\hat{\lambda}\approx4$ and $\hat{\lambda}\approx16$,
marked by open squares), $\Delta$ is merely \emph{tangent} to $+2$
from below and does not cross it --- the signature of coexistence,
which collapses the even instability zones to points ($\lambda_{m}^{-}=\lambda_{m}^{+}$).
The exact discriminant is computed numerically as in Figure~\ref{fig:Hill-discriminant}.
The dash-dot curve shows the MW leading-order approximation~\eqref{eq:Delta-LC},
which captures the qualitative structure away from resonances. Compare
with Figure~\ref{fig:Hill-discriminant} for the Mathieu equation,
where $\Delta$ crosses $+2$ at all even resonances, producing finite-width
instability zones.}
\label{fig:LC-discriminant} 
\end{figure}

Figure~\ref{fig:LC-discriminant} shows the discriminant of the LC
circuit equation for $\delta=0.25$, contrasting directly with the
Mathieu case of Figure~\ref{fig:Hill-discriminant}.

\subsection{Connection to the Magnus--Winkler expansion}

Magnus and Winkler~\cite[\S\,2.1]{MagWin} write Hill's equation
in the form 
\begin{equation}
y''+[\lambda+Q(t)]\,y=0,\label{eq:Hill-MW}
\end{equation}
where $Q$ has zero mean ($\frac{1}{\pi}\int_{0}^{\pi}Q\,dt=0$) and
$\lambda$ is the spectral parameter. The discriminant $\Delta(\lambda)=y_{1}(\pi,\lambda)+y_{2}'(\pi,\lambda)$
is then an entire function of $\lambda$, \cite[Thm.~2.2]{MagWin}.
Its expansion in powers of the Fourier coefficients $g_{n}$ of $Q$
(the MW discriminant expansion, equation~\eqref{eq:D0} of Chapter~\ref{sec:MWInce})
is the principal computational tool. The stability criterion of Theorem~\ref{thm:Hill-stability}
reads $|\Delta(\lambda)|\leq2$ (stable) and $|\Delta(\lambda)|>2$
(unstable) in this notation, and the instability zone boundaries are
where $\Delta=\pm2$ exactly.

\medskip{}
\emph{Leading-order approximation.} Writing the Fourier series in
MW's complex two-sided form~\cite[eq.~2.40]{MagWin} 
\[
Q(x)=\sum_{n\neq0}g_{n}\,e^{2inx},\qquad g_{-n}=\bar{g}_{n},
\]
the MW expansion in~\cite[Cor.~2.6]{MagWin} gives 
\begin{equation}
\Delta(\lambda)=2\cos\pi\sqrt{\lambda}+\frac{\pi\sin\pi\sqrt{\lambda}}{2\sqrt{\lambda}}\sum_{n=1}^{\infty}\frac{|g_{n}|^{2}}{\lambda-n^{2}}+O(g^{4}),\quad g\to0.\label{eq:Delta-MW}
\end{equation}
The zeroth term $\Delta_{0}=2\cos\pi\sqrt{\lambda}$ is the free discriminant
(no perturbation): it oscillates between $\pm2$, with $\Delta_{0}=+2$
at $\lambda=(2k)^{2}$ ($\sqrt{\lambda}$ an even integer, $k=0,1,2,\ldots$)
and $\Delta_{0}=-2$ at $\lambda=(2k-1)^{2}$ ($\sqrt{\lambda}$ an
odd integer, $k=1,2,\ldots$). These are the points from which the
instability zones open as $|g_{n}|$ increase from zero.

\medskip{}
\noindent\emph{Mathieu equation ($Q(x)=2q\cos2x$, $g_{\pm1}=q$, $g_{n}=0$
for $|n|\geq2$).} Equation~\eqref{eq:Delta-MW} reduces to~\cite[Cor.~2.6]{MagWin}
\begin{equation}
\Delta^{{\rm Math}}(\lambda)=2\cos\pi\sqrt{\lambda}+\frac{\pi q^{2}\sin\pi\sqrt{\lambda}}{2\sqrt{\lambda}\,(\lambda-1)}+O(q^{4}),\quad q\to0.\label{eq:Delta-Math}
\end{equation}
computed numerically with $q=0.8$ in Figure~\ref{fig:Hill-discriminant}.
The MW approximation~\eqref{eq:Delta-Math} is shown as the dash-dot
curve in the figure; it tracks the exact discriminant closely for
$\lambda\gtrsim2$ but fails near $\lambda=1$ due to the resonance
singularity $({\lambda-1})^{-1}$, which is resolved by the $O(q^{4})$
terms.

\medskip{}
\noindent\emph{LC circuit ($Q(x)=c/(1+a\cos2x)-c$, all harmonics present).}
The inverse capacitance\index{capacitance sensing} coefficient $1/(1+a\cos2x)$
of the LC circuit is closely related to the classical \emph{Poisson
kernel} of the unit disk. The \emph{Poisson kernel Fourier series}~\cite[Vol.~I, Ch.~III, \S\,3.3]{Zygmund}:
\begin{equation}
P(r,\theta)=\frac{1-r^{2}}{1-2r\cos\theta+r^{2}}=1+2\sum_{n=1}^{\infty}r^{n}\cos n\theta,\quad|r|<1,\label{eq:Poisson-kernel}
\end{equation}
arises naturally in potential theory (the Dirichlet problem for the
unit disk) and in Fourier analysis (Abel summability of trigonometric
series). Evaluating at $r=\delta$, $\theta=2x$, and using $a=-2\delta/(1+\delta^{2})$
gives (GR~\cite[Sec.~1.4, eq.~(1.421.2)]{GraRyzh}): 
\begin{multline}
\frac{1}{1+a\cos2x}=\frac{1+\delta^{2}}{1-\delta^{2}}\,P(\delta,2x)=\frac{1+\delta^{2}}{1-\delta^{2}}\Bigl[1+2\sum_{n=1}^{\infty}\delta^{n}\cos2nx\Bigr],\\
a=-\frac{2\delta}{1+\delta^{2}},\quad|\delta|<1,\label{eq:Poisson-LC}
\end{multline}
so the LC coefficient is a constant multiple $(1+\delta^{2})/(1-\delta^{2})$
of the Poisson kernel, not the Poisson kernel itself. The geometrically
decaying Fourier coefficients $\delta^{n}$ are the key reason $\delta$
is the natural parameter for the LC circuit. The zero-mean perturbation
is $Q(x)=c/(1+a\cos2x)-c-\bar{Q}$, where $\bar{Q}=2c\delta^{2}/(1-\delta^{2})$
is the mean, and $\lambda$ in~\eqref{eq:Hill-MW} is the shifted
eigenvalue $\hat{\lambda}=c(1+\delta^{2})/(1-\delta^{2})$. The Fourier
coefficients of the zero-mean $Q$ are 
\begin{equation}
g_{n}=\frac{c(1+\delta^{2})}{1-\delta^{2}}\,\delta^{|n|}=\hat{\lambda}\,\delta^{|n|},\quad\hat{\lambda}=c\frac{1+\delta^{2}}{1-\delta^{2}},\quad n=\pm1,\pm2,\ldots.\label{eq:gn-LC}
\end{equation}
Substituting into~\eqref{eq:Delta-MW}: 
\begin{equation}
\Delta^{{\rm LC}}(\hat{\lambda})=2\cos\pi\sqrt{\hat{\lambda}}+\frac{\pi\hat{\lambda}^{2}\delta^{2}\sin\pi\sqrt{\hat{\lambda}}}{2\sqrt{\hat{\lambda}}}\sum_{n=1}^{\infty}\frac{\delta^{2(n-1)}}{\hat{\lambda}-n^{2}}+O(\delta^{4}),\quad\delta\to0.\label{eq:Delta-LC}
\end{equation}
The MW approximation~\eqref{eq:Delta-LC} is shown as the dash-dot
curve in Figure~\ref{fig:LC-discriminant}. The key distinction from
the Mathieu case is that \emph{all harmonics} $n\geq1$ contribute
with geometrically decaying weights $\delta^{2(n-1)}$. Near $\hat{\lambda}\approx m^{2}$
for odd $m$, the sum is dominated by the $n=m$ term (the resonant
harmonic), which drives $\Delta$ through $-2$ to create the instability
zone. For even $m$, the contributions from all harmonics combine
to produce only a tangency of $\Delta$ with $+2$ rather than a crossing
--- the algebraic mechanism behind the coexistence and zone collapse
proved by the Ince theorems of Section~\ref{sec:InceGeneral}.

\section{Parametric resonance}

\label{app:ParRes}

The following are the primary references for parametric resonance\index{resonance}\index{parametric resonance}
theory used in this work. Yakubovich and Starzhinskii~\cite[Chap.~V]{YakSta2}
give a very detailed and insightful presentation of the theory of
parametric resonance and its relation to the stability of differential
equations with periodic coefficients, including the general formulas
for critical (resonance) frequencies stated below. Arnold~\cite[\S\S\,25,~42]{ArnMech}
gives a concise, qualitative analysis of parametric resonance and
the influence of friction. Seyranian and Mailybaev~\cite[Chap.~9]{SeyMai}
treat stability of periodic systems dependent on parameters, the Mathieu\index{Mathieu equation}
equation, and effective perturbation theory. Landau and Lifshitz~\cite[\S27]{LanLif}
provide an especially clear physical introduction to parametric resonance
and energy pumping. Nayfeh and Mook~\cite[Chap.~5]{NayfMook} treat
parametric excitation systematically for both single- and multi-degree-of-freedom
systems, including combination resonances and the energy input from
parameter modulation. For a modern overview connecting the Mathieu
equation to parametric resonance phenomena across several physical
settings, see~\cite[\S\,1]{ChenG}.

\bigskip{}

\emph{Historical note.} Humanity practiced parametric resonance long
before naming it: every child pumping a swing modulates the effective
pendulum length at twice the natural frequency. The grandest such
machine is medieval --- O Botafumeiro, the giant censer of the cathedral
of Santiago de Compostela, swung to amplitudes above eighty degrees
by men rhythmically shortening and lengthening its rope; its mechanics,
analyzed by Sanmartin (\emph{Amer.\ J. Phys.}~\textbf{52} (1984),
pp.~937--945), is parametric pumping executed to near-optimal precision
centuries before the theory existed. The laboratory and theoretical
lineage --- Faraday, Melde, Rayleigh, and the Mathieu--Hill machinery
--- was traced in the note of Chapter~\ref{app:genLin}; here we
follow the subject into electricity, where this book lives. FitzGerald
attempted as early as 1892 to excite oscillations in an LC circuit
by driving a variable inductance with a dynamo; iron-core parametric
amplifiers carried radio telephony between Berlin, Vienna, and Moscow
in 1913--1915; and in the early 1930s the school of Mandelstam and
Papaleksi turned parametric excitation of electric oscillations into
a systematic theory and into working machinery --- parametric generators
producing power from a periodically varying reactance, with the instability
zones charted by their nonlinear-dynamics school (Mandelstam and Papaleksi,
\emph{On the parametric excitation of electric oscillations}, Zh.\ Tekh.\
Fiz., 1934). Parametric driving then revealed its paradoxical other
face: it can \emph{stabilize} as well as destabilize. Stephenson predicted
in 1908 that a pendulum could stand upright if its pivot vibrates
fast enough (\emph{Mem.\ Proc.\ Manch.\ Lit.\ Phil.\
Soc.}~\textbf{52} (1908), pp.~1--10), and Kapitza explained the
phenomenon and made it an experimental art (1951), founding what is
now called vibrational stabilization. The electrical line was reborn
when van der Ziel observed (1948) that amplification by a modulated
\emph{reactance}, dissipating nothing, is inherently quiet; the microwave
parametric amplifiers of the 1950s exploited this (for the early history
see Mumford, \emph{Proc.\ IRE}~\textbf{48} (1960), pp.~848--853),
and its present-day descendants --- Josephson parametric amplifiers,
modulated LC circuits in the most literal sense --- operate at the
quantum noise limit as the readout stage of superconducting quantum
processors. The EPD-based sensing of Chapter~\ref{sec:EPD-sensor}
belongs to this same unbroken line, in which the modulated LC circuit
has been protagonist for over a century.

\subsection{Contrast with ordinary (forced) resonance}

Parametric resonance differs from ordinary resonance in several fundamental
ways~\cite[Chap.~V, \S\,1.1]{YakSta2}, \cite[\S27]{LanLif}: 
\begin{enumerate}
\item \emph{Mechanism of excitation.} In ordinary (forced) resonance, energy
enters the oscillator via an explicit driving term $F(t)$ on the
right-hand side of the equation of motion. In parametric resonance
the right-hand side is identically zero; instead, one of the system's
own parameters --- spring constant, inductance, capacitance\index{capacitance sensing},
length --- is varied periodically. The oscillator equation has the
homogeneous form $\ddot{x}+\omega^{2}(t)x=0$, yet solutions may grow
without bound. The energy delivered to the system comes from the external
agent that maintains the parameter variation, not from a direct forcing
term. 
\item \emph{Spectrum of dangerous frequencies.} In ordinary resonance, unbounded
growth occurs only at the discrete natural frequencies of the system.
In parametric resonance the dangerous frequencies form a union of
open intervals (instability tongue\index{instability tongue}s) in
parameter space. The widths of these intervals depend on the amplitude
of the parameter modulation and shrink to zero as the modulation amplitude
vanishes. The frequency values to which the tongues collapse in the
zero-amplitude limit are called the critical frequencies~\cite[Chap.~V, \S1.1]{YakSta2}. 
\item \emph{Growth law.} In ordinary resonance, amplitude growth is algebraic
--- typically linear in time. In parametric resonance, growth is
exponential: within an instability tongue the amplitude behaves as
$e^{\mu t}$ with Floquet\index{Floquet theory} exponent\index{Floquet theory!Floquet exponent}
$\mu>0$, a qualitatively stronger instability~\cite[Chap.~V, \S1.2]{YakSta2},
\cite[\S27]{LanLif}. 
\end{enumerate}

\subsection{Energy extraction from the parameter source}

A physically illuminating way to understand parametric growth is through
the work done by the agent controlling the varying parameter~\cite[\S27]{LanLif},
\cite[Chap.~5]{NayfMook}. Consider the prototypical parametric oscillator
$\ddot{x}+\omega^{2}(t)x=0$ with $\omega^{2}(t)=\omega_{0}^{2}(1+\varepsilon f(t))$
and $f$ periodic of period $T=2\pi/\vartheta$. The total energy
of the oscillator is $E=\frac{1}{2}\dot{x}^{2}+\frac{1}{2}\omega^{2}(t)x^{2}$.
Because the equation of motion gives $\ddot{x}=-\omega^{2}(t)x$,
its time derivative along solutions reduces to 
\begin{equation}
\frac{\dd E}{\dd t}=\frac{1}{2}\frac{\dd\omega^{2}}{\dd t}(t)\,x^{2}(t)=\frac{1}{2}\omega_{0}^{2}\varepsilon\dot{f}(t)\,x^{2}(t).\label{eq:energy-rate}
\end{equation}
The net work delivered to the oscillator over one period is therefore
\begin{equation}
\Delta E=\frac{1}{2}\int_{0}^{T}\frac{\dd\omega^{2}}{\dd t}(t)\,x^{2}(t)\,\dd t.\label{eq:energy-cycle}
\end{equation}
This integral is positive whenever $\omega^{2}(t)$ is \emph{increased}
at moments of large displacement $|x(t)|$ (large potential energy)
and \emph{decreased} at moments of small displacement~\cite[\S27]{LanLif}.
For an oscillation near frequency $\omega_{0}$, with $x(t)\approx A\cos(\omega_{0}t+\phi)$,
the squared displacement $x^{2}(t)=\tfrac{1}{2}A^{2}(1+\cos(2\omega_{0}t+2\phi))$
oscillates at frequency $2\omega_{0}$. The integrand $\frac{\dd\omega^{2}}{\dd t}\cdot x^{2}(t)$
therefore has a non-zero period-average precisely when $\frac{\dd\omega^{2}}{\dd t}$
also contains frequency $2\omega_{0}$, i.e.\ when $\vartheta=2\omega_{0}$
(primary parametric resonance). More generally, the $m$-th harmonic
of $\frac{\dd\omega^{2}}{\dd t}$ couples to the $m$-th subharmonic
of $x^{2}$, yielding the family of critical frequencies $\vartheta_{0}=2\omega_{0}/m$~\cite[\S27]{LanLif},
\cite[Chap.~V, \S1.10]{YakSta2}. Within an instability tongue, this
net work per cycle is positive, $\Delta E>0$, and the oscillation
amplitude grows geometrically each period: $E(t+T)=\Lambda E(t)$
with $\Lambda>1$~\cite[\S27]{LanLif}.

For the LC circuit, the role of $\omega^{2}(t)$ is played by $1/[L\,C(t)]$,
and the agent doing the work is the external mechanism that varies
the capacitance $C(t)$. At parametric resonance the circuit extracts
energy from the capacitance-driving mechanism each period, and the
stored electromagnetic energy grows exponentially.

\subsection{Critical (resonance) frequencies: general formula}

We now state and sketch the argument behind the general formula for
critical frequencies~\cite[Chap.~V, Thm.~II]{YakSta2}.

\medskip{}
\emph{Setting.} Consider a Hamiltonian (canonical) system with $k$
degrees of freedom whose Hamiltonian $H(\mathbf{q},\mathbf{p},t;\varepsilon)$
depends smoothly on a small perturbation amplitude $\varepsilon\geq0$
and is periodic in $t$ with period $T=2\pi/\vartheta$. At $\varepsilon=0$
the system reduces to a linear oscillator with constant positive definite
Hamiltonian; its $2k$ characteristic multipliers lie on the unit
circle in conjugate pairs $e^{\pm i\omega_{j}T}$, $j=1,\ldots,k$,
where $\omega_{j}$ are the natural frequencies. All multipliers are
assumed to be of \emph{definite Krein\index{Krein collision theory}
kind}~\cite[Chap.~V, \S\,1.2]{YakSta2} (no mixed-kind pairs at $\varepsilon=0$),
which is the generic situation for a Hamiltonian system with distinct
frequencies.

\medskip{}
\noindent\emph{Floquet perturbation argument.} By the Floquet theorem (Chapter~\ref{app:genLin}),
the monodromy matrix\index{monodromy matrix} $M(\varepsilon)$ of
the perturbed system is an analytic function of $\varepsilon$. Its
eigenvalues $\rho_{j}(\varepsilon)$ (characteristic multipliers)
are therefore analytic branches. At $\varepsilon=0$ they lie on the
unit circle in conjugate pairs $\{\rho_{j},\bar{\rho}_{j}\}=\{e^{i\omega_{j}T},e^{-i\omega_{j}T}\}$;
instability requires them to leave the unit circle, which for a symplectic\index{symplectic system}
matrix can happen only when two multipliers collide on the unit circle
and then bifurcate off it. A collision of the multiplier $e^{i\omega_{j}T}$
(from the $j$-th pair) with the conjugate $e^{-i\omega_{h}T}$ (from
the $h$-th pair) occurs when their product equals $1$: 
\begin{equation}
e^{i(\omega_{j}+\omega_{h})T}=1,\quad\text{i.e.}\quad(\omega_{j}+\omega_{h})\frac{2\pi}{\vartheta}=2\pi m,\quad m=1,2,3,\ldots,\label{eq:YS-collision}
\end{equation}
which gives $\vartheta=(\omega_{j}+\omega_{h})/m$. (The case $j=h$
gives $e^{2i\omega_{j}T}=1$, the collision of the two elements of
the $j$-th conjugate pair with each other.) However, for a Hamiltonian
system the Krein theory~\cite[Ch.~III, \S\,1]{YakSta1}, \cite[Chap.~V, \S\,1.5--1.6]{YakSta2}
(see also Chapter~\ref{app:Krein}) shows that two multipliers can
leave the unit circle only if they collide \emph{and} carry opposite
Krein signature\index{Krein signature}s. Multipliers of the same
Krein kind can collide but then remain on the unit circle (a ``strong''
or ``safe'' collision); they do not produce instability. Under the
assumption of no mixed-kind pairs at $\varepsilon=0$, each conjugate
pair $\{e^{i\omega_{j}T},e^{-i\omega_{j}T}\}$ has opposite signatures
(one positive, one negative), so any collision of $e^{i\omega_{j}T}$
with $e^{-i\omega_{h}T}$ involves multipliers of opposite kind and
is therefore potentially destabilizing. This singles out exactly the
collision conditions~\eqref{eq:YS-collision}, which are~\cite[Chap.~V, \S\,1.1, Thm.~II]{YakSta2}
\begin{equation}
\vartheta_{0}=\frac{\omega_{j}+\omega_{h}}{m},\qquad j,h=1,\ldots,k;\quad m=1,2,3,\ldots\label{eq:YS-critical-freq}
\end{equation}
Two cases arise: 
\begin{itemize}
\item \emph{Principal resonance} ($j=h$): $\vartheta_{0}=2\omega_{j}/m$.
The excitation frequency is a rational multiple of a single natural
frequency. The instability tongue is attached to $\vartheta_{0}$
with width $O(\varepsilon^{m})$ for small modulation amplitude~\cite[Chap.~V, \S1.10]{YakSta2}. 
\item \emph{Combination resonance} ($j\neq h$): $\vartheta_{0}=(\omega_{j}+\omega_{h})/m$.
The excitation frequency couples two distinct natural frequencies.
This type is absent for scalar ($k=1$) equations~\cite[Chap.~V, \S1.11]{YakSta2}. 
\end{itemize}

\subsection{Scalar Hill equation: only principal resonance}

\noindent For a scalar second-order equation with a single natural
frequency $\omega_{0}$ (one degree of freedom, $k=1$), combination
resonance is absent and the critical frequencies reduce to 
\begin{equation}
\vartheta_{0}=\frac{2\omega_{0}}{m},\qquad m=1,2,3,\ldots\label{eq:YS-Hill-critical}
\end{equation}
This is the fundamental resonance condition for Hill's equation\index{Hill equation}
\cite[Chap.~V, \S1.10--1.11]{YakSta2}. Fixing the excitation frequency
$\mu=\vartheta_{0}$, instability occurs when the natural frequency
satisfies $\omega_{0}=m\mu/2$ for some positive integer $m$. This
formula is confirmed by the rigorous Floquet theory of Hill's equation:
within each instability zone\index{instability zone} the discriminant\index{discriminant}
satisfies $|\Delta|>2$, and the zone boundaries (where $|\Delta|=2$)
accumulate at $\omega_{0}=m\mu/2$ as the modulation amplitude $\varepsilon\to0$~\cite[Thm.~2.1]{MagWin},
\cite[Chap.~V, \S1.10]{YakSta2}.

\subsection{On the term \textquotedblleft negative energy\textquotedblright{}
in parametric resonance}

As established by the work integral~\eqref{eq:energy-rate}--\eqref{eq:energy-cycle},
within each instability tongue the oscillator gains energy $\Delta E_{\mathrm{osc}}>0$
per period, while by conservation of energy the external agent maintaining
the parameter variation simultaneously loses exactly that amount:
$\Delta E_{\mathrm{source}}=-\Delta E_{\mathrm{osc}}<0$. We use the
phrase \emph{negative energy} in this context to refer to this negative
energy balance of the parametric source --- the energy it surrenders
to the resonating system each period. This usage is consistent with
the physical picture of Landau and Lifshitz~\cite[\S27]{LanLif},
whose description of the mechanism --- increase the spring constant
at large displacement, decrease it at small --- is precisely the
condition under which the source runs such a deficit, though they
do not use that label.
\begin{rem}[On the term ``negative energy'']
\label{rem:neg-energy}Neither Landau--Lifshitz~\cite[\S27]{LanLif}
nor Arnold~\cite[\S\S25,42]{ArnMech} nor Yakubovich--Starzhinskii\index{Yakubovich--Starzhinskii series}~\cite[Ch.~III]{YakSta1},
\cite[Ch.~V]{YakSta2} uses the term ``negative energy'' in the
context of parametric resonance. The same phrase appears in Kirillov~\cite[Sec.~3.3.8]{Kiril}
with a different meaning: there it refers to the sign of the modal
energy $\mathcal{E}_{j}=(\mathbf{H}\mathbf{u}_{j},\mathbf{u}_{j})$
of a normal mode of a multi-degree-of-freedom Hamiltonian system with
indefinite Hessian $\mathbf{H}$ --- a structural, frame-independent
property of the linearized system at equilibrium, unrelated to the
source's energy budget during excitation. That usage, due to MacKay~\cite[Sec.~3]{MacKay87}
and systematized by Kirillov, applies to systems with at least two
degrees of freedom and an indefinite Hessian; it does not apply to
the single-mode LC oscillator of this work, whose Hessian is always
positive definite. The two usages should not be conflated. For a concrete
physical realization of Kirillov's negative-energy modes in electric
circuits --- via negative capacitance and a gyrator producing an
EPD\index{exceptional point of degeneracy (EPD)} point --- see Remark~\ref{rem:Krein-energy}
and~\cite[\S\,1]{FigPert,FigSynbJ}. 
\end{rem}

The same energy-transfer idea --- a source losing energy to a resonating
system --- appears naturally in the theory of traveling-wave tube
(TWT) amplifiers~\cite[\S\,1]{FigTWT,RabTrub89,LouPie55,ChuKPT51,Sturrock60},
where it goes under the same ``negative energy'' label. There the
source is the kinetic energy of the electron beam, which ``drags''
the slow space-charge wave on the helix, decelerating slightly to
pass energy to it: the beam runs a negative energy balance while the
slow wave grows. The physical mechanism differs from parametric resonance
--- in the TWT the energy comes from the flow's kinetic energy continuously,
with no periodic parameter variation --- but the energy bookkeeping
has the same sign structure. For a detailed treatment of the negative-energy
slow wave and its role in TWT instability and amplification, see~\cite[\S\,1]{FigTWT}
and references therein.

\subsection{Application to the LC circuit}

The LC circuit equation~\eqref{eq:LC-Hill} is a scalar Hill equation
with natural frequency $\omega_{0}=1/\sqrt{LC}$ and excitation frequency
$\mu$ (the capacitance modulation frequency; see the two-frequency
convention at~\eqref{eq:LC-tau}). Formula~\eqref{eq:YS-Hill-critical}
gives critical frequencies $\mu=2\omega_{0}/m$, $m=1,2,3,\ldots$,
equivalently $\omega_{0}=m\mu/2$. However, as established in this
work (Corollaries~\ref{cor:even}--\ref{cor:odd}), the rational
structure of the LC coefficient causes the even-$m$ instability domains
to collapse to zero width for all $0<\delta<1$: only the odd sub-harmonics
$\omega_{0}=(2k-1)\mu/2$ ($m=2k-1$, $k=1,2,3,\ldots$) produce genuine
instability tongues. This selective suppression of even resonances
is invisible to the Mathieu approximation, which incorrectly predicts
nonzero-width instability at all $m$. The energy viewpoint of~\eqref{eq:energy-cycle}
reinforces this: at even critical frequencies the net work integral
$\Delta E$ vanishes identically --- at a closed gap every solution
is periodic, a consequence of the coexistence established by the Ince
analysis --- so no energy can accumulate and no instability tongue
opens.

\section{Basics of finite difference equations}

\label{app:FDE}

Chapters~\ref{app:FDE}--\ref{app:Moeb} form a coherent unit whose
connecting thread is this: three-term recurrences generate sequences
of Mobius\index{Mobius transformation} transformations whose compositions
are exactly continued fractions (Chapter~\ref{app:CF}, Lemma~\ref{lem:Moebpr});
convergence of those fractions is equivalent to the existence of a
minimal solution\index{continued fraction!minimal solution} of the
recurrence (Pincherle\index{continued fraction!Pincherle theorem}'s
theorem, Chapter~\ref{app:CF}); and the geometry of iterated Mobius
maps (Chapter~\ref{app:Moeb}) explains why convergence holds or
fails. Together these tools underpin the continued-fraction\index{continued fraction}
eigenvalue method of Chapter~\ref{sec:LCInce}. We follow~\cite[Sec.~4.1]{KelPet}
and~\cite[Ch.~2]{Elay}.

A system of linear difference equations has the general vector form
\begin{equation}
\begin{aligned}\begin{alignedat}{1}u_{1}(n+1) & =a_{11}(n)u_{1}(n)+\cdots+a_{1m}(n)u_{m}(n)+f_{1}(n)\\
u_{2}(n+1) & =a_{21}(n)u_{1}(n)+\cdots+a_{2m}(n)u_{m}(n)+f_{2}(n)\\
 & \vdots\\
u_{m}(n+1) & =a_{m1}(n)u_{1}(n)+\cdots+a_{mm}(n)u_{m}(n)+f_{m}(n)
\end{alignedat}
\\
n=n_{0},n_{0}+1,\ldots
\end{aligned}
\label{eq:fdifeq1a}
\end{equation}
This system can be written as an equivalent vector equation 
\begin{equation}
U\left(n+1\right)=A\left(n\right)U\left(n\right)+F\left(n\right),\quad U\left(n_{0}\right)=U_{0},\quad n=n_{0},n_{0}+1,\cdots,\label{eq:fdifeq1b}
\end{equation}
where 
\begin{multline}
U\left(n\right)=\left[\begin{array}{r}
u_{1}\left(n\right)\\
u_{2}\left(n\right)\\
\vdots\\
u_{m}\left(n\right)
\end{array}\right],\quad A\left(n\right)=\left[\begin{array}{rcr}
a_{11}\left(n\right) & \cdots & a_{1m}\left(n\right)\\
\vdots & \ddots & \vdots\\
a_{m1}\left(n\right) & \cdots & a_{mm}\left(n\right)
\end{array}\right],\\
F\left(n\right)=\left[\begin{array}{r}
f_{1}\left(n\right)\\
f_{2}\left(n\right)\\
\vdots\\
f_{m}\left(n\right)
\end{array}\right],\label{eq:fdifeq1c}
\end{multline}
and vector $U_{0}$ is the so-called initial value.

We also consider a homogeneous version of equation~\eqref{eq:fdifeq1b}
when $F\left(n\right)=0$, that is 
\begin{equation}
U\left(n+1\right)=A\left(n\right)U\left(n\right),\quad U\left(n_{0}\right)=U_{0},\quad n=n_{0},n_{0}+1,\cdots.\label{eq:fdifeq1d}
\end{equation}
Equation~\eqref{eq:fdifeq1d} can be solved iteratively yielding
\begin{equation}
U\left(N\right)=A\left(N-1\right)A\left(N-2\right)\cdots A\left(n_{0}\right)U_{0},\quad N>n_{0}.\label{eq:fdifeq1e}
\end{equation}
In particular, if $A\left(n\right)=A$ independent of $n$ then equation
\eqref{eq:fdifeq1e} turns into 
\begin{equation}
U\left(N\right)=A^{N-n_{0}}U_{0},\quad N\geq n_{0}.\label{eq:fdifeq1f}
\end{equation}

Let us consider the $m$-th order scalar difference equation 
\begin{multline}
y\left(n+m\right)+p_{m-1}\left(n\right)y\left(n+m-1\right)+\cdots+p_{0}\left(n\right)y\left(n\right)\\
=r\left(n\right),\quad n=n_{0},n_{0}+1,\cdots,\label{eq:fdifeq2a}
\end{multline}
where $p_{j}\left(n\right)$ and $r\left(n\right)$ are real (or complex)
valued functions defined for $n\geq n_{0}$ and $p_{0}\left(n\right)\neq0$
for all $n\geq n_{0}$. If $r\left(n\right)$ is identically zero
equation~\eqref{eq:fdifeq2a} is called \emph{homogeneous}. Equation
\eqref{eq:fdifeq2a} can be written in the form 
\begin{multline}
y\left(n+m\right)=-p_{m-1}\left(n\right)y\left(n+m-1\right)-\cdots\\
-p_{0}\left(n\right)y\left(n\right)+r\left(n\right),\quad n=n_{0},n_{0}+1,\cdots.\label{eq:fdifeq2aa}
\end{multline}

Equations~\eqref{eq:fdifeq2a},~\eqref{eq:fdifeq2aa} can be reduced
to the vector equation~\eqref{eq:fdifeq1b}. To see this, let $y\left(n\right)$
solve equation~\eqref{eq:fdifeq2a} and define 
\begin{equation}
U\left(n\right)=\left[\begin{array}{r}
y\left(n\right)\\
y\left(n+1\right)\\
\vdots\\
y\left(n+m-1\right)
\end{array}\right],\quad n=n_{0},n_{0}+1,\cdots.\label{eq:fdifeq2b}
\end{equation}
Then the vector sequence $U\left(n\right)$ defined by equation~\eqref{eq:fdifeq2a}
satisfies vector equations~\eqref{eq:fdifeq1b} if 
\begin{multline}
A\left(n\right)=\left[\begin{array}{rcrcc}
0 & 1 & 0 & \cdots & 0\\
0 & 0 & 1 & \cdots & 0\\
\vdots & \vdots & \ddots & \ddots & \vdots\\
0 & 0 & \cdots & 0 & 1\\
-p_{0}\left(n\right) & -p_{1}\left(n\right) & \cdots & -p_{m-2}\left(n\right) & -p_{m-1}\left(n\right)
\end{array}\right],\\
F\left(n\right)=\left[\begin{array}{r}
0\\
\vdots\\
0\\
r\left(n\right)
\end{array}\right].\label{eq:fdifeq2c}
\end{multline}
The matrix $A\left(n\right)$ in equation~\eqref{eq:fdifeq2c} is
called \emph{companion matrix} of the scalar equation~\eqref{eq:fdifeq2a}.

\bigskip{}

\emph{Historical note.} The calculus of finite differences is the
mathematics of the table maker, and it is as old as the tables themselves.
Newton created interpolation by differences for the subtabulation
of astronomical data (the interpolation lemma of the \emph{Principia};
\emph{Methodus Differentialis}, 1711), and Taylor's \emph{Methodus
Incrementorum} (1715) --- the book in which Taylor's series first
appeared --- founded the subject as a calculus in its own right.
Euler and Lagrange developed its symbolic operators, Boole's \emph{Treatise
on the Calculus of Finite Differences} (1860) taught it to generations,
and in the Petersburg school it was cultivated by Markov, whose computational
credo we met in the note of Chapter~\ref{app:CF}; the mature analytic
theory was consolidated in Nörlund's \emph{Vorlesungen über Differenzenrechnung}
(Springer, 1924). The part of the theory on which this book turns,
however, is asymptotic. Poincaré proved in 1885 (\emph{Amer.\ J.
Math.} \textbf{7}, pp.~203--258) that for a linear recurrence whose
coefficients converge, the ratio of successive terms of a solution
tends to a root of the limiting characteristic equation, provided
the roots differ in modulus; Perron closed the gaps in the statement
(\emph{J. reine angew.\ Math.} \textbf{136} (1909), pp.~17--37;
\textbf{137} (1910), pp.~6--64), giving the Poincaré--Perron theory
presented below. Pincherle had meanwhile isolated the decisive concept
(\emph{Giorn.\ Mat.\ Battaglini} \textbf{32} (1894), pp.~209--291):
among the solutions of a three-term recurrence there may exist a \emph{minimal}
one, vanishing relative to every other, and its existence is equivalent
to the convergence of the associated continued fraction --- the theorem
that welds this chapter to Chapter~\ref{app:CF}, and the reason
the recurrences of this book are solved backward, by the algorithm
of Chapter~\ref{app:Miller}.

\subsection{Linear homogeneous finite difference equations with constant coefficients}

Consider a special case of homogeneous finite difference equations
\eqref{eq:fdifeq2aa} when coefficients $p_{j}\left(n\right)=p_{j}$
are constants independent of $n.$ In this case equations~\eqref{eq:fdifeq2a}
turn into 
\begin{equation}
y\left(n+m\right)+p_{m-1}y\left(n+m-1\right)+\cdots+p_{0}y\left(n\right)=0.\label{eq:fdifeq4a}
\end{equation}
To find a fundamental set of solutions we consider solutions of the
form $y\left(n\right)=\lambda^{n}$ where $\lambda$ is a complex
number. Substituting this form of solution into~\eqref{eq:fdifeq4a}
we obtain 
\begin{equation}
P\left(\lambda\right)\stackrel{\mathrm{def}}{=}\lambda^{m}+p_{m-1}\lambda^{m-1}+\cdots+p_{0}=0.\label{eq:fdifeq4b}
\end{equation}
Polynomial $P\left(\lambda\right)$ defined in equation~\eqref{eq:fdifeq4b}
is called the characteristic polynomial of finite difference equation
\eqref{eq:fdifeq4a} and equation $P\left(\lambda\right)=0$ is called
the characteristic equation of finite difference equation~\eqref{eq:fdifeq4a}.
The companion matrix $A\left(n\right)$ of equation~\eqref{eq:fdifeq4a}
does not depend on $n$ and it equals 
\begin{equation}
A=\left[\begin{array}{rcrcc}
0 & 1 & 0 & \cdots & 0\\
0 & 0 & 1 & \cdots & 0\\
\vdots & \vdots & \ddots & \ddots & \vdots\\
0 & 0 & \cdots & 0 & 1\\
-p_{0} & -p_{1} & \cdots & -p_{m-2} & -p_{m-1}
\end{array}\right].\label{eq:fdifeq4c}
\end{equation}

The eigenvalues $\zeta_{j}$ of $A$ (equivalently, the roots of the
characteristic polynomial $P(\lambda)=0$) determine a fundamental
set of solutions. If $\zeta_{1},\ldots,\zeta_{m}$ are \emph{distinct},
then $y_{j}(n)=\zeta_{j}^{n}$ for $j=1,\ldots,m$ form a fundamental
set and the general solution is 
\begin{equation}
y(n)=c_{1}\,\zeta_{1}^{n}+c_{2}\,\zeta_{2}^{n}+\cdots+c_{m}\,\zeta_{m}^{n},\quad c_{j}\in\mathbb{C}.\label{eq:fdifeq4d}
\end{equation}
If $\zeta_{j}$ is a root of $P(\lambda)=0$ of multiplicity $r$,
it contributes $r$ linearly independent solutions $\zeta_{j}^{n},\,n\zeta_{j}^{n},\,\ldots,\,n^{r-1}\zeta_{j}^{n}$
to the fundamental set \cite[Sec.~4.1]{KelPet}, \cite[Sec.~2.3]{Elay}.

\subsection{Three-term recurrences and asymptotic behavior of solutions}

A three-term recurrence relation is a second-order ($m=2$) homogeneous
difference equation of the form 
\begin{equation}
a_{n}\,y(n+1)+b_{n}\,y(n)+c_{n}\,y(n-1)=0,\quad a_{n}\neq0,\quad n=1,2,3,\ldots,\label{eq:ttrec-gen}
\end{equation}
where the coefficients $a_{n},b_{n},c_{n}$ may depend on $n$. This
is the form that arises from the eigenvalue conditions for Ince's
equation (Chapter~\ref{sec:LCInce}): the recurrences~\eqref{eq:bulk-rec}
for the Fourier coefficients $A_{2n+1}$ are precisely of the form~\eqref{eq:ttrec-gen}
with 
\begin{equation}
a_{n}=a(2n+3)^{2},\quad b_{n}=2[(2n+1)^{2}-c],\quad c_{n}=a(2n-1)^{2}.\label{eq:ttrec-gen1}
\end{equation}

The asymptotic behavior of the solutions of~\eqref{eq:ttrec-gen}
as $n\to\infty$ is governed by the \emph{Poincaré--Perron\index{Poincaré--Perron theorem}
theorem}, which we state here in the form most useful for our applications
\cite[Sec.~8.2, Thm.~8.9--8.10]{Elay}, \cite[Sec.~5.3, Thm.~5.1--5.2]{KelPet}.

Equation~\eqref{eq:ttrec-gen} is said to be of \emph{Poincaré type}
if the coefficient ratios $c_{n}/a_{n}$ and $b_{n}/a_{n}$ converge
to finite limits as $n\to\infty$. In that limit the recurrence~\eqref{eq:ttrec-gen}
approaches the constant-coefficient equation 
\begin{equation}
\zeta^{2}+p\,\zeta+q=0,\quad p=\lim_{n\to\infty}\frac{b_{n}}{a_{n}},\quad q=\lim_{n\to\infty}\frac{c_{n}}{a_{n}},\label{eq:char-asymp}
\end{equation}
whose roots $\zeta_{1},\zeta_{2}$ are called the \emph{characteristic
roots} of the recurrence. 
\begin{thm}[{Poincaré--Perron; see {\cite[Sec.~8.2]{Elay}}}]
\index{Poincaré--Perron theorem} \label{thm:PP} Suppose that equation~\eqref{eq:ttrec-gen}
is of Poincaré type, that $c_{n}\neq0$ for all $n$, and that the
characteristic roots $\zeta_{1},\zeta_{2}$ satisfy $|\zeta_{1}|>|\zeta_{2}|>0$. 
\begin{enumerate}
\item \emph{(Poincaré)} Every nontrivial solution $y(n)$ satisfies 
\begin{equation}
\lim_{n\to\infty}\frac{y(n+1)}{y(n)}=\zeta_{i}\label{eq:PP-ratio}
\end{equation}
for some $i\in\{1,2\}$. 
\item \emph{(Perron)} There exist two linearly independent solutions $y_{1}(n)$
and $y_{2}(n)$ such that 
\begin{equation}
\lim_{n\to\infty}\frac{y_{1}(n+1)}{y_{1}(n)}=\zeta_{1},\qquad\lim_{n\to\infty}\frac{y_{2}(n+1)}{y_{2}(n)}=\zeta_{2}.\label{eq:Perron-fund}
\end{equation}
\end{enumerate}
The solution $y_{2}(n)$ associated with the smaller characteristic
root $|\zeta_{2}|<|\zeta_{1}|$ is called the \emph{minimal solution}:
it satisfies $y_{2}(n)/y_{1}(n)\to0$ as $n\to\infty$, and any solution
not proportional to $y_{2}$ satisfies the same ratio with $y_{2}$
in the numerator tending to zero. The solution $y_{1}(n)$ associated
with the larger root is called \emph{dominant}. 
\end{thm}

\begin{rem}[Connection to continued fractions and the Ince eigenvalue problem]
\label{rem:PP-CF} The Poincaré--Perron theorem is the key to understanding
why the continued-fraction eigenvalue method of Chapter~\ref{sec:LCInce}
is valid. By Pincherle's theorem (Theorem~\ref{thm:Pinch} of Chapter~\ref{app:CF}),
\emph{the continued fraction associated with recurrence~\eqref{eq:ttrec-gen}
converges if and only if the recurrence has a minimal solution}. By
the Poincaré--Perron theorem, this is guaranteed whenever $|\zeta_{1}|\neq|\zeta_{2}|$,
i.e., whenever the two characteristic roots have distinct moduli. 
\end{rem}

\begin{rem}[Poincaré--Perron applied to the Ince recurrence]
\label{rem:PP-Ince} For the Ince recurrence~\eqref{eq:bulk-rec},
the coefficient ratios $c_{n}/a_{n}=(2n-1)^{2}/(2n+3)^{2}\to1$ and
$b_{n}/a_{n}=2[(2n+1)^{2}-c]/[a(2n+3)^{2}]\to2/a$ as $n\to\infty$
(for fixed $c$ and $a\neq0$). The limiting characteristic equation
is $\zeta^{2}+(2/a)\zeta+1=0$, with roots 
\[
\zeta_{1}=\frac{1}{\delta},\qquad\zeta_{2}=\delta,
\]
obtained by substituting $a=-2\delta/(1+\delta^{2})$ and computing
$-1/a\pm\sqrt{1/a^{2}-1}=(1+\delta^{2})/(2\delta)\pm(1-\delta^{2})/(2\delta)$.
Both roots are real, nonzero, and satisfy $|\zeta_{1}|=1/\delta>1>\delta=|\zeta_{2}|>0$
for all $\delta\in(0,1)$; neither depends on $c$. Hence the Poincaré--Perron
hypotheses hold unconditionally for all $c$: the trailing coefficient
$c_{n}=a(2n-1)^{2}$ never vanishes, and $|\zeta_{1}|>|\zeta_{2}|>0$,
guaranteeing a minimal solution (with ratio $y(n+1)/y(n)\to\delta$)
and a dominant solution (with ratio $y(n+1)/y(n)\to1/\delta$) regardless
of which zone $c$ lies in. This gives the rigorous foundation for
the continued-fraction eigenvalue condition $F_{\mathrm{even/odd}}(c)=0$
derived in Section~\ref{subsec:Widths}. 
\end{rem}

\section{Continued fractions}

\label{app:CF}

A continued fraction\index{continued fraction} is the value to which
a sequence of nested Mobius\index{Mobius transformation} transformations
converges. This chapter collects the definitions and convergence theorems
needed for the eigenvalue method of Chapter~\ref{sec:LCInce}; the
geometric picture behind convergence is developed in Chapter~\ref{app:Moeb}.
The conceptual role of three-term recurrences as the structural foundation
for continued fractions --- and of Pincherle\index{continued fraction!Pincherle theorem}'s
theorem (Theorem~\ref{thm:Pinch} below) as the bridge to minimal
solution\index{continued fraction!minimal solution}s --- is discussed
in Remark~\ref{rem:CF-three-term}; see~\cite[\S1]{Gautschi67}
and~\cite[Sec.~5.2--5.3]{JonThr}. We follow mainly~\cite[Ch.~1]{JonThr},
\cite[Ch.~1]{LorWaad}, \cite[\S\,12]{Henrici2}, \cite[Ch.~9]{Elay}.

A continued fraction is an ordered pair $\left\langle \left\langle \left\{ a_{n}\right\} ,\left\{ b_{n}\right\} \right\rangle ,\left\{ f_{n}\right\} \right\rangle $,
where $a_{1},a_{2},\ldots$ and $b_{0},b_{1},b_{2},\ldots$ are complex
numbers with $a_{n}\neq0$ and where $f_{n}$ is a sequence in extended
complex plane defined as follows~\cite[\S\,2.1]{JonThr}, \cite[\S\,1.1.2]{LorWaad}:
\begin{equation}
f_{n}=S_{n}\left(0\right),\quad n=1,2,3,\ldots,\label{eq:abfS1a}
\end{equation}
where 
\begin{equation}
S_{0}\left(w\right)=s_{0}\left(w\right);\qquad S_{n}\left(w\right)=S_{n-1}\left(s_{n}\left(w\right)\right),\quad n=1,2,3,\ldots,\label{eq:abfS1b}
\end{equation}
and 
\begin{equation}
s_{0}\left(w\right)=b_{0}+w;\qquad s_{n}\left(w\right)=\frac{a_{n}}{b_{n}+w},\quad n=1,2,3,\ldots.\label{eq:abfS1c}
\end{equation}
The continued fraction algorithm is a function $\mathbf{K}$ which
assigns to each ordered pair $\left\langle \left\{ a_{n}\right\} ,\left\{ b_{n}\right\} \right\rangle $
the sequence $\left\{ f_{n}\right\} $ defined by~\eqref{eq:abfS1a}--\eqref{eq:abfS1c}.
The numbers $a_{n}$ and $b_{n}$ are called \emph{partial numerator}
and \emph{denominator} of the continued fraction, respectively. Sometimes
they are simply called the elements; $f_{n}$ is called the \emph{(classical)
$n$-th approximant}. Thus a continued fraction $\left\langle \left\langle \left\{ a_{n}\right\} ,\left\{ b_{n}\right\} \right\rangle ,\left\{ f_{n}\right\} \right\rangle $
is often denoted by symbol 
\begin{equation}
b_{0}+\frac{a_{1}}{b_{1}+\cfrac{a_{2}}{b_{2}+\cfrac{a_{3}}{b_{3}+\cdots}}}.\label{eq:abfS2a}
\end{equation}
For convenience the continued fraction is commonly represented by
the following more concise notations 
\begin{equation}
b_{0}+\mathbf{K}\left[\frac{a_{m}}{b_{m}}\right]=b_{0}+\mathbf{K}_{m=1}^{\infty}\left[\frac{a_{m}}{b_{m}}\right]=b_{0}+\frac{a_{1}}{b_{1}+}\:\frac{a_{2}}{b_{2}+}\:\frac{a_{3}}{b_{3}+\cdots}.\label{eq:abfS2b}
\end{equation}
Similarly the $n$-th \emph{approximant} $f_{n}=S_{n}\left(0\right)$
may be denoted by 
\begin{equation}
f_{n}=b_{0}+\frac{a_{1}}{b_{1}+}\:\frac{a_{2}}{b_{2}+}\cdots\frac{a_{n}}{b_{n}}=b_{0}+\mathbf{K}_{m=1}^{n}\left[\frac{a_{m}}{b_{m}}\right].\label{eq:abfS2c}
\end{equation}

The Mobius transformations $s_{n}\left(w\right)$, $n=1,2,3,\ldots$
are associated with the following matrices, see \cite[Sec.~2.1]{JonThr}:
\begin{equation}
s_{0}\left(w\right)=b_{0}+w:M_{0}=\left[\begin{array}{rr}
1 & b_{0}\\
0 & 1
\end{array}\right],\label{eq:abfS3a}
\end{equation}
\begin{multline}
s_{n}(w)=\frac{a_{n}}{b_{n}+w},\quad M_{n}=\begin{bmatrix}0 & a_{n}\\
1 & b_{n}
\end{bmatrix},\\
\tilde{M}_{n}=\frac{1}{\sqrt{-a_{n}}}M_{n}=\begin{bmatrix}0 & -\sqrt{-a_{n}}\\
\frac{1}{\sqrt{-a_{n}}} & \frac{b_{n}}{\sqrt{-a_{n}}}
\end{bmatrix},\quad\det\tilde{M}_{n}=1.\label{eq:abfS3b}
\end{multline}

Two different \emph{continued fractions can be equivalent} in the
following sense, \cite[\S\,2.3.1]{JonThr}, \cite[\S\,2.2.2]{LorWaad},
\cite[\S\,9.2]{Elay}, \cite[\S\,12.1]{Henrici2} 
\begin{defn}[equivalence of continued fractions]
\label{def:equi} We say that two continued fractions are equivalent
if they have the same sequence of classical approximants $\left\{ f_{n}\right\} $
defined by~\eqref{eq:abfS1a}--\eqref{eq:abfS1c}. 
\end{defn}

There is an important equivalency criterion for two continued fractions
which is as follows, \cite[\S\,2.3.1]{JonThr}, \cite[\S\,2.2.2]{LorWaad},
\cite[\S\,9.2]{Elay}, \cite[\S\,12.1]{Henrici2}. 
\begin{thm}[criterion of equivalency]
\label{thm:equi} Continued fractions $b_{0}+\mathbf{K}\left[\frac{a_{m}}{b_{m}}\right]$
and $b_{0}^{\prime}+\mathbf{K}\left[\frac{a_{m}^{\prime}}{b_{m}^{\prime}}\right]$
are equivalent, denoted 
\begin{equation}
b_{0}^{\prime}+\mathbf{K}\left[\frac{a_{m}^{\prime}}{b_{m}^{\prime}}\right]\cong b_{0}+\mathbf{K}\left[\frac{a_{m}}{b_{m}}\right],\label{eq:abfS3c}
\end{equation}
if and only if there exists a sequence of nonzero complex numbers
$\left\{ c_{m}:m=0,1,2,\ldots\right\} $ such that 
\begin{equation}
b_{0}^{\prime}=b_{0},\quad a_{m}^{\prime}=c_{m}c_{m-1}a_{m},\quad b_{m}^{\prime}=c_{m}b_{m};\quad c_{0}=1,\quad c_{m}\neq0,\quad m\geq1.\label{eq:abfS3d}
\end{equation}
\end{thm}

Note that formula~\eqref{eq:abfS3c} is manifestly consistent with
the continued fraction representation~\eqref{eq:abfS2a}. Indeed,
let us multiply the numerator and the denominator of the first fraction
by $c_{1},$ then multiply the numerator and the denominator of the
second fraction by $c_{2}$ and so on. These actions preserve classical
approximants $\left\{ f_{n}\right\} $ defined by~\eqref{eq:abfS1a}--\eqref{eq:abfS1c}.
As a result of the sequence of actions each $b_{m}$ is multiplied
by $c_{m}$ whereas each $a_{m}$ is multiplied by first by $c_{m-1}$
and then by $c_{m}$ for $m=1,2,\ldots$ which is consistent with
equations~\eqref{eq:abfS3d}.

Let us turn now to Mobius transformations $S_{n}\left(w\right)$ defined
recurrently by equations~\eqref{eq:abfS1b}--\eqref{eq:abfS1c}.
They satisfy the following representations, \cite[\S\,2.3.1]{JonThr},
\cite[\S\,1.1.2]{LorWaad}, \cite[\S\,9.1]{Elay}. 
\begin{lem}[Mobius transformation representation]
\index{continued fraction!Mobius representation} \label{lem:Moebpr}
Let $S_{n}$ be defined by~\eqref{eq:abfS1b}, \eqref{eq:abfS1c}.
Then 
\begin{equation}
S_{n}\left(w\right)=\frac{A_{n-1}w+A_{n}}{B_{n-1}w+B_{n}},\quad n=1,2,3,\ldots,\label{eq:abfS4a}
\end{equation}
where 
\begin{equation}
A_{n}=b_{n}A_{n-1}+a_{n}A_{n-2},\quad B_{n}=b_{n}B_{n-1}+a_{n}B_{n-2}\label{eq:abfS4b}
\end{equation}
with initial values 
\begin{equation}
A_{-1}=1,\quad A_{0}=b_{0},\quad B_{-1}=0,\quad B_{0}=1.\label{eq:abfS4c}
\end{equation}
\end{lem}

Using Lemma \ref{lem:Moebpr} one can derive the \emph{Euler-Minding
formula} which is useful for the computation of classical approximants
$f_{n}$, \cite[\S\,1.1.2]{LorWaad}, \cite[\S\,2.1.4]{JonThr} 
\begin{equation}
f_{n}=\frac{A_{n}}{B_{n}}=\frac{A_{0}}{B_{0}}+\sum_{k=1}^{n}\left(\frac{A_{k}}{B_{k}}-\frac{A_{k-1}}{B_{k-1}}\right)=b_{0}-\sum_{k=1}^{n}\frac{\left(-1\right)^{k}a_{1}a_{2}\cdots a_{k}}{B_{k}B_{k-1}}.\label{eq:abfS4d}
\end{equation}

\bigskip{}

\emph{Historical note.} The subject is older than the calculus. Bombelli
used a continued-fraction scheme to approximate square roots in 1572,
and Cataldi introduced the first notation in 1613; the term itself
was coined by Wallis in his \emph{Arithmetica Infinitorum} (1655),
in his commentary on Brouncker's expansion of $4/\pi$. Huygens chose
the tooth counts for the gears of his planetarium from continued-fraction
convergents (\emph{Descriptio automati planetarii}, 1703) --- the
first engineering application of the theory, and in spirit a precursor
of every rational approximation in this book. The analytic theory
begins with Euler (\emph{De fractionibus continuis dissertatio}, 1737),
whose expansions connected continued fractions with series and with
differential equations; Lambert used continued fractions to prove
the irrationality of $\pi$ (1768), and Lagrange proved that the quadratic
irrationals are exactly the numbers with eventually periodic expansions
(1770). The first published paper of Galois (1829) concerned periodic
continued fractions. In the Petersburg school the subject was a thread
of continuity: Markov won the university's gold medal in 1877 for
a study of the solution of differential equations by continued fractions
--- precisely the circle of ideas of the present book --- took his
1884 doctorate on applications of algebraic continued fractions to
limiting values of integrals, and lectured on the theory to the end
of his career. His credo suits the computational spirit of the subject:
``Many mathematicians apparently believe that going beyond the field
of abstract reasoning into the sphere of effective calculations would
be humiliating'' (A.~A. Markov, 1899; quoted in translation in G.~P.
Basharin, A.~N. Langville, and V.~A. Naumov, \emph{Linear Algebra
and its Applications} \textbf{386} (2004), p.~19) --- an opinion
Markov refuted by example. The line from Chebyshev and Markov leads
onward to Stieltjes, whose \emph{Recherches sur les fractions continues}
(1894) created the moment problem, and to the convergence theory of
Pincherle, Poincaré, and Perron on which the present chapter rests.

\bigskip{}

\subsection{Tail sequences}

\label{subsec:cftails}

We follow \cite[\S\,2.1.4--2.1.5]{LorWaad}, \cite[\S\,1.9]{Cuyt}.
When the classical approximant $f_{n}$ of the continued fraction
$b_{0}+\mathbf{K}\left[\frac{a_{m}}{b_{m}}\right]$ is obtained by
truncating the continued fraction after $n$ fraction terms the part
that was cut out is the so-called $n$-th \emph{tail} defined by 
\begin{equation}
f^{\left(n\right)}=\frac{a_{n+1}}{b_{n+1}+}\:\frac{a_{n+2}}{b_{n+2}+}\:\frac{a_{n+3}}{b_{n+3}+\cdots}.\label{eq:abfS5a}
\end{equation}
Indeed, expression~\eqref{eq:abfS5a} converges if and only if $b_{0}+\mathbf{K}\left[\frac{a_{m}}{b_{m}}\right]$
converges to 
\begin{equation}
f=S_{n}\left(f^{\left(n\right)}\right).\label{eq:abfS5b}
\end{equation}
The sequence $\left\{ f^{\left(n\right)}\right\} $ is then called
the sequence of tail values for $b_{0}+\mathbf{K}\left[\frac{a_{m}}{b_{m}}\right]$.

\subsection{Convergence of continued fractions}

\label{subsec:conv}

Consider a system of \emph{three-term recurrence} relations, \cite[\S\,5.3]{JonThr}
\begin{equation}
y_{n+1}=b_{n}y_{n}+a_{n}y_{n-1},\quad a_{n}\neq0,\quad n=1,2,3,\ldots,\label{eq:ttrec1a}
\end{equation}
where $a_{n}$, $b_{n}$ and $y_{n}$ are elements of a normed field
$\mathbb{F}$~\cite[\S\,2.1.1]{JonThr} with $\left\Vert y\right\Vert $
being the norm of $y\in\mathbb{F}$. Field $\mathbb{F}$ in applications
of our interest is often the set of meromorphic functions of complex
$z$ over a domain in $\mathbb{C}$. The set of all solutions $\left\{ y_{n}\right\} $
of~\eqref{eq:ttrec1a} forms a linear vector space $V$ of dimension
two over the field $\mathbb{F}$. If there exists a non-trivial solution
$\left\{ h_{n}\right\} $, that is $h_{n}\neq0$ for some $n$, and
another solution $\left\{ g_{n}\right\} $ such that 
\begin{equation}
\lim_{n\rightarrow\infty}\frac{h_{n}}{g_{n}}=0,\label{eq:ttrec1b}
\end{equation}
then $\left\{ h_{n}\right\} $ is called \emph{minimal solution} of
\eqref{eq:ttrec1a}. One can readily verify that for any solution
$\left\{ y_{n}\right\} $ of~\eqref{eq:ttrec1a} not proportional
to $\left\{ h_{n}\right\} $ 
\begin{equation}
\lim_{n\rightarrow\infty}\frac{h_{n}}{y_{n}}=0.\label{eq:ttrec1c}
\end{equation}
If the set of all solutions $\left\{ h_{n}\right\} $ of~\eqref{eq:ttrec1a}
having the property~\eqref{eq:ttrec1b} is not empty, then it is
a one-dimensional subspace of $V$. A solution of~\eqref{eq:ttrec1a}
which is not minimal is called \emph{dominant}. In general a system
of recurrence relations~\eqref{eq:ttrec1a} may or may not have a
minimal solution.

The following important statement relates converging continued fraction
to the minimal solution of the corresponding three-term recurrence
relation, \cite[Thm.~5.7]{JonThr}, \cite[Thm.~4.5]{Wimp}, \cite[Thm.~9.5]{Elay},
\cite[Ch.~2, \S\,2.1]{AhlPet}. 
\begin{thm}[Pincherle generalized, convergence]
\index{continued fraction!Pincherle theorem} \label{thm:Pinch}
For each $n=1,2,3,\ldots,$ let $a_{n}$ and $b_{n}$ be elements
of a normed field $\mathbb{F}$~\cite[\S\,2.1.1]{JonThr}, with 
\begin{equation}
a_{n}\neq0,\quad n=1,2,3,\ldots.\label{eq:confr1a}
\end{equation}
\begin{enumerate}
\item The system of three-term recurrence relations 
\begin{equation}
y_{n+1}=b_{n}y_{n}+a_{n}y_{n-1},\quad n=1,2,3,\ldots,\label{eq:confr1b}
\end{equation}
has a minimal solution $\left\{ h_{n}\right\} ,\;h_{n}\in\mathbb{F}$,
if and only if the continued fraction over the normed field $\mathbb{F}$
\begin{equation}
\frac{a_{1}}{b_{1}+}\:\frac{a_{2}}{b_{2}+}\:\frac{a_{3}}{b_{3}+\cdots}\label{eq:confr1c}
\end{equation}
converges to a finite value in $\mathbb{F}$ or to infinity. 
\item Suppose that~\eqref{eq:confr1b} has a minimal solution $\left\{ h_{n}\right\} ,\;h_{n}\in\mathbb{F}$.
Then 
\begin{equation}
\frac{h_{m}}{h_{m-1}}=-\frac{a_{m}}{b_{m}+}\:\frac{a_{m+1}}{b_{m+1}+}\:\frac{a_{m+2}}{b_{m+2}+\cdots},\quad m=1,2,3,\ldots.\label{eq:confr1d}
\end{equation}
Note that according to formula~\eqref{eq:abfS5a} the expression
on the right in equation~\eqref{eq:confr1d} is minus of $\left(m-1\right)$-th
tail, that is $-f^{\left(m-1\right)}$, of continued fraction~\eqref{eq:confr1c}. 
\item By~\eqref{eq:confr1d} we mean the following: If $h_{m-1}=0$, then
$h_{m}\neq0$ and the continued fraction~\eqref{eq:confr1d} converges
to $\infty=\frac{h_{m}}{h_{m-1}}$. If $h_{m-1}\neq0$, then the continued
fraction~\eqref{eq:confr1d} converges to the finite value $\frac{h_{m}}{h_{m-1}}\in\mathbb{F}$. 
\end{enumerate}
\end{thm}

In the case of our applications the field $\mathbb{F}$ is the set
of meromorphic functions over a domain $D$ or the extended complex
plane $\hat{\mathbb{C}}$ furnished with the spherical (chordal) metric~\eqref{eq:Moeb1b},
see, e.g., \cite[\S\,2.1.1]{JonThr}.

Here is another classical convergence statement, obtained via the
Euler-Minding formula~\eqref{eq:abfS4d}, \cite[Thm.~4.35]{JonThr},
\cite[\S\,3.2.4]{LorWaad}, 
\begin{thm}[Sleszynski-Pringsheim, convergence]
\index{Sleszynski--Pringsheim theorem} \label{thm:sLePrin} Suppose
that the coefficients of the continued fraction $\mathbf{K}_{n=1}^{\infty}\left[\frac{a_{n}}{b_{n}}\right]$
satisfy 
\begin{equation}
\left|b_{n}\right|\geq\left|a_{n}\right|+1.\label{eq:confr2a}
\end{equation}
Then $\mathbf{K}_{n=1}^{\infty}\left[\frac{a_{n}}{b_{n}}\right]$
converges to a finite value $f$ and its $n$-th approximant $f_{n}$
satisfies 
\begin{equation}
\left|f_{n}\right|<1,\quad n=1,2,3,\ldots.\label{eq:confr2b}
\end{equation}
Moreover, $\left\{ \left|A_{n}\right|\right\} $ and $\left\{ \left|B_{n}\right|\right\} $
defined by equations~\eqref{eq:abfS4a}--\eqref{eq:abfS4c} are
strictly increasing sequences, and $0<\left|S_{n}\left(w\right)\right|\leq1$
for all $\left|w\right|\leq1$. 
\end{thm}

There are some more statements on convergence of continued fractions,
\cite[Cors.~4.34,~4.36, Thm.~4.37]{JonThr}. 
\begin{thm}[convergence]
\label{thm:JoThr1} The continued fraction $\mathbf{K}_{n=1}^{\infty}\left[\frac{1}{b_{n}}\right]$
converges if for a real number $c$ the following inequalities hold
\begin{equation}
\left|b_{n}+2c\right|\geq2\sqrt{1+c^{2}},\quad n=1,2,3,\ldots.\label{eq:confr2c}
\end{equation}
\end{thm}

\begin{thm}[convergence]
\label{thm:JoThr2} 
\begin{enumerate}
\item $\mathbf{K}_{n=1}^{\infty}\left[\frac{a_{n}}{b_{n}}\right]$ converges
to a finite value if 
\begin{equation}
\left|\frac{a_{1}}{b_{1}}\right|\leq\frac{p_{1}-1}{p_{1}},\quad\left|\frac{a_{n}}{b_{n-1}b_{n}}\right|\leq\frac{p_{n}-1}{p_{n-1}p_{n}},\quad n=2,3,\ldots,\label{eq:confr3a}
\end{equation}
where $p_{n}$ are real and $p_{n}>1$. 
\item (Worpitzky) $\mathbf{K}_{n=1}^{\infty}\left[\frac{a_{n}}{1}\right]$
converges to a finite value if 
\begin{equation}
\left|a_{n}\right|\leq\frac{1}{4},\quad n=1,2,3,\ldots.\label{eq:confr3b}
\end{equation}
\item $\mathbf{K}_{n=1}^{\infty}\left[\frac{1}{b_{n}}\right]$ converges
to a finite value if 
\begin{equation}
\left|\frac{1}{b_{2n-1}}\right|+\left|\frac{1}{b_{2n}}\right|\leq1,\quad n=1,2,3,\ldots.\label{eq:confr3c}
\end{equation}
\end{enumerate}
\end{thm}

\begin{thm}[convergence]
\label{thm:JoThr3}Let $D$ be the circular region defined by 
\begin{equation}
D=\left\{ w:\left|w-c\right|\leq r\right\} ,\text{ where }\left|c\right|<r.\label{eq:confr3d}
\end{equation}
Let continued fraction $\mathbf{K}_{n=1}^{\infty}\left[\frac{a_{n}}{b_{n}}\right]$
be such that 
\begin{equation}
s_{n}\left(D\right)\subseteq D,\text{ where }s_{n}\left(w\right)=\frac{a_{n}}{b_{n}+w},\quad n=1,2,3,\ldots.\label{eq:confr3e}
\end{equation}
Then the continued fraction converges to a finite value. 
\end{thm}

\subsection{Convergence of meromorphic functions}

\label{subsec:meromo}

We review here concisely some statements on the convergence of meromorphic
functions, using the spherical (chordal) metric defined in \S\,\ref{subsec:Moeb-def}
of Chapter~\ref{app:Moeb} (eq.~\eqref{eq:Moeb1b}).
\begin{thm}[Weierstrass; spherically uniform limit of meromorphic functions]
\index{Weierstrass convergence theorem} \label{thm:Weierstrass-uniform}
Let $\{f_{n}\}$ be a sequence of meromorphic functions on a region
$G\subseteq\mathbb{C}$ that converges to a function $f$ in the spherical
metric, uniformly on every compact subset of $G$. Then $f$ is either
meromorphic on $G$ or identically $\infty$. If each $f_{n}$ is
analytic on $G$, then $f$ is either analytic on $G$ or identically
$\infty$. 
\end{thm}

\begin{proof}[References]
\cite[Ch.~5, \S\,5, Lemma]{Ahlf} (stated as a lemma extending Theorems~1
and~2 to meromorphic functions; the proof reduces to the ordinary
Weierstrass theorem near points where $f\neq\infty$ and to $1/f_{n}\to1/f$
near poles); \cite[Ch.~VII, Thm.~3.4]{Conway} (convergence in $C(G,\mathbb{C}_{\infty})$,
i.e.\ spherically uniformly on compact sets; same two-case conclusion). 
\end{proof}
\begin{thm}[Hurwitz; poles and zeros of spherically uniform limits]
\index{Hurwitz theorem} \label{thm:Hurwitz} Let $G\subseteq\mathbb{C}$
be a region and $\{f_{n}\}\subset M(G)$ a sequence of meromorphic
functions converging to $f$ in the spherical metric, uniformly on
every compact subset of $G$. 
\begin{enumerate}
\item[(a)] \emph{Non-vanishing limit (\cite[Ch.~5, \S\,5, Lemma]{Ahlf}; \cite[Ch.~VII, Thm.~3.4]{Conway})}.
If each $f_{n}$ is analytic and nowhere zero on $G$, then either
$f\equiv0$, $f\equiv\infty$, or $f$ is analytic and nowhere zero
on $G$.
\item[(b)] \emph{Zero and pole count in a disk (\cite[Ch.~VII, Thm.~2.5]{Conway};
\cite[Thm.~6.4.1]{SimBCA2A})}. If $f\not\equiv0$ and $f\not\equiv\infty$,
$\bar{B}(a;R)\subset G$, and $f(z)\neq0,\infty$ for $|z-a|=R$,
then there exists $N$ such that for all $n\geq N$ the functions
$f$ and $f_{n}$ have the same number of zeros and the same number
of poles (counting multiplicity) in $B(a;R)$.
\item[(c)] \emph{Convergence of zeros and poles (\cite[Ch.~4, \S\,3.3, Thm.~11]{Ahlf})}.
If $f(z_{0})=0$ (respectively $=\infty$) with multiplicity $m$,
then for every sufficiently small $r>0$ and all sufficiently large
$n$, $f_{n}$ has exactly $m$ zeros (resp.\ poles) in $|z-z_{0}|<r$,
and these converge to $z_{0}$ as $n\to\infty$. 
\end{enumerate}
\end{thm}

\begin{proof}[Remarks on sources]
Part~(a) follows from Theorem~\ref{thm:Weierstrass-uniform} applied
to $1/f_{n}$: since $f_{n}$ is analytic and nowhere zero, $1/f_{n}$
is analytic, and its spherical limit $1/f$ is analytic or $\equiv\infty$.
The three alternatives correspond to $1/f$ analytic and nowhere zero,
$1/f\equiv0$ (i.e.\ $f\equiv\infty$), or $1/f$ analytic with a
zero (impossible if $f_{n}$ nowhere vanishes, by Hurwitz applied
to $1/f_{n}$). This is the conclusion of~\cite[Ch.~VII, Cor.~3.6]{Conway}.
Part~(b) extends Conway's Theorem~2.5 to the meromorphic setting
by applying it to $f_{n}/f$ near zeros and to $f_{n}\cdot g$ near
poles for a local uniformizing factor $g$. Part~(c) follows from
the local mapping theorem~\cite[Ch.~4, \S\,3.3, Thm.~11]{Ahlf} applied
to both $f$ near zeros and $1/f$ near poles. 
\end{proof}
\begin{prop}[Convergence of meromorphic functions]
\index{meromorphic function!convergence} \label{prop:mero-conv}
(\cite[Thm.~15.2.1]{HilleA2}) A sequence of functions $f_{n}\left(z\right)$
meromorphic in a domain $D$, which converges spherically uniformly
on each compact subset of $D$, converges to a meromorphic function
$f\left(z\right)$. It is not excluded that $f\left(z\right)$ is
actually holomorphic in $D$ or reduces to a finite or infinite constant. 
\end{prop}

\begin{thm}[Montel]
\index{Montel theorem} \label{thm:Montel} (\cite[Thm.~15.3.1]{HilleA2})
Let the functions $f_{n}\left(z\right)$ be meromorphic and spherically
equicontinuous in a domain $D$, and let the sequence $\left\{ f_{n}\left(z\right)\right\} $
converge spherically on a point set $\left\{ z_{k}\right\} $ which
has a limit point $z_{0}$ in $D$. Then $f_{n}\left(z\right)$ converges
spherically everywhere on $D$, uniformly on every compact subset
of $D$. 
\end{thm}

Here is a statement on convergence of continued fraction with coefficients
depending on a complex variable, \cite[Thm.~4.54 and Remarks]{JonThr} 
\begin{thm}[Meromorphic convergence]
\index{meromorphic function!spherically uniform convergence} \label{thm:sLconfmero1}
(\cite[Thm.~4.54]{JonThr}) Let $\mathbf{K}_{n=1}^{\infty}\left[\frac{a_{n}\left(z\right)}{b_{n}\left(z\right)}\right]$
be a continued fraction such that, for each $n=1,2,3,\ldots$, $a_{n}\left(z\right)$
and $b_{n}\left(z\right)$ are holomorphic functions of a complex
variable $z$ in a domain $D$. Further assume that there exist distinct
numbers $\alpha$ and $\beta$ such that for each $n=1,2,3,\ldots$
the $n$-th approximant $f_{n}\left(z\right)$ of the continued fraction
satisfies the conditions 
\begin{equation}
f_{n}\left(z\right)\neq\alpha,\beta,\infty\text{ for all }z\in D.\label{eq:confr4a}
\end{equation}
If the continued fraction converges to a finite value $f\left(z\right)$
for each $z\in\varDelta$, where $\varDelta$ is an infinite set with
at least one limit point in $D$, then it converges uniformly on every
compact subset of $D$, and hence its limit function is holomorphic
in $D$.

In addition, if this theorem is valid for a ``tail'' of the continued
fraction, then the continued fraction converges to a function which
is meromorphic in $D$ or identically $\infty$. 
\end{thm}

\begin{rem}[Continued fractions and the Ince eigenvalue problem]
\label{rem:CF-Ince} Theorems~\ref{thm:Pinch}--\ref{thm:sLconfmero1}
are the rigorous foundation for the continued-fraction eigenvalue
conditions used in Chapter~\ref{sec:LCInce} and in Cambi's solution
(Chapter~\ref{sec:Cambi}). In both cases the Fourier-coefficient
recurrence~\eqref{eq:bulk-rec} is of three-term form~\eqref{eq:ttrec-gen}
with coefficients that are rational functions of the eigenvalue parameter
$c$ (equivalently $\hat{\lambda}$). By the Pincherle theorem (Theorem~\ref{thm:Pinch}),
the associated continued fraction converges if and only if the recurrence
has a minimal solution; by the Poincaré--Perron analysis of Remark~\ref{rem:PP-Ince},
a minimal solution exists for \emph{every} value of $c$, so the continued
fractions converge unconditionally and their values --- the tail
functions --- are well defined. The eigenvalue condition $F_{\mathrm{even/odd}}(c)=0$
of Section~\ref{subsec:Widths} of Chapter~\ref{sec:LCInce} expresses
the further requirement that the solution singled out by the lowest-index
(boundary) equation coincide with the minimal solution; this matching
occurs precisely when $c$ is an eigenvalue --- i.e., when the Floquet\index{Floquet theory}
solution is periodic or antiperiodic. The meromorphic convergence
theorem (Theorem~\ref{thm:sLconfmero1}) guarantees that the continued-fraction
value is meromorphic in $c$, which underpins the analytic structure
of the instability boundary\index{stability boundary} curves derived
in Chapters~\ref{sec:LCInce} and~\ref{sec:Boundaries}. 
\end{rem}

The geometric underpinning of continued fraction convergence via iterated
Mobius transformations is developed in Chapter~\ref{app:Moeb}.

\section{Linear fractional (Mobius) transformations}

\label{app:Moeb}

A \emph{Mobius\index{Mobius transformation} transformation} (also
called a linear fractional\index{linear fractional transformation}
transformation) is a map $T(z)=(az+b)/(cz+d)$ on the Riemann sphere.
Mobius transformations are the building blocks of both continued fraction\index{continued fraction}s
and transfer-matrix methods: the $n$-th approximant of a continued
fraction is the composition $S_{n}=s_{1}\circ\cdots\circ s_{n}$ of
elementary Mobius maps (Lemma~\ref{lem:Moebpr}), and the transfer
matrices of Chapter~\ref{sec:LCInce} act on the ratio of successive
Fourier coefficients by a Mobius map. The type of the limiting Mobius
transformation --- elliptic, parabolic, or hyperbolic --- determines
whether the continued fraction converges and whether the instability
boundary\index{stability boundary} is a genuine EPD\index{exceptional point of degeneracy (EPD)}
or a coexistence point. Additional references: \cite[Sec.~5.3--5.4]{Henrici1}
and \cite[Ch.~3]{Ahlf}.

A remark on notation: this chapter follows the standard symbols of
the Mobius literature, which are local to the chapter --- here $T(z)$
denotes a Mobius transformation (not the period), $G(z)$ a conjugating
map (not the recurrence coefficient $G(u)$ of Chapter~\ref{sec:expanding-Cambi}),
$\mu$ the multiplier of a transformation (not the driving frequency),
and $c(\rho)$, $r(\rho)$ the center and radius of an invariant circle
(not the Hill spectral parameter).

\subsection{Definition and basic properties}

\label{subsec:Moeb-def} We review the basic structure of a Mobius
transformation $T$ and the limiting behavior of its iterates $T^{n}$
as $n\to\infty$, following~\cite[Sec.~3.1--3.2]{HilleA1}, \cite[Ch.~1]{FordA},
\cite[Sec.~1--2]{JonSin}, \cite[Ch.~3]{Need}, \cite[Sec.~7.3]{SimBCA2A}.

A linear fractional (Mobius) transformation $T:\mathbb{C}\rightarrow\mathbb{C}$
is defined by 
\begin{equation}
T\left(z\right)=\frac{az+b}{cz+d},\quad ad-bc\neq0,\label{eq:Moeb1a}
\end{equation}
where $a$, $b$, $c$ and $d$ are given complex numbers. The Mobius
transformation defined by~\eqref{eq:Moeb1a} can be naturally extended
as a one-to-one mapping in the \emph{extended complex plane} $\hat{\mathbb{C}}=\mathbb{C}\cup\left\{ \infty\right\} $.
We furnish the extended complex plane $\hat{\mathbb{C}}$ with the
spherical metric, which is based on the so-called \emph{stereographic
projection} with ``chordal distance'' $\chi$ defined as follows,
\cite[\S\,2.5]{HilleA1}, \cite[\S\,2.1.1]{LorWaad}: 
\begin{equation}
\chi\left(z_{1},z_{2}\right)\stackrel{\mathrm{def}}{=}\left\{ \begin{array}{rcr}
\frac{\left|z_{1}-z_{2}\right|}{\sqrt{\left(1+\left|z_{1}\right|^{2}\right)\left(1+\left|z_{2}\right|^{2}\right)}}\leq1 & \text{for } & z_{1},z_{2}\in\mathbb{C},\\
\frac{1}{\sqrt{\left(1+\left|z_{1}\right|^{2}\right)}}\leq1 & \text{for } & z_{1}\in\mathbb{C},\;z_{2}=\infty,\\
0 & \text{for } & z_{1}=z_{2}=\infty.
\end{array}\right.\label{eq:Moeb1b}
\end{equation}
We call a sequence $\left\{ z_{k}\right\} $ \emph{spherically convergent}
if it converges in the chordal distance $\chi$ defined by equation
\eqref{eq:Moeb1b}. The extended complex plane $\hat{\mathbb{C}}$
is a complete metric space with respect to the chordal metric~\eqref{eq:Moeb1b}.
The cross-ratio identity~\eqref{eq:crosrat2d} satisfied by any Mobius
transformation $T$ and formula~\eqref{eq:Moeb1b} for $\chi$ readily
imply, \cite[\S\,2.1.1]{LorWaad}: 
\begin{equation}
\frac{\chi\left(T\left(u\right),T\left(z\right)\right)}{\chi\left(T\left(u\right),T\left(w\right)\right)}\cdot\frac{\chi\left(T\left(v\right),T\left(w\right)\right)}{\chi\left(T\left(v\right),T\left(z\right)\right)}=\frac{\chi\left(u,z\right)}{\chi\left(u,w\right)}\cdot\frac{\chi\left(v,w\right)}{\chi\left(v,z\right)}.\label{eq:Moeb1ba}
\end{equation}

Let us denote the set of all Mobius transformations by $\mathbb{M}$.
The composition of any two Mobius transformations is a Mobius transformation,
so the set $\mathbb{M}$ is naturally a group.

There is a strong connection between Mobius transformations and matrices.
More precisely, let $\mathbb{GL}\left(2,\mathbb{C}\right)$ be the
\emph{general linear group} of $2\times2$ complex matrices 
\begin{equation}
M=\left[\begin{array}{cc}
a & b\\
c & d
\end{array}\right],\quad\det\left\{ M\right\} =ad-bc\neq0.\label{eq:Moeb1c}
\end{equation}
For any such matrix $M\in\mathbb{GL}\left(2,\mathbb{C}\right)$ we
define the corresponding Mobius transformation 
\begin{equation}
T_{M}=\frac{az+b}{cz+d}.\label{eq:Moeb1d}
\end{equation}
Then the following identities hold for these transformations, \cite[\S\,2.1]{JonSin},
\cite[\S\,3.6.1]{Need}, \cite[\S\,7.3]{SimBCA2A}: 
\begin{equation}
T_{M_{1}}\circ T_{M_{2}}\left(z\right)\stackrel{\mathrm{def}}{=}T_{M_{1}}\left(T_{M_{2}}\left(z\right)\right)=T_{M_{1}M_{2}}\left(z\right),\quad T_{M}^{-1}=T_{M^{-1}}.\label{eq:Moeb1e}
\end{equation}
Consequently, $T:\mathbb{GL}\left(2,\mathbb{C}\right)\rightarrow\mathbb{M}$
is a group homomorphism. Note that the mapping $T:M\rightarrow T_{M}$
is evidently not one-to-one since for any nonzero complex number $\lambda$
we have $T_{M}=T_{\lambda M}$ indicating that matrices $M$ and $\lambda M$
correspond to one and the same Mobius transformation. In fact, the
kernel of the mapping $T:\mathbb{GL}\left(2,\mathbb{C}\right)\rightarrow\mathbb{M}$
is exactly the set of all matrices $\lambda\mathbb{I}$ where $\mathbb{I}$
is the identity matrix and $\lambda\in\mathbb{C}\setminus\left\{ 0\right\} $,
\cite[\S\,2.1]{JonSin}. In view of that, it is natural to introduce
the special linear group of normalized matrices, namely 
\begin{equation}
\mathbb{SL}\left(2,\mathbb{C}\right)=\left\{ M\in\mathbb{GL}\left(2,\mathbb{C}\right):\det\left\{ M\right\} =ad-bc=1\right\} .\label{eq:Moeb1f}
\end{equation}
Now the correspondence between $\mathbb{M}$ and $\mathbb{SL}\left(2,\mathbb{C}\right)$
is ``almost'' one-to-one up to the sign, namely if $M_{1},M_{2}\in\mathbb{SL}\left(2,\mathbb{C}\right)$
\begin{equation}
T_{M_{1}}=T_{M_{2}}\Rightarrow M_{1}=\pm M_{2}.\label{eq:Moeb1g}
\end{equation}
It is often convenient to assume that the matrix $M$ associated with
the Mobius transformations satisfies 
\begin{equation}
\det\left\{ M\right\} =ad-bc=1,\label{eq:Moeb1h}
\end{equation}
and then we refer to $M$ as a \emph{normalized matrix}.

The following statement describes an important geometric property
of Mobius transformations, \cite[\S\,3.1-3.2]{HilleA1}, \cite[\S\,I.4]{FordA},
\cite[\S\,2.4]{JonSin}, \cite[\S\,7.3]{SimBCA2A}. 
\begin{thm}[circles and straight lines]
\index{Mobius transformation!circle-line preservation} \label{thm:Moeb1}A
Mobius transformation maps the family $F$ of straight lines and circles
in the plane onto itself. If $L_{1}$ and $L_{2}$ are any two elements
of $F$, then there exists a Mobius transformation that maps $L_{1}$
onto $L_{2}$. 
\end{thm}

We say that $\zeta_{0}$ is a \emph{fixed point} of the Mobius transformation
$T$ defined by~\eqref{eq:Moeb1a} if $z=\zeta_{0}$ solves the equation
\begin{equation}
T\left(z\right)=\frac{az+b}{cz+d}=z,\quad ad-bc\neq0,\label{eq:Moeb2a}
\end{equation}
that is 
\begin{equation}
cz^{2}+\left(d-a\right)z-b=0.\label{eq:Moeb2b}
\end{equation}
Consequently, transformation $T$ always has two fixed points $\zeta_{1}$
and $\zeta_{2}$ in the extended complex plane $\hat{\mathbb{C}}=\mathbb{C}\cup\left\{ \infty\right\} $,
namely, \cite[\S\,3.5.4]{Need} 
\begin{equation}
\zeta_{1},\zeta_{2}=\frac{a-d\pm\sqrt{D}}{2c},\quad D=\left(d-a\right)^{2}+4bc,\label{eq:Moeb2c}
\end{equation}
which may coincide if the \emph{discriminant\index{discriminant}}
$D$ in equation~\eqref{eq:Moeb2c} is zero.

\subsection{Normal form}

The fixed points of a Mobius transformation turn out to be important
geometric characteristics. One can use them to parametrize Mobius
transformation as follows, \cite[\S\,3.2]{HilleA1}, \cite[\S\,1.6]{FordA},
\cite[\S\,3.7]{Need}. Suppose that $c\neq0$ and that fixed points
$\zeta_{1}$ and $\zeta_{2}$ of $T$ are finite and distinct. Then
the transformation $T\left(z\right)$ defined by~\eqref{eq:Moeb1a}
can be recast in terms of $\mu$, $\zeta_{1}$ and $\zeta_{2}$ viewed
as parameters as follows: 
\begin{equation}
T\left(z\right)=w:\quad\frac{w-\zeta_{1}}{w-\zeta_{2}}=\mu\frac{z-\zeta_{1}}{z-\zeta_{2}},\quad\mu=\frac{a-c\zeta_{1}}{a-c\zeta_{2}},\quad T=T\left(\mu;\zeta_{1},\zeta_{2}\right).\label{eq:Moeb2d}
\end{equation}
Representation~\eqref{eq:Moeb2d} is often called the \emph{normal
form of the Mobius transformation}, \cite[\S\,3.7.3]{Need}. Parameter
$\mu$ defined by the last equation in~\eqref{eq:Moeb2d} is called
\emph{multiplier} of the transformation. It is straightforward to
verify that multiplier $\mu$ satisfies the following identities,
\cite[\S\,1.6]{FordA}, \cite[\S\,3.7.1, 3.7.5]{Need}: 
\begin{equation}
\mu+\frac{1}{\mu}=\frac{\left(a+d\right)^{2}-2\left(ad-bc\right)}{ad-bc}=\left(a+d\right)^{2}-2,\text{ if }ad-bc=1.\label{eq:Moeb2e}
\end{equation}
\begin{equation}
\sqrt{\mu}+\frac{1}{\sqrt{\mu}}=a+d,\text{ if }ad-bc=1.\label{eq:Moeb2f}
\end{equation}

Equation~\eqref{eq:Moeb2d} suggests introducing the following auxiliary
Mobius transformation 
\begin{equation}
G\left(z\right)=\frac{z-\zeta_{1}}{z-\zeta_{2}}\Rightarrow G\left(\zeta_{1}\right)=0,\quad G\left(\zeta_{2}\right)=\infty.\label{eq:fMoeb2g}
\end{equation}
Using this Mobius transformation we can readily recast the normal
form~\eqref{eq:Moeb2d} of Mobius transformation $T\left(z\right)$
as follows 
\begin{equation}
T\left(z\right)=w:G\left(w\right)=\mu G\left(z\right),\quad G\left(z\right)=\frac{z-\zeta_{1}}{z-\zeta_{2}}.\label{eq:Moeb2da}
\end{equation}
Then transformation $T\left(\mu;\zeta_{1},\zeta_{2}\right)$ defined
by equations~\eqref{eq:Moeb2d},~\eqref{eq:Moeb2da} can be recast
into the following \emph{normal form}, \cite[\S\,3.2]{HilleA1}, \cite[\S\,1.6]{FordA},
\cite[\S\,3.7.1, 3.7.5]{Need} 
\begin{equation}
T\left(z\right)=T\left(\mu;\zeta_{1},\zeta_{2}\right)\left(z\right)=\left[G^{-1}\circ U_{\mu}\circ G\right]\left(z\right),\quad G\left(z\right)=\frac{z-\zeta_{1}}{z-\zeta_{2}},\label{eq:Moeb2h}
\end{equation}
where the complex-valued multiplier $\mu$ and the corresponding Mobius
transformation $U_{\mu}$ are defined as follows: 
\begin{equation}
\mu=G\left(\frac{a}{c}\right)=\frac{a-c\zeta_{1}}{a-c\zeta_{2}},\quad U_{\mu}:z\rightarrow\mu z.\label{eq:Moeb2hmu}
\end{equation}

In view of equations~\eqref{eq:Moeb2h} the Mobius transformations
$T\left(\mu;\zeta_{1},\zeta_{2}\right)$ and $U_{\mu}:z\rightarrow\mu z$
are \emph{conjugate}. Note that transformation $T\left(z\right)$
defined by~\eqref{eq:Moeb1a} or alternatively by~\eqref{eq:Moeb2d}
carries points $\zeta_{1}$, $\zeta_{2}$ and $\infty$ to $\zeta_{1}$,
$\zeta_{2}$ and $\frac{a}{c}$ respectively. Note also that equations~\eqref{eq:Moeb2d}
and~\eqref{eq:Moeb2h} for the Mobius transformation $T\left(z\right)$
readily imply, \cite[\S\,3.7.3]{Need}, \cite[\S\,3.2]{JonThr}: 
\begin{multline}
\frac{dT\left(z\right)}{dz}=\frac{\mu\left(\zeta_{2}-\zeta_{1}\right)^{2}}{\left[\zeta_{2}-z+\mu\left(z-\zeta_{1}\right)\right]^{2}}\\
\Rightarrow\left.\frac{dT\left(z\right)}{dz}\right|_{z=\zeta_{1}}=\mu,\quad\left.\frac{dT\left(z\right)}{dz}\right|_{z=\zeta_{2}}=\frac{1}{\mu},\quad\mu=\frac{a-c\zeta_{1}}{a-c\zeta_{2}}.\label{eq:Moeb2i}
\end{multline}
The association of the multiplier and its reciprocal with the two
fixed points implied by these derivatives is developed in Remark~\ref{rem:mulfix}
below.

In the case when the discriminant $D$ in equation~\eqref{eq:Moeb2c}
is zero the Mobius transformation defined by~\eqref{eq:Moeb2a} has
a single fixed point $\zeta_{0}$. Then the Mobius transformation
can be recast into the following normalized form, \cite[\S\,1.6]{FordA}
\begin{gather}
T\left(c;\zeta_{0}\right)\left(z\right)=w:\quad\frac{1}{w-\zeta_{0}}=\frac{1}{z-\zeta_{0}}\pm c,\nonumber \\
\text{if }ad-bc=1,\quad a+d=\pm2,\quad c\neq0,\label{eq:Moeb2ha}
\end{gather}
which in turn can be recast as follows 
\begin{equation}
T\left(z\right)=T\left(c;\zeta_{0}\right)\left(z\right)=\left[G^{-1}\circ\left(z\pm c\right)\circ G\right]\left(z\right),\quad G\left(z\right)=\frac{1}{z-\zeta_{0}},\quad c\neq0,\label{eq:Moeb2hb}
\end{equation}
where evidently transformation $G\left(z\right)$ carries $\zeta_{0}$
to $0$. In the case when $c=0$ and $T\left(z\right)\neq\pm z$ we
have 
\begin{equation}
T\left(c;\zeta_{0}\right)\left(z\right)=z\pm b,\quad\text{ if }ad-bc=1,\quad a=d=\pm1,\quad b\neq0.\label{eq:Moeb2hc}
\end{equation}

Recall that if two transformations $T$ and $U$ are related by $T=G^{-1}\circ U\circ G$
where $G$ is yet another transformation we call $T$ and $U$ \emph{conjugate}.
Conjugacy is an equivalence relation on the group $\mathbb{M}$, and
the equivalence classes are called \emph{conjugacy classes}. The following
general statement holds for the conjugacy classes of Mobius transformations,
\cite[\S\,2.9]{JonSin}, \cite[\S\,3.7.4, 3.7.5, 3.7.6]{Need}, \cite[\S\,7.3]{SimBCA2A},
\cite[\S\,3.2]{JonThr}. 
\begin{thm}[conjugacy classes]
\index{Mobius transformation!conjugacy classes} \label{thm:conj}
If $T$ is a non-identity Mobius transformation, then there exist
some $\mu\in\mathbb{C}\setminus\left\{ 0\right\} $ such that $T$
is conjugate to 
\begin{equation}
U_{\mu}\left(z\right)=\left\{ \begin{array}{rcr}
\mu z & \text{if } & \mu\neq0,1,\\
z+1 & \text{if } & \mu=1.
\end{array}\right.,\label{eq:Moeb3ab}
\end{equation}
Transformations $U_{\mu}\left(z\right)$ correspond to the following
normalized matrices $M_{\mu}$ from $\mathbb{SL}\left(2,\mathbb{C}\right)$,
\cite[\S\,3.7.5]{Need}: 
\begin{equation}
M_{\mu}=\pm\left[\begin{array}{rr}
\sqrt{\mu} & 0\\
0 & \frac{1}{\sqrt{\mu}}
\end{array}\right],\quad\mathrm{Tr}\,\left\{ M_{\mu}\right\} =\pm\left(\sqrt{\mu}+\frac{1}{\sqrt{\mu}}\right),\quad\mu\neq0,1;\label{eq:Moeb3ac}
\end{equation}
\begin{equation}
z\rightarrow z+1:\quad M_{1}=\pm\left[\begin{array}{rr}
1 & 1\\
0 & 1
\end{array}\right],\quad\mathrm{Tr}\,\left\{ M_{1}\right\} =2.\label{eq:Moeb3ad}
\end{equation}
\end{thm}

\subsection{Geometric classification}

\label{subsec:geomclas}

Geometric classification of Mobius transformations is based on their
normal form~\eqref{eq:Moeb2d} and Theorem \ref{thm:conj} describing
all possible conjugacy classes of Mobius transformations, \cite[\S\,3.2]{HilleA1},
\cite[\S\,I.7-I.9]{FordA}, \cite[\S\,2.10]{JonSin}, \cite[\S\,3.7]{Need},
\cite[\S\,7.3]{SimBCA2A}, \cite[\S\,3.2]{JonThr}.

Suppose that the fixed points $\zeta_{1}$ and $\zeta_{2}$ in the
normalized form~\eqref{eq:Moeb2d} of a Mobius transformation $T\left(z\right)$
are distinct. Then using equations~\eqref{eq:Moeb2h} we readily
obtain 
\begin{equation}
T\left(\mu;\zeta_{1},\zeta_{2}\right)\circ T\left(\nu;\zeta_{1},\zeta_{2}\right)=T\left(\mu\nu;\zeta_{1},\zeta_{2}\right),\quad T\left(1;\zeta_{1},\zeta_{2}\right)\left(z\right)\equiv z,\label{eq:Moeb3a}
\end{equation}
showing that the transformations $T\left(\mu;\zeta_{1},\zeta_{2}\right)$
when $\zeta_{1}$ and $\zeta_{2}$ are fixed and the parameter $\mu$
runs over the complex plane $\mathbb{C}$ form a one-parameter group
$\mathbb{GL}\left(\zeta_{1},\zeta_{2}\right)$ with respect to $\mu$.
In particular, it follows from~\eqref{eq:Moeb3a} that for any integer
$n$ we have 
\begin{gather}
T^{n}\left(z\right)=\left[G^{-1}U_{\mu}^{n}G\right]\left(z\right),\quad n=0,\pm1,\pm2,\ldots,\nonumber \\
U_{\mu}\left(z\right)=\mu z,\qquad G\left(z\right)=\frac{z-\zeta_{1}}{z-\zeta_{2}}.\label{eq:Moeb3aa}
\end{gather}
Equations~\eqref{eq:Moeb3aa} readily imply that for any $z\neq\zeta_{1},\zeta_{2}$
the following statements hold, \cite[Thm.~3.3]{JonThr}: 
\begin{gather}
\text{if }\left|\mu\right|>1\text{ then}:\lim_{n\rightarrow+\infty}T^{n}\left(z\right)=\zeta_{2},\quad\lim_{n\rightarrow-\infty}T^{n}\left(z\right)=\zeta_{1},\label{eq:Moeb5a}\\
\text{if }\left|\mu\right|<1\text{ then}:\lim_{n\rightarrow+\infty}T^{n}\left(z\right)=\zeta_{1},\quad\lim_{n\rightarrow-\infty}T^{n}\left(z\right)=\zeta_{2}.\label{eq:Moeb5b}
\end{gather}
Equations~\eqref{eq:Moeb5a} and~\eqref{eq:Moeb5b} suggest that
if $\left|\mu\right|>1$ and we apply the Mobius transformation $T$
repeatedly to any point $z\neq\zeta_{1},\zeta_{2}$, then $T^{n}\left(z\right)$
moves away from the fixed point $\zeta_{1}$ and approaches the fixed
point $\zeta_{2}$ exponentially fast. We refer then to $\zeta_{1}$
as the \emph{repulsive fixed point} and $\zeta_{2}$ as the \emph{attractive
fixed point}. If $\left|\mu\right|<1$ the roles of $\zeta_{1}$ and
$\zeta_{2}$ are reversed, \cite[\S\,3.7.2]{Need}, \cite[\S\,3.2]{JonThr}. 
\begin{rem}[Association of multipliers with fixed points]
\label{rem:mulfix} Equations~\eqref{eq:Moeb2i} imply that for
the transformation $T$ defined by equation~\eqref{eq:Moeb2d} its
multiplier $\mu=\frac{a-c\zeta_{1}}{a-c\zeta_{2}}$, as in equations
\eqref{eq:Moeb2hmu}, is associated with fixed point $\zeta_{1}$
whereas its reciprocal $\frac{1}{\mu}$ is associated with fixed point
$\zeta_{2}$. This is consistent with the prior statements that if
$\left|\mu\right|>1$, then $\zeta_{1}$ is a repulsive fixed point
and $\zeta_{2}$ is an attractive one. If $\left|\mu\right|<1$ the
roles of $\zeta_{1}$ and $\zeta_{2}$ are reversed. 
\end{rem}

The one-parameter group $\mathbb{GL}\left(\zeta_{1},\zeta_{2}\right)$
associated with representations~\eqref{eq:Moeb3a} has important
subgroups which are as follows, \cite[\S\,3.2]{HilleA1}.

\medskip{}
\emph{Loxodromic transformations.}

\noindent Let us introduce a multiplicative one-parameter subgroup
of the complex plane $\mathbb{C}$ defined by the following set $\left\{ \mu\left(\sigma\right)\right\} $
of multipliers, \cite[\S\,3.2]{HilleA1}: 
\begin{equation}
\mu^{\sigma}=e^{\zeta\sigma}=e^{\alpha\sigma}\left[\cos\left(\beta\sigma\right)+\mathrm{i}\sin\left(\beta\sigma\right)\right],\;\;\mu=e^{\zeta},\;\;\zeta=\alpha+\mathrm{i}\beta,\;\;\alpha,\beta,\sigma\in\mathbb{R},\label{eq:Moeb3b}
\end{equation}
where $\zeta\in\mathbb{C}$ and $\alpha,\beta\in\mathbb{R}$ are fixed,
whereas real-valued $\sigma$ varies over $\left(-\infty,\infty\right)$.
Let us introduce also the following Mobius transformations 
\begin{equation}
S\left(\sigma;\alpha,\beta;\zeta_{1},\zeta_{2}\right)\stackrel{\mathrm{def}}{=}T\left(\exp\left\{ \left[\alpha+\mathrm{i}\beta\right]\sigma\right\} ;\zeta_{1},\zeta_{2}\right),\label{eq:Moeb3c}
\end{equation}
where (i) $T$ is defined by equations~\eqref{eq:Moeb2h},~\eqref{eq:Moeb2hmu};
(ii) parameters $\alpha$, $\beta$, $\zeta_{1}$ and $\zeta_{2}$
are fixed whereas $\sigma$ varies over $\left(-\infty,\infty\right)$.
Equation~\eqref{eq:Moeb3a} implies that the transformations $S\left(\sigma;\alpha,\beta;\zeta_{1},\zeta_{2}\right)$
form a one-parameter group, called \emph{loxodromic}, with the law
of composition being 
\begin{equation}
S\left(\sigma;\alpha,\beta;\zeta_{1},\zeta_{2}\right)\circ S\left(\tau;\alpha,\beta;\zeta_{1},\zeta_{2}\right)=S\left(\sigma+\tau;\alpha,\beta;\zeta_{1},\zeta_{2}\right).\label{eq:Moeb3d}
\end{equation}

\begin{figure}[htbp]
\begin{centering}
\includegraphics[width=12cm]{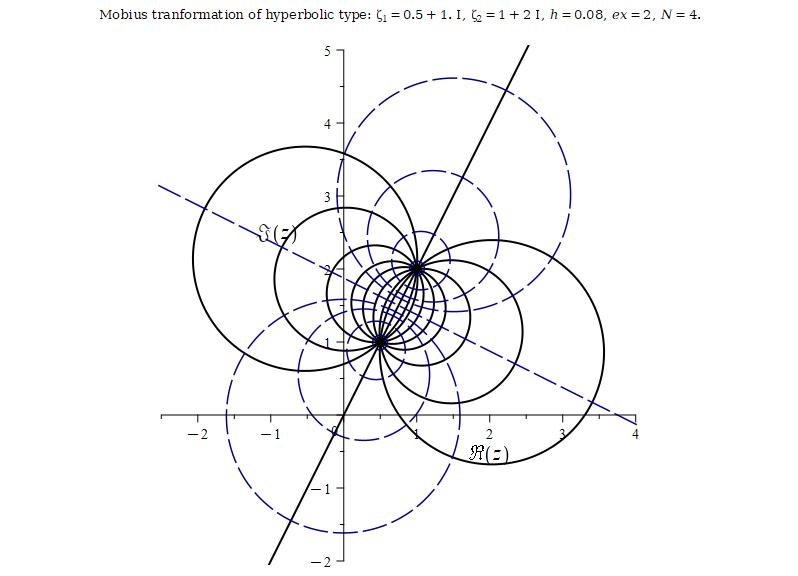} 
\par\end{centering}
\centering{}\caption{Plot of two orthogonal pencils of circles $P_{1}\left(\zeta_{1},\zeta_{2}\right)$
(circles of Apollonius with limit points $\zeta_{1}$ and $\zeta_{2}$)
and $P_{2}\left(\zeta_{1},\zeta_{2}\right)$ defined by equations
\eqref{eq:Moeb2a},~\eqref{eq:Moeb4b} for a Mobius transformation
of hyperbolic type. The limit points $\zeta_{1}$ and $\zeta_{2}$
are the points of intersection of all circles of pencils of circles
$P_{2}\left(\zeta_{1},\zeta_{2}\right)$. They are also the limit
points of the circles of Apollonius associated with pencil $P_{1}\left(\zeta_{1},\zeta_{2}\right)$.}
\label{fig:Mob-hyp} 
\end{figure}

To better understand the properties of the group transformations $T\left(\mu;\zeta_{1},\zeta_{2}\right)$
and $S\left(\sigma;\alpha,\beta;\zeta_{1},\zeta_{2}\right)$ we introduce
two \emph{orthogonal (conjugate) pencils of circles} $P_{1}\left(\zeta_{1},\zeta_{2}\right)$
and $P_{2}\left(\zeta_{1},\zeta_{2}\right)$ defined as follows: 
\begin{equation}
P_{1}:\left\{ C\left(\rho\right):\rho>0\right\} \text{ where }C\left(\rho\right)=\left\{ z:\left|G\left(z\right)\right|=\left|\frac{z-\zeta_{1}}{z-\zeta_{2}}\right|=\rho\right\} ,\label{eq:Moeb4a}
\end{equation}
\begin{gather}
P_{2}:\left\{ \varGamma\left(\theta\right):-\infty<\theta<\infty\right\} ,\nonumber \\
\text{where }\varGamma\left(\theta\right)=\left\{ z:\arg G\left(z\right)=\arg\left\{ \frac{z-\zeta_{1}}{z-\zeta_{2}}\right\} =\theta\right\} .\label{eq:Moeb4b}
\end{gather}
In fact $\varGamma\left(\theta\right)$ is a circular arc from $\zeta_{1}$
to $\zeta_{2}$, and the complementary to it arc is $\varGamma\left(\theta+\pi\right)$.
Evidently $\varGamma\left(\theta\right)=\varGamma\left(\theta+2\pi\right)$
for each $\theta$. The two sets $P_{1}\left(\zeta_{1},\zeta_{2}\right)$
and $P_{2}\left(\zeta_{1},\zeta_{2}\right)$ of circles are orthogonal
to each other, see Fig. \ref{fig:Mob-hyp} . The set of circles $P_{1}\left(\zeta_{1},\zeta_{2}\right)$
is known as the \emph{circles of Apollonius} with limit points $\zeta_{1}$
and $\zeta_{2}$, \cite[\S\,3.8.3]{Need}.

Let us define now a complex number $z$ and introduce its \emph{orbit
under the transformation group} $\left\{ S\left(\sigma\right)\right\} $:
\begin{equation}
w=w\left(\sigma\right)=S\left(\sigma;\alpha,\beta;\zeta_{1},\zeta_{2}\right)\left(z\right),\quad-\infty<\sigma<\infty,\quad w\left(0\right)=z.\label{eq:Moeb4c}
\end{equation}
One can infer from the above formulas that if $z=w\left(0\right)$
lies at the intersection of $C\left(\rho\right)$ and $\varGamma\left(\theta\right)$
where $w\left(\sigma\right)$ lies on the intersection of $C\left(\rho e^{\alpha\sigma}\right)$
with $\varGamma\left(\theta+\beta\sigma\right)$.

There are two important special cases of loxodromic transformation
$\left\{ S\left(\sigma\right)\right\} $ called hyperbolic and elliptic
transformations.

\medskip{}
\emph{Hyperbolic transformations.} A loxodromic transformation $T$
defined by equation~\eqref{eq:Moeb3a} is called \emph{hyperbolic}
if $\beta=0$ and $\alpha\neq0$, that is the transformation multiplier
$\mu=e^{\alpha}>0$ and $\mu\neq1$, \cite[\S\,3.2]{HilleA1}. In
this case the orbit $w\left(\sigma\right)$ coincides with circular
arc $\varGamma\left(\theta\right)$ defined by~\eqref{eq:Moeb4b}.
To understand better how $w\left(\sigma\right)$ moves along circular
arc $\varGamma\left(\theta\right)$ as $\sigma$ varies from $0$
to $\infty$ let us suppose that $\alpha>0$, that is $\mu=e^{\alpha}>1$.
Then as $\sigma$ varies from $0$ to $\infty$ point $w\left(\sigma\right)$
moves along $\varGamma\left(\theta\right)$ from $w\left(0\right)=z$
to $\zeta_{2}$ and when $\sigma$ varies from $0$ to $-\infty$
point $w\left(\sigma\right)$ moves along $\varGamma\left(\theta\right)$
from $w\left(0\right)=z$ to $\zeta_{1}$. Consequently, one may think
of the fixed points $\zeta_{1}$ and $\zeta_{2}$ as \emph{centers
of attraction or repulsion}. For a fixed $\sigma>0$, $w\left(\sigma\right)$
is closer to $\zeta_{2}$ and farther away from $\zeta_{1}$ than
$w\left(0\right)=z$ was, distances being taken along arc $\varGamma\left(\theta\right)$.
Consequently, if $\sigma$ varies from $0$ to $+\infty$ point $w\left(\sigma\right)$
moves from $\zeta_{1}$ toward $\zeta_{2}$, that is fixed point $\zeta_{1}$
\emph{is a center of repulsion} and fixed point $\zeta_{2}$ \emph{is
the center of attraction}. We refer then to $\zeta_{1}$ as \emph{repulsive
fixed point} and $\zeta_{2}$ as \emph{attractive fixed point}. But
if $\sigma$ varies from $0$ to $-\infty$ then point $w\left(\sigma\right)$
moves from $\zeta_{2}$ toward $\zeta_{1}$, that is now $\zeta_{2}$
is a center of repulsion whereas $\zeta_{1}$ becomes the center of
attraction. Hence the designations of fixed points $\zeta_{1}$ and
$\zeta_{2}$ as centers of repulsion and attraction are switched when
$\sigma$ approaches $+\infty$ or $-\infty$.

\noindent Fig. \ref{fig:Mob-hyp} illustrates geometric properties
of a hyperbolic Mobius transformation $T\left(z\right)$. The set
of circles there shown in black solid lines corresponds to pencil
$P_{2}\left(\zeta_{1},\zeta_{2}\right)$. All these circles can be
viewed as orbits of $T\left(z\right)$ passing through limit points
$\zeta_{1}$ and $\zeta_{2}$. Depending on the value of the relevant
multiplier $\mu$ of the hyperbolic Mobius transformation one of the
limit points is the attractive fixed point whereas another one is
the repulsive fixed point. The orthogonal to pencil $P_{2}\left(\zeta_{1},\zeta_{2}\right)$
is another nested set of circles $P_{1}\left(\zeta_{1},\zeta_{2}\right)$
shown in Fig. \ref{fig:Mob-hyp} in dashed blue lines. These circles
are known as the \emph{circles of Apollonius} with limit points $\zeta_{1}$
and $\zeta_{2}$, \cite[\S\,3.8.3]{Need}.

\emph{A Mobius transformation $T\left(z\right)$ is hyperbolic if
and only if it maps each circle of pencil $P_{2}\left(\zeta_{1},\zeta_{2}\right)$
into itself}, \cite[\S\,3.7.2]{Need}. In addition to that a hyperbolic
transformation $T\left(z\right)$ maps each circle of pencil $P_{1}\left(\zeta_{1},\zeta_{2}\right)$
into another circle of this very pencil away from the repulsive fixed
point and closer to the attractive fixed point.\emph{ More precisely,
each Apollonius circle $C\left(\rho\right)$ of pencil $P_{1}\left(\zeta_{1},\zeta_{2}\right)$
is mapped by powers $T^{n}\left(z\right)$ of $T\left(z\right)$ onto
corresponding Apollonius circles $C\left(\mu^{n}\rho\right)$ as described
by formula~\eqref{eq:Moeb4bb}, where $0<\mu\neq1$ is the multiplier
of hyperbolic transformation $T\left(z\right)$.} 

\medskip{}
\emph{Elliptic transformations.} A loxodromic transformation defined
by equation~\eqref{eq:Moeb3a} is called \emph{elliptic} if $\alpha=0$
and $\beta\neq0$, that is the transformation multiplier $\mu=e^{\mathrm{i}\beta}\neq1$
and $\left|\mu\right|=1$, \cite[\S\,3.2]{HilleA1}. In this case
the roles of pencils of circles $P_{1}\left(\zeta_{1},\zeta_{2}\right)$
and $P_{2}\left(\zeta_{1},\zeta_{2}\right)$ compare to the hyperbolic
case are reversed. In the elliptic case $C\left(\rho\right)$ is the
path curve and point $w\left(\sigma\right)$ passes infinitely many
times through the points of $C\left(\rho\right)$ as $\sigma$ varies
from $0$ to $\pm\infty$. In this case one can think of the fixed
points $\zeta_{1}$ and $\zeta_{2}$ as \emph{centers of rotation}.
The same two orthogonal pencils $P_{1}\left(\zeta_{1},\zeta_{2}\right)$
and $P_{2}\left(\zeta_{1},\zeta_{2}\right)$ shown in Fig.~\ref{fig:Mob-hyp}
apply to the elliptic case as well, with their roles interchanged
relative to the hyperbolic case. \emph{A Mobius transformation $T\left(z\right)$
is elliptic if and only if it maps each circle of pencil $P_{1}\left(\zeta_{1},\zeta_{2}\right)$
into itself}, \cite[\S\,3.7.2]{Need}. An elliptic transformation
$T\left(z\right)$ also maps each circle of pencil $P_{2}\left(\zeta_{1},\zeta_{2}\right)$
into another circle of this very pencil.

\medskip{}
\noindent\emph{Parabolic transformations.} The parabolic case is the case of
a Mobius transformation with a single fixed point $\zeta_{0}$ with
the normalized form described by equations~\eqref{eq:Moeb2ha}--\eqref{eq:Moeb2hc}
and Theorem \ref{thm:conj}, \cite[\S\,3.2]{HilleA1}. In this case
$\mu=1$ and the transformation is of the form 
\begin{equation}
T\left(\alpha;\zeta_{0}\right)\left(z\right)=w:\quad\frac{1}{w-\zeta_{0}}=\frac{1}{z-\zeta_{0}}+\alpha,\label{eq:Moebpar1a}
\end{equation}
where $\alpha$ can be any finite complex number. Here is an alternative
form of transformation $T\left(\alpha;\zeta_{0}\right)$: 
\begin{equation}
T\left(\alpha;\zeta_{0}\right)\left(z\right)=\zeta_{0}+\frac{z-\zeta_{0}}{1+\alpha\left(z-\zeta_{0}\right)}.\label{eq:Moebpar1b}
\end{equation}
The normalized form of transformation $T\left(\alpha;\zeta_{0}\right)$
is 
\begin{equation}
T\left(\alpha;\zeta_{0}\right)\left(z\right)=\left[G^{-1}\circ\left(z+\alpha\right)\circ G\right]\left(z\right),\quad G\left(z\right)=\frac{1}{z-\zeta_{0}},\quad\alpha\neq0.\label{eq:Moebpar1c}
\end{equation}
Note that 
\begin{equation}
G\left(\zeta_{0}\right)=\infty.\label{eq:Moebpar1ca}
\end{equation}

\begin{figure}[htbp]
\begin{centering}
\includegraphics[width=12cm]{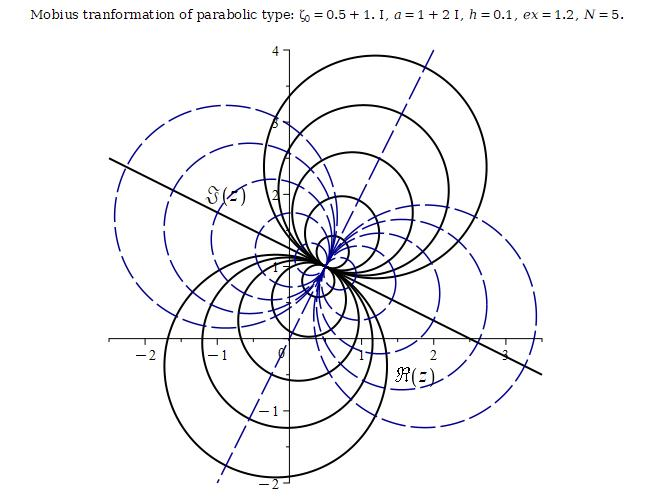} 
\par\end{centering}
\centering{}\caption{The plot of circles in the complex plane associated with a Mobius
transformation of parabolic type. The transformation orbits are a
set of nested circles (black solid lines); an orthogonal set of nested
circles is shown in blue dashed lines. The fixed point is at the intersection
of two straight lines, the solid and the dashed ones.}
\label{fig:Mob-par} 
\end{figure}

\noindent The family of transformations $\left\{ T\left(\alpha;\zeta_{0}\right):\alpha\in\mathbb{C}\right\} $
is a one-parameter transformation group $GP\left(\zeta_{0}\right)$
with the law of composition 
\begin{equation}
T\left(\alpha;\zeta_{0}\right)\circ T\left(\beta;\zeta_{0}\right)=T\left(\alpha+\beta;\zeta_{0}\right),\quad T\left(0;\zeta_{0}\right)\left(z\right)=z,\label{eq:Moebpar1d}
\end{equation}
and $T\left(-\alpha;\zeta_{0}\right)$ is the inverse of $T\left(\alpha;\zeta_{0}\right)$.
Acting similarly to the hyperbolic and elliptic cases, we fix complex
$\alpha\neq0$ and $\zeta_{0}$, and set 
\begin{equation}
w\left(\sigma\right)=\left[T\left(\sigma\alpha;\zeta_{0}\right)\right]\left(z\right),\quad-\infty<\sigma<\infty,\quad w\left(0\right)=z.\label{eq:Moebpar1e}
\end{equation}
The orbit described by $w\left(\sigma\right)$ as $\sigma$ runs from
$-\infty$ to $+\infty$ is a circle passing through $z$ and $\zeta_{0}$,
and its tangent at point $\zeta_{0}$ has direction $-\arg\left\{ \alpha\right\} $.
Note that equations~\eqref{eq:Moebpar1c} readily imply 
\begin{equation}
T\left(\sigma\alpha;\zeta_{0}\right)\left(z\right)=\left[G^{-1}\circ\left(z+\sigma\alpha\right)\circ G\right]\left(z\right),\quad G\left(z\right)=\frac{1}{z-\zeta_{0}},\quad\alpha\neq0,\label{eq:Moebpar2a}
\end{equation}
\begin{equation}
\left[T\left(\alpha;\zeta_{0}\right)\right]^{n}=T\left(n\alpha;\zeta_{0}\right),\quad n=0,\pm1,\pm2,\cdots.\label{eq:Moebpar2b}
\end{equation}
Fig. \ref{fig:Mob-par} illustrates geometric properties of a parabolic
Mobius transformation $T\left(z\right)$ with a single fixed point
$\zeta_{0}$. 

Equations~\eqref{eq:Moebpar1ca} and~\eqref{eq:Moebpar2b} imply
in turn 
\begin{equation}
\lim_{n\rightarrow+\infty}T^{n}\left(z\right)=\lim_{n\rightarrow-\infty}T^{n}\left(z\right)=\zeta_{0},\quad T\left(z\right)=T\left(\sigma\alpha;\zeta_{0}\right)\left(z\right),\quad\alpha\neq0.\label{eq:Moebpar2c}
\end{equation}

\subsection{Classification summary}

The following general statements classifying all four types of Mobius
transformations hold, \cite[\S\,I.7-I.9]{FordA}, \cite[\S\,2.10]{JonSin},
\cite[\S\,3.7.5]{Need}, \cite[\S\,7.3]{SimBCA2A}. 
\begin{thm}[Mobius transformations types]
\index{Mobius transformation!classification} \label{thm:mobtype}
Any Mobius transformation 
\begin{equation}
T\left(z\right)=\frac{az+b}{cz+d},\quad ad-bc=1,\label{eq:Moeb6a}
\end{equation}
is of the type stated below if and only if the following conditions
on $a+d$ hold: 
\begin{equation}
\begin{array}{rcr}
\text{hyperbolic } & \text{if } & a+d\text{ is real and }\left|a+d\right|>2,\\
\text{elliptic } & \text{if } & a+d\text{ is real and }\left|a+d\right|<2,\\
\text{parabolic } & \text{if } & a+d=\pm2,\\
\text{loxodromic } & \text{if } & a+d\text{ is complex. }
\end{array}\label{eq:Moeb6b}
\end{equation}
Consequently, the type of the Mobius transformation~\eqref{eq:Moeb6a}
is completely determined by the value of $a+d$\emph{ provided $ad-bc=1$}. 
\end{thm}

\begin{thm}[{Mobius transformation types and dynamics; see {\cite[\S\,7.3]{SimBCA2A}}}]
\label{thm:mobtype1} Let $T\left(z\right)$ be a Mobius transformation
of the form~\eqref{eq:Moeb6a} with $ad-bc=1$. 
\begin{enumerate}
\item If $T$ is hyperbolic and $a+d=\pm2\cosh\left(\alpha\right)$ with
$\alpha>0$, then $T$ has two fixed points $\zeta_{\pm}$, it is
conjugate to 
\begin{equation}
T_{\alpha}\left(z\right)=\exp\left\{ -2\alpha\right\} z,\label{eq:Moeb7a}
\end{equation}
and 
\begin{equation}
\text{if }z\neq\zeta_{-}:\lim_{n\rightarrow+\infty}T^{n}\left(z\right)=\zeta_{+},\quad\text{if }z\neq\zeta_{+}:\lim_{n\rightarrow-\infty}T^{n}\left(z\right)=\zeta_{-}.\label{eq:Moeb7b}
\end{equation}
\item If $T$ is parabolic, then $T$ has a single fixed point $\zeta_{0}$,
it is conjugate to 
\begin{equation}
U\left(z\right)=z+1,\text{ and for any }z:\lim_{n\rightarrow\pm\infty}T^{n}\left(z\right)=\zeta_{0}.\label{eq:Moeb7c}
\end{equation}
\item If $T$ is elliptic and $a+d=\pm2\cos\left(\beta\right)$ with $0<\beta<\frac{\pi}{2}$,
then $T$ has two fixed points $\zeta_{\pm}$, it is conjugate to
\begin{equation}
T_{\beta}\left(z\right)=\exp\left\{ 2\mathrm{i}\beta\right\} z.\label{eq:Moeb7d}
\end{equation}
If $\frac{\beta}{\pi}$ is rational, the orbit $\left\{ T^{n}\left(z\right):n=0,\pm1,\pm2,\ldots\right\} $
for any fixed $z$ is a finite set. If $\frac{\beta}{\pi}$ is irrational,
then for any fixed $z$ distinct from the fixed points the closure
of the orbit is a circle or straight line. 
\item If $T$ is loxodromic and $a+d=\pm2\cosh\left(\alpha+\mathrm{i}\beta\right)$
with $\alpha>0$ and $0<\beta<\pi$, then $T$ is conjugate to 
\begin{equation}
T_{\beta}\left(z\right)=\exp\left\{ -2\left(\alpha+\mathrm{i}\beta\right)\right\} z,\label{eq:Moeb7e}
\end{equation}
with two fixed points $\zeta_{\pm}$ having the same attracting properties
as in the hyperbolic case. 
\end{enumerate}
\end{thm}

\subsection{The circles of Apollonius and their properties}

\label{subsec:Apollo}

The circles of Apollonius introduced in Section \ref{subsec:geomclas}
are associated with pencil $P_{1}\left(\zeta_{1},\zeta_{2}\right)$,
see Fig. \ref{fig:Mob-hyp}. They are very helpful in quantifying
the action of the relevant Mobius transformation particularly in the
case of hyperbolic transformations. Indeed the normal form~\eqref{eq:Moeb2da}
of Mobius transformation $T$ implies it maps the Apollonius circle
$C\left(\rho\right)$, $\rho>0$ onto another Apollonius circle $C\left(\left|\mu\right|\rho\right)$,
namely 
\begin{gather}
\text{if }\left|\frac{z-\zeta_{1}}{z-\zeta_{2}}\right|=\rho\;\text{ then }\;\left|\frac{T\left(z\right)-\zeta_{1}}{T\left(z\right)-\zeta_{2}}\right|=\rho\left|\mu\right|,\nonumber \\
\text{implying }\;T\!\left[C\left(\rho\right)\right]=C\!\left(\left|\mu\right|\rho\right).\label{eq:Moeb4ba}
\end{gather}
Relations~\eqref{eq:Moeb4ba} combined with~\eqref{eq:Moeb3aa}
yields the following simple formula for the action of powers $T^{n}$
of transformation $T$ applied to Apollonius circles $C\left(\rho\right)$,
$\rho>0$: 
\begin{equation}
T^{n}\left[C\left(\rho\right)\right]=C\left(\left|\mu\right|^{n}\rho\right),\quad n=0,\pm1,\pm2,\ldots.\label{eq:Moeb4bb}
\end{equation}

The radius $r\left(\rho\right)$ of Apollonius circles $C\left(\rho\right)$
and its center $c\left(\rho\right)$ can be found from equations~\eqref{eq:Moeb4ba}
and they are as follows 
\begin{gather}
r\left(\rho\right)=\frac{\rho\left|\zeta_{2}-\zeta_{1}\right|}{\left|\rho^{2}-1\right|},\qquad\rho>0,\nonumber \\
c\left(\rho\right)=\zeta_{1}+\frac{\rho^{2}}{\rho^{2}-1}\left(\zeta_{2}-\zeta_{1}\right)=-\frac{1}{\rho^{2}-1}\zeta_{1}+\frac{\rho^{2}}{\rho^{2}-1}\zeta_{2}.\label{eq:Moeb4bc}
\end{gather}
Equations~\eqref{eq:Moeb4bc} readily imply the following limit relations
\begin{equation}
\lim_{\rho\rightarrow0}r\left(\rho\right)=\lim_{\rho\rightarrow\infty}r\left(\rho\right)=0;\quad\lim_{\rho\rightarrow0}c\left(\rho\right)=\zeta_{1},\quad\lim_{\rho\rightarrow\infty}c\left(\rho\right)=\zeta_{2},\label{eq:Moeb4bd}
\end{equation}
as well as the series expansions: 
\begin{equation}
c\left(\rho\right)=\zeta_{1}-\left(\zeta_{2}-\zeta_{1}\right)\left[\rho^{2}+\rho^{4}+\cdots\right],\quad\rho\rightarrow0,\label{eq:Moeb4be}
\end{equation}
\begin{equation}
c\left(\rho\right)=\zeta_{2}+\left(\zeta_{2}-\zeta_{1}\right)\left[\frac{1}{\rho^{2}}+\frac{1}{\rho^{4}}+\cdots\right],\quad\rho\rightarrow\infty.\label{eq:Moeb4bf}
\end{equation}

\subsection{Eigenvectors and eigenvalues}

\label{subsec:Moebeig}

The relation between the fixed points of a Mobius transformation and
the eigenvectors and eigenvalues of the relevant matrix is described
as follows, \cite[\S\,3.6.3, 3.7.6]{Need}, \cite[\S\,7.3]{SimBCA2A}.
Keeping in mind the relations~\eqref{eq:Moeb1c} and~\eqref{eq:Moeb1d}
between Mobius transformations and $2\times2$ matrices one can verify
the following relation between fixed points of the Mobius transformation
$T=T_{M}$ and eigenvectors and eigenvalues of the corresponding normalized
matrix $M$ ($\det\left\{ M\right\} =1$), \cite[\S\,3.6.3, 3.7.6]{Need},
\cite[\S\,7.3]{SimBCA2A}: 
\begin{equation}
T\left(z\right)=z\Longleftrightarrow\left[\begin{array}{c}
z\\
1
\end{array}\right]\text{ is an eigenvector of matrix }M:T=T_{M}.\label{eq:Moebeig1a}
\end{equation}
In the case when the fixed point of $T$ is infinity, that is $z=\infty$,
the relation~\eqref{eq:Moebeig1a} is interpreted as 
\begin{equation}
T\left(\infty\right)=\infty\Longleftrightarrow\left[\begin{array}{c}
1\\
0
\end{array}\right]\text{ is an eigenvector of matrix }M:T=T_{M}.\label{eq:Moebeig1b}
\end{equation}
Let us assume that the Mobius transformation has the normal form~\eqref{eq:Moeb2d},
that is 
\begin{equation}
T\left(z\right)=w:\quad\frac{w-\zeta_{1}}{w-\zeta_{2}}=\mu\frac{z-\zeta_{1}}{z-\zeta_{2}}.\label{eq:Moebeig1c}
\end{equation}
where $\zeta_{1}$ and $\zeta_{2}$ are distinct fixed points of $T$
and $\mu$ is its multiplier. Then in accordance with Theorem \ref{thm:conj}
and equations~\eqref{eq:Moeb2h},~\eqref{eq:Moeb2hmu} (see also
Remark \ref{rem:mulfix}) we have

\begin{multline}
e_{1}=\left[\begin{array}{c}
\zeta_{1}\\
1
\end{array}\right],\quad e_{2}=\left[\begin{array}{c}
\zeta_{2}\\
1
\end{array}\right]\text{ are eigenvectors of }M\\
\text{with eigenvalues }\lambda_{1}=\frac{1}{\sqrt{\mu}},\quad\lambda_{2}=\sqrt{\mu}.\label{eq:Moebeig1d}
\end{multline}
Note that if $\left|\mu\right|>1$, then $\left|\lambda_{2}\right|>1>\left|\lambda_{1}\right|$
implying that eigenvector $e_{2}$ is ``asymptotically dominant''
for matrix $M^{n}$ as $n\rightarrow+\infty$ which is consistent
with Remark \ref{rem:mulfix} stating that in this case $\zeta_{2}$
is an attractive fixed point. If $\left|\mu\right|<1$ then $e_{1}$
is ``asymptotically dominant'' for matrix $M^{n}$ as $n\rightarrow+\infty$
and $\zeta_{1}$ is an attractive fixed point.

\subsection{Cross-ratios}

\label{subsec:crosrat}

If $\left(z,z_{0},z_{1},z_{\infty}\right)$ is a quadruple of distinct
elements of the extended complex plane $\hat{\mathbb{C}}=\mathbb{C}\cup\left\{ \infty\right\} $
we can consider the following quantity called the \emph{cross-ratio},
\cite[\S\,2.5]{JonSin}, \cite[\S\,7.3]{SimBCA2A}, \cite[\S\,3.5.6]{Need}
\begin{equation}
\left[z,z_{0},z_{1},z_{\infty}\right]\stackrel{\mathrm{def}}{=}\frac{z-z_{0}}{z-z_{\infty}}\frac{z_{1}-z_{\infty}}{z_{1}-z_{0}}.\label{eq:crosrat1a}
\end{equation}
The fundamental importance of the cross-ratio~\eqref{eq:crosrat1a}
as well as the particular choice of indexing of elements $z$ are
explained by the following statements, \cite[\S\,2.5]{JonSin}, \cite[\S\,7.3]{SimBCA2A},
\cite[\S\,3.5.6]{Need}. 
\begin{thm}[cross-ratio: mapping a triple to $(0,1,\infty)$]
\label{thm:crosrat1}If $\left(z_{0},z_{1},z_{\infty}\right)$ is
a triple of distinct elements of the extended complex plane $\hat{\mathbb{C}}$,
then there exists a unique Mobius transformation $T_{z_{0},z_{1},z_{\infty}}\in PGL\left(2,\mathbb{C}\right)$
such that 
\begin{equation}
T_{z_{0},z_{1},z_{\infty}}\left(z_{0}\right)=0,\quad T_{z_{0},z_{1},z_{\infty}}\left(z_{1}\right)=1,\quad T_{z_{0},z_{1},z_{\infty}}\left(z_{\infty}\right)=\infty,\label{eq:crosrat1b}
\end{equation}
and this transformation is represented by the cross-ratio~\eqref{eq:crosrat1a},
that is 
\begin{equation}
T_{z_{0},z_{1},z_{\infty}}\left(z\right)=\left[z,z_{0},z_{1},z_{\infty}\right]=\frac{z-z_{0}}{z-z_{\infty}}\frac{z_{1}-z_{\infty}}{z_{1}-z_{0}}.\label{eq:crosrat1c}
\end{equation}
Note that the transformation $w=T_{z_{0},z_{1},z_{\infty}}\left(z\right)$
maps the oriented circle $C$ through the three points $z_{0},z_{1},z_{\infty}$
to the real axis $\Im\left\{ w\right\} =0$, and consequently the
point $z$ lies on the circle $C$ if and only if $\Im\left\{ \left[z,z_{0},z_{1},z_{\infty}\right]\right\} =0$. 
\end{thm}

\begin{thm}[{cross-ratio invariance and triple-mapping uniqueness; see {\cite[\S\,3.5.6]{Need}}}]
\index{cross-ratio} \label{thm:crosrat2}If $\left(z_{1},z_{2},z_{3}\right)$
and $\left(w_{1},w_{2},w_{3}\right)$ are triples of distinct points
in $\hat{\mathbb{C}}$, then there exists a unique $T\in PGL\left(2,\mathbb{C}\right)$
such that 
\begin{equation}
T\left(z_{1}\right)=w_{1},\quad T\left(z_{2}\right)=w_{2},\quad T\left(z_{3}\right)=w_{3}.\label{eq:crosrat2a}
\end{equation}
Let $\left(z_{0},z_{1},z_{2},z_{3}\right)$ and $\left(w_{0},w_{1},w_{2},w_{3}\right)$
be quadruples of distinct points in $\hat{\mathbb{C}}$. Then there
exists some $T\in PGL\left(2,\mathbb{C}\right)$ with 
\begin{equation}
T\left(z_{j}\right)=w_{j},\quad j=0,1,2,3,\label{eq:crosrat2b}
\end{equation}
if and only if 
\begin{equation}
\left[z_{0},z_{1},z_{2},z_{3}\right]=\left[w_{0},w_{1},w_{2},w_{3}\right].\label{eq:crosrat2c}
\end{equation}
Consequently, for any triple $\left(z_{1},z_{2},z_{3}\right)$ of
distinct points in $\hat{\mathbb{C}}$ and any $T\in PGL\left(2,\mathbb{C}\right)$
we have for any $z$ 
\begin{equation}
\left[T\left(z\right),T\left(z_{1}\right),T\left(z_{2}\right),T\left(z_{3}\right)\right]\;=\;\left[z,z_{1},z_{2},z_{3}\right].\label{eq:crosrat2d}
\end{equation}
\end{thm}

\subsection{Connection to the LC circuit analysis}

\label{subsec:MoebConnect}

The Mobius transformation framework of this chapter connects directly
to the instability analysis of the LC circuit in Chapter~\ref{sec:LCInce}.
We summarize the correspondence here for the reader's convenience.

\medskip{}
\emph{The ratio recurrence.} The transfer-matrix recurrence $V_{n+1}=M_{n}V_{n}$
of Section~\ref{subsec:VecForm} induces a Mobius recurrence on the
ratio $z_{n}=A_{2n+1}/A_{2n-1}$: 
\begin{equation}
z_{n+1}=T_{M_{n}}(z_{n}),\qquad T_{M_{n}}(z)=\frac{N_{n}\,z-(2n-1)^{2}/(2n+3)^{2}}{z},\label{eq:TMn-ratio}
\end{equation}
where $N_{n}=2(c-(2n+1)^{2})/[a(2n+3)^{2}]$, which converges asymptotically
to $T_{M_{\infty}}$ given in~\eqref{eq:TMinfz}.

\medskip{}
\noindent\emph{Classification of $T_{M_{\infty}}$.} By Theorem~\ref{thm:mobtype},
the type of $T_{M_{\infty}}$ is determined by its trace $\operatorname{Tr}M_{\infty}=(1+\delta^{2})/\delta$.
Since $(1+\delta^{2})/\delta>2$ for all $0<\delta<1$ (AM--GM inequality,
with equality only at $\delta=1$), $T_{M_{\infty}}$ is always \emph{hyperbolic}
in the interior of the modulation range. Its two fixed points are
$\zeta_{1}=\delta$ (repulsive, multiplier $1/\delta^{2}>1$) and
$\zeta_{2}=1/\delta$ (attractive, multiplier $\delta^{2}<1$).

\medskip{}
\noindent\emph{Generic $c$: dominant and minimal solution\index{continued fraction!minimal solution}s.}
For generic $c$, every solution vector $V_{n}$ is a linear combination
of the dominant solution (ratio $z_{n}\to1/\delta$) and the minimal
solution (ratio $z_{n}\to\delta$), by the Poincaré--Perron\index{Poincaré--Perron theorem}
theorem (Theorem~\ref{thm:PP}). The characteristic roots of the
Ince recurrence are $\zeta_{1}=1/\delta$ and $\zeta_{2}=\delta$,
both real, with $|\zeta_{1}|>|\zeta_{2}|>0$ for all $\delta\in(0,1)$,
and neither depends on $c$. The PP condition therefore holds unconditionally
for all $c$. For generic $c$ the coefficient of the dominant solution
in any initial condition is nonzero, so $z_{n}\to1/\delta$ in every
generic case; the minimal solution is present but swamped. The exceptional
convergence $z_{n}\to\delta$ --- in which the dominant coefficient
is forced to zero --- is enforced by the eigenvalue condition $F_{\mathrm{even/odd}}(c)=0$,
which selects precisely those $c$ at which this occurs.

\medskip{}
\noindent\emph{Parabolic regime (boundary eigenvalue $c=c_{k}^{\pm}$).} At
a boundary eigenvalue the Mobius map induced by the monodromy matrix
$X(\pi)$ degenerates from hyperbolic to \emph{parabolic}: its two
fixed points --- the eigenvector directions of $X(\pi)$ --- merge
into one. By Theorem~\ref{thm:mobtype1}, the orbit under a parabolic
transformation converges to the single fixed point from both directions.
In the monodromy matrix\index{monodromy matrix} this corresponds
to the Jordan form~\eqref{eq:Jordan-EPD}: both Floquet\index{Floquet theory}
multiplier\index{Floquet theory!Floquet multiplier}s equal $-1$,
and the monodromy has a non-trivial nilpotent part. This is the exceptional
point of degeneracy (EPD) --- the boundary between stability and
instability.

\medskip{}
\noindent\emph{Identity regime (even resonance\index{resonance}s).} At even
resonances $c\approx(2k)^{2}$, the coexistence theorem forces period-$\pi$
coexistence (Floquet multiplier $\rho=+1$): the monodromy is $X(\pi)=+\mathbb{I}$,
not a Jordan block\index{Jordan block}. In Mobius terms, the Mobius
map induced by the monodromy is the \emph{identity} --- every point
is a fixed point. This is qualitatively different from the parabolic
regime: there is no EPD, no instability, and the two independent periodic
solutions both survive. The instability interval has zero width for
all $\delta\neq0$.

\section{The discriminant as an entire function: removable resonances}\index{resonance}

\label{app:entire}

\subsection{Motivation}

The MW expansion of the discriminant\index{discriminant} (eq.~\eqref{eq:Delta-MW},
from MW~\cite[Cor.~2.6]{MagWin}) contains terms of the form 
\begin{equation}
\frac{\pi\sin\pi\sqrt{\lambda}}{2\sqrt{\lambda}}\cdot\frac{|g_{n}|^{2}}{\lambda-n^{2}},\label{eq:MW-term}
\end{equation}
one for each Fourier harmonic $g_{n}$ of $Q$. To a reader familiar
with perturbation theory, the factor $(\lambda-n^{2})^{-1}$ is immediately
recognizable as a \emph{resonance denominator}: it signals a near-degeneracy
when the spectral parameter $\lambda$ approaches the square of the
harmonic index $n$. Such denominators are ubiquitous in quantum-mechanical
perturbation theory, where they produce genuine poles and divergences
at resonance.

Yet Magnus and Winkler prove~\cite[Thm.~2.2]{MagWin} that $\Delta(\lambda)$
is an \emph{entire} function of $\lambda$ --- it has no poles, no
branch points, no singularities anywhere in $\mathbb{C}$. The two
facts appear to be in direct conflict. This chapter resolves the conflict,
explains the mechanism by which $\Delta$ remains entire despite the
resonance denominators, and describes the deeper consequences of this
structure for the LC circuit equation.

\subsection{The discriminant is entire}

The proof of entireness is by Picard iteration~\cite[Thm.~2.2]{MagWin}.
Writing $y_{1}$ and $y_{2}$ as uniformly convergent Neumann series
in $Q$, each term $u_{n}$ and $v_{n}'$ in the expansion $\Delta=\sum_{n\geq0}(u_{n}(\pi)+v_{n}'(\pi))$
is an entire function of $\lambda$, and the series converges uniformly
on every bounded set in $\mathbb{C}$. Hence $\Delta$ is entire of
order of growth exactly $\tfrac{1}{2}$~\cite[Thm.~2.2]{MagWin},
meaning $|\Delta(\lambda)|\leq K_{0}\exp(M_{0}|\lambda|^{1/2})$ for
constants $K_{0},M_{0}>0$. This is the sharpest possible order for
a function whose zeros (the spectrum) grow like $n^{2}$.

\subsection{The resonance denominators are removable}

The apparent contradiction resolves as follows. The $O(g^{2})$ term
in the MW expansion is, in full, 
\[
\Delta_{2}(\lambda)=\frac{\pi\sin\pi\sqrt{\lambda}}{2\sqrt{\lambda}}\sum_{n=1}^{\infty}\frac{|g_{n}|^{2}}{\lambda-n^{2}}.
\]
Setting $\omega=\sqrt{\lambda}$, the generic factor in the sum is
\[
\frac{\sin\pi\omega}{\omega^{2}-n^{2}}=\frac{\sin\pi\omega}{(\omega-n)(\omega+n)}.
\]
As $\omega\to n$ (i.e.\ $\lambda\to n^{2}$), L'Hôpital's rule gives
\[
\lim_{\omega\to n}\frac{\sin\pi\omega}{\omega-n}=\pi\cos\pi n=\pi(-1)^{n},
\]
so 
\[
\lim_{\lambda\to n^{2}}\frac{\sin\pi\sqrt{\lambda}}{\lambda-n^{2}}=\frac{\pi(-1)^{n}}{2n}.
\]
The limit is finite and nonzero. Every apparent pole of $\Delta_{2}$
at $\lambda=n^{2}$ is therefore a \emph{removable singularity}: the
zero of $\sin\pi\sqrt{\lambda}$ at each integer $\sqrt{\lambda}=n$
cancels the zero of $\lambda-n^{2}$ exactly. After removing these
singularities, $\Delta_{2}$ is itself entire.

The same cancellation mechanism operates at every order: Magnus and
Winkler show~\cite[Thm.~2.5]{MagWin} that each coefficient $c(l_{1},\ldots,l_{n})$
in the expansion has the form $A(\omega)\cos\pi\omega+B(\omega)\sin\pi\omega$,
where $A$ and $B$ are even rational functions of $\omega$ whose
poles lie at the partial sums $\omega=\pm(l_{r}+\cdots+l_{s})$. These
poles are canceled by the zeros of $\cos\pi\omega$ or $\sin\pi\omega$
at the same points, rendering every individual term $\Delta_{n}$
entire. The full discriminant $\Delta=\sum\Delta_{n}$, being a uniformly
convergent sum of entire functions, is therefore entire.

\subsection{Physical interpretation}

The resonance denominators $(\lambda-n^{2})^{-1}$ in the MW expansion
are not warnings of true singularities but \emph{bookkeeping devices}.
They record which Fourier harmonic $g_{n}$ is responsible for the
opening of the $m=n$ instability zone\index{instability zone}, and
encode how wide that zone is (through the residue of the apparent
pole, which equals the half-width of the zone to leading order). The
cancellation of the pole by $\sin\pi\sqrt{\lambda}$ reflects a deep
constraint: because $\Delta$ must be entire, the perturbation expansion
is \emph{not free} to produce genuine singularities at $\lambda=n^{2}$.
Any singularity that appears at one order must be canceled by the
next.

A useful analogy is the product expansion $\sin\pi z/(\pi z)=\prod_{n=1}^{\infty}(1-z^{2}/n^{2})$:
each factor in the product introduces zeros at $z=\pm n$, but the
zeros conspire to produce an entire function with prescribed growth.
Similarly, the discriminant is a canonical entire function of order
$\tfrac{1}{2}$ whose zeros are exactly the spectrum, and the MW expansion
is the perturbative unfolding of this canonical structure.

\subsection{Consequences for the LC circuit}

For the LC circuit the Fourier coefficients are $g_{n}=\hat{\lambda}\delta^{n}$
(equation~\eqref{eq:gn-LC}), so every harmonic is present. Here
$\hat{\lambda}$ is the MW spectral parameter (eq.~\eqref{eq:c-lMW-relation});
the resonances $\hat{\lambda}=n^{2}$ in MW space correspond to instability
zones near the physical parameter values $c\approx n^{2}(1-\delta^{2})/(1+\delta^{2})$.
The removable-singularity mechanism operates at every resonance $\hat{\lambda}=n^{2}$,
but with a crucial asymmetry between odd and even $n$: 
\begin{itemize}
\item At \emph{odd} resonances $\hat{\lambda}=(2k-1)^{2}$, the contributions
from all harmonics add constructively, driving $\Delta$ through $-2$
and creating a genuine instability zone of finite width $\sim\delta^{2k-1}$. 
\item At \emph{even} resonances $\hat{\lambda}=(2k)^{2}$, the contributions
from all harmonics interfere \emph{destructively}, so that $\Delta$
approaches $+2$ from below but does not cross it. The apparent pole
is canceled not only by the zero of $\sin\pi\sqrt{\hat{\lambda}}$
(which makes the singularity removable) but also by the Ince coexistence
condition (which forces $\lambda_{2k}^{-}=\lambda_{2k}^{+}$, collapsing
the zone to a point). The two cancellation mechanisms --- analytic
and algebraic --- are not independent: both are consequences of the
same structural symmetry of the LC Ince equation\index{Ince equation}
proved in Section~\ref{sec:InceGeneral}. 
\end{itemize}
Thus the entire function property of $\Delta$ and the Ince coexistence
of even zones are two faces of the same phenomenon. The absence of
poles in $\Delta(\hat{\lambda})$ is the global (entire-function)
statement; the coexistence is its local (zone-by-zone) consequence.

\subsection{\texorpdfstring{Polynomial expansion in $\hat{\lambda}$ and $\delta$}{Polynomial
expansion in lambda-hat and delta}}

\label{subsec:poly-expansion}

The analysis of the removable singularities suggests a natural reorganization
of the LC discriminant approximation. Let $\hat{\lambda}$ denote
the MW spectral parameter treated as an independent variable, decoupled
from $\delta$. The connection to the circuit parameters is $\hat{\lambda}=c(1+\delta^{2})/(1-\delta^{2})$
with $c=4\omega_{0}^{2}/\mu^{2}$, but this relation is deferred to
Stage~2 of the analysis. In Stage~1, $\hat{\lambda}$ and $\delta$
are treated as independent.

The Fourier coefficients of the zero-mean perturbation in MW complex
notation are $g_{n}=\hat{\lambda}\delta^{n}$ ($n\geq1$), so $|g_{n}|^{2}=\hat{\lambda}^{2}\delta^{2n}$.
Substituting into~\cite[Cor.~2.6]{MagWin}: 
\begin{gather}
\Delta^{{\rm LC}}(\hat{\lambda},\delta)=2\cos\pi\sqrt{\hat{\lambda}}+\frac{\pi\hat{\lambda}^{3/2}\delta^{2}}{2}\sin\pi\sqrt{\hat{\lambda}}\,\sum_{n=1}^{\infty}\frac{\delta^{2(n-1)}}{\hat{\lambda}-n^{2}}+O(\delta^{4}),\quad\delta\to0.\label{eq:Delta-LC-poly}
\end{gather}
Equivalently, in the universal $\psi$-basis of Definition~\ref{defn:phi-main}
(using eq.~\eqref{eq:Delta-univ-alt} with $g_{n}=\hat{\lambda}\delta^{n}$,
$\varepsilon=1$, $\omega=\sqrt{\hat{\lambda}}$), 
\begin{equation}
\Delta^{{\rm LC}}(\hat{\lambda},\delta)=2\cos\pi\sqrt{\hat{\lambda}}+\frac{\pi\hat{\lambda}}{2\sqrt{2}}\sum_{n=1}^{\infty}(-1)^{n}\delta^{2n}\,\psi_{n}(\sqrt{\hat{\lambda}})+O(\delta^{4}),\quad\delta\to0.\label{eq:Delta-LC-psi}
\end{equation}
Both forms display the same content: equation~\eqref{eq:Delta-LC-poly}
exhibits the MW resonance denominators that get canceled by $\sin\pi\sqrt{\hat{\lambda}}$,
while~\eqref{eq:Delta-LC-psi} makes the entireness of each $O(\delta^{2n})$
contribution manifest, since each $\psi_{n}$ is an entire function
of $\omega=\sqrt{\hat{\lambda}}$ (Definition~\ref{defn:phi-main}).
Truncating at order $\delta^{2N}$, equation~\eqref{eq:Delta-LC-psi}
is an exact polynomial of degree $N$ in $\delta^{2}$ with entire
$\hat{\lambda}$-coefficients; the error is $O(\delta^{2N+2})$ uniformly
on compact subsets of $\mathbb{C}$.

\subsection{Boundary curves from the polynomial discriminant}

The boundary curve\index{boundary curves}s $\hat{\lambda}_{m}^{\pm}(\delta)$
are the solutions of $\Delta^{{\rm LC}}(\hat{\lambda},\delta)=\pm2$
near $\hat{\lambda}=m^{2}$. Since $\Delta_{0}(m^{2})=-2$ for odd
$m$ (and $+2$ for even $m$), the unperturbed equation $\Delta_{0}(\hat{\lambda})=\pm2$
is already satisfied at $\hat{\lambda}=m^{2}$, and the perturbation
opens a zone of width $O(\delta^{m})$ around $\hat{\lambda}=m^{2}$.

For the \emph{first} instability zone ($m=1$), write $\hat{\lambda}=1+h$
and expand in powers of $\delta$. The equation $\Delta^{{\rm LC}}(1+h,\delta)+2=0$
gives, to leading order, 
\begin{equation}
\frac{\pi^{2}}{4}h^{2}-\frac{\pi^{2}}{4}\delta^{2}+O(\delta^{2}h,\delta^{4})=0,\quad\delta\to0.\label{eq:bdry-eq-m1}
\end{equation}
Here the first term arises from $\Delta_{0}(1+h)+2=\frac{\pi^{2}}{4}h^{2}+O(h^{4})$
and the second from the $n=1$ term of~\eqref{eq:Delta-LC-psi},
which at the resonance $\hat{\lambda}=1$ evaluates to $-\frac{\pi\hat{\lambda}}{2\sqrt{2}}\psi_{1}(1)\,\delta^{2}=-\frac{\pi}{2\sqrt{2}}\cdot\frac{\pi}{\sqrt{2}}\,\delta^{2}=-\frac{\pi^{2}}{4}\delta^{2}$
(using $\psi_{1}(1)=\pi/\sqrt{2}$~\eqref{eq:psi-resonance}). Solving
order by order yields the Puiseux expansions 
\begin{equation}
\hat{\lambda}_{1}^{\pm}=1\pm\delta+\frac{7}{8}\delta^{2}\pm\!\left(\frac{93}{128}-\frac{\pi^{2}}{96}\right)\!\delta^{3}+O(\delta^{4}),\quad\delta\to0.\label{eq:lh-bdry-m1}
\end{equation}
with width 
\begin{equation}
\hat{\lambda}_{1}^{+}-\hat{\lambda}_{1}^{-}=2\delta+O(\delta^{3}),\quad\delta\to0.\label{eq:lh-width-m1}
\end{equation}
and center $(\hat{\lambda}_{1}^{+}+\hat{\lambda}_{1}^{-})/2=1+\frac{7}{8}\delta^{2}+O(\delta^{4})$.

\medskip{}
\emph{Stage~2: converting to physical variables.} Using $\hat{\lambda}=c(1+\delta^{2})/(1-\delta^{2})$,
the boundary values $c_{1}^{\pm}$ satisfy $c=\hat{\lambda}(1-\delta^{2})/(1+\delta^{2})$,
which at $\hat{\lambda}=1\pm\delta+O(\delta^{2})$ gives 
\begin{equation}
c_{1}^{\pm}=(1\pm\delta+\tfrac{7}{8}\delta^{2})(1-2\delta^{2}+O(\delta^{4}))=1\pm\delta-\tfrac{9}{8}\delta^{2}+O(\delta^{3}),\quad\delta\to0.\label{eq:c-bdry-from-lh}
\end{equation}
in agreement with the boundary table values of Chapter~\ref{sec:Boundaries}
(after converting $\delta\to\varepsilon/2+O(\varepsilon^{3})$).

\noindent Table~\ref{tab:lh-bdry} compares the polynomial-$\hat{\lambda}$
boundary formulas with the exact numerics and the CF results of Chapter~\ref{sec:Boundaries}
for representative values of $\delta$.

\begin{table}[ht]
\centering %
\begin{tabular}{ccccc}
\toprule 
$\delta$  & $\hat{\lambda}_{1}^{-}$ (poly)  & $\hat{\lambda}_{1}^{-}$ (num.)  & $\hat{\lambda}_{1}^{+}$ (poly)  & $\hat{\lambda}_{1}^{+}$ (num.)\tabularnewline
\midrule 
$0.05$  & $0.9521$  & $0.9521$  & $1.0523$  & $1.0523$\tabularnewline
$0.10$  & $0.9081$  & $0.9080$  & $1.1094$  & $1.1097$\tabularnewline
$0.15$  & $0.8676$  & $0.8671$  & $1.1718$  & $1.1731$\tabularnewline
$0.20$  & $0.8300$  & $0.8292$  & $1.2400$  & $1.2435$\tabularnewline
$0.25$  & $0.7949$  & $0.7938$  & $1.3144$  & $1.3224$\tabularnewline
\bottomrule
\end{tabular}\caption{First instability zone boundaries $\hat{\lambda}_{1}^{\pm}(\delta)$
in $\hat{\lambda}$-space: polynomial expansion~\eqref{eq:lh-bdry-m1}
truncated at $O(\delta^{3})$ vs.\ numerical values. Agreement is
excellent for $\delta\protect\leq0.15$ and degrades gracefully for
larger $\delta$.}
\label{tab:lh-bdry} 
\end{table}

\subsection{Literature}

The entire-function property and its implications are treated in depth
in Eastham~\cite[Chap.~1--2]{Eastham73}, Reed--Simon~\cite[Sec.~XIII.16]{ReeSim4},
and the inverse-spectral work of Trubowitz~\cite[Sec.~1--2]{Trub77}
and McKean--Trubowitz~\cite[\S\S\,1--2]{McKTrub76}. Hochstadt~\cite[Thm.~1]{Hochst65}
gives the sharpest function-theoretic results on the discriminant.
The connection between the Riemann surface of $\sqrt{\Delta(\lambda)^{2}-4}$
and the hyperelliptic geometry underlying Hill's equation\index{Hill equation}
is developed in McKean--van Moerbeke~\cite[Sec.~1]{McKVanM}.

\section{Floquet theory: constructive determination of the periodic Lyapunov
transformation}

\label{app:YS-Floquet}

The Floquet\index{Floquet theory}--Lyapunov theorem guarantees the
existence of the decomposition $X(t,\varepsilon)=F(t,\varepsilon)e^{tK(\varepsilon)}$
but provides no constructive procedure for computing $F(t,\varepsilon)$.
Yakubovich and Starzhinskii~\cite[Ch.~IV]{YakSta1} developed a systematic
method --- which we call the \emph{Yakubovich--Starzhinskii\index{Yakubovich--Starzhinskii series}
(YS) method} --- for computing both $F(t,\varepsilon)$ and $K(\varepsilon)$
as convergent power series in a small parameter $\varepsilon$, with
coefficients determined by explicit Fourier series formulas. To our
knowledge, this is the only systematic constructive procedure for
the periodic Lyapunov transformation available in the literature.
This chapter presents the essential structure of the YS method, following~\cite[Ch.~IV, \S\,4]{YakSta1}.
The application to the LC circuit is in Chapter~\ref{sec:YS-LC}.

\subsection{Setup and the Yakubovich--Starzhinskii series}

\label{subsec:YS-general-setup}

Consider a system with $T$-periodic coefficients analytic in a small
parameter $\varepsilon$: 
\begin{equation}
\mathbf{x}'=\mathbf{A}(t,\varepsilon)\,\mathbf{x},\qquad\mathbf{A}(t,\varepsilon)=\mathbf{C}+\varepsilon\mathbf{B}_{1}(t)+\varepsilon^{2}\mathbf{B}_{2}(t)+\cdots,\label{eq:YS-gen-system}
\end{equation}
where $\mathbf{C}$ is a constant $n\times n$ matrix and $\mathbf{B}_{j}(t)=\mathbf{B}_{j}(t+T)$
are $T$-periodic with absolutely convergent Fourier series. The parameter
$\varepsilon$ may enter linearly ($\mathbf{B}_{j}=0$ for $j\geq2$,
the Mathieu\index{Mathieu equation}/Hill linear case) or as a full
power series (the general case, which covers the exact LC circuit
with $\varepsilon=\delta$). The YS method constructs the Floquet
decomposition $X(t,\varepsilon)=F(t,\varepsilon)e^{tK(\varepsilon)}$
by expanding: 
\begin{align}
F(t,\varepsilon) & =\mathbf{I}+\varepsilon\mathbf{F}_{1}(t)+\varepsilon^{2}\mathbf{F}_{2}(t)+\cdots,\label{eq:YS-F-series}\\
K(\varepsilon) & =\mathbf{K}_{0}+\varepsilon\mathbf{K}_{1}+\varepsilon^{2}\mathbf{K}_{2}+\cdots,\label{eq:YS-K-series}
\end{align}
where each $\mathbf{F}_{j}(t)$ is $T$-periodic and each $\mathbf{K}_{j}$
is a constant matrix; whether $F(0,\varepsilon)=\mathbf{I}$ holds
depends on the normalization convention chosen for the constant parts
of the $\mathbf{F}_{j}$ (\S\,\ref{subsec:YS-conventions}). We
call~\eqref{eq:YS-F-series}--\eqref{eq:YS-K-series} the \emph{Yakubovich--Starzhinskii
(YS) series}.

\subsection{\texorpdfstring{General recursion for $\mathbf{K}_{j}$ and $\mathbf{F}_{j}$}{General
recursion for K-j and F-j}}

\label{subsec:YS-recursion}

After the preliminary transformation that handles the singular case
(see below), the system takes the form~\eqref{eq:YS-gen-system}
with $T$-periodic coefficient matrices 
\begin{equation}
\mathbf{D}_{j}(t)=\mathbf{D}_{j}(t+T),\qquad j=1,2,3,\ldots,\label{eq:YS-Dj}
\end{equation}
which in the nonsingular case coincide with $\mathbf{B}_{j}(t)$,
and in the singular case are the \emph{transformed} coefficient matrices
obtained after the preliminary change of variables (see Nonsingular
and singular cases below). Define the $l$-th order source~\cite[\S\,IV.4.6, eq.~(4.49)]{YakSta1}:
\begin{equation}
\boldsymbol{\Phi}_{l}(t)=\mathbf{D}_{l}(t)+\sum_{j=1}^{l-1}\mathbf{D}_{j}(t)\,\mathbf{F}_{l-j}(t),\qquad l=1,2,3,\ldots\label{eq:YS-source}
\end{equation}
(for the linear case $\mathbf{D}_{j}=\mathbf{B}_{j}\equiv0$ for $j\geq2$;
for the LC circuit $\mathbf{D}_{j}$ is computed from $\mathbf{B}_{j}$
via the preliminary transformation). The exponent and Lyapunov factor
coefficients then satisfy: 
\begin{align}
\mathbf{K}_{l} & =\Bigl[\boldsymbol{\Phi}_{l}(t)-\sum_{j=1}^{l-1}\mathbf{F}_{j}(t)\,\mathbf{K}_{l-j}\Bigr]_{{\rm av}},\label{eq:YS-Kl}\\
\mathbf{F}_{l}' & =\mathbf{K}_{0}\mathbf{F}_{l}-\mathbf{F}_{l}\mathbf{K}_{0}+\boldsymbol{\Phi}_{l}(t)-\sum_{j=1}^{l-1}\mathbf{F}_{j}(t)\,\mathbf{K}_{l-j}-\mathbf{K}_{l},\label{eq:YS-Fl}
\end{align}
where $[\cdot]_{{\rm av}}=T^{-1}\int_{0}^{T}(\cdot)\,\dd t$ denotes
the time average over one period. Equation~\eqref{eq:YS-Kl} is exact
under the zero-mean normalization $[\mathbf{F}_{j}]_{{\rm av}}=\mathbf{0}$
of \S\,\ref{subsec:YS-conventions}, under which the averages of
the commutator term and of the sums $\mathbf{F}_{j}\mathbf{K}_{l-j}$
vanish --- in agreement with the second equality of YS1's recursion~(4.49).
When $\operatorname{ad}_{\mathbf{K}_{0}}=0$\index{adjoint action}
(for the adjoint action $\operatorname{ad}_{\mathbf{K}_{0}}$, its
meaning and role, see \S\,\ref{subsec:YS-singular}) --- e.g.\
when $\mathbf{K}_{0}$ is a multiple of the identity, as in the Hill
singular case of \S\,\ref{subsec:YS-Hill-formulas} --- the commutator
term in~\eqref{eq:YS-Fl} drops and the equation integrates directly
to $\mathbf{F}_{l}(t)=\int_{0}^{t}[\boldsymbol{\Phi}_{l}-\sum_{j}\mathbf{F}_{j}\mathbf{K}_{l-j}-\mathbf{K}_{l}]\,\dd s$;
in general $\mathbf{F}_{l}$ is obtained from the Fourier analysis
of the Sylvester equation~\eqref{eq:YS-Sylvester} below.
\begin{rem}[The averaging operator and the scalar Floquet prototype]
\label{rem:av-scalar} The meaning of $[\cdot]_{{\rm av}}$ and its
role in the recursion are perfectly transparent in the scalar case
of Chapter~\ref{app:genLin}, \S\,\ref{subsec:YS-scalar} (``The
Floquet theory for a scalar''). For the scalar equation $x'=a(t)x$
with $T$-periodic $a(t)$, the Floquet decomposition $x(t)=f(t)e^{kt}$
gives (eq.~\eqref{eq:FLsca2a}--\eqref{eq:FLsca2b}): 
\begin{equation}
k=[a]_{{\rm av}}=\frac{1}{T}\int_{0}^{T}a(s)\,\dd s,\qquad f(t)=\exp\!\int_{0}^{t}(a(s)-k)\,\dd s.\label{eq:scalar-Floquet-av}
\end{equation}
The Floquet exponent\index{Floquet theory!Floquet exponent} $k$
is the \emph{average} of $a(t)$; the periodic factor $f(t)$ integrates
the zero-mean remainder $a(t)-k$. The YS matrix recursion is the
exact matrix generalization of this splitting: $\mathbf{K}_{l}=[\boldsymbol{\Phi}_{l}-\sum_{j}\mathbf{F}_{j}\mathbf{K}_{l-j}]_{{\rm av}}$
extracts the \emph{constant} (DC) part of the source that contributes
to the Floquet exponent matrix $K$, while $\mathbf{F}_{l}(t)$ in~\eqref{eq:YS-Fl}
integrates the zero-mean oscillatory remainder that builds the periodic
Lyapunov factor $F(t,\varepsilon)$. This is the fundamental organizing
principle of the YS method: at every order, averaging isolates the
secular (exponential) behavior from the oscillatory (periodic) behavior,
exactly as in the scalar model~\eqref{eq:scalar-Floquet-av}. Yakubovich
and Starzhinskii make this connection explicit in~\cite[Ch.~IV, \S\S\,4.5--4.6]{YakSta1}. 
\end{rem}

Each $\mathbf{F}_{l}(t)$ is determined as an explicit trigonometric
sum with $\mathbf{F}_{l}(0)=\mathbf{0}$; no truncation of the Fourier
series is made.

\emph{Application to the exact LC circuit.} For the LC Hill equation~\eqref{eq:LC-Hill-firstorder}
with small parameter $\varepsilon=\delta$, the preliminary singular-case
substitution $\mathbf{x}=e^{x\mathbf{C}_{0}}\mathbf{y}$ with $\mathbf{C}_{0}=\bigl[\begin{smallmatrix}0 & 1\\
-1 & 0
\end{smallmatrix}\bigr]$ (so that $e^{x\mathbf{C}_{0}}=\bigl[\begin{smallmatrix}\cos x & \sin x\\
-\sin x & \cos x
\end{smallmatrix}\bigr]$ is exactly $\pi$-periodic) leaves the residual constant matrix $\mathbf{K}_{0}=\mathbf{C}-\mathbf{C}_{0}=\bigl[\begin{smallmatrix}0 & 0\\
-(\hat{\lambda}-1) & 0
\end{smallmatrix}\bigr]$ and transforms $\mathbf{B}_{n}(x)=\hat{\lambda}\tilde{\mathbf{B}}_{n}(x)$
to~\cite[\S\S\,IV.4.2--4.3, eq.~(4.10)]{YakSta1} 
\begin{equation}
\mathbf{D}_{n}(x)=e^{-x\mathbf{C}_{0}}\mathbf{B}_{n}(x)e^{x\mathbf{C}_{0}}=\hat{\lambda}\begin{bmatrix}d_{11}(x) & d_{12}(x)\\[6pt]
d_{21}(x) & -d_{11}(x)
\end{bmatrix},\label{eq:LC-Dn-exact}
\end{equation}
where 
\begin{align*}
d_{11}(x) & =\frac{\sin2(n+1)x-\sin2(n-1)x}{2},\\
d_{12}(x) & =\cos2nx-\frac{\cos2(n-1)x+\cos2(n+1)x}{2},\\
d_{21}(x) & =-\cos2nx-\frac{\cos2(n-1)x+\cos2(n+1)x}{2}.
\end{align*}

\begin{rem}[Choice of $\mathbf{C}_{0}$ and its relation to YS1]
\label{rem:C0-convention} The reduction $\mathbf{x}=e^{x\mathbf{C}_{0}}\mathbf{y}$
of the singular case requires a choice of the constant generator $\mathbf{C}_{0}$,
and the YS1 construction~\cite[\S\S\,IV.4.2--4.3]{YakSta1} leaves
this choice to the user subject to one requirement: $e^{x\mathbf{C}_{0}}$
must be (anti)periodic with the period of the perturbation, so that
the transformed coefficients $\mathbf{D}_{n}(x)=e^{-x\mathbf{C}_{0}}\mathbf{B}_{n}(x)e^{x\mathbf{C}_{0}}$
remain periodic and the residual system is nonsingular. Two choices
are natural for the LC Hill equation, and it is worth stating which
one is in force here to avoid ambiguity. 
\begin{itemize}
\item \emph{The choice used throughout this work} is the $\pi$-periodic
generator $\mathbf{C}_{0}=\bigl[\begin{smallmatrix}0 & 1\\
-1 & 0
\end{smallmatrix}\bigr]$, for which $e^{x\mathbf{C}_{0}}$ has period $\pi$ for \emph{every}
$\hat{\lambda}$. It leaves the constant residual $\mathbf{K}_{0}=\mathbf{C}-\mathbf{C}_{0}=\bigl[\begin{smallmatrix}0 & 0\\
-(\hat{\lambda}-1) & 0
\end{smallmatrix}\bigr]$ (vanishing at the resonance\index{resonance} $\hat{\lambda}=1$),
and the transformed $\mathbf{D}_{n}(x)$ above are $\pi$-periodic
with integer Fourier modes $k=n-1,n,n+1$. This is the choice that
maps directly onto YS1 equations~(4.9)--(4.10): system~(4.9) carries
the residual exponent $\mathbf{K}_{0}$ as its $\varepsilon^{0}$
term, and the recursion $\mathbf{K}_{1}=[\mathbf{D}_{1}]_{\mathrm{av}}$
together with the Sylvester step~\eqref{eq:YS-Sylvester} is applied
with this nonzero $\mathbf{K}_{0}$. It is also the convention used
in the main-text computation of Chapter~\ref{sec:YS-exact-LC} (\S\,\ref{subsubsec:YS-LC-F1F2}). 
\item \emph{The alternative}, $\mathbf{C}_{0}=\mathbf{C}=\bigl[\begin{smallmatrix}0 & 1\\
-\hat{\lambda} & 0
\end{smallmatrix}\bigr]$, absorbs the entire unperturbed matrix, so that $\mathbf{K}_{0}=\mathbf{0}$
and $e^{x\mathbf{C}_{0}}$ has period $2\pi/\sqrt{\hat{\lambda}}$.
This choice is \emph{not} used here: its $e^{x\mathbf{C}_{0}}$ is
not $\pi$-periodic except at $\hat{\lambda}=1$, so the transformed
coefficients are not periodic in $x$ at general $\hat{\lambda}$
and the clean polynomial-in-$\hat{\lambda}$ structure of the $\mathbf{D}_{n}$,
$\mathbf{K}_{m}$, $\mathbf{F}_{m}$ would be lost. 
\end{itemize}
The two choices yield the same instability boundaries (the boundary
curves depend only on the conjugacy class of the monodromy matrix\index{monodromy matrix},
which is invariant under the choice of $\mathbf{C}_{0}$); they differ
only in the bookkeeping of the intermediate series. The companion
question of how the periodic factors $\mathbf{F}_{j}$ themselves
are normalized once $\mathbf{C}_{0}$ is fixed is addressed separately
in \S\,\ref{subsec:YS-conventions}. 
\end{rem}

Each $\mathbf{D}_{n}$ carries \emph{exactly one factor} of $\hat{\lambda}$
(from $\mathbf{B}_{n}$), confirming that $\mathbf{K}_{m}$ and $\mathbf{F}_{m}$
are polynomials in $\hat{\lambda}$ (eq.~\eqref{eq:YS-lam-poly}).
Since $\mathbf{D}_{n}(x,\hat{\lambda})$ is analytic (in fact polynomial)
in $\hat{\lambda}$, Theorem~\ref{thm:joint-analytic} of Chapter~\ref{app:genLin}
applies with parameter vector $z=(\delta,\hat{\lambda})$: the YS
coefficients $\mathbf{K}_{m}(\delta,\hat{\lambda})$ and $\mathbf{F}_{m}(x,\delta,\hat{\lambda})$
are jointly analytic in $(\delta,\hat{\lambda})$ for $|\delta|<r_{0}$
and $\hat{\lambda}$ near any fixed positive value. At primary resonance
($\omega=1$, $n=1$), $\mathbf{D}_{1}$ has Fourier modes $k=0$,
$1$, and $2$ (the $k=0$ term is the constant part that enters the
average $[\mathbf{D}_{1}]_{{\rm av}}$); for general $n\geq2$, $\mathbf{D}_{n}$
has modes $k=n-1$, $n$, and $n+1$. The source $\boldsymbol{\Phi}_{l}$
at each order $l$ is therefore a finite trigonometric polynomial
whose highest Fourier mode is $k=l+1$ (proved inductively in Remark~\ref{rem:YS-freq-conj}
of Chapter~\ref{sec:YS-LC}), and $\mathbf{F}_{l}(x)$, $\mathbf{K}_{l}$
are obtained by finite Fourier operations --- no truncation error.

\emph{$\hat{\lambda}$-linearity of the YS series.} Since every $\mathbf{D}_{n}=\hat{\lambda}\cdot(\hat{\lambda}$-free
matrix), and the recursion~\eqref{eq:YS-Kl}--\eqref{eq:YS-Fl}
is linear in the source $\boldsymbol{\Phi}_{l}$ (which is built from
$\mathbf{D}_{j}$ and previous $\mathbf{F}_{r}$), every $\mathbf{K}_{m}$
and $\mathbf{F}_{m}$ is a polynomial of degree $m$ in $\hat{\lambda}$
(eq.~\eqref{eq:YS-lam-poly}). The instability boundary\index{stability boundary}
width is therefore $L_{1}=\hat{\lambda}\,\ell(\delta)$ where $\ell(\delta)$
is computed from the recursion at $\hat{\lambda}=1$ and $\hat{\lambda}=c(1+\delta^{2})/(1-\delta^{2})$
(with $c=1$ at the primary resonance center) is substituted at the
end. For example, the exact $\mathbf{K}_{1}$ at $\mu_{0}=0$ satisfies
the remarkable identity 
\begin{equation}
\mathbf{K}_{1}[2,1]=\hat{\lambda}\cdot\mathbf{K}_{1}[1,2]\label{eq:K1-lam-identity}
\end{equation}
(verified analytically from~\eqref{eq:LC-Dn-exact} and confirmed
numerically for all $\delta\in(0,1)$), which is a direct manifestation
of the $\hat{\lambda}$-linearity.

\subsection{Nonsingular and singular cases}

\label{subsec:YS-singular}

The determination of the YS series coefficients depends critically
on whether the system is in the \emph{nonsingular} or \emph{singular}
case~\cite[\S\,IV.4.1]{YakSta1}.

\emph{The role of $\mathbf{K}_{0}$: resonance frequencies as eigenvalues
of $\operatorname{ad}_{\mathbf{K}_{0}}$.} The matrix $\mathbf{K}_{0}$
is the Floquet exponent of the \emph{unperturbed} system ($\varepsilon=0$),
whose eigenvalues $\lambda_{j}$ are the characteristic exponents
of $\mathbf{x}'=\mathbf{C}\mathbf{x}$~\cite[\S\,IV.2.1]{YakSta1}.
A central role is played by what in modern Lie-algebraic notation
is the \emph{adjoint action}\index{adjoint action} of $\mathbf{K}_{0}$
on the space $M_{n}(\mathbb{C})$ of $n\times n$ matrices: 
\begin{equation}
\operatorname{ad}_{\mathbf{K}_{0}}\colon\mathbf{W}\mapsto\mathbf{K}_{0}\mathbf{W}-\mathbf{W}\mathbf{K}_{0}.\label{eq:ad-K0}
\end{equation}
If $\mathbf{K}_{0}$ has eigenvalues $\lambda_{1},\ldots,\lambda_{n}$
with corresponding right eigenvectors $v_{j}$ and left eigenvectors
$w_{k}^{\top}$, then $\operatorname{ad}_{\mathbf{K}_{0}}$ acts as
a linear map on the $n^{2}$-dimensional space $M_{n}(\mathbb{C})$
with eigenvalues 
\begin{equation}
\lambda_{j}-\lambda_{k},\qquad j,k=1,\ldots,n,\label{eq:ad-eigenvalues}
\end{equation}
and corresponding rank-one eigenvectors $v_{j}w_{k}^{\top}$ (the
verification $\operatorname{ad}_{\mathbf{K}_{0}}(v_{j}w_{k}^{\top})=(\lambda_{j}-\lambda_{k})\,v_{j}w_{k}^{\top}$
is immediate, and for diagonalizable $\mathbf{K}_{0}$ these $n^{2}$
rank-one matrices form a basis of $M_{n}(\mathbb{C})$). The notation
$\operatorname{ad}_{\mathbf{K}_{0}}$ is ours, not Yakubovich--Starzhinskii's:
YS carry out the equivalent solvability analysis by means of their
operators $S$ and $P$~\cite[\S\,IV.2.2]{YakSta1} and the Fourier
analysis of the Sylvester equation~\eqref{eq:YS-Sylvester} (\cite[\S\,IV.4.5, eq.~(4.31)]{YakSta1}).
The commutator form is adopted here because it makes the resonance
structure most transparent. In the language of Lie theory\index{Lie algebra},
the space $M_{n}(\mathbb{C})$ equipped with the commutator $[\mathbf{A},\mathbf{B}]=\mathbf{A}\mathbf{B}-\mathbf{B}\mathbf{A}$
is the Lie algebra $\mathfrak{gl}(n,\mathbb{C})$ of the general linear
group $GL(n,\mathbb{C})$, and $\operatorname{ad}_{\mathbf{K}_{0}}=[\mathbf{K}_{0},\cdot\,]$
is the value at $\mathbf{K}_{0}$ of the \emph{adjoint representation}\index{adjoint action}
$\operatorname{ad}\colon\mathfrak{gl}(n,\mathbb{C})\to\operatorname{End}(M_{n}(\mathbb{C}))$.
The group acts by conjugation through the adjoint representation $\operatorname{Ad}_{\mathbf{S}}\mathbf{W}=\mathbf{S}\mathbf{W}\mathbf{S}^{-1}$,
and the two are linked by the basic identity $\operatorname{Ad}_{e^{\mathbf{X}}}=e^{\operatorname{ad}_{\mathbf{X}}}$;
in this language the preliminary transformation $\mathbf{D}_{n}=e^{-x\mathbf{C}_{0}}\mathbf{B}_{n}e^{x\mathbf{C}_{0}}$
of \S\,\ref{subsec:YS-recursion} is precisely the group adjoint
action, $\mathbf{D}_{n}=\operatorname{Ad}_{e^{-x\mathbf{C}_{0}}}\mathbf{B}_{n}$.
For the Lie-theoretic background the reader may consult Humphreys~\cite[\S\S\,1.2--1.3]{Humph},
Hall~\cite[Chs.~1--3]{HallLie}, and Knapp~\cite[Ch.~I]{KnappLie}.
\begin{rem}[$\mathbf{K}_{0}$ encodes all parametric resonance\index{parametric resonance}s]
\label{rem:K0-resonance} The differences $\lambda_{j}-\lambda_{k}$
are precisely the \emph{resonance frequencies} of the unperturbed
system. For a single oscillator with natural frequency $\omega_{0}$,
$\mathbf{K}_{0}$ has eigenvalues $\pm i\omega_{0}$, so $\operatorname{ad}_{\mathbf{K}_{0}}$
has eigenvalues $0,+2i\omega_{0},-2i\omega_{0}$. Primary parametric
resonance ($\omega_{{\rm forcing}}=2\omega_{0}$) occurs when the
forcing frequency $2\pi/T$ equals $2\omega_{0}$, i.e., when $2i\omega_{0}=2\pi i\cdot1/T$
--- exactly when $2\pi i/T$ is an eigenvalue of $\operatorname{ad}_{\mathbf{K}_{0}}$.
Thus $\mathbf{K}_{0}$ encodes in a single matrix all information
about which forcing harmonics produce resonance: \emph{order-$m$
parametric resonance occurs if and only if $2\pi im/T$ is an eigenvalue
of $\operatorname{ad}_{\mathbf{K}_{0}}$}. Yakubovich and Starzhinskii~\cite[\S\S\,IV.4.1--4.3]{YakSta1}
make this the organizing principle of their theory: the singularity/non-singularity
of the YS system is completely determined by whether any Fourier frequency
falls in the spectrum of $\operatorname{ad}_{\mathbf{K}_{0}}$ (in
their formulation: whether the eigenvalues of $\mathbf{C}$ are congruent
modulo $2\pi i/T$). The classical resonance condition --- ``modulation
frequency commensurate with natural frequency of the unperturbed system''
--- thus takes the clean algebraic form 
\begin{equation}
\operatorname{spec}\bigl(\operatorname{ad}_{\mathbf{K}_{0}}\bigr)\cap\frac{2\pi i}{T}\mathbb{Z}^{\times}\neq\emptyset.\label{eq:resonance-algebraic}
\end{equation}
The physical mechanism underlying this resonance condition --- the
net work done by the parameter source on the oscillator per period,
and its sign within vs.\ outside instability tongue\index{instability tongue}s
--- is developed in Chapter~\ref{app:ParRes}, equations~\eqref{eq:energy-rate}--\eqref{eq:energy-cycle}. 
\end{rem}

\emph{Nonsingular case.} The system is nonsingular when~\eqref{eq:resonance-algebraic}
fails: 
\begin{equation}
\operatorname{spec}\bigl(\operatorname{ad}_{\mathbf{K}_{0}}\bigr)\cap\frac{2\pi i}{T}\mathbb{Z}^{\times}=\emptyset,\label{eq:nonsingular-cond}
\end{equation}
i.e., no nonzero Fourier frequency $2\pi im/T$ is an eigenvalue of
$\operatorname{ad}_{\mathbf{K}_{0}}$, equivalently the eigenvalues
$\lambda_{j}$ of $\mathbf{C}$ are incongruent modulo $2\pi i/T$~\cite[\S\,IV.4.1]{YakSta1}.
In this case the coefficients $\mathbf{F}_{j}(t)$ and $\mathbf{K}_{j}$
are determined inductively by the linear Sylvester equation~\cite[\S\,IV.4.5, eq.~(4.31)]{YakSta1}:
\begin{equation}
\dot{\mathbf{Z}}=\mathbf{K}_{0}\mathbf{Z}-\mathbf{Z}\mathbf{K}_{0}+\mathbf{\Phi}(t)-\mathbf{L},\label{eq:YS-Sylvester}
\end{equation}
where $\mathbf{Z}(t)=\mathbf{F}_{j}(t)$, $\mathbf{L}=\mathbf{K}_{j}$,
and $\mathbf{\Phi}(t)$ is known from the previous step. Fourier expanding~\eqref{eq:YS-Sylvester},
the $m$-th coefficient $\mathbf{Z}^{(m)}$ satisfies $\bigl(\operatorname{ad}_{\mathbf{K}_{0}}-\frac{2\pi im}{T}\bigr)\mathbf{W}=-\mathbf{\Phi}^{(m)}$,
uniquely solvable precisely when $2\pi im/T\notin\operatorname{spec}(\operatorname{ad}_{\mathbf{K}_{0}})$
--- the condition~\eqref{eq:nonsingular-cond}. The average $\mathbf{K}_{j}=[\mathbf{\Phi}(t)]_{{\rm av}}$
(zeroth Fourier coefficient, $m=0$).

\emph{Singular case (parametric resonance).} The system is singular
when condition~\eqref{eq:nonsingular-cond} fails, i.e., when~\eqref{eq:resonance-algebraic}
holds: 
\begin{equation}
\frac{2\pi i\,p}{T}=\lambda_{j}-\lambda_{k}\quad\text{for some }j\neq k\text{ and }p\in\mathbb{Z}^{\times}.\label{eq:YS-singular-cond}
\end{equation}
The positive integer $p$ is called the \emph{resonance order}: it
labels which harmonic of the forcing is in resonance with the unperturbed
beat frequency $\lambda_{j}-\lambda_{k}$. Order $p=1$ (primary resonance,
the primary instability tongue) means the forcing period $T$ itself
matches the beat period; order $p=m$ means the $m$-th forcing harmonic
is in resonance. This is the same resonance order used in \S\,\ref{subsec:YS-Hill-formulas}
and Chapters~\ref{sec:YS-LC}--\ref{sec:YS-exact-LC} (written there
as $m$ to avoid confusion with Cambi's $p=\omega_{0}/\mu$; see the
note in Chapter~\ref{subsec:YS-Mathieu}). In the singular case the
preliminary substitution $\mathbf{x}=e^{t\mathbf{C}_{0}}\mathbf{y}$,
where $\mathbf{C}_{0}$ is chosen so that $\mathbf{K}_{0}=\mathbf{C}-\mathbf{C}_{0}$
has no resonant Fourier modes~\cite[\S\S\,IV.4.2--4.3]{YakSta1},
reduces the system to the nonsingular case in the new variable $\mathbf{y}$.
The transformed coefficient matrices are $\mathbf{D}_{j}(t)=e^{-t\mathbf{C}_{0}}\mathbf{B}_{j}(t)e^{t\mathbf{C}_{0}}$
~\cite[\S\S\,IV.4.2--4.3, eq.~(4.10)]{YakSta1}, and the YS recursion~\eqref{eq:YS-Kl}--\eqref{eq:YS-Fl}
applies to the transformed system. For the exact LC circuit at primary
resonance, the correct choice is $\mathbf{C}_{0}=\bigl[\begin{smallmatrix}0 & 1\\
-1 & 0
\end{smallmatrix}\bigr]$ (so that $e^{\tau\mathbf{C}_{0}}$ is exactly $\pi$-periodic), with
residual $\mathbf{K}_{0}=\mathbf{C}-\mathbf{C}_{0}=\bigl[\begin{smallmatrix}0 & 0\\
-(\hat{\lambda}-1) & 0
\end{smallmatrix}\bigr]$ (a constant matrix, vanishing at $\hat{\lambda}=1$). The transformed
matrices $\mathbf{D}_{n}$ are then $\pi$-periodic for all $\hat{\lambda}$,
placing the system in the nonsingular case. See \S\,\ref{subsubsec:YS-LC-F1F2}
for the explicit computation.

\subsection{\texorpdfstring{Normalization conventions for $\mathbf{F}_{j}$}{Normalization
conventions for F-j}}

\label{subsec:YS-conventions}

The YS series $F(t,\varepsilon)=\mathbf{I}+\sum_{j=1}^{\infty}\varepsilon^{j}\mathbf{F}_{j}(t)$
determines $\mathbf{F}_{j}$ from the equation $\mathbf{F}_{j}'=\mathbf{K}_{0}\mathbf{F}_{j}-\mathbf{F}_{j}\mathbf{K}_{0}+\boldsymbol{\Phi}_{j}-\mathbf{K}_{j}$
only up to an additive constant matrix (the kernel of the adjoint
action $\operatorname{ad}_{\mathbf{K}_{0}}$ on constants). A normalization
condition must be imposed to fix $\mathbf{F}_{j}$ uniquely. YakSta
use two standard conventions, each optimized for a different situation,
plus a third special prescription for the primary resonance $m=1$
(the $\varphi'_{ij}$ prescription, see below and Chapter~\ref{subsec:YS-Mathieu}).

\medskip{}
\emph{Convention~1 (zero-mean): $[\mathbf{F}_{j}]_{{\rm av}}=\mathbf{0}$.}
In the \emph{singular case} after the preliminary transformation,
YakSta impose~\cite[\S\,IV.4.6, eq.~(4.48)]{YakSta1}: 
\begin{equation}
[\tilde{\mathbf{F}}_{j}(t)]_{{\rm av}}=\frac{1}{T}\int_{0}^{T}\tilde{\mathbf{F}}_{j}(t)\,\dd t=\mathbf{0}.\label{eq:YS-norm-zeromean}
\end{equation}
\emph{Motivation:} $\tilde{\mathbf{F}}_{j}$ is then a pure Fourier
series with no constant term, and the Fourier coefficients $\mathbf{R}_{m}(\boldsymbol{\Phi}_{j}^{(m)})$
are obtained by direct Fourier inversion~\cite[\S\,IV.4.6, eq.~(4.50)]{YakSta1}
without integrating a differential equation. This gives the simplest
computational procedure and $\operatorname{tr}\tilde{\mathbf{F}}_{j}=0$
follows automatically from $\operatorname{tr}\mathbf{A}=0$. The cost
is that $\tilde{\mathbf{F}}_{j}(0)=\mathbf{V}_{j}\neq\mathbf{0}$
in general, so $F(0,\varepsilon)\neq\mathbf{I}$: the matriciant normalized
by $X(0)=\mathbf{I}$ then carries the constant right factor $F(0,\varepsilon)^{-1}$,
and the monodromy $X(T)=F(0)e^{TK}F(0)^{-1}$ is conjugation-equivalent
to $e^{TK}$, leaving all spectral data (exponents, boundaries) unchanged.

\medskip{}
\noindent\emph{Convention~2 (initial condition): $\mathbf{F}_{j}(0)=\mathbf{0}$.}
In the \emph{nonsingular case}, YakSta integrate from $t=0$~\cite[\S\,IV.5.3]{YakSta1}:
\begin{equation}
\mathbf{F}_{j}(t)=\int_{0}^{t}\bigl[\boldsymbol{\Phi}_{j}(t_{1})-\mathbf{K}_{j}\bigr]\,\dd t_{1},\label{eq:YS-norm-initial}
\end{equation}
giving $\mathbf{F}_{j}(0)=\mathbf{0}$ for all $j\geq1$. (When $\operatorname{ad}_{\mathbf{K}_{0}}\neq\mathbf{0}$
the integrand acquires the commutator term of~\eqref{eq:YS-Fl},
and the integral is replaced by the solution of the Sylvester equation~\eqref{eq:YS-Sylvester}
with $\mathbf{F}_{j}(0)=\mathbf{0}$.) \emph{Motivation:} Since $\mathbf{F}_{0}(t)=\mathbf{I}$
and $X(t,0)=\mathbf{I}$, the condition $\mathbf{F}_{j}(0)=\mathbf{0}$
ensures $F(0,\varepsilon)=\mathbf{I}$ at every order in $\varepsilon$,
so $X(t,\varepsilon)=F(t,\varepsilon)e^{tK(\varepsilon)}$ is the
standard matrizant with $X(0,\varepsilon)=\mathbf{I}$. This is physically
transparent: $\mathbf{F}_{j}$ is the $j$-th-order perturbation of
the identity, starting from zero. The cost is that $[\mathbf{F}_{j}]_{{\rm av}}\neq\mathbf{0}$
for $j\geq2$ in general.

\medskip{}
\noindent\emph{The $\varphi'_{ij}$ prescription (resonance order $m=1$).}
For the Mathieu equation at primary resonance, the resonance order
is $m=1$ (primary instability tongue). Yakubovich and Starzhinskii~\cite[\S\,IV.5.4]{YakSta1}
define functions $\varphi_{ij}(t)$ (the entries of $\mathbf{F}_{1}(t)$)
for general resonance order $p\geq2$ by the explicit formulas~\cite[\S\,IV.5.3, eqs.~after~(5.9)]{YakSta1}:
\begin{align}
\varphi_{11}(t) & =-\frac{\mu_{0}}{4p^{3}}\cos2pt+\frac{\mu_{0}}{4p^{3}}-\frac{\cos2(p+1)t}{8p^{2}(p+1)}+\frac{1}{4(p^{2}-1)}-\frac{\cos2(p-1)t}{8p^{2}(p-1)},\nonumber \\
\varphi_{12}(t) & =-\frac{\mu_{0}}{4p^{3}}\sin2pt+\frac{\sin2pt}{4p^{2}}-\frac{\sin2(p+1)t}{8p^{2}(p+1)}-\frac{\sin2(p-1)t}{8p^{2}(p-1)},\nonumber \\
\varphi_{21}(t) & =-\frac{\mu_{0}}{4p}\sin2pt-\frac{\sin2t}{4}-\frac{\sin2(p+1)t}{8(p+1)}-\frac{\sin2(p-1)t}{8(p-1)},\nonumber \\
\mathbf{F}_{1}(t) & =\begin{bmatrix}\varphi_{11}(t) & \varphi_{12}(t)\\
\varphi_{21}(t) & -\varphi_{11}(t)
\end{bmatrix}\quad(p\geq2).\label{eq:YS-phi-general}
\end{align}
At $p=1$ the formulas for $\varphi_{11}$, $\varphi_{12}$, $\varphi_{21}$
contain the factor $1/(p-1)$ or $1/(p^{2}-1)$ which is singular.
YS1 resolves this by using the \emph{time derivatives}: 
\begin{equation}
\mathbf{F}_{1}(t)=\begin{bmatrix}\varphi'_{11}(t) & \varphi'_{12}(t)\\
\varphi'_{21}(t) & -\varphi'_{11}(t)
\end{bmatrix}\quad(p=1),\qquad\varphi'_{ij}(t)=\frac{d\varphi_{ij}}{dt}.\label{eq:YS-phi-prime}
\end{equation}
Differentiating~\eqref{eq:YS-phi-general} and setting $p=1$ removes
the $1/(p-1)$ singularity (since $d/dt[\cos2(p-1)t/(p-1)]=-2\sin2(p-1)t$
remains finite at $p=1$), giving the explicit Mathieu result eq.~\eqref{eq:YS-F1}
of Section~\ref{subsec:YS-F1}. The $\varphi'$ prescription applies
only to $\mathbf{F}_{1}$ at $m=1$; $\mathbf{F}_{2}$ and higher
use Convention~1 (zero-mean).

\medskip{}
\noindent\emph{Relation between the two standard conventions.} At each order
$j\geq1$: 
\begin{equation}
\mathbf{F}_{j}^{{\rm zm}}(t)=\mathbf{F}_{j}^{{\rm ic}}(t)-[\mathbf{F}_{j}^{{\rm ic}}]_{{\rm av}},\label{eq:YS-conv-relation}
\end{equation}
where superscripts ``zm'' and ``ic'' denote zero-mean and initial-condition
respectively. The difference is the constant matrix $[\mathbf{F}_{j}^{{\rm ic}}]_{{\rm av}}$,
and $\mathbf{F}_{j}^{{\rm zm}}(0)=-[\mathbf{F}_{j}^{{\rm ic}}]_{{\rm av}}$.
For $\mathbf{F}_{1}$ of the exact LC circuit at $\hat{\lambda}=1$
(eq.~\eqref{eq:F1-omega1}, zero-mean convention): $[\mathbf{F}_{1}^{{\rm ic}}]_{{\rm av}}=\bigl[\begin{smallmatrix}1/8 & 0\\
0 & -1/8
\end{smallmatrix}\bigr]$, so $\mathbf{F}_{1}^{{\rm zm}}(0)=\bigl[\begin{smallmatrix}-1/8 & 0\\
0 & 1/8
\end{smallmatrix}\bigr]$. Note: the Mathieu $\mathbf{F}_{1}$ (eq.~\eqref{eq:YS-F1}) uses
the $\varphi'_{ij}$ singular-case prescription (neither zero-mean
nor IC), and does not fit into the two-convention framework above;
see the discussion in Chapter~\ref{subsec:YS-Mathieu}.

\medskip{}
\noindent\emph{Which convention is used in this work.} 
\begin{itemize}
\item \emph{Chapter~\ref{subsec:YS-Mathieu} (Mathieu equation, singular
case)}: The $\varphi'_{ij}$ prescription from YakSta~\cite[\S\,IV.5.4]{YakSta1},
which is the singular-case $p=1$ result. This is \emph{not} Convention~1
or Convention~2: $\mathbf{F}_{1}$ (eq.~\eqref{eq:YS-F1}) has non-zero
off-diagonal mean ($-1/4$) and non-zero initial value, but satisfies
$\operatorname{tr}\mathbf{F}_{1}=0$ and is $\pi$-periodic. $\mathbf{F}_{2}$
(eq.~\eqref{eq:YS-F2}) uses Convention~1 (zero-mean, $[\mathbf{F}_{2}]_{{\rm av}}=\mathbf{0}$). 
\item \emph{\S\,\ref{subsubsec:YS-LC-F1F2} (exact LC circuit, after preliminary
transformation, nonsingular case)}: Convention~2 (initial-condition,
$\mathbf{F}_{j}(0,\hat{\lambda})=\mathbf{0}$), giving $\mathbf{F}_{1}(0,\hat{\lambda})=\mathbf{F}_{1}(\pi,\hat{\lambda})=\mathbf{0}$
and globally $\pi$-periodic entries polynomial in $\hat{\lambda}$.
The intermediate derivation in \S\,\ref{subsubsec:YS-Km-exact}
(at $\hat{\lambda}=1$) uses Convention~1 (zero-mean). 
\end{itemize}

\subsection{Convergence}

\label{subsec:YS-convergence}

Yakubovich and Starzhinskii prove that the series~\eqref{eq:YS-F-series}
and~\eqref{eq:YS-K-series} converge for $|\varepsilon|<r$, where
the radius of convergence $r>0$ depends on the system data~\cite[\S\,IV.4.4]{YakSta1}.
Explicit lower bounds for $r$ in terms of the Fourier coefficient
norms $|\mathbf{D}_{j}(t)|\leq\delta_{j}$ and the spectral gap of
$\operatorname{ad}_{\mathbf{K}_{0}}$ are given in~\cite[\S\,IV.4.6, eq.~(4.60)]{YakSta1}.
For Hill's equation\index{Hill equation} with analytic $f(t)$, the
series converge in a neighborhood of $\varepsilon=0$ explicitly estimated
in~\cite[\S\S\,IV.5.3--5.4]{YakSta1}.

For the exact LC circuit, the norms satisfy $|\mathbf{D}_{n}|\leq\kappa\hat{\lambda}\delta^{n-1}$
(geometric decay, from~\eqref{eq:LC-Dn-exact}), so the YS series
converge for $\delta<r_{0}$ where $r_{0}>0$ depends only on $\hat{\lambda}$.
Since $\hat{\lambda}=c(1+\delta^{2})/(1-\delta^{2})$ is bounded for
$c$ fixed and $\delta\leq\delta_{0}<1$ (e.g.\ $\hat{\lambda}\leq5/3$
for $c=1$, $\delta\leq1/2$), the convergence is uniform in $\hat{\lambda}$
on any compact subset $\delta\in[0,\delta_{0}]$. In particular, the
$O(\cdot)$ error terms in the truncated YS series are bounded by
$C(\delta_{0})\cdot\hat{\lambda}^{k}\delta^{m}$ for explicit constants
$C(\delta_{0})$ and integers $k,m$, making the $\hat{\lambda}$-dependence
of the remainder fully transparent.

\emph{Joint analyticity and exact substitution.} By Theorem~\ref{thm:joint-analytic}
of Chapter~\ref{app:genLin} (Picard iteration in the parameter vector
$z=(\delta,\hat{\lambda})$ plus Hartogs'~theorem~\cite[Ch.~2]{FuksBVA}),
the YS series coefficients $\mathbf{K}_{m}(\delta,\hat{\lambda})$
and $\mathbf{F}_{m}(x,\delta,\hat{\lambda})$ are jointly analytic
in $(\delta,\hat{\lambda})$ for $|\delta|<r_{0}$, $\hat{\lambda}$
near any fixed positive value. Consequently, the instability boundary
curve\index{boundary curves}s computed from the truncated YS series
$K(\delta,\hat{\lambda})=\sum_{m=1}^{p}\delta^{m}\mathbf{K}_{m}(\hat{\lambda})$
to order $O(\delta^{p})$ at fixed $\hat{\lambda}$, and then evaluated
at the \emph{exact} value $\hat{\lambda}=c(1+\delta^{2})/(1-\delta^{2})$
(with $c$ the spectral parameter, fixed by the tongue) without any
further expansion in $\delta$, retain $O(\delta^{p+1})$ accuracy
uniformly on $[0,\delta_{0}]$. This is because $\hat{\lambda}$ is
bounded on $[0,\delta_{0}]$, the truncation error is $O(\hat{\lambda}^{p}\delta^{p+1})$
at fixed $\hat{\lambda}$, and after substituting $\hat{\lambda}(\delta)$
this remains $O(\delta^{p+1})$. The result is a simple closed-form
expression in $\delta$ --- not a polynomial but an exact function
--- which can be compared directly with the CF or MW boundary formulas.

\subsection{Two-parameter YS framework and separation principle}

\label{subsec:YS-two-param}

The YS construction applies to systems depending on \emph{two} parameters:
a \emph{bounded} parameter $\xi$ (playing the role of $\hat{\lambda}$
for the LC circuit) and a \emph{small} parameter $\delta$ (the amplitude
of the periodic perturbation). The system matrix takes the form 
\begin{equation}
\mathbf{A}(x,\xi,\delta)=\mathbf{C}(\xi)+\sum_{n=1}^{\infty}\delta^{n}\,\mathbf{B}_{n}(x,\xi),\label{eq:YS-two-param-system}
\end{equation}
where $\mathbf{C}(\xi)$ and each $\mathbf{B}_{n}(x,\xi)$ are analytic
in $\xi$ near a reference value $\xi_{0}>0$. For the exact LC circuit:
$\xi=\hat{\lambda}$, $\xi_{0}=1$, and $\mathbf{B}_{n}=\hat{\lambda}\tilde{\mathbf{B}}_{n}$
(eq.~\eqref{eq:LC-Hill-factored}).

\emph{Boundedness and the $O(\Lambda_{0},\delta^{p})$ notation.}
Fix $\delta_{0}\in(0,1)$ and set $\Lambda_{0}=\sup_{\delta\in[0,\delta_{0}]}\xi(\delta)<\infty$;
for the LC circuit $\Lambda_{0}=(1+\delta_{0}^{2})/(1-\delta_{0}^{2})$.
We write $O(\Lambda_{0},\delta^{p})$ for a remainder $R$ satisfying
\begin{equation}
|R|\leq C(\Lambda_{0})\,\delta^{p}\label{eq:big-O-two-param}
\end{equation}
for a finite constant $C(\Lambda_{0})<\infty$, uniformly in $\delta\in[0,\delta_{0}]$.
This makes the $\xi$-dependence of the error explicit.

\emph{Joint analyticity.} By Theorem~\ref{thm:joint-analytic} of
Chapter~\ref{app:genLin} combined with Hartogs'~theorem~\cite[Ch.~2]{FuksBVA},
the YS coefficients $\mathbf{K}_{m}(\xi)$ and $\mathbf{F}_{m}(x,\xi)$
are analytic in $\xi$ at each order $m$, with truncation error 
\begin{equation}
K(\xi,\delta)-{\textstyle \sum_{m=0}^{p}}\delta^{m}\mathbf{K}_{m}(\xi)=O(\Lambda_{0},\delta^{p+1}).\label{eq:YS-truncation-error}
\end{equation}

\emph{Separation principle.} The instability boundary curves depend
\emph{only} on $K(\xi,\delta)$, not on $F(x,\xi,\delta)$. The primary
boundary condition is 
\begin{equation}
K_{12}(\xi,\delta,\mu_{0})\cdot K_{21}(\xi,\delta,\mu_{0})=0.\label{eq:YS-sep-principle}
\end{equation}
To determine the boundaries to $O(\Lambda_{0},\delta^{p})$ one needs
only $\mathbf{K}_{0},\ldots,\mathbf{K}_{p}$, not the full $F$ series.

\emph{Exact substitution preserves accuracy.} If $\xi=\xi(\delta)$
is itself analytic (as is $\hat{\lambda}=c(1+\delta^{2})/(1-\delta^{2})$
for the LC circuit, with $c$ fixed at the tongue center), then the
boundary curves obtained by solving $K_{12}=0$, $K_{21}=0$ with
$\xi$ free, truncated at order $p$, and evaluated at the exact $\xi=\xi(\delta)$
retain $O(\Lambda_{0},\delta^{p+1})$ accuracy. The result is a closed-form
expression in $\delta$ --- not a polynomial, but an exact smooth
function --- comparable directly with MW or CF formulas.

\subsection{Specialization: Hill's equation, linear parameter dependence}

\label{subsec:YS-Hill-formulas}

When $\mathbf{B}_{j}\equiv0$ for all $j\geq2$ (parameter enters
linearly, i.e.\ the Mathieu equation), the source~\eqref{eq:YS-source}
reduces to $\boldsymbol{\Phi}_{l}=\mathbf{B}_{1}\mathbf{F}_{l-1}$
for $l\geq2$, and YakSta give closed-form formulas for all coefficients.
For the scalar Hill equation $\xi''+\bigl\{(p\pi/T)^{2}+\varepsilon[\mu_{0}+f(t)]\bigr\}\xi=0$
(system~\eqref{eq:YS-gen-system} with $n=2$), where $p=1,2,\ldots$
is the resonance order (positive integer labeling the instability
tongue; $p=1$ is the primary tongue), YakSta derive explicit formulas
for all coefficients~\cite[\S\S\,IV.5.3--5.4]{YakSta1}. Let $f^{(k)}$
denote the complex Fourier coefficients of $f(t)$, and $a_{p}=2\operatorname{Re}f^{(p)}$,
$b_{p}=-2\operatorname{Im}f^{(p)}$ the real cosine and sine coefficients
at the resonant frequency~\cite[\S\S\,IV.5.3--5.4, eq.~(5.13)]{YakSta1}.
The following formulas apply for general $p$ in the singular case
($\alpha=-p\pi i/T$, $\mathbf{K}_{0}=-p\pi i\mathbf{I}/T$): 
\begin{align}
\mathbf{K}_{1}(\mu_{0}) & =\frac{T}{4p\pi}\begin{bmatrix}b_{p} & \dfrac{T}{p\pi}(2f_{{\rm av}}+2\mu_{0}-a_{p})\\[8pt]
-\dfrac{p\pi}{T}(2f_{{\rm av}}+2\mu_{0}+a_{p}) & -b_{p}
\end{bmatrix},\label{eq:YS-K1-general}\\[6pt]
\mathbf{F}_{1}(t) & =\frac{T^{2}}{8p\pi^{2}}\sum_{k\neq0}\frac{1}{k}\Bigl(e^{2k\pi it/T}-1\Bigr)\begin{bmatrix}M_{11}^{(k)} & M_{12}^{(k)}\\[6pt]
M_{21}^{(k)} & M_{22}^{(k)}
\end{bmatrix},\label{eq:YS-F1-general}
\end{align}
with matrix entries 
\begin{align*}
M_{11}^{(k)} & =\beta^{(k+p)}-\beta^{(k-p)}, & M_{22}^{(k)} & =\beta^{(k-p)}-\beta^{(k+p)},\\
M_{12}^{(k)} & =-\frac{T}{p\pi i}\bigl(\beta^{(k+p)}+\beta^{(k-p)}-2\beta^{(k)}\bigr),\\
M_{21}^{(k)} & =\frac{p\pi i}{T}\bigl(\beta^{(k+p)}+\beta^{(k-p)}+2\beta^{(k)}\bigr).
\end{align*}
where $\beta^{(k)}=f^{(k)}$ for $k\neq0$ and $\beta^{(0)}=\mu_{0}+f_{{\rm av}}$~\cite[\S\S\,IV.5.3--5.4, eq.~(5.15)]{YakSta1}.
For $p=1$ (primary resonance), the $\mathbf{F}_{1}$ formula above
uses Convention~1 (zero-mean, $[\mathbf{F}_{1}]_{{\rm av}}=\mathbf{0}$),
which gives eqs.~\eqref{eq:YS-phi-prime} and~\eqref{eq:YS-F1}
after specializing to the Mathieu forcing $f(t)=\cos2t$, $T=\pi$.
The formula for $\mathbf{K}_{2}$ in terms of the Fourier coefficients
$a_{m}$, $b_{m}$ of $f(t)$ is given in~\cite[\S\S\,IV.5.3--5.4, eq.~(5.14)]{YakSta1};
for $p=1$ and $\mu_{0}=0$ it simplifies to $\mathbf{K}_{2}^{{\rm zm}}=\bigl(\begin{smallmatrix}0 & -\frac{1}{16}\\
\frac{1}{16} & 0
\end{smallmatrix}\bigr)$ (zero-mean convention, unit-amplitude $f(t)=\cos2t$; see also Remark~\ref{rem:k2k3-derivation}
and eq.~\eqref{eq:K2-omega1}). These series converge and yield the
YS series for any Hill equation with $T$-periodic analytic perturbation.

\medskip{}
\emph{Scope in this work.} This book treats only the primary instability
tongue ($p=m=1$) for both the Mathieu equation (Chapter~\ref{sec:YS-LC})
and the exact LC circuit (Chapter~\ref{sec:YS-exact-LC}). The general-$p$
(resonance order) formulas (eqs.~\eqref{eq:YS-K1-general}--\eqref{eq:YS-F1-general})
are given here for completeness; only their $p=1$ specializations
are used in the computations. Higher tongues ($m=3,5,7,\ldots$) require
the odd-$p$ formulas from YS1~\cite[\S\,IV.5.4]{YakSta1} and are
not pursued here (Remark~\ref{rem:YS-higher-tongues}). To our knowledge,
the systematic analytic construction of the periodic Lyapunov transformation
$F(t,\varepsilon)$ via convergent Fourier series with explicitly
computable coefficients was developed by Yakubovich and Starzhinskii~\cite[Ch.~IV]{YakSta1}
and does not appear elsewhere in the literature in this form. We refer
to the resulting expansions as the \emph{Yakubovich--Starzhinskii
series}.

\noindent\phantomsection

\section{Krein signature theory and strong stability}

\label{app:Krein}

This chapter gives a self-contained treatment of Krein's signature
theory for canonical Hamiltonian systems with periodic coefficients,
following Yakubovich--Starzhinskii\index{Yakubovich--Starzhinskii series}~\cite[Ch.~III]{YakSta1}
and~\cite[Ch.~VIII]{YakSta2}, with additional perspective from Kirillov~\cite[Sec.~3.3]{Kiril}.
The theory provides the symplectic\index{symplectic system} explanation
of the alternating even/odd instability structure of the LC circuit
(Theorem~\ref{thm:EPD}): it gives the precise condition under which
a collision of Floquet\index{Floquet theory} multiplier\index{Floquet theory!Floquet multiplier}s
on the unit circle produces a Jordan block\index{Jordan block} (EPD\index{exceptional point of degeneracy (EPD)},
instability tongue\index{instability tongue}) rather than a semisimple
double eigenvalue (coexistence, zero-width interval).

\bigskip{}

\emph{Historical note.} The theory of this chapter descends from Lyapunov
by an unusually short and explicit line: Krein announced it in a Doklady
note whose very title is a bow to the master --- \emph{A generalization
of some investigations of A.~M. Lyapunov on linear differential equations
with periodic coefficients} (\emph{Dokl.\
Akad.\ Nauk SSSR} \textbf{73} (1950), pp.~445--448). The discovery
was that for canonical systems the multipliers on the unit circle
are not all alike: each carries a \emph{kind} --- a signature ---
and collisions of multipliers of the same kind are harmless, while
only a collision of opposite kinds can throw multipliers off the circle.
These traffic rules for multipliers, developed further in Krein's
papers of 1951 and above all in his great memoir on $\lambda$-zones
of stability, published in the volume in memory of A.~A. Andronov
(USSR Academy of Sciences, 1955, pp.~413--498) --- the same Andronov
of the Mandelstam school met in the note of Chapter~\ref{app:ParRes}
--- turned Lyapunov's stability question into structural, symplectic
mathematics. In the same year Gelfand and Lidskii described the topology
of the stability regions of canonical systems (\emph{Uspekhi Mat.\ Nauk}
\textbf{10} (1955), no.~1, pp.~3--40), and in 1958 Moser, unaware
of the Soviet work, rediscovered the signature criterion in the West
(\emph{Comm.\ Pure Appl.\ Math.} \textbf{11} (1958), pp.~81--114)
--- whence the name Krein--Moser theorem and the theory's entry
into celestial mechanics and Hamiltonian dynamics. The school of Yakubovich
consolidated and extended the theory into the treatise this chapter
follows, and its modern life runs through the multiparameter perturbation
theory of Seyranian and Mailybaev and the spectral analysis of Kirillov
cited above, where Krein signature governs the unfolding of degeneracies
across mechanics and physics. For this book the theory supplies the
final word of explanation: an EPD of the modulated LC circuit is precisely
a Krein collision --- multipliers of opposite kind meeting at $\rho=\pm1$
--- and the alternating even/odd pattern of its instability tongues
is the signature calculus in action.

\subsection{Canonical systems and symplectic structure}

\label{subsec:Krein-setup}

A \emph{canonical} (Hamiltonian) system of linear differential equations
with $T$-periodic coefficients has the form 
\begin{equation}
\mathbf{J}\,\frac{d\mathbf{x}}{dt}=\mathbf{H}(t)\,\mathbf{x},\qquad\mathbf{H}(t+T)=\mathbf{H}(t)=\mathbf{H}(t)^{*},\label{eq:Krein-canonical}
\end{equation}
where $\mathbf{J}=\bigl[\begin{smallmatrix}0 & -\mathbf{I}_{k}\\
\mathbf{I}_{k} & 0
\end{smallmatrix}\bigr]$ is the standard $2k\times2k$ symplectic matrix ($\mathbf{J}^{2}=-\mathbf{I}_{2k}$,
$\mathbf{J}^{*}=-\mathbf{J}$) and $\mathbf{H}(t)$ is the real symmetric
Hamiltonian matrix. State vectors are ordered (coordinate, momentum),
as in Section~\ref{subsec:Hamiltonian}, so that $\mathbf{J}$ and
$\mathbf{H}$ here agree with the canonical form~\eqref{eq:canonical-form}.
The monodromy matrix\index{monodromy matrix} $\mathbf{X}(T)$ of~\eqref{eq:Krein-canonical}
satisfies~\cite[Ch.~III, \S\,1.1]{YakSta1} 
\begin{equation}
\mathbf{X}(T)^{*}\mathbf{J}\,\mathbf{X}(T)=\mathbf{J},\label{eq:Krein-symplectic}
\end{equation}
i.e., $\mathbf{X}(T)$ is \emph{symplectic}, or equivalently $(i\mathbf{J})$-unitary
(since $\mathbf{X}^{*}(i\mathbf{J})\mathbf{X}=i\mathbf{J}$ follows
from~\eqref{eq:Krein-symplectic}). By the Lyapunov--Poincaré theorem,
the Floquet multipliers (eigenvalues of $\mathbf{X}(T)$) are symmetric
about the unit circle: if $\rho$ is a multiplier then so is $\bar{\rho}^{-1}$.
Consequently, multipliers on the unit circle always appear in conjugate
pairs $\rho,\bar{\rho}$.

\subsection{Elementary divisors and Jordan structure}

\label{subsec:Krein-elementary-divisors}

We recall the relevant linear-algebraic notions~\cite[\S\S\,I.1.10--I.1.12]{YakSta1}.
The \emph{elementary divisors} of a matrix $\mathbf{X}$ are the binomials
$(\lambda-\lambda^{(\sigma)})^{m_{\sigma}}$, one for each Jordan
block $Q_{\sigma}(\lambda^{(\sigma)})$ in the Jordan canonical form.
An elementary divisor is \emph{simple} if $m_{\sigma}=1$ (corresponding
to a $1\times1$ Jordan block, i.e.\ a true eigenvector). The elementary
divisors of $\mathbf{X}$ are simple if and only if $\mathbf{X}$
is diagonalizable (semisimple).

An eigenvalue $\rho$ is said to have a \emph{multiple elementary
divisor} (equivalently, a \emph{non-trivial Jordan block}) if at least
one $m_{\sigma}\geq2$. In that case the root subspace $\mathfrak{L}_{\rho}$
is strictly larger than the eigenspace $\mathfrak{E}_{\rho}$: besides
true eigenvectors $\mathbf{a}$ (satisfying $\mathbf{X}\mathbf{a}=\rho\mathbf{a}$)
there are \emph{root vectors} (generalized eigenvectors) $\mathbf{f}$
satisfying $(\mathbf{X}-\rho\mathbf{I})\mathbf{f}=\mathbf{a}$, and
so on.

\subsection{The indefinite scalar product}

\label{subsec:Krein-indef-product}

The symplectic structure endows the state space $\mathbb{C}^{2k}$
with an \emph{indefinite scalar product}~\cite[Ch.~III, \S\,2.1]{YakSta1}
\begin{equation}
[\mathbf{x},\mathbf{y}]\stackrel{{\rm def}}{=}(i\mathbf{J}\,\mathbf{x},\,\mathbf{y})=\mathbf{y}^{*}\,(i\mathbf{J})\,\mathbf{x}.\label{eq:Krein-indef}
\end{equation}
The matrix $i\mathbf{J}$ is Hermitian ($(i\mathbf{J})^{*}=i\mathbf{J}$)
and nonsingular, but indefinite (its eigenvalues are $\pm1$). Hence
$[\cdot,\cdot]$ is a non-degenerate Hermitian form on $\mathbb{C}^{2k}$: 
\begin{enumerate}
\item[(1)] $[\mathbf{x},\mathbf{y}]=\overline{[\mathbf{y},\mathbf{x}]}$ (Hermitian
symmetry); 
\item[(2)] $[\mathbf{x},\mathbf{x}]$ is real for all $\mathbf{x}$ (but can
be positive, negative, or zero); 
\item[(3)] if $[\mathbf{x}_{0},\mathbf{x}]=0$ for all $\mathbf{x}$, then $\mathbf{x}_{0}=\mathbf{0}$
(nondegeneracy). 
\end{enumerate}
A vector $\mathbf{x}$ with $[\mathbf{x},\mathbf{x}]=0$ is called
\emph{isotropic}~\cite[Ch.~III, \S\,2.1]{YakSta1}. Because the scalar
product is indefinite, isotropic vectors can be nonzero; this is in
contrast with the positive-definite inner product where $(\mathbf{x},\mathbf{x})=0$
implies $\mathbf{x}=\mathbf{0}$.

The monodromy $\mathbf{X}(T)$ is $(i\mathbf{J})$-unitary in the
sense that 
\[
[\mathbf{X}(T)\mathbf{x},\mathbf{X}(T)\mathbf{y}]=[\mathbf{x},\mathbf{y}]\quad\text{for all }\mathbf{x},\mathbf{y}.
\]

\subsection{Eigenvalues of the first and second kinds: the Krein definition}

\label{subsec:Krein-signature}

The following definition is due to Yakubovich and Starzhinskii; we
follow~\cite[Ch.~III, \S\,1.2]{YakSta1} exactly.
\begin{defn}[Krein signature\index{Krein signature}]
\label{defn:Krein-sig} (\cite[Ch.~III, \S\,1.2]{YakSta1}) Let $\mathbf{X}$
be a $(i\mathbf{J})$-unitary ($2k\times2k$ symplectic) matrix. 
\begin{enumerate}
\item[(a)] \emph{Simple eigenvalue on the unit circle.} Let $\rho'$ ($|\rho'|=1$)
be a simple eigenvalue of $\mathbf{X}$ with eigenvector $\mathbf{a}$,
$\mathbf{X}\mathbf{a}=\rho'\mathbf{a}$. Then $[\mathbf{a},\mathbf{a}]\neq0$
(see Remark~\ref{rem:Krein-noniso} below), and $\rho'$ is called
an eigenvalue of the \emph{first kind} if $[\mathbf{a},\mathbf{a}]>0$,
and of the \emph{second kind} if $[\mathbf{a},\mathbf{a}]<0$. 
\item[(b)] \emph{Multiple definite eigenvalue on the unit circle.} Let $\rho'$
($|\rho'|=1$) be an $r$-fold eigenvalue of $\mathbf{X}$ and $\mathfrak{E}_{\rho'}$
its eigensubspace. If $[\mathbf{x},\mathbf{x}]>0$ for all $\mathbf{x}\neq\mathbf{0}$
in $\mathfrak{E}_{\rho'}$, then $\rho'$ is an $r$-fold eigenvalue
of the \emph{first kind}; if $[\mathbf{x},\mathbf{x}]<0$ for all
$\mathbf{x}\neq\mathbf{0}$ in $\mathfrak{E}_{\rho'}$, it is of the
\emph{second kind}. Eigenvalues of the first or second kind are called
\emph{definite}. 
\item[(c)] \emph{Indefinite (mixed-kind) eigenvalue on the unit circle.} Let
$\rho'$ ($|\rho'|=1$) be an $r$-fold eigenvalue on the unit circle
such that $[\mathbf{x},\mathbf{x}]$ is \emph{not} of fixed sign on
$\mathfrak{E}_{\rho'}$, i.e., there exists a nonzero $\mathbf{x}\in\mathfrak{E}_{\rho'}$
with $[\mathbf{x},\mathbf{x}]=0$. Then $\rho'$ is called \emph{indefinite}
or of \emph{mixed kind}. 
\item[(d)] \emph{Eigenvalue off the unit circle.} Let $\rho$ ($|\rho|\neq1$)
be an $r$-fold eigenvalue of $\mathbf{X}$. Then $\rho$ is called
an eigenvalue of the \emph{first kind} if $|\rho|<1$, and of the
\emph{second kind} if $|\rho|>1$.

\emph{Motivation.} This labeling is consistent with the on-circle
classification under continuous Hamiltonian deformation. Every canonical
system has exactly $k$ multipliers of the first kind and $k$ of
the second kind, and this partition is preserved under continuous
deformation of the Hamiltonian~\cite[Ch.~III, \S\,2]{YakSta1}. If
a pair of multipliers leaves the unit circle, the symplectic symmetry
$\rho\mapsto\bar{\rho}^{-1}$ forces one of them into the open unit
disk and the other out of it; the labeling of part~(d) --- first
kind inside, second kind outside --- is the unique one that preserves
the $k$/$k$ partition along the deformation. Such an excursion can
occur only through a collision of multipliers of \emph{different}
kinds (Theorem~\ref{thm:Krein} below): one enters the disk and one
leaves it --- this is the mixed-kind collision responsible for the
EPD and the opening of instability tongues (see also Figure~III.8
of~\cite[Ch.~III, \S\,1]{YakSta1}). 

\end{enumerate}
\end{defn}

\begin{rem}[Non-isotropicity of simple-eigenvalue eigenvectors on the unit circle]
\label{rem:Krein-noniso} In part~(a), the fact that $[\mathbf{a},\mathbf{a}]\neq0$
is \emph{not} obvious and requires proof. It follows from the theory
of $G$-orthogonality of root subspaces~\cite[Ch.~III, \S\,2.4]{YakSta1}:
each root subspace $\mathfrak{L}_{\rho'}$ (for $|\rho'|=1$) is a
nondegenerate subspace for the indefinite scalar product, meaning
the Gram matrix of $\mathfrak{L}_{\rho'}$ is nonsingular. In particular,
if $\rho'$ is a simple eigenvalue on the unit circle, its one-dimensional
eigenspace is nondegenerate, so $[\mathbf{a},\mathbf{a}]\neq0$~\cite[Ch.~III, \S\,2.4, end of proof]{YakSta1}.
This nondegeneracy fails for eigenvalues \emph{off} the unit circle
(part~(d) and Lemma~\ref{lem:Krein-offcircle} below). 
\end{rem}

\subsection{Two key lemmas}

\label{subsec:Krein-lemmas}

The following two lemmas are the technical core of the theory~\cite[Ch.~III, \S\,1]{YakSta1}.
\begin{lem}[Jordan block $\Rightarrow$ isotropic eigenvector]
\index{Jordan block!isotropic eigenvector} \label{lem:Krein-Jordan}
(\cite[Lemma~III.1.1]{YakSta1}) Let $\mathbf{X}$ be a $2k\times2k$
$(i\mathbf{J})$-unitary matrix with a multiple eigenvalue $\rho'$
on the unit circle ($|\rho'|=1$) having at least one non-trivial
Jordan block. Then there exists an eigenvector $\mathbf{a}\neq\mathbf{0}$
belonging to $\rho'$ such that $[\mathbf{a},\mathbf{a}]=0$, i.e.,
$\rho'$ is an eigenvalue of mixed kind. 
\end{lem}

\begin{proof}
Since $\mathbf{X}$ has a Jordan block at $\rho'$, the root subspace
$\mathfrak{L}_{\rho'}$ has a cyclic basis $\mathbf{f}_{1}=\mathbf{a}$,
$\mathbf{f}_{2}$, \ldots\ of dimension $\geq2$, with 
\[
\mathbf{X}\mathbf{a}=\rho'\mathbf{a},\quad\mathbf{X}\mathbf{f}_{2}=\rho'\mathbf{f}_{2}+\mathbf{a},\quad\ldots
\]
By $(i\mathbf{J})$-unitarity of $\mathbf{X}$: $[\mathbf{X}\mathbf{a},\mathbf{X}\mathbf{f}_{2}]=[\mathbf{a},\mathbf{f}_{2}]$.
Computing directly using $|\rho'|=1$: 
\[
[\mathbf{X}\mathbf{a},\mathbf{X}\mathbf{f}_{2}]=[\rho'\mathbf{a},\,\rho'\mathbf{f}_{2}+\mathbf{a}]=|\rho'|^{2}[\mathbf{a},\mathbf{f}_{2}]+\rho'[\mathbf{a},\mathbf{a}]=[\mathbf{a},\mathbf{f}_{2}]+\rho'[\mathbf{a},\mathbf{a}].
\]
Equating the two expressions gives $\rho'[\mathbf{a},\mathbf{a}]=0$,
and since $\rho'\neq0$ we conclude $[\mathbf{a},\mathbf{a}]=0$. 
\end{proof}
\begin{lem}[Off-circle eigenvalue $\Rightarrow$ isotropic eigenvector]
\label{lem:Krein-offcircle} (\cite[Lemma~III.1.2]{YakSta1}) Let
$\rho$ be an eigenvalue \emph{not} on the unit circle ($|\rho|\neq1$)
of an $(i\mathbf{J})$-unitary matrix $\mathbf{X}$. Then every corresponding
eigenvector $\mathbf{a}$ is isotropic: $[\mathbf{a},\mathbf{a}]=0$. 
\end{lem}

\begin{proof}
From $\mathbf{X}\mathbf{a}=\rho\mathbf{a}$ and $(i\mathbf{J})$-unitarity:
$[\mathbf{X}\mathbf{a},\mathbf{X}\mathbf{a}]=|\rho|^{2}[\mathbf{a},\mathbf{a}]=[\mathbf{a},\mathbf{a}]$.
Since $|\rho|^{2}\neq1$, this forces $[\mathbf{a},\mathbf{a}]=0$. 
\end{proof}
\begin{rem}[Complete equivalence]
\label{rem:Krein-converse} The converse of Lemma~\ref{lem:Krein-Jordan}
is also true (it is part~(a) of Theorem~\ref{thm:Krein} below):
a \emph{definite} multiple multiplier on the unit circle has only
simple elementary divisors. Together, the two directions give: 
\begin{equation}
\text{Jordan block at }\rho'\text{ on unit circle}\iff\rho'\text{ is of mixed kind (indefinite).}\label{eq:Krein-equiv}
\end{equation}
Lemma~\ref{lem:Krein-offcircle} shows that the notion of Krein kind
is intrinsically tied to the unit circle: eigenvectors of off-circle
multipliers are always isotropic, so parts~(a)--(c) of Definition~\ref{defn:Krein-sig}
genuinely classify only eigenvalues on the unit circle. 
\end{rem}

\subsection{Krein's theorem on perturbation of multipliers}

\label{subsec:Krein-theorem}
\begin{thm}[Krein]
\index{Krein collision theory!Krein theorem} \label{thm:Krein}
(\cite[Ch.~III, \S\,1.3]{YakSta1}) Let $\mathbf{X}_{0}$ be a $2k\times2k$
$(i\mathbf{J})$-unitary matrix with a \emph{definite} $r$-fold multiplier
$\rho^{(0)}$ on the unit circle. Then: 
\begin{enumerate}
\item[(a)] All elementary divisors of $\mathbf{X}_{0}$ belonging to $\rho^{(0)}$
are simple (no Jordan blocks). 
\item[(b)] There exist $\gamma>0$ and $\eta>0$ such that for any $(i\mathbf{J})$-unitary
matrix $\mathbf{X}$ with $|\mathbf{X}-\mathbf{X}_{0}|<\eta$, all
eigenvalues of $\mathbf{X}$ in the disk $|\rho-\rho^{(0)}|<\gamma$
lie on the unit circle and have simple elementary divisors. 
\end{enumerate}
In particular: \emph{a definite multiple multiplier on the unit circle
cannot leave the unit circle under any perturbation of the $(i\mathbf{J})$-unitary
(symplectic) matrix.} 
\end{thm}

\begin{proof}[Proof sketch]
Part~(a) follows directly from Lemma~\ref{lem:Krein-Jordan}: if
there were a Jordan block, the eigenvector $\mathbf{a}$ would satisfy
$[\mathbf{a},\mathbf{a}]=0$, contradicting definiteness. Part~(b)
follows by a compactness argument using (a) and the continuity of
eigenvalues; see~\cite[Ch.~III, \S\,1.3]{YakSta1} for the full proof. 
\end{proof}
The physical content is clear: same-kind multipliers meeting on the
unit circle are ``frozen'' there by the symplectic structure ---
they cannot open an instability gap. Opposite-kind multipliers meeting
on the unit circle form a mixed-kind double multiplier (by~\eqref{eq:Krein-equiv}),
which can and generically does leave the unit circle, opening an instability
tongue.
\begin{rem}[Energy-sign interpretation of Krein signature]
\label{rem:Krein-energy} For a canonical Hamiltonian system~\eqref{eq:Krein-canonical}
with a simple pure imaginary eigenvalue $\lambda=i\omega$ ($\omega\in\mathbb{R}\setminus\{0\}$)
and eigenvector $\mathbf{u}$, the Krein signature $\kappa(\lambda):=\operatorname{sign}([\mathbf{u},\mathbf{u}])$
is related to the sign of the modal energy $\mathcal{E}:=(\mathbf{H}\mathbf{u},\mathbf{u})$
by 
\begin{equation}
\operatorname{sign}(\mathcal{E})=\kappa(\lambda)\operatorname{sign}(\omega),\label{eq:Krein-energy-sign}
\end{equation}
see~\cite[Sec.~3.3.8, eq.~(3.57)]{Kiril}. Consequently, modes of
the \emph{first kind} ($\kappa=+1$, $\omega>0$) carry positive energy,
while modes of the \emph{second kind} ($\kappa=-1$, $\omega>0$)
carry negative energy. Theorem~\ref{thm:Krein} then acquires a transparent
physical meaning: two modes of the \emph{same} energy sign cannot
exchange energy to produce exponential growth, so their collision
is harmless (an avoided crossing); two modes of \emph{opposite} energy
sign can do so, and their collision generically opens an instability
tongue. This energy-sign picture is developed in the autonomous Hamiltonian
setting (equilibria of $\dot{z}=\mathbf{J}\cdot DH$, eigenvalues
of $\mathbf{J}D^{2}H$) by MacKay~\cite[Sec.~3]{MacKay87}, who shows
that stability can be lost only by collision of eigenvalues with opposite
sign of energy or by collision at zero. For the periodic-coefficient
Floquet setting of the present work the complete proofs are in~\cite[Ch.~III]{YakSta1};
Kirillov~\cite[Sec.~3.3.8]{Kiril} gives a detailed account including
the connection to negative energy waves and dispersive wave propagation.
A concrete physical realization of negative-energy modes in electric
circuits is provided by circuits containing a \emph{negative capacitance\index{capacitance sensing}}
element: since the potential energy stored in a capacitor is $q^{2}/2C$,
a negative capacitance $C<0$ contributes a negative term to the circuit
Hamiltonian, rendering the Hessian $\mathbf{H}$ indefinite and producing
a mode of the second kind ($\kappa=-1$) in the sense of Krein. The
papers~\cite[\S\,1]{FigPert,FigSynbJ} show that the combination
of a negative capacitance and a gyrator in a lossless circuit can
be synthesized to produce precisely an exceptional point of degeneracy
(EPD) --- the point at which two modes of opposite energy sign coalesce,
their Floquet multipliers merge on the unit circle, and the monodromy
matrix acquires a non-trivial Jordan block. This is the circuit-theoretic
incarnation of the Krein collision\index{Krein collision theory}
mechanism, with the negative-capacitance mode playing the role of
the negative-energy (second-kind) mode. 
\end{rem}

\subsection{The Krein--Gel'fand--Lidski\u{\i} theorem}

\label{subsec:KGL}
\begin{thm}[Krein--Gel'fand--Lidski\u{\i}]
\index{Krein--Gel'fand--Lidskii theorem} \label{thm:KGL} (\cite[Ch.~III, \S\,3]{YakSta1};
\cite[Sec.~3.3.6]{Kiril}) The canonical system~\eqref{eq:Krein-canonical}
is \emph{strongly stable} --- meaning all solutions of every sufficiently
nearby canonical system (in the $L^{1}$-norm of the Hamiltonian)
are bounded on $(-\infty,+\infty)$ --- \emph{if and only if} all
Floquet multipliers lie on the unit circle and all are definite (of
pure kind, no mixed-kind multipliers). 
\end{thm}

Strong stability characterizes the interior of the stability zones
in Hamiltonian parameter space. Weak stability (stable but not strongly
stable) occurs exactly when all multipliers are on the unit circle
but at least one multiple multiplier is of mixed kind --- i.e., at
an EPD curve\index{exceptional point of degeneracy (EPD)!EPD curve}.

\subsection{\texorpdfstring{The $2\times2$ case: Hamiltonian space partition}{The
2x2 case: Hamiltonian space partition}}

\label{subsec:Krein-2x2}

For canonical systems of two equations ($k=1$, $2\times2$ matrices),
the theory specializes to a complete three-way partition of Hamiltonian
space $\mathcal{L}^{3}$ (the space of real symmetric $T$-periodic
$2\times2$ matrix functions with the $L^{1}$ norm) into stability
and instability domains~\cite[Ch.~VIII, \S\S\,1.4--1.9]{YakSta2}:

\begin{equation}
\mathcal{L}^{3}=\mathcal{H}\cup\mathcal{G}\cup\Pi,\qquad\Pi=\Pi^{**}\cup\Pi^{*+}\cup\Pi^{*-},\label{eq:Krein-partition}
\end{equation}
where the behavior of the matrizant $\mathbf{X}(t)$ in each region
is: 
\begin{itemize}
\item $\mathcal{H}$ (\emph{instability}): $\mathbf{X}(T)$ has two real
multipliers $\rho_{1},\rho_{2}$ with $|\rho_{1}|>1>|\rho_{2}|$;
one solution grows exponentially. 
\item $\mathcal{G}$ (\emph{strong stability}): all multipliers on the unit
circle, distinct, definite --- all solutions bounded and almost-periodic. 
\item $\Pi^{*+}$ or $\Pi^{*-}$ (\emph{EPD curve}): double multiplier $\rho=-1$
or $\rho=+1$ of \emph{mixed kind}, with non-trivial Jordan block;
exactly one periodic or antiperiodic solution; the other solution
grows linearly in $t$. 
\[
\mathbf{X}(T)=\pm\,\mathbf{S}\begin{bmatrix}1 & 1\\
0 & 1
\end{bmatrix}\mathbf{S}^{-1}.
\]
\item $\Pi^{**}$ (\emph{coexistence}): double multiplier of \emph{definite
kind}; all solutions periodic or antiperiodic; $\mathbf{X}(T)=\pm\mathbf{I}_{2}$. 
\end{itemize}
The regions $\mathcal{H}$ and $\mathcal{G}$ are open; $\Pi^{*\pm}$
and $\Pi^{**}$ are the boundary sets $\partial\mathcal{H}\cup\partial\mathcal{G}$.
\begin{rem}[Transition at a boundary]
\label{rem:Krein-transition} As a parameter varies, two multipliers
on the unit circle move toward each other. By Theorem~\ref{thm:Krein},
if they meet as a definite double multiplier ($\Pi^{**}$), they cannot
leave the unit circle: the system passes through the boundary without
opening an instability gap. If they meet as a mixed-kind double multiplier
($\Pi^{*\pm}$), they leave the unit circle generically, and an instability
tongue opens. The condition for the first scenario is that both approaching
multipliers carry the \emph{same} Krein signature; for the second,
\emph{opposite} signatures. This is Krein's collision theory in its
simplest form~\cite[Sec.~3.3.2]{Kiril}. 
\end{rem}

\subsection{Application to the LC circuit}

\label{subsec:Krein-LC}

The LC circuit Hill equation\index{Hill equation}~\eqref{eq:LC-Hill}
is a scalar second-order equation with real coefficients; its monodromy
matrix is $2\times2$ symplectic (in the class of the preceding subsection).
The Hamiltonian space partition~\eqref{eq:Krein-partition} applies
directly.

At the $m$-th resonance\index{resonance} ($\hat{\lambda}\approx m^{2}$,
$\delta\to0$): the unperturbed multiplier is $\rho_{0}=(-1)^{m}$
(double). By the Floquet analysis of the unperturbed system $q''+m^{2}q=0$:
two independent solutions of period $2\pi/m$ coexist, so the unperturbed
monodromy is $(-1)^{m}\mathbf{I}_{2}$, a definite double multiplier
(both solutions carry the same energy sign, $\Pi^{**}$ configuration).

Under perturbation ($\delta>0$): 
\begin{itemize}
\item At \emph{even} resonances ($m=2,4,6,\ldots$): the Ince coexistence
theorem (Propositions~\ref{prop:even-vanish}, \ref{prop:odd-survive})
forces the two independent periodic solutions to persist for all $0<\delta<1$.
The monodromy remains $+\mathbf{I}_{2}$ --- a $\Pi^{**}$ configuration.
By Lemma~\ref{lem:Krein-Jordan}, a definite multiple multiplier
has only simple elementary divisors. By Theorem~\ref{thm:Krein},
it cannot leave the unit circle. The instability interval collapses
to zero width.
\item At \emph{odd} resonances ($m=1,3,5,\ldots$): the two multipliers
approach $-1$ from opposite sides of the unit circle, carrying \emph{opposite}
Krein signatures ($\Pi^{*\pm}$ configuration). By Lemma~\ref{lem:Krein-Jordan},
the resulting double multiplier on the unit circle has a non-trivial
Jordan block (EPD). Generically the multipliers leave the unit circle,
opening an instability tongue of width $O(\delta^{m})$ (Theorem~\ref{thm:EPD-bdry}). 
\end{itemize}
This is the complete symplectic explanation of Theorem~\ref{thm:EPD}:
the alternating structure of the LC circuit instability zone\index{instability zone}s
is encoded in the Krein signatures of the colliding multipliers. Lemma~\ref{lem:Krein-Jordan}
provides the mechanism by which opposite-signature collisions force
a Jordan block (EPD), while Theorem~\ref{thm:Krein} guarantees that
same-signature collisions cannot open a tongue.

% =============================================================================

\section{The Miller algorithm for minimal solutions}

\label{app:Miller} % =============================================================================

This chapter collects the definitions and results from Wimp~\cite[\S\S\,2.6,\,4]{Wimp}
that are used in \S\,\ref{subsec:Cambi-notes2} and \S\,\ref{subsec:Cambi-notes2-res}.
We follow Wimp's notation throughout; our mapping to the notation
of the main text is given at the end.

\bigskip{}

\emph{Historical note.} Miller's algorithm was born in the last great
age of mathematical table making. The computers of that age --- the
word then meant people --- knew from hard experience that a three-term
recurrence run forward destroys a decreasing solution: rounding errors
plant a seed of the dominant solution, which then outgrows everything
(in the language of Chapter~\ref{app:FDE}, forward recursion is
unstable for the minimal solution). The cure appeared in the British
Association tables of Bessel functions (\emph{Mathematical Tables},
vol.~X: \emph{Bessel Functions, Part~II}, Cambridge, 1952), where
J.~C.~P. Miller introduced the now-classical device: run the recurrence
\emph{backward} from arbitrary trial values set far beyond the range
of interest, then normalize --- in that direction the dominant solution
extinguishes itself, and only the minimal solution survives. What
began as a table maker's trick became theory: Olver gave the error
analysis (\emph{Math.\ Comp.} \textbf{18} (1964), pp.~65--74) and
a refined algorithm (\emph{J. Res.\ Nat.\ Bur.\
Standards} \textbf{71B} (1967), pp.~111--129), and Gautschi's survey
(\emph{SIAM Review} \textbf{9} (1967), pp.~24--82) connected the
algorithm to its true foundation, Pincherle's theorem of 1894: backward
recurrence computes precisely the minimal solution whose existence
is equivalent to the convergence of the associated continued fraction.
The circle of ideas thus closes across three chapters --- finite
difference equations (Chapter~\ref{app:FDE}), continued fractions
(Chapter~\ref{app:CF}), and the present algorithm --- and it is
no museum piece: the continued-fraction evaluations behind the boundary
curves of Part~I are Miller's backward recurrence at work on the
LC recurrence.

\subsection*{Three-term recurrences and minimal solutions}

Write the three-term recurrence in the form~\cite[\S\,4.1]{Wimp}:
\begin{equation}
y(n)+a(n)\,y(n+1)+b(n)\,y(n+2)=0,\qquad n\geq0,\quad b(n)\neq0.\label{eq:Wimp-rec}
\end{equation}

\begin{defn}[Minimal solution]
\index{minimal solution} \label{defn:Wimp-minimal} (\cite[\S\,2.6, Def.~2.7]{Wimp})
A nontrivial solution $\mu(n)$ of~\eqref{eq:Wimp-rec} is called
a \emph{minimal solution\index{continued fraction!minimal solution}}
if $\mu(n)$ is dominated by each of the other solutions in some fundamental
set containing $\mu(n)$: 
\[
\lim_{n\to\infty}\frac{\mu(n)}{y(n)}=0
\]
for any solution $y(n)$ not proportional to $\mu(n)$. A minimal
solution, if it exists, is unique up to a constant multiple. 
\end{defn}

Wimp notes~\cite[\S\,4.1]{Wimp}: \emph{``{[}A{]} minimal solution
is uniquely determined by one initial value. These observations suggest
that minimal solutions are ideal candidates for the application of
the Miller algorithm\index{Miller algorithm}.''}

\subsection*{The Miller algorithm}

Let $w(n)$ be the desired minimal solution of~\eqref{eq:Wimp-rec}.
Assume a convergent normalizing condition is known: 
\begin{equation}
S:=\sum_{k=0}^{\infty}c(k)\,w(k)\neq0.\label{eq:Wimp-norm}
\end{equation}
For integer $N\geq0$ define the backward recursion~\cite[\S\,4.1, eqs.~(4.3)--(4.5)]{Wimp}:
\begin{multline}
y_{N}(N+1):=0,\quad y_{N}(N):=1;\\
y_{N}(n)+a(n)\,y_{N}(n+1)+b(n)\,y_{N}(n+2)=0,\quad n=N-1,\ldots,0.\label{eq:Wimp-Miller}
\end{multline}
The $N$-th approximant to $w(n)$ is: 
\begin{equation}
w_{N}(n):=\frac{S\,y_{N}(n)}{S_{N}},\qquad S_{N}:=\sum_{k=0}^{N}c(k)\,y_{N}(k).\label{eq:Wimp-approx}
\end{equation}

\begin{thm}[Convergence, Wimp]
\index{Miller algorithm!convergence} \label{thm:Wimp-conv} (\cite[\S\,4.1, Thm.~4.2]{Wimp})
Let~\eqref{eq:Wimp-rec} have a minimal solution $w(n)$ with $S\neq0$
as in~\eqref{eq:Wimp-norm}. Then the Miller algorithm~\eqref{eq:Wimp-Miller}--\eqref{eq:Wimp-approx}
converges, $w_{N}(n)\to w(n)$ as $N\to\infty$, if and only if 
\begin{equation}
\lim_{N\to\infty}\tau_{N}S_{N}^{*}=0,\qquad\tau_{N}:=\frac{w(N+1)}{y(N+1)},\quad S_{N}^{*}:=\sum_{k=0}^{N}c(k)\,y(k),\label{eq:Wimp-conv-cond}
\end{equation}
where $y(n)$ is any dominant solution of~\eqref{eq:Wimp-rec}. 
\end{thm}

\begin{cor}[Simplified algorithm]
\label{cor:Wimp-simplified} (\cite[\S\,4.1, Cor.~to Thm.~4.2 and Thm.~4.4]{Wimp})
If $w(0)\neq0$ is known, replace~\eqref{eq:Wimp-approx} by: 
\begin{equation}
w_{N}(n):=\frac{w(0)\,y_{N}(n)}{y_{N}(0)},\qquad y_{N}(0)\neq0.\label{eq:Wimp-simplified}
\end{equation}
Then $w_{N}(n)\to w(n)$ as $N\to\infty$. Conversely, if $y_{N}(n)/y_{N}(0)$
converges as $N\to\infty$, its limit is a minimal solution. 
\end{cor}

\subsection*{Connection to continued fractions}

The ratio $r(n):=w(n+1)/w(n)$ of consecutive minimal solution values
satisfies~\cite[\S\,4.5, eq.~(4.41)]{Wimp}: 
\begin{equation}
r(n)=\frac{-1}{a(n)+b(n)\,r(n+1)},\label{eq:Wimp-ratio-rec}
\end{equation}
and hence equals the continued fraction\index{continued fraction}~\cite[Thm.~4.5]{Wimp}:
\begin{equation}
r(n)=\cfrac{-1}{a(n)-\cfrac{b(n)}{a(n+1)-\cfrac{b(n+1)}{a(n+2)-\cdots}}}.\label{eq:Wimp-CF}
\end{equation}
The CF-based algorithm~\cite[\S\,4.5, eq.~(4.44)]{Wimp} computes
$r_{N}(n)\to r(n)$ by backward recursion: 
\begin{equation}
r_{N}(N):=0;\quad r_{N}(n):=\frac{-1}{a(n)+b(n)\,r_{N}(n+1)},\quad n=N-1,\ldots,0.\label{eq:Wimp-CF-algo}
\end{equation}
This is equivalent to the Miller algorithm but avoids overflow since
the ratios $r_{N}(n)$ are typically moderate in size.

\subsection*{Application to our setting}

Our recurrence~\eqref{eq:notes2-ttrec}, $A_{n+1}=-G(u+n)/\gamma\cdot A_{n}-A_{n-1}$,
is equivalently written as 
\[
A_{n}\;+\;\frac{G(u+n+1)}{\gamma}\,A_{n+1}\;+\;A_{n+2}\;=\;0,
\]
which matches Wimp's form~\eqref{eq:Wimp-rec} with $y(n)=A_{n}$
and 
\begin{equation}
a(n)\;=\;\frac{G(u+n+1)}{\gamma},\qquad b(n)\;=\;1.\label{eq:Wimp-our-ab}
\end{equation}
The minimal solution is $\{h_{n}(u)\}$ with ratio $r(n)=h_{n+1}/h_{n}=w(u+n)$
(eq.~\eqref{eq:notes2-hm-ratio}). The CF formula~\eqref{eq:Wimp-CF}
applied to~\eqref{eq:Wimp-our-ab} gives: 
\begin{equation}
\frac{h_{n+1}}{h_{n}}\;=\;w(u+n)\;=\;\cfrac{-1}{G(u+n+1)/\gamma-\cfrac{1}{G(u+n+2)/\gamma-\cfrac{1}{\ddots}}},\label{eq:Wimp-our-CF}
\end{equation}
which is the continued fraction~\eqref{eq:notes2-CF-orig} evaluated
at $u+n$. The CF-based backward recursion~\eqref{eq:Wimp-CF-algo}
reads: 
\[
r_{N}(N):=0;\quad r_{N}(n):=\frac{-1}{G(u+n+1)/\gamma+r_{N}(n+1)},\quad n=N-1,\ldots,0,
\]
and is equivalent to the backward recursion~\eqref{eq:notes2-wdef}
used to compute $w(u)$ (with the index shift $n\mapsto n-1$).

The convergence condition of Theorem~\ref{thm:Wimp-conv} holds because:
with $c(k)=1$, the normalizing sum $S=\sum_{k=0}^{\infty}h_{k}(u)$
converges geometrically ($|h_{k}|\sim|\zeta_{+}|^{k}$, $|\zeta_{+}|<1$)
and, being analytic in $u$ and not identically zero, is nonzero outside
a discrete set; for the dominant solution $y(k)\sim\zeta_{-}^{k}$
($|\zeta_{-}|>1$), the partial sum $S_{N}^{*}=\sum_{k=0}^{N}y(k)$
grows as $|S_{N}^{*}|\sim|\zeta_{-}|^{N}$, while $\tau_{N}=h_{N+1}/y(N+1)\sim|\zeta_{+}/\zeta_{-}|^{N}$.
The product $|\tau_{N}S_{N}^{*}|\sim|\zeta_{+}/\zeta_{-}|^{N}\cdot|\zeta_{-}|^{N}=|\zeta_{+}|^{N}\to0$
exponentially, verifying the convergence condition~\eqref{eq:Wimp-conv-cond}.

The \emph{CF-based algorithm}~\eqref{eq:Wimp-CF-algo} applied to
our setting gives $r_{N}(n)\to w(u+n)=h_{n+1}/h_{n}$, the ratios
of the minimal solution, independently of any scalar normalization
of $\{h_{m}\}$. These ratios are then used to compute the product
$C(u)=\prod_{k=0}^{\infty}w(u+k)/\zeta_{+}$ and hence $H(u)=\tilde{h}_{0}(u)$
(eq.~\eqref{eq:notes2-H-new-def}--\eqref{eq:notes2-H-new-functional}).

\section{Notation and Abbreviations}

\label{app:notation}

The following table collects the principal symbols and abbreviations
used throughout, together with their meaning and the section or equation
where each is first defined or introduced.

\medskip{}

\begin{longtable}[c]{@{}p{0.24\textwidth}p{0.53\textwidth}p{0.16\textwidth}@{}}
\toprule 
\textbf{Symbol}  & \textbf{Meaning}  & \textbf{Defined}\tabularnewline
\midrule
\endfirsthead
\midrule 
\textbf{Symbol}  & \textbf{Meaning}  & \textbf{Defined}\tabularnewline
\midrule
\endhead
\midrule 
\multicolumn{3}{@{}l@{}}{\emph{Physical variables and circuit parameters}}\tabularnewline
$q(t)$  & charge on the capacitor  & Chapter~\ref{sec:Intro}\tabularnewline
$L$  & inductance  & Chapter~\ref{sec:Intro}\tabularnewline
$C$  & mean capacitance\index{capacitance sensing}  & Chapter~\ref{sec:Intro}\tabularnewline
$\varepsilon$  & modulation amplitude, $0<\varepsilon<1$; $\varepsilon=2\delta/(1+\delta^{2})$  & \eqref{eq:LC-original}\tabularnewline
$\mu$  & modulation (driving) frequency  & Chapter~\ref{sec:Intro}\tabularnewline
$\omega_{0}=1/\sqrt{LC}$  & natural frequency of the unperturbed circuit  & Chapter~\ref{sec:Intro}\tabularnewline
$r=\mu/\omega_{0}$  & frequency ratio  & Chapter~\ref{sec:Intro}\tabularnewline
$\tau=\mu t$  & rescaled time  & Chapter~\ref{sec:Intro}\tabularnewline
$x=\tau/2=\mu t/2$  & Hill/Ince independent variable; $x=\tau/2$, period $\pi$  & Chapter~\ref{sec:LCHill}\tabularnewline
\addlinespace
\multicolumn{3}{@{}l@{}}{\emph{Modulation parameters}}\tabularnewline
$\delta=\dfrac{1-\sqrt{1-\varepsilon^{2}}}{\varepsilon}$  & Mobius\index{Mobius transformation} modulation parameter, $0<\delta<1$;
inverse: $\varepsilon=2\delta/(1+\delta^{2})$; $\delta\approx\varepsilon/2$
for small $\varepsilon$; repulsive fixed point of Mobius ratio recurrence
$T_{M_{\infty}}$; see Table~\ref{tab:aux-params}  & \eqref{eq:eps-delta}, \eqref{eq:Ince-MR}\tabularnewline
$T_{M_{\infty}}(z)=\dfrac{1+\delta^{2}}{\delta}-\dfrac{1}{z}$  & Mobius ratio recurrence associated with the limiting transfer matrix
$M_{\infty}$ of the Ince recurrence; fixed points $\zeta_{1}=\delta$
(repulsive) and $\zeta_{2}=1/\delta$ (attractive), simultaneously
the eigenvalues of $M_{\infty}$ (Remark~\ref{rem:eig-fixpt})  & \eqref{eq:TMinfz}, \eqref{eq:Minf-eigs}\tabularnewline
$c=\dfrac{4\omega_{0}^{2}}{\mu^{2}}=\dfrac{4}{r^{2}}$  & Ince spectral parameter (primary dimensionless form eq.~\eqref{eq:Ince-MR});
$c=m^{2}$ at $m$-th resonance\index{resonance} center ($m=1,2,3,\ldots$);
see also \emph{Ince's equation\index{Ince equation} and its parameters}
below; Table~\ref{tab:aux-params}  & \eqref{eq:Ince-MR}, \eqref{eq:c-vs-mu}\tabularnewline
\addlinespace
\multicolumn{3}{@{}l@{}}{\emph{Hill's equation\index{Hill equation} and spectral variables}}\tabularnewline
$\lambda$  & spectral (eigenvalue) parameter of Hill's equation  & Chapter~\ref{sec:MWInce}\tabularnewline
$\hat{\lambda}=\dfrac{c(1+\delta^{2})}{1-\delta^{2}}$,\quad{}$\omega=\sqrt{\hat{\lambda}}$  & Magnus--Winkler\index{Magnus--Winkler theory} spectral parameter
and its square root; $\hat{\lambda}=c$ at $\delta=0$; $\hat{\lambda}>c$
for $\delta>0$  & Chapter~\ref{sec:MWInce}\tabularnewline
$Q(x)$  & zero-mean periodic perturbation in Hill's equation  & \eqref{eq:Hill-MW}\tabularnewline
$Q_{0}(x)$  & normalized perturbation profile in two-parameter family~\eqref{eq:Hill-family}  & \eqref{eq:Hill-family}\tabularnewline
\addlinespace
\multicolumn{3}{@{}l@{}}{\emph{Floquet\index{Floquet theory} theory}}\tabularnewline
$\mathbf{z}=(q,p)^{T}$  & state vector ordered (coordinate, momentum) throughout this work,
following Yakubovich--Starzhinskii\index{Yakubovich--Starzhinskii series}~\cite[Ch.~III, \S\,3]{YakSta1};
fixes the form of $\tilde{\mathbf{J}}=\bigl[\begin{smallmatrix}0 & -1\\
1 & 0
\end{smallmatrix}\bigr]$ and $\mathbf{H}=\mathrm{diag}(C^{-1},L^{-1})$ in the canonical form~\eqref{eq:canonical-form}  & \S\ref{subsec:Hamiltonian}\tabularnewline
$X(\pi)$  & monodromy matrix\index{monodromy matrix} of Hill's equation (period
$T=\pi$ throughout this work); general case: $X(T)$  & Ch.~\ref{app:genLin}\tabularnewline
$\rho_{1},\rho_{2}=e^{\pm i\pi\alpha}$  & Floquet multiplier\index{Floquet theory!Floquet multiplier}s (eigenvalues
of $X(\pi)$); $\rho_{1}\rho_{2}=1$; $\rho=+1$: coexistence (two
independent period-$\pi$ or $2\pi$ solutions, instability zone\index{instability zone}
collapses); $\rho=-1$: EPD curve\index{exceptional point of degeneracy (EPD)!EPD curve}
(non-trivial Jordan block\index{Jordan block}, instability tongue\index{instability tongue}
opens)  & Ch.~\ref{app:Hill}\tabularnewline
$\nu$  & Floquet exponent\index{Floquet theory!Floquet exponent}; $\rho=e^{2\pi i\nu}$;
period-$2\pi$ convention; $\nu=u_{0}$ in Cambi's notation  & Ch.~\ref{app:Hill}\tabularnewline
$\alpha$  & characteristic exponent; $\rho=e^{i\pi\alpha}$; period-$\pi$ convention;
$\alpha=2\nu$; at primary EPD: $\alpha=1$; the relation $\rho=e^{i\pi\alpha}=e^{2\pi i\nu}$
shows $\alpha=2\nu$; this work uses $\alpha$ throughout except where
Cambi's $u_{0}=\nu$ is referenced  & \eqref{eq:Floquet-mode-intro}\tabularnewline
\addlinespace
\multicolumn{3}{@{}l@{}}{\emph{EPD sensor quantities (Chapter~\ref{sec:EPD-sensor})}}\tabularnewline
$\xi=\omega_{0}-\omega_{0}'$  & frequency shift of natural frequency due to $\Delta C$; $\xi\approx\omega_{0}\Delta C/(2C_{0})$  & \eqref{eq:xi-DeltaC}\tabularnewline
$\epsilon_{w}$  & \emph{work-point offset}: design detuning from the EPD boundary at
which the sensor is biased; sets the baseline split $\sqrt{2\delta\,\epsilon_{w}}$
and the sensitivity coefficient $\sqrt{2\delta}/(2\sqrt{\epsilon_{w}})$  & \eqref{eq:work-point-total}\tabularnewline
$\hat{\xi}=\xi/\mu$  & dimensionless signal; $\hat{\xi}=\omega_{0}\Delta C/(2C_{0}\mu)$;
physically $|\hat{\xi}|\ll1$  & \eqref{eq:xi-hat-def}\tabularnewline
$\Delta\alpha=\alpha_{+}-\alpha_{-}$  & frequency split; $\Delta\alpha=2\sqrt{2\delta}\sqrt{\hat{\xi}(1-\hat{\xi})}\cdot(1+O(\delta))$  & \eqref{eq:Deltaalpha-m1}\tabularnewline
$\hat{\xi}^{*}=2\delta$  & EPD/conventional crossover: EPD wins for $\hat{\xi}<\hat{\xi}^{*}$  & \eqref{eq:sensitivity-ratio}\tabularnewline
\addlinespace
\multicolumn{3}{@{}l@{}}{\emph{Fourier coefficients}}\tabularnewline
$g_{n}$, $Q=\sum_{n\neq0}g_{n}e^{2inx}$  & MW complex Fourier coefficients of $Q$; $g_{-n}=\overline{g_{n}}$
for real $Q$; for LC: $g_{\pm n}=\hat{\lambda}\delta^{|n|}$  & Chapter~\ref{sec:MWInce}\tabularnewline
$A_{n}$  & cosine Fourier amplitudes; $A_{n}=2g_{n}$ for real even $Q$  & Chapter~\ref{sec:LCHill}\tabularnewline
\addlinespace
\multicolumn{3}{@{}l@{}}{\emph{Discriminant and its expansion}}\tabularnewline
$\Delta(\lambda)$  & Hill's discriminant\index{discriminant}; $\Delta=y_{1}(\pi)+y_{2}'(\pi)$  & Chapter~\ref{sec:MWInce}, Ch.~\ref{app:Hill}\tabularnewline
$\Delta_{0}$  & unperturbed discriminant; $\Delta_{0}=2\cos\pi\sqrt{\hat{\lambda}}$  & \eqref{eq:D0}\tabularnewline
$\Delta_{2}$  & leading-order $O(\varepsilon^{2})$ MW correction to $\Delta$  & \eqref{eq:D0}\tabularnewline
$\Delta^{(N)}$  & partial sum of the $\psi$-basis expansion with $N$ terms  & Chapter~\ref{sec:UnivExp}\tabularnewline
\addlinespace
\multicolumn{3}{@{}l@{}}{\emph{Universal $\psi$-basis (Chapter~\ref{sec:UnivExp})}}\tabularnewline
$\psi_{n}(\hat{\lambda})$  & $n$-th universal basis function (entire); see eq.~\eqref{eq:psi-def}  & Def.~\ref{def:psi-basis}\tabularnewline
$\psi_{n}(n)=\pi/\sqrt{2}$ (eq.~\eqref{eq:psi-resonance})  & resonance value of $\psi_{n}$  & \eqref{eq:psi-resonance}\tabularnewline
$\mathcal{K}(s,t;\lambda)$  & Volterra kernel; $\sin[\sqrt{\lambda}(s-t)]/\sqrt{\lambda}$, $\;0\leq t\leq s$  & \eqref{eq:Volterra-kernel}\tabularnewline
$\mathcal{L}_{\mathrm{SL}}$  & Sturm--Liouville\index{Sturm--Liouville theory} operator $-(\sqrt{\hat{\lambda}}\,y')'$,
weight $1/\sqrt{\hat{\lambda}}$  & \eqref{eq:SL-formal}\tabularnewline
\addlinespace
\multicolumn{3}{@{}l@{}}{\emph{Instability zones and boundary curve\index{boundary curves}s}}\tabularnewline
$\lambda_{m}^{\pm}$  & boundaries of $m$-th instability zone in $\hat{\lambda}$-space  & Chapter~\ref{sec:Boundaries}\tabularnewline
$L_{m}=\lambda_{m}^{+}-\lambda_{m}^{-}$  & width of $m$-th instability interval  & \eqref{eq:Lm-main}\tabularnewline
$c_{\pm}^{(m)}(\delta)$  & EPD curves of the $m$-th instability tongue in $c$-space ($m=1,3,5,\ldots$
odd); $c_{-}^{(m)}(\delta)<m^{2}<c_{+}^{(m)}(\delta)$; tongue width
$L_{m}^{(c)}=c_{+}^{(m)}-c_{-}^{(m)}$; equivalently written $c_{k}^{\pm}$
with $m=2k-1$ in the CF method of Chapter~\ref{sec:LCInce}  & \eqref{eq:EPD-curves-def}, Thm~\ref{thm:EPD-bdry}\tabularnewline
\multicolumn{3}{@{}l@{}}{%
\parbox[c]{0.92\textwidth}{%
\emph{Terminological note: ``EPD curves'' $\equiv$ ``stability
boundaries.'' Every point on $c=c_{\pm}^{(m)}(\delta)$ is an EPD
point (exceptional point of degeneracy\index{exceptional point of degeneracy (EPD)},
$\rho=\pm1$, non-trivial Jordan block), so the curve is simultaneously
the stability boundary\index{stability boundary}} of the $m$-th
tongue and a \emph{curve of EPD points}. Both terms refer to the same
locus; this work uses ``EPD curve'' throughout.%
}}\tabularnewline
$\mu_{\pm}^{(m)}$  & boundaries of $m$-th instability zone in $\mu$-space  & \eqref{eq:primary-domain}\tabularnewline
\addlinespace
\multicolumn{3}{@{}l@{}}{\emph{Cambi's notation (Chapter~\ref{sec:Cambi})}}\tabularnewline
$p=\omega_{0}/\mu=1/r$  & Cambi's inverse frequency ratio; $p=\frac{1}{2}$ at primary resonance
($\mu=2\omega_{0}$); related to our spectral parameter by $c=4p^{2}$;
resonant (critical) values $p=m/2$ for $m=1,2,3,\ldots$ correspond
to $c=m^{2}$, with only odd $m$ giving genuine instability tongues
for the LC circuit (Corollaries~\ref{cor:even}--\ref{cor:odd});
see Ch.~\ref{app:ParRes}  & \eqref{eq:Cambi-res}, \eqref{eq:gamma-delta1}\tabularnewline
$\eta=p-\tfrac{1}{2}$  & detuning of $p$ from the primary resonance center; both $\eta$ and
$\gamma$ are small near the first tongue tip; $|\eta/\gamma|\approx\tfrac{1}{4}$
on the boundary curves; near-boundary stable strip: $3<|\gamma/\eta|<4$
(eq.~\eqref{eq:stable-near-bdry-CF}); see eq.~\eqref{eq:eta-def}  & \eqref{eq:eta-def}\tabularnewline
$\gamma=\varepsilon/2$  & Cambi's half-modulation amplitude; $\gamma=\delta/(1+\delta^{2})$;
$\gamma<\frac{1}{2}$ for $|\varepsilon|<1$  & \eqref{eq:gamma-delta}\tabularnewline
$u_{0}$  & Cambi's Floquet parameter (a third convention for the Floquet exponent);
defined by $\rho=e^{2\pi iu_{0}}$ with period $2\pi$ for Cambi's
equation; taken in $(0,\tfrac{1}{2})$ as the fundamental exponent;
at the $m$-th resonance center $u_{0}=m/2$; related to $\alpha$
by $u_{0}=\alpha/2$ and to $\nu$ by $u_{0}=\nu$; generates the
Floquet comb\index{Floquet theory!Floquet comb} $Z_{F}=\{\pm u_{0}+n\}$,
with $u^{*}\equiv-u_{0}\pmod{\mathbb{Z}}$ (see $u^{*}$ and $Z_{F}$)  & \eqref{eq:Cambi-ansatz}, \eqref{eq:ZF-def}\tabularnewline
\addlinespace
\multicolumn{3}{@{}l@{}}{\emph{Floquet exponent computation (Chapter~\ref{sec:expanding-Cambi})}}\tabularnewline
$G(u)=1-p^{2}/u^{2}$  & \emph{spectral ratio function}; coefficient in the three-term recurrence~\eqref{eq:notes2-ttrec};
$G(u)=0$ iff $u=\pm p$ (resonance); $G(u)\to1$ as $|u|\to\infty$  & \eqref{eq:Cambi-res}\tabularnewline
$\zeta_{\pm}=\dfrac{-1\pm\sqrt{1-4\gamma^{2}}}{2\gamma}$  & \emph{characteristic roots} of the constant-coefficient limit of the
recurrence~\eqref{eq:notes2-ttrec} (roots of $\gamma\zeta^{2}+\zeta+\gamma=0$);
$\zeta_{+}\zeta_{-}=1$ and $|\zeta_{+}|<1<|\zeta_{-}|$ for $0<2\gamma<1$;
in the modulation parameter, $\zeta_{+}=-\delta$, $\zeta_{-}=-1/\delta$
(via $\gamma=\delta/(1+\delta^{2})$, eq.~\eqref{eq:gamma-delta};
distinct from the Möbius fixed points $\zeta_{1},\zeta_{2}$ of $T_{M_{\infty}}$
above); $w\to\zeta_{+}$ at infinity; $|\zeta_{+}|^{2n}$ is the geometric
rate of the pole ladders  & \eqref{eq:Cambi-char-roots}\tabularnewline
$F_{w}(u)=w(-u)-1/w(u-1)$  & \emph{double minimality function}; meromorphic; zeros at Floquet exponents
$u^{*}$ and their integer translates; poles at poles of $w$; satisfies
$F_{w}(u)=\widetilde{F}_{w}(u)/\gamma$; $-F_{w}$ is the Casoratian
$K[h,h^{-}]$ of the two minimal solutions, the discrete Evans function
(see $K_{m}[A,B]$ below)  & \eqref{eq:Fw-def}\tabularnewline
$\widetilde{F}_{w}(u,\gamma)=\gamma w(u)+G(u)+\gamma w(-u)$  & \emph{symmetrized double minimality function}; $\widetilde{F}_{w}=0$
is the double minimality equation for $u^{*}$; equivalent to $F_{w}=0$;
symmetric: $\widetilde{F}_{w}(u)=\widetilde{F}_{w}(-u)$  & \eqref{eq:Ftilde-def}\tabularnewline
$u^{*}$  & \emph{primary Floquet exponent}; the zero of $\widetilde{F}_{w}$
nearest to $p$; satisfies $u^{*}\to p$ as $\gamma\to0$; computed
to arbitrary precision via the $\gamma^{2}$-series~\eqref{eq:ustar-c2}--\eqref{eq:ustar-c8};
the representative of $Z_{F}^{-}$ near $p$, with $u^{*}\equiv-u_{0}\pmod{\mathbb{Z}}$,
so $\{\pm u^{*}+n\}=Z_{F}$ (see $u_{0}$ above and $Z_{F}$ below)  & \eqref{eq:ustar-c2}, \eqref{eq:ZF-def}\tabularnewline
$u^{**}$  & \emph{base comb point}, $u^{**}\in\{-u^{*},\,u^{*}-1\}$: one ladder
anchor per sub-comb ($-u^{*}\in Z_{F}^{+}$, the reflection of the
primary exponent and host of the second Floquet solution; $u^{*}-1\in Z_{F}^{-}$,
its unit translate); the leftward stealthy-pole ladders of Theorem~\ref{thm:pole-ladders}
descend from $u^{**}$, and the two values are exchanged by the companion
map $u^{**}\mapsto-u^{**}-1$  & \eqref{eq:ustarstar-def}\tabularnewline
$Z_{F}^{+}=\{u_{0}+n\}$  & one of the two sub-combs of the Floquet comb $Z_{F}=\{\pm u_{0}+n\}$
($u_{0}\in(0,\tfrac{1}{2})$ the fundamental exponent); equals $\{-u^{*}+n\}$
in the $u^{*}$ form; associated with the multiplier $\rho_{+}=e^{+2\pi iu_{0}}$
and hosting the secondary exponent $u^{**}\approx-p$; its zeros of
$F_{w}$ are stealthy at double precision  & \eqref{eq:family-def}\tabularnewline
$Z_{F}^{-}=\{-u_{0}+n\}$  & the other sub-comb of $Z_{F}$; equals $\{u^{*}+n\}$ in the $u^{*}$
form; associated with the multiplier $\rho_{-}=e^{-2\pi iu_{0}}$
and hosting the primary exponent $u^{*}\approx+p$; its zeros of $F_{w}$
are accessible at any precision  & \eqref{eq:family-def}\tabularnewline
$Z_{F}$  & \emph{Floquet comb}: the zero set of the double minimality function
$F_{w}$; the union of $Z_{F}^{+}$ and $Z_{F}^{-}$, $Z_{F}=\{u_{0}+n\}\cup\{-u_{0}+n\}=\{\pm u_{0}+n\}=\{\pm u^{*}+n\}$
(the two forms agree since $u^{*}\equiv-u_{0}\pmod{\mathbb{Z}}$);
two interleaved unit-spaced sub-combs whose consecutive points are
separated by alternating gaps $s$ and $1-s$, with $s=2u^{*}\bmod1$
(see $s$ below)  & \eqref{eq:ZF-def}\tabularnewline
$s$ (smallest gap)  & \emph{smallest gap of the Floquet comb}; $s=\min(2u^{*}\bmod1,\,1-(2u^{*}\bmod1))=\operatorname{dist}(2u^{*},\mathbb{Z})\in(0,\tfrac{1}{2}]$;
the smaller of the two alternating gaps between adjacent points of
$Z_{F}^{+}$ and $Z_{F}^{-}$ (the larger being $1-s$); $s\to0$
as $2u^{*}\to\mathbb{Z}$ (sub-combs merge, period-doubling boundary),
$s=\tfrac{1}{2}$ at maximal separation; $s\approx0.484$ at $p=0.70,\gamma=0.30$  & \eqref{eq:s-offset}\tabularnewline
$c_{2k}$  & \emph{series coefficients} of $u^{*}=p+\sum_{k\geq1}c_{2k}\gamma^{2k}$;
closed-form rational functions of $p$; poles at the odd resonances
$p=\tfrac{1}{2},\tfrac{3}{2},\ldots$ of increasing order; computed
by the recursion~\eqref{eq:c2k-recursion}  & \eqref{eq:ustar-c2}--\eqref{eq:ustar-c8}\tabularnewline
$[S]_{k}$  & \emph{coefficient extraction}: for a power series $S(\gamma)=\sum_{n\geq0}s_{n}\gamma^{n}$,
$[S]_{k}=s_{k}$ denotes the coefficient of $\gamma^{k}$; e.g.\ $[S]_{2k}=s_{2k}$
for even-index extraction; used in the recursion~\eqref{eq:c2k-recursion}  & \eqref{eq:c2k-recursion}\tabularnewline
$b_{k}=-\sum_{j=1}^{k}a_{j}b_{k-j}$  & \emph{inversion-series coefficients}; $b_{0}=1$; give $1/(1+\sum a_{j}\xi^{j})=\sum b_{k}\xi^{k}$;
precomputed to fixed depth and used in the recursion~\eqref{eq:c2k-recursion}
to invert CF denominators efficiently  & \eqref{eq:inv-series-table}\tabularnewline
Stealthy zero  & A zero $u_{0}$ of a meromorphic function $F$ is \emph{stealthy at
precision $D$} if evaluating $F(u_{0})$ with $D$ significant digits
yields $|F(u_{0})|\gg1$ rather than $\approx0$, due to catastrophic
cancellation between large terms; the $Z_{F}^{+}$ zeros of $F_{w}$
are stealthy at double precision but accessible with sufficient arithmetic
(Remark~\ref{rem:double-precision-misses}, Theorem~\ref{thm:precision-cost})  & \S\,\ref{subsec:Cambi-notes2-precision}\tabularnewline
$\varepsilon_{\mathrm{series}}=\gamma^{2}/(p^{2}-\tfrac{1}{4})$  & \emph{natural small parameter} for the $u^{*}$ series; equals $4\gamma^{2}/(c-1)=\varepsilon_{\mathrm{LC}}^{2}/(c-1)$;
measures proximity to the primary stability boundary $c=1$ ($p=\tfrac{1}{2}$);
series converges when $\varepsilon_{\mathrm{series}}\ll1$, i.e.\ when
the system is far from the primary instability tongue  & \S\,\ref{subsec:Cambi-notes2-precision}\tabularnewline
$v(u)$  & \emph{Cambi continued fraction}~\eqref{eq:Cambi-v-CF}, the central
object of Cambi's method; related to the minimal solution ratio by
$w(u)=-\gamma\,v(u+1)$ (Lemma~\ref{lem:notes2-w-v}); approaches
$-\zeta_{+}/\gamma$ as $|u|\to\infty$~\eqref{eq:Cambi-asymp-v}  & \eqref{eq:Cambi-v-CF}\tabularnewline
$w(u)=h_{1}(u)/h_{0}(u)$  & \emph{minimal solution\index{continued fraction!minimal solution}
ratio}: ratio of consecutive values of the minimal solution of the
three-term recurrence~\eqref{eq:notes2-ttrec}; equals the CF~\eqref{eq:notes2-wdef};
satisfies $w(u)\to\zeta_{+}$ as $u\to\infty$; meromorphic, with
poles at $u<0$ near the Floquet comb; satisfies the discrete Riccati
recurrence~\eqref{eq:notes2-wrec} (Theorem~\ref{thm:notes2-wrec-exact}\index{Riccati recurrence}),
with residues and pole propagation governed by \S\,\ref{subsec:Riccati-structure}  & \eqref{eq:notes2-wdef}\tabularnewline
$\operatorname*{Res}_{u=u_{p}}w=-1/w'(u_{p}-1)$  & residue of $w$ at a pole $u_{p}$, via the slope at its source zero
$z_{0}=u_{p}-1$; negative on the first sheet, sign set by $w'(z_{0})$
(Proposition~\ref{prop:Riccati-residues})  & \eqref{eq:Riccati-residue}\tabularnewline
$H(u)=1/\prod_{m\geq0}(w(u+m)/\zeta_{+})$  & \emph{minimal solution generator}: meromorphic function defined by
the infinite product~\eqref{eq:notes2-H-new-def}; satisfies $H(u+1)/H(u)=w(u)/\zeta_{+}$
and $H(u)\to1$ as $u\to+\infty$; zeros at poles of $w$ shifted
left, second-order poles at negative integers  & \eqref{eq:notes2-H-new-def}\tabularnewline
\addlinespace
\multicolumn{3}{@{}l@{}}{\emph{Casoratian--Weyl and Nevanlinna-type structure (\S\,\ref{subsec:imported-Jacobi},
\S\,\ref{subsec:Casoratian-Weyl})}}\tabularnewline
$K_{m}[A,B]=A_{m}B_{m+1}-A_{m+1}B_{m}$  & \emph{Casoratian} of two solutions of the recurrence~\eqref{eq:notes2-rec};
constant in $m$ by Abel's lemma; $K[h,h^{-}]=-F_{w}$, exhibiting
$F_{w}$ as the discrete Evans function  & \eqref{eq:CW-Casoratian-def}, \eqref{eq:CW-Casoratian}\tabularnewline
$q_{m}(u)=G(u+m)-1$  & perturbation sequence of the recurrence, $q_{m}(u)=-p^{2}/(u+m)^{2}$;
absolutely summable in $m$, with poles swept by the shifts of the
single pole of $G$ at the origin  & Lem.~\ref{lem:Jost-right}\tabularnewline
$\widehat{Q}_{k}(u)$, $S_{k}(u)$  & tail sums of the perturbation: $\widehat{Q}_{k}=\sum_{m\leq k-1}\sup_{D}|G(u+m)-1|$
(leftward; explicitly $p^{2}\sum_{m\leq k-1}\sup_{D}|u+m|^{-2}=O(1/|k|)$)
and $S_{k}=\sum_{m\geq k+1}|q_{m}|$ (rightward); they control the
Jost errors  & \eqref{eq:CW-Qhat}, \eqref{eq:CW-Qhat-explicit}, Lem.~\ref{lem:Jost-right}\tabularnewline
$\Gamma(k,m)$, $C_{\gamma}$  & Green kernel of the free recurrence, $\Gamma(k,m)=\zeta_{+}\bigl(1-\zeta_{+}^{2(k-m)}\bigr)/\bigl(\gamma(1-\zeta_{+}^{2})\bigr)$
for $k\geq m$, with $|\Gamma|\leq C_{\gamma}=|\zeta_{+}|/\bigl(\gamma(1-\zeta_{+}^{2})\bigr)$;
the transposed kernel $\Gamma(m,k)$ serves the rightward construction  & \eqref{eq:CW-Green}, \eqref{eq:CW-Jost-bound}\tabularnewline
$\hat{h}_{k}(u)$, $\hat{h}_{k}^{+}(u)$  & left and right \emph{Jost solutions}: solutions of~\eqref{eq:notes2-rec}
with $\hat{h}_{k}\sim\zeta_{+}^{-k}$ as $k\to-\infty$ and $\hat{h}_{k}^{+}\sim\zeta_{+}^{\,k}$
as $k\to+\infty$, with uniform error bounds; $w=\hat{h}_{1}^{+}/\hat{h}_{0}^{+}$  & Lem.~\ref{lem:Jost-LC}, Lem.~\ref{lem:Jost-right}\tabularnewline
$\mathbb{C}_{\pm}$  & open upper and lower half-planes $\{z\in\mathbb{C}:\pm\operatorname{Im}z>0\}$  & \S\,\ref{subsec:imported-Jacobi}\tabularnewline
$Q$ (quadrant)  & the quadrant $\{\operatorname{Im}u>0,\ \operatorname{Re}u>-1\}$ on
which $w$ has the Nevanlinna-like positivity $\operatorname{Im}w>0$
(maps $Q$ into $\mathbb{C}_{+}$)  & Lem.~\ref{lem:w-Nevanlinna}\tabularnewline
$g(z)=w(\sqrt{z}-1)$  & \emph{Herglotz transform} of $w$: a standard Herglotz (Nevanlinna)
function $\mathbb{C}_{+}\to\mathbb{C}_{+}$ in the variable $z=(u+1)^{2}$,
with representation $g=\zeta_{+}+\int_{\mathbb{R}}d\rho(\lambda)/(\lambda-z)$  & \eqref{eq:NV-g-def}, \eqref{eq:NV-g-rep}\tabularnewline
$\rho$ (Herglotz measure)  & finite positive measure representing $g$: atoms at $P=\{(u_{p}+1)^{2}\}$
for the poles $u_{p}$ of $w$ in $(-1,\infty)$, absolutely continuous
part on $(-\infty,0]$ with density $\pi^{-1}\operatorname{Im}w(-1+\mathrm{i}\sqrt{|\lambda|})$;
not to be confused with the Floquet multipliers $\rho_{1},\rho_{2}$
above  & Thm.~\ref{thm:w-Herglotz}\tabularnewline
$\hat{\mathbb{C}}=\mathbb{C}\cup\{\infty\}$, $\chi$  & extended complex plane (Riemann sphere) and chordal metric, a complete
metric space; the setting of the spherical convergence theory of Chapter~\ref{app:CF}  & \eqref{eq:Moeb1b}, Ch.~\ref{app:Moeb}\tabularnewline
$\tau$, $z$, $m_{p}^{\pm}(z)$, $\varphi_{p,\pm}(z,\cdot)$  & imported Jacobi notation (Teschl): difference expression, spectral
parameter, Floquet multipliers and Floquet solutions of the periodic
background; $\varphi$ replaces Teschl's $u$, reserved here for the
Floquet-exponent parameter  & \eqref{eq:imported-tau}--\eqref{eq:imported-Floquet}\tabularnewline
\addlinespace
\multicolumn{3}{@{}l@{}}{\emph{Yakubovich--Starzhinskii series (Chapters~\ref{sec:YS-LC},
\ref{sec:YS-exact-LC}, \ref{app:YS-Floquet})}}\tabularnewline
$\mathbf{K}(\varepsilon)$, $\mathbf{K}_{m}$  & Floquet exponent matrix and its YS series coefficients, $\mathbf{K}(\varepsilon)=\mathbf{K}_{0}+\varepsilon\mathbf{K}_{1}+\varepsilon^{2}\mathbf{K}_{2}+\cdots$;
for the LC circuit each $\mathbf{K}_{m}(\hat{\lambda})$ is a polynomial
of degree $m$ in $\hat{\lambda}$  & \eqref{eq:YS-K-series}, \eqref{eq:YS-lam-poly}\tabularnewline
$\mathbf{F}(t,\varepsilon)$, $\mathbf{F}_{m}$  & periodic Lyapunov factor and its YS series coefficients, $\mathbf{F}=\mathbf{I}+\varepsilon\mathbf{F}_{1}+\cdots$;
normalization conventions in Chapter~\ref{app:YS-Floquet}  & \eqref{eq:YS-F-series}\tabularnewline
$\mathbf{K}_{0}$  & residual exponent of the unperturbed system after the preliminary
transformation; in the LC convention $\mathbf{K}_{0}=\bigl[\begin{smallmatrix}0 & 0\\
-(\hat{\lambda}-1) & 0
\end{smallmatrix}\bigr]$, vanishing at $\hat{\lambda}=1$  & Rem.~\ref{rem:C0-convention}\tabularnewline
$\mathbf{D}_{j}(t)$  & transformed coefficient matrices of the reduced system, $\mathbf{D}_{j}=e^{-t\mathbf{C}_{0}}\mathbf{B}_{j}e^{t\mathbf{C}_{0}}=\operatorname{Ad}_{e^{-t\mathbf{C}_{0}}}\mathbf{B}_{j}$  & \eqref{eq:YS-Dj}, \eqref{eq:LC-Dn-exact}\tabularnewline
$\boldsymbol{\Phi}_{l}(t)$  & $l$-th order source of the YS recursion  & \eqref{eq:YS-source}\tabularnewline
$[\cdot]_{{\rm av}}$  & time average over one period, $T^{-1}\int_{0}^{T}(\cdot)\,\dd t$;
extracts the secular (exponent) part of the source at each order  & \eqref{eq:YS-Kl}\tabularnewline
$\operatorname{ad}_{\mathbf{K}_{0}}$  & adjoint action $\mathbf{W}\mapsto\mathbf{K}_{0}\mathbf{W}-\mathbf{W}\mathbf{K}_{0}$
on the Lie algebra $\mathfrak{gl}(n,\mathbb{C})=M_{n}(\mathbb{C})$;
eigenvalues $\lambda_{j}-\lambda_{k}$ encode all parametric resonances;
our notation for the solvability analysis that YS perform via their
operators $S$ and $P$  & \eqref{eq:ad-K0}\tabularnewline
\addlinespace
\multicolumn{3}{@{}l@{}}{\emph{Mathieu\index{Mathieu equation} equation}}\tabularnewline
$y''+(\lambda-2q\cos2x)y=0$  & Mathieu equation (degenerate Ince case $a=0$, $b=d=0$, $c=\lambda$,
$d=-2q$)  & \eqref{eq:Mathieu-intro}\tabularnewline
$q=\dfrac{2\varepsilon\omega_{0}^{2}}{\mu^{2}}=\dfrac{2\varepsilon}{r^{2}}$  & Mathieu parameter in LC circuit variables; $q\approx\varepsilon/2$
at primary resonance $\mu\approx2\omega_{0}$  & \eqref{eq:q-LC-params}\tabularnewline
\addlinespace
\multicolumn{3}{@{}l@{}}{\emph{Ince's equation and its parameters}}\tabularnewline
$(1+a\cos2x)y''+(b\sin2x)y'+(c+d\cos2x)y=0$  & Ince's four-parameter equation; $|a|<1$  & \eqref{eq:Ince-intro}, \eqref{eq:Ince}\tabularnewline
$a=-\varepsilon,\;b=0,\;c=4/r^{2},\;d=0$  & Ince parameters for the LC circuit; general Ince eq.\ uses $a,b,c,d$
freely with $|a|<1$  & \eqref{eq:Ince-intro}, \eqref{eq:IncePars}\tabularnewline
\addlinespace
\multicolumn{3}{@{}l@{}}{\emph{Ince equation polynomials}}\tabularnewline
$Q(\mu)$  & coexistence polynomial for period-$\pi$ solutions; $\mu$ is the
polynomial variable (a real indeterminate), distinct from the modulation
frequency $\mu$; coexistence requires $Q$ to have an integer root
(MW Theorem~\ref{thm:MW71})  & \S\ref{sec:InceGeneral}\tabularnewline
$Q^{*}(\mu)$  & coexistence polynomial for period-$2\pi$ solutions; same note on
the variable $\mu$  & \S\ref{sec:InceGeneral}\tabularnewline
$F_{\mathrm{even}}(c),\;F_{\mathrm{odd}}(c)$  & continued-fraction\index{continued fraction} eigenvalue functions  & Chapter~\ref{sec:LCInce}\tabularnewline
\addlinespace
\multicolumn{3}{@{}l@{}}{\emph{Special functions}}\tabularnewline
$\operatorname{sinc}(x)=\sin(\pi x)/(\pi x)$  & normalized sinc function  & \eqref{eq:psi-sinc}\tabularnewline
$j_{0}(z)=\sin(z)/z$  & zeroth spherical Bessel function; $\operatorname{sinc}(x)=j_{0}(\pi x)$  & \eqref{eq:psi-sinc}\tabularnewline
\addlinespace
\multicolumn{3}{@{}l@{}}{\emph{Asymptotic notation}}\tabularnewline
$O(x)$  & $f=O(x)$ means $|f|\leq C|x|$ for some constant $C>0$ independent
of all parameters  & Chapter~\ref{sec:Intro}\tabularnewline
$O_{m}(x)$  & $f=O_{m}(x)$ means $|f|\leq C(m)|x|$ for some constant $C(m)>0$
depending on the tongue index $m$  & \eqref{eq:Lm-main}\tabularnewline
\addlinespace
\multicolumn{3}{@{}l@{}}{\emph{Yakubovich--Starzhinskii (YS) series}}\tabularnewline
$\mu_{0}$  & \emph{YS detuning parameter}; $\mu_{0}=(c-1)/\varepsilon$, measures
how far the driving frequency deviates from the primary resonance
$c=1$ (i.e.\ $\omega_{0}=\mu/2$), in units of the perturbation
strength $\varepsilon$. Although $c=4/r^{2}>0$ always, $\mu_{0}$
is \emph{unrestricted in sign}: $\mu_{0}>0$ when $r<2$ (driving
below resonance), $\mu_{0}=0$ when $r=2$ (exact resonance), $\mu_{0}<0$
when $r>2$ (driving above resonance); the constraint $c>0$ only
requires $\mu_{0}>-1/\varepsilon$. Key values: $\mu_{0}=0$ (tongue
center, exact resonance, maximum instability); $|\mu_{0}|=\tfrac{1}{2}$
(EPD curves, instability tongue edges, $\mathbf{K}_{1}$ eigenvalues
$=0$); $|\mu_{0}|<\tfrac{1}{2}$ (inside tongue, unstable); $|\mu_{0}|>\tfrac{1}{2}$
(outside tongue, stable); $\mu_{0}=\pm\tfrac{1}{2}$ also mark the
exact Steiner-deltoid degenerations of the Floquet-factor column portraits
(second column at $+\tfrac{1}{2}$, first at $-\tfrac{1}{2}$, Theorem~\ref{thm:F1-topology});
further portrait transitions (not stability boundaries) at $\mu_{0}=0$
and $|\mu_{0}|=\tfrac{3}{2}$.  & \eqref{eq:YS-LC-params}\tabularnewline
YS$\leftrightarrow$this work  & \emph{Parameter dictionary} for reading YS1~\cite[Ch.~IV]{YakSta1}
against this book. YS writes Hill's equation as $\ddot{\xi}+\{(p\pi/T)^{2}+\varepsilon[\mu_{0}+f(t)]\}\xi=0$;
the identifications are: YS resonance order $p\leftrightarrow$ our
tongue index $m$ ($p=m=1$ for the primary tongue); YS period $T=\pi$;
YS small parameter $\varepsilon=2\delta/(1+\delta^{2})$ (our modulation,
not YS's $a$); YS harmonic $f(t)=\cos2\tau$ (single retained harmonic
of $f_{{\rm LC}}$); YS detuning $\mu_{0}=(c-1)/\varepsilon$ with
$c=1+\varepsilon\mu_{0}$. Caution: YS's $(\varepsilon,\mu_{0})$
parametrization is \emph{not} the literal Mathieu $(\lambda,q)$ coordinate
(Remark~\ref{rem:YS-bdry-derivation}).  & \eqref{eq:YS-LC-params}, \eqref{eq:YS-Hill-eq}\tabularnewline
IC  & Initial-condition convention for $\mathbf{F}_{j}$: $\mathbf{F}_{j}(0)=\mathbf{0}$
(Convention~2 of \S\,\ref{subsec:YS-conventions}, Chapter~\ref{app:YS-Floquet});
used in \S\,\ref{subsubsec:YS-LC-F1F2} for the exact LC closed-form
formulas  & \eqref{eq:LC-F1-lam}\tabularnewline
ZM  & Zero-mean convention for $\mathbf{F}_{j}$: $[\mathbf{F}_{j}]_{{\rm av}}=\mathbf{0}$
(Convention~1 of \S\,\ref{subsec:YS-conventions}, Chapter~\ref{app:YS-Floquet});
used in \S\,\ref{subsubsec:YS-Km-exact} for the intermediate derivation  & \eqref{eq:YS-norm-zeromean}\tabularnewline
$\mathbf{K}_{m},\;\mathbf{F}_{m}(\tau)$  & $m$-th order YS exponent matrix and Lyapunov factor; $K(\varepsilon)=\sum_{m}\varepsilon^{m}\mathbf{K}_{m}$
(constant), $F(\tau,\varepsilon)=\mathbf{I}+\sum_{m}\varepsilon^{m}\mathbf{F}_{m}(\tau)$
($\pi$-periodic); $\mathbf{K}_{1}$ governs EPD curves  & Chapter~\ref{sec:YS-LC}\tabularnewline
$\Delta\alpha_{{\rm YS}}=2\sqrt{\alpha\beta}$  & YS characteristic exponent split (full difference $\varkappa_{1}-\varkappa_{2}$);
$\alpha,\beta$ are the off-diagonal entries of $\varepsilon\mathbf{K}_{1}+\varepsilon^{2}\mathbf{K}_{2}$~\eqref{eq:YS-Dalpha};
$\alpha\beta>0$ (inside tongue): $\Delta\alpha_{{\rm YS}}$ real
(unstable); $\alpha\beta<0$ (outside): imaginary (stable); $\alpha\beta=0$
at EPD curve  & \eqref{eq:YS-Dalpha}\tabularnewline
YS  & Yakubovich--Starzhinskii \cite{YakSta1,YakSta2}  & Chapter~\ref{sec:YS-LC}\tabularnewline
CF  & continued fraction  & Chapter~\ref{sec:Intro}\tabularnewline
EPD  & exceptional point of degeneracy (Jordan-block monodromy)  & Chapter~\ref{sec:LCInce}\tabularnewline
LC  & inductor--capacitor (circuit)  & Chapter~\ref{sec:Intro}\tabularnewline
MW  & Magnus--Winkler \cite[Ch.~2--8]{MagWin}  & Chapter~\ref{sec:Intro}\tabularnewline
SL  & Sturm--Liouville  & Chapter~\ref{sec:UnivExp}\tabularnewline
\bottomrule
\end{longtable}

\end{document}